\documentclass[biblatex,twoside]{cernyrep2018}

\usepackage{float}
\usepackage{authblk}

\usepackage{graphicx}

\usepackage{booktabs}  
\usepackage{refstyle}

\usepackage{amssymb}
\usepackage{amsmath}
\usepackage{wasysym}
\usepackage[detect-all,separate-uncertainty=true,multi-part-units=single,detect-weight=true,alsoload=hep,per=slash,eVcorrb=0.4ex]{siunitx}

\usepackage{xcolor}

\usepackage{silence}

\usepackage{placeins}

\usepackage[subrefformat=parens]{subcaption}

\usepackage{verbatim}
\usepackage{tabularx}
\usepackage{multirow}

\usepackage{makecell}
\usepackage{lscape}
\usepackage{textgreek}

\newcolumntype{L}{>{\raggedright\arraybackslash}X}

\usepackage{array}

\usepackage{enumitem}
\setlist{noitemsep} 

\usepackage{longtable}
\usepackage{tabu}
\usepackage{adjustbox}

\DeclareSIUnit\protons{\text{p$^\text{+}$}}
\newcommand*\rot{\rotatebox{90}} 

\usepackage[toc,page]{appendix}

\newcommand{\BLKP}{
\ifthenelse{\isodd{\value{page}}}{\relax}{\mbox{}\thispagestyle{empty}\newpage}}

\begin{document}

\thispagestyle{empty}
\setlength{\unitlength}{1mm}

\vskip 5cm
\begin{center}
{\huge\bfseries
SPS Beam Dump Facility \\}

  \vspace{5mm}
  
  {\LARGE\bfseries Comprehensive Design Study}
  
\end{center}

 \vspace{15mm}



{\large
\noindent
  C.C.~Ahdida$^1$,
  R.G. Alia$^1$, 
  G.~Arduini$^1$,
  A.~Arnalich$^1$,
  P.~Avigni$^1$,
  F.~Bardou$^1$,
  M.~Battistin$^1$,
  J.~Bauche$^1$,
  M.~Brugger$^1$,
  J.~Busom$^1$,
  M.~Calviani$^1$,
  M.~Casolino$^1$,
  N. Colonna $^1$, 
  L.~Dougherty$^1$,
Y.~Dutheil$^1$,
E. Fornasiere$^1$, 
M.A.~Fraser$^1$,
L.~Gatignon$^1$,
  J.~Gall$^1$,
  S.~Gilardoni$^1$,
  B.~Goddard$^1$,
  J-L.~Grenard$^1$,
 D.~Grenier$^1$,
  C.~Hessler$^1$,
  R.~Jacobsson$^1$,
  V.~Kain$^1$,
  K.~Kershaw$^1$, 
  E.~Koukovini~Platia$^1$,
  M.~Lamont$^1$,
  E.~Lopez~Sola$^1$,
  S.~Marsh$^1$,
  R.~Morton$^1$,
  Y.~Muttoni$^1$,
  P.~Ninin$^1$,
  J.A.~Osborne$^1$,
  A.~Perillo~Marcone$^1$,
  J.~Prieto~Prieto$^1$,
  F.~Sanchez~Galan$^1$,
  P.~Santos~Diaz$^1$,
  S.~Schadegg$^1$,
  L.~Stoel$^1$,
   C. Torregrosa Martin$^1$, 
  Heinz~Vincke$^1$,
  Helmut~Vincke$^1$,
  F.M.~Velotti$^1$,
  P.~Vojtyla$^1$,
  T.~Wijnands$^1$,
  O.~Williams$^1$.
}

\vspace{5mm}

\hspace{2mm}
\begin{minipage}{15cm}
{\em\footnotesize

${}^{1}$  European Organization for Nuclear Research, CERN CH-1211 Gen\'eve 23, Switzerland \\
${}^{2}$ INFN-Bari, Italy
}
\end{minipage}


\vspace{10mm}


\begin{quotation}

{\bf Abstract:} The proposed Beam Dump Facility (BDF) is foreseen to be located at the North Area of the SPS.
It is designed to be able to serve both beam dump like and fixed target experiments.
The SPS and the new facility would offer unique possibilities to enter a new era of exploration at the intensity frontier.
Possible options include searches for very weakly interacting particles predicted by Hidden Sector models, and flavour physics measurements.
In the first instance, exploitation of the facility, in beam dump mode, is
envisaged to be for the Search for Hidden Particle (SHiP) experiment.


Following the first evaluation of the BDF in 2014-2016,
CERN management launched a Comprehensive Design Study over three years for the facility. 
The BDF study team has since executed an in-depth feasibility study of proton delivery to target, the target complex, and the underground experimental area,
including prototyping of key sub-systems and evaluations of the radiological aspects and safety. 
A first iteration of detailed integration and civil engineering studies have been performed in order to produce a realistic schedule and cost. 
This document gives a detailed overview of the proposed facility together with the results of the studies, 
and draws up a possible road map for a three-year Technical Design Report phase, 
followed by a 5 to 6 year construction phase. 

\end{quotation}

\vskip 8.em
\begin{center}
Geneva  \\
	December 2019\\
\end{center}
\newpage

\thispagestyle{empty}
\vspace{18mm}
\hspace{2mm}
\newpage


\pagestyle{plain}
\pagenumbering{roman}
\setcounter{page}{1}

\begingroup\baselineskip.99\baselineskip
\tableofcontents
\endgroup



\BLKP

\pagestyle{fancy}
\pagenumbering{arabic}
\setcounter{page}{1}
\renewcommand{\floatpagefraction}{0.9}
\renewcommand*\thesection{\thechapter.\arabic{section}}


 \chapter{Introduction}
\label{Chap:Introduction}

\section{Context and motivation}

The proposed Beam Dump Facility is foreseen to be located at the North Area of the SPS.
It is designed to be able to serve both beam dump like and fixed target experiments.
Beam dump in this context implies a target which aims at provoking hard interactions of all of the incident protons
and the containment of most of the associated cascade. 
In the first instance, exploitation of the facility, in beam dump mode, is envisaged to be for the Search for Hidden Particle (SHiP) experiment~\cite{2013arXiv1310.1762B, Anelli:2007512, 2016RPPh...79l4201A}.

Several recently proposed experiments~\cite{2019arXiv190200260A, 2019arXiv190109966B}
highlight that the SPS operating in beam dump mode or in fixed-target mode would be an excellent 
way to go beyond the current SPS program and enter a new era of physics studies at the 
intensity frontier.
These studies would complement the exploration of the high-energy frontier at the Large 
Hadron Collider (LHC) after 2026. Papers have been submitted~\cite{SHiP_EPPSU, TauFV, Moulson:2018mlx}
to demonstrate the unique potential in searches for particles predicted by Hidden 
Sector models and in flavour physics measurements.

The multi-user design
and full exploitation of the SPS accelerator with its 
present performance could allow the delivery of an annual yield of up to \num{4e19} 
protons on target with a beam momentum of \SI{400}{GeV/c} while respecting the beam requirements of
the HL-LHC, and while maintaining the operation of the existing SPS beam facilities. 
Currently, CERN has no experimental facility which is compatible with this beam power.
The consolidation and upgrades of the CERN injector complex and the 
continued operation of the SPS with the unique combination of the high-intensity 
proton beam and slow beam extraction, motivates the construction of a new 
high-intensity experimental facility which is capable of fully exploiting its 
capacity in parallel to the operation of the HL-LHC. 
CERN's North Area has a large area next to the SPS beam transfer lines which is for the most part free of 
structures and underground galleries, and which could accommodate the proposed 
facility. In addition, the facility may be designed with future extensions in 
mind. On a longer time scale, the possible future large-scale programs at CERN 
could also pave the way for upgrading the facility to take advantage of new or upgraded 
injectors.


Following first evaluation of the beam dump facility (BDF) in 2014-2016, submitted together with the SHiP Technical Proposal and Physics Case, 
the CERN management launched a Comprehensive Design Study over three years for SHiP and BDF. 
The BDF study team has executed an in-depth feasibility study of proton delivery to target, the target complex, and the underground experimental area, 
including prototyping of key sub-systems and evaluations of the radiological aspects and safety. 
A first iteration of detailed integration and civil engineering studies have been performed in order to produce a realistic schedule and cost. 
In complement to the SHiP physics and detector studies, the following document gives a detailed overview of the proposed facility together with the results of the in-depth studies, 
and draws up the road map and project plan for a three-year Technical Design Report phase, 
and a 5-6 year construction phase. 
The document shows that the feasibility is proven, 
the technologies and techniques, although challenging, appear to be within CERN’s established competencies, 
and that the project, given the resources, is ready to move towards the detailed design and execution phase.

\section{Objectives}

The lack of firm hints to the mass scale of new particles calls for a concerted
effort by direct searches and precision measurements. At the same time the absence of new 
particles is not necessarily due to their high scale of masses but could equivalently 
be due to their weak scale of couplings with the Standard Model particles. This motivates 
investing in an ambitious complementary programme to investigate the possibility 
of a light Hidden Sector coupled to the Standard Model.

Beam dump experiments are potentially superior to collider experiments in the 
sensitivity to GeV-scale hidden particles given potential luminosities several 
orders of magnitude larger than those at colliders. The large forward boost for light 
states, gives good acceptance despite the smaller angular coverage and allows 
efficient use of filters against background between the target and the detector, 
making the beam dump configuration ideal to search for new particles with long 
lifetimes.

The detailed specification of the BDF is, in the first instance, 
mainly driven by the requirements of the SHiP experiment. 
SHiP has been optimised to search for light long-lived particles 
produced in decays of charm and beauty hadrons and radiative processes, and 
consists of two complementary apparatuses which are sensitive to both decay 
of hidden particles and scattering signatures of light dark matter. The combination 
of the intensity and the energy of the SPS proton beam allows the production of a very 
large yield of the processes potentially capable of giving rise to the different 
Hidden Sector particles. 
During five years of operation with \num{4e19}\,protons on 
a high-density target per year, it is expected to produce ${\cal O}(10^{18})$ 
charmed hadrons and more than $10^{21}$\,photons above \SI{100}{MeV}. In addition, 
it has been found that the \SI{400}{GeV/c} proton beam at the SPS provides a good 
compromise between the large yield of heavy hadrons and photons, and a 
manageable background. Furthermore, the unique feature of slow extraction of 
a de-bunched beam on a timescale of around a second allows a tight control of 
combinatorial background.

The new beam dump facility would also allow the performance of unprecedented measurements 
with tau neutrinos. Five years of operation on the BDF target at \SI{400}{GeV/c} would yield 
${\cal O}(10^{16})$ tau and anti-tau neutrinos. The first direct observation of the anti-tau 
neutrino and the measurement of tau neutrino and anti-tau neutrino cross-sections 
are among the goals of the SHiP experiment. 
As charm hadron decays are also a source of electron and muon neutrinos, it will also
be possible to study neutrino-induced charm production from all flavours with a 
data-set which is more than one order of magnitude larger than those collected by 
previous experiments, solely performed with muon neutrino interactions. 

The BDF beam-line offers a potential opportunity to host and operate 
in parallel an experiment~\cite{TauFV} to search for lepton flavour violation and rare decays 
with the very large production of tau leptons and D mesons. Intercepting 
about 2\% of the intensity delivered to SHiP with a thin target, the experiment would 
have access to about \num{8e13} tau leptons and \num{e16} $D^0$ and $D_s$ meson decays.

In the medium term, extensions of the long-lived particle search programme at the 
BDF could be possible with a large volume light dark matter and neutrino 
scattering detector located downstream of the SHiP detector. Assuming the same angular acceptance
and a liquid argon target, the equivalent mass to the current proposal with a 
\SI{10}{tonne} scattering target of lead at \SI{30}{m} from the target, would be \SI{\sim 450}{tonne} detector at \SI{120}{m}.

In the longer term, beyond, say, 2035, depending on the future large-scale project at CERN 
after the HL-LHC, upgraded or new injectors could open a new search region.
Continued operation of the SPS leaves several options open
for the BDF depending on the development of the physics landscape. 
Findings at SHiP could motivate continued operation with protons to further 
establish or measure properties of new particles, but also potentially operation with electrons.
An upgrade of the SPS to a superconducting machine at \SI{0.9}{}--\SI{1.3}{TeV} in conjunction
with an upgrade of the slow extraction and the transfer lines, could open a new mass 
window in the searches for weakly coupled particles.
Alternatively, with a reconfigured target system the facility could also host a 
kaon physics experiment in the future, such as the proposed experiment KLEVER~\cite{Moulson:2018mlx} 
which aims at making a measurement of the branching ratio for the 
$K_L\rightarrow \pi^0\nu\overline{\nu}$ decay.

\section{Interested Community}

The physics community's interest in Hidden Sector physics has grown rapidly in recent years, with an important increase in the activity of operational experiments,
both in the domain of direct and indirect searches, as well as new experiment proposals. 
SHiP formed an official collaboration with 45 institutes shortly after the submission of an Expression of Interest~\cite{2013arXiv1310.1762B}
in 2013 as part of the recommendation from the CERN management to prepare a Technical Proposal. 
Today the collaboration consists of 50 institutes and 4 associated institutes from 18 countries, CERN, and JINR. 
Several institutes in already associated, and new, countries are showing interest, and the collaboration is expected to grow significantly upon an approval to proceed with TDRs.
Cooperation is already established with the NA62 and LHCb collaborations, both on physics and detector developments, and in particular with LHCb detector upgrade programmes.
It is expected that this synergy will further increase. 
The proposed light dark matter/neutrino scattering detector is also expected to stimulate further interest in the light dark matter search community and could lead to extensions of the current concepts.

Other potential users of the BDF include the TauFV and KLEVER proposals, both of which are in the domain of flavour physics.
A dedicated TauFV study group has been active since the end of 2017, and the proposal, as developed so far, is presented in Appendix C of this document.
The flavour physics community is large and worldwide.
Violation of lepton universality has sustained particular interest of the community in Europe, US, and Japan for several decades and has raised even more interest in recent years.
The TauFV proposal is complementary to LHCb and BELLE-II~\cite{2010arXiv1011.0352A} in this respect.
It is expected that the proposal will rapidly stimulate interest and attract collaborators. 
The project is technologically very challenging, requiring detectors with performances similar to those of the HL-LHC and FCC~\cite{Benedikt:2651299} detectors. 
A collaboration to draw up a plan for detector R\&D is already underway. 
The KLEVER experiment draws experience from NA62 and is expected to receive strong support from the kaon physics community.
Its potential realisation at ECN3 of the SPS North Area is also under consideration.

The development required for the target/dump design and construction, and for the Target Complex technologies,
could have important implications for facilities worldwide. 
Collaboration is currently being explored and could be expanded upon project approval in order to optimise resources and costs, and to increase the impact of any technological breakthroughs. 

The target/dump assembly is based on diffusion bonding via the Hot Isostatic Pressing technique, employing refractory metals such as pure tungsten, molybdenum alloys, tantalum and tantalum alloys.
New developments such as the 
Second Target Station of the Oak Ridge National Laboratories Spallation Neutron Source~\cite{Kustom:2000rj} might benefit from such R\&D and experience. 
Similarly, upgrade projects at the UK-based ISIS spallation neutron source might benefit from the technologies developed for the CERN's Beam Dump Facility. 
Developments of W-based technologies and radiation damage studies would create natural links with the European Spallation Source currently under construction.

The Target Complex, operating with a system with a purification technology for the helium vessel loop, currently under design, 
might be very beneficial for the upgrade of the T2K JPARC neutrino beam line, both in the framework of the current operation as well as for the Tokai to Hyper-Kamiokande project (T2HK)~\cite{Ishida:2013kba} project presently under consideration. 
Similar interest and collaborative opportunities exist in the framework of the LBNF/DUNE\cite{2016arXiv160105471A} neutrino project in the US. 
The radiation levels close the the production target would be extremely useful for irradiation purposes in the framework of the RaDIATE Collaboration~\cite{Ammigan:2018gfb}
and would open unprecedented equipment and material test possibilities for CERN related to the Radiation to Electronics (R2E) Project.

\section{Overview of the BDF facility}

\begin{figure}[htb]
\centering
\includegraphics[width=0.7\linewidth]{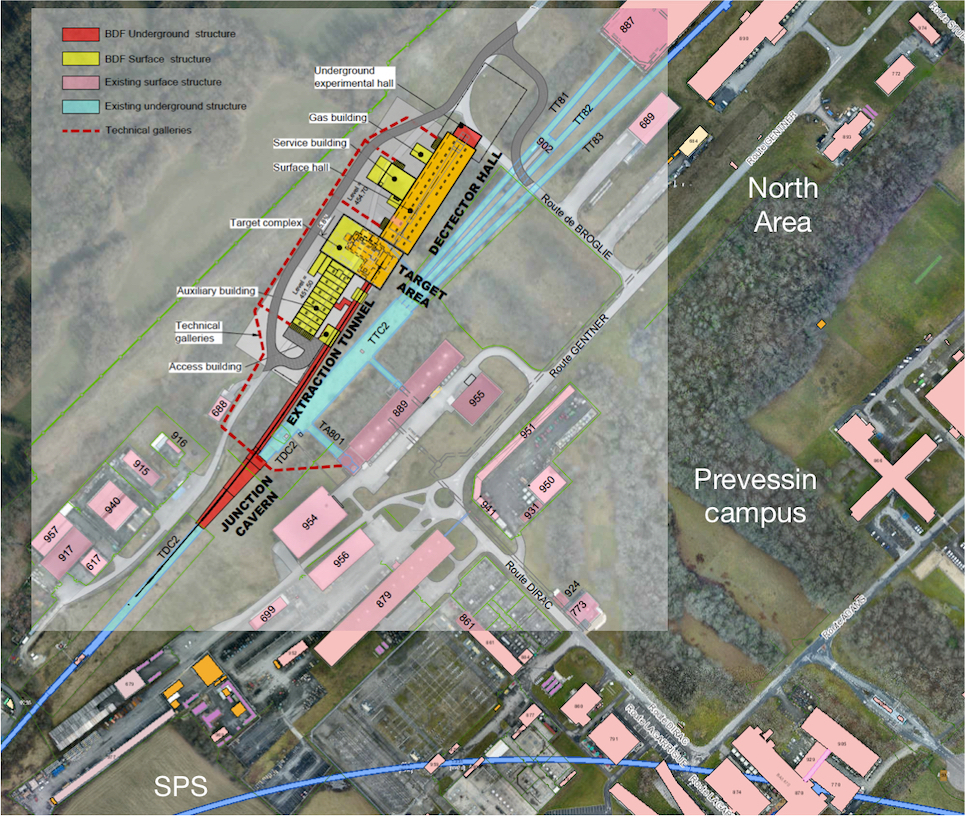}
\caption{Overview of the proposed implementation of the BDF at the SPS North Area at the CERN Pr\'evessin campus.}
\label{fig:Intro-location}
\end{figure}

The BDF relies on slow extraction of protons from the SPS, from an upgraded extraction channel in Long Straight Section 2 (LSS2).
After about \SI{600}{m} of the present TT20 transfer line,
a new splitter/switch magnet system, which will maintain compatibility with the existing NA operation, 
will deflect the beam into a dedicated new transfer line. 
The new line will be connected by a modified junction cavern to the existing tunnel. 
The beam is dumped on a high power target housed in a purpose-built, heavily shielded, target complex.
Beyond the target complex, a muon shield is followed by an experimental hall. 
The scope of the facility studies includes these elements plus the associated CE and service infrastructure. 
The proposed location and overall layout of the facility is shown in Fig.~\ref{fig:Intro-location}. 
A description of the key features of the new facility and a brief overview of the progress made by the study on the technical aspects follows.

\subsection{Extraction from SPS}
Third-integer slow extraction of \SI{400}{GeV/c} protons from the SPS is well established. A few percent of beam losses on the septum wires is intrinsic to the process and results in machine activation, reduces component lifetime and places severe limitations on personnel access and maintenance. Significant operational effort already goes into minimising these losses and ensuring a high quality, uniform spill. The proton intensity on target (PoT) requested by SHiP poses significant challenges~\cite{Fraser:2017ndi} and improvements will need to come from a combination of lower beam loss per extracted proton, reduced activation per lost proton, and improved or remote interventions. The most effective solution is to reduce the beam loss per extracted proton, since this also reduces the radiation dose to cables and high voltage feedthroughs. A factor of four reduction is needed.

Methods to reduce the losses concerning the extraction process, hardware and controls have been developed and tested in 2018, both with a SHiP cycle and the longer NA cycle.

The Q-sweep method used for slow extraction in the SPS since its construction has been replaced by a new type of Constant Optics Slow Extraction (COSE), 
where the optics are kept constant in normalised strength while the whole machine momentum is ramped. 
COSE has several advantages over the Q-sweep, since it keeps the orbit and separatrix angle at the electrostatic septum (ZS) fixed through the spill. 
Since mid-2018 COSE has been systematically deployed for NA operation.

For loss reduction, both passive scatterers and bent silicon crystals have been developed and tested as diffusers to locally reduce the proton density at the ZS wires. For the passive diffuser~\cite{Goddard:2017kbl}, a 240~\si{\um} wide, 30~\si{\mm} long array of Ta wires achieved a loss reduction of 15\%, consistent with an effective ZS width of 500 -- 600~\si{\um}. For the bent silicon crystal \cite{Velotti:2017oey} a 780~\si{\um} width and 2.5~\si{\mm} long crystal with a large channelling angle of 150~\si{\micro\radian} gave a loss reduction of slightly over 40\%, again consistent with a ZS width of around 500~\si{\um}. Both diffuser types were tested for 12 hour periods with the full beam intensity of \num{3e13} protons per spill, and demonstrated that ZS shadowing was stable and reproducible.

A separate technique of loss reduction by separatrix folding\,\cite{Stoel:2017xbm} was also tested successfully. The extraction sextupoles which govern the speed of diffusion across the ZS wires were increased in strength to reduce the particle density and losses, while octupole fields slow the diffusion speed at higher amplitude to avoid increased losses on the ZS cathodes. In beam tests this method also reduced the losses by slightly over 40\%. Importantly, it was successfully tested in combination with the crystal ZS shadowing. The combination of the two methods gave a loss reduction of slightly more than a factor three.

The alignment of the five ZS anodes is a crucial factor in the overall beam loss, since it determines both the absolute beam loss and the potential gain from the shadowing methods. The control of the alignment was improved to a resolution of below 50~\si{\um}, while numerical optimisers were simulated and deployed to align the five anodes to the extracted beam. The alignment time (with 9 degrees of freedom) was reduced from 8~hours to 40~minutes. 

\subsection{Transfer lines, switching and dilution}

The location of the BDF target complex allows the re-use of about \SI{600}{m} of the present transfer line TT20, which is already operated with slow-extracted beam at \SI{400}{GeV/c}. After the upgraded switch/splitter element, a new \SI{380}{m} long section of beam line (TT90) is required to deliver the beam to the target.

The powering scheme for the TT20 transfer line will remain largely unchanged. The design of TT90 uses 23 standard bending magnets at a field of up to \SI{1.9}{T} in a FODO structure to ensure adequate separation between the new and existing beam lines, and to minimise the longitudinal extent of the CE works in the new junction region. Six standard focusing quadrupole magnets provide flexibility and tunability of the beam spot size and dispersion at the proton target. The line optics have been finalised with the completion of trajectory correction and aperture studies, and the parameter feasibility at the target has been demonstrated with the use of existing magnet designs. 

The transfer line design replaces the three existing splitter magnets by laminated versions with dual functionality: either splitting the beam destined for the NA targets as today; or deflecting the entire beam into TT90 for transport to the BDF target. For the BDF beam sent to TT90, the entire beam will be steered through the switch/splitter aperture without losses. This solution maintains full compatibility with the present NA operation. 

The present splitter magnet is an in-vacuum Lambertson septum with a yoke machined from solid iron, with the coil based on a water-cooled lead of copper with an insulation of compacted MgO powder. For the new magnets a laminated yoke is required to perform the polarity switch between SPS cycles in about \SI{2}{s}.

A magnetic and mechanical design has been made and the magnet performance simulated, complicated by the details of the possible mechanical errors and effect on the beam losses. Prototyping of the laminated yoke design is underway to evaluate the feasibility of the very tight mechanical tolerances required to maintain low beam losses. MgO coils will provide the required radiation resistance. Procurement of parts for the construction of a short magnet prototype will start in early 2019.

The beam dilution sweep will be implemented with two sets of two orthogonal kicker magnets with Lissajous powering functions to produce a circular sweep at \SI{4}{Hz}. With a  free drift length for the beam of about \SI{120}{m} and a bending angle of \SI{0.5}{mrad} per plane, the sweep radius will be \SI{50}{mm}. Since the survival of the proton target relies critically on the beam dilution, the SPS beam will be interlocked with the beam dilution system and the instantaneous loss rate at the target. New concepts for interlocking of the slow extraction have been developed.

A straightforward reconfiguration of the existing beam elements in the BDF extraction channel would allow the accommodation of 
the drift space required to implement the in-line target and experimental zone for the tau lepton flavour experiment.


\subsection{Production target/dump}

The BDF/SHiP target can be considered as a beam dump, as it has to safely absorb the full \SI{400}{GeV/c} SPS primary beam every \SI{7.2}{s}. The target is required to maximise the production of charm and beauty hadrons, and to maximise the re-absorption of pions and kaons, which implies a high-Z material with a short nuclear interaction length, contrary to a neutrino-producing target. The high deposited power is the most challenging aspect, with up to 355~kW average power deposited on target and \SI{2.56}{MW} over the \SI{1}{s} spill. To produce sufficient dilution of the energy density in the target, the slow extraction needs to be combined with a beam spot of at least 8~\si{mm} root-mean square in both planes and a \SI{300}{mm} long sweep of the beam over the target surface.

Detailed energy deposition and thermo-mechanical design studies have been performed. The required performance may be achieved with a longitudinally segmented hybrid target consisting of blocks of four nuclear interaction lengths (58~\si{cm}) of titanium-zirconium doped molybdenum alloy (TZM, density 10.22~\si{g/cm^3}) in the core of the proton shower followed by six nuclear interaction lengths (58~\si{cm}) of pure tungsten (density 19.3~\si{g/cm^3}).

A medium-density material is required in the first half of the target to reduce the energy density and resulting thermal-induced stresses. The thickness of each block and location of each cooling slot has been optimised to provide uniform energy deposition and sufficient energy extraction. The blocks are interleaved with 5~\si{mm} wide slots for water cooling. Tantalum alloy cladding of the TZM and the tungsten blocks - by means of diffusion bonding via Hot Isostatic Pressing - will prevent corrosion and erosion of the core material by the high water flow rate. The design limits the peak power density in the target to below 850~\si{J/cm^3}/spill and compressive stresses to below 130~\si{MPa}. 

The target blocks will be assembled in a double-walled helium vessel. The inner vessel will enforce the high-flow 35~\si{m^3/h} water circulation between the proton target blocks at 20~\si{bar} to avoid water boiling. The outer vessel acts as a safety  hull to contain hypothetical leaks, and is filled with He gas to prevent corrosion.

A prototype target built to the BDF/SHiP design was installed and tested with beam in the North Area target area (TCC2) in 2018. Lower intensity 1\,s spills of \num{4e12}\,protons without dilution sweep was used to achieve stress levels comparable to the final target. The prototype was successfully operated with beam for over 14\,h, accumulating \num{2.4e16}\,PoT. Online measurements of strains and temperature on instrumented target blocks showed a very good agreement with simulation. In 2019 the target blocks will be disassembled and analysed in a dedicated facility to quantify the target material behaviour under irradiation conditions.

\subsection{Target complex}
\label{sec:tc}

The target will be subject to severe radiological constraints, and will  be located in a shielded bunker around 15 metres below ground level. Remote handling for manipulation of the target and surrounding shielding will be mandatory due to the high residual dose rates expected after operation. The target complex has been designed to house the target and its shielding in a helium vessel, along with the cooling, ventilation and helium purification services below ground level. The SPS beam will enter the surrounding helium vessel through a removable beam window, then pass through a collimator which serves to protect the target and adjacent equipment from misalignment of the incident SPS beam and to protect the equipment in the extraction tunnel from particles (essentially neutrons generated by the target) travelling backwards relative to the incident beam. 

The target will be surrounded by approximately 3700 tonnes of cast iron and steel shielding (part of it water-cooled to dissipate the deposited power) with outer dimensions of around \num{6.8 x 7.9 x 11.2}~\si{\cubic\metre} (the so-called bunker/hadron absorber) to reduce the prompt dose rate during operation and the residual dose rate around the target during shutdown. The target and its surrounding shielding will be housed in the vessel containing gaseous helium slightly above atmospheric pressure in order to reduce air activation and reduce the radiation accelerated corrosion of the target and surrounding equipment. The design allows for removal and temporary storage of the target and shielding blocks in the cool-down area below ground level and includes dedicated shielded pits for storage of the highest dose rate equipment.

In 2018 the target complex design has been developed in detail with full definition of the handling and remote handling operations required throughout the life of the facility. This work demonstrated the feasibility of the construction, operation, maintenance of the BDF target complex along with decommissioning of the key elements. The remote handling of highly activated radioactive objects, such as target, beam window, collimator, shielding blocks and magnetic coil, along with their connection and disconnection within the target complex building were studied in detail, including foreseen remote handling operations such as target exchange as well as unforeseen operations needed to recover from failures or damage to equipment.

The study included the conceptual design of lifting, handling and remote handling equipment for the highly activated objects along with the necessary water, helium and electrical connections compatible with the radiation environment and remote handling constraints. These have been integrated in the overall target complex design \cite{1748-0221-13-10-P10011}.

\subsection{Muon shield}

The total flux of muons emerging from the proton target with a momentum larger than \SI{1}{GeV/c} amounts to ${\cal O}(10^{11})$ muons per spill of \num{4e13} protons.
To control the background from random combinations of muons producing fake decay vertices in the detector decay volume,
and from muon deep inelastic scattering producing long-lived neutral particles in the surrounding material, 
and to respect the occupancy limits of the sub-detectors, 
the muon flux in the detector acceptance must be reduced by at least six orders of magnitude over the shortest possible distance. 
To this end, a muon shield entirely based on magnetic deflection has been developed~\cite{Akmete:2017bpl}. The first section of the muon shield starts within the target complex shielding assembly, one meter downstream of the target, with a magnetic coil which magnetises the hadron stopper made of US1010 steel with a field of \SI{1.6}{T} over \SI{4.5}{m}. The rest of the muon shield consists of 6 free-standing magnets, each \SI{5}{m} long, located in the upstream part of the experimental hall.

\subsection{Experimental Area}

\begin{figure}[th]
\centering
\includegraphics[width=0.99\columnwidth]{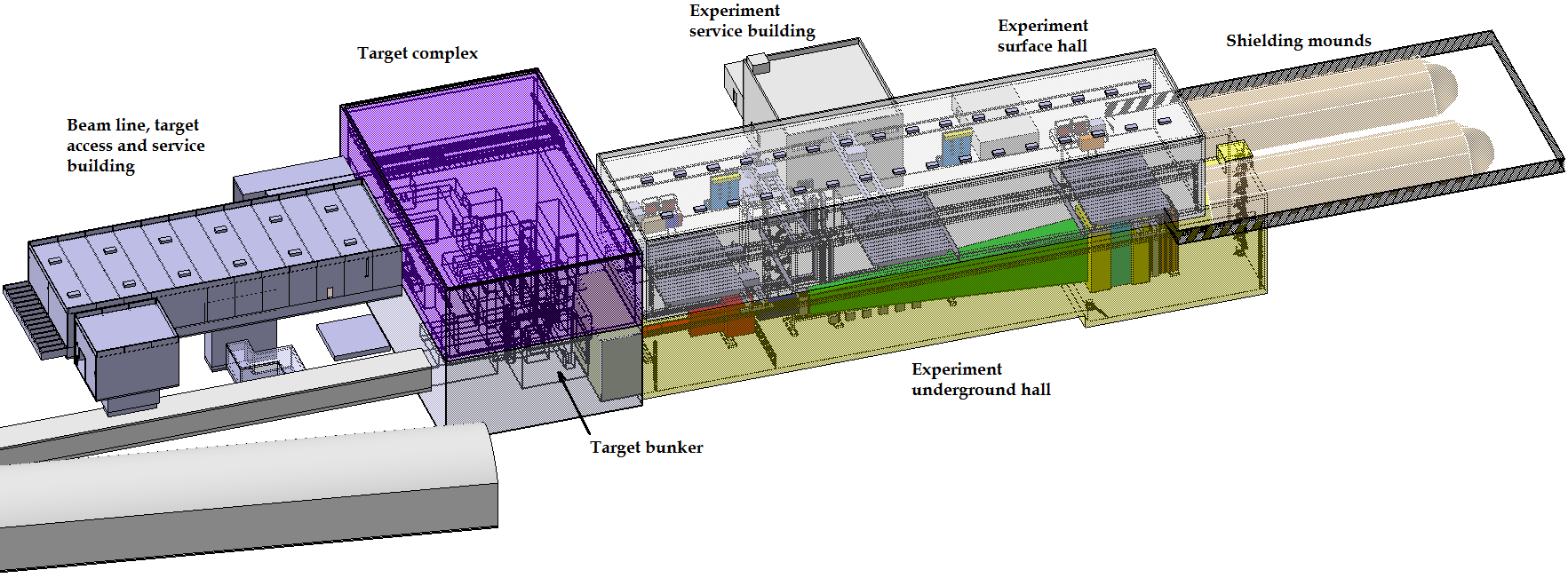}
\caption{Overview of the target complex and experimental area.}
\label{fig:ExperimentalAreaOverview}
\end{figure}

BDF is designed to house a multi-purpose large-scale experimental program using a single beam line and a single main target station. The initial design of the experimental area (Fig.~\ref{fig:ExperimentalAreaOverview}) has been dictated to a large extent by the requirements of the SHiP experiment. All phases of the experiment, including assembly, construction and installation, as well as operation, have been taken into consideration. The complex consists of a \SI{120}{m} long underground experimental hall, centred on the beam axis. In order to reduce background from particle scattering in the walls, the underground hall is \SI{20}{m} wide along its entire length. A \num{100 x 26}~\si{\square\metre} surface hall is located on top of the underground hall. The installation plan foresees pre-assembly in the surface hall in three principal work zones with the help of a \SI{40}{tonne} and a \SI{10}{tonne} crane, in parallel to final assembly in the underground hall using a dual \SI{40}{tonne} hoist crane and a single  \SI{40}{tonne} hoist crane. Three \num{14.5 x 18}~\si{\square\metre} access openings between the surface hall and the underground hall provide direct access to the principal detector installation areas. For shielding purposes each opening would be covered by 18 concrete beams during operation. A \num{20 x 34}~\si{\square\metre} three-storey service building, adjacent to the service hall, houses all services related to the infrastructure and the detector, control room, workshop, labs, and offices. 

\subsection{Overview Summary}

The above outline presents a succinct overview of the proposed facility.
Since the inception of the BDF feasibility study in 2016, the BDF team has attempted to
address all pertinent technological challenges, 
along with a detailed look at the deployment of such a facility on the proposed site.

By end 2018, in-depth studies and prototyping had been performed or are already well underway for all critical components. 
Through a mixture of novel hardware development, beam physics and technology, the study and prototype validations have shown
that SPS can deliver the beam with the required characteristics and with acceptable losses, to a robust target housed in a suitable target complex. 
The detailed results of these studies are presented in the following chapters.

With the delivery of this report, and the corresponding document from the SHiP collaboration,
it may be argued that BDF and SHiP are mature proposals, and are ready, with appropriate approval, to move into the TDR phase. 

\FloatBarrier

\printbibliography[heading=subbibliography]

 \chapter{Delivery of beam from the SPS}
\label{Chap:Beam}

\section{Introduction}

\begin{figure}[ht]
\centering
\includegraphics[width=0.98\textwidth]{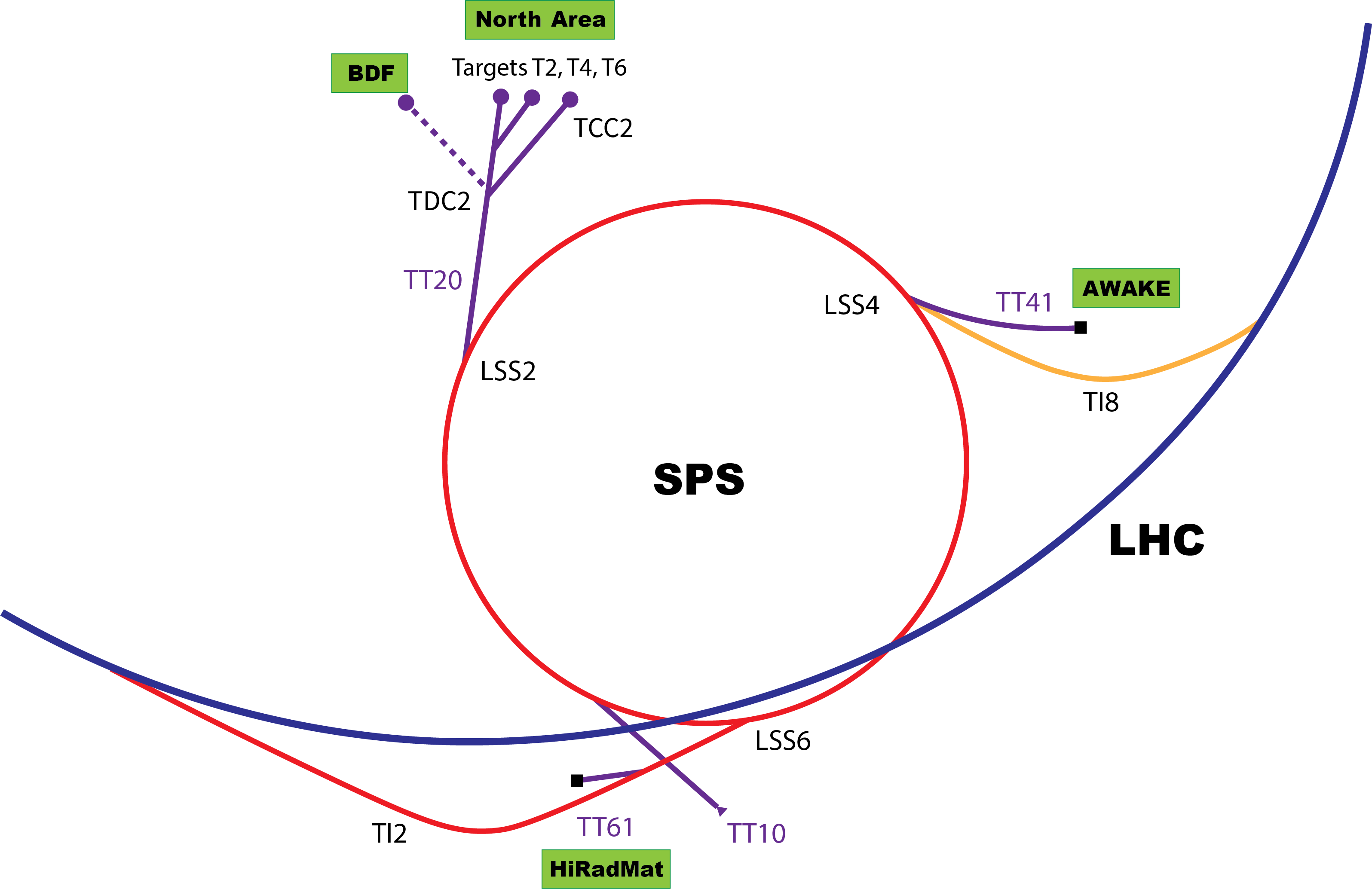}
\caption{\label{fig:SPSlayout}Schematic layout of the SPS and its associated facilities. Abbreviations used extensively in this report are shown. These include: 
Long Straight Section (LSS); 
TT20 -- the beam lines to the North Area targets;
TCC2 -- the target hall of the SPS North Area; 
TDC2 -- the tunnel/cavern before the target hall containing the splitters and downstream beam lines leading to TCC2.
}
\end{figure}

The Super Proton Synchrotron (SPS) serves several clients in sequence via a repetitive operational sequence (``Supercycle") of a length of a few tens of seconds. 
Users can include the LHC, the Advanced Proton Driven Plasma Wakefield Acceleration Experiment (AWAKE), the High-Radiation to Materials (HiRadMat), and the North Area (NA). 
The NA receives a primary proton beam at 400 GeV/c wherein the full SPS beam is slowly extracted over a flat top of typically several seconds.
The recent maximum slow-extracted proton intensity was about \num{3.5e13} protons over 4.8~s. 
This proton flux is transported and shared through two series of splitter magnets onto three primary targets, T2, T4 and T6, from which the NA secondary beam lines are served. The targets are housed in TCC2, the target hall of the SPS North Area. 
Important limitations on the intensities extracted to the NA are the losses inherent to slow extraction, and to the splitting process.
A schematic overview of the SPS and its associated facilities is shown inf Fig. \ref{fig:SPSlayout}.

The beam and SPS cycle parameters foreseen for SHiP \cite{SHIP} at BDF are shown in Table \ref{tab:SHiP-P}. 
Of note is the total beam intensity, the relatively short slow extraction spill length, and high beam power on target. 

\begin{table}[ht]
\begin{center}
\caption{Key SPS beam and cycle parameters foreseen for SHiP}
\label{tab:SHiP-P}
\begin{tabular}{ll}
\hline
Momentum 	   & 400   GeV/c            \\
Beam Intensity per cycle  & 	\num{4.2e13}              \\
Beam Intensity on target  & 	\num{4.0e13}              \\
Cycle length     & 	7.2   \si{s}            \\
Spill duration     & 	1.0   \si{s}            \\
Avg. beam power on target 	   & 355   \si{kW}           \\
Avg. beam power on target during spill  	   & 2560   \si{kW}             \\
Protons on target (PoT)/year	   & \num{4.0e19}             \\
Total PoT in 5 years data taking	   &   \num{2.0e20}           \\
\hline
\end{tabular}
\end{center}
\end{table}

A detailed study of possible future proton sharing between potential SPS users was performed
previously and was reported in \cite{SHIP_protonsharing_note}. 
This chapter summarises a new study, taking into account the operational experience from the recent years in terms of machine time usage and SPS users,
supercycle compositions, limitations from activation and recent progress in loss reduction for the SPS slow extraction on the third order resonance achieved in machine development (MD) studies. 
Furthermore, new considerations on the machine interlocking strategy for the operation of the NA and the SHiP experiments are provided. 
Finally, future operational scenarios with varying spill lengths to the NA targets are discussed.

\section{SPS intensity reach for slow extracted beams}

\subsection{SPS intensity reach}

The maximum beam intensities accelerated so far and extracted from the SPS (peak values) in the last 20 years together with the peak operational parameters from the 2018 run are listed below (the duration quoted corresponds to the cycle length):

\begin{itemize}
    \item $4.8\times10^{13}$~p/cycle (1997 - slow and fast slow extraction - 9.6~s - 450~GeV/c)
    \item $4.5\times10^{13}$~p/cycle (2008 - CNGS - fast extraction - 6.0~s - 400~GeV/c)
    \item $4.0\times10^{13}$~p/cycle (2009 - slow extraction - 15.6~s - 400~GeV/c)
    \item $3.5\times10^{13}$~p/cycle (2018 - slow extraction - 10.8~s - 400~GeV/c)
\end{itemize}

The intensity reach for fixed target beams from the SPS was studied in the past in preparation of the CERN accelerator complex serving the CERN Neutrinos to Gran Sasso (CNGS) facility \cite{CNGS_proton_flux}. The maximum intensity accelerated in the SPS during MDs (but not extracted) has been $5.3\times10^{13}$~p/cycle (2004) \cite{PAC2005}. 
The main intensity limitation for these beams was identified to come from losses in the PS and SPS. 
In particular, losses are encountered at PS-to-SPS transfer due to the extraction process and in the SPS itself due to losses at the vertical aperture and due to radiofrequency (RF) power limitations. 

To mitigate the losses at PS extraction, the ``Continuous Transfer'' (CT) \cite{CT} scheme in which the beam was  split by the extraction septum was replaced by the ``Multi-Turn-Extraction'' (MTE) \cite{MTE2001} scheme in which the beam is split magnetically. This scheme has been used operationally since 2015 and has reduced activation levels in the PS significantly \cite{MTE}. The maximum intensity used operationally with MTE has been about $3.5\times10^{13}$ p/cycle in the SPS (during the 2018 run). A beam intensity of $4.0\times10^{13}$~p/cycle could be accelerated in the SPS in a recent high-intensity test (2017) \cite{HighIntens}. During this test no particular issues were encountered related to the high beam intensity in combination with the MTE. 

\begin{figure}[ht]
\centering
\includegraphics[trim=0 10 0 0, clip,width=0.98\textwidth]{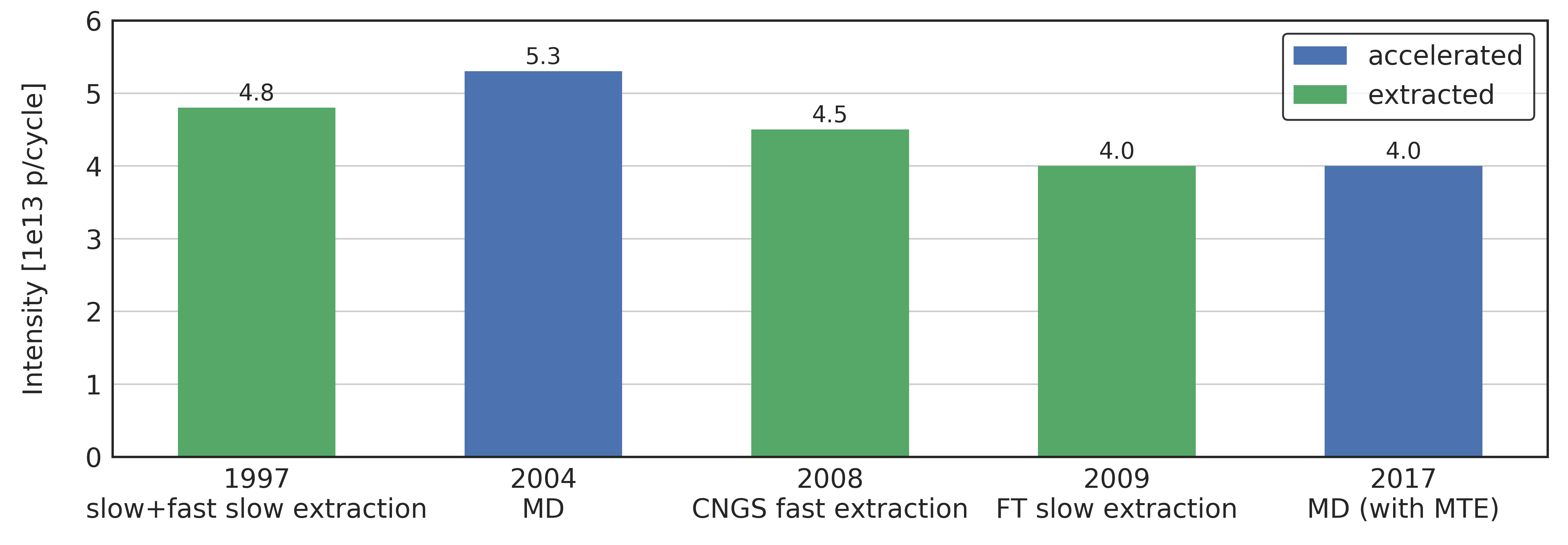}
\caption{\label{fig:SPSintensityrecords}Intensity per cycle achieved in the SPS.}
\end{figure}

The fact that the beam transmission in the SPS is degrading with intensity is mostly related to the increase of the vertical emittance proportional to intensity (due to the beam production in the PSB) and particles lost at the vertical aperture of the SPS. 
With the connection of Linac4 to the PSB during the second long shutdown (LS2) as part of the Large Hadron Collider (LHC) injectors upgrade (LIU) project \cite{LIU, LIU2}, the vertical emittance of the fixed target beam is expected to be reduced by up to a factor two, which would improve transmission in the SPS. 
In addition, it should be mentioned that the SPS RF system will receive a major upgrade as part of the LIU project, hence more RF power will become available after LS2. 
This could allow to further increase the intensity reach for fixed target beams in the SPS. 
A beam intensity of $4.2\times10^{13}$~p/cycle can therefore be safely assumed as future average intensity accelerated in the SPS with the MTE deployed in the PS. 
A summary of the SPS intensity records is shown in Fig.~\ref{fig:SPSintensityrecords}.

\subsection{Limitations from slow extraction and splitting losses}

\begin{figure}[t]
\centering
\includegraphics[width=0.98\textwidth]{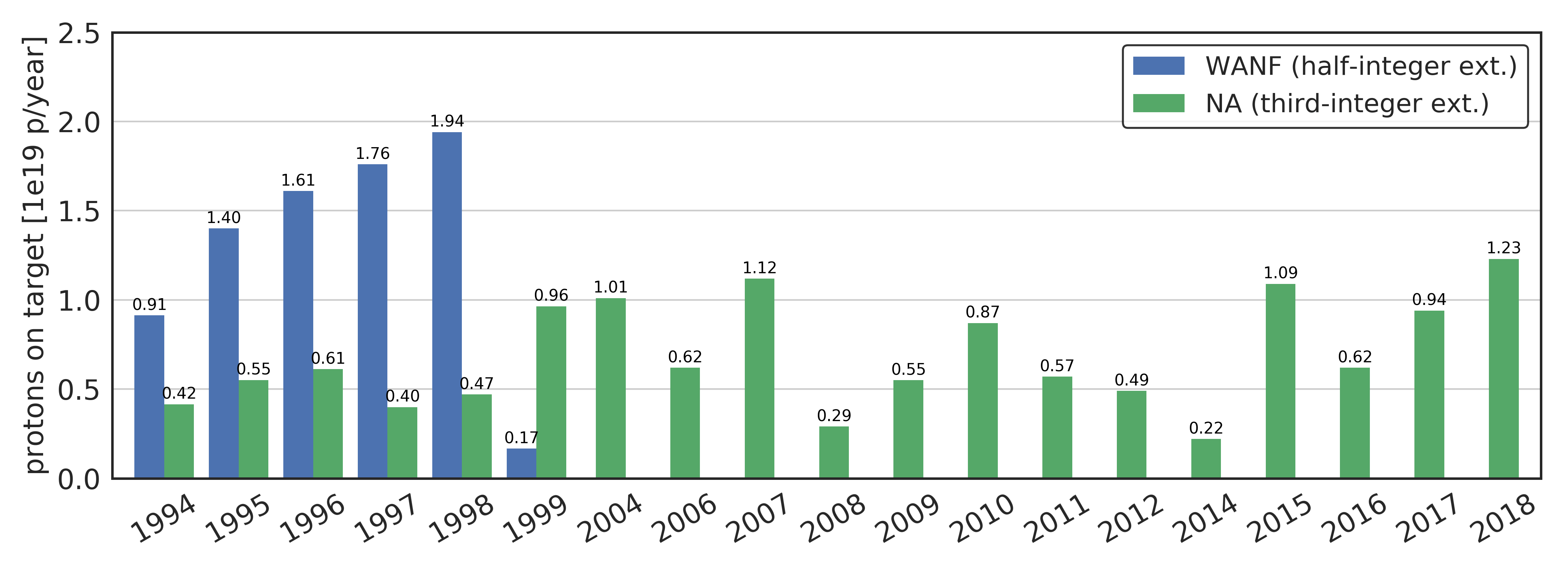}
\caption{\label{fig:SPSpots}Protons on target delivered by the SPS in the last years (no single infallible reference source for data, cross-checked as best as possible with SPS operation logging data, SPS reports/minutes, radiation protection reports etc.). Intensity data taken using secondary emission foils (BSI) located in the NA transfer lines and in front of the targets might be affected by calibration uncertainties~\cite{cali}.}
\end{figure}

An overview of the PoT delivered by the SPS in the last years to different experiments using resonant extraction schemes is shown in Fig.~\ref{fig:SPSpots}. For SHiP, the beam will be extracted on the third-integer resonance using the long straight section (LSS2) extraction channel as for the NA beams. 

Slow extraction using third-integer resonance and thin electrostatic septa (ES) is a process with inherent beam loss. For the 400~GeV protons required for the present NA operation this beam loss already results in machine activation, reduced component lifetime, and severe limitations on personnel access and maintenance. With the request from SHiP of an additional  $4.0\times10^{19}$ PoT annually, the instantaneous and integrated loss levels become key limitations. 

The activation and radiation doses in LSS2 and also at the splitters in TT20 are correlated to the total PoT and also to other less tangible aspects of the SPS operation and extraction channel setup, such as the beam stability, the alignment of the extraction septa, any hardware faults, and the beam quality. There is therefore a significant scatter in the activation levels per proton, depending on the specific combination of conditions and also the level and quality of the follow-up both of the hardware and of the operation.

\begin{figure}[ht!]
\centering
\includegraphics[trim=0 5 0 0, clip, width=0.85\textwidth]
{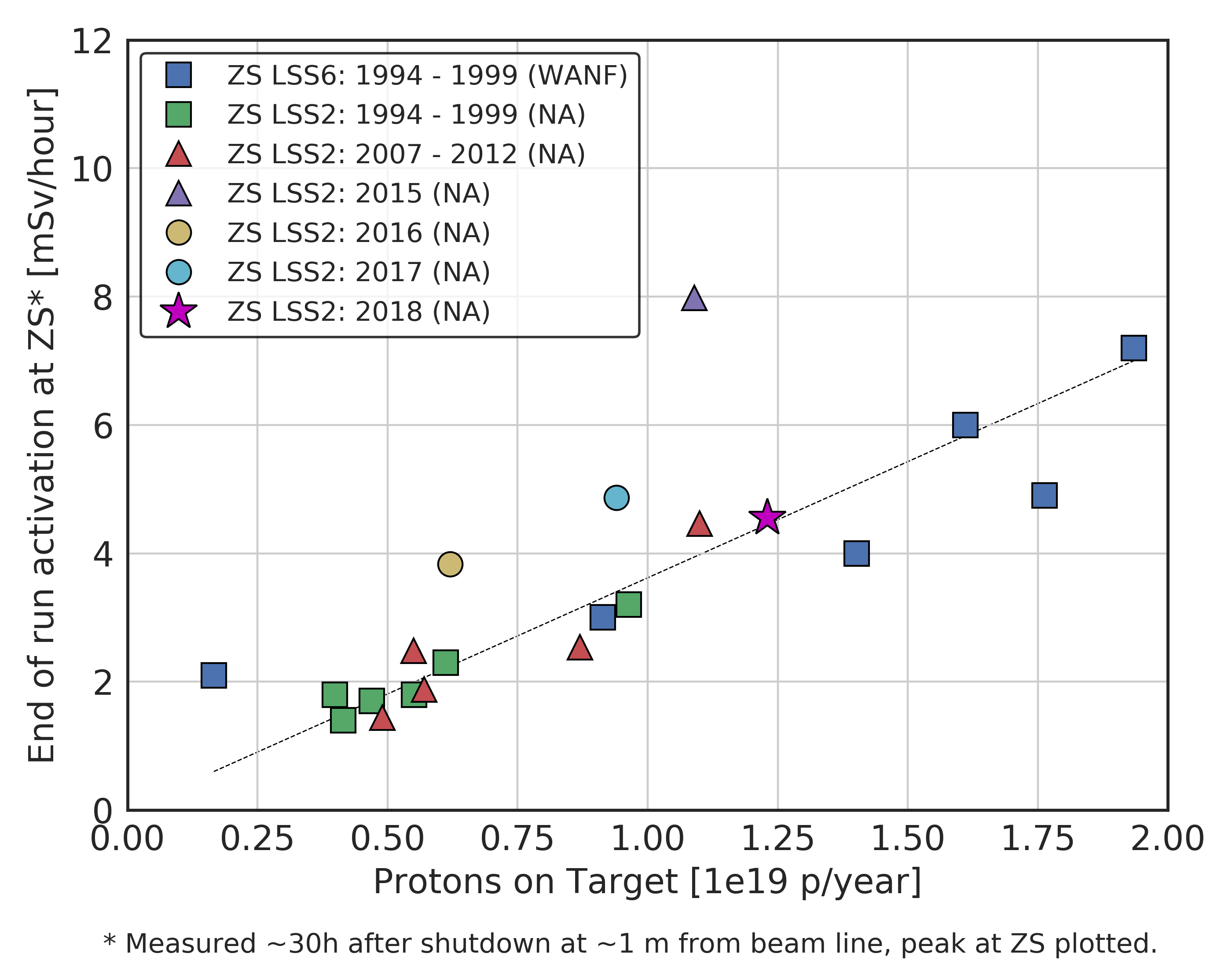}
\caption{\label{fig:SPSactivation}Comparison of activation levels in the SPS extraction channels.}
\end{figure}

The residual dose after the end of proton operation has been measured over many years. Figure~\ref{fig:SPSactivation} shows the results for both LSS2 and LSS6. 
To keep the residual activation levels at around 5~mSv/h while extracting around $5.0\times 10^{19}$ PoT will require a reduction in the activation per proton of about a factor four\cite{fraser_ipac17, shadowing, brunner2, SPS_diffuser} compared to the results obtained with around $1.0\times10^{19}$ PoT annually.

The SHiP study has developed and tested methods to reduce the losses per proton at extraction by a factor of four. 
Since the beam for SHiP will pass through the gap of the splitter system in an essentially loss-free transport, this gain will mean that the present loss levels both in the extraction channel and in the splitter region can be maintained for the simultaneous delivery of $4.0\times10^{19}$ to SHiP and $1.0\times10^{19}$ to the NA (via the lossy splitting). Studies will be launched in 2019 to investigate whether there are ways to reduce the losses per proton during the splitting process, but before these conclude the present situation needs to be assumed.

The radiation dose to the cables and extraction equipment depends on the beam loss at extraction and the total number of protons extracted.
If the factor four loss reduction can be applied to both SHiP and NA extraction, then the limitation for the total number of protons extracted per year (to whatever target) will be $5.0\times10^{19}$. At these dose levels, a full re-cabling of the extraction channel region (control, high voltage and DC cables) is required after about 8 years of operation, or every other long shutdown.

\begin{figure}[ht]
\centering
\includegraphics[trim=-20 0 20 0, clip, width=0.9\textwidth]{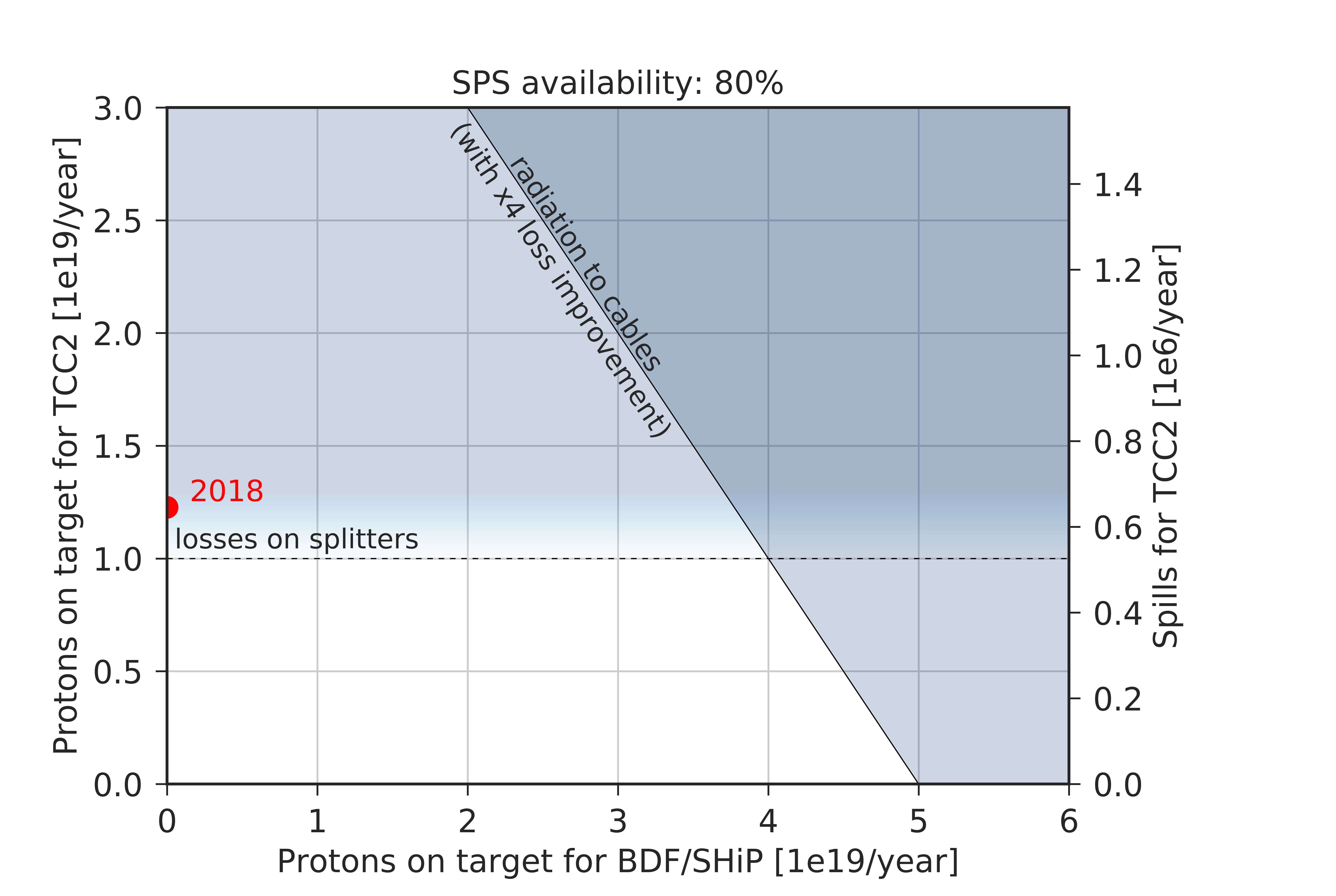}
\caption{\label{fig:SPS_protonsharing_2018}Intensity limitations (shaded areas) from SPS slow extraction and operationally achieved protons on target in 2018. The secondary axis shows the number of spills.}
\end{figure}

The slow extracted beam is distributed via a network of transfer lines in the NA and delivered simultaneously to multiple experimental targets by splitting the beam directly on steel Lambertson septa magnets \cite{Evans}. This process, when combined with the extraction and transfer efficiency, introduces a transmission from ring to targets that is typically measured at approximately 70\%. The splitting efficiency also depends on the splitting ratio and intensities demanded by each target. Due to systematic errors on the absolute calibration of the intensity monitors in the NA this number is intrinsically uncertain and should be taken with caution. The limitations discussed above are plotted graphically in Fig.~\ref{fig:SPS_protonsharing_2018}. It should be emphasised that, in this plot, the beam loss per proton in the LSS2 extraction channel of the SPS is assumed to be reduced by a factor 4 as compared to present operational values. 

\subsection{Beam parameters considered for SHiP}
The slow extraction on the third integer to TT20 in the time scale of one second is retained as baseline for the SHiP beam parameters, taking into account the following considerations:
\begin{itemize}
    \item maximum acceptable instantaneous particle flux at the detector
    \item calculations of the power and power density deposition on the target.
\end{itemize}

SHiP assumes a proton beam of 400~GeV/c. This beam momentum corresponds to the momentum of the proton beam extracted to the NA targets T2, T4 and T6 during proton fixed target operation. In the past it was considered to define different extraction momenta for the beams circulating in the SPS according to their destination to allow proper interlocking. In the case of SHiP, the present machine interlocking approach will not rely on different beam momenta. Instead both the beam to the NA targets as well as the beam for the SHiP experiment will be extracted at 400~GeV/c. The programmed timing destination will be used to distinguish between beam to SHiP or the NA targets in the extraction interlock system. 

The minimum cycle length that is compatible with the above parameters is 7.2~s provided that it can be proven that an extraction of $4.2\times10^{13}$ p/cycle over 1~s can be performed without damaging the electrostatic septa or significantly increasing the spark rate. In the WANF era in LSS6, intensities of $1.5\times10^{13}$ p were extracted in a spill length of under $10$~ms, twice per cycle. Although there were relatively frequent stops due to damage of the electrostatic septum ZS, many improvements were made and these instantaneous rates are considered acceptable. During machine development studies in 2017 and 2018 up to $1.0\times10^{13}$ p/cycle slow extracted over 1~s were demonstrated on a SHiP cycle. Extracting higher intensities on such a cycle has not been attempted yet. 

\section{Proton sharing}
\subsection{SPS cycles and sharing of machine time in 2018}

In its present configuration, the SPS delivers beam to the LHC, the NA, the AWAKE, and the HiRadMat. In addition, a rich program of MD studies are carried out in order to improve the machine performance and prepare for future beam requests. Table~\ref{tab:SPScycles} shows a representative selection of cycles used during SPS operation in 2018. 
\begin{table}[ht!]
    \begin{center}
    \caption{SPS cycles from 2018.}
    \begin{tabular}{lcc}
\toprule
{} & Length [s] & Power [MW] \\
\midrule
AWAKE        &        7.2 &      31.23 \\
BDF/SHiP     &        7.2 &      32.50 \\
HiRadMat     &       22.8 &      16.83 \\
LHC filling  &       22.8 &      16.83 \\
LHC pilot    &       12.0 &      31.67 \\
MD dedicated &       22.8 &      16.83 \\
MD parallel  &        7.2 &       2.98 \\
Scrubbing    &       22.8 &      16.83 \\
TCC2         &       10.8 &      52.79 \\
Zero         &        1.2 &       0.10 \\
deGauss      &        3.6 &       4.77 \\
\bottomrule
\end{tabular}

    \label{tab:SPScycles}
    \end{center}
\end{table}

Note that on the ``Zero'' and the ```Degauss'' cycles it is not possible to inject any beam. The latter is typically placed in front of the fixed target cycle (here called the ``TCC2'' cycle) to achieve reproducible magnetic behaviour of the machine for optimising transmission and slow extraction conditions. The parallel MD cycle has a short flat bottom for measurements (about 3~s) and a short ramp to 200~GeV/c to establish the magnetic reference in the main magnets for the cycle after (typically the TCC2 cycle).

The maximum acceptable average resistive power dissipated in the main dipole magnets for the SPS is 37.9~MW, while the total average power for the main dipoles and main quadrupoles is 44~MW. This constraint limits the possible supercycle combinations. A list of the supercycles used during 2018 operation is shown in Table~\ref{tab:2018supercycles}. In both tables the quoted power corresponds to the sum of the power in the main dipoles and main quadrupoles. 

\begin{table}[ht]
    \begin{center}
    \caption{SPS supercycles used in 2018.}
    \begin{tabular}{l|ccccccccccc|cc}
\toprule
{} & \rot{AWAKE} & \rot{BDF/SHiP} & \rot{HiRadMat} & \rot{LHC filling} & \rot{LHC pilot} & \rot{MD dedicated} & \rot{MD parallel} & \rot{Scrubbing} & \rot{TCC2} & \rot{Zero} & \rot{Degauss} & \rot{Length [s]} & \rot{Power [MW]} \\
\midrule
AWAKE                    &           2 &              - &              - &                 - &               - &                  - &                 - &               - &          2 &          - &             2 &             43.2 &            37.60 \\
AWAKE with parallel MD   &           2 &              - &              - &                 - &               - &                  - &                 2 &               - &          2 &          - &             - &             50.4 &            32.40 \\
Dedicated MD             &           - &              - &              - &                 - &               - &                  1 &                 - &               - &          - &          - &             - &             22.8 &            16.83 \\
HiRadMat                 &           - &              - &              1 &                 - &               - &                  - &                 - &               - &          1 &          - &             1 &             37.2 &            26.10 \\
LHC filling              &           - &              - &              - &                 1 &               - &                  - &                 - &               - &          1 &          - &             1 &             37.2 &            26.10 \\
LHC setup                &           - &              - &              - &                 - &               1 &                  - &                 - &               - &          1 &          6 &             2 &             37.2 &            26.48 \\
Physics                  &           - &              - &              - &                 - &               - &                  - &                 - &               - &          2 &          - &             2 &             28.8 &            40.79 \\
Physics with parallel MD &           - &              - &              - &                 - &               - &                  - &                 2 &               - &          2 &          - &             - &             36.0 &            32.87 \\
Scrubbing                &           - &              - &              - &                 - &               - &                  - &                 - &               1 &          1 &          - &             - &             33.6 &            28.39 \\
Thursday MD              &           - &              - &              - &                 - &               - &                  1 &                 - &               - &          1 &          - &             - &             33.6 &            28.39 \\
\bottomrule
\end{tabular}

    \label{tab:2018supercycles}
    \end{center}
\end{table}

To calculate the number of cycles and number of protons to the NA targets, the time sharing between the different supercycle configurations during the operational run has to be taken into account. The 2018 proton run was scheduled over 31 weeks with two planned technical stops of 30 hours. This corresponds to a total of 5148 hours machine time allocated for operation. The time sharing between the different SPS users as obtained from the 2018 injector schedule is summarised in Table~\ref{tab:2018hours}. 

\begin{table}[ht]
    \begin{center}
    \caption{SPS supercycle sharing from schedule of 2018 proton run.}
    \begin{tabular}{l|ccccccccccc|cc}
\toprule
{} & Scheduled [hours] &  Effective [hours] \\
\midrule
Physics                    &          1,852.00 &           1,389.00 \\
Physics with parallel MD   &          1,356.00 &           1,017.00 \\
HiRadMat                   &            240.00 &             180.00 \\
AWAKE with parallel MD     &            265.51 &             199.13 \\
AWAKE                      &            742.49 &             556.87 \\
Dedicated MD               &            370.00 &             277.50 \\
Thursday MD                &            250.00 &             187.50 \\
Scrubbing                  &             72.00 &              54.00 \\
LHC setup (10\% of time)   &                 - &             514.80 \\
LHC filling (15\% of time) &                 - &             772.20 \\
Total                      &          5,148.00 &           5,148.00 \\
\bottomrule
\end{tabular}

    \label{tab:2018hours}
    \end{center}
\end{table}

\begin{figure}[ht!]
\centering
\includegraphics[trim=53 20 50 0, clip,width=0.49\textwidth]{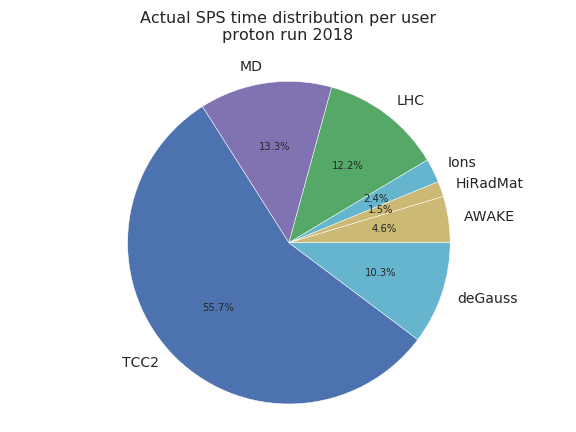}
\includegraphics[trim=53 20 50 0, clip,width=0.49\textwidth]{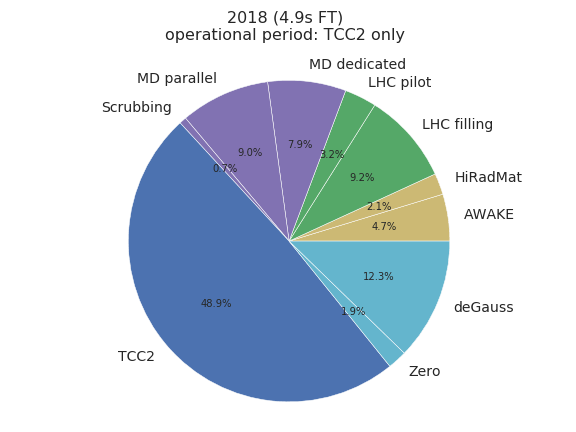}
\caption{\label{fig:SPS_timesharing_2018}Distribution of the actual machine time sharing in 2018. Data, provided by J.~Dalla-Costa (left), are compared to the expectation from schedule (right).}
\end{figure}

The left column shows the bare hours for which the corresponding supercycle (cf.~Table~\ref{tab:2018supercycles}) was scheduled. The right column shows the effective hours expected taking into account the LHC filling and LHC setup periods. Based on these numbers, the total number of cycles per user over the entire run is obtained. Figure~\ref{fig:SPS_timesharing_2018} shows the sharing of the SPS machine time per user comparing the actual numbers from 2018 (left) with the expected values obtained from the analysis described above (right). 

The agreement is very good, thus validating the approach. Small differences are explained by the fact that the ion setting up was not included explicitly in the schedule and the TCC2 cycle was played slightly more frequently compared to schedule as it was present in the supercycle during some dedicated MDs (without taking beam). This kind of modelling will be used for the projection of future proton sharing presented in what follows.  

\subsection{Future proton sharing scenarios}
The proton sharing scenarios in the SHiP era are generated in the following way: the maximum number of protons to the TCC2 experiments is obtained assuming an operational year without SHiP, similar to the analysis performed for the 2018 run. On the other hand, the maximum number of protons for SHiP is obtained assuming that the SPS serves both SHiP and the TCC2 targets throughout the entire run using supercycle configurations with a high duty cycle for SHiP. Any intermediate scenario can be obtained by splitting the operational run in periods with only TCC2 experiments and periods with both TCC2 and SHiP experiments with according duration. 

\subsubsection{Scenarios with 4.9~s flat top for TCC2 experiments}
The SHiP baseline scenario considers operational runs with protons only, i.e.~there is no operational period with ions. Instead the duration of the proton run is extended to a total of 245 days including two times 30 hours for technical stops. The considered supercycle time sharing is summarised in Table~\ref{tab:future_hours_baseline}. 
\WarningFilter{latex}{Text page 10 contains only floats}
\begin{table}[ht!]
    \begin{center}
    \caption{Considered supercycle sharing for a run with protons only.}
    \begin{tabular}{l|ccccccccccc|cc}
\toprule
{} & Scheduled [hours] &  Effective [hours] \\
\midrule
Physics                    &           2,332.0 &            1,749.0 \\
Physics with parallel MD   &           1,548.0 &            1,161.0 \\
HiRadMat                   &             240.0 &              180.0 \\
AWAKE with parallel MD     &             268.1 &              201.1 \\
AWAKE                      &             739.9 &              554.9 \\
Dedicated MD               &             370.0 &              277.5 \\
Thursday MD                &             250.0 &              187.5 \\
Scrubbing                  &              72.0 &               54.0 \\
LHC setup (10\% of time)   &                 - &              582.0 \\
LHC filling (15\% of time) &                 - &              873.0 \\
Total                      &           5,820.0 &            5,820.0 \\
\bottomrule
\end{tabular}

    \label{tab:future_hours_baseline}
    \end{center}
\end{table}
The corresponding supercycle compositions for the two possible running periods are summarised in Table~\ref{tab:futuresupercycles_TCC2_only} (serving only the TCC2 experiments) and in Table~\ref{tab:futuresupercycles_TCC2_and_SHiP} (serving both TCC2 and SHiP experiments). 
\begin{table}[htp!]
    \begin{center}
    \caption{Future SPS supercycles (running period without BDF/SHiP).}
    \begin{tabular}{l|ccccccccccc|cc}
\toprule
{} & \rot{AWAKE} & \rot{BDF/SHiP} & \rot{HiRadMat} & \rot{LHC filling} & \rot{LHC pilot} & \rot{MD dedicated} & \rot{MD parallel} & \rot{Scrubbing} & \rot{TCC2} & \rot{Zero} & \rot{Degauss} & \rot{Length [s]} & \rot{Power [MW]} \\
\midrule
AWAKE                    &           2 &              - &              - &                 - &               - &                  - &                 - &               - &          2 &          - &             2 &             43.2 &            37.60 \\
AWAKE with parallel MD   &           2 &              - &              - &                 - &               - &                  - &                 2 &               - &          2 &          - &             - &             50.4 &            32.40 \\
Dedicated MD             &           - &              - &              - &                 - &               - &                  1 &                 - &               - &          - &          - &             - &             22.8 &            16.83 \\
HiRadMat                 &           - &              - &              1 &                 - &               - &                  - &                 - &               - &          1 &          - &             1 &             37.2 &            26.10 \\
LHC filling              &           - &              - &              - &                 1 &               - &                  - &                 - &               - &          1 &          - &             1 &             37.2 &            26.10 \\
LHC setup                &           - &              - &              - &                 - &               1 &                  - &                 - &               - &          1 &          - &             1 &             26.4 &            36.64 \\
Physics                  &           - &              - &              - &                 - &               - &                  - &                 - &               - &          2 &          - &             2 &             28.8 &            40.79 \\
Physics with parallel MD &           - &              - &              - &                 - &               - &                  - &                 2 &               - &          2 &          - &             - &             36.0 &            32.87 \\
Scrubbing                &           - &              - &              - &                 - &               - &                  - &                 - &               1 &          1 &          - &             - &             33.6 &            28.39 \\
Thursday MD              &           - &              - &              - &                 - &               - &                  1 &                 - &               - &          1 &          - &             - &             33.6 &            28.39 \\
\bottomrule
\end{tabular}

    \label{tab:futuresupercycles_TCC2_only}
    \end{center}
\end{table}
\begin{table}[ht!]
    \begin{center}
    \caption{Future SPS supercycles  (running period with BDF/SHiP).}
    \begin{tabular}{l|ccccccccccc|cc}
\toprule
{} & \rot{AWAKE} & \rot{BDF/SHiP} & \rot{HiRadMat} & \rot{LHC filling} & \rot{LHC pilot} & \rot{MD dedicated} & \rot{MD parallel} & \rot{Scrubbing} & \rot{TCC2} & \rot{Zero} & \rot{Degauss} & \rot{Length [s]} & \rot{Power [MW]} \\
\midrule
AWAKE                    &           2 &              3 &              - &                 - &               - &                  - &                 - &               - &          1 &          - &             1 &             50.4 &            34.50 \\
AWAKE with parallel MD   &           2 &              3 &              - &                 - &               - &                  - &                 1 &               - &          1 &          - &             1 &             57.6 &            30.56 \\
Dedicated MD             &           - &              - &              - &                 - &               - &                  1 &                 - &               - &          - &          - &             - &             22.8 &            16.83 \\
HiRadMat                 &           - &              4 &              1 &                 - &               - &                  - &                 - &               - &          - &          - &             - &             51.6 &            25.58 \\
LHC filling              &           - &              1 &              - &                 1 &               - &                  - &                 - &               - &          - &          - &             - &             30.0 &            20.59 \\
LHC setup                &           - &              4 &              - &                 - &               1 &                  - &                 - &               - &          - &          - &             - &             40.8 &            32.26 \\
Physics                  &           - &              4 &              - &                 - &               - &                  - &                 - &               - &          1 &          - &             1 &             43.2 &            35.26 \\
Physics with parallel MD &           - &              4 &              - &                 - &               - &                  - &                 1 &               - &          1 &          - &             - &             46.8 &            32.64 \\
Scrubbing                &           - &              - &              - &                 - &               - &                  - &                 - &               1 &          1 &          - &             - &             33.6 &            28.39 \\
Thursday MD              &           - &              2 &              - &                 - &               - &                  1 &                 - &               - &          1 &          - &             1 &             51.6 &            27.89 \\
\bottomrule
\end{tabular}

    \label{tab:futuresupercycles_TCC2_and_SHiP}
    \end{center}
\end{table}
In both cases, it is assumed that the SPS serves in addition the LHC, AWAKE, HiRadMat and MD users similarly to the operational run in 2018. 

\begin{figure}[ht!]
\centering
\includegraphics[trim=53 0 50 0, clip, width=0.48\textwidth]{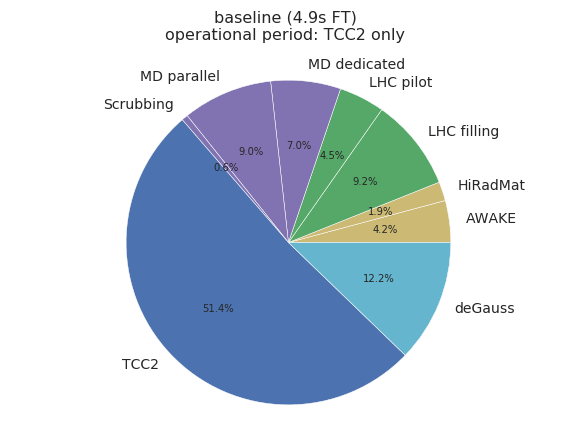}
\includegraphics[trim=53 0 50 0, clip, width=0.48\textwidth]{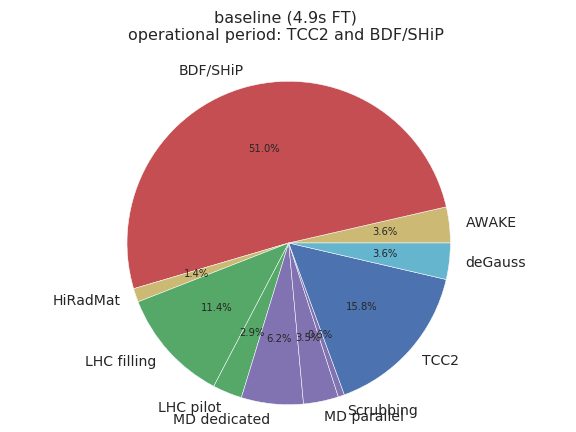}
\caption{\label{fig:SPS_timesharing_baseline}Projected future distribution of machine time usage for a running period without BDF/SHiP (left) compared to a running period with BDF/SHiP (right).}
\end{figure}

The resulting machine time sharing for the two running periods is shown in Fig.~\ref{fig:SPS_timesharing_baseline}. As expected, without SHiP the main user of SPS machine time remains the NA with slightly more than 50\% overall. On the other hand, during operational periods with both NA and SHiP more than 50\% of the machine time would be used for the SHiP cycles while the TCC2 cycles would use only about 15\% of machine time. To be noted that the amount of ``degauss'' cycles is also reduced as they are mostly needed to reduce the RMS power of the SPS main magnets when playing the NA cycle. Furthermore, it should be noted that the amount of parallel MD time in this running period would also be reduced due to the lower duty cycle in the supercycle configurations considered for SHiP.
\begin{figure}[ht!]
\centering
\includegraphics[trim=53 0 50 0, clip, width=0.49\textwidth]{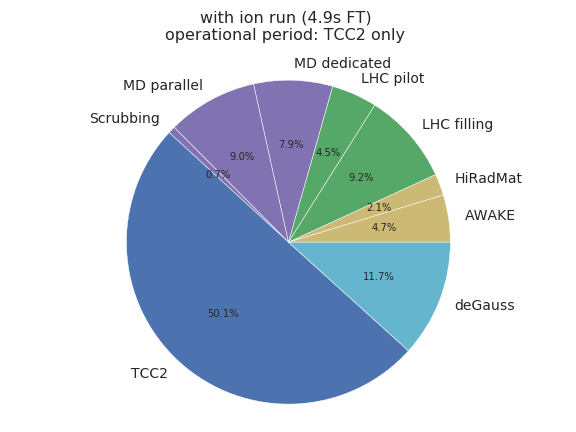}
\includegraphics[trim=53 0 50 0, clip, width=0.49\textwidth]{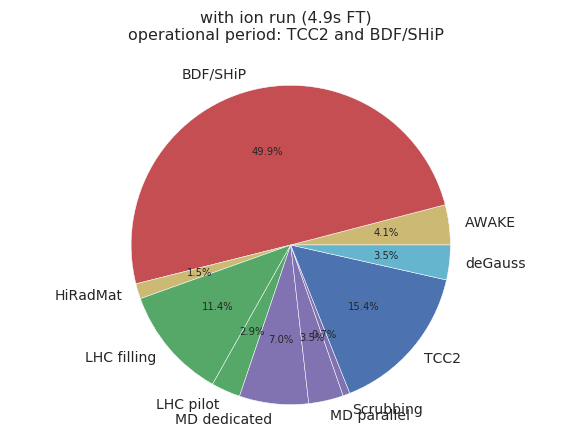}
\caption{\label{fig:SPS_timesharing_ions4p9}Projected future distribution of machine time usage for a running period without BDF/SHiP (left) compared to a running period with BDF/SHiP (right) with ion run as in 2018.}
\end{figure}
Figure~\ref{fig:SPS_timesharing_ions4p9} shows a similar comparison of the machine time sharing during the proton run for an operational year with both protons and ion physics like operation in 2018. 

\begin{figure}[ht!]
\centering
\includegraphics[trim=-20 0 20 0, clip, width=0.99\textwidth]{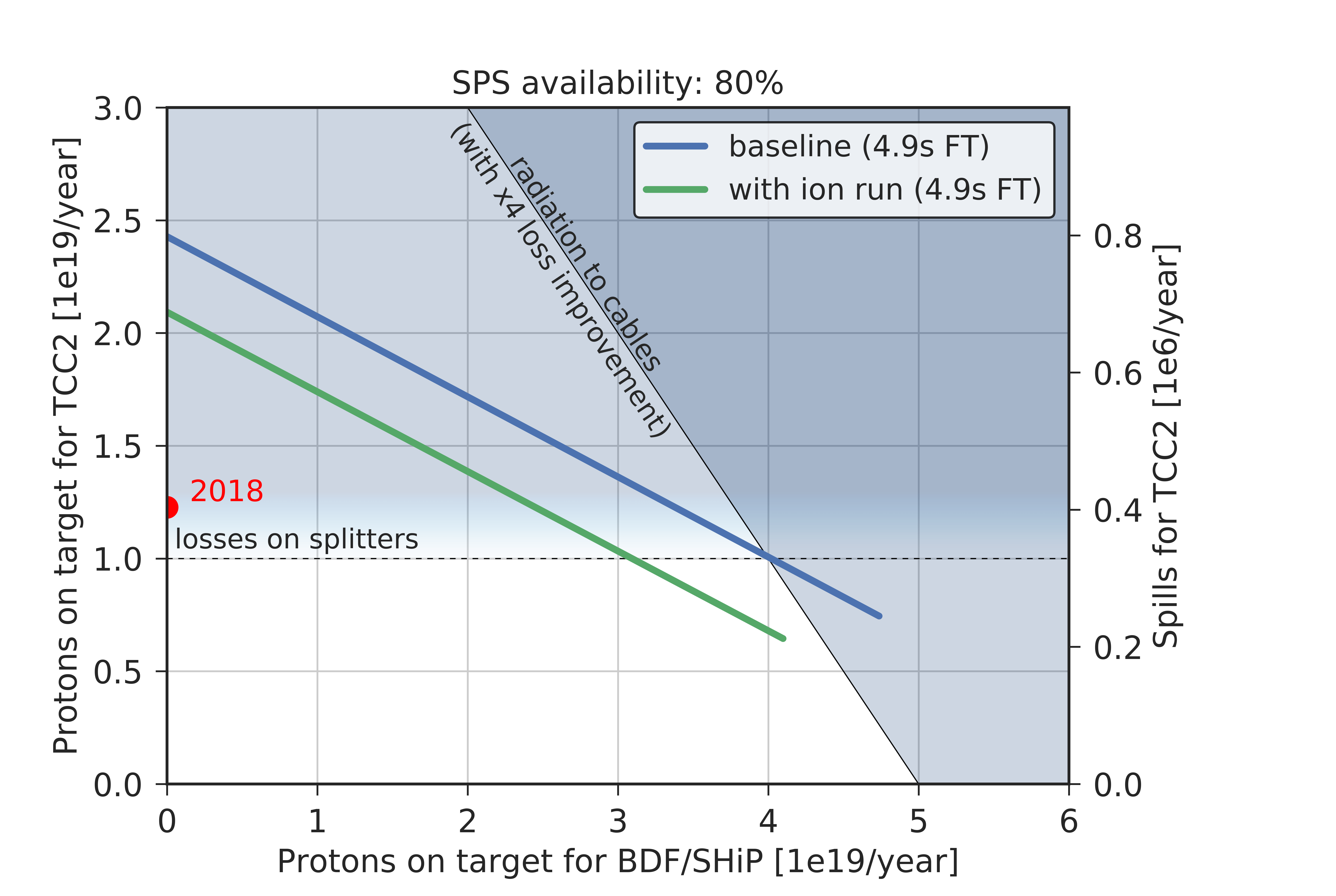}
\caption{\label{fig:SPS_protonsharing_future}Future proton sharing scenarios with (green) and without (blue) ion operation. A 4.9~s flat top (FT) length is considered for the TCC2 experiments.}
\end{figure}

With the considered transmission values and intensity per cycle as summarised in Table~\ref{tab:efficiency_future}, the resulting proton sharing is shown in Fig.~\ref{fig:SPS_protonsharing_future}. Furthermore, a machine availability of 80\% is considered, which is a realistic but conservative number (in 2018 the SPS availability was 80\%, slightly lower than previous years). The curve labelled ``with ion run'' corresponds to a schedule as in 2018, i.e.~with the supercycle sharing given in Table~\ref{tab:2018hours}.
\begin{table}[ht!]
    \begin{center}
    \caption{Transmission and intensity per cycle considered.}
    \begin{tabular}{lcc}
\toprule
{} &  TCC2 &  BDF/SHiP \\
\midrule
Extraction transmission            &  0.99 &      0.99 \\
TT20 transmission                  &  0.96 &      0.96 \\
Splitting transmission             &  0.80 &      1.00 \\
Total transmission                 &  0.76 &      0.95 \\
Protons per cycle [1e13]           &  4.00 &      4.20 \\
Effective protons per spill [1e13] &  3.04 &      3.99 \\
\bottomrule
\end{tabular}

    \label{tab:efficiency_future}
    \end{center}
\end{table}
The 4.9~s flat top for the TCC2 experiments maximises the total proton flux to the NA and the BDF/SHiP experiment. This would be optimum in case there are experiments in the NA that require large proton flux (e.g.~KLEVER). 

\subsubsection{ Scenarios with 9.7~s flat top for TCC2 experiments}
\begin{figure}[ht!]
\centering
\includegraphics[trim=40 0 50 0, clip, width=0.49\textwidth]{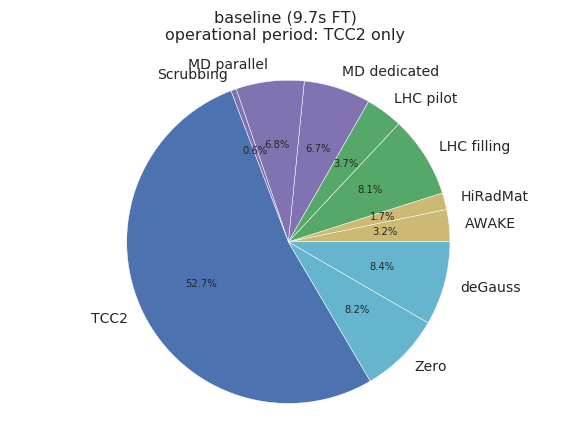}
\includegraphics[trim=40 0 50 0, clip, width=0.49\textwidth]{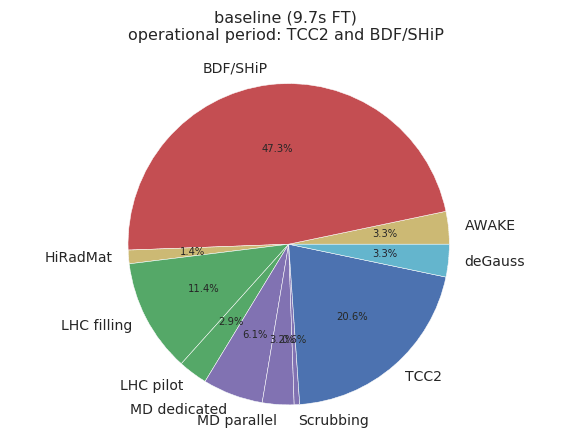}
\caption{\label{fig:SPS_timesharing_baseline9p7}Projected future distribution of machine time usage for a running period without BDF/SHiP (left) compared to a running period with BDF/SHiP (right) with a 9.7~s flat top duration.}
\end{figure}

\begin{figure}[ht!]
\centering
\includegraphics[trim=40 0 50 0, clip, width=0.49\textwidth]{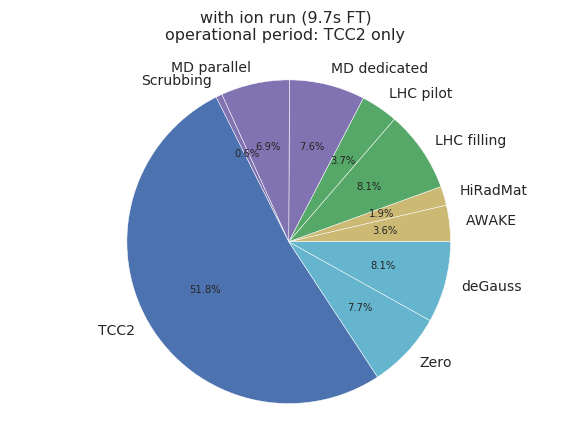}
\includegraphics[trim=40 0 50 0, clip, width=0.49\textwidth]{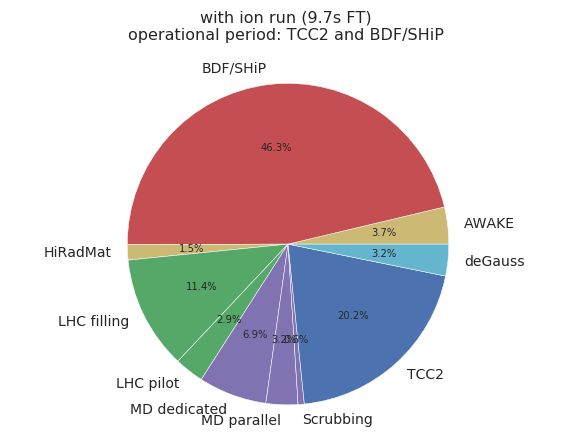}
\caption{\label{fig:SPS_timesharing_ions9p7}Projected future distribution of machine time usage for a running period without BDF/SHiP (left) compared to a running period with BDF/SHiP (right), with ion run and with a 9.7~s flat top duration.}
\end{figure}

If the main interest of the NA experiments would be in test beams with limitations on the instantaneous proton flux, a longer flat top cycle is more advantageous (similar to what was done during the operation of the CNGS experiment). Proton sharing scenarios with 9.7~s flat top for the TCC2 experiments have been calculated as presented below. The total length of the TCC2 cycle with 9.7~s flat top is 15.6~s with an average power (main bends and quadrupoles) of 63.6~MW. Assuming again the supercycle sharing of Table~\ref{tab:future_hours_baseline} for operational runs with only protons and Table~\ref{tab:2018hours} for operational runs with protons and ions. The resulting time sharing between the different users is given in Figs.~\ref{fig:SPS_timesharing_baseline9p7} and  \ref{fig:SPS_timesharing_ions9p7}. The considered supercycle compositions are summarised in Table~\ref{tab:futuresupercycles_9p7s_TCC2_only} and Table~\ref{tab:futuresupercycles_9p7s_TCC2_and_SHiP}. 

\begin{table}[ht!]
    \begin{center}
    \caption{Future SPS supercycles (running period without BDF/SHiP).}
    \begin{tabular}{l|ccccccccccc|cc}
\toprule
{} & \rot{AWAKE} & \rot{BDF/SHiP} & \rot{HiRadMat} & \rot{LHC filling} & \rot{LHC pilot} & \rot{MD dedicated} & \rot{MD parallel} & \rot{Scrubbing} & \rot{TCC2} & \rot{Zero} & \rot{Degauss} & \rot{Length [s]} & \rot{Power [MW]} \\
\midrule
AWAKE                    &           1 &              - &              - &                 - &               - &                  - &                 - &               - &          1 &          2 &             1 &             28.8 &            42.86 \\
AWAKE with parallel MD   &           1 &              - &              - &                 - &               - &                  - &                 1 &               - &          1 &          - &             - &             30.0 &            41.28 \\
Dedicated MD             &           - &              - &              - &                 - &               - &                  1 &                 - &               - &          - &          - &             - &             22.8 &            16.83 \\
HiRadMat                 &           - &              - &              1 &                 - &               - &                  - &                 - &               - &          1 &          - &             1 &             42.0 &            33.17 \\
LHC filling              &           - &              - &              - &                 1 &               - &                  - &                 - &               - &          1 &          - &             1 &             42.0 &            33.17 \\
LHC setup                &           - &              - &              - &                 - &               1 &                  - &                 - &               - &          1 &          1 &             1 &             32.4 &            42.89 \\
Physics                  &           - &              - &              - &                 - &               - &                  - &                 - &               - &          1 &          4 &             1 &             24.0 &            42.08 \\
Physics with parallel MD &           - &              - &              - &                 - &               - &                  - &                 2 &               - &          2 &          1 &             - &             46.8 &            43.32 \\
Scrubbing                &           - &              - &              - &                 - &               - &                  - &                 - &               1 &          1 &          - &             - &             38.4 &            35.83 \\
Thursday MD              &           - &              - &              - &                 - &               - &                  1 &                 - &               - &          1 &          - &             - &             38.4 &            35.83 \\
\bottomrule
\end{tabular}

    \label{tab:futuresupercycles_9p7s_TCC2_only}
    \end{center}
\end{table}
\begin{table}[ht]
    \begin{center}
    \caption{Future SPS supercycles  (running period with BDF/SHiP).}
    \begin{tabular}{l|ccccccccccc|cc}
\toprule
{} & \rot{AWAKE} & \rot{BDF/SHiP} & \rot{HiRadMat} & \rot{LHC filling} & \rot{LHC pilot} & \rot{MD dedicated} & \rot{MD parallel} & \rot{Scrubbing} & \rot{TCC2} & \rot{Zero} & \rot{Degauss} & \rot{Length [s]} & \rot{Power [MW]} \\
\midrule
AWAKE                    &           2 &              3 &              - &                 - &               - &                  - &                 - &               - &          1 &          - &             1 &             55.2 &            39.15 \\
AWAKE with parallel MD   &           2 &              3 &              - &                 - &               - &                  - &                 1 &               - &          1 &          - &             1 &             62.4 &            34.98 \\
Dedicated MD             &           - &              - &              - &                 - &               - &                  1 &                 - &               - &          - &          - &             - &             22.8 &            16.83 \\
HiRadMat                 &           - &              4 &              1 &                 - &               - &                  - &                 - &               - &          - &          - &             - &             51.6 &            25.58 \\
LHC filling              &           - &              1 &              - &                 1 &               - &                  - &                 - &               - &          - &          - &             - &             30.0 &            20.59 \\
LHC setup                &           - &              4 &              - &                 - &               1 &                  - &                 - &               - &          - &          - &             - &             40.8 &            32.26 \\
Physics                  &           - &              4 &              - &                 - &               - &                  - &                 - &               - &          1 &          - &             1 &             48.0 &            40.53 \\
Physics with parallel MD &           - &              4 &              - &                 - &               - &                  - &                 1 &               - &          1 &          - &             - &             51.6 &            37.78 \\
Scrubbing                &           - &              - &              - &                 - &               - &                  - &                 - &               1 &          1 &          - &             - &             38.4 &            35.83 \\
Thursday MD              &           - &              2 &              - &                 - &               - &                  1 &                 - &               - &          1 &          - &             1 &             56.4 &            33.00 \\
\bottomrule
\end{tabular}

    \label{tab:futuresupercycles_9p7s_TCC2_and_SHiP}
    \end{center}
\end{table}

\begin{figure}[htb]
\centering
\includegraphics[trim=0 0 0 0, clip, width=0.95\textwidth]{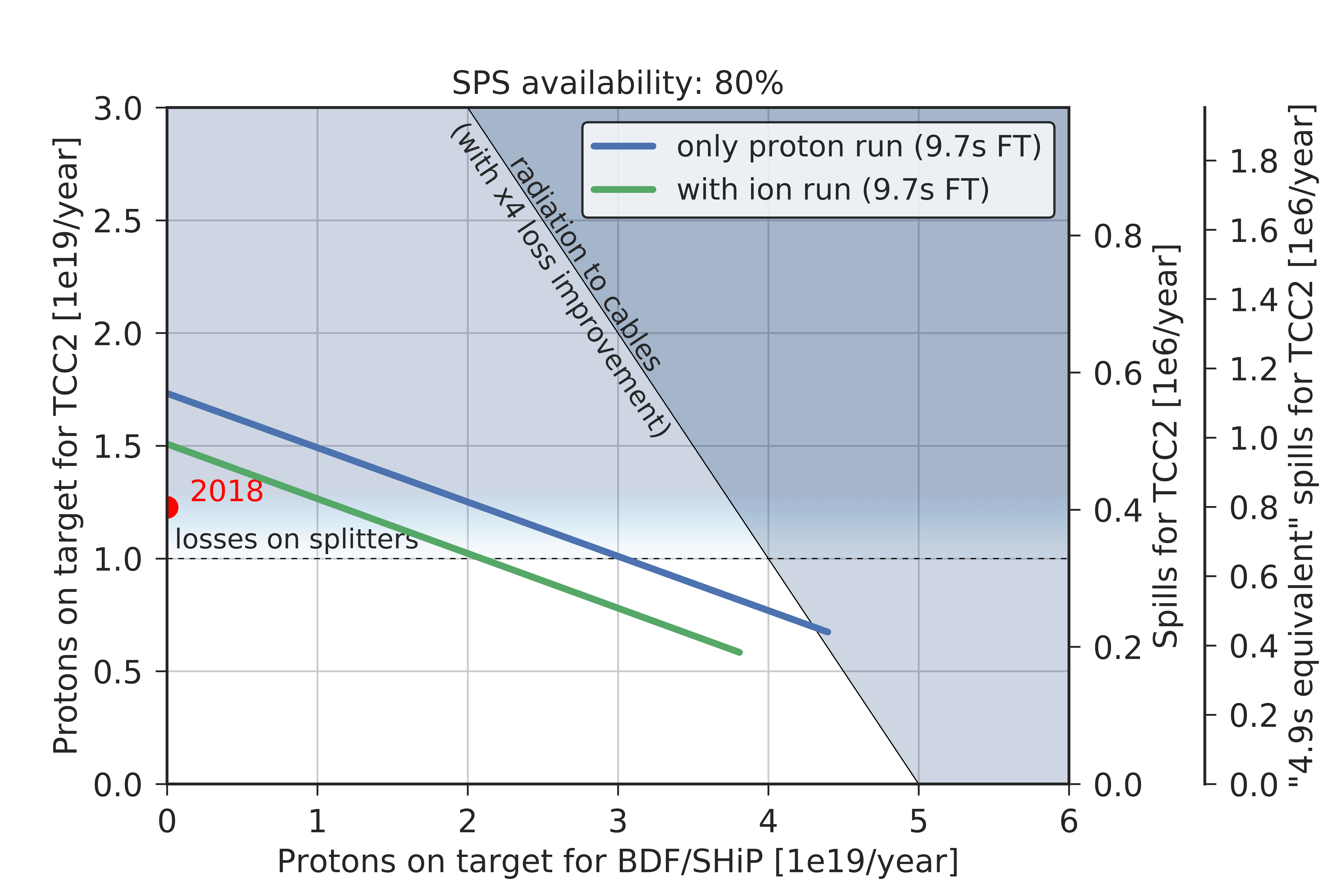}
\caption{\label{fig:SPS_protonsharing_futurenoKLEVER}Future proton sharing scenarios with (green) and without (blue) ion operation. A 9.7~s flat top length is considered for the TCC2 experiments.}
\end{figure}

The obtained scenarios of proton sharing between TCC2 experiments and the BDF/SHiP experiment are shown in Fig.~\ref{fig:SPS_protonsharing_futurenoKLEVER}. The secondary vertical axis on the right side of the plot shows the number of spills for the TCC2 experiments in units of ``4.9~s equivalent'' spills, which is obtained by multiplying the number of actual spills by the ratio of the flat top lengths (9.7/4.9). This represents the situation where TCC2 experiments are limited in the instantaneous proton flux and therefore would profit from longer spills. In other words, a flat top of 9.7 s gives double the data-taking time compared to 4.9 s flat top, at a constant spill rate.

\section{Summary and conclusions}

The SPS serves a large variety of physics users. A detailed analysis has been performed to analyse the compatibility and possible proton sharing scenarios between the TCC2 experiments and proposed future experiments such as BDF/SHiP, taking into account the parallel operation of the LHC, AWAKE, HiRadMat and MDs. The analysis is based on the actual operational conditions and constraints of the 2018 proton run, in order to be as realistic as possible. 
The methodology presented in this note has been benchmarked with the operational period of 2018: the operationally obtained number of protons for the TCC2 targets of the 2018 run is correctly reproduced for the corresponding parameters. For the future proton sharing scenarios, operational periods with and without dedicated ion physics have been considered. Two different flat top lengths for the slow extraction to the TCC2 experiments have been analysed taking into account realistic supercycle compositions and respecting the SPS limits on power dissipation in the magnets. Intensities considered for the TCC2 cycle and for the BDF/SHiP cycle are based on operationally achieved values during the CNGS era of the SPS. The limitations from losses at the splitters to the TCC2 targets as well as the intensity limitation from activation of the slow extraction equipment are also taken into account. With all these assumptions, the target of $4.0\times10^{19}$ PoT for the BDF/SHiP experiment can be achieved in different scenarios. The PoT for the TCC2 experiments then depends on the chosen flat top length and the total days of physics (i.e.~if there is an ion run or not). However, it should be stressed that the PoT numbers would be reduced in case of more frequent LHC fillings as compared to today's operation (which might be required if one of the backup operational scenarios with reduced LHC fill lengths is needed operationally during the HL-LHC era).

\FloatBarrier

\printbibliography[heading=subbibliography]

 \chapter{Slow Extraction from the SPS}
\label{Chap:Extraction}

\section{Introduction}\label{sec:ext:intro}

\subsection{Slow extraction in the SPS}\label{sec:ext:slowExt}

Third integer slow extraction from SPS uses a set of suitably located extraction sextupoles to create a stable area in horizontal phase space. In the standard scheme (Q-sweep) the beam is debunched with chromaticity set to a large negative value, and the machine tune moved toward the third-integer. The extraction is made in combined momentum and betatron space, with lowest momentum particles coming into resonance and being extracted first across the electrostatic septum (ES); ZS is used synonymously when discussing the installed hardware. There is a momentum ramp of the extracted beam during the spill as it sweeps through the intrinsic momentum spread of the circulating beam, which couples into separatrix position and angle changes in time at the ES. For the newly-developed Constant Optics Slow Extraction (COSE), the normalised machine optics are set constant and the momentum of the entire beam is trimmed to move into the resonance, which has the advantage of keeping the separatrix presentation to the ES constant.

\begin{figure}[htbp]
   \centering
   \includegraphics[trim={0 20 0 300}, clip,width=\linewidth]{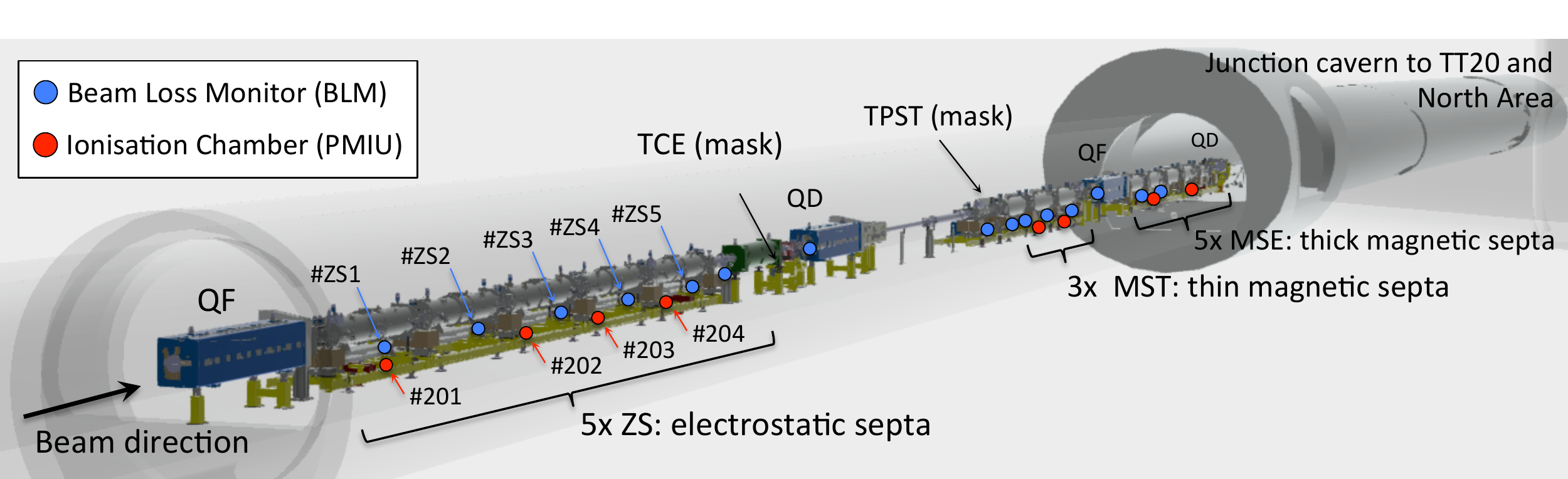}
  \caption{LSS2 slow extraction region towards the TT20 transfer line and the NA~\cite{bjorkman}.~\label{fig:layout}}
\end{figure}

The SPS presently provides beam to the North Area (NA) Fixed Target physics programme with spill lengths of several seconds; in recent years spills were extracted on a flat-top length of \SI{4.8}{s}. The extraction system is located in Long Straight Section (LSS) 2 and is composed of an ES (ZS) upstream of magnetic septa (MST and MSE), as shown in Fig.~\ref{fig:layout}. Dedicated protection devices, e.g. TCE and TPST, are strategically positioned to intercept and absorb a large part of the beam energy that would otherwise be deposited in the septa and quadrupoles. The largest amplitude particles jumping into the ES are deflected off into the extraction channel, leaving the SPS via the coil window of the wide-aperture quadrupole QDA.219.

The ES is divided into 5 separate tanks, 3.15 m in length, each containing an array of 2080 Tungsten-Rhenium (WRe) wires that create the boundary delimiting the low and high electric field regions. The wires are made as thin as possible (ZS1-2: $\diameter$\SI{60}{\micro\meter}, ZS3-5: $\diameter$\SI{100}{\micro\meter}) in order to reduce beam losses, but a small fraction of beam is unavoidably scattered on the wires, inducing radioactivity in LSS2. Five such units are needed to extract the beam at 400 GeV, individually aligned with the beam to minimise overall losses. Primary proton scattering with these wires is the dominant source of the beam loss. In order to minimise the effective thickness of the septum seen by the beam, the anode supporting the wires is made as straight as possible with a very accurate positioning system, but nevertheless several percent of the beam is lost in the extraction process.

\subsection{Implications of protons on target request}

\begin{figure}[htbp]
\centering\includegraphics[width=0.7\linewidth]{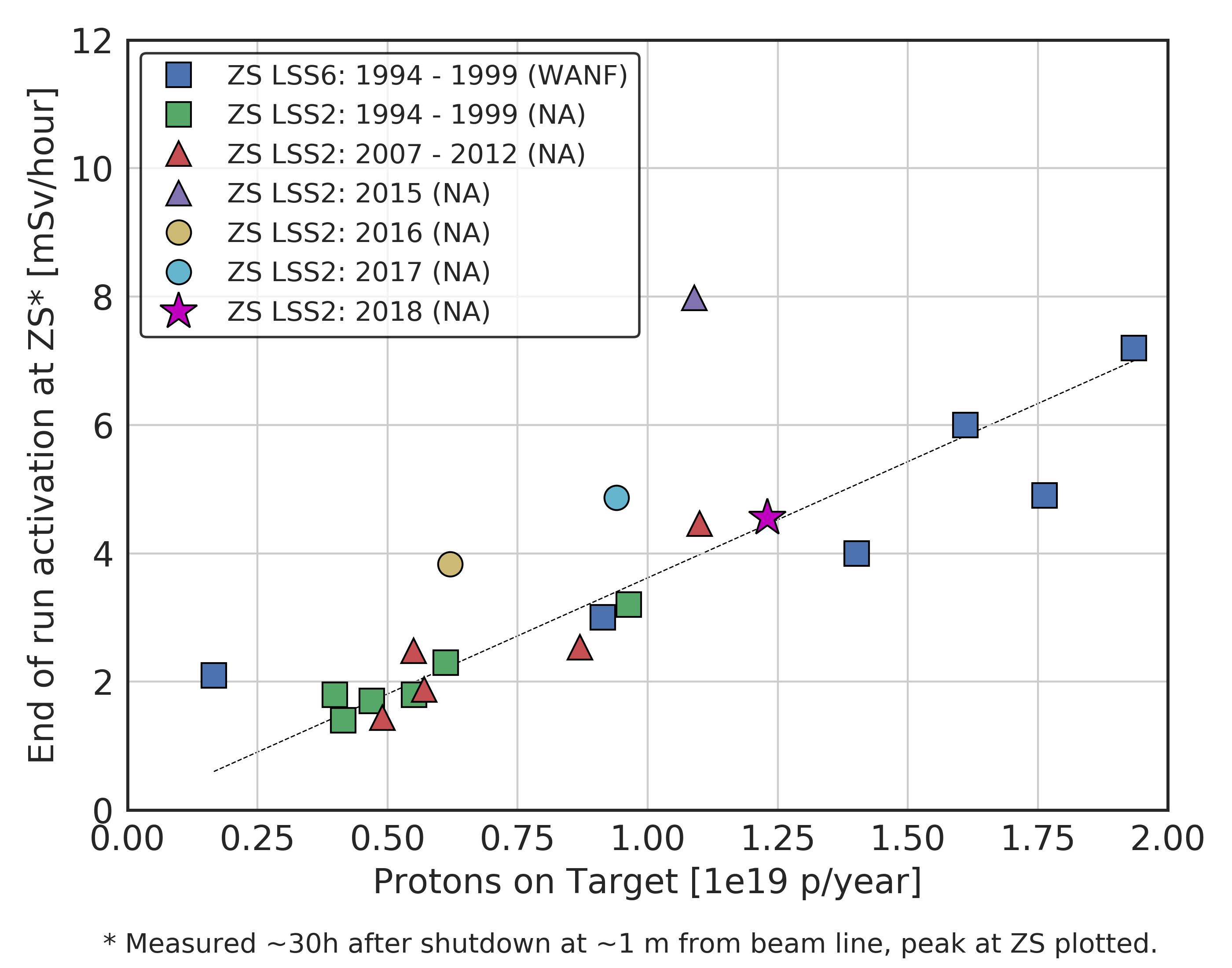}
\caption{Induced activation as a function of PoT per year, for the ES septa in the slow extraction channels LSS2 and LSS6.}
\label{fig:ACTvsPOT}
\end{figure}

The radiation dose and induced radio-activation in LSS2, and also at the splitters in TT20, are correlated to the total Protons On Target (POT) as well as to other less tangible aspects of the SPS operation and extraction channel setup, like the beam stability, the alignment of the extraction septa, any hardware faults and the beam quality. There is therefore a significant scatter in the activation levels per proton, depending on the specific combination of conditions and also the level and quality of the follow-up both of the hardware and of the operation. The residual dose after the end of proton operation has been measured over many years, summarised in Fig.~\ref{fig:ACTvsPOT}, showing the results for both LSS2 and LSS6. To keep the residual activation levels at around \SI{5}{\milli\sievert\per\hour} after 30 hours of cool-down while extracting around $5 \times 10^{19}$ POT per year will require a reduction in the activation per proton of approximately a factor four. 

The high losses and activation observed in 2015 were problematic for the hands-on maintenance of the systems, and also were likely contributors to the severe failures seen in 2018, where both a High Voltage (HV) feedthrough broke on one ZS, and the main HV cable powering all the ZS failed. Accumulated dose is known to be a major factor in the lifetime of both these critical elements, and since changes to ZS tank or anode materials will not help with the dose to equipment, the reduction in the induced radioactivity per extracted proton is critical for the equipment performance.

The transfer line TT20 to the NA presents several zones of rather high activation but the highest occurs at the radiation-resistant Lambertson splitter septa, which are protected by an upstream collimator. The fraction of beam lost on the splitters is about 3\%, as estimated by intensity monitors in the NA. For the future BDF, one major advantage of the concept is that the splitters will be used in a ‘switch’ mode, where the beam is not split, nor shared, but steered toward the new beam line and target. Since the beam for BDF will pass through the gap of the splitter system in an essentially loss-free transport, this gain will mean that the present loss levels both in the extraction channel and in the splitter region can be maintained for the simultaneous delivery of \num{4e19} to the BDF and \num{\sim 1e19} to the NA. 

Further, the radiation dose to the cables and extraction equipment depends on the beam loss at extraction and the total number of protons extracted. If the factor 4 loss reduction can be applied to both SHiP and NA extraction, then again the limitation for the total number of protons extracted per year (to whatever target) will be $5.0\times10^{19}$. At these dose levels, a full recabling of the extraction channel region, e.g. control, High Voltage (HV) and DC cables, is required after about 8 years of operation, or every other long shutdown based on experience.

\subsection{Overview of challenges, possibilities for improvement and associated R\&D efforts}

In 2018 the highest integrated number of protons in the history of the NA was slow extracted from the SPS for the FT physics programme. At well over $1.2\times 10^{19}$ POT, this represented the highest annual figure for almost two decades, since the West Area Neutrino Facility was operational. The high intensity POT requests look set to continue in the foreseeable future, especially in view of the proposed SPS Beam Dump Facility (BDF) and experiments, e.g. SHiP~\cite{bdf}. Without significant improvements, the attainable annual POT will be limited to well below the total the SPS machine could deliver, due to activation of accelerator equipment and associated personnel dose limitations. The instantaneous and integrated loss levels are therefore strong limitations on the annual attainable POT.

Over the past three years a significant effort has been made on several fronts to conceive, design, deploy and test where possible methods to reduce the extraction beam losses or to mitigate their effects through different material choice or handling. In this chapter the requirements, constraints and potential improvements to the slow extraction system in the SPS are described. The present and future challenges are numerous, including machine stability and reproducibility, monitoring and surveillance of the extraction, carrying out interventions and remote handling of the equipment in the activated extraction region and improving the extraction inefficiency. Several possibilities to reduce dose to personnel during hands-on maintenance, which is considered as one of the most important figures of merit alongside machine availability are being investigated, spanning improved spill quality, manipulation and control of the extracted separatrix, alternative or upgraded extraction hardware concepts, including diffuser devices (passive and active) upstream of the ZS, low-activation materials and the extended use of remote handling techniques. 

It is clear that there will not be one single solution but rather a combination of improvements, which together may achieve the factor of 4 reduction in extraction beam loss desired. Interestingly, some of the methods for loss reduction, activation and personnel dose reduction can be accumulated directly, with a multiplicative gain. The techniques evaluated are compared in terms of what has been learned from simulation and recent measurements with beam, and what can be expected operationally from a realistic extrapolation of the results.

\subsection{Dose rate and cool-down predictions in BDF era}

In order to understand how the extracted proton flux affects the build-up and subsequent cool-down of the induced radio-activation of the extraction straight section and make predictions for the BDF scenario, a model based on a simple empirical relationship has been developed~\cite{IPAC_philipp} and shown to predict the measured radioactive decay on ionisation chambers in LSS2. The empirical model was first developed in the mid-1990's with dedicated measurements of the induced radioactivity (IR) of the LSS6 extraction region to understand the build-up of activation during the high intensity operation of the West Area Neutrino Facility (WANF)~\cite{keizer1,keizer2,keizer3}. To put the relative intensities in perspective, consider that the BDF is requesting $20\times10^{19}$ POT over 5 years, compared to the $7\times10^{19}$ POT delivered to WANF over a similar period. During this period the model was developed to fit the measured data as a function of the extracted proton flux, allowing predictions of cool-down times to be made during operation. More recently, and with far improved software and analysis tools, the model was revived and exploited as a predictive tool for estimating cool-down times as a function of extracted proton flux. 

The instrumentation relevant to this study is shown in Fig. \ref{fig:layout}. Beam Loss Monitors (BLMs) placed along the extraction region are used to measure the prompt beam loss induced by the small fraction of the beam that impacts the wires of the ZS septum during extraction. In addition to the BLMs, a series of ionisation chambers (PMIUs) are used to measure the IR. Unlike the BLMs, the PMIU detectors have a fixed gain and saturate during extraction due to the high prompt loss signal but give a reliable signal during periods of cool-down.

The objective of this work was to estimate the cool-down times required for the POT requested annually by the NA experiments and to estimate the improvement in the extraction efficiency required to keep cool-down times reasonable for the future BDF.

\subsubsection{Empirical models of IR($t$)}
 
One of the most challenging aspects of predicting the evolution of the IR is the non-linear time dependence of its effective half-life arising from the mixture of different radionuclides produced both during the initial irradiation and in the resulting chains of radioactive decay. After irradiation, the exponential decay of the IR can be expressed as,
\begin{equation}
\textrm{IR}(t) \propto \exp\Bigg({-\frac{t}{\tau(t)}}\Bigg)\label{model_form}
\end{equation}
where different functional forms for the time evolution have been proposed~\cite{keizer1}. The most suitable model at the SPS was shown empirically to take the above exponential form with,
\begin{equation}
\tau(t) = \frac{t}{k_1\ln(t)^{k_2}} \label{model}
\end{equation}
where $k_1$ and $k_2$ are decay constants. By differentiating Eq.~\ref{model_form} with respect to time and re-writing the result as a first-order linear ordinary differential equation one can write the effective half-life of the empirical model by inspection as,
\begin{equation}
t_\textrm{$1/2$}(t) = \frac{t}{k_1k_2\ln(t)^{k_2-1}}\ln(2) \label{model_eff}
\end{equation}
A similar derivation based on another empirical model developed by Sullivan and Overton can be found in~\cite{stevenson}. As one would expect for a physical effective half-life describing a mixture of different radioisotopes, each with different populations and half-lives, the expression continually increases at an exponentially slower rate towards stability, i.e. $\lim_{t \to \infty}t_\textrm{$1/2$}(t) = \infty$ and $\lim_{t \to \infty} \frac{\partial t_\textrm{$1/2$}(t)}{\partial t} = 0$, for $k_1,k_2 \in \mathbb{R}_{>0}$ and $k_2 > 1$, which is not the case for many of the empirical models that have been proposed to date~\cite{sullivan,sullivan2}. The model fits the empirical data well, as well as simulated data generated by activating different materials with a 400 GeV proton beam for 200 days using the ActiWiz code~\cite{actiwiz}. The fit constants showed good agreement with the simulated activation of stainless steel, which is the dominant material component of the ZS, and also with the fit constants attained empirically.

A predictive model of the build-up and decay of the IR was developed as a function of time by introducing the measured extracted proton flux $P_\textrm{ex}(t)$ and the prompt normalised loss per proton $N_L(t)$. The introduction of $N_L$, as measured on the BLMs next to the ZS, was an attempt to introduce changes in the extraction efficiency over time into the model. The model was discretised in time, using bins of duration $\Delta t$, and an exponential decay function generated at every bin with a starting value proportional to both $P_\textrm{ex}$ and $N_L$. At the $n^\textrm{th}$ bin the IR can be expressed as a sum of exponentials arising from all previous bins according to,
\begin{equation*}
\textrm{IR}_n = G\sum\limits_{i=1}^n N_{L,n+1-i} P_{\textrm{ex},n+1-i}\exp{\Big(-k_1\ln(\Delta t (i-1))^{k_2}\Big)}
\end{equation*}
where $G$ is a constant conversion factor that depends on the primary beam energy, material composition and geometry of the machine, including the relative positions of the detectors and their calibrations. This relatively simple analytic function depends on only three constants determined empirically by applying a non-linear least-squares fitting routine on logged measurement data taken during past operational years.
\begin{figure}[htbp]
   \centering
   \includegraphics[width=0.7\linewidth]{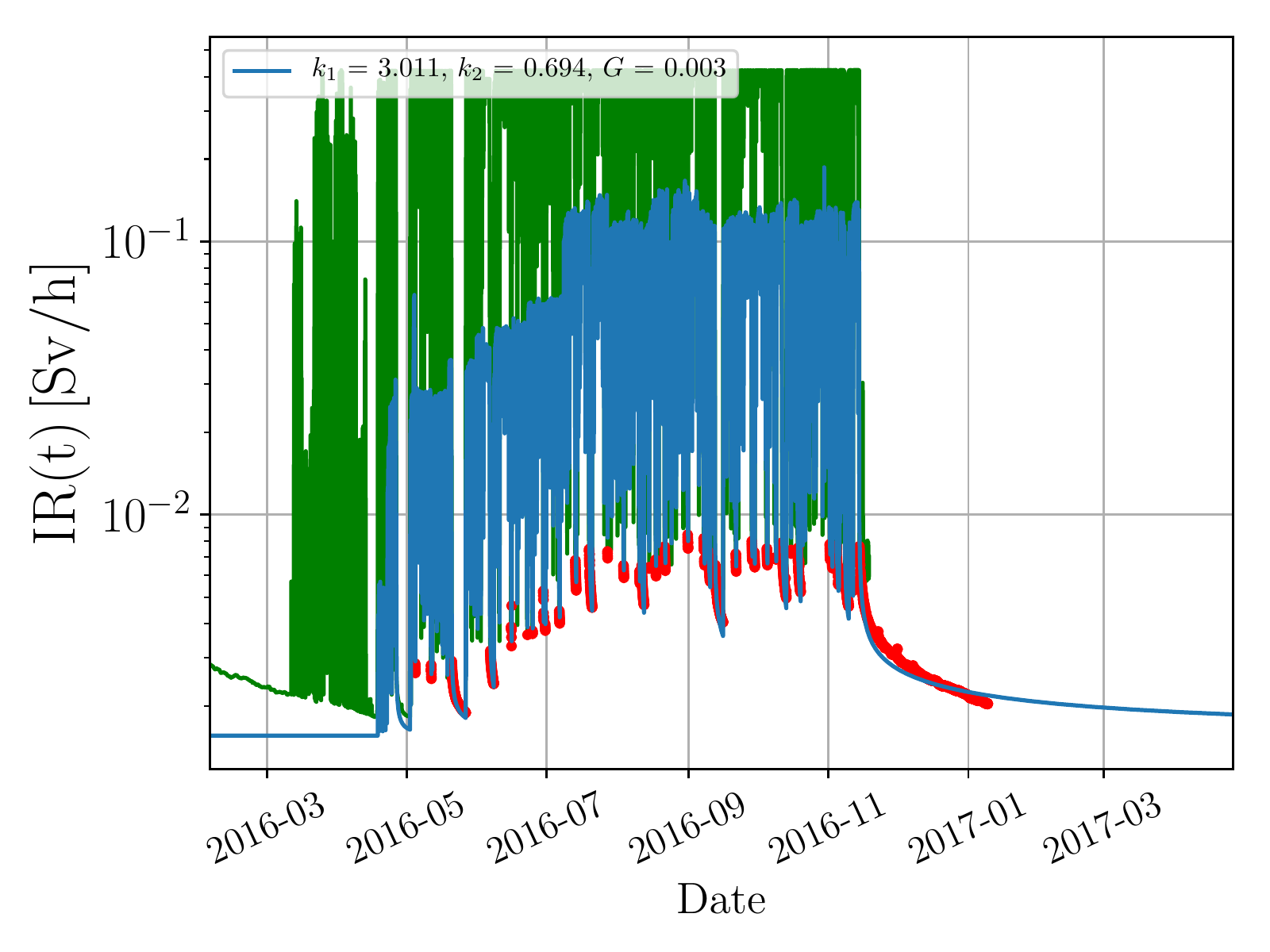}
  \caption{IR model (blue) fitted to measured filtered data (red): first \SI{5}{h} of data after beam stops and saturated PMIU data are filtered from the raw data (green), $\Delta t = \SI{0.5}{h}$.\label{fit_2016}}
  \end{figure}
An example is shown in Fig.~\ref{fit_2016} using data logged during 2016. In this example, PMIU.202 was paired with its closest BLM on ZS2 in order to account for variations in the extraction efficiency. The agreement between measurement and the fit is better than 10\% from a few days to 2 months of cool-down. The IR measured on the PMIU detectors has to be filtered to remove saturated values acquired in the presence of beam, as well as data taken immediately after a stop in operation. The quality of the fitting over longer time periods is limited by the fast initial variation of the half-life caused by the rapid decay of short-living radioisotopes and the length of the stops during the operational year over which the fitting is made. The fit convergence is improved significantly if the first few hours of data is removed; the amount of data removed is a free parameter chosen to improve the fitting in the time range of interest.

The fit constants determined from the logged 2016 data are shown in Table~\ref{constants} by pairing each PMIU with its nearest BLM. The variation in $G$ can be attributed to the relative positions and differences in calibration of the PMIUs and BLMs. The variations in the decay constants $k_1$ and $k_2$ indicate localised differences in the decay rate. A spatial dependence of the cool-down rate is indeed observed along LSS2 and is likely caused by local differences in the material composition of the equipment in proximity to the detectors. The effect is most prevalent in the first days of cool-down. The fitting was also carried out on the average of all detectors next to the ZS to give a more global description of the IR, from which similar fit constants were also attained.
\begin{table}[htbp]
   \centering
   \caption{Empirical constants determined from 2016 data\label{constants}}
   \begin{tabular}{ccccc}
       \hline
       \textbf{BLM.ZS \#} & \textbf{PMIU \#}  & \textbf{$k_1$} & \textbf{$k_2$} & \textbf{$G$}  \\ 
              $N_L(t)$~[Gy p$^{-1}$] & $\textrm{IR}(t)$ [Sv h$^{-1}$]  &  & & [Sv h$^{-1}$ Gy$^{-1}$]  \\ 
       \midrule
           1& 201 & 3.14 & 0.69&0.0062\\
           2& 202 & 3.01 &0.69&0.0027\\
	  3& 203 & 2.73 &0.66 &0.0007\\
	  4 & 204 & 2.44 & 0.76 &0.0008\\
 \midrule
 average & average & 3.31 & 0.67 &0.0016\\
       \hline
   \end{tabular}
   \label{specs}
\end{table}

\subsubsection{Predictive power of the IR($t$) model}

The predictive power of the model was tested on data logged in 2011 and 2015 by applying the 2016 fit constants. The discrepancy with the model was tested using PMIU.202 and BLM.ZS2, and was accurate to within 10\%. The model has been used in recent years to predict the end-of-year activation for different operational scenarios. An example during the 2017 physics run is shown below in Fig.~\ref{fig:Model_prediction_YETS17}, where it was required to assess the impact on the cool-down time during the Year End Technical Stop (YETS) in view of a possible exchange of the second ZS tank. Using the planned intensity and the Injector Accelerator Schedule as input, the model was able to accurately predict the measured induced radioactivity during the YETS.
\begin{figure}[htbp]
   \centering
   \includegraphics[width=0.7\linewidth]{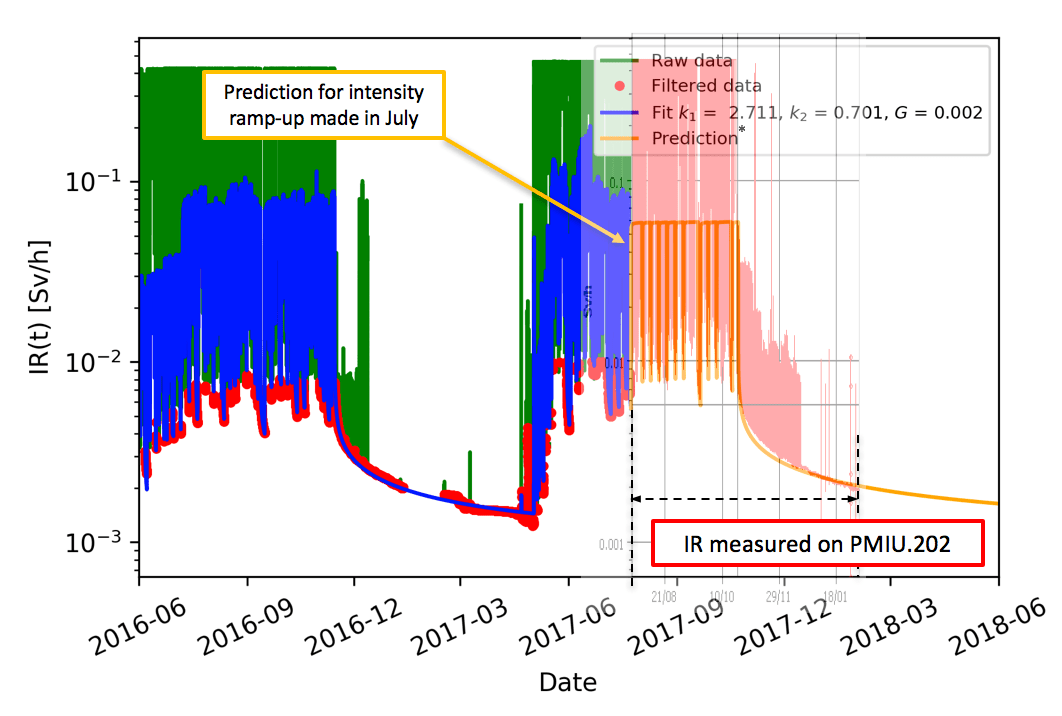}
  \caption{IR model (blue) fitted to measured filtered data (red dots) and extrapolated (orange) in view of the requested increase in POT: the prediction is compared to the measured IR (red shaded line) on PMIU202.}
  \label{fig:Model_prediction_YETS17}
  \end{figure}
  
\begin{figure}[htbp]
   \centering
   \includegraphics[width=0.7\linewidth]{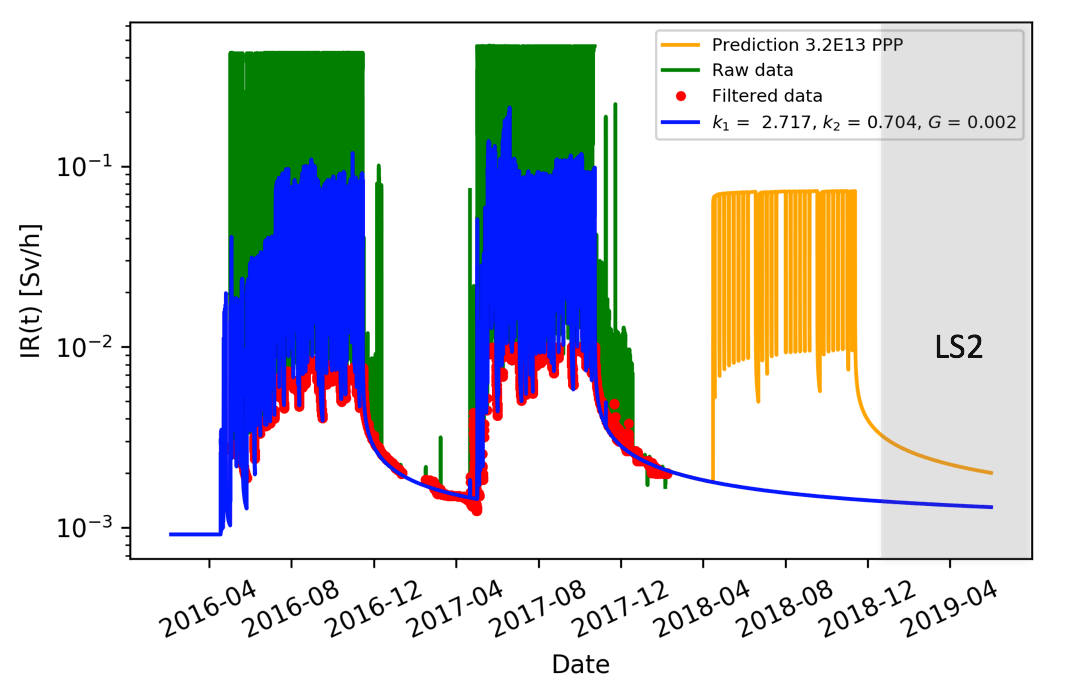}
  \caption{IR model (blue) fitted to measured filtered data (red dots) and extrapolated (orange) in view of the 2018 POT request and cool-down into LS2.}
  \label{fig:Model_prediction_LS2}
  \end{figure}

With increasing amounts of good quality, precisely logged data, the model is reaching its full potential and several years can be strung together to consider the build-up of longer term induced radio-activation. Fig.~\ref{fig:Model_prediction_LS2} demonstrates this with both the 2016 and 2017 years included in the fitting algorithm and extrapolated to give prediction of the end of run activation and cool-down into Long Shutdown 2 (LS2) based on the 2018 request for proton flux by the NA experiments.

In 2018, the model was also used to predict the cool-down time and expected dose to personnel in the event that the tank on which the HV feed-through broke would need replacing.

\subsubsection{Impact of future BDF operational scenario on IR}

During the shutdown period early in 2016, several ZS tanks had to be exchanged for preventative maintenance due to problems observed in 2015. Using the actual dose taken by the personnel involved in these interventions and the measured IR at PMIU.202 as a reference, the dose for given cool-down times was estimated for future operational scenarios. In the following estimates, the cool-down times are quoted at the end of an operational year for a 5\,mSv collective dose using the exchange of ZS tank 2 on 19$^\textrm{th}$ February 2016 as the reference; the collective dose taken was 1.7 mSv after 100 days of cool-down.

\begin{figure}[htbp]
   \centering
   \includegraphics[width=0.7\linewidth]{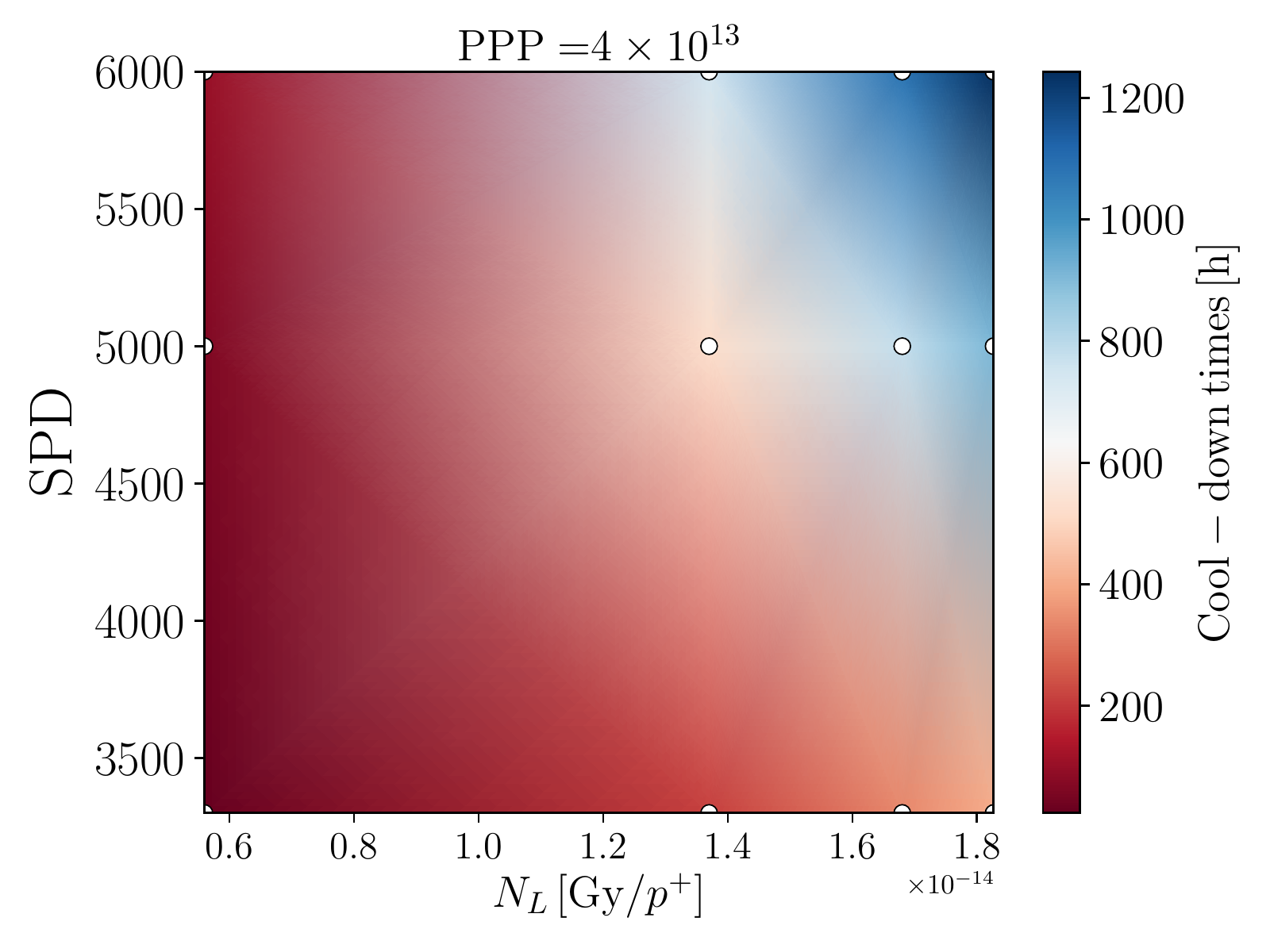}
  \caption{Parametric study of waiting times for the reference intervention: ZS2 tank exchange at 5 mSv collective dose.\label{cooldown}}
\end{figure}

An intensity profile $P_\textrm{ex}(t)$ based on the draft 2017 CERN Injector Schedule was assumed and parameterised by the number of Spills Per Day (SPD), Protons Per Pulse (PPP) and $N_L$, where the extraction efficiency is inversely proportional to $N_L$. The cool-down times parameterised in terms of SPD and $N_L$ are shown in Fig.~\ref{cooldown} for an intensity of $4\times10^{13}$ PPP, as requested by both the NA in 2017 and the future SPS BDF. In this case a model pairing PMIU.202 with BLM.ZS2 was used. The cool-down times scale almost quadratically with $N_L$ and intensity. In 2017, an average of 3300 SPD was predicted, whereas for the BDF over 6000 SPD would be needed to meet the requested POT. Considering the same average extraction efficiency as measured on BLM.ZS2 in 2015 ($N_L = 1.8\times10^{-14}$ Gy/p) one can consider cool-down times of approximately 17 days. For the SPS BDF the cool-down times would extend to over 7 weeks with today's extraction efficiency. An improvement of at least a factor 3 is required in order to keep future waiting times below a week during operation of the BDF.

The estimates assume that the shape of the activation profile along LSS2 does not change significantly and that no local hot spots arise. As observed in recent operation, localised hot spots could significantly affect the dose taken during interventions and the conclusions made with the aforementioned assumptions should be taken with care. The model provides a powerful tool to understand changes in the activation levels as a function of the extracted proton flux and extraction efficiency, which could permit the identification of hot spots before the end of year radio-protection survey.

Further work is needed to understand the build-up of the induced radioactivity from longer-living radioisotopes over extended periods of operation and to tune the model to the these timescales. To this end, the LSS2 geometry has been implemented into the FLUKA code to generate loss and activation maps.


\section{Understanding today's extraction efficiency}

Since careful attention was given to the operational alignment of the ZS after Long Shutdown 1 (LS1) due to the increasing proton flux to the North Area and induced radio-activation of LSS2, it has become evident that the measured extraction efficiency is lower than what one would expect from simulations of the extraction process with scattering, including even relatively conservative mechanical and alignment tolerances. The measured extraction inefficiency indicates values a little over 3\% \cite{mf_ipac_ee} improving over recent years, but simulations of the nominal extraction system suggest values should be closer to 1\%. Recent FLUKA simulations do indeed confirm that the measured prompt beam loss is higher than expected, consistent with a thicker septum. The increased effective thickness is expected to arise from a combination of electro- and thermo-mechanical deformation as well as relative alignment of the separate ZS tanks. The studies summarised in this section focus on recent work to mitigate and reduce the extraction losses as well as to understand the present limitations and identify where future effort is needed to improve the extraction efficiency. In particular, the tests of the prototype diffusers with beam provided important information on the effective ZS width and angular spread of the beam at the ZS, amongst other things.

\subsection{Slow extraction efficiency measurements at the SPS}

The high efficiency of most slow extraction systems makes quantifying the exact amount of beam lost in the process extremely challenging. This is compounded by the lack of time structure in the extracted beam, as is typically requested by the high-energy physics experiments, and the difficulty in accurately calibrating D.C. intensity monitors in the extraction line at count rates of $\approx \SI{1e13}{Hz}$. As a result, it is common for the extraction inefficiency to be measured by calibrating the beam loss signal induced by the slow extraction process itself. The ability to accurately measure the extraction efficiency is becoming more relevant at a time when global research efforts~\cite{seWS} are intensifying to find loss mitigation methods  to meet the demanding requests for higher intensity slow extracted beams. Accurate efficiency measurements are also important in order to compare (i) the expected performance of different extraction techniques with simulation and (ii) the state of the art performance achieved at different laboratories.

The efficiency ($\epsilon$) of the extraction process is a very important figure of merit because the IR is directly proportional to the number of protons lost in the extraction process,
\begin{equation}
\textrm{IR} \propto 1 - \epsilon
\end{equation}
As $\epsilon \rightarrow 1$, it is more accurate to measure the extraction inefficiency ($\bar{\epsilon}$) and to infer $\epsilon$ from the relation,
\begin{equation}
\epsilon + \bar{\epsilon} = 1 \label{Eqeff}
\end{equation}
Even relatively large systematic errors on $\bar{\epsilon}$ result in small absolute errors on $\epsilon$. To illustrate this point, assume $\epsilon = 0.99$; a systematic error of 10\% on a measurement of $\bar{\epsilon}$ yields a systematic error of only 0.1\% on an indirect measurement of $\epsilon$. The measured beam loss during extraction, which is proportional to the number of protons lost, can be calibrated and used to measure the inefficiency. The challenge is to carefully calibrate the beam loss measured on a BLM system to the number of protons lost. Most laboratories use beam loss measurements to quantify their slow extraction efficiency~\cite{baconnier,gvozdev, tanaka,tanaka2,tanaka3,andrews,tomizawa}.

\begin{figure}[htbp]
   \centering
   \includegraphics[trim={150 130 120 136}, clip,width=0.7\linewidth]{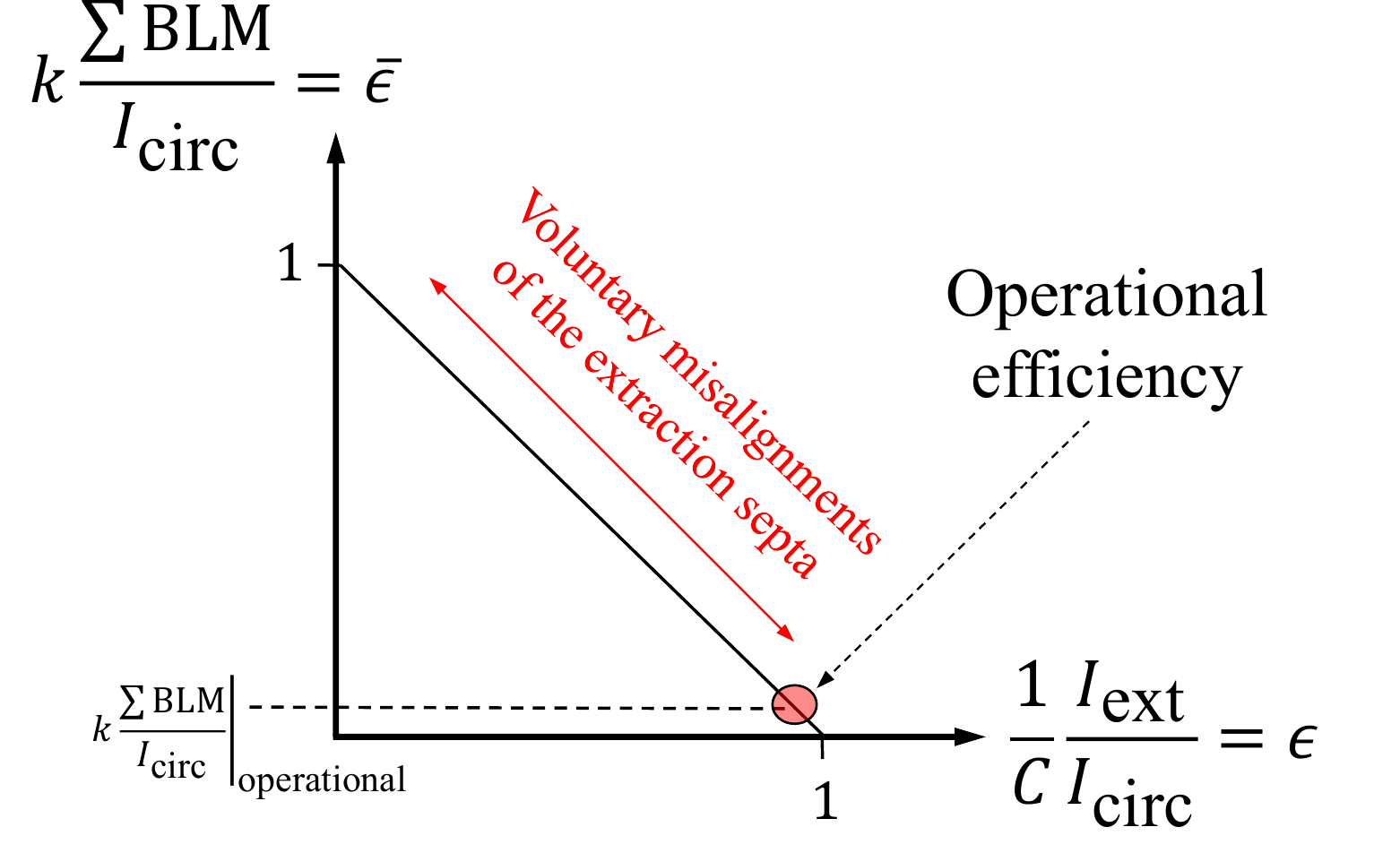}
  \caption{FNAL efficiency measurement concept~\cite{hornstra}~\label{concept}.}
\end{figure}

At the CERN SPS, a technique developed at FNAL's Main Ring (MR) in the 1970's was applied to calibrate the response of the BLM system as a function of the extraction efficiency by gently skewing the ZS~\cite{hornstra,hornstra2}. The measurement concept is described schematically in Fig.~\ref{concept}. Equation~\ref{Eqeff} can be expressed in terms of the intensity measured in the external transfer line $I_\textrm{ext}$, the extracted intensity inferred from measurements of the circulating beam intensity in the machine $I_\textrm{circ}$ and the total beam loss signal summed on the BLM system,
\begin{equation}
\underbrace{k\frac{\sum{\textrm{BLM}}}{I_\textrm{circ}}}_{\bar{\epsilon}} = 1 - \underbrace{\frac{1}{C}\frac{I_\textrm{ext}}{I_\textrm{circ}}}_\epsilon
\end{equation}
where $k$ and $C$ are calibration constants. Once the calibration constants are determined empirically, the extraction efficiency can be measured online using the relationship,
\begin{equation}
\epsilon \approx 1- kC\frac{\sum\textrm{BLM}}{I_\textrm{ext}}
\end{equation}

The alignment of the ES was voluntarily skewed during dedicated measurement sessions with low intensity extractions of approximately $2\times10^{12}$ protons. The downstream end of the girder on which all 5 of the ES tanks sit was moved in steps to maximum excursions of $\approx\pm \SI{1.5}{mm}$, therefore rotating the septum by up to $\approx\pm \SI{100}{\micro\radian}$. The beam intensity in the extraction beam line was measured on a secondary emission monitor (BSI) composed of titanium foils and placed into the beam approximately \SI{200}{m} downstream of the ES. The results of the measurement campaigns are summarised in Table \ref{results}. More recent measurements were also made in 2018 with similar results to the 2017 data, however, the results were less conclusive due to several problems such as missing BLM signals during start-up due to a hardware electronics failure, incorrect gain settings of the BSI foils and the limited Machine Development (MD) time available with other higher priority topics such as tests of the diffusers and extraction with octupoles taking precedence. These results and the difficulties encountered are documented in~\cite{fraser_ee_2018}.
\begin{table}[htbp]
   \centering
   \caption{SPS slow extraction efficiency measurement results using BSI.210279.}
   \begin{tabular}{lccccc}
       \hline
       \textbf{Year} & \textbf{BSI} & \textbf{Girder Scan} & \textbf{$k$} & \textbf{$C$} &\textbf{$\epsilon = 1 -\bar{\epsilon}$} \\ 
        & \textbf{Plate} & \textbf{Direction} & \textbf{[$\mathbf{10^{13}}~\textrm{p}^+~\textrm{mGy}^{-1}$]} &\textbf{[$\mathbf{\frac{I_\textrm{ext}}{I_\textrm{circ}}}$]}  & \textbf{ [$\mathbf{\%\pm\delta_{\epsilon}}$]} \\      
       \midrule
           2016 & \textbf{A} & \textbf{All data} & $\mathbf{24.0\pm 1.2}$ & $\mathbf{0.66\pm0.005}$ & $\mathbf{95.7 \pm 0.8}$\\
       \midrule
	 & A & Towards outside ring & 21.7 & 0.78 & $97.0 \pm 0.6$\\
	2017	 & A & Towards inside ring & 26.3 & 0.79 & $96.4 \pm 0.7$\\
	         &  \textbf{A} & \textbf{All data} &  $\mathbf{23.8\pm 0.9}$&  $\mathbf{0.78\pm0.005}$ & $\mathbf{96.6 \pm 0.7}$\\
	 \midrule
	 & B & Towards outside ring & 21.3 & 0.94 & $97.1 \pm 0.6$\\
	2017	 & B & Towards inside ring & 27.0 & 0.93 & $96.4 \pm 0.7$\\
	         & \textbf{B} & \textbf{All data} & $\mathbf{25.9\pm1.0}$ & $\mathbf{0.93\pm0.005}$ & $\mathbf{96.6 \pm 0.7}$\\
       \hline
   \end{tabular}
   \label{results}
\end{table}

\begin{figure}[htbp]
    \centering
        \begin{subfigure}{0.48\linewidth}
            \centering
            \includegraphics[width=\textwidth,clip=true,trim =  150 75 170 125]{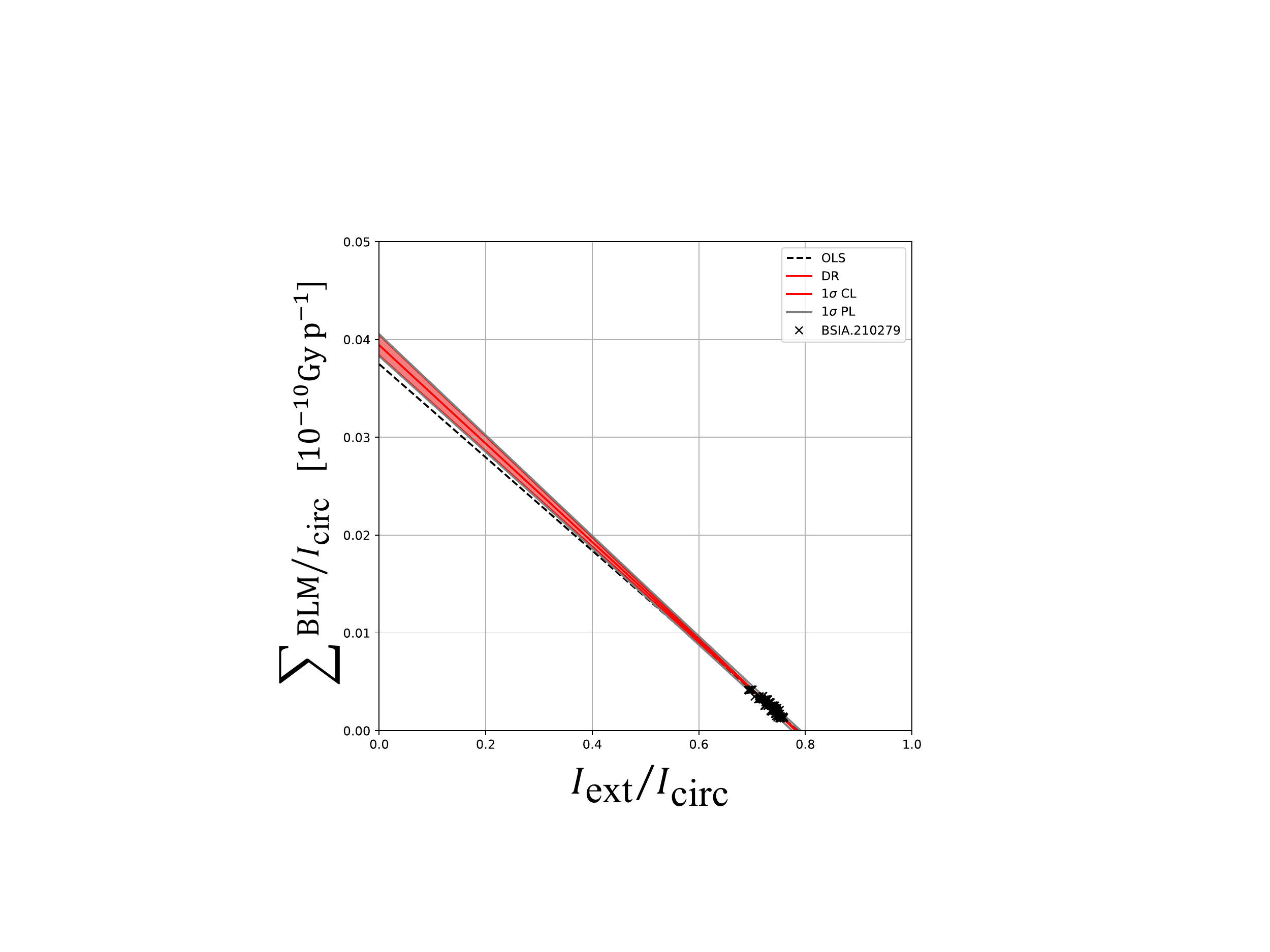}
            \caption{Large extrapolation to determine $k = (23.8\pm0.9)\times10^{13}~\textrm{p}^+~\textrm{mGy}^{-1}$.}
            \label{plot1}
        \end{subfigure}
        \begin{subfigure}{0.48\linewidth}
            \includegraphics[width=\textwidth,clip=true,trim = 150 75 170 125]{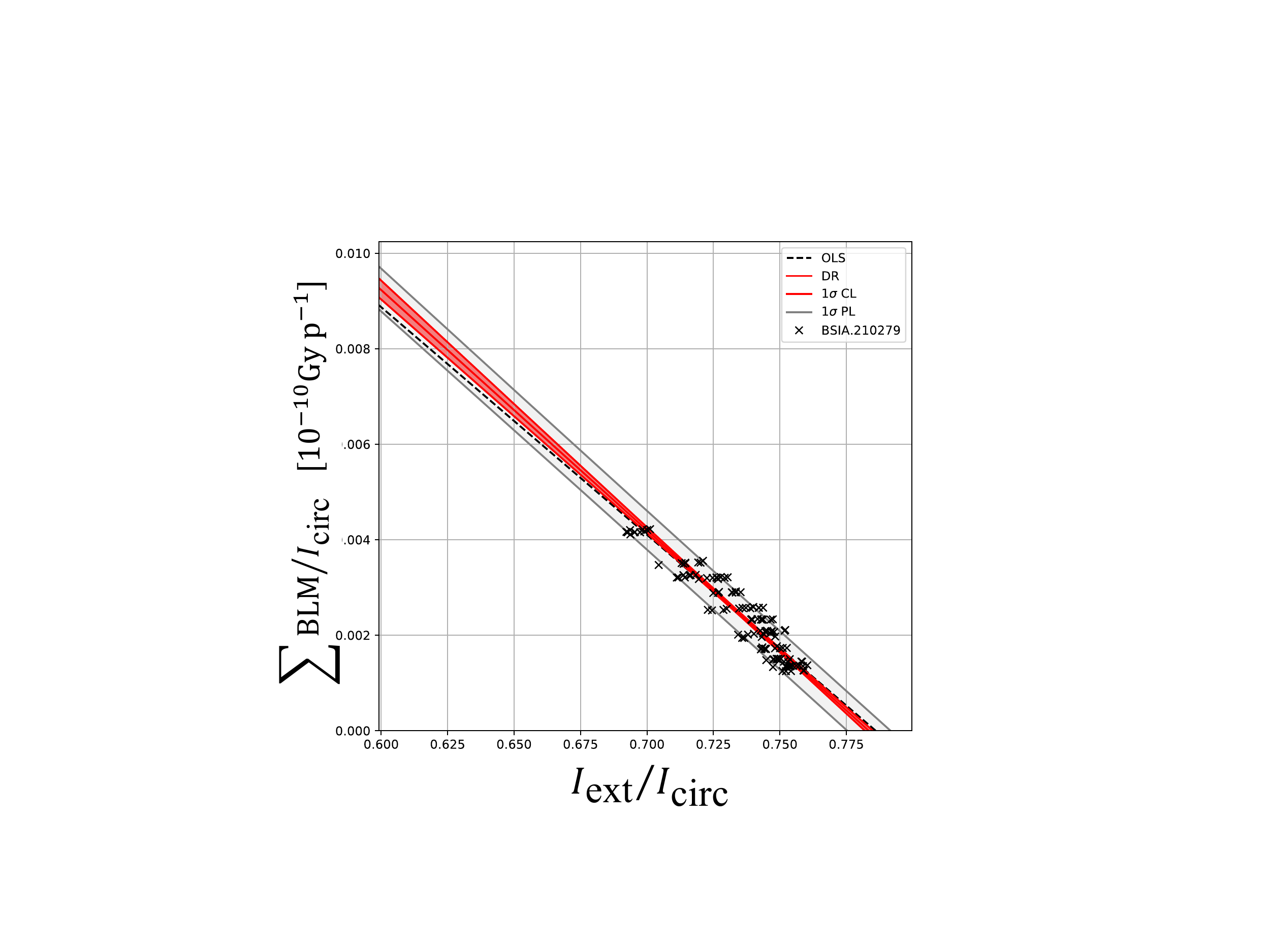}
            \caption{Extrapolation to determine $C = 0.78\pm0.005$\\$\,$.}
            \label{plot2}
        \end{subfigure}
        \caption{Measurement data: BSIA.210279 taken in 2017\label{plot_data}.}
\end{figure}

An example dataset in 2017 is shown in Fig.~\ref{plot_data} where an Ordinary Least Squares (OLS) regression analysis was performed alongside a Deming Regression (DR) (to account for errors in both scattered variables) plotted with the corresponding $1\sigma$ Confidence (CL) and Prediction Levels (PL).

The inefficiency in 2016 was measured at $4.3\pm0.8$\,\% and improved by 20\,\% in 2017 and 2018. The measured inefficiency should be compared to the theoretical value of approximately 1.2\,\% computed using MAD-X, \textit{pycollimate}~\cite{linda,velotti_thesis} and FLUKA~\cite{Fluka1,Fluka2,Fluka3,esposito} simulations with an ES set to an effective thickness of \SI{200}{\micro\meter}.  The empirical determination of $k$ showed a strong dependence on the direction of the movement of the ZS girder, which is an indicator that the changing loss profile as measured on the BLM system is a source of non-linearity and systematic error. The quoted error on $\epsilon$ is an estimation based on the systematic variations observed in the girder scan direction, including a propagation of the errors from the regression analysis.

Unlike at FNAL, where a dedicated longitudinal (co-axial) BLM was installed on the ceiling of the accelerator tunnel, relatively far and vertically above the beam line, no dedicated longitudinal BLM is presently available in LSS2. The BLMs are well distributed but located relatively close to the beam line, in the plane of extraction (horizontal) and biased by their position on the inside of the ring. FLUKA simulations have been launched to understand the dependence of the systematic errors in the measurements on the location of BLMs in LSS2~\cite{esposito2}. There is a BLM on main quadrupole 218 in LSS2 that failed in 2017. Although it was replaced for the 2018 run the BLM is saturated during high intensity operation, affecting the measurement of the efficiency. These issues may systematically affect the efficiency measurements. An online measurement of the extraction efficiency will be implemented as part of the SPS Quality Control application.

\subsection{Calibration of beam intensity monitors in the extraction and transfer lines}

As a result of the slow extraction efficiency measurements, large calibration errors on the transfer line intensity monitors were identified, which are far from guaranteeing a few \% accuracy. The first results in 2016 identified a large discrepancy between the calibration of the BSI.210279 monitor located at the upstream end of the TT20 extraction line and the Beam Current Transformer (BCT) in the ring, i.e. $C = 0.66\pm0.005$. To complicate issues the BSI assembly is composed of a stack of two measurement plates (A upstream of B) with a bias plate in between~\cite{bernier}.

\begin{figure}[htbp]
   \centering
   \includegraphics[width=0.7\linewidth,trim={2 2.25 2 0}, clip]{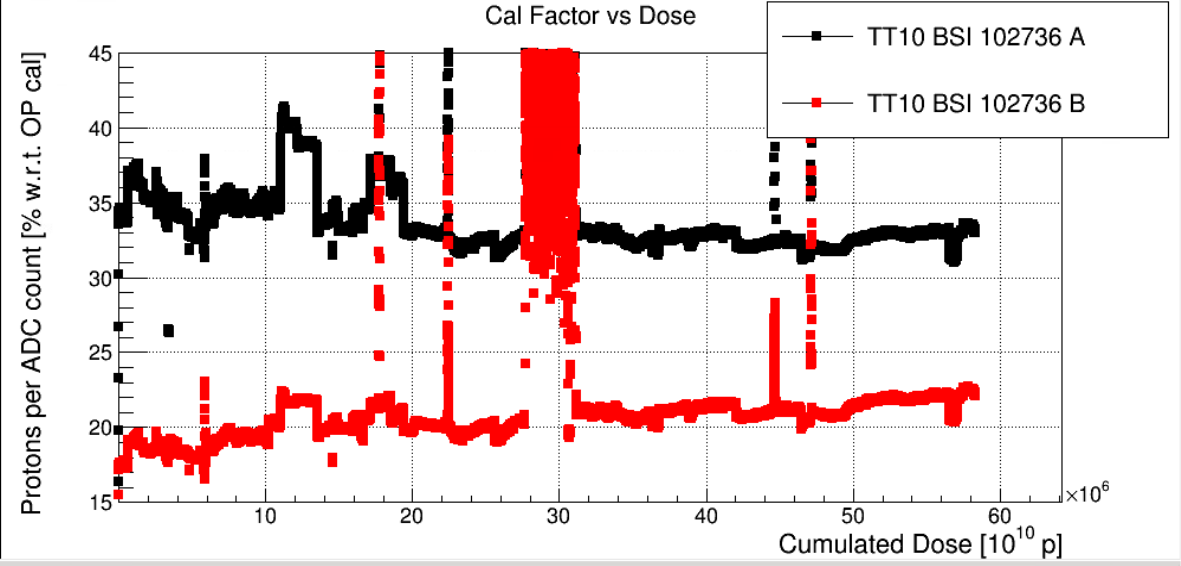}
  \caption{Measured BSI calibration factors in 2017~\cite{roncarolo}\label{bsi}.}
\end{figure}

To further understand the discrepancy and behaviour of the BSI, plate A was removed at the end of the 2016 physics run and installed in the TT10 transfer line between the Proton Synchrotron (PS) and SPS, where a fast-BCT is available for cross-calibration purposes. The difference in the Secondary Emission Yield (SEY) due to the lower beam energy, which is a factor of $\sim30$ below the SPS extraction energy, is expected to be negligible. The BSI plates in TT20 were replaced by two new titanium plates with a third new plate installed at position B in TT10. The measurements in TT10 indicate that exposing the old plate to air during its displacement from TT20 to TT10 affected its SEY, which increased by about 10\%. The calibration constant determined using the extraction efficiency measurements in LSS2 was confirmed with the TT10 measurements in 2017, although drifting of the SEY throughout the year was observed, as shown in Fig.~\ref{bsi}. The measurements also confirmed a systematically higher signal of $\sim15\%$ measured on plate B compared to A, consistent with the LSS2 measurements, with the likely explanation that secondary electrons generated on plate A reach plate B.

Work is actively on-going to provide accurately calibrated intensity measurements in the extraction transfer lines and a summary of the analysis can be found in~\cite{roncarolo}. For now one must take numbers recorded by the BSI's with caution and assume systematic errors on the order of 10 - 20\%.

\subsection{ZS alignment}

The alignment of the ZS has improved iteratively over the last few years as operational knowledge of the system has improved along with the required tools, both operational and simulation. An important improvement was to check the size of the extracted beam and its impact of beam loss on the cathodes of the ZS. For most of the last two years of operation, the first two cathodes were retracted slightly to help reduce losses coming from this source, which helped alignment and especially to reduce the loss contribution from particles scattered back into the machine and returning at large amplitude after three more turns of the synchrotron. The improvement is best illustrated in Fig.~\ref{fig:ACTvsPOT} where, since the 2015 physics run, the end-of-year RP survey measured close to the ZS 30 hours after operation has improved year-on-year, and is now lying back on the historical trend as a function of total extracted flux.

\subsubsection{Simulation of alignment procedures}

\begin{figure}[htbp]
    \centering
    \begin{subfigure}{0.38\linewidth}
        \centering
        \includegraphics[width=\textwidth]{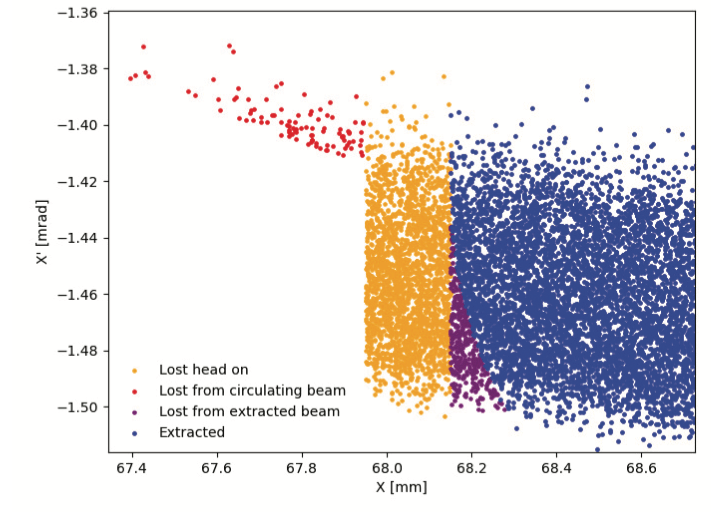}
        \caption{Phase space upstream ZS and losses (blackbody absorber model).}
    \end{subfigure}
    \begin{subfigure}{0.6\linewidth}
        \includegraphics[width=\textwidth]{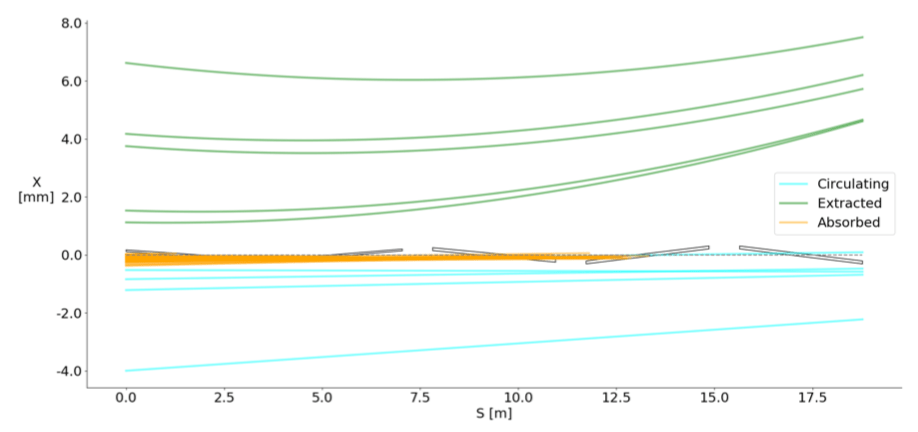}
        \caption{Particle trajectories (\textit{pycollimate} scattering model) along the ZS\\$\,$.}
    \end{subfigure}
    \caption{(a) Particles lost or extracted are shown for the nominal alignment with simplified ZS geometry (effective thickness \SI{200}{\micro\meter}, no gaps between tanks) in the black absorber approximation. Most particles are lost due to non-zero wire thickness at the ES upstream (orange). The remaining losses stem from the angular spread of the beam, which occur downstream, both from the inside (purple) and outside (red) of the ZS. (b) Particle trajectories along the ZS in the presence of scattering.}
    \label{fig:particles_at_ZS}
\end{figure}

The active length of the ZS is over \SI{16}{m} and composed of 5 separate units containing separate wire-arrays that can be moved independently. The tanks are all mounted on a single support structure that can move the ensemble coherently. As a result, the large number of positional degrees of freedom complicates the alignment procedure in operation. Obtaining and maintaining accurate alignment with the beam is therefore crucial for minimising beam loss. The particles impinging the ZS are shown in phase space in Fig.~\ref{fig:particles_at_ZS}(a) along with a sample of trajectories with misaligned anode wires and scattering in Fig.~\ref{fig:particles_at_ZS}(b).

To understand and investigate the efficacy of different alignment procedures, particle tracking simulations were carried out and the beam loss along the extraction straight section computed as a function of the relative positions of each of the 5 separate ZS units. An important aspect of the study was to understand the required alignment tolerance and motor precision to achieve optimum extraction efficiency for a given configuration of wire-array thicknesses. The anode wires were taken as nominally designed, without imperfections, between \SI{60}{\micro\meter} (tanks 1 and 2) and \SI{100}{\micro\meter} (tanks 3-5) thick. The upstream and downstream ends of the anodes and the girder can be moved independently, yielding a total of 12 degrees of freedom. By fixing the upstream end of the girder and the first tank, together with the extraction bump amplitude, the spiral step (transverse extracted beam size) is kept constant, reducing the dimensionality of the problem to 10 degrees of freedom. Random misalignments between the tanks mean the effective thickness of the anode wires is increased and the probability of particles being scattered and lost increases.

Simulations, such as FLUKA or even multi-turn tracking simulations with MAD-X coupled to \textit{pycollimate}, demand a level of detail that is computationally very expensive. Given the relatively high dimensionality of our problem, simplified models are necessary in order to cut down simulation time to reasonable values when running large batches of error seeds. To reduce the simulation time, a fixed particle distribution at the upstream end of the ZS was pre-computed by MAD-X and then sampled for every alignment error seed tested. Sampled particles are then tracked along the ZS taking into account wire misalignments. Only the last three turns before extraction are simulated, giving a large speed-up in the simulation time. If a particle hits the wires, it is handed over to \textit{pycollimate} to simulate the scattering process. Particles that are scattered back to the circulating side are then tracked around the ring until they are either extracted, absorbed by the ZS in an inelastic nuclear interaction or lost somewhere on an aperture restriction. The tracking around the ring is done with a simplified lattice containing only linear elements and non-linear thin lens extraction sextupoles. This scheme allows us to reduce the number of turns the particles are tracked by a factor of $\sim10^{4}$, since the bulk of the work only has to be done once in MAD-X. The  simulation code can be found on Gitlab \cite{pyTRACK} and has been made in such a way that different custom or off-the-shelf optimisers can be easily plugged into the library.

Currently, the ZS tanks are aligned manually in a few 8 hour shifts over the course of the commissioning period at the beginning of the operational year. In 2017 and 2018, once the alignment was adjusted and losses acceptable for high intensity operation the ZS could be kept in the same position through the rest of the year's operation.

\begin{figure}[htbp]
\centering\includegraphics[width=\linewidth]{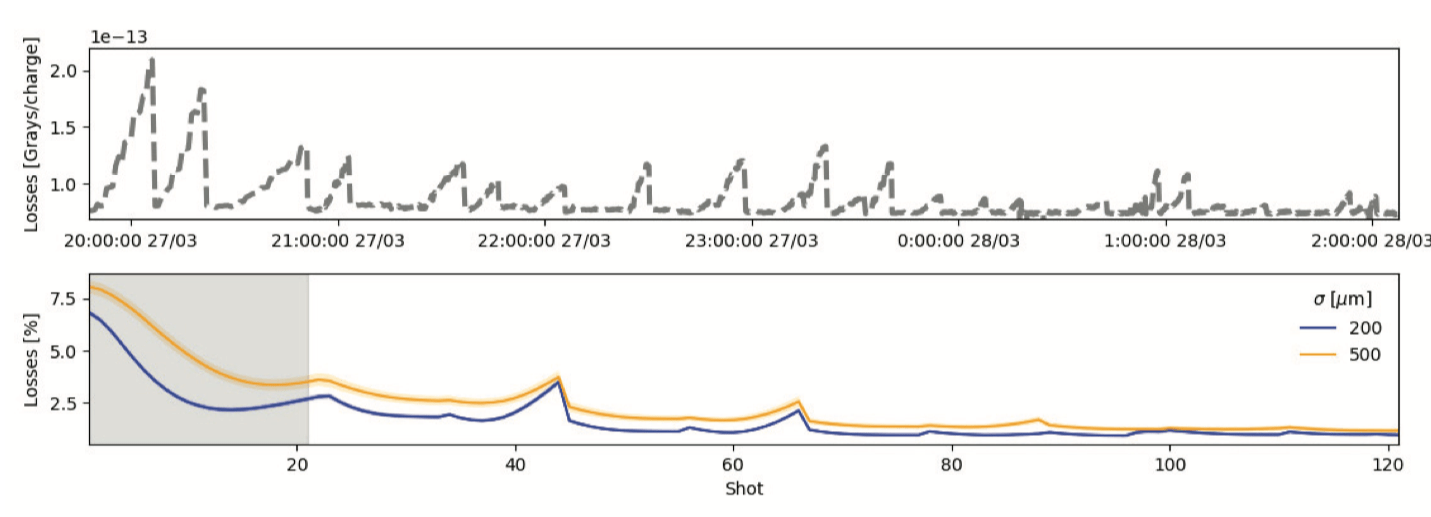}
\caption{Top: Normalised BLM losses measured during the last alignment procedure over 6 hours for a single iteration of alignment during 2018 re-commissioning. Overall improvement was 31\%. Bottom: 500 anode scan simulations with normally distributed initial misalignments. Average and 95\% confidence interval of the losses is shown. Shaded area indicates girder scan with subsequent scans following ZS tank anodes 1 to 5.}
\label{fig:OP_alignment_vs_sim}
\end{figure}

The procedure is as follows: once the slow extraction has been correctly set-up with the orbit flattened in LSS2, and the extraction bump and sextupoles correctly scaled, the angle of the girder is scanned, whilst holding the upstream position constant. The position that yields the lowest losses, as measured by the LSS2 BLMs, is chosen. The alignment set-up of the ZS is carried out at low intensity to minimise risk of damaging the septum and to minimise the induced radio-activation. Next, similar scans are repeated for all the anode motors, starting from the downstream end of ZS1, before moving onto the upstream end of ZS2 and moving downstream sequentially. The anode scans are then repeated until no improvement in beam loss is observed. The sum of the extraction losses in LSS2 during the alignment procedure over 6 hours for a single iteration during the 2018 re-commissioning is shown in Fig.~\ref{fig:OP_alignment_vs_sim} and compared to large-scale error study simulations with two different tolerances for the starting misalignment of the anodes.

Since the procedure is labour intensive, requiring several dedicated sessions of about 8 hours each, and the loss profiles for each scan are reproducible, the automation of the procedure was investigated.  In addition, these studies were motivated by the pressure for physics time that severely limits the number of iterations that can be carried out. For example, it was evident that an algorithm exploiting the global structure of the problem instead of locally optimising each degree of freedom would be of interest. Different optimisation and alignment algorithms capable of yielding similar performance with faster convergence were investigated. A full summary of the study can be found here~\cite{javier_slawg}. 

\begin{figure}[htbp]
\centering\includegraphics[width=\linewidth]{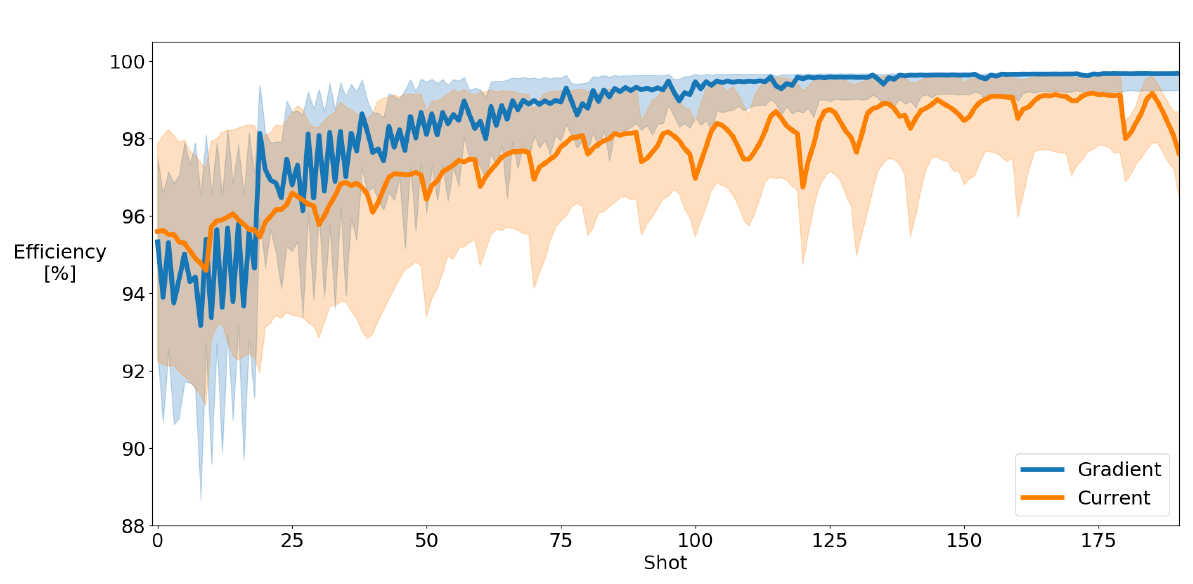}
\caption{Simulated evolution of the current operational alignment procedure compared to a gradient descent optimiser algorithm starting from random initial anode alignments with $\sigma = \SI{500}{\micro\meter}$. The median and  90\% confidence interval of the error seeds are plotted.}
\label{fig:OP_vs_grad}
\end{figure}

\begin{figure}[htbp]
\centering\includegraphics[width=\linewidth]{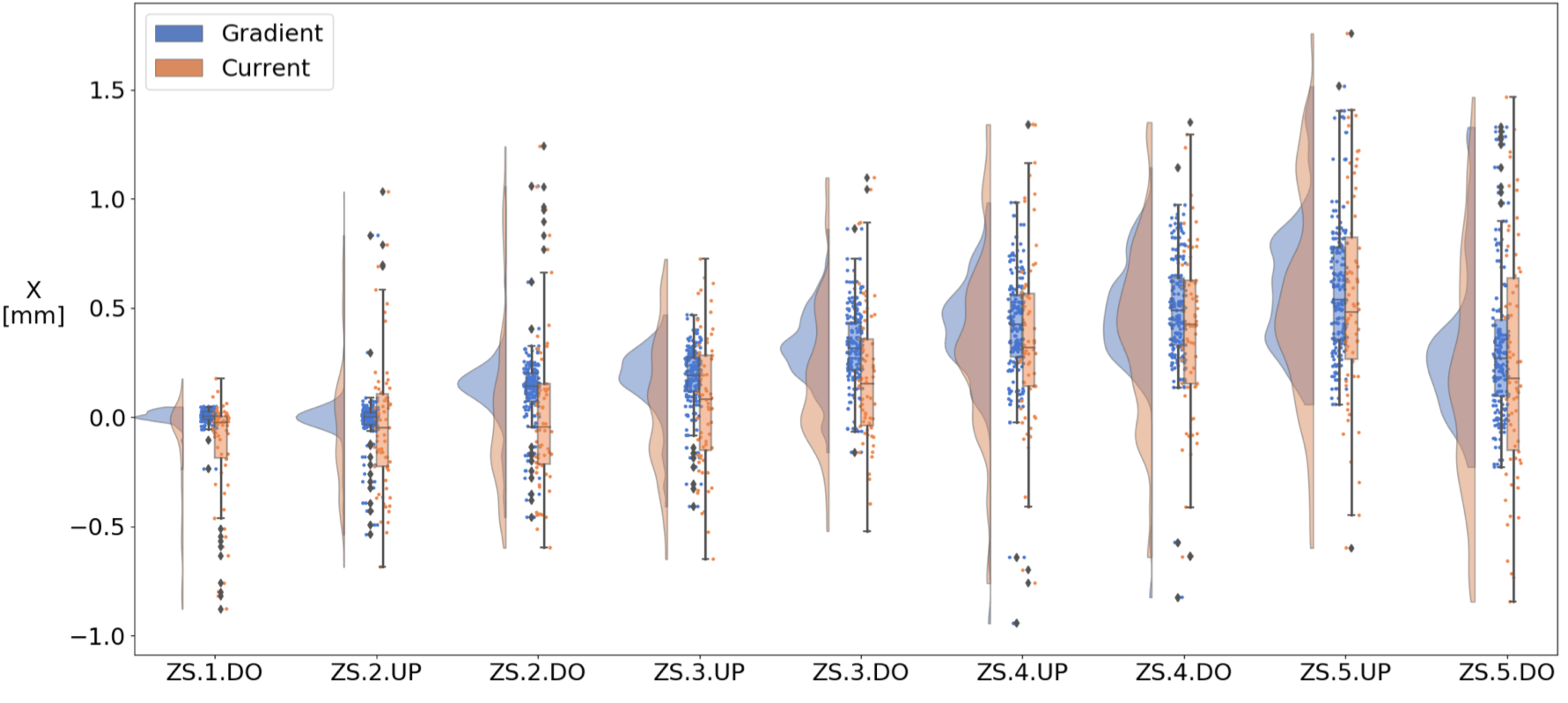}
\caption{Simulated ZS anode positions at the end of one iteration of the current operational alignment procedure compared to a gradient descent optimiser algorithm starting from random initial anode alignments with $\sigma = \SI{500}{\micro\meter}$. }
\label{fig:ZS_final_positions}
\end{figure}

The current operational algorithm was simulated and compared to a gradient descent algorithm that computed the gradient of the loss function in 9 degrees of freedom by moving each ZS anode motor left and right by a given step, with the upstream end fixed, as described in detail here~\cite{azsawg_marcell}. At each iteration all anodes were moved in the direction of greatest descent of the total loss function before repeating with the gradient computation and an exponentially shrinking step size.  Each error seed drew its initial misalignments from a random normal distribution with $\sigma = \SI{500}{\micro\meter}$. When studying motor errors, a random shift drawn from a normal distribution was added on top of the requested anode position. The results summarised in Fig.~\ref{fig:OP_vs_grad} show the stark difference in the efficiency of the current operational procedure compared to the gradient descent algorithm, both in the number of shots needed for convergence but also in the end scatter of the data. The results are highlighted by the improved clustering of the final positions of the anodes summarised in Fig.~\ref{fig:ZS_final_positions}.

\subsubsection{Simulation with motor accuracy error}

\begin{figure}[htbp]
    \begin{subfigure}{\linewidth}
        \centering
        \includegraphics[width=\textwidth]{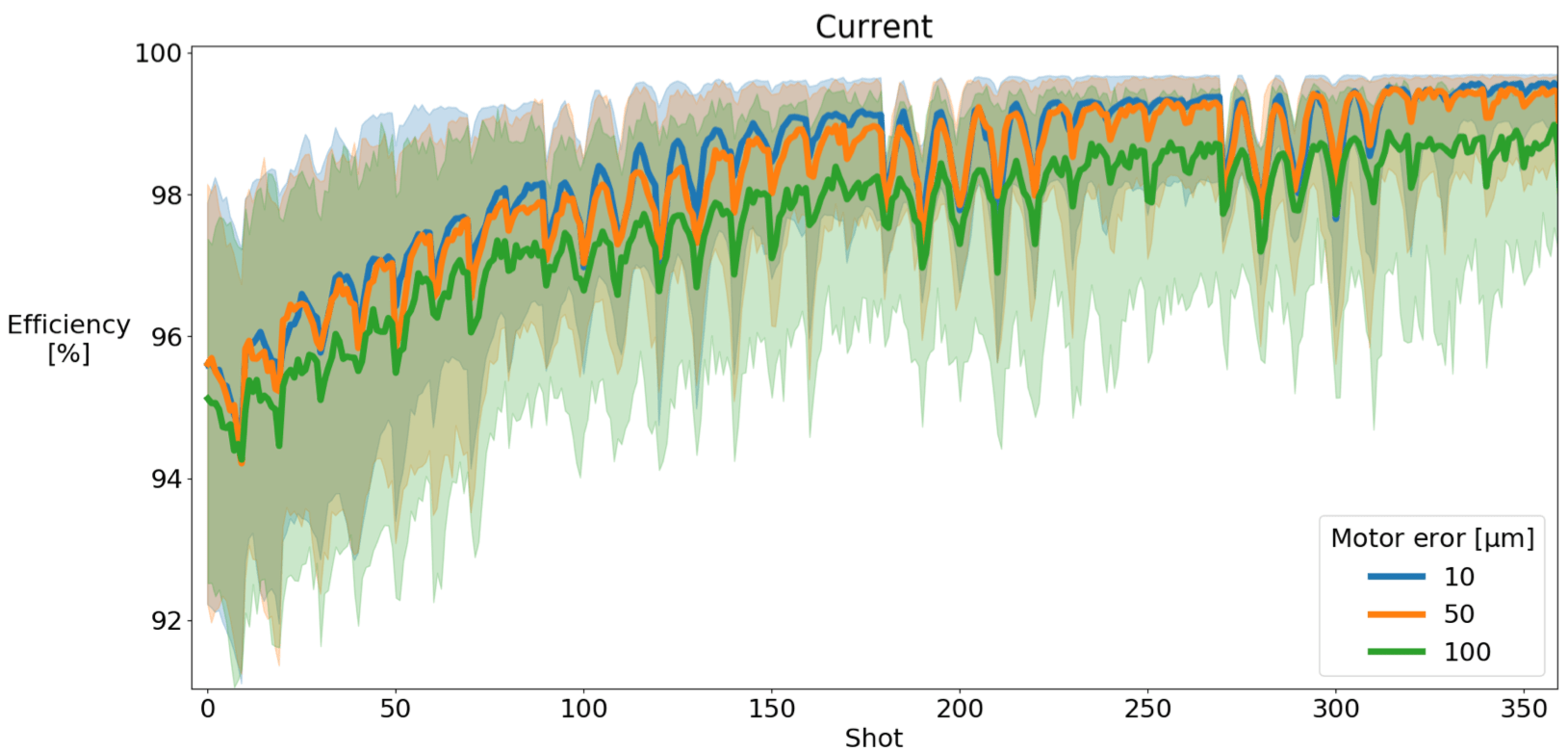}
        \caption{Simulated evolution of the current operational alignment procedure including random motor position errors of different magnitude.}
    \end{subfigure}
    \begin{subfigure}{\linewidth}
        \centering
        \includegraphics[width=\textwidth]{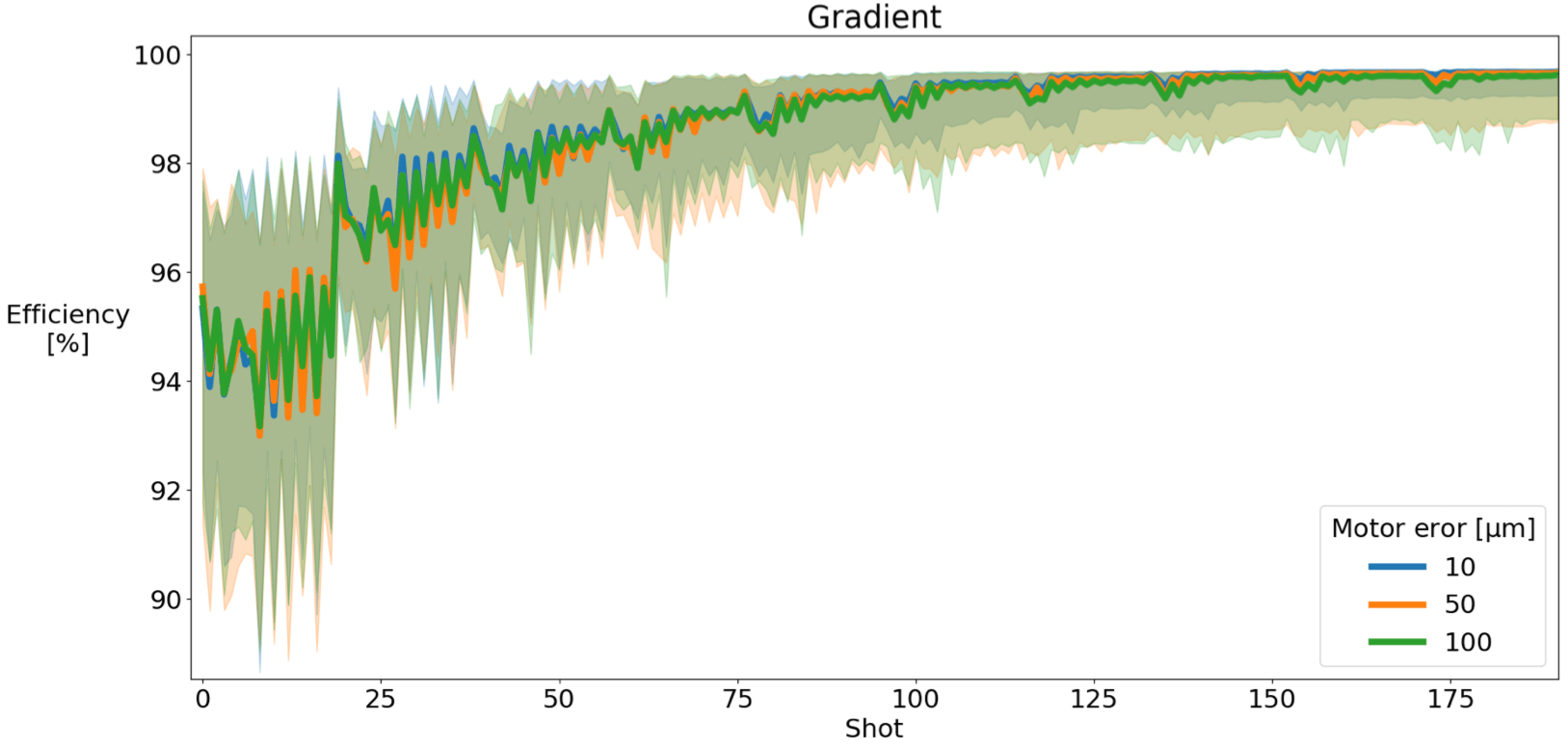}
        \caption{Simulated gradient descent optimiser including random motor position errors of different magnitude.}
    \end{subfigure}
    \caption{Simulations of the alignment procedure in the presence of different random motor errors.}
    \label{fig:align_motor_error}
\end{figure}

The simulations, presented in Fig.~\ref{fig:align_motor_error}, indicate that the required motor precision depends heavily on the type and efficacy of the alignment algorithm employed. The simulations indicate that for today's scenario a motor precision of \SI{50}{\micro\meter} is sufficient, whereas the accuracy could be relaxed in the future should improved alignment algorithms be successfully deployed.

\subsubsection{Simulation of higher voltage ZS}

It goes without saying that the fewer independent septum tanks that are needed to align, the simpler and more efficient the alignment procedure. In addition, if higher voltages could be applied less material will be presented to the beam and the extraction efficiency would be significantly higher. This was demonstrated in simulation by increasing the voltage and extracting with just two ZS tanks, as is the case at J-PARC~\cite{javier_slawg}. Although the improvement is dramatic, it is unfeasible when one considers that one would need electric fields surpassing 11 MV/m with voltages of over 500 kV. The electric field is limited by today's technology used for the large electrodes and gap sizes.

\subsubsection{Operational implementation of an automatic ZS alignment procedure}

In a recent beam test, an operational optimiser based on a Powell optimisation algorithm was applied to the ZS alignment problem. The septum was successfully aligned within about 40 minutes of beam time and about 130 shots, reducing the time needed by over an order of magnitude. The optimiser acted on the 9 degrees of freedom whilst observing and minimising the normalised losses of all BLMs in LSS2. Future studies will concentrate on algorithms which memorise (speed up full alignment to a few cycles) and possibly generalise the automatisation via reinforcement learning. These first promising results, reported in~\cite{hirlander_slawg}, give us confidence that the septum can be aligned far quicker in the future and potentially with a better extraction efficiency. Further beam tests are needed to understand the improvement of the extraction efficiency with a dedicated optimisation algorithm, as well as to assess the role of noise in the optimisation.

\subsection{Impact of bunched beam extraction on extraction efficiency}

\begin{figure}[htbp]
\centering\includegraphics[width=0.7\linewidth]{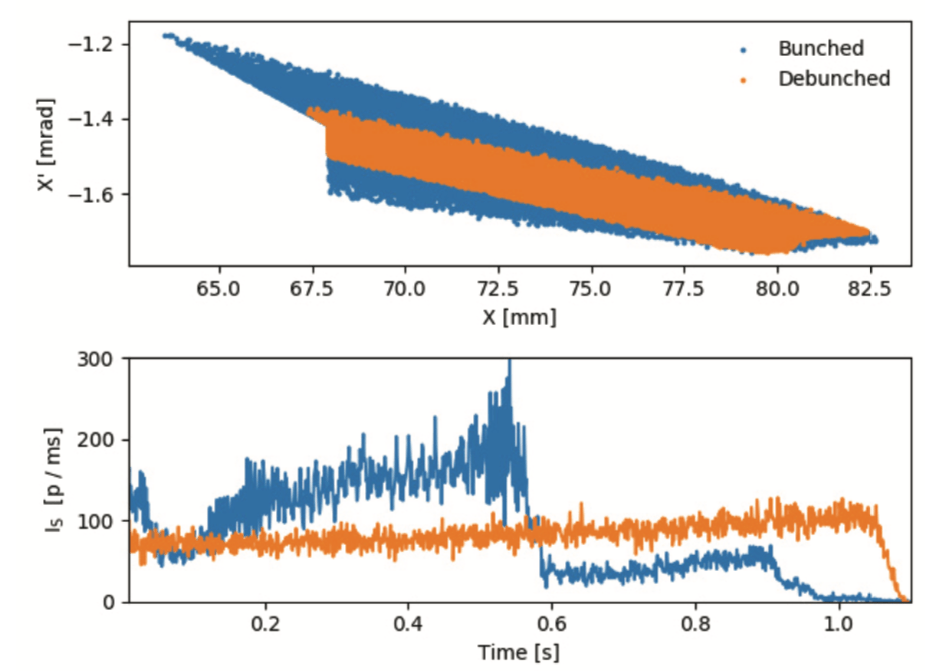}
\caption{Comparison of debunched and bunched ($V_{RF}$ = 7 MV) chromatic extraction. Top: horizontal phase space at the extraction point showing particles lost on the ZS or extracted. Bottom: Spill rate.}
\label{fig:bunched_beam}
\end{figure}

In order to provide slow-extracted beams with a time structure on the order 5 - 100 of MHz, the RF system of the SPS must remain on to keep the beam bunched during the extraction. Due to the non-zero chromaticity used in the present extraction scheme, the resulting synchrotron motion induces a tune modulation in a way similar to that of a strong ripple on the current of the main power converters, see~\cite{javier_ripple}. The difference in this case is that the tune modulation is not coherent; the modulation frequency and phase of each particle depend on its initial location in the bunch (longitudinal action and angle). In recent simulations of the extraction efficiency, the extracted beam emittance and spill quality for a 5 ns bunch structure  with an RF voltage of 7 MV was investigated and summarised in Fig.~\ref{fig:bunched_beam}. The spill quality degrades, with a duty factor of $F = 0.55$, with the spill length effectively halving. This is due to the fact that particles undergoing synchrotron motion visit the lower half of the bucket every $\sim40$ turns. As the tune sweep progresses in momentum from the lower tip to the centre of the RF bucket, virtually every particle will have the chance to become resonant and be extracted. Thus, when the sweep reaches half-way and matches on-momentum particles, the extracted intensity drops as all particles have been pushed through resonance by their synchrotron motion. The tune sweep can be easily adjusted for the bunched case.

A threefold increase in the extraction inefficiency was observed. This can be attributed to the higher angular spread of the extracted beam at the ES. As in the case of ripple, particles coming in and out of resonance don’t follow clean trajectories along the separatrix.

Slow extracted beams have been provided to the NA in the past for irradiation tests of LHC components with a \SI{25}{ns} bunch structure. However, the extraction inefficiency was reported as much as 10 times worse, limiting the intensity that could be delivered. For experimental requests with a larger bunch spacing than \SI{5}{ns}, e.g. 25 - \SI{100}{ns}, the extra cycle time required in the PS to carry out the required RF manipulations will have a severe impact on the duty cycle in the SPS.

These results highlight the need to fundamentally rethink the extraction scheme in the face of experimental requests for beam with a MHz time structure. Other extraction schemes are a matter of ongoing study, but they have their own short-comings. For instance, a similar scheme with near zero chromaticity has been shown to reduce the angular spread of the extracted beam, however, this comes at the expense of having to make a slower tune sweep, thus increasing the sensitivity to ripple. Possibilities of smoothing the spill with zero chromaticity by applying transverse RF noise were successful in beam tests in 2018 and will be investigated further.

\subsection{FLUKA modelling of LSS2 beam loss profile} \label{FlukaModel}
\begin{figure}[htbp]
\centering\includegraphics[width=0.7\linewidth]{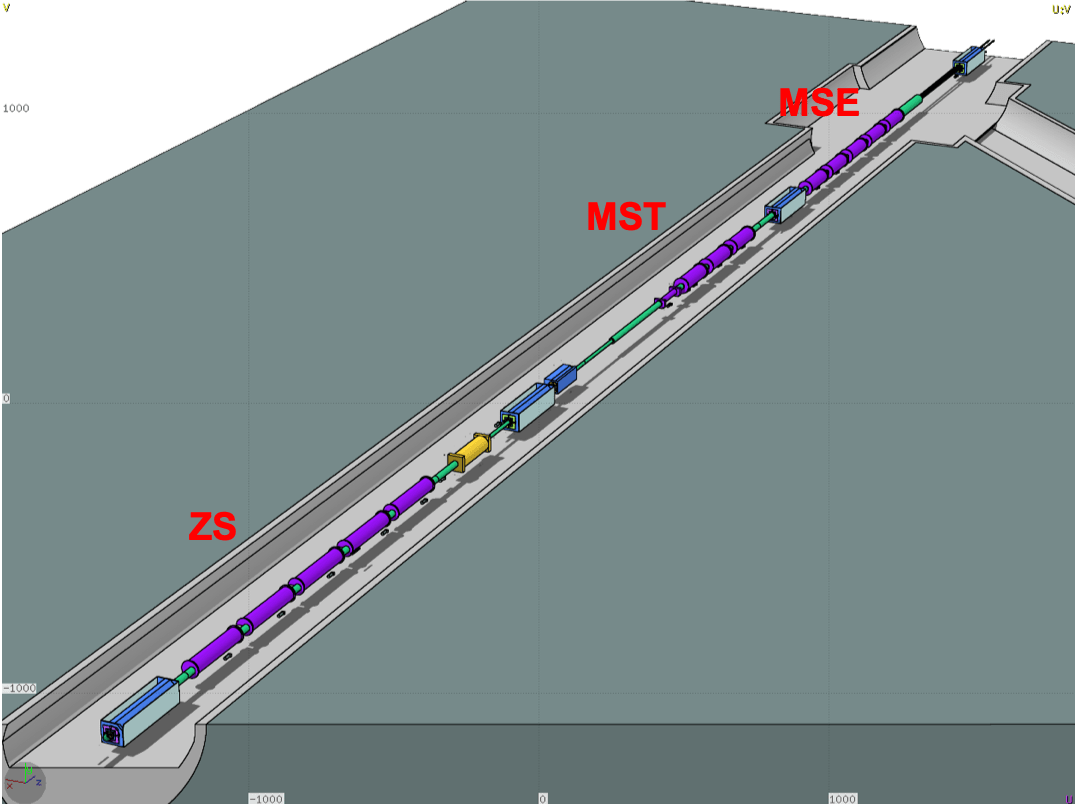}
\caption{FLUKA model of the SPS-LSS2. The geometry extends from QFA.216 to QDA.219.}
\label{fig:fluka_model}
\end{figure}

An extensive FLUKA model of the SPS LSS2 was developed to study the energy deposition in the beam line elements  
and evaluate the response of the BLMs as consequence of the protons lost during the slow extraction process.
The model, shown in Fig~\ref{fig:fluka_model}, consists of:
\begin{itemize}
\item all beam elements from QFA.216 to QDA.219, including the electrostatic septa (ZS), the magnetic septa (MST and MSE), the bumper magnets and the beam intercepting devices (TCE and TPST);
\item the SPS BLMs;
\item a detailed description of the aperture of the beam-line elements;
\item the time-constant magnetic fields of the quadrupoles, MST and MSE;
\item the electric field of the ZS emulated by an equivalent magnetic field.
\end{itemize}

The ZS wires are modelled as a \SI{200}{\micro\meter} thick WRe ribbon with scaled density (to maintain the same mass of the real wires) and are perfectly aligned along the 5 ZS tanks. It was checked that the modelling of the wire array as a scaled-density ribbon does not change significantly the loss pattern with respect to a case where each single wire is modelled.

The proton distribution used as source term for the FLUKA calculations is computed by MAD-X tracking and the last 4 turns for each particles are given at a position upstream the ZS. Two distributions have been used: 

\begin{itemize}
\item the first, referred to as \emph{black} distribution, where the ZS wires are treated as black absorbers, i.e. any proton hitting the ZS wire is not transported anymore;
\item the second, referred to as \emph{full} distribution, where the MAD-X tracking is linked to \textit{pycollimate} \cite{velotti_thesis}. In this case,  a proton scattered off by the ZS wires is tracked for additional turns until it is extracted, hits the vacuum chamber aperture, or undergoes a nuclear interaction in the ZS.
\end{itemize}

\begin{figure}[htbp]
\centering\includegraphics[width=0.7\linewidth]{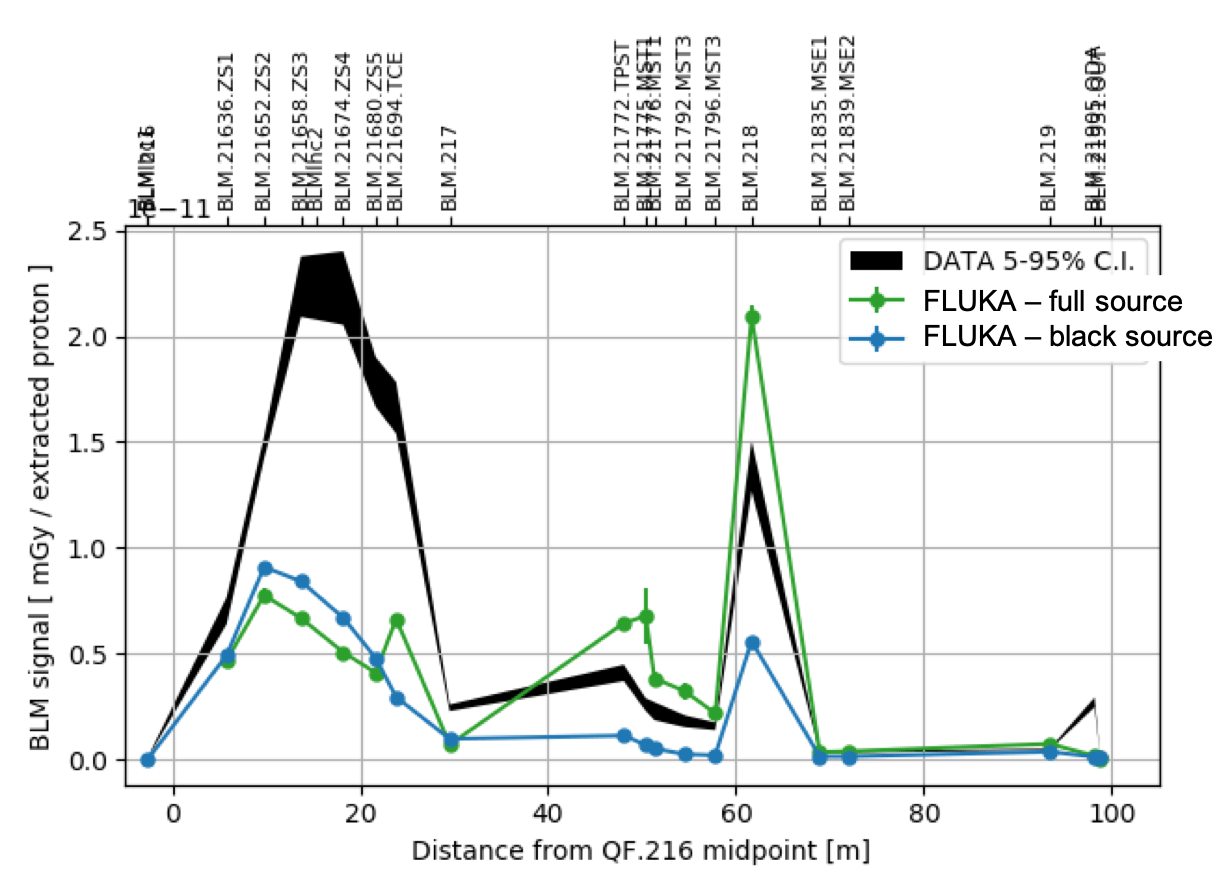}
\caption{ Normalised loss pattern of the SPS LSS2 BLMs. (black) Measured data from the 15$^{\textrm{th}}$ to the 18$^{\textrm{th}}$ of April 2018. 
BLM response computed by FLUKA using the \emph{black} distribution (blue) and the \emph{full} distribution (green) are also shown. 
The BLM.21694.TCE was misplaced in the geometry model used with  \emph{black} distribution. 
For this reason its signal is increased by a factor of 2. }
\label{fig:fluka_blm_pattern}
\end{figure}

Figure \ref{fig:fluka_blm_pattern} compares the measured pattern of the BLMs installed in the SPS LSS2 with the one simulated by FLUKA.
The simulated losses along the ZS are mainly due to the single-pass effect and are a factor 3 below measured values. The losses at TPST and QFA.218 are instead due to protons scattered by the ZS that make additional turns in the SPS.

The inefficiency of the slow extraction process, defined as the number of protons lost divided by the number of protons extracted toward TT20, is 0.7\% and 1.2\% for the \emph{black} and \emph{full} distribution respectively.

\begin{figure}[htbp]
\centering\includegraphics[width=0.7\linewidth]{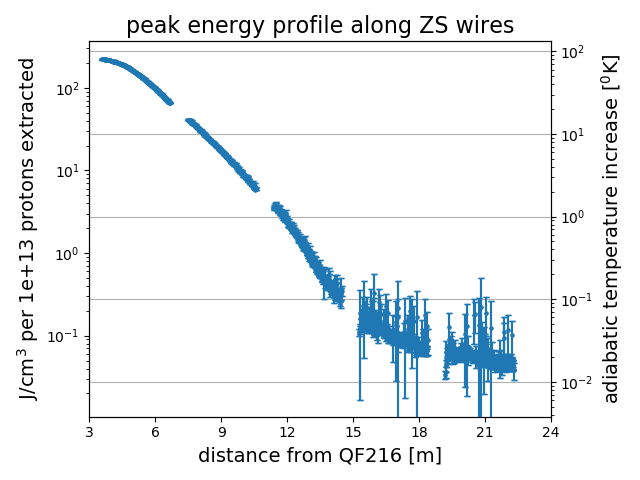}
\caption{ Energy density profile along the ZS wire array normalised to $10^{13}$ protons extracted. 
The right vertical axis indicates the corresponding adiabatic temperature increase.}
\label{fig:fluka_endep_ZSwires}
\end{figure}

The relative energy deposition in the ZS wires is shown in Fig.~\ref{fig:fluka_endep_ZSwires}, which reports also the corresponding temperature increase. The latter is computed in a first approximation by assuming a constant specific heat capacity and considering no mechanism of heat transfer. For a perfectly aligned geometry, the peak is located at the beginning of the upstream end of the wire array, while the downstream wires are in their shadow.

\begin{figure}[htbp]
\centering\includegraphics[width=0.7\linewidth]{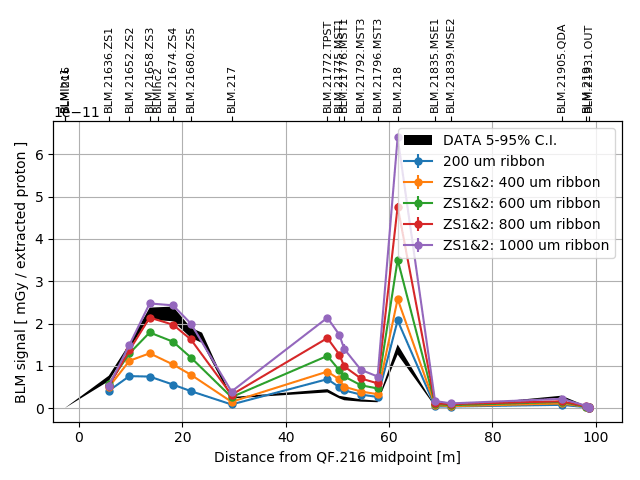}
\caption{Energy density profile along the ZS wire array for different ribbon thickness .}
\label{fig:fluka_blm_vs_ZS_thickness}
\end{figure}

As an illustration of the effect of the wire misalignment, the loss pattern with different effective thicknesses of the wire ribbon (the uniform ribbon density is scaled accordingly) are compared in Fig.~\ref{fig:fluka_blm_vs_ZS_thickness}. 
The measured loss pattern along the ZS matches the simulation when the wire ribbon is $\sim 800-\SI{1000}{\micro\meter}$ thick. At the same time, the losses at TPST and QFA.218, due to the protons scattered-off by the ZS, are clearly overestimated. It should be pointed out that ribbon is assumed as straight and therefore aligned. Recent studies have shown that coherent deformations of the anode support  holding the wires worsens the inefficiency compared to the ribbon approximation. The computed inefficiency of the slow extraction process is 3.1\% (3.9\%) for the 800 (1000)~\si{\micro\meter} thick wire ribbon, consistent with recent measurements giving 3.4~$\pm$~0.7 \%~\cite{fraser_ee_2018}. Further investigations are on going to assess potential coherent misalignments of the wires and the interplay between the scattering of protons off the ZS and the multi-turn effects.

\newpage
\section{Extraction loss reduction techniques}

\subsection{Constant Optics Slow Extraction (COSE)}

\begin{figure}[htbp]
  \centering
  \includegraphics*[width=0.55\linewidth]{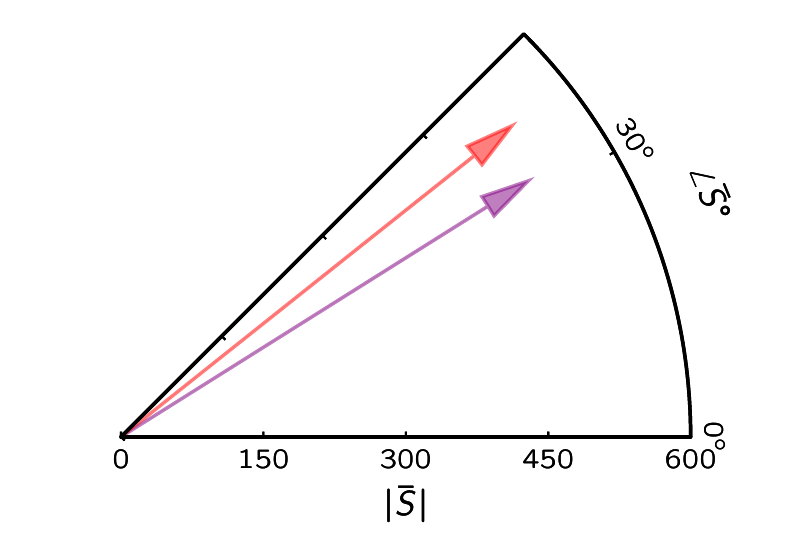}
  \caption{Schematic view of the resonance driving term evolution in time (violet is $t=0$ and red is $t = \SI{4.8}{s}$.}
  \label{fig:ps_zs}
\end{figure}

As described above, the SPS slow extraction is a quadrupole-driven chromatic extraction. The machine tune is changed in time during the spill to extract particles with increasing momentum as a function of time. In the SPS this was done by ramping the main quadrupoles to change the horizontal tune and is referred to as the "Q-sweep" method in this document. As the rest of the machine's elements are stationary during the spill, the optics seen by the particles with the resonant tune is different from the nominal one, changing the presentation of the separatrix as a function of momentum, i.e. time. The non-perfect overlap in time of the extraction separatrix actually results in an increase of the beam angular spread at the ZS entrance, which also results in a higher number of particles that intercept the ZS, i.e. losses. The change of optics in time also means that the phase advance between the ZS and the resonance driving term are different, see Fig.~\ref{fig:ps_zs}, as well as beta-beating and variation of natural chromaticity. The concept and results briefly discussed in the section can be found published in more detail  in~\cite{PRSTAB_kain}.   

\subsubsection{Dynamic bump}
In order to keep the extracted separatrix at the same phase space location at the ZS, the first idea was to directly act on position and angle of the beam. This could be done using orthogonal independent steering bumps, for angle and position, with time varying amplitude to follow the beam variations, as implemented already at the JPARC Main Ring \cite{jparc}. Theoretical studies and experimental tests were carried out showing the ability to act on the beam presentation at the entrance of the ZS wires over time. Although a possible reduction in the order of 10\% was foreseen in simulations, it was never experimentally confirmed~\cite{stoel_thesis}. The machine tests only highlighted the operational complexity of the dynamic bump concept at the SPS.  The beam presentation change at the electrostatic septum was identified as coming from the variation of the machine optics during the spill. Eventually, it was realised that the SPS control system offers the possibility to eliminate the optics change without applying dynamic bumps.

\subsubsection{Implementation at SPS}
The SPS is operated by controlling high-level parameters thanks to its control infrastructure, the so called LHC Software Architecture (LSA) \cite{LSA}. LSA offers the ability to work in physics parameters, e.g. tune ($Q$), machine momentum ($p$), charge ($q$), etc. The framework takes care of translating these into hardware settings like voltage or current. The spill structure of the slow extraction in Q-sweep mode is optimised by acting directly on the horizontal tune parameter. All magnetic parameters in the SPS are defined normalised to $p/q$ in the control room tools, where $ q = 1$ for protons. The $p$-parameter links the defined normalised magnetic strength with the required magnetic field and current (applying the calibration function). 

\begin{figure}[htbp]
  \centering
  \includegraphics*[width=240pt]{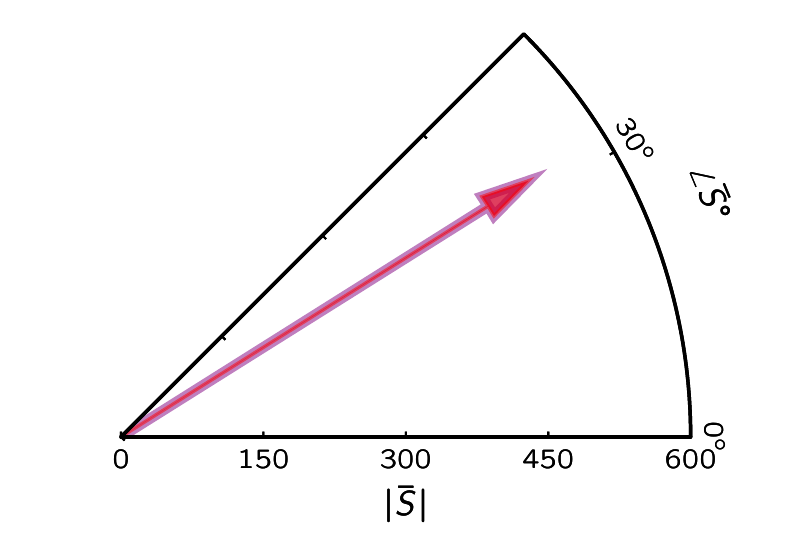}
  \caption{Schematic view of the resonance driving term evolution in time (violet is $t=0$ and red is $t = \SI{4.8}{s}$ for COSE. The two arrows are overlapping.}
  \label{fig:driving_term_cose}
\end{figure}

\begin{figure}[htbp]
    \begin{subfigure}{0.45\linewidth}
        \centering
        \includegraphics[width=\textwidth]{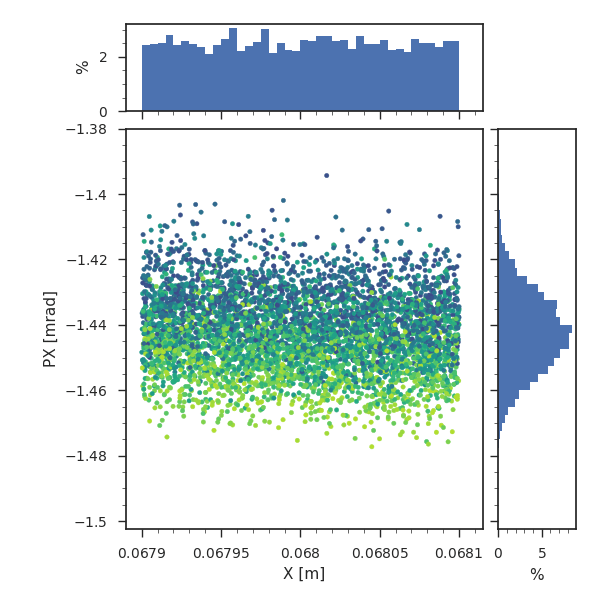}
        \caption{Q-sweep ($\sigma_{x'} = \SI{12}{\micro\radian}$).}
    \end{subfigure}
    \begin{subfigure}{0.45\linewidth}
        \centering
        \includegraphics[width=\textwidth]{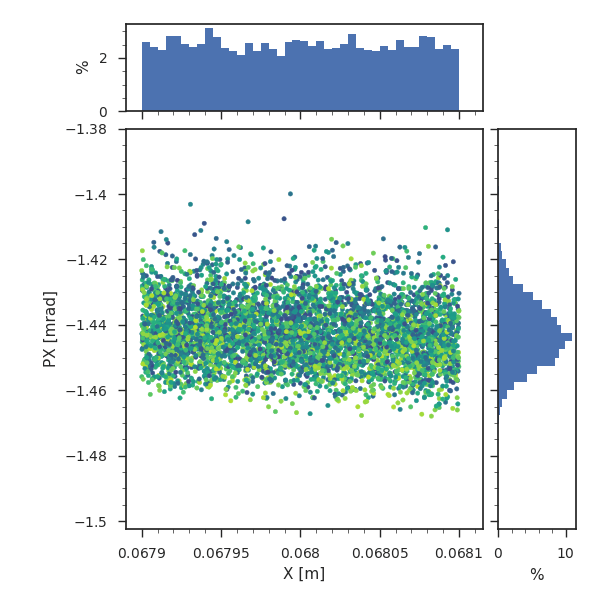}
        \caption{COSE $\sigma_{x'} = \SI{9.5}{\micro\radian}$.}
    \end{subfigure}
    \caption{Simulated angular spread of the beam presented to the ZS wires across a \SI{200}{\micro\meter} thickness.}
    \label{fig:cose_q_sweep_angular spread}
\end{figure}

To drive chromatic slow extraction either the machine tune is varied to select the resonant particles according to their tune (Q-sweep) or the machine tune is adjusted to be the resonant one for the particle with the lowest momentum and then the $p$-function is increased accordingly to extract all other momenta across the beam's momentum distribution. For the latter, the machine tune and the entire machine optics stay constant for the particles on resonance. This is the concept of Constant Optics Slow Extraction (COSE). Fig.~\ref{fig:driving_term_cose} shows the resonant driving term with COSE. It does not change anymore with time as the optics is frozen during the extraction process. Locking the optics during the spill results in an overall reduction of the angular spread of the separatrix at the ZS wires of about 20\% in simulation, from 12 to \SI{9.5}{\micro\radian} as shown in Fig.~\ref{fig:cose_q_sweep_angular spread}. This should also result in a loss reduction at the ZS.

Several low intensity MDs were dedicated to detailed studies of COSE and  the preparation of an operational procedure for its deployment during the 2018 run. During these studies no clear loss reduction was observed when compared to the Q-sweep method, but due to its operational simplicity, elegance and as a prerequisite for other loss reduction schemes, it was deployed on the operational cycle for beam to the North Area targets in September 2018. A comparison between 10 days of operation with Q-sweep and COSE in terms of extraction losses was carried out. A slight reduction of losses of approximately 5\% was observed when comparing periods with similar POT delivered and duty cycle. Also, no increase of overall losses around the machine was measured. More systematic studies will be required to conclude if the apparent loss reduction was as a direct consequence of the implementation of COSE. For example, changes in the relative alignment between the ZS could account for loss differences before and after the implementation of COSE after the Technical Stop; the ZS girder was not aligned before the switch and girder alignment was performed right after the deployment of COSE. 

\subsection{Passive diffuser (wire array)}

A passive diffuser, or pre-scatterer, in a suitable configuration has been predicted to reduce beam loss on the ES significantly, with the actual gain factor depending on the parameters and details of the extraction process and hardware. In this section, the optimisation of diffuser configurations is investigated for the SPS, and the sensitivity to the available parameters explored via simulation results. The design, construction and installation of a prototype diffuser are described, together with the experimental results obtained from its use in the SPS. The expected performance gain for an optimised design is discussed and quantified by simulation, in view of the obtained experimental results.

\subsubsection{Concept and theoretical potential} 

A diffuser, as initially postulated by Durand \cite{diffuser}, is a promising approach to reduce the performance-limiting beam loss at the ES in the SPS \cite{ref:SPS_diffuser}. The scatterer generates an angular spread in the particle distribution, which reduces the transverse density at the septum wires and can result in an overall beam loss reduction. This technique has been used at CERN in the PS and is under study for SHiP \cite{ref:SPS_diffuser,ref:PS_diffuser}, and also for the Mu2e beam from the Fermilab debuncher \cite{ref:FNAL_Mu2e_diffuser, ref:IPAC18_diffuser}.

The distance of the diffuser from the septum entrance determines the extent to which the angular scattering is translated into a positional spread at the ES. Small distances corresponding to a few degrees only in betatron phase have been tested, in a local configuration with the diffuser in the extraction bump. Much larger phase advances may be advantageous in a non-local configuration, with the added complexity of needing a second bump system at the diffuser's location elsewhere in the SPS.

\begin{figure}[htbp]
\centering\includegraphics[width=0.49\linewidth]{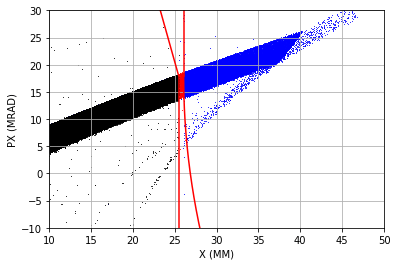}
\centering\includegraphics[width=0.49\linewidth]{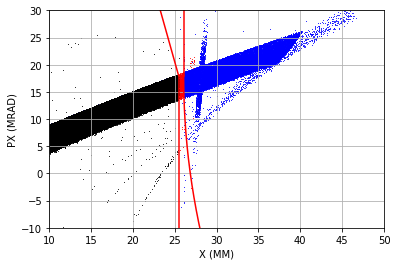}
\caption{Separatrix (in normalised phase space) with lost (red) and extracted particles (blue) at ZS without (left) and with (right) diffuser placed at a phase advance 4 degrees upstream. The diffuser here is not at the optimum alignment to shadow the ZS. The coordinates of particles lost at the diffuser are also plotted.}
\label{fig:diffuser_01}
\end{figure}

The diffuser works because the main loss source is particles traversing the ES with small impact angle. A small scattering angle upstream of the ES produces a spread in the particle positions at the ES, illustrated in Fig.~\ref{fig:diffuser_01}. If this spread is large enough, an overall reduction in beam loss is produced, provided that the additional losses induced by the diffuser itself remain small.

The required RMS Multiple Coulomb (MC) scattering angle $\theta_{MC}$ can be estimated by assuming that the diffuser has negligible length and that the ES losses are proportional to the ES hits.  The position spread at the septum is:

\begin{equation}
X_{ES} \approx \beta_{x} \sqrt{(\mu_{x}^{2}\theta_{MC}^{2} + \text{w}_{s}^{2} / 12)}
\label{eq:diffuser_X_ES}
\end{equation}

for phase advance $\mu_{x}$, diffuser width w$_{s}$ and beta function at ES and diffuser $\beta_{x}$.

With the 4 degrees of phase advance available for a realistic diffuser location in the SPS, a factor 2 loss reduction is possible for a scattering angle of $\approx \SI{30}{\micro\radian}$. Clearly, compared to a crystal, the incoherent diffuser suffers because the peak density of the scattered particle distribution is always at zero scattering angle and aligned with the ES, so that a large scattering angle is needed to produce a significant loss reduction factor. The material length of the diffuser needed to produce the given scattering angle is crucial for the overall loss reduction factor, which depends on the number of nuclear interaction lengths of material. 

For the diffuser, the ratio of radiation length $X_{0}$ to the nuclear interaction length $\lambda_{I}$ should be small, since large $\lambda_{I}$ minimises loss through nuclear scattering, while short $X_{0}$ maximises the MC scattering angle. 

A comparison of considered materials and lengths needed to achieve \SI{30}{\micro\meter} RMS scattering angle with a momentum of \SI{400}{\giga\eVperc} is shown Table \ref{tab:1}. The total loss includes all protons scattered inelastically, and those scattered elastically by more than 0.5 mrad.

Somewhat counter-intuitively, denser materials are actually significantly better for the SPS. The use of a dense diffuser like W$_{75}$Re$_{25}$ (widely used for ES wires) can provide over a factor 10 gain in the loss per impacting proton at the diffuser itself (for the length needed to generate a specific MC scattering angle), when compared to materials with a lower atomic number such as carbon. 

\begin{table}[htbp]
\centering
\caption{Diffuser Length and Loss Fraction for $\theta_{\text{MC}} = \SI{30}{\micro\meter}$.}
\begin{tabular}{lrrrrrc}
\hline
\textbf{Parameter} & \textbf{$^{9}_{4}$Be} & \textbf{$^{12}_{6}$C} & \textbf{$^{28}_{14}$Si} & \textbf{$^{96}_{42}$Mo} & \textbf{$^{181}_{73}$Ta} & \textbf{$^{184}_{74.3}($W$_{75}$Re$_{25}$) alloy} \\
\midrule
$\rho$~[g/cm$^{3}]$            & 1.8 & 2.0     & 2.3  & 10.2 & 16.7 & 19.7 \\ 
$\lambda_{n}$~total [cm]       & 29.9 & 29.6   & 30.2 & 9.1 & 6.6 & 5.6 \\ 
$\lambda_{i}$~inelastic [cm]   & 42.1 & 42.9   & 46.5 & 15.3 & 11.5 & 9.8 \\ 
$X_{0}$~[cm]       		     & 35.3 & 21.4   & 9.4 & 0.96 & 0.41 & 0.35 \\ 
Length [cm]		             & 26   & 16     & 7.0 & 0.70 & 0.32 & 0.26 \\ 
$\theta_{e}$ [\si{\micro\radian}]        & 237 & 215    & 162 & 108 & 87 & 87 \\ 
Inelastic loss [\%]          & 46 & 31 & 14  & 4.5 & 2.5  & 2.6 \\ 
Total loss [\%]                & 56  & 40 & 19 & 6.4 & 3.6 & 3.7 \\ 
\hline
\end{tabular}
\label{tab:1}
\end{table}

\subsubsection{Prototype diffuser mechanical design}

For the SPS prototype diffuser demonstration tests, the local shadowing option with the diffuser installed close to the ES was selected as being the most straightforward to integrate and to test with beam. A suitable location in the SPS lattice was identified and the diffuser specification developed with the help of particle tracking simulations.
\begin{figure}[htbp]
\centering\includegraphics[width=0.95\linewidth]{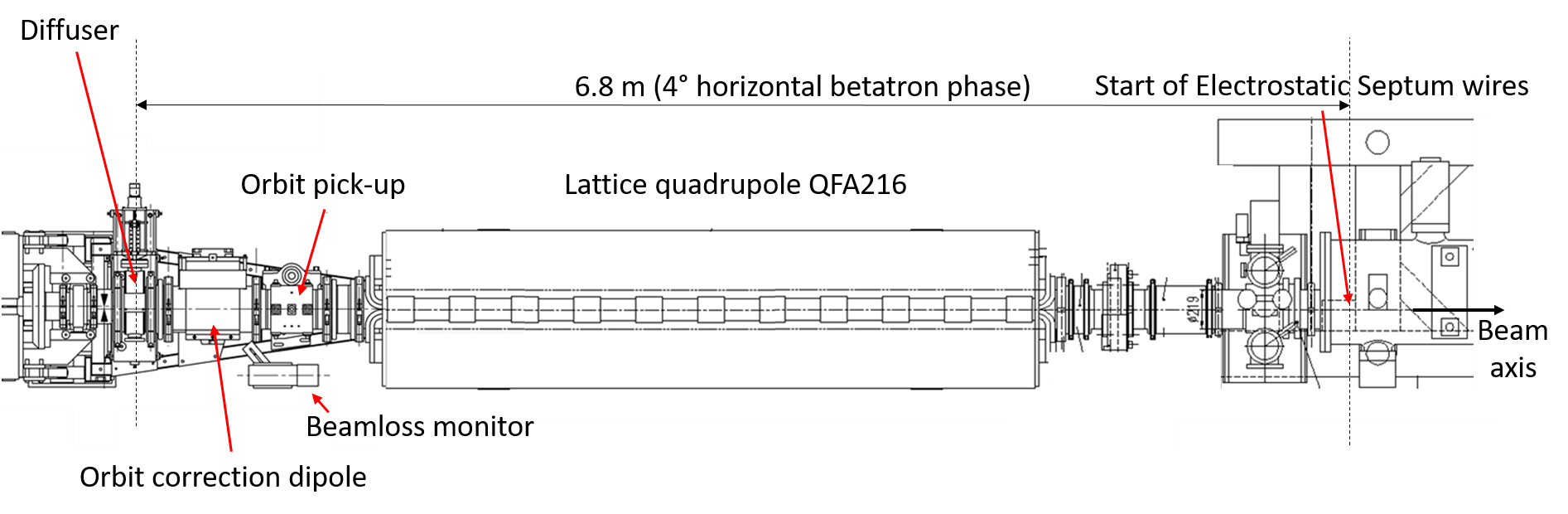}
\caption{Integration of prototype diffuser in SPS, upstream the electrostatic septum and focussing lattice quadrupole (QFA.21610).}
\label{fig:diffuser_02}
\end{figure}

\begin{figure}[htbp]
\centering\includegraphics[width=0.7\linewidth]{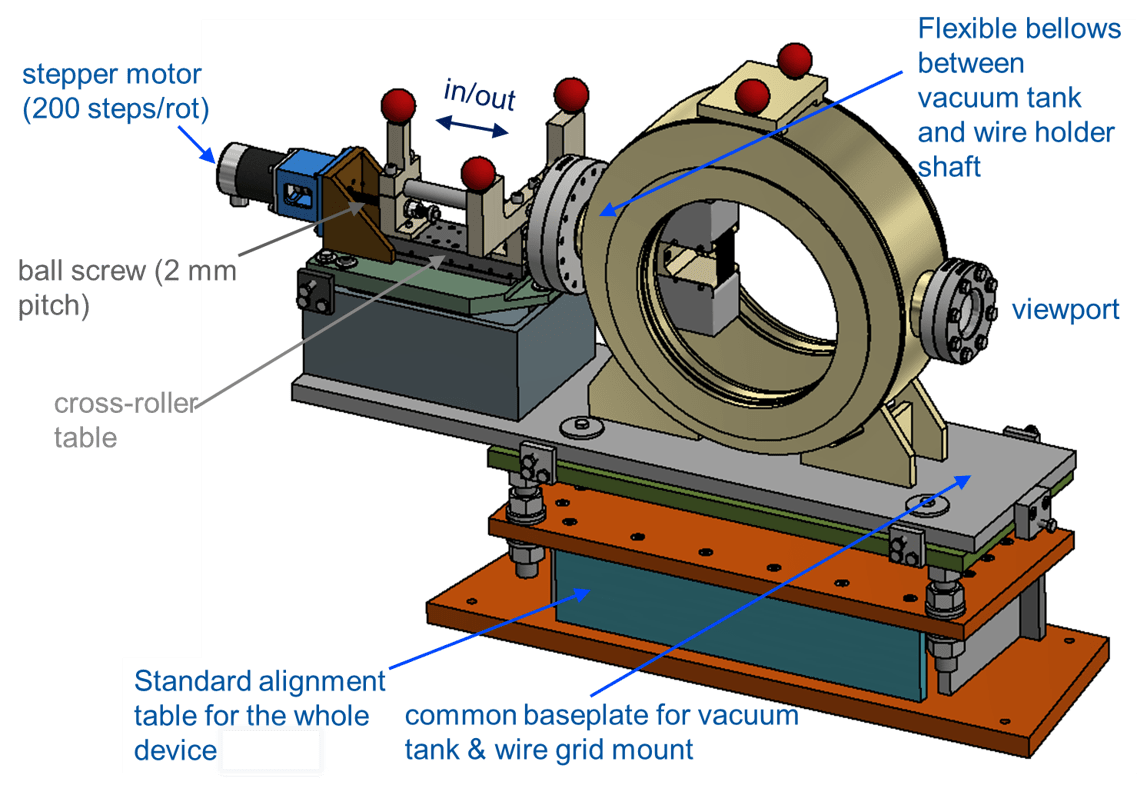}
\caption{Prototype diffuser installed in SPS. The device has 20 Ta wires of $\diameter$\SI{200}{\micro\meter}, aligned in two halves with a \SI{60}{\micro\meter} offset, and spaced by \SI{1.5}{mm} for a total length of \SI{30}{mm}.}
\label{fig:diffuser_03}
\end{figure}

A prototype diffuser was built in collaboration with the Wigner Institute, Hungary. A single degree of freedom was used, for translation of the wires in and out of the beam, using a precision screw driven by a stepping motor permitting a step size of \SI{10}{\micro\meter}. Tantalum wire of $\diameter$\SI{0.2}{mm} was chosen for the diffuser material; the $\diameter$\SI{0.2}{mm} wire is much more malleable than similar thickness WRe and could be mounted with good straightness. Its performance, as specified in Table~\ref{tab:1}, is almost identical to WRe in terms of the loss produced for a given scattering angle. The mounted wire array had a \SI{0.26}{mm} effective thickness, which was obtained by offsetting half of the wires. The mechanical design of the device is shown in Fig.~\ref{fig:diffuser_03}. 
 
\subsubsection{Summary of main results of machine development and operational tests}

\begin{figure}[htbp]
\centering\includegraphics[width=0.7\linewidth]{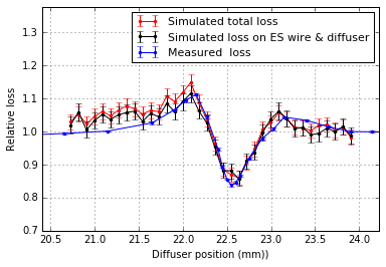}
\caption{Measured diffuser response (total loss in extraction region) compared to simulation using \SI{0.6}{mm} ZS thickness.}
\label{fig:diffuser_04}
\end{figure}

A series of tests were made in 2018 with simple scans of the diffuser position in front of the ZS wires, initially with a low intensity beam of $2\times 10^{12}$ protons per spill. The results were highly reproducible with a reduction in the overall beam loss on the ZS (summed over all extraction BLMs) of 15\%, compared to extraction with no diffuser. A typical loss response profile is shown in Fig.~\ref{fig:diffuser_04}, compared to simulation results. There is a very good agreement with the simulation in terms of the simulated loss reduction and the side features of the profile, for a ZS width of \SI{0.6}{mm}, which is significantly larger than expected (the wires of the first ZS are only \SI{0.06}{mm} in diameter). The specified positioning accuracy of the diffuser's wire array of $\pm \SI{50}{\micro\meter}$ was confirmed to be necessary, and attainable.

\begin{figure}[htbp]
\centering\includegraphics[width=0.7\linewidth]{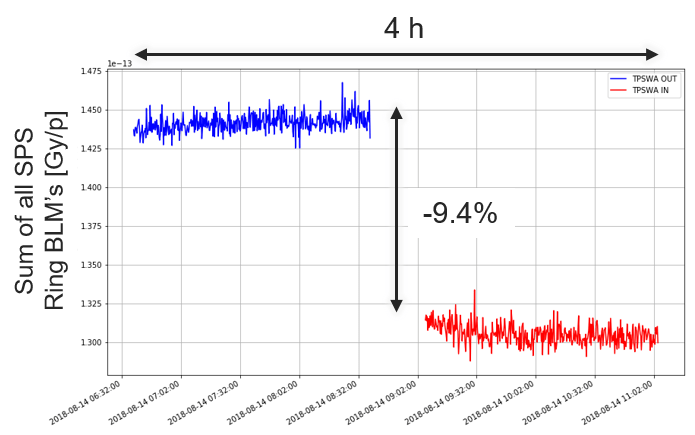}
\caption{Total SPS beam loss (integral on BLMs around the full ring) for a 4 hour period spanning the insertion of the prototype diffuser into the operational beam.}
\label{fig:diffuser_05}
\end{figure}

The diffuser was also deployed for a 23~hour period on the operational beam to the NA, with an intensity of $3\times 10^{13}$ protons per spill. After the beam was inhibited, it took 3~minutes to move the diffuser to the operational position and a further 22 minutes to optimise the position with respect to the beam. No issues were seen during this longer test period, although the extraction loss reduction factor was slightly lower, at 10\%. A total of $1.2\times 10^{17}$ protons were extracted with the diffuser in beam. A period of 2 hours each side of the insertion where the SPS was in stable operation with no super-cycle changes or LHC filling was analysed, see Fig.~\ref{fig:diffuser_05}. There was a short initial period of outgassing where the vacuum pressure increased from $1\times 10^{-8}$ to $2\times 10^{-7}$ mbar, but it recovered within about 15 minutes.  No re-alignment of the ZS girder was possible and further optimisation to the 15\% gain may have been possible.

\subsubsection{FLUKA simulations of the diffuser}

\begin{figure}[ht!]
\centering\includegraphics[width=0.9\linewidth]{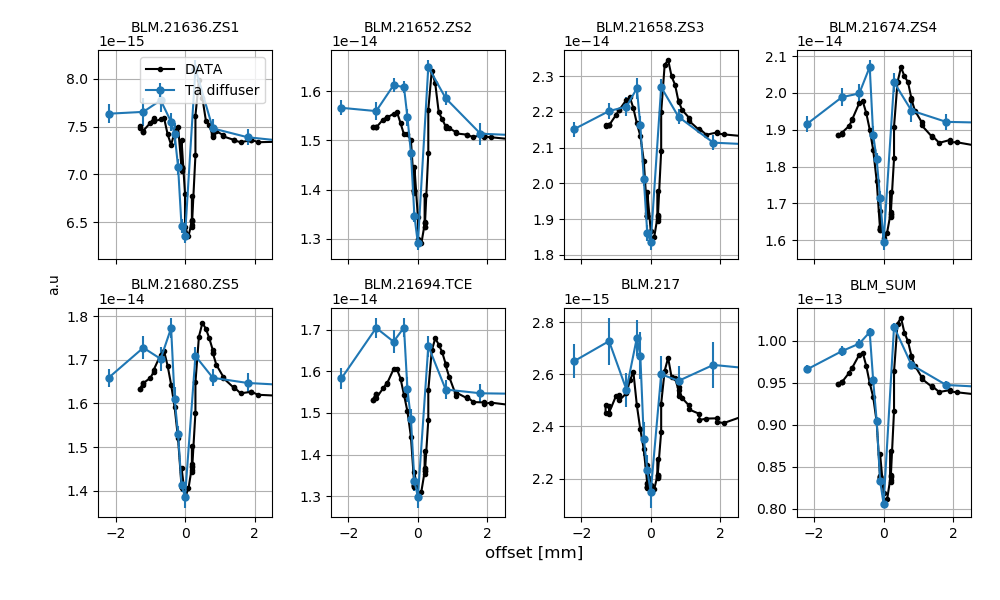}
\caption{ Relative comparison of the signal of the BLMs along the ZS, TCE and QF.217 measured as a function of the diffuser position with the result of the FLUKA simulation.  The bottom right plot displays the sum of the signal of the five BLMs along the ZS and the TCE. The simulation points are normalised to the minimum value of the experimental data.}
\label{fig:fluka_blm_vs_diffuser_scan}
\end{figure}

The response of the BLMs to the diffuser linear scan was measured during MD time and compared with the FLUKA model, discussed in more detail below. Despite the difference in the absolute comparison, as was indicated in Fig.~\ref{fig:fluka_blm_pattern}, the functional pattern is fairly well reproduced and, in particular, the height and the width of the loss dip are in a reasonable agreement with the experimental observations. Fig.~\ref{fig:fluka_blm_vs_diffuser_scan} compares the measured loss response with the FLUKA simulation and the sum of the six BLMs along the ZS and TCE, where the data is normalised to the same minimum value found empirically.

\begin{figure}[ht!]
\centering\includegraphics[width=0.7\linewidth]{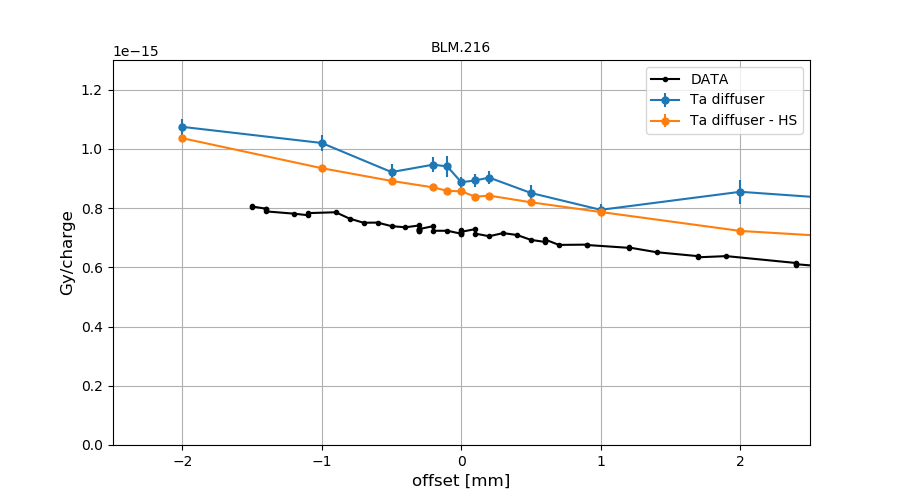}
\caption{ Absolute comparison of the signal of the BLM.216 measured as a function of the diffuser position with the FLUKA simulation. 
(black) Measured data.  (blue) Monte Carlo points from the same simulation of Fig.~\ref{fig:fluka_blm_vs_ZS_thickness}.
(orange) Monte Carlo points from a high statistics run with a geometry model including only the diffuser and the BLM.216.
 All curves are normalized to the extracted current.
}
\label{fig:fluka_blm216_vs_diffuser_scan}
\end{figure}

The response of the BLM.216, which is located near the diffuser, is sensitive to the linear density of the extracted beam.
For this BLM, the FLUKA estimation of the signal per extracted proton is $\sim15\%$ systematically higher than the measured signal.
Fig.~\ref{fig:fluka_blm216_vs_diffuser_scan} shows the experimental data and the results of the simulation.  This figure shows also a curve obtained with a high statistics run (about 100 times the number of the events) but with the model geometry including only the diffuser and the BLM.216 in order to speed up the computing time and  investigate the statistical convergence of the observable. The systematic difference for the signal of the BLM.216 can stem from an incorrect calibration factor. In the simulation, the conversion from the energy deposited in the active gas volume of the BLM to dose is computed using the mass of the gas and the the same factor is applied to all BLMs. Therefore, the factor of 3 in the underestimation of the losses at the ZS by FLUKA simulations is not compatible with a systematic error, but it is likely originated by effects that increase the effective thickness of the septum that are not taken into account in the simulation model yet.

\subsubsection{Issues and challenges observed}
The increased beam loss at the location of the diffuser is a concern since this zone is a ``low-dose'' area used for access during interventions on the much more radioactive ZS. The dose at this location with the diffuser in beam was measured on the BLM on the quadrupole 216 was indeed observed to increase by a factor of 7.5. Although the expected remnant dose at the diffuser is expected to be more than an order of magnitude less than the ZS, this increase in this previously low radiation area needs to be taken into consideration in the overall optimisation.

\subsubsection{Outlook for design of operational device}

The key parameter for an operational device is the effective ZS width, as seen from Fig.~\ref{fig:diffuser_02}, and estimated at approximately \SI{0.6}{mm}, while from the results with the crystal a width of \SI{0.5}{mm} was derived. The diffuser width needs to be matched precisely to this value, to obtain the optimum loss reduction; too narrow and unscattered particles will directly impact the ZS, too wide and the losses from nuclear scattering in the diffuser are unnecessarily large. A Ta diffuser with a width of \SI{0.6}{mm} would allow approximately 30\% loss reduction in regular operation, depending on the separatrix angular width that can be achieved. This reduction could be improved for a thinner ZS, to 50\% - so for the SPS the diffuser is significantly less effective that the crystal, albeit with expected simpler operation.

\subsubsection{Conclusions}

The Wigner prototype passive diffuser (TPSWA) in LSS2 was tested successfully in dedicated MD and for almost 24 hours with operational beam. A total of $1.2\times 10^{17}$ PoT were extracted to the NA during the test with just 3 minutes downtime caused by the deployment. A 9.4\% reduction of slow extraction beam loss was recorded on the SPS BLM system, with no impact on TT20 losses or at experiments. For a well-aligned ZS the loss reduction of 15\% is consistent with a ZS width of \SI{0.6}{mm}. A redesigned diffuser for this ZS effective width could reduce the extraction losses by approximately 30\%.

\subsection{Active diffuser (thin bent crystal)}\label{sec:crystal}

\subsubsection{A novel extraction scheme}


The application of bent silicon crystals has been pursued at CERN for various applications involving the deflection of high-energy particles in the SPS and LHC~\cite{scandale1, scandale2, scandale3}. The beam loss on a thin electrostatic extraction septum can be reduced by aligning a thin, bent crystal to both the beam and septum to deplete the beam density impinging the blade of the septum in a scheme we term as ``shadowing''. The particle that would have otherwise hit the septum blade are deflected into the extraction channel by the crystal. The concept and results briefly discussed in the section can be found published in more detail  in~\cite{PRSTAB_velotti}.

\subsubsection{Concept and theoretical potential}

\begin{figure}[htbp]
\centering
\includegraphics[width=0.7\linewidth]{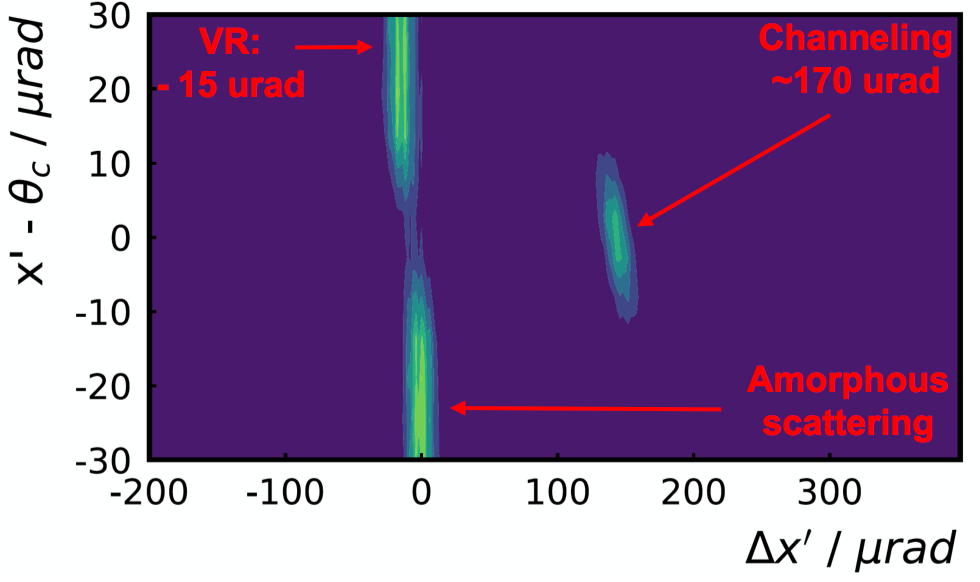}
\caption{Probability density function describing particle interaction with a bent crystal as implemented in \textit{pycollimate} from data obtained by the UA9 collaboration taken in H8.}
\label{fig:2d_pdf}
\end{figure}

Due to the highly non-linear regime of the beam dynamics during the third-integer slow extraction process, particle tracking simulations were carried out to assess the performance reach of this novel extraction concept. In order to simulate the effect of the crystal on the beam, the thin-tracking module of MADX was used and coupled with \textit{pycollimate}~\cite{velotti_thesis} to simulated the interation with the crystal and the ES. The different processes that generate deflections of charged particles interacting with the crystal lattice can be divided into different regimes. The probability of a charged particle undergoing each process occurs as a function of the incidence angle of the particle with respect to the crystal. To design and specify the extraction scheme, measurements performed in 2014 on a UA9 silicon bent crystal (SFT45) \cite{rossi} were used to map the relation between input and output angles of particles impinging the crystal. A 2D probability density function (Fig.~\ref{fig:2d_pdf}) was implemented in \textit{pycollimate} to simulate the effect of the crystal throughout the slow extraction process. The results refer to single-pass effects of the crystal measured on an experimental transfer line (H8) in the NA. In MADX, particles that interact with the crystal are given a thin kick assigned randomly from the probability density function. For an incoming angular range of $\pm\SI{10}{\micro\radian}$ the single-pass channelling efficiency in the simulation is approximately 55\%.

To avoid restricting the aperture of the synchrotron at injection, a fast-actuating crystal or a series of high-energy magnetic bumpers are needed to move the beam close to the crystal. The latter option is preferable for mechanical reasons. There are two ways to shadow the ES wires: (i) locally, by installing a crystal immediately upstream of the ZS inside the extraction bump or, (ii) non-locally, by installing it at a favourable optics location in the ring that is equipped with another set of bumpers. Both options have been studied in detail in the SPS and reported in~\cite{shadowing}. The first option was chosen for the prototyping of the concept because of its apparent advantages for ease of operation and optimisation, although the performance reach of a non-local system is expected to be better than the local system due to the flexibility in choosing the phase advance.

\begin{figure}[htbp]
    \begin{subfigure}{0.48\linewidth}
        \centering
        \includegraphics[width=\textwidth]{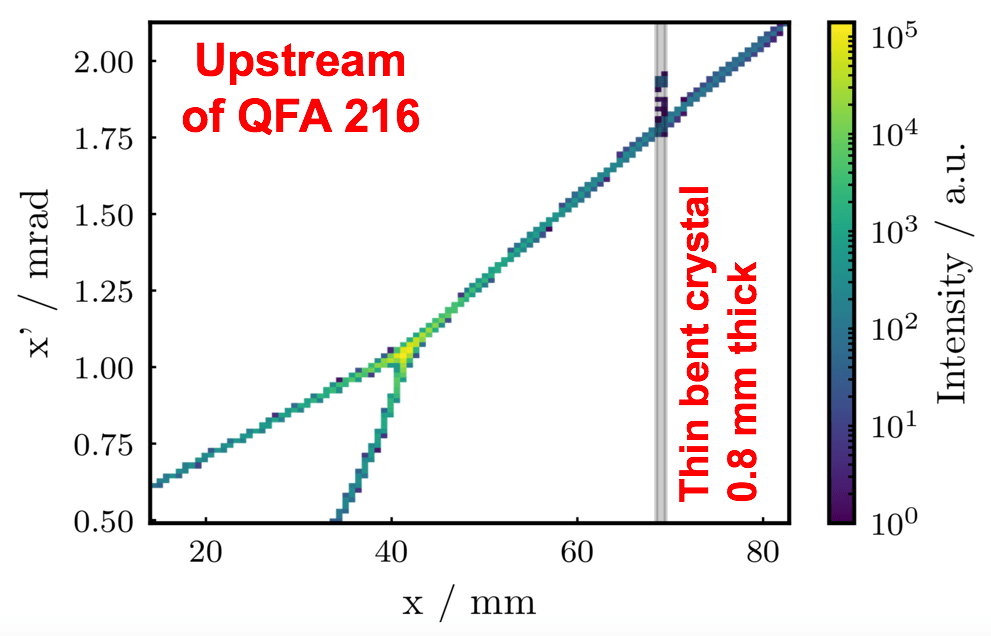}
        \caption{Phase space presentation at the crystal.}
    \end{subfigure}~
    \begin{subfigure}{0.48\linewidth}
        \centering
        \includegraphics[width=\textwidth]{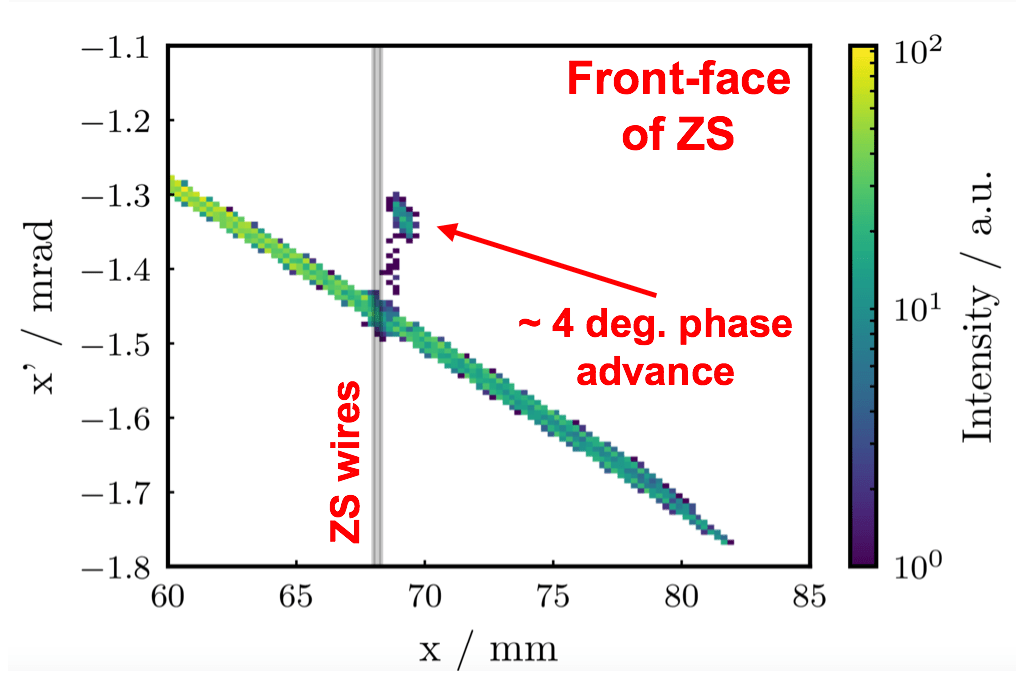}
        \caption{Phase space presentation at the ZS.}
    \end{subfigure}
    \caption{Phase space presentation of the resonantly extracted beam at the crystal and close to the wires of the ZS.}
    \label{fig:crsytal_phase_space}
\end{figure}

\begin{figure}[htbp]
  \centering
  \includegraphics*[width=0.7\textwidth]{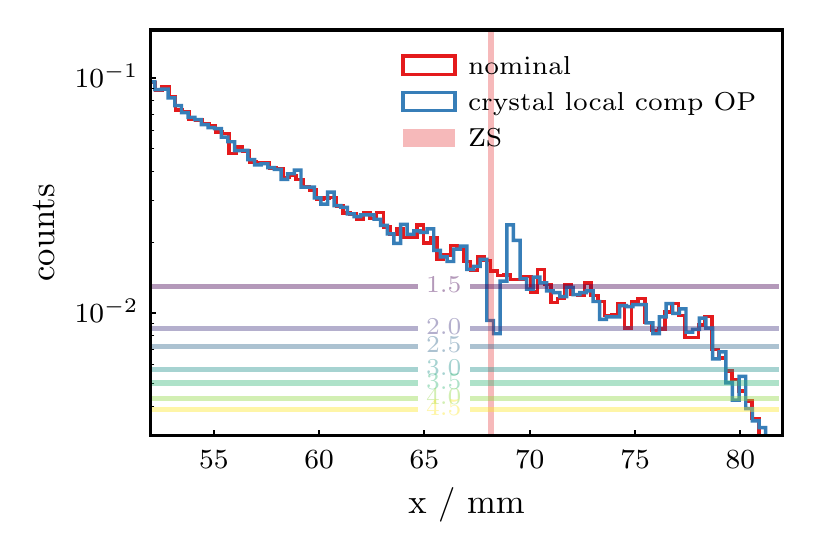}
  \caption{Histogram of the horizontal particle distribution at the ZS. In red is shown the case for a nominal SPS FT extraction and in blue the density for ZS local shadowing.}
  \label{fig:zs_density}
\end{figure}

The results of the MADX-\textit{pycollimate} tracking simulations are shown Fig.~\ref{fig:crsytal_phase_space}, where the beam presentation to the upstream crystal and the downstream electrostatic septum is shown in phase space. The crystal thickness was specified at \SI{0.6}{mm} with a deflection angle of \SI{170}{\micro\radian}~\cite{edms_crsytal_FS}. A clear region of intensity depletion can be observed at the location of the ZS wires with a consequent increase in the number of extracted particles. The simulated loss reduction at the ES is a factor 2 for the local case, as shown by the transverse beam density in Fig.~\ref{fig:zs_density}.

In a similar way as a passive diffuser, the relative thickness of the crystal and ES plays an important role on the overall loss reduction potential of the concept. In order to maximise the efficiency of the scheme and improve the loss reduction, the angular spread of the beam at the crystal must be minimised such that it fits well inside the channelling acceptance. In the original concept it was thought that this could be achieved with a dynamic extraction bump, which is varied together with the horizontal tune to compensate the beam movement throughout the spill. As discussed previously, the implementation and setup of the dynamic bump is operationally complex and the same problem was instead solved by employing the COSE extraction technique.

The prototype crystal was installed in LSS2 of the SPS, just upstream the ZS (\SI{0.6}{m} upstream the QFA.21610). In this case, a crystal with a large channelling angle is required because the phase-advance between the crystal and the ZS is only a few degrees. Also, the crystal should be installed such that the channelled particles are deflected towards the outside of the ring. The expected loss reduction using this extraction configuration is about a factor 2. When the crystal is aligned in volume reflection a density depleted region is also formed with particles being kicked back into the machine to circulate another three turns, before being extracted as part of the tail at large amplitude on the extracted beam. The expected loss reduction factor is 3.4 for the non-local case with a crystal located in LSS4 of the SPS.

\subsubsection{Prototype mechanical design}

\begin{figure}[htbp]
\centering\includegraphics[width=0.6\linewidth]{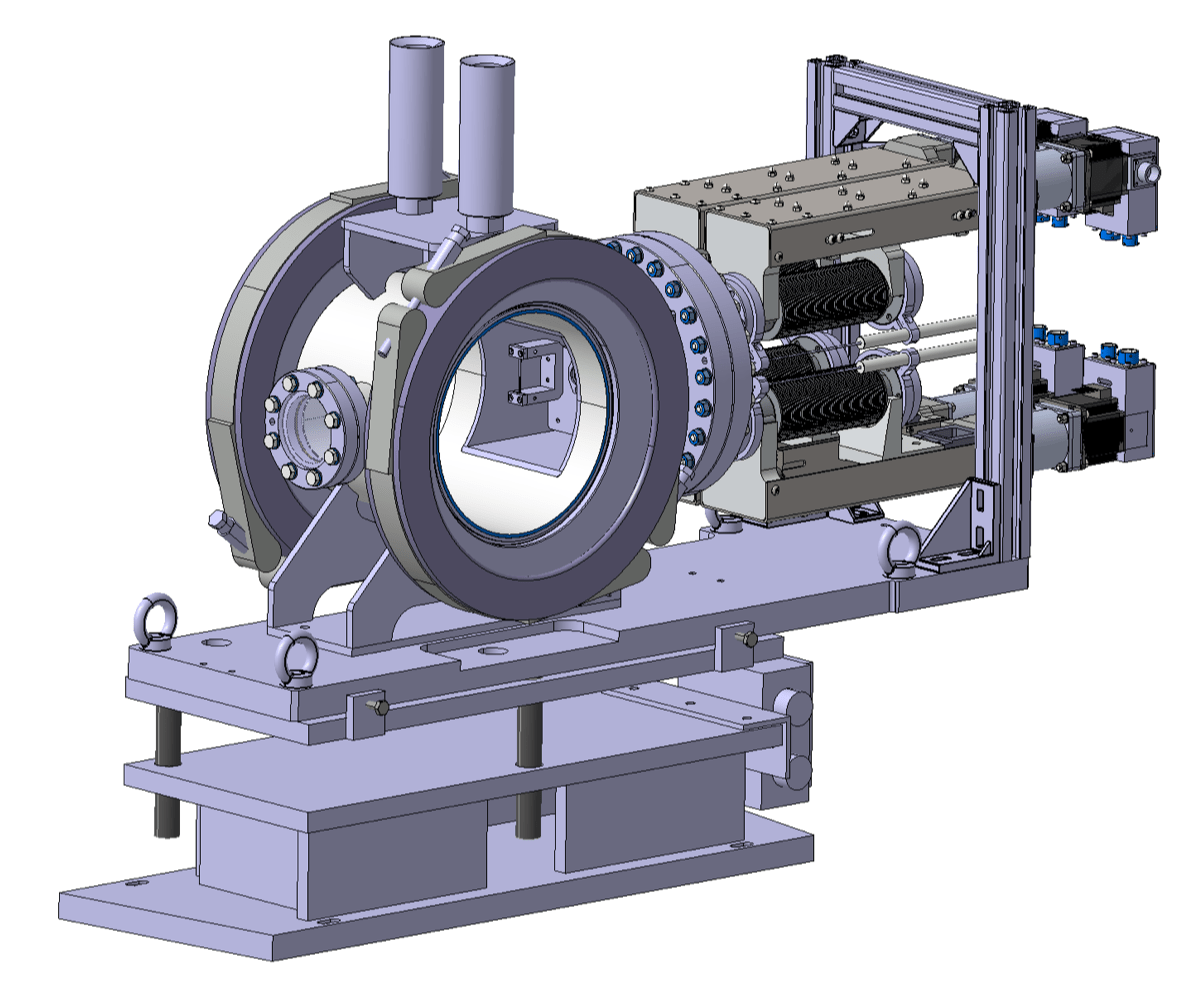}
\caption{Prototype of the compact vacuum tank TECS.21602 installed in the SPS. It contains the goniometer and the crystal.
}
\label{fig:TECS_model}
\end{figure}

The UA9 prototype crystal diffuser tank (TECS.21602) is installed in the same location as the passive diffuser, see Fig.~\ref{fig:diffuser_02}, which was swapped out during the second Injectors Technical Stop in September 2018. The crystal is \SI{2.5}{mm} long, \SI{0.8}{mm} wide with a \SI{175}{\micro\radian} bending angle. It is mounted on a holder with a large vertical clearance and a 35.8 mm horizontal clearance to allow the resonantly extracted beam to jump the crystal. The holder is mounted inside a tank equipped with a goniometer, which is needed to allow the correct angular alignment of the crystal with respect to the incoming beam. Fig.~\ref{fig:TECS_model} shows the prototype of the crystal installed in the SPS.

The alignment of the crystal is accomplished by two linear stage motors: the first one moves transversely the crystal and the second one changes the orientation of the crystal with respect to the beam. The lever arm between the two motors is \SI{10}{cm}. The measurement of the position of each stage motor is performed by means of a Linear Variable Differential Transformer (LVDT) sensor.
The motor gears experience backlash when the direction of the movement is reverted. The backlash of the two motors was measured in vacuum in SPS and it is \SI{62}{\micro\m} and \SI{37}{\micro\m}. In terms of the crystal angle, the backlash affects the absolute value of crystal orientation up to a few hundreds of \si{\micro\radian}. 

For the prototype goniometer, this can only be partially corrected using the LVDT measurements as they were affected by a variation of the pulsed currents in the nearby main magnets. The design of a future operational crystal device would implement appropriate measures to minimise these effects in view of the lessons learnt.

\subsubsection{Issues and challenges observed}

The particles channelled by the  aligned crystal and entering the transfer line as a coherent beamlet caused losses measured at a specific location at the beginning of TT20. With the new BDF optics developed for TT20, it was possible to transport the channelled beamlet to the targets with no additional losses recorded along the line by applying local trajectory bumps to avoid the aperture restrictions. This is also expected to be the case for the transport along the new transfer line to the BDF target. It should be noted that the trajectory excursions predicted for the channelled beamlet in the TT20 operational optics for the NA are more severe and require further investigation to understand if the beamlet can also pass the splitters and be transferred to the targets, or if it will need collimating in the transfer line.

\subsubsection{Summary of main results of machine development and operational tests}

The majority of the machine tests took place in parallel to the dedicated prototype BDF target test, where a \SI{1}{s} spill was used to slowly extract \SI{400}{\giga\eVperc} protons towards the T6 target in the NA. The advantage of such a parasitic test was the high duty-cycle achieved and the large number of position and angular scans that could be made with the crystal actuating to its demanded position and angle in-between cycles, before staying fixed and not moving during each extraction. A significant amount of data was also taken  using the crystal with ions in parallel to operation in 2018 at different beam energies, which is not discussed here.

\begin{figure}[htbp]
\centering
    \begin{subfigure}{0.35\linewidth}
        \centering
        \includegraphics[width=\textwidth]{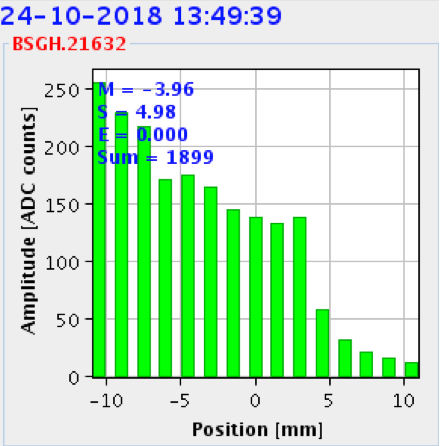}
        \caption{Crystal as amorphous scatterer.}
    \end{subfigure}~
    \begin{subfigure}{0.35\linewidth}
        \centering
        \includegraphics[width=\textwidth]{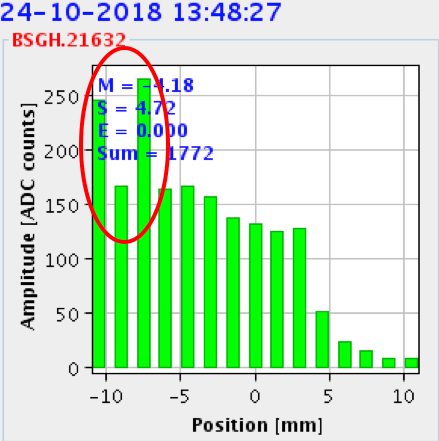}
        \caption{Crystal aligned and channelling.}
    \end{subfigure}
    \caption{Horizontal beam density measured on a wire-grid profile monitor directly upstream of the ZS, with the observed depletion of intensity at the ZS wires circled in red.}
    \label{fig:BSG216_crystal}
\end{figure}

The creation of a channelled beamlet was directly observed on screens at different locations in the transfer line and, most impressively, the expected manipulation of the horizontal beam density was clearly observed on the wire-grid profile monitor located directly upstream of the ZS, as shown in Fig~\ref{fig:BSG216_crystal}.

\begin{figure}[htbp]
\centering\includegraphics[width=0.7\textwidth]{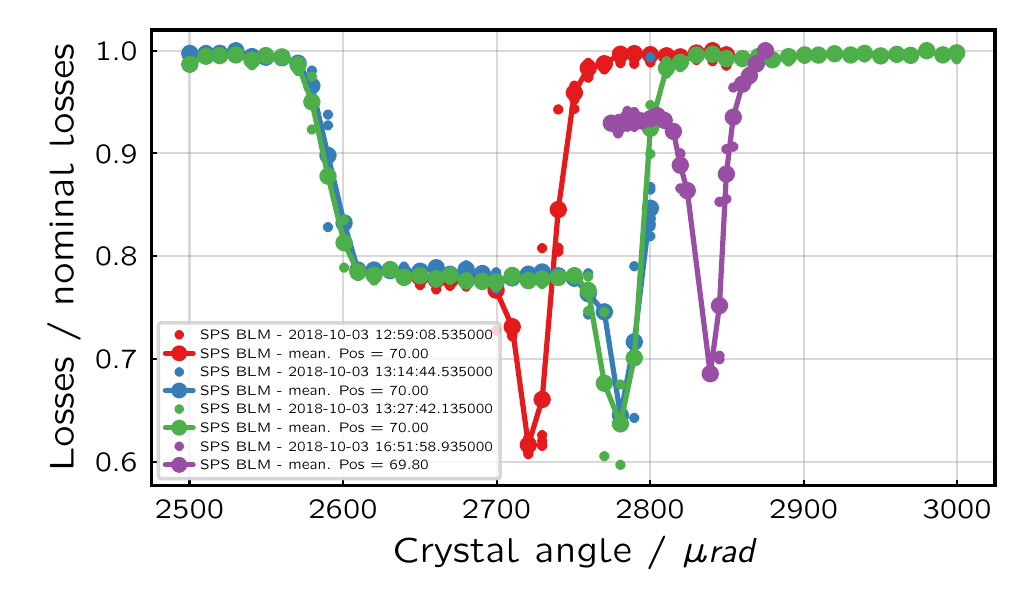}
\caption{Measured relative total loss evolution in the LSS2 extraction channel as a function of the crystal angle.}
\label{fig:shadowing_data}
\end{figure}

\begin{figure}[htbp]
\centering\includegraphics[width=0.6\textwidth]{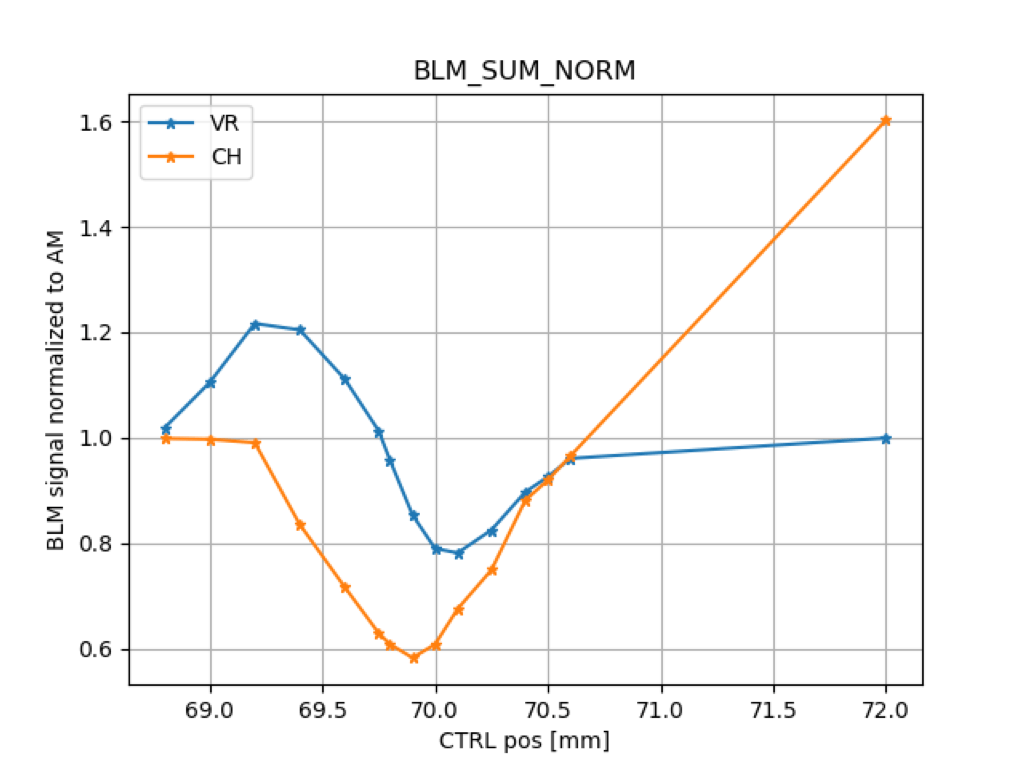}
\caption{Measured loss reduction factor when the crystal is aligned either in CH or VR as a function of the crystal position. The losses are normalised to the amorphous (AM) crystal orientation.}
\label{fig:loss_reduction_CH_VR}
\end{figure}

In order to find the optimum settings for the shadowing, i.e. to maximise the loss reduction, many angular scans were performed at different transverse positions of the crystal in front of the ZS. The results from the first measurement campaign are shown in Fig.~\ref{fig:shadowing_data}. 
For the best position and angle, a loss reduction of about 40\% was observed. This is obtained when the crystal is correctly aligned with the beam angle permitting the protons impinging the crystal to be channelled. Another interesting regime for loss reduction was observed to be volume reflection \cite{vr_crystal}. In these conditions, the particles on the separatrix are instead coherently deflected towards the inside of the machine with a small angle ($\approx$\SI{-15}{\micro\radian}) and extracted three turns later. This leads to a loss reduction of $\sim20$\% when the crystal is oriented in volume reflection. The tests are well summarised in Fig.~\ref{fig:loss_reduction_CH_VR}, where the loss reduction factor is shown through linear scans of the crystal position aligned either in channelling or volume reflection.

\begin{figure}[htbp]
\centering\includegraphics[width=0.6\textwidth]{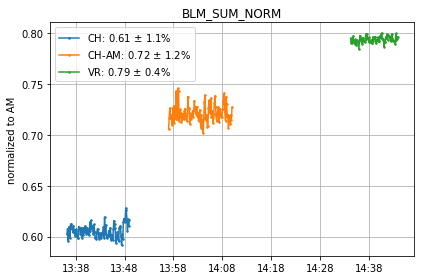}
\caption{Loss level normalised to amorphous (AM) during the stability checks. The crystal was fixed in channelling (CH), in the channelling-amorphous (CH-AM) and volume reflection (VM) regimes. The loss reduction factors and the RMS's are indicated in the legend.}
\label{fig:stability_check}
\end{figure}

One of the main initial concerns was regarding the beam stability because the channelling angular acceptance of the crystal is limited to about $\pm$\SI{10}{\micro\radian} for \SI{400}{GeV} protons. The tests demonstrated that the loss reduction could be achieved with remarkable stability. The stability was tested in channelling and volume reflection, resulting in an RMS stability of 1.1\% and 0.4\%, respectively, as shown in Fig.~\ref{fig:stability_check}. The stability was even acceptable at an angle between channelling and amorphous scattering, where the slope of the loss with angle was steepest.

\begin{figure}[htbp]
\centering\includegraphics[width=0.6\textwidth]{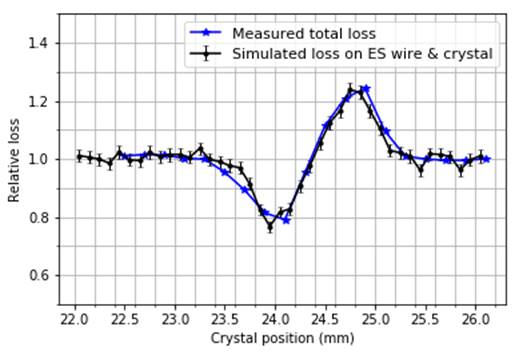}
\caption{Linear scan data in volume reflection fitted to simulation with an effective ZS thickness of \SI{0.5}{mm}.}
\label{fig:linear_scan_simulation}
\end{figure}

The angular scans performed at different transverse positions were used to map the phase space presentation of the extracted beam separatrix to the crystal. Analysis of the data is on-going to include corrections using the LVDT measurements of the motor position. The data, when compared to simulation, also provides information on the effective thickness of the ES, as shown in Fig.~\ref{fig:linear_scan_simulation}, where values of close to \SI{500}{\micro\meter} are found, consistent with the passive diffuser results.

\begin{figure}[htbp]
\centering\includegraphics[width=0.65\linewidth]{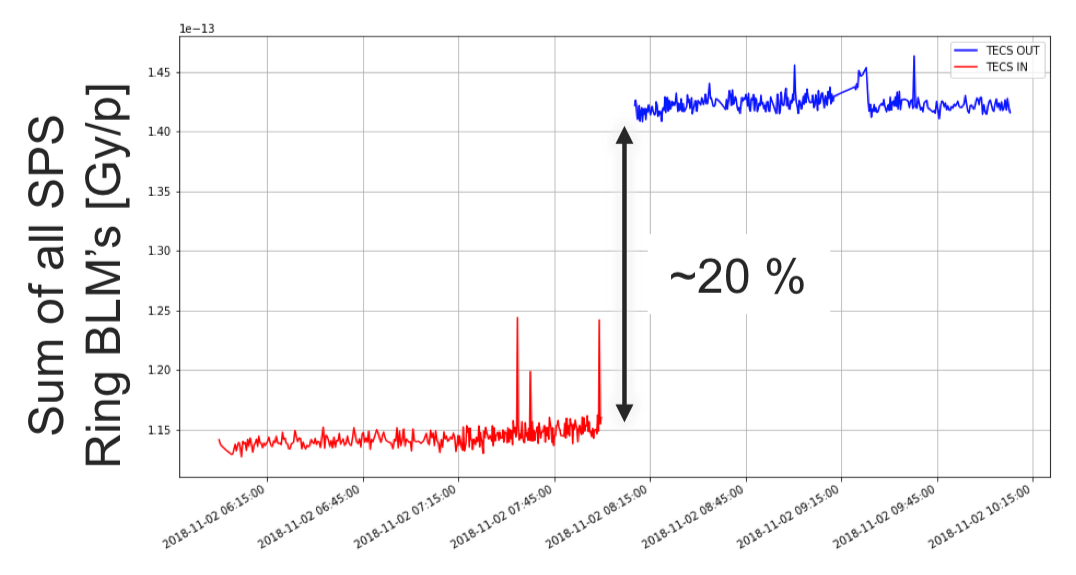}
\caption{Total SPS beam loss (integral around the full ring) for the 4 hour period spanning the retraction of the prototype crystal after 13 h in the operational beam.}
\label{fig:crystal_op_test}
\end{figure}

The crystal was aligned in volume reflection and tested for 13 hours on the operational beam to the NA, with an intensity of $2.8\times 10^{13}$ protons per spill. A total of $\sim6\times 10^{16}$ protons were extracted. The test was carried out in volume reflection to avoid causing downtime to operation from losing the channelled beamlet on an aperture restriction in TT20. The operational transfer line optics is different to that used during the MD tests and could not be tested in advance. The crystal was aligned with only 10 minutes of downtime and the performance shown in Fig.~\ref{fig:crystal_op_test} for a period of 2 hours each side of the removal of the crystal at the end of the 13 hours.
The expected loss reduction of 20\% was demonstrated with evidence  that the extracted beam intensity was increased with the crystal inserted. A negligible amount of outgassing was observed when the crystal was inserted, an order of magnitude lower than observed for the passive diffuser. The absolute prompt dose level observed upstream of the QDA.216 with the crystal inserted was a factor two lower than for the passive diffuser, with only 1\% of the total dose saved around the SPS ring being increased locally at this location.

\subsubsection{Conclusions}

The prototype thin bent crystal provided by the UA9 collaboration for ZS shadowing tests showed huge promise for significant extraction loss reduction factors being sought with up to 40\% loss reduction being demonstrated. The tests made on the operational beam showed the ease of alignment and excellent stability, even in operation at high intensity and duty cycle with a stable loss reduction of approximately 20\%. Further improvements and optimisation of the shadowing concept could yield even more significant gains and are being studied for future implementation. Important lessons were also learnt regarding the crystal hardware itself. 

\subsection{Phase space folding with octupoles}

Higher order multipoles can be added to the sextupole driven extraction in order to manipulate the phase space trajectories in such a way that a `folded' separatrix can be extracted. This can lower the particle density at the ZS wires and increase the density in the extraction aperture, leading to reduced beam loss per proton extracted. Theoretical and experimental studies at SPS have shown that the normalised loss can be reduced by over 40\% \cite{stoel_thesis,stoel_ipac18,stoel_ipac19}.

\subsubsection{Kobayashi Hamiltonian}
The Kobayashi Hamiltonian, upon which the theory of slow extraction is based, is conventionally shown with only a sextupole field. When we add octupoles as well, the Hamiltonian in non-dimensionalised coordinates $(\hat{X},\hat{P}) = K_2\cdot(X,P)$ reads
\[
 \hat{H} = \frac{\epsilon}{2}\left(\hat{X}^2+\hat{P}^2\right)+\frac{1}{4}\left(\hat{X}^3-3\hat{X}\hat{P}^2\right)+\frac{9}{32}\kappa_3\left(\hat{X}^2+\hat{P}^2\right)^2
\]
where $\epsilon=6\pi\left(Q-Q_\mathrm{res}\right)$ is a measure of the tune distance to resonance, $\kappa_3=K_3/K_2^2$ and
\[K_n = \frac{1}{n!}\frac{L}{B\rho}\left[\frac{\partial^nB_y}{\partial x^n}\right]_{x=y=0}\beta_x^{(n+1)/2}\]
are the normalised multipole strengths. Its equivalent in polar coordinates reads
\[
 \hat{H} = \frac{\epsilon}{2}\hat{A}^2+\frac{1}{4}\hat{A}^3\cos(3\hat{\theta})+\frac{9}{32}\kappa_3 \hat{A}^4
\]
showing the threefold symmetry more clearly. For zero octupole strength, the Kobayashi Hamiltonian shows a stable triangle with its size dependent on the distance to resonance and separatrix arms extending toward infinity. With the addition of octupoles, we still see a stable triangle on one side of the resonance, but with the separatrix arms coming out of a stable triangle and bending around three new stable points returning towards the stable triangle. At resonance, the stable triangle shrinks to a point and on the other side of the resonance the stable triangle exists only in a narrow range of tunes, as the corners of the stable triangle merge with the other stable points to leave a fully stable phase-space. 

It is also theoretically possible to reduce the normalised extraction losses by adding decapoles to the extraction instead of octupoles. However, simulations have shown that in the case of the SPS the required decapole strengths would be unfeasible~\cite{stoel_ipac17}.

\subsubsection{Driving term rotation}
The strong curvature of the separatrix with added octupoles means that the angle of particles on the separatrix at the position of the ZS wires changes significantly. Therefore, a knob was designed that can change the angle of the effective sextupole driving term, without changing the effective sextupole strength \cite{stoel_thesis}. This knob makes use of the extraction sextupoles that are already present in SPS, but are not normally used in operation. Since this knob uses sextupoles, but does not affect the effective sextupole strength, it is referred to as the `orthogonal sextupole knob'.

\begin{figure}[htbp]
    \centering
    \includegraphics[width=0.6\textwidth]{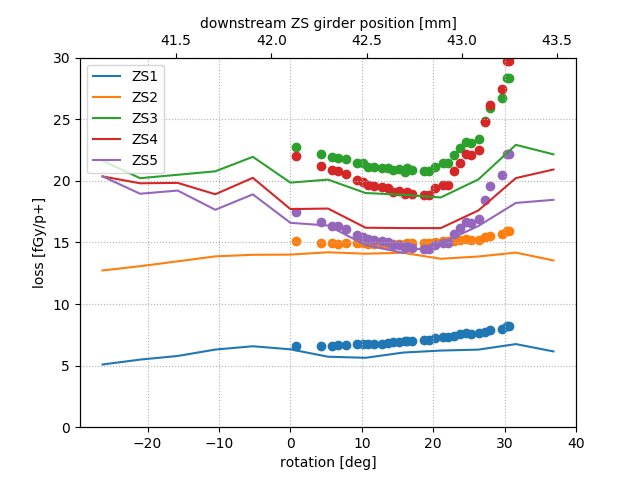}
    \caption{Measured normalised losses on the ZS BLMs during a scan of the rotational knob (solid lines, lower horizontal scale) compared to those during an earlier scan of the downstream ZS girder position (dots, upper horizontal scale). The two horizontal scales were aligned so that the minima correspond, and stretched in accordance to simulation results for the scale equivalence.}\label{fig:rotation_is_girder}
\end{figure}

The knob was tested in MD. First observations showed minor changes in the extracted beam profile at the upstream end of the ZS, but a movement of several \si{\mm}  at the grid \SI{90}{\degree} downstream, as expected for a rotation of the separatrix arm. Furthermore, the normalised losses on the ZS BLMs were recorded and compared to those for a scan of the girder on which the ZS sits. Since in one case we are changing the angle of the ZS with respect to the beam, and in the other we are changing the beam angle with respect to a stationary ZS, the scan results were expected to be similar. Fig. \ref{fig:rotation_is_girder} shows this is indeed the case, confirming that the knob works. The shift in the baseline losses is explained by the fact that the girder scan had been performed several months earlier, and the relative alignment of the ZS tanks has changed in the meantime.

\subsubsection{Simulated extraction}
The LOF octupoles in SPS are nominally unused during the slow extraction and are installed at high $\beta_x$, making them good candidates to use for separatrix folding. These octupoles were used in simulation to curve the separatrix and extract a 'folded' separatrix in combination with the orthogonal sextupole knob to correct the angle of the beam at the ZS wires back to the nominal. In order to eliminate changes in optics with momentum, all simulations used the COSE extraction method.

\begin{figure}[htbp]
    \centering
    \begin{subfigure}{\textwidth}
        \centering
        \begin{subfigure}{0.325\textwidth}
    	\centering
            \includegraphics[width=\textwidth]{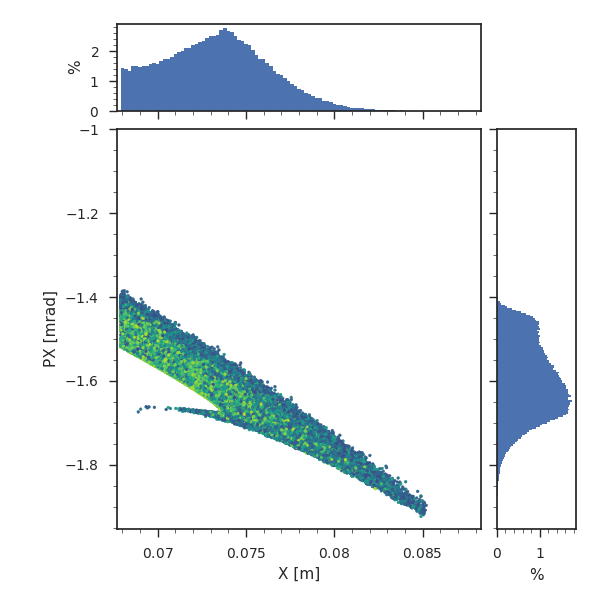}
        \end{subfigure}~
        \begin{subfigure}{0.325\textwidth}
    	\centering
            \includegraphics[width=\textwidth]{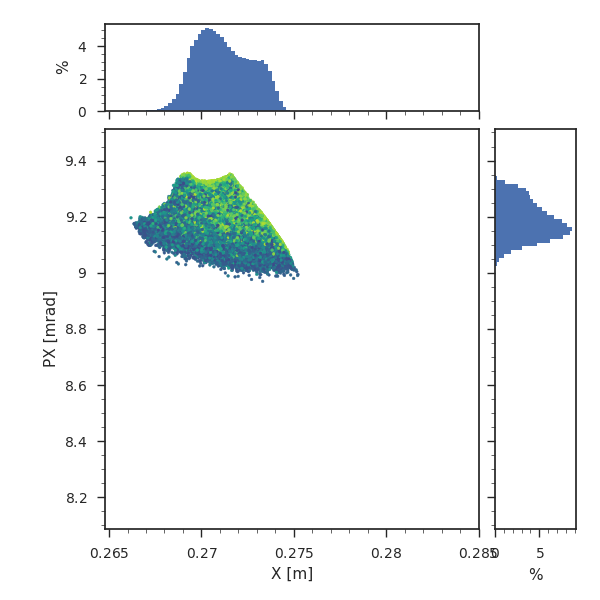}
        \end{subfigure}
        \caption{Extraction with nominal sextupole strength, $k_3L=\SI{4.5}{\meter^{-3}}$ per octupole and \SI{-85}{\degree} driving term rotation, resulting in 1.74\% of beam impacting the ZS wires.}
        \label{fig:oct_neg}
    \end{subfigure}

    \begin{subfigure}{\textwidth}
        \centering
        \begin{subfigure}{0.325\textwidth}
    	\centering
            \includegraphics[width=\textwidth]{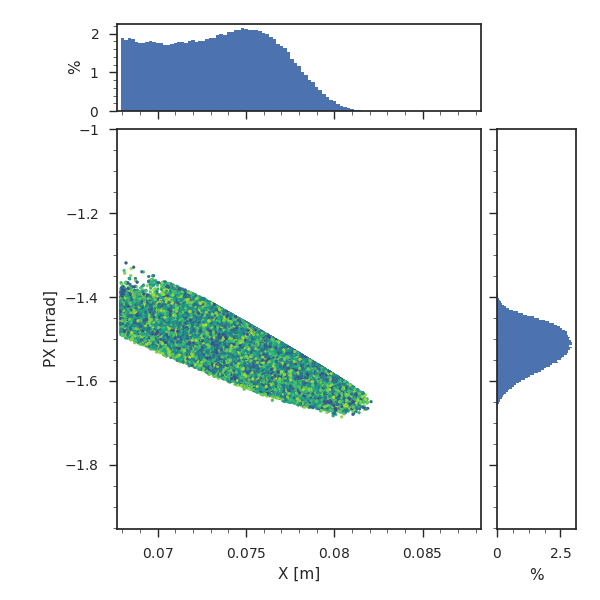}
        \end{subfigure}~
        \begin{subfigure}{0.325\textwidth}
    	\centering
            \includegraphics[width=\textwidth]{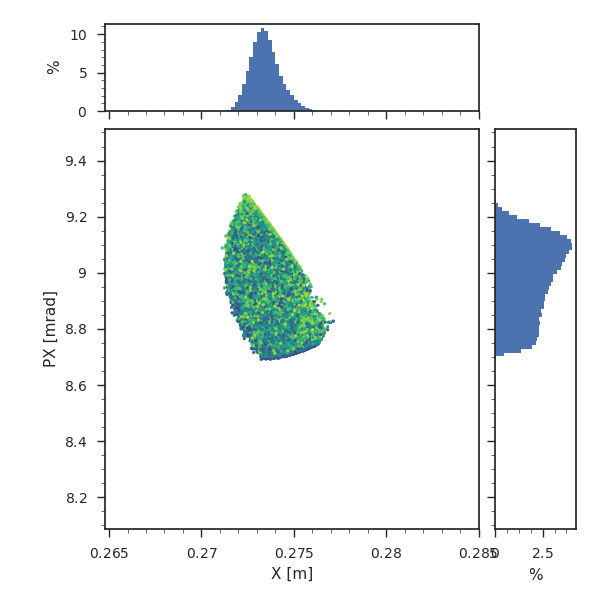}
        \end{subfigure}
        \caption{Extraction with 1.4x nominal sextupole strength, $k_3L=\SI{-2.5}{\meter^{-3}}$ per octupole and \SI{50}{\degree} driving term rotation, resulting in 2.35\% of beam impacting the ZS wires.}
        \label{fig:oct_pos_lo}
    \end{subfigure}

    \begin{subfigure}{\textwidth}
        \centering
        \begin{subfigure}{0.325\textwidth}
    	\centering
            \includegraphics[width=\textwidth]{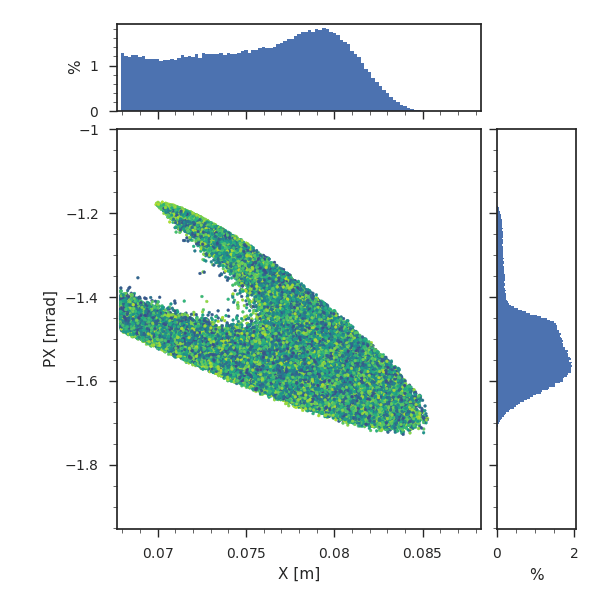}
        \end{subfigure}~
        \begin{subfigure}{0.325\textwidth}
    	\centering
            \includegraphics[width=\textwidth]{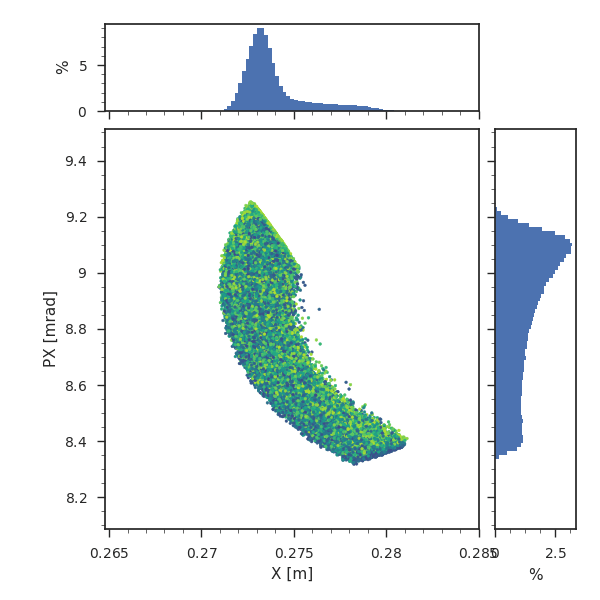}
        \end{subfigure}
        \caption{Extraction with 2.0x nominal sextupole strength, $k_3L=\SI{-2.2}{\meter^{-3}}$ per octupole and \SI{30}{\degree} driving term rotation, resulting in 1.52\% of beam impacting the ZS wires.}
        \label{fig:oct_pos_hi}
    \end{subfigure}
    \caption{Extracted beam in phase space at the upstream end of the ZS (left)and at the LSS2-TT20 handover point (right) for several multipole settings. Simulated particles are coloured by momentum from blue (low) to yellow (high). Nominally, 2.66\% of beam impacts the ZS wires.}\label{fig:oct_sim}
\end{figure}

The simulated extracted beams are shown in Fig. \ref{fig:oct_sim}. In simulation, a reduction of up to 43\% in the number of particles impinging on the ZS wires was observed, shown in Fig.~\ref{fig:oct_pos_hi}. The shape of the extracted beam varies greatly with the multipole settings. Therefore, the optics in TT20 would need to be rematched in order to transport the beam to the target. In these simulations, the orthogonal sextupole knob was used to rotate the separatrix arm in such a way the beam angle at the ZS wires was nominal. The advantage is that the ZS girder alignment would not have to be changed, but the drawback is that the beam centre does move, both at the upstream ZS and the handover point, which would require a steering correction in the extraction channel in LSS2 and the upstream part of TT20.

\newpage


\newpage
\subsubsection{Machine protection}
A procedure for the MD tests was discussed and approved by the SPS and LHC Machine Protection Panel~\cite{octupole_procedure}. Even though the combination of octupoles and sextupoles can be used to reduce normalised losses, a wrong configuration could also increase normalised losses and potentially damage the ZS wires. When the octupoles are too strong compared to the sextupoles, the horizontal extent of the beam in the extraction aperture becomes very small. For extreme cases, the separatrix can even be bent so strongly that particles become trapped at large amplitudes without being extracted/ When the beam is trapped at large amplitudes, near the ZS wires, one risks to collimate the circulating beam directly on the ZS wires, which is to be avoided at all cost.


This trapping effect was confirmed in machine studies (under safe conditions), and safe octupole limits for several sextupole strengths were identified. Since the safe sextupole and octupole limits are inter-dependent, interlocking for the general case is not possible with the current SPS Interlock System (SIS). If extraction with octupoles were to be used operationally, one would either need to define safe settings around the new operational parameters, or require an updated SIS with functionality of inter-dependent interlock settings.

\subsubsection{Summary of main results of Machine Development (MD)}
Several combinations of octupole and sextupole strengths  listed in Table~\ref{tab:octu_md} were tested in the machine, and a scan of the angular rotation knob carried out for each. In the best case, the summed losses were reduced by 42\%. For this best-case scenario, unlike for the other configurations the loss signal at BLM 219 increased. The most likely cause for the increased losses at BLM 219 would be beam losses on the downstream end of the MSE. These losses can probably be avoided, either by moving the downstream end of the MSE, or by correcting the trajectory of the beam through LSS2.

\begin{table}[htbp]
  \centering
  \caption{Optimised ZS losses for several combinations of octupole and sextupole strengths. For these settings the rotation could not be used to fully optimise the ZS losses, since losses at BLM 219 increased as the ZS losses decreased.}
  \label{tab:octu_md}
  \begin{tabular}{ccccc}
    \hline
    \textbf{$\mathbf{k_2L/(k_2L)_\mathrm{ref}}$} & \textbf{$\mathbf{k_3L}$ per LOF} & \textbf{Rotation} & \textbf{Loss ZS1-5} & \textbf{Loss BLM 219}\\
    \textbf{} & \textbf{[\si{\meter^{-3}}]} & \textbf{[deg.]} & \textbf{[\si{\femto\gray/\protons}]} & \textbf{[\si{\femto\gray/\protons}]}\\\midrule
    0.95 & 0.0 & 15.8 & 69.4 & 2.7\\
    0.95 & 1.7 & -35.8 & 67.8 & 3.6\\
    1.4 & -2.25 & 50.0 & 47.0 & 10.0\\
    2.1 & -2.25 & 14.2 & 43.7 & 9.9\\
    2.1 & -2.45 & 28.5$^*$ & 40.2 & 18.5\\
    \hline
  \end{tabular}
\end{table}



\subsection{Phase space folding with a massless septum}
An alternate approach to folding the separatrix in phase space using the fringe field region of a massless septum to deliver a kick to part of the separatrix arm has also been studied~\cite{brunner,brunner2}. Like the octupole method, this allows the use of higher sextupole strengths. Initial studies showed a loss reduction of 50\%. The simulated optimum used the sextupoles at 2.3 times their nominal strength and with a massless septum placed in between the final sextupole and the ZS. Although the preliminary studies are promising, the aperture requirements in the ring have not yet been taken into account in the simplified simulation code used, which is likely to be a limitation. It has not yet been determined whether folding with octupoles and folding with massless septa can be combined to achieve an even greater loss reduction.

\subsection{Combining loss reduction techniques: phase space folding applied with crystal shadowing}

\begin{figure}[htbp]
    \centering
    \includegraphics[width=0.8\linewidth]{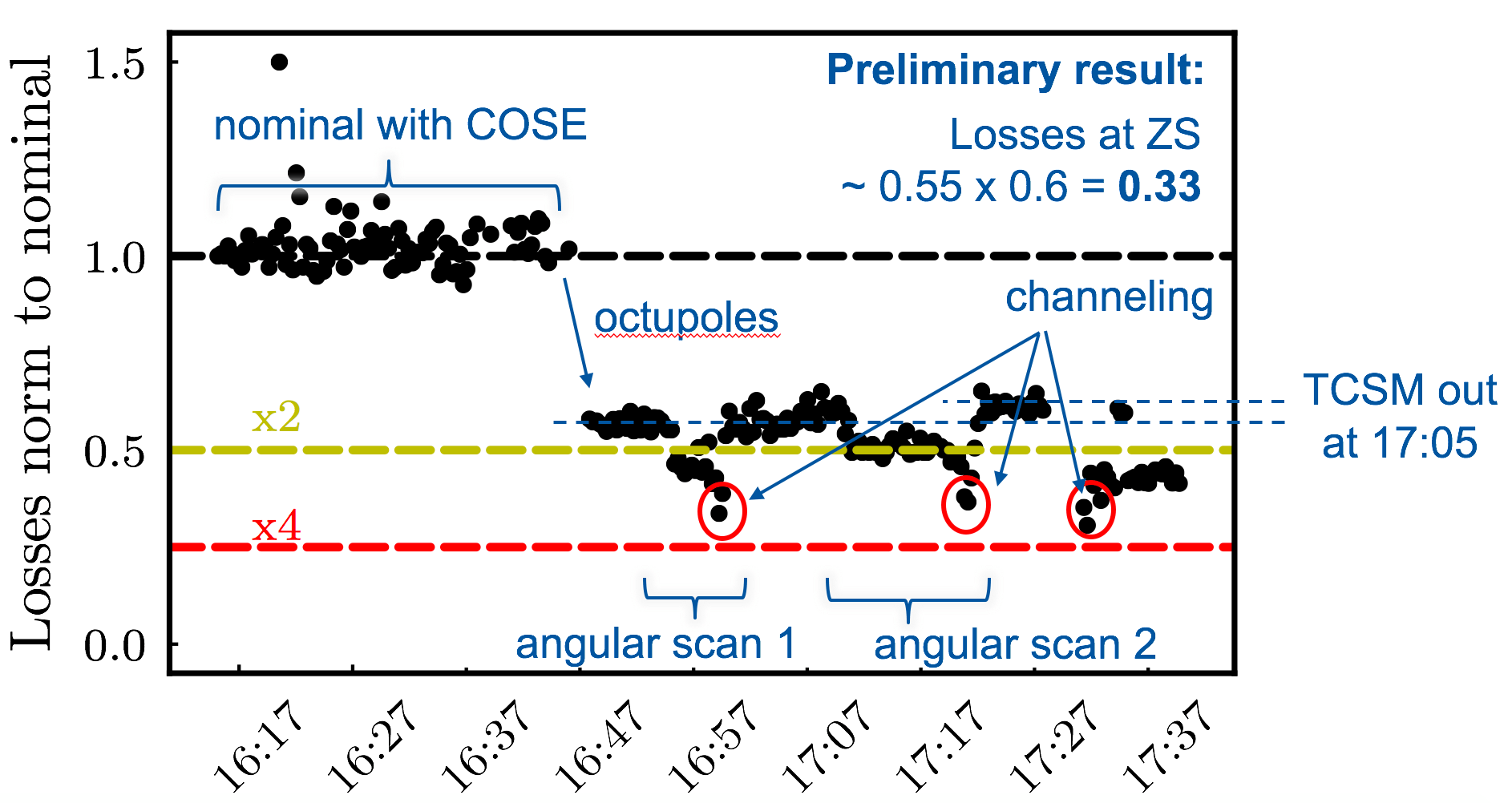}
    \caption{Relative loss reduction as COSE, phase space folding with octupoles and shadowing with the crystal were combined on 1$^\text{st}$ November 2018. In this case, the octupoles were powered as mentioned above, i.e. $k_2L/(k_2L)_\mathrm{ref} = 2.1$ and $k_3L = -2.45$ m$^{-3}$ per LOF.\label{fig:combination}}
\end{figure}

In a final test with protons, the crystal was aligned with the octupoles powered and COSE implemented. During the crystal alignment scans a loss reduction factor at the ZS BLMs of over 3 was achieved when the crystal was channelling, see~Fig.~\ref{fig:combination}. More dedicated tests will be needed in the future to assess the stability of the combined extraction technique as time was limited to the data collected in~Fig.~\ref{fig:combination}. The angular spread of the beam at the crystal in the presence of octupoles is an important consideration for the specification of a future extraction system combining the two techniques. The same can be said about transporting or collimating the channelled beamlet in the transfer line in the presence of a larger horizontal emittance created by the octupoles

\section{Machine stability, reproducibility and operation}

The SPS provides different beams to serve multiple users on different magnetic cycles played in a sequence, the so-called Super-Cycle (SC). The composition of the SC is changed several tens of times every day. Proton and ion beams are accelerated for fixed target physics in the North Area hall, the SPS assembles and accelerates the LHC injection trains, beams are also provided to the HiRadMat irradiation facility \cite{HiRadMat}, the AWAKE plasma wakefield facility \cite{AWAKE} and to Machine Development
(MD) users who often fill any remaining space in the SC to push the performance and prepare for future running scenarios. In addition, a dynamic economy system has been introduced to save energy by ramping the main power supplies only if beam is received on the injection plateau. The frequent changes to the magnetic cycling of the machine impacts the reproducibility, the effects of which are most noticeable on the resonant fixed target cycle with degradation of the uniformity of the slow extracted spill. There is evidence to suggest that hysteresis, which amounts to just a few Gauss at flat-top, is responsible for these variations on timescales of hours.

In addition to the hysteresis effects, the longer term stability of the closed-orbit in the bending plane of the synchrotron has been shown to drift by over a mm on timescales of several weeks in the LHC extraction regions of the SPS [8]. This makes it very difficult to maintain the relative alignment between the ZS and the beam, demanding regular realignment of the septa. The source of the drift has not yet been identified.

\begin{figure}[htbp]
   \centering
   \includegraphics*[width=250pt]{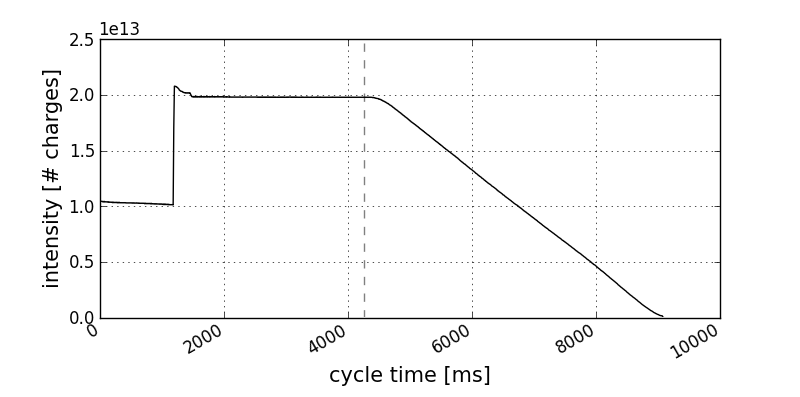}
   \caption{Evolution of the intensity through the fixed target cycle in the SPS in 2016. The dashed line indicates the moment at which the slow extraction starts.}
   \label{fig:bct_good}
\end{figure}
\begin{figure}[htbp]
   \centering
   \includegraphics*[width=250pt]{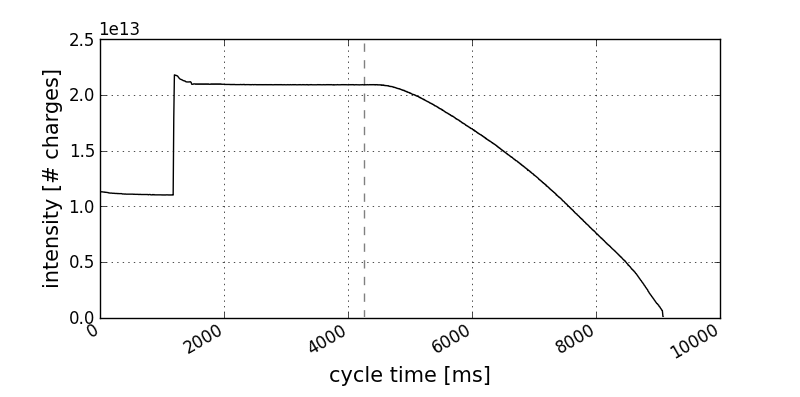}
   \caption{Evolution of the intensity through the fixed target cycle in the SPS in 2016 after a super cycle change. The dashed line indicates the moment at which the slow extraction starts. The beam is not extracted with a constant rate.}
   \label{fig:bct_bad}
\end{figure}
\begin{figure}[htbp]
   \centering
   \includegraphics*[width=250pt]{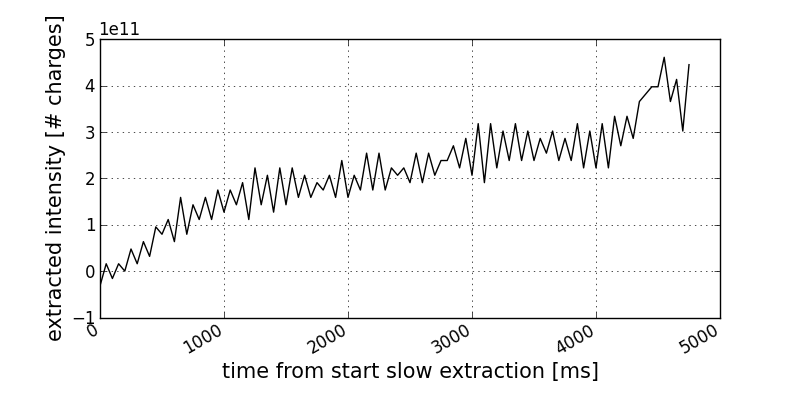}
   \caption{Extracted intensity as function of time on the extraction plateau of the fixed target cycle calculated from the intensity evolution in Fig. \ref{fig:bct_bad}.}
   \label{fig:bct_spill_bad}
\end{figure}

\subsection{Spill quality and effect of spill length}

For the NA experiments and BDF, the slow extracted spill quality is important for several reasons. The experiments are sensitive to combinatorial background, and large spikes in the extracted proton rate have an impact on the sensitivity.  In addition the target is designed for a certain maximum transverse proton density, which could be exceeded if the spill shape departs too far from the ideal trapezoid. A measure of the uniformity of the spill is the "effective spill length" \cite{verena_spill} which is defined as:
\begin{equation}
t_\textrm{efs}=\frac{[\int^{t_2}_{t_1} f(t) dt]^2}{\int^{t_2}_{t_1}[f(t)]^2 dt}
\label{eq:effective_spill_length}
\end{equation}
where $f(t)$ is the extracted intensity as a function of time. 
Figures \ref{fig:bct_good} and \ref{fig:bct_bad} show the evolution of the circulating intensity and extracted intensity calculated from the decay of the intensity during the extraction flat-top. The extracted intensity ramp-up at the beginning of the spill is introduced on purpose as the RF structure takes roughly \SI{500}{ms} to diminish to an acceptable level for the North Area experiments. Any events during this time are not taken into account. The situation in Fig. \ref{fig:bct_good} corresponds to a well adjusted spill and the effective spill length calculated according to Eq.~\ref{eq:effective_spill_length} is $t_\textrm{efs}\approx \SI{4500}{ms}$. If LHC \SI{450}{\giga\eVperc} cycles are added to the SC or the dynamic economy mode is enabled, where the cycles are not fully played in case no beam is injected, the beam parameters will change. The effect on the extracted intensity before running the feed-forward algorithm is shown in Fig. \ref{fig:bct_bad} and \ref{fig:bct_spill_bad}. The effective spill length is reduced to 3800 ms for that case. 

\begin{figure}[htbp]
   \centering
   \includegraphics*[width=270pt]{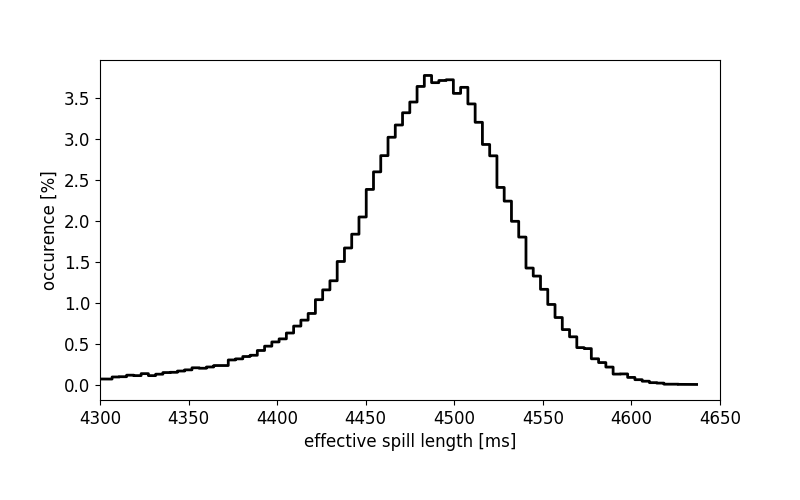}
   \caption{Distribution of effective spill length during 4 weeks in July 2018. The spread is roughly 4250 to 4600 ms.}
   \label{fig:effective_spill_length_distribution}
\end{figure}

In the SPS the quoted effective spill length is derived from the Beam Current Transformer (BCT) measurement sampling at \SI{200}{Hz} during the extraction flat-top, as indicated in the figures above. The effective spill length derived in this manner is a measure of the macro-structure of the spill. Often also the spill duty factor is quoted which corresponds to $t_\textrm{efs}/(t_2-t_1)$ where $t_2-t_1$ is the flat-top length in the SPS. Typical values for the SPS spill duty factor are 95 \%. Fig. \ref{fig:effective_spill_length_distribution} shows the distribution of the effective spill length over 4 weeks in July 2018. The mean effective spill length was 4450 ms.  

\subsection{Possible origin of spill macro-structure changes}

\begin{figure}[htbp]
   \centering
   \includegraphics*[width=0.7\textwidth]{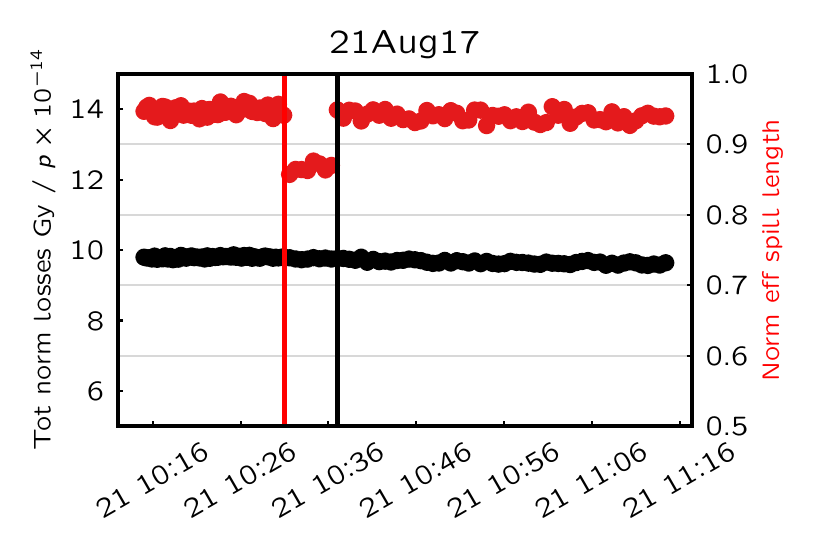}
   \caption{Time evolution of normalised effective spill length and total extraction losses in the slow extraction channel. The vertical red line represents a change in SC and the black vertical line the correction applied on the tune to readjust the spill quality.}
   \label{fig:losses}
\end{figure}

The various SPS cycles not only differ in the maximum energy reached, but also in the optics used, see e.g.~\cite{hannes_q20}. This implies that, following a SC change, not only the magnetic history of the main bends will be different, but also the main quadrupoles. It was observed that variations of the SC reduces the spill duty factor (red curve in Fig.~\ref{fig:losses}) by a few percent. Interestingly, the losses in the slow extraction channel are not affected (black curve in Fig.~\ref{fig:losses}). This can be explained by the fact that the beam position at the electromagnetic septum (ZS) is not significantly perturbed by SC changes. 

\begin{figure}[htbp]
   \centering
   \includegraphics*[width=0.7\textwidth]{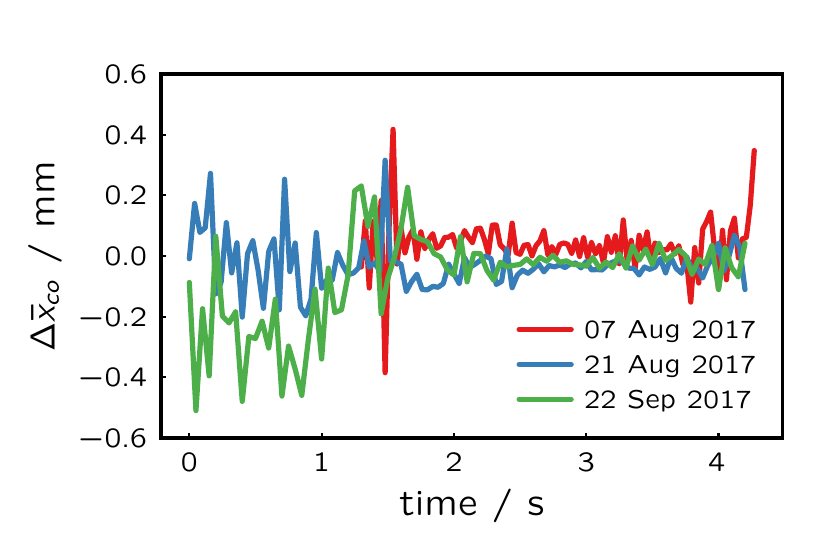}
   \caption{Difference of mean horizontal beam closed orbit along FT cycle before and after a SC change on three different dates.}
   \label{fig:mean}
\end{figure}

In Fig.~\ref{fig:mean}, the difference of the measured mean horizontal position of the beam along the slow extraction cycle, before and after a SC change, is shown. It can be seen that, except for the difference at flat bottom, which can be explained as the radial loop is activated at the start of the ramp, the mean of the horizontal orbit does not change. 

\begin{figure}[htbp]
\centering
   \includegraphics*[width=0.8\textwidth]{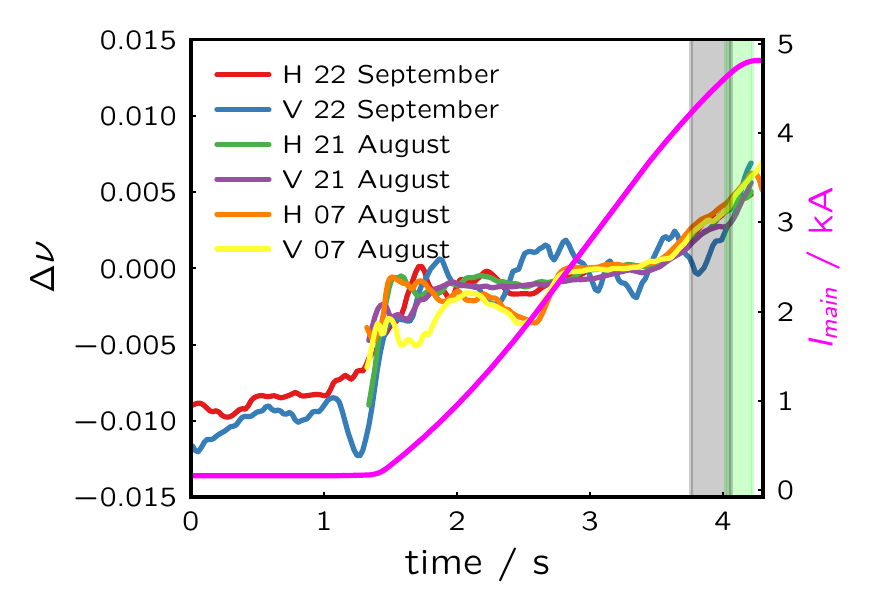}
   \caption{Difference of horizontal and vertical tune along FT cycle before and after a SC change. The magenta line represents the current time evolution of the main SPS power supplies.}
   \label{fig:delta_tune}
\end{figure}

During the same measurement period as in Fig.~\ref{fig:losses}, the transverse tunes were also measured (Fig.~\ref{fig:delta_tune}). At injection, the SC change shifts both horizontal and vertical tune by 0.04\%. Then, thanks to the radial loop, the difference in tune from one cycle to another is kept zero for more than half of the acceleration ramp. When approaching the flat-top (starting from around \SI{310}{\giga\eVperc}), the difference diverges from zero reaching about 0.02\% at flat top. Such a difference, mainly on the horizontal tune, provokes the above mentioned spill quality degradation. Earlier investigations of the response errors of the main dipoles and quadrupoles, as well as their compensation, are detailed in \cite{jorg}. It was already observed that, starting from \SI{300}{\giga\eVperc} and until the flat top is reached, an error between the demanded and measured field is present on all SPS cycles. This effect is more critical for the FT cycle due to its faster ramp. The difference in the tune functions could originate from the different behaviour between main bends and quadrupoles when the saturation levels are approached. Also, the radial loop is still active at that moment, which could have an additional effect. 

The good reproducibility of the tune variation following a SC change shown here could be used to eliminate the observed spill quality degradation by feed-forward compensation. Further analysis and measurements are still needed to fully conclude on the origin of the observed effect to possibly cure it at the source. 

Insufficient tracking of the demanded current function by the main quadrupole power converter leads to another possible spill quality degradation factor. At the round off to the extraction flat-top of the current function, the power supplies tend to overshoot. These current variations lead to spikes of extracted intensity before slow extraction is supposed to start.

\subsection{Effect of power supply noise on spill quality and correction of spill ripple}

\begin{figure}[htbp]
   \centering
   \includegraphics*[width=0.48\textwidth]{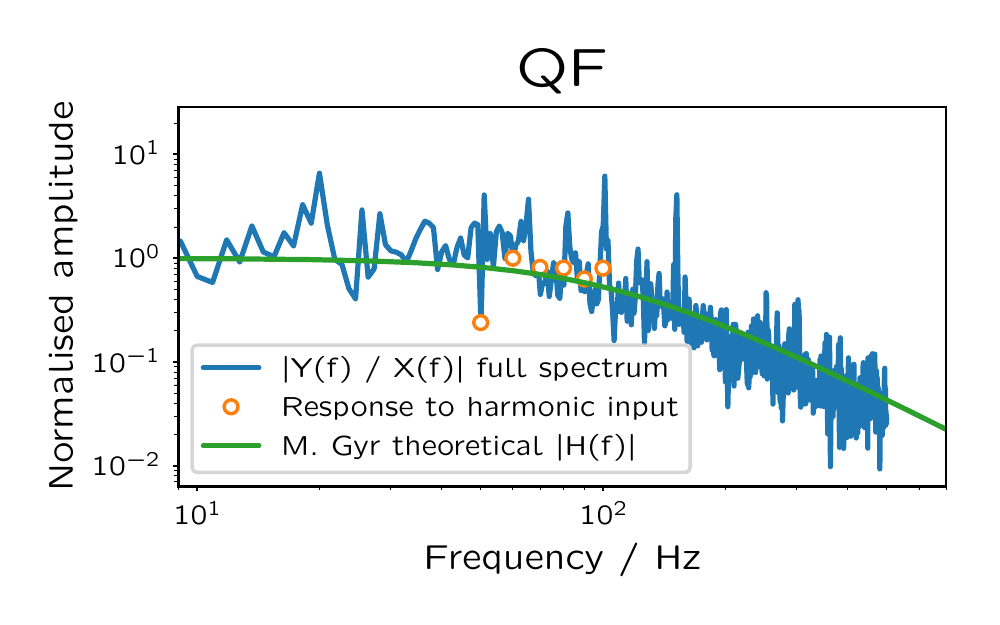}
   \includegraphics*[width=0.48\textwidth]{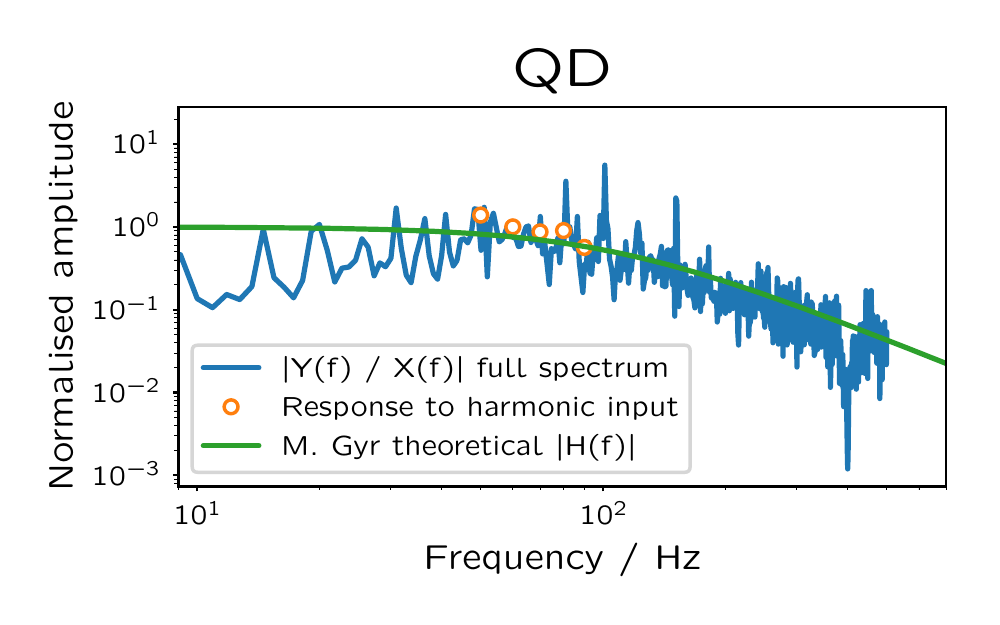}
   \caption{Transfer function between QF (left) / QD (right) PS and extracted intensity.}
   \label{fig:tf_qf}
\end{figure}

Spill quality is also significantly influenced by noise on the machine power supplies. The spill harmonic content can be measured using the \SI{2.5}{kHz} bandwidth intensity monitor installed in the extraction transfer line to the NA. Without correction, the intensity is almost 100~\% modulated along the flat-top. The main source of the noise has been identified to originate from the main power supplies, especially the focusing quadrupoles (QF). As already shown in literature \cite{marcel, jorg_servo}, the figure of merit of the sensitivity of the spill to power supply noise is the transfer function (TF) between each individual converter and the extracted intensity. In 2017, a campaign of measurements to characterise the TF for all SPS main power supplies was carried out. The results for the main focusing and defocusing quadrupoles (QD) are shown in Fig.~\ref{fig:tf_qf}. The observed behaviour is the one explained and measured in \cite{marcel, jorg_servo}, i.e. the dynamics of the slow extraction as well as the machine with its vacuum chambers act as low pass filter for noise~\cite{javier_ripple}. At about \SI{300}{Hz} a reduction of about one order of magnitude in the passing amplitudes can be observed. The same measurements were carried out in two different ways. The blue curves in Fig.~\ref{fig:tf_qf} shows the normalised ratio between the Fourier spectra of measured current of the QF and QD circuits with the extracted intensity over time. The yellow markers, instead, are obtained by injecting external noise into the magnet circuits at each individual frequency and measuring the effect on the extracted spill. Both measurements are in good agreement with the theoretical prediction presented in~\cite{marcel}. 

\subsubsection{Simulations of spill with power supply noise}

\begin{figure}[htbp]
   \centering
   \includegraphics*[width=220pt]{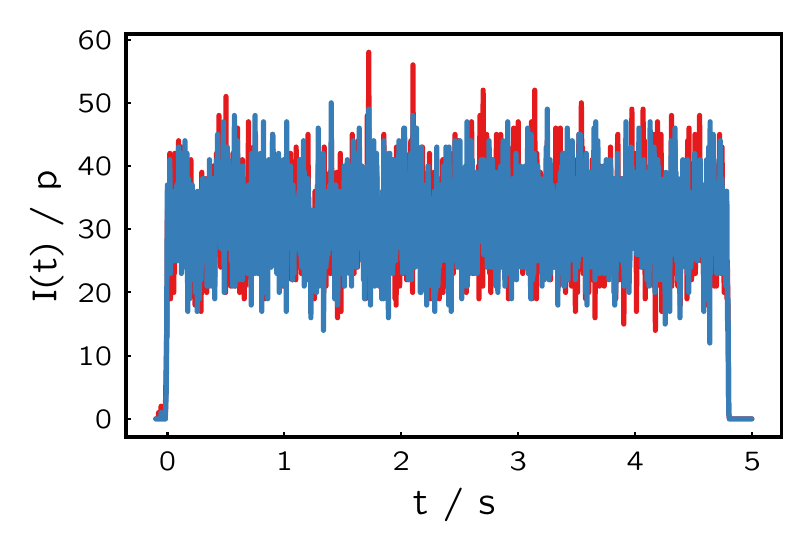}
   \caption{Example of spill structure simulations using the described semi-analytic model. In red, the expected spill structure when 50, 100 and \SI{150}{Hz} noise on the horizontal tune is injected. In blue, the same noise is used, but the extraction speed was increased by \SI{50}{\%}, together with the extraction sextupole strength. }
   \label{fig:simulation}
\end{figure}

In order to predict the expected spill quality from a measured current waveform, a semi-analytic model was built. The driving idea is to simplify the slow extraction simulations and make them less time consuming. The simplified code was benchmarked with more sophisticated simulation tools such as MADX \cite{javier_ripple, me_se}. This tool will be used to optimise slow extraction parameters and, in the best case, propose ways to reduce the impact of noise, e.g. increasing the absolute value of chromaticity and hence increase the speed of the tune sweep during extraction.

The first version of this model is built parametrising the probability that a particle can be extracted judging by its transverse amplitude, its momentum and the distance from resonance. Such a probability is parametrised as Gaussian in momentum space and exponential in amplitude space. Then, it is shaped using the classic stop-bandwidth definition \cite{slow_extr} according to the instantaneous machine tune, hence the distance from resonance, accounting for chromaticity and individual momentum.

The input current is fed into the model decomposing it into its main harmonics and adding it to the main quadrupole current function which is driving the slow extraction. This is used as input to evaluate at each step in time the exact machine tune and the size of the stable area for each particle. 

\begin{figure}[htbp]
   \centering
   \includegraphics*[width=220pt]{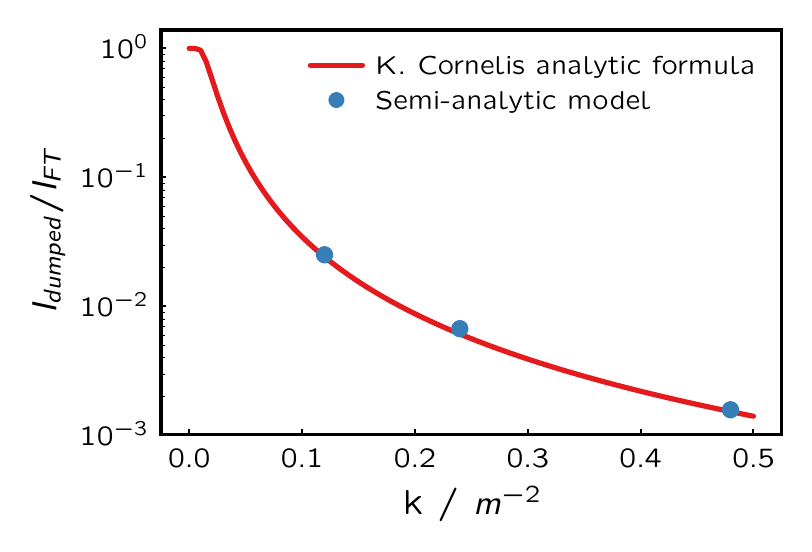}
   \caption{Evolution of non-extracted particles, normalised by the total intensity, as a function of the sextupole strength. In red, the analytic formula presented in \cite{karel} is shown. The blue markers represents the results obtained with the described semi-analytic model.}
   \label{fig:karel}
\end{figure}

The main SPS machine and beam parameters that play a role in the slow extraction process are taken into account, i.e. chromaticity, sextupole strength, emittance and momentum distribution. An example of two simulated spills (using two different machine configurations) is shown in Fig.~\ref{fig:simulation}. The gain in the spill noise level achievable from increasing the machine chromaticity (hence the slow extraction speed) by \SI{50}{\%} was simulated to be about \SI{30}{\%} for the main harmonics (50, 100 and \SI{150}{Hz}). Measurements are still needed to benchmark these predictions. As a first benchmark, the results from this model have been compared with an analytical description \cite{karel} of the effect of faster tune sweep on the non-extracted intensity, see Fig.~\ref{fig:karel}. 

\begin{figure}[htbp]
   \centering
   \includegraphics*[width=300pt]{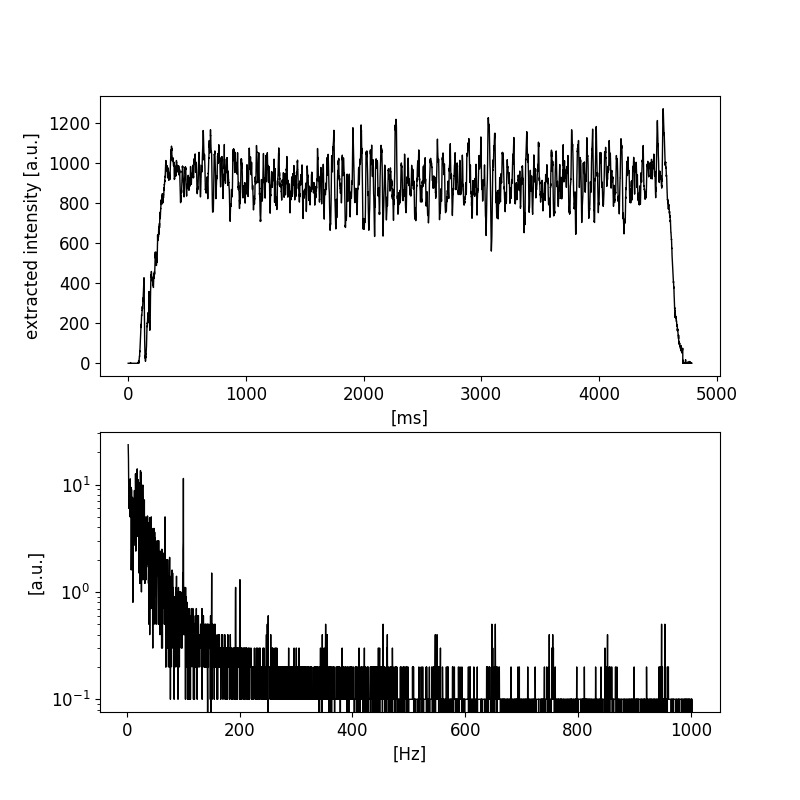}
   \caption{Typical spill measured by the BSI foil in the transfer line TT20 for \textasciitilde $2.7\times 10^{13}$ protons extracted at \SI{400}{\giga\eVperc} on 4$\textrm{th}$ November 2018 at 12:29 on the upper plot and FFT of spill on the lower plot. The 100 Hz component was not well corrected. The 50 Hz amplitude is within tolerance of amplitude 50 a.u. }
   \label{fig:typical_spill}
\end{figure}

\begin{figure}[htbp]
   \centering
   \includegraphics*[width=350pt]{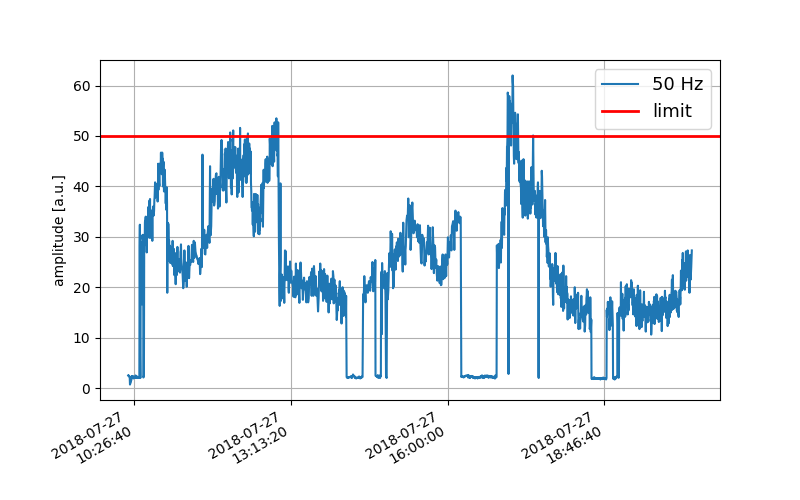}
   \caption{Typical evolution of the 50 Hz amplitude in the spill during a duration of 10 h. Data was obtained on the $27^\textrm{th}$ of July, 2018.}
   \label{fig:50Hz_evolution}
\end{figure}

\begin{figure}[htbp]
   \centering
   \includegraphics*[width=300pt]{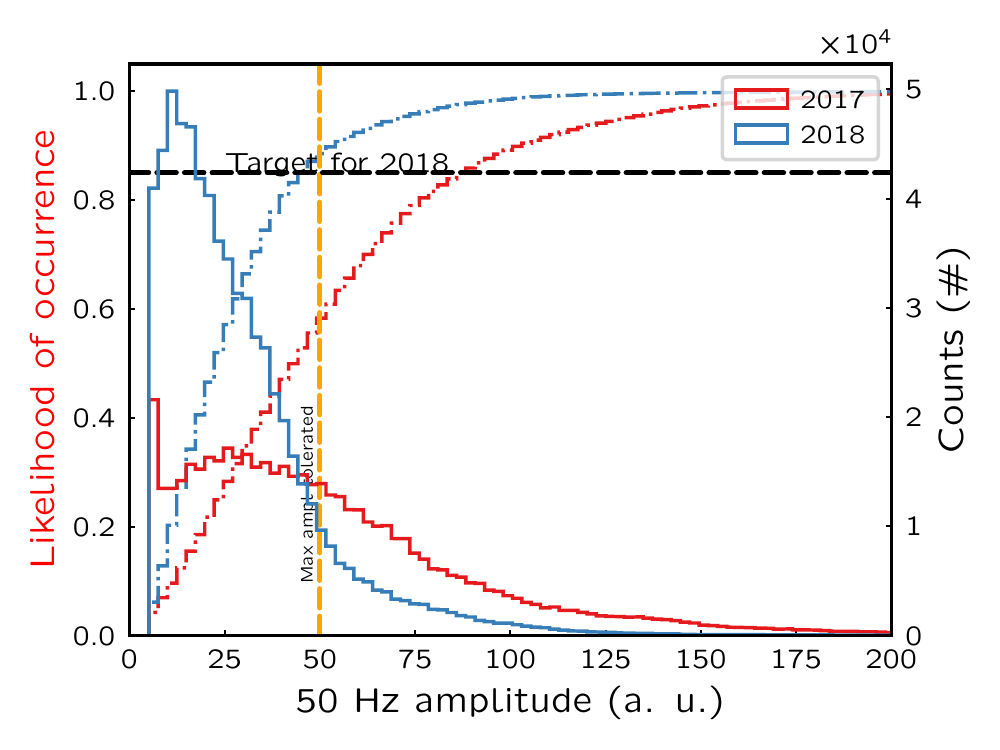}
   \caption{Distribution of spill \SI{50}{Hz} amplitude for both 2017 (red) and 2018 (blue). The dashed curves represent the cumulative distribution function for both set of data. }
   \label{fig:50Hz_distribution}
\end{figure}

Until the 2018 proton run, the spill harmonic content was corrected with the dedicated servo-quadrupole system consisting of 4 short QMS quadrupoles installed in cell 116. It is equipped with a power converter capable of providing modulation in current of 50, 100, 150 and \SI{300}{Hz} with adjustable phase and amplitude on top of its reference function, which is zero nowadays. A measurement of the network 50 Hz signal was made available in the control room to deterministically find the correct phase with respect to the 50 Hz modulation of the spill. Figure~\ref{fig:typical_spill} shows a typical spill of the 2018 proton run corresponding to a total extracted intensity of $2.7 \times 10^{13}$ protons with servo-quadrupole harmonic correction on. Unfortunately the correction settings do not stay valid for long because the amplitude and phase drift. The time-scale of the drift for the 50 Hz amplitude is illustrated in Fig.~\ref{fig:50Hz_evolution}. The SPS quality check (SPS QC) software monitors the effective spill length and harmonic content and issues a warning if the spill quality has deteriorated below a given threshold. Increased awareness of the operations crew, better tools and automatic algorithms for spill corrections have much improved the spill quality and stability of correction over the years. Figure \ref{fig:50Hz_distribution} shows the improved control of the 50 Hz amplitude in the slow extracted spill between 2017 and 2018. For about 85 \% of the time the 50 Hz content was well corrected in 2018 with respect to only 60 \% of the time in 2017. 

The electronics of the servo-quadrupole $n\times 50$~Hz injection system is obsolete and a new way of correcting the spill ripple has been proposed. It will be based on directly injecting the current modulation at $n\times 50$~Hz into the QF main power supply circuit consisting of 108 quadrupoles. Amplitude and phase can be adjusted. First tests in 2018 looked promising and the SPS will start up with correcting the harmonic content of the spill via the QF circuit in 2021. Additional improvements will come from the spill measurement system. Currently the acquisition system of the spill monitoring system (BSI) is in the surface building BA1 close to the servo-quad system. The long cables from the surface buildings BA2 to BA1 add noise. After LS2 the acquisition system will be directly placed in BA2 to improve the signal-to-noise ratio, such that even low intensity ion spills will be measurable.

\subsection{Spill quality at higher frequency}

\subsubsection{Medium frequencies (synchrotron revolution period)  -- \SI{46}{\kilo\hertz}}

The fixed-target beam spectrum in the SPS has a pronounced structure at the revolution frequency of 43 kHz and its second harmonic of 86 kHz, due to the filling of the SPS from the PS, where the \SI{2.1}{\micro\second} PS single turn is injected into the SPS in 5 turns, leading to a pattern of two trains of \SI{10.5}{\micro\second} of beam with two gaps of \SI{1}{\micro\second}. The time needed for the beam to completely debunch to wash out this structure is much longer than the $\sim1$~s time available for the extraction. 

\subsubsection{Higher frequencies (RF time structure) -- 200 MHz}

 The main SPS accelerating RF system imposes a 200 MHz structure on the proton beam. When the RF is switched off just prior to extraction the protons start to debunch, helped by the large (increased by RF gymnastics) momentum spread. The particles with negative momentum offset move forward in phase while those with positive offset move backwards. As measured by a pickup in the SPS ring, after some \SI{10}{ms} the debunching is complete, as the particles with negative offset in one bunch overlap fully with the positive offset particles from adjacent bunches. However, as the slow extraction effectively selects a narrow band of momentum (via the large chromaticity), the 200 MHz structure is still present in the extracted beam for several 100 ms, until the debunching and phase space mixing is complete. This structure and how it varies throughout the spill is important to understand for experiments at the BDF.
 
 In this context, research has been launched to provide a higher bandwidth spill measurement using a detector based on Cherenkov radiation, allowing to measure the decay of the SPS main RF 200 MHz component and other components induced by the machine's impedance during  extraction~\cite{cpfm_fs,cpfm_addesa,addesa_thesis}. Options for suppressing higher frequency spill components are being investigated, including lowering the chromaticity (extracting using amplitude detuning instead of chromaticity via momentum) and appliying longitudinal noise (stochastic extraction, see e.g.~\cite{marcel_stochastic}).
 
\section{Extraction hardware, activation and interventions}

\subsection{Low-Z septa}
To extract the beam from the accelerator, electrostatic septa (ZS) and magnetic septa (MS) are being used. The magnetic septa are protected from accidental beam impact by a diluter (TPST) installed in front of the first magnetic septum (MST1). A fraction of the slow extracted beam impacts the ZS (wire array) of the electrostatic septa, provoking secondary particle showers that develop and activate equipment further downstream. Hence, there is a strong interest in septa fabricated from materials with low atomic number (Z), provoking only the minimal amount of loss via scattering, and causing less activation per proton lost. Low-Z septa therefore serve two purposes: 
\begin{itemize}
	\item to reduce the interaction of the septum wires with the beam, thus reducing the production of secondaries;
	\item to limit its activation by minimising interactions of the equipment itself with stray particles and secondaries.
\end{itemize}

\subsubsection{Low-Z septum wires}
Low-Z electrostatic septa wires yield an immediate gain in prompt extraction losses by reducing the production of secondary particle showers~\cite{azsawg_dora}. The quantity of secondaries produced by the beam scattering from the septum wires is obviously also dependant on the wire thickness and anode straightness (see Sec.~\ref{sub:AnodeStraightness}), in particular in the upstream part of the septum where the beam is not yet deflected significantly.

\begin{figure}[htbp]
    \centering
    \includegraphics[width=\linewidth]{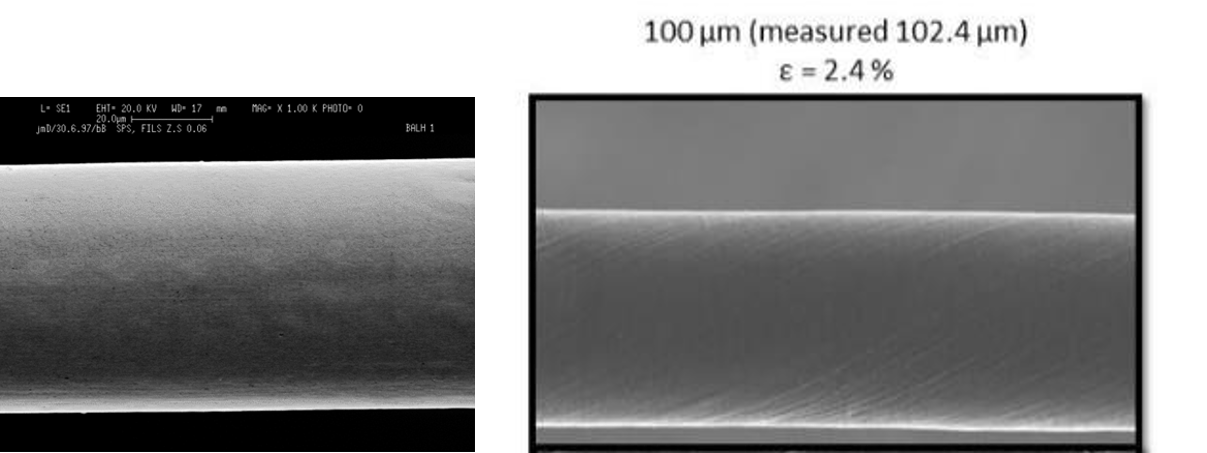}
    \caption{Surface finish of \SI{60}{\micro\meter} Polished WRe wire (left) and \SI{100}{\micro\meter} CNT wire (right), as observed under an electron microscope.}
    \label{WRe_CNT_wire}
\end{figure}

Historically, the septum wires are made of Tungsten Rhenium (WRe) due to its high melting temperature and ductility. Encouraged by the initial experience with carbon nanotube (CNT) wires at KEK~\cite{Tomizawa2}, a comparison with a microscope of the surface finish between polished WRe wire and CNT wire was made, see Fig. \ref{WRe_CNT_wire}.  It can be concluded that the surfaces of these two wires are comparable. The yield stress of the CNT wire was tested at CERN~\cite{Mariet}, and it is higher than the yield stress of WRe wire of the same cross-section. WRe wire is difficult to fix mechanically and for the ZS they are clamped with rods onto the ZS anode. CNT wire is delivered on reels, and it is anticipated that a similar fixation system can be used. CERN has ordered $\diameter$\SI{100}{mm} CNT wires from Hitachi-Zosen (Japan) with the aim of testing the behaviour of these wires as an anode with high electric fields, since the behaviour of the CNT under HV is a key issue to the feasibility of its deployment as septum (anode) material.

\subsubsection{Low-Z components}
The activation of the equipment due to direct beam loss and secondary beam showers can be reduced by carefully selecting low-Z materials or low density geometries for equipment components. 

In the ZS the WRe wires and their anode supports also heat up when the beam is extracted. To avoid deformation of the anode support, it is made of INVAR (FeNi36, an alloy with a very low thermal expansion coefficient) for the first 3 (of total 5) ZS's. Changing the wires to low-Z materials will result in less heating of the wires and their support. In this case, it would be of interest to study the possibility of using low-Z materials for the anode support, such as aluminium. Aluminium anode supports are used in the CERN PS electrostatic septa. Most notably, it was used for the beam slicing septum PE.SEH31 for the Continuous Transfer extraction scheme up to 2017. 

\begin{figure}[htbp]
    \centering
    \includegraphics[width=\linewidth]{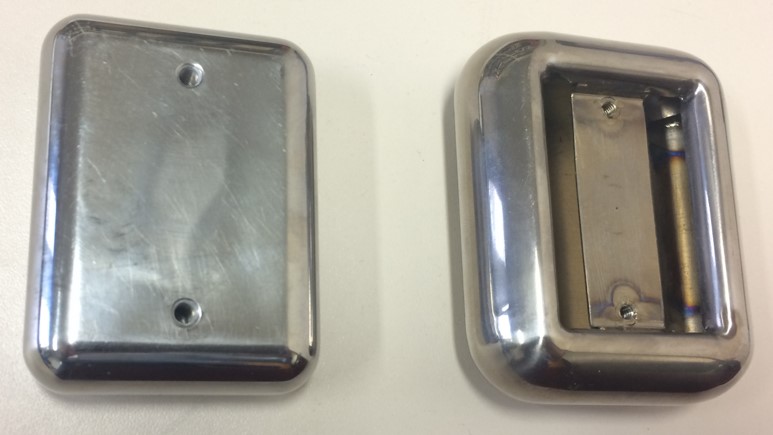}
    \caption{Stainless steel electrode machined from solid, \SI{432}{g}, (left) and hollow titanium electrode, \SI{66}{g}, (right) made using sheet metal working techniques.}
    \label{Hollowelectrodes}
\end{figure}

To achieve low-Z septa, one could also think of different manufacturing methods. The electrodes, i.e. the anode support and cathode, are presently made of solid materials.  Alternative approaches that could be explored are the manufacture of hollow electrodes. This could be achieved by either the conventional sheet metal approach, such as shown in Fig. \ref{Hollowelectrodes}, where a test electrode was made of thin titanium sheet instead of solid material. An alternative approach to explore if components such as an anode support can be manufactured by additive layer manufacturing. This may allow designs using material only where needed for mechanical strength and stability, reducing the weight and hence chance of particles to interact with the anode support. Also, the solid HV deflectors installed on the insulating support rods and HV feedthroughs could be made lighter using one of these techniques. 

\subsection{Anode straightness and positioning control}
\label{sub:AnodeStraightness}

The ZS wires are attached to precision machined anode supports. The straightness of these anode supports prior to wire installation was measured to be better than \SI{20}{\micro\meter} at ambient temperature \cite{Keizer}. Of the 5 successive ZS's, the first 3 anode supports are made of Invar and the last 2 made of stainless steel. This was done to preserve as much as possible the straightness during extraction, when beam is lost on the wires and the anode supports are heated. To limit the beam loss on the first ZS's, the first 2 ZS's employ $\diameter$\SI{60}{\micro\meter} wires, while the last 3 are equipped with $\diameter$\SI{100}{\micro\meter} wires. 

The wires are tensioned using springs. These springs ensure a tensile force of approximately \SI{1}{\newton} for the $\diameter$\SI{60}{\micro\meter} wires and \SI{5}{\newton} for $\diameter$\SI{100}{\micro\meter} wires. This equates to roughly 20\% of their breaking strength. The wires are subjected to a force due to the electric field and are deflected towards the cathode. This deflection is proportional to the tension applied to the wire, to the wire diameter and to the electric field. Consequently, the $\diameter$\SI{60}{\micro\meter} wire will be deflected by more than the $\diameter$\SI{100}{\micro\meter} wire. A simple analytic calculation, neglecting friction forces, predicts a deflection of \SI{75}{\micro\meter} and \SI{27}{\micro\meter}, for the $\diameter$\SI{60}{\micro\meter} and $\diameter$\SI{100}{\micro\meter} wires respectively, at an electric field of \SI{11}{MV\per\meter}~\cite{balhan}.

A measurement of the wire deflection due to the electric field was carried out using the following setup. A ZS equipped with a DN150 viewport on the upstream flange was put under vacuum. An optical system was installed to allow the observation of the radial displacement of a wire due to the electrical field. The cross hair of a theodolite was aligned with a wire. Once this wire was subjected to the full electric field the wire displacement could not be measured directly due to the angle between the wire array and the line of sight. Therefore, using the radial displacement system of the anode, the anode position was adjusted to line up the wire with the previously set cross hair of the theodolite. This radial displacement then yields the displacement of the wire.

These measurements yielded a radial displacement of \SI{105}{\micro\meter} for the \SI{60}{\micro\meter} wire, and \SI{45}{\micro\meter} for the \SI{100}{\micro\meter} wire, both for a main field of \SI{11}{MV\per\meter}.  Additional measurements on wires on the extremities of the anode support (not subject to the field) confirmed that they did not move, i.e. their displacement was \SI{0}{\micro\meter}, and as such validated the measurement method. 

\begin{figure}[htbp]
    \centering
   \includegraphics[width=0.7\linewidth]{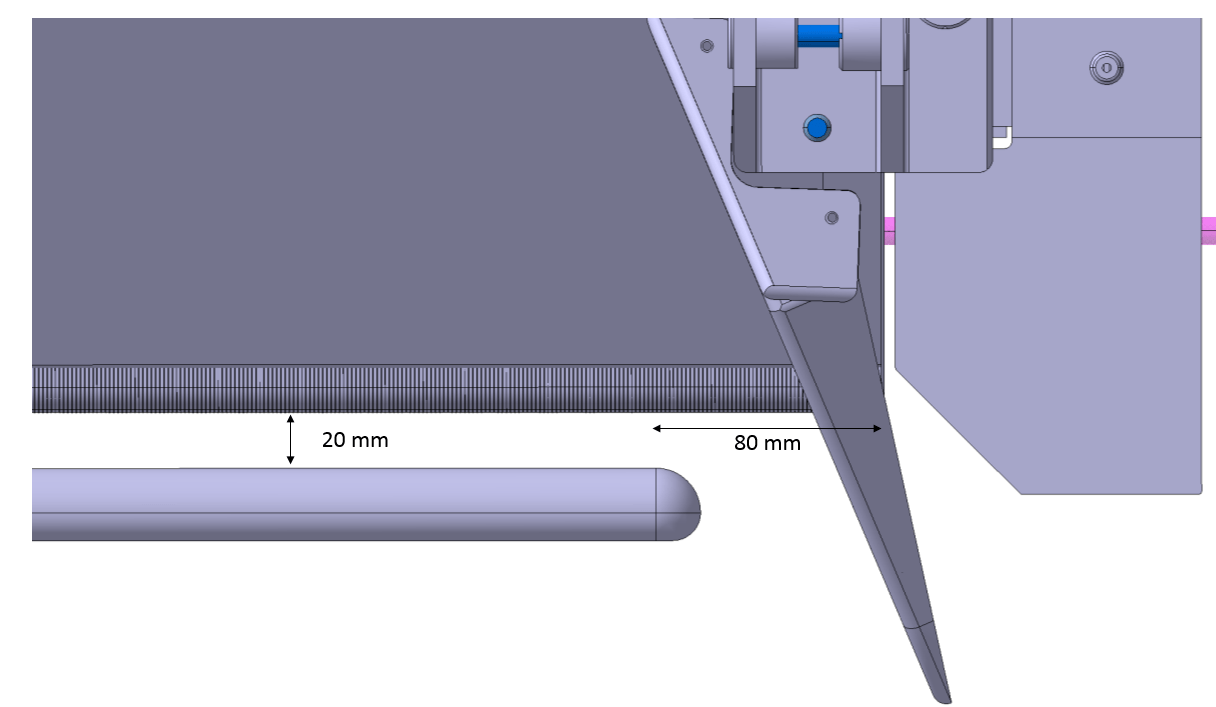}
    \caption{Top view of the ZS electrode extremities, showing the cathode (bottom), the edge of the anode support (top) and a field deflector (diagonally placed, on the right).}
    \label{Fig:CatiaZStopview}
\end{figure}

\begin{figure}[htbp]
    \centering
   \includegraphics[width=0.7\linewidth]{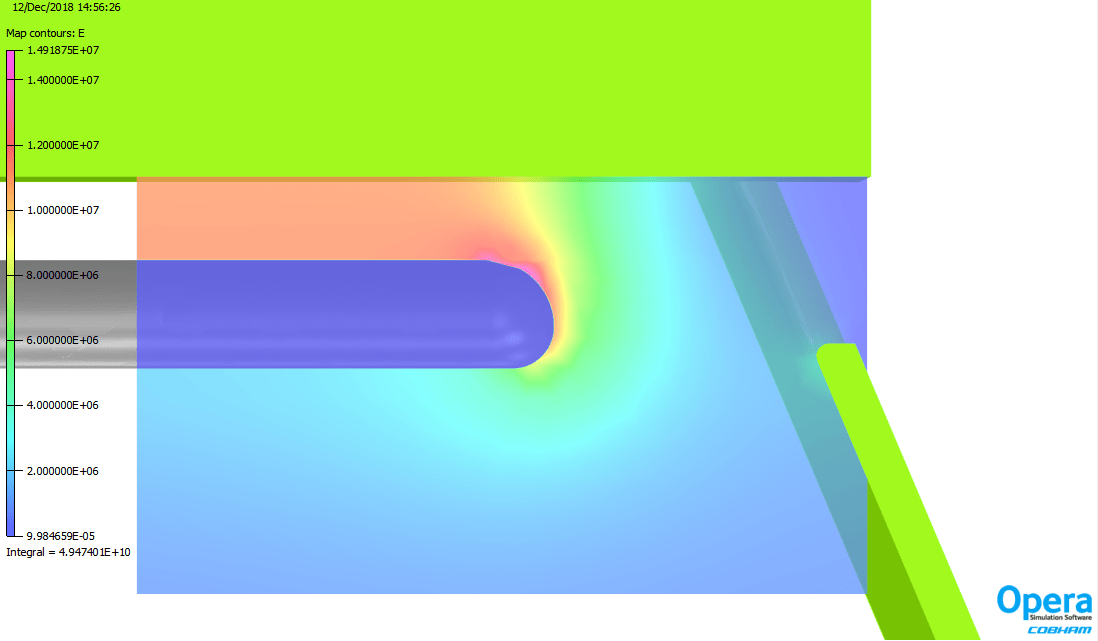}
    \caption{The radial electric field in the mid plane near the extremity of the ZS anode.}
    \label{Fig:ZSendField}
\end{figure}

\begin{figure}[htbp]
    \centering
   \includegraphics[width=0.8\linewidth]{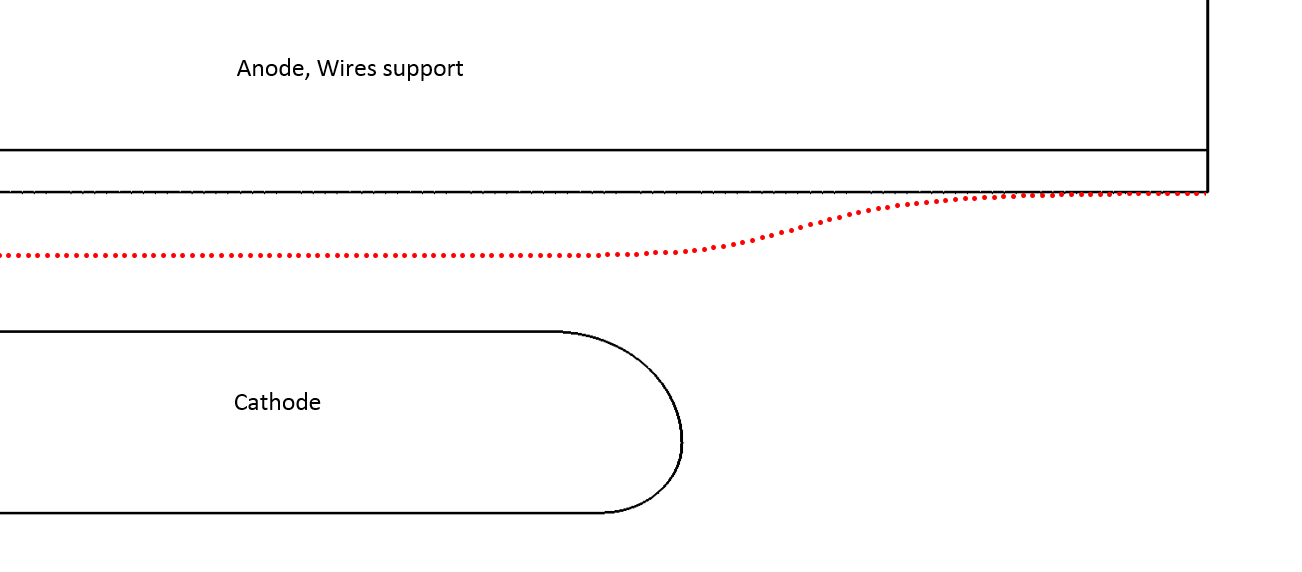}
    \caption{The wire displacement near the extremities of the anode support (not to scale).}
    \label{Fig:WireDisplacement}
\end{figure}

As can be seen in Fig.~\ref{Fig:CatiaZStopview} the wire array is longer than the cathode. The field on the central portion of the anode supports is uniform, but the wires installed close to the extremities of the anode support are subject to a gradually decreasing field (see Fig.~\ref{Fig:ZSendField}). These regions are about 80 mm upstream and downstream of each ZS. The subsequent wire displacement is shown in Fig.~\ref{Fig:WireDisplacement}. Here it can be observed that the effective septum thickness is increased by the wire deflection due to the electric field.   

Since the end field shape is well known at the nominal position, a deliberate anode wire offset (by machining the anode support) could be considered to compensate this effect. It must be noted that the perfect wire alignment can only be obtained for a predetermined (nominal) cathode voltage and gap width.
In addition, the leak, or stray, field penetrating the anode wires is being investigated as another source of the large measured effective thickness.

The actual motorisation system of the anodes is based on DC motors with position feedback using potentiometers. This allows for a positioning accuracy and reproducibility of about \SI{25}{\micro\meter}. Since the power is switched off after displacement, even if a drift is observed on the potentiometer, the anode position remains unchanged.

To obtain a more accurate positioning system, one could consider replacing the present system with a new system using radiation-hard stepping motors and encoders. It is expected that this could improve the positioning accuracy to \SI{5}{\micro\meter}. To allow the verification of the anode positions when the ZS's are installed in the tunnel and are under vacuum, external alignment references have been added directly on the anode shafts. By installing targets on these supports an accurate measurement referring directly the upstream and downstream mechanical wire position can be performed. The advantage of this method is that it allows a local measurement of the wire position independent of the tilt error of the installed magnet and the tank deformation due to the vacuum forces. Unfortunately, this approach will not allow an online read out, but will require the help of surveyors.

\subsection{Interventions}
\label{sub:Interventions}

\subsubsection{Present case study}

Since the lessons learnt from the high dose levels in LSS2 during 2015, it has become policy that no major in-situ repairs are carried out on the ZS. It is now preferable to replace the ZS tanks with an operational spare to limit dose and to preserve HV performance. During the Extended YETS of 2016 three ZS tank exchanges were carried out with dose to personnel improving each time from the experience gained, which was used to improve training and procedures. Ultimately, the goal is to respect CERN's As Low As Reasonably Allowable (ALARA) guidelines with hard-limits presently set for Level 2 at a collective dose of \SI{5}{\milli\sievert} for all participants, per intervention, with far stricter limits on individuals. As explained earlier, the exchange of the second ZS tank on 19 February 2016 has been used as a case study to estimate doses and cool-down times required in future operational scenarios using the dose measured on the ionisation chamber PMIU.202 and the \SI{1.7}{\milli\sievert} collective dose to personnel that was recorded for the intervention. Dose measurements at PMIU.202 are used to scale the predicted dose to an intervening team using this reference.

\subsubsection{Remote handling}

Remote handling during the exchange of a ZS tank offers a significant reduction in dose to personnel and potentially to less machine downtime. First steps have already been made in this direction with the recent ZS tank exchanges being partially carried out with remote handling techniques. It will be important to include the required modifications in the design of new devices to fully benefit from remote handling techniques in the future.

\subsection{Radiation effects to equipment}
The Replacement of Irradiated and Ageing Cables (RIAC) working group was established in 2009 to establish a cable replacement strategy based on quantifiable ageing of cables. Cable samples were placed in the CERN accelerators and a plan was defined for their collection and testing during the following years, up to the Long Shutdown 5. The working group was discontinued in 2014, but cable samples installed in LSS2 were analysed in 2017 \cite{DorisReport} and the results indicated clearly that the cables in LSS2 were severely degraded by ageing and radiation. In particular, cross-linked polyethylene (XLPE) cables used to supply the ZS tanks with HV from the surface are prone to radiation damage. This cable is split in 2 main parts: 
\begin{itemize}
\item Two cables (of which 1 spare) in the shaft from BA2 (surface building where the HV generator is installed) to a junction box where the cable enters the tunnel. These 2 cables were installed during the construction of the SPS in the 1970’s;
\item Two (of which 1 spare) cables in the tunnel from the junction box (downstream of the MSEs in LSS2) along the extraction equipment in a cable tray towards a distribution box upstream of the ZS. 
\end{itemize}

After 10 years of operation, in 2008, the operational cable in the tunnel broke down and operation resumed after having switched to the spare cable. In the subsequent Technical Stop both the operational cable in the tunnel and the spare were replaced by new cables from the same production batch as the previous ones (manufactured around 1987). In 2018 the operational cable in the tunnel broke again, and switching to the spare cable permitted finishing the run with only 21 hours of down-time for this fault. Since this cable was from the same manufacturing batch as the cables installed previously, this indicates clearly that the cable lifetime is reduced to approximately 10 years due to the radiation in LSS2. This is even more evident when taking into account the cable that has been installed in the shaft for around 50 years and not subject to radiation from the accelerator except for its final few metres.

Historically, the extraction regions were re-cabled every 10 years. The cable sample campaigns launched by the RIAC WG have not been able to demonstrate that this frequency is unsuitable. More concretely, the fact that the XLPE HV cable failed in 2008 and 2018 show clearly that little margin exists for recabling the extraction region. However, due to lack of resources, during LS2 only the ZS region will be recabled since new upgraded ZS tanks (in the context of LIU-SPS) need to be installed. The cables in the remainder of the extraction region (for example around the MS tanks) will only be recabled at a later date (likely LS3). The consequence of this situation is that no equipment will be changed in these other regions, nor other major works will be done, to limit the risk of damaging the degraded cables. 

Concerning radiation damage to feedthroughs, one can observe that the number of feedthrough failures has dropped significantly since the early 2000s. This coincides with the disappearance of leptons in the SPS, hence the reduced amount of synchrotron radiation to which the feedthroughs were subjected. 

\subsection{Induced radioactivity and dose to personnel}

\begin{figure}[htbp]
    \centering
   \includegraphics[width=0.9\linewidth]{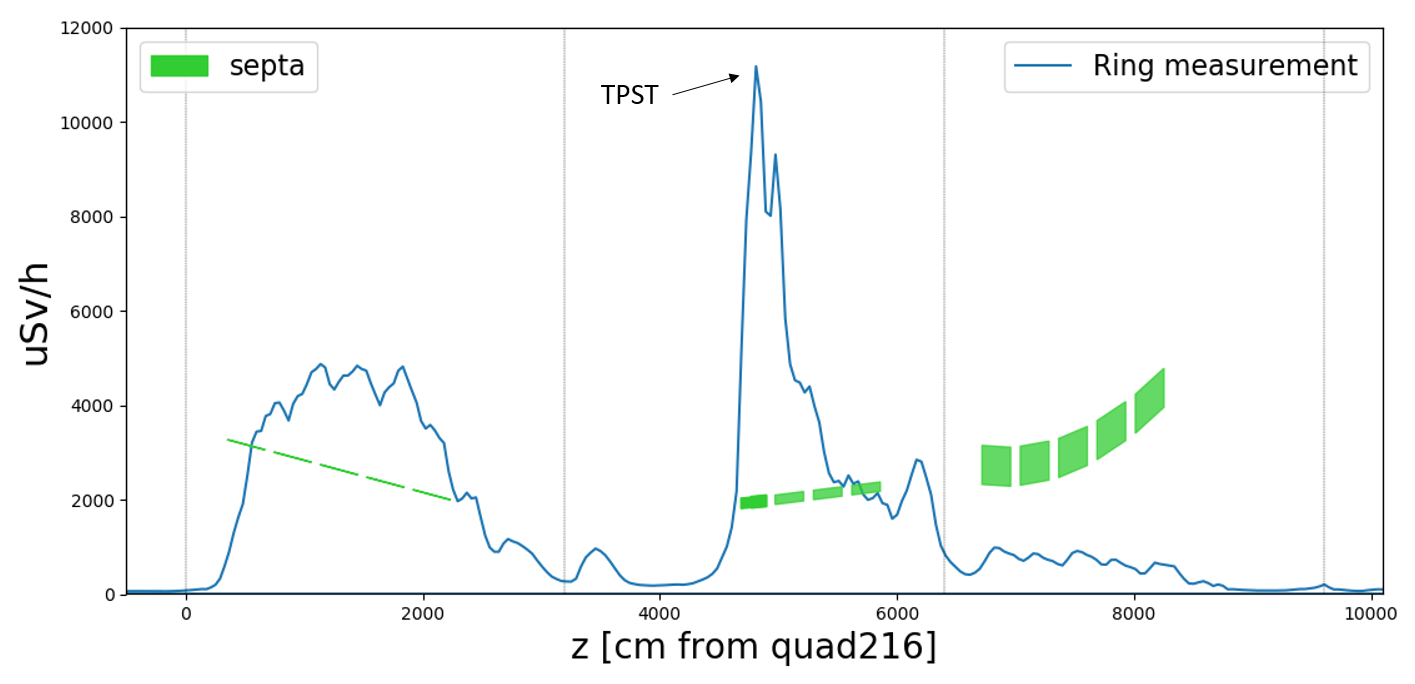}
    \caption{Dose rate profile of LSS2 measured 30 hours after beam stop on the 24th of October 2017. Radiation hazard most prominent at ZS's, TPST and MST tanks.}
    \label{LSS2profile}
\end{figure}

The slow extraction process in LSS2 provokes losses and induces radio-activation most prominently at locations where the septa present an aperture restriction for the beam. Figure \ref{LSS2profile} shows the residual dose rate profile measured 30 hours after beam stop 1 metre from beam axis at the end of the 2017 run. The measured dose rate profile highlights that the most critical elements for residual activation hazard of the extraction equipment are the ZS tanks, the TPST and the MST tanks.

\subsubsection{The electrostatic septum materials}
\label{subsub:ZSmaterials}

\begin{figure}[htbp]
    \centering
   \includegraphics[width=0.5\linewidth]{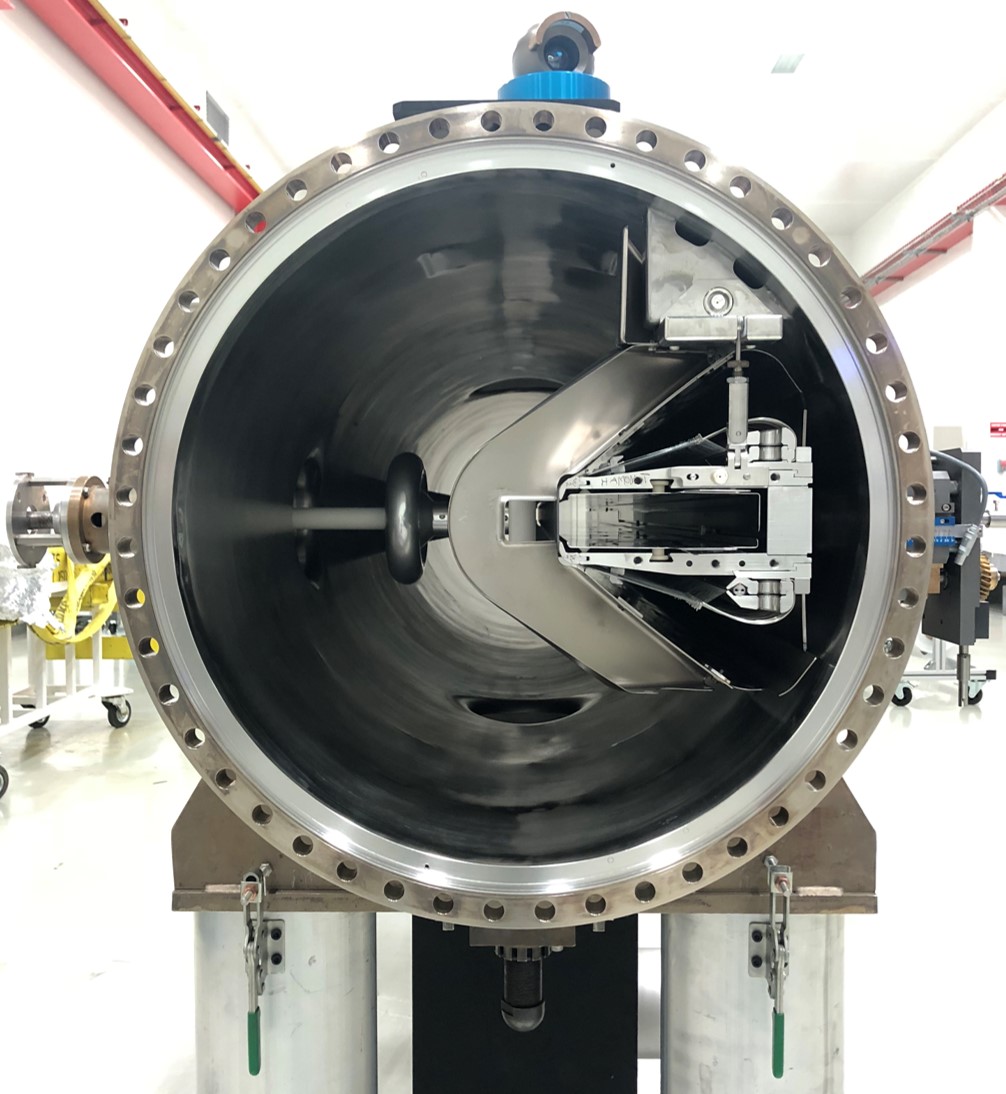}
    \caption{ZS septum, showing circular stainless steel vacuum vessel and the INVAR C-shaped anode support on the right inside the vacuum vessel.}
    \label{ZStank}
\end{figure}

Presently at CERN, the electrostatic septa (ZS for example, see Fig. \ref{ZStank}) use mostly stainless steel (for vacuum vessels and mechanical supports), some aluminium (for the cathodes and HV deflectors) and Invar (as anode supports in the first ZS's).
These materials activate considerably, and the remnant radio-activation imposes severe constraints on the interventions as described above. Instead, low-Z materials could be selected for the vacuum vessel or even anode support to improve this. Experience with alternative materials exists, such as in the following examples:
\begin{itemize}
    \item J-PARC built and installed a septum with a titanium vacuum vessel, anode support and end-plates to reduce residual radioactivity \cite{Tomizawa2};
    \item LEP used aluminium vacuum chambers;
    \item Electrostatic septa in the PS (PESEH23/31) use aluminium anode supports.
\end{itemize}

To evaluate the potential of changing some of the ZS materials, the software package Actiwiz~\cite{actiwiz} was used to assess the impact on the radiological hazards. This software package developed at CERN allows a fast and simple analysis, and as such is very suitable for a first look into alternative materials for the different ZS components. Actiwiz provides hazard values (risk factors) allowing the linear comparison of activation triggered consequences for different materials and different irradiation cases. The higher the risk factor, the higher the activation. For the Global Radiation Risk an average value is provided of the material after activation of 1 day, 1 week, 200 days and 20 years. The operational values presented consider the sum of weighted dose rate contributions for 10 cooling times between 1 hour and 20 years, while for the waste only the cooling time of 20 years is taken into account. 

To this end a very simplistic model of the ZS was used. This model for the presently installed ZS is made up of:
\begin{itemize}
	\item Tank: 3.2m $\times$ $\diameter$\SI{0.6}{m} tank made of \SI{6}{mm} thick stainless steel 304L,  with 30 mm thick covers;
	\item Anode support: 3.1 m $\times$ 200 mm $\times$ 94 mm C-shaped support made of Invar or stainless steel.
\end{itemize}
Looking into alternative materials, the following assumptions were made for the tank:
\begin{itemize}
   \item Stainless steel: baseline with dimensions as stated above;
   \item Aluminium, but using 67 \% thicker material (i.e. 10 mm tank body, 50 mm covers), to take into account the reduced yield stress of aluminium compared to stainless steel;
   \item Titanium, but assuming 33 \% thinner material (i.e. 4 mm tank body, 25 mm covers). This choice probably gives an optimistic indication, since it does not take into account that Ti has a much lower Young’s modulus, and may need additional reinforcements to maintain the required rigidity.
   \end{itemize}
For the anode support, the dimensions are kept constant for all material alternatives, since these are driven by the space requirements, not the mechanical stress. The material properties assumed in the calculations are shown in Table \ref{Table:MatProperties}. In this table also the Global Radiation Risk in the vicinity of a 400 GeV/$c$ beam (at 10 cm lateral distance to the target) are listed for the operational case as well as for consideration of the disposal of the device as waste.

\begin{table}[htbp]
\centering
\caption{Material properties and radiation risk factors for different materials considered.}
 \label{Table:MatProperties}
 \begin{tabular}{l c c c c}
 \hline
  & INOX 304L & Al (6061) & Ti-A6-V & INVAR  \\
 \hline
Young’s modulus [GPa] &	200 & 69 & 120 & 148\\
Yield strength [MPa] &	200 & 120 & 760 & 483\\
Density [Mg/m$^3$] & 7.85 & 2.7 & 4.43 & 8.05\\
Thermal expansion [\num{e-6}/\si{\kelvin}] & 17 & 23 & 9 & 1.5\\
Global Radiation risk at \SI{400}{\giga\eVperc} (operation) & 1.6 & 0.27 & 0.94 & 2.13\\
Global Radiation risk at \SI{400}{\giga\eVperc} (waste) & 0.83 & 0.35 & 1.16 & 0.81\\
 \hline
 \end{tabular}
 \end{table}
   
To establish the radiological hazard factors for each tank/anode support material combination, the following approach was used: 
\begin{itemize}
    \item To calculate the Operational Hazard Factor: the volume of each topology (tank volume) is multiplied by the compound risk factor of each material;
    \item To determine the Waste Hazard Factor: a normalised hazard factor is calculated based on the weight of the anode and the tank ratio. 
\end{itemize}
   
\begin{figure}[htbp]
    \centering
   \includegraphics[width=0.95\linewidth]{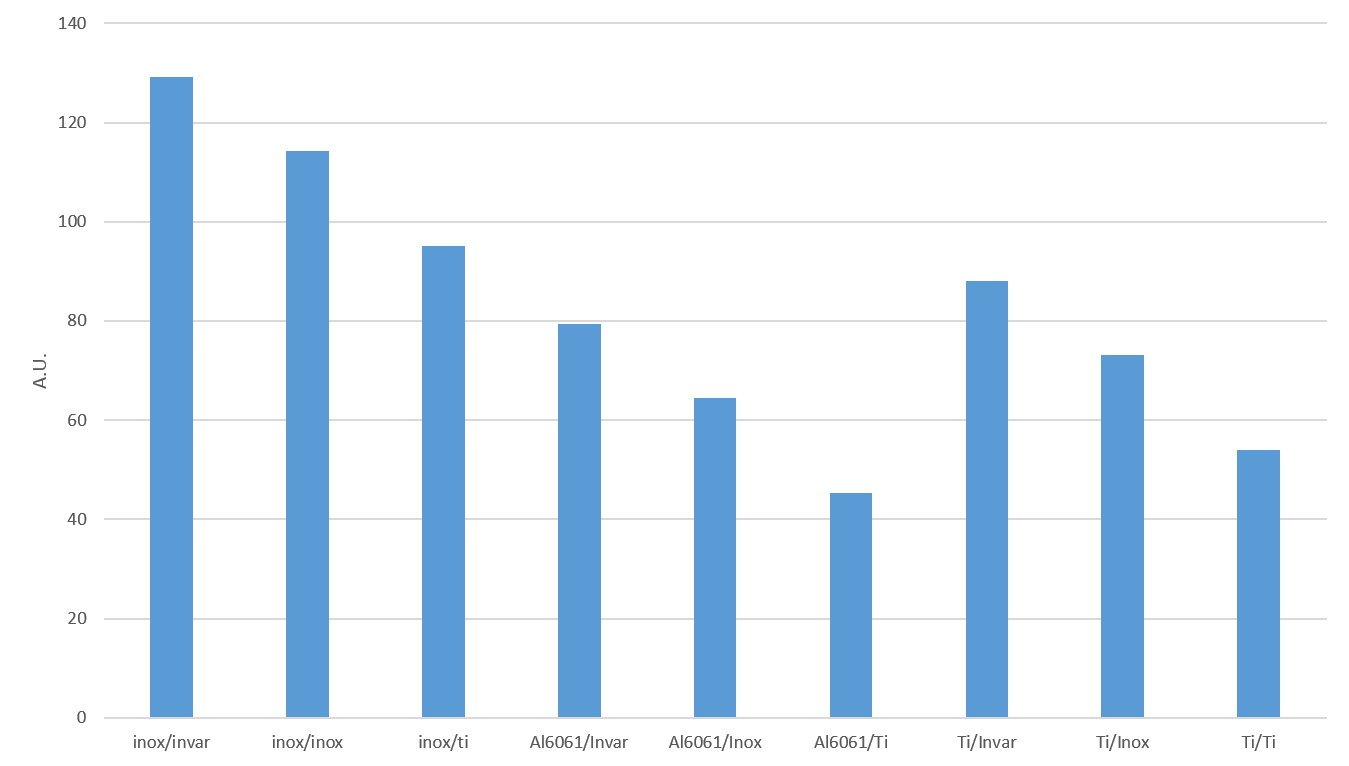}
    \caption{Global radiation hazard factor for a ZS tank in operation for different tank/anode support material combinations. Presently, ZS1-3: Inox/Invar, ZS4-5: Inox/Inox.}
    \label{Fig:OperationHazard}
\end{figure}

\begin{figure}[htbp]
    \centering
   \includegraphics[width=0.95\linewidth]{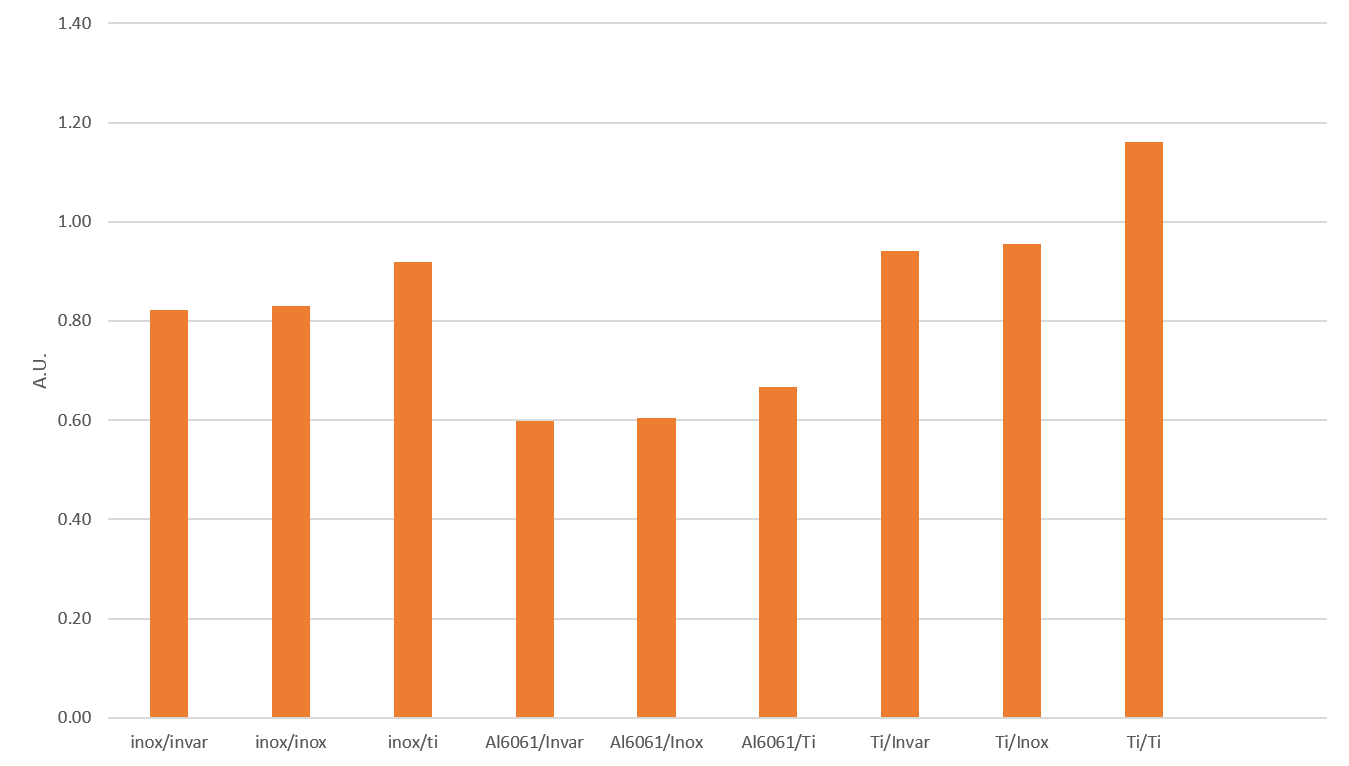}
    \caption{Normalised global radiation hazard factor for a ZS tank as waste for different tank/anode support material combinations. Presently, ZS1-3: Inox/Invar, ZS4-5: Inox/Inox.}
    \label{Fig:WasteHazard}
\end{figure}

The hazard factors for the different tank/anode support material combinations are shown in Figs.~\ref{Fig:OperationHazard} and \ref{Fig:WasteHazard}. This preliminary analysis shows that changing materials could be beneficial for the activation of the ZS's and further, more detailed studies were undertaken. 

\subsubsection{Material alternatives to reduce the radiation hazard in the extraction region}

Encouraged by the potential radiation hazard reduction of alternative materials, the newly developed LSS2 FLUKA model~\cite{bjorkman,Fluka1,Fluka2,Fluka3,bjorkman2} has been used to evaluate alternative materials while taking into account both the spatial complexity of the extraction beamline and the specific radiation fields produced during and after the extraction process. This section presents the material exchanges that were found to reduce residual dose related hazards of the radioactive elements of the extraction beamline. The lower Z material suggestions of Table \ref{tab:material} are either directly found to be advantageous through the FLUKA model, or indirectly by obtained particle fluence spectra with the model to be used as input for Actiwiz \cite{actiwiz}. Estimated hazard magnitude differences with Actiwiz are without geometrical considerations, such as component shadowing or the internal distribution of the radioisotopes within the components themselves.

\begin{table}[htbp]
\centering
\caption{Radiological beneficial component material exchanges for extraction equipment.}
\begin{tabular}{lrrrrr}
\hline
\textbf{Element} &\textbf{Component} & \textbf{Current material} & \textbf{Beneficial material} & \textbf{Dose rate reduction factor} \vspace{2mm} \\
\toprule
ZS & Anode wires          & Rhenium/ & Titanium or  & 0.2-0.6 integrated reduction  \\ 
 & &Tungsten & Graphite & after 30 hours. 0.1-0.001 anode   \\
  & & & &   contribution at 1 metre distance \\ 
   & & & &   (Figs. \ref{ZSseptamaterial} and \ref{ZSseptaTotalActivation})\vspace{1mm}\\ 
\midrule
ZS & Anode support       & Stainless steel & Titanium  &  0.25-0.33 after 1 week,\\ 
 & & or Invar& &   1 metre distance (Fig. \ref{ZSseptaSupport}) \vspace{1mm}\\ 
\midrule
ZS & Vacuum tank   & Stainless steel & AL6061   &  0.6 after 1 week, 1 metre  \\ 
 & & & &   distance (Fig. \ref{AluContainers}) \vspace{1mm} \\
\midrule
TPST & Vacuum tank        		     & Stainless steel & AL6061    &  0.8 after 1 week, 1 metre  \\ 
 & & & &  distance (Fig. \ref{AluContainers}) \vspace{1mm}\\
\midrule
MST & Vacuum tank		             & Stainless steel   & AL6061  &  0.5 after 1 week, 1 metre  \\ 
& & & &   distance (Fig. \ref{AluContainers}) \vspace{1mm} \\
\midrule
MSE & Vacuum tank        & Stainless steel & AL6061  &  0.7 after 1 week, 1 metre  \\ 
& & & &  distance (Fig. \ref{AluContainers}) \vspace{1mm}\\
\midrule
& Beam pipe          & Stainless steel & AL6061 &  Up to 0.1 upon contact \\ 
& & & & after 1 week (Fig. \ref{AluPipe}) \\
\hline
\\
\end{tabular}
\label{tab:material}
\vspace*{-1\baselineskip}
\end{table}

\begin{figure}[htbp]
    \centering
   \includegraphics[width=\linewidth]{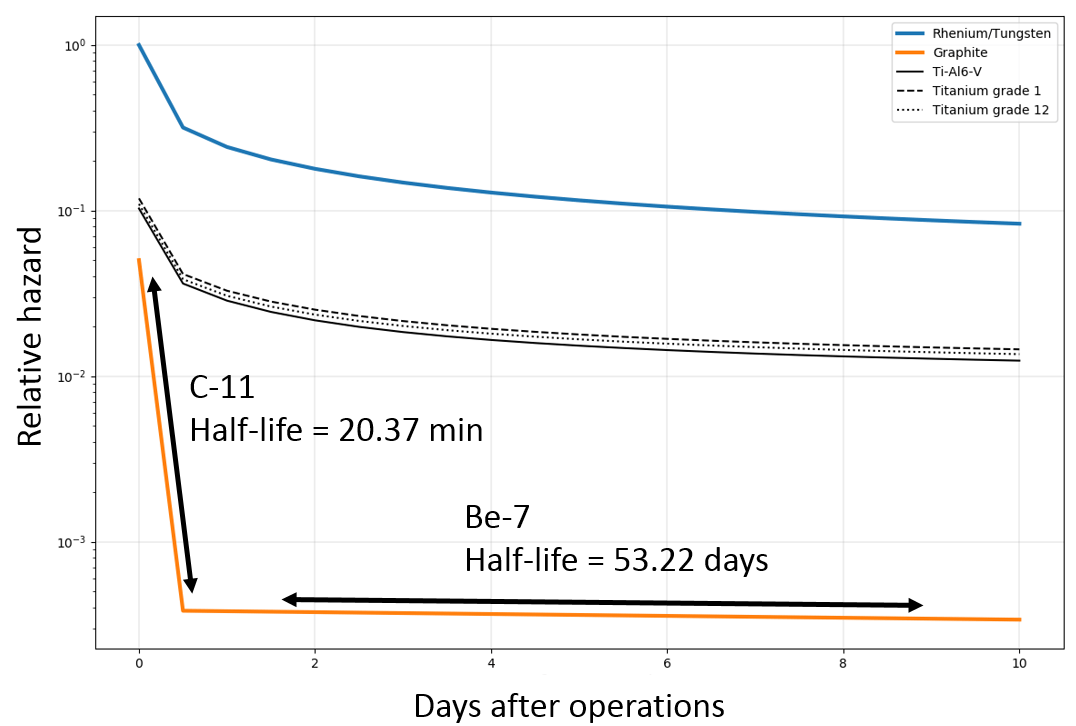}
    \caption{Actiwiz comparison of residual dose rate contribution for different ZS septa material at 1 metre distance assuming identical wire volume for each material choice.}
    \label{ZSseptamaterial}
\end{figure}

\begin{figure}[htbp]
    \centering
   \includegraphics[width=\linewidth]{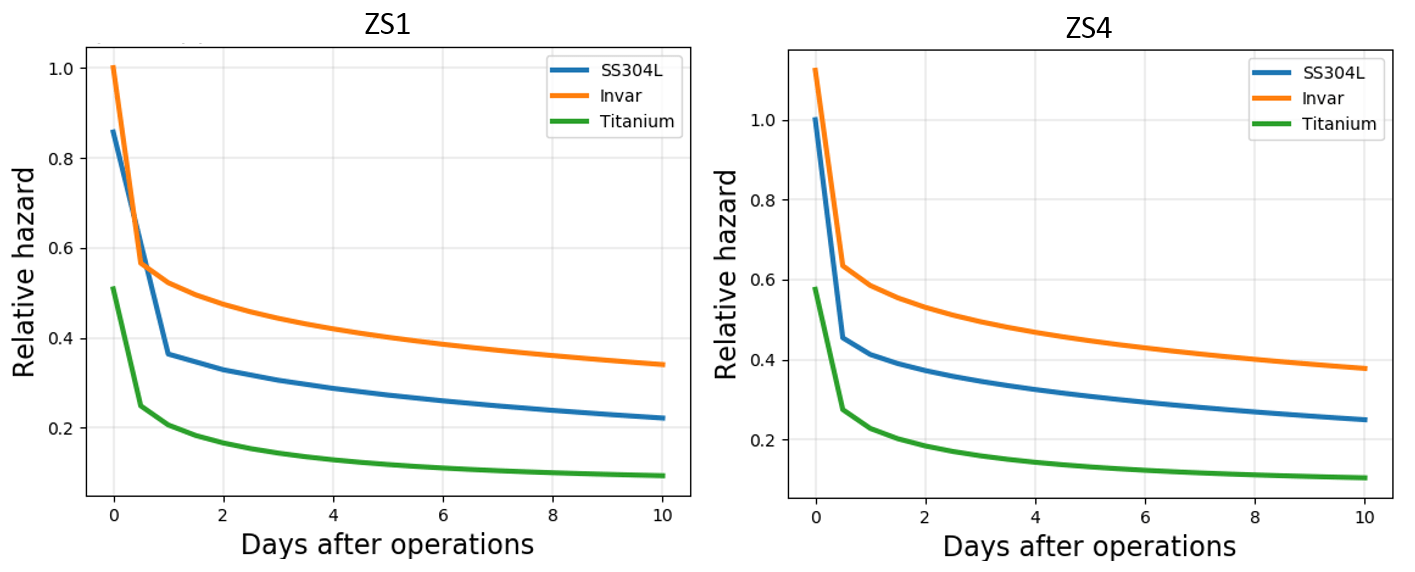}
    \caption{Actiwiz comparison of residual dose rate contribution for different ZS anode support material at 1 metre distance.}
    \label{ZSseptaSupport}
\end{figure}

Figures \ref{ZSseptamaterial} and \ref{ZSseptaSupport} show Actiwiz estimates of the impact that different material choices have on residual dose rates one metre from the ZS anode wire and ZS anode support. The septa material estimation are evaluated with particle fluences from separate FLUKA simulations, all of which consider the same volume wires averaged over the full ribbon volume as explained Sec. \ref{FlukaModel}. 

\begin{figure}[htbp]
    \centering
   \includegraphics[width=1\linewidth]{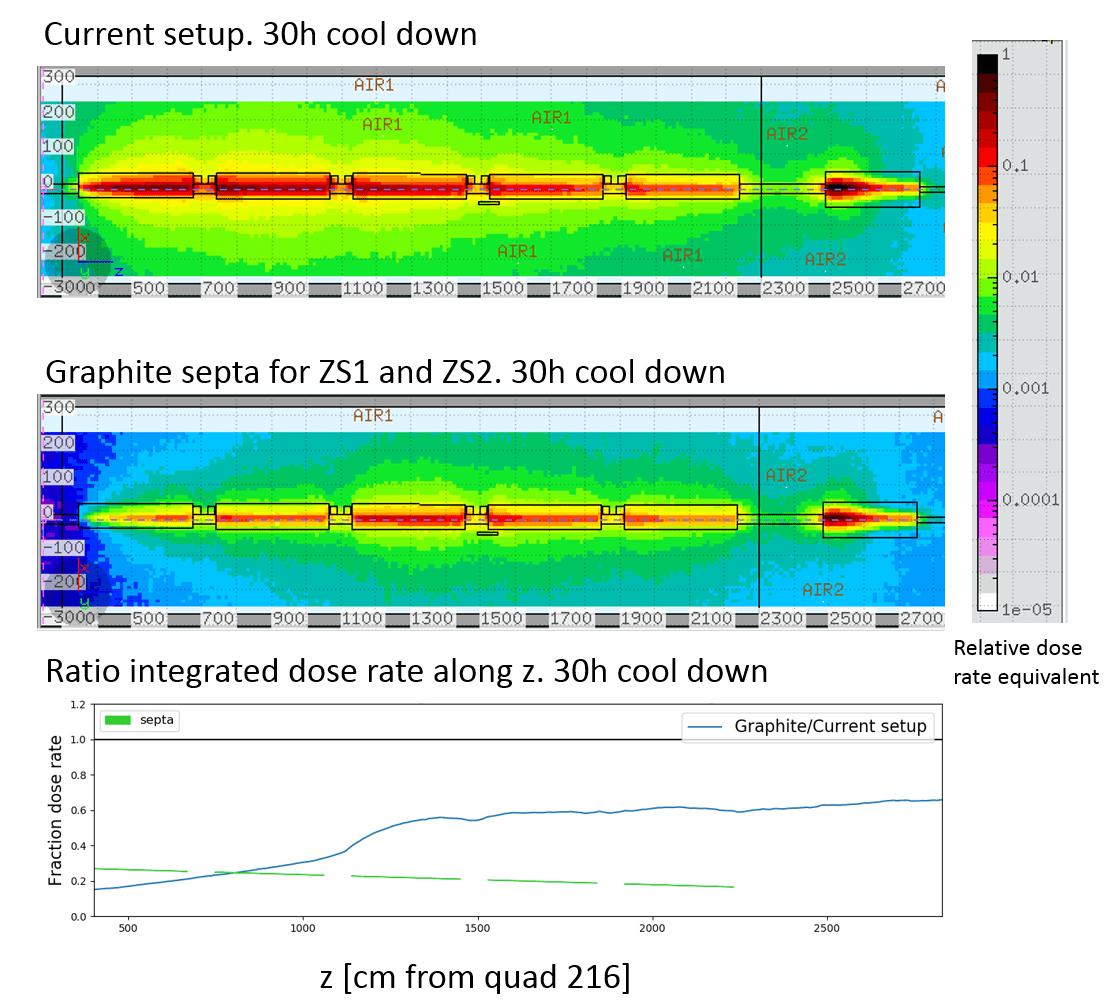}
    \caption{FLUKA residual dose rate comparison for exchanging wire septa material to graphite for ZS1 and ZS2 only. Reduced activation extends outside of exchanged wire septa. }
    \label{ZSseptaTotalActivation}
\end{figure}

Constructing the ZS anode wire septa of lower-Z material will both change its radiological contribution by a factor of 0.001 - 0.1 (Fig.~\ref{ZSseptamaterial}) and the overall residual dose rate in the ZS region by at least a factor of 0.6 in the case of carbon-based wires (in this case simulated as graphite), as shown in Fig.~\ref{ZSseptaTotalActivation}. The overall radiation hazard in the region will be decreased since fewer secondary particles are produced from the reduced probability of hadronic interactions between the beam and the nuclei of the wire septa. The produced nuclear fragments from nuclear interactions of the beam with lower-Z materials will be of decreased variety and less likely to further activate the machine by the release of fewer neutrons from the process of nuclear evaporation. The fact that the residual activation is reduced outside the wire septa itself is shown in Fig.~\ref{ZSseptaTotalActivation}, which compares FLUKA simulations of the current setup to a setup where the wire septa of ZS1 and ZS2 are replaced with graphite of the same volume. The comparison shows that all ZS elements and the downstream TCE element are activating less from the lower-Z septa material exchange of the first two ZS elements.

\begin{figure}[htbp]
    \centering
   \includegraphics[width=0.8\linewidth]{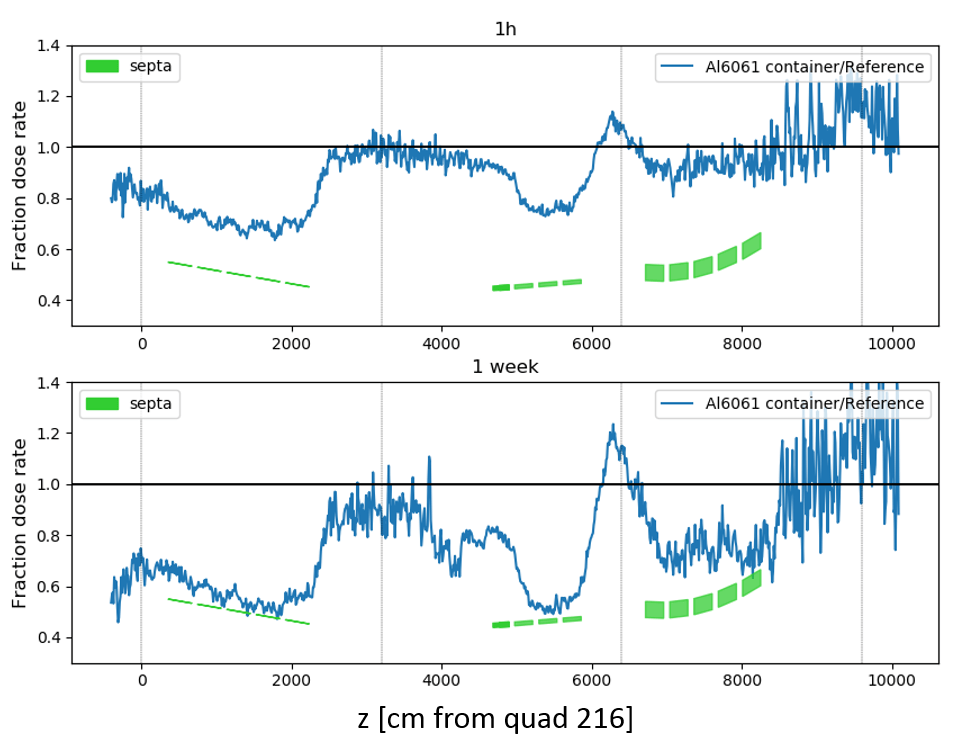}
    \caption{FLUKA estimate of dose rate reduction for Al6061 vacuum tank exchange for the ZS's, TPST, MSTs and the MSEs.}
    \label{AluContainers}
\end{figure}

\begin{figure}[htbp]
    \centering
   \includegraphics[width=0.8\linewidth]{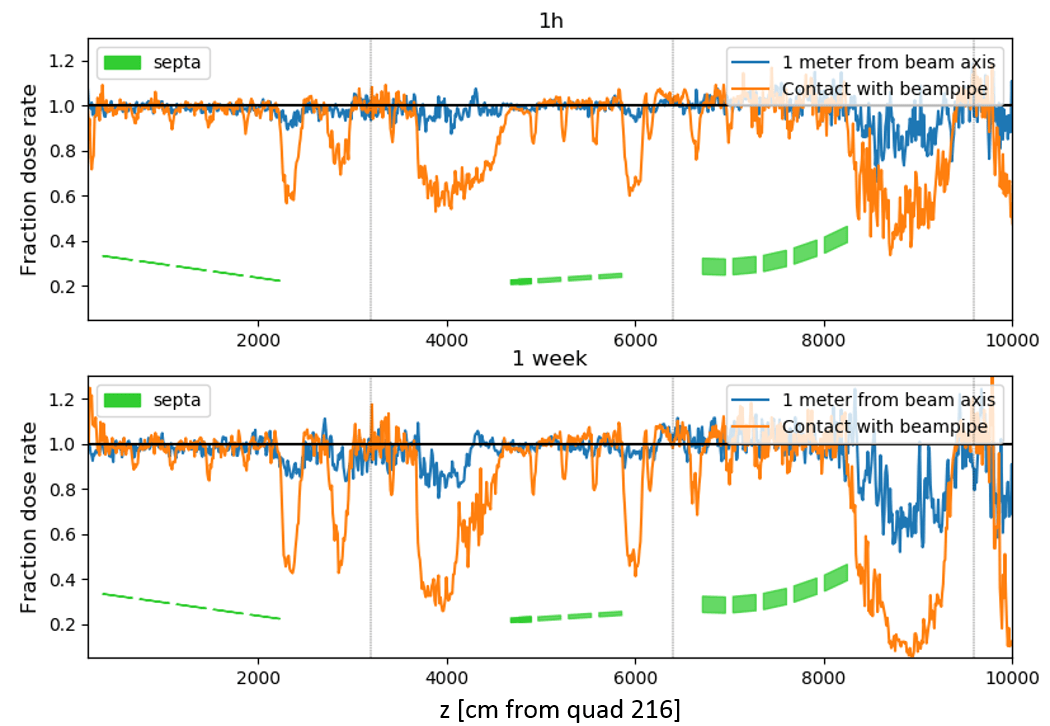}
    \caption{FLUKA estimate of dose rate reduction for Al6061 beampipe upon contact and 1 metre distance from beam axis.}
    \label{AluPipe}
\end{figure}

The aluminium alloy Al6061 was found to be favourable to stainless steel for the vacuum tank containers of the extraction elements and for the beam pipe in the region, see Figs.~\ref{AluContainers} and \ref{AluPipe}. The activation reduction of the Al6061 vacuum tanks is particularly useful since it can be used to reduce the residual activation hazard of the most critical elements of the extraction region without interfering with their functionality. Exchanging the beam pipe material to Al6061 will be the most radiologically beneficial in close proximity to the beampipe at locations where frequent work is performed in their vicinity.

\subsubsection{Shielding design and impact on radiation hazard}
\begin{figure}[htbp]
    \centering
   \includegraphics[width=1\linewidth]{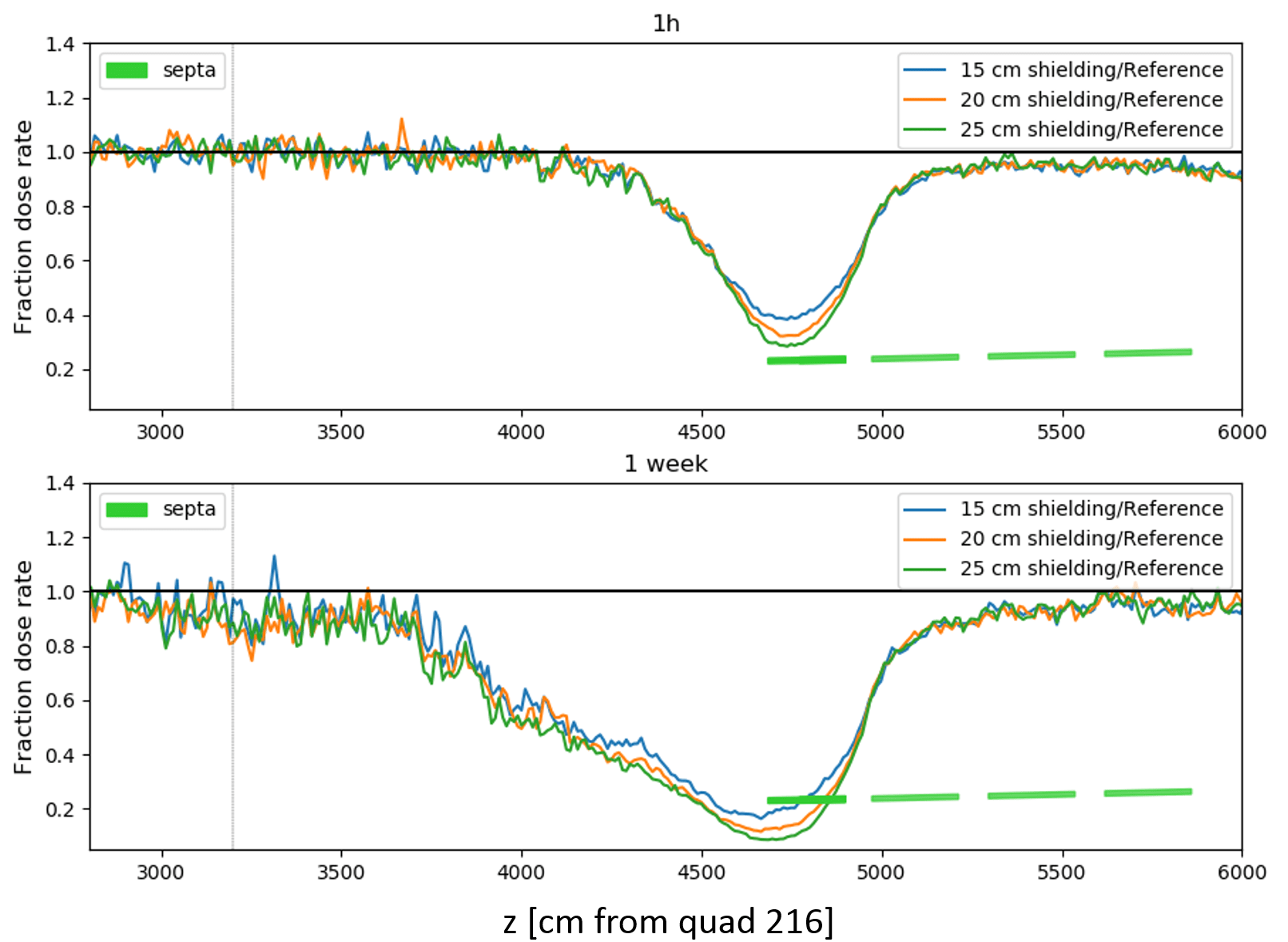}
    \caption{FLUKA estimate of residual dose rate reduction from marble shielding of TPST at 1 metre from beam axis for different cooling times.}
    \label{TPSTshielding}
\end{figure}

The extraction protection element TPST is currently protecting the downstream magnetic septa from stray particle showers provoked by scattering or interactions with the upstream ZS anode wires. The number of hadronic interactions with the TPST blade are sufficient to make this protection element the hottest part of the extraction beamline, however recent operational improvements in the set-up of the extraction are seeing this device cool year-on-year. Figure~\ref{TPSTshielding} shows that shielding the TPST within a marble encasement can reduce the residual dose rate 1 metre from beam axis by a factor of 10 after 1 week of cool down. Shielding the highly activated TPST addresses the most radioactive part of the extraction region.

\subsubsection{Conclusion}
The typical residual dose rate profile of the extraction region shows peaks for the first few ZS's, and around the TPST and first MST. Studies have shown that changing the materials used for the septa can be beneficial to reduce this dose rate. As such, changing all vacuum tanks and beam pipes from stainless steel to aluminium would be favourable, and could potentially reduce the activation of the septa by around 25\%, with the subsequent reduction in dose to personnel intervening on this equipment for maintenance and or repairs. Before aluminium vacuum vessels can be designed however, the following aspects need to be considered: 
\begin{itemize}
  \item Aluminium Conflat Flanges (CF) made are being developed at CERN but are not yet commercially available;
  \item Bi-metal CF flanges (stainless steel/ aluminium) are commercially available;
  \item Larger flange diameters or aluminium flanges to replace ‘Suchet’ or ‘Wheeler’ flanges will have to be developed. 
\end{itemize}
Replacing the tungsten-rhenium wires of the first two ZS septa by carbon wires appears very promising, and will reduce the activation of all ZS's due to the reduced production of nuclear fragments during the beam interaction with the wires. Before this can be considered, the compatibility of carbon wires as anode material for the ZS septa needs to be demonstrated. To this end Carbon Nano Tube (CNT) wires have been ordered, and will be tested in the foreseeable future. 

Using titanium instead of Invar or stainless steel as an anode support could be a good alternative, if combined with carbon wires and after further studies have demonstrated that the energy deposition in the titanium anode supports is sufficiently reduced compared to the present Invar/Tungsten anode to avoid the anode support from deforming in operation. 

To reduce the residual dose rate for personnel near the extraction protection element TPST the best option is to shield the device using 15 to 25~cm of marble, depending on the limitations of the girder support on which the TPST is installed together with the MSTs. 

\section{Conclusion}

The SPS slow extraction study for the BDF has actively developed and tested methods to reduce the prompt beam loss per proton during extraction. A variety of approaches have been applied to the extraction process, including hardware and controls, and deployed on the SPS for tests and put into operation, both with the shorter SPS BDF (SHiP) cycle and the longer NA cycle. 

The Q-sweep method used for slow extraction in the SPS since its construction has the disadvantage that the machine optics changes through the spill as the machine tune is varied. A new type of slow extraction was developed and deployed operationally, where the optics is kept constant in normalised strength while the whole machine momentum is ramped. This Constant Optics Slow Extraction (COSE) has several advantages over the Q-sweep and is now systematically used for regular NA operation. It is also a pre-requisite for exploiting the full potential of the diffusers by minimising the angular spread of the beam presented to the septum. In addition, the recent improvements in the quality of the spill have been important, allowing the experiments to better exploit the extracted flux. Looking to future high intensity operation, further improvements in the spill quality will be important to ensure every extracted proton is exploited to its full potential.

The alignment of the ZS anodes is also a crucial factor in the overall beam loss, since the effective width determines both the absolute beam loss and the potential gain factor from the shadowing method. The control of the alignment was improved to a resolution below \SI{50}{\micro\meter} and numerical optimisers have been simulated and tested with beam to align the 5 anodes to the extracted beam, bringing the alignment time down from over 8 hours to 40 minutes. Measurements coupled to simulations of the extraction efficiency and tests with the diffusers have provided strong evidence that the effective thickness of the ZS is far larger than expected. Investigations are underway to understand the source of this discrepancy.

Both passive scatterers and thin, bent silicon crystals have been been developed and prototypes tested to reduce the proton density at the ZS wires during extraction. For the passive diffuser, a \SI{260}{\micro\meter} wide, \SI{30}{\mm} long array of Ta wires achieved a loss reduction of 15\%, consistent with an effective ZS width of 500-\SI{600}{\micro\meter}. For the bent silicon crystal of \SI{780}{\micro\meter} width and \SI{2.5}{\mm} length, the large channelling angle of \SI{175}{\micro\radian} allowed a loss reduction of slightly over 40\%, again consistent with a ZS width of around \SI{500}{\micro\meter} and with the angular spread in the beam expected with the COSE extraction. Both diffuser types were tested with the full operational beam intensity of approximately $3 \times 10^{13}$ protons per spill, demonstrating that the concept of ZS shadowing is stable and reproducible.

In addition, a separatrix folding technique was tested successfully to reduce the beam loss at the ZS. In this method, the extraction sextupoles that govern the speed of diffusion across the ZS wires are increased in strength to reduce the particle density and losses, while octupole magnets are used to slow the diffusion speed at higher amplitude, folding the extracted beam back to avoid beam losses on the ZS cathodes presented by the limited ZS gap size. This method demonstrated a beam loss reduction of slightly over 40\%. Most importantly, it was also tested in combination with the crystal aligned in channelling and shadowing the ZS. The combination of methods gave a loss reduction factor at the ZS BLMs of up to $\sim$3.1 and demonstrated that some of the methods for loss reduction, activation and personnel dose reduction can be accumulated directly, with a multiplicative gain.

It is expected that after further optimisation of the different concepts presented in this study, a factor of 4 reduction of the prompt extraction losses in LSS2 is within reach for operational scenarios. Since the beam for BDF will pass through the gap of the splitter system in an essentially loss-free transport, this gain will mean that the present loss levels both in the extraction channel and in the splitter region can be maintained for the simultaneous delivery of $4.0\times10^{19}$ protons to SHiP and $1.0\times10^{19}$ protons to the NA (via the lossy splitting). Nevertheless, studies will be launched in 2019 to investigate whether there are ways to reduce the losses per proton during the splitting process to the NA.

The radiation dose to cables, extraction equipment and to personnel carrying-out interventions depends on both the beam loss at extraction and the total number of protons extracted, and remains a limitation. In addition to developing the loss reduction concepts, a significant effort is on-going in order to investigate new low-Z hardware to reduce losses and minimise the radiation hazard in the extraction region, as well as ensuring interventions can be carried out with a minimum dose to personnel.

The reduced radiation hazard from material exchanges and shielding designs looks promising for the future. Implementing these improvements will allow the SPS to increase beam intensities
delivered to the North Area without losing availability from increased activation of the machine.

\FloatBarrier
\printbibliography[heading=subbibliography]

 \chapter{Transfer to target}
\label{Chap:Transfer}

\section{Introduction}

\begin{figure}[htbp]
\centering
\includegraphics[width=16cm]{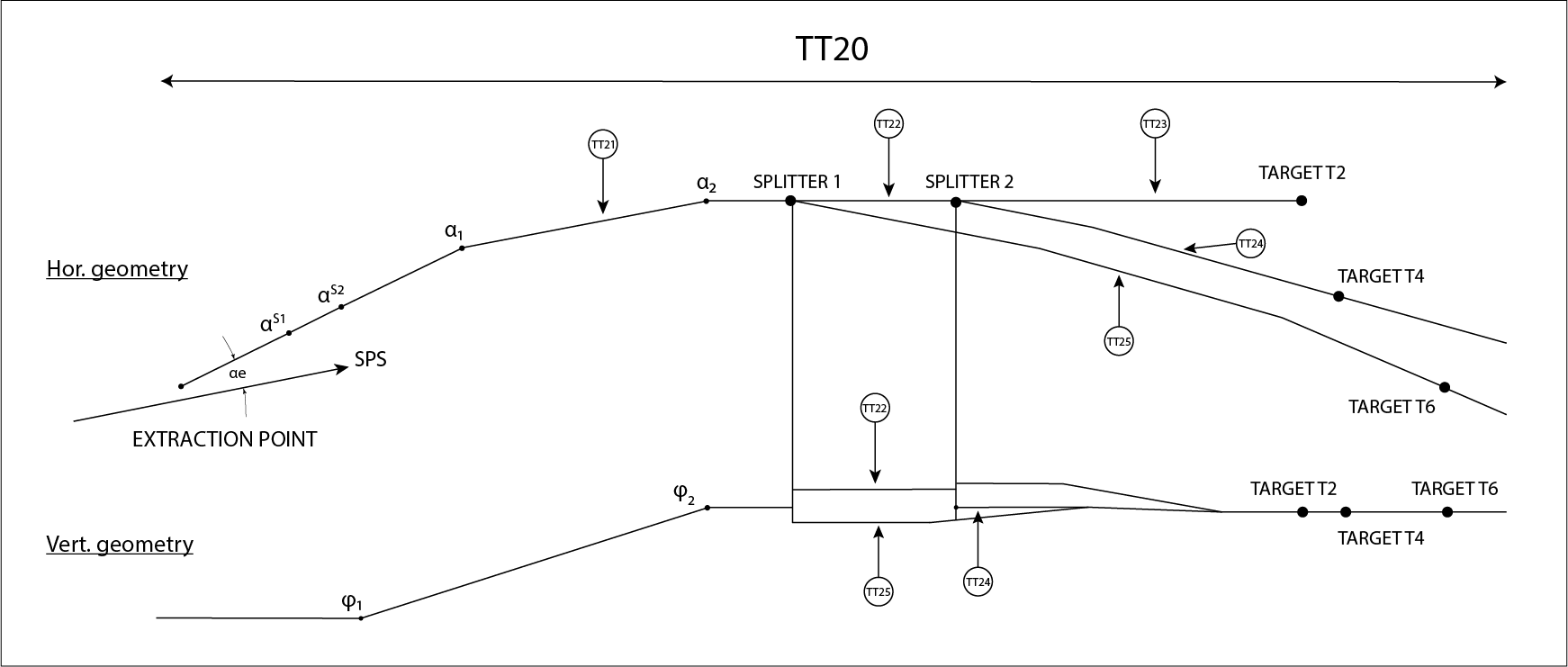}
\caption{Present TT20 layout showing the position of the splitters and North Area targets -- the proposed new beam line would branch off towards the BDF at splitter 1}
\label{transf:fig:TT20-layout}
\end{figure}

The Beam Dump Facility is foreseen to be located on the side of the present SPS North Area towards the Jura mountains (Fig. \ref{fig:Intro-location}). 
This position allows the re-use of the existing slow extraction channel from the SPS and the TT20 transfer line to the North Area up to the first splitter (MSSB2117), 
which is in total about 600\,m of existing beamline. 
The present layout of the TT20 line is shown in Fig. \ref{transf:fig:TT20-layout}.

The existing splitter, consisting of three individual magnets, has to be replaced by a new laminated and bipolar design, which will permit the deflection of the entire beam to the left into a new beam line leading to the BDF target complex while maintaining the possibility to operate the North Area in the present mode by splitting the beam into two parts sent towards the T2 and T6 targets. 
The new splitter will be able to switch between the destinations BDF and North Area on a cycle-by-cycle basis. 
The new beamline is approximately 360 m in length and will transport, enlarge, and dilute the slow extracted beam onto the target.


\section{Extraction line optics}

\subsection{Beam parameters} \label{transf:sec:beamparms}
\begin{figure}[htbp]
    \begin{subfigure}[t]{.48\linewidth}
        \centering
        \includegraphics[width=\linewidth]{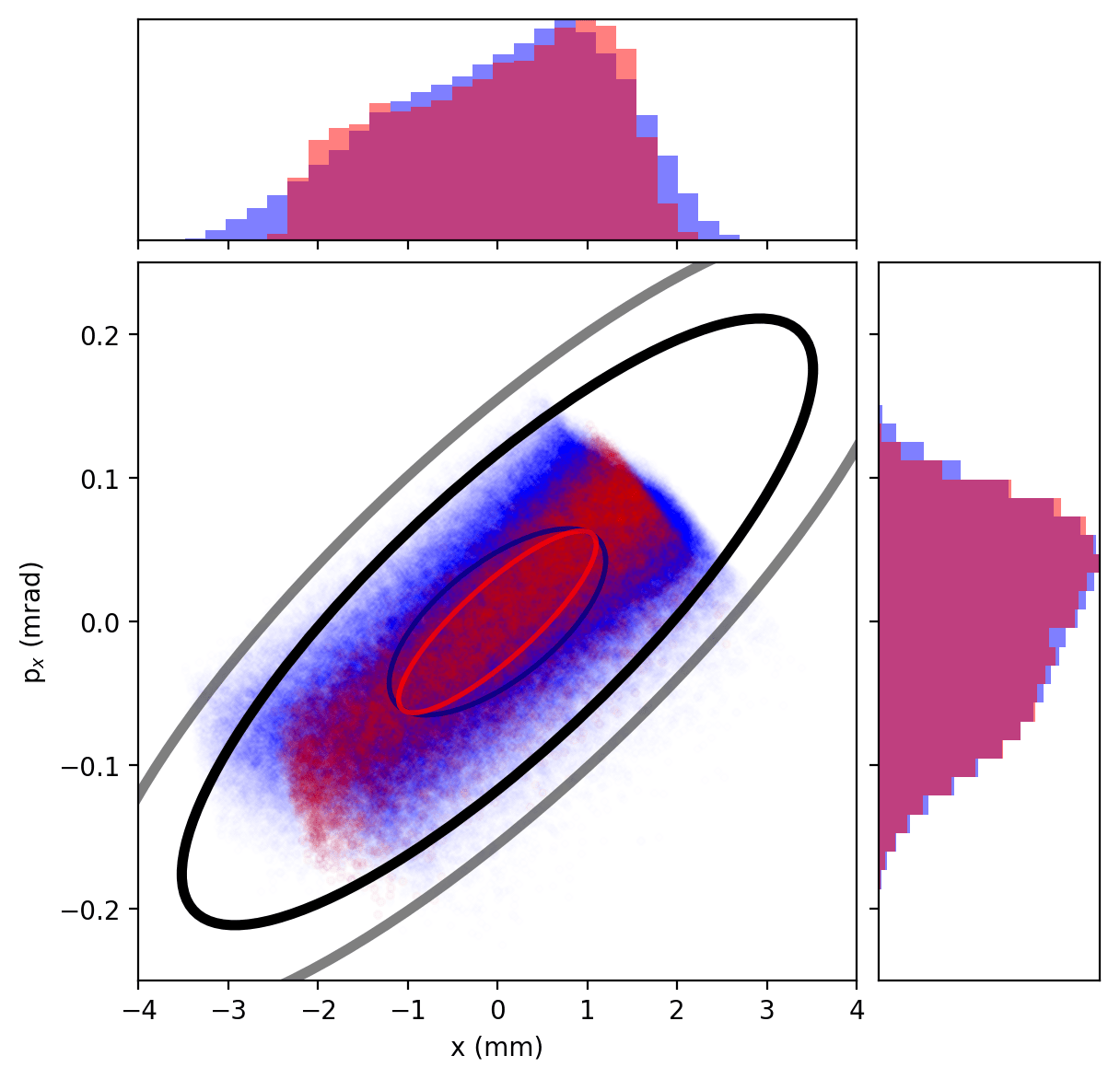}
        \caption{} \label{transf:fig:simPS:qsweep}
    \end{subfigure}
    \begin{subfigure}[t]{.48\linewidth}
        \centering
        \includegraphics[width=\linewidth]{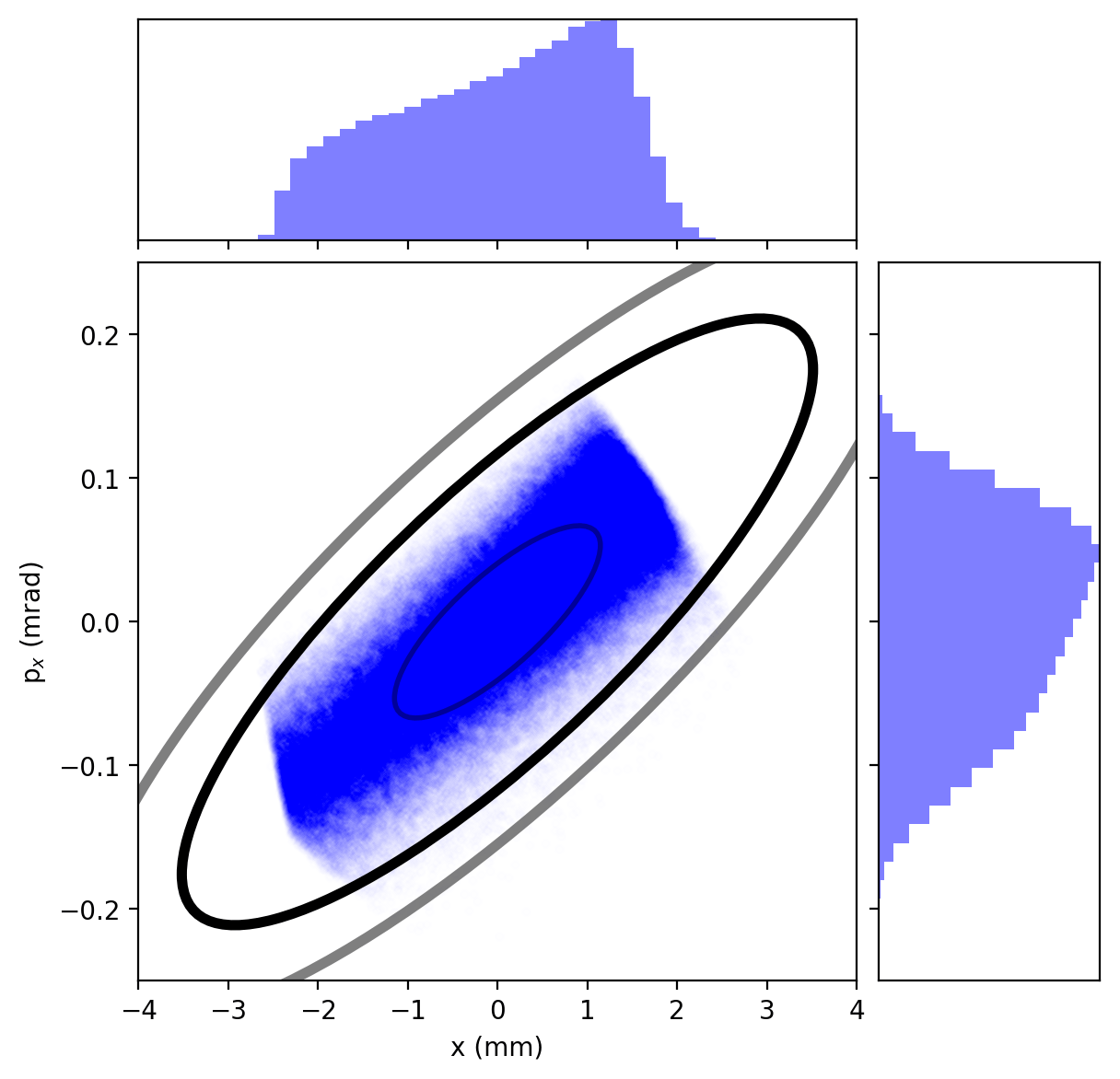}
        \caption{} \label{transf:fig:simPS:cose}\
    \end{subfigure}
    \caption{Horizontal phase space and projections of the extracted beam at the start of the extraction line for the current extraction method \subref{transf:fig:simPS:qsweep} and for the constant optics method \subref{transf:fig:simPS:cose}.}
    \label{transf:fig:simPS}
\end{figure}

Extraction methods were discussed Chap. \ref{Chap:Extraction} and Fig.\ref{transf:fig:simPS} shows resulting horizontal distributions at the start of the extraction line. The vertical plane is not shown as it remains unchanged by the extraction process and considered Gaussian.

Figure \ref{transf:fig:simPS} shows all extracted particles in blue. A relative cut around the central momentum of $\pm \num{1.5e-4}$ leads to the particles in red and better represents the phase space transported in the line since dipoles ramp during extraction to follow the extracted momentum. The constant optics extraction method provides a constant phase space during extraction and Fig. \ref{transf:fig:simPS:cose} only shows the total extracted phase space. Ellipses in red and blue are defined using the covariance matrix of the associated distributions and such that their projections cover the space between $\pm 1\sigma$. 

Table \ref{transf:tab:beamparms} lists the simulated beam parameters associated with the ellipses discussed earlier. In the horizontal plane the horizontal phase space distribution depend strongly on the extraction method. Therefore we decided to use approximate beam parameters as reference in this document. A visual representation of those reference beam parameters in the horizontal plane is shown Fig.\ref{transf:fig:simPS} as the black and grey ellipse which projections cover respectively $\pm 3\sigma$ and $\pm 4\sigma$ of the matched Gaussian distribution. Measurements conducted during the SPS BDF extraction cycle in 2017 are also listed in Table \ref{transf:tab:beamparms}. The reference beam parameter considered rely on simulations, measurements and the fact that a more precise instrumentation (see Sec. \ref{transf:sec:instrumentation}) will allow a more accurate setting of the extraction line to minimise the emittance and effective momentum spread transported during operation.

\begin{table}[htbp]
\begin{center}
\caption{Beam parameters at the start of the extraction line.}
\label{transf:tab:beamparms}
\begin{tabular}{lcccccc}
\hline
\textbf{Condition}          & \textbf{$\alpha_x$}   	& \textbf{$\beta_x$}   & \textbf{$\epsilon_x^N$} & \textbf{$\alpha_y$}   	& \textbf{$\beta_y$}   & \textbf{$\epsilon_y^N$} \\
~ & ~ & (\si{m}) &(\si{mm.mrad})& ~ & (\si{m}) &(\si{mm.mrad}) \\
\hline
\multicolumn{7}{r}{\textbf{from simulations}} \\
Current method & -0.87 & 24.7 & 26.0 & -3.79 & 140.4 & 5.0\\
Current method, w. cut & -1.62 & 33.1 & 16.2 & -3.79 & 140.4 & 5.0\\
COSE method & -1.30 &  28.2 & 20.6 & -3.79 & 140.4 & 5.0\\
\hline 
\multicolumn{7}{r}{\textbf{from 2017 measurement \cite{Hessler:2018snj}}} \\
\SI{1}{s} spill  &  \num[separate-uncertainty=false]{-0.85(8)}  & \num[separate-uncertainty=false]{25.8(9)}  &  \num[separate-uncertainty=false]{37.0(1)} & \num[separate-uncertainty=false]{-4.2(4)} & \num[separate-uncertainty=false]{162(18)} & \num[separate-uncertainty=false]{4.5(4)}  \\
\hline
\multicolumn{7}{r}{\textbf{selected reference}} \\
reference & -1.50  & 30.0 & 20.0 & -3.79 & 140.4 & 5.0\\
\hline
\end{tabular}
\end{center}
\end{table}

In the vertical plane the beam is essentially unperturbed by the extraction process and has the circulated beam parameters. Table \ref{transf:tab:beamparms} shows measurements taken in 2017 in good agreement with the simulated quantities. Simulated beam parameters are therefore set as reference and used thereafter.

The last beam characteristic is the correlation between the particle relative momentum and their position or angle in both horizontal and vertical planes. We refer to those correlations as dispersion (D) and dispersion prime (D') in a given plane. Depending on the extraction method the horizontal dispersion is null or small. In the vertical plane the dispersion is the same as for the circulating beam and null or very small since the SPS is planar horizontal. Furthermore, the evolution of the dispersion in the extraction line is dominated by long sections of bending magnets that bring the beam from the extraction point to the North Area. Therefore D and D' in both planes are set to zero at the extraction point.

Due to the ramp of the dipole in the line during the extraction process, the effective momentum spread of the beam is much lower than the extracted momentum spread from the ring (see Sec. \ref{sec:ext:slowExt}). We use a value primarily based on experience of the transported relative momentum spread of $\sigma _{dp/p} = \num{1e-4}$.

\subsection{Trajectory and optics} \label{transf:sec:optics}
The extracted beam needs to be transported from the SPS to the BDF target. Unlike for North Area operation the beam will not undergo transverse splitting during its transport. At the target the beam is required to be circular with transverse size of $\sigma = \SI{8}{mm}$. Additionally we impose the condition that the dispersion in both planes be zero at the target.

\begin{figure}[htbp]
\centering
\includegraphics[width=16cm]{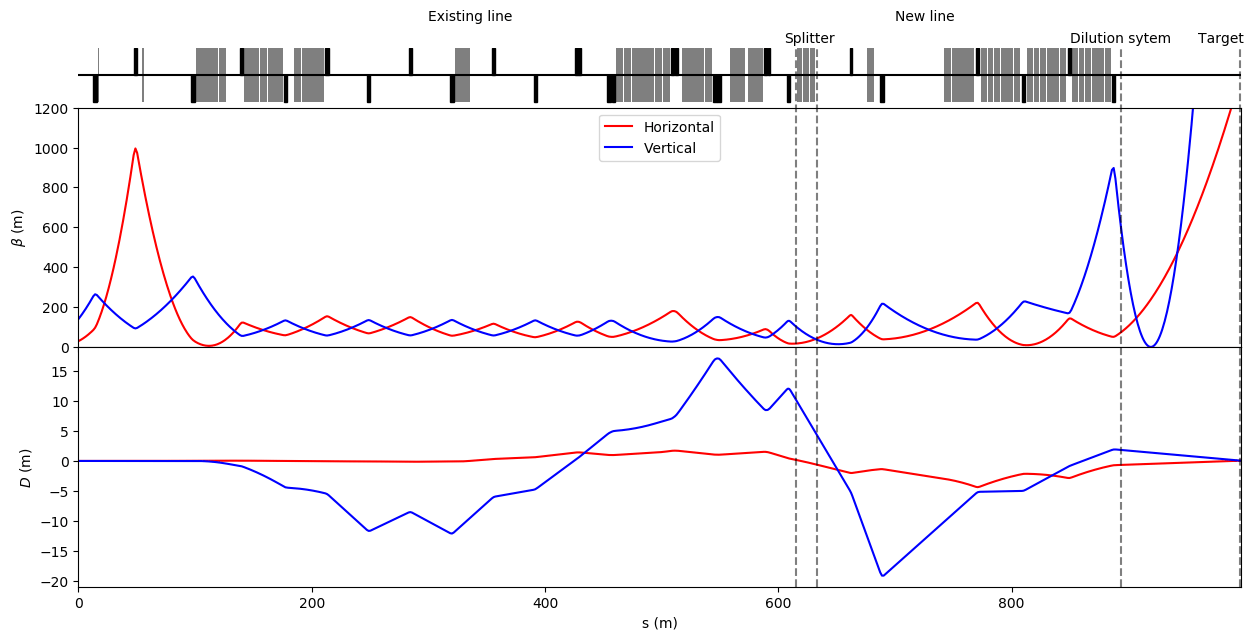}
\caption{Synoptic and optical functions along the transfer line from the SPS to the BDF target. Dipoles are represented as grey bars while quadrupoles use black squares above the centreline for focusing, and below for defocusing.}
\label{transf:fig:optics}
\end{figure}
The beam optics for the BDF transfer line have been studied in detail using the MAD-X simulation code. 
Figure \ref{transf:fig:optics} shows the resulting optics, taking into account input from the integration work detailed Chap. \ref{Chap:Integration}. Optics in the existing part of the line is essentially the same as for current North Area operation. However, contrary to the existing optics, the optical functions are kept small at the splitter, as no splitting is performed for the BDF beam. 
Positions and powering of all quadrupoles were optimised to use available magnets and to allow integration in the available space near the TT22 line.

Complete cancellation of the dispersion at the target is achieved using carefully tuned phase advances and quadrupole settings in both the existing and the new line. Optical functions of respectively \SI{1364}{m} and \SI{5456}{m} in the horizontal and vertical planes allow to reach round beam size on the target of $\sigma = \SI{8}{mm}$ using the reference emittances defined in Sec. \ref{transf:sec:beamparms}. Optical functions in Fig. \ref{transf:fig:optics} are clipped above \SI{1200}{m} for easier visualisation but follow a typical quadratic evolution. Magnet choices as well as powering scheme will be discussed Sec. \ref{transf:sec:magnets}.

\subsection{Aperture and correction scheme} \label{transf:sec:apertures}
It is critical to consider that magnetic characteristics as well as device positioning cannot be perfect. Both lead to offsets between ideal and effective beam trajectory. In this section we study the effects of those errors with a realistic modelling of the beam that includes apertures and beam size. 

Static errors, which comprise the alignment of magnets and monitor as well as systematic field errors were studied here. Using \cite{Herr:2004ng} we decided to consider the following realistic errors:
\begin{itemize}
\item to the dipole field of the \verb+RBEND+ elements a relative error of $\mathcal{N}\left(0, (\num{2.5e-4})^2\right)$ \footnote{Where $\mathcal{N}\left(\mu, (\sigma)^2\right)$ describes a normal distribution of median $\mu$ and standard deviation $\sigma$.} truncated at $\pm 2 \sigma$
\item to the quadrupole field of the \verb+QUADRUPOLE+ elements a relative error of $\mathcal{N}\left(0, (\num{2.5e-4})^2\right)$ truncated at $\pm 2  \sigma$
\item to the tilt around the longitudinal axis of the  \verb+RBEND+ elements an angle of $\mathcal{N}\left(0, (\num{1.6e-6})^2\right) \text{rad}$ truncated at $\pm 4\sigma$
\item to the transverse position of the \verb+QUADRUPOLE+ elements a vertical and horizontal misalignment of  $\mathcal{N}\left(0, (\num{.2e-3})^2\right) \text{m}$ truncated at $\pm 3\sigma$
\item to the transverse position of the \verb+MONITOR+ elements a vertical and horizontal misalignment of  $\mathcal{U}\left( \num{-0.5e-3}, \num{+0.5e-3}\right) \text{m}$ \footnote{Where $\mathcal{U}\left( a, b\right)$ describes a uniform distribution bounded by $a$ and $b$.} 
\item to the transverse position at the start of the line a  vertical and horizontal misalignment of \\ $\mathcal{N}\left(0, (\num{.5e-3})^2\right) \text{m}$ truncated at $\pm 2\sigma$
\item to the transverse angles at the start of the line a  vertical and horizontal angle of  \\ $\mathcal{N}\left(0, (\num{.05e-3})^2\right) \text{rad}$ truncated at $\pm 2\sigma$
\end{itemize}

Those errors are introduced in the MADX model of the line and 500 different sets of errors are generated. 
For each set of errors, correction is simulated using the beam centroid position at each monitor and finding the set of corrector strengths that minimises the offset at those monitors.
Here a total of 13 horizontal monitors, 16 vertical monitors, 5 horizontal correctors and 11 vertical correctors are used. 
New correctors are discussed in detail Sec. \ref{transf:sec:corrmag} and new monitors Sec. \ref{transf:sec:instrumentation}.

\begin{figure}[htbp]
\centering
\includegraphics[width=16cm]{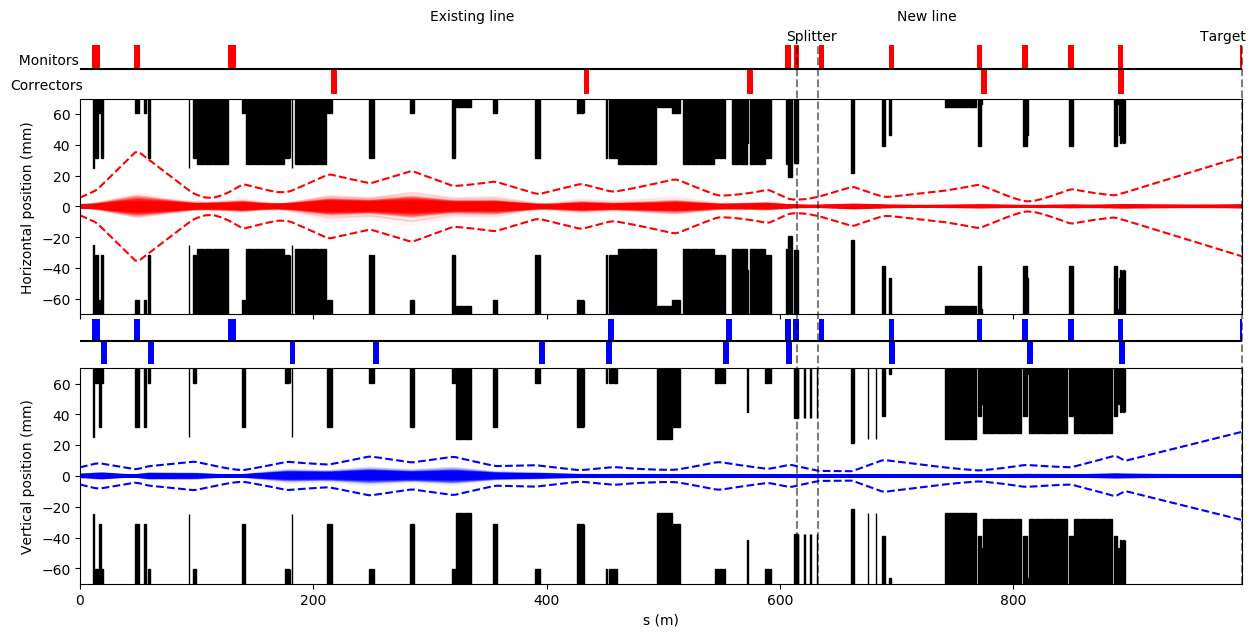}
\caption{Beam size along the line after correction of the trajectory in the horizontal and vertical planes with element apertures in black. A synoptic of the line shows the position, for each plane, of each monitor and corrector magnets respectively above and below the centreline.}
\label{transf:fig:all_apers}
\end{figure} 

Figure \ref{transf:fig:all_apers} shows the maximum expected envelopes of the beam from the correction scheme modelling. Each of the 500 corrected trajectories are plotted along the line and visible in the centre around the centreline for each plane. An envelope's maximum $Z$ is defined as:
\begin{equation}
    \pm Z = N \times \sqrt{\epsilon \beta_z + \left(\sigma_{dp/p} \times D_z \right)^2} + z_{max} \label{transf:eq:beamsize}
\end{equation}
where $z$ stands for either $x$ in the horizontal plane or $y$ in the vertical one. The absolute maximum excursion of the trajectory after correction is $z_{max}$. The Twiss parameters are $\beta$ and $D$ while the beam parameters $\epsilon$ and $\sigma_{dp/p}$ were discussed Sec. \ref{transf:sec:beamparms}. The envelopes shown Fig \ref{transf:fig:all_apers} in dashed lines use $N=4$ and refer to the transport, with error and after application of the correction scheme, of the grey ellipse in Fig. \ref{transf:fig:simPS}.

It is clear that the correction scheme allows transport of the beam along the line and within the aperture of both existing and new elements. In the vertical plane the envelope remains well within the aperture of all elements along the line. In the horizontal plane Fig. \ref{transf:fig:all_apers} omits some specific apertures in the first few tens of meters of the new line, but are discussed below.

\begin{figure}[htbp]
\centering
\includegraphics[width=11cm]{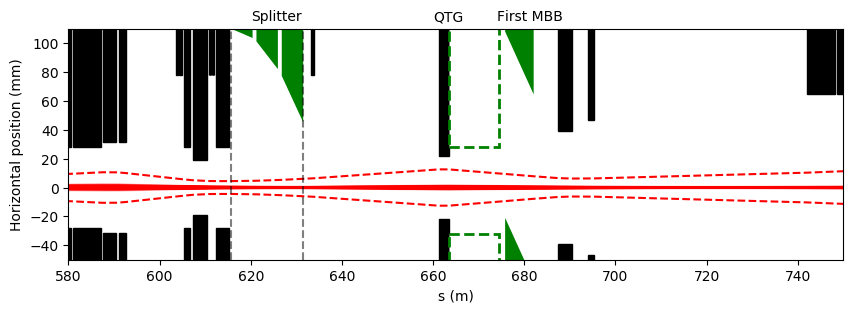}
\caption{Horizontal beam size after correction in the early part of the new line with specific aperture limitations in green.}
\label{transf:fig:zoom_H_apers}
\end{figure}

At the splitter the BDF beam travels in the lower field region of the 3 magnets. 
The splitter magnets are aligned and symmetric as discussed in more detail in Sec. \ref{transf:sec:splitter}. 
Therefore as the BDF beam is bent towards the target, the left side of the aperture is closing with the reference trajectory. 
Figure \ref{transf:fig:zoom_H_apers} shows a zoom of the envelope in the horizontal plane with the splitter aperture closing with the beam envelope. 
This representation is relative to the ideal reference trajectory but quickly allows to confirm the size and aperture of the new splitter is compatible with the transported beam.

Due to limited space immediately downstream of the splitter (see Sec. \ref{integ:sec:vac}) the first quadrupole of the new line, of QTG type~\cite{norma:QTG}, features a smaller aperture than the other quadrupoles used in the extraction line. The smaller aperture of the quadrupole allows also tighter transverse external sizes as well as stronger maximum gradient, both useful here. 
The aperture presented in Fig. \ref{transf:fig:zoom_H_apers} refers to the largest inscribed radius fitting within the magnet poles and after accounting for the thickness of the vacuum pipe.
This is a very conservative choice as the beam is not round in this location and the real beam pipe has a lozenge section. 
Nevertheless, the envelope shown Fig. \ref{transf:fig:zoom_H_apers} fits well within the aperture of the quadrupole. 

Again due to size constraints the upstream side of the first dipole is shifted away from the TT22 line by \SI{40}{mm}, as discussed in Sec. \ref{integ:sec:magnets}.
Due to the curvature of the beam within the magnet as well as the angle between the two lines, the downstream side of the magnet is aligned with the reference trajectory of the BDF beam.
Figure \ref{transf:fig:zoom_H_apers} shows the resulting aperture relative to the reference trajectory and the aperture reduction caused by the shift of the upstream side of the magnet. 
The modelled beam envelope remains away from that limiting aperture and confirms the validity of the solution.

Lastly, the vacuum chamber used between the QTG quadrupole and the first dipole had to be reduced from the usual circular of internal diameter \SI{156}{mm}, to a rectangular vacuum chamber of internal dimensions under vacuum of \SI{154}{mm} in vertical and \SI{60.4}{mm} in horizontal. 
In the vertical direction the size is large enough to be ignored and well beyond the scale in Fig. \ref{transf:fig:all_apers}. 
Figure \ref{transf:fig:zoom_H_apers} shows in dashed green lines the position of this aperture limitation in the horizontal plane. The aperture is rather close to the beam but still far enough to pose no risk to clip the BDF beam.

\subsection{Modification to the TT22 line}
As discussed Sec. \ref{integ:sec:magnets} and \ref{integ:sec:vac}, the existing TT22 line has to be slightly modified to accommodate the new BDF line. 

\begin{figure}[htbp]
    \begin{subfigure}[b]{.36\linewidth}
        \centering
        \includegraphics[width=\linewidth]{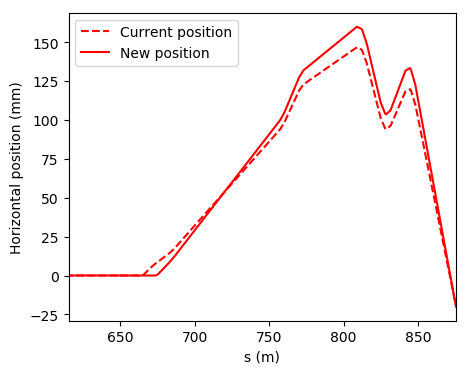}
        \caption{} \label{transf:fig:oldLine:traj}
    \end{subfigure}
    \begin{subfigure}[b]{.60\linewidth}
        \centering
        \includegraphics[width=\linewidth]{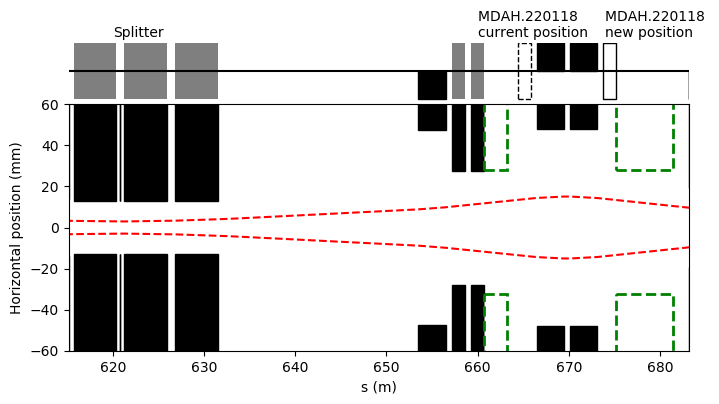}
        \caption{} \label{transf:fig:oldLine:aper}
    \end{subfigure}
    \caption{Trajectory offset in TT22 created by a kick of \SI{1}{mrad} at MDAH.220118 at its current location in dashed line and at its new location in plain line \subref{transf:fig:oldLine:traj}. Apertures and beam size downstream of the splitter in the TT22 line with synoptic representation of the lattice \subref{transf:fig:oldLine:aper}.}
    \label{transf:fig:oldLine}
\end{figure}

The first modification consists of the relocation of the first horizontal dipole corrector in the TT22 line: MDAH.220118. 
Figure \ref{transf:fig:oldLine:aper} shows that the dipole corrector is moved from upstream to downstream of the two QTAF.2202 and QTAF.2203 focusing quadrupoles, at around $s=\SI{670}{m}$. 
This change is required to integrate the new BDF line, as discussed in Sec. \ref{integ:sec:magnets}. 
The full correction scheme of the line has not been studied here but Fig. \ref{transf:fig:oldLine:traj} shows that the effect of the shift on the trajectory offset created by this dipole is minimal. 
Furthermore the operating current of the correct is always around \SI{-20}{A}, much below the rating of \SI{330}{A} of the magnet. 
Therefore the relocation of that dipoles corrector would only marginally modify the behaviour of the TT22 line.

The second modification required on the TT22 line is a reduction of the dimensions of the vacuum chamber (see Sec. \ref{integ:sec:vac}).
Figure \ref{transf:fig:oldLine:aper} shows the evolution of the $N=4$ envelope in dashed red lines according to equation \ref{transf:eq:beamsize} and following the horizontal beam parameters discussed Sec. \ref{transf:sec:beamparms} but does not include trajectory correction. 
The new vacuum chamber is rectangular of internal dimensions under vacuum of \SI{154}{mm} in vertical and \SI{60.4}{mm} in horizontal. 
The chamber aperture in dashed green remains well away from the beam envelope. 
Even though the correction scheme and trajectory offsets were not considered here, the large distance that remains between the new aperture and the beam envelope validates this change.

Other changes not impacting beam dynamics also have to be made. 
In particular the Secondary Emission Grid monitor (SEM) BSGV.220075 has to be redesigned to accommodate the new BDF line. 
This modification is discussed Sec. \ref{integ:sec:instrumentation} and Sec. \ref{trasf:sec:instr:upgrade}.

%

\section{Magnets} \label{transf:sec:magnets}

\subsection{Main dipoles and quadrupoles}
The new part of the line was designed with the intention of using magnets already available at CERN.
Table \ref{transf:tab:magTypes} shows the magnets used in the current design of the BDF line. 
In particular, the apertures shown here were used in Sec. \ref{transf:sec:apertures} (Fig. \ref{transf:fig:all_apers}) and refer to the total internal size, 
accounting for chamber compression effects from vacuum and magnet fitting. 
In the case of quadrupoles, the aperture used corresponds to the diameter of the largest inscribed circle fitting within the poles,
and reduced by the thickness of the vacuum chamber.
This definition of the aperture underestimates the available space when the beam size is not circular but has proved to be sufficient for our study.

\begin{table}[htbp]
\begin{center}
\caption{Main magnet types used for the new BDF line and their specifications.}
\label{transf:tab:magTypes}
\begin{tabular}{cccccccc}
\hline
\textbf{Function}  & \textbf{Type} & \textbf{Number} & \textbf{Magnetic} & \multicolumn{2}{c}{\textbf{Aperture}} & \textbf{Max.} & \textbf{Max. Field/}\\
 & & &\textbf{length} &  &  &\textbf{current} & \textbf{Gradient} \\
 & & & & x & y & & \\
  & & & \si{m} & \multicolumn{2}{c}{\si{mm}} & \si{A} & \si{T} or \si{T.m^{-1}} \\
\hline
Dipole & MBB~\cite{norma:MBB} & 5 & 6.26 & 129 & 48.5 & 5750 & 2.02   \\
Dipole & MBN~\cite{norma:MBN} & 18 & 5.0 & 152 & 55.5 & 1340 & 1.8 \\
Quadrupole & QTG~\cite{norma:QTG} & 1 & 2.2 & 43 & 43 & 530 & 34 \\
Quadrupole & QTL~\cite{norma:QTL} & 5 & 2.99 & 78 & 78 & 416 & 24 \\
\hline
\end{tabular}
\end{center}
\end{table}

\subsubsection{Powering scheme}
A detailed powering scheme was investigated and provides a comprehensive input for the costing of the project.
It was developed to provide a \SI{7.2}{s} repetition cycle with a \SI{1.2}{s} magnetic flat-top, coherent with the most demanding scenarios discussed Chap. \ref{Chap:Beam}.
Present standards for new beam lines at CERN require that all magnetic elements be pulsed and that the power supply be able to recover the inductive energy in order to minimise power consumption.
All the power converters identified to power the magnets of the new BDF line are part of the new SIRIUS family~\cite{EPS:Sirius}.

\begin{table}[htbp]
\begin{center}
\caption{Powering schemes with nominal currents and indicative maximum fields achievable. Negative values refer to quadrupoles used for defocusing.}
\label{transf:tab:magPS}
\begin{tabular}{ccSSSc}
\hline
 \textbf{Type} & \textbf{Number of}   & \multicolumn{2}{c}{\textbf{Nominal}} & \multicolumn{2}{c}{\textbf{Power supply maximum}} \\
 & \textbf{power circuits} & \multicolumn{1}{c}{\textbf{Current}} & \multicolumn{1}{c}{\textbf{Field}} &  \multicolumn{1}{c}{\textbf{Current}} & \multicolumn{1}{c}{\textbf{Field}} \\
 & & \multicolumn{1}{c}{\si{A}} & \multicolumn{1}{c}{\si{T} or \si{T.m^{-1}}} & \multicolumn{1}{c}{\si{A}} & \multicolumn{1}{c}{\si{T} or \si{T.m^{-1}}} \\
\hline
MBB & 1 & 5493 & 1.92 & 5750 & 2.02 \\
MBN & 6 & 1070 & 1.46 & 1800 & > 1.8 \\
QTG.01 & 1 & 297 & 23.5 & 450 & 28.9 \\
QTL.02 & 1 & -287 & -16.6 & 900 & > 24.0 \\
QTL.03 & 1 & 251 & 14.5 & 450 & > 24.0 \\
QTL.04 & 1 & -160 & -9.2 & 450 & > 24.0 \\
QTL.05 & 1 & 248 & 17.3 & 450 & > 24.0 \\
QTL.06 & 1 & -396 & -21.3 & 900 & > 24.0 \\
\hline
\end{tabular}
\end{center}
\end{table}

Table \ref{transf:tab:magPS} shows the result of the power supply study and presents some important parameters. 
Nominal currents are associated with the optics discussed Sec. \ref{transf:sec:optics} and will be achievable in the most stringent cycle considered.
The strongest constraint to this powering scheme is that every circuit needs enough internal energy storage capacity to limit their electrical consumption to the compensation of the ohmic losses in each circuit.

The powering scheme is sized to provide up to 5\% and 20\% current beyond nominal values for respectively the dipoles and quadrupoles. 
The values for maximum power supply current and associated field in Tab. \ref{transf:tab:magPS} are only indicative and may not be reachable by the system with the nominal repetition rate or associated ramp rate. Furthermore, the quadrupole power supplies are designed for 4 quadrant operation and thus will be capable of changing polarity.

\subsection{Dipole corrector magnets} \label{transf:sec:corrmag}
The new BDF line requires dipole correctors with integrated field $Bl \approx \SI{0.5}{\tesla\meter}$ and a laminated core to handle the pulsing of the BDF cycle. 
Unlike the main dipoles and quadrupoles, no existing dipole corrector magnets in storage at CERN could be used for the design of the new line. 
However, an existing short dipole corrector based on a solid core exists: the MDX corrector magnet~\cite{norma:MDX}. 
A laminated version of the MDX magnet is already being developed in the context of another project~\cite{EDMS:MDXl}. 
Table \ref{transf:tab:MDXspecs} summarises the preliminary specifications for the dipole correctors used in our study.

\begin{table}[htbp]
\begin{center}
\caption{Preliminary specifications for the laminated MDX magnet design.}
\label{transf:tab:MDXspecs}
\begin{tabular}{ll}
\hline
Aperture in bending plane & \SI{140}{mm} \\
Aperture in non bending plane & \SI{100}{mm} \\
Total length & \SI{630}{mm} \\
Maximum integrated field & \SI{0.509}{\tesla\meter} \\
Maximum current & \SI{240}{A} \\
Resistance & 320 m$\Omega$  \\ 
Inductance & \SI{221}{\milli\henry} \\
\hline
\end{tabular}
\end{center}
\end{table}

A total of 5 of those correctors are used along the new line to correct the trajectory. 
An additional single dipole corrector referred to as VREF is provisioned to impart a small vertical angle to the BDF line if required 
because of incoherence in the current 3D integrated model related to change in definition of verticality between the Meyrin and Pr\'evessin sites.
A further 4 of those correctors will be required for the dilution system that will be discussed Sec. \ref{transf:sec:dill}. 
In total, 11 of the new laminated MDX-type dipole corrector will be used in the new BDF line.

Each of the 6 correctors plus the single VREF will be each powered by a single four quadrant power supply capable of recovering and storing the inductive energy between cycles. 
Voltage specifications will also allow small changes during the spill in order to compensate for any systematic motion of the beam along the \SI{1}{s} spill.

\subsection{Dilution system}
As discussed Sec. \ref{Sec:TGT:Simus:accident:dilution} the target cannot handle the extracted beam for a full \SI{1}{s} spill, despite the relatively large round beam of $\sigma = \SI{8}{mm}$.
A dilution system with the aim of moving the beam spot across the target during the spill is necessary. Simulations presented Sec. \ref{Sec:TGT:Simus:accident:dilution} show that a circular pattern of radius \SI{50}{mm} repeated 4 times during the spill is sufficient. This dilution scenario is considered the nominal case.

\subsubsection{Dilution system design} \label{transf:sec:dill}
The last \SI{100}{m} of the BDF line is empty of magnetic elements.
It is natural that the dilution system would make use of this drift space by imparting a small angle to the beam that translates into the required offset at the target. 

The required angle imparted to the beam \SI{100}{m} upstream the target to reach an offset $R=\SI{50}{mm}$ is $\theta_{max} = \SI{500}{\micro \radian}$. 
Using the magnetic rigidity of the \SI{400}{\giga \eVperc} proton beam we find that the maximum required integrated field is $Bl_{max} \sim \SI{0.667}{T m}$. 
This integrated gradient can be achieved using two laminated MDX (see Tab. \ref{transf:tab:MDXspecs}) powered at $I_{max} \sim \SI{160}{A}$.

\begin{figure}[htbp]
    \begin{subfigure}[b]{.48\linewidth}
        \centering
        \includegraphics[width=7.0cm]{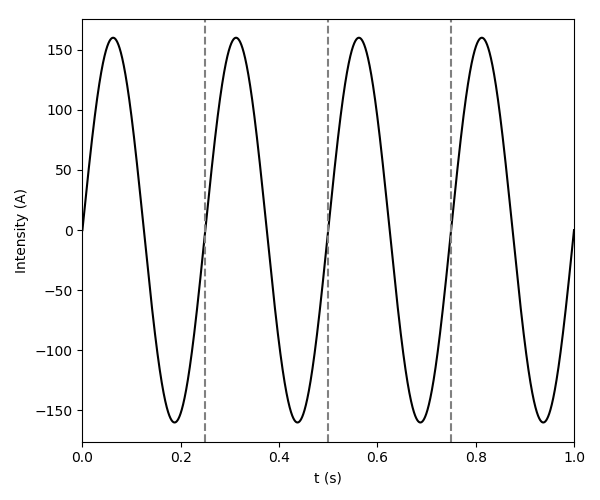}
        \caption{} \label{transf:fig:dil:I}
    \end{subfigure}
    \begin{subfigure}[b]{.48\linewidth}
        \centering
        \includegraphics[width=7.0cm]{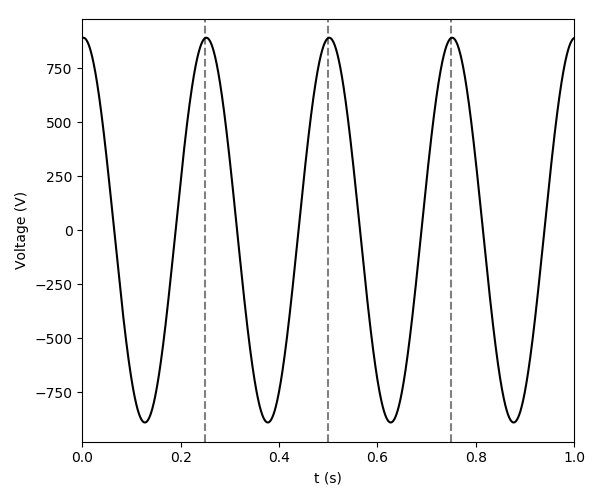}
        \caption{} \label{transf:fig:dil:U}
    \end{subfigure}
    \caption{Required intensity \subref{transf:fig:dil:I} and voltage \subref{transf:fig:dil:U} at the MDX magnet for the dilution system and as a function of time.}
    \label{transf:fig:dil}
\end{figure}

The required pattern can be obtained by driving each quadrupole with a sinusoidal current at a frequency of $f_{dil} = \SI{4}{\hertz}$. 
We need to consider both voltage $U(t)$ and current $I(t)$ as a function of time $t$ for this application:
\begin{align}
   I(t) &= I_{max} \times \sin \left( 2\pi f_{dil} t \right) \label{transf:eq:dilI} \\
   U(t) &= I_{max} \times \left( 2\pi f_{dil} t \times L \times \cos \left( 2\pi f_{dil} t \right) + R\times \sin\left(2\pi f_{dil}\right)\right) \label{transf:eq:dilU}
\end{align}
where $R$ and $L$ are respectively the magnet resistance and inductance, listed in Table \ref{transf:tab:MDXspecs}.
Figure \ref{transf:fig:dil} shows the evolution of the intensity and voltages during a full spill. 
The large required maximum voltage per magnet of $U_{max} \sim \SI{890}{V}$ prevents the use of more than one magnet per power supply.
Therefore each of the two magnets needed to reach the offset $R$ is to be powered by a dedicated power supply and through a dedicated pair of cables. 

The power supply selected for this application may reach up to \SI{900}{V} and \SI{450}{A}.
Here the frequency directly translates into strong voltages due to the inductance of the magnet. 
The margin between between the required maximum voltage  $U_{max}$ and the power supply rating is particularly small. 
However a more complex voltage function such as a clipped sinusoidal function may be applied to increase the maximum current reached and with only a small effect on the shape of the intensity function.

Since we need two independent circuits to reach the integrated field $Bl_{max}$, we need 4 independent circuits to obtain a circular pattern of the beam on the target. 
This can be achieved by splitting the 4 circuits into two groups, referred to as $0$ and $\pi/2$, in which magnets are oriented horizontally and vertically respectively.
Each group is driven by the current functions:
\begin{align}
I_{0}(t) &= I_{max} \times \sin \left( \omega t \right)  \\
I_{\pi/2}(t) &= I_{max} \times \sin \left( \omega t + \pi/2 \right) \nonumber
\end{align}
where $\omega = 2 \pi f_{dil}$. 
We refer to this scheme as the $pi/2$ dilution scheme.
This scheme provides the required circular pattern with radius of \SI{50}{mm} on the target.

Given that there are 4 independently powered circuits, we can investigate a more complex scheme. 
By rotating each magnet by $\pi/4$ and powering them with the correct dephasing, we achieve the same circular dilution pattern. 
Each of the four magnet is then oriented along a direction $\theta$ and powered by a current function $I_{\theta}$:
\begin{align}
I_{0}(t) &= I_{max} \times \sin \left( \omega t \right)  \\
I_{\pi/4}(t) &= I_{max} \times \sin \left( \omega t + \pi/4 \right) \nonumber \\
I_{\pi/2}(t) &= I_{max} \times \sin \left( \omega t + \pi/2 \right) \nonumber \\
I_{3\pi/4}(t) &= I_{max} \times \sin \left( \omega t + 3\pi/4 \right) \nonumber \\
\end{align}
This arrangement is called $pi/4$.

The two schemes detailed here provide the same dilution pattern on the target. 
Therefore the simpler $\pi/2$ scheme would be favoured. 
However, we will go into more detail about the interests of the $\pi/4$ scheme when studying failure cases in Sec. \ref{transf:sec:dillFailure}.

\subsection{Splitter} \label{transf:sec:splitter}
In order to allow switching of the BDF beam in the opposite direction to the existing North Area beams,
three MSSB splitters of a new design are planned to be installed in place of the existing first splitter triplet (MSSB.2117).
In the present configuration, the beam can be sent straight towards T2 or bent right towards T6.
The new MSSB design will add the capability to switch the beam towards the BDF target complex while maintaining full compatibility with the operation of existing experiments in the North Area.

To allow the switching of the beam, the magnets must have the possibility of reversing their polarity in a relatively short time between cycles of the machine. 
The main way to achieve this is to replace the present solid iron yoke with laminations, so as to mitigate the eddy currents induced by the field variation associated with the switch in polarity.

The high radiation levels and the fact that the iron yoke is enclosed in a vacuum tank, imply that the insulating coating on the laminations has to be inorganic and vacuum-compatible.
Another radiation-hard measure is the use of mineral insulated copper conductors (MICC) to build the coils of the magnets.
These conductors use compacted magnesium oxide (MgO) powder as insulator instead of epoxy resin-glass fibre composite. 
A water cooled MICC has been successfully employed in the existing MSSB design and is still in operation. 
To reduce the inductance of the magnet and improve the compatibility with the powering circuit, the number of turns in the coil will be reduced from 48 to 35 and the current correspondingly increased with simultaneous increase of the conductor cross-section.
To allow for a quick exchange of a magnet in case of a fault, all connections, both hydraulic and electrical, would be of a plug-in type.
This will also allow the automatic alignment of the magnet without the necessity to enter the high radiation area.

The construction of a short prototype of the laminated top yoke is currently under study. 
The aim of the prototype is to validate the proposed manufacturing technology and prove that the mechanical tolerances required are achievable over \SI{5}{m} long laminated yoke.
The main area of interest is the septum blade region, in which even small imperfections can drastically increase beam losses.

\section{Instrumentation} \label{transf:sec:instrumentation}

\subsection{Beam position and profile monitors}

Beam position and beam size will have to be measured in several locations along the line. 
These systems will also be used to measure the evolution of the beam position and beam size during the extraction process, with a temporal resolution of \SI{100}{ms}. 
The system should also be able to function with a wide range of intensity i.e. for beams of \num{1e11} to \num{4e13} protons per second.

\subsubsection{Existing transfer line upgrade} \label{trasf:sec:instr:upgrade}

The existing TT21 line up to the first splitter is shared by all beams. 
The instrumentation of these lines is being reviewed as some systems no longer fulfil present and foreseen operational requirements. 
The expected time line for renovation is during Long Shutdown 3 (LS3) but funding has yet to be found.
However, in the scope of the current BDF work, a few specific monitors and their specifications have been identified. 

\begin{table}[htpb]
  \center
  \caption{Required modifications for the TT21 beam line instrumentation. BSP are split foils used to measure position. }
  \label{transf:tab:beam_pos_size_existingTT20}
\begin{tabular}{ccSSSS}
\hline 
\textbf{s}  & \textbf{Current Name} & \multicolumn{2}{c}{\textbf{Horizontal}} & \multicolumn{2}{c}{\textbf{Vertical}}  \\ 
~              &                  &  \textbf{Resolution} & \textbf{Coverage} & \textbf{Resolution} & \textbf{Coverage} \\
\si{m} & & \multicolumn{4}{c}{\si{mm}} \\
\hline
181            &  BSP.210508      &  1.5             &   22.5        &  1.5            &  22.5 \\
318            &  BSP.210858      &  2.5             &   37.5        &  2.5            &  37.5 \\
397            &  None          & \multicolumn{2}{c}{None}        &   2.5           &  37.5 \\
515            &  BSPH.211411     &  5.0             &   75.0        &  \multicolumn{2}{c}{None} \\
  \hline 
\end{tabular} 
\end{table}

Table \ref{transf:tab:beam_pos_size_existingTT20} summarises the specifications of those requested monitors requiring replacement. 
The proposed solution is to provide 16 wires per grid per plane, consistent with the above
requirements, while keeping cabling and electronics costs reasonable.

The new beam line layout impacts an existing SEM grid tank (BSGV.220075) that must be modified mechanically as the present tank is too wide.
A new tank where the external diameter does not exceed \SI{225}{mm} is under study, with the grid movement operated from below. 
All other parameters including the wire number and position for this SEM grid are expected to remain unchanged.

\subsubsection{New BDF line}

In the new beam line which branches off downstream of the first splitter (MSSB.2117) a total of 5 beam position (centroid) and beam size measurement devices should be provided.
The specifications in terms of coverage and resolution are summarised in Table \ref{transf:tab:beam_pos_size_newBDF}. 

\begin{table}[htpb]
  \center
  \caption{Specifications for the beam instrumentation of the new BDF line.}
  \label{transf:tab:beam_pos_size_newBDF}
\begin{tabular}{ccSSSS}
\hline
\textbf{s}  & \textbf{Provisional Name} & \multicolumn{2}{c}{\textbf{Horizontal}} & \multicolumn{2}{c}{\textbf{Vertical}}  \\ 
~              &                  &  \textbf{Resolution} & \textbf{Coverage} & \textbf{Resolution} & \textbf{Coverage} \\
\si{m} & & \multicolumn{4}{c}{\si{mm}} \\
\hline
  692 & BDF.02 & 1.5 &22.5& 1.5 &22.5\\
  769 & BDF.03 & 2.5 &37.5& 0.8 & 12\\
  808 & BDF.04 & 0.8 & 12 & 1.5 &22.5\\
  847 & BDF.05 & 1.5 &22.5& 1.5 &22.5\\
  889 & BDF.06 & 1.5 &22.5& 2.5 &37.5\\
  \hline
\end{tabular} 
\end{table}

The proposed solution is here also to provide SEM grids with 16 wires per grid per plane, consistent
with the above requirements, while keeping cabling and electronics costs reasonable.

\subsubsection{Target beam monitoring} \label{transf:sec:dil_live_mon}

A new target screen monitor (BTV) is requested at a location approximately \SI{50}{m} upstream of the
target where the beam RMS size is expected to be around \SI{5}{mm} in both planes.
It should provide the 2D image of the beam allowing beam position and size to be extracted. 
The total area to be covered is $\num{70} \times \SI{70}{mm}$ with images taken every \SI{100}{ms} for off-line analysis.

\subsection{Other instruments}

\subsubsection{Intensity measurement}
It is requested to measure the beam intensity at one specific location, close to the monitor BDF.06 in the new beam line.
The intensity should be provided throughout the slow extraction with a time resolution of \SI{100}{ms}. 
The intensity range to be covered ranges from \num{1e11} to \num{4e13} charges with a resolution of \num{5E10} charges and an accuracy on the level of a \%.

\subsubsection{High frequency spill structure monitor}
The high frequency time structure of the beam is relevant for the facility up to \SI{10}{GHz}. 
R\&D efforts are to start after LS2 to provide a spill monitor covering frequencies up to and including \SI{200}{MHz}.

\subsubsection{Beam loss monitors} \label{transf:sec:lossmon}
A total of 9 new ionisation monitors are requested along the new beam line at locations yet to be defined. 
In case losses exceed operational thresholds during the slow extraction a hardware
interlock should trigger the beam dump with a reaction time not exceeding \SI{10}{ms}.

\section{Failure scenarios and machine protection} \label{transf:sec:Failure}
As discussed Tab. \ref{tab:SHiP-P}, both instantaneous and average beam power are very high. This calls for a specific approach of studying failure scenarios in order to prevent catastrophic machine damage.
The results of preliminary investigations are presented below. A rigorous study should be conducted in the preparatory stage of the technical design report.

\subsection{Dilution system failure} \label{transf:sec:dillFailure}
The dilution system is required for the target to handle the full spill from the SPS. Uninterrupted extraction without nominal dilution pattern is potentially a very serious incident for the target.
For instance, a single full spill without dilution would lead to a complete failure of the BDF target (see Sec. \ref{Sec:TGT:Simus:accident:dilution}).

\subsubsection{Dilution failure patterns}
\begin{figure}[htbp]
    \begin{subfigure}[b]{.48\linewidth}
        \centering
        \includegraphics[width=7cm]{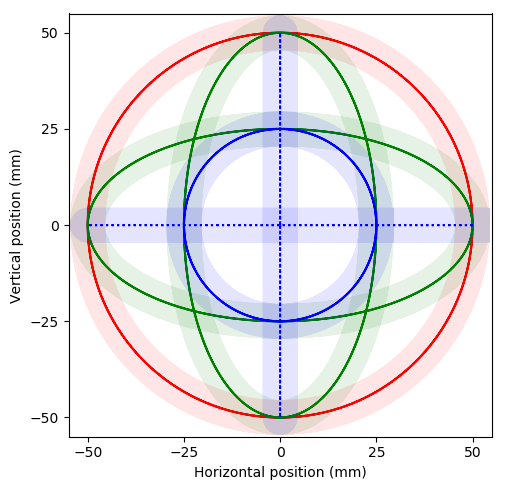}
        \caption{} \label{transf:fig:dilF:pi2}
    \end{subfigure}
    \begin{subfigure}[b]{.48\linewidth}
        \centering
        \includegraphics[width=7cm]{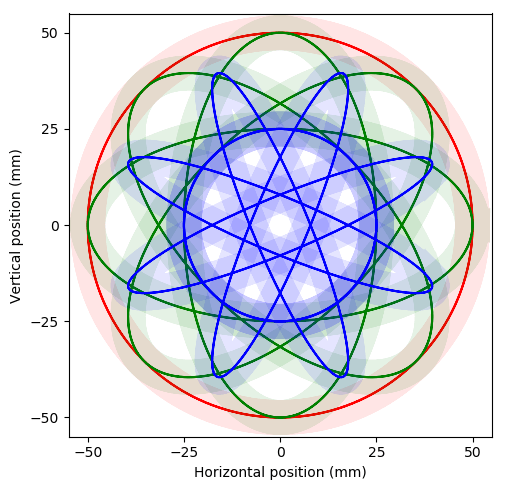}
        \caption{} \label{transf:fig:dilF:pi4}
    \end{subfigure}
    \caption{Possible dilution patterns on the target with all four circuits in red, 3 circuits in green and only two circuits in blue. Figure \subref{transf:fig:dilF:pi2} shows patterns for the $\pi/2$ scheme while the $\pi/4$ possible patterns are shown in \subref{transf:fig:dilF:pi4}.}
  \label{transf:fig:dilF}
\end{figure}

Due to the 4 fully independent dilution kickers, a wide range of patterns are possible depending on the scheme and the power supply failure. 
Figure \ref{transf:fig:dilF} shows the different dilution patterns possible on the target and represents the extent in transparency of the round beam of $\sigma = \SI{8}{mm}$.

The loss of one single circuit causes the dilution pattern to take a oblong shape, in green in Fig. \ref{transf:fig:dilF}. 
This scenario is referred to as case 1 in Sec. \ref{Sec:TGT:Simus:accident:dilution} and would not require a replacement of the target.

In the eventuality of the loss of two dilution circuits, multiple different patterns may be considered. 
We can see here the effect of the $\pi/4$ magnet scheme introduced Sec. \ref{transf:sec:dill}. 
In the case of the most simple $\pi/2$ scheme, the simultaneous loss of two circuits may lead to flat dilution patterns in 33\% of the cases and is represented in dotted lines Fig \ref{transf:fig:dilF:pi4}. 
In the case of the $\pi/4$ scheme no two circuits share the same axis. 
This prevents the loss of two power supplies to lead to a flat dilution pattern. 
In this $\pi/4$ scheme the loss of two circuits leads to a flattened elliptical pattern in 66\% of the cases.

The loss of two circuits with the $\pi/2$ scheme and flat pattern or $\pi/4$ schemes and elliptical pattern are respectively referred to as case 3 and case 4 in Sec. \ref{Sec:TGT:Simus:accident:dilution}.
It was established that both scenarios would require a replacement of the target. 
Case 2, that refers to the smaller circular patter of \SI{25}{mm} radius would likely not necessitate a replacement of the target. 
Therefore the $\pi/2$ scheme is favoured since the loss of two circuits would require a replacement of the target in 33\% of occurrences while that probability would be 66\% for the  $\pi/4$ scheme. 
It is the $\pi/2$ scheme, with two horizontal and two vertical dilution circuits that is implemented in the drawings of the line presented Sec. \ref{Sec:TGT:Simus:accident:dilution}.

Finally, the simultaneous loss of 3 circuits would lead to a \SI{50}{mm} long and flat dilution pattern. 
This scenario would also require a replacement of the target.

\subsubsection{Mitigation systems and interlocks} \label{transf:sec:XPOC}
The currents in the power supplies (dilution system or else where in the transfer line) can be monitored continuously during their execution of the programmed function. 
The front-end power converter control system (FGC) provides the possibility to interlock on regulation error (i.e. power converter not following programmed function within a predefined tolerance) ~\cite{EPC:PS_mon} at a frequency of \SI{1}{\kilo \hertz}.
The output of the FGC needs to be connected to a dedicated slow extraction beam interlock system (BIS), the details of which will be discussed below. 
The trigger to the BIS can occur within \SI{200}{\micro s}. 
The time it takes to stop slow extraction will have to be evaluated taking cable lengths all the way to the SPS beam dump system into account.
The total BIS reaction time will however be significantly below \SI{100}{ms}, as was studied in case 6 Sec. \ref{Sec:TGT:Simus:accident:dilution}.
As additional protection to ensure that the correct functions are loaded to the power converters, a rough absolute current value interlocking should be provided by the SPS software interlock system (SIS) for all converters except the ones used in the dilution system. 
The loaded flattop current function should be compared to a reference current value at the SPS start of acceleration. 
The tolerances have to be sufficiently large to allow for COSE. 
This consistency check will ensure that not only the power converters are following the programmed functions, but that they are all consistent with each other for a given mode of operation - either BDF or NA.
For the dilution system another software application/makerule will have to be prepared that allows to only load functions fulfilling certain criteria. 
The dilution system current settings should be managed by the Management of Critical Settings (MCS).  

The monitoring system discussed Sec. \ref{transf:sec:dil_live_mon} will also be used to validate the dilution system performance. After every extraction cycle, data from the system will be automatically analysed and compared to the nominal dilution pattern. The system can use other information sources such as power supply current or radiation levels in the tunnel. Only once the normality of the previous cycle will have been confirmed, a new extraction cycle will be permitted. Such automatic post operation checking system is already used for the LHC dump~\cite{Gallet:2008zz}, but would here need to be completed within no more than $\sim \SI{4}{s}$.

The methods for interlocking the dilution system presented in this section so far have in common that they are software based solutions. Taking into account the criticality of the beam dilution onto the target, an additional hardware-based system is advisable for interlocking the dilution system. Such a hardware interlock could be realised by obtaining a signal proportional to the sweep velocity by analogue signal processing of the current measurement signals of the horizontal and vertical dilution magnets, which are proportional to the horizontal and vertical beam position on the target. This sweep velocity should be constant during the flat top in regular operation. By comparing the sweep velocity signal with a reference, a deviation can be detected and the beam interlock triggered.

\subsubsection{Failure probability listing}
We established that it is not possible to absolutely prevent failure of the dilution system. Therefore, a statistical approach of the failure cases can give us an objective expectation of the performance of the system and direct studies to more complex protection systems if needed.

The SIRIUS power converter expected for the dilution system have not yet been operated at CERN. Hence no measured failure probabilities are available. Furthermore, the complexity of the power converters prevent an accurate estimation of the MTBF (Mean Time Between Failure) to be undertaken. Only the design goal lifetime of \SI{50000}{h} is known. 
As the SIRIUS system will be operated for powering of the TT10 line magnets, experimental numbers will become available after LS2.

\subsection{Interlocking slow extraction at the CERN SPS}

Beam interlock controllers (BIC) are used to gather the interlock signals (True, False) from different systems such as beam loss monitors, power supplies, \emph{etc}. 
The inputs are combined by a logical AND and the resultant input to a so-called master BIC. 

The TT21 BIC will combine extraction equipment (ZS voltage and magnetic septa) as well as power converters, SIS software and BLM interlocks before the first splitter. 
The NA BIC will have inputs from the first splitter current, beam loss monitors and TT22/TT25 power converter interlocks. 
Finally the BDF BIC will cover the power converters in the new BDF line, the muon shield, dilution kickers, \emph{etc}. 

Faults from BDF systems should not stop the NA or any other beams in the SPS and vice versa.
For this purpose a concept of arbitration between slow extraction interlocking for NA or BDF has to be provided. 
Timing destinations are a useful concept for this purpose. 
The master BIC would have as additional inputs BDF destination or NA destination (provided by the timing system) and only take the relevant BICs into account for its output.
For example, if the destination is BDF only the TT21 and BDF BICs have to give a green light.
At the same time if none of the destinations BDF or NA are provided by the timing system, none of the TT21, BDF or NA BICs are taken into account.
The output of the master BIC is connected to an unmaskable input of the SPS ring permit BA2 BIC. 
If any input of a ring permit BIC goes false, the SPS beam dump is triggered. 
The simplified master BIC equation would be:
\begin{multline}
[\lnot \text{destination}_{BDF} \land \lnot \text{destination}_{NA}] \lor \\ [\text{BIC}_{TT21} \land [(\text{destination}_{BDF} \land \text{interlocks}_{BDF})\lor (\text{destination}_{NA}\land \text{interlocks}_{NA})]] 
\end{multline}

As timing destinations are not rigorously enforced, and BDF interlocks are only taken into account if the correct timing destination is set, a mechanism has to be provided so that no slow extraction is possible without correct timing destination. 
A way to achieve this is to modify the power converter controls of the SPS extraction bumper as well as the SPS extraction sextupole circuits,
such that they only pulse if either NA or BDF timing destination is set for the given cycle.

\FloatBarrier
\printbibliography[heading=subbibliography]

\chapter{Production target design and R\&D} 
\label{Chap:Target}



\section{Introduction and target requirements}
\label{Sec:TGT:Intro}
The Beam Dump Facility target assembly is the core of the proposed Beam Dump Facility. The BDF target sits at the core of the installation with a double function: on one side, it must absorb safely and reliably the full SPS primary beam. On the other side, its design has been optimized from a physics perspective point of view (in terms of geometry, material,gaps, etc.) to maximize the production of charmed mesons. It can be considered as a beam dump/absorber, since it will contain most of the cascade generated by the interaction with the primary beam. The high power deposited is one of the main challenges of the BDF target design, with 320 kW of average power deposited and 2.56 MW over the 1-second slowly extracted spill.

Such high beam power expected on target requires beam dilution, as well as a large beam spot diameter, in order to avoid target failure. The dilution pattern generated by the upstream magnets and the beam size have been optimized taking into account the aperture restrictions from the extraction line magnets and the mechanical performance of the target. As a result, the SPS primary beam will be diluted in 4 turns over a 50 mm radius circle for each 1-second pulse, with a beam spot size of 8 mm 1$\sigma$.

The materials sought for the BDF target are high-Z materials with a short nuclear interaction length, in order to increase the re-absorption of pions and kaons produced in the intra-nuclear cascade process. Details of the design will be discussed in Section~\ref{Sec:TGT:Design} and later paragraph, which will include energy deposition \& radiation damage considerations (Section~\ref{Sec:TGT:EneRad}), thermo-mechanical simulations \& CFD (Sections~\ref{Sec:TGT:Simus} and \ref{Sec:TGT:CoolingCFD}) as well as preliminary mechanical design of the assembly (Section~\ref{Sec:TGT:MechDesign}). Discussion of the target material selection will be developed in Section~\ref{Sec:TGT:Materials} and the detailed presentation of the BDF prototype target which have been irradiated during 2018 will be presented in Section~\ref{Sec:TGT:Proto}.

\section{Target design introduction}
\label{Sec:TGT:Design}
\subsection{Production target material selection}
\label{Sec:TGT:Design:material}

\begin{figure}[ht!]
\centering %
\includegraphics[width=0.8\linewidth]{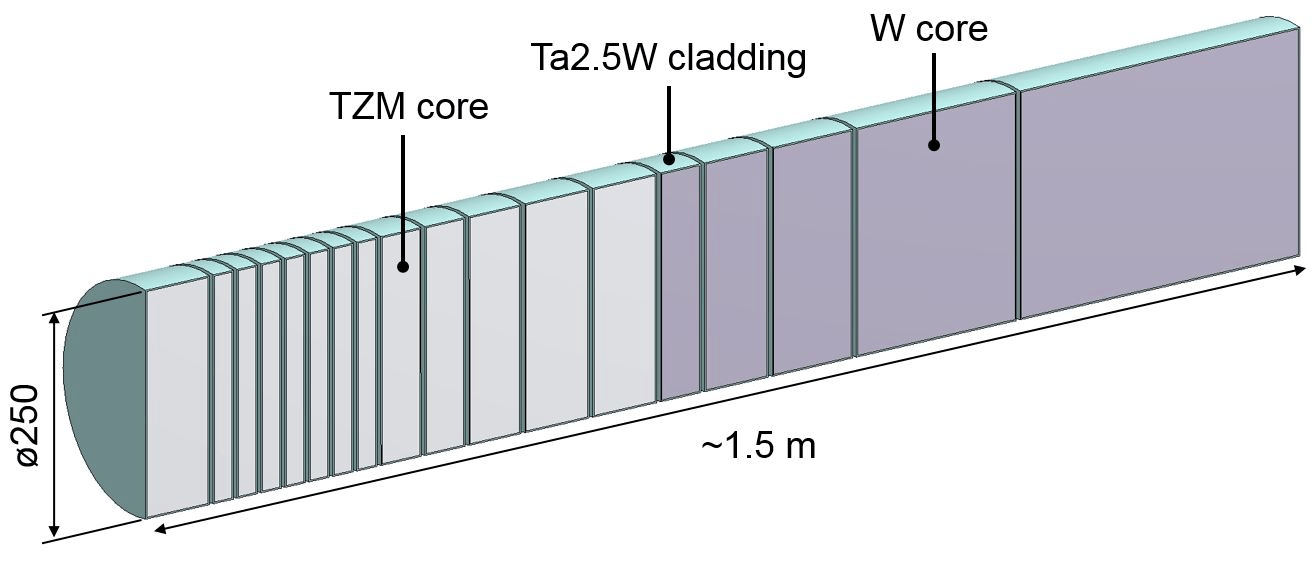}
\caption{\label{fig:TGT:BDFtarget} Layout of the Beam Dump Facility target core, showing the first part of TZM and the second one of pure W, both clad with Ta2.5W.}
\end{figure}

The proposed Beam Dump Facility production target is a hybrid mechanical assembly, consisting of several collinear cylinders of TZM ((0.08\%) titanium - (0.05\%) zirconium - molybdenum alloy) and pure tungsten (W), clad with a W-containing Ta-alloy, Ta2.5W ((2.5\%) tungsten - tantalum alloy), as depicted in Figure~\ref{fig:TGT:BDFtarget}. The function of the target is to produce particles for the downstream experiment, but at the same time also to fully absorb the primary energy beam from SPS. For this reason, the assembly is often referred as the target/dump device of the BDF infrastructure.

In order to match the physics requirements for the BDF facility, the target/dump assembly should be characterized by the highest possible density and shortest nuclear interaction length. However thermo-mechanical and technical constraints limit the possibility of potential materials. 
\begin{itemize}
    \item For the first part of the target core, which will absorb a large fraction of the total beam power deposited on target, TZM has been considered as the absorbing material. The material has a sufficiently high density to fulfill the experiment requirements, but it leads to energy deposition and stresses lower than those encountered if materials with higher density - such as pure tungsten - would be chosen. This Mo-based alloy has been selected because of its higher strength, better creep resistance and higher recrystallization temperature compared to pure molybdenum~\cite{TantalumW2}.
    \item  For the second part of the target, that will receive much less power deposition from the primary beam, pure W has been chosen, since it fulfills the physics requirements and has proven good performance under irradiation~\cite{tungstenprops}. 
\end{itemize}

The target needs to be actively cooled by water, given the high energy deposited and the high temperatures reached during operation. The cooling system design is based on high velocity water flowing through 5 mm gaps foreseen between the different target blocks. The high-speed water flow over the blocks surface allows for an effective heat transfer coefficient (HTC) between the cooling medium and the target, dissipating the power deposited by the SPS primary beam. More details about the cooling system design are given in Section~\ref{Sec:TGT:CoolingCFD}. 

The high velocity water flow in contact with the pure W and TZM blocks could induce undesired corrosion-erosion effects. Therefore, all the target core blocks will be clad via diffusion bonding achieved by means of the Hot Isostatic Pressing (HIPing) method with Ta2.5W \cite{HIP1,HIP2}, a Ta compound alloyed with W. This material is selected as cladding material due to its high corrosion resistance and its convenience as high-Z material with short interaction length. Ta2.5W has enough ductility to plastify and allow the diffusion bonding with the target core materials, and is soluble with molybdenum and tungsten, reducing risks of forming any intermetallic layers during the HIP diffusion bonding.

In the preliminary target design phase, pure tantalum was considered as cladding material for the target core blocks, given the vast experience with tantalum-clad targets in other operating facilities such as the ISIS spallation neutron source at the Rutherford Appleton Laboratory~\cite{ISISclad} in UK. However, the structural calculations performed on the BDF target blocks (which are detailed in Section~\ref{Sec:TGT:Simus:struct}) have shown that the maximum stresses reached in the tantalum layer may be critical for the target operation, limiting its lifetime significantly. For that reason, Ta2.5W has been considered as alternative cladding material, with the advantage of a considerably higher strength at high temperatures and a similar corrosion-erosion resistance \cite{TantalumW,TantalumW2}. 

An exhaustive R\&D study has been carried out in the framework of the BDF Project in order to test the bonding quality of Ta2.5W with TZM and pure tungsten after the HIP process. The interface mechanical tests performed have proven that the intermetallic bonding strength of TZM or W with Ta2.5W is comparable to the one with pure Tantalum, validating the selection of Ta2.5W as target cladding material. The procedures, results and conclusions of the R\&D studies carried out have been published in Ref.~\cite{HIP_Busom}. The target materials selection and properties will be further detailed in Section \ref{Sec:TGT:Materials}.

As a further validation of the use of Ta2.5W in the BDF target blocks cladding, a prototype of the BDF target has been tested under beam in the North Area of CERN~\cite{EDMS-T6}. The BDF target prototype consists in a scale replica of the BDF target, with identical length and reduced diameter. Pure tantalum and Ta2.5W have been used as cladding materials for the BDF target prototype, in order to compare the performance of both materials under beam irradiation. A Post Irradiation Examination (PIE) campaign is foreseen during the course of 2019-2020 on several blocks of the target prototype, to characterize the mechanical bonding of the cladding and core materials after the irradiation. The description and results of the BDF target prototype beam tests in 2018 will be published on a separate paper.

\subsection{Beam parameters and target design optimisation}
\label{Sec:TGT:Design:optimisation}

The 400~GeV/c proton beam pulse at \num{4e13} protons/pulse will be extracted from the SPS to impact the target during 1 second, delivering an average power of 2.56~MW, followed by a cooling of 6.2 seconds. Out of the 355~kW average beam power impacting on the target, roughly 305~kW will be deposited inside the target assembly, while the rest will be lost in the surrounding (water cooled) shielding of the BDF target complex. A detailed list of the BDF operation beam parameters is specified in Table~\ref{tab:TGT:beamparameters}.

\begin{table}[htbp]
\centering
\caption{\label{tab:TGT:beamparameters} Baseline beam parameters of the BDF target operation}
\smallskip
\begin{tabular}{lc}
\toprule
\multicolumn{2}{c}{\textbf{Baseline characteristics}}\\
\midrule
Proton momentum {[}GeV/c{]}              & 400                        \\
Beam intensity {[}p+/cycle{]}            & \num{4e13} \\
Cycle length {[}s{]}                     & 7.2                        \\
Spill duration {[}s{]}                   & 1.0                        \\
Beam dilution pattern [-] & Circular \\
Beam sweep frequency [turns/s] & 4\\
Dilution circle radius [mm] & 50 \\
Beam sigma (H,V) [mm] & (8,8) \\
Average beam power  {[}kW{]}    & 356                       \\
Average beam power deposited on target {[}kW{]}    & 305                        \\
Average beam power during spill {[}MW{]} & 2.3                       \\
\bottomrule
\end{tabular}
\end{table}

The BDF target core is constituted by 18 collinear cylinders with a diameter of 250~mm and variable thicknesses, from 25 to 80~mm for the TZM blocks and from 50 to 350~mm for the pure tungsten blocks, giving a total effective target length of around 1300 mm. The target cylinders length has been iteratively adjusted to reduce the level of temperatures and stresses reached in the different materials. Table~\ref{tab:TGT:targetblocks} summarizes the target core materials and the blocks longitudinal thickness.

\begin{table}[ht!]
\centering
\caption{\label{tab:TGT:targetblocks} Summary of the BDF final target cylinders longitudinal thickness and material configuration, including those of the core and the cladding.}
\smallskip
\begin{tabular}{cccc}
\toprule
\textbf{Block number} & \textbf{Core material} & \textbf{Cladding material} & \textbf{Length (mm)} \\
\midrule
1            & TZM           & Ta2.5W            & 80          \\
2            & TZM           & Ta2.5W            & 25          \\
3            & TZM           & Ta2.5W            & 25          \\
4            & TZM           & Ta2.5W            & 25          \\
5            & TZM           & Ta2.5W            & 25          \\
6            & TZM           & Ta2.5W            & 25          \\
7            & TZM           & Ta2.5W            & 25          \\
8            & TZM           & Ta2.5W            & 25          \\
9            & TZM           & Ta2.5W            & 50          \\
10           & TZM           & Ta2.5W            & 50          \\
11           & TZM           & Ta2.5W            & 65          \\
12           & TZM           & Ta2.5W            & 80          \\
13           & TZM           & Ta2.5W            & 80          \\
14           & W             & Ta2.5W            & 50          \\
15           & W             & Ta2.5W            & 80          \\
16           & W             & Ta2.5W            & 100         \\
17           & W             & Ta2.5W            & 200         \\
18           & W             & Ta2.5W            & 350         \\
\bottomrule
\end{tabular}
\end{table}

\begin{figure}[htbp!]
\centering %
\includegraphics[width=1\linewidth]{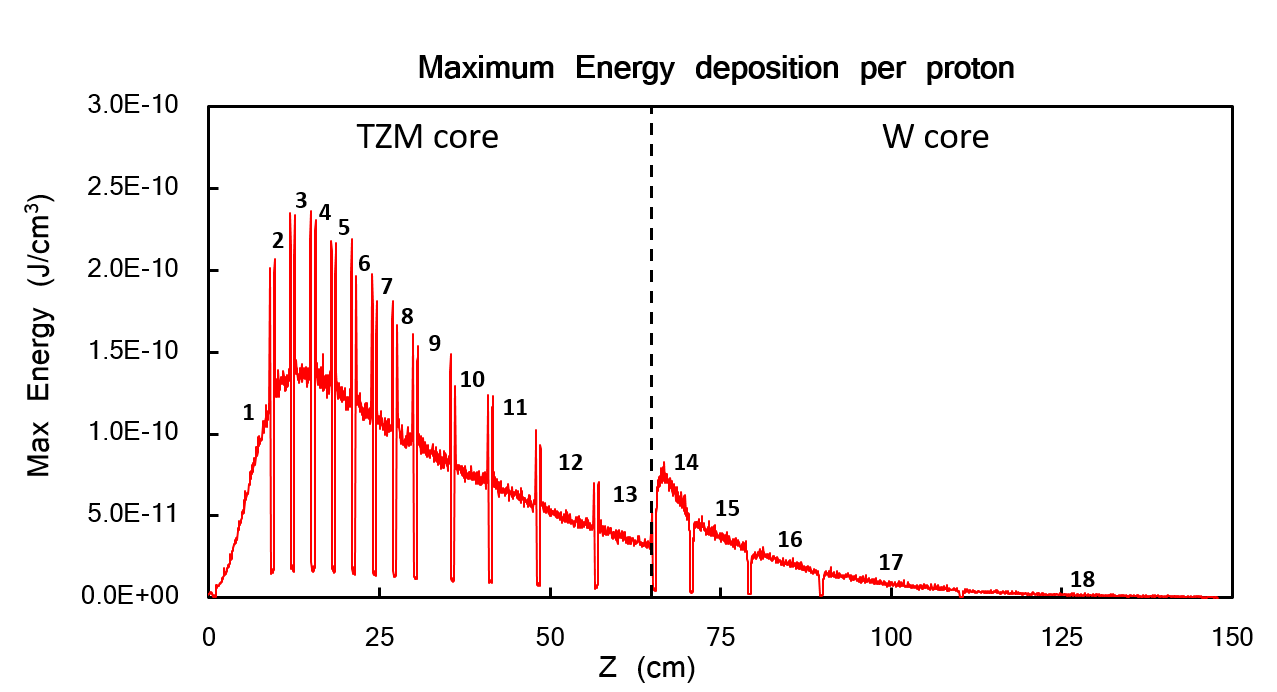}
\caption{\label{fig:TGT:FLUKAenergy} The figure shows the maximum energy deposition (per unit volume and per proton) in the BDF target blocks along the longitudinal axis, obtained via FLUKA Monte Carlo simulations.}
\end{figure}

Figure~\ref{fig:TGT:FLUKAenergy} shows the maximum energy deposition in the longitudinal direction after one proton impact on target, obtained via FLUKA Monte Carlo simulations~\cite{FLUKA_Code}. Owing to the fact that Ta2.5W has a higher density than TZM, the peak energy deposition is reached in the target blocks cladding. The maximum energy values for the TZM core part (first half of the target) are concentrated in blocks 2 to 8, which have been segmented to have the shortest length (25 mm). The reduced thickness of these blocks allows for a more effective heat dissipation by the water flowing through the 5 mm gaps between the target cylinders, aiming for a reduction of the temperatures reached in the core and on the surface. 
The thickness of the following blocks is gradually increased as the total energy deposition is decreasing. A similar approach is used for the second - higher density - part of the target made of pure tungsten: the first block is the most loaded one in terms of heat deposition, similarly to blocks 9 and 10 (TZM core), hence the thickness of this block has been set to 50 mm. The length of the following tungsten core blocks is then increased progressively as the energy deposited decreases. 

The optimization of target blocks thickness is essentially based in the distribution of maximum energy deposition per block (shown in Figure~\ref{fig:TGT:FLUKAenergy}). As will be discussed in Section~\ref{Sec:TGT:Simus}, the most critical stresses and cladding temperatures will be reached in the upstream thin blocks (i.e. blocks 3 to 6), where the interaction with the primary beam leads to the highest values of peak energy deposition and power density.

A further improvement of the target design aiming at increasing its operational lifetime could be possible through an additional segmentation of the low-thickness target blocks. However, the insertion of supplementary water gaps in the target design is not desirable from the physics perspective, since the presence of water in the longitudinal direction reduces the production of charmed mesons for the experiment leading to higher background. 

The advanced optimization of the target blocks thickness would require an evaluation of the temperatures and stresses in parallel with the variation of the different cylinders length. These calculations would be very costly in terms of time and computational power, given the high complexity of the thermal and structural Finite Element Model (FEM) simulations carried out for the target performance evaluation (see Section~\ref{Sec:TGT:Simus}). The target blocks thickness optimization could be the subject of future studies if the operational conditions would require it.

\begin{figure}[htbp]
\centering %
\includegraphics[width=1\linewidth]{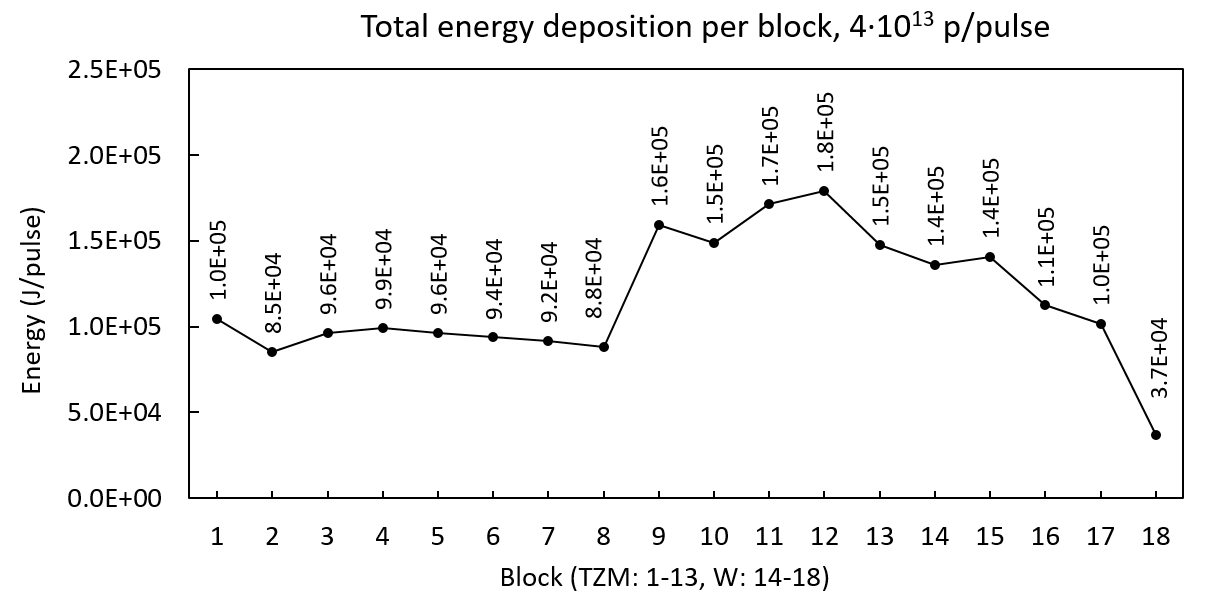}
\caption{\label{fig:TGT:totalenergy} Total energy in every block of the BDF target, assuming  a nominal pulse intensity of \num{4e13} protons.}
\end{figure}

The total energy deposited in each target block is reported in Figure~\ref{fig:TGT:totalenergy}; it can be observed that the total energy distribution for blocks 1 to 8 is rather uniform, with an energy deposition per block around 100 kJ/pulse, lower than the more downstream cylinders. This is justified by the fact that the volume of the first target blocks is much smaller than for the rest of the blocks, the thickness of blocks 2 to 8 being the shortest (25 mm). Blocks 9 to 18 also present a quite uniform distribution of total energy, around 150 kJ/pulse, excepting for the last 3 tungsten blocks where the energy deposition is considerably low, due to the dump/absorber nature of the target (roughly a total of 12 nuclear interaction lengths). 

\subsection{Dilution system for target}
\label{Sec:TGT:Design:dilution}

\begin{figure}[htbp!]
\centering %
\includegraphics[width=1\linewidth]{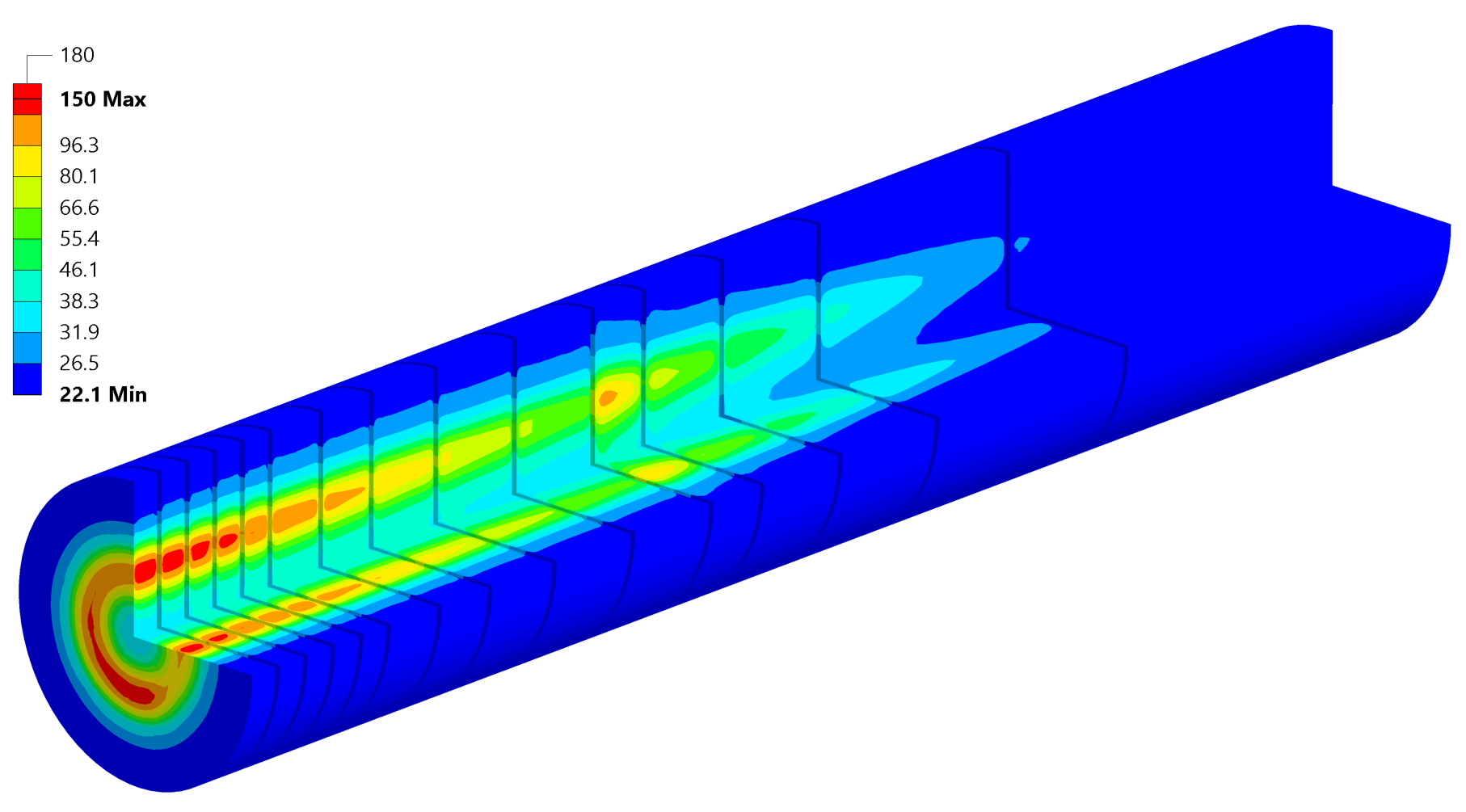}
\caption{\label{fig:TGT:dilution_temp} The figure shows the temperature distribution (in degrees Celsius) in the BDF target/dump during the beam dilution. The BDF beam dilution scheme currently consists in 4 beam sweep turns over a 50 mm radius circular pattern during 1 second (i.e. 250 ms per circle).}
\end{figure}

The high energy deposited on target requires beam dilution by the upstream magnets in the TDC2, as detailed in Chapter~\ref{Chap:Transfer}, since the impact of a non-diluted beam would lead to the premature target failure by fracture of the core and/or cladding (melting is not likely to occur given the high melting point of the target materials). The pattern for the beam dilution has been optimized taking into account the mechanical performance of the target and the different restrictions imposed by the upstream magnets. As a result, the SPS primary beam is foreseen to be diluted following a circular pattern, with 4 turns over a 50 mm radius circle for each 1-second pulse (Figure~\ref{fig:TGT:dilution_temp}).

The dilution system design has been improved with respect to the initial design, which consisted in an idealised Archimedean spiral with a radius from 5 to 35 mm, and 5 turns in 1 second. In order to evaluate the performance of the target under different dilution systems, thermo-mechanical calculations have been carried out in one of the most loaded target blocks, to compare the maximum temperatures reached after several pulses for each dilution scenario. 

\begin{figure}[htbp!]
\centering %
\includegraphics[width=1\linewidth]{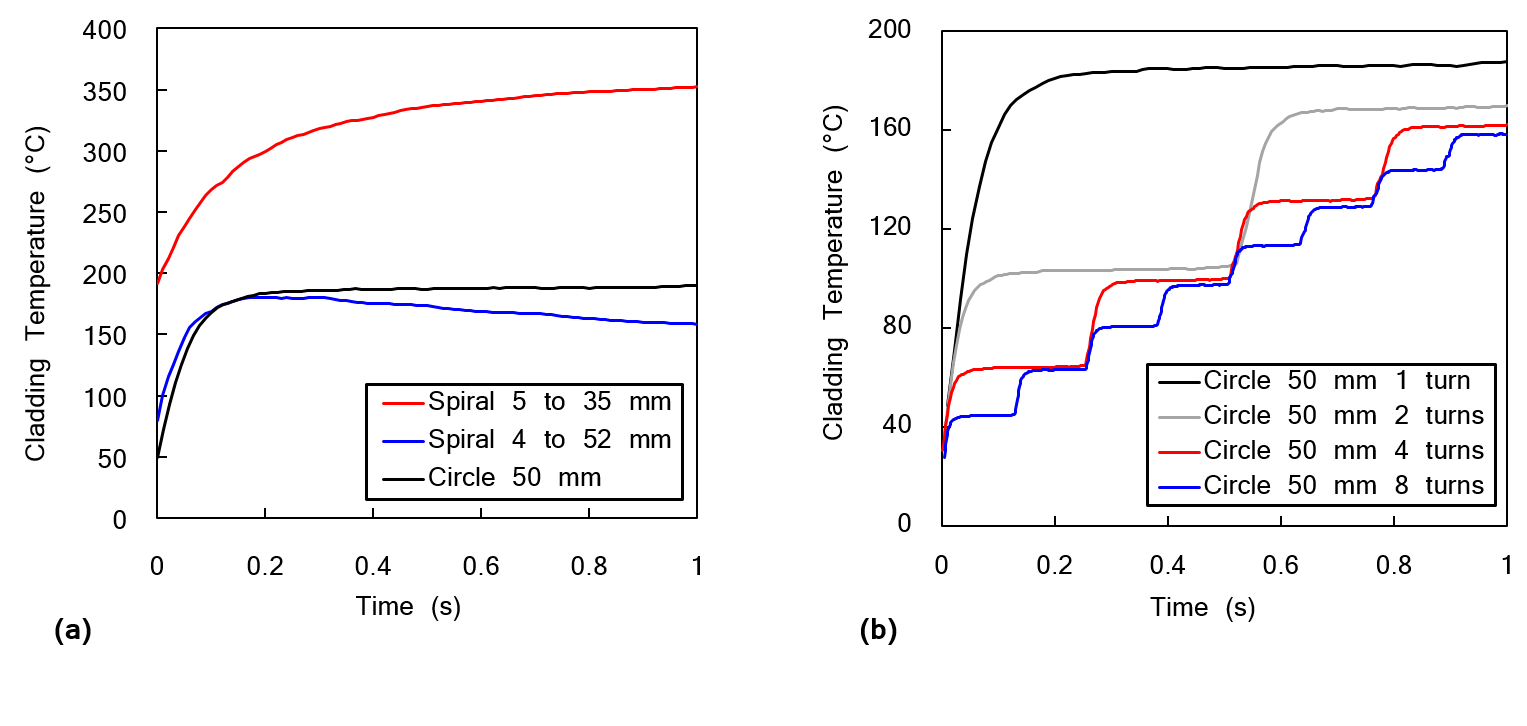}
\caption{\label{fig:TGT:dilution} The figure shows a comparison of the maximum cladding temperature between (a) spiral and circular dilution patterns and (b) several circular turns.}
\end{figure}

Figure \ref{fig:TGT:dilution}a presents the evolution of the maximum temperature in the Ta2.5W cladding during a one second beam pulse (after several beam pulses) for different dilution scenarios. A similar trend is observed for both the TZM and W cores. The increase of the sweep radius has a considerable impact in the reduction of the maximum temperatures; it can be seen that the temperatures reached for a beam sweep over a 4 to 52 mm radius spiral are $200\,^{\circ}\mathrm{C}$ lower than for the first dilution design over a 5 to 35~mm radius spiral. The dilution amplitude was therefore increased to the limits accepted by the upstream magnets (around 50~mm). 

Furthermore, it can be observed that for an almost equivalent dilution amplitude of 50 mm, the circular sweep of the upstream magnets leads to a comparable level of temperatures in the cladding with respect to the spiral dilution pattern. The dilution system complexity is substantially reduced by considering a circular pattern dilution instead of a spiral pattern: the spiral pattern requires a controlled variation in frequency and amplitude from the dilution magnets, while the circular sweep of the beam requires constant amplitude and frequency in the dilution. Therefore, a circular dilution of radius 50 mm is chosen as baseline sweep pattern for the magnet dilution system.

Finally, Figure~\ref{fig:TGT:dilution}b presents a comparison between the maximum temperature build-up for a circular dilution over a 50~mm radius circle with several turns for each 1-second pulse. It can be seen that the increase in the number of turns reduces the maximum temperatures reached in the cladding to $160\,^{\circ}\mathrm{C}$ approximately. 4 circular turns were selected as an optimal compromise between reducing the temperature and stress levels and keeping the dilution frequency within reasonable limits for the dilution magnets. More details about the beam dilution system can be found in Chapter~\ref{Chap:Transfer}.

As a further improvement of the dilution system design, the beam transverse size at the target/dump position has been maximized taking into account the limitations imposed by the aperture of the upstream magnets. It has been proven that a larger beam spot size leads to lower temperatures and stresses on target, since the energy deposition is more distributed in the materials volume. The compromise between maximizing the spot size of the round beam and the aperture restrictions of the transfer line magnets (see Chapter~\ref{Chap:Transfer}), concluded with the selection of a beam size on target of 8 mm 1$\sigma$.


\section{Target energy deposition \& radiation damage studies}
\label{Sec:TGT:EneRad}

\subsection{Introduction}

FLUKA Monte Carlo simulations~\cite{FLUKA_Code,Ferrari2005} have been performed in order to evaluate the energy deposited on the target blocks as well as radiation damage studies around the target.

\begin{figure}[ht!]
\center
\includegraphics[width=0.45\textwidth]{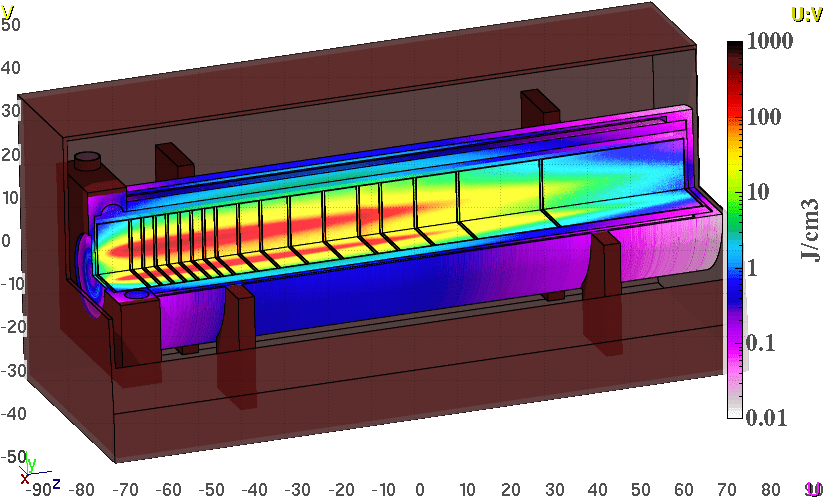}
\includegraphics[width=0.495\textwidth]{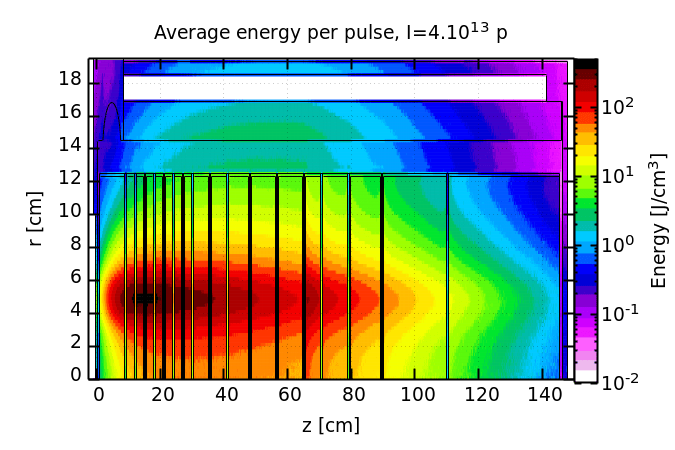}
\caption{The figure shows the FLUKA model of the BDF target, including the target, container and the energy depositions itself (left). On the right, the figure shows the radial energy deposition (in J/cm$^{3}$) of the BDF target in 1 pulse of \num{4e13} protons.
\label{fig:TGT:EnDep}}
\end{figure}

The target core is composed of 18 cylinder blocks with 125 mm radius and different thicknesses. The thickness have been iteratively adjusted as discussed in the previous sections.
The first 13 blocks are made of TZM while the remaining blocks are pure tungsten (W). Every block is cladded with a tantalum alloy of 1.5 mm thickness. The FLUKA model of the BDF target assembly could be seen in the left part of Figure \ref{fig:TGT:EnDep}, visualized using the FLAIR \cite{FLAIR} interface. The blocks and the target vessels are displayed (including the energy deposition on the blocks normalized to 1 pulse with
\num{4e13} proton). On the right, the Figure shows the radial energy deposition (in J/cm$^{3}$) of the BDF target in 1 pulse of \num{4e13} protons.

The FLUKA simulations with version 2011.2x.6 have been performed with a 400 GeV/c proton beam, having a Gaussian shape with 8 mm sigma on target, in a circular sweep with 5 cm radius, to reproduce the shape referred in Section~\ref{Sec:TGT:Design:optimisation}. The sweep beam helps to dissipate destructive effect of the beam in the material, such as high peak of energy deposition or damage to the material structure itself. The energy deposition have been normalized to 1 pulse with \num{4e13} protons, while DPA and radiation to electronics (R2E) parameters to 1 year of irradiation assuming \num{4e19} protons.

\subsection{Energy deposition}

\begin{figure}[!ht]\center
\includegraphics[width=0.495\textwidth]{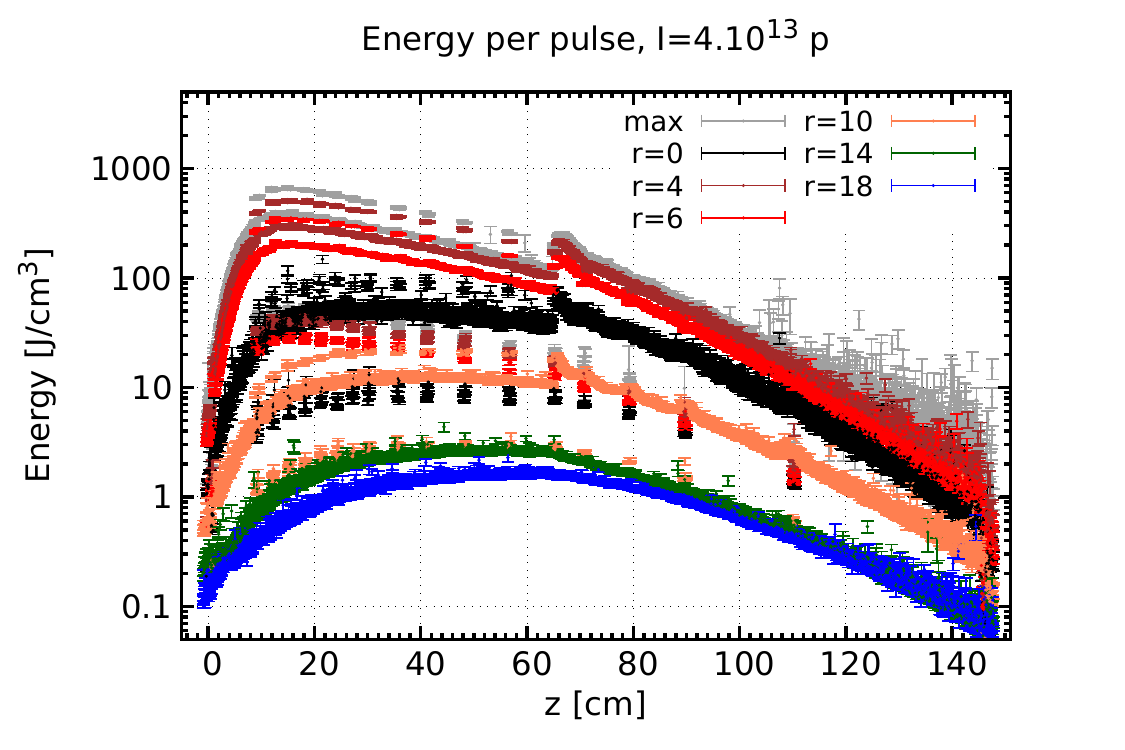}
\includegraphics[width=0.495\textwidth]{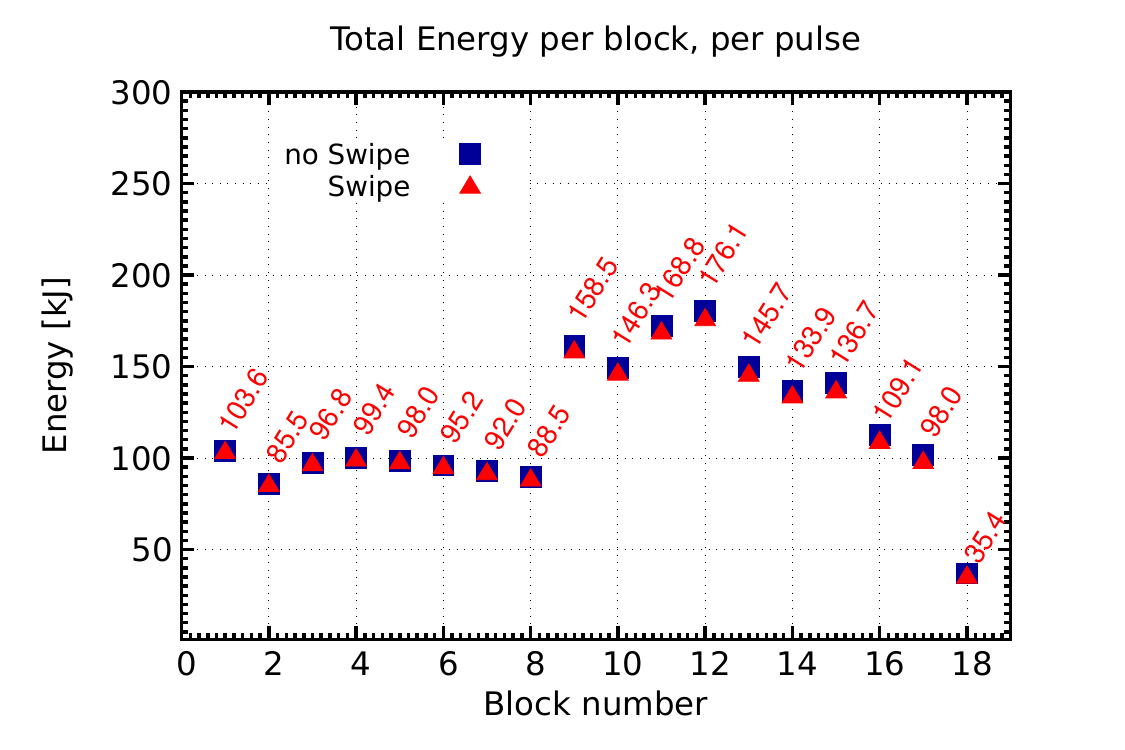}
\caption{(Left) The figure shows the average deposited energy density along the beam axis (z) for several radius for 1 pulse at \num{4e13} p as well as (Right) the total energy deposited per block and per pulse. For the latter, a comparison between the case with and without sweep is shown.
\label{fig:TGT:EnDep2}}
\end{figure}

Figure~\ref{fig:TGT:EnDep2} shows the average energy deposited per normalized pulse along the beam axis (z axis) for several radius. The target energy density can reach values between 0.1 J/cm$^3$ up to a maximum of 500 J/cm$^3$. From the distribution is clear to see the transition between the TZM and tungsten blocks. Similarly, the right part of the figure shows the total energy deposited in each of the blocks. From block 8 to 9, the total energy increases since the thickness changes from 25~mm to 50~mm. The energy deposited on the 18 blocks is approximately 2.2 MJ per pulse, which means they absorb about 78\% of the total proton beam energy.

\begin{figure}[ht!]\center
\includegraphics[width=0.495\textwidth]{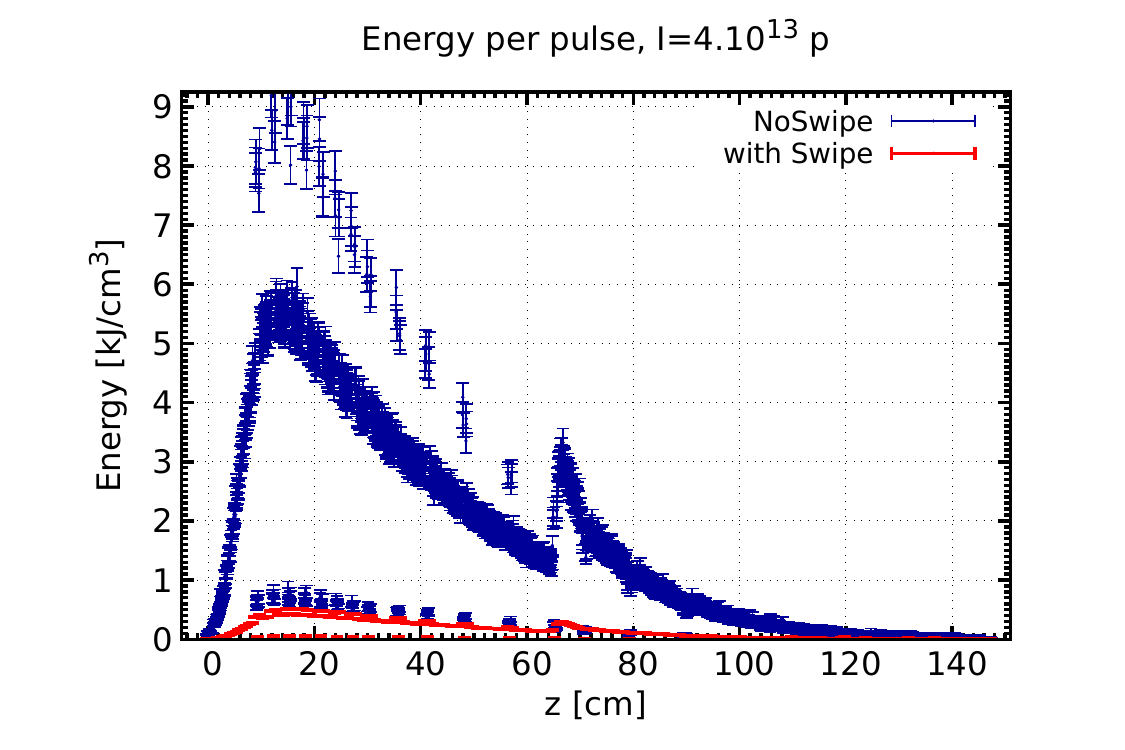}
\caption{The figure shows the maximum deposited energy density along the beam axis (z) for the same beam, concentrated in the center (blue) or swept in a circular path (red). The sweep version is the one considered for the target.
\label{fig:TGT:EnDep3}}
\end{figure}

Using a beam with same shape but in a sweep path or concentrated in the center gives approximately the same energy deposition per block, as can be seen in the right part of Figure~\ref{fig:TGT:EnDep2}. However, if the beam is not dissipated with the circular path, the maximum energy density deposited in the target per pulse is about 10 times higher (as shown in figure \ref{fig:TGT:EnDep3}). This would significantly increase the temperature and pressures gradients inside the material. It would rise the risk of failure, although the total energy per block remain similar.  

\subsection{Dose around the target}
The BDF target is subjected to a considerable high amount of protons per year, expected in the order of \num{4e19} protons/year. Owing to the high average beam power and due to the high density of the production target, extremely high radiation levels are expected for the components around the target assembly. The information of the cumulative dose is important to determine the radiation hardness of the target and proximity shielding, water-cooling plugin system, for example, or other equipment placed nearby. To this aim, the same FLUKA Monte Carlo code model was employed. 

\begin{figure}[!ht]\center
\includegraphics[width=0.495\textwidth]{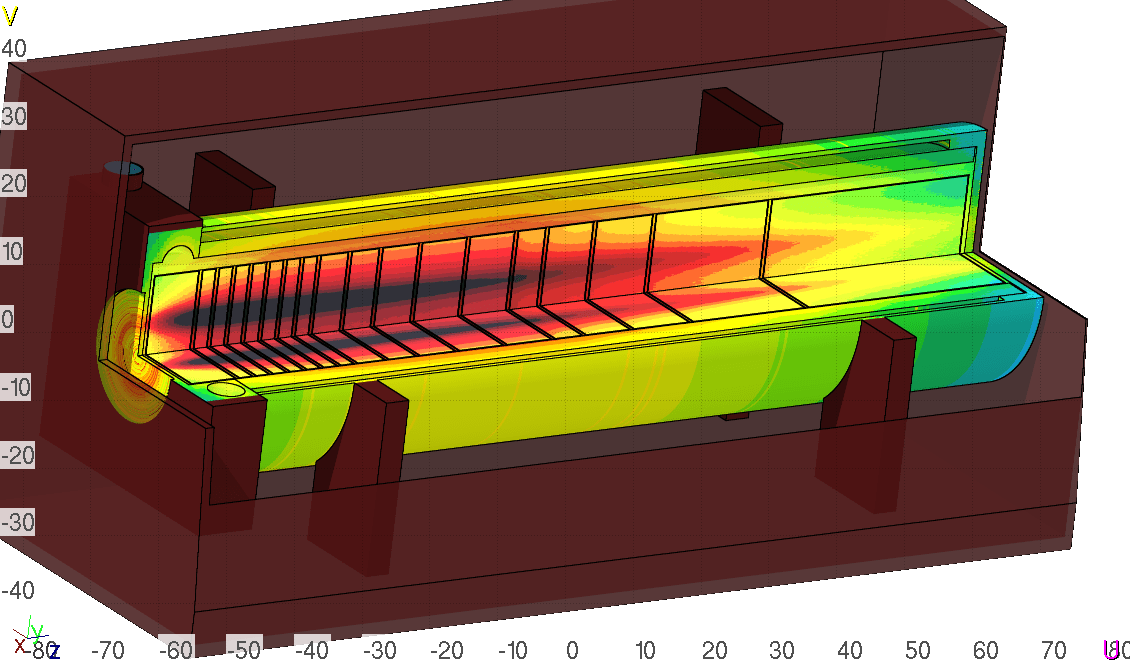}
\caption{The figure shows an isometric view of the FLUKA model including the dose on the target per 1 year assuming \num{4e19} p. 
\label{fig:TGT:Dosebdf}}
\end{figure}

\begin{figure}[!ht]\center
\includegraphics[width=0.495\textwidth]{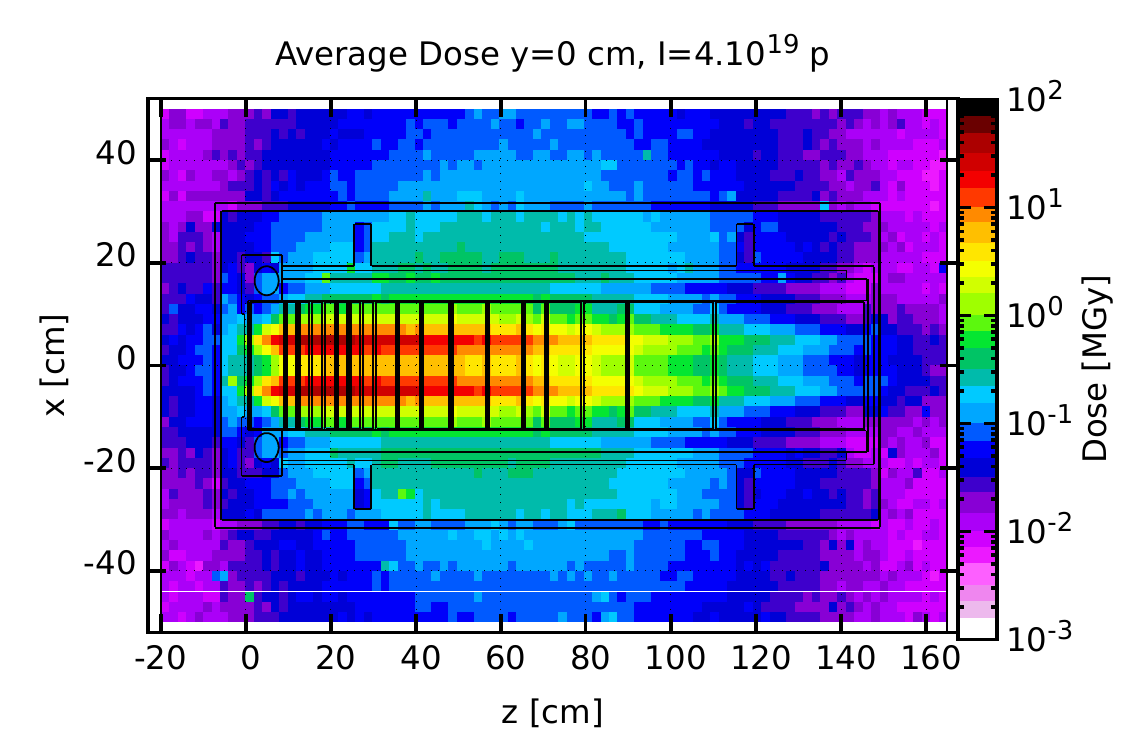}
\includegraphics[width=0.495\textwidth]{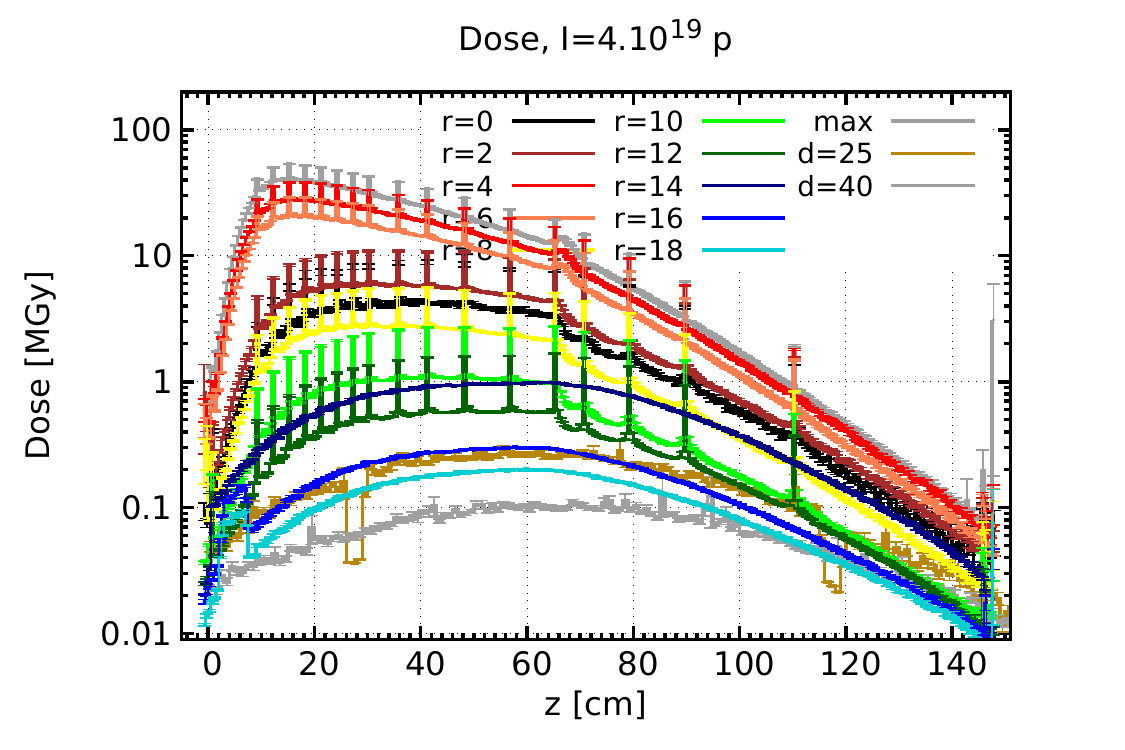}
\caption{(Left) The figure shows the yearly average dose in $y=[-1,1]$ cm around the target center. (Right) The figure shows the dose along the beam axis (z) for several radius, and outside the target at 25 and 40 cm from the target axis. Outside the target, the cylindrical geometry is lost, so the positions d=25 and d=40 cm corresponds to the average in y=[-1,1] cm and x= 25 or 40 cm respectively. Results normalized for 1 year, with \num{4e19} p.
\label{fig:TGT:dosebdf2}}
\end{figure}

In Figure~\ref{fig:TGT:Dosebdf}, the FLUKA model cut out, to display the center of the target is shown. Similarly, the left part of Figure~\ref{fig:TGT:dosebdf2} shows the yearly average dose averaged around the target center, while the right part shows the dose value at different distances from the target center axis. At 25 cm from the target center it can reach doses around $\sim300$ MGy per year and at about 40 cm from the target center it reaches $\sim100$ MGy per year.

\subsection{Neutron and protons fluence}
\begin{figure}[!ht]\center
\includegraphics[width=0.495\textwidth]{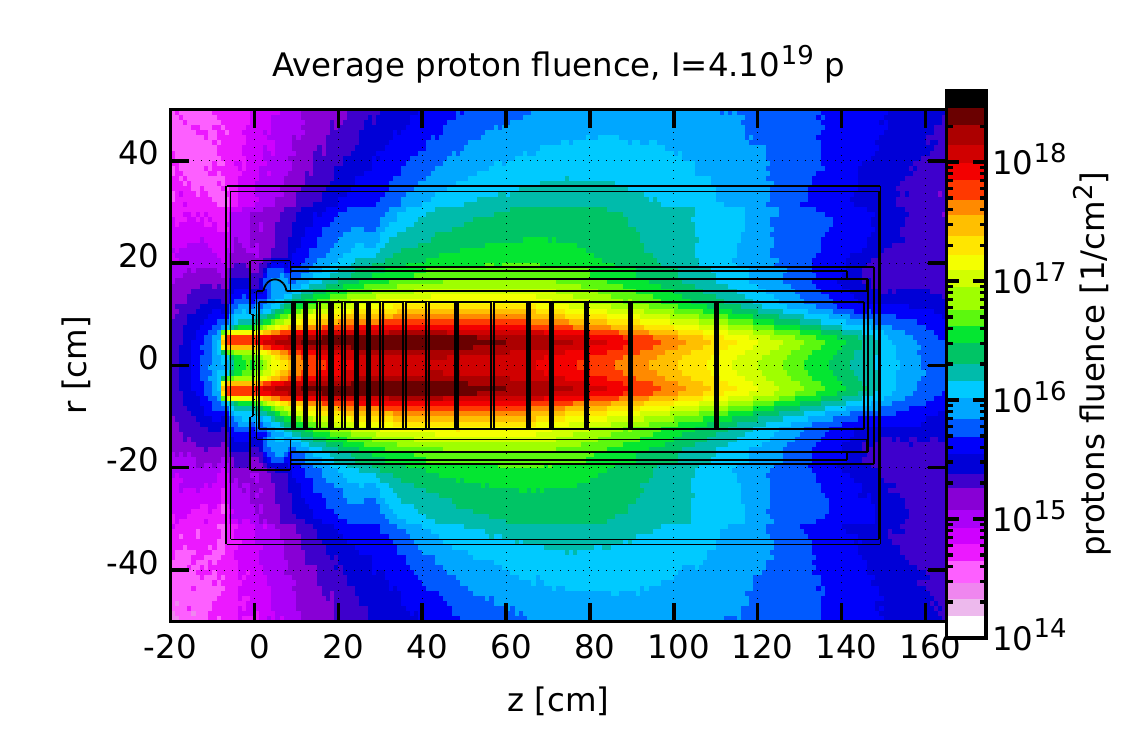}\includegraphics[width=0.495\textwidth]{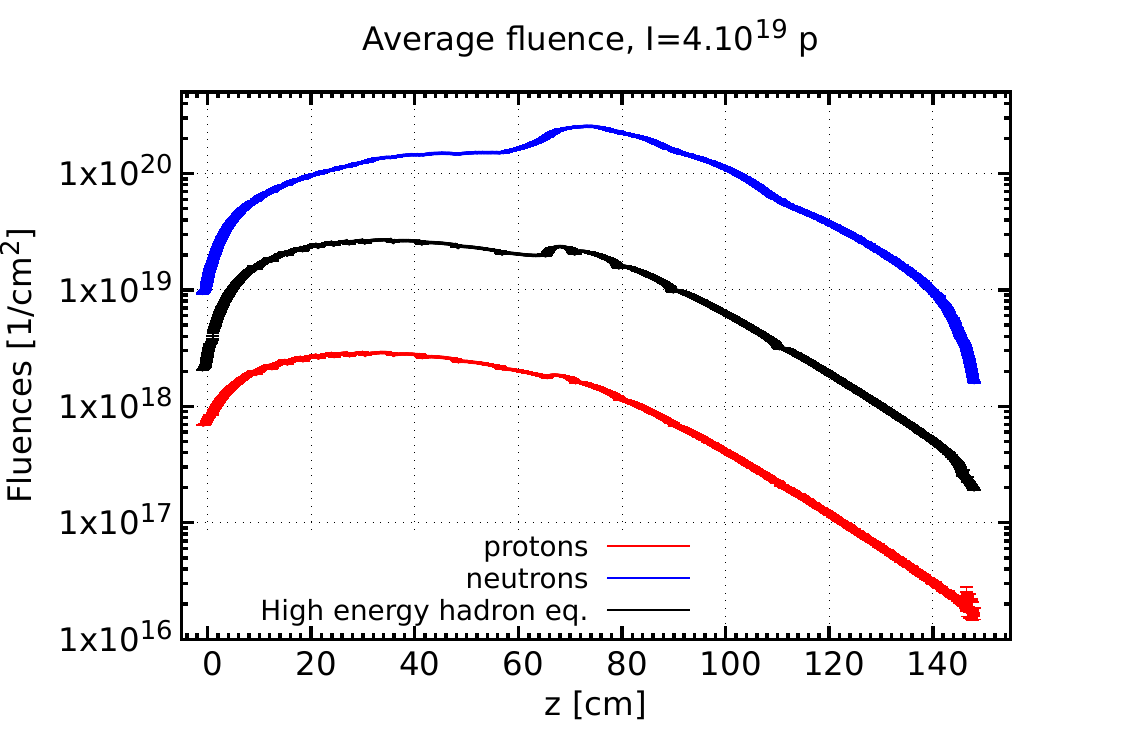}
\caption{(Left) Average proton fluence in $y=[-1,1]$ cm around the target center for 1 year of irradiation. The effect of the beam dilution is evident by the presence of an apparent double impact beam around the target center. (Right) Maximum protons and neutron density and high energy hadrons equivalent along the beam axis (z). Results normalized for 1 year, assuming \num{4e19} p.
\label{fig:TGT:Fluences}}
\end{figure}

The fluence of particles in the material is also important to understand the potential materials damage. In Figure~\ref{fig:TGT:Fluences} the yearly proton fluence (p/cm$^{2}$) is shown on the left, while the yearly maximum fluence of protons, neutrons and high energy hadron equivalent is shown on the right. The neutrons thresholds were considered at thermal energies, while the protons thresholds were set to 100 keV.
The target will see a maximum fluence of $\sim$\num{3e18} protons/cm$^2$ per year around the block 8 to 9 (TZM, $\sim$30 cm).
The yearly maximum fluence of neutrons is around the blocks 15 (W, $\sim$75 cm), 
with \num{2.5e20} n/cm$^2$. More discussion about the materials properties and properties variation with irradiation are shown in Section~\ref{sec:TGT:mats:radiation}.

\subsection{Radiation damage (DPA and gas production)}
The materials on the target are expected to suffer significant variations on the mechanical properties, due to the high cumulative number of protons colliding. More precisely, the Displacement per Atom (DPA) as well as the H and He gas production inside the core provides an indication on the extent of the effects on thermo-physical and mechanical properties. The same FLUKA model was used to simulate these values per year of operation (\num{4e19} p). 
The energy required to permanently displace an atom (damage energy threshold) for TZM has been assumed equal to the one for pure molybdenum (i.e. 60 eV), while 90 eV have been assumed for pure tungsten. The value for pure Ta was assumed at 53 eV and the one for stainless steel (316L) at 40 eV.

\begin{figure}[!ht]
\center
\includegraphics[width=0.495\textwidth]{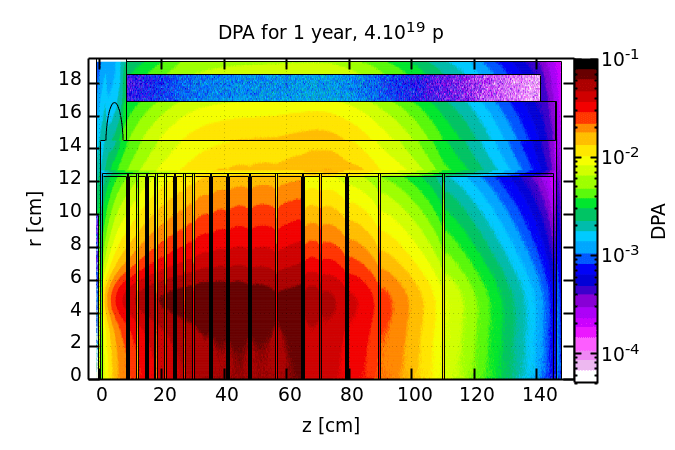}
\includegraphics[width=0.495\textwidth]{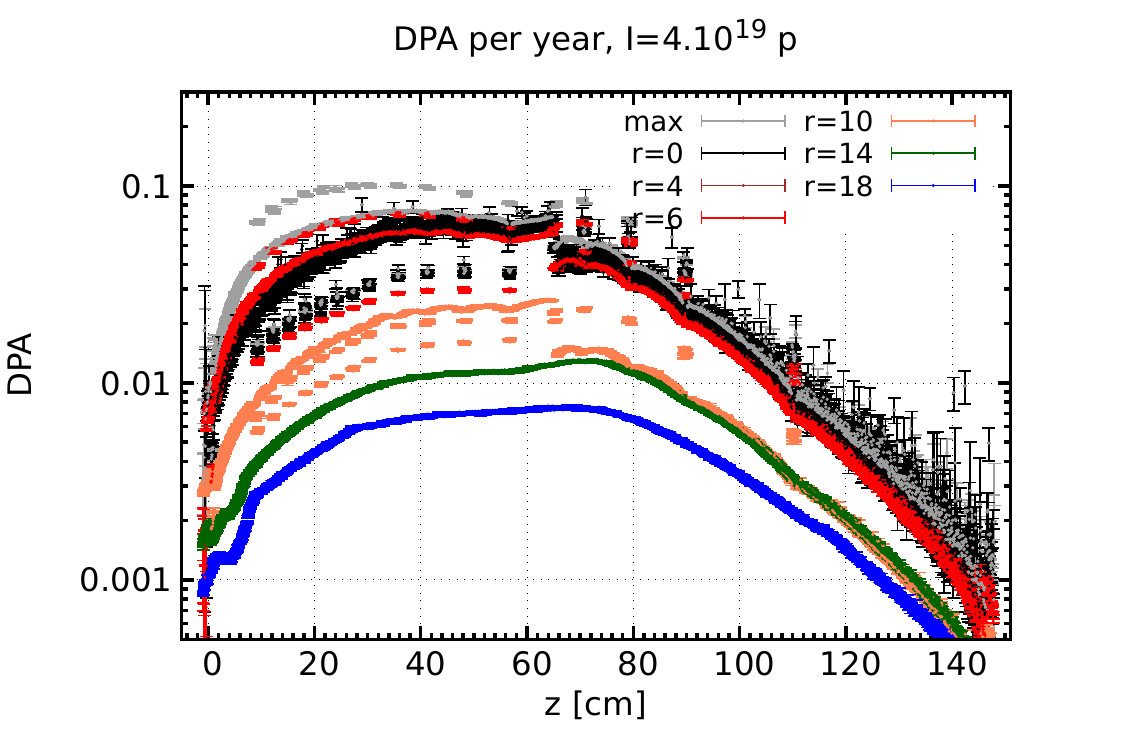}
\caption{The figure shows the DPA for the BDF target in 1 year of operation (\num{4e19} p). (Left) Radial average versus z (beam direction) and (Right) Maximum DPA along the beam axis (z) for several radius on the right.
\label{fig:TGT:DPA}}
\end{figure}

\begin{figure}[!ht]
\center
\includegraphics[width=0.48\textwidth]{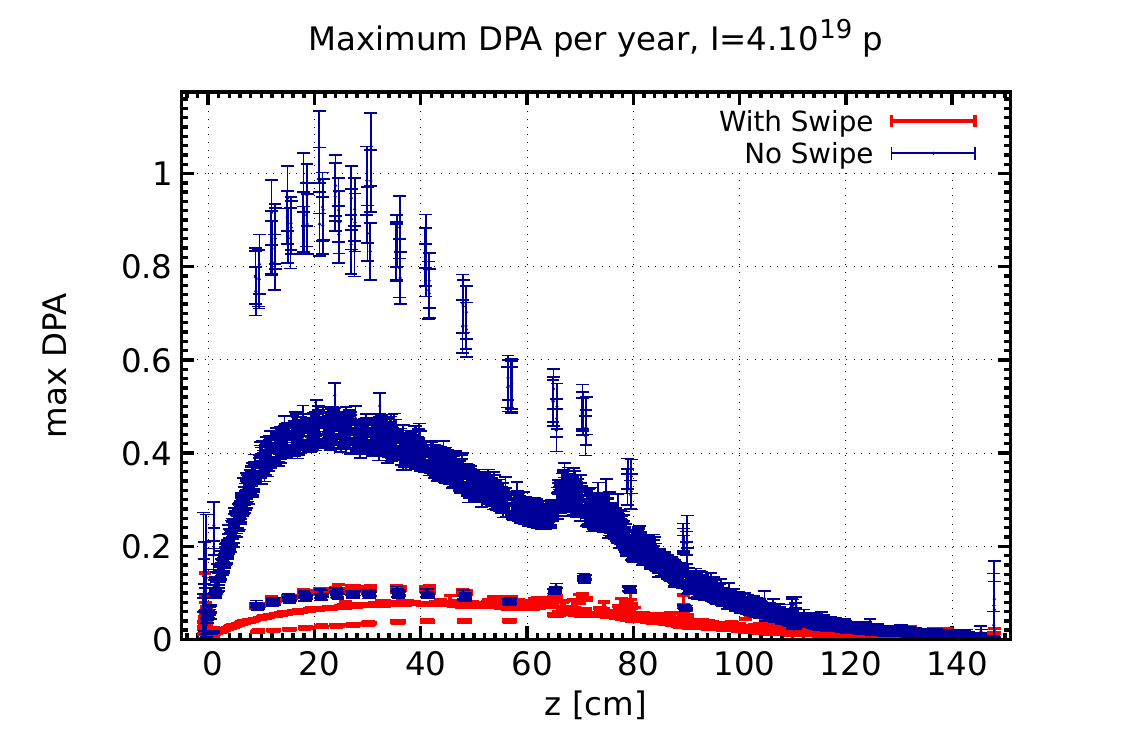}
\caption{Maximum DPA per year along the beam axis (z) for the same beam, concentrated in the center (red) or swept in a circular path (blue).
\label{fig:TGT:DPA2}}
\end{figure}

Figure~\ref{fig:TGT:DPA} shows the DPA radial distribution, as well as the maximum DPA for different radius.
The maximum DPA occurs at 5 cm from the center, where the beam is focused, and larger between the TZM blocks 9 and 12, where it could reach values about 0.1 per year. 
If the beam was focused on the center instead of a circular sweep, considerably higher values would be expected. In Figure~\ref{fig:TGT:DPA2}, a comparison between the DPA reached in case of no beam sweep and the baseline configuration case can be seen. The material, considering an unswept beam, would suffer about 5 times more DPA than the sweeped version. 

The gas production, hydrogen and helium are very important for the metal materials properties in operation, since gas particles can have severe life-limiting consequences for materials, even at lower concentrations. Moreover, this is special important in the helium case due to its low solubility in the crystal lattice.
The gas production can accumulate producing bubbles and producing defects in the metal, which could lead to swelling and/or grain-boundary embrittlement of the materials. 

\begin{figure}[ht]\center
\includegraphics[width=0.495\textwidth]{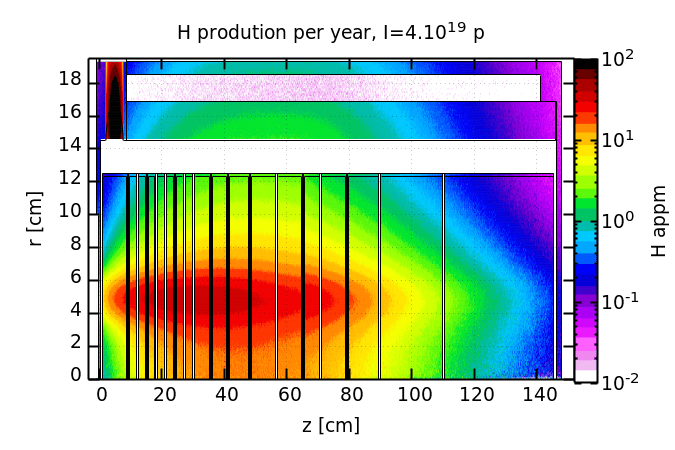}
\includegraphics[width=0.495\textwidth]{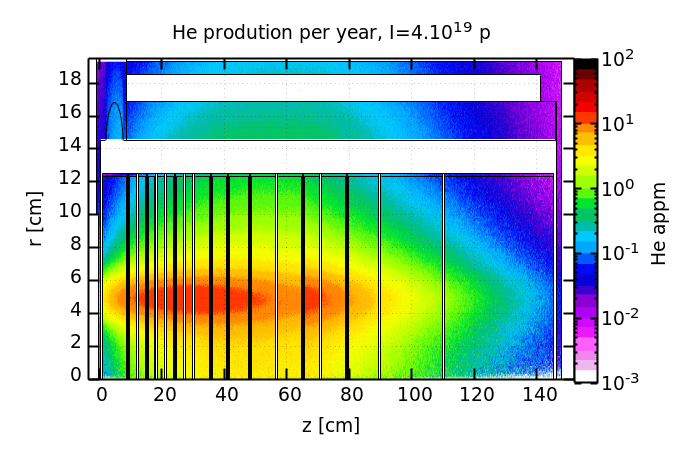}
\includegraphics[width=0.495\textwidth]{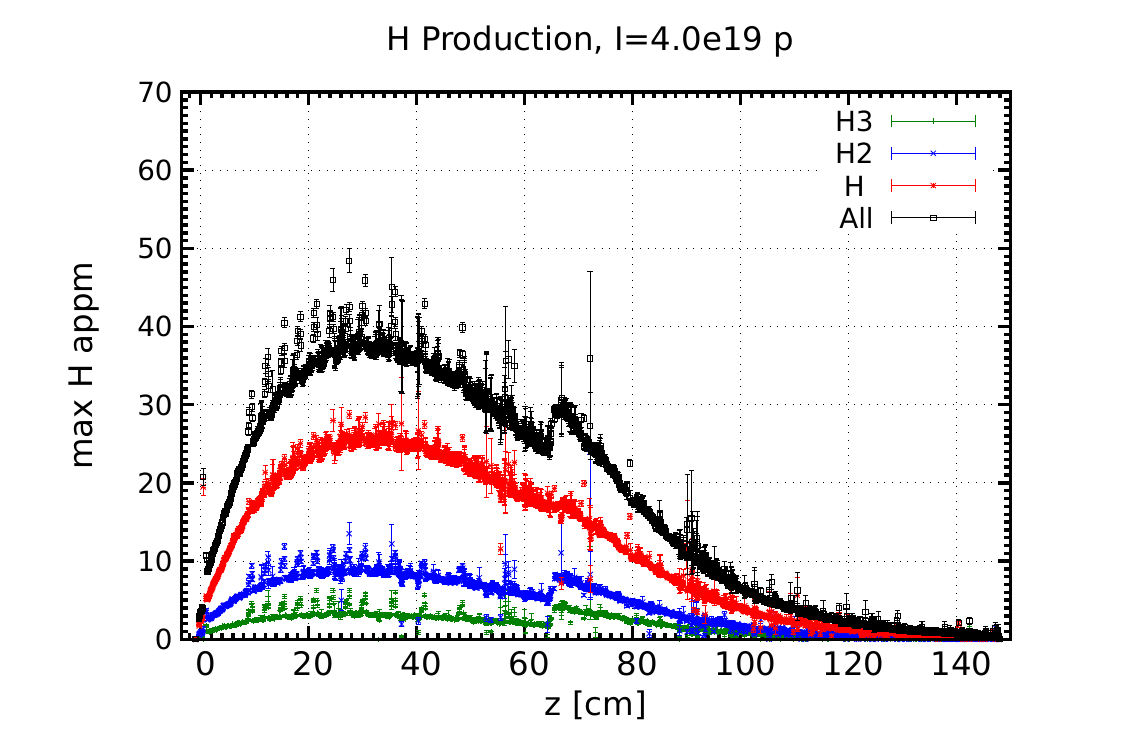}
\includegraphics[width=0.495\textwidth]{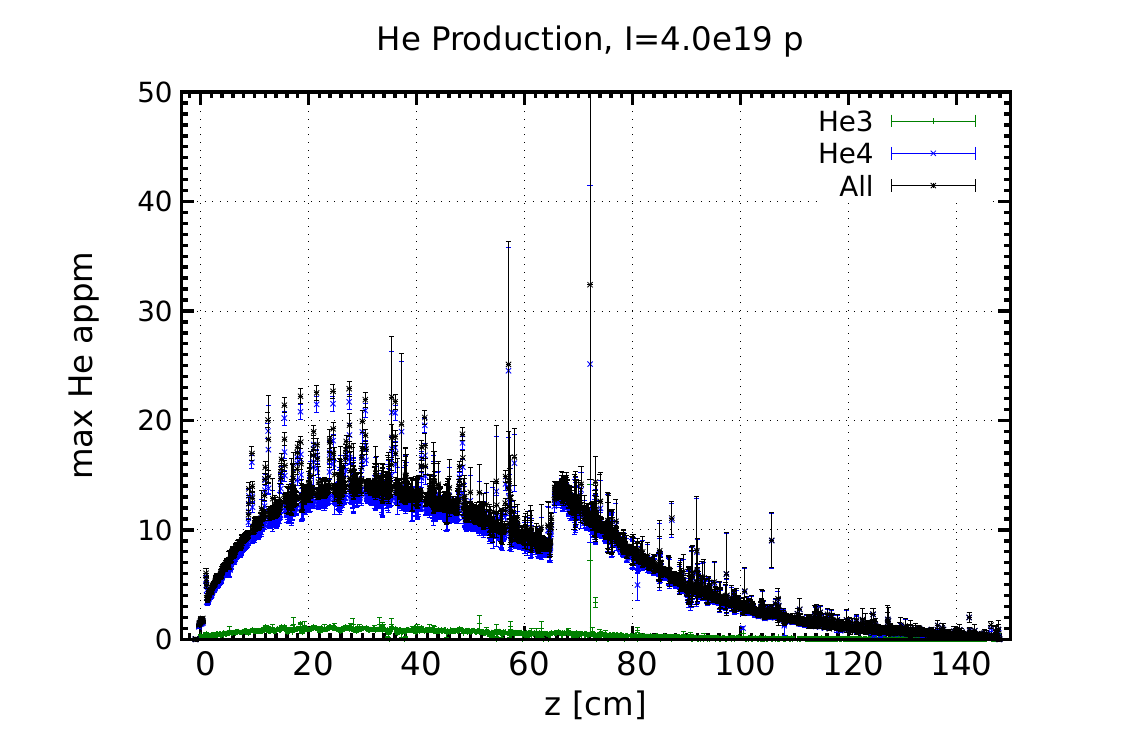}
\caption{Gas production per 1 year of operation in the BDF target: hydrogen production on the left and helium production on the right. On top is the average DPA around the target core and on the bottom is maximum along the beam axis (z).
\label{fig:TGT:GasProd}}
\end{figure}

In one year of operation, a maximum of about 40 hydrogen ppm and about 15 helium ppm can be produced in the target core, as shown in Figure~\ref{fig:TGT:GasProd}.
  
\begin{figure}[ht]\center
\includegraphics[width=0.495\textwidth]{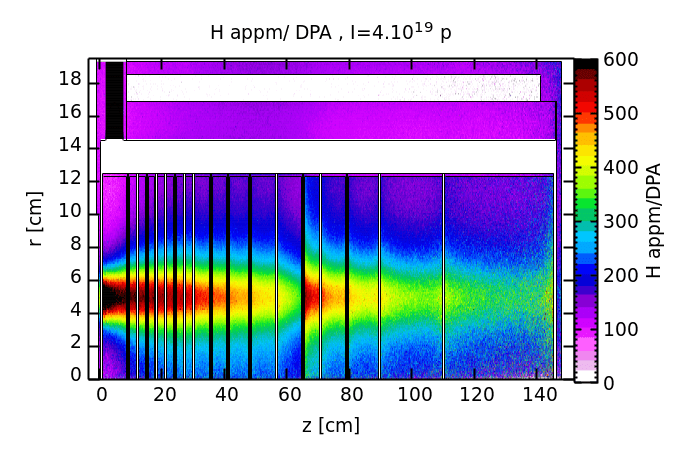}
\includegraphics[width=0.495\textwidth]{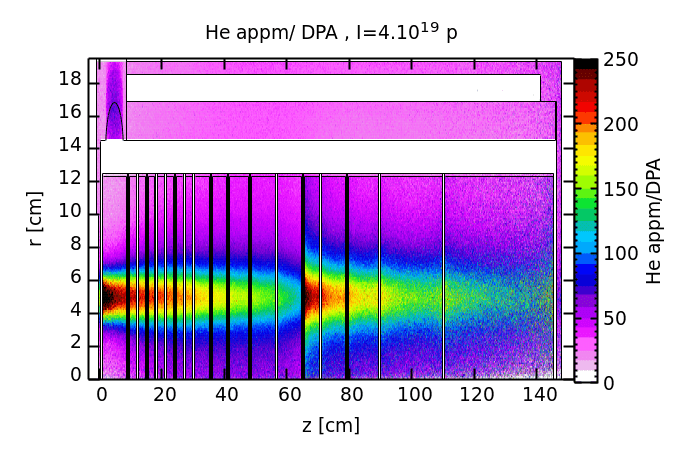}
\includegraphics[width=0.495\textwidth]{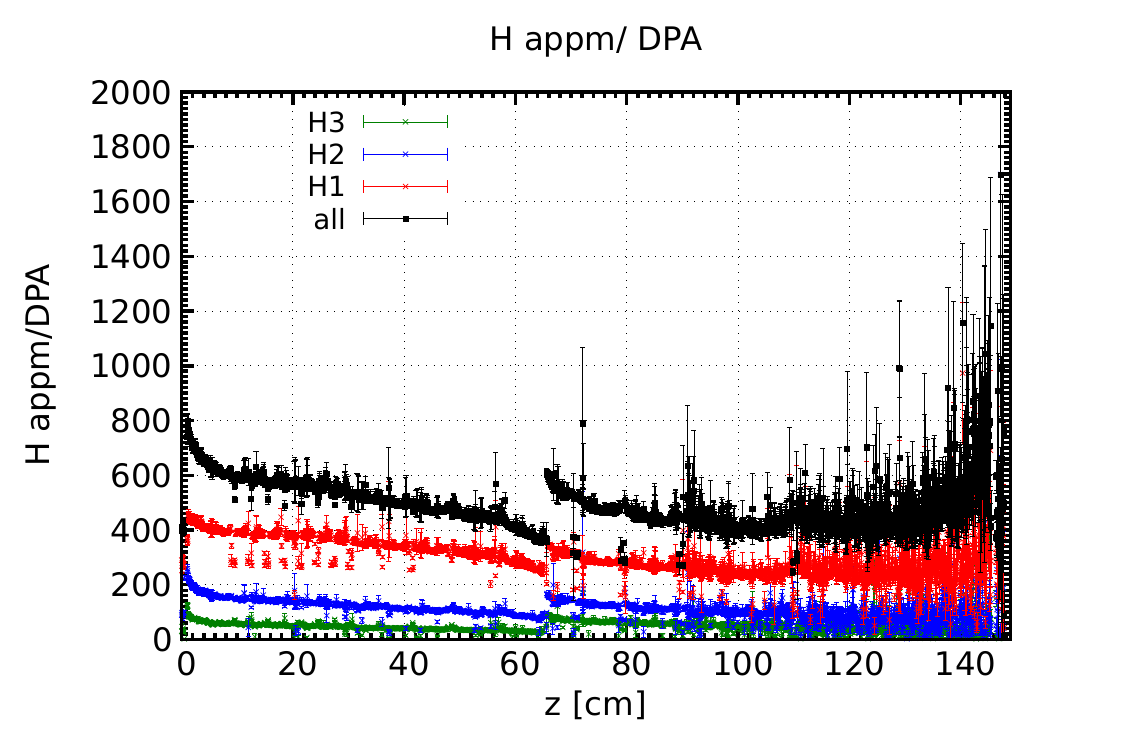}
\includegraphics[width=0.495\textwidth]{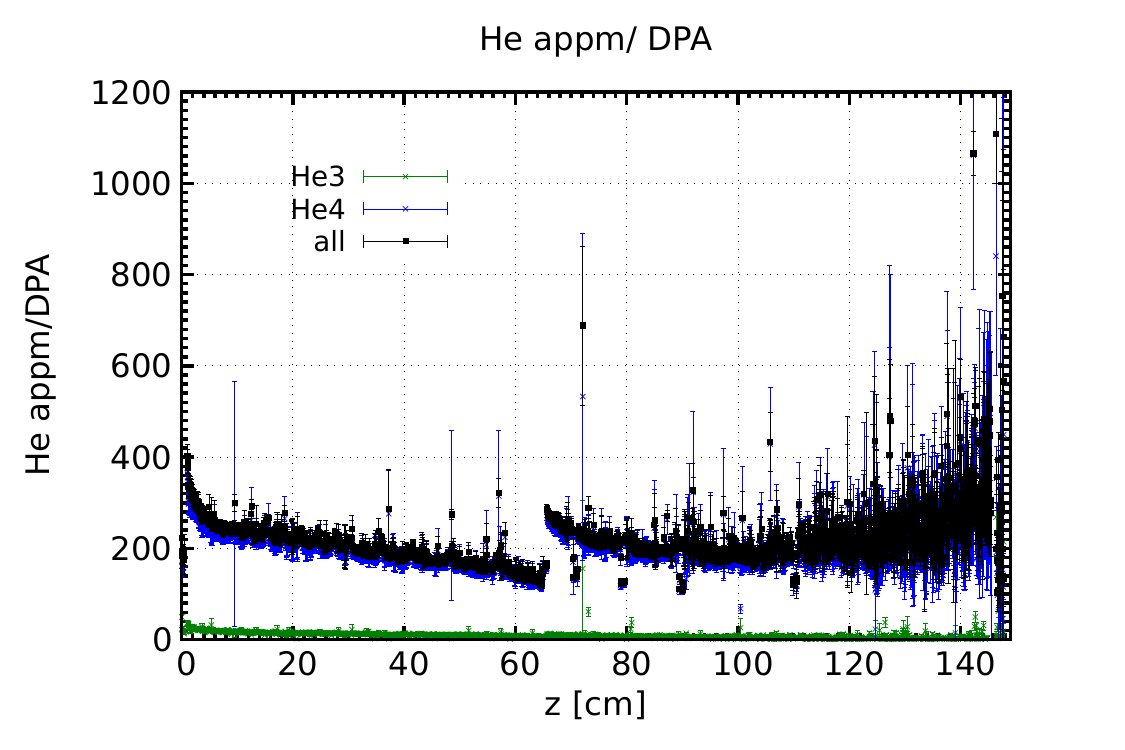}
\caption{Ratio of gas (H or He) production over DPA: hydrogen production on the left and helium production on the right. On top is the average ratio in 2D and on the bottom is maximum along the beam axis (z).
\label{fig:TGT:GasDPA}}
\end{figure}

Another important factor to be considered is the ratio of gas content over DPA. Figure~\ref{fig:TGT:GasDPA} shows the ratio for hydrogen (left) and for helium (right). It can be seen that the ratio is about 600 and 200 for hydrogen and helium respectively. 

For a summary of the results, the average maximums along z are reported in Table~\ref{tab:TGT:tableSum}. To avoid statistical fluctuations, the values correspond to the maximum of the average of 3 points. The Ta cladding receives slightly more dose, DPA and H/He production since it is in the border of the blocks and have a high density.  

\begin{table}[ht] \center \small
\begin{tabular}{lrrrrrrrrr} \hline 
\textbf{material/}&\textbf{Dose}&\textbf{pr fluence}&\textbf{n fluence}&\textbf{HeHad}&\textbf{DPA}&\textbf{H}&\textbf{He}&\textbf{H/DPA}&\textbf{He/DPA} \\
position&  [GGy] & [1/cm$^2$] & [1/cm$^2$]& [1/cm$^2$]&     & [appm] & [appm] &       &        \\ \hline

TZM  & $\sim$40 & \num{2.9e18} & \num{1.7e20} & \num{2.7e19} & 
0.074 & 37.9 & 15 &  $\sim650$ & $\sim275$     \\
block &  4    &    9     &     13      &   9    & 
10    &  8   & 8    &   1 & 1        \\ \hline

W &  $\sim$14 &  \num{1.8e18} & \num{2.5e20}   & \num{2.3e19} & 
0.053 & 30 & 13 &  $\sim560$ & $\sim300$     \\
block &  14    &   14     &     15      &   14    & 
14    &  14   & 14    &   14 & 18        \\ \hline

Ta2.5W &  $\sim$54 &  \num{2.9e18} & \num{2.5e20}    &\num{2.7e19} & 
0.053 & 51 & 30 &  $\sim$900 & $\sim$280     \\
block &  3-4    &   8 \text{and} 9     &     15      &   9    & 
9    &  9-10   & 9-10    &   1 & 1 \text{and} 2        \\ \hline
\end{tabular}
\caption{Summary of the maximum values for TZM, W and Ta2.5W cladding, considering each parameter reported in this section. HeHad corresponds to high energy hadron equivalent. For the Ta cladding, there are 3 bins in z per slice of cladding. The maximum here corresponds to the maximum of the average of the 3 bins per slice (to reduce fluctuations).  
\label{tab:TGT:tableSum}}
\end{table}

\FloatBarrier

\section{Summary of BDF target radiation protection considerations}

This section summarizes the radiological assessment for the final BDF target. The high intensity beam power deposited on the target poses challenges to the radiation protection for all the maintenance interventions. 
The studies include expected residual dose rates, storage and transport considerations. Finally, studies on target disposal were conducted. The studies are based extensive simulations with the FLUKA Monte Carlo particle transport code~\cite{FLUKA_Code,Ferrari2005} and ActiWiz3~\cite{Actiwiz}. Figure~\ref{fig:TGT:Dosebdf} shows the target as implemented in FLUKA. All studies assume $4\times10^{13}$ protons on target per spill (duration of 7.2 s) and an integrated total of $2\times10^{20}$ protons on target over five years operation, each with 83 days of operation followed by 272 days of shutdown.

\begin{figure}[!htb]

 
    \begin{subfigure}[b]{0.5\linewidth}
            \centering
            \includegraphics[width=\linewidth]{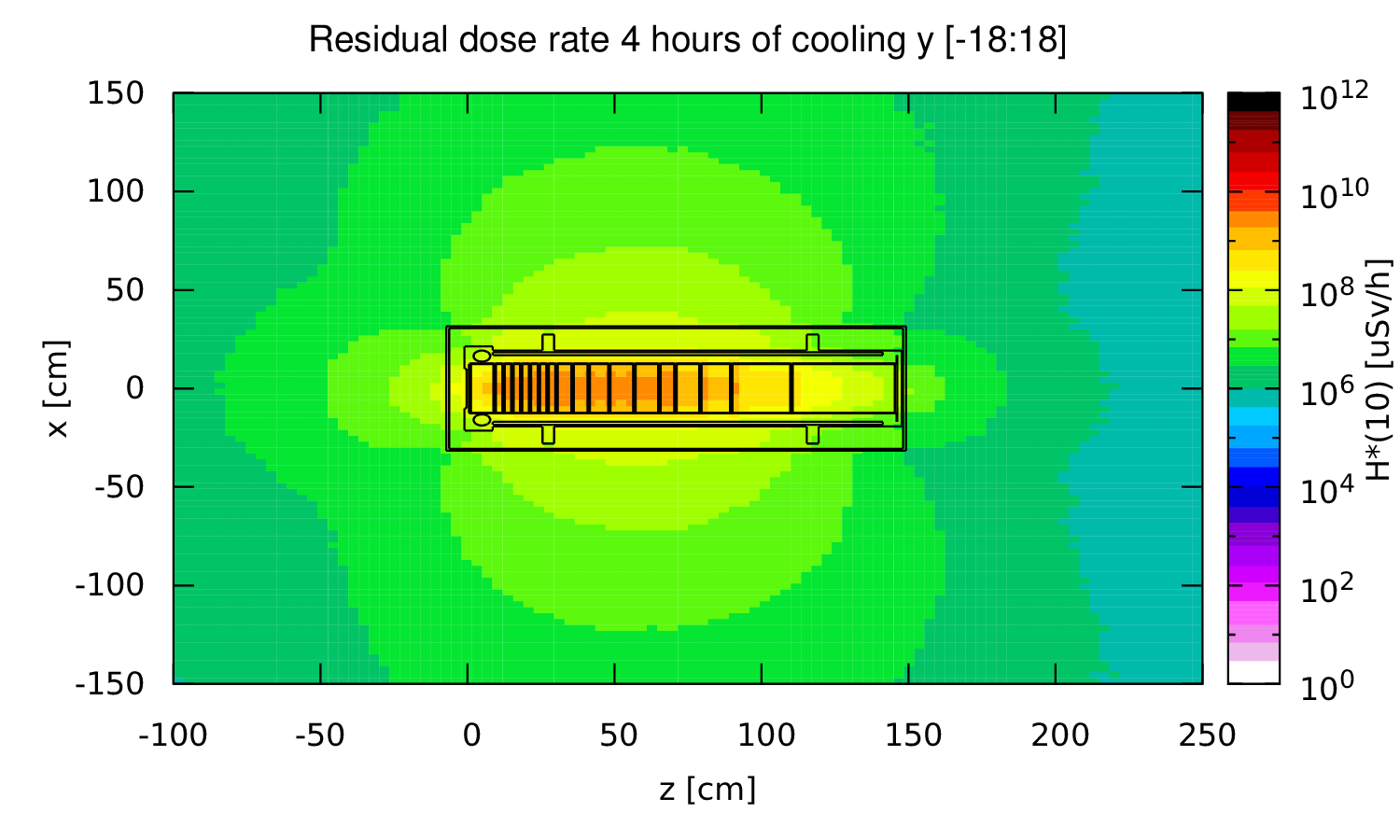}
    \label{fig:fred1}
    \end{subfigure}
   \begin{subfigure}[b]{0.5\linewidth}
            \centering
            \includegraphics[width=\linewidth]{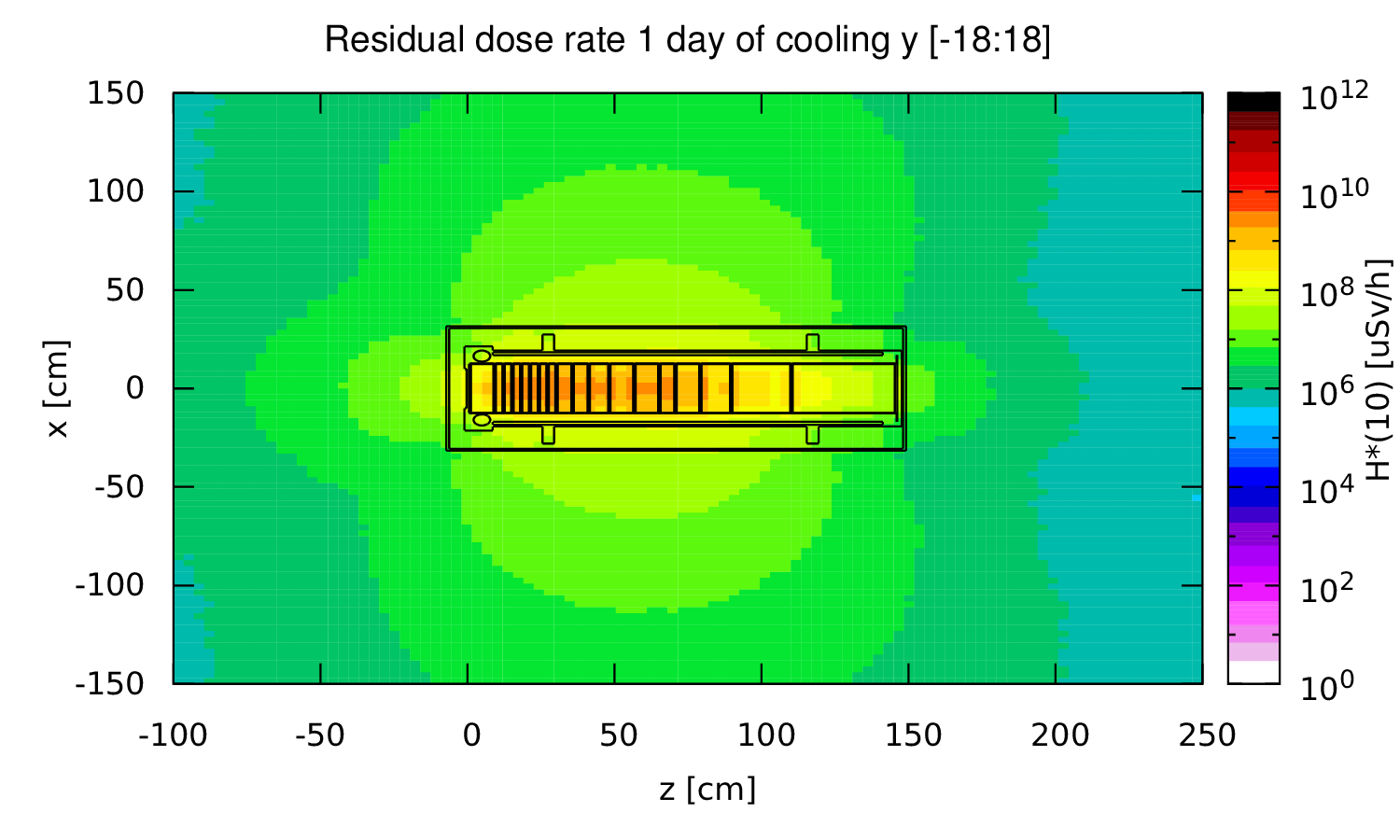}
    \label{fig:fred2}
    \end{subfigure}
    \begin{subfigure}[b]{0.5\linewidth}
            \centering
            \includegraphics[width=\linewidth]{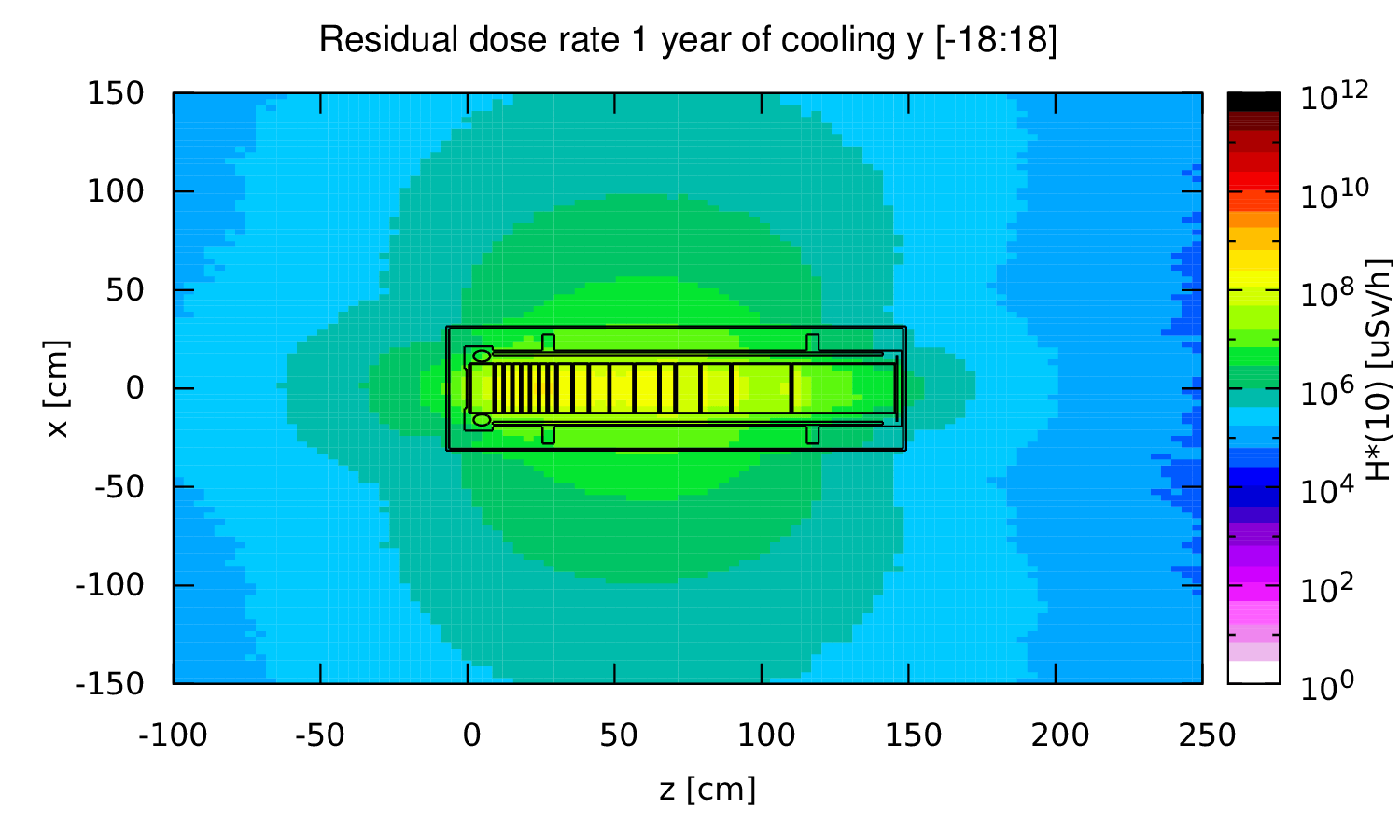}
    \label{fig:fred3}
    \end{subfigure}
    \begin{subfigure}[b]{0.5\linewidth}
            \centering
            \includegraphics[width=\linewidth]{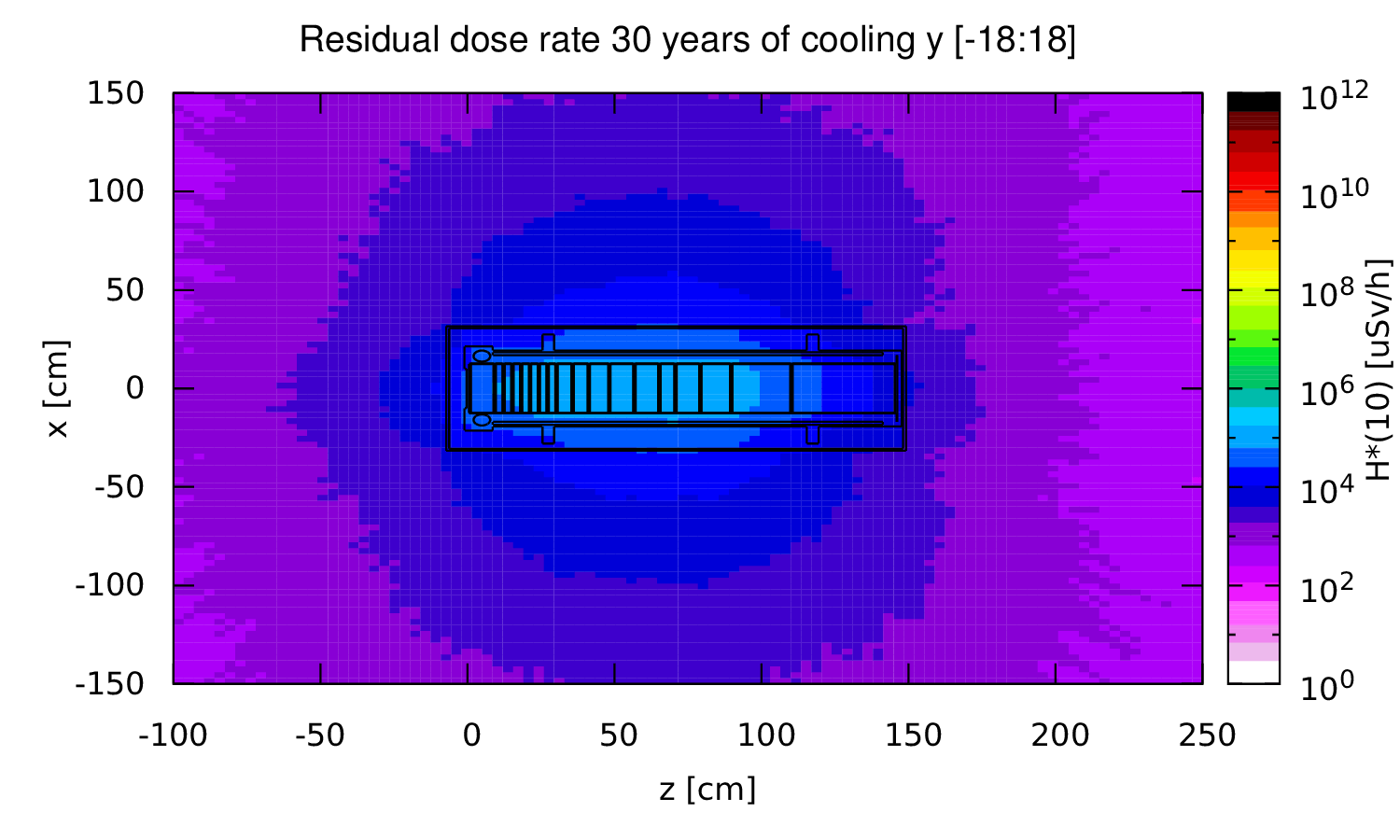}
    \label{fig:fred4}
    \end{subfigure}
 
\captionsetup{width=0.85\textwidth} \caption{\small Residual dose rates  in $\mu$Sv/h after (a) 4 hours, (b) 1 day, (c) 1 year and (d) 30 years cooling.}
\label{fig:TargetRD}

\end{figure}

Figure~\ref{fig:TargetRD} shows the expected residual dose rates for the target for different cooling times. The highest dose rates are in the order of $10^{8}$ $\mu$Sv/h after 4 hours of cooling. For this reason, the target exchange is a delicate procedure that was carefully addressed in Chapter~\ref{Chap:TargetComplex}.

Even after 30 years of cooling, the dose rates at 40 cm will still be in the order of a few mSv/h, thus dedicated storage place was designed in the facility in order to safely store the target.  For transportation of radioactive materials, as indicated in Ref.~\cite{Transport}, the maximum dose rate level at any point on the external surface of a package shall not exceed 2 mSv/h. Therefore, the container will be made out of iron and with the thickness of 30 cm for all sides. This transportation cask is used as well during the exchange procedure in order not to expose the target.

\begin{figure}[!htb]
 \centering
\includegraphics[width=0.9\textwidth]{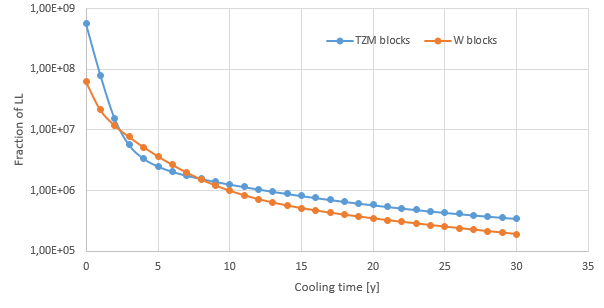}
\captionsetup{width=0.85\textwidth} \caption{\small Time evolution of the LL for TZM and W blocks.}
\label{fig:LLevolution}
\end{figure}

Some preliminary studies were performed for the disposal of the BDF target after irradiation. The results of ActiWiz 3 calculations are reported in terms of the Swiss Liberation Limits (LL) (see Ref. \cite{ORAP}). In the calculation the same irradiation profile was assumed, but 1 year cooling time. The total LL is $5.6\times10^{8}$ and $6.3\times10^{7}$ for the TZM and W block respectively, with Y-88 ($\approx87\% $LL) and Ta-182 ($\approx42\%$ LL) as top contributors for TZM and W, respectively. The time evolution of the LL value is shown in Figure \ref{fig:LLevolution}, even after 30 years cool down period the LL is still above the liberation limit. The target should be treated as radioactive waste in case of disposal. A proper assessment of the disposal will be performed in the future.

\FloatBarrier

\label{Sec:TGT:RP}

\section{Target thermo-mechanical simulations}
\label{Sec:TGT:Simus}

\subsection{Thermal calculations}
\label{Sec:TGT:Simus:thermal}
As discussed in the previous Section (Section~\ref{Sec:TGT:EneRad}), the energy deposited by the primary proton beam on the target is obtained via FLUKA Monte Carlo simulations~\cite{FLUKA_Code,Ferrari2005}. The energy induced by beam-matter interactions is imported into ANSYS$\copyright$ Mechanical in order to perform FEM thermo-mechanical calculations, to evaluate the target performance during operation.

Forced convection has been applied on the target blocks surface as boundary condition for the FEM thermal simulations, with an estimated film coefficient value of 20000 W/m\textsuperscript{2}K and a water temperature of $30\,^{\circ}\mathrm{C}$. This value is consistent with the average heat transfer coefficient (HTC) calculated via CFD simulations, as will be shown in Section~\ref{Sec:TGT:CoolingCFD}.

One of the most challenging aspects of the thermal calculations performed is the implementation of the proton beam dilution into the FEM software. An ANSYS$\copyright$ APDL code has been developed to simulate the beam sweep trajectory, where the FLUKA data corresponding to one single proton impact is imported into a thermal analysis of ANSYS$\copyright$ Mechanical and moved along the target following a circular pattern identical to the dilution system design. As a result, the temperature distribution in the target blocks is obtained as a function of time. Table~\ref{tab:TGT:maxtemps} summarizes the maximum temperatures reached in the different target materials for the most critical blocks in terms of thermal loads for each material. 

\begin{table}[htbp]
\centering
\caption{\label{tab:TGT:maxtemps} The table shows the maximum temperatures reached in the different BDF target materials for the most critical blocks.}
\smallskip
\begin{tabular}{lcc}
\toprule
\textbf{Material} & \textbf{Block number} & \textbf{Maximum temperature} \\
\midrule
Ta2.5W   & 4            & $160\,^{\circ}\mathrm{C}$              \\
TZM      & 9            & $180\,^{\circ}\mathrm{C}$              \\
W        & 14           & $150\,^{\circ}\mathrm{C}$        \\
\bottomrule
\end{tabular}
\end{table}

\begin{figure}[htbp]
\centering %
\includegraphics[width=1\linewidth]{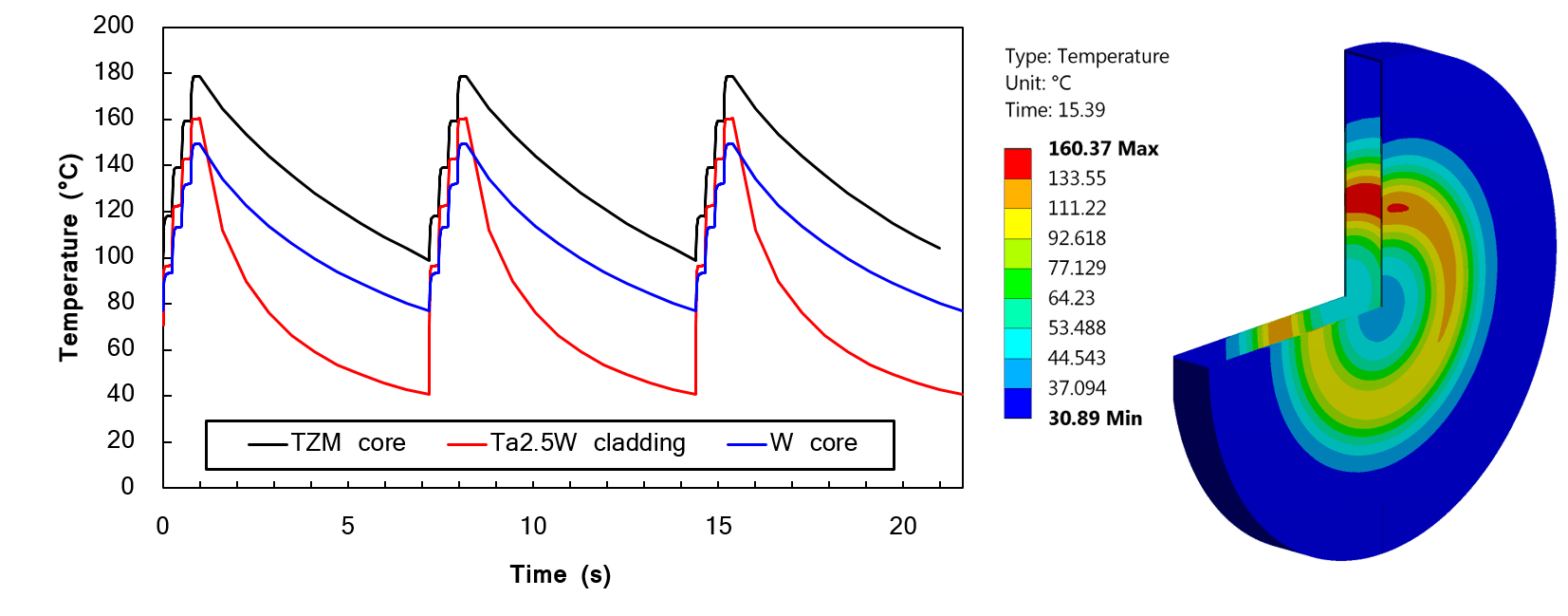}
\caption{\label{fig:TGT:thermalplot} (a) Maximum temperature evolution during 3 beam pulses after long-time operation for the BDF target materials, results for the most loaded target blocks. (b) Temperature distribution in the Ta2.5W cladding of block number 4 at the time of maximum temperature.}
\end{figure}

Figure~\ref{fig:TGT:thermalplot}a shows the maximum temperature evolution during 3 pulses for TZM, W and Ta2.5W, while Figure~\ref{fig:TGT:thermalplot}b shows the temperature distribution in the Ta2.5W cladding of block number 4 after the beam impact. The beneficial effect of the beam dilution in reducing the maximum temperature can be observed in both figures.

The temperature reached in the target core and cladding materials is around 0.1 T$_{\text{m}}$ (melting temperature), and these materials do not present any allotropic transformation in this range of temperatures. As a consequence, the evolution of the physical and mechanical properties with temperature is expected to evolve gradually, without abrupt changes. However, there are several limitations associated with the high temperatures expected during operation. First, the degradation of the material properties at high temperatures, especially for Ta2.5W, leads to a reduction of the strength of the material (see Section~\ref{Sec:TGT:Simus:struct}). This is the main reason for the selection of a tantalum alloy as cladding material instead of pure tantalum, the latter having a reported yield strength reduction from 185 MPa at room temperature to 70 MPa at $200\,^{\circ}\mathrm{C}$~\cite{Tantalum_Schmidt}. 

Furthermore, it is undesirable to reach temperatures above the boiling point of water in the target surface, which could lead to localized boiling of the cooling water, inducing a severe degradation of the heat dissipation from the blocks. This issue will be reported in more detail in Section~\ref{Sec:TGT:CoolingCFD}. Finally, the thermal loads applied to the target will be responsible for high levels of stresses, in particular for the cladding material where the temperature increase after each proton beam impact can reach $120\,^{\circ}\mathrm{C}$, as shown in Figure~\ref{fig:TGT:thermalplot}.

\subsection{Structural calculations}
\label{Sec:TGT:Simus:struct}

\subsubsection{Transient structural simulations}
\label{sec:TGT:Simus:struct:transient}

\subsubsubsection{Target materials properties review}

\begin{figure}[htbp]
\centering %
\includegraphics[width=0.65\linewidth]{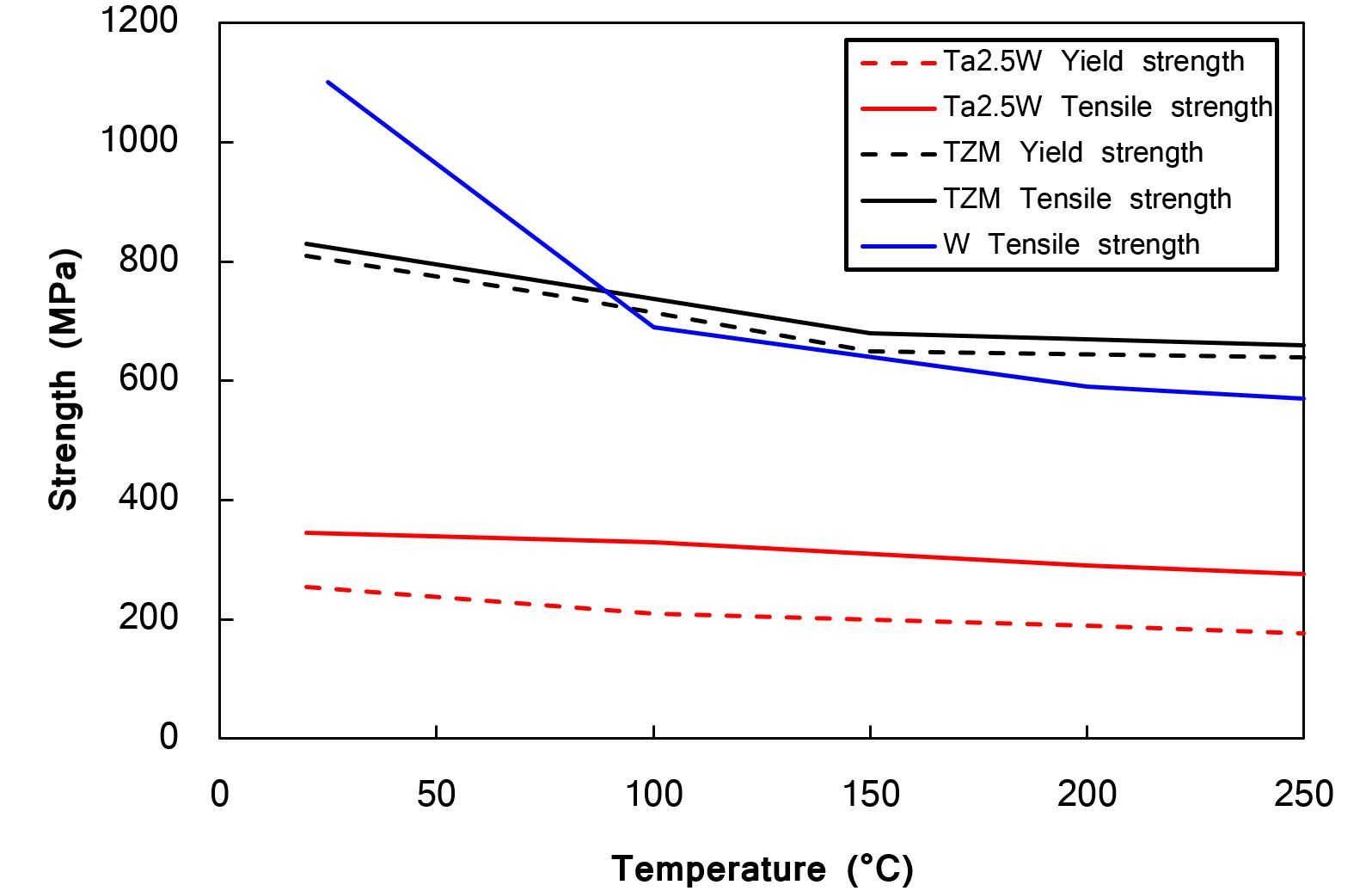}
\caption{\label{fig:TGT:Yieldtensile} The figure reports the yield strength and tensile strength as a function of temperature for TZM~\cite{TZM_Filacchioni}, tungsten~\cite{Tungsten_Schmidt} and Ta2.5W~\cite{TaW_HCStarck}.}
\end{figure}

As mentioned in the previous section, considering the level of thermal-induced stresses reached in the target materials, it is important to consider the reduction of the material properties with temperature, since the target materials are expected to reach temperatures around $150\,^{\circ}\mathrm{C}$ during operation. The evolution of the yield and tensile strength of Ta2.5W, TZM and pure tungsten as obtained in literature is shown in Figure~\ref{fig:TGT:Yieldtensile}, as an indication of the strength reduction with temperature. The yield strength of tungsten is not plotted given the brittle nature of this material below $250\,^{\circ}\mathrm{C}$, as will be clarified in the next paragraph.

As a general trend, for the three materials, the yield strength and tensile strength decrease with temperature. Ta2.5W shows much lower strength values than TZM and W in the entire range of temperatures. It shall be remarked that the differences in mechanical properties between products of the same material can be significant and they highly depend on the geometry and manufacturing process applied. 

The strength of the target materials is a crucial parameter to estimate the target lifetime, and thus, to perform the target design. A material characterization campaign is currently ongoing within the framework of the BDF Project, and will help elucidating the properties of Ta2.5W, TZM and W for a more robust target design. In the following paragraphs, the most relevant values of yield and tensile strength that have been measured or found in literature will be presented:

\begin{itemize}
    \item The yield and tensile strength of Ta2.5W was measured at room temperature for specimens obtained from Ta2.5W disks treated with a HIP cycle identical to the one used to produce the BDF target blocks. The measured yield strength was 270 MPa, and the tensile strength 360 MPa~\cite{TaW_yieldFH}. These results are compatible with the ones plotted in Figure~\ref{fig:TGT:Yieldtensile} at room temperature, even if the material supplier is different in both cases. For comparison purposes, pure tantalum specimens with identical geometry have also been tested after the HIP process, showing much lower yield strength (170 MPa) and tensile strength (220 MPa). This low strength exhibited by pure tantalum would probably compromise the target lifetime as mentioned in Section~\ref{Sec:TGT:Design:material}.
    \item Regarding the TZM core material, samples obtained from a 200 mm diameter, 100 mm length rod were tested at 20 and $700\,^{\circ}\mathrm{C}$~\cite{TZM_yieldPlansee}. At room temperature, the measured yield strength and mean tensile strength are 480 and 525 MPa respectively, which are significantly lower values than the ones shown in Figure~\ref{fig:TGT:Yieldtensile}. This is probably due to the bigger grain size obtained in the production of rods with such a large diameter and length, which is also the case of the TZM rods produced for the BDF target. Considering this reduction of material strength, the estimated yield strength at $200\,^{\circ}\mathrm{C}$ is around 370 MPa.
    \item Pure tungsten presents brittle behaviour at room temperature, and for common commercial tungsten products the ductile-to-brittle-transition-temperature (DBTT) is around $300\,^{\circ}\mathrm{C}$ in most cases \cite{W_DBTT_Zhang,W_DBTT_2}. Furthermore, the tungsten DBTT can be strongly affected by the manufacturing process and by radiation damage on the material; studies have shown an increase of the DBTT of tungsten with low levels of radiation damage (around 0.1 dpa) to values above $400\,^{\circ}\mathrm{C}$~\cite{W_radiation}. It is then foreseen that the BDF target tungsten will behave in a brittle manner during operation, when the maximum temperatures reached in tungsten are expected to be around $150\,^{\circ}\mathrm{C}$. The tensile strength of tungsten specimens produced via sintering and HIPing has been reported in Ref.~\cite{Wfatigue}, showing much lower strength (567 MPa) than in Figure~\ref{fig:TGT:Yieldtensile}. This value is more relevant to the BDF target material properties study, given that the large diameter and length of the BDF tungsten blocks will most probably constrain the tungsten cylinders production to the sintering and HIPing method instead of forging or rolling (see Section~\ref{Sec:TGT:MechDesign}). Taking into account the reduction of tensile strength with temperature shown in Figure~\ref{fig:TGT:Yieldtensile}, the estimated Ultimate Tensile Strength (UTS) at $150\,^{\circ}\mathrm{C}$ is around 330 MPa.
\end{itemize}

Table~\ref{tab:TGT:strength_data} summarizes the strength data found in literature and via mechanical testing relevant for the BDF operational conditions. The data for the different materials is presented at room temperature and high temperature, in order to show meaningful data at the target operational temperatures.

\begin{table}[htbp]
\centering
\caption{\label{tab:TGT:strength_data} BDF target materials yield and tensile strength at room temperature as obtained from literature and mechanical testing~\cite{TaW_yieldFH,TZM_yieldPlansee,Wfatigue}. Estimated material strength at high temperatures according to the trend shown in Figure~\ref{fig:TGT:Yieldtensile}.}
\smallskip
\begin{tabular}{cccc}
\toprule
\textbf{Material} & \textbf{\begin{tabular}[c]{@{}c@{}}Yield strength  \\at RT $\left[\text{MPa}\right]$ \end{tabular}} & \textbf{\begin{tabular}[c]{@{}c@{}}UTS at RT \\ $\left[\text{MPa}\right]$\end{tabular}}& \textbf{\begin{tabular}[c]{@{}c@{}}Yield strength\\ at $\boldsymbol{200\,^{\circ}\mathrm{C}}$ $\left[\text{MPa}\right]$\end{tabular}} \\
\midrule
\textbf{Ta2.5W} & 270 &  360 & 190 \\
\textbf{TZM}& 480 & 525 & 370 \\
\bottomrule
\end{tabular}
\begin{tabular}{ccc}
\toprule
\textbf{Material} & \textbf{UTS at RT $\left[\text{MPa}\right]$}& \textbf{UTS at $\boldsymbol{150\,^{\circ}\mathrm{C}}$ $\left[\text{MPa}\right]$} \\
\midrule
\textbf{W}& 567 & 330  \\
\bottomrule
\end{tabular}
\end{table}

\subsubsubsection{Simulation results}

The stresses induced by the high temperatures reached in the target materials are estimated by means of FEM calculations. The preliminary simulations performed have shown that the level of stresses is substantially reduced if the target blocks present free body expansion after the temperature raise generated by the beam impact. Therefore, in the analysis the target cylinders are considered to be resting on the support surface (not fixed or clamped to the support). A detailed description of the target assembly will be given in Section~\ref{Sec:TGT:MechDesign}. 

The temperature distribution for each of the blocks is imported as a thermal load for the structural analysis. A transient structural analysis is performed, in order to evaluate the stress evolution over time given the temperature distribution at each time step. However, the pulse duration of one second is considered sufficiently long for the beam-induced dynamic effects to be negligible, for that reason the structural analysis can be regarded as quasi-static.

Ta2.5W and TZM present ductile behaviour at the operational temperatures, they are expected to deform elastically under the thermal loading induced by the beam; if the yield point of the materials is reached, the blocks will start deforming in the plastic regime. However, the cyclic plastic deformation of the core or the cladding during a long period could lead to premature fracture of the material and/or to detachment of the cladding with respect to the base refractory metal, reducing or blocking the heat dissipation through the cladding material. Therefore, the von Mises yield criterion was used to evaluate the safety margin of the stresses reached in the Ta2.5W cladding and TZM core with respect to the yield strength of these materials. 

For pure tungsten, that is considered as a brittle material at the target operational temperatures, the Christensen criterion~\cite{Christensen} is considered the most suited failure criterion for this case. However, due to the low availability of compressive strength data for tungsten under the target operational conditions, the maximum normal stress criterion was used to assess if the maximum stresses reached in the tungsten core are within the safety limits of the material.

Table~\ref{tab:TGT:maxstress} summarizes the maximum stresses found in each material, as well as the safety margin with respect to the yield strength or ultimate tensile strength of the material at the operational temperatures.

\begin{table}[htbp]
\centering
\caption{\label{tab:TGT:maxstress} Maximum stresses reached in the BDF target materials and safety factor with respect to the material limits presented in Table~\ref{tab:TGT:strength_data}.}
\smallskip
\begin{tabular}{cccc}
\toprule
\textbf{Material} & \textbf{\begin{tabular}[c]{@{}c@{}}Maximum von Mises\\equivalent stress $\left[\text{MPa}\right]$\end{tabular}}& \textbf{\begin{tabular}[c]{@{}c@{}}Yield strength\\ at $\boldsymbol{200\,^{\circ}\mathrm{C}}$ $\left[\text{MPa}\right]$\end{tabular}} & \textbf{\begin{tabular}[c]{@{}c@{}}Safety \\ factor\end{tabular}} \\
\midrule
\textbf{TZM}& 128 & 370 & 3 \\
\textbf{Ta2.5W} & 95 &  190 & 2 \\
\bottomrule
\toprule
\textbf{Material} & \textbf{\begin{tabular}[c]{@{}c@{}}Maximum normal\\stress $\left[\text{MPa}\right]$\end{tabular}}& \textbf{\begin{tabular}[c]{@{}c@{}}UTS at $\boldsymbol{150\,^{\circ}\mathrm{C}}$\\ $\left[\text{MPa}\right]$\end{tabular}} & \textbf{\begin{tabular}[c]{@{}c@{}}Safety \\ factor\end{tabular}} \\
\midrule
\textbf{W}& 80 & 330 & 4 \\
\bottomrule
\end{tabular}
\end{table}

\begin{figure}[htbp]
\centering %
\includegraphics[width=1\linewidth]{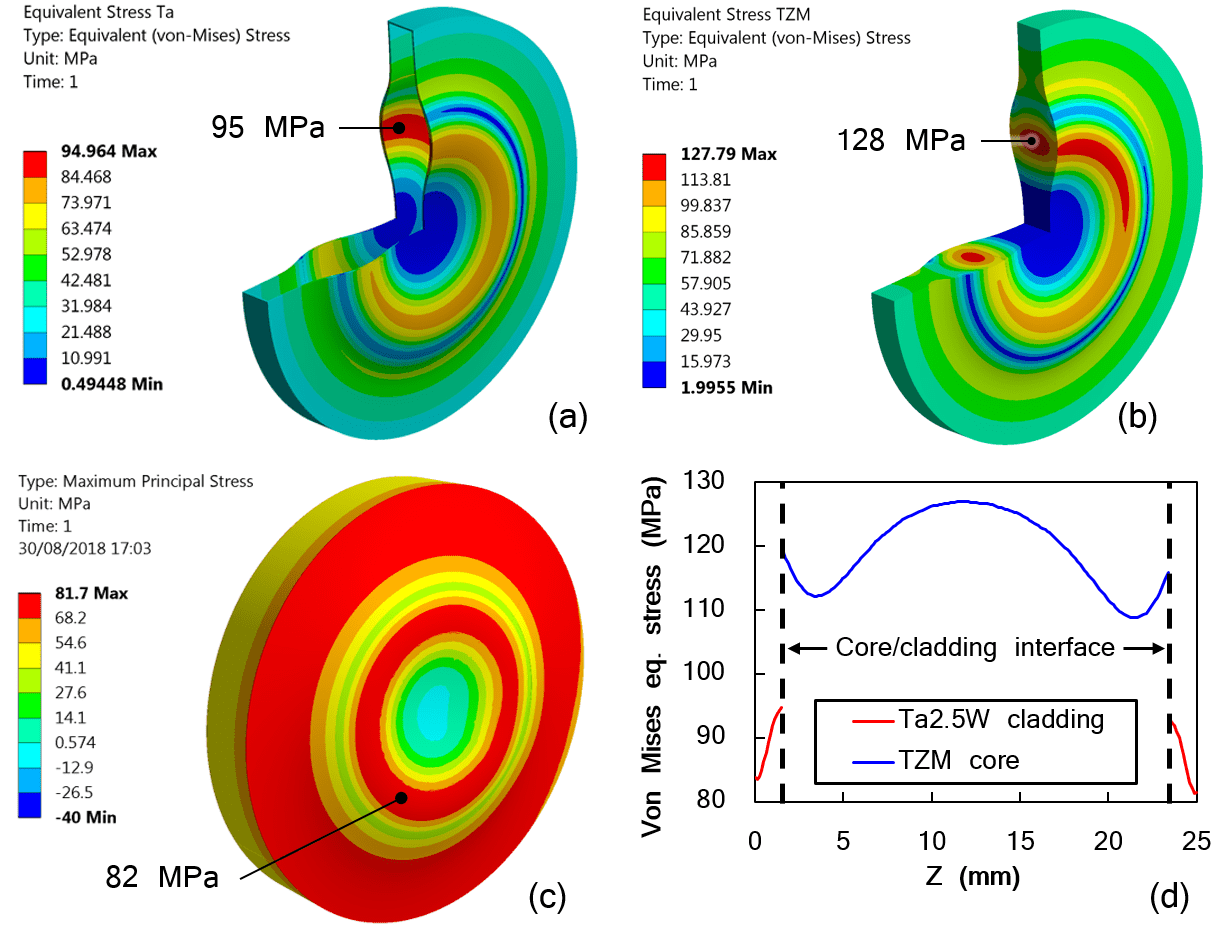}
\caption{\label{fig:TGT:vmstresses} The figure shows the von Mises equivalent stress distribution after one beam pulse in the most loaded blocks for each target material: Block \#4 for Ta2.5W (a) and TZM (b), Block \#14 for W (c). Stress distribution in the longitudinal axis at the beam dilution position (Y = 50 mm) for block \#4 (d).}
\end{figure}

The beam impact on the target blocks leads to high stresses that follow the geometrical distribution beam dilution, as can be seen in Figure~\ref{fig:TGT:vmstresses}. The highest von Mises equivalent stress in the cladding is reached in block number 4, in the beam impact region at the end of the pulse, the maximum stress being located in the interface with the block core. As for the TZM, the highest von Mises equivalent stress is found in the core center of block 4, where high compressive stresses develop, as well as in the interface with the tantalum cladding. The maximum principal stress for the pure tungsten blocks is reached in the upstream face of block 14, following the beam dilution pattern.

As discussed in Section~\ref{Sec:TGT:Design:optimisation}, the largest values of stress are found in the most upstream blocks (3 to 6). The stress distribution inside one of the most critical target blocks is displayed in Figure~\ref{fig:TGT:vmstresses}(d), showing an abrupt stress variation in the interface between the target cladding and the core.

\begin{figure}[htbp]
\centering %
\includegraphics[width=1\linewidth]{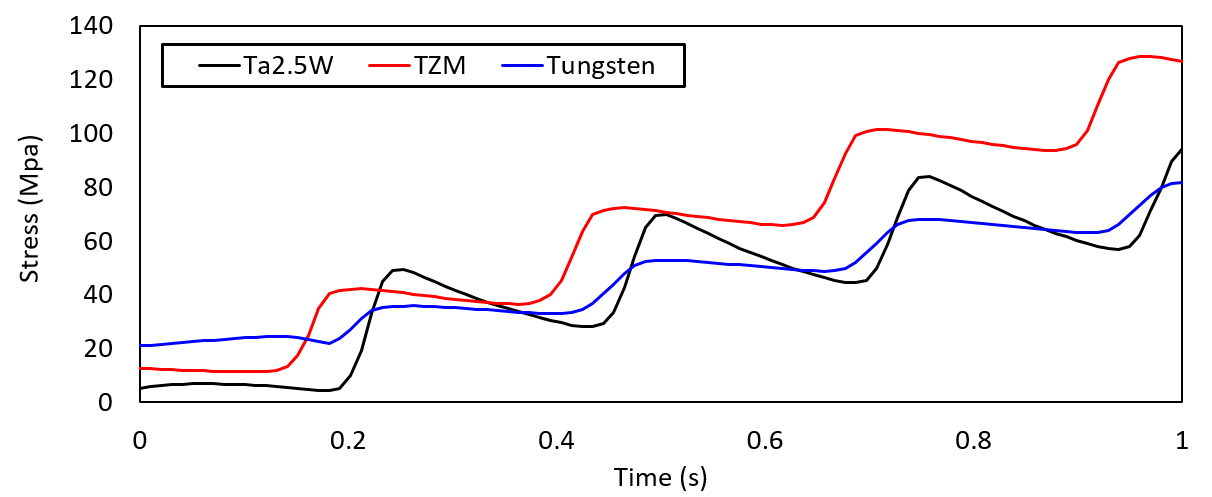}
\caption{\label{fig:TGT:stress_evolution_sweep} Effect of the beam dilution in the stress evolution in the target materials. Results plotted for the locations of maximum stress shown in Figure \ref{fig:TGT:vmstresses}. The stress values correspond to the von Mises equivalent stress for TZM and Ta2.5W, and to the maximum principal stress for tungsten.}
\end{figure}

The stress increase in the target blocks during the beam impact is highly influenced by the beam dilution pattern. Figure~\ref{fig:TGT:stress_evolution_sweep} presents the stress evolution in the maximum stress locations during the 1-second SPS beam pulse. The maximum stress locations studied can be seen in Figure~\ref{fig:TGT:vmstresses}. The beam sweep following four circular turns can be clearly appreciated, leading to a stress increase when the beam approaches the specific location plotted, and to a stress decrease as the beam moves away.

For TZM, the maximum von Mises equivalent stress is well below the yield strength of the material, then the TZM core is not expected to present any plastic deformation during operation. In the case of pure W, the maximum principal stress is considerably lower than the tensile strength of the material at $150\,^{\circ}\mathrm{C}$, leading to a high safety margin even considering the brittle behavior of tungsten at the target operational temperatures. The target Ta2.5W cladding is the most critical part of the target operation, with an acceptable safety factor with respect to the design criteria limits, but lower than for the TZM and tungsten cores.

\subsubsection{Fatigue analysis}
\label{sec:simus:struct:fatigue}

\subsubsubsection{Fatigue literature review}

The SPS primary beam is foreseen to impact the BDF target during 1 second every 7.2 seconds, and the total number of cycles expected during the target lifetime is of the order of 10\textsuperscript{7}. The target will operate under cyclic structural loads, it is then necessary to evaluate the fatigue life of the target materials under high-cycle loading.

A literature study has been carried out to obtain fatigue data for the target materials under loading conditions similar to the BDF target ones. Literature for fatigue life is very limited, specially for fatigue behavior at high temperatures and under irradiation. This makes the estimation of fatigue properties for the BDF target materials specifically difficult, as the different material and test conditions (geometry, heat treatment, purity, test mode, test stress ratio, test temperature, test frequency...) can significantly change the resulting fatigue strength value. A material characterization campaign is foreseen in the framework of the BDF Project in order to acquire experimental data of the target materials fatigue life at room and high temperatures and allow for a robust BDF target engineering design.

Fatigue data for TZM is available in Ref.~\cite{TZMfatigue}, with the special interest that the fatigue limit at 10\textsuperscript{7} cycles is analyzed at room and high temperatures. The most relevant source for pure tungsten fatigue is found in recent studies from the European Spallation Source (ESS), where the fatigue limit of several tungsten specimens produced from different manufacturing routes is evaluated for \num{2e6} cycles~\cite{Wfatigue}. For Ta2.5W, fatigue data is even more limited. It can be found in the Material Database of a tantalum supplier~\cite{TantalumW2}, that includes data for fully-reversed bending fatigue at 10\textsuperscript{7} cycles for 1 mm plates. 

Table~\ref{tab:TGT:fatiguedata} summarizes the fatigue data reported in the studied papers, which are considered to present the closest conditions to the BDF target operation. Nevertheless, it is worth mentioning that some of the testing conditions are not identical to the ones of the BDF target materials (e.g. fatigue data given at room temperature, different material production route, etc.).

\begin{table}[htbp]
\centering
\caption{\label{tab:TGT:fatiguedata} Summary of the reviewed high-cycle fatigue data relevant for the BDF target operational conditions. Sources: TZM~\cite{TZMfatigue}, Tungsten~\cite{Wfatigue}, Ta2.5W~\cite{TantalumW2}. P/M: Powder Metallurgy, Aw: As worked, HIP: Hot Isostatic Pressing, Rxx: Recrystallized, RT: Room Temperature}
\smallskip
{\renewcommand{\arraystretch}{1.2}%
\begin{tabular}{ccccccc}
\toprule

\textbf{Material} & \begin{tabular}[c]{@{}c@{}}\textbf{Production} \\ \textbf{process}\end{tabular} & \textbf{Dimensions} & \textbf{Test mode} & \begin{tabular}[c]{@{}l@{}}\textbf{Number} \\ \textbf{of cycles}\end{tabular} & \begin{tabular}[c]{@{}c@{}} \textbf{Stress ratio,} \\ \textbf{temperature,} \\ \textbf{frequency (Hz)}\end{tabular} & \begin{tabular}[c]{@{}c@{}}\textbf{Fatigue limit}\\ \textbf{(MPa)}\end{tabular}\\
\midrule
\multirow{2}{*}{\textbf{TZM}} & \multirow{2}{*}{P/M, Aw} & \multirow{2}{*}{$\varnothing$50 mm bar} & \multirow{2}{*}{Push-pull} & \multirow{2}{*}{10\textsuperscript{7}} & -1, RT, 25 & \textbf{440} \\
 &  &  &  &  & -1, $850^{\circ}\mathrm{C}$, 25 & \textbf{250} \\
\midrule
\multirow{2}{*}{\textbf{W}} & \begin{tabular}[c]{@{}c@{}}Sintered \\ + HIP\end{tabular} & $\varnothing$5 mm bar  & Push-pull & \num{2e6} & 0, RT, 25 & \textbf{180}  \\
 & \begin{tabular}[c]{@{}c@{}}Rolled \\ + Annealed\end{tabular} & $\varnothing$5 mm bar & Push-pull &  \num{2e6} & 0, RT, 25 & \textbf{350}\\
\midrule
\textbf{Ta2.5W} & P/M, Rxx & Plate 1 mm  & \begin{tabular}[c]{@{}c@{}}Bending \\ fatigue\end{tabular} & \begin{tabular}[c]{@{}c@{}}10\textsuperscript{7}, 50\% \\ fracture\end{tabular} & -1, RT, 25 & \textbf{310} \\
\bottomrule
\end{tabular}}
\end{table}

It can be seen that TZM presents the highest fatigue strength at $N=10\textsuperscript{7}$ with 440~MPa at room temperature, and the fatigue limit decreases at high temperatures as expected. 
Pure tungsten presents the lowest fatigue strength of 180~MPa for the sintered and HIPed manufacturing route, which is considered closer to the BDF target case. It is worth mentioning that the measured fatigue strength of 180~MPa can be considered as conservative because it does not correspond to fully reverse loading, since all the tests presented in~\cite{Wfatigue} are carried out in the tensile regime. Ta2.5W presents a fatigue strength of 310~MPa at room temperature, which is relatively close to the tensile strength of the material. 

\subsubsubsection{Simulation results}

As a result of the high temperatures reached during operation for every beam impact on target, the target materials are subjected to significant mean stresses and stress amplitudes, which may be critical for the target operation. The fatigue data found in literature is usually obtained from uni-axial fully reversed tests (with a mean stress equal to zero). However, the state of stresses in the target blocks is multi-axial and with a non-zero mean stress. In order to correlate the stresses calculated for the BDF target with the available fatigue strength, a two-step approach is necessary:

\begin{itemize}
    \item First, an equivalent mean stress and an equivalent stress amplitude must be calculated, that are expected to give the same fatigue life in uni-axial loading as the multi-axial stress-state found in the target. 
    \item Then, it is needed to compute an equivalent fully-reversed stress from the values of equivalent mean stress and equivalent stress amplitude. This equivalent fully-reversed stress must take into account the contribution of the mean stress to the fatigue life of the target materials, in order to compare it with the fatigue strength under fully-reversed loading found in literature.
\end{itemize}

For TZM and Ta2.5W, which are considered as ductile materials, the Sines method~\cite{Fatemi} is considered the most suited for this analysis. However, given the low availability of fatigue strength data under loading with non-zero mean stresses for the target materials, a similar approach requiring only data for fully-reversed fatigue has been adopted. 

The octaedral shear stress (von Mises) theory applied to a multi-axial state of stresses is used to calculate the equivalent stress amplitude $\sigma_{q,a}$ and the equivalent mean stress $\sigma_{q,m}$ from the evolution of principal stresses in the target materials \cite{Fatemi}, following:

\begin{equation}
\label{eq:TGT:VMfatigue2}
\centering
\smallskip
\sigma _{q,a}={\sqrt  {{\frac  {(\sigma _{1,a}-\sigma _{2,a})^{2}+(\sigma _{2,a}-\sigma _{3,a})^{2}+(\sigma _{3,a}-\sigma _{1,a})^{2}}{2}}}}\!\hspace{5pt},
\smallskip
\end{equation}
and
\begin{equation}
\label{eq:TGT:VMfatigue}
\centering
\smallskip
\sigma _{q,m}={\sqrt  {{\frac  {(\sigma _{1,m}-\sigma _{2,m})^{2}+(\sigma _{2,m}-\sigma _{3,m})^{2}+(\sigma _{3,m}-\sigma _{1,m})^{2}}{2}}}}
\smallskip
\end{equation}

where $\sigma _{i,a}$ and $\sigma _{i,m}$ are the stress amplitude and mean stress of the maximum, middle and minimum principal stresses.

This equivalent stress method is limited to proportional loading conditions, where the principal directions remain unchanged during the loading cycle, which is the case for the BDF target blocks. The von Mises equivalent stress approach is also limited to the case in which the principal stresses are in-phase. In some areas of the target materials, the maximum and minimum principal stresses are in opposition of phase (180$^\circ$ out-of-phase). The methodology applied takes into account this situation by adding a negative sign to the out-of-phase component, as described in \cite{Socie}.

It is worth noting that the value of $\sigma_{q,m}$ obtained from von Mises equation is always positive, and does not take into account the possible beneficial effects of compressive mean stresses. Given that the mean principal stresses of the target materials are compressive in most of the cases, the applied method can be regarded as conservative.

For pure tungsten, which is assumed as brittle material, the maximum stress criterion is considered \cite{Fatemi}. It is expected that the fatigue endurance of the tungsten core under the multi-axial state of stresses during operation is principally influenced by the tensile contribution of the stress tensor. Therefore, the equivalent mean stress and the equivalent stress amplitude for pure tungsten are taken from the maximum principal stress cyclic variation:

\begin{equation}
\centering
\smallskip
\label{eq:TGT:Maxnormalfatigue}
\sigma _{q,a}={{\sigma _{1,a}}}\!\hspace{10pt}
\text{and} \hspace{10pt}
\sigma _{q,m}={{\sigma _{1,m}}}\!\hspace{5pt} ,
\smallskip
\end{equation}

\begin{figure}[htbp]
\centering %
\includegraphics[width=0.6\linewidth]{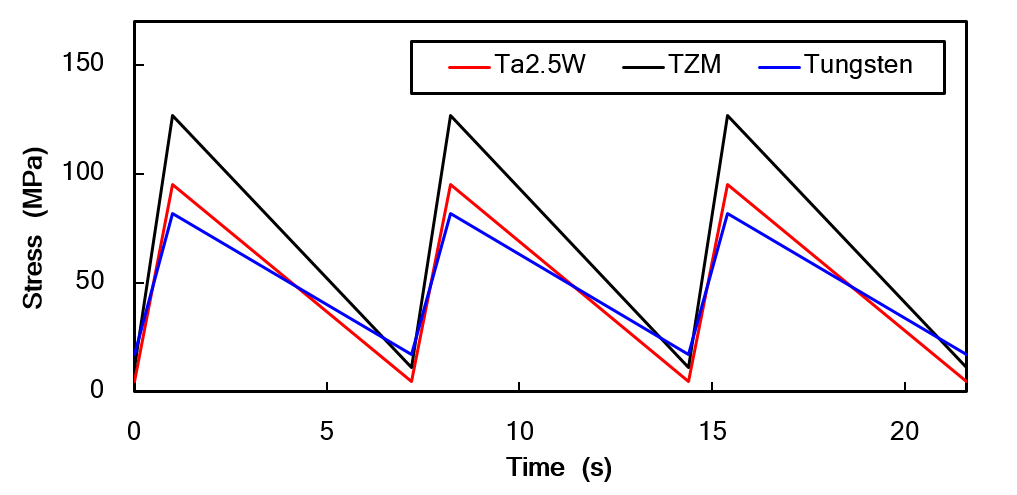}
\caption{\label{fig:TGT:stressfatigueplot} Evolution during 3 beam pulses of the equivalent stresses in the most loaded areas of the BDF target materials.}
\end{figure}

Figure \ref{fig:TGT:stressfatigueplot} shows the equivalent mean stress and stress amplitude calculated at the locations of maximum stress for the three target materials and illustrates the equivalent stress evolution during three SPS beam pulses. The stress evolution presented corresponds to uni-axial loading and is expected to be equivalent in terms of fatigue life than the actual multi-axial stress evolution in the target materials.

TZM displays the highest equivalent stress amplitude and mean stress, followed by Ta2.5W. The results displayed do not take into account the stress variations during the circular sweep of the beam presented in Figure~\ref{fig:TGT:stress_evolution_sweep}. The effect of the beam dilution on the fatigue life of the target materials shall be further studied in the future, but the preliminary calculations performed have shown that it is not expected to significantly affect the target lifetime. 

The influence of the mean stress in the fatigue life has been deeply studied~\cite{failurebook}, concluding that in general non-zero mean stresses have a detrimental effect in the fatigue behavior of materials. The modified Goodman equation shown in \ref{eq:TGT:Goodman} is used as design criterion to compare the stresses calculated for the BDF target materials which have a non-zero mean stress, with the fully reversed stress amplitude found in literature. This equation relates the mean stress $\sigma_{q,m}$ and equivalent stress amplitude $\sigma_{q,a}$ obtained from \ref{eq:TGT:VMfatigue}, \ref{eq:TGT:VMfatigue2} and \ref{eq:TGT:Maxnormalfatigue} with an equivalent fully reversed stress amplitude $\sigma_{q,f}$ which is assumed to give the same fatigue life:

\begin{equation}
\label{eq:TGT:Goodman}
\centering
\smallskip
\sigma _{q,f}=\dfrac{{\sigma _{q,a}}}{\left({1-\dfrac{\sigma_{q,m}}{\sigma_{UTS}}}\right)}\leqslant\sigma_{fat}\!\hspace{5pt} ,
\smallskip
\end{equation}

where $\sigma_{UTS}$ is the Ultimate Tensile Strength of the material, which is taken for the present calculations from the measured values at room temperature cited in Section~\ref{Sec:TGT:Simus:struct}, and $\sigma_{fat}$ is the fatigue limit of the material.

Table \ref{tab:TGT:fatigueresults} presents the equivalent mean stress, stress amplitude and Goodman equivalent stresses calculated for the target materials, as well as the fatigue endurance for each material and the safety margin achieved.

\begin{table}[htbp]
\centering
\caption{\label{tab:TGT:fatigueresults}Stress summary of the BDF target under high-cycle fatigue loading. Comparison with the fatigue strength of the target materials for 10\textsuperscript{7} cycles, obtained from \cite{Wfatigue,TZMfatigue,TantalumW2}. The values of UTS used for the calculation of the Goodman equivalent stress are found in \cite{TaW_yieldFH,TZM_yieldPlansee,Wfatigue}.}
\smallskip
{\renewcommand{\arraystretch}{1.2}%
\begin{tabular}{ccccccc}
\toprule
\textbf{Material} & 
\begin{tabular}[c]{@{}c@{}}$\boldsymbol{\sigma_{q,\ m}}$\\ \textbf{(MPa)}\end{tabular} & \begin{tabular}[c]{@{}c@{}}$\boldsymbol{\sigma_{q,\ a}}$\\ \textbf{(MPa)}\end{tabular} & \begin{tabular}[c]{@{}c@{}}$\boldsymbol{\sigma_{UTS}}$\\ \textbf{(MPa)}\end{tabular} & \begin{tabular}[c]{@{}c@{}}$\boldsymbol{\sigma_{q,\ f}}$\\ \textbf{(MPa)}\end{tabular} & 
\begin{tabular}[c]{@{}c@{}}\textbf{Fatigue limit}\\ \textbf{(MPa)}\end{tabular} & \begin{tabular}[c]{@{}c@{}}\textbf{Safety}\\ \textbf{margin}\end{tabular} \\
\midrule
\textbf{TZM} & 68 & 58 & 525 & \textbf{66} & 440 & 6.7\\
\textbf{W} & 49.5 & 32.5 & 567 & \textbf{36} & 180 & 5 \\
\textbf{Ta2.5W} & 50 & 45 & 360 &\textbf{53} & 310 & 5.8\\
\bottomrule
\end{tabular}}
\end{table}

The stresses found in the BDF target are well within the endurance limits found in literature. The performed thermo-mechanical calculations have proven that it is unlikely that the target blocks will fail due to high-cycle fatigue. In this case the Ta2.5W cladding is not a critical aspect for target operation, due to the fact that the fatigue limit of the material is relatively close to its tensile strength (85\% of the UTS at room temperature).

Further studies are required to evaluate the effects of radiation damage on the target materials, that could eventually reduce the fatigue life of the target, even though the safety margin appears to be sufficient. Additional fatigue data under more representative loading conditions is necessary for an accurate estimation of the target materials fatigue life under operation.

\subsection{Residual stresses}
\label{Sec:TGT:Simus:residual}
\begin{figure}[htbp]
\centering %
\includegraphics[width=0.8\linewidth]{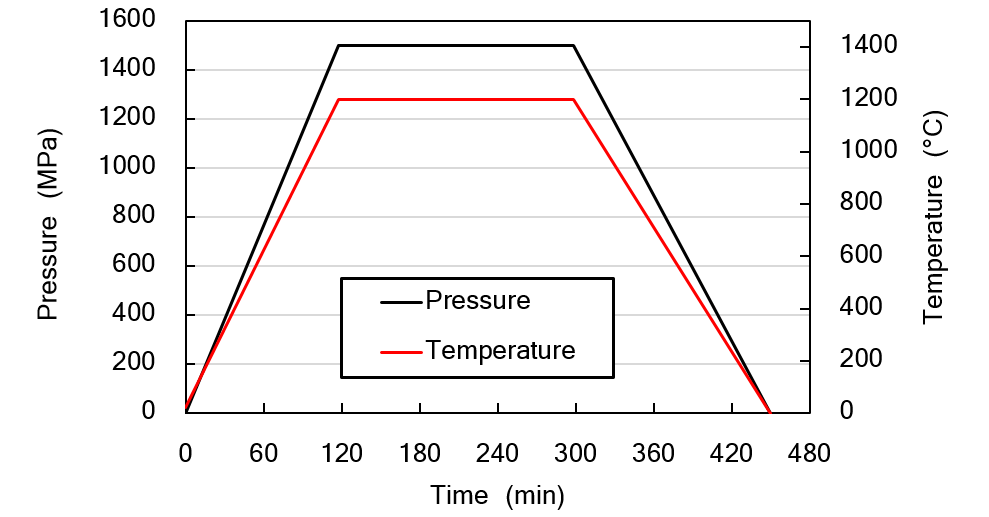}
\caption{\label{fig:TGT:HIPcycle} Hot Isostatic Pressing cycle applied for the target blocks production.}
\end{figure}

The HIP process carried out for the BDF target blocks production is crucial to ensure the mechanical and chemical bonding between the cladding and core materials. The TZM and tungsten cylindrical parts are inserted into a Ta2.5W tube with a small gap (around 0.1 mm in radius), and then closed by two Ta2.5W disks. The top and bottom Ta2.5W disks are welded under vacuum to the tube by means of electron beam welding (EBM), in order to close the assembly leak-tight. Finally the target block assembly undergoes a HIP cycle, reaching a temperature of $1200\,^{\circ}\mathrm{C}$ and a pressure of 150 MPa for 2 hours. Figure~\ref{fig:TGT:HIPcycle} describes the proposed HIP cycle applied for the production of the BDF target blocks. It is worth noting that other HIP cycles have been investigated as presented in Section~\ref{Sec:TGT:mat:HIP}.

During the cool-down phase of the HIP process, some residual stresses appear in the target blocks due to the different coefficient of thermal expansion (CTE) of the target materials. At high temperatures, this stresses are expected to be relieved, but as the temperature decreases below a certain temperature, residual stresses appear in the blocks due to the differential thermal contraction. 

This so-called 'lock-in' temperature, as thoroughly described in Ref.~\cite{ISIScladding}, has a strong influence in the level of residual stresses reached. Literature regarding the lock-in temperature is very limited and shows large scattering. An experimental setup has been tested at Rutherford Appleton Laboratory (RAL), measuring a lock-in temperature for tantalum-cladded tungsten blocks in the range of 350 to 500$\,^{\circ}\mathrm{C}$~\cite{ISISlocking}. Concerning the BDF target blocks case, the effect of the residual stresses in the expected target lifetime has been studied through FEM calculations, by calculating the residual stresses appearing in the target for an assumed lock-in temperature of 500$\,^{\circ}\mathrm{C}$. 

The maximum stresses reached after several beam pulses considering the residual stresses in the blocks has also been computed, as well as the Goodman equivalent stress, calculated using Equations \ref{eq:TGT:VMfatigue} to \ref{eq:TGT:Goodman} in order to evaluate the fatigue life of the target materials. 

\subsubsubsection{Ta2.5W-cladded TZM blocks}

\begin{figure}[htbp]
\centering %
\includegraphics[width=1\linewidth]{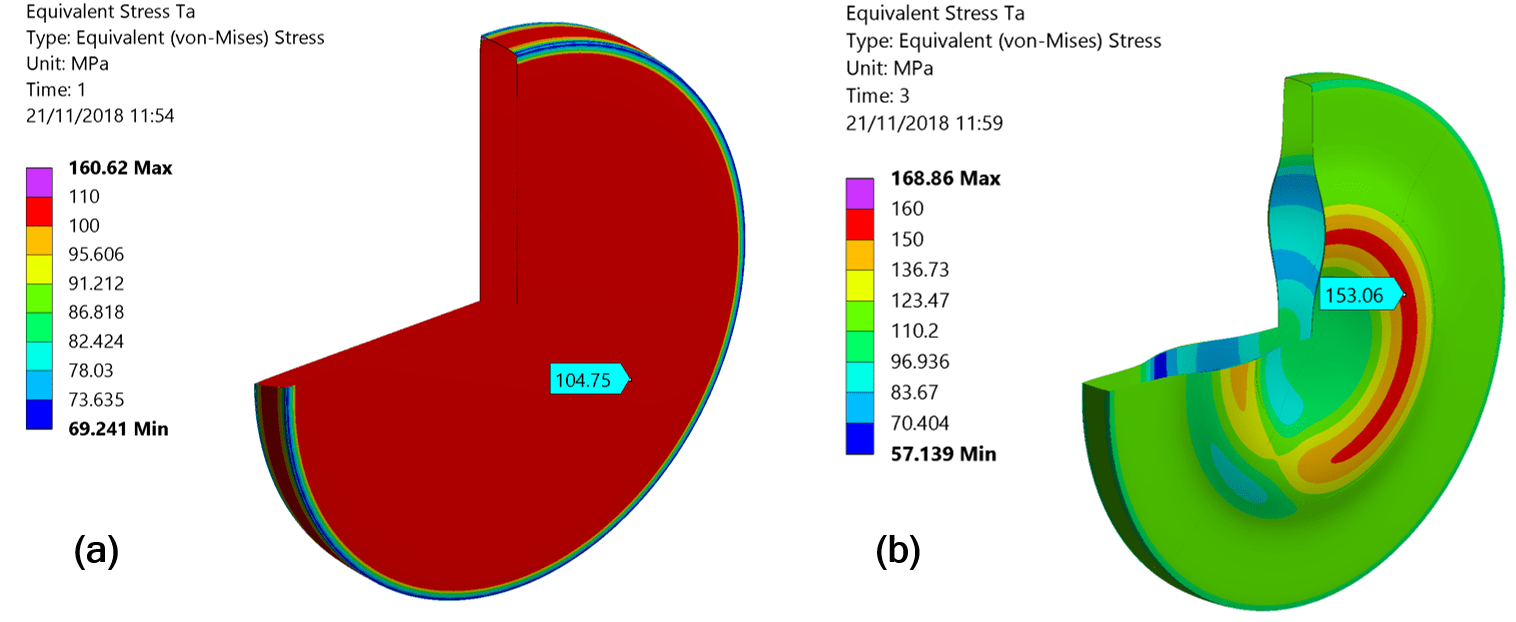}
\caption{\label{fig:TGT:Residual4TaW} Von Mises equivalent stress distribution for the Ta2.5W cladding of block \#4: (a) after HIP (residual stress); (b) after beam impact.}
\end{figure}

For the Ta2.5W-cladded TZM blocks, the residual stresses are higher in the Ta2.5W cladding, reaching around 105 MPa for the most beam-loaded block. This residual stress is purely tensile in the Ta2.5W layer. After the beam impact, the maximum von Mises equivalent stress obtained is higher than in the previously studied case, reaching around 150 MPa. The reason for that is that the beam-induced stresses in the external part of the cladding are highly tensile in the radial direction, and add up to the tensile residual stresses. However, this value is still below the yield strength of the material at $200\,^{\circ}\mathrm{C}$. Figure \ref{fig:TGT:Residual4TaW} illustrates the von Mises equivalent stress distribution in the Ta2.5W cladding of block 4 after the HIP process and after one beam impact on target (after steady-state).

In terms of fatigue life, the equivalent stress amplitude is reduced with respect to the values without residual stresses, and the mean stress is considerably increased by the residual stresses. The reduction in stress amplitude turns out to be apparently beneficial for the fatigue life estimation of the target, as the calculated Goodman equivalent stress is slightly lower than the previous scenario. 

\begin{figure}[htbp]
\centering %
\includegraphics[width=1\linewidth]{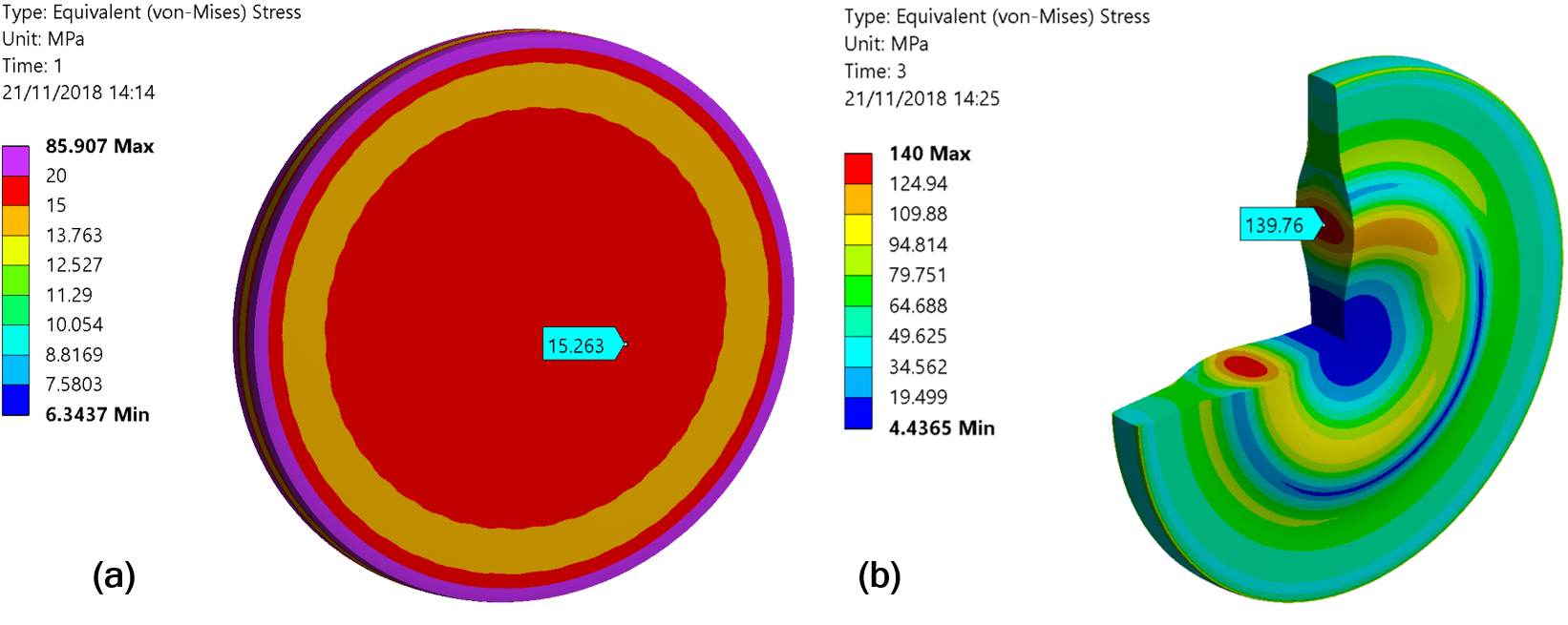}
\caption{\label{fig:TGT:Residual4TZM} Von Mises equivalent stress distribution in the TZM core of block \#4, after HIP (residual stress) (a) and after beam impact (b).}
\end{figure}

For the TZM part, the residual stresses are quite low, around 15 MPa, and purely compressive. The equivalent stress amplitude after the beam irradiation is similar to the case without residual stress, but the equivalent mean stress is higher (15 MPa more). This is explained by the fact that the stresses induced by the beam interaction in the TZM core are mainly compressive in the circumferential direction, and are added to the compressive residual stress already present in the material. As a result, the maximum von Mises equivalent stress is higher, but well within the material limits, reaching 140 MPa (Figure \ref{fig:TGT:Residual4TZM}). 

Regarding fatigue life, the Goodman equivalent stress is slightly higher than without considering residual stresses, mainly due to the increase of mean stress. Due to the small difference between both scenarios it is considered that the target lifetime won't be reduced. 

\begin{figure}[htbp]
\centering %
\includegraphics[width=1\linewidth]{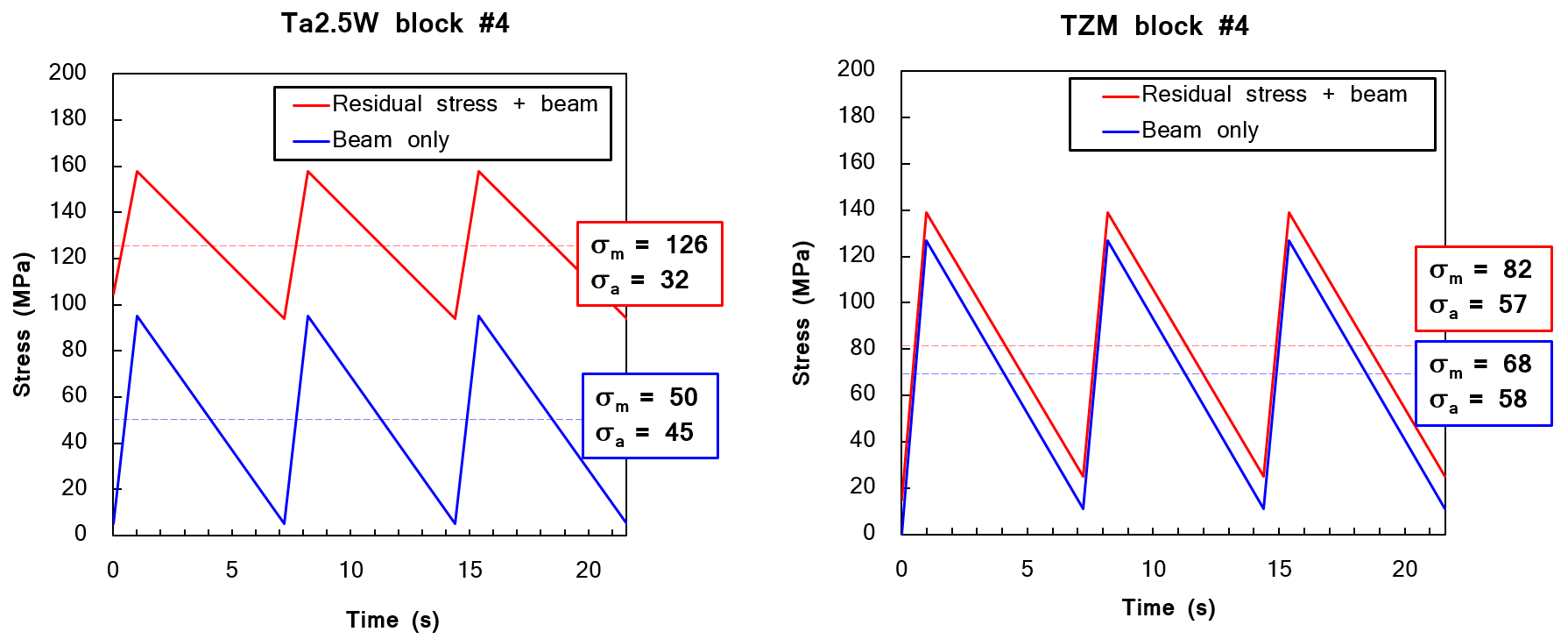}
\caption{\label{fig:TGT:ResidualPlot} Equivalent mean stress and stress amplitude for the Ta2.5W cladding and TZM core of block \#4. Comparison between the simulation results for a beam impact on the block with or without residual stresses due to HIP.}
\end{figure}

 Table~\ref{tab:TGT:residual} summarizes the most relevant calculations results for the Ta2.5W-cladded TZM blocks. The values obtained from the calculations without residual stresses shown in Section~\ref{Sec:TGT:Simus:struct} are displayed in parentheses for comparison purposes. Figure~\ref{fig:TGT:ResidualPlot} presents the fatigue behavior for the cladding and core materials of block number 4 as described previously. 

\begin{table}[htbp]
\centering
\caption{\label{tab:TGT:residual} Summary of the FEM calculations performed to evaluate the influence of the residual stresses in the behavior of the target materials for the most loaded TZM block (number 4). The values obtained Section \ref{Sec:TGT:Simus:struct} without residual stresses are presented in parentheses and italic font for comparison. Yield strength values obtained from \cite{TaW_yieldFH,TZM_yieldPlansee} and fatigue limit from \cite{TZMfatigue,TantalumW2}.}
\smallskip
{\renewcommand{\arraystretch}{1.2}%
\begin{tabular}{cccccc}
\toprule
\textbf{Material} & \textbf{\begin{tabular}[c]{@{}c@{}}Maximum \\ residual stress \\ {[}MPa{]}\end{tabular}} & \textbf{\begin{tabular}[c]{@{}c@{}}Maximum stress \\ after beam impact \\ {[}MPa{]}\end{tabular}} & \textbf{\begin{tabular}[c]{@{}c@{}}Yield \\ strength \\ {[}MPa{]}\end{tabular}} & \textbf{\begin{tabular}[c]{@{}c@{}}Goodman \\ equivalent stress \\ {[}MPa{]}\end{tabular}} & \textbf{\begin{tabular}[c]{@{}c@{}}Fatigue\\ limit \\ {[}MPa{]}\end{tabular}} \\
\midrule
\textbf{Ta2.5W}   & 105 & 153 \textit{(95)}& 190  & 50 \textit{(53) } & 310 \\
\textbf{TZM}  & 15 & 140 \textit{(128)}  & 370  & 68 \textit{(66)} & 440 \\
\bottomrule  
\end{tabular}}
\end{table}

\subsubsubsection{Ta2.5W-cladded tungsten blocks}

\begin{figure}[htbp]
\centering %
\includegraphics[width=1\linewidth]{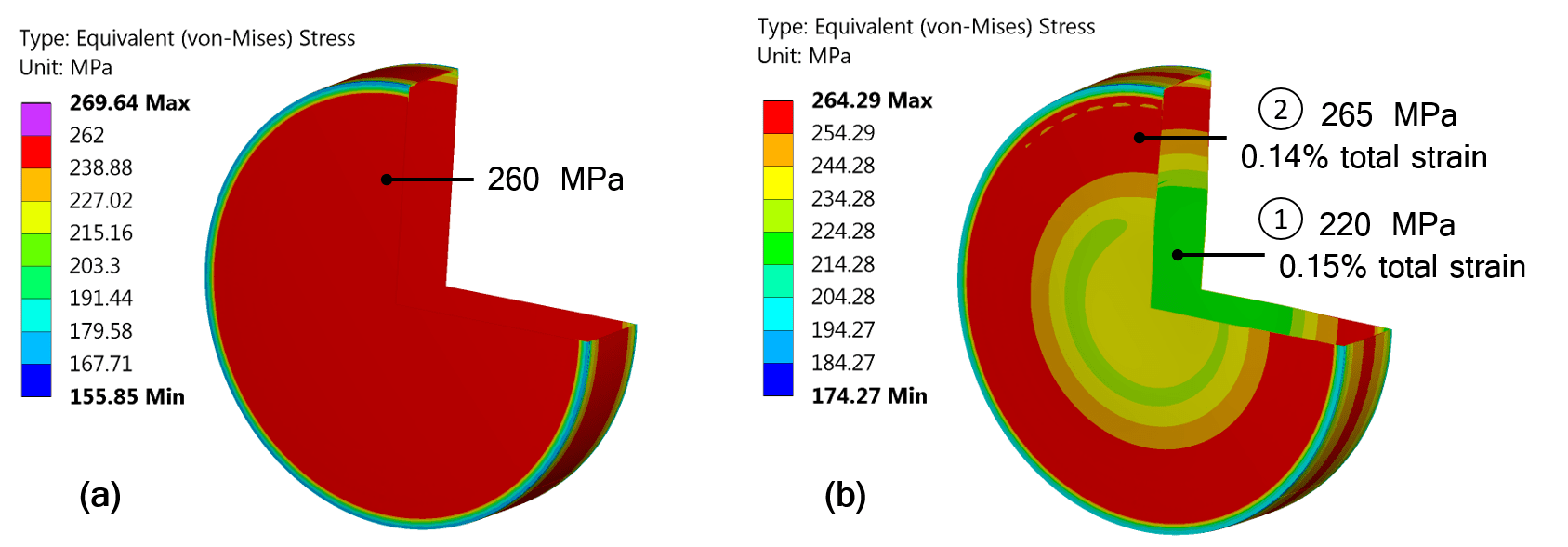}
\caption{\label{fig:TGT:ResidualPlastic} Von Mises equivalent stress corresponding to the residual stress after HIP (a), and after beam impact (b) in the cladding of block \#14 (Ta2.5W cladding, tungsten core).}
\end{figure}

Concerning the Ta2.5W-cladded tungsten blocks, the residual stresses that appear in the Ta2.5W cladding are very high, given the high difference in coefficient of thermal expansion (CTE) between Ta2.5W and pure tungsten. A deeper investigation is needed in this case to evaluate the amount of residual stresses in the cladding. A temperature-dependent plasticity model has been used in order to predict the cladding behavior, taking into account the yield strength evolution presented in Figure~\ref{fig:TGT:Yieldtensile}. The calculations performed have shown that the cladding is not expected to suffer any plastic deformation during the HIP cool-down phase, but the residual stresses appearing in the Ta2.5W layer are very close to the yield strength of the material. The remaining residual stress in Ta2.5W after the HIP cycle is around 250 MPa overall the cladding, leading to a low safety margin with respect to the yield strength of the material that is assumed to be 270 MPa at room temperature, as shown in Section~\ref{Sec:TGT:Simus:struct}. Figure~\ref{fig:TGT:ResidualPlastic} displays the von Mises equivalent stress distribution in the Ta2.5W cladding after the HIP process and after the beam impact.

During and right after the proton beam impact, the target blocks temperature raises significantly, with the subsequent reduction of the yield strength of Ta2.5W. Given that the residual stresses in the Ta2.5W layer are already close to the yield strength at room temperature, the cladding is expected to deform plastically after the beam pulse. Two different regions have been studied in detail to evaluate the effect of the residual stresses in the cladding:

\begin{itemize}
\item In the beam impact region and the internal part of the cladding (point 1 of Figure~\ref{fig:TGT:ResidualPlastic}), the highest temperature raise is found, reaching $125\,^{\circ}\mathrm{C}$. The stresses induced by the beam interaction are highly compressive in the circumferential direction in this region, and they relieve the existing tensile residual stresses. The maximum von Mises equivalent stress found in the beam dilution area is 220 MPa. This stress is slightly above the yield strength of the material at $125\,^{\circ}\mathrm{C}$, producing plastic deformation of the cladding. The deformation in the plastic regime is quite low though, thanks to the fact that the residual stresses have been relieved, and the total strain obtained is 0.15\% after the beam pulse. The cladding is not expected to fail since the total strain is well below the elongation to failure of Ta2.5W that is around 10\% at $150\,^{\circ}\mathrm{C}$~\cite{TaW_HCStarck}.

\item In the external part of the cladding (point 2 of Figure ~\ref{fig:TGT:ResidualPlastic}), far from the beam dilution path, some of the stresses generated by the beam impact are tensile in the circumferential direction, and they add up to the tensile residual stress. However, the beam-induced tensile stresses are considerably small in that area, and the temperatures reached are of the order of 50$\,^{\circ}\mathrm{C}$, much lower than in the beam impact region. The von Mises equivalent stress obtained is 265 MPa, slightly above the yield strength of Ta2.5W at 50$\,^{\circ}\mathrm{C}$, which leads to plastic deformation in the cladding. The calculated total strain is 0.14\% after the beam pulse, largely below the elongation limit. It can be concluded that, as expected, the whole cladding will present plastic behaviour, but the amount of plastic deformation is quite low and well within the material limits. 
\end{itemize}

It can be concluded that in several areas of some of the target blocks cladding, plastic deformation is expected as a consequence of the first beam pulse after the HIP process. The cyclic plastic deformation of the cladding might eventually lead to detachment of the latter, which is undesirable for operation as it would reduce heat dissipation. However, the calculations performed have shown that the cladding will always operate in the elastic regime during the steady-state beam pulses (elastic shakedown), after the first plastic deformation produced by the first beam impact post-HIP. 

\begin{figure}[htbp]
\centering %
\includegraphics[width=1\linewidth]{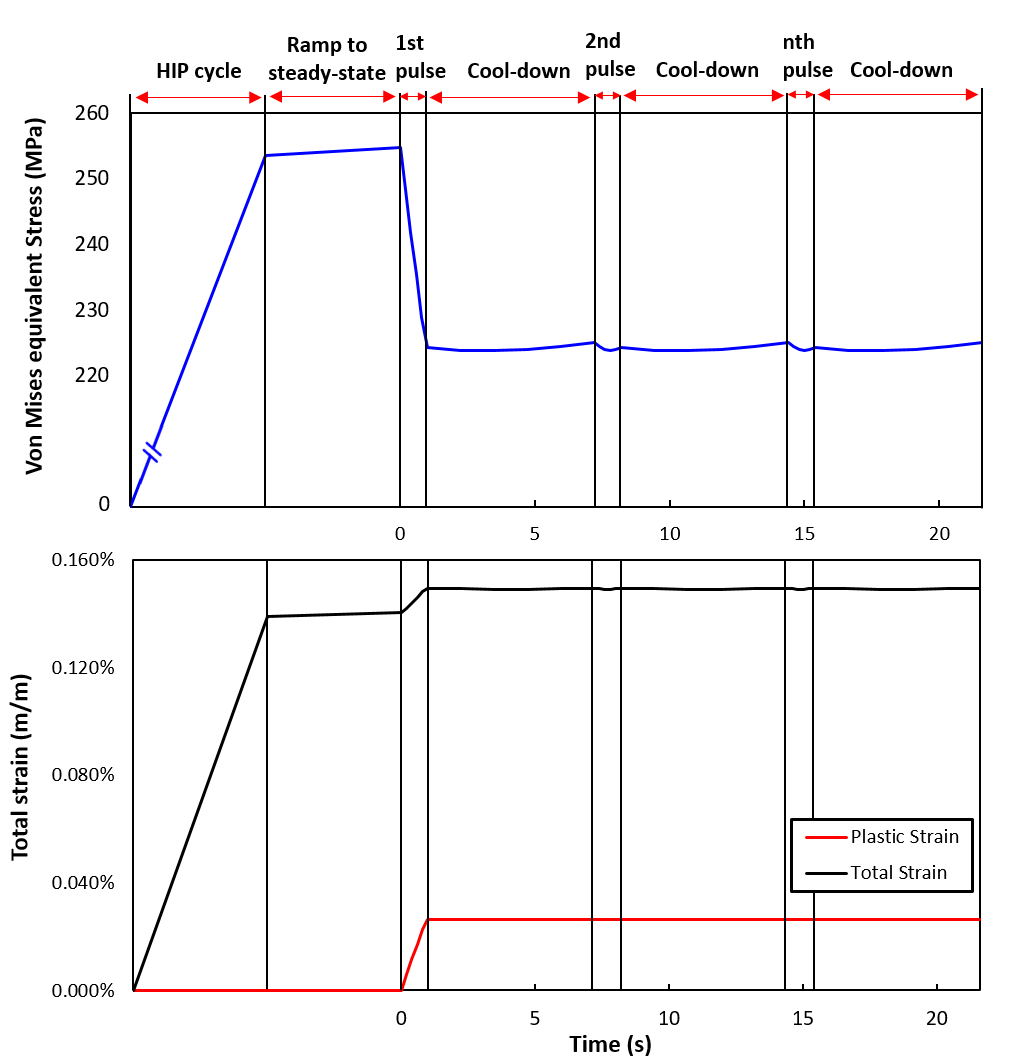}
\caption{\label{fig:TGT:Residual_evolution} Evolution of the von Mises equivalent stress, total and plastic strain in the external part of the cladding (point 1) of Figure \ref{fig:TGT:ResidualPlastic}.}
\end{figure}

This behavior can be observed in Figure~\ref{fig:TGT:Residual_evolution}, where the evolution of the von Mises equivalent stress, total and plastic strain at point 1 of the cladding is plotted. The residual stress builds up during the HIP process, always in the elastic regime, and increases slightly until reaching steady-state operation. With the first beam impact after reaching steady-state, the residual stresses are relieved, and the cladding deforms plastically due to the raise of temperature (and subsequent decrease of yield strength). The maximum plastic strain reached is around 0.03\%. During the next pulses, no further plastic deformation occurs, the deformation after the following beam impacts takes place in the elastic regime (elastic shakedown). For that reason, it is considered that there is no risk of further plastic deformation or ratcheting, which could be potentially harmful for the target lifetime.
 
Regarding the fatigue life of the Ta2.5W cladding in the tungsten core blocks, the maximum Goodman equivalent stress calculated is 50 MPa, and is found in the beam dilution path. This value is higher than the one obtained without considering residual stresses for the same location in block 14, mainly due to the substantial increase of mean stresses. In any case, this value is similar to the Goodman equivalent stress obtained for block 4 (TZM core, Ta2.5W cladding) without residual stresses, and is very far from the fatigue limit; this means that the residual stresses are not expected to have a detrimental effect in the target lifetime from this point of view.

\begin{figure}[htbp]
\centering %
\includegraphics[width=1\linewidth]{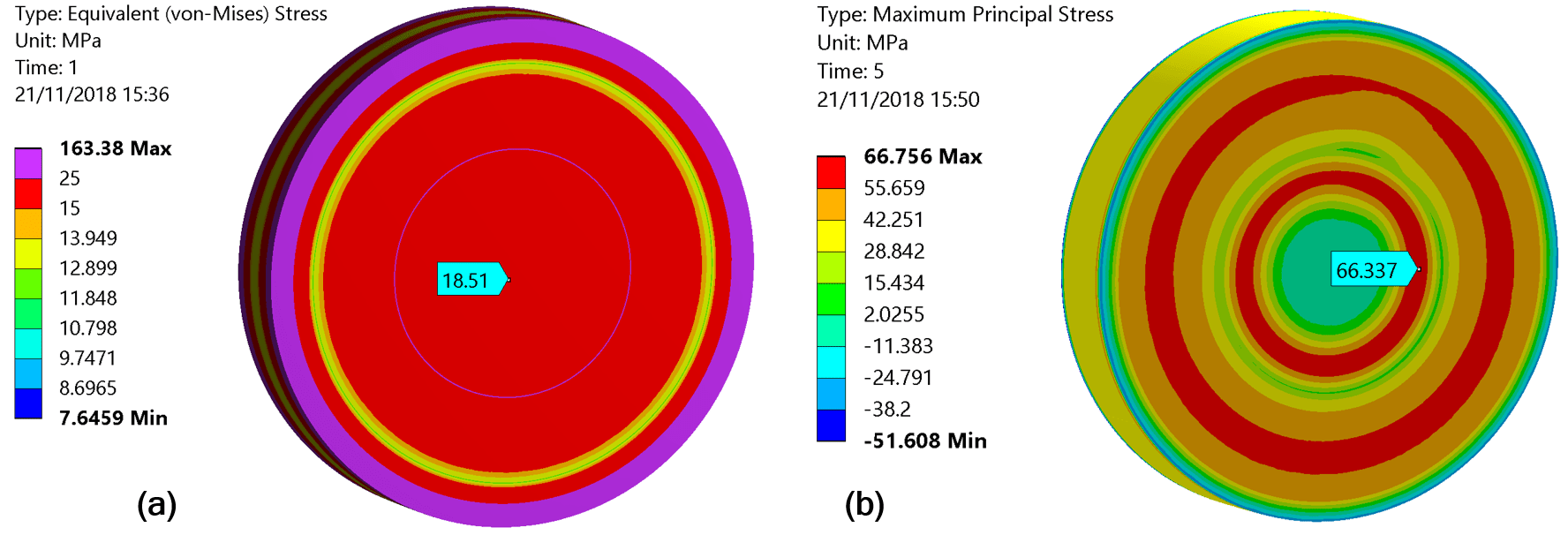}
\caption{\label{fig:TGT:Residual14W} Von Mises equivalent stress distribution for the Ta2.5W cladding of block \#14 after HIP (residual stress) (a) and maximum principal stress after beam impact (b).}
\end{figure}

In the pure tungsten core, similarly to the TZM blocks, the residual stress is compressive, reaching around 18 MPa. After the  beam impact, the values of maximum principal stress reached are lower than without residual stresses, since the beam-induced tensile stresses are applied to a stress field which is purely compressive. Therefore, the safety margin with respect to the UTS of the material is higher than for a non-stressed block. For the same reason, the maximum compressive stresses are higher than in the previous scenario, but since the compressive strength of tungsten is much superior to its tensile strength, the maximum principal stress criteria is used as exposed in Section~\ref{Sec:TGT:Simus:struct}. Figure~\ref{fig:TGT:Residual14W} shows the residual stress distribution after HIP and the maximum principal stress distribution at the pulse peak in the pure tungsten core of block number 14.

The calculated Goodman equivalent stress is similar to the previous case without residual stresses (around 35 MPa), therefore it is considered that the residual stress won't have a detrimental influence in the target lifetime from this point of view.

Table~\ref{tab:TGT:residualW} summarizes the main results obtained from the FEM simulations for the Ta2.5W-cladded tungsten blocks. 
 
\begin{table}[htbp]
\centering
\caption{\label{tab:TGT:residualW} Summary of the FEM calculations performed to evaluate the influence of the residual stresses in the behavior of the target materials. Results shown for the most loaded W core block (number 14). The values obtained in Section \ref{Sec:TGT:Simus:struct} without residual stresses are presented in parentheses and italic font for comparison. Strength limit values obtained from \cite{TaW_yieldFH,Wfatigue} and fatigue strength from \cite{Wfatigue,TantalumW2}.}
\smallskip
{\renewcommand{\arraystretch}{1.2}%
\begin{tabular}{cccccc}
\toprule
\textbf{Material} & \textbf{\begin{tabular}[c]{@{}c@{}}Maximum \\ residual stress \\ {[}MPa{]}\end{tabular}} & \textbf{\begin{tabular}[c]{@{}c@{}}Maximum stress \\ after beam impact \\ {[}MPa{]}\end{tabular}} & \textbf{\begin{tabular}[c]{@{}c@{}}Strength\\ limit \\ {[}MPa{]} \end{tabular}} & \textbf{\begin{tabular}[c]{@{}c@{}}Goodman \\ equivalent stress \\ {[}MPa{]}\end{tabular}} & \textbf{\begin{tabular}[c]{@{}c@{}}Fatigue\\ limit\\ {[}MPa{]}  \end{tabular}} \\
\midrule
\smallskip
\textbf{Ta2.5W}  & 260  & \begin{tabular}[c]{@{}c@{}}265 (plastic) \textit{(63)}\\ Total strain: 0.15\%\end{tabular}  & \begin{tabular}[c]{@{}c@{}}Elongation to\\ failure: 10\%\end{tabular}  & 50 \textit{(25)}  & 310   \\
\textbf{W} & 18  & 67 \textit{(80)}  & 370 (UTS)     & 34 \textit{(36)}  & 180 \\
\bottomrule  
\end{tabular}}
\end{table}

It can be concluded that in general the residual stresses post-HIP are not likely to reduce the target lifetime considerably, given that the maximum stress and Goodman equivalent stress values are close to the calculated values without residual stress. The residual stresses can be even beneficial in some cases, as in the case of pure tungsten where the compressive residual stress leads to an increase of the safety margin between the maximum tensile stress and the UTS of the material. However, the fact that the cladding is expected to present plastic deformation in some of the target tungsten blocks requires further investigation on this subject. In order to evaluate the residual stresses in the HIPed blocks, experimental measurements are foreseen with different techniques (hole drilling, ring core technique, neutron diffraction,...). The experimental setup and results of this measurements will be presented in a separate paper.

\subsection{Accidental scenarios and target survivability}
\label{Sec:TGT:Simus:accident}

\subsubsection{Cooling system failure}
\label{Sec:TGT:Simus:accident:LOCA}

The target cooling circuit is critical for the successful operation of the BDF target. A disconnection or rupture of the cooling pipes or a failure of the cooling system equipment could lead to the sudden stop of the water circulation or to the loss of cooling water in the circuit. Even if the circuit pressure and flow rate is continuously monitored and interlocked in order to stop the target irradiation in case of cooling system failure, there are several operational (and not only) risks associated to the loss of active cooling in the target.

In the event of a cooling accident with beam stop, the heat produced by the decay of the long-time irradiated target materials will dissipate at a very slow rate, due to the absence of forced cooling, increasing the temperature of the target materials for the hours and days following the accident. Two different scenarios have been considered:
\begin{itemize}
    \item In case of a sudden water circulation failure, where the water remains inside the target inner tank, the heat will mainly dissipate through the stagnant water via natural convection with a low convection coefficient. The main risks associated to this scenario are the target melting or cracking due to the high temperatures reached, and the possible over-pressure in the tank in case of water evaporation.
    \item In the event of a Loss Of Coolant Accident (LOCA), where all the cooling water is lost from the circuit, the decay heat can only be dissipated by natural convection with air or helium, with a much lower convection coefficient than water. In this case, the main risk is the possible melting of the target materials given the high concentration of heat in the target or the production of tungsten trioxide (WO$_3$) species, which are highly volatile.
\end{itemize} 

\begin{figure}[htbp]
\centering %
\includegraphics[width=0.7\linewidth]{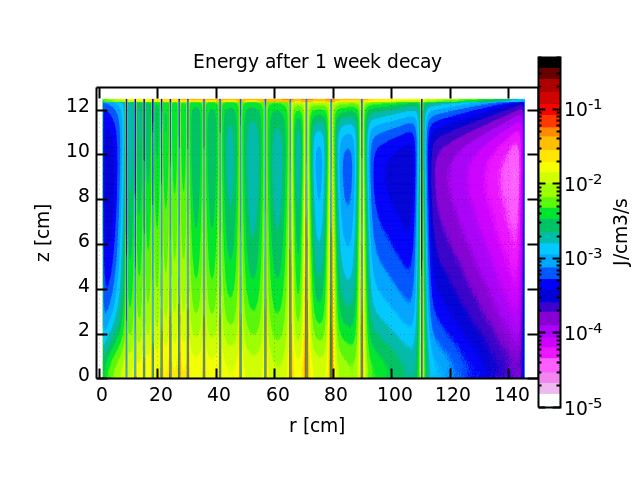}
\caption{\label{fig:TGT:Decay1w} Decay heat in the BDF target blocks after 1 week. FLUKA calculation results considering 5 years operation with a total of \num{2e20} POT.}
\end{figure}

The decay heat produced in the target materials for different periods has been estimated via FLUKA simulations. Figure~\ref{fig:TGT:Decay1w} displays the heat distribution in all the target blocks one week after the event obtained in FLUKA. It can be seen that the decay heat is dominated by the decay of tantalum isotopes.

\begin{figure}[htbp]
\centering %
\includegraphics[width=0.7\linewidth]{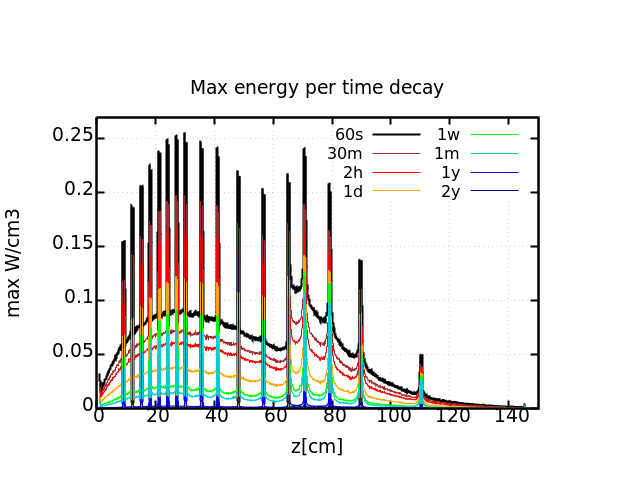}
\caption{\label{fig:TGT:Decay_all} Longitudinal plot of the maximum decay heat in the BDF target blocks for different time-steps. FLUKA calculation results are conservatively considering 5 years operation with a total of \num{2e20} POT before failure of the cooling system.}
\end{figure}

Figure~\ref{fig:TGT:Decay_all} presents the maximum decay heat in the target blocks along the Z-axis for different time-steps, from 60 seconds to 2 years; the power generated is of the order of 600 W one minute after the failure. 

\begin{figure}[htbp]
\centering %
\includegraphics[width=1\linewidth]{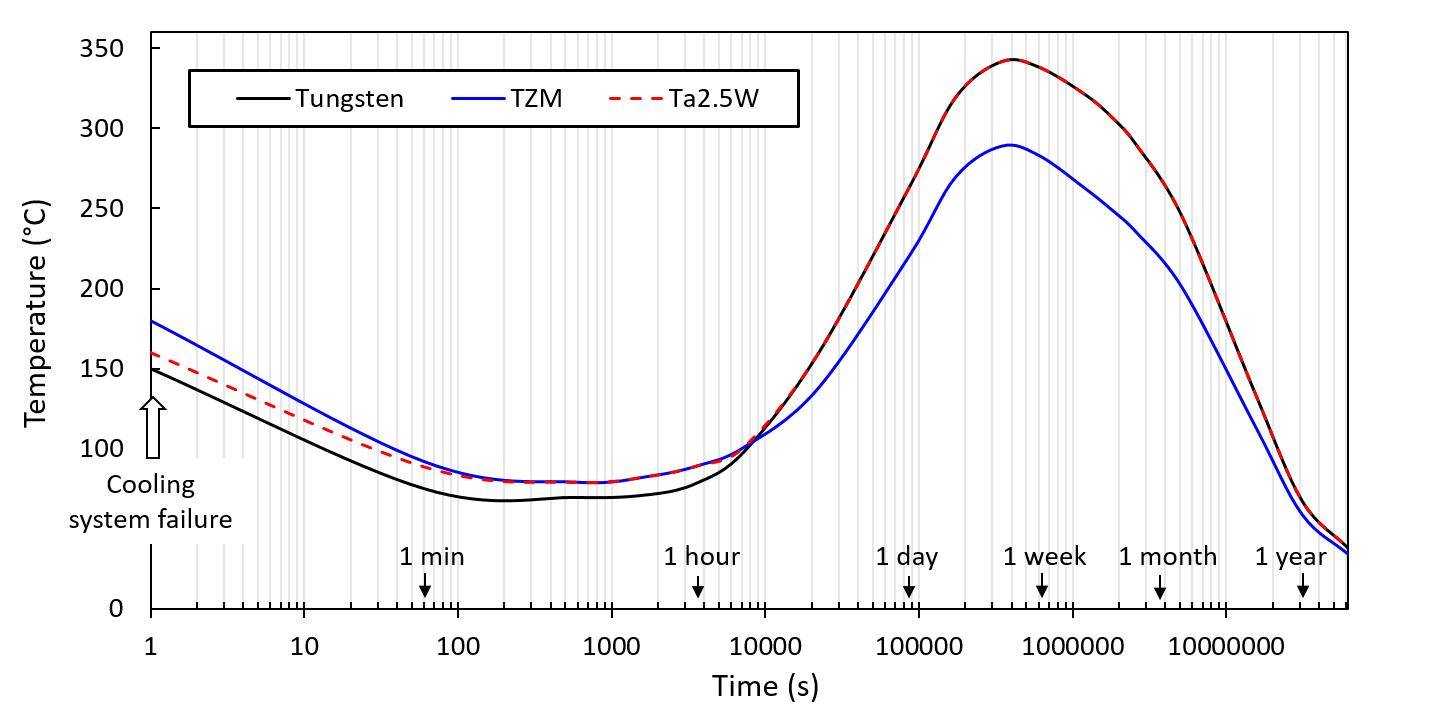}
\caption{\label{fig:TGT:Decay_temp_2y} Evolution of the maximum temperature in the different target materials after a cooling system failure due to the decay heat generated inside the target.}
\end{figure}

Thermal calculations have been carried out to evaluate the effects on target of a cooling system failure. As a preliminary approach, a low convection coefficient of 1 W/m$^{2}$K has been selected as boundary condition for the calculations. Such a film convection coefficient could be considered as conservative for natural convection with air, helium or water. Under these conditions, the temperature evolution in the target blocks has been estimated. Figure~\ref{fig:TGT:Decay_temp_2y} presents the maximum temperature evolution in the target blocks after the cooling system failure during 2 years. The maximum temperature in the target blocks is around 350$^{\circ}$C and is reached 1 week after the accident, which is sufficiently low as to avoid oxidation even in the case of breached tungsten. Further studies are required in order to understand the possibility for producing oxidation at these temperatures in TZM.

\begin{figure}[htbp]
\centering %
\includegraphics[width=1\linewidth]{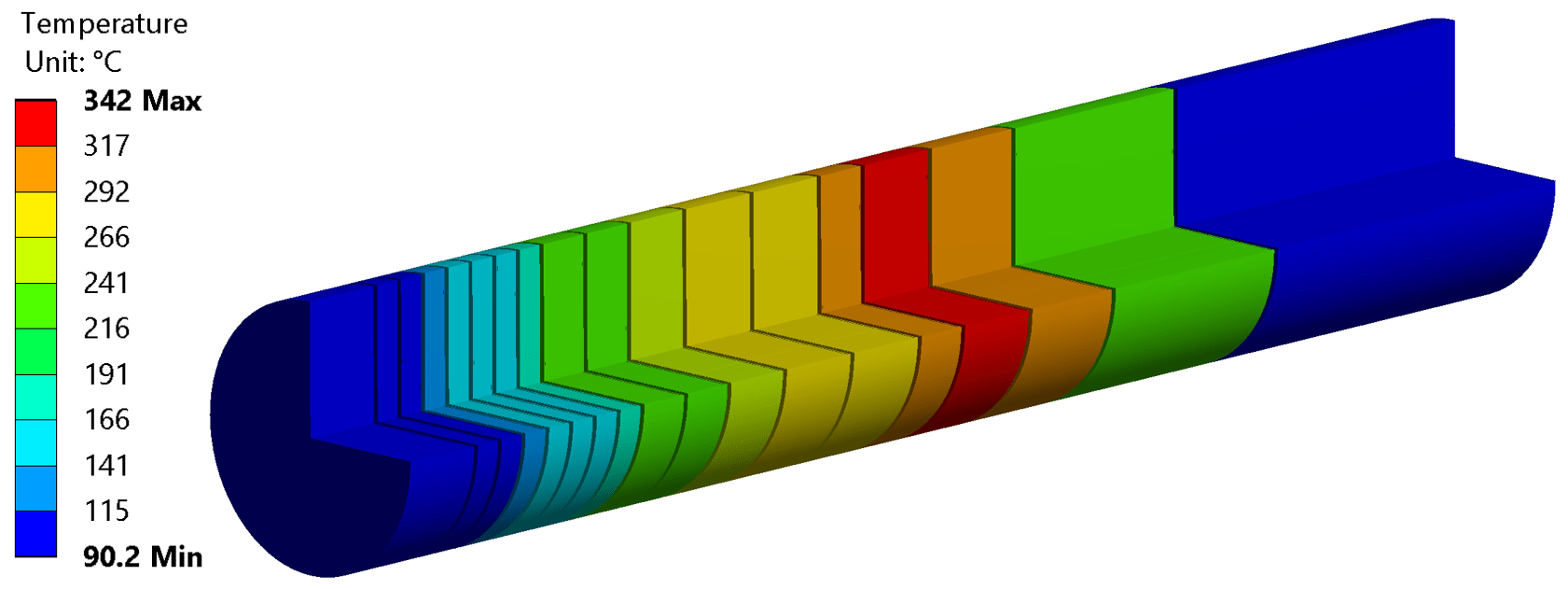}
\caption{\label{fig:TGT:Decay_temp_1w} Temperature distribution in the target blocks 1 week after the cooling system failure. Maximum temperature $\approx$ 350$^{\circ}$C, located in block 15.}
\end{figure}

Figure \ref{fig:TGT:Decay_temp_1w} presents the temperature distribution in the target blocks one week after the failure. It can be observed that the temperatures found are different from one block to another, and that the peak temperature is reached in block number 15 (tungsten core), coincident with the position of maximum energy in the core materials shown in Figure \ref{fig:TGT:Decay_all}.

The temperatures expected on target for this accidental scenario are well below the melting point of the materials. The stresses induced by the thermal loads have also been calculated. The equivalent von Mises stress and maximum principal stress in the core materials (TZM and tungsten) are below 20 MPa for all the blocks during the whole period considered (2 years), according to the structural calculations carried out. Hence, the target core materials are not expected to compromise the target survival after the cooling system failure.

\begin{figure}[htbp]
\centering %
\includegraphics[width=1\linewidth]{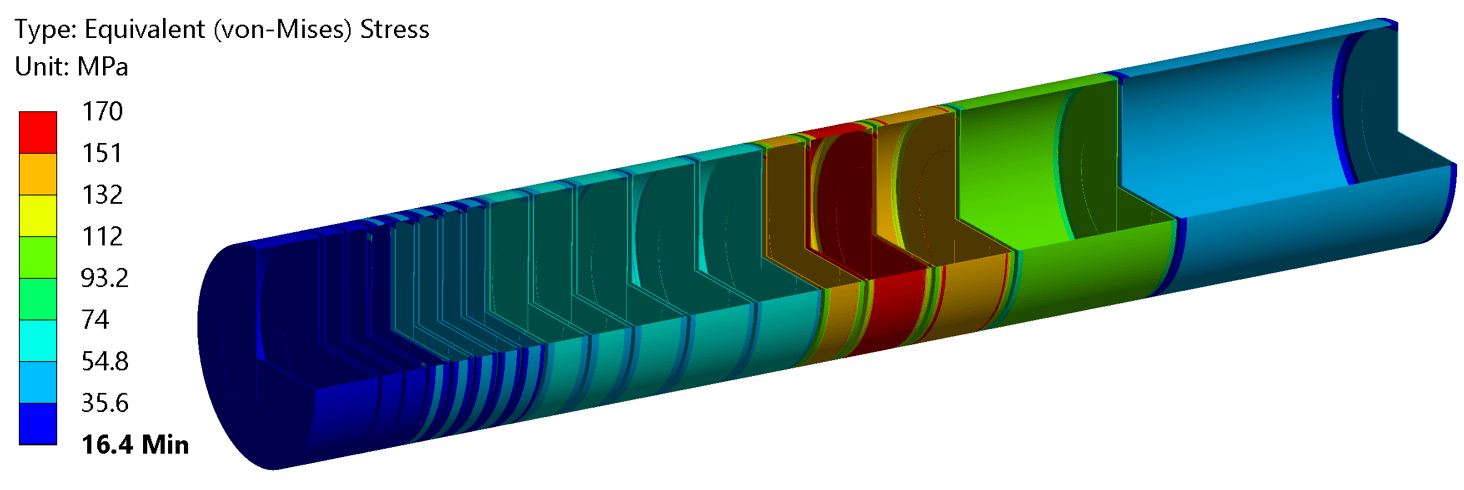}
\caption{\label{fig:TGT:Decay_stress} Von Mises equivalent stress distribution in the cladding of the target blocks 1 week after the cooling system failure. Maximum stress $\approx$ 170 MPa, located in block 15.}
\end{figure}

In the Ta2.5W cladding, higher stresses are expected, given the high amount of heat generated by the decay of tantalum isotopes. Figure~\ref{fig:TGT:Decay_stress} presents the von Mises equivalent stress distribution in the Ta2.5W cladding of the target blocks at the moment of maximum temperature, i.e. one week after the cooling system failure. 

The maximum stress is around 170 MPa, found in the cladding of block 15. This value is above the yield strength of the material at 350$^{\circ}$C, which is estimated around 140 MPa. Therefore, a certain plastic deformation is expected on the Ta2.5W cladding if the cooling system is not restarted after several days. Fracture of the cladding is not foreseen, since the maximum stress is below the tensile strength of the material and the total strain is very low compared with the elongation to failure of Ta2.5W. 

As a further study, it should be evaluated if the plastic deformation produced has a detrimental effect in the target behavior if operation is resumed after more than one week. Given that the plastic strain is quite low (0.02\%) and localized, it is expected that the plastic deformation provoked by the decay heat will not be impacting the target lifetime.

\begin{figure}[htbp]
\centering %
\includegraphics[width=1\linewidth]{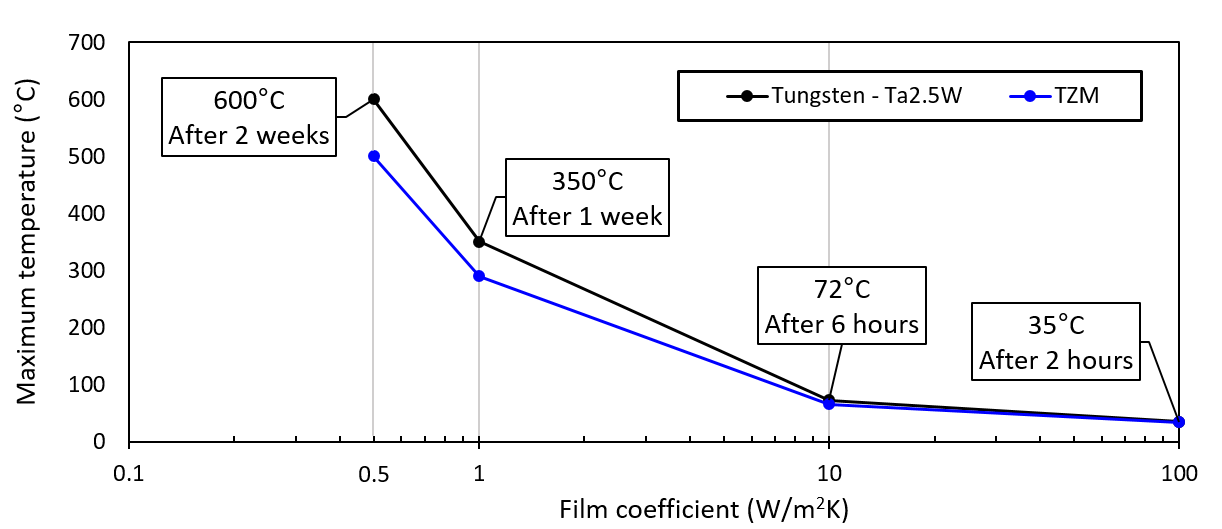}
\caption{\label{fig:TGT:Decay_temp_comp} Maximum temperature in the target materials obtained via FEM calculations for 4 different heat transfer coefficients used as boundary condition to model the natural convection. The value used for the detailed thermal simulations is 1 W/m$^{2}$K, which is considered to be realistic and conservative.}
\end{figure}

The FEM calculations performed have shown that the cooling system failure could lead to high temperatures in the target materials, which are not critical in terms of target melting or fracture, but could induce plastic deformation in some of the blocks cladding. However, the preliminary calculations performed present several limitations to fully understand the consequences of a cooling circuit failure in the target:

\begin{itemize}
    \item First, they do not take into account the water temperature increase as the heat is dissipating into the stagnant water, in the case of a sudden stop of the cooling circuit. Therefore, it is not possible to estimate the risk of water evaporation and tank over-pressure.
    \item Then, further studies are required to evaluate the possibility of releasing volatile isotopes at high temperatures. This is the case for pure tungsten, that has been reported to form oxide layers above 400$^\circ$C~\cite{WO3_oxi} and release tungsten trioxide species (WO$_3$) at 800$^\circ$C~\cite{WO3}. Additional studies are required to investigate the potential volatile isotope production for TZM and Ta2.5W.
    \item Finally, the convection coefficient used has a considerable influence in the temperatures reached on target. Figure \ref{fig:TGT:Decay_temp_comp} shows a comparison between the maximum temperatures expected on target for 4 different heat transfer coefficients used as boundary condition for the FEM calculations: 0.5 W/m$^{2}$K, 1 W/m$^{2}$K (used for the thermal simulations presented), 10 W/m$^{2}$K and 100 W/m$^{2}$K. It can be seen that the variation of the film coefficient leads to very different scenarios for the target materials. Further dedicated studies are required to be able to provide a more firm assessment of the situation.
\end{itemize} 

In order to achieve an accurate modelling of the air, helium or water behavior after a cooling system failure accident, CFD studies are currently ongoing. These will allow calculating precisely the natural convection coefficient and the temperatures reached on target, thereby assessing the target survivability for the different scenarios considered.


\subsubsection{Dilution system failure}
\label{Sec:TGT:Simus:accident:dilution}

The beam dilution system is critical for the target to absorb safely the full SPS primary beam without damage. The transfer line design considers 4 independent dilution magnets powered by 4 different power supplies. Several dilution failure patterns are possible depending on the number of failing power supplies and the dilution scheme used. Two different dilution schemes have been considered, one with two pairs of magnets rotated by $\pi/2$ and another one with 4 magnets rotated by $\pi/4$. More details on the transfer line dilution systems and the possible failure scenarios are given in Chapter~\ref{Chap:Transfer}. 

\begin{figure}[ht]
  \centering
  \begin{subfigure}[b]{0.5\linewidth}
    \centering\includegraphics[width=200pt]{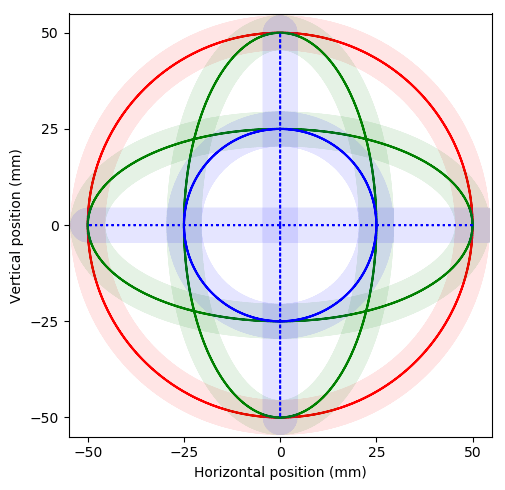}
    \caption{\label{fig:TGT:dilution_fail_2}}
  \end{subfigure}%
  \begin{subfigure}[b]{0.5\linewidth}
    \centering\includegraphics[width=200pt]{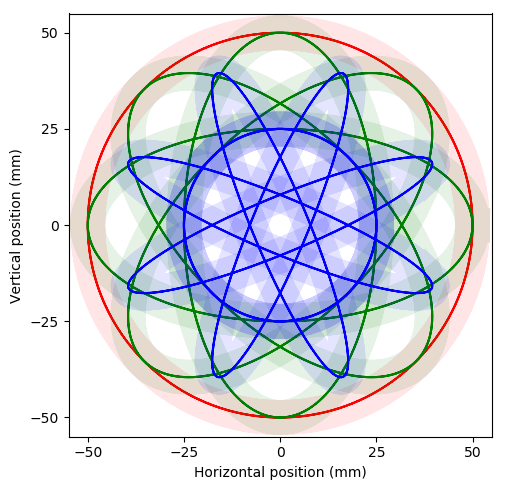}
    \caption{\label{fig:TGT:dilution_fail_4}}
  \end{subfigure}
  \caption{Possible dilution patterns on the target with all four circuits in red, 3 circuits in green and only two circuits in blue. Figure~\subref{fig:TGT:dilution_fail_2} shows the patterns for the $\pi/2$ scheme while the $\pi/4$ possible patterns are shown in \subref{fig:TGT:dilution_fail_4}. See Chapter 5 for more details about the beam dilution system.}
    \label{fig:TGT:dilution_fail_all}
\end{figure}

Figure~\ref{fig:TGT:dilution_fail_all} displays the possible dilution failure patterns that could take place during operation. The beam size of $\sigma$ = 8 mm is also represented in transparency.

It is necessary to evaluate the thermal-induced effects on target for the different dilution failure scenarios foreseen, in order to estimate the target failure probability and decide whether a continuous monitoring system of the dilution system is required or not. The results presented in this section summarize the thermo-mechanical calculations performed for that purpose.

The survivability of the target under the different failure cases has been evaluated by studying two parameters that are considered to be among the most critical for the target operation: 1) the maximum von Mises equivalent stress in the Ta2.5W cladding, which shall be lower than the yield strength of Ta2.5W at the operational temperature (as discussed in Section \ref{Sec:TGT:Simus}) and 2) the blocks surface temperature, which has to be maintained below the boiling temperature of water (as will be shown in Section~\ref{Sec:TGT:CoolingCFD}).

\smallskip
\underline{Scenario 1 - loss of one single dilution power supply}
\smallskip

The loss of one single power supply leads to an oval dilution pattern (for both $\pi/2$ and $\pi/4$ schemes), shown in green in Figure~\ref{fig:TGT:dilution_fail_all}. This scenario induces a maximum temperature of 310$^{\circ}$C in the blocks and a maximum surface temperature of 280$^{\circ}$C, which can lead to very localized water boiling for a short period of time. The maximum von Mises equivalent stress in the Ta2.5W cladding is found to be 140 MPa, below the yield strength at 300$^{\circ}$C which is estimated around 160 MPa. 

Given that the amount of boiling in this case can be negligible, and that the probability of plastic deformation in the cladding is low, it is considered that for the failure of one single power supply, the target would survive and no replacement would be needed.

\smallskip
\underline{Scenario 2 - loss of two dilution power supplies (circular dilution pattern)}
\smallskip

The loss of two dilution circuits may lead to different dilution patterns. The first one considered is a small circular pattern of 25 mm radius, in blue in figure \ref{fig:TGT:dilution_fail_all}.

This failure scenario leads to a maximum temperature of 340$^{\circ}$C in the blocks and a maximum surface temperature of 300$^{\circ}$C, which again could induce very localized water boiling for a short period of time. The maximum von Mises equivalent stress in the Ta2.5W cladding is found to be 170 MPa, above the yield strength at 350$^{\circ}$C which is estimated around 140 MPa. This means that local plastic deformation is expected, and further analysis was performed as described hereafter.

\begin{figure}[htbp]
\centering %
\includegraphics[width=1\linewidth]{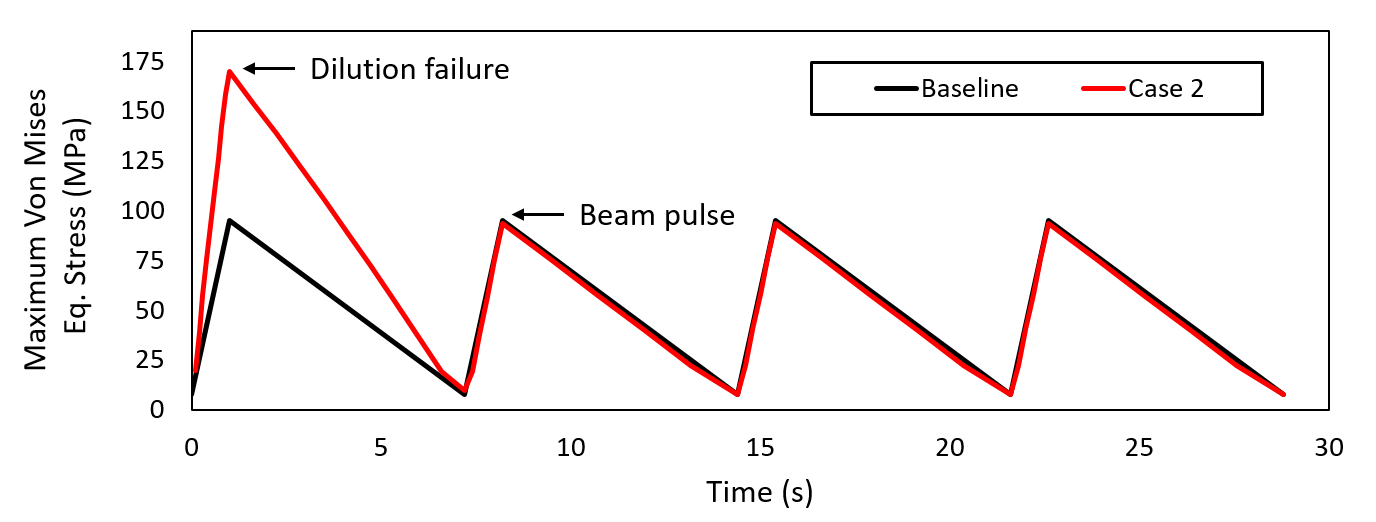}
\caption{\label{fig:TGT:Dilution_case2} Stress evolution during 3 pulses after the beam dilution failure presented in case 2. Comparison between the results of the baseline case and case 2 for the maximum von Mises equivalent stress found in the Ta2.5W cladding.}
\end{figure}

The target behavior after the dilution failure has been simulated in order to understand the effects on target of such a failure scenario. A plastic model with temperature-dependent hardening has been used to simulate the materials behavior above the yield point and after plastic deformation. The maximum von Mises equivalent stress reached after the beam impact subsequent to the dilution failure is 95 MPa, exactly as for the baseline case (see Section~\ref{Sec:TGT:Simus:struct}), and no further plastic deformation takes place. The situation described is illustrated in Figure~\ref{fig:TGT:Dilution_case2}. 

It can be concluded that a residual stress has been created by the plastic deformation induced after the beam dilution failure impact, but this residual stress is negligible and is not expected to affect the target behavior for normal operation. Taking into account these considerations, it is considered that - according to the current knowledge - for this failure scenario the target would survive and no replacement would be needed.

\smallskip
\underline{Scenario 3 - loss of two dilution power supplies (flat dilution pattern)}
\smallskip
 
Another possible dilution pattern produced by the failure of two power supplies is a flat line of 50 mm, represented in blue dotted lines in Figure \ref{fig:TGT:dilution_fail_all}a, obtained only for the $\pi/2$ scheme.

In this case, the maximum temperature reached is 450$^{\circ}$C in the blocks and a maximum surface temperature of 400$^{\circ}$C is found, with temperatures over 250$^{\circ}$C in large areas of the block surface. Water boiling is not considered to be negligible under these operational conditions and could pose a risk for target operation. 

The maximum von Mises equivalent stress in the Ta2.5W cladding is found to be 215 MPa, well above the yield strength at 400$^{\circ}$C which is estimated around 125 MPa. This means that local plastic deformation is expected. The maximum von Mises equivalent stress is lower than the UTS of the material at 400$^{\circ}$C, and the calculated total strain is well below the elongation limit of Ta2.5W, hence the risk of fracture in the cladding is considered to be very low. 

The plastic deformation created by the beam impact under dilution failure could lead to residual stresses that affect the target lifetime. Similarly to Scenario 2, the maximum von Mises stress reached in the Ta2.5W cladding after the first beam impact subsequent to the beam dilution failure has been estimated at a value of 145 MPa. This stress is below the yield strength of Ta2.5W at the operational temperatures, and during the next pulses no further plastic deformation is expected to occur. 

However, the stress level is around 1.5 times higher than the one of the baseline case, and it is considered that the dilution failure will affect the target behavior by creating residual stresses, modifying the stress state in the target and increasing the maximum stress during operation. It is concluded that the target shall be replaced for this failure scenario.

\smallskip
\underline{Scenario 4 - loss of 2 dilution power supplies (elliptical dilution pattern)}
\smallskip

Scenario 4 has effects on target very similar to Scenario 3. In this configuration, the failure of two power supplies leads to a flattened elliptical pattern represented by a blue ellipse in Figure~\ref{fig:TGT:dilution_fail_all}b. The maximum temperature reached is 470$^{\circ}$C in the blocks and a maximum surface temperature of 415$^{\circ}$C is found, with temperatures over 250$^{\circ}$C in large areas of the block surface. Water boiling is not considered to be negligible under this conditions, and could be a risk for the target operation. 

The maximum von Mises equivalent stress in the Ta2.5W cladding is found to be 215 MPa, well above the yield strength at 400$^{\circ}$C which is estimated around 140 MPa. As for Case 3, local plastic deformation is expected. The risk of cladding fracture is again considered to be very low, given that the maximum von Mises equivalent stress lower than the UTS of the material at 400$^{\circ}$C, and the calculated total strain is well below the elongation limit of Ta2.5W. 

\begin{figure}[htbp]
\centering %
\includegraphics[width=0.9\linewidth]{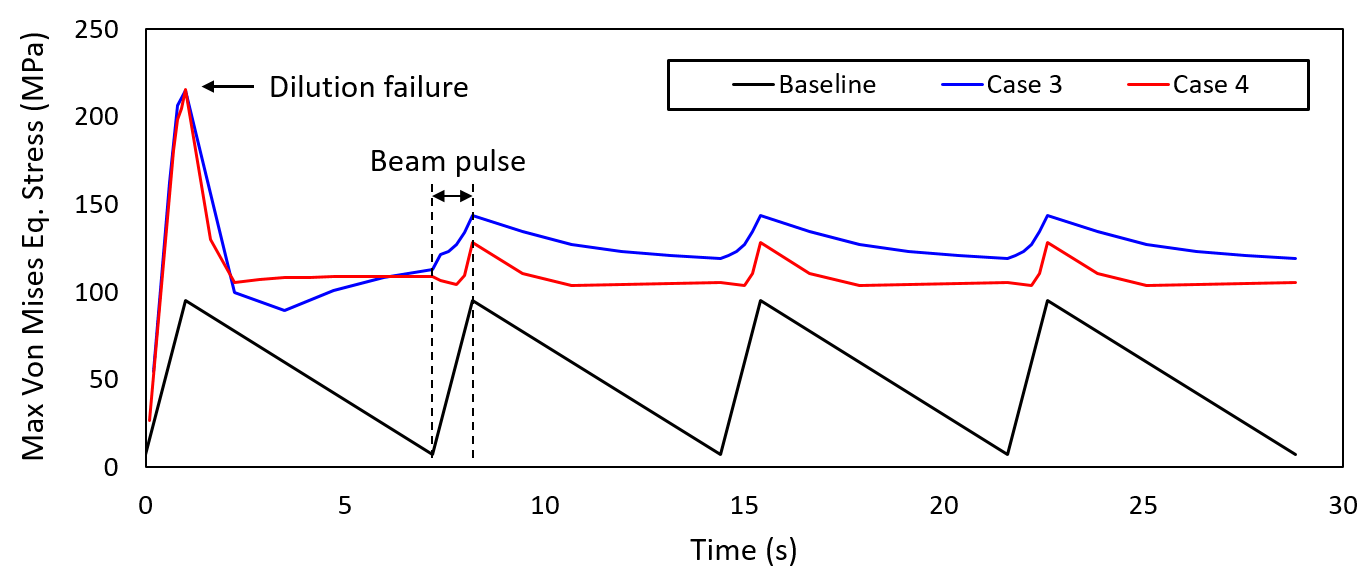}
\caption{\label{fig:TGT:Dilution_case3} Stress evolution during 3 pulses after the beam dilution failure presented in cases 3 and 4. Comparison between the results of the baseline case, case 3 and case 4 for the maximum von Mises equivalent stress found in the Ta2.5W cladding.}
\end{figure}

The maximum von Mises stress reached in the Ta2.5W cladding after the first beam impact subsequent to the beam dilution failure is around 120 MPa, 1.25 times higher than the maximum stress for the baseline case. Figure~\ref{fig:TGT:Dilution_case3} illustrates the aforementioned situation, comparing the results of the baseline case and scenarios 3 and 4. Same as for scenario 3, it is considered that the dilution failure will affect the target lifetime, and that the target replacement shall be performed in this case too.

\smallskip
\underline{Scenario 5 - loss of 3 dilution power supplies}
\smallskip

The simultaneous failure of 3 dilution circuits would lead to a flat line pattern of length 25 mm. The calculations performed have shown that the temperatures reached on target would be around 550$^{\circ}$C, with large areas of the target surface exposed to water boiling. The maximum von Mises equivalent stress obtained is 225 MPa, above the UTS of Ta2.5W at 500$^{\circ}$C (estimated around 200 MPa), leading to a high risk of fracture in the cladding. Therefore, it is concluded that this scenario would require the replacement of the target.

\smallskip
\underline{Conclusions}
\smallskip

Table~\ref{tab:TGT:dilution_fail} summarizes the effects on the BDF target for the different dilution failure scenarios considered up to now in the study. Further studies will be required at the TDR in order to validate the technical solution to be implemented in the final Project.

\begin{table}[htbp]
\centering
\caption{\label{tab:TGT:dilution_fail} The table presents a summary of the effects on target foreseen for the 5 different cases of dilution system failure considered in the present study, evaluated via FEM calculations.}
\smallskip
{\renewcommand{\arraystretch}{1.2}%
\begin{tabular}{cclcc}
\toprule
\textbf{\begin{tabular}[c]{@{}c@{}}Dilution \\ failure\end{tabular}} & \textbf{\begin{tabular}[c]{@{}c@{}}Number of failing\\power supplies\end{tabular}} & \multicolumn{1}{c}{\textbf{Effects on target}} & \textbf{\begin{tabular}[c]{@{}c@{}}Additional \\ risks\end{tabular}} & \textbf{\begin{tabular}[c]{@{}c@{}}Target \\ replacement\end{tabular}} \\ \midrule
Case 1 & 1 & \begin{tabular}[c]{@{}l@{}}Low risk of plastic deformation \\ in the cladding\end{tabular} & \begin{tabular}[c]{@{}c@{}}Localized \\ water boiling\end{tabular} & No \\ \midrule
\multirow{2}{*}{Case 2} & \multirow{2}{*}{2} & \begin{tabular}[c]{@{}l@{}}Localized plastic deformation \\ in the cladding\end{tabular} & \multirow{2}{*}{\begin{tabular}[c]{@{}c@{}}Localized\\  water boiling\end{tabular}} & \multirow{2}{*}{No} \\
 &  & No influence in target lifetime &  &  \\ \midrule
\multirow{3}{*}{Case 3} & \multirow{3}{*}{2} & Plastic deformation in the cladding & \multirow{3}{*}{Water boiling} & \multirow{3}{*}{Yes} \\
 &  & Increased stress after pulse (x1.5) &  &  \\
 &  & Influence in target lifetime &  &  \\ \midrule
\multirow{3}{*}{Case 4} & \multirow{3}{*}{2} & Plastic deformation in the cladding & \multirow{3}{*}{Water boiling} & \multirow{3}{*}{Yes} \\
 &  & Increased stress after pulse (x1.25) &  &  \\
 &  & Influence in target lifetime &  &  \\ \midrule
Case 5 & 3 & \begin{tabular}[c]{@{}l@{}}Possible fracture of the Ta2.5W\\ cladding\end{tabular} & Water boiling & Yes \\
\bottomrule  
\end{tabular}}
\end{table}

\FloatBarrier

\section{Target cooling system and CFD simulations}
\label{Sec:TGT:CoolingCFD}

\subsection{Cooling system design}
\label{Sec:TGT:CoolingCFD:design}

The target cooling system is one of the most critical parts of the BDF target design, given the high energy and average power deposited on target during operation. Pressurized water has been chosen as cooling medium, other coolants such air or helium would require a much higher flow rate to dissipate such a high amount of power. A rotating target, sought to be employed for other similar spallation targets, has been for the moment excluded due to the additional technical complexity. Several requirements have been considered for the cooling system design: first, a high water velocity is necessary to obtain an effective HTC between the cooling medium and the target blocks. Then, the pressure of the circuit should be high, and the pressure drop minimized, in order to ensure that the water in contact with the solid blocks is always below the boiling temperature. Furthermore, the cooling system design has been optimized in order to limit the increase of temperature in the circulating water whilst minimizing the necessary flow rate.

\begin{figure}[htbp]
\centering %
\includegraphics[width=0.9\linewidth]{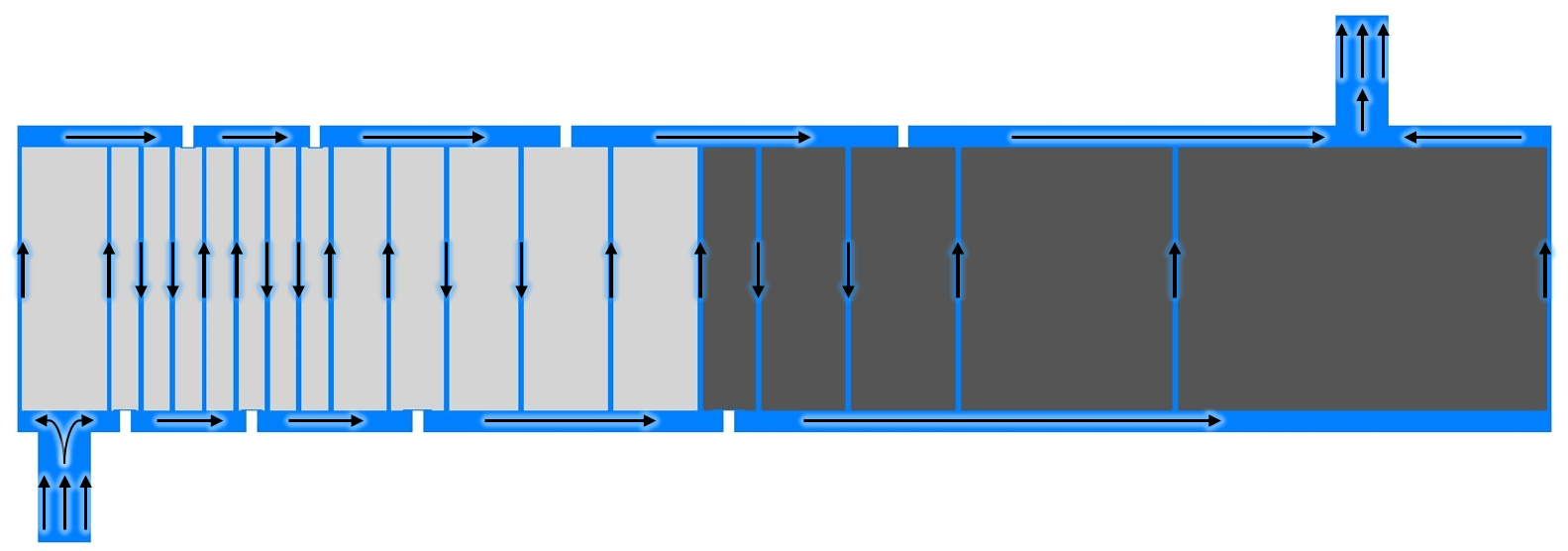}
\caption{\label{fig:TGT:2Dcooling} BDF target longitudinal cross-section. Top view of the cooling circulation path. The beam is coming from the left side; the blocks in light grey are the TZM ones, while the blocks in dark grey are those made of pure W.}
\end{figure}

\begin{figure}[htbp]
\smallskip
\centering %
\includegraphics[width=0.6\linewidth]{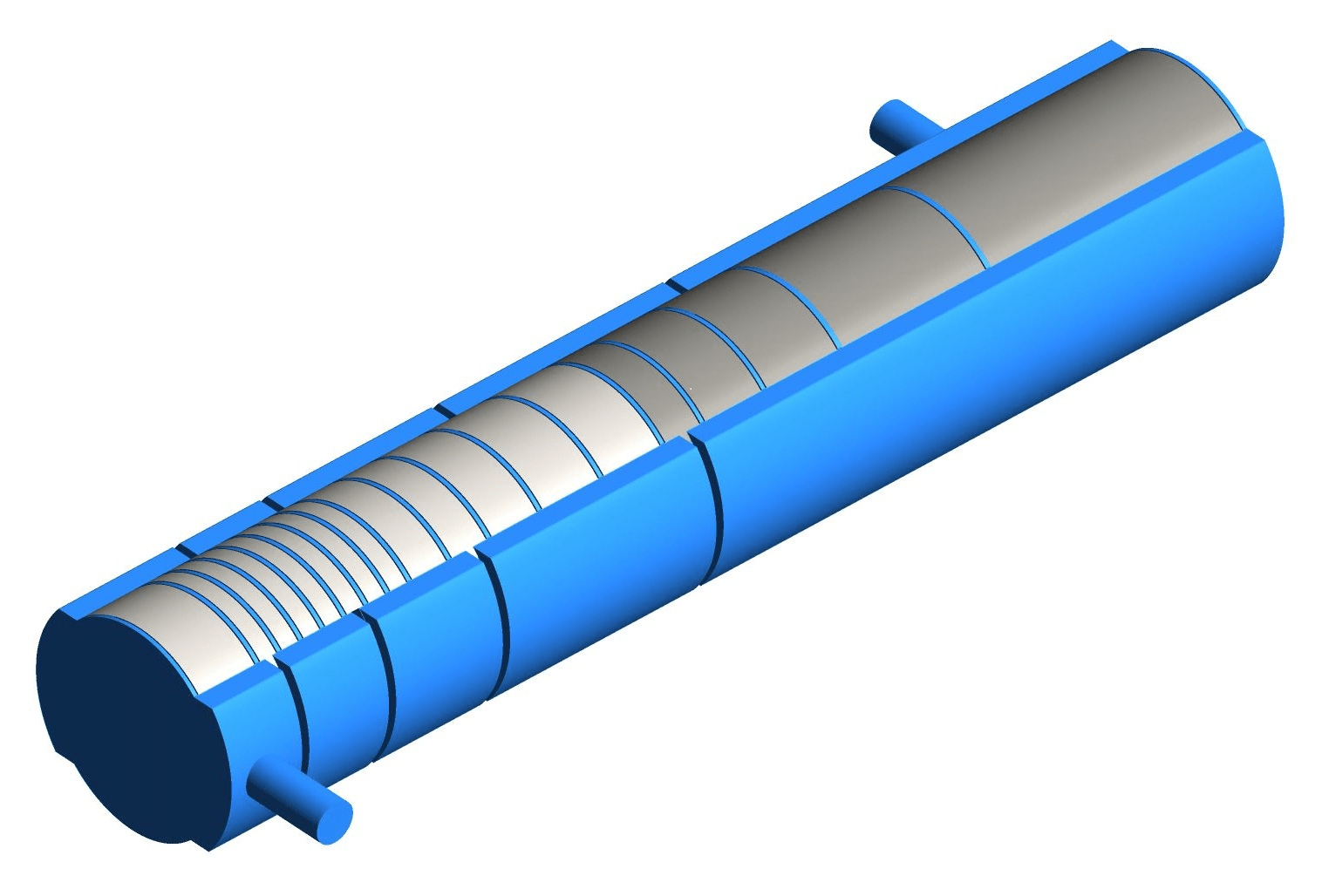}
\caption{\label{fig:TGT:3Dcooling3} 3D model of the BDF target cooling system, showing in blue the water volumes around the target core.}
\end{figure}

Figures~\ref{fig:TGT:2Dcooling} and \ref{fig:TGT:3Dcooling3} describe the cooling system circulation path around the target blocks. The target cylinders are separated by 5 mm channels that allow the water passage between the blocks. The water circulation is designed to cool down effectively the flat faces of the target cylinders, that are impacted by the diluted beam and will reach the highest temperatures during operation. The proposed cooling system design consists in a serpentine configuration (series flow) with two parallel streams. The serpentine circulation can provide high water speeds in the channels with a quite low flow rate, maintaining a reasonable pressure drop in the entire assembly. 

The target cooling circuit has been designed to allow the circulation of two streams in parallel. This configuration aims to reduce the total pressure drop of the circuit as well as the temperature increase in the water from inlet to outlet. Another reason for this arrangement is to avoid a complete cooling water circuit failure in the case that one of the channels is blocked, because of potential swelling of the blocks due to thermal expansion or debris in the circuit. If one of the channels is blocked, the water flow can continue through the other parallel channel, improving the circuit reliability. The last three channels are set in parallel, given that the number of channels is odd and the last tungsten blocks are the ones receiving the lowest amount of energy and having the lowest raise of temperature during operation.

The channels are connected by "manifolds" that receive the water from two channels and distribute it to the following two. The manifold size is reduced to minimize the total water volume of the cooling system, and is constrained by the target support design, that will be described in Section~\ref{Sec:TGT:MechDesign}. At the same time, the manifold cross-section has been designed large enough to avoid increasing the circuit pressure drop. The water velocity in the manifold is not a critical parameter for the cooling system design, given that the side circular faces of the target core blocks will not receive direct beam impact. 

The serpentine circulation is horizontal, in order to avoid the formation of air pockets during the filling process, as well as to prevent having stagnant water after draining. If a vertical configuration was chosen, the manifolds would be placed on top and below the channels, and it would be very likely that air bubbles would remain in the top manifolds during the filling process, and stagnant water could stay in the bottom manifolds during the circuit draining. The BDF target prototype design, construction and operation (see Section ~\ref{Sec:TGT:Proto}), that will be detailed in a separate publication, has improved the understanding of the requirements and limitations of the final BDF target cooling design.

\begin{figure}[htbp]
\centering %
\includegraphics[width=0.9\linewidth]{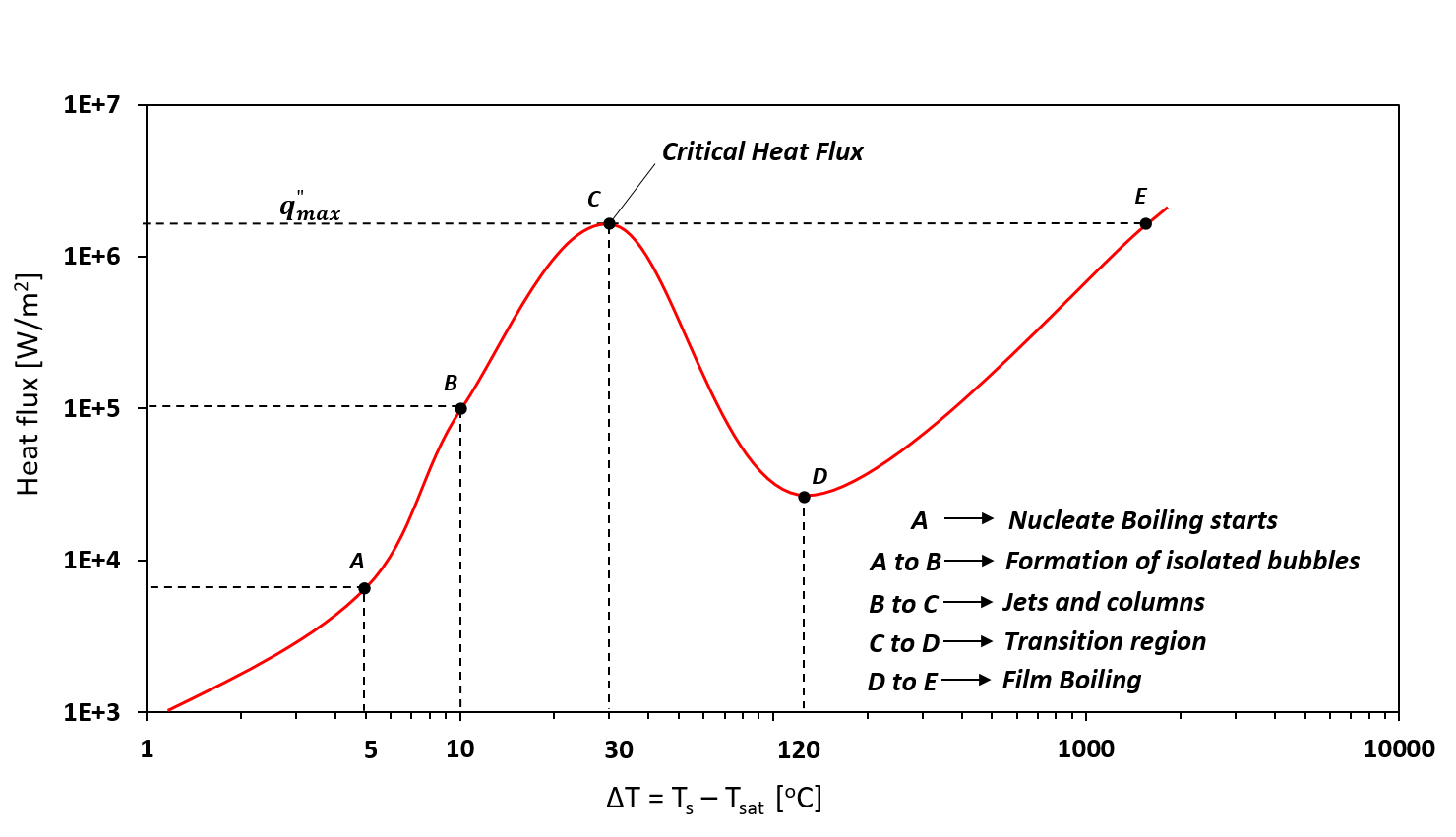}
\caption{\label{fig:TGT:boiling_curve} Typical boiling curve for water at 1 atm. Surface heat flux as a function of temperature difference with the saturation temperature \cite{CFD_Incropera}.}
\end{figure}

The chosen operational pressure of the water cooling circuit is 22 bar. The thermal-fluid simulations carried out in the study and detailed in the next Sections have shown that the maximum surface temperature in the target blocks is around $110\,^{\circ}\mathrm{C}$ at the pulse peak, well below the boiling point of water at 22 bar that is 212$\,^{\circ}\mathrm{C}$. The main risk of having localized boiling is the loss of heat transfer between the solid blocks and the water, which would prevent the heat dissipation by convection. It is worth mentioning that the safety margin between the expected water temperature and the damaging effects of boiling is considered sufficiently large, given that the first boiling phase occurring beyond the boiling point is the so-called nucleate boiling. During nucleate boiling, the heat flux is increased with respect to the one achieved by convection with liquid water and therefore, the effects on heat dissipation are not detrimental, but rather beneficial. This phase takes place after reaching the boiling temperature until this one is exceeded by around 30$\,^{\circ}\mathrm{C}$, at that point the critical heat flux is reached and the heat flux decreases drastically, leading to a significant reduction in the heat dissipation that could be harmful for the target behavior. At 22 bar, the critical heat flux is reached at around 240$\,^{\circ}\mathrm{C}$, much higher than the expected maximum water temperatures in the circuit. Figure~\ref{fig:TGT:boiling_curve} illustrates the before-mentioned effect.

The design water flow rate is 9 kg/s, leading to a rather uniform water velocity in the channels of around 5 m/s, as will be shown in the following section. The water velocity has been limited by design to 5 m/s in the channels, in order to avoid potential undesired erosion effects on the tantalum-tungsten surface. Other spallation sources such as ISIS at RAL (UK) currently use high water velocities up to 10 m/s in operational tantalum-cladded targets (TS1)~\cite{ISISlocking}. At the present conceptual design stage of the BDF target, it has been considered to apply a safety factor of 2 with respect to this value in order to ensure the safe operation of the target. As will be shown in the following sections, the average HTC obtained with a velocity of 5 m/s in the channels is sufficiently high to ensure temperatures and stresses well within the operational limits of the target.


\subsection{Analytical calculations}
\label{Sec:TGT:CoolingCFD:analytical}

Based on the fundamental knowledge of flow dynamics and heat transfer, analytical calculations were performed in order to have a preliminary idea of the water flow behavior in the cooling channels and the manifold. The average fluid-solid interface temperature and the average heat transfer coefficient on the wall of the cooling channel were also calculated. The properties of water were conservatively taken at 20 bar and 30$^\circ$C.

\begin{figure}[htbp]
\centering %
\includegraphics[width=0.7\linewidth]{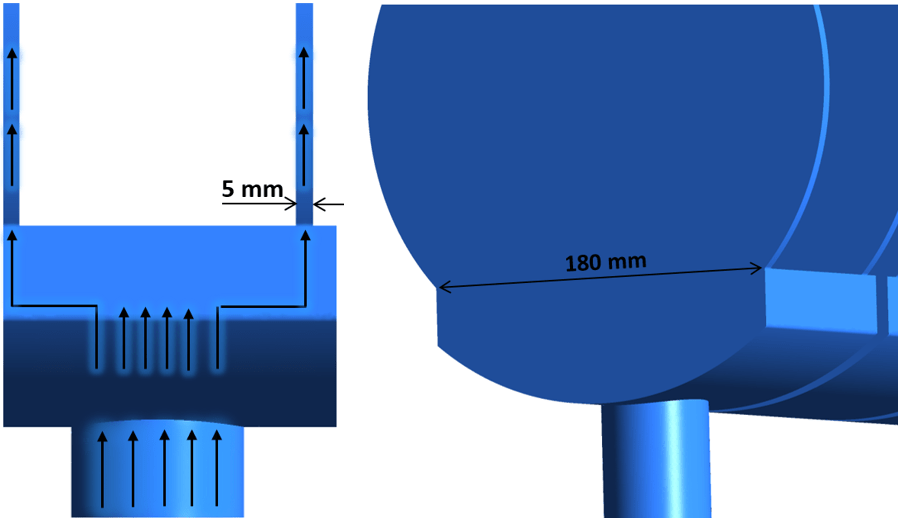}
\caption{\label{fig:TGT:CFD_inlet} The figure shows the BDF target cooling circuit channel dimensions and flow bifurcation.}
\end{figure}

As discussed earlier, a channel velocity of 5~m/s was considered as initial design parameter. A comparative study will be done later in this section with channel velocities of 1, 2.5, 7.5 and 10 m/s. From Figure~\ref{fig:TGT:2Dcooling} it can be appreciated that after the water enters the cooling domain from the inlet, it gets bifurcated into two streams of parallel channels passing adjacent to the first target blocks. A zoomed view of Figure~\ref{fig:TGT:2Dcooling} including two cooling channels and the inlet is shown in Figure~\ref{fig:TGT:CFD_inlet}. 

Due to the serpentine nature of the flow, the same pattern shown in Figure~\ref{fig:TGT:CFD_inlet} is found in the following sections of the domain, which means that the sum of mass flow rate in any couple of parallel cooling channels is identical and is also equivalent to the inlet mass flow rate. However, in the last part of the cooling domain, three channels are in parallel and therefore the average velocities in these channels will correspondingly be less than 5~m/s. 

Using the principle of mass conservation, the inlet mass flow rate can be calculated as follows: 

\begin{equation}
    \dot{m} = 2\rho A_c v_{channel}
\end{equation}

where $\rho$ is the density of water, $v_{channel}$ is the velocity of water in the channel and $A_c$ is the cross-section area of the cooling channels which can be simplified as a rectangular cross-section with a channel thickness of 5~mm and a width of 180~mm. Using this information, the mass flow rate can be calculated and corresponds to $\dot{m}\approx 9\text{kg/s}$.

Using the energy balance equation (Equation~\ref{eqn:TGT:CFD_tempraise}), the temperature rise of water from the inlet to the outlet of the flow configuration can be evaluated: 

\begin{gather}
\label{eqn:TGT:CFD_tempraise}
    Q = \dot{m} C_P \Delta T \\
    \Delta T = \frac{Q}{\dot{m}C_p}\approx \boldsymbol{8.1^{\circ} C},
\end{gather}

where $Q$ is the average beam power deposited on the target core (305 kW) and $Cp$ is the specific heat of water.

The primary focus is to resolve the flow and heat transfer in the channels (and not in the manifolds) because of two reasons: 

\begin{itemize}
    \item The beam impacts along the longitudinal direction of the target blocks and hence the maximum energy deposition in all the blocks will be mainly located along the beam dilution path, and not in the cylindrical surface of the blocks.
    \item The conjugate heat transfer is primarily dependent on the flow speed and the contact area (between solid and fluid) and since the average velocity of the water in the channels is approximately 5 times more than in the manifold, it can be safely assumed that the main heat dissipation will take place in the channels.
\end{itemize}

Using the velocity of water in channel, the average Reynolds number inside the channel can be calculated as follows:

\begin{gather}
    \label{eqn:TGT:cfd5}
    Re_{c} = \frac{\rho\times v_{channel}\times D_H}{\mu}
\end{gather}

where $\mu$ is the dynamic viscosity of water and $D_H$ is the hydraulic diameter of the channel, which can be calculated as:

\begin{gather}
    \label{eqn:TGT:cfd6}
    D_H = \frac{4\times A_c}{P_c} = 0.00973 \: m
\end{gather}

where $P_c$ is the wetted perimeter of the channel cross-section.
Using this value in equation \ref{eqn:TGT:cfd5}, the average Reynolds number $Re_D$ inside the channels is around 60000, showing that the flow is highly turbulent in the cooling circuit channels.

The HTC at the  fluid-solid interface can be analytically calculated using the following expression:

\begin{gather}
    \label{eqn:TGT:cfd8}
    h = \frac{Nu\times k}{D_H}
\end{gather}

where $k$ is thermal conductivity of the water, $D_H$ is the hydraulic diameter of the channel and $Nu$ is the Nusselt number. Several convection correlations can be used to calculate the value of $Nu$ for turbulent flow in a channel. Assuming water flow over a smooth surface, either Dittus-Boelter's or Gnielinski's equation can be used~\cite{CFD_Incropera} (however Gnielinski's equation gives higher level of accuracy, which will be employed in the following sections). Dittus-Boelter's equation can be written as:

\begin{gather}
\label{eqn:TGT:cfd9}
    Nu_D = 0.023 \, Re_D^{4/5}Pr^n
\end{gather}

which is valid for, $0.6\leq Pr\leq 160 $, $Re_D \geq 10,000$ and $\frac{L}{D}\geq 10$. $n$ in equation \ref{eqn:TGT:cfd9} is 0.4 if the solid temperature is higher than the liquid temperature and 0.3 if the liquid temperature is higher than the solid temperature. 

Gnielinski's equation \cite{CFD_Gnielinski} can be written as:

\begin{gather}
\label{eqn:TGT:cfd10}
    Nu_D = \frac{(f/8)(Re_D-1000)Pr}{1+12.7(f/8)^{1/2}(Pr^{2/3}-1)}
\end{gather}

which is valid for $0.5\leq Pr\leq 2000$ and $3000\leq Re_D\leq 10^6$. $f$ is the friction factor which can be calculated by the Petukhov relationship~\cite{CFD_Petukhov} as given in equation \ref{eqn:TGT:cfd11}.

\begin{gather}
\label{eqn:TGT:cfd11}
    f = \big(0.79 \, \text{ln} (Re_D) - 1.64\big)^{-2} \:\:\:\: \text{valid for} \:\:\:\: 3000\leq Re_D\leq 5 \times 10^6
\end{gather}

Using the above relationship and the value of $Re_D$, the friction factor in the channel is found out to be $f=0.02$. The Prandtl number considered for water at $30\,^{\circ}\mathrm{C}$ is 5.4 \cite{CFD_Incropera}, and is assumed to be constant for the flow. Using the values of $f$, $Re_D$ and $Pr$, $Nu_D$ can be calculated in Equations \ref{eqn:TGT:cfd9} and \ref{eqn:TGT:cfd10}:

\begin{gather}
\label{eqn:TGT:cfd13}
  Nu_D = 300\:\:\:\:\:  (\text{using Dittus-Boelter equation}) \\
  \label{eqn:TGT:cfd14}
   Nu_D = 350\:\:\:\:\: (\text{using Gnielinski equation}) 
\end{gather}

It is recommended to use the value of $Nu_D$ from Gnielinski equation if a higher level of accuracy is desired~\cite{CFD_Incropera}. Using the value of $Nu_D$ from Equation \ref{eqn:TGT:cfd14} in Equation \ref{eqn:TGT:cfd8}, the HTC can be estimated as:

\begin{gather}
    \label{eqn:TGT:cfd15}
    h = \frac{350 \times 0.61}{0.00973} \:\:\: \Rightarrow \:\:\: \boldsymbol{h \approx 22000 \:W/(m^2K)}
\end{gather}

The high HTC value calculated using the above expression is attributed to the large velocities reached in the channels. A comparative analytical study has been performed considering 1 m/s, 2.5 m/s, 7.5 m/s and 10 m/s as average channel velocity. Table~\ref{tab:TGT:CFD_analytical} reports a comparison between the results for different channel velocities. 

\begin{table}[htbp]
\centering
\caption{Comparison of various analytical results with the channel velocity}
\label{tab:TGT:CFD_analytical}
\begin{tabular}{ccccccc}
\toprule
$\boldsymbol{Case \: No.}$ & $\boldsymbol{v_c}$ & $\boldsymbol{\dot{m}}$ & $\boldsymbol{\Delta T}$ & $\boldsymbol{f}$    & $\boldsymbol{Nu}$ (using $\ref{eqn:TGT:cfd10}$)  & $\boldsymbol{HTC}$ (using $\ref{eqn:TGT:cfd10}$) \\
\midrule
1       & 1     & 1.8       & 41       & 0.03                          & 86                                               & 5400                      \\ 
2       & 2.5   & 4.5       & 16      & 0.024                         & 190                                             & 12000                    \\ 
$\boldsymbol{3}$       & $\boldsymbol{5}$     & $\boldsymbol{9}$         & $\boldsymbol{8.1}$        & $\boldsymbol{0.02}$                         & $\boldsymbol{350}$                                             & $\boldsymbol{22000}$                        \\ 
4       & 7.5   & 13.4     & 5.4        & 0.018                       & 490                                             & 31000                    \\ 
5       & 10    & 18        & 4.1        & 0.017                       & 630                                             & 40000                      \\
\bottomrule
\end{tabular}
\end{table}

As it can be seen, the temperature rise from inlet to outlet of the cooling domain is around 41$\,^{\circ}\mathrm{C}$ and 16$\,^{\circ}\mathrm{C}$ for cases 1 and 2 respectively. However, it has been considered by design of the target cooling stations that the temperature rise shall be limited to 10$\,^{\circ}\mathrm{C}$. Furthermore, it can be seen that the HTC values for cases 1 and 2 are much lower (four and two times less respectively) than the ones obtained with 5~m/s velocity.

In cases 4 and 5, even though the temperature rise is within the design limits and the HTC values are above the ones obtained for 5~m/s, the high speed of water (7.5 and 10 m/s) in the channels may induce long-term erosion effects in the cladding, which is undesired for reliable target operation. In addition to that, higher water velocities lead to a correspondigly higher pressure drop in the circuit, as will be shown in the following section. 

The analytic calculations have given a broad idea about the flow and heat transfer characteristics in the cooling circuit and specifically in the cooling channels. In spite of that, an extensive CFD study of the full scale cooling system is needed because of the complexity of the flow in the turbulent regime, as well as to resolve the flow and heat transfer in the boundary layer of the channels. 

\subsection{CFD calculations}
\label{Sec:TGT:CoolingCFD:CFD}

\subsubsection{Simulation setup}

A commercial CFD code, ANSYS Fluent, has been used to perform the extensive 3D turbulence modelling of the flow configuration. In conjugate heat transfer problems (like the present one), the heat transmitted from the solid body to the liquid is highly dependent on the thermal boundary layer. Therefore, sophisticated turbulence models (like $k-\omega$ SST or realizable $k-\epsilon$ with enhanced wall treatment) must be used to resolve the boundary layers. 

A schematic model of the full scale 3D flow domain along with the target blocks is shown in Figure~\ref{fig:TGT:3Dcooling3}. A velocity of~5 m/s in the cooling channels was selected as design parameter, and hence a velocity of 4.58 m/s is given as inlet boundary condition, the inlet diameter being 50 mm. A homogeneous Neumann boundary condition for pressure is given at the outlet (constant outlet pressure). A no-slip boundary condition is given at the walls of the cooling circuit and the target blocks. A non-uniform energy deposition is given on all the target blocks, the energy deposition map is imported from the FLUKA Monte Carlo~\cite{FLUKA_Code,Ferrari2005} simulation results to the ANSYS Fluent model with help of a User Defined Function (UDF). The turbulence model used for all the simulations presented in this section is $k\omega$-SST, unless mentioned explicitly.

\begin{figure}[htbp]
\centering %
\includegraphics[width=1\linewidth]{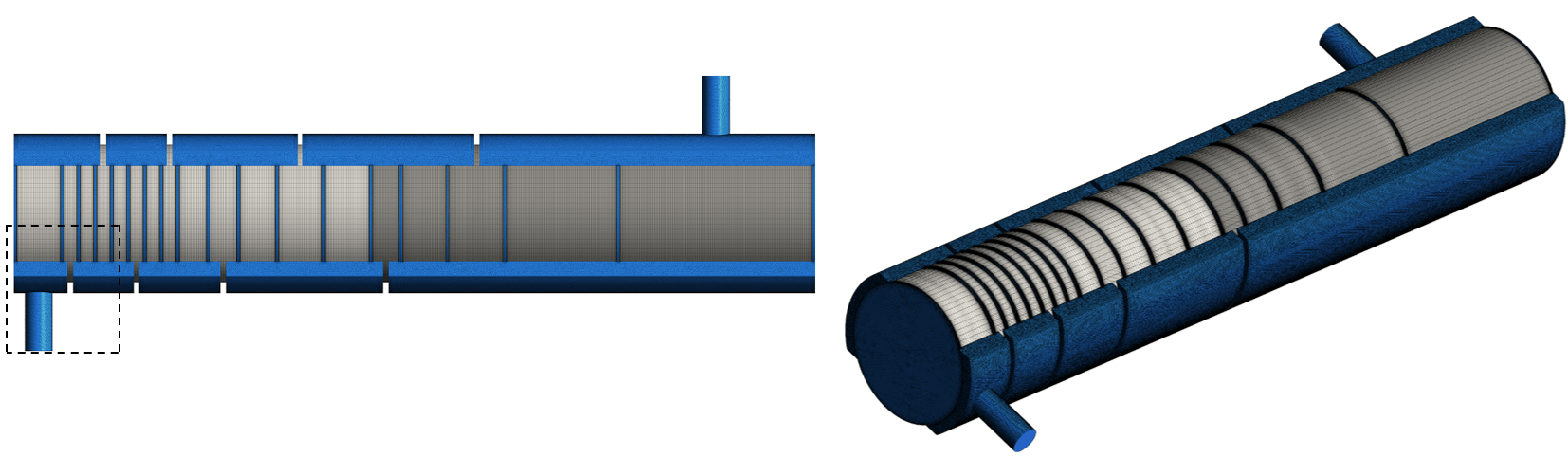}
\caption{\label{fig:TGT:CFD_Mesh} Computational hybrid grid used for the 3D CFD calculations (top view).}
\end{figure}

\begin{figure}[htbp]
\centering %
\includegraphics[width=0.8\linewidth]{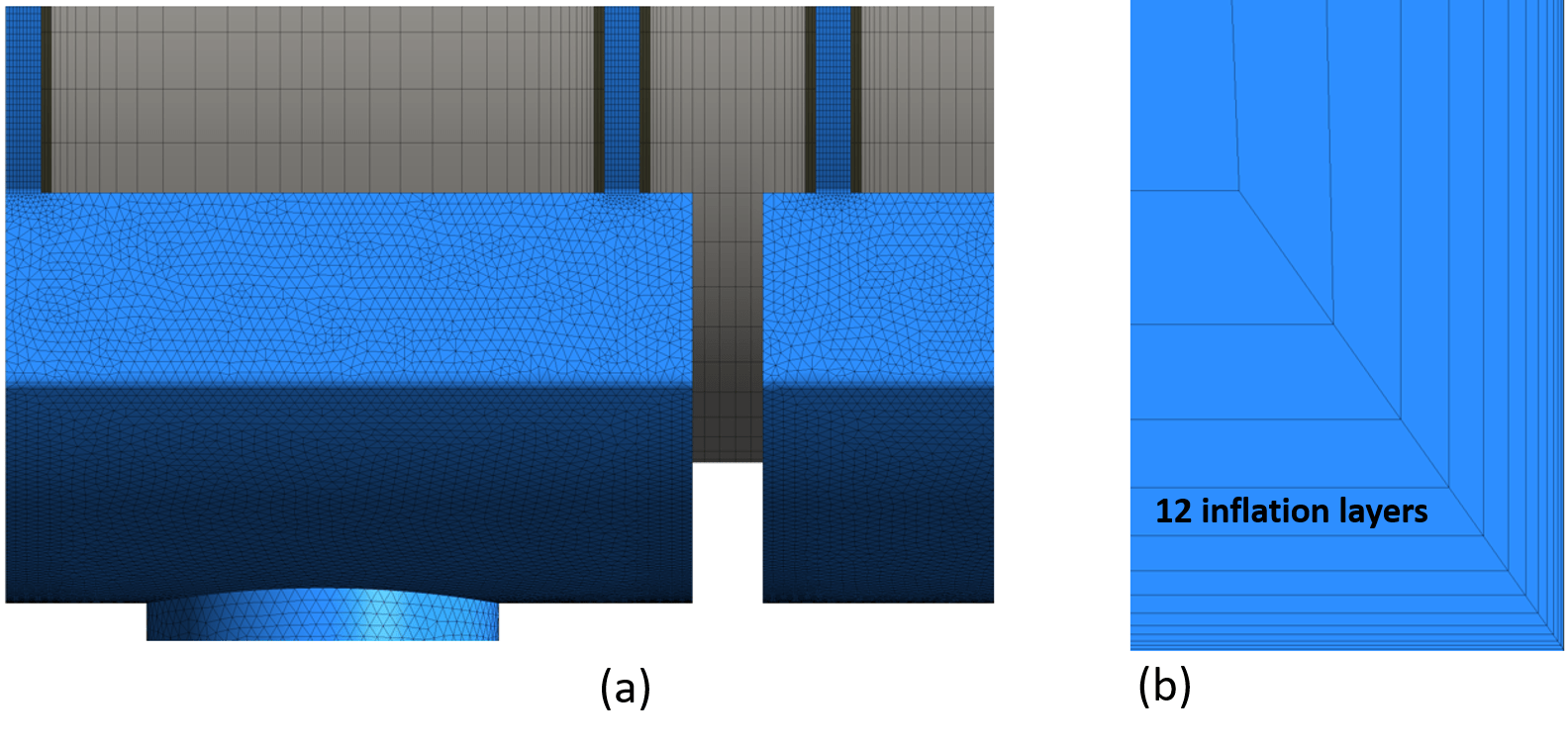}
\caption{\label{fig:TGT:CFD_Mesh2} Zoomed part of the computational grid: (a) of Figure~\ref{fig:TGT:CFD_Mesh} (rectangular dashed box) and (b) near the solid-liquid interface in the cooling channels.}
\end{figure}

Figure~\ref{fig:TGT:CFD_Mesh} illustrates the mesh used for the present model, that consists of both structured and unstructured elements. In all the target blocks and the cooling channels, hexahedral elements are used, which is imperative to improve the mesh quality (skewness, orthogonal quality etc.) and achieve a faster convergence of the simulation. Wedge elements are used in all the boundary layer region which is constituted by several inflation layers, as shown in Figure~\ref{fig:TGT:CFD_Mesh2}. The inflation layers are necessary in turbulent flow simulations near the wall region in order to resolve the flow in the boundary layer region.

\subsubsection{Steady-state results}

Steady state simulations were carried out in order to investigate the flow and heat transfer behaviour in the cooling domain and the target.

\begin{figure}[htbp]
\centering %
\includegraphics[width=0.8
\linewidth]{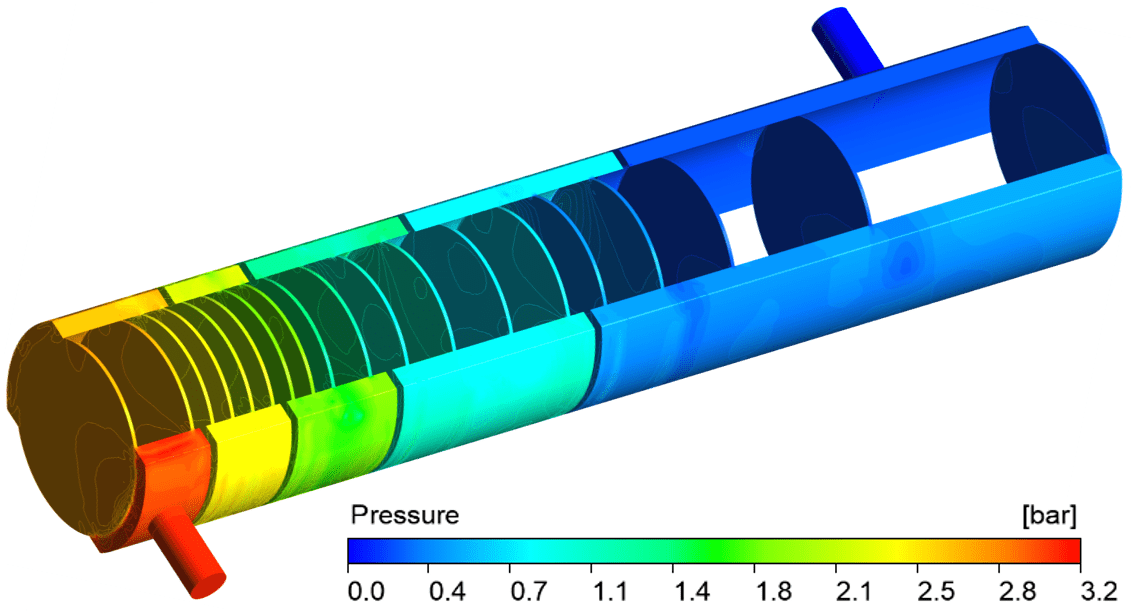}
\caption{\label{fig:TGT:CFD_pressure} Relative pressure variation map in the target cooling system, assuming an outlet pressure set to 0 bar.}
\end{figure}

Figure~\ref{fig:TGT:CFD_pressure} shows the contour for the static pressure distribution in the cooling system. As is illustrated in the figure, the pressure drop in the domain from inlet to outlet is around 3.2~bar, which means that the absolute pressure at the outlet would be $(22-3.2) = 18.8\:\text{bar}$. This pressure is acceptable since the boiling temperature at this pressure is over $200^{\circ} \text{C}$.

\begin{figure}[htbp]
\centering %
\includegraphics[width=1
\linewidth]{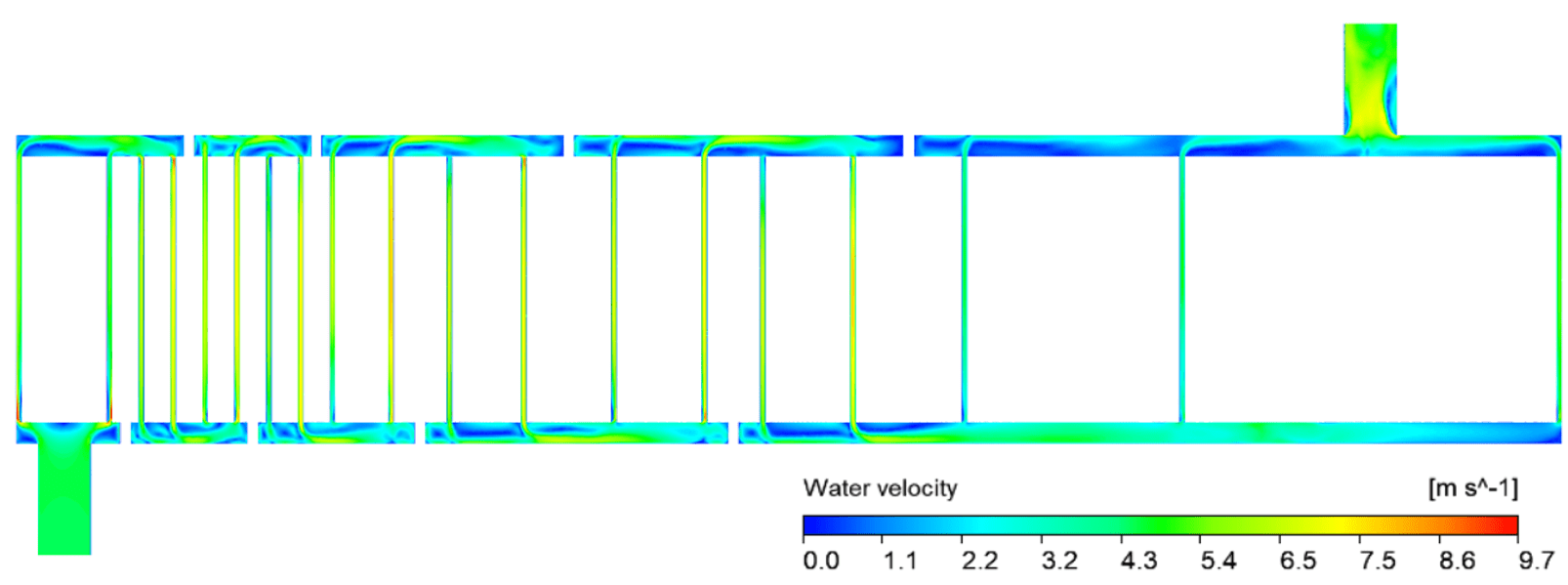}
\caption{\label{fig:TGT:CFD_vel} Velocity contour in the cooling system along the central longitudinal plane of the BDF target.}
\end{figure}

Figure~\ref{fig:TGT:CFD_vel} illustrates the velocity contour in the cooling domain along the longitudinal plane of the BDF target. It can be seen that the average velocity in all the cooling channels vary approximately between 4 to 6~m/s except for the last three channels where the average velocity is around 3 to 4 m/s. In the manifolds and in the outlet pipe, the blue region denotes the re-circulation zones due to flow separation.

\begin{figure}[htbp]
\centering %
\includegraphics[width=0.8
\linewidth]{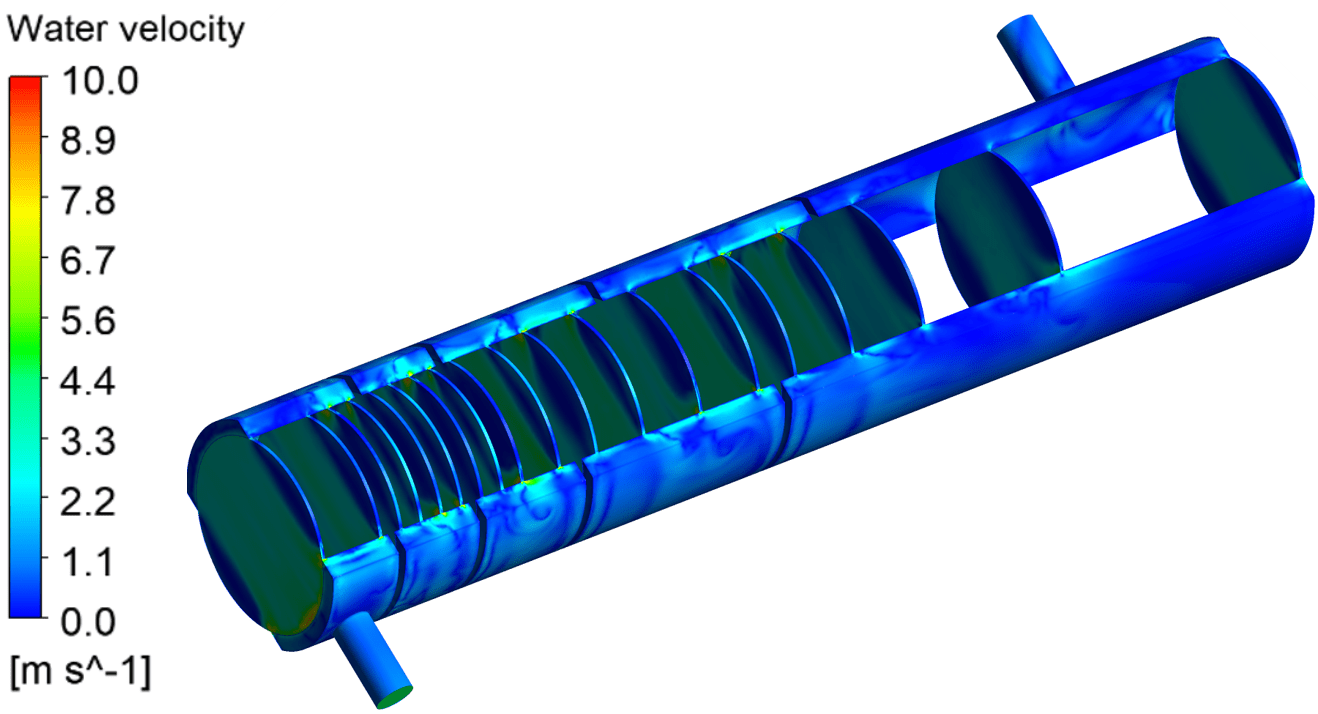}
\caption{\label{fig:TGT:CFD_vel3D} 3D velocity distribution in the BDF target cooling system. The re-circulation zones in the channels and manifolds can be appreciated in dark blue.}
\end{figure}

The re-circulation zone was also observed in the cooling channels because of their circular shape. Figure~\ref{fig:TGT:CFD_vel3D} illustrates the 3D velocity distribution in the cooling system. 

\begin{figure}[htbp]
\centering %
\includegraphics[width=1
\linewidth]{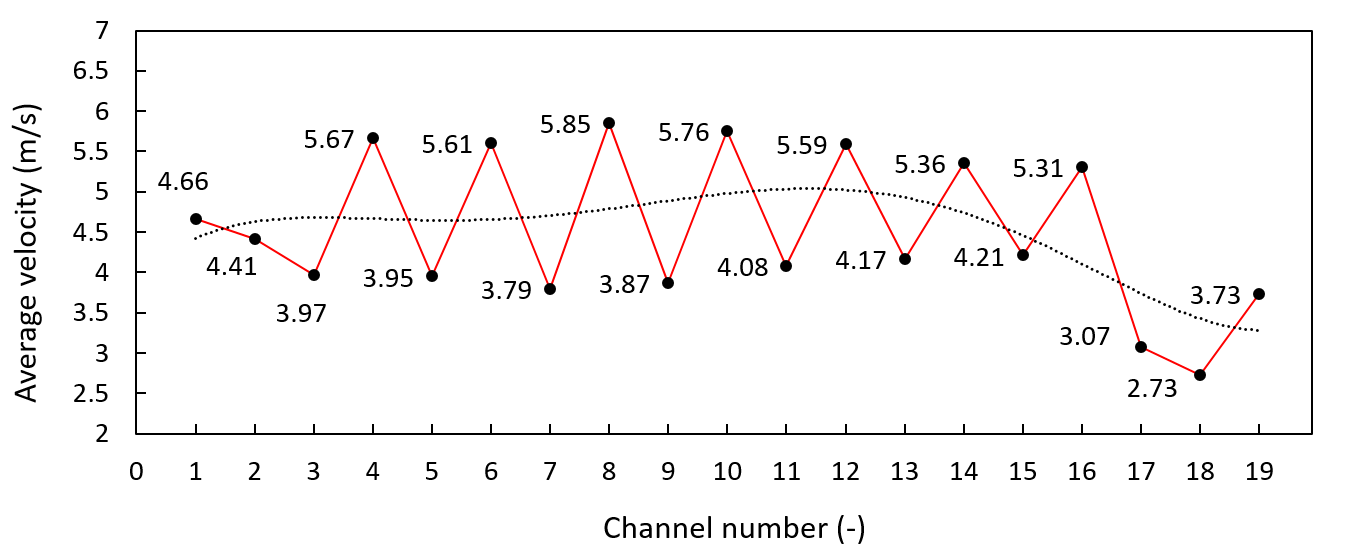}
\caption{\label{fig:TGT:CFD_plot_velo} Variation of average velocities in the target water channels. The trend-line of the plotted values is represented by a dotted line.}
\end{figure}

Figure~\ref{fig:TGT:CFD_plot_velo} displays the average velocity of the water in all the 19 channels. As can be appreciated from the figure, the average velocity in the channels is around 5 m/s for the first 16 channels and around 3 m/s for the last 3 channels. As the water stream enters the cooling system from the inlet, the flow is bifurcated into two streams that enter the channels with similar mass flow rate. Hence, the average velocity in the first two cooling channels is almost identical. However, due to the presence of blockers (which make the serpentine design possible), the mass flow rate in the even channels is higher than the one of the odd channels. This effect can be highlighted in Figure~\ref{fig:TGT:CFD_plot_velo} with the ''zig-zag'' pattern from channel 3 to 16.

\begin{figure}[htbp]
\centering %
\includegraphics[width=1
\linewidth]{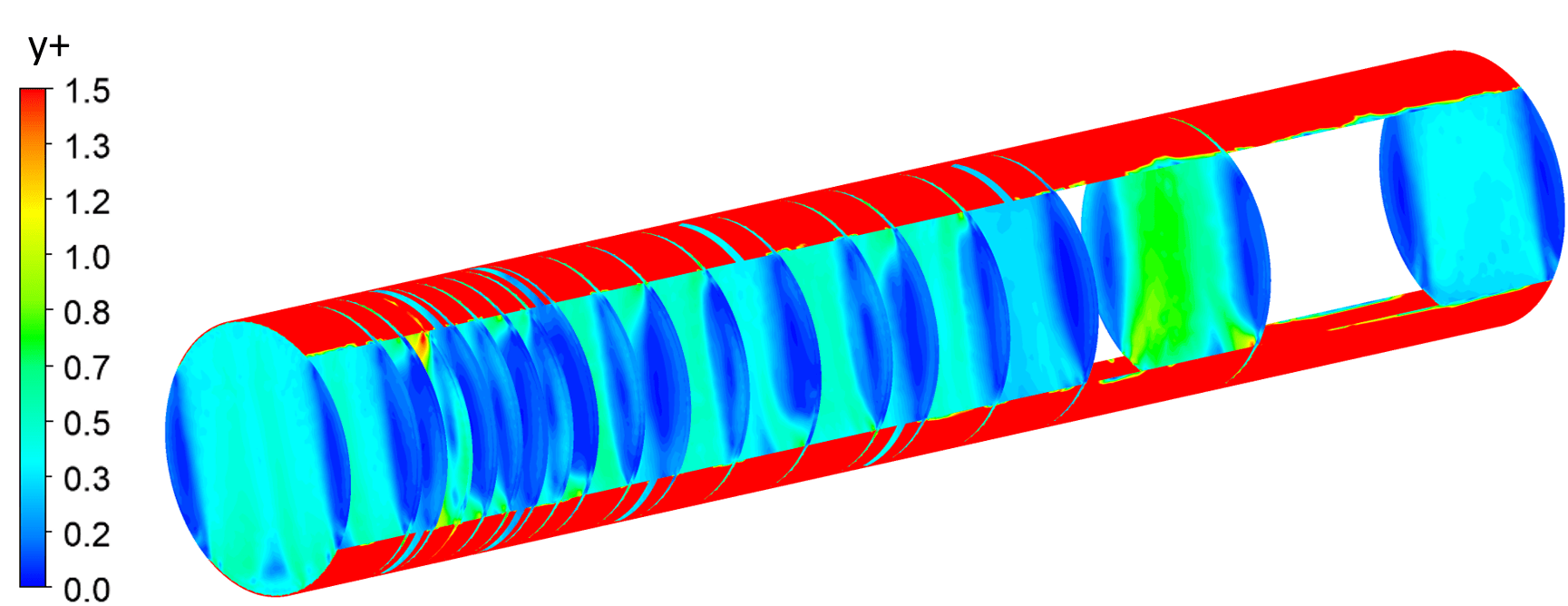}
\caption{\label{fig:TGT:CFD_yplus} $y^+$ contour on the BDF target cooling channels surface.}
\end{figure}

Figure~\ref{fig:TGT:CFD_yplus} shows the $y^+$ distribution plot near the walls of the channels. It can seen that the value of $y^+$ in all the walls of the channels is near 1 and hence it can be concluded that the boundary layer is sufficiently resolved in the cooling channels using the $k-w$ SST turbulence model. The top and bottom part corresponds to the manifolds and shows higher $y^+$ values (above 30).

\begin{figure}[htbp]
\centering %
\includegraphics[width=1
\linewidth]{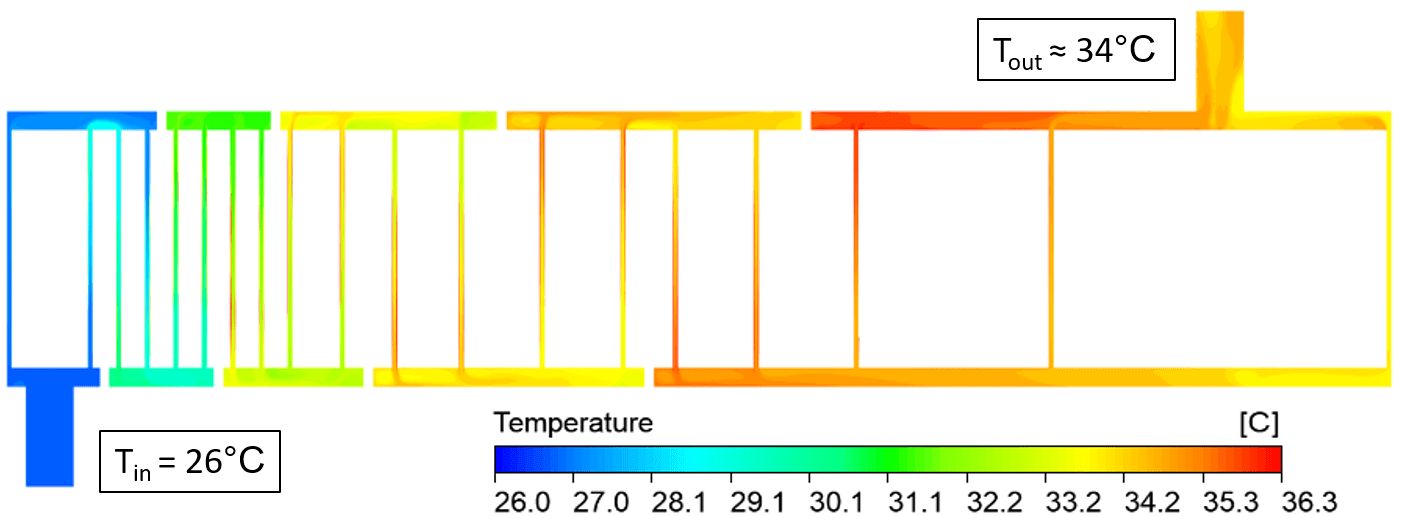}
\caption{\label{fig:TGT:CFD_2D_temp} Temperature contour in the cooling circuit along the YZ plane. Results from steady-state thermal CFD simulations.}
\end{figure}

The total temperature rise at the outlet obtained from the steady-state CFD simulations performed is reported to be around 8$^\circ$C. This behavior is illustrated in Figure~\ref{fig:TGT:CFD_2D_temp} which shows the temperature contour in the cooling circuit along the YZ plane. This is found to be in good comparison with the temperature rise calculated from the energy balance (Equation~\ref{eqn:TGT:CFD_tempraise}). However, in the transient case, the temperature at the outlet will fluctuate influenced by the beam impact on target, as will be shown in the following section.

\begin{figure}[htbp]
\centering %
\includegraphics[width=1
\linewidth]{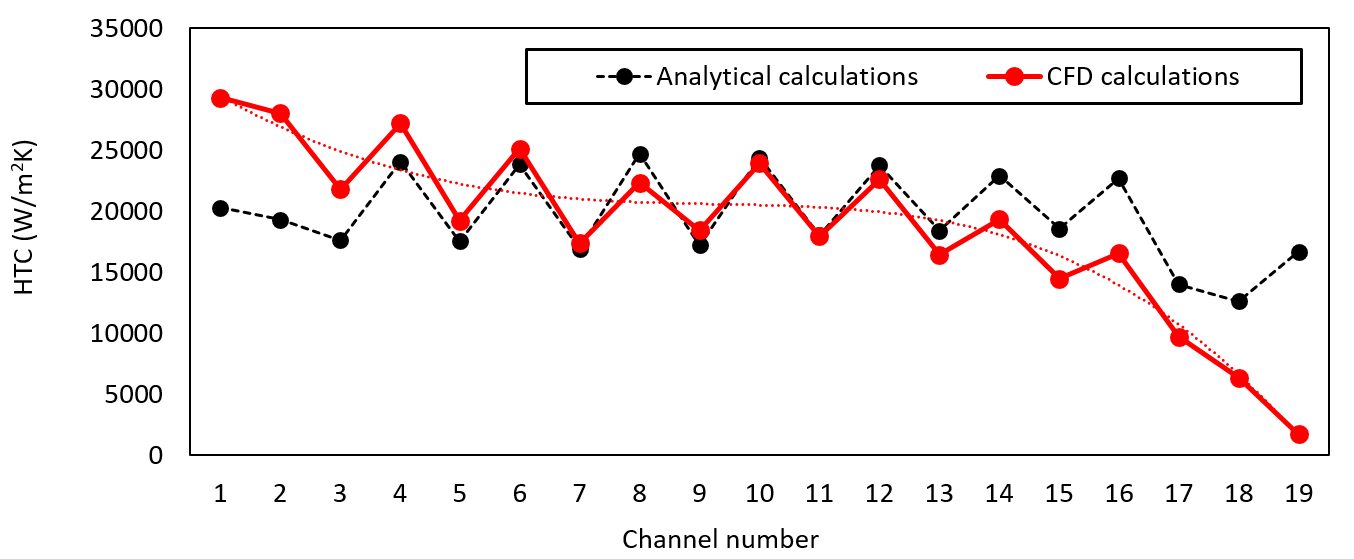}
\caption{\label{fig:TGT:CFD_HTC_plot} Average HTC in the cooling channels. Comparison between the values obtained from analytical calculations and numerical simulations ($k\omega$ SST turbulence model). Trend-line of the HTC average values obtained from CFD simulations shown as a dotted line.}
\end{figure}

Figure~\ref{fig:TGT:CFD_HTC_plot} displays the average heat transfer coefficient for all the 19 channels of the cooling domain. Analytical values obtained as in Equation~\ref{eqn:TGT:cfd15} using the average channel velocities shown in Figure~\ref{fig:TGT:CFD_plot_velo} are also plotted for comparison purposes. As shown by the trend-line of the simulated HTC average values, the HTC in the target cooling channels oscillates between 15000 and 25000~W/(m$^{2}$K) in the first 16 cooling channels where the velocity is around 5 m/s, as predicted by the analytical calculations. From this plot it can also be concluded that the HTC values obtained using the $k\omega$-SST turbulence model are in good comparison with the analytical solution.

\begin{figure}[htbp]
\centering %
\includegraphics[width=1
\linewidth]{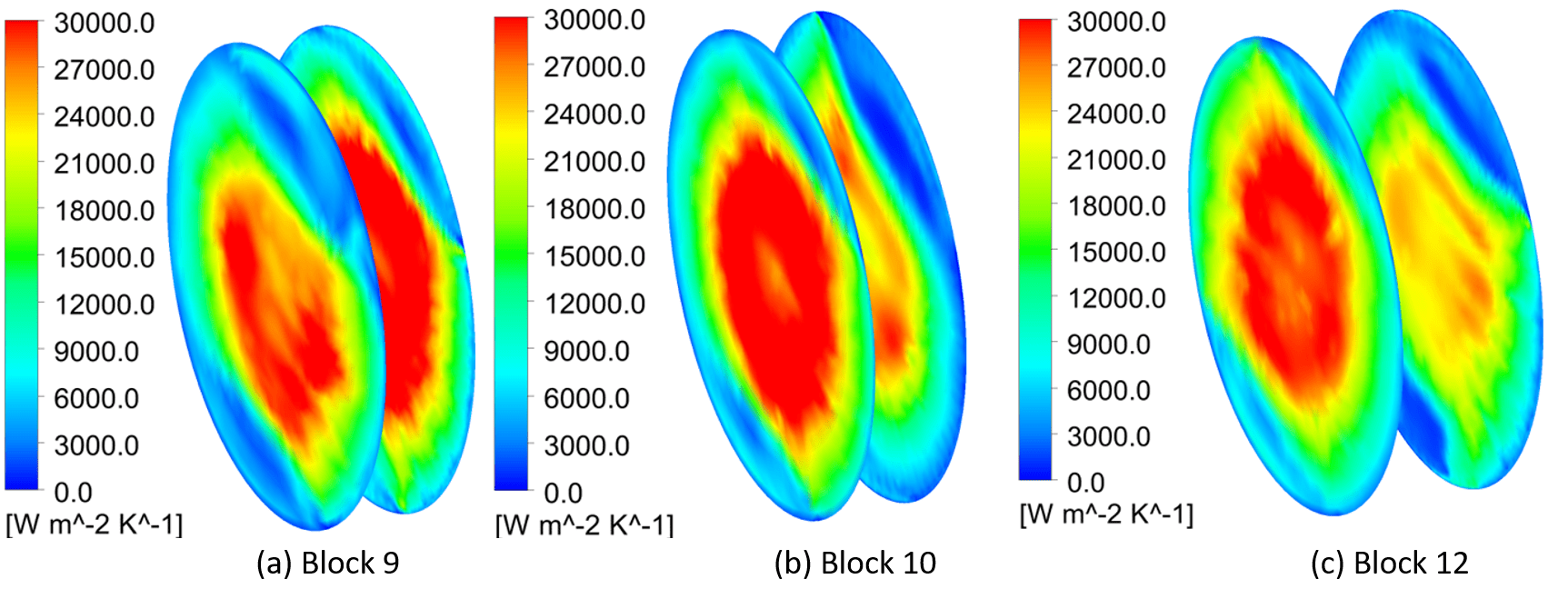}
\caption{\label{fig:TGT:CFD_HTC_map} Convective heat transfer coefficient distribution on the water-target interface of different blocks in the channel.}
\end{figure}

Figure~\ref{fig:TGT:CFD_HTC_map} illustrates the HTC distribution on the solid-liquid interface for the $9^{th}$, $10^{th}$, $11^{th}$ and $12^{th}$ blocks, respectively, where it can be seen that, locally, higher values of HTC (above 30000 W/(m$^{2}$K)) are reached on the surface of the blocks. It can be observed that for each block one of the two interfaces presents higher values of HTC, which is attributed to the higher velocities reached in that particular channel (as shown in Figure~\ref{fig:TGT:CFD_plot_velo}).

\begin{figure}[htbp]
\centering %
\includegraphics[width=1
\linewidth]{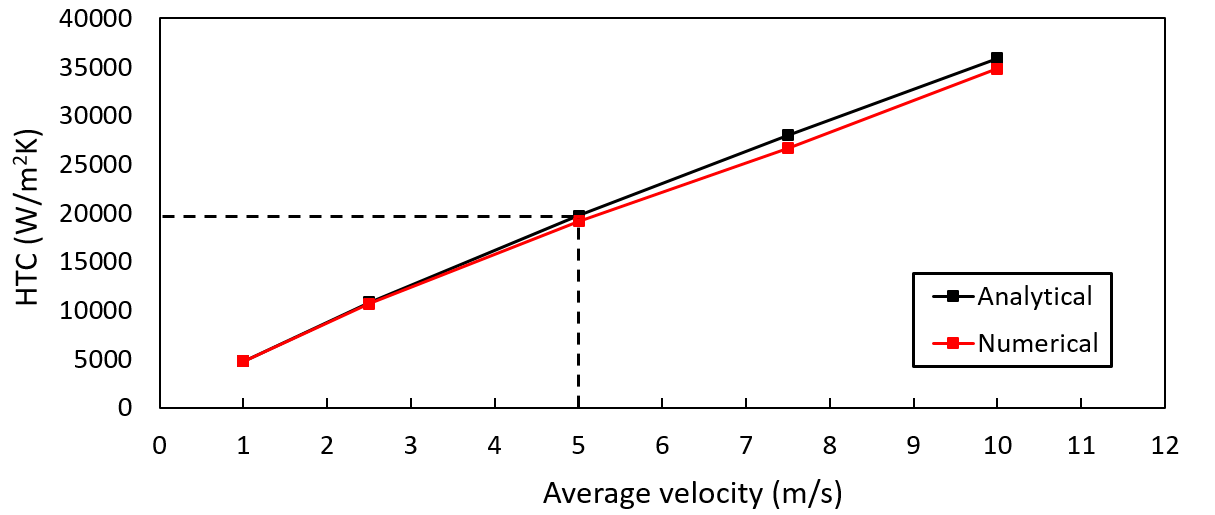}
\caption{\label{fig:TGT:CFD_HTC_comparison} Variation of convective heat transfer coefficient with channel velocity. Comparison between the analytical and CFD calculations results, showing the good agreement between the two.}
\end{figure}

The HTC, as explained earlier, is highly dependent on the flow velocity and contact area of the interface. A comparative study to investigate the effect of different channel velocities (1, 2.5, 5, 7.5 and 10 m/s) on the average HTC value in the cooling channels has been performed. The results are shown in Figure~\ref{fig:TGT:CFD_HTC_comparison}. A good comparison was found between the average value of HTC obtained numerically and analytically (see Table~\ref{tab:TGT:CFD_analytical}). From this figure it can be concluded that the heat transfer coefficient on the solid-liquid interface increases as the velocity of water is increased. On the other hand, high velocities in the channel can lead to erosion of the solid surface, as well as higher pressure drop in the circuit. A compromise between both effects is found in the selected value of 5 m/s, leading to a considerably high HTC (20000 W/(m$^{2}$K)) and reducing the risk of erosion on tantalum-tungsten.

\subsubsection{Transient results}

Transient simulations were performed for the model described in the previous section with identical beam size, computational grid and boundary conditions. In this case, the User-Defined Function responsible for importing the energy deposition map on the target blocks was modified such that the energy deposited by the primary beam was applied to all the target blocks for 1 second, and then removed for 6.2 seconds, thereby simulating the 7.2 seconds BDF/SHiP cycle. The transient simulations were carried out by providing the results of the steady-state simulations as initial condition.

\begin{figure}[htbp]
\centering %
\includegraphics[width=1
\linewidth]{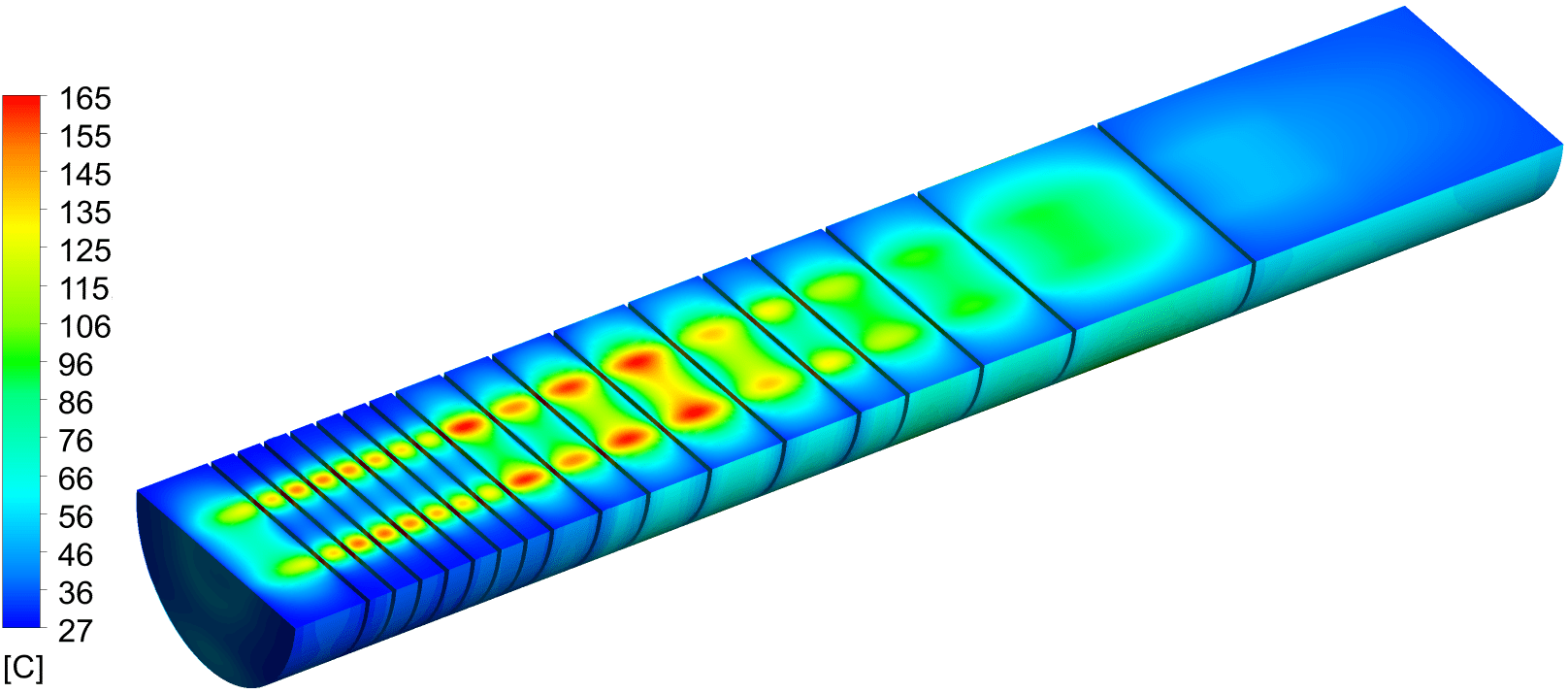}
\caption{\label{fig:TGT:CFD_3D_temp} Temperature contour in the target blocks at the end of the 1 second proton pulse.}
\end{figure}

\begin{figure}[htbp]
\centering %
\includegraphics[width=1
\linewidth]{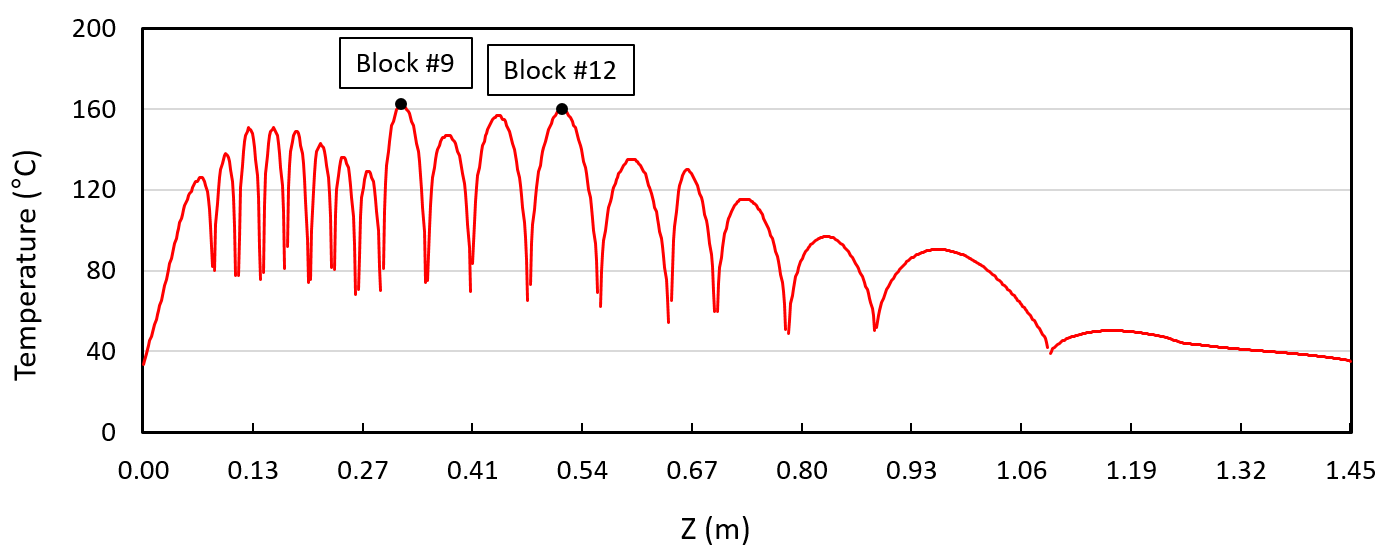}
\caption{\label{fig:TGT:CFD_temp_transient} Temperature variation in all the target blocks at the end of the 1 second pulse along the longitudinal axis at Y = 50 mm.}
\end{figure}

Figure~\ref{fig:TGT:CFD_3D_temp} illustrates the temperature distribution contour in a longitudinal cut at mid plane of all the target blocks at the peak of the pulse (1 second). The corresponding temperature profile in the different blocks following a line parallel to the Z-axis and at a distance of 50 mm from the axis (coincident with the beam impact location) is plotted in Figure~\ref{fig:TGT:CFD_temp_transient}. From this figure, it can be noticed that the maximum temperatures are found in blocks 9 and 12 with similar temperatures (around 160$^{\circ}\text{C}$) in the core of the block due to heat diffusion during the 1 second-long pulse. 

\begin{figure}[htbp]
\centering %
\includegraphics[width=1
\linewidth]{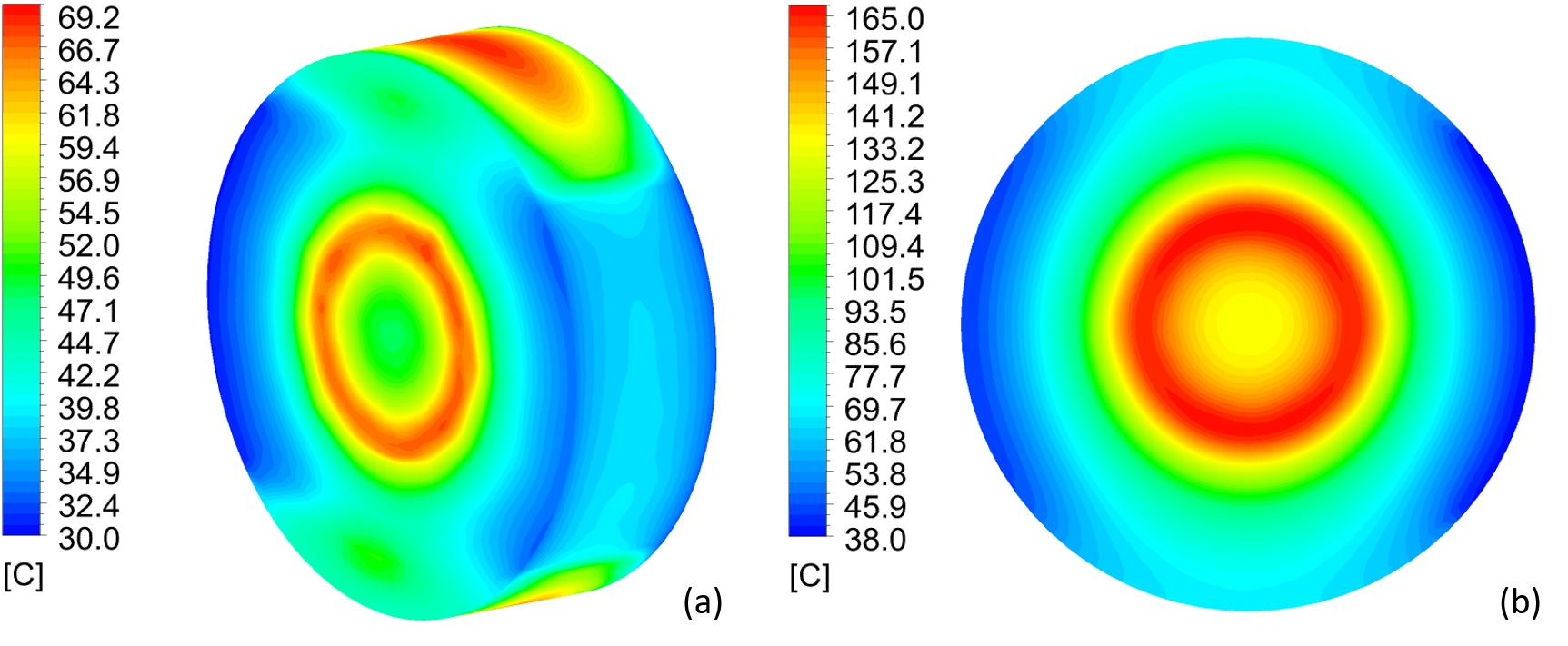}
\caption{\label{fig:TGT:CFD_block12}Temperature contour in the $12^{th}$ block at the peak of the pulse (1 second). (a) Side view, (b) Mid section}
\end{figure}

Figure~\ref{fig:TGT:CFD_block12} shows the temperature distribution in block number 12 at the end of the proton pulse, the maximum temperature in the TZM core of the block being 165$^{\circ}$C. It can be observed that the circular surfaces of the block which are not in direct contact with water (top and bottom surface), are nearly $30^{\circ}$C warmer than the side surfaces which are cooled by water through the manifold.

\begin{figure}[htbp]
\centering %
\includegraphics[width=0.8
\linewidth]{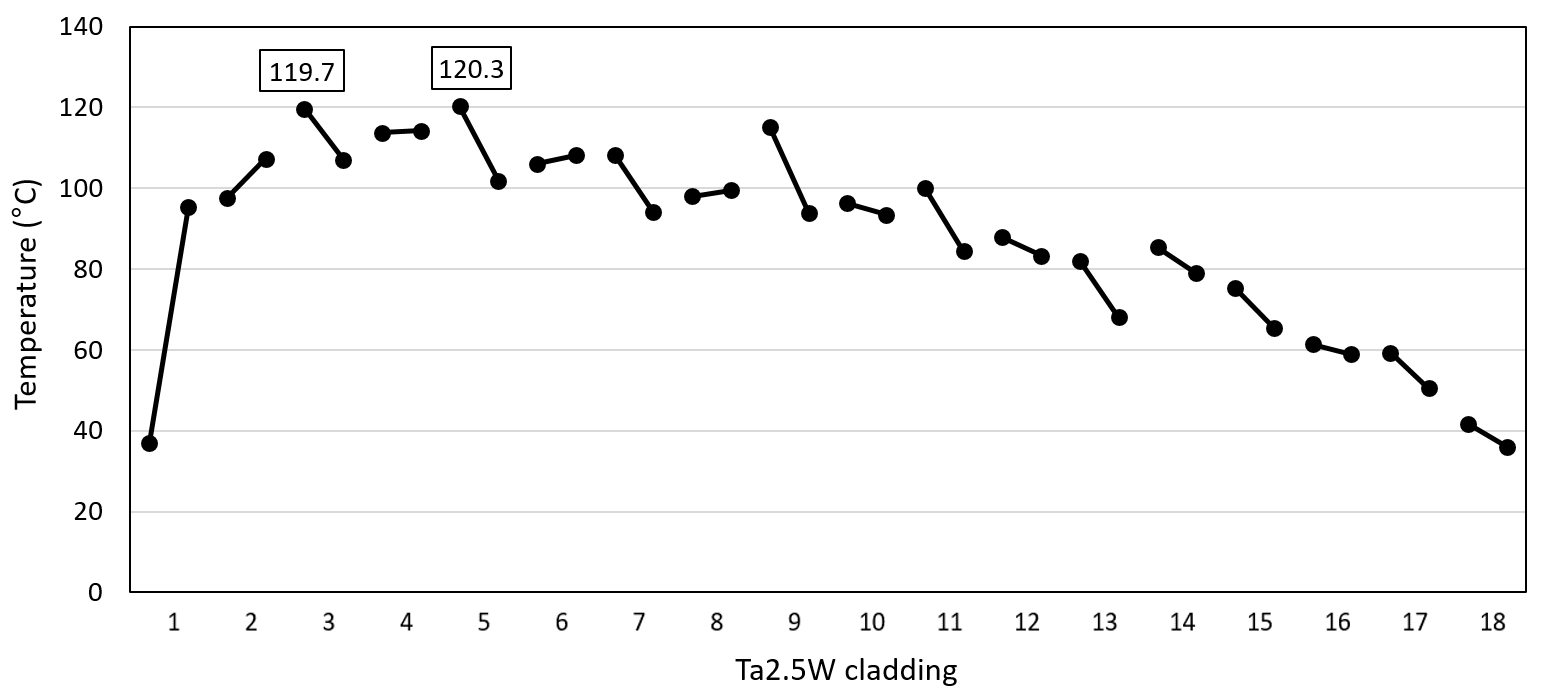}
\caption{\label{fig:TGT:CFD_Ta_temp}Maximum temperature in the left and right Ta2.5W layers of the target blocks.}
\end{figure}

As presented in Section~\ref{Sec:TGT:Simus}, the Ta2.5W cladding is one of the most critical aspect of the BDF target design, due to the high intensity nature of the SPS pulse. Hence, it is important to investigate the maximum temperature reached in the Ta2.5W layer in order to ensure that mechanical failure does not occur due to excessive thermal stresses. Figure~\ref{fig:TGT:CFD_Ta_temp} shows the maximum temperature in the left and right layers of the Ta2.5W cladding of each block at the peak of the pulse. The maximum temperature is obtained in the left side layer of blocks 3 and 5 with a value of around $120^{\circ}$C.

It is worth mentioning that the maximum temperatures obtained in the core and cladding materials via CFD calculations differ from the results of the FEM simulations shown in Section~\ref{Sec:TGT:Simus}. The maximum temperature in the Ta2.5W cladding is not obtained in block number 4 as in the FEM calculations, but in blocks 3 and 5 (see Figure~\ref{fig:TGT:CFD_Ta_temp}).

\begin{figure}[htbp]
\centering %
\includegraphics[width=1
\linewidth]{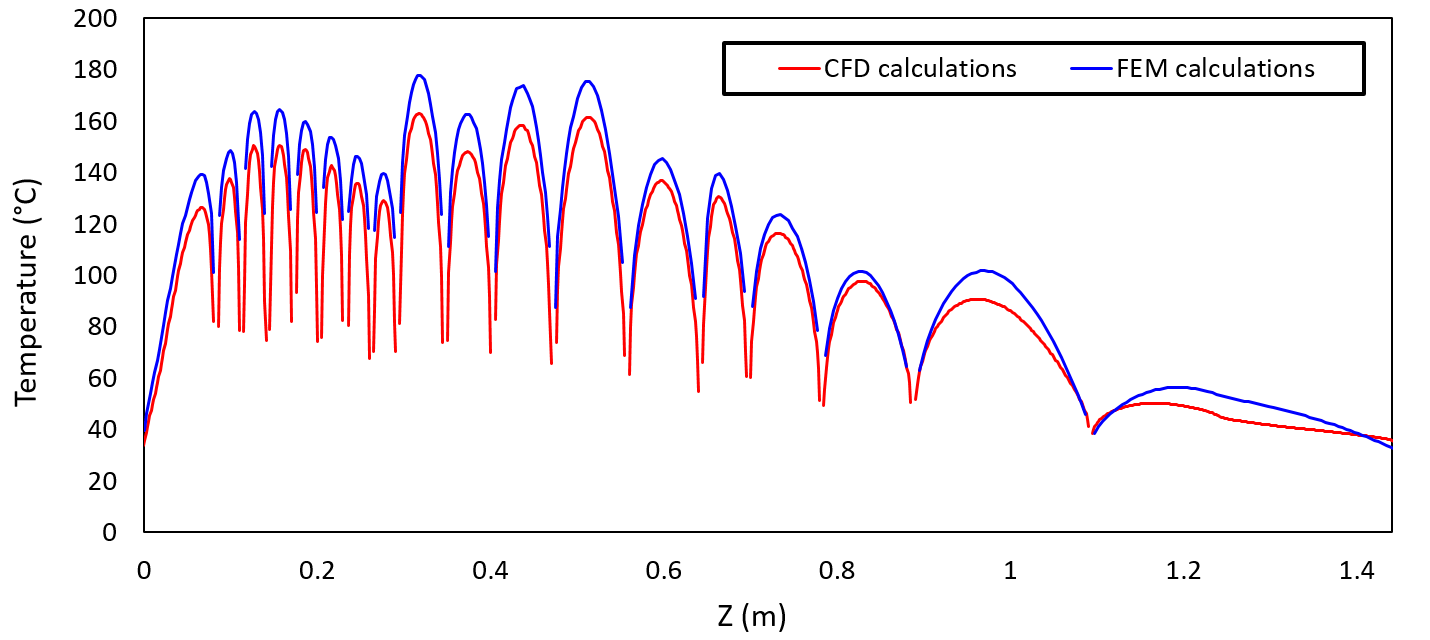}
\caption{\label{fig:TGT:CFD_FEM_comparison}Comparison between FEM and CFD calculation results for the temperature profile along the longitudinal axis 50 mm far from the target center. Results obtained for t = 1 second (end of pulse).}
\end{figure}

Figure~\ref{fig:TGT:CFD_FEM_comparison} displays a comparison of the temperature longitudinal profile in the target blocks at 50 mm from the target axis for both cases. It can be seen that the maximum temperatures obtained in the core of the target blocks is around 20$^{\circ}$C lower in the CFD simulations, and the surface temperature of the blocks is also substantially reduced. This difference can be explained by the fact that the HTC used as boundary condition in the FEM simulations was constant with a value of 20000 W/(m$^{2}$K), while for the CFD calculations the HTC is iteratively calculated and can reach values over 30'000 W/(m$^{2}$K), as shown in Figure~\ref{fig:TGT:CFD_HTC_map}.

Therefore, the stresses calculated in Section~\ref{Sec:TGT:Simus:struct} by importing the temperatures obtained via FEM calculations are conservative, and the safety margin with respect to the material limits is expected to be higher. For an accurate modelling of the state of stresses in the BDF target during operation, the temperature map obtained through CFD calculations shall be imported into the FEM structural code. This activity remains as part of a future study, even if the results of the present calculations are conservative and nevertheless appears to be well within the target material limits.

\begin{figure}[htbp]
\centering %
\includegraphics[width=1
\linewidth]{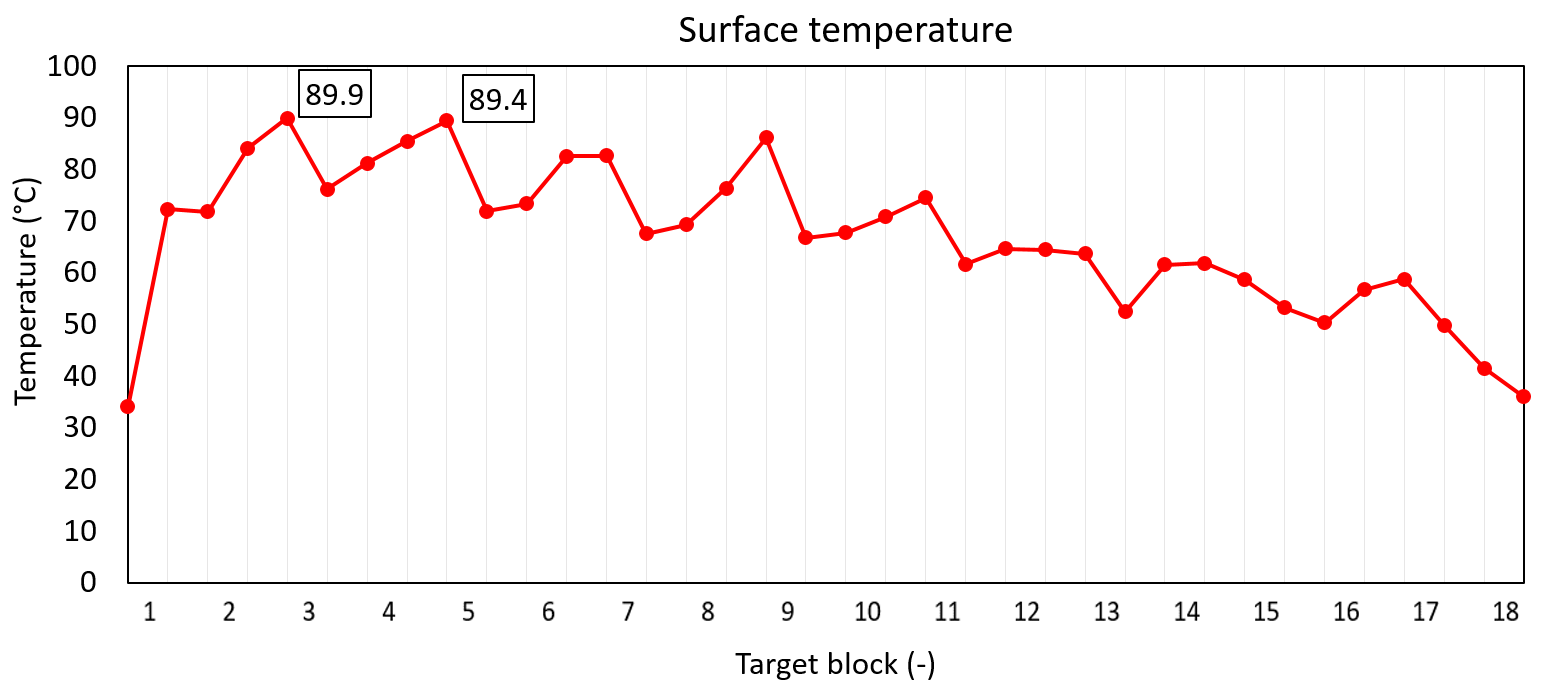}
\caption{\label{fig:TGT:CFD_surface_temp} Maximum temperature in the left and right surfaces of the target blocks. Maximum surface temperature $\approx$ 90$^{\circ}$C.}
\end{figure}

Figure~\ref{fig:TGT:CFD_surface_temp} presents the maximum temperature found on the surface of the target blocks, displaying the maximum value for the left and right fluid-solid interface in every block. It can be observed that the maximum surface temperature is around 90$^{\circ}$C, reached in the left interface of target blocks 3 and 5. This temperature is more than 2 times lower than the boiling temperature of water at 22 bar which is 212$^{\circ}$C. It can be concluded from the results of the CFD calculations that water boiling is not likely to occur during normal operation of the BDF target.

\FloatBarrier

\section{Mechanical design of the BDF target and target assembly}
\label{Sec:TGT:MechDesign}

\subsection{Target assembly design}

\begin{figure}[htbp]
\centering %
\includegraphics[width=1\linewidth]{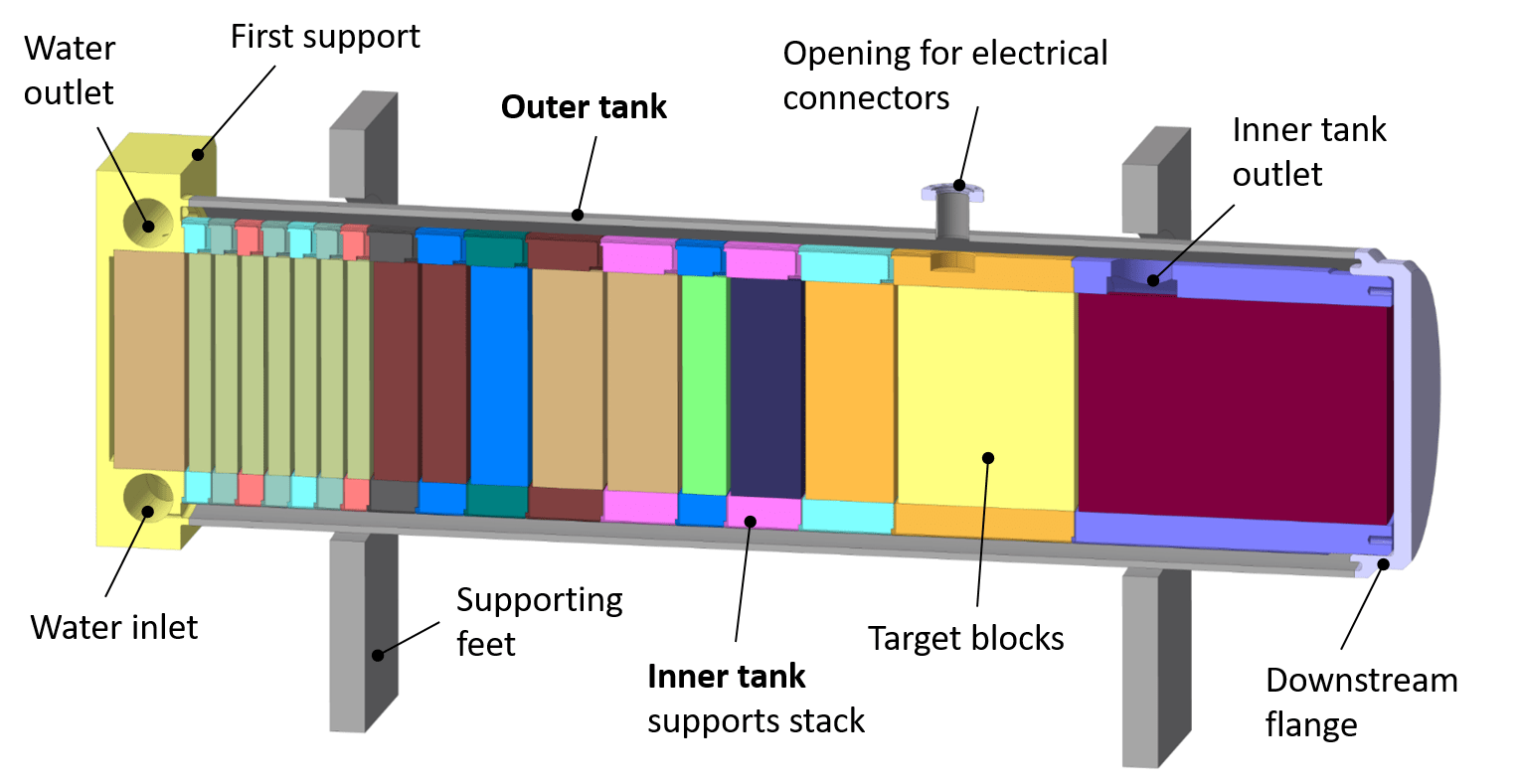}
\caption{\label{fig:TGT:assembly_cut} Longitudinal section of the BDF target inner and outer tank structure. The target core blocks supported by the inner tank can be seen, as well as many functional elements of the cooling circuit and the supporting structure of the target itself.}
\end{figure}

\begin{figure}[htbp]
\centering %
\includegraphics[width=0.8\linewidth]{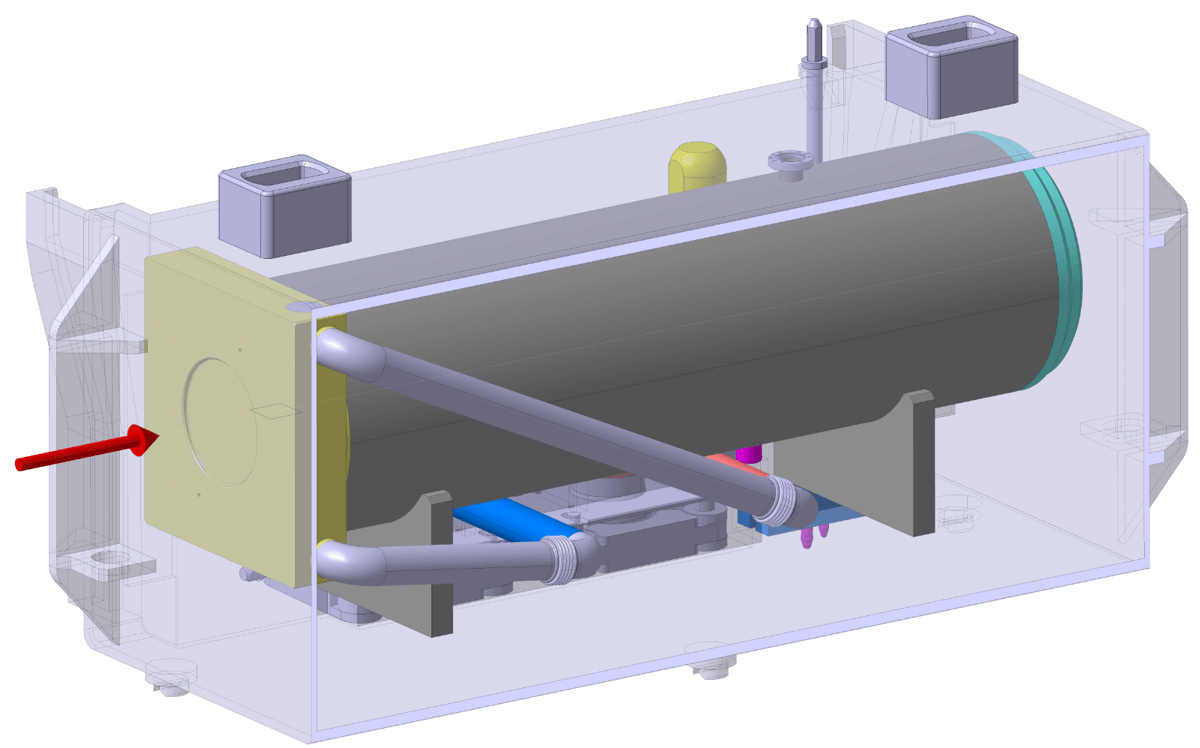}
\caption{\label{fig:TGT:He_box} BDF target full assembly: view of the Helium container, outer tank, upstream and downstream flanges and inlet and outlet pipes.}
\end{figure}

The BDF target assembly consists essentially of 4 main parts: the target core blocks, an inner tank that supports the target blocks, a leak-tight outer tank that encloses the inner tank, and a helium container that holds the whole assembly. A section of the inner and outer tank supporting the target blocks can be seen in Figure~\ref{fig:TGT:assembly_cut}. The helium container enclosing the full target assembly, which is also the interface for the handling system in the Target Complex design, is shown in Figure~\ref{fig:TGT:He_box}.

\subsubsection{Target blocks production}

In order to assess the feasibility of the target assembly, a preliminary mechanical design of the target as well as of its assembly procedure has been executed. The BDF target core blocks consist of two different parts:
\begin{itemize}
    \item A TZM or W cylinder with different length according to the block position in the target core. Preliminary investigations have shown that all the TZM cylinders can be manufactured via multi-axial forging, while not all the pure tungsten cylinders can be obtained by this method~\cite{Production_plansee}. The length of some of the tungsten cylinders, that reach up to 350 mm long, is a limiting factor to apply both longitudinal and radial forging. For that reason, it is foreseen to produce the W cylinders via sintering and subsequent HIPing, this process leading to an isotropic material structure and an acceptable density of around 97\%. 
    \item The Ta2.5W cladding, which encloses the TZM or W cylinder, and consists of a tube with variable length and two disks. The Ta2.5W tubes can be rolled, and must be seamless as a this is a requirement for the HIP process that will be described later on. The Ta2.5W disks can be obtained via forging.
\end{itemize}

For the production of the target blocks, the TZM or W cylinder is inserted into the Ta2.5W tube and closed above and below by the two Ta2.5W disks. The disks and the cylinder have to be precisely machined to ensure a gap of around 0.1 mm between the disk and cylinder diameters and the inner diameter of the Ta2.5W tube. This gap should be sufficient for the disks and cylinder insertion into the Ta2.5W tube, and tight enough to achieve the diffusion bonding between the materials during the HIP process. 

\begin{figure}[htbp]
\centering %
\includegraphics[width=0.6\linewidth]{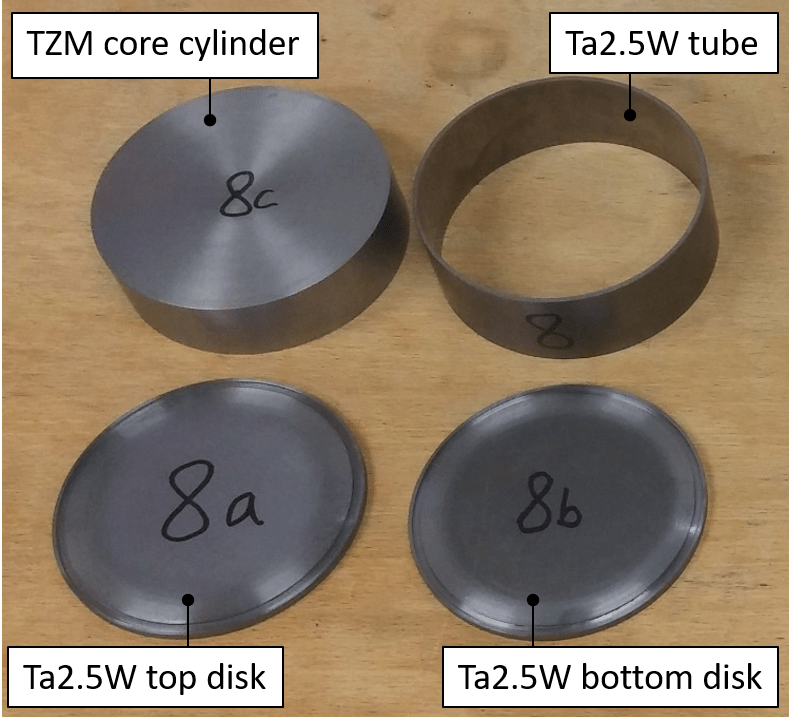}
\caption{\label{fig:TGT:Block_exploded} Refractory metals parts required for the production of a 80 mm diameter, 25 mm thick Ta2.5W-clad TZM target block for the BDF target prototype test in the North Area of CERN (see Section~\ref{Sec:TGT:Proto}).}
\end{figure}

As an example, Figure~\ref{fig:TGT:Block_exploded} shows the different parts necessary for the target blocks production, in this case for the production of a reduced scale target block (25 mm thickness, 80 mm diameter) for the BDF target prototype tested in the North Area of CERN~\ref{Sec:TGT:Proto}.

Before the HIP run, the top and bottom Ta2.5W disks are electron-beam (EB) welded to the Ta2.5W tube, and the whole assembly is tested under vacuum to guarantee the capsules leak-tightness. After that, the capsules are covered with a Zirconium foil to prevent oxidation. Then, every assembled target block undergoes a HIP cycle, reaching a temperature of $1200\,^{\circ}\mathrm{C}$ and a pressure of 150 MPa for 2 hours. Figure~\ref{fig:TGT:HIPcycle} illustrates the HIP cycle applied for the production of the BDF target blocks. 

Once the HIP cycle has been completed, the target blocks have to be machined to ensure that the design dimensions are respected. The HIP process carried out for the BDF target blocks production is crucial to ensure the mechanical and chemical bonding between the cladding and core materials. A separate publication will detail the studies carried out to investigate different HIP cycles and evaluate the bonding quality achieved. Further studies are currently ongoing to explore other cladding cycles that could be similarly adopted to improve the cladding process.

\subsubsection{Target vessel inner stainless steel tank}

\begin{figure}[htbp]
\centering %
\includegraphics[width=1\linewidth]{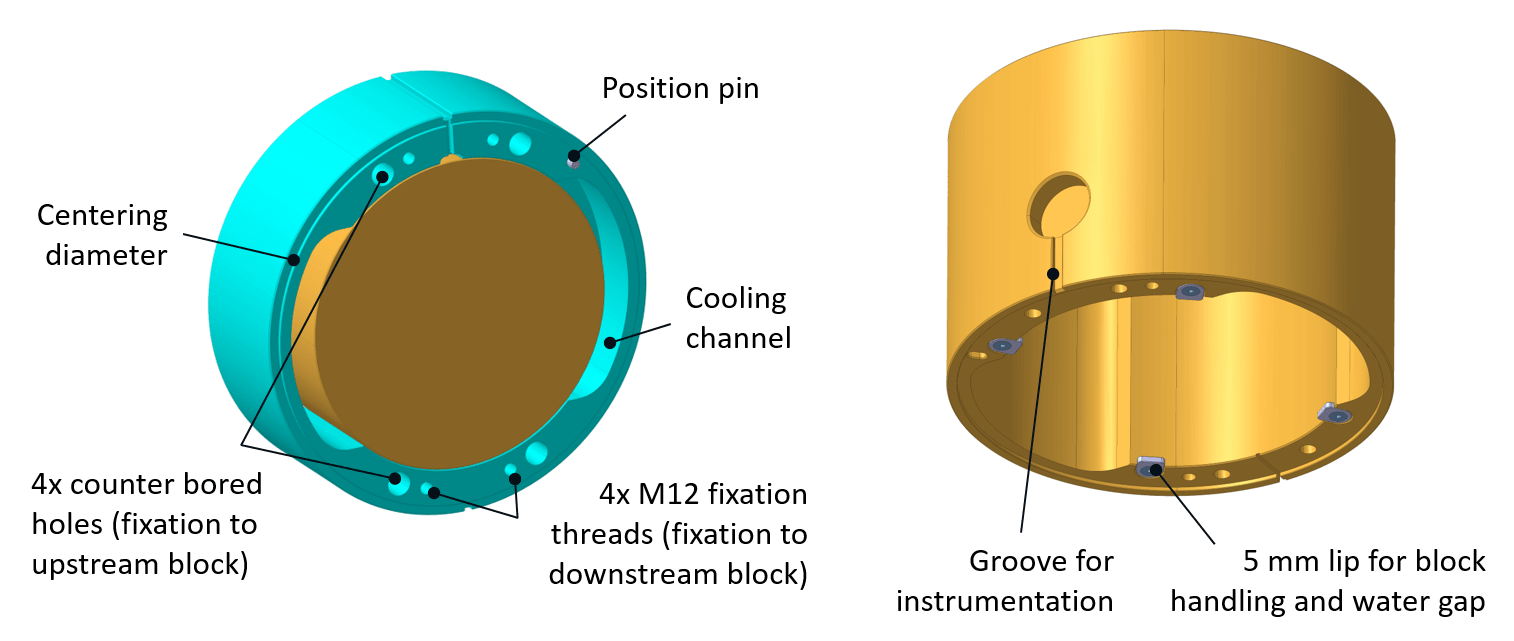}
\caption{\label{fig:TGT:support_description} Description of the target inner tank supports. Each support integrates different elements that permit the assembly with the previous and following supports, as well as the water circulation around the target blocks.}
\end{figure}

\begin{figure}[htbp]
\centering %
\includegraphics[width=0.8\linewidth]{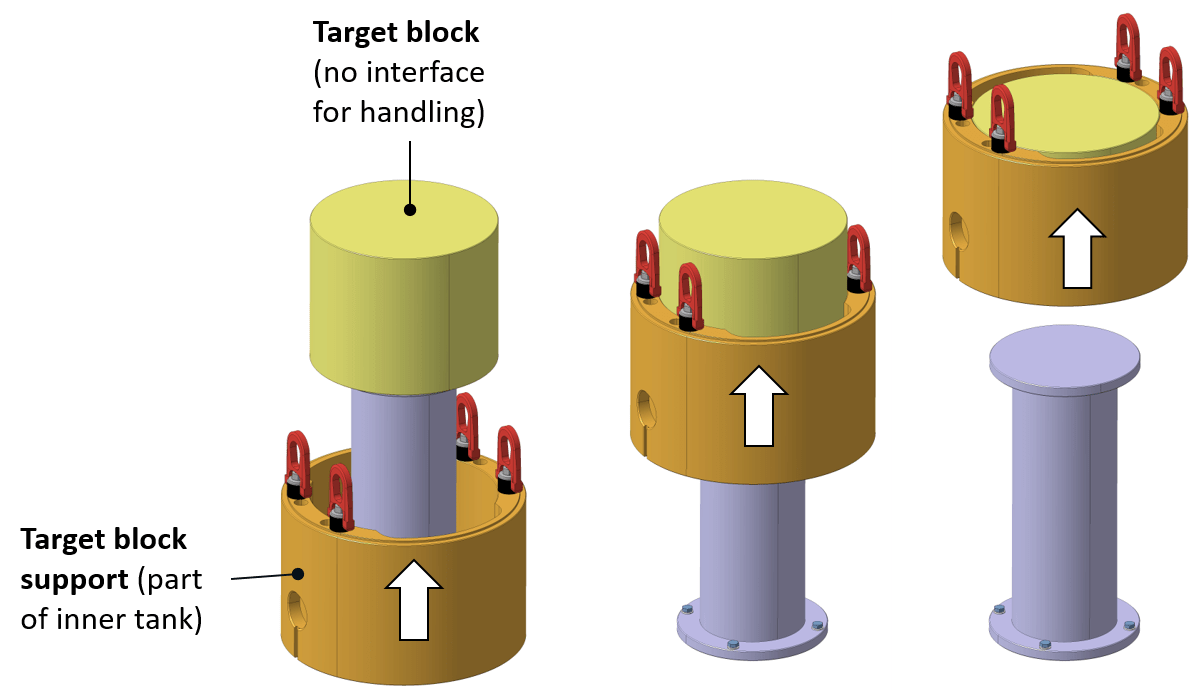}
\caption{\label{fig:TGT:handling_blocks} Handling process of the target blocks: each target block is lifted by the corresponding support, which includes a dedicated interface for the target blocks handling operation.}
\end{figure}

\begin{figure}[htbp]
\centering %
\includegraphics[width=0.8\linewidth]{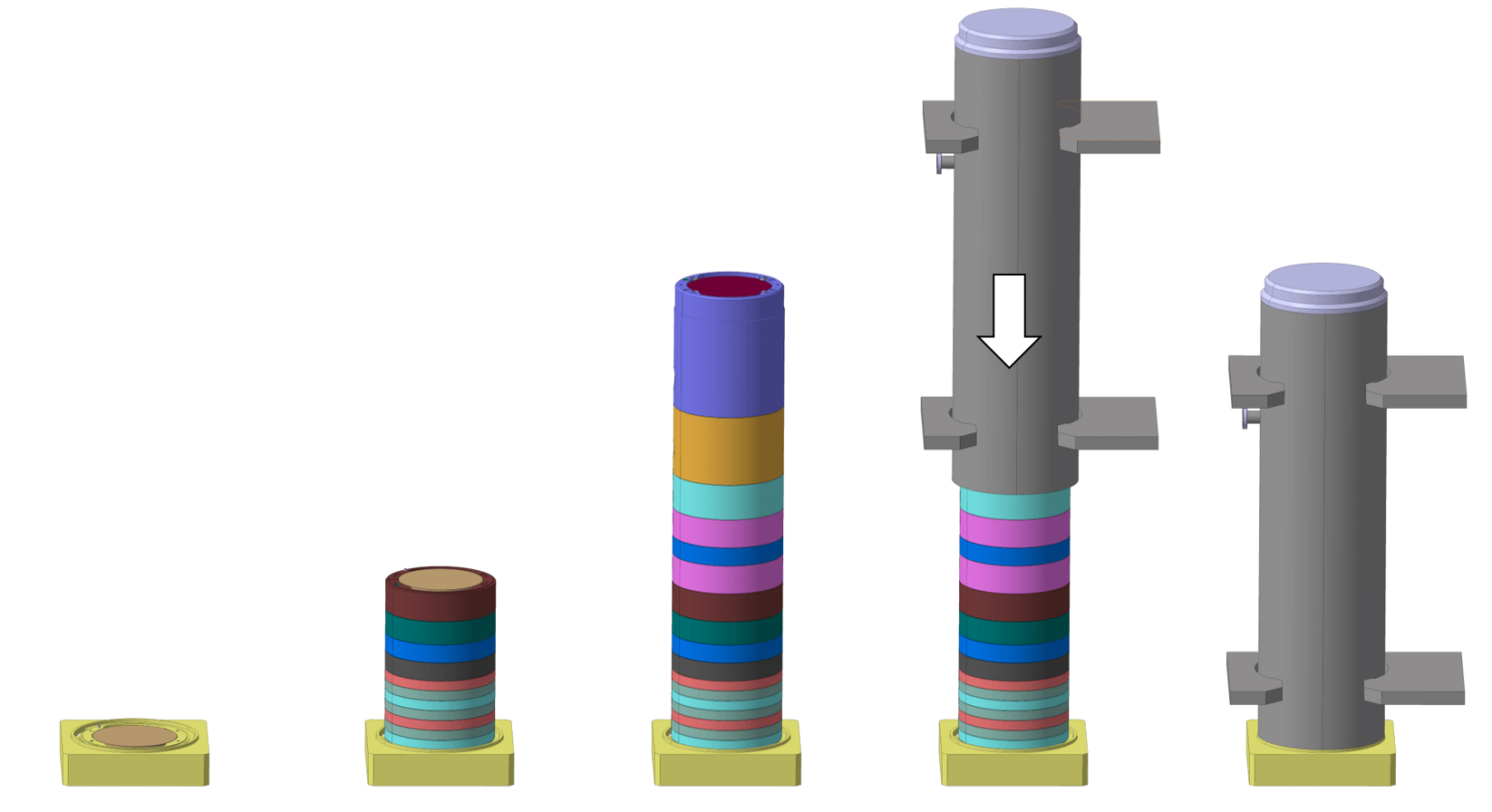}
\caption{\label{fig:TGT:assembly_blocks} Target inner tank assembly process and outer tank fitting: (i) each support is mounted on top of the previous one in vertical position; (ii) once all the supports have been assembled, the outer tank is installed and welded to the first support, enclosing the whole inner tank ; (iii) the whole assembly is placed in horizontal position (Figure~\ref{fig:TGT:assembly_cut}).}
\end{figure}

The inner tank is composed of several "supports" that are assembled together, made out of SS316LN (see Table~\ref{Tab:TC:Material_properties} for more info). Each target block has its own support, that is acting at the same time as handling tool. The target blocks can be very heavy, up to 300 kg for the heaviest tungsten cylinder, making necessary a handling mechanism for their assembly. Each support holds the corresponding target block in vertical position during the assembly, and all the supports are stacked on top of each other starting from the first support. Figure~\ref{fig:TGT:support_description} shows a description of the supports that make up the inner tank, and Figures~\ref{fig:TGT:handling_blocks} and \ref{fig:TGT:assembly_blocks} depict the handling and assembly process of the inner tank. This is a preliminary assessment of the constructability of the assembly, which will be further developed during the Technical Design (TDR) phase.

The supports are progressively screwed to each other, making one whole supporting structure. Once the assembly is completed, the target blocks supporting structure (hereafter, inner tank) is placed in horizontal position for subsequent operation. The inner tank structure is held by the first support in one side and by the outer tank downstream flange in the other side. The structure has been designed to safely withstand the weight of all the target blocks, with an acceptable vertical deformation in the center of around 0.1 mm, as will be shown in Section~\ref{Sec:TGT:MechDesign:struct}. Figure~\ref{fig:TGT:assembly_cut} presents a longitudinal cross-section of the inner tank structure in horizontal position, the target core blocks can be seen enclosed by the different supports, as well as the outer tank containing the whole assembly.

Another function of the inner tank is to enclose the target cooling system. The different supports include dedicated grooves in order to provide the foreseen circulation path for the water cooling, compatible with the cooling system design presented in Section~\ref{Sec:TGT:CoolingCFD}. The 5 mm gap between the different blocks necessary for the water channels is ensured by a 5 mm "lip" added to the supports, that also allows the blocks to be held in vertical position for the handling process (Figure~\ref{fig:TGT:handling_blocks}). The inlet of the target cooling system is located in the first support, and the outlet in the last one. The inlet is placed at the bottom of the first support, making it the lowest point of the whole cooling circuit. This arrangement grants a more effective draining of the water in case it would be necessary to empty the circuit, avoiding water stagnation points. Finally, the inner tank includes also grooves for the instrumentation cabling, since it is foreseen to monitor the temperature and strain of several blocks during operation.

\subsubsection{Target vessel outer stainless steel tank}

The outer tank - also made out of SS316LN (see Table~\ref{Tab:TC:Material_properties} for more info) - is responsible for providing leak-tightness to the target assembly and structural stability to the equipment. The cooling circuit is enclosed by the inner tank, but the inner tank structure is - by design - not sealed against water leaks. Given the high number of connections between the supports that constitute the inner tank, and the complexity of the water circulation path, an external tank has been chosen as the most reliable barrier to prevent any leaks from the cooling system. The outer tank is welded to the first support in one side, and to the downstream flange in the other side (see Figure~\ref{fig:TGT:assembly_cut}). Regarding the assembly procedure, once all the inner tank supports are stacked and screwed one on top of each other, the outer tank is inserted and welded to the first inner tank support. The fact of having only two welds adds simplicity and robustness to the target manufacturing and assembly process, while ensuring a good reliability of the system. 

\begin{figure}[htbp]
\centering %
\includegraphics[width=1\linewidth]{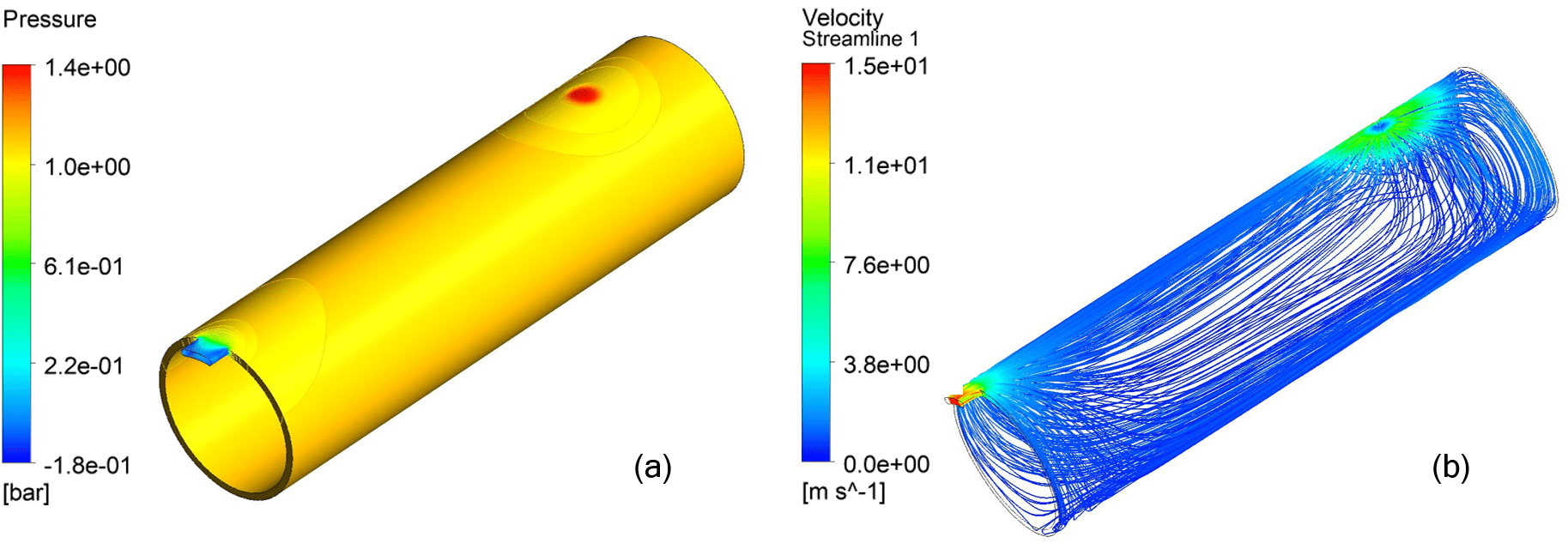}
\caption{\label{fig:TGT:CFD_outer} The figure shows the pressure (a) and velocity (b) distribution in the external cooling circuit obtained via CFD calculations. Maximum pressure drop $\approx$~1.5 bar, average velocity $\approx$ 1 m/s.}
\end{figure}

In order to avoid any stagnation of water between the inner and outer tank, a sufficiently large gap is foreseen to force the water circulation in the volume between both tanks, creating an "external" cooling loop. The outer tank diameter has been optimized taking into account the pressure drop, that should be minimized, and the water velocity, that should be high enough to have continuous water circulation. CFD calculations have been carried out, concluding that the optimal inner diameter of the external tank should be of the order of 360 mm, leading to a pressure drop of 1.5 bar in the volume enclosed between the two tanks and an average water velocity of 1 m/s (see Figure~\ref{fig:TGT:CFD_outer}). The outer tank is laying on two rectangular feet and is 8 mm thick, it has been designed to withstand an internal pressure of 31 bar (22 bar with the safety factor of 1.43), an external pressure of 1 bar, and to support the whole weight of the inner tank and the blocks. 

\begin{figure}[htbp]
\centering %
\includegraphics[width=0.9\linewidth]{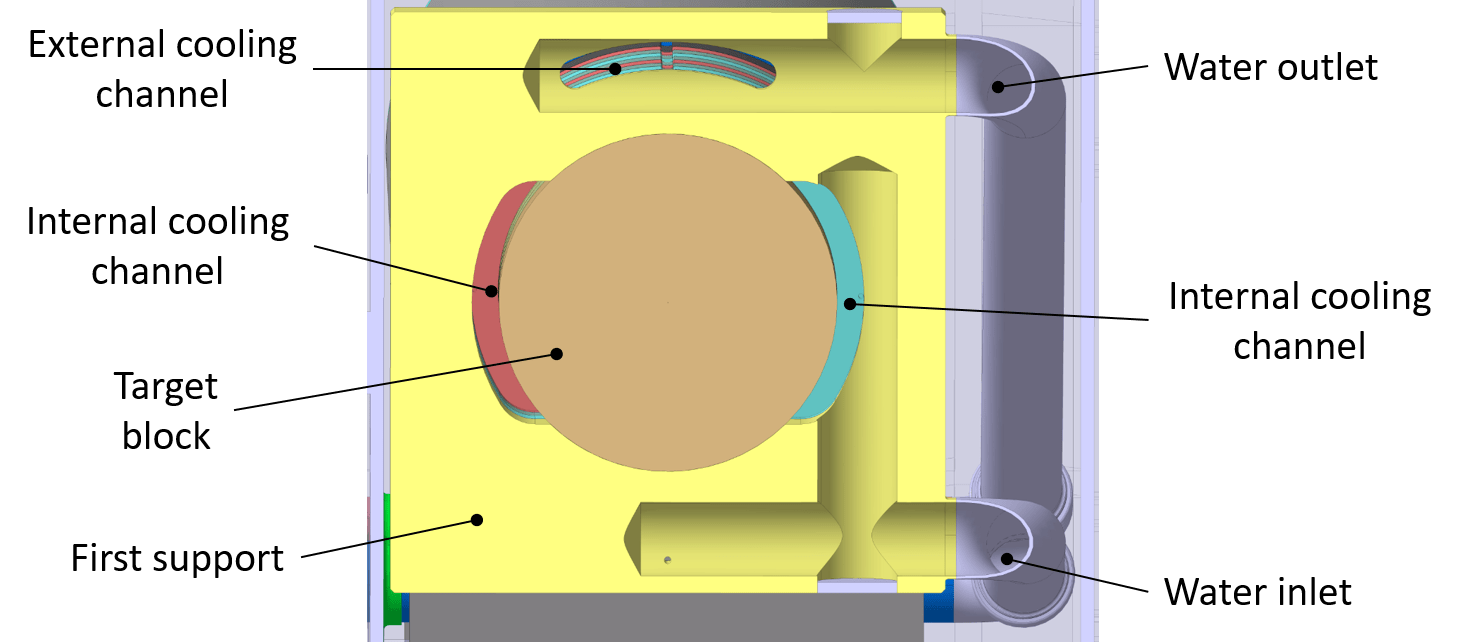}
\caption{\label{fig:TGT:First_support} Layout of the first support of the target inner tank, also upstream flange of the outer tank.}
\end{figure}

Figure~\ref{fig:TGT:First_support} details the layout of the first support of the inner tank, that is one of the key elements of the target assembly. This support acts as upstream flange of the outer tank, and includes both the internal cooling circuit inlet and the external cooling circuit outlet. The water outlet is the highest point of the circuit, preventing the formation of air pockets during the filling process, since the presence of air bubbles could be harmful for the cooling system operation. The cooling pipes providing the water circulation are welded to the sides of the first support, and are routed below the outer tank, facilitating the water drainage for the target replacement, as shown in Figure~\ref{fig:TGT:assembly_cut}. 

The first support assembly includes also the proton beam window. The thickness of the central part of the first support, where the diluted beam impact will take place, has been reduced to 15 mm. This dimension has been optimized to obtain a beam window thick enough for a good mechanical reliability, taking into account the internal pressure of 22 bar; and thin enough to avoid critical levels of energy deposition by the beam impact.

The outer tank provides also an interface with the electrical connections of the target. As mentioned in the previous section, a groove is foreseen in the inner tank supports for the instrumentation cabling. When the outer tank is inserted, the cabling can be collected in a larger groove machined in one of the target supports (see Figure~\ref{fig:TGT:assembly_cut}), and the electrical connectors can be extracted from an opening in the outer tank. The connectors will be then connected to a feed-through in order to ensure the leak-tightness of the outer tank. The number of target blocks to be instrumented, as well as the physical parameters to monitor and the type of instrumentation to use is yet to be defined. The aforementioned target prototype test under beam, where some of the target blocks have been instrumented employing different technologies (Pt100, strain gauges, optical fibers), will be of particular interest to clarify this subject, coupled with the operating experience of spallation neutron sources such as ISIS (UK), SNS (US) and MLF (JP).

\subsubsection{Target helium tank}

The whole target assembly is contained inside a square-section tank filled with inert gas (He at the current stage), as shown in Figure~\ref{fig:TGT:He_box}. The presence of helium gas ensures a dry and controlled environment for the target operation, reducing the corrosion risks on the target assembly components. Additionally, the closed circulation of helium allows the monitoring of possible water leaks from the target vessel. 

\begin{figure}[htbp]
\centering %
\includegraphics[width=1\linewidth]{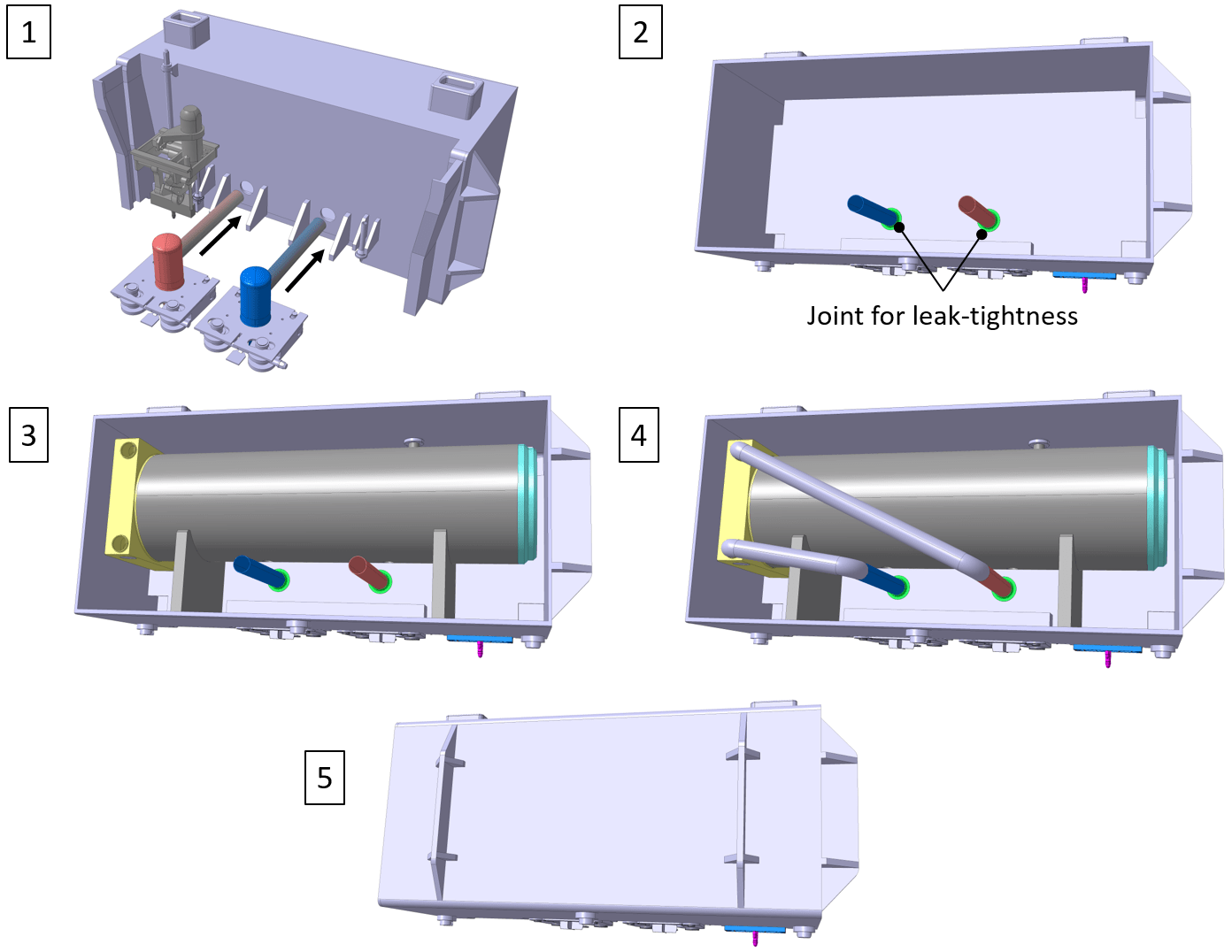}
\caption{\label{fig:TGT:He_box_mounting} The figure shows a simplified assembly process of the BDF target outer tank in the helium containing tank.}
\end{figure}

The mounting process of the target outer tank in the helium container is summarized in Figure~\ref{fig:TGT:He_box_mounting}. In order to ensure the leak-tightness of the helium tank and the feasibility of the welds inside the container, it is necessary to define several steps in the assembly procedure. First, the water inlet and outlet attached to the remote handling connections are inserted into the helium tank through two dedicated holes in the box side. Then, the leak-tightness is achieved by inserting two joints around the inlet and outlet pipes that avoid any possible passage of helium or water to the target complex environment. The next step is to install the target outer tank inside the helium tank, and attach the support of the tank to the container to ensure a proper alignment of the target assembly. Then, the connection between the pipes coming from the target complex water supply and the inlet and outlet of the BDF target cooling circuit is performed. Two water pipes are welded to the first support of the target and to the external pipes for that purpose. Finally, the side cover of the helium container is welded to the tank, ensuring the leak-tightness and reliability of the whole assembly.

In case of target failure, it is foreseen to replace the whole helium tank with the internal components. A comprehensive study of the target complex handling and integration has been performed, and reported in detail in a separate publication~\cite{BDFcomplex} and in Chapter~\ref{Chap:TargetComplex}. The target assembly design is fully compatible with the target complex integration, that foresees the target helium container remote disconnection and exchange. All the electrical, water and helium interfaces of the helium tank that can be seen in Figure~\ref{fig:TGT:He_box} have been designed be disconnected via remote handling, and are detailed in~\cite{BDFcomplex} and in Chapter~\ref{Chap:TargetComplex}. 

\subsection{Structural calculations for the target vessel}
\label{Sec:TGT:MechDesign:struct}

\subsubsection{Scope of the analysis}
The model under investigation is the BDF target assembly, $\varnothing$ 373 mm, approximately 1500 mm long and 1.5 t heavy, which will be subject to a cooling with pressurized water at 22 bar in operational conditions, but it should ramp up to 31 bar during test phase.
The goal of the analysis is to check the structural resistance of the inner tank supports stack and determine the size and number of required screws. An additional scope of the analysis is to assess the vertical deflection of the structure, with an assigned threshold of 1 mm. The relative displacement between the end of the stack and the supporting housing structure, consequent to the self-weight induced bending deformation of the assembly, is also to be checked to ensure that no locking occurs.
Detailed results are described in Ref.~\cite{EDMS-BDF-Tank-Struct}.

\subsubsection{Material selection}
The mechanical properties of the different materials used in the target structure are reported in Table \ref{Tab:TC:Material_properties}.

\begin{table}[htbp]
\centering
\caption{\label{Tab:TC:Material_properties} Material properties.}
\smallskip
\begin{tabular}{l|c|c|c|c|c}
\hline
\textbf{Property} & \textbf{St.Steel 1.4404} & \textbf{St.Steel 1.4435} & \textbf{TZM} & \textbf{Tungsten} & \textbf{St.Steel A4-100}\\
\hline
{$\rho$ [kg/m$^3$]} & 8000 & 8000 & 10220 & 19300 & 8000\\
\hline
{$E$ [GPa]} & 200 & 200 & 288 & 405 & 200\\
\hline
{$\nu$ [*]} & 0.3 & 0.3 & 0.28 & 0.28 & 0.3\\
\hline
{$R_{p0.2}$ [MPa]} & 190 & 200 & 764 & 1350 & 940\\
\hline
{$R_{p1.0}$ [MPa]} & 225 & 235 & - & - & -\\
\hline
{$R_m$ [MPa]} & 490 & 520 & 965 & 1670 & 1000\\
\hline
A & 35\% & 45\% & 10\% & 2\% & 9\%\\
\hline
\end{tabular}
\end{table}

\subsubsection{Failure criteria}

\subsubsubsection{Deformations and stresses}
It is required that the vertical displacement does not exceed 1 mm. Also, the relative motion between the end of the stack and the supporting structure is checked to ensure that no locking occurs. The stresses in the model are assessed against the standard for pressure vessels EN-13445-3~\cite{EN-13445}. This standard defines the maximum allowable values of the nominal design stress for pressure parts, in operating conditions and during testing cases, according to the steel specification and its minimum rupture elongation.
As a conservative approach, the properties of the stainless steel 1.4404, austenitic steel with a minimum elongation after fracture larger than 35\%, are considered for the calculation of the maximum allowable stresses.
The maximum allowable value of the nominal design stress for the present vessel in operating conditions is:

\begin{gather}
\label{eqn:TGT:MaxNomDesignStress}
f_d = max\left[\left(\frac{R_{p1.0/T}}{1.5}\right);min\left(\frac{R_{p1.0/T}}{1.2};\frac{R_{m/T}}{3}\right)\right] = 163.3 \: \text{MPa}
\end{gather}

The maximum allowable stress during testing is:

\begin{gather}
\label{eqn:TGT:MaxTestDesignStress}
f_{test} = max\left[\left(\frac{R_{p1.0/Ttest}}{1.05}\right);\left(\frac{R_{m/Ttest}}{2}\right)\right] = 245 \: \text{MPa}
\end{gather}

\subsubsubsection{Bolt assessment methodology}
Bolts have been assessed against VDI 2230-1~\cite{VDI-2230}. In so doing, the pre-load to be applied to the screws has been estimated taking into account the effects of external forces, so as to guarantee that loads acting on the planes orthogonal to the screws main axes are equilibrated by the friction forces arising between the jointed surfaces (the number of screws acting on the same joint has been also considered): a conservative friction coefficient of 0.1 has been considered. The necessary tightening torque to be applied to each screw has been assessed, verifying that it does not exceed the prescribed limit value. Finally, the die has also been verified, estimating the minimum required engagement length.

\subsubsection{Boundaries and load conditions}
\begin{figure}[ht!]
\centering %
\includegraphics[width=0.8\linewidth]{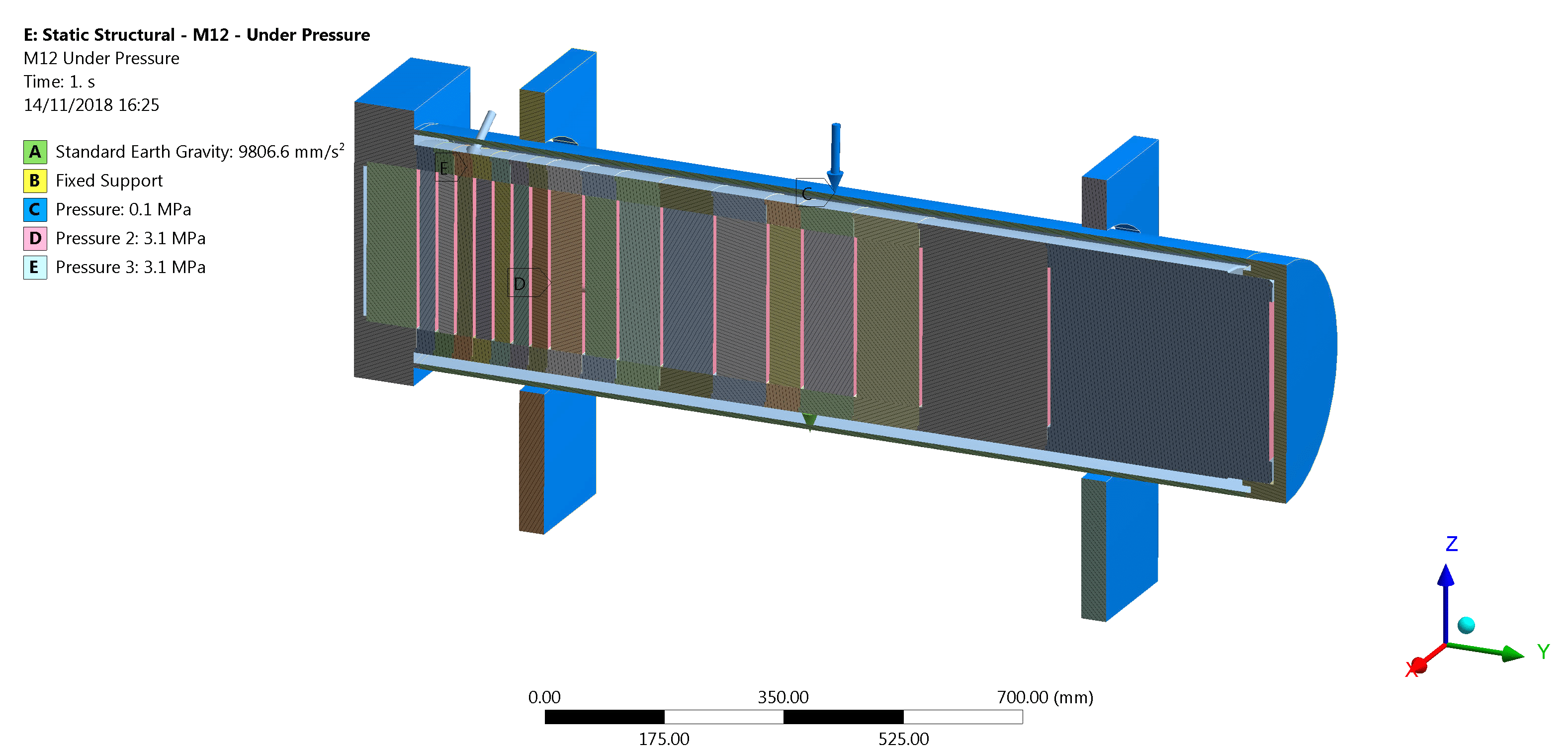}
\caption{\label{fig:TGT:MA_Target_Boundaries} The figure shows the 
initial conditions of the numerical model required to assess the structural robustness of the BDF target assembly.}
\end{figure}

In all analyses, the whole model is fixed on the ground at the bottom faces of the two vertical supports and standard earth gravity applies. 
As shown on Figure~\ref{fig:TGT:MA_Target_Boundaries}, a pressure of 0.1 MPa mimicking atmospheric pressure is applied on all the external faces, while a pressure of 3.1 MPa is applied on the surfaces inside the structure to simulate the maximum pressure that the target will suffer during the test phase.

\subsubsection{Results}
Numerical simulations have been used to perform a static structural analysis of the target support, in order to evaluate its deformations, stress levels, and the reaction forces in the screws.

\begin{figure}[htbp]
\centering %
\includegraphics[width=0.8\linewidth]{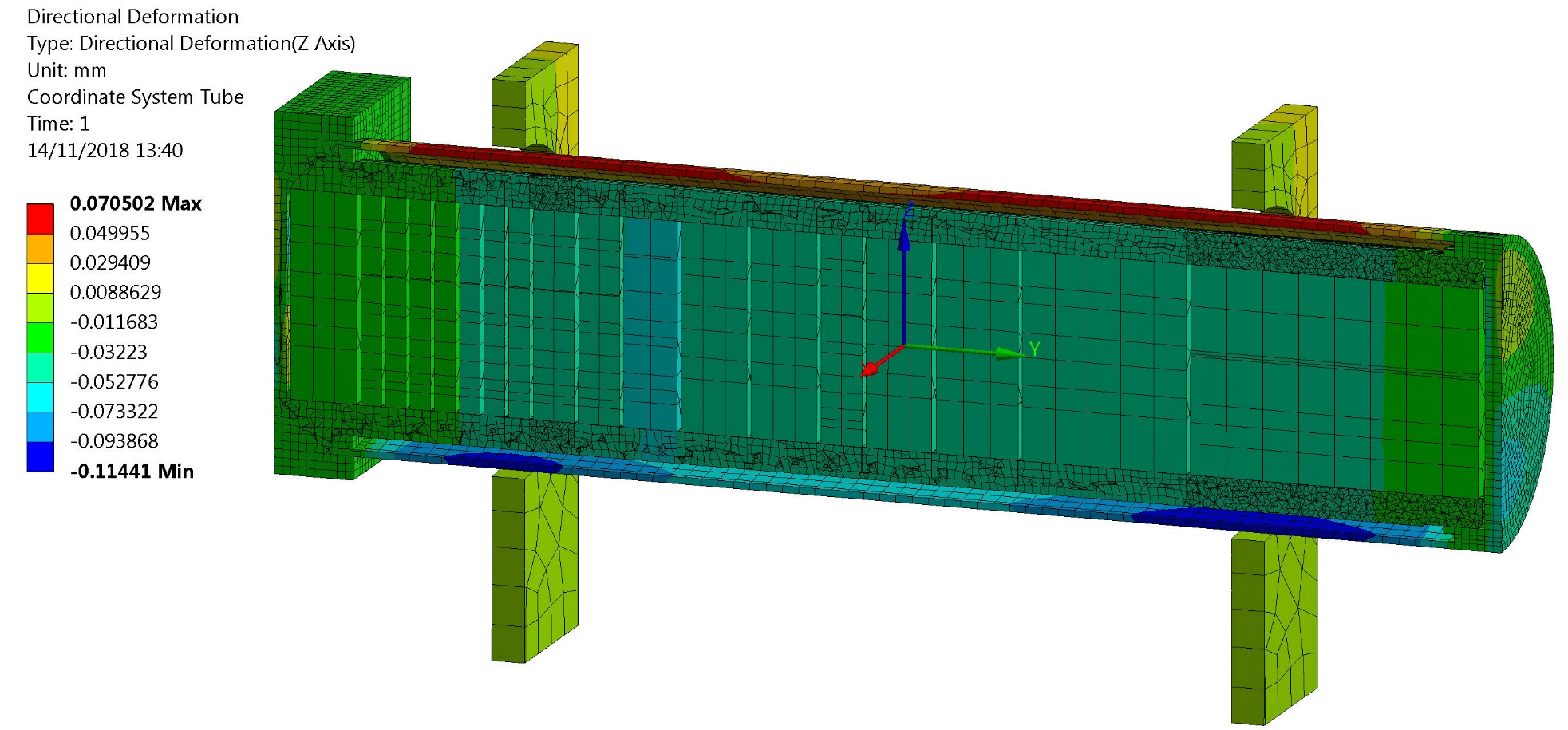}
\caption{\label{fig:TGT:MA_Target_Displacement} The figure shows the vertical displacement of the BDF target assembly.}
\end{figure}

\subsubsubsection{Displacements}
The vertical deformation of the model has been calculated. Figure~\ref{fig:TGT:MA_Target_Displacement} shows the vertical deformation of the assembly. It can be seen that the maximum displacement 0.11 mm. Hence, the criteria of a maximum displacement of 1 mm is fulfilled.

\begin{figure}[htbp]
\centering %
\includegraphics[width=0.9\linewidth]{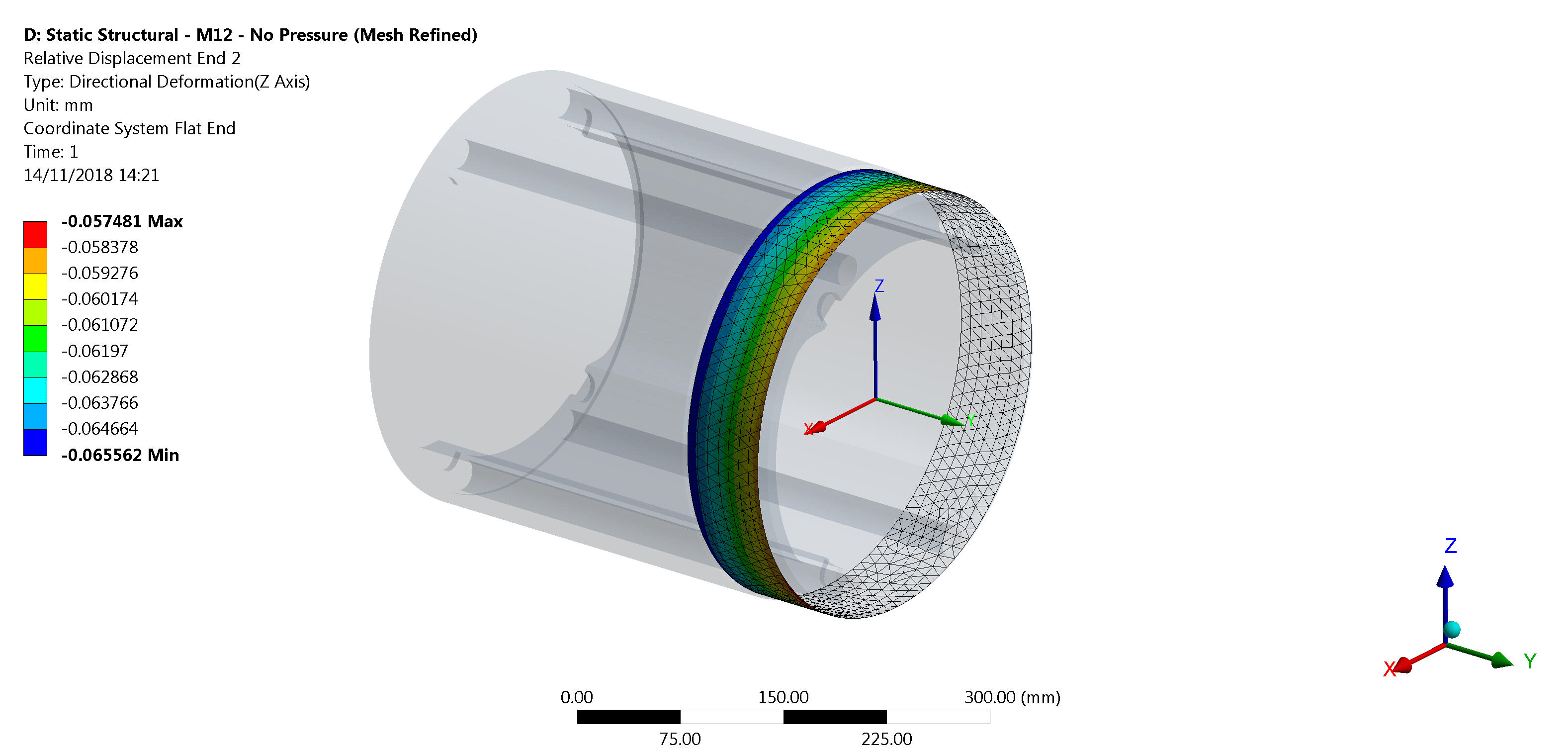}
\caption{\label{fig:TGT:MA_Target_Relative_displacement} Relative displacement at the end of the BDF target stack.}
\end{figure}

\subsubsubsection{Relative displacement between the end of the stack and the supporting housing structure}
The annular support at the end of the stack is embedded in the closing flange. The relative displacement between these 2 parts needs to be evaluate to ensure that no locking occurs.
The reference system is fixed to the closing flange and the displacement of the annular support are assessed in this reference.
The most critical case is when the assembly is only submitted to its own weight.
As shown on Figure~\ref{fig:TGT:MA_Target_Relative_displacement}, the maximum displacement in the end of the stack relatively to the supporting housing structure is 0.066 mm. Consequently, a minimum radial gap of 0.1 mm between the 2 parts should prevent the locking.

\begin{figure}[htbp]
\centering %
\includegraphics[width=0.9\linewidth]{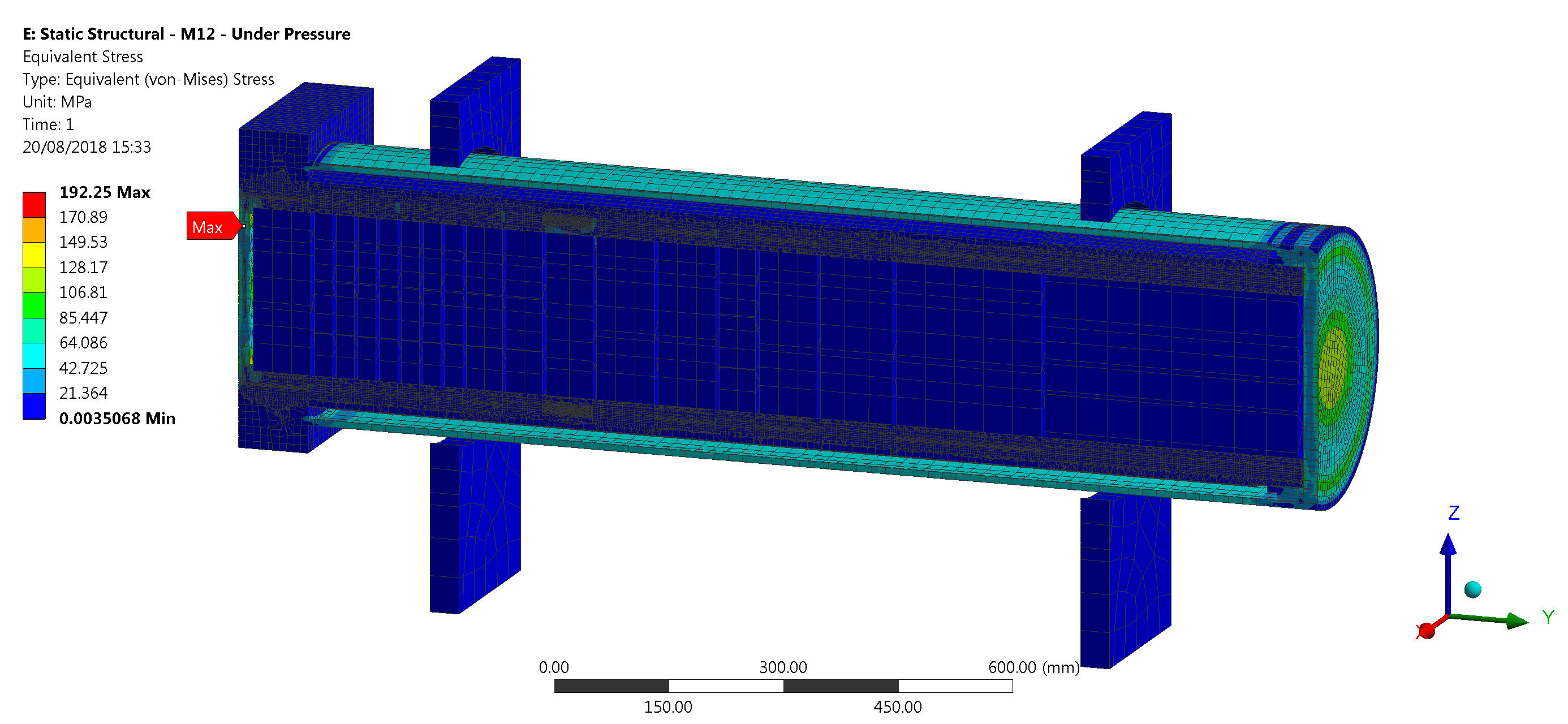}
\caption{\label{fig:TGT:MA_Target_Stress} Equivalent (structural) stresses in the BDF target assembly (Von Mises).}
\end{figure}

\subsubsubsection{Equivalent stress}
The stresses in the model are evaluated according to EN-13445-3. According to this standard, the maximum design limit in the simulation should not exceed $f_{test}$ = 245 MPa.
In so doing, the stresses arising over the contacts simulating the screws cross sections have been excluded from this evaluation, as the analysis of the reaction forces associated to the contacts simulating the screws is carried out according to VDI 2230-1.
Figure~\ref{fig:TGT:MA_Target_Stress} shows the equivalent stress. The stress levels in the model are globally below 200 MPa. Only a few points are getting close to the maximum allowable stress during testing. As they are very localized and in regions that are non-critical, they could be tolerated. 

\begin{figure}[htbp]
\centering %
\includegraphics[width=1\linewidth]{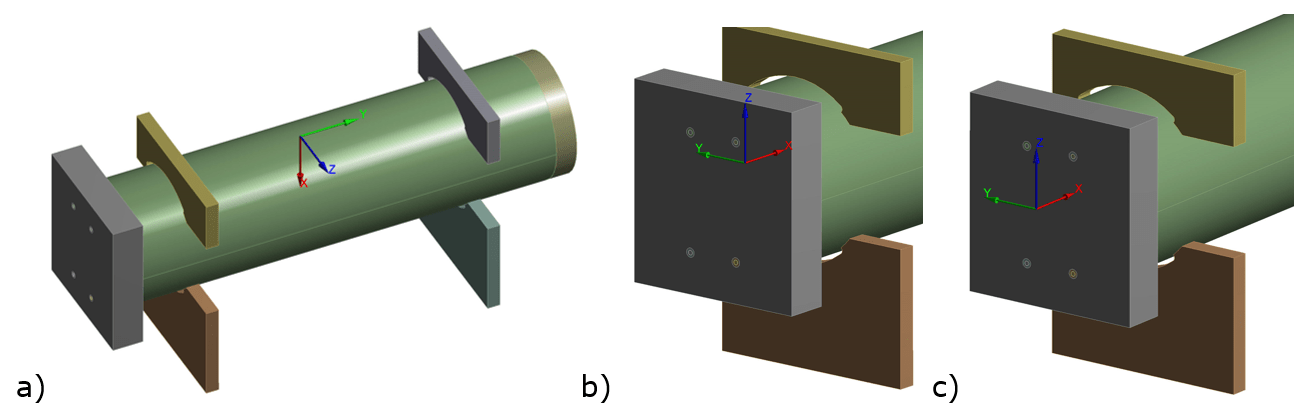}
\caption{\label{fig:TGT:MA_Target_Linearization} Locations of: (a) maximum equivalent principal membrane stress ${\sigma_{Pm}^{eq}}$; (b) maximum equivalent local membrane stress ${\sigma_{Pl}^{eq}}$; and (c) maximum equivalent total stress ${\sigma_{P+Q}^{eq}}$.}
\end{figure}

A stress linearization in the through-the-thickness direction is performed according to EN-13445-3: in particular, given the static nature of the loads acting on the structure, the following condition has been conservatively considered as design limit: ${\sigma_{P+Q}^{eq}<1.5\times f_{test}}$. The maximum equivalent principal membrane stress has the value ${\sigma_{Pm}^{eq}=68.5 \: \text{MPa}}$ and is found in the middle of the major tube, while the maximum equivalent local membrane stress has the value ${\sigma_{Pl}^{eq}=45.5 \: \text{MPa}}$ and is obtained in the 1st block support, at an extremity of the window. The equivalent total stress ${\sigma_{P+Q}^{eq}=140 \: \text{MPa}}$ is found in the middle of the first block support, as shown in Figure~\ref{fig:TGT:MA_Target_Linearization}. The target stress distribution appears to be compliant with EN-13445-3.

\begin{gather}
\label{eqn:TGT:Linearization_a}
{\sigma_{Pm}^{eq}} < f_{test}
\end{gather}

\begin{gather}
\label{eqn:TGT:Linearization_b}
{\sigma_{Pl}^{eq}} < f_{test}
\end{gather}

\begin{gather}
\label{eqn:TGT:Linearization_c}
{\sigma_{P+Q}^{eq}} \leqslant 1.5 \times f_{test}
\end{gather}

\subsubsubsection{Screws assessment}
For each screw, the reaction force has been calculated numerically and has been used to estimate analytically the necessary preload.
In parallel, the admissible preload corresponding to the screw type has been estimated, as well as the tightening torque, according to VDI 2230-1.
Table~\ref{Tab:TC:Calculated_Parameters} summarizes the values of the design and the values of reference calculated according to VDI 2230-1.

\begin{table}[htbp]
\centering
\caption{\label{Tab:TC:Calculated_Parameters} Results of the calculated parameters for the bolts required for the BDF target inner tank assembly.}
\smallskip
\begin{tabular}{l|c}
\hline
\textbf{Parameter} & \textbf{Value}\\
\hline
Maximum necessary preload & 33.5 kN\\
Admissible preload & 42.5 kN\\
\hline
Maximum necessary tightening torque & 100 N.m\\
Admissible tightening torque & 107.7 N.m\\
\hline
Actual length of engagement & 17.2 mm\\
Required length of engagement & 11.1 mm\\
\hline
\end{tabular}
\end{table}

The admissible preload for the M12 screw in stainless steel A4-100 is 42.5 kN, and the necessary preloads calculated for the screws of the design are all below this value (the maximum required preload is 33.5 kN).
An estimation of the tightening torque shows that 100 Nm is enough to guarantee the necessary preload.
The length of engagement of the screws has also been calculated according to VDI 2230-1, and it can be noticed that the actual length of engagement is larger than the minimum recommended.
Therefore, the design parameters for the bolts are compliant with the standard VDI 2230-1.

\subsubsection{Conclusion on the structural calculations for the BDF inner tank}
The BDF target inner and outer tank structural strength has been reviewed considering that the atmospheric pressure and an internal pressure of 31 bar are applied.
The vertical displacement does not exceed the limit of 1 mm.
The analysis of the relative displacement between the end of the inner tank stack and the supporting housing structure revealed that a minimum gap of 0.1 mm between the 2 parts will ensure that no locking occurs.
The simulations showed that the stress levels are below the acceptability limits, in accordance with EN-13445-3.
Analytical calculations, based on numerical results and following the standard VDI 2230-1, have demonstrated the compliance of the fixations of the supports in terms of bolt area, length of engagement and necessary preload for each screw.

\FloatBarrier

\section{Considerations on BDF target material selection}
\label{Sec:TGT:Materials}

\subsection{Target materials selection} \label{Sec:TGT:Mat:matselec}
Based on simulations involving a spallation target split into a Mo based and a W based portion, respectively, the suitability of these metals and of their alloys has been considered. Starting from the available physical and mechanical properties, it has been checked whether the above materials can guarantee sufficient strength. The most severe stresses are in tensile regime, up to $\sim$140 MPa for Mo  (see Figure~\ref{tab:TGT:residual}) and $\sim$70 MPa for W (see Figure~\ref{tab:TGT:residualW}),  in a temperature range between  100 $^{\circ}$C and 180 $^{\circ}$C for Mo and between 80$^{\circ}$C and 150 $^{\circ}$C for W (see Figure~\ref{fig:TGT:thermalplot} for both materials). The above figures arise from the results of the simulations summarized in Section~\ref{Sec:TGT:Simus}. 

The spallation materials will experience each year $10^6$ cycles of thermo-mechanical fatigue in irradiation conditions, hence fatigue strength and irradiation resistance have to be considered. 

\subsubsection{Mo and Mo alloys}

The mechanical properties of pure Mo are highly dependent of its temper state. Stress relieved Mo shows sufficient yield strength and tensile strength at room temperature. However, tensile properties for recrystallized Mo are quite limited \cite{Tungsten_Schmidt}. Independently of the temper state, yield and tensile strength values are diminished by 10 \% - 15 \% at 200 $^{\circ}$C respect to room temperature~\cite{TantalumW2}. In all cases, stress relieved temper should be preferred for Mo and as well as for its alloys.

\begin{figure}[htb]
    \centering
    \includegraphics[scale=0.65]{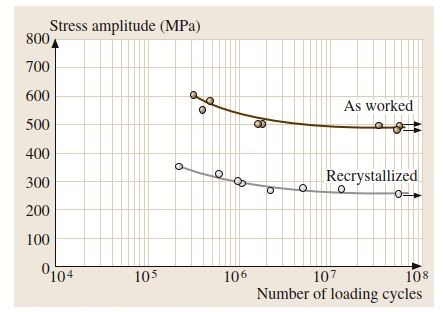}
    \caption{Rotatory-bending fatigue test results for as-received and recrystallized Mo 25 mm diameter rods, tested at room temperature \cite{TantalumW2}}
    \label{fig:TGT:mats:Fat_mo}
\end{figure}

Mo features an elevated fatigue strength at room temperature for high cycle fatigue regime (up to 10$^8$ cycles) which is also highly dependent of the Mo temper state. As an example, fatigue strength values of 500 MPa (\% 68 UTS) and 300 MPa (\% 44 UTS) are reported for as-worked and recrystallized 25 mm rods respectively, appreciable in Figure~\ref{fig:TGT:mats:Fat_mo} (see Ref.~\cite{TantalumW2}).

Irradiation of Mo leads to general loss of tensile ductility compared to the unirradiated state in the case of irradiation temperatures below 700 $^{\circ}$C. These phenomena was homogeneously observed for two types of irradiation at representative fluences: protons (2.0x10$^{20}$ p/cm$^2$) and neutrons (1.5x10$^{22}$ n/cm$^2$) (Ref.~\cite{SIMOS2017,GORYNIN1992}).

\begin{figure}[htb]
    \centering
    \includegraphics[scale=0.8]{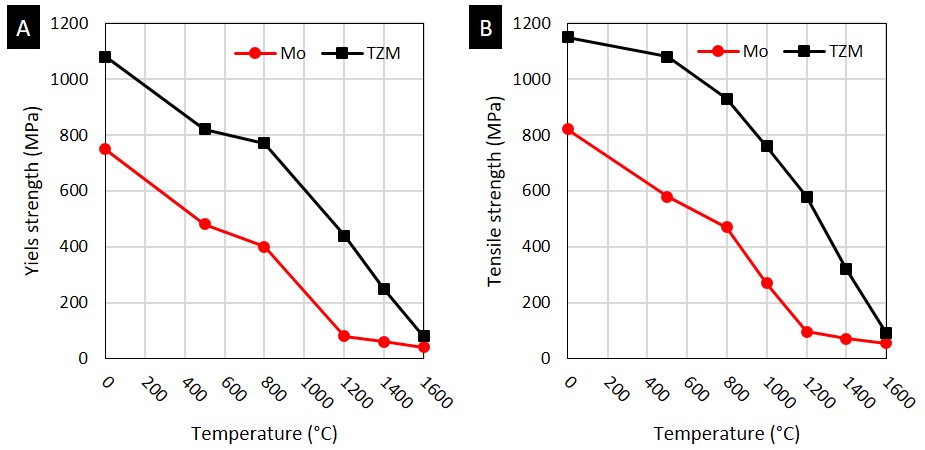}
    \caption{a) Yield strength and b) Ultimate tensile strength versus temperature for Mo and TZM 1 mm thick stress-relieved sheets \cite{TantalumW2}.}
    \label{fig:TGT:mats:UTSMoTZM}
\end{figure}

Among the Mo-alloys, one of the most available commercially and largely tested is TZM 
((0.5 \%) titanium - (0.08 \%) zirconium - (0.03 \%) carbon - molybdenum alloy). In the unirradiated state, TZM features higher strength and ductility than pure Mo for both stress-relieved and recrystallized tempers in a wide temperature range (up to 1600 $^{\circ}$C, see Figure~\ref{fig:TGT:mats:UTSMoTZM}) \cite{TantalumW2}.

TZM exhibits equal sensitivity to temper state than pure Mo. Tensile strength for stress-relieved 1 mm sheets at room temperature is 820 MPa whilst for the recrystallized temper is 550 MPa. For the same material grade, fatigue strength at representative cycles (10$^7$) was measured in 620 MPa and 390 MPa for stress-relieved and recrystallized tempers respectively, values representing 70 \% of the tensile strengths \cite{FURUYA1984}

The alloying elements in TZM develop carbides that precipitate during
the manufacturing process that act as re-crystallization inhibitors. In consequence, re-crystallization temperature is rise by roughly 400 K with respect to pure Mo. Full re-crystallization temperatures for 1 h annealing in 1 mm sheets with similar deformation degree was found in 1000 $^{\circ}$C for Mo and 1450 $^{\circ}$C for TZM~\cite{TantalumW2}. Re-crystallization temperature becomes of importance when considering the thermal cycling that the target materials will be subject during the bonding to the cladding material, as described in Section~\ref{Sec:TGT:mat:HIP}. As exposed here-above, mechanical properties are significantly enhanced in the stress-relieved tempers, thus re-crystallization shall be ideally avoided during materials processing. 

\begin{figure}[htb]
    \centering
    \includegraphics[scale=0.9]{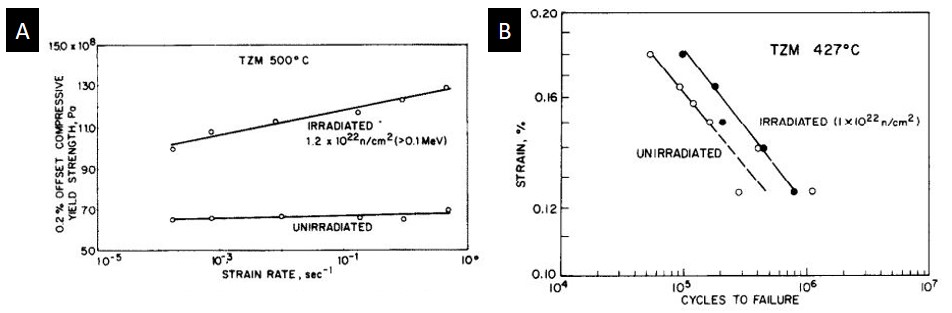}
    \caption{a) Effect of strain rate on the compressive yield strength of irradiated and unirradiated TZM at 500 $^{\circ}$C and b) Effect of total strain per half cycle on the fatigue life of notched specimens of irradiated and unirradiated TZM at 427 $^{\circ}$C in vacuum~\cite{SMITH1977}.}
    \label{fig:TGT:mats:YSIrrTZM}
\end{figure}

Irradiation data for TZM is currently mainly available from neutron irradiation. With representative neutron fluences (0.9x10$^{22}$ n/cm$^2$), significant strength increase was found when irradiating below 650 $^{\circ}$C in representative strain rates (10$^{-3}$s$^{-1}$ to 10$^{-1}$s$^{-1}$) as appreciable in Figure~\ref{fig:TGT:mats:YSIrrTZM}a. Moreover, the ductile to brittle transition (DBTT) increased from -85 $^{\circ}$C to 150 $^{\circ}$C~\cite{tungstenprops}, resulting in a potential loss of ductility in the target operational conditions. Figure~\ref{fig:TGT:mats:YSIrrTZM}b shows TZM fatigue behaviour for similar neutron fluence, resulting in a fatigue life increased by a factor of two in the irradiated state \cite{SMITH1977}.

\subsubsection{W and W alloys}

Pure W is available in different product forms including rolled plates (thickness up to 20 mm), billets, rods (diameter up to 80 mm) and forgings that are relevant forms for BDF target application. Rolled plates can feature a fully dense microstructure, but which becomes layered hence strongly anisotropic. Flat grains parallel to the rolled surface affect the mechanical properties and result in a potential risk of delamination. Double-forged (radially 
and axially) W blanks feature a more isotropic microstructure than unidirectional forged 
shapes. It is considered by the fusion community that double-forged (multi-directional) W should act as 
a reference grade for establishing a reliable materials database for finite element 
calculations~\cite{PINTSUK2012}.

\begin{figure}[htb]
    \centering
    \includegraphics[scale=0.6]{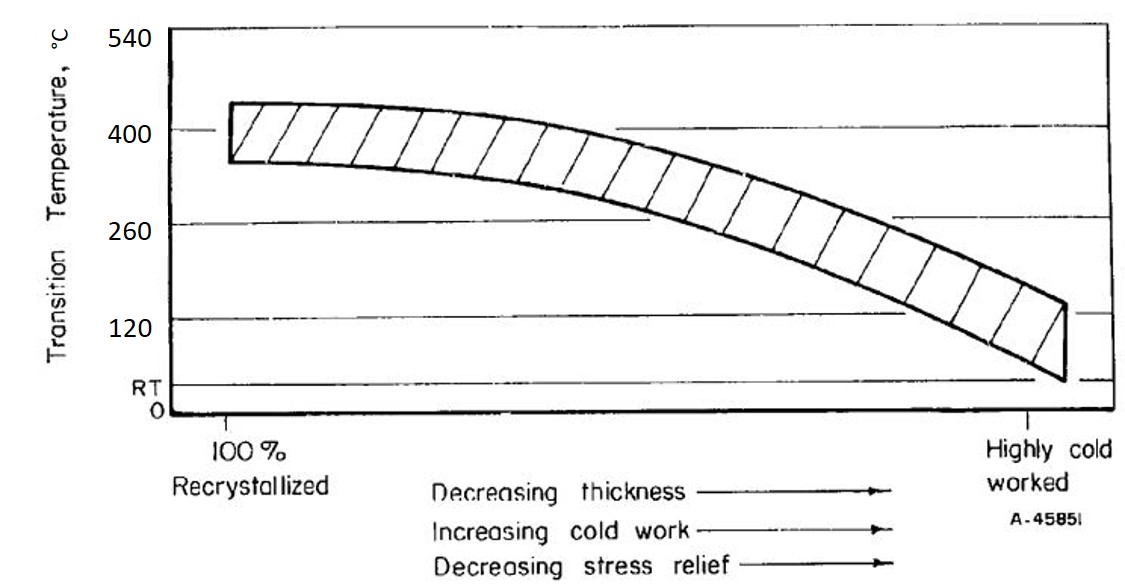}
    \caption{Influence of processing parameters on the DBTT of flat-rolled powder-metallurgy W products \cite{Tungsten_Schmidt}.}
    \label{fig:TGT:mats:DBTTW}
\end{figure}

The ductile to brittle transition (DBTT) for pure W is highly influenced by the deformation degree applied below the re-crystallization temperature, as shown in Figure~\ref{fig:TGT:mats:DBTTW}. With a sufficient degree of deformation, DBTT can be lowered to 200 $^{\circ}$C - 400 $^{\circ}$C. With annealing, DBTT raises until attaining that of recrystallized metal. Thus, multi-directionally forged products in the stress-relieved state are the preferred W products for offering the maximal strength and potential ductility~\cite{Tungsten_Schmidt}.

Sufficient tensile strength is exhibited for the latter W products in the target operational temperatures, with values above 1100 MPa at room temperature and 600 MPa at 200 $^{\circ}$C for rolled sheets. These strength values are halved for the same product in the recrystallized state~\cite{Tungsten_Schmidt}.

Nevertheless, due to W brittle nature, the use of any W grade in structural components at low temperatures ($<$ 500 $^{\circ}$C) is subject to a suitable design to limit the thermal stress~\cite{BARABASH1999}. Even in the case of low DBTT achieved with severe deformation, at irradiation fluences of 0.3 - 0.5 dpa (representative of the target material) any achieved ductility will be already vanished~\cite{BARABASH1999}. Therefore, a W grade with the best available mechanical properties shall be employed in order to maintain high safety factor respect to the operational stress field.

\begin{figure}[htb]
    \centering
    \includegraphics[scale=0.9]{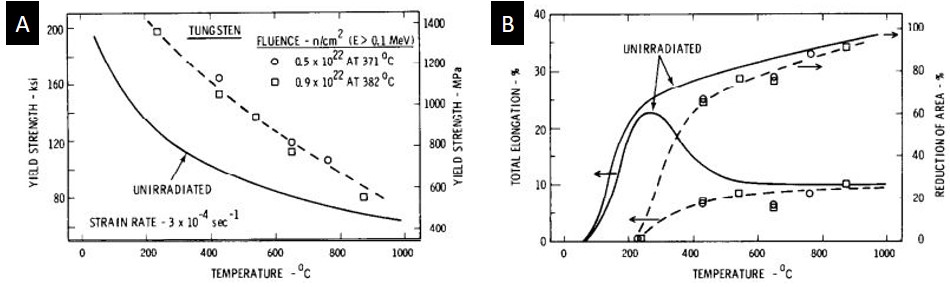}
    \caption{Temperature dependence of the strength (a) and ductility (b) of unirradiated and irradiated W \cite{tungstenprops}.}
    \label{fig:TGT:mats:UTSIRRW}
\end{figure}

Relevant data in the literature on mechanical properties of neutron-irradiated W 
is available. The effect of irradiation at low temperatures (300 $^{\circ}$C and below) is considered minor for fluences around 5x10$^{19}$ n/cm$^2$ (with E $>$ 0.1 MeV). Nevertheless, at neutron fluences of 0.9x10$^{22}$ n/cm$^2$ it results in a yield strength increase and a diminution of ductility (see Figure~\ref{fig:TGT:mats:UTSIRRW}) \cite{MAKIN1957,tungstenprops}. Analogous behaviour is also observed in representative proton irradiation (0.5 - 1 dpa) at target operational temperatures (50 $^{\circ}$C - 150 $^{\circ}$C), with increase of in the compressive yield strength and decrease of ductility \cite{MALOY2005}.

Fatigue data for unirradiated W have been recently reported for two grades: rolled plates and sintered + Hot Isostatic Pressed (HIP) plates. Data is reported in the tensile regime and for high cycle fatigue (2x10$^6$ cycles). Material grade raised as key factor in the fatigue performance, with fatigue strengths of 350 MPa and 180 MPa for rolled W and sintered+HIP W respectively. Both values represent a percentage of 30 \% - 35 \%  respect to the static tensile strength values~\cite{Wfatigue}.

Concerning W alloys, W-Re alloys have the advantage of featuring higher 
recrystallization T and lower DBTT compared to pure W and preserving ductility after 
recrystallization \cite{KLOPP1966}. The effect of Re additions is significant above 3\% and towards the lower end range (3\% to 5\%), it provides a favourable effect on the ductility in the non-irradiated state. However, under neutron irradiation it results in more rapid and severe 
embrittlement than it is observed for pure W~\cite{PINTSUK2012}. High Re activation is also expected under neutron (and also proton) irradiation~\cite{BARABASH1996}, owing to its high thermal neutron cross-section.

Similarly, less mechanical strength and an increased loss of ductility compared to pure W was 
found for particle strengthened W alloys such as W - 1\% La$_2$O$_3$ "when tested up to 
700 $^{\circ}$C~\cite{PINTSUK2012}.

An exception between all explored W alloys is the mechanically alloyed W - TiC  which offers better machinability and improved ductility compared to pure W (for the latter, both characteristics are a major concern). The addition of TiC allows to form an isotropic and fine grain structure that is maintained even in the recrystallized condition. Moreover, the finer dispersoids of TiC particles improve the low temperature impact toughness~\cite{PINTSUK2012}.

Ultra-fine grained W-TiC (additions in the range 0.25-0.8 \% TiC) has been developed for use in irradiation environments. It has been observed that, contrarily than for pure W, this alloys presents no hardening when irradiated with neutrons at a fluence of 2x10$^{24}$ n/cm$^2$ at 600 $^{\circ}$C. However, W-TiC is not yet available on industrial scale~\cite{KURISHITA2007,KURISHITA2013}. Such material have been tested with high intensity proton beam at the HiRadMat facility in the HRMT48-PROTAD experiment during 2018.

Between the different available W alloys, unalloyed W has been the target material choice in several neutron spallation facilities (ISIS~\cite{BROOME1997}, LANSCE~\cite{LANSCEcladding}, KENS~\cite{HIP2}) over decades. Operational conditions of those targets are relevant for BDF target in terms of target configuration, magnitude of the cyclic thermal loads, and irradiation conditions. Extensive operational experience as target material has been built up with no major issues reported, therefore pure W remains the reference grade as spallation target material for future facilities (such as ESS~\cite{PEGGS2013}).

\subsubsection{Target material selection conclusion}

In conclusion, for Mo-based components a Mo-alloy should be preferred to pure Mo. TZM, in the stress relieved temper, is a promising candidate because of better mechanical properties and lower recrystallisation temperature. Preliminary investigations indicated that this material can fulfill the target material requirements in terms of physical, mechanical properties and considering the irradiation environment. Accordingly, TZM was selected as baseline material for the first half of the BDF target. 

Concerning the W-based components, available literature from plasma facing applications recommends to avoid the use of W in heavy cyclic thermal loads at temperatures below 500 $^{\circ}$C. However, BDF target conditions are much less critical in terms of T gradients, strain rates, flux heat factor... Conditions in neutron spallation targets are more representative of the BDF target conditions, and many facilities have successfully operated W based targets. Literature on material properties combined with facilities operational feedback indicated that pure W can fulfill the physical and mechanical properties considering the irradiation environment. Thus, pure W in forged and stress-relieved temper was chosen as the baseline material for the second half of the BDF target. Nevertheless, pure W brittle nature should be taken in account in the target design, to avoid issues related with machinability and residual stresses, which can be accentuated in irradiation conditions. Yet, W alloys such W-TiC could be considered as alternative, pending industrial availability.

Radiation damage effects on the target material and the respective variation of thermo-physical and mechanical properties with representative proton irradiation remain still to be addressed by specific tests in order to have a robust target design.


\subsubsection{Target protection: protective claddings}

W and TZM feature limited corrosion resistance in high temperature flowing water (up to $320\,^{\circ}\mathrm{C}$), even when maintaining stringent control of the water oxygen content ($<$1 ppm) \cite{ISHIJIMA2002}. In these conditions, phenomena such as exfoliation has been observed for W. Furthermore, corrosion rates of these materials are expected to be enhanced in irradiation environments due to the build up of radiolysis products, effect already reported for pure W~\cite{LILIARD2002,MALOY2012}. 
As consequence, high corrosion rates can lead to significant amounts of activated material in the cooling circuit and generating operational issues, like it has been reported in facilities such as LANSCE~\cite{LANSCEcladding}.

Alternative materials such as Zircaloys, stainless steels, titanium alloys, tantalum alloys exhibit excellent corrosion and erosion-corrosion resistance in aqueous environments~\cite{BERMUDEZ2005}. However, the target physics performance can be severely penalized because of longer interaction lengths and lower Z, therefore limiting their application as target materials. 
Covering the target material with a layer of the latter materials, while maintaining the maximum target volume in W and TZM raised as an elegant solution to protect the target materials without limiting the physics requirements \cite{BROOME1997,HIP2,LANSCEcladding} .

\subsubsection{Cladding materials selection}
\label{Sec:TGT:mat:cladselec}
Between several options, tantalum (Ta) was selected as protective material because of good bondability to W and Mo alloys. Ta features perfect solubility with W and Mo, avoiding brittle crystalline phases development at the interfaces when bonding to the target. Tantalum also presents similar thermal expansion than W and Mo, required to minimize stresses at the interface during thermal cycles \cite{METCALFE1963,DOBROWSKI2009,FENGLI2003}.

The BDF target will face approximately $10^6$ beam pulses par year of operation, and therefore the same number of thermal cycles. Mechanical stresses are expected in the target-protective layer interface due to the unavoidable thermal expansion mismatch. Hence, strong and reliable bonding between the Ta and the target is mandatory to avoid any detachment during target lifetime.
Any loss of contact could originate a loss of heat evacuation to cooling water, cooling channels blocking or accelerate cladding/target failure.

Thank to extensive R\&D studies (see Section~\ref{Sec:TGT:mat:RandD}), the applicability of a protective Ta layer was fully validated. The optimal configuration was determined as a Ta layer of 1 mm - 2 mm thickness which is diffusion bonded to the target materials by means of HIP (see Section~\ref{Sec:TGT:mat:HIP}).

Thermal and structural simulations revealed that, in some of the most loaded blocks, the stress and temperature field can reach up to 100 MPa and 150$^{\circ}$C in some points of the cladding (see Section~\ref{Sec:TGT:Simus}). Considering that the yield stress of the unalloyed Ta is approximately 80 MPa at this temperature, the accumulation of cyclic plastic strain could lead to clad premature failure. Alternative cladding materials which could offer higher strength but also offer good erosion-corrosion resistance and HIP assisted diffusion bonding compatibility to the target materials were sought. 
The explored material was a commercial Ta alloy, solution strengthened with 2.5\% of W. This material presented the advantage of a close chemistry to unalloyed Ta but with enhanced mechanical properties. Theoretical yield strength of this material at the operational BDF temperatures is over 200 MPa, offering sufficient operational safety margin. The alloy can withstand the conditions of a HIP cycle and both the Ta and the W present perfect solubility with the target materials, resulting in interfaces without undesirable formation of brittle intermetallic phases. The alloy exhibits improved corrosion resistance respect to Ta, proven in severe conditions such as with hot H$_2$SO$_4$ and HCl, both of the environments were employing Ta is required. Furthermore, hydrogen embrittlement resistance is enhanced~\cite{GYPEN1984}.

\begin{figure}[htb]
    \centering
    \includegraphics[scale=0.85]{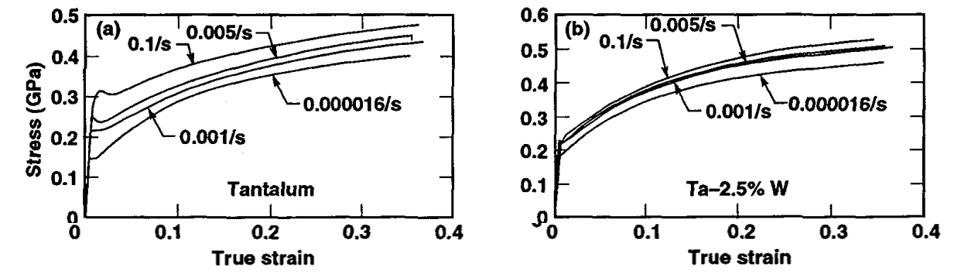}
    \caption{Stress strain compression curves at different strain rates for a) unalloyed Ta and b) Ta-2.5\%W \cite{GOURDIN1994}.}
    \label{fig:TGT:mats:UTSTaTa25W}
\end{figure}

Unalloyed Ta sheet features sufficient yield and tensile strength of 185 MPa and 200 MPa respectively at room temperature. However, the same properties are reduced to 70 MPa and 180 MPa at 200$^{\circ}$C. Recrystallized temper is considered in the materials selection since the HIP thermal cycle will certainly induce full recrystallisation of the Ta microstructure. Ta2.5\%W sheet features greater yield and tensile stress in the target operational temperatures, with values of 255 MPa and 345 MPa at room temperature and 190 MPa and 290 MPa at 200 $^{\circ}$C respectively~\cite{TaW_HCStarck,Tantalum_Schmidt}. Solution strengthen Ta with W results also in a decrease  of the material sensibility to the strain rate as appreciable from Figure~\ref{fig:TGT:mats:UTSTaTa25W} in several compression tests~\cite{GOURDIN1994}.

Fatigue data for both materials is available but limited to certain conditions, i.e. fully reversed loading and at room temperature. Recrystallised Ta and Ta2.5W sheets feature fatigue strength of 200 MPa and 270 MPa respectively at high cycle fatigue regimes ($\sim$10$^{7}$ cycles) \cite{TantalumW2,KONG1994}.

Extensive irradiation data is available for unalloyed Ta in representative conditions. Irradiation with 800 MeV protons at temperatures of 25 $^{\circ}$C - 250 $^{\circ}$C resulted in a progressive hardening up to 11 dpa. For the same specimens strain to necking was fast reduced from 30\% to 10\% at 0.6 dpa but maintained constant in 10\% up to 11 dpa~\cite{CHEN2001}. Alloying effects in the mechanical properties evolution after irradiation do not show clear trends: in Ta-10\%W alloy, increased irradiation hardening compared to Ta was observed when irradiated with protons; contrarily, similar hardening than for Ta was observed for the alloy T111 (Ta-8\%W-2\%Hf) for comparable neutron irradiation conditions~\cite{CLAUDSON1965, WIFFEN1973}. 

\begin{figure}[htb]
    \centering
    \includegraphics[scale=0.9]{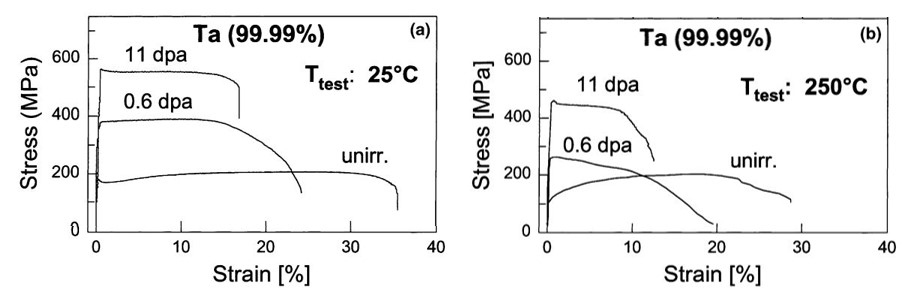}
    \caption{Stress-strain curves of Ta specimens tensile tested with a strain rate of 10$^{-3}$s$^{-1}$ at a) RT and b) 250$^{\circ}$C~\cite{CHEN2001}.}
    \label{fig:TGT:mats:UTSTaIrr2}
\end{figure}

Few specific irradiation data exists for the Ta2.5W alloy. Ref.~\cite{IPATOVA2017} reported that alloying Ta with W delays the radiation induced lattice damage, as observed by irradiating several Ta alloys (Ta, Ta2.5\%W and Ta-10\%W) with a proton beam at 3 MeV. Further studies are required in order to explore the radiation damage with BDF representative beam conditions and activities are ongoing in the framework of the RaDIATE Collaboration \cite{Rad2018,Rad2018_2}. More details on the present activities are given in Section~\ref{Sec:TGT:Mat:BLIP}.

To our knowledge, there is no reported experience in literature corresponding to diffusion bonding Ta2.5W by means of HIP and thus Ta2.5W it was included in the R\&D studies as alternative cladding material to unalloyed Ta. Successful results of the R\&D studies brought Ta2.5W as the baseline cladding material for the BDF target over unalloyed Ta (see section~\ref{Sec:TGT:mat:RandD}).

\subsection{Target blocks manufacturing}\label{Sec:TGT:mat:manufac}
\subsubsection{Target materials manufacturing}
The baseline material grade for the target production is multi-directionally forged products. This material grade is basic to ensure isotropic microstructure since the target stresses are equally severe in the three axis, as presented in Section~\ref{Sec:TGT:Simus}. Forging can also ensure the maximal materials density and the best mechanical properties. 

\begin{figure}[htb]
    \centering
    \includegraphics[scale=0.8]{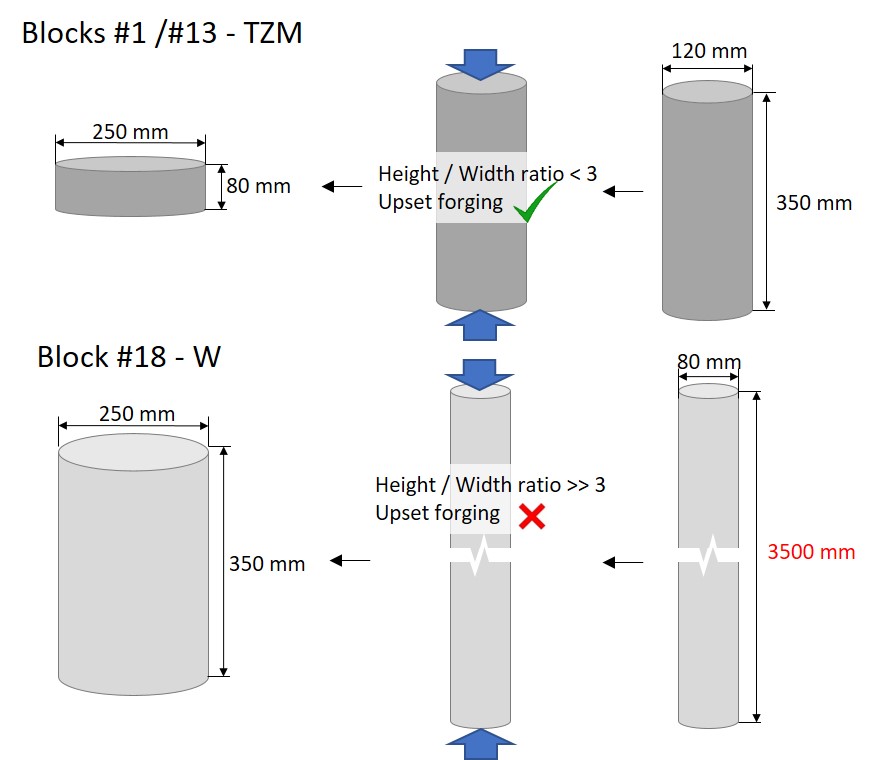}
    \caption{Schema of the upset forging process for the longest target blocks of TZM and W, applicable for the BDF target.}
    \label{fig:TGT:mats:upset}
\end{figure}

Since the W and TZM blocks are conceived in cylindrical shapes, the most indicated product available commercially to build the target blocks are from 2D forged (radially) rods which are then cut to the respective target block lengths. The maximal available rod diameters from known providers are 80 mm for W and 120 mm for TZM, much smaller than the final target blocks diameter (250 mm). One potential solution proposed by a vendor to obtain the target blocks is to cut rods from the maximal available diameter with a certain length and afterwards upset forge the rods to increase the diameter to 250 mm. This procedure would additionally improve the mechanical properties in the axial direction and the target blocks would had been forged in the 3 directions (2D radially + 1D axially from the upset forging). Nevertheless, upset forging operation is limited by the height/diameter (H/D) ratio, which shall be smaller than a certain value in order to avoid buckling and excessive accumulated strain in the materials. Maximal aspect ratio for forging is normally considered between 2 and 3. This limitation can represent an issue in the longest blocks of the BDF target. In the case of TZM, the longest blocks are \#1 and \#13 with a length of 80 mm. To obtain this geometry from 120 mm diameter rods, upset forging should start with a 350 mm long rod. The radio H/D in this case is 2.9 and therefore upset forging possible for all the TZM blocks (see Figure~\ref{fig:TGT:mats:upset}). For the W blocks, the longest block is \#18 with 350 mm length. From 80 mm diameter rod, upset forging shall start on a 3500 mm long rod. The ratio H/W in this case is 27.5 much grater than the recommendable. 

\begin{figure}[htb]
    \centering
    \includegraphics[scale=1]{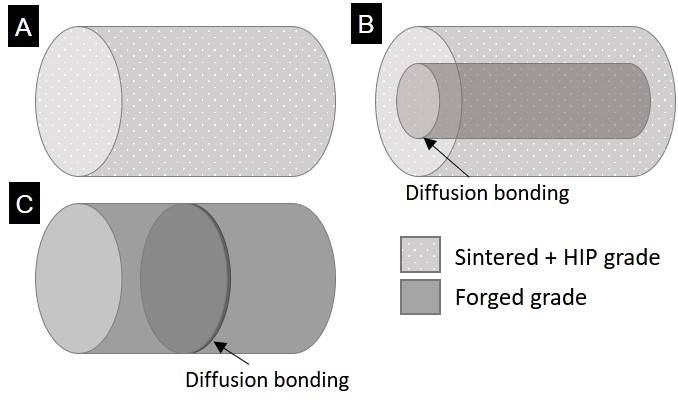}
    \caption{Schema of the target W blocks different fabrication possibilities}
    \label{fig:TGT:mats:Wfabr}
\end{figure}

As a consequence, the production of the longest W target blocks by upset forging route was discarded. Several alternative manufacturing routes were explored:
\begin{itemize}
    \item Sintered W pre-shapes with analogous aspect ratio than the target blocks, which are then densified by means of Hot Isostatic Pressing (see Figure \ref{fig:TGT:mats:Wfabr}a). The main advantage of this production route is the flexibility in terms of shapes and dimensions that can be achieved. As discussed before, the mechanical properties of the W are highly dependent on the material density and deformation grade below the recrystallisation temperature \cite{Tungsten_Schmidt}. The density and deformation achieved by this production route is significantly lower than for forging, hence the mechanical properties are inferior than W forged grades (see Section~\ref{Sec:TGT:mat:litprops}) \cite{Wfatigue}
    \item To employ two different W grades in a same target block. A multidirectionally forged rod with the maximal available diameter would take the central part of the target block. This "core" would be afterwards covered by a "shell" of sintered + HIP W grade to reach the 250 mm diameter. The intimate union between both parts would be carried out afterwards by means of HIP assisted diffusion bonding (see Figure~\ref{fig:TGT:mats:Wfabr}b).
    \item To employ target blocks composed of two or several 3D forged W cylinders bonded coaxially. Since the length of the target blocks is the main limitation for the upset forging, cylinders with reduced length would be individually upset forged, piled up coaxially to reach the desired target block length and finally diffusion bonded by means of HIP (see Figure~\ref{fig:TGT:mats:Wfabr}c).
\end{itemize}

The FEM simulations (see Section~\ref{Sec:TGT:Simus}) predict moderate mechanical requirements in the W target blocks, which will face maximal cyclic stresses below 100 MPa. Commercial forged W presents tensile strengths in the order of 1000 MPa whilst sintered + HIP W presents tensile strengths in the order of 500 MPa - 600 MPa~\cite{Wfatigue}. Despite the significant strength reduction, strength for sintered + HIP W appears to fulfill the target requirements. The combination of satisfactory mechanical properties and the production flexibility prompted sintered + HIP W grade as baseline over other options.

In one hand, the option of using two different W grades in the same target block would allow to employ forged W grade in the target core, resulting in improved mechanical properties. Additionally, ductility might be enhanced, but soon vanished with irradiation as described in Section~\ref{Sec:TGT:Mat:matselec}. In the other hand, ensuring reliable thermal conductivity and therefore mechanically strong bonding between the core and the external would be mandatory condition to consider the applicability. Preliminary studies such are presented afterwards (see Section~\ref{Sec:TGT:mat:RandD}) revealed difficulties to diffusion bond W to W even in simplified geometries (flat to flat surfaces). Diffusion bonding two concentric cylindrical surfaces would also represent a challenge in terms tolerances due to the difficulties in machining the W. This option was abandoned in views of the added complexity and the important prototyping and additional development which would require.

A third option was considered and it is currently pending results from prototyping. As said, diffusion bonding two flat surfaces revealed difficulties during preliminary R\&D activities (see Section~\ref{Sec:TGT:mat:RandD}). Nevertheless, promising results were obtained when diffusion interfacial aids such are interfoils placed in between two target blocks (see Section~\ref{Sec:TGT:mat:RandD}). With this work, the solution appears to be validated but the reliability of the interface with cyclic thermal fatigue remains unclear.

\subsubsection{Target cladding operation} \label{Sec:TGT:mat:HIP}
As introduced in section \ref{Sec:TGT:mat:cladselec}, the target blocks will be protected with an external 1 mm - 2 mm thick Ta based layer. The Ta layer is diffusion bonded on the target blocks by means of HIP. The principles of the bonding process are highlighted this part.

\subsubsubsection{Principles of diffusion bonding}

\begin{figure}[htb]
    \centering
    \includegraphics[scale=1]{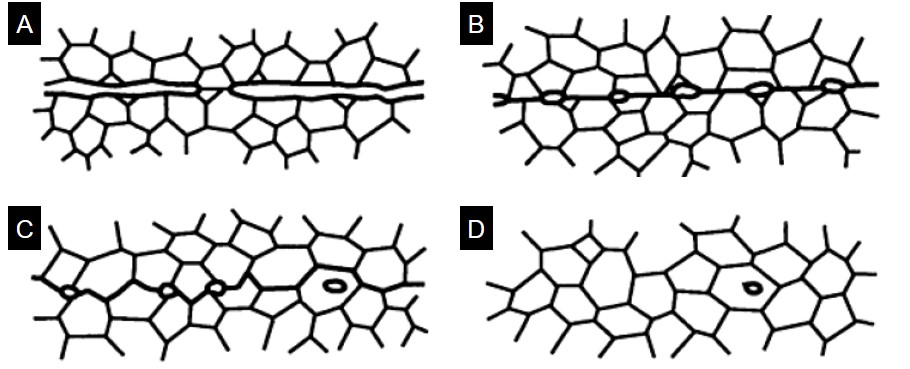}
    \caption{Stages of diffusion bonding: a) Initial contact, limited to few asperities, b) first stage: deformation of surface asperities by plastic flow and creep, c) second stage: diffusion of atoms to the voids and grain boundary migration and d) third stage: volume diffusion of atoms to the voids~\cite{OLSON1993}.}
    \label{fig:TGT:mats:Difbon}
\end{figure}

In diffusion bonding, the nature of the joining process is essentially the coalescence of two atomically clean solid surfaces. The process can be divided in three stages, listed below and schematized in Figure~\ref{fig:TGT:mats:Difbon} \cite{OLSON1993}.

\begin{itemize}
    \item Firstly, due to surface inherent roughness, contact between surface is limited to a small fraction. With pressure, time and temperature, the contact area increases through plastic deformation of asperities and creep;
    \item Atoms diffuse to the remaining voids in order to reduce the surface free energy. In parallel interfacial grain boundary migrates out of the plane of the joint to a lower energy equilibrium;
    \item The remaining intra-granular voids are eliminated through atom diffusion though the volume of the grains;
\end{itemize}

\subsubsubsection{HIP assisted diffusion bonding}

When intending to diffusion bond complicated geometries, like a continuous Ta layer over a cylindrical geometry, conventional diffusion bonding techniques (uni-axial, rolling) are not applicable. This is due to the difficulties to apply uniform pressures between the base materials. HIP allows applying isostatic pressure even on complicated geometries thanks to a pressurized gas.  

In order to bring the cladding in intimate contact with the target and allow subsequent diffusion bonding, isostatic pressure only on the external side of the cladding material is required. To do so, a continuous and hermetic cladding material capsule is fabricated around the target by welding the several cladding parts together (e.g. two plates and a tube to surround the target cylinder in our case). Previous to the welding, the air between the cladding capsule and target material shall be removed. Any infiltration of gas between between the cladding material and the target cylinder would equalize pressure in both sides of the cladding material and cancel the applied pressure towards the target. 

\begin{figure}[htb]
    \centering
    \includegraphics[scale=1]{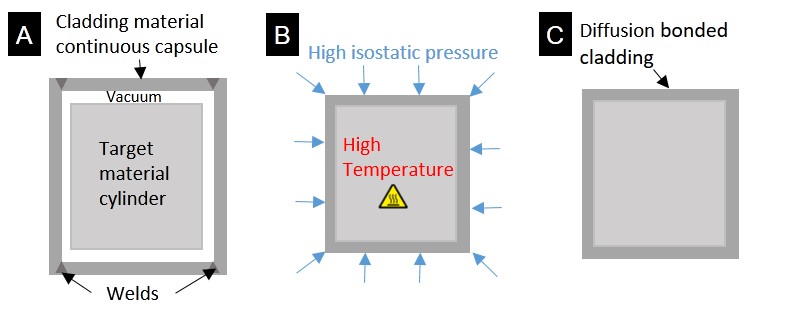}
    \caption{Schema of the different stages of the HIP assisted diffusion bonding: a) welding a continuous and hermetic clad material capsule over the target material, b) application of isostatic pressure and temperature to bring two surfaces together and start diffusion bonding c) resulting geometry after the HIP cycle, with a diffusion bonded clad.}
    \label{fig:TGT:mats:HIP}
\end{figure}

HIP is carried out in a chamber which allows continuous and controlled application of high gas pressures (up to 2000 bar) and high temperatures (up to 1500 $^{\circ}$C - 2000 $^{\circ}$C). During the HIP cycle, gas pressure and temperature are simultaneously ramp up and hold for a certain time. The cladding internal surface and the target are brought in contact and diffusion bonding phenomena (described here above) takes place between them. Further details on the preparation procedure and HIP parameters are given in Section~\ref{Sec:TGT:mat:RandD}). Different steps of the HIP assisted diffusion bonding are schematized in Figure~\ref{fig:TGT:mats:HIP}.

\subsection{Target materials properties}

After materials selection validation (see Sections~\ref{Sec:TGT:Mat:matselec} and \ref{Sec:TGT:mat:manufac}), it was necessary to gather the available material properties from literature and internal studies. Careful analysis of the available data was performed to determine its applicability for the BDF target. Finally, the necessary materials testing should be carried out in order to complete the available data and provide a reliable input for the FEM simulations and target design. All these steps are presented in the current section.

\subsubsection{Relevant material properties}

The energy deposition from the pulsed proton beam will be translated into cyclic temperature increase in the target materials. Knowledge of materials physical properties like heat capacity (C$_p$) and thermal conductivity is required to determine the target temperature field evolution. At the same time, the temperature evolution in the blocks will induce thermal expansion. Physical properties such are the coefficient of thermal expansion and young modulus are required to determine the derived stress field. Yield strength, tensile strength, elongation at failure and fatigue strength are crucial properties to determine the target integrity when subject to the cyclic stress field, either during normal operation or in failure scenarios. Many of these properties are dependent on temperature, thus it is relevant to know how they evolve within the target operational temperature range. 
 
In parallel to withstanding the mechanical solicitations, cladding material resistance to corrosion in demineralised water is of extreme importance to assess the reliability of the target. Furthermore, all the target materials will be subject to irradiation conditions which can affect material properties. Both factors are discussed in separated sections (see Section~\ref{Sec:TGT:mat:corrosion} for corrosion considerations and Section~\ref{sec:TGT:mats:radiation} for irradiation considerations). 

\subsubsection{BDF target material properties literature review} 
\label{Sec:TGT:mat:litprops}

\subsubsubsection{Physical properties}

\begin{figure}[htb]
    \centering
    \includegraphics[scale=0.9]{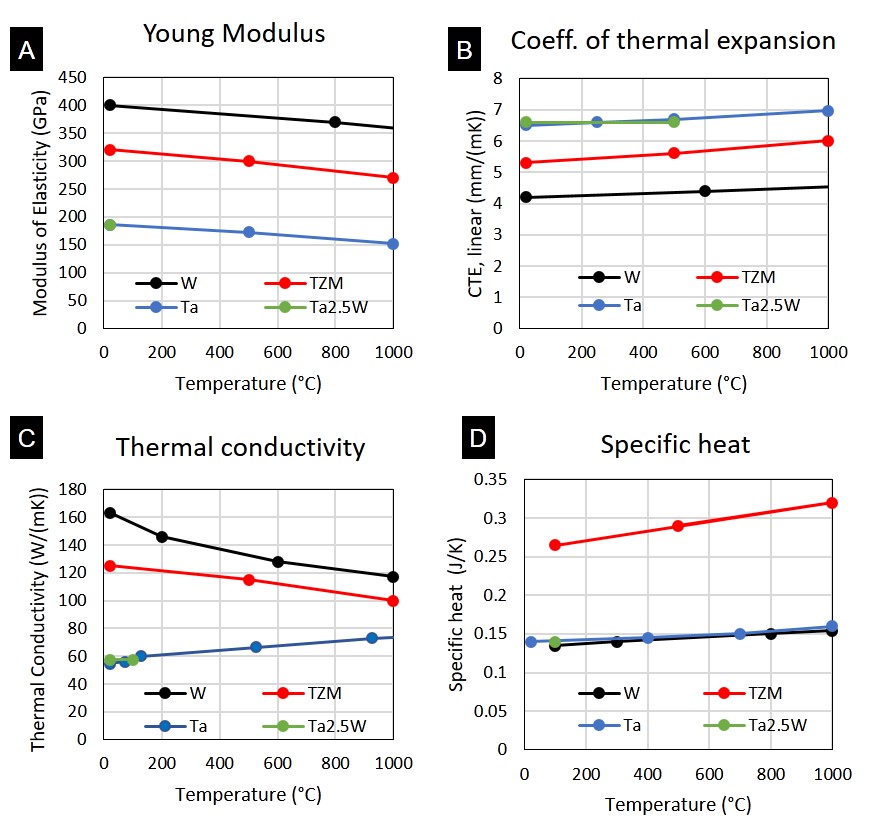}
    \caption{Physical properties of the target materials, a) Young Modulus, b) Coefficient of thermal expansion, c) Thermal conductivity and d) Specific heat~\cite{TantalumW2,TaW_HCStarck,SHI1998}.}
    \label{fig:TGT:mats:physic}
\end{figure}

Physical properties of the four target materials an their evolution with temperature are reported in Figure~\ref{fig:TGT:mats:physic}. Physical properties of these materials are not sensitive to the material product, microstructure, purity or thermal history. Generic values are available for W, TZM and Ta for the entire operational temperatures while data for Ta2.5W is mostly limited to room temperature. 

\subsubsubsection{Static mechanical properties}

Static mechanical properties are highly dependent on the material product, microstructure, purity and thermal history. Literature values are presented therefore together with a discussion on their applicability for the BDF target material selection. Target material stress field present similar intensities in both tensile and compressive behaviour. Tensile values are presented (if available) because of the more conservative approach.

\begin{figure}[htb]
    \centering
    \includegraphics[scale=0.8]{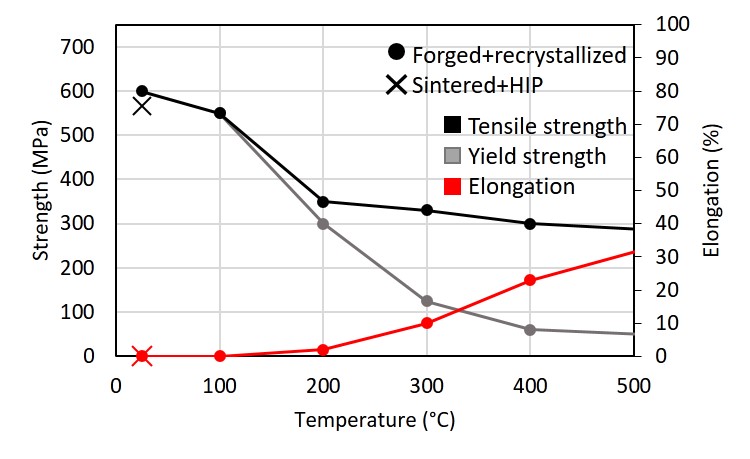}
    \caption{Tensile properties of the two W grades forged+recrystallized~\cite{Tungsten_Schmidt} and sintered+HIP~\cite{Wfatigue}.}
    \label{fig:TGT:mats:UTSW}
\end{figure}

For the BDF target, two W grades are considered from two production routes, forged material and sintered + HIP material. For the forged W grade, the most representative (and conservative) data is for recrystallized forged rods~\cite{Tungsten_Schmidt}. The HIP diffusion bonding cycle will bring the W close to the recrystallisation temperature for 3 hours and therefore recrystallization can occur. Specimens were tested at representative strain rate (10$^{-3}$s$^{-1}$) but the diameter and grain size of the rods was not reported. The obtained values and their evolution with temperature are plotted in Figure~\ref{fig:TGT:mats:UTSW}. Tensile strength values are 600 MPa at 20 $^{\circ}$C and 350 MPa at 200 $^{\circ}$C. It is remarkable the brittle nature of the material in the entire temperature operational range. Data for sintered + HIP W grade is only available at room temperature~\cite{Wfatigue}. Tensile strength value is similar to the forged grades, with 560 MPa at RT and no plastic deformation was either observed. Samples were tested at representative strain rates (10$^{-2}$ s$^{-1}$) but no references to the material product were given. 

Extensive data on the static mechanical properties is available for TZM. Most of the data is obtained from rolled sheet products up to 20 mm thickness, not representative of the target TZM cylindrical forgings. Nevertheless, data from TZM sheet is useful to appreciate the evolution of the properties with temperature. Data for stress relieved sheet from Plansee, with grain size of 190 microns (longitudinal) x 35 microns (transverse) tested at strain rate of 10$^{-3}$ s$^{-1}$ is given in Figure~\ref{fig:TGT:mats:UTSMoTZM}~\cite{TZM_Filacchioni}. This material is characterized by the high tensile strength and elongation at failure, which at room temperature reach values of 830 MPa and 21\% respectively. Representative data is available from TZM cylindrical forgings of 200 mm diameter and 100 mm length, with similar production route and dimensions of the BDF target but tested only at two temperatures (20 $^{\circ}$C, 700 $^{\circ}$C). In this case tensile strength (axial directions) is much lower with 525 MPa at 20 $^{\circ}$C and 320 MPa at 700 $^{\circ}$C. It is also remarkable the low ductility offered by this product at room temperature. 
The significant difference on the mechanical properties between both material products is due to the applied deformation degree. The high strength of TZM limits the deformation degree that can be achieved with conventional forging equipment, which becomes relevant for big cross-sections (e.g. forgings). For small sections (e.g. sheets or plates), lower load is required and greater  deformation degree can be achieved, resulting in superior mechanical properties.  

\begin{figure}[htb]
    \centering
    \includegraphics[scale=0.8]{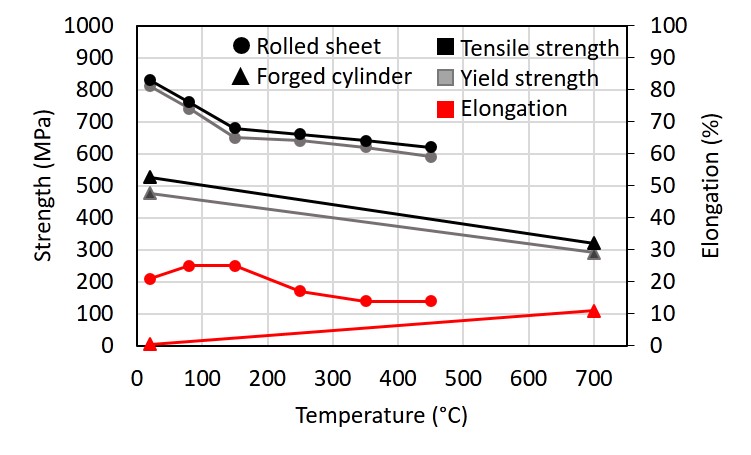}
    \caption{Tensile properties of two TZM grades, stress relieved rolled sheet~\cite{TZM_Filacchioni} and forged cylinder (internal data).}
    \label{fig:TGT:mats:UTSTZM}
\end{figure}

\begin{figure}[htb]
    \centering
    \includegraphics[scale=0.8]{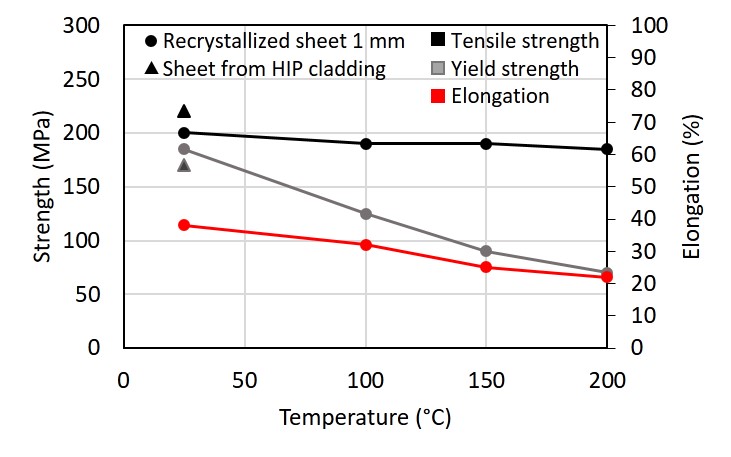}
    \caption{Tensile properties of Ta, recrystallized rolled sheet (1 mm thickness)~\cite{Tantalum_Schmidt} and sheet from HIP cladding (internal data).}
    \label{fig:TGT:mats:UTSTa}
\end{figure}

Evolution of tensile properties with temperature for Ta is available in literature in quite representative material: cold rolled and recrystallized 1 mm thick sheet, tested at strain rate of 1.5x10$^{-3}$ s$^{-1}$ (see Figure~\ref{fig:TGT:mats:UTSTa}~\cite{Tantalum_Schmidt}). Some measurements on Ta sheet specimens extracted from Ta clad W/TZM prototypes were carried out internally and resulted in similar values at room temperature, with 170 MPa yield strength and 220 MPa tensile strength (see Figure~\ref{fig:TGT:mats:UTSTa}).

In the case of Ta2.5W, tensile properties evolution with temperature are only available from the minimal guaranteed properties from HC Starck Ta2.5W products~\cite{TaW_HCStarck}. Even though no information of the tested product is given, values are similar than in-house Ta2.5W sheet specimens extracted from Ta2.5W clad W/TZM prototypes, tested at room temperature. For the latter, values of 270 MPa and 360 MPa were measured. 

\begin{figure}[htb]
    \centering
    \includegraphics[scale=0.8]{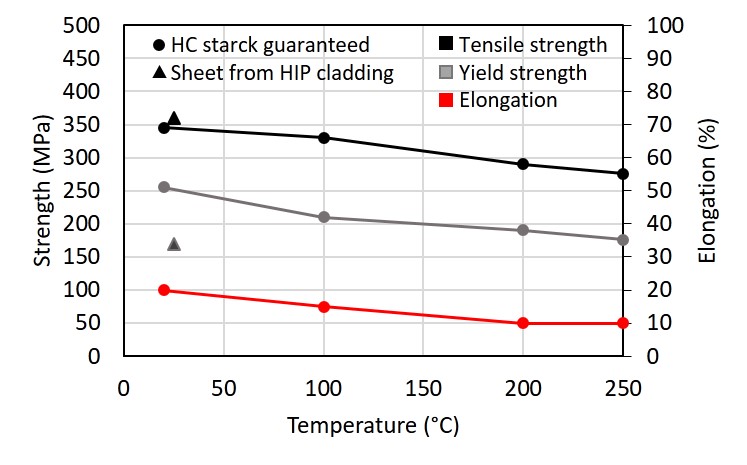}
    \caption{Tensile properties of Ta2.5W, guaranteed values from HC Starck products~\cite{TaW_HCStarck} and sheet from HIP cladding (internal data).}
    \label{fig:TGT:mats:UTSTa25W}
\end{figure}

Summary of the reviewed data is given in Table~\ref{tab:TGT:mats:strengthdata} which compiles the available static mechanical strength data, from literature or obtained via mechanical testing, relevant for the BDF operational conditions and source for the FEM simulations.

\begin{table}[htb]
\centering
\caption{BDF target materials yield and tensile strength at room temperature as obtained from available literature and mechanical testing. Estimated material strength at high temperatures according to the trend shown in Figures \ref{fig:TGT:mats:UTSW}-\ref{fig:TGT:mats:UTSTa25W}.}
\label{tab:TGT:mats:strengthdata}
\smallskip
\begin{tabular}{lllllll}
\hline
\multirow{2}{*}{\textbf{Material}} & \multicolumn{2}{l}{\textbf{Yield strength (MPa)}} & \multicolumn{2}{l}{Tensile strength (MPa)} & \multicolumn{2}{l}{Total elongation (\%)} \\ \cline{2-7} 
 & RT & 200 °C & RT & 200 °C & RT & 200 °C \\ \hline
W & - & 300 & 560 & 350 & 0 & 2 \\ \hline
TZM & 480 & 425 & 525 & 475 & 0.5 & 3 \\ \hline
Ta & 180 & 70 & 200 & 180 & 35 & 22 \\ \hline
Ta2.5W & 270 & 190 & 360 & 290 & 20 & 10 \\ \hline
\end{tabular}
\end{table}

\subsubsubsection{Dynamic mechanical properties}

Dynamic mechanical properties are highly dependent on two factors, both the material grade (product, microstructure, purity, thermal history...) and the testing setup (specimen dimensions, specimen surface preparation, testing mode, testing frequency...). Literature values are presented together with discussion on the representativity for the BDF target materials. 
Representative data is considered as values obtained for equal products than the target (forged rods for TZM and W, recrystallized sheet for Ta and Ta2.5W) and tested at similar conditions than during operation (push-pull mode, 10$^7$ cycles, operational temperatures, stress ratio $\sim0$, frequency $<$100 Hz).

Fatigue data for sintered + HIP W is limited to the work of Ref.~\cite{Wfatigue}, although it is highly representative for the BDF target application. Specimens were tested in push-pull configuration for 2x10$^6$ cycles, stress ratio $\sim$0, testing frequency 25 Hz and at RT. Fatigue strength with these conditions is 175 MPa, 31\% of the UTS. The same author studied the fatigue behaviour of forged plates with 90 mm thickness, representative of alternative W grade. Fatigue strength for the same conditions is similar to Sintered + HIP grade, with 170 MPa - 200 MPa (34\%-38\% of the UTS). In this work, high temperature tests were also carried out resulting in fatigue strengths of 250 MPa at 280 $^{\circ}$C and 190 MPa at 480 $^{\circ}$C employing the same testing conditions~\cite{HABAINY2015}.

\begin{figure}[htb]
\centering
\includegraphics[scale=0.8]{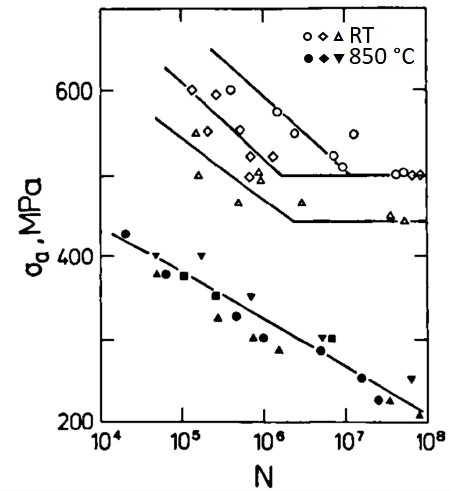}
\caption{Fatigue strength for different number of cycles (N) for TZM at two temperatures, RT and 850 $^{\circ}$C \cite{TZMfatigue}.}
\label{fig:TGT:mats:TZMFati}
\end{figure}

For TZM, fatigue properties are extensively reported in literature, but the data in the most representative conditions for the BDF target is from Ref.~\cite{TZMfatigue}. Samples were extracted from 50 mm bar, as-worked, and tested with push-pull fatigue up to 10$^8$ cycles. Testing conditions were stress ratio -1, testing frequency 25 Hz and temperatures RT and 850 $^{\circ}$C. Fatigue strength at 10$^7$ cycles were measured in 440 MPa and 250 MPa respectively. The obtained Wohler curves are given in Figure~\ref{fig:TGT:mats:TZMFati}.

\begin{figure}[htb]
\centering
\includegraphics[scale=0.4]{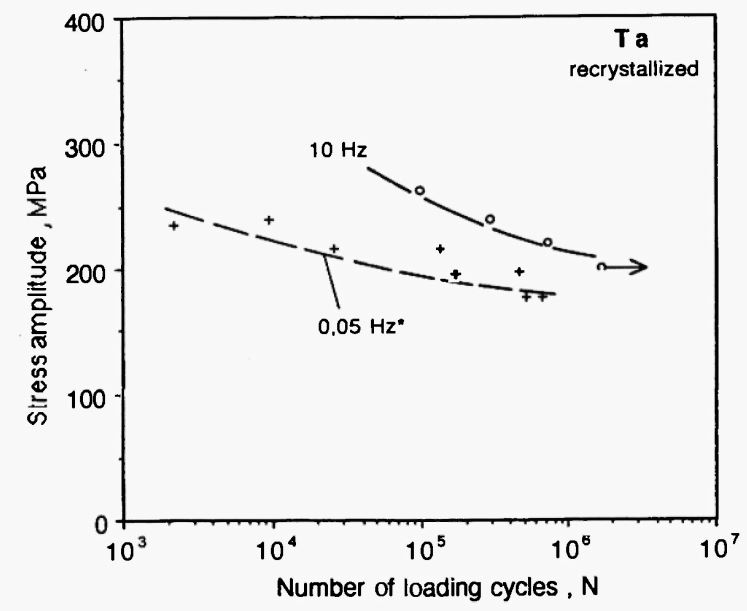}
\caption{Fatigue strength for different number of cycles (N) for Ta at room temperature and two different frequencies 0.05 Hz and 10 Hz~\cite{KONG1994}}
\label{fig:TGT:mats:TaFati}
\end{figure}

For Ta and Ta2.5W representative fatigue data exists but limited to room temperature. For Ta, fatigue strengths of 180 MPa and 210 MPa were found for 2 mm thick recrystallized plate at 10$^7$ cycles, tested in push-pull mode, at stress ratio -1 and frequencies of 0.05 Hz and 10 Hz respectively~\cite{KONG1994}. The Wholer curves are given in Figure~\ref{fig:TGT:mats:TaFati}. For Ta2.5W, fatigue strengths between 270 MPa and 310 MPa were found for bending fatigue mode, at 10$^7$ cycles, stress ratio -1 and 25 Hz. This values represented between 60\% and 70\% of the material UTS~\cite{TantalumW2}. 

Table~\ref{tab:TGT:mats:fatiguedata} summarizes the available fatigue data most relevant for the BDF operational conditions and source for the FEM simulations.

\begin{table}[htb]
\caption{Summary of the reviewed high-cycle fatigue data relevant for the BDF target operational conditions. Sources: TZM 
\label{tab:TGT:mats:fatiguedata}
\cite{TZMfatigue}, W \cite{Wfatigue, HABAINY2015}, Ta \cite{KONG1994} and Ta2.5W \cite{TantalumW2}. P/M: Powder Metallurgy, Aw: As worked, Rxx: Recrystallized}
\small
\begin{tabular}{ccccccc}
\toprule
\textbf{Material} & \makecell{\textbf{Production}\\ \textbf{process}} & \textbf{Dimensions} & \textbf{Test mode} & \textbf{Number of cycles} & \makecell{\textbf{Stress ratio,}\\ \textbf{temperature,}\\ \textbf{frequency (Hz)}} & \makecell{\textbf{Fatigue}\\ \textbf{limit}\\ \textbf{(MPa)}} \\ \midrule
\multirow{2}{*}{\textbf{TZM}} & \multirow{2}{*}{P/M,  Aw} & \multirow{2}{*}{50 mm bar} & \multirow{2}{*}{Push-pull} & \multirow{2}{*}{10$^7$} & -1, RT, 25 & \textbf{440} \\
 &  &  &  &  & -1, 850 $^{\circ}$C, 25 & \textbf{250} \\ \midrule
\multirow{4}{*}{\textbf{W}} & \makecell{Sintered +\\HIP} & 5 mm bar & Push-pull & 2x10$^6$ & $\sim$0, RT, 25 & \textbf{180} \\
 & \multirow{3}{*}{Forged} & \multirow{3}{*}{90 mm thick plate} & \multirow{3}{*}{Push-pull} & \multirow{3}{*}{2x10$^6$} & $\sim$0, RT, 25 & \textbf{170-200} \\
 &  &  &  &  & $\sim$0, 280 $^{\circ}$C, 25 & \textbf{250} \\
 &  &  &  &  & $\sim$0, 480 $^{\circ}$C, 25 & \textbf{190} \\ \midrule
\multirow{2}{*}{\textbf{Ta}} & \multirow{2}{*}{Rxx} & \multirow{2}{*}{2 mm thick plate} & \multirow{2}{*}{Push-pull} & \multirow{2}{*}{1x10$^7$} & -1, RT, 0.05 & \textbf{180} \\
 &  &  &  &  & -1, RT, 10 & \textbf{210} \\ \midrule
\textbf{Ta2.5W} & Rxx & 1 mm thick plate & \makecell{Bending \\ fatigue} & \makecell{1x10$^7$ \\ 50 \% fracture} & -1, RT, 25 & \textbf{270-310} \\ \bottomrule
\end{tabular}
\end{table}

\subsubsection{Materials characterization campaign}

 Literature review was convenient to validate the target materials selection and define a preliminary materials database for the FEM simulations. However, specific materials properties obtained from representative material is required to validate the target design. Thus, a dedicated characterization campaign was launched for the target materials within the framework of the BDF Project. 
 
 The justifications and objectives of a specific characterization campaign were the following:
 
\begin{itemize}
    \item The material products for W and TZM, which are big forged cylinders, are not conventional products. Mechanical properties in refractory metals highly depend on the applied deformation below recrystallisation temperature. Deformation application is limited (in quantity and homogeneity) due to the big dimensions of the forgings. Hence, microstructure and mechanical properties highly differ between big forged cylinders and conventional rolled sheet, being the latter the main source of literature data. One clear example of these differences is the static mechanical properties for TZM presented in Section~\ref{Sec:TGT:mat:litprops};
    \item Target materials will be subject to the HIP cycle necessary to bond the cladding material to the target material. It represents a thermal treatment of roughly 3 h at temperatures which might exceed the recrystallisation temperatures of the target materials (most certainly for Ta and Ta2.5W). Material properties might be affected due to several factors: grain growth, recrystallisation, segregations, gas absorption... Literature of the material properties after these thermal treatments is scarcely available and not representative;
    \item There are material properties, sampling procedure or testing parameters in literature that are not reported or that differ from the target application conditions. The effect of these differences on the properties is unknown, but can eventually become significant. One example is dynamic mechanical testing, which is extremely sensible to specimen geometry, surface preparation and testing parameters. Ref.~\cite{HABAINY2015} reported endurance limit in 150 MPa in sintered + HIP W, which increased up to 240 MPa only with specimen polishing (electro-polishing) before testing. Another example is the sensibility of BCC metals to strain rate; in Ref.~\cite{PAPAKYRIACOU2001}  several refractories materials were tested for fatigue at high number of cycles, reporting endurance values of 250 MPa and 350 MPa for Ta at low (100 Hz) and high (20 kHz) frequencies respectively. Only the frequency change affected significantly the endurance value and even changed the fracture mode of the samples from ductile to brittle when increasing testing frequency.
\end{itemize}

\subsubsubsection{Testing matrix}

The material characterization was carried out on the same materials than the used for the BDF target prototype (see Section~\ref{Sec:TGT:Proto}), with the same characteristics and production route, foreseen to be highly representative of the final target. The characteristics for each material are given in Table~\ref{tab:TGT:mats:T6mat}.

\begin{table}[htb]
\caption{List of materials employed for the characterization with the respective productions routes, product form, densification method and product dimensions. These corresponds to the blocks acquired for the execution of the BDF target prototype tests reported in Section~\ref{Sec:TGT:Proto}. P/M: Powder Metallurgy, EB: Electron Beam melting}
\label{tab:TGT:mats:T6mat}
\center
\begin{tabular}{@{}lllll@{}}
\toprule
\textbf{Material} & \textbf{Production route} & \textbf{Product} & \textbf{Dimensions} & \textbf{Densification route} \\ \midrule
TZM & P/M & Rod & 80 mm diameter & Radial forging (2D) \\ \midrule
W & P/M & Rod & 80 mm diameter & HIP \\ \midrule
Ta & EB & Sheet &  1.5 mm thickness& Rolling \\ \midrule
Ta2.5W & EB & Sheet & 1.5 mm thickness & Rolling \\ \bottomrule
\end{tabular}
\end{table}

\begin{figure}[htb]
    \centering
    \includegraphics{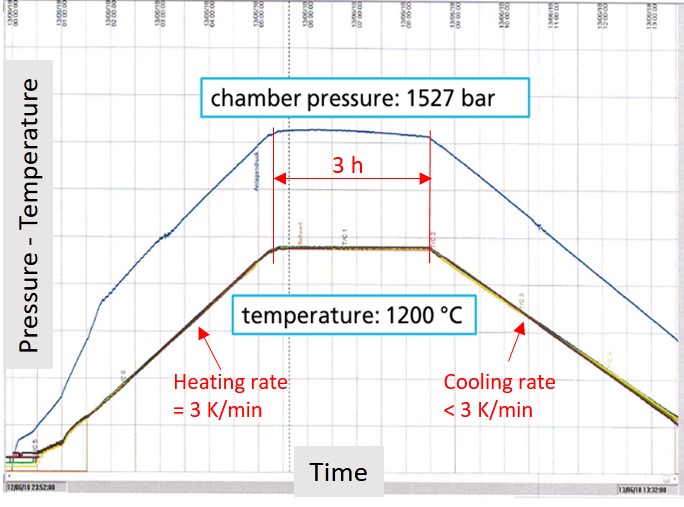}
    \caption{Pressure and temperature evolution during the HIP cycle.}
    \label{fig:TGT:mats:HIPchara}
\end{figure}

\begin{figure}[htb]
    \centering
    \includegraphics{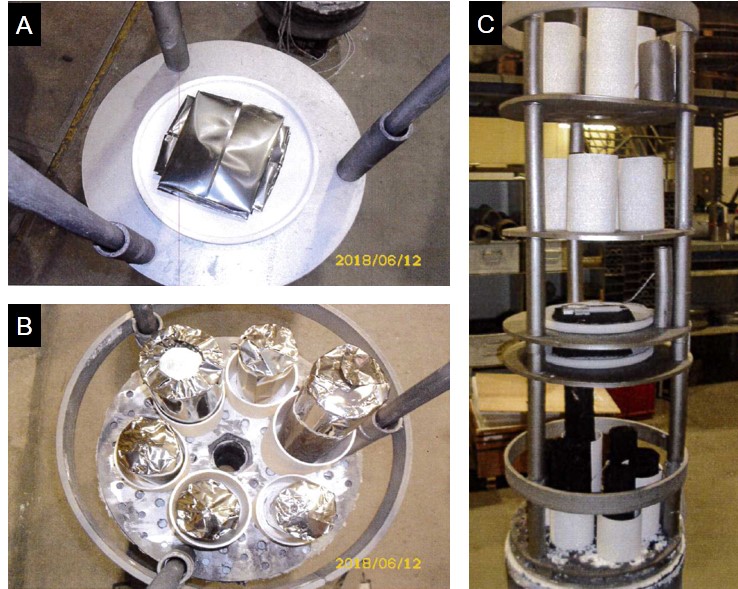}
    \caption{Images of the setup for the HIP. a) Ta and Ta2.5W sheet wrapped with getter foils before the HIP, b) W and TZM cylinders wrapped with getter foils before the HIP, c) full set of samples after the HIP, d) Pressure and temperature evolution during the HIP cycle.}
    \label{fig:TGT:mats:HIPsetup}
\end{figure}

All the materials followed afterwards a HIP cycle analogous than the one followed by the target blocks to bond the target and cladding material, in order to obtain fully representative microstructure. Pressure and temperature were ramped up to 1500 bar and 1200 °C respectively and hold for 3 hours. Evolution of pressure and temperature during the HIP cycle is also given in Figure~\ref{fig:TGT:mats:HIPchara}. Images of the materials, wrapped in getter foils, before and after the HIP cycle are given in Figure~\ref{fig:TGT:mats:HIPsetup}.

The characterization was focused in measuring the relevant properties for the four target materials in the entire operational temperature range. Three groups of properties were measured, the thermo-physical (specific heat capacity, linear thermal expansion and thermal diffusivity or thermal conductivity), the static mechanical (tensile testing) and dynamic mechanical properties (high cycle fatigue testing). The testing conditions for the characterization are given in Table~\ref{tab:TGT:mats:thermo}, Table~\ref{tab:TGT:mats:meca} and Table~\ref{tab:TGT:mats:fatiguedata} respectively. The characterization results were not yet available at the time of writing this document, but will be part of a dedicated publication.

\begin{table}[htb]
\centering
\caption{Characterized thermo-physical properties of the BDF target materials with the corresponding testing conditions.}
\label{tab:TGT:mats:thermo}
\begin{tabular}{@{}cc@{}}
\toprule
\textbf{Properties} & \textbf{Test conditions} \\ \midrule
Specific heat capacity (Cp) & \multirow{3}{*}{RT - 500 $^{\circ}$C} \\ \cmidrule(r){1-1}
Thermal Diffusivity or thermal conductivity &  \\ \cmidrule(r){1-1}
Linear thermal expansion (CTE) &  \\ \bottomrule
\end{tabular}
\end{table}

\begin{table}[htb]
\caption{Characterized static mechanical properties of the BDF target materials with the corresponding testing conditions.}
\label{tab:TGT:mats:meca}

\begin{tabular}{cccccc}
\toprule
\multirow{2}{*}{\textbf{Properties}} & \multicolumn{5}{c}{\textbf{Test conditions}} \\ \cmidrule(l){2-6} 
 & Standard & Temperature ($^{\circ}$C) & \makecell{Strain \\ rate (s$^{-1}$)} & \makecell{Sampling \\direction} & Atmosphere \\ \midrule
\multirow{10}{*}{\makecell{-Young Modulus\\-Yield strength\\-Tensile strength\\ -Elongation at\\ failure}} & \multirow{10}{*}{\makecell{-ASTM E8 \\(RT)\\-ASTM E21\\(HT)}} & \makecell{TZM: RT-125-\\200-450-800\\ W: RT-125-200-\\350-450-800\\ Ta and Ta2.5W:\\ RT-125-200-\\300-450} & 1x10$^{-3}$ & \makecell{TZM: Axial\\ W: Axial\\ Ta: Long.\\ Ta2.5W: Long.} & \multirow{10}{*}{\makecell{-Inert gas for\\ Ta and Ta2.5W\\ at T \textgreater 250 $^{\circ}$C\\-Inert gas for\\ TZM and W \\ 
at T \textgreater 400 $^{\circ}$C}} \\ \cmidrule(lr){3-5}
 &  & \makecell{TZM: 200\\ W: 200\\ Ta: 200\\ Ta2.5W: 200} & 1x10$^{-1}$ & \makecell{TZM: Axial\\ W: Axial\\ Ta: Long.\\ Ta2.5W: Long.} &  \\ \cmidrule(lr){3-5}
 &  & \makecell{TZM: 200\\ W: 200\\ Ta: 200\\ Ta2.5W: 200} & 1x10$^{-3}$ & \makecell{TZM: Radial\\ W: Radial\\ Ta: Trans.\\ Ta2.5W: Trans.} &  \\ \bottomrule
\end{tabular}
\end{table}

\begin{table}[htb]
\small

\centering
\caption{Characterized dynamic mechanical properties of the BDF target materials with the corresponding  testing conditions.}
\label{tab:TGT:fati}
\begin{tabular}{@{}llllllll@{}}
\toprule
\multirow{2}{*}{\textbf{Properties}} & \multicolumn{7}{l}{\textbf{Testing conditions}} \\ \cmidrule(l){2-8} 
 & \begin{tabular}[c]{@{}l@{}}Temperatures\\ ($^{\circ}$C)\end{tabular} & \begin{tabular}[c]{@{}l@{}}Stress \\ ratio\end{tabular} & \begin{tabular}[c]{@{}l@{}}Frequency\\ (Hz)\end{tabular} & \begin{tabular}[c]{@{}l@{}}Number \\ of cycles\end{tabular} & \begin{tabular}[c]{@{}l@{}}Sampling \\ direction\end{tabular} & Atmosphere & \begin{tabular}[c]{@{}l@{}}Test \\ method\end{tabular} \\ \midrule
\begin{tabular}[c]{@{}l@{}}Fatigue \\ strength\end{tabular} & \begin{tabular}[c]{@{}l@{}}TZM: RT-200\\ W: RT-200\\ Ta: RT-200\\ Ta2.5W: RT-200\end{tabular} & $~$0 & <100 & 10$^7$ & \begin{tabular}[c]{@{}l@{}}TZM: Axial\\ W: Axial\\ Ta: Long.\\ Ta2.5W: Long.\end{tabular} & \begin{tabular}[c]{@{}l@{}}Inert gas \\ for T=200 $^{\circ}$C\end{tabular} & \begin{tabular}[c]{@{}l@{}}Staircase \\ method\end{tabular} \\ \bottomrule
\end{tabular}
\end{table}

\subsection{Studies and development of target cladding via HIPing process} 
\label{Sec:TGT:mat:RandD}

\subsubsection{Cladding technique selection}
During the preceding years (2015-2018), strong efforts were already dedicated to obtain a reliable bonding between W and Mo cylindrical geometries and surrounding Ta external protective layers.

\begin{figure}[htb]
\includegraphics[width=\linewidth]{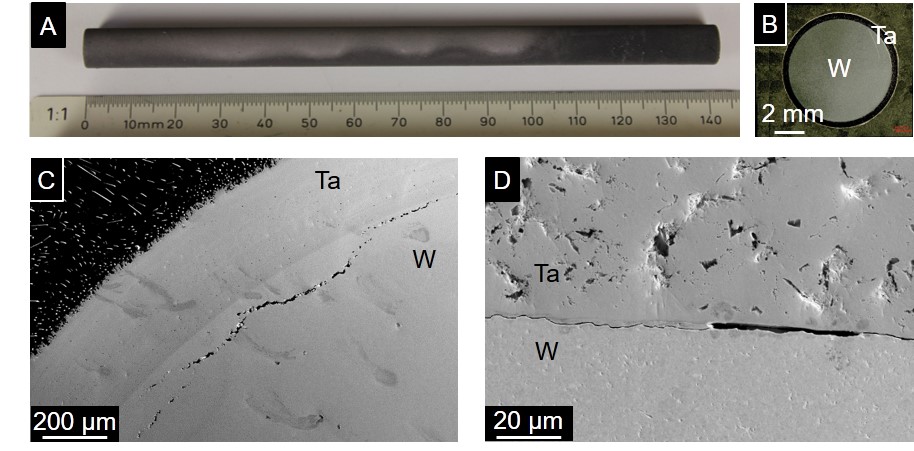}
\caption{\label{fig:TGT:mats:Tantaline} a) Image of one Ta coated W rod sample, b) optical microscope image of a rod cross-section, c)-d) Secondary Electron Microscope (SEM) images of a rod cross section at the Ta-W interface.}
\end{figure}

First investigations explored the application of a Ta layer via coating. Chemical vapour deposition (CVD) technique was employed to grow Ta layers with thickness ranging from 0.25 mm to 0.5 mm directly on W rods of 20 mm diameter. Several sample rods were coated as shown in Figure~\ref{fig:TGT:mats:Tantaline}A. Metallographic inspections at the interface level revealed punctual detachments and cracks which eventually propagated into the bulk W (see Figure~\ref{fig:TGT:mats:Tantaline}B-D)~\cite{EDMS1513398}. 

\begin{figure}[htb]
\includegraphics[width=\linewidth]{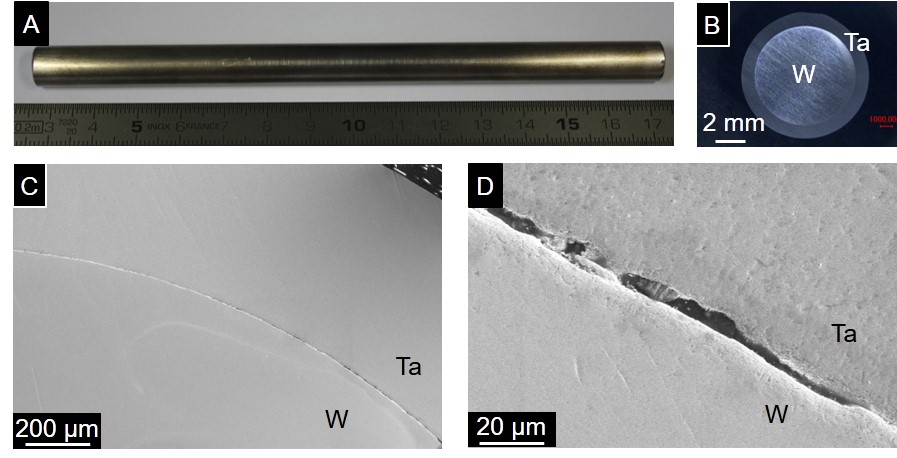}
\caption{\label{fig:TGT:mats:Hempel}a) Image of one W rod sample fitted inside a Ta tube, b) optical microscope image of a rod cross-section, c)-d) Secondary Electron Microscope (SEM) images of a rod cross section at the Ta-W interface}
\end{figure}

The second attempt was undertaken from an opposite approach. W rods of 20 mm diameter were fitted inside 1 mm thick Ta tubes and the contact between the two materials maintained by Ta elasticity. Various samples were produced as shown in Figure~\ref{fig:TGT:mats:Hempel}A. Metallographic inspections at the interface level pointed out recurrent detachments up to few microns width (see Figure~\ref{fig:TGT:mats:Hempel}B-D)~\cite{EDMS1538148}.

Following these trials, solid state diffusion bonding was considered as a potential via, since it was already applied for refractory metals in the 1960s as explained in Ref.~\cite{TARR1965} and \cite{METCALFE1963}. Simultaneous application of elevated pressure and temperature maintained over a certain time would eventually lead to surface diffusion processes between the Ta and and W or TZM. 

The candidate technique was Hot Isostatic Pressing (HIP). The utilization of this technique to clad Ta to W was first reported in Ref.~\cite{BROOME1997} to protect a water cooled W target at ISIS (Didcot, United Kingdom). The construction of successive spallation neutron targets at KENS (KEK, Japan) in 2000 and at LANSCE (LANL, United States) in 2010 based on this concept significantly increased the knowledge on this application. On the contrary, no literature reported the HIP diffusion bonding between Ta and Ta-alloys and Mo alloys.

\begin{figure}[htb]
\includegraphics[width=\linewidth]{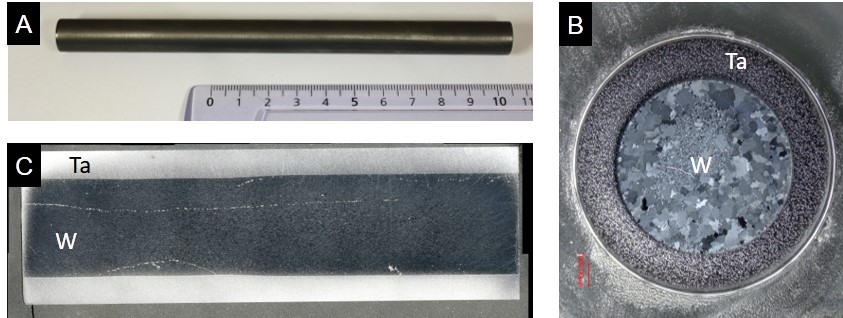}
\caption{\label{fig:TGT:mats:ATM}a) Image of one HIP assisted diffusion bonded Ta on a W rod, b) optical microscope image of a rod cross-section, showing the significant W grain growth observed after the HIP cycle and c) optical microscope image of a rod longitudinal cross-section.}
\end{figure}

Several actors in the refractory metals industry were asked to provide W, Mo and TZM cylinders of different diameters clad with a Ta layer of 2 mm diffusion bonded by means of HIP. ATM (Beijing, China) provided two Ta clad W prototypes (see Figure~\ref{fig:TGT:mats:ATM}A) in which diffusion bonding was successful, with homogeneous and defect-free interfaces. However, the cladding thickness was found variable along the prototypes and the W material presented heterogeneous grain growth and some cracking (see Figure~\ref{fig:TGT:mats:ATM}B-C) \cite{EDMS1566868}.

\begin{figure}[htb]
\includegraphics[width=\linewidth]{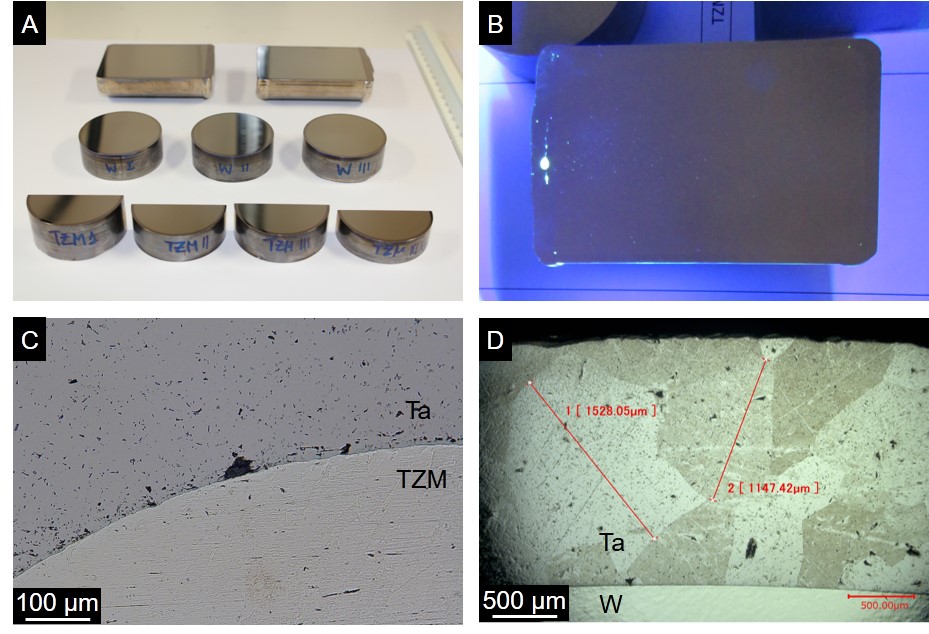}
\caption{\label{fig:TGT:mats:Plansee}a) Image of some HIP assisted diffusion bonded Ta on W cylinders after cutting, b) image of one prototype longitudinal cross-section after fluorescent dye penetrant testing, which revealed some indications at the interface level, c) SEM image at the Ta-TZM interface level, d) optical microscope image of the Ta layer microstructure.}
\end{figure}

Plansee (Reutte, Austria) supplied eight prototypes of W, Mo and TZM~\cite{TORREGROSSA2016}, some of them visible in  Figure~\ref{fig:TGT:mats:Plansee}A after cutting. The diffusion bonding with Ta was successful and additionally all the prototypes presented homogeneous interfaces and cladding thickness. Ta layer was first applied by using a welded Ta tube covering the cylindrical surface and two Ta plates covering the cylinder flat covers. The two plates were welded to the tube by Tungsten Inert Gas (TIG) welding before the HIP to obtain an hermetic seal. Interfacial defects were observed at the vicinity of the weld (see Figure~\ref{fig:TGT:mats:Plansee}B-C). Furthermore, since Ta tube was welded, significant Ta grain growth was observed in the weld surroundings (see Figure \ref{fig:TGT:mats:Plansee}D) \cite{EDMS1749441}.

 Moreover, in order to solve the issues observed in the prototypes some actions were taken: subsequent prototypes would employ seamless Ta cladding tubes, the welds previous to the HIP operation would be carried out at the cylinder edges and with an external lip not to interfere with the diffusion bonded interface and welds would be carried out employing electron beam welding to minimize the energy deposition and consequent Ta grain growth. 
 
\begin{figure}[htb]
\includegraphics[width=\linewidth]{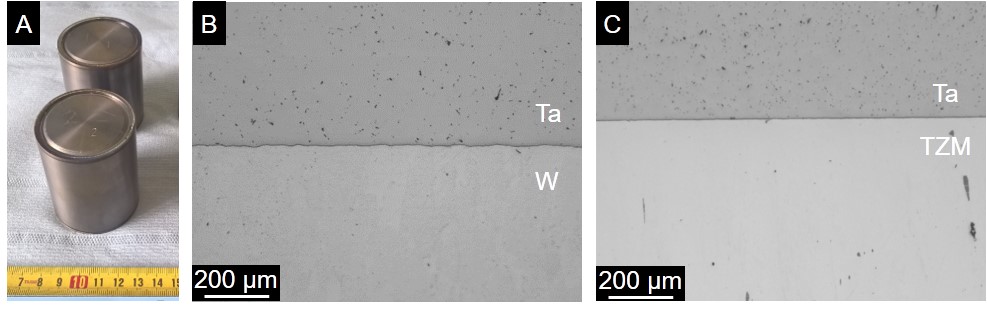}
\caption{\label{fig:TGT:mats:FH}a) Image of HIP assisted diffusion bonded Ta on W and TZM cylinders, b) SEM image at the Ta-W interface level, C) SEM image at the Ta-TZM interface level.}
\end{figure}

 In views of the promising results HIP was fixed as baseline technique to bond a Ta layer of 1 mm - 2 mm to the target materials. Numerous reduced-scale target prototypes were produced and characterized since then with continuous improvement in the framework of a collaboration with Fraunhofer - IFAM (Dresden, Germany) (see Figure~\ref{fig:TGT:mats:FH}). Successful diffusion bonding with homogeneous interfaces free of defects were achieved.

\begin{figure}[htb]
\includegraphics[width=\linewidth]{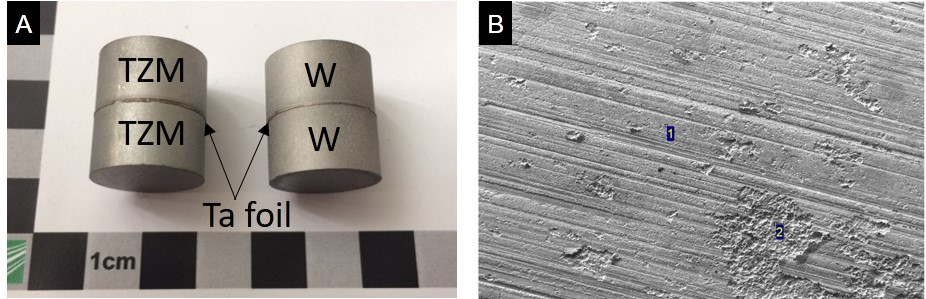}
\caption{\label{fig:TGT:mats:SPS} a) Image of two samples where the bonding between target materials TZM-TZM and W-W was explored, b) SEM image of a W cylinder surface after bonding operation. Bonding was limited to the areas with "machining like" lines.}
\end{figure}

In parallel, following the fabrication potential issues for the biggest W blocks required for the BDF target (see Section~\ref{Sec:TGT:mat:manufac}), development of the diffusion bonding of two forged blocks coaxially was explored with Fraunhofer - IFAM.
Preliminary exploration of this option was carried out employing Spark Plasma Sintering (SPS) to diffusion bond small cylinders co-axially (see Figure~\ref{fig:TGT:mats:SPS}A). The bonding operation between cylinder couples of W-W and TZM-TZM was carried out at 1200 $^{\circ}$C with an axial pressure of 45 MPa for 1 h. Diffusion phenomena was limited to few regions and it was concluded that the pressure applied was too low (see Figure~\ref{fig:TGT:mats:SPS}B). Hence, it was decided to explore HIP assisted diffusion bonding because of the higher applicable pressure (up to 200 MPa).

Successful bonding of TZM -TZM and Mo-Mo was already reported in Refs.~\cite{METCALFE1963} and \cite {HAMMOND1973} but all employing interfacial aids such are metal and refractory metal foils (Ta, Re, V, Ni, Ti....) of few tens of microns thickness. The effectiveness and/or the need of interfacial aids to diffusion bond W to W and TZM to TZM couples required specific development.
 
\subsubsection{Studies for HIP assisted diffusion bonding development}

Several points related to the materials bonding still remained open. Therefore an ambitious specific R\&D study was launched in collaboration with Fraunhofer - IFAM. The results presented hereafter have been published in Ref.~\cite{HIP_Busom}.

The objectives of the study were multiple. First, to validate the diffusion bonding by HIP for all the cladding and target candidate materials for BDF, specially for the alternative cladding material Ta2.5W. Second, to study the potential bonding between several target cylinders by HIP assisted diffusion bonding. Third, to study the potential use of interfacial aids (as Ta foil) in the two previous cases. In parallel there was also an interest to explore different HIP parameters in order to optimize the bonding properties and to study how the HIP cycle affects the bulk materials. 

To this aim, several down-scaled target block prototypes were built in order to extract bonding specimens for the subsequent characterization at three levels: micro-structural, mechanical and thermal.

Several similar studies, specially those from Ref.~\cite {HIP2} and \cite{LANSCEcladding} dedicated efforts on improving prototype preparation routines, ameliorate the interface cladding-target and solving surface state related problems. Prototype preparation routine was already optimized in previous internal studies (as described here-above) and no surface related problems were observed in this work, thanks to the application of vast prior experience from \cite {HIP2,LANSCEcladding} as well as private communications with RAL/ISIS (UK). Hence, the study was specifically addressed to the cladding-target interface. 

\subsubsubsection{Materials for prototyping}

Four materials were employed in the fabrication of the prototypes, unalloyed W (W) and molybdenum alloy "TZM" as target materials and unalloyed tantalum (Ta) and tantalum alloy "Ta2.5W" as cladding materials. W was supplied by Plansee in form of forged and annealed 25 mm - 50 mm diameter rods, with a minimal density of 99.97 \% and a hardness of HV 420-480. TZM was also supplied by Plansee in form of forged and annealed 25 mm - 50 mm diameter rods, with a hardness of HV 250-310. Ta and Ta alloy with 2.5\% W (Ta2.5W) were supplied by Plansee and WHS Sondermetalle respectively, both in form of plates, seamless tubes and foils and all in the annealed state. The surface of all the materials employed in the study did not present specific surface preparation further than conventional machining. Henceforth, the materials are referred as "W", "TZM", "Ta" and "Ta2.5W".

\subsubsubsection{Prototypes configuration}

For this study, prototypes with geometries equivalent to the final target configuration were considered to already anticipate any geometrical issue in the conception of the target.

The prototypes were built using of one or two cylinders of the target material (W or TZM) with diameters between 25~mm and 50~mm and a total lengths between 50~mm and 100~mm. The cylinders were fitted inside a tube of cladding material (Ta or Ta2.5W) and closed from the two sides by two covers also from cladding material. The cladding material thickness was 1.5 mm in the cylindrical part but it was increased to 10 mm for the two covers to allow the extraction of bigger bonding specimens, required for mechanical testing. The bonding of the cladding material to the target material was created by HIP, technique already introduced in section~\ref{Sec:TGT:mat:HIP}.

\begin{figure}[htb]
\includegraphics[scale=0.95]{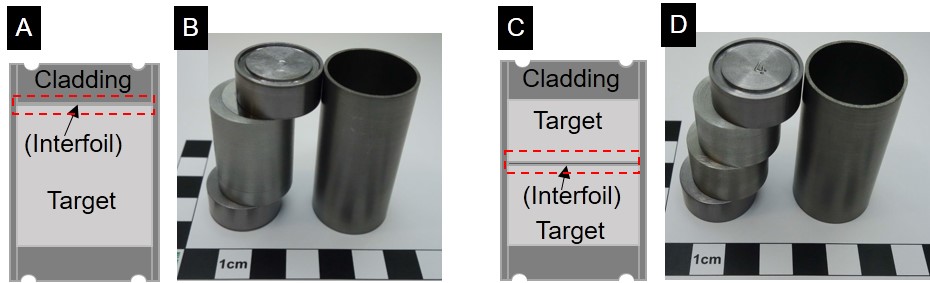}
\caption{\label{fig:TGT:mats:protos} a) cross-sectional schema of the prototypes to study the target to cladding materials bonding, b) image of the target to cladding materials prototype components before assembly, c) cross-sectional schema of prototypes to study the target to target materials bonding and d) image of the target to target materials prototype components before assembly~\cite{HIP_Busom}.}
\end{figure}

Two types of prototypes were built, either with a single or double target cylinders. The single cylinder prototypes were used to study the target to cladding materials bonding whilst the double cylinder prototypes were intended to study the target to target materials bonding.  The two types of prototypes are schematized in Figure~\ref{fig:TGT:mats:protos} with the respective components. Ta interfoils of 50\textmu m thickness were eventually introduced between the two materials (cladding/target or target/target) as bonding interfacial aids.

\subsubsubsection{Prototypes fabrication}

Before assembly of each prototype, the target material cylinder and the two cladding material covers were machined to a diameter 100\textmu m smaller than the tube. This tolerance requirement aimed to ensure sufficient spacing for the introduction of the target material cylinder inside the cladding material tube but narrow enough to avoid cladding corrugation during the HIP cycle.

\begin{figure}[ht]
    \centering
    \includegraphics[width=\linewidth]{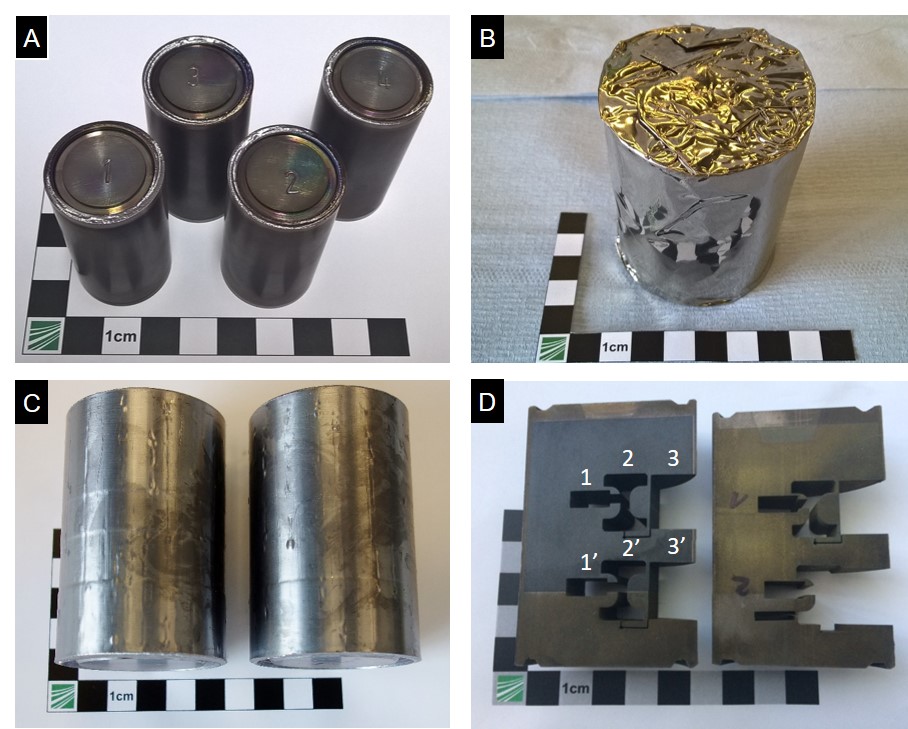}
    \caption{a) Photograph of four prototypes after welding of the Ta / Ta2.5W capsule, b) Photograph of one prototype wrapped with the getter foils before the HIP cycle, c) Photograph of two prototypes after the HIP cycle and removal of the getter foils. Cladding deformation and corrugation is appreciable d) Photograph of two half prototypes after cross-sectional cut with EDM . Specimen extraction can be appreciated in 1,2,3 for target to target materials bonding and 1', 2', 3' for cladding to target materials bonding. Specimens "1" are for thermal conductivity measurements, "2" for tensile testing and "3" for interface microscopy  ~\cite{HIP_Busom}.}
    \label{fig:TGT:mats:protofab}
\end{figure}

A continuous and hermetic (gas tight) cladding material capsule covering the target material cylinder was required to ensure application of isostatic pressure between the cladding and target materials during the HIP cycle. To this aim, electron beam welding was employed to weld the tube and both covers of each prototype. The covers presented a machined lip in order to facilitate the welding operation and to maintain the heat affected zone far from the cladding-target materials interface. Electron beam welding was performed at STD Strahltechnologie (Dresden, Germany). The prototypes were evacuated for one night before welding to 10$^{-2}$ mbar to remove the entrapped air between the cladding material and the target material. After the welding, each prototype was helium leak tested in order to ensure the tightness of the cladding material capsules around the target material cylinders. Images of the capsules production are reported in Figure~\ref{fig:TGT:mats:protofab}.

HIP cycle was carried out for the prototypes to diffusion bond the cladding material to the target material and eventually also the target to target materials. HIP cycle was carried out by HIP PM Volker (Dorfen, Germany). Especial attention was dedicated to the HIP furnace atmosphere to avoid the maximal of Ar impurities which could potentially be absorbed by the cladding surface. Ref.\cite{HIP2} studied the effect different atmosphere purity of HIP furnace during the HIP cycle to bond Ta cladding to W. The growth of TaC and Ta$_2$O$_5$ fragile layers was observed on the Ta surface due to atmosphere gas and furnace impurities. From this valuable experience, the HIP atmosphere for the study was chosen as Argon 5.0 purity grade in a furnace with Mo heater. Additionally all the prototypes were wrapped with 1 Ta and 1 Zr foil to getter all the remaining impurities in the furnace atmosphere.

The heating rate during the HIP cycle was fixed in 10 K/min and the dwell time at the nominal temperature and pressure was 3 h. HIP cycle temperature and time are determining parameters in the plastic deformation, inter-diffusion kinetics and microstructure of the prototype materials. In this work, two HIP cycles parameters were employed: 1200 \textdegree{}C and 150 MPa (from now on referred as "L") in which temperature was retained below the recrystallization temperatures of the W and the TZM \cite{TantalumW2}, in order to achieve diffusion bonding and at the same time intending to preserve the refined microstructure in the target materials; 1400 \textdegree{}C and 200 MPa (from now on referred as "H") in which temperature was raised above the recrystallization temperature of W and TZM target materials with the objective of obtaining the best diffusion bonding even if partially sacrificing target material properties due to recrystallization or grain growth.

After the HIP cycle, electro-discharge machining (EDM) was employed to extract specimens from the prototypes. The specimens were extracted from the areas marked in red in Figure~\ref{fig:TGT:mats:protos}. The target to cladding materials interface specimens were extracted from the prototypes shown in Figure~\ref{fig:TGT:mats:protos}A and the target to target material interface specimens from the prototypes shown in Figure~\ref{fig:TGT:mats:protos}C.

Interface microstructure inspections were carried out with SEM. Samples were prepared by conventional mechanical grinding and polishing. Interface strength was measured with tensile testing, carried out on miniaturised tensile specimens of 5 mm gauge length and squared section of 4 mm$^2$. When measuring the bonded interfaces strength, the interface was coincident with the minimal specimen section. Due to the small specimen gauge length, elongation could not be precisely measured, therefore only values of tensile strength are reported. Testing was carried out at a strain rate of 2x10$^{-4}$s$^{-1}$. Hardness measurements were carried out on bulk materials employing a Falcon 500 system from INNOVATEST with a vickers indenter and a load of 5 N (HV5).

The thermal conductivity was indirectly measured by employing the following relation:
\[\lambda = \alpha * C_{p} * \rho\]
where $\alpha$ is the measured thermal diffusivity, $C_{p}$ is heat capacity and $\rho$ is the density. 

\subsubsubsection{Testing matrix}

The number of prototypes built for the study was determined in order to obtain all the combinations of target material to cladding materials, target to target materials, use of Ta interfoil, and HIP cycles. The list of the specimens under study resulting from all these combinations is given in Table~\ref{fig:TGT:mats:speci}.

\begin{table}[htb]
\caption{\label{fig:TGT:mats:speci} List of the BDF target-related specimens studied in this work with the corresponding materials, use of interfoil and HIP cycle~\cite{HIP_Busom}.
}
\begin{tabular}{|l|l|l|l|l|}
\hline
\multirow{2}{*}{Bonding specimen number} & \multicolumn{4}{l|}{Target to cladding} \\ \cline{2-5} 
 & Target material & Cladding material & Interfoil & HIP cycle \\ \hline
1 & \multirow{3}{*}{TZM} & \multirow{2}{*}{Ta2.5W} & o & \multirow{6}{*}{L} \\ \cline{1-1} \cline{4-4}
2 & & & - & \\ \cline{1-1} \cline{3-4}
3 & & Ta & - & \\ \cline{1-4}
4 & \multirow{3}{*}{W} & \multirow{2}{*}{Ta2.5W} & o & \\ \cline{1-1} \cline{4-4}
5 & & & - & \\ \cline{1-1} \cline{3-4}
6 & & Ta & - & \\ \hline
7 & \multirow{3}{*}{TZM} & \multirow{2}{*}{Ta2.5W} & o & \multirow{6}{*}{\begin{tabular}[c]{@{}l@{}}H\\ \end{tabular}} \\ \cline{1-1} \cline{4-4}
8 & & & - & \\ \cline{1-1} \cline{3-4}
9 & & Ta & - & \\ \cline{1-4}
10 & \multirow{3}{*}{W} & \multirow{2}{*}{Ta2.5W} & o & \\ \cline{1-1} \cline{4-4}
11 & & & - & \\ \cline{1-1} \cline{3-4}
12 & & Ta & - & \\ \hline
\multirow{2}{*}{} & \multicolumn{4}{l|}{Target to target} \\ \cline{2-5} 
 & Target material I & Target material II & Interfoil & HIP cycle \\ \hline
13 & \multirow{2}{*}{TZM} & \multirow{2}{*}{TZM} & - & \multirow{4}{*}{L} \\ \cline{1-1} \cline{4-4}
14 & & & o & \\ \cline{1-4}
15 & \multirow{2}{*}{W} & \multirow{2}{*}{W} & - & \\ \cline{1-1} \cline{4-4}
16 & & & o & \\ \hline
17 & TZM & TZM & o & \multirow{2}{*}{H} \\ \cline{1-4}
18 & W & W & o & \\ \hline

\end{tabular}
\end{table}

\subsubsection{Results of the interface study}

\subsubsubsection{Target-cladding materials bonding }

\begin{figure}[htb]
\includegraphics[width=\textwidth]{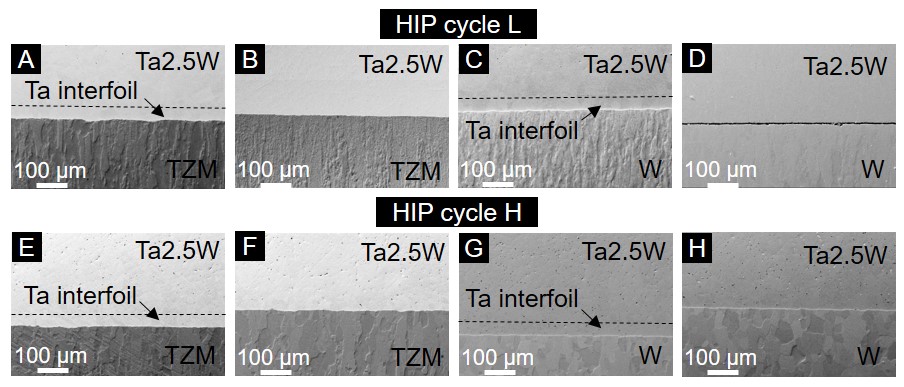}
\caption{\label{fig:TGT:mats:SEM1}Secondary electron micrographs at the interface level for the specimens under study at 150x magnifications, for the two mentioned HIP cycles~\cite{HIP_Busom}.}
\end{figure}

The bonding interfaces between target and cladding materials are observable from the SEM micrographs given in Figure~\ref{fig:TGT:mats:SEM1}. Apart from the specimens between W and Ta2.5W which did not show any apparent bonding for the HIP cycle "L" (Figure~\ref{fig:TGT:mats:SEM1}D), all the rest specimens revealed interfaces in which the two materials are in perfect contact indicating potential diffusion bonding. Difficulties were encountered to image the diffusion layer due to specimen preparation artifacts and the small layer thickness, which was measured in approximately \( 1\ \mu \)m when achievable. This value was in concordance with Ref.~\cite {HIP2}, that estimated a diffusion layer thickness between 0.5 \textmu m and 2 \textmu m for the HIP parameters employed in the study.
For the interfaces between target and cladding materials no features such as heterogeneities, retained porosity or observed. However, in the interfaces between the Ta interfoil and Ta2.5W barely visible micro-porosities aligned at the interface level were appreciable for the specimens with HIP cycle "L". Such features were not observed in the specimens with HIP cycle "H".

\begin{figure}[htb]
\centering
\includegraphics[width=\linewidth]{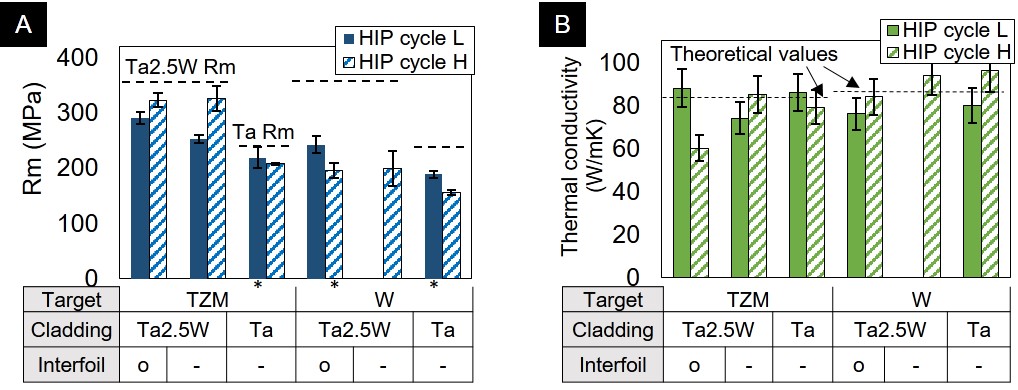}
\caption{\label{fig:TGT:mats:Tens1} a) Measured average tensile strength for each type of bonding between cladding and target materials. Ta2.5W and Ta tensile strength reported as reference. Specimens marked with * failed outside the gauge length b) measured thermal conductivity for each type of bonding between cladding and target materials. Theoretical thermal conductivity for bonding without looses at the interface are reported as reference~\cite{HIP_Busom}.}
\end{figure}

For the tensile testing, three tensile specimens were tested for each bonding combination and the averaged values are given in Figure~\ref{fig:TGT:mats:Tens1}. Both cladding materials present lower strength than the target materials, thus, all the bonding tensile strengths were theoretically limited by the latter. The tensile strengths of the cladding materials were measured in 360 MPa for Ta2.5W and 220 MPa for Ta, employing the same setup than for the interface specimens of the study. Most part of the bonding specimens failed at the interface level therefore measuring the interface strength. However, some of them failed (tests marked with *) at the base materials and therefore the interface strength might be greater. 

In general terms, strong bonding was achieved for the combinations between TZM-Ta2.5W (330 MPa), TZM-Ta (215 MPa) and W-Ta (185 MPa), all values close to the tensile strength of the respective cladding material. Nevertheless, for the couple W-Ta2.5W (230 MPa), the highest mechanical strength was 60 \% - 70 \% of the Ta2.5W tensile strength.

For TZM as target material, the strongest bonding was reached with Ta2.5W as cladding material and the HIP cycle "H", independently of the use of Ta foil. For W as target material, the highest bonding strength was achieved with Ta2.5W as cladding material but requiring the use of the Ta interfoil. Stronger bonding was always achieved for the HIP cycle "L".

Higher bonding strengths were achieved for Ta2.5W, most probably due to the higher strength of this cladding material compared to Ta. However, in certain cases (specimen W-Ta - HIP cycle "L") Ta2.5W required the use of a Ta interfoil in order to create bonding with W. 

Interfoils increased the bonding strength only in the case of Ta2.5W bonding to W and for the "L" cycle. In all the other combinations, the Ta interfoil affected negatively the bonding strengths. Nevertheless, the use of the Ta interfoil was the only option to obtain a successful bonding W to Ta2.5W with the "L" HIP cycle.

On one hand, the HIP cycle with higher temperature was useful to increase the bonding strength for the Ta2.5W claddings. At the same time, the higher temperature cycle reduced the bonding strength for the Ta claddings.

Thermal conductivity of the specimens was mainly limited by the Ta and Ta2.5W bulk material conductivity (55 W/mK), much lower compared to TZM and W (120 W/mK and 179 W/mK respectively). Thermal conductivity was measured across the cladding to target materials interface and the obtained values are given in Figure~\ref{fig:TGT:mats:Tens1}. Values are compared to the theoretical thermal conductivities, calculated with the thermal resistance law and assuming a perfect bonding without loses at the interface. The theoretical thermal conductivity for the couples W-Ta and W-Ta2.5W was calculated in 85 W/mK and for the couples TZM-Ta and TZM-Ta2.5W was calculated in 82 W/mK. Thermal conductivity for combination W-Ta2.5W without foil and at low temperature cycle was considered 0 due to the absence of mechanical bonding.
Only one specimen, TZM-Ta2.5W with foil and HIP cycle "H" showed significant reduced thermal conductivity with 60 W/mK, 73 \% respect to the theoretical. For all the other specimens, thermal conductivity values close or equal to the theoretical conductivity were observed independently of the type of bonding, ranging from 74 W/mK (89 \% of the theoretical) for specimen TZM-Ta2.5W and HIP cycle "L" to 96 W/mK (113 \% of the theoretical) for specimen W-Ta and HIP cycle "H". No clear tendency was observed between the thermal conductivity with either the target and cladding materials, the use of interfoil or the HIP cycle. 

\subsubsubsection{Target-target materials bonding}

\begin{figure}[htb]
    \centering
    \includegraphics[width=\linewidth]{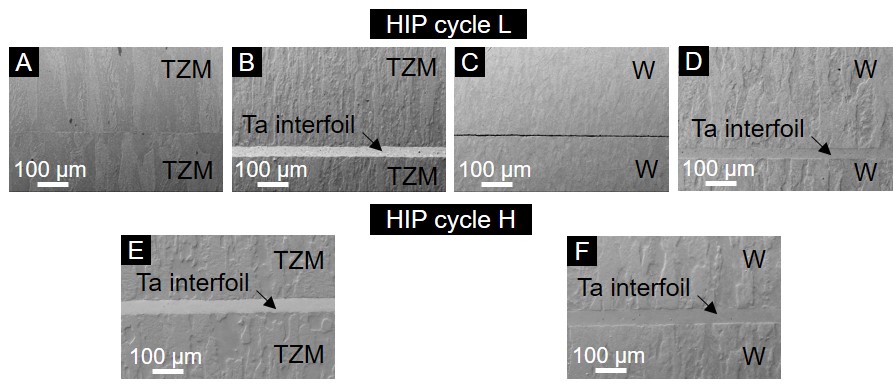}
        \caption{\label{fig:TGT:mats:SEM2tts}Secondary electron micrographs at the interface level for the specimens under study at 150x magnifications, for the two mentioned HIP cycles~\cite{HIP_Busom}.}
  \end{figure}

The bonding interfaces between target materials are observable from the SEM micrographs given in Figure~\ref{fig:TGT:mats:SEM2tts}. The interfaces between target materials presented incomplete diffusion bonding when no interfoil was used: the interface between TZM and TZM presented some bonding, but limited to some areas and leaving recurrent gaps; the interface between W and W did not presented any bonding. The interfaces between target materials, when containing an interfoil, indicated potential complete diffusion bonding. Interfaces were homogeneous with no features such pores, voids, segregation or any other heterogeneity. Diffusion layer was not discernible as for the target to cladding materials bondings.

\begin{figure}[htb]
    \centering
    \includegraphics[width=\linewidth]{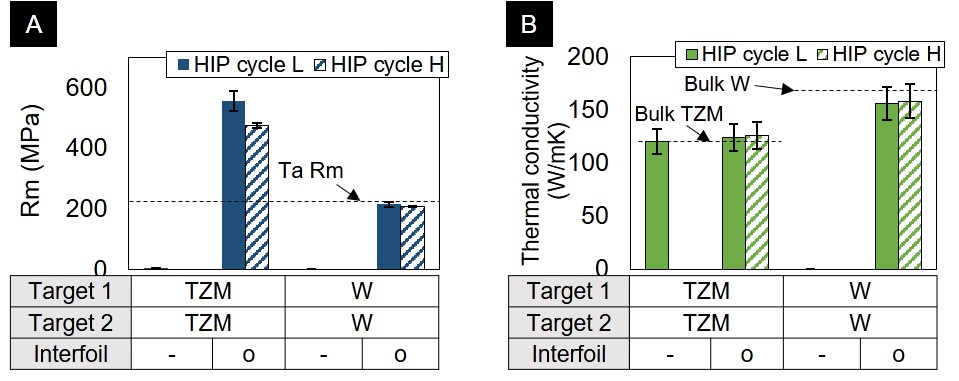}
    \caption{a) Measured average tensile strength for each type of bonding between target and target materials. Ta tensile strength reported as reference b) Measured thermal conductivity for each type of bonding between target and target materials. Bulk materials thermal conductivity are reported as reference~\cite{HIP_Busom}.}
    \label{fig:TGT:mats:Tens2}
\end{figure}

For the tensile testing of the target to target materials interface, three tensile specimens were tested for each interface combination and the averaged values are given in Figure~\ref{fig:TGT:mats:Tens2}A.  The Ta tensile strength is given in the figure for ease interpretation. All the tested specimens failed at the bonded interfaces. The bonding strength for the interfaces without interfoil presented extremely low values for TZM-TZM (5 MPa) and no mechanical bonding for W-W. However strong bonding was observed when using an intefoil. The interface W-W showed tensile strengths of 200 MPa when using an interfoil and independently of the HIP cycle, strength which is close to the Ta tensile strength. The interface TZM-TZM presents much higher tensile strengths of 550 MPa (cycle L) and 475 MPa (cycle H) when using a Ta interfoil, much stronger than the Ta tensile strength. 

Thermal conductivity was measured across the target to target materials interface and the obtained values are presented in Figure~\ref{fig:TGT:mats:Tens2}B. In the case of TZM-TZM bonding, the thermal conductivity presented values ranging between 120 W/mK and 126 W/mK, values equal to bulk TZM thermal conductivity, indicating perfect conductivity across the interface independently of the bonding parameters. The W-W bonding presented thermal conductivity values slightly lower than bulk W, from values of 126 W/mK in the bonding without interfoil to values of 158 W/mK when employing interfoils and HIP cycle H. In general terms, slight increase in the thermal conductivity was observed when using interfoil and for the HIP cycle H.

\subsubsubsection{Bulk material properties}

Temperatures of HIP cycles can be equal or higher than recrystallization temperatures of the materials under study. Subsequently, materials can present recrystallisation, grain growth, softening and degradation of thermal properties. In order to evaluate these effects, hardness and thermal conductivity measurements were additionally carried out on bulk material specimens from the prototypes.

The microstructure of the TZM and W was appreciable from the micrographs in Figure~\ref{fig:TGT:mats:SEM1} and Figure~\ref{fig:TGT:mats:SEM2tts} TZM and W showed deformed and elongated grains after HIP cycle "L" indicating a microstructure as worked, without indications of recrystallization. Similar microstructure but with slight larger grains was observed after the HIP cycle ("H"), most probably due to slight grain growth. This growth was more pronounced for W and even equiaxial microstructure was observed in Figure~\ref{fig:TGT:mats:SEM1}H, indicating potential recrystallization. The microstructure of Ta and Ta2.5W was not appreciable from the micrographs but no significant differences were expected, since the temperature and time of the two HIP cycles was significantly above from the recrystallization temperatures of the materials.

\begin{figure}[htb]
\centering
\includegraphics[width=\linewidth]{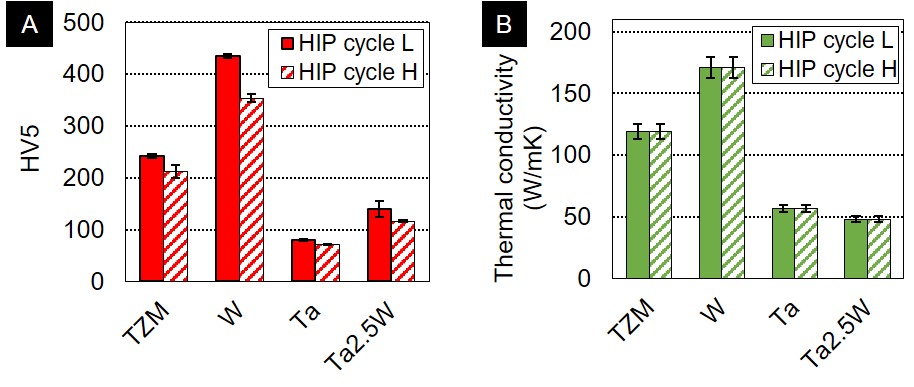}
\caption{\label{fig:TGT:mats:Tens3} a) HV5 micro-hardness and b) thermal conductivity values for the four materials after the respective HIP cycles~\cite{HIP_Busom}.}
\end{figure}

Vickers hardness values were measured for the four materials and after each HIP cycle. The measured values are plotted in Figure \ref{fig:TGT:mats:Tens3}A. The four materials presented lower hardness values for HIP cycle "H" respect to the hardness values after the HIP cycle "L". Reduction was of similar levels for the four materials: 13\% in TZM, 19\% in W, 10\% in Ta and 17\% in Ta2.5W.

Thermal conductivity of the four materials was measured after each HIP cycle. The measured values values are given in Figure \ref{fig:TGT:mats:Tens3}. No significant changes of the thermal conductivity were observed in the materials due to the different HIP cycles.

\subsubsection{Discussion}

Successful diffusion bonding was achieved for all the combinations of target materials (TZM and W) and cladding materials (Ta and Ta2.5W), only some difficulties were encountered when bonding W and Ta2.5W at HIP cycle "L". Microscopy inspection of the interfaces did not bring exploitable information of the diffusion bonding extent since the diffusion layers were extremely thin (around 1 \textmu m). However, microscopy proved useful to appreciate the morphology of the interfaces, which apart from the above-mentioned exception, were homogeneous and did not present any defect such as voids, pores or segregations. Only in the case of Ta to Ta2.5W interfaces and for the cycle "L" slight retained porosity was observed at the interface level.

Diffusion bonding evaluation at the interfaces was mainly assessed by parameters which are critical for the application and performance of the joints such are the interfacial mechanical strength and thermal conductivity.

Ta as cladding material provided satisfactory results either with TZM or W as target materials. With TZM, tensile strength of 100 \% Ta yield strength and thermal conductivity of 100 \% of the theoretical interfacial thermal conductivity were achieved. For W, slightly lower performance was achieved, with tensile strength in the range of 70 \% - 90 \% of Ta yield strength but maintaining 100 \% of the theoretical interfacial thermal conductivity. 

Ta2.5W alloy was chosen in views of increasing cladding strength at BDF target operational temperatures (up to 200 \textdegree{}C). Moreover, it could additionally improve corrosion resistance to demineralised water and resistance to proton irradiation. Ta2.5W alloy was successfully validated as an alternative cladding material to the widely used Ta. Diffusion bonding to TZM was successful in all the cases, obtaining tensile strengths of 330 MPa (92\% of Ta2.5W tensile strength) and thermal conductivity of 88 W/mK (107\% of the theoretical interface thermal conductivity). Diffusion bonding to W required to use either higher temperature and pressure HIP parameters (cycle "H") or the use of a Ta interfoil. With these conditions an interface tensile strength of 240 MPa (67 \% of Ta2.5W tensile strength) and thermal conductivity of 95 W/mK (110 \% of the theoretical interface thermal conductivity) were achieved. This difficulties to bond directly W to Ta2.5W could be due to the higher strength of the Ta2.5W compared to Ta. This enhanced strength could limit the cladding plastic deformation to adapt to the W and therefore reducing the contact area where diffusion could take place. This phenomena was not observed with the TZM as target material, most probably because the increase of strength of the cladding was compensated by the TZM higher ductility.

The parameters of the HIP cycle such are the temperature and pressure were found to be of high influence in the bonding properties. For the Ta2.5W as cladding material, the use of higher T and P (cycle "H") was beneficial for generating bonding and to increase the mechanical strength. For the Ta as cladding material, good mechanical bonding was already achieved with lower T and P (cycle "L"). Lower strength was found when using higher T and P (cycle "H") most probably due to the materials softening due to recrystallisation or grain growth.

The effect of using an interfacial aid such an interfoil presented strong synergies with other factors. As example, when bonding Ta2.5W to TZM, using an interfoil increased the thermal conductivity from 74 W/mK to 88 W/mK when using only the HIP "L". Contrarily if using the HIP cycle "H" the interfoil reduced the thermal conductivity from 85 W/mK to 60 W/mK. Thus, the influence of an interfoil required assessment case by case.

Successful diffusion bonding was also accomplished in the case of two target to target material combinations (between TZM-TZM and W-W) but only when using interfacial aids (Ta interfoil). In these cases, interface microscopy could not image diffusion layers but revealed homogeneous and defect-free interfaces. Diffusion bonding extent in the interfaces was evaluated in the same manner than the target to cladding materials bonding, by  measuring the interfacial mechanical strength and thermal conductivity. In the case of bonding between W target material, high quality bonding was achieved, with a mechanical strength of 215 MPa (90\% Ta Rm) and a thermal conductivity 158 W/mK (93\% of bulk W). In the case of bonding between TZM target material, the results were outstanding, with mechanical strength of 550 MPa (250\% Ta Rm) and thermal conductivity of 126 W/mK (105\% of bulk TZM).
Contrarily to the target to cladding materials bonding, the HIP parameters did not play an important role, with only small reduction of bonding strengh which was observed for the HIP cycle "H", most probably due to bulk material softening. Nevertheless the use of interfacial aids was of extreme importance in the mechanical strength of the bondings: the bonding strength for TZM-TZM bonding increased from 5 MPa to 550 MPa and for the W-W bonding from 0 MPa to 215 MPa. Even if thermal conductivity might attain high values, the mechanical strength of the bonding without interfoil was found residual. 

In both types of bonding (cladding to target materials and target to target materials) it was found that incipient diffusion bonding is enough to obtain high values of thermal conductivity. Contrarily, tensile strength of the bondings was more indicated to assess the extent of diffusion bonding.

Decrease of 10\% - 20\% in the hardness values was found for the four studied materials after the HIP cycle "H" compared to HIP cycle "L". For the target materials higher decrease was found for W most probably due to the lower recrystallisation temperature of this materials compared to TZM, confirmed by the micrograph in Figure~\ref{fig:TGT:mats:SEM2tts}H. Significant hardness decrease was also found for both Ta and Ta2.5W. This softening phenomena could be the cause of the interfacial strength decrease for specimens with Ta as cladding material, since the decrease in the tensile strength for the specimens with cycle "H" corresponds also to 10 \% - 20 \%. The diffusion bonding for Ta cladding could be already completed with the HIP cycle "L" and the increase of T and P for cycle "H" could act only in detrimental of the Ta bulk properties. Contrarily, the bonding of Ta2.5W would not be complete for HIP cycle "L" and therefore strength is increased with the higher T and P from HIP cycle "H". Same reasoning can be expressed for the target to target materials: diffusion bonding with interfoil is already complete with HIP cycle "L" and increasing T and P leads to bonding with less strength due to bulk material softening.

\subsubsection{Conclusions}

Several reduced scale BDF target prototypes were successfully built and clad with Ta based materials by HIP. The microstructure, mechanical properties and thermal properties of the resulting bondings were studied. HIP has been proven as a valid technique to bond the candidate target materials for BDF (W and TZM) to Ta based erosion-corrosion resistant claddings in representative geometries of the final target. Several previous studies with different techniques led to not reliable bondings. In the study, bondings with homogeneous interfaces, stronger than the cladding material and with thermal conductivities equal than the theoretical were achieved in the combinations Ta-TZM, Ta-W and Ta2.5W-TZM. The same was valid for the combination Ta2.5W-W only with a slight lower strength (70\% of the cladding material).
In order to comply with the desired target geometry and at the same time employ multi-directionally forged material, the division of a target block into smaller multidirectionally forged blocks coaxially bonded inside the Ta based cladding was explored. Diffusion bonding assisted by HIP was successfully achieved to bond W to W and TZM to TZM cylindrical geometries. Homogeneous and defect-free interfaces, bonding strengths of 550 MPa and 215 MPa (for TZM-TZM and W-W respectively) and thermal conductivities equal to the theoretical were achieved. However the use of interfacial aids was mandatory in both cases.  
Mastering of key parameters such are the interfacial aids, HIP parameters revealed to be of special consideration for the final target to optimize the interface and bulk material properties.

\subsection{Target corrosion considerations} 
\label{Sec:TGT:mat:corrosion}

In this section, an assessment of the potential corrosion phenomena for the Ta and Ta2.5W, the target materials which will be exposed to the target coolant environment.

\subsubsection{Target conditions}
The target blocks will be actively cooled by demineralised water, at a temperature between RT and 110 $^{\circ}$C and at high pressures (20 bar) and velocity (5 m/s). Detailed design of the target cooling is already given in Section~\ref{Sec:TGT:CoolingCFD}. In these conditions several corrosion phenomena are potentially applicable:

\begin{itemize}
    \item General/localized corrosion of the Ta and Ta2.5W due to direct contact between target and coolant;
    \item Galvanic corrosion of the Ta and Ta2.5W due to direct contact and through-coolant contact between the target and the vessel materials;
    \item Hydrogen embrittlement of the Ta and Ta2.5W, inherent to both materials;
    \item Erosion-corrosion of the Ta and Ta2.5W due to the high velocity of the coolant and eventual particles dissolved in coolant;
    \item Irradiation-enhanced corrosion;
\end{itemize}

The applicability of each corrosion phenomena is discussed hereafter.

\subsubsection{Literature review}

\subsubsubsection{General/localized corrosion}

According to the ASM Metals Handbook, Ta shows no corrosion in de-ionized water at 40 $^{\circ}$C. Moreover, failures due to exposure to steam condensate have not been recorded, with many cases operating at temperatures up to 250 $^{\circ}$C. Only a general recommendation on maintaining the pH below 8 is given~\cite{CRAMER2005}. 

The corrosion behaviour or Ta at high temperature water was studied in Ref.~\cite{ISHIJIMA2002}. The corrosion rate of Ta was measured at in flowing water at 180 $^{\circ}$C (maximal target water temperature),  260 $^{\circ}$C, and 320 $^{\circ}$C, with stringent control of dissolved oxygen ($<$ 1 ppm). Extremely low corrosion rates in the order of 0.001 mm/year - 0.004 mm/year were reported. In addition, no occurrence of exfoliation or spallation was observed~\cite{ISHIJIMA2002}.

No specific data is available for general/localised corrosion in water environments for Ta2.5W. Nevertheless, as mentioned here-above, Ta2.5W has proven equal or improved corrosion resistance compared to Ta, in critical environments such are hot nitric and sulfuric acids, environments where Ta and alloys are generally employed \cite{GYPEN1984,AIMONE2001}.

\subsubsubsection{Galvanic corrosion}

The target blocks will be supported in a stainless steel vessel (AISI 316L) and the cooling water will circulate in between the inner and outer tank (see Section~\ref{Sec:TGT:Design}). Hence, it is of interest to assess the potential galvanic corrosion between both materials. Data about the galvanic coupling between Ta and 316L is available from internal studies in representative conditions, in demineralized water at 20$^{\circ}$C and 80$^{\circ}$C, with two oxygen levels, 0.1 ppm and 10 ppm. In all the cases, the measured galvanic currents were negligible, below 2 \textmu A/cm$^2$~\cite{EDMS1698449}.
No specific data is available for Ta2.5W but due to the close chemistry to Ta, no big differences in the corrosion potential are expected for this alloy. Further tests will be however required in the BDF Project follow-up.

\subsubsubsection{Hydrogen embrittlement}

In presence of stray currents or when coupled to a less noble material (e.g. low-carbon steel) in acid media, tantalum can become the cathode in the created galvanic cell and absorb hydrogen, with the subsequent embrittlement. Although stray currents and coupling might be transient, hydrogen absorption is cumulative in its effect. Even if no stray currents are expected, and the Ta will be coupled to a more noble metal (AISI 316L) Ta might eventually become cathodic if given sufficient time. Therefore it is desirable to prevent galvanic cell formation by providing adequate electrical insulation from the other metals~\cite{CRAMER2005}.

Substitutional alloying of Ta can reduce the hydrogen absorption. Refs.~\cite{GYPEN1984,AIMONE2001} reported reduced hydrogen absorption in hot sulfuric acid environment particularly for Ta2.5W. 

\subsubsubsection{Erosion corrosion aspects}

Erosion-corrosion considers simultaneous interactions between erosion and corrosion, with synergies which can increase significantly the metal loss further than the addition of both phenomena independently. 

Erosion-corrosion studies were reported in Ref.~\cite{BERMUDEZ2005} for several metals (AISI 304L, AISI 316L, Ti, Ta, Ta2.5W and Zr). In the areas of the specimens protected from impinging alumina particles, no alteration of the Ta or Ta2.5W surface was observed, even in aggressive media (10\% HCl). Nevertheless, due to the lower hardness of Ta compared to stainless steels, this metal was found susceptible to failure by impinging particles. Even with the additions of alloying elements such as W, the presence of abrasive particles in the coolant induced severe erosion-corrosion.

Erosion-corrosion test was carried out on a Ta cladding at RAL~\cite{CARETTA2017}. A water jet was impinging the Ta surface at 34 m/s and at an angle of 25$^{\circ}$. In this test, the water was filtered before impinging the surface and therefore erosion-corrosion was limited to liquid erosion. After more than 3000 h of testing Ta did not show any sign of pitting or erosion.

\subsubsubsection{Irradiation enhanced corrosion}
\%smallskip

 Corrosion behaviour in water and in irradiation environments presents specific conditions. The interaction of high energy protons with water produces the decomposition of water molecules into different species (radiolysis) which can attack the metal surface. Published literature data on Ta corrosion behaviour in irradiation environments is limited to Ref.~\cite{LILIARD2002}. Ta specimens were introduced in a corrosion loop with de-ionized water at 30 $^{\circ}$C. The Ta specimens were continuously irradiated by 800 MeV protons inside the corrosion loop. Thanks to careful pre-cleaning of the circuit and employing hydrogen water chemistry, the formation of radiolysis products was mitigated, and the Ta corrosion rate was extremely low, less than 0.12 microns / year.

\subsubsubsection{Remarks}

From the previous literature review on corrosion behaviour of Ta and Ta2.5W it can be considered that no significant potential issues related to corrosion are expected for Ta and Ta2.5W if several points are properly addressed in the design of the water circuits. Still, dedicated R\&D will be pursue further to quantitatively clarify this point.

\begin{itemize}
\item Water chemistry shall be carefully monitored and controlled (oxygen and hydrogen concentrations, pH, conductivity...);
\item Water shall be continuously filtered to avoid re-circulation of spallation products, either coming from the target or other parts of the cooling circuit;
\item Circuit cleaning shall be performed before initiation of target operation;
\item If feasible, isolate electrically the target blocks from the other metals;
\end{itemize}

Dedicated corrosion studies are planned to be performed in the future. 

\subsection{Radiation damage considerations and R\&D studies} 
\label{sec:TGT:mats:radiation}

In this section, a literature review of the irradiation effects on the target materials is presented with discussion about its representativity. Afterwards, on-going and foreseen specific irradiation studies are presented.   

\subsubsection{Literature review}

Irradiation of the target materials will affect the materials properties. Irradiation causes lattice damage through atom displacements (dpa) and atomic gas generation (mainly hydrogen and helium) in the material. dpa and gas generation introduce microscopic defects in the material lattice which translate into macroscopic changes in the material properties. The extent of this damage is directly related to the irradiation type and fluence. Irradiation temperature plays also an important role since high temperatures promote defects annihilation by thermal recovery. 

The irradiation conditions that the target will be subject during one year of operation are summarized in Table~\ref{tab:TGT:irrcond} for the four target materials (from Section~\ref{Sec:TGT:EneRad} and Section~\ref{Sec:TGT:Simus}).

\begin{table}[htb]
\center
\caption{Irradiation conditions for each BDF target material with the aproximate maximal annual fluences, dpa and gas production, as presented in Section~\ref{Sec:TGT:EneRad}.}
\label{tab:TGT:irrcond}
\begin{tabular}{cccccc}
\hline
\textbf{Material} & \textbf{Irradiation type} & \textbf{\makecell{Irradiation\\ T ($^{\circ}$C)}} & \textbf{Fluence} & \textbf{dpa} & \textbf{Gas production} \\ \hline
TZM & proton + neutron & 35-180 & \makecell{3x10$^{18}$p/cm$^2$\\ 2x10$^{20}$n/cm$^2$} & 0.1 & \makecell{38 appm H,\\ 15 appm He} \\ \hline
W & proton + neutron & 35-150 & \makecell{2x10$^{18}$p/cm$^2$\\ 3x10$^{20}$n/cm$^2$} & 0.1 & \makecell{30 appm H,\\ 16 appm He} \\ \hline
Ta/Ta.25W & proton + neutron & 35-160 & \makecell{3x10$^{18}$p/cm$^2$\\ 3x10$^{20}$n/cm$^2$} & 0.06 & \makecell{51 appm H,\\ 30 appm He} \\ \hline
\end{tabular}
\end{table}

It should be highlighted that most irradiation damage will be concentrated in areas in the center of the target due to the beam impact and damage will decrease gradually to the external diameters. Irradiation microscopic effects are not discussed in this part, but only the macroscopic variation, which will impact directly the thermo-physical and mechanical properties of the BDF target assembly.

Swelling due to irradiation is a concern for the target materials. Since most of the target volume is occupied either by W or TZM, even small swelling of 1\%-2\% in these materials could drastically reduce the spacing between target blocks and reduce the cooling water flow. Literature on both materials at representative neutron fluences reported swelling which starts above 300 $^{\circ}$C of irradiation temperature and becomes critical at 600 $^{\circ}$C with swelling values up to 1.5\% for W and 2.5\% for TZM. Therefore, materials swelling under neutron irradiation appears not to represent a significant concern in within target operational temperatures~\cite{COCKERAM2011, MATOLICH1974}. However further studies will be requested in order to assess the situation under a representative proton irradiation, as gas production - contributing to swelling - is significantly larger for proton irradiation as compared to (thermal) neutron irradiation.

Literature review of the available mechanical properties after irradiation for the materials of the target are summarized in Table~\ref{tab:TGT:radlit}. The minimal temperature for thermal recovery initiation in the target materials is estimated as 600 $^{\circ}$C. Therefore, data from irradiation temperatures between RT and 600 $^{\circ}$C is considered representative and presented in this review.

\begin{landscape}
\begin{table}
\footnotesize
\caption{Literature review of the irradiation effects on the mechanical properties for the four materials. UTS: Ultimate tensile strength, YS: Yield strength, FS: Fracture strength, FT: Fracture toughness, E: Elongation at failure, MH: Microhardness, NH: Nanohardness, SR:Strain rate.}
\label{tab:TGT:radlit}
\begin{tabular}{lllllllll}
\hline
\textbf{Product form}\ & \textbf{\makecell{Irradiation\\ type}} & \textbf{\makecell{Irradiation \\T ($^{\circ}$C)}} & \textbf{Fluence} & \textbf{dpa} & \textbf{Gas} & \textbf{Strength} & \textbf{Ductility} & \textbf{Ref.} \\ \hline
\multicolumn{9}{l}{\textbf{TZM}} \\ \hline
Bar 6 mm & Proton & 35-400 & - & 0.4-0.7 & 50-70 He appm & UTS +20\% & E: 8\% to 0\% & \citeyear{MULLER1994} \\ \hline
Plate 6.35 mm & Neutron & 300-560 & 0.7-2.5x10$^{22}$ n/cm$^2$ & 3.9-12.3 & - & YS: +5\%-100 \% &  & \citeyear{BYUN2008} \\ \hline
\multirow{2}{*}{Sheet 1.52 - 6.35 mm} & \multirow{2}{*}{Neutron} & \multirow{2}{*}{500-600} & \multirow{2}{*}{1-5.5x10$^{22}$ n/cm$^2$} & \multirow{2}{*}{-} & - & YS: +50 \%, SR: + sensible & DBTT: 430 $^{\circ}$C-500 $^{\circ}$C & \multirow{2}{*}{\citeyear{SMITH1977}} \\ \cline{6-8}
 &  &  &  &  & 10 He appm & FS: + &  &  \\ \hline
Bar 105 mm diam. & Neutron & 380 & 0.5-0.9x10$^{22}$ n/cm$^2$ & - & - & YS: 60-70 \% at 200 $^{\circ}$C & E: 25\% to 0\% at 200 $^{\circ}$C & \citeyear{tungstenprops} \\ \hline
Sheet 0.3 mm thickness & Neutron & 50-100 & 1.5x10$^{20}$ n/cm$^2$ & - & - & UTS: +50\%-120\% & E: 3\%-11\% to 0\% & \citeyear{SINGH1995} \\ \hline
Rods 3.1 mm diam. & Neutron & 40-475 & 2.9-3.5x10$^{20}$ n/cm$^2$ & 0.29-0.35 & - & YS: +50\%, FT: - & \makecell{E: 30\%-50\% to\\ 0\% at RT-200 $^{\circ}$C }& \citeyear{SCIBETTA2000} \\ \hline
\multicolumn{9}{l}{\textbf{W}} \\ \hline
Rod 3 mm diam. & Proton & 50-270 & - & 0.6-23.3 & \makecell{200-11000 H appm,\\ 40-2020 He appm} & YS: 30\% increase at 1 dpa & E:+ & \makecell{\citeyear{MALOY2002} \\ \citeyear{MALOY2005}} \\ \hline
Rod 3 mm - 6 mm diam. & Proton & $<$300 & 3.1x10$^{14}$ p/cm$^2$ & - & - & MH: +18\% & E: 0\% & \citeyear{SOMMER1995} \\ \hline
Plate & Self-ion & 500 & - & 0.5-5 & - & \makecell{NH: +18\% at 0.2 dpa,\\ +40\% at 5 dpa} &  & \citeyear{HWANG2016} \\ \hline
Wire  1 mm diam. & Neutron & 100 & 5x10$^{19}$ n/cm$^2$ & - & - & No significant differences &  & \citeyear{MAKIN1957} \\ \hline
- & Neutron & 330-450 & 1-1.5x10$^{21}$ n/cm$^2$ & - & - & YS: - (ann), + (def) & E: 0\% & \citeyear{GORYNIN1992} \\ \hline
Bar 105 mm diam. & Neutron & 380 & 0.5-0.9x10$^{22}$ n/cm$^2$ & - & - & YS: +55\% at 200 $^{\circ}$C & DBTT: 60$^{\circ}$C to 240$^{\circ}$C & \citeyear{tungstenprops} \\ \hline
\multicolumn{9}{l}{\textbf{Ta}} \\ \hline
Plate & Proton & $<$200 $^{\circ}$C & 0-1.7x10$^{21}$ p/cm$^2$ & 0-11.3 & 0-580 appm He & \makecell{MH: +80\% at 1 dpa,\\ +150\% at 11 dpa} & E: 15\% at 11 dpa & \citeyear{CHEN2001} \\ \hline
- & Proton & 150-260 & - & 6.5-16.3 & - & UTS: +170\% at 7.3 dpa & E: 2\% - 10\% at 15 dpa & \citeyear{SAITO2012} \\ \hline
Sheet 0.5 mm thickness & Proton & 385-400 & \makecell{4.8x10$^{19}$ p/cm$^2$-\\5.3x10$^{20}$ p/cm$^2$} & 0.26-2.9 & - & TS: up to +85\%  & E: 22\% to 2\% & \citeyear{BROWN1987} \\ \hline
- & neutron & 60-100 & \makecell{1.1x10$^{17}$ n/cm$^2$-\\ 4.2x10$^{20}$ n/cm$^2$} & \makecell{0.00004-\\0.14} & - & UTS: +200\% at 0.14 dpa & E: 2\%-12\% for 0.14 dpa & \citeyear{BYUN2008Ta} \\ \hline
\multicolumn{9}{l}{\textbf{Ta2.5W}} \\ \hline
Ta-1.2W & neutron & 60-100 & \makecell{1.1x10$^{17}$ n/cm$^2$-\\ 4.2x10$^{20}$ n/cm$^2$} & \makecell{0.00004-\\0.14} & - & UTS: +100\% at 0.14 dpa & E: 10\% at 0.14 dpa & \citeyear{BYUN2008Ta} \\ \hline
Ta-1.2W & \makecell{Proton+\\neutron} & 50-160 & \makecell{8.3x10$^{20}$ p/cm$^2$-\\ 3.6x10$^{21}$ p/cm$^2$ \\ and 2.9x10$^{20}$n/cm$^2$-\\ 1.1x10$^{21}$ n/cm$^2$\}} & 0.7-7.5 & - & UTS: +120\% at 7.5 dpa & E: 10\% at 7.5 dpa & \citeyear{BYUN2008Ta} \\ \hline
\end{tabular}
\end{table}

\end{landscape}

\begin{figure}[htb]
    \centering
    \includegraphics[scale=0.35]{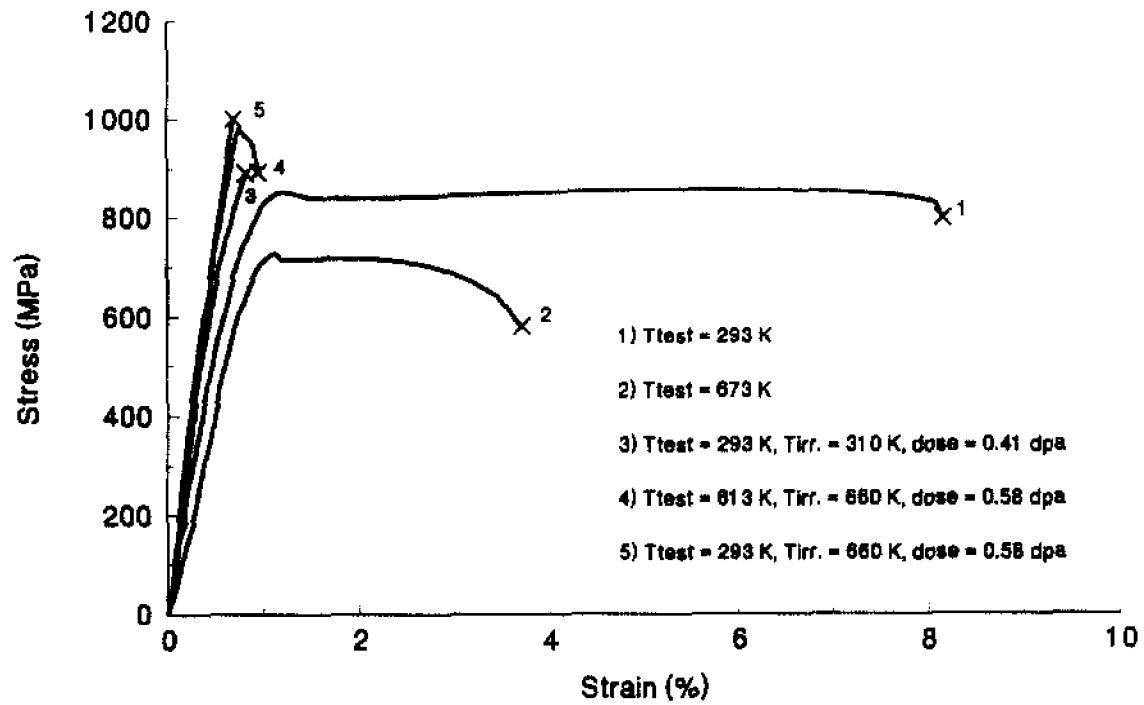}
    \caption{Tensile deformation curves for TZM, irradiated at different temperatures and doses, as presented in Ref.~\cite{MULLER1994}.}
    \label{fig:TGT:mats:TZMIrrProt}
\end{figure}

For TZM, quite representative data is reported in Ref.~\cite{MULLER1994}, with irradiation temperature, type, dpa and He generation representative of 3-5 years of target operation. In these conditions, TZM exhibits higher UTS (+20\% respect to the unirradiated) even if fracturing before yielding, as appreciable in Figure~\ref{fig:TGT:mats:TZMIrrProt}. This underlines the embrittlement of the base materials, which shows a negligible ductility. Data with neutron irradiation is less representative of the target conditions but effects on the mechanical properties provides a general indication. Moreover, studies with neutron irradiation reported additional irradiation effects: slight increase in strain rate sensibility, fatigue life and fracture strength reported by~\cite{SMITH1977} or the slight decrease in fracture toughness reported by~\cite{SCIBETTA2000}.

\begin{figure}[htb]
    \centering
    \includegraphics[scale=0.4]{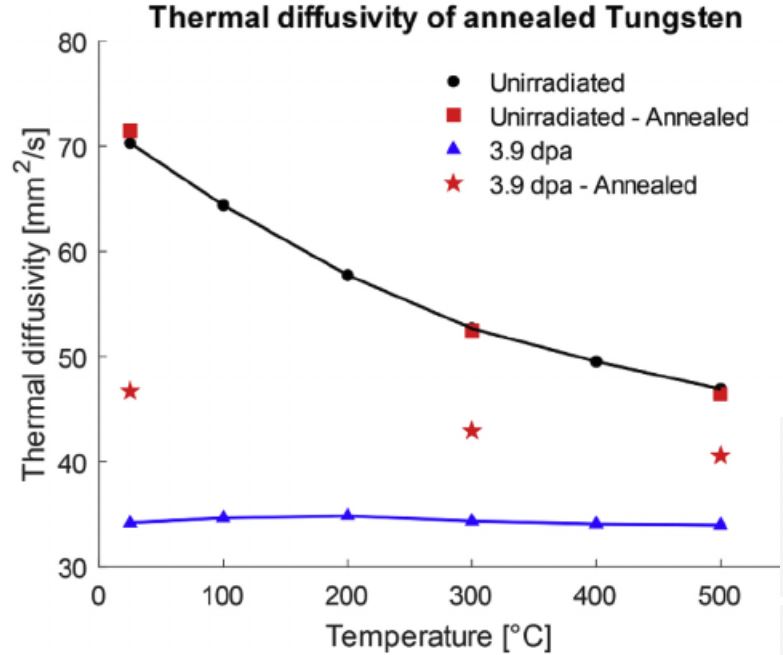}
    \caption{Thermal diffusivity of irradiated W at 25$^{\circ}$C-500$^{\circ}$C, as presented by Ref.~\cite{Wfatigue}.}
    \label{fig:TGT:mats:Wdiff}
\end{figure}

Data with equivalent irradiation conditions and for a wide range of dpa and He appm is available for pure W from Ref.~\cite{MALOY2002,MALOY2005}. Increase in the compressive yield strength of 30 \% was measured already at 0.6 dpa (40 He appm), equivalent to 5-6 years of target operation. This value stayed constant up to 4 dpa (290 He appm). Cracking on the sides of the irradiated specimens indicated decrease in ductility which was confirmed by Ref.~\cite{SOMMER1995} in bending tests for proton irradiated W. Less attention is commonly given to physical properties Ref.~\cite{Wfatigue} recently reported significant decrease in the thermal diffusivity for proton irradiated W, as depicted in Figure~\ref{fig:TGT:mats:Wdiff}. 

\begin{figure}[htb]
    \centering
    \includegraphics[scale=0.7]{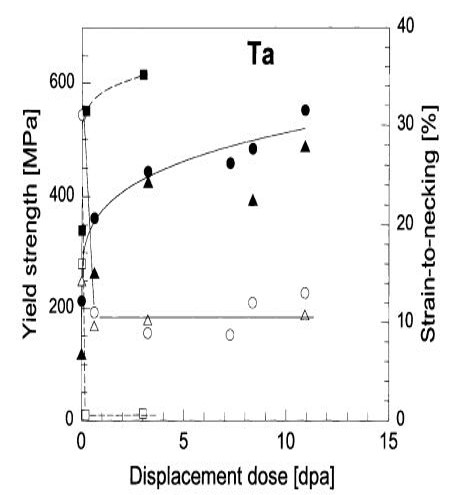}
    \caption{Yield strength (filled symbols) and strain-to-necking (empty symbols) as a function of the displacement dose for Ta. The circles and triangles indicate the results for high pure Ta tested at RT and 250 $^{\circ}$C respectively (from Ref.~\cite{CHEN2001}), the squares indicate data for less pure Ta from Ref.~\cite{BROWN1987}.}
    \label{fig:TGT:mats:UTSTaIrr}
\end{figure}

Several studies on pure Ta proton irradiation are available and applicable to the target conditions. Literature agree in reporting fast Ta strength increase from low dpa values (0.1-0.3) which is reduced but constant up to several tens of dpa (see Figure~\ref{fig:TGT:mats:UTSTaIrr}). Less agreement is shown in the embrittlement rate of Ta with increasing dpa values. While Ref.~\cite{BROWN1987} reported severe elongation reduction already at 0.3 dpa, Ref.~\cite{CHEN2001} reported 15 \% elongation up to 8.4 dpa, as appreciable in Figure~\ref{fig:TGT:mats:UTSTaIrr}. This different behaviour was attributed to Ta purity level. In accordance, Ref.~\cite{BYUN2008Ta} also reported Ta severe embrittlement in Ta which was attributed to oxygen pick-up during the annealing. 

\begin{figure}[htb]
    \centering
    \includegraphics[scale=0.8]{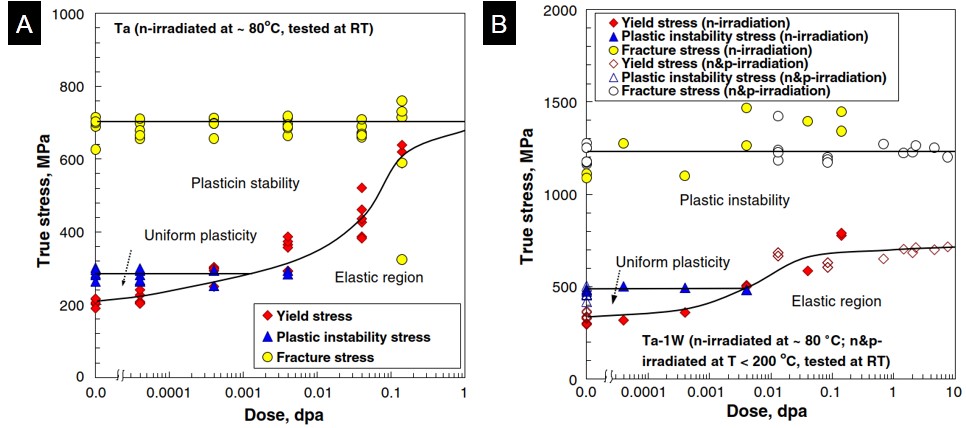}
    \caption{Macroscopic deformation map for a) pure Ta and b) Ta-1W alloy, from Ref.~\cite{BYUN2008Ta}.}
    \label{fig:TGT:mats:Ta1W}
\end{figure}

No specific data is available for the  irradiation affectation on the mechanical properties of Ta2.5W. Nevertheless, Ref.~\cite{BYUN2008Ta} studied the effects of combined irradiation (neutron+proton) of two different Ta alloys with W in solid solution: Ta-1\%W and Ta-10\%W. It was found that Ta-1\%W exhibits significant plastic region up to 10 dpa, much greater compared to Ta or Ta-10\%W at high dpa, appreciable in Figure~\ref{fig:TGT:mats:Ta1W}.

For all the target materials data for proton irradiation is limited and most of the data is available for neutron irradiation. Neutron data is difficult to correlate to proton data since the effects on the material are different: He gas production can reach higher appm per dpa for high energy proton irradiation compared to fission reactor neutron irradiation~\cite{JUNG2002}. Moreover, there are many difficulties to correlate the existing data, since many parameters can affect the irradiation consequences apart from the irradiation conditions: material impurities, heat treatment, grain size, cold work... These parameters are not always reported or considered during the experiments. 

In general terms, many of the irradiation works converge in the irradiation consequences on the materials, by showing and increase in tensile strength combined with reduced ductility (i.e. DBTT significant increase). It is widely accepted that low irradiation temperature embrittlement is the major concern in the working temperatures below 0.3T$_m$ for these materials~\cite{LEONARD2012}.

By putting together the estimated radiation damage on the target materials, it can be considered that W and TZM will become slightly harder but brittle already at few years of operation. Nevertheless is expected that Ta and Ta2.5W will become slightly stronger but keep significant ductility and if high purity material is employed. 

There are several material properties, which are relevant for the target design, for which no irradiation effects have been studied or correlation is difficult. Data for fatigue behaviour, fracture toughness and thermal properties after irradiation are scarcely available for refractory metals. This is a critical aspects that requires to be addressed in order to produce a robust target design with sufficient safety margins in terms of radiation damage. A plan in the framework of the RaDIATE Collaboration~\cite{Rad2018,Rad2018_2} is being defined. Further discussions are reported in Section~\ref{Sec:TGT:Mat:BLIP}.

\subsubsection{BLIP studies}
\label{Sec:TGT:Mat:BLIP}
\subsubsubsection{Introduction to RaDIATE collaboration}

With the increase of power on proton accelerator particle sources such as target facilities, there is a pressing need to better understand and predict long-term radiation damage effects of structural window and target materials. In this framework, CERN joined in~2017 the "Radiation Damage In Accelerator Target Environments" (RaDIATE) collaboration~\cite{radiate2019}. Many research institutes  in the accelerators domain, such as Fermilab, Argonne, FRIB, Brookhaven National Laboratory (BNL), Pacific Northwest National Laboratory (PNNL), etc. were already members of the RaDIATE collaboration. Its final goal is to be able to predict as best as possible operating lifetimes for materials in uses in target facilities in terms of integrated proton fluence for the high energy proton accelerator parameter space (e.g. temperature, dose rate, duty factor, dynamic stress). 

In order to achieve is final goal~\cite{radiate2019}, "the RaDIATE Collaboration draws on existing expertise in related fields in fission and fusion research [..] to formulate and implement a research program that will apply the unique combination of facilities and expertise at participating institutions to a broad range of high power accelerator projects of interest to the collaboration. The broad aims are threefold:

\begin{itemize}
\item 	to generate new and useful materials data for application within the 			accelerator and fission/fusion communities;
\item 	to recruit and develop new scientific and engineering experts who can 		cross the boundaries between these communities;
\item	to initiate and coordinate a continuing synergy between research in 			these currently disparate communities, benefiting both proton 					accelerator applications in science and industry and carbon-free 				energy technologies."
\end{itemize}

\subsubsubsection{Introduction to BLIP facility and experiment}

The BLIP facility at BNL has for mission to produce medical isotope using 116~MeV primary proton beams~\cite{blip2019,Ammigan:2017qws}. It delivers a rastered beam with peak current of 165~{\textmu}A and fluence of $7 \times 10^{13}$~p/cm\textsuperscript{2}~$\cdot$~s (3~cm diameter footprint) to isotope targets. However, the BNL Linac is capable to deliver protons up to 200 MeV. Therefore, it is possible to operate BLIP at higher energies and in tandem with material targets upstream of the isotope targets. In this operating mode, the optimal beam energy and proton flux still needs to be delivered to the downstream isotope targets preserving isotope yield. This can be achieved by \emph{selecting} some target materials with defined parameters (such as the thickness, the Z number of the materials, etc. through which the proton beam is passing) responsible for the degradation of the initial beam energy. Taking also into account the type of tests to be performed after irradiation, significant fine tuning and multiple sensitivity studies were performed to optimize and configure samples size, the sample configuration inside the capsule and the final target array composition.

One big advantage of the BLIP experiment is that is offers a unique opportunity to irradiate materials with protons with energies above 100~MeV. Indeed, reactor materials studies with neutron flux with energies between 1 to 14~MeV are limited in relevant to target systems. Effects from low energy neutron irradiations do not equal effects from high-energy proton irradiations. Typical obtained irradiation parameters are summarized in table~\ref{tab:TGT:mats:diff_proton_neutron} between proton irradiation~($E_{p^{+}} \geq 100 MeV$) and neutron irradiation~($1 \leq E_{n} \leq 14 MeV$). For proton irradiation, the obtained DPA rate and gas production are higher by a factor two at least~\cite{Hurh_radiate2018}.

\begin{table}[htb]
\centering
\caption{Typical irradiation parameters between proton irradiation~($E_{p^{+}} \geq 100 MeV$) and neutron irradiation~($1 \leq E_{n} \leq 14 MeV$) obtained in irradiated materials.}
\label{tab:TGT:mats:diff_proton_neutron}
\begin{tabular}{cccc}
\hline 
\textbf{\begin{tabular}[c]{@{}c@{}}Irradiation\\ source\end{tabular}} & \textbf{\begin{tabular}[c]{@{}c@{}}DPA rate\\ (DPA/s)\end{tabular}} & \textbf{\begin{tabular}[c]{@{}c@{}}He gas production\\ (appm/DPA)\end{tabular}} & \textbf{\begin{tabular}[c]{@{}c@{}}Irradiation Temp.\\ ({\textdegree}C)\end{tabular}} \\ \hline
\begin{tabular}[c]{@{}c@{}}Mixed spectrum\\ fission reactor\end{tabular} & $3 \times 10^{-7}$ & $1 \times 10^{-1}$ & $200-600$ \\
Fusion reactor & $1 \times 10^{-6}$ & $1 \times 10^{1}$ & $400-1000$ \\
\begin{tabular}[c]{@{}c@{}}High energy\\ proton beam\end{tabular} & $6 \times 10^{-3}$ & $1 \times 10^{3}$ & $100-800$ \\ \hline
\end{tabular}
\end{table}

\subsubsubsection{Framework and Objectives}

In the context explained above, CERN had the possibility to irradiate materials in the proton line of the "Brookhaven Linac Isotope Producer" (BLIP) facility at BNL~\cite{blip2019}. The BLIP experiment will complement the long-term radiation damage effect on materials (such as iridium, TZM, Ta-2.5\%W, etc.) used for targets and dumps application like the BDF facility for example. Results obtained from the BLIP irradiation campaign will provide important knowledge on the mechanical and thermal behavior of materials of interest. 

The main objectives for CERN were to re-design a new irradiation capsule in order to accommodate the samples and to send the capsules at BNL for irradiation. Additionally, Post-Irradiation Examinations (PIE) works had to be defined prior to the irradiation. A complete PIE procedure had to be written taking into account several aspects: capsules reception and handling of activated capsules, disassembly, mechanical and thermal tests and metallographic observations.

All the information presented here is well documented and summarized in CERN EDMS~\cite{Canhoto:BLIP:2019,Fornasiere:BLIP:2017,Fornasiere:BLIP:2019}. The reader can refer to these documents for more details.

\subsubsubsection{RaDIATE irradiation run}

\begin{figure}[htb] 
\centering 
\includegraphics[width=0.8\columnwidth]{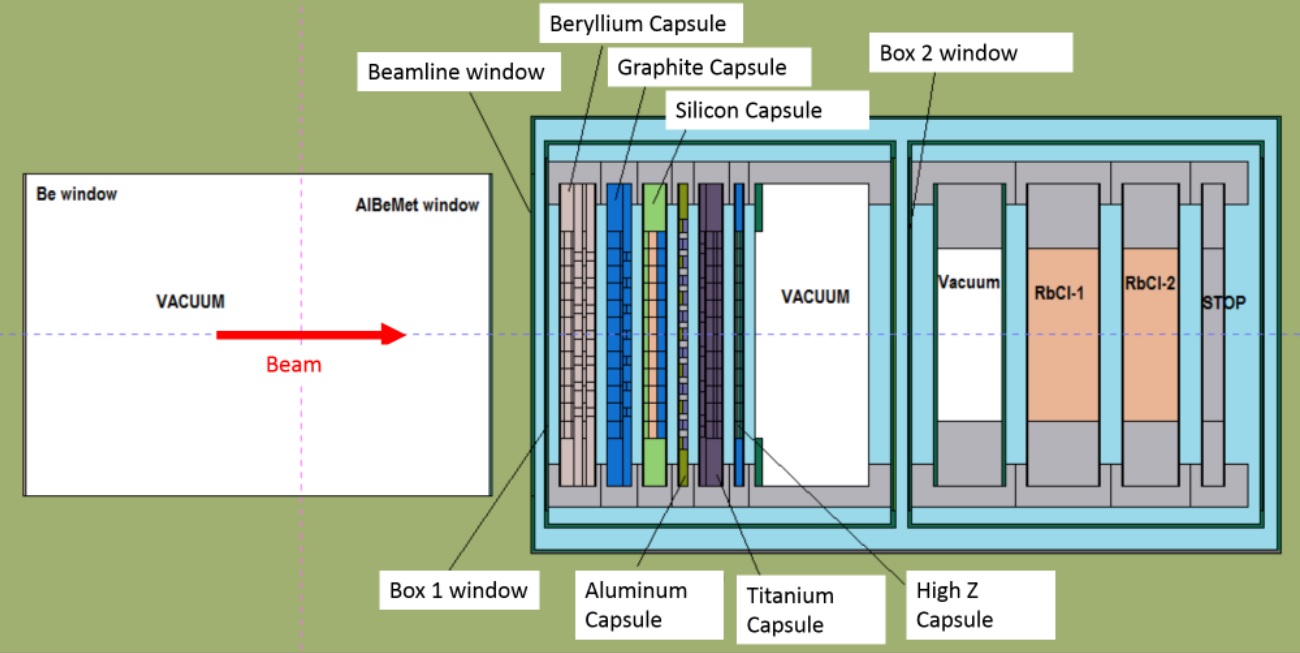} 
\caption[Targets arrangement in BLIP beam line]{Targets arrangement in BLIP beam line~\cite{blip2019}.}
\label{fig:TGT:targets_arrangement} 
\end{figure}

Various materials relevant to the participating institutions have been irradiated up to 8 weeks in the framework of the RaDIATE collaboration. The target box arrangement in the BLIP beamline, containing various materials just upstream of the isotope target box, is depicted in figure~\ref{fig:TGT:targets_arrangement}~\cite{blip2019}. The target box was configured in order to degrade the 181 MeV incoming Linac beam to the exact energy required for optimal isotope production. Each material type is enclosed in their individual stainless steel capsule, separated in series by a 2.5 mm wide gap of flowing cooling water.

Up to three capsules have been built shared between two runs. Several samples have been irradiated, such as iridium, TZM, copper-alloy, Ta-2.5W (tantalum with 2.5{\%} tungsten alloy) or Si samples provided insight for the behavior of materials under proton irradiation used for future Beam Intercepting Devices (BIDs) at CERN. The irradiation of the first two capsules successfully took place in July 2017 at BNL. Finally, the success of the first BLIP test campaign led to the realization of a second irradiation test campaign completed in March 2018. The different runs, capsules configurations, time of irradiation and Protons On Target (POT) are summarized in %
table~\ref{tab:TGT:mats:radiate_run}~\cite{Ammigan_radiate2018}. 
The BLIP run \#1 comprises two irradiated phases where capsules could be swapped.

\begin{table}[htb]
\centering
\caption{Capsules configurations, different runs, time of irradiation and Protons On Target (POT) for the BLIP irradiation campaign.}
\label{tab:TGT:mats:radiate_run}
\resizebox{\textwidth}{!}{%
\begin{tabular}{lccc}
\hline 
\textbf{Capsule name} & \begin{tabular}[c]{@{}c@{}}CERN High-Z Capsule\\ (High-Z CERN)\end{tabular} & \begin{tabular}[c]{@{}c@{}}CERN Silicon Capsule\\ (Si CERN)\end{tabular} & \begin{tabular}[c]{@{}c@{}}CERN High-Z Capsule II\\ (High-Z II CERN)\end{tabular} \\ \hline
\textbf{\begin{tabular}[c]{@{}l@{}}Irradiated materials\\ and \\ associated experiments\end{tabular}} & \begin{tabular}[c]{@{}c@{}}Iridium (AD)\\ TZM (BDF)\\ CuCrZr \\ (SPS internal dump)\\ PGS graphite\end{tabular} & \begin{tabular}[c]{@{}c@{}}Silicon (SPS internal dump)\\ SiC-coated graphite \\ (KEK muon production target)\\ Sigraflex{\textregistered} graphite\end{tabular} & \begin{tabular}[c]{@{}c@{}}Ta-2.5W (BDF)\\ Mo-coated CFC and\\ Mo-coated MoGr.\\ (materials for collimators)\\ Monocrystalline Si \\ (Crystal collimation)\\ PGS graphite\end{tabular} \\ \hline
\textbf{Run} & \multicolumn{2}{c}{BLIP Run \#1} & BLIP Run \#2 \\ \hline
\textbf{\begin{tabular}[c]{@{}l@{}}Capsule position\\ in the array\end{tabular}} & 6 & 3 & 3 \\ \hline
\textbf{Irradiation phases} & 2 & 1,2 & 3 \\
\textbf{Weeks of irradiaiton} & 1.8 & 3.15 & 4.7 \\
\textbf{POT} & 1.03E+21 & 1.76E+21 & 2.81E+21 \\ \hline
\textbf{Completed irradiation} & \multicolumn{2}{c}{July 2017} & March 2018 \\ \hline 
\end{tabular}%
}
\end{table}

\subsubsubsection{Capsule configuration}

\textit{\textbf{High-Z Capsule}}
\label{Sec:TGT:mat:High-Z}

\begin{figure}[htb]
\centering
\includegraphics[width=.9\columnwidth]{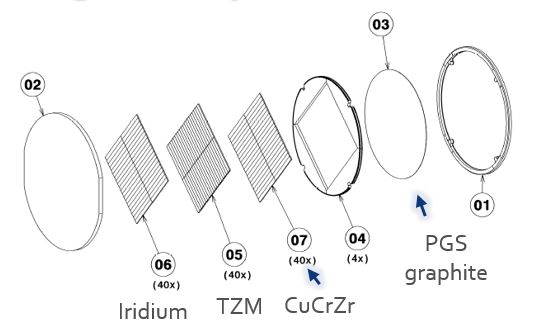}
\caption{Sketch of the bend specimen layers arrangement for the different materials of the high Z capsule}
\label{fig:TGT:high-z_capsule}
\end{figure}

The High-Z capsule, enclosed in vacuum, contained 40 samples of TZM (candidate material for BDF targets) as well as other Iridium and CuCrZr specimens. Figure~\ref{fig:TGT:high-z_capsule} shows a sketch of the bend specimen layers for the different materials and a picture of the capsule during its assembly. Due to the high density of the materials, this capsule was only irradiated for two weeks in order to minimize the residual dose rate and stay below limits for handling and transportation purposes. In addition, the proton budget allowed for our materials (<10 MeV/capsule) was a limiting factor for the thickness of the samples to be irradiated. Dimensions of samples were minimized to $20 \times 2 \times 0.5$ mm for micromechanical testing. All samples were manufactured to be tested in 4-points bending. Samples dimensions together with reached peak temperatures during irradiation are summarized in table~\ref{tab:TGT:mats:blip_samples}. More information about the obtained temperatures can be found in section~\ref{Sec:TGT:mat:Irradiation_results}.

\textit{\textbf{High-Z II Capsule}}

The High-Z II capsule is also vacuum-sealed and was irradiated during run \#2 and which received the highest number of POT (2.81E+21 total POT, see table~\ref{tab:TGT:mats:radiate_run}). Estimated peak irradiation temperatures were calculated to be reasonably low (around 330 {\textdegree}C, see table~\ref{tab:TGT:mats:blip_samples}) thanks to the systematic use of PGS in between the materials arrays. Between other materials, the capsule contains 40 samples of Ta-2.5\%W (used as cladding material for TZM and W blocks for BDF target) for 4-points bend tests.

\begin{figure}[htb]
		{\label{fig:TGT:high-z_2_capsule_configuration}%
		\includegraphics[width=.9\columnwidth]{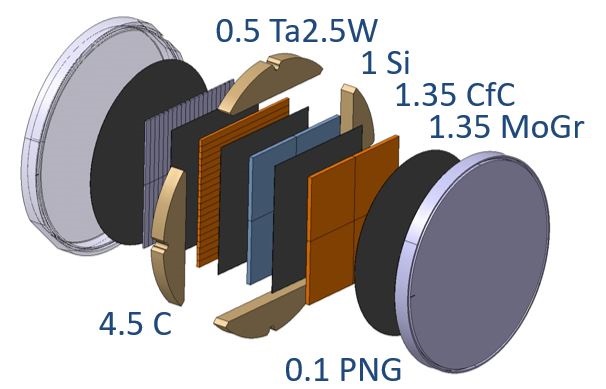}}
		\caption{%
		3D view of the bend specimen layers arrangement for the different 				materials of the High-Z II capsule
	}
\end{figure}

\subsubsubsection{Irradiation results}
\label{Sec:TGT:mat:Irradiation_results}

To calculate the temperatures and stressed reached in each capsules and their sample materials, thermo-mechanical calculations using the commercial \uppercase{ANSYS}{\textregistered} Mechanical~\cite{Ansys:Guide:2013} Finite-Element (FE) software have been performed. For these calculations, FLUKA Monte-Carlo code~\cite{Ferrari2005} was first used to generate an input thermal heat load file coming from the irradiation. The generated file was then mapped to the \uppercase{ANSYS}{\textregistered} mesh and applied as an internal heat generation source. The transfer from one code to the other one was realized through an \uppercase{APDL} script~\cite{Ansys:APDL:2013}. The obtained values could fluctuate depending on the used models and therefore here are presented the most reasonable retained estimations. Unfortunately, estimated values by simulations could not be confirmed by measurements during irradiation due to the obvious difficulty of the operation.

\begin{figure}[htb] 
	\centering 
	\includegraphics[width=0.8\columnwidth]{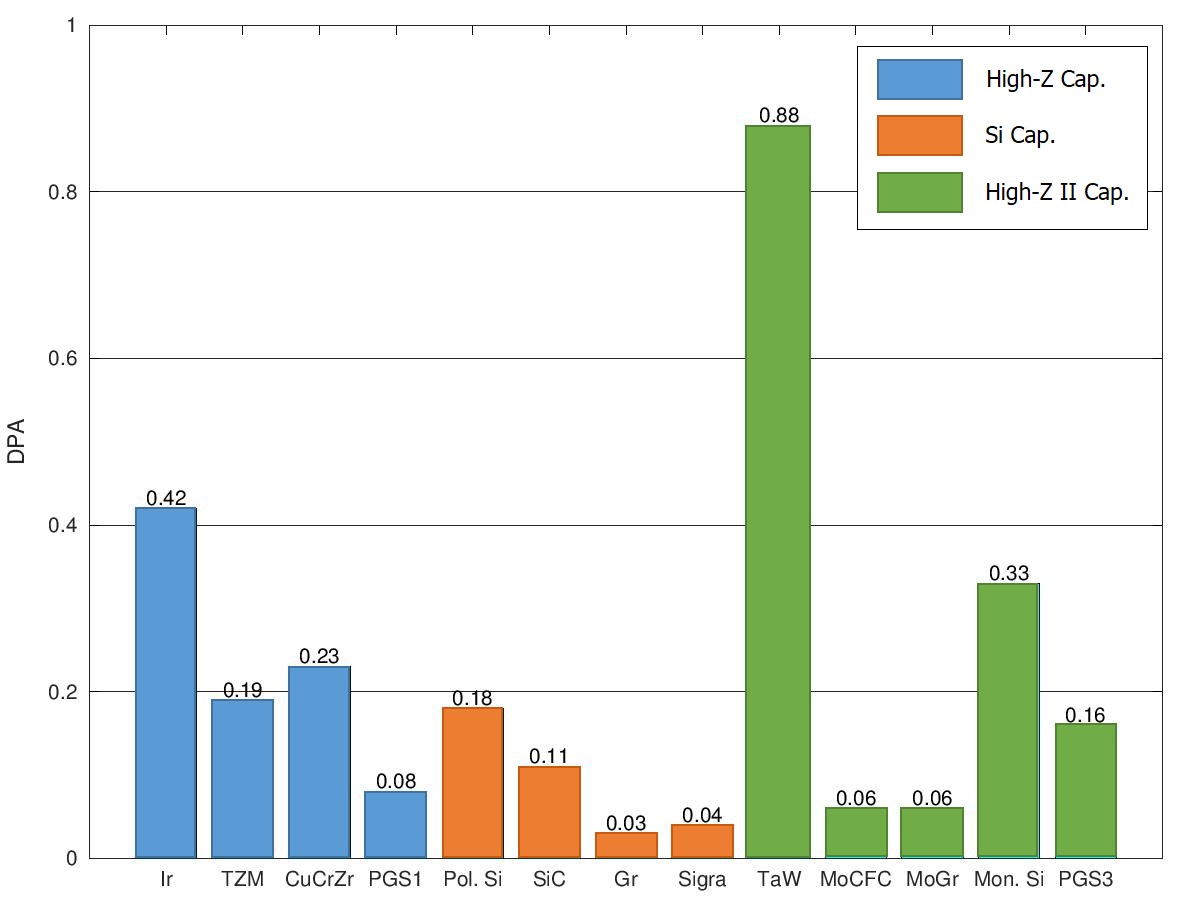} 
	\caption{Reached DPA in irradiated CERN BLIP materials.}
	\label{fig:TGT:dpa} 
\end{figure}

\begin{figure}[htb] 
	\centering 
	\includegraphics[width=0.8\columnwidth]{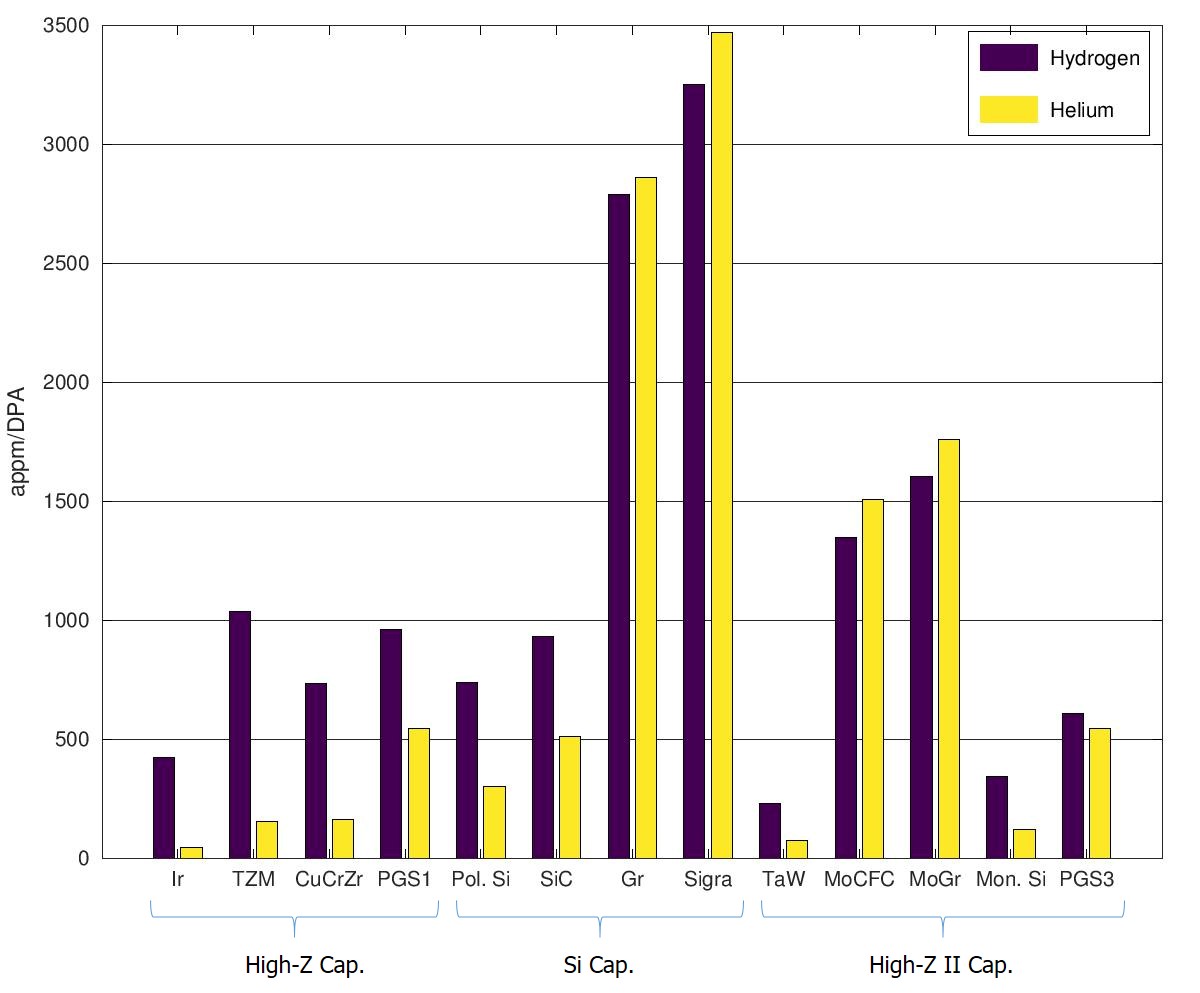} 
	\caption{Peak gas production rate in irradiated CERN BLIP materials.}
	\label{fig:TGT:gas_production} 
\end{figure}

DPA calculations and gas productions were performed by means of \uppercase{FLUKA} Monte-Carlo simulations~\cite{Canhoto:BLIP:2019}. Reached peak DPA obtained during irradiation and gas productions are depicted in figures~\ref{fig:TGT:dpa} and~\ref{fig:TGT:gas_production}, respectively. Materials are presented following the order of table~\ref{tab:TGT:mats:blip_samples}. For the High-Z capsule with a POT of 1.03E+21 TZM reached 0.19 DPA. Ta-2.5W reached the highest obtained DPA of~0.88 for a POT of 2.81E+21 on the High-Z II capsule. 

\begin{table}[htb]
\centering
\caption{Samples configuration, samples size and thickness, reached DPA and temperature during the BLIP irradiation campaigns.}
\label{tab:TGT:mats:blip_samples}
\begin{tabular}{ccccc}
\hline 
\multicolumn{1}{l}{\textbf{Capsule}} & \textbf{Material} & \textbf{\begin{tabular}[c]{@{}c@{}}Sample size \\ $l \times w$\\ (mm $\times$ mm)\end{tabular}} & \textbf{\begin{tabular}[c]{@{}c@{}}Sample \\ thickness\\ (mm)\end{tabular}} & \textbf{\begin{tabular}[c]{@{}c@{}}Peak \\ Temp. (\textdegree C)\end{tabular}} \\ \hline
High-Z & Iridium & $20 \times 2$ & 0.5 & 860 \\
High-Z & TZM & $20 \times 2$ & 0.5 & 820 \\
High-Z & CuCrZr & $20 \times 2$ & 0.5 & 230 \\
High-Z & PGS graphite & $\varnothing$ 55.5 & 0.1 & 165 \\ \hline
Si & Silicon & $40 \times 2$ & 1 & 220 \\
Si & SiC & $\varnothing$ 9.8 & 0.2 & 220 \\
Si & Graphite & $\varnothing$ 9.8 & 0.8 & 220 \\
Si & Sigraflex{\textregistered} graphite & $\varnothing$ 55.5 & 2 & 190 \\ \hline
High-Z II & Ta-2.5W & $40 \times 2$ & 1 & 270 \\
High-Z II & Mo-CFC & $20 \times 20$ & 1.35 & 330 \\
High-Z II & Mo-MoGr & $20 \times 20$ & 1.35 & 250 \\
High-Z II & Single crystal Si & $20 \times 2$ & 0.5 & 320 \\
High-Z II & PGS graphite & $\varnothing$ 55.5 & 0.1 & 290 \\ 
\hline 
\end{tabular}
\end{table}

\subsubsubsection{BLIP PIE}
\label{Sec:TGT:mat:PIE}

During the irradiation, the interaction of the proton beam with targets and dumps creates radioactive isotopes inside the materials through nuclear fragmentation processes, resulting in their activation. After irradiation, a sufficient cooling time is necessary in order to proceed with the treatment, handling and process of the samples. The capsules were shipped to the various institutions of the RaDIATE collaboration for the Post-Irradiation Examinations (PIE) works. Thus, a cooling time of the capsules to meet Type A radioactive shipment was required. PIE will characterize strength, thermal and microstructural material property changes due to radiation damage effects.

Together with the fact of coping with brittle materials and the little available space inside the capsule, micromechanical methods was retained for material testing. For 4-points bending tests, in accordance with the available proton budget (see section~\ref{Sec:TGT:mat:High-Z}), the design and dimensions of the samples were chosen to accommodate the ASTM E-1820 and ASTM E-399 standards as guideline for the test campaign.

Required mechanical tests to be performed of the irradiated samples, samples preparation for metallographic observations, metallography investigation techniques to be used (such as optical microscopy and scanning electron microscopy SEM), thermal tests on flexible graphite and required analysis of results have been all documented in details in a PIE technical specification procedure~\cite{Fornasiere:BLIP:PIE:2017}.  For CERN, all PIE works of the specimens will be performed at PNNL. The High-Z capsule, PIE works will start current of the year 2019, while for the High-Z II capsule, PIE works are foreseen during 2020.

\FloatBarrier


\section{Development of a BDF prototype target for beam tests}
\label{Sec:TGT:Proto}

\subsection{Introduction and motivations to the BDF target prototype construction}

Given the unprecedented regime of temperature and stresses in the BDF target, a beam test of a prototype target during 2018 had been proposed back in 2017 in order to provide a timely feedback for the European Strategy for Particle Physics process. In order to reach similar operational scenarios, representative beam scenarios had to be reproduced.

Since BDF conditions require a slow extraction (SX) process, the HiRadMat facility could not be used. Considering that high intensity SX is available in the North Area target zone (TCC2), the experiment had been proposed to take place upstream the T6 primary target (thanks also to the absence of a vacuum beam pipe until roughly 15 meters upstream the T6 target assembly). 
This is allowed by the lack of wobbling magnets upstream T6, contrary to T2 and T4. In addition, intensities up to 10$^{13}$ protons per pulse are regularly sent on the T6 beryllium targets for the COMPASS experiment.

The experiment has been designed in order to reproduce similar levels of temperature and stress in the core and in the cladding as it will be expected in the final BDF application, despite the different beam configuration.

The main objectives of the prototype target tests are summarized hereafter:

\begin{itemize}
    \item Reproduce experimentally the temperatures and magnitude of thermal-induced stresses of the final target despite the lack of dilution;
    \item Evaluate the response under thermal shock of refractory cladded materials, that will be subjected to temperature gradients of the order of 50 to 100$^\circ$C per pulse;
    \item Cross-check the Finite Element Model simulations performed. To that purpose, several blocks have been instrumented in order to execute on-line measurements;
    \item Further explore the instrumentation survivability in challenging environments, including high levels of accumulated dose, high water speed and pressure;
    \item Validate the performance of the target assembly cooling system and specifically the high velocity in contact with the blocks; 
    \item Execute tritium out-diffusion experiments in order to quantitatively assess the diffusion coefficient that will be employed for the environmental impact study for the final installation;
    \item Perform detailed post-irradiation experiments (PIE) after irradiation
\end{itemize}

More details on the different objectives presented will be given in the following sections.

\subsection{Design of the prototype target}
\label{Sec:TGT:Proto:design}

\subsubsection{Prototype target mechanical design}

The target replica that has been tested in the North Area consists in a reduced scale prototype of the BDF final target. The target prototype cylinders have a diameter of 80 mm (instead of 250 mm), and the same length distribution as the BDF final target. The core materials used for the target replica are equivalent to the ones of the final target, the first 13 blocks are made out of forged TZM and the last 5 are made out of sintered+HIP pure tungsten. The cladding materials used are pure tantalum and Ta2.5W, in order to compare the performance of both materials under beam irradiation conditions.

The pure tungsten blocks are all cladded with pure tantalum and not with Ta2.5W. This is due to the fact that at the design stage of the target prototype, a good mechanical and chemical bonding between tungsten and Ta2.5W could not be achieved via HIPing. A successful bonding was later on accomplished by changing the HIP parameters (see Section~\ref{Sec:TGT:Materials}), justifying the use of Ta2.5W as cladding material for all the target blocks~\cite{HIP_Busom}. 

A summary of the materials and dimensions of the target prototype cylinders is given in Table~\ref{tab:TGT:proto_dimensions}.

\begin{table}[htbp]
\centering
\caption{\label{tab:TGT:proto_dimensions} BDF target prototype blocks description, with core and cladding material as well as total length and weight of each block.}
\smallskip
\begin{tabular}{ccccc}
\toprule
\textbf{Block number} & \textbf{Core material} & \textbf{Cladding material} & \textbf{Length (mm)}&\textbf{Weight (kg)}  \\
\midrule
1                     & TZM                    & Ta                         & 80 & 4.1                  \\
2                     & TZM                    & Ta2.5W                     & 25 &1.3                  \\
3                     & TZM                    & Ta2.5W                     & 25 &1.3                  \\
4                     & TZM                    & Ta2.5W                     & 25&1.3                   \\
5                     & TZM                    & Ta2.5W                     & 25 &1.3                  \\
6                     & TZM                    & Ta2.5W                     & 25&1.3                   \\
7                     & TZM                    & Ta2.5W                     & 25&1.3                   \\
8                     & TZM                    & Ta                         & 25 &1.3                  \\
9                     & TZM                    & Ta2.5W                     & 50 &2.6                  \\
10                    & TZM                    & Ta                         & 50 &2.6                     \\
11                    & TZM                    & Ta                         & 65 &3.3                  \\
12                    & TZM                    & Ta                         & 80 &4.1                  \\
13                    & TZM                    & Ta                         & 80      &4.1             \\
14                    & W                      & Ta                         & 50 &4.7                  \\
15                    & W                      & Ta                         & 80      &7.5             \\
16                    & W                      & Ta                         & 100         &9.4         \\
17                    & W                      & Ta                         & 200             &18.8     \\
18                    & W                      & Ta                         & 350       &32.9          \\
\bottomrule
\end{tabular}
\end{table}

Similarly to the BDF final target, the prototype assembly includes two concentric stainless steel tanks. The outer tank ensures the leak-tightness of the assembly, compatible with an operational pressure of 22 bar, provides an interface for the electrical and water connections, and encloses the inner stainless steel tank. The inner tank supports the target blocks and encloses the prototype cooling circuit. 

The prototype inner tank differs significantly from the final target inner tank design. In this case, the target prototype blocks are much lighter, and don't require a specific handling tool for the assembly. The inner tank consists of two half-shells, clamped together with screws. The target blocks are sitting on the inner tank lower shell, only constrained in the Z direction by 2 pins that allow free-body expansion of the blocks within 50 \textmu m, but ensure a gap of 5 mm between the blocks, necessary for the water cooling acting on the blocks surfaces. 

\begin{figure}[htbp]
\centering %
\includegraphics[width=0.8\linewidth]{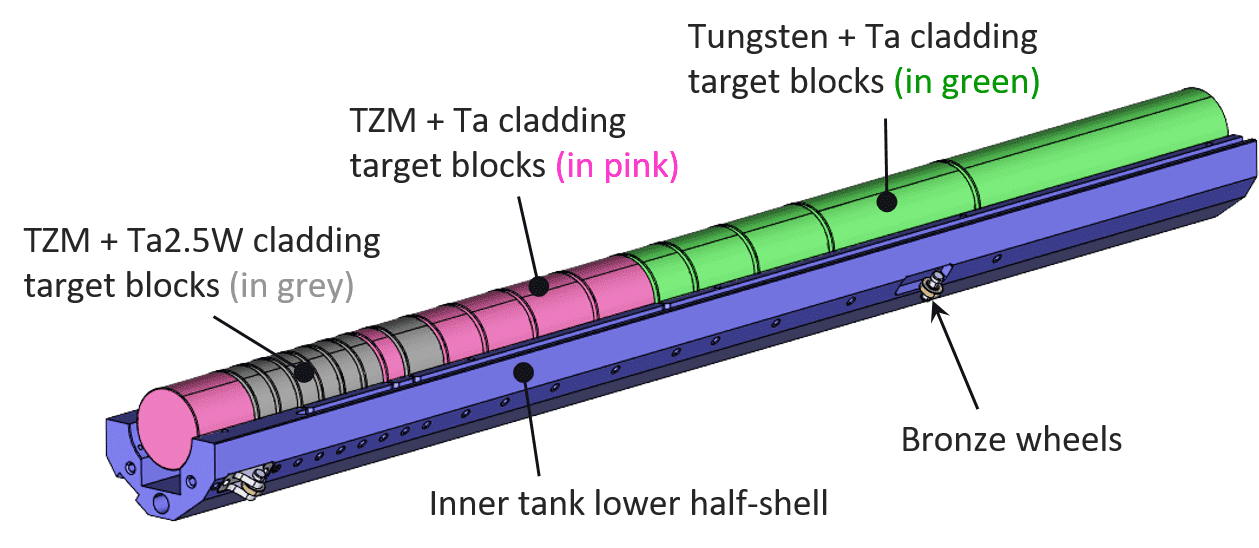}
\caption{\label{fig:TGT:proto_inner_tank2} BDF target prototype inner tank lower shell and target blocks description.}
\end{figure}

\begin{figure}[htbp]
\centering %
\includegraphics[width=0.9\linewidth]{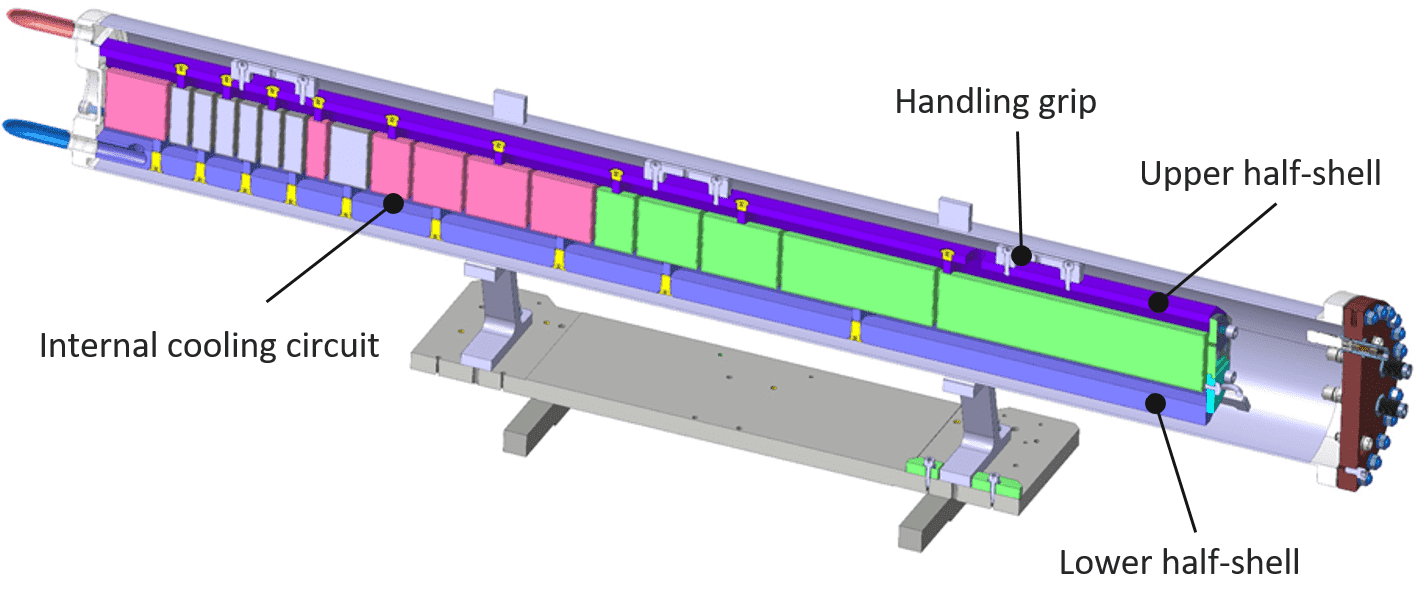}
\caption{\label{fig:TGT:proto_inner_tank} Longitudinal cross-section of the target prototype inner tank inserted into the outer tank. The upstream and downstream flanges are also represented.}
\end{figure}

The inner tank shells have been designed taking into account the water circulation foreseen for the target prototype cooling system. A more detailed description of the cooling circuit is given in the following sections. The inner tank upper shell includes several grips to allow the removal of the half-shell via remote handling, which is necessary for the remote extraction of the target blocks for PIE (see Section~\ref{Sec:TGT:Proto:PIE}). A scheme of the inner tank can be seen in Figure~\ref{fig:TGT:proto_inner_tank} and \ref{fig:TGT:proto_inner_tank2}, where the prototype target blocks and the different materials used are also shown.

The target inner tank is mounted on four bronze wheels, that allow the tank insertion into the outer tank. Once inserted, the inner tank is bolted to the upstream flange of the outer tank. The outer tank has two guides in its inner surface that serve as alignment guides for the inner tank, ensuring the target correct positioning with respect to the beam axis. 

\begin{figure}[htbp]
\centering %
\includegraphics[width=0.9\linewidth]{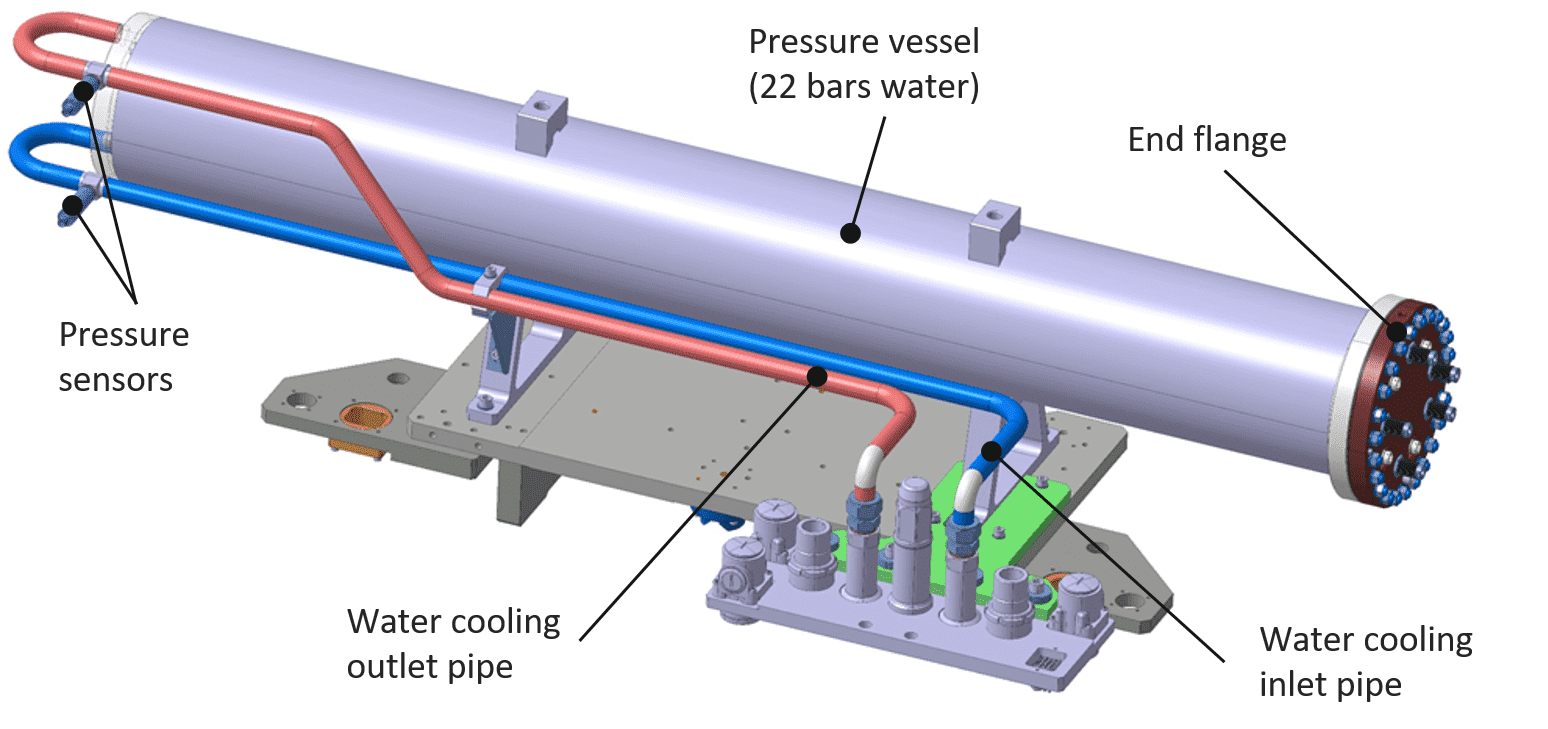}
\caption{\label{fig:TGT:proto_outer_tank} BDF target prototype outer tank description. The modified LHC collimators plugin, the water cooling pipes and the intermediate connection plate are visible.}
\end{figure}

The downstream part of the tank is closed by an end flange that includes all the feedthroughs for the target instrumentation (see Section~\ref{Sec:TGT:Proto:Instru}). The water connections are located in the upstream flange of the outer tank, and the pressure drop in the whole target assembly can be measured via two pressure sensors installed in the water connection pipes. The outer tank is considered as a pressure vessel capable of safely withstanding the 22 bar operational pressure of the cooling system (31 bar tested with a 1.43 safety factor~\cite{EN-13445}). Figure~\ref{fig:TGT:proto_outer_tank} shows the outer tank assembly with the inlet and outlet of the cooling system.

The whole target assembly is placed on a motorized table with horizontal movement that permits to align the target with the beam axis for the prototype beam tests, and remove it from the beam after operation. As intermediate support between the target outer tank and the motorized table, a modified LHC collimator plugin table is used. The collimator standard support is made out of two plates designed in such a way that the upper plate can be installed on top of the lower plate by means of remote handling. This setup allows the remote installation, replacement and removal of the whole target assembly (inner tank, outer tank and collimator top plate) by only using the remotely manipulated crane in the TCC2 target complex zone. This is a critical implementation due to the limited access to TCC2 during installation, and the high dose rate expected at the end of the beam irradiation period.

\begin{figure}[htbp]
\centering %
\includegraphics[width=1\linewidth]{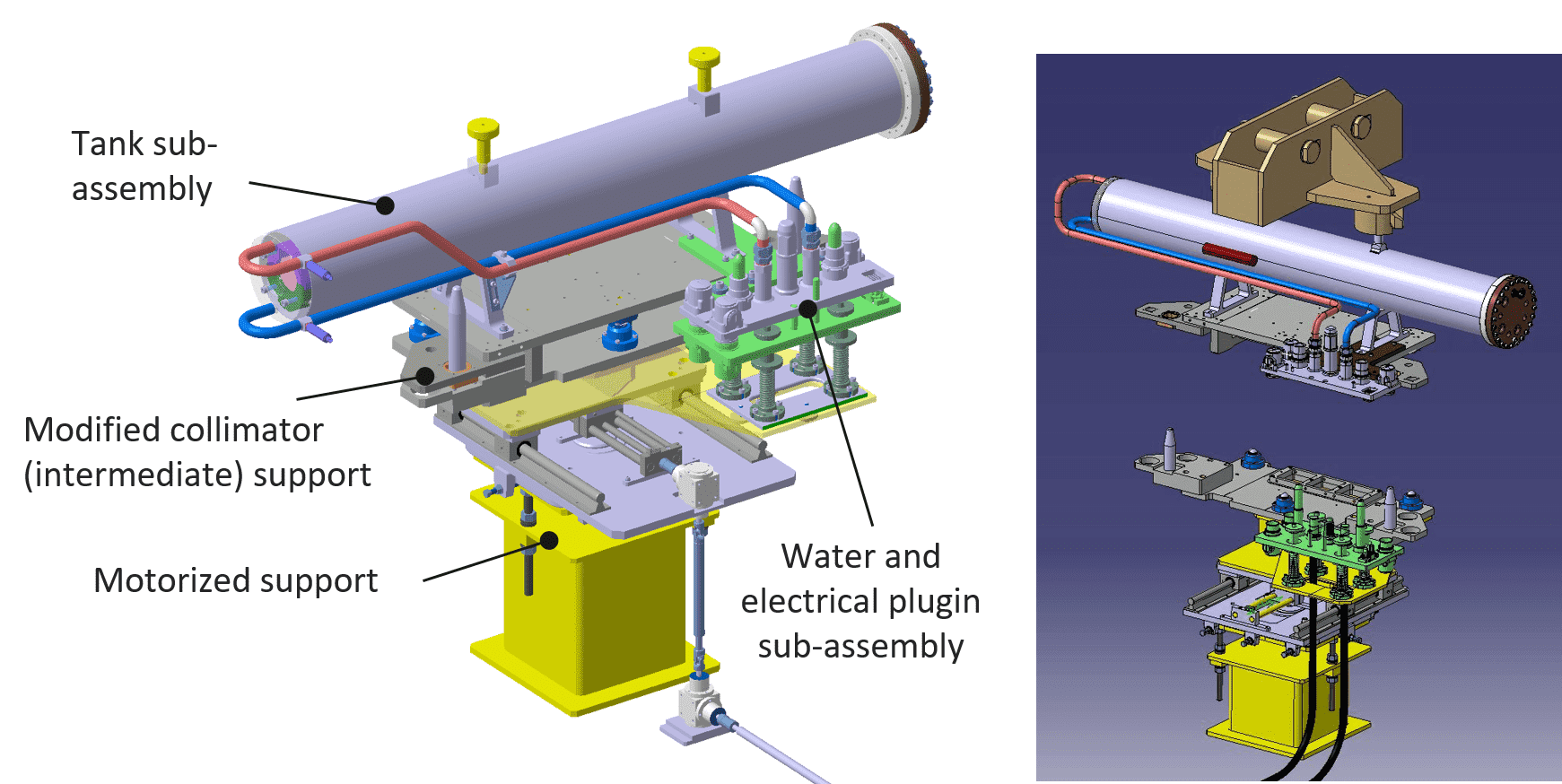}
\caption{\label{fig:TGT:proto_support} (a) BDF target prototype motorized support, plugin system and intermediate support. (b) Representation of the remote installation/removal of the target assembly.}
\end{figure}

Additionally, a specific interface is needed to perform the remote connection and disconnection of the water and electrical connectors. Due to the specific requirements of high water flow, high pressure and high radiation levels, the LHC collimator plugin connection had to be significantly modified. Dedicated R\&D has been carried out in order to develop a plugin system for the water and signal connections suited for the prototype target application, leading to a fully metallic system. Figure~\ref{fig:TGT:proto_support} describes the different sub-assemblies of the whole prototype target and support assembly (a), as well as the remote installation or dismantling procedure (b).

\subsubsection{Beam parameters and thermo-mechanical calculations}

\subsubsubsection{}{Target prototype beam parameters}

The target prototype has been tested in TCC2 using a non-diluted primary proton beam (no diluter is available in the TDC2/TCC2 area) and the same cycle as for the BDF final target (i.e. spill length of 1 second and repetition rate of 7.2 seconds). Consequently, the required beam intensity to reach representative temperatures and stresses with respect to the final target is lower and is expected to be in the range of 3-4$\times 10^{12}$p\textsuperscript{+}/cycle. 
Table~\ref{tab:TGT:proto_beamparam} shows a comparison between the BDF final target beam and BDF target prototype beam parameters.

\begin{table}[htbp]
\centering
\caption{\label{tab:TGT:proto_beamparam} The table summarizes the BDF final target and BDF target prototype beam parameters.}
\smallskip
\begin{tabular}{lcc}
\toprule
\textbf{Baseline characteristics}                  & \textbf{BDF final target} & \textbf{BDF target prototype} \\
\midrule
Proton momentum {[}GeV/c{]}                        & 400                       & 400                           \\
Beam intensity {[}p\textsuperscript{+}/cycle{]}                      & \num{4e13}                    & 3-4$\time 10^{12}$                    \\
Beam dilution                                      & 4 circular sweeps / s     & No                            \\
Expected beam spot size (H/V) {[}mm{]}             & 8/8                       & 3/3                           \\
Cycle length {[}s{]}                               & 7.2                       & 7.2                           \\
Spill duration {[}s{]}                             & 1.0                       & 1.0                           \\
Average beam power  {[}kW{]}              & 350                       & 35                            \\
Average beam power on target  {[}kW{]}              & 305                       & 30                            \\
Average beam power during spill {[}MW{]} & 2.56                       & 0.26                                        \\
\bottomrule
\end{tabular}
\end{table}

\subsubsubsection{Thermal calculations}

The energy deposited on target by the SPS 400 GeV/c beam was calculated via FLUKA simulations, and imported into ANSYS Mechanical for thermo-structural analysis. Detailed CFD calculations have also been performed, defining the water cooling parameters of the target prototype in order to obtain a homogeneous water velocity distribution in the block channels (see next section). The estimated average heat transfer coefficient (HTC) value on the blocks surface is around 16000 W/(m$^2$K), which was used as boundary condition for the thermal analysis.

\begin{figure}[htbp]
\centering %
\includegraphics[width=0.7\linewidth]{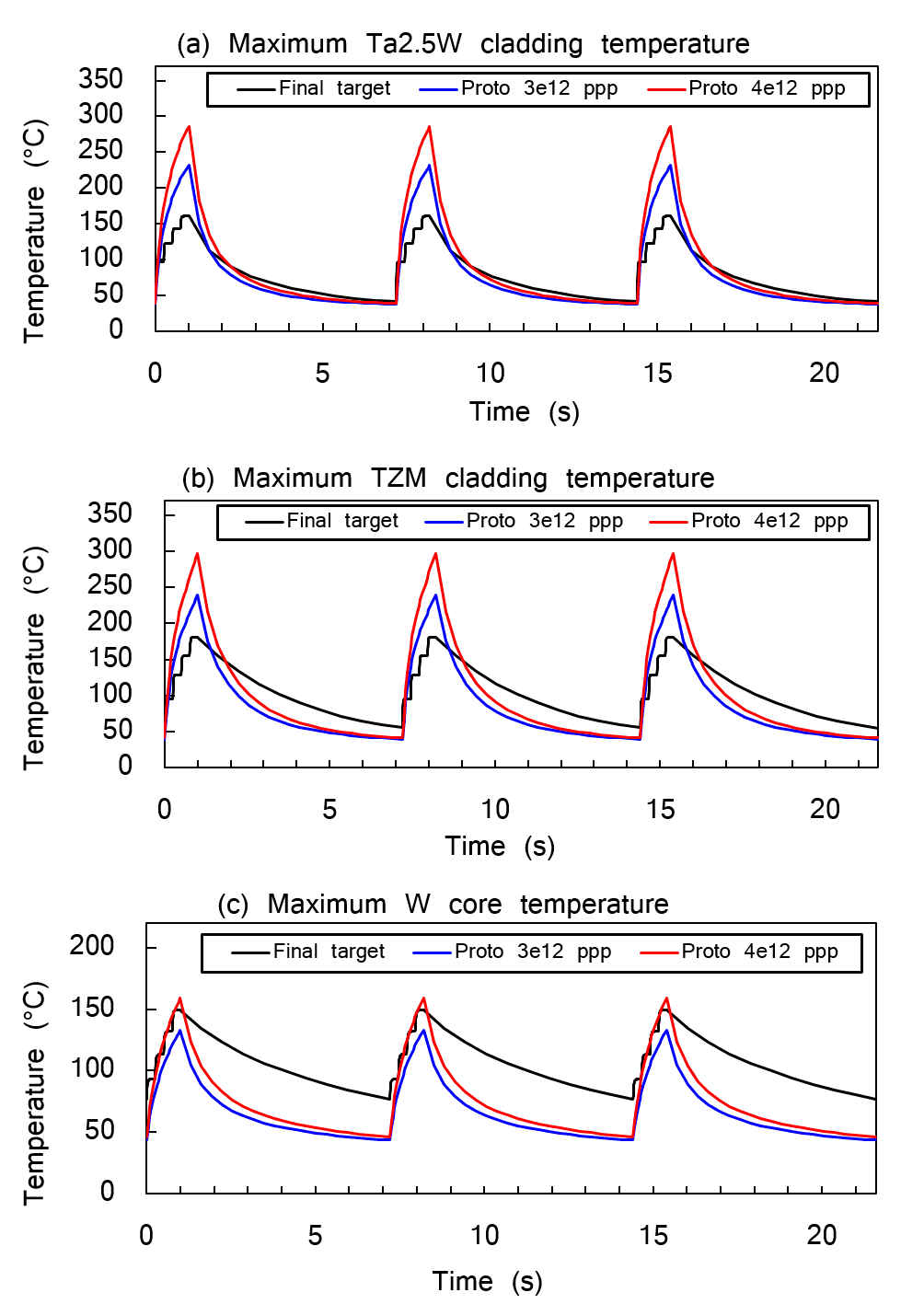}
\caption{\label{fig:TGT:proto_temps} Maximum temperature evolution during three beam pulses after steady-state for the different target materials: (a) Ta2.5W cladding of block \#4 for the final and prototype target, (b) TZM core of block \#9 for the final target and TZM core of block \#4 for the prototype target, (c) W core of block \#14 for the final and prototype target. Results comparison for the final target operation and the prototype operation under \num{3e12} or \num{4e12} ppp (protons per pulse).}
\end{figure}

A comparison between the maximum temperatures estimated in the different materials for the final target and the prototype target at two different beam intensities is shown in Figure~\ref{fig:TGT:proto_temps}(a), (b) and (c). 

The temperatures reached at \num{3e12} p\textsuperscript{+}/pulse and \num{4e12} p\textsuperscript{+}/pulse in the TZM core and the cladding materials are higher than the ones expected in the final target. For the pure W core, the temperature reached at \num{3e12} p\textsuperscript{+}/pulse is 20\% lower than for the final application, which justifies the use of a higher intensity beam for the prototype test. At \num{4e12} p\textsuperscript{+}/pulse, the temperatures reached in the pure W core are comparable to the ones of the final target.

\begin{figure}[htbp]
\centering %
\includegraphics[width=0.9\linewidth]{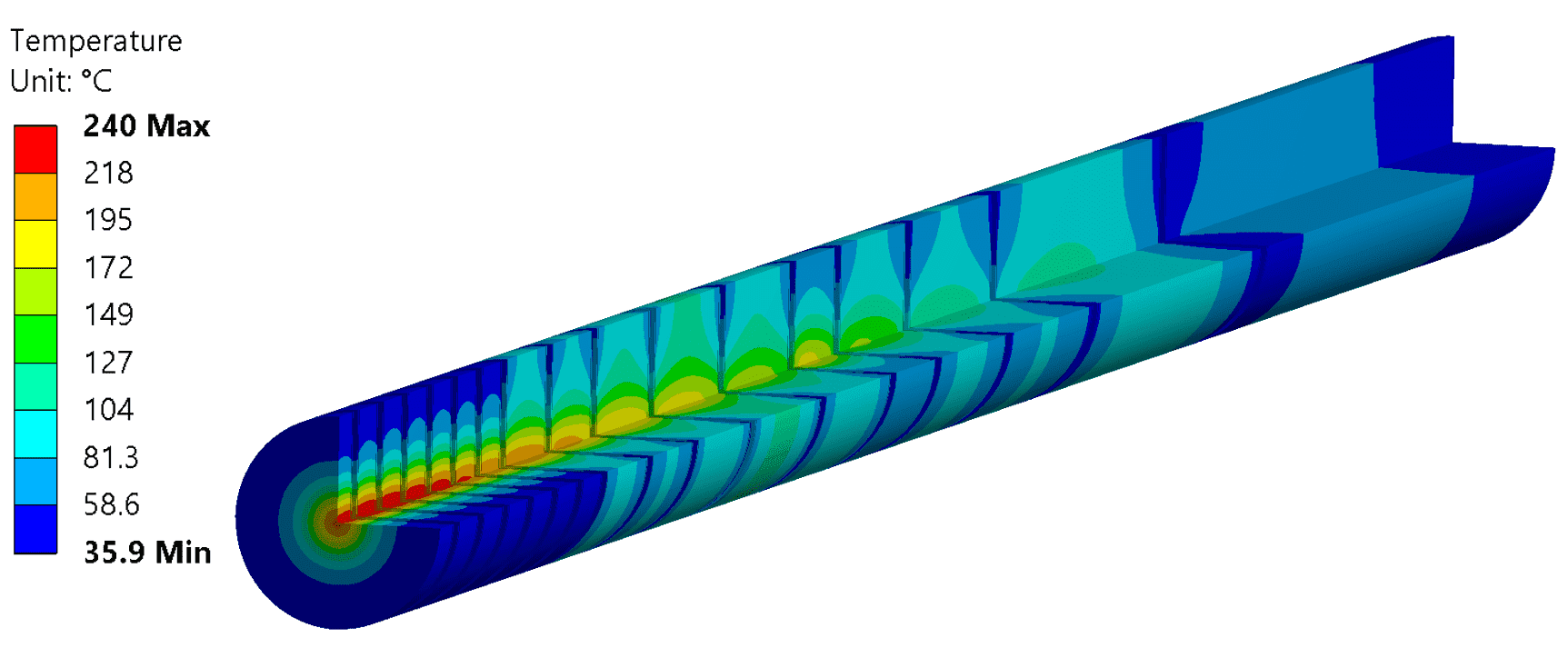}
\caption{\label{fig:TGT:proto_temp_all} Temperature distribution in the BDF target prototype after the beam impact. the maximum temperature reached is around $240\,^{\circ}\mathrm{C}$, found in the TZM core of block \#4.}
\end{figure}

Figure~\ref{fig:TGT:proto_temp_all} presents the temperature distribution in the target prototype at the end of the beam impact. It can be seen that the temperature distribution differs from the one obtained in the BDF final target (see Figure~\ref{fig:TGT:dilution_temp}), which can be mainly explained by the fact that the beam is diluted on a circular trajectory in the BDF final target operation, contrary to the prototype test where the beam is impacting directly on the target center. 

Despite the impacted area being different, the thermal-induced stresses are expected to have the same effect on the interface between the refractory metal core block and the respective cladding material, which is one of the main objectives of the beam test. 

\subsubsubsection{Structural calculations}

The temperature distribution evolution with time is imported as an input for the structural analysis. The stresses induced by the temperature increase are considered as quasi-static (similarly to the BDF final target case), since the pulse duration of 1 second is considered long enough for the dynamic effects on the materials to be negligible.

A comparison between the maximum stress in the different target materials obtained for the BDF final target and for the prototype target test at two different intensities is shown in Table~\ref{tab:TGT:proto_stress}. 

\begin{table}[htbp]
\centering
\caption{\label{tab:TGT:proto_stress} Maximum stress expected in the target materials for the final target and the target prototype at two different intensities. For TZM and Ta2.5W, the maximum von Mises equivalent stress is shown, for pure tungsten the maximum principal stress is presented.}
\smallskip
\begin{tabular}{cccc}
\toprule
\multirow{3}{*}{\textbf{Material}} & \multicolumn{3}{c}{\textbf{Maximum stress {[}MPa{]}}} \\ \cmidrule(l){2-4} 
 & Final target & \begin{tabular}[c]{@{}c@{}}Target prototype\\ \num{3e12} p+/pulse\end{tabular} & \begin{tabular}[c]{@{}c@{}}Target prototype\\ \num{4e12} p+/pulse\end{tabular} \\ \midrule
TZM & 128 & 116 & 160 \\
W & 80 & 70 & 85 \\
Ta2.5W & 95 & 90 & 120 \\ \bottomrule                                                         
\end{tabular}
\end{table}

It can be observed that for an intensity of \num{3e12} p+/pulse on the BDF target prototype, a similar level of stresses that the one on the final BDF target is reached. Higher intensities such as \num{4e12} have been considered for the target prototype, which will be subjected to slightly more challenging conditions than the BDF final target. It can be seen that the level of stresses in this case is higher than for the final target, specially for the TZM core and the Ta2.5W cladding (20\% higher stress), but still within the material limits. Higher intensities were sought also to compensate for expected slightly larger beams than the one requested (3 mm 1$\sigma$).

\begin{figure}[htbp]
\centering %
\includegraphics[width=0.7\linewidth]{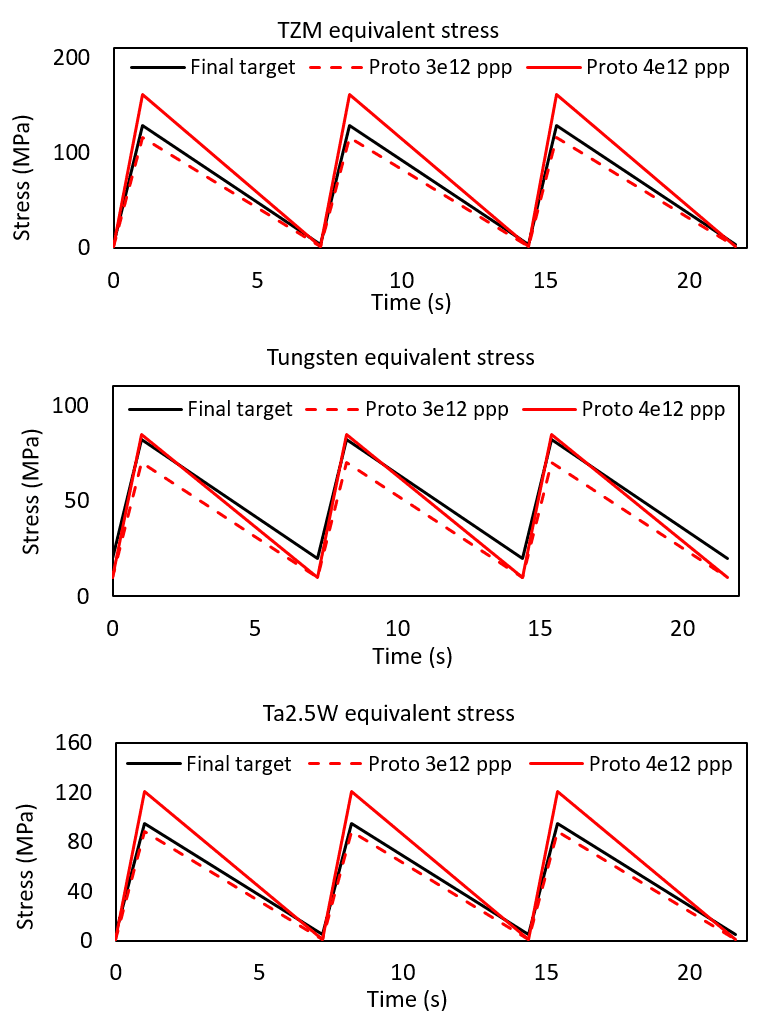}
\caption{\label{fig:TGT:proto_fatigue} Evolution of the equivalent stress in the different target materials during 3 beam pulses (once steady-state conditions have been reached). The plots show a comparison between the BDF final target and the BDF target prototype at two different intensities.}
\end{figure}

The aim of the experiment has been to reach around $10^{4}$ pulses on target. The number of cycles is substantially lower to the one expected in the final BDF target ($N=10^{7}$ over the lifetime of the assembly), but will help understanding the materials response to high-cycle fatigue and the robustness of the core/cladding interface. Figure~\ref{fig:TGT:proto_fatigue} shows a comparison between the maximum equivalent stress evolution in TZM, W and Ta2.5W for the final BDF target and the target prototype at the two different intensities. The plotted results correspond to the position where the maximum stresses are reached in the most loaded target blocks (block \#4 for TZM and Ta2.5W, block \#14 for tungsten).

\begin{figure}[htbp]
\centering %
\includegraphics[width=1\linewidth]{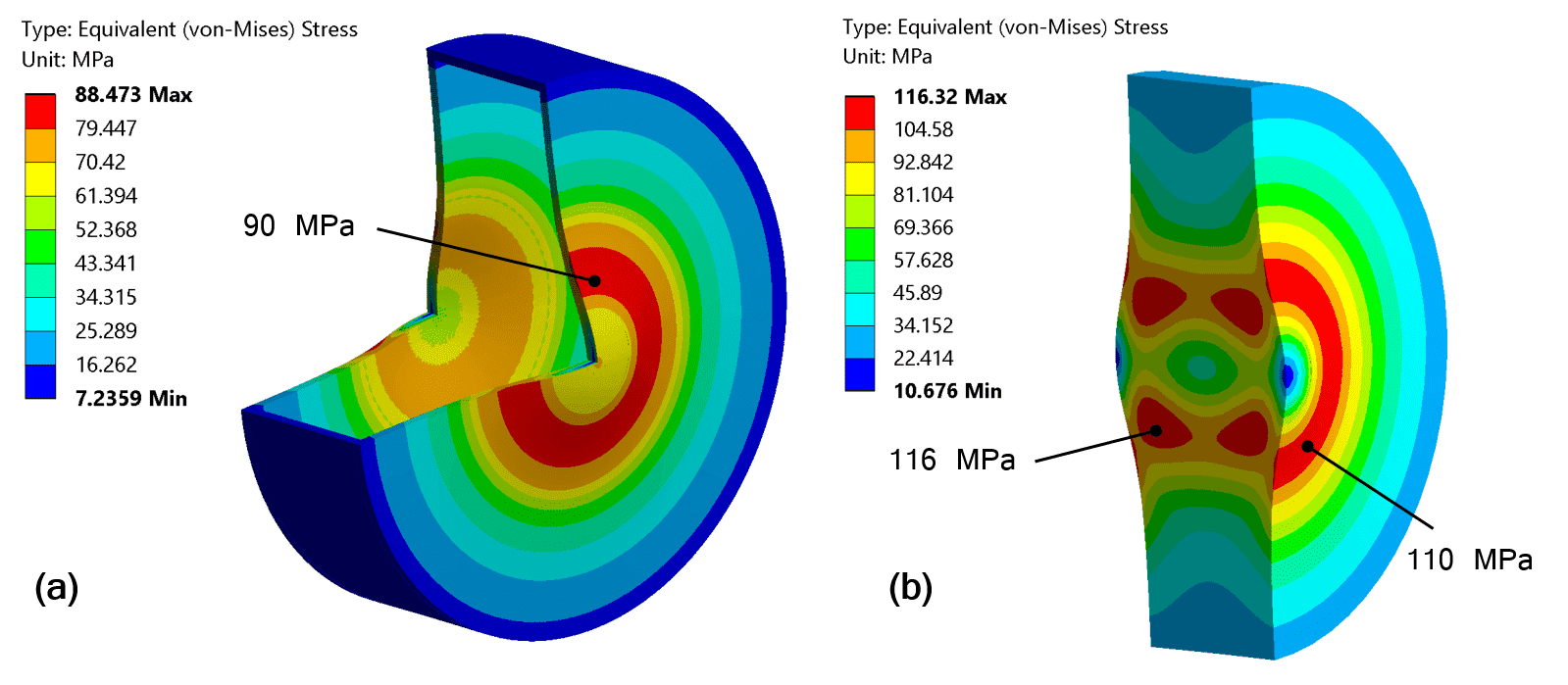}
\caption{\label{fig:TGT:proto_stress_map} Von Mises equivalent stress distribution in the target Ta2.5W cladding (a) and TZM core (b) of block \#4 of the target prototype. Results for \num{3e12} p+/pulse.}
\end{figure}

The performed calculations have shown that the mean stress and stress amplitude are expected to be well reproduced in the target prototype tests for the two different intensities. However, there are some limitations to recreate the stress state of the final target blocks. As shown in Figure~\ref{fig:TGT:proto_stress_map}(a), the maximum von Mises equivalent stress for the cladding materials is reached in the external surface of the cladding, and about 10 mm far from the beam axis. As for the core materials, the maximum stresses are reached in the core center and the bonding interface with the cladding layer, 5-10 mm far from the beam axis as well (Figure~\ref{fig:TGT:proto_stress_map}(b)).

This stress distribution differs from the final BDF target one, where the maximum stresses are expected to follow the beam impact trajectory (Figure~\ref{fig:TGT:vmstresses}). The difference between both stress distributions can be explained by the fact that the beam is diluted in one case and not in the other, leading to a different temperature distribution and therefore, to a different stress field.

\begin{figure}[htbp]
\centering %
\includegraphics[width=1\linewidth]{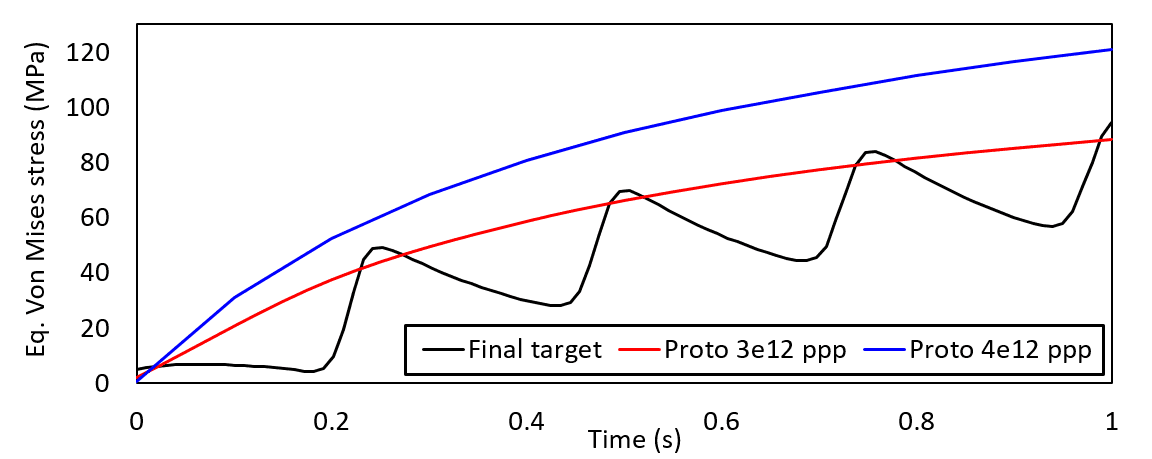}
\caption{\label{fig:TGT:proto_stress_evolution} von Mises equivalent stress evolution in the point of maximum stress of the Ta2.5W cladding (block \#4) during the beam impact of 1 second. Comparison between the final target and the target prototype at two different intensities.}
\end{figure}

The stress evolution for a given point of the target is also different depending on the beam dilution presence or not. As shown in Figure~\ref{fig:TGT:stress_evolution_sweep}, the von Mises equivalent stress in a given point of the BDF final target increases with a clear influence of the beam dilution. In the BDF target prototype the situation is different, since the beam impact occurs always in the target center. A comparison between the evolution of the von Mises equivalent stress in the points of maximum stress for the Ta2.5W cladding of block \#4 for the final target and the target prototype at \num{3e12} and \num{4e12} p+/pulse is presented in Figure~\ref{fig:TGT:proto_stress_evolution}.

The absence of beam dilution in the target prototype tests leads to some limitations in reproducing an identical stress state in the prototype blocks. However, it has been shown that the maximum level of stress foreseen in the final target was expected to be recreated and even surpassed in the target prototype operation, which will provide - when the collected data will be fully exploited - fundamental information on the robustness of the BDF final target design.

\subsubsection{Prototype cooling system and CFD analysis}

The target prototype cooling system design intended to prove an initial validation of the design of the BDF final target cooling. The BDF target prototype cooling design replicates the major features of the BDF final target cooling system:

\begin{itemize}
    \item Water cooling at a high pressure of around 22 bar;
    \item 5 mm channels (allowing the placement of sensors on the surface of the blocks);
    \item Significant water speed between plates (up to 4 m/s);
    \item Serpentine configuration of the water flow;
\end{itemize}

The prototype cooling design also includes some differences with respect to the final target, due to the specific design of the experiments and the available SPS beams:

\begin{itemize}
    \item Reduced size channels, due to the blocks diameter being only 80 mm instead of 250 mm;
    \item The channel flow velocity is lower than the one expected for the BDF final target, i.e. 4 m/s respect to 5 m/s;
    \item One stream circulating in series instead of 2 parallel streams following the serpentine path;
\end{itemize}

\begin{figure}[htbp]
\centering %
\includegraphics[width=1\linewidth]{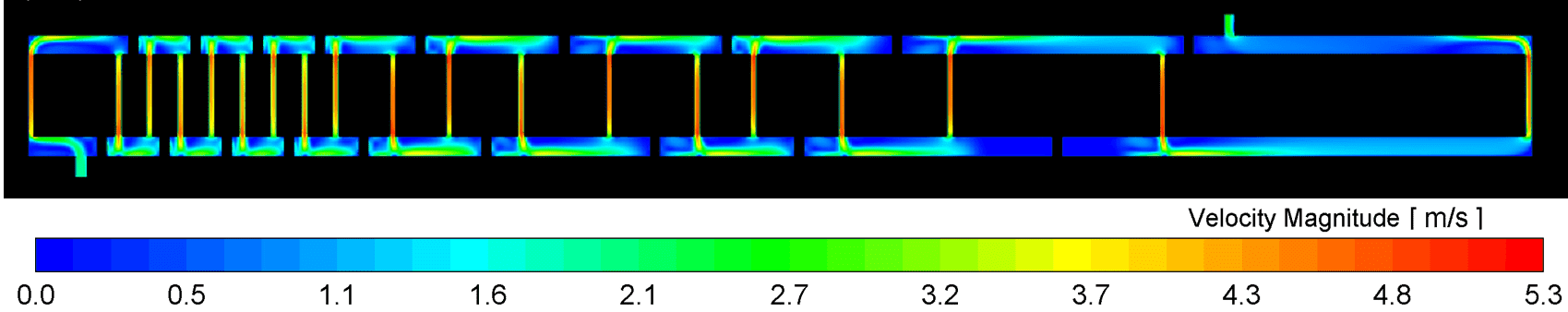}
\caption{\label{fig:TGT:proto_CFD_vel} 2D contour of velocity magnitude [m/s] in the target prototype cooling system; serpentine configuration.}
\end{figure}

Several flow configurations have been investigated in order to minimize the required mass flow rate and at the same time obtain uniform fluid velocity and high HTC in the channels. Each configuration has a specific combination of number of parallel channels and number of turns along the target body. For the specific set of parameters of the BDF target prototype, the single-channel (serpentine) configuration presented in Figure~\ref{fig:TGT:proto_CFD_vel} results to be the optimal one in terms of flow velocity uniformity and overall mass flow rate.

\begin{figure}[htbp]
\centering %
\includegraphics[width=0.8\linewidth]{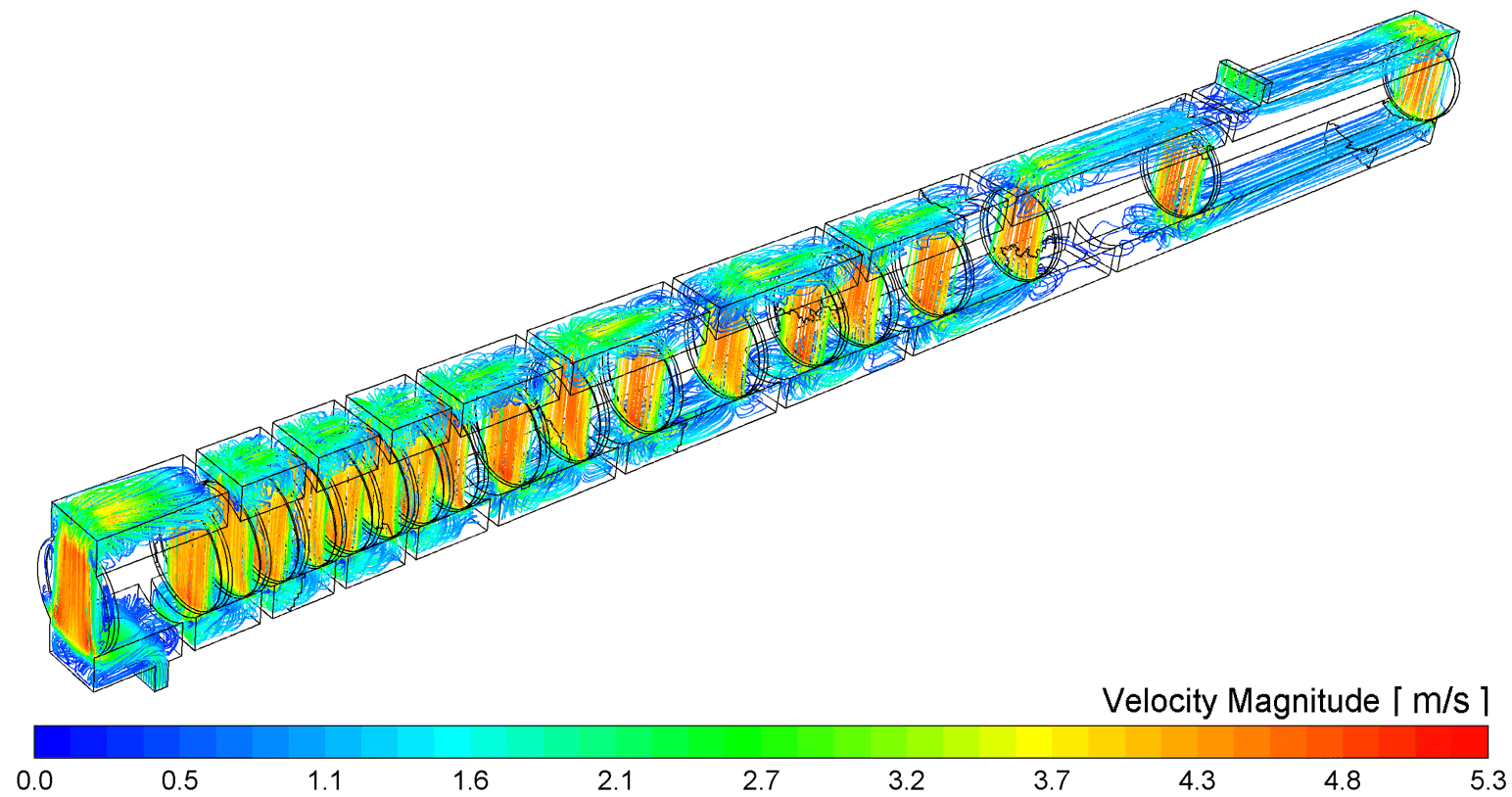}
\caption{\label{fig:TGT:proto_CFD_vel3D} Path lines colored by velocity magnitude [m/s] from a 3D CFD simulation of the BDF target cooling system.}
\end{figure}

\begin{figure}[htbp]
\centering %
\includegraphics[width=0.8\linewidth]{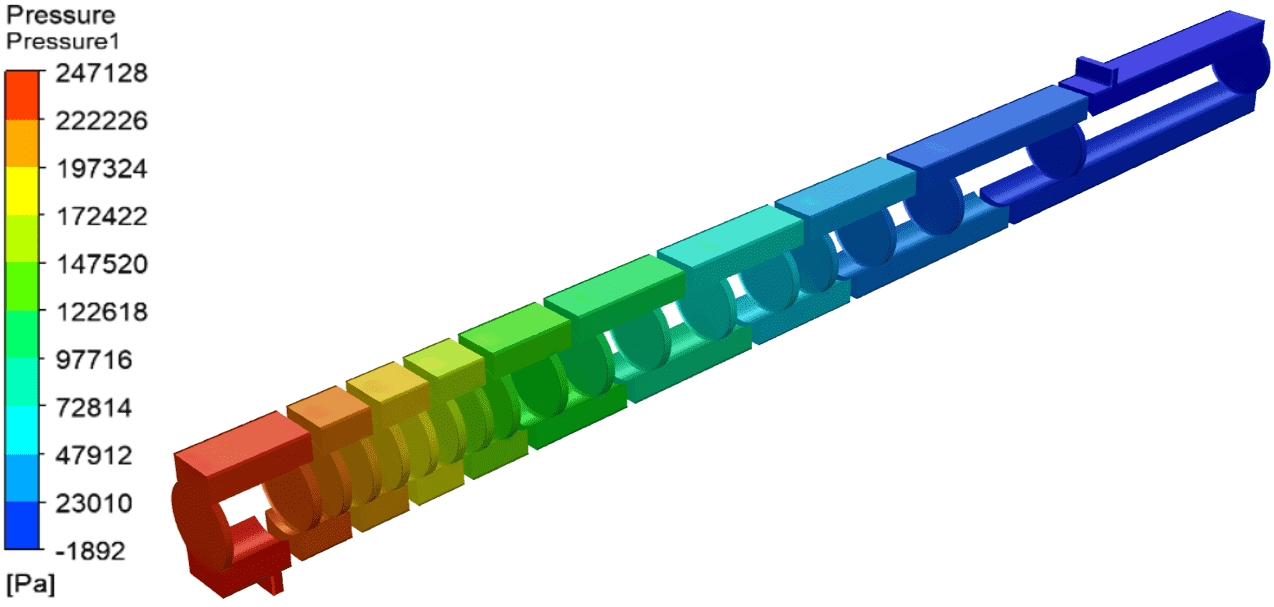}
\caption{\label{fig:TGT:proto_CFD_pres} Static pressure [Pa] distribution along the target cooling circuit. The total pressure drop is $\approx$ 2.5 bar.}
\end{figure}

Figure \ref{fig:TGT:proto_CFD_vel3D} shows the 3D flow distribution in the water volume of the inner vessel; the high velocity in the gap between the plates is very clear. Figure~\ref{fig:TGT:proto_CFD_pres} illustrates the static pressure distribution on the surfaces of the fluid volume, and provides an estimate of about 2.5 bar for the pressure drop along the target.

\begin{figure}[htbp]
\centering %
\includegraphics[width=0.5\linewidth]{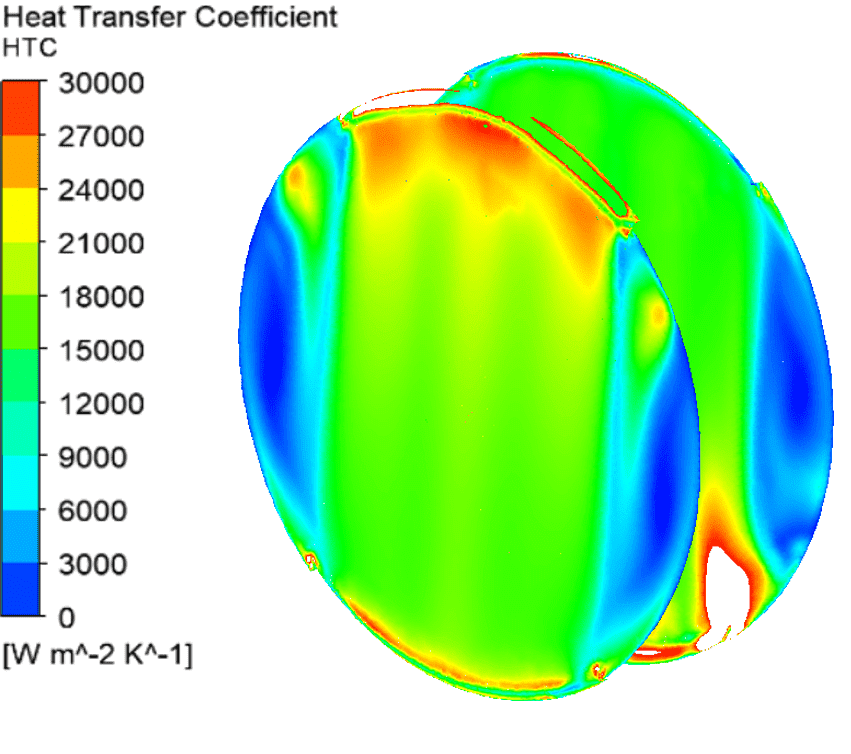}
\caption{\label{fig:TGT:proto_CFD_HTC} Heat transfer coefficient distribution in the front surface of block \#4.}
\end{figure}

The thermal simulations results described in the previous section show that, due to a combination of low thermal conductivity, high power deposition and potential localized boiling, the cladding temperature is a critical parameter. In order to limit the cladding temperature, a surface heat transfer coefficient of about 16000 W/m$^2$K has been assumed for the FEM calculations by selecting the average flow velocity in the vertical gaps at 4 m/s. In reality, the actual value of HTC predicted by the simulations is slightly higher and non-uniform, as shown in Figure~\ref{fig:TGT:proto_CFD_HTC}, providing conservative values. 

\begin{figure}[htbp]
\centering %
\includegraphics[width=1\linewidth]{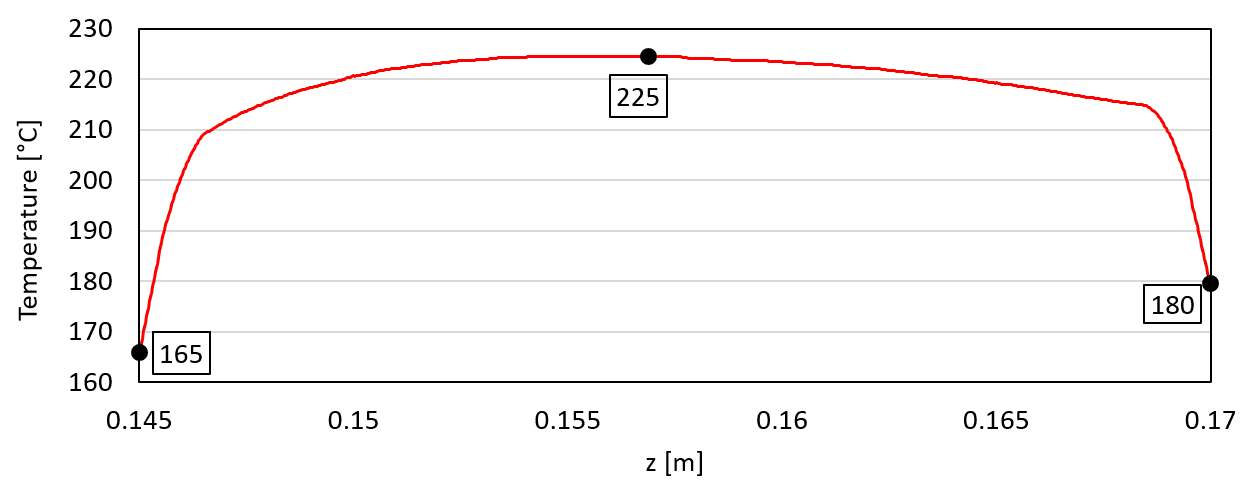}
\caption{\label{fig:TGT:proto_CFD_surface} Temperature distribution along z-axis of block \#4 at the end of the beam pulse (peak profile). Maximum temperature in the cladding surface $\approx$ $180^{\circ}\mathrm{C}$, peak temperature in the block $\approx$ $225^{\circ}\mathrm{C}$.}
\end{figure}

Transient simulations in this configuration have shown that the predicted peak surface temperature at \num{3e12} p+/pulse for the cladding of block \#4 (most critical block from a thermal standpoint) is about $180^{\circ}\mathrm{C}$, well below the boiling temperature of water at 22 bar ($\sim212^{\circ}\mathrm{C}$). The temperature distribution along the longitudinal axis in block \#4 is shown in Figure \ref{fig:TGT:proto_CFD_surface}. The peak temperature after beam impact is about $225^{\circ}\mathrm{C}$, found in the center of the TZM core. 

It shall be noted that the results of the CFD (Figure~\ref{fig:TGT:proto_CFD_surface}) and FEM (Figure~\ref{fig:TGT:proto_temp_all}) simulations show a discrepancy of maximum 15$\,^{\circ}\mathrm{C}$ on the temperature field. The peak temperature in the TZM core of block 4 obtained from the calculations in ANSYS Mechanical is around $240^{\circ}\mathrm{C}$. This difference is due to the fact that FEM assumes a constant HTC value on all wetted surfaces (calculated from a correlation for fully-developed flow), while the CFD simulation accounts for the actual HTC distribution. This discrepancy is negligible in the scope of the respective analyses.

\begin{figure}[htbp]
\centering %
\includegraphics[width=1\linewidth]{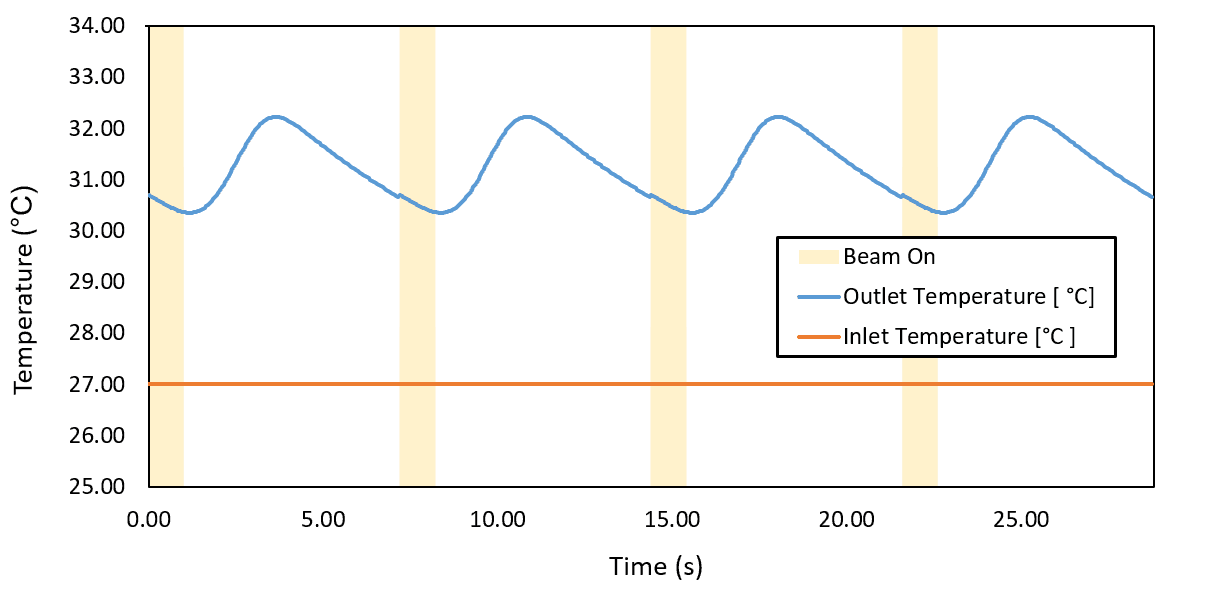}
\caption{\label{fig:TGT:proto_CFD_temp} Target inlet and outlet water temperature as a function of time. Evolution during 4 beam pulses after long-time operation.}
\end{figure}

With respect to the water temperature distribution, the thermal inertia of the blocks smooths out the rapid variations of the power versus time, and the resulting outlet water temperature only slightly oscillates around its average power, as shown in Figure~\ref{fig:TGT:proto_CFD_temp}.

\subsection{Target prototype construction and assembly}
\label{Sec:TGT:Proto:assembly}

\subsubsection{Prototype construction}

\begin{figure}[htbp]
\centering %
\includegraphics[width=1\linewidth]{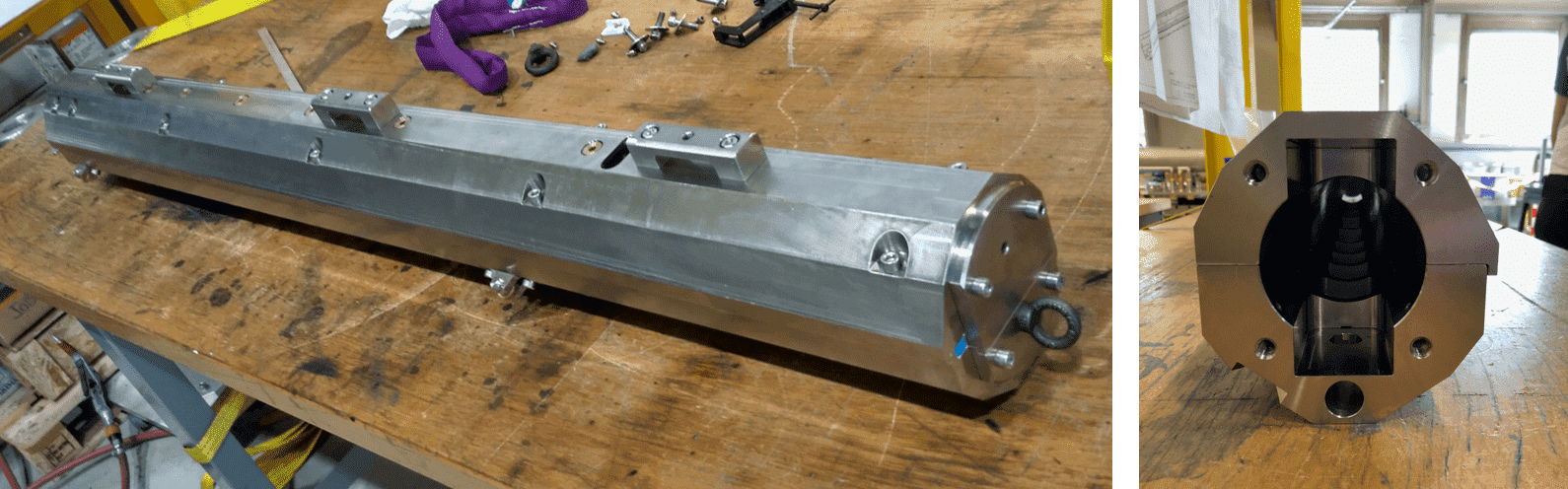}
\caption{\label{fig:TGT:proto_assy_inner} BDF target prototype inner tank side and front view. Lower and upper half shells mounted without target blocks. The right picture shows clearly the housing of the target blocks.}
\end{figure}

\begin{figure}[htbp]
\centering %
\includegraphics[width=1\linewidth]{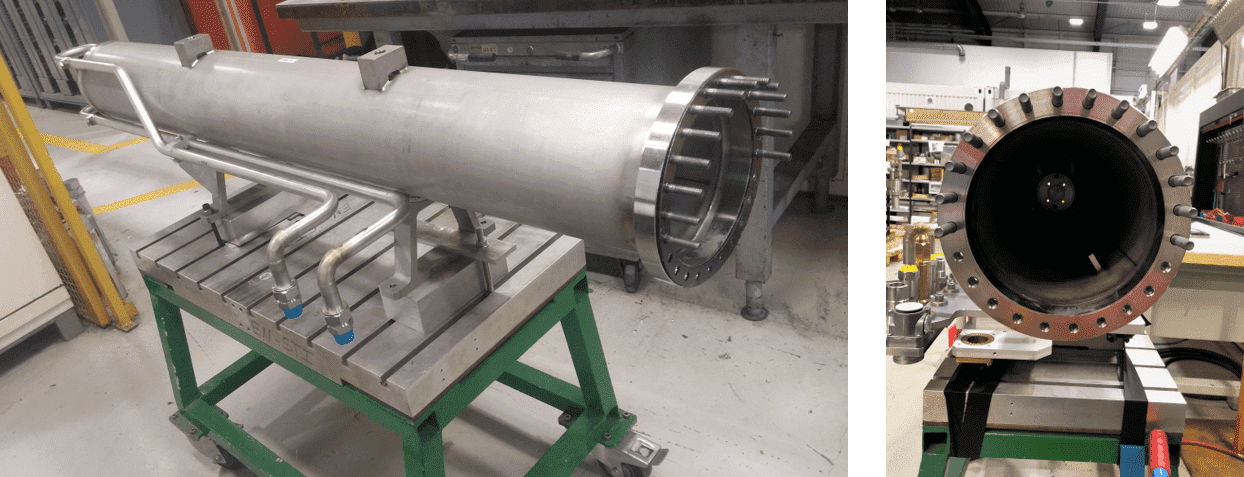}
\caption{\label{fig:TGT:proto_assy_outer} BDF target prototype outer tank side and front view.}
\end{figure}

The target inner tank manufacture was carried out by an external contractor. In parallel, the outer tank was manufactured at CERN. Figures~\ref{fig:TGT:proto_assy_inner} and \ref{fig:TGT:proto_assy_outer} present several views of the inner and outer tank.

The production of the target core blocks has been one of the most challenging aspect of the prototype construction. The prototype blocks are made out of different parts, following a similar process than the one foreseen for the final target blocks production (described in Section~\ref{Sec:TGT:MechDesign}). For the prototype blocks manufacturing via HIP process, several parts of different materials are needed. Each target block is made out of 4 different parts:
\begin{itemize}
    \item 1 TZM or W cylinder with diameter 77 mm and different length according to the block position in the target
    \item 1 Ta2.5W or Ta tube with inner diameter 77 mm, outer diameter 80 mm, and variable length
    \item 2 Ta2.5W or Ta disks with diameter 77 mm and thickness 3 mm.
\end{itemize}

In a first stage, the different parts were produced at the premises of an external supplier and received at CERN. All the parts were tested with ultrasounds, mechanical and chemical testing at the contractor's premises. Furthermore, vacuum leak testing was performed at CERN once the parts were received. The TZM and W cylinders, as well as the Ta2.5W and Ta disks were requested to have larger diameters than the ones aforementioned. Then, precise machining was performed at CERN to ensure the required 0.1 mm gap between the cylinders/disks and the Ta or Ta2.5W tubes. The machined parts prior to the HIP process can be seen in Figure~\ref{fig:TGT:Block_exploded}.

\begin{figure}[htbp]
\centering %
\includegraphics[width=0.6\linewidth]{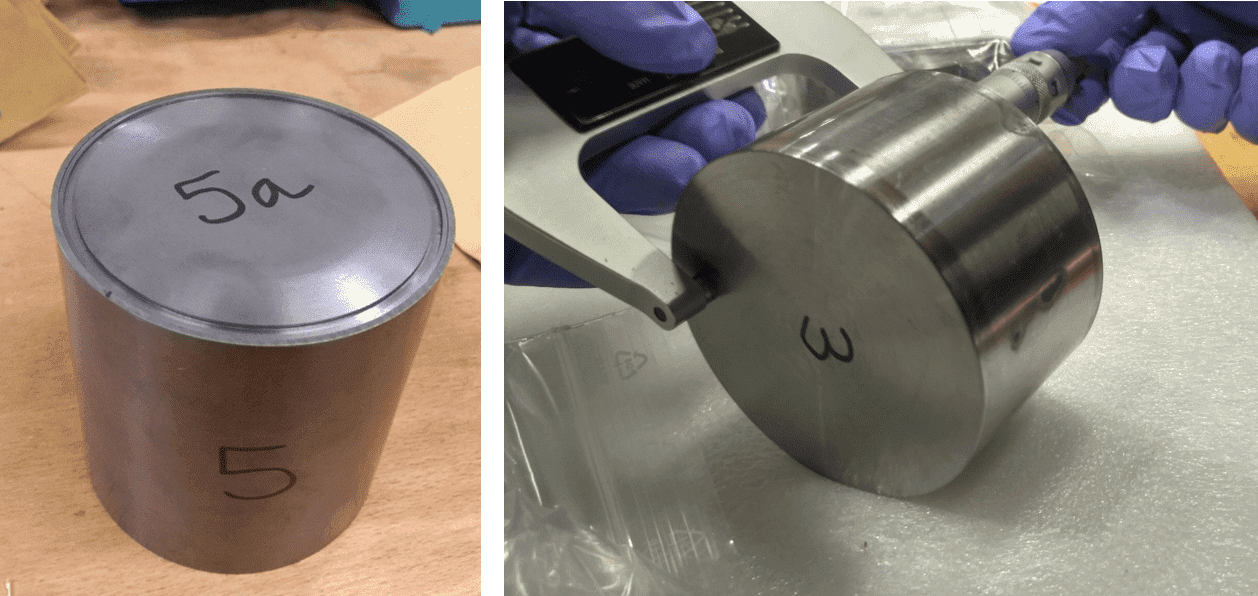}
\caption{\label{fig:TGT:proto_assy_block} Target block parts mounted before the HIP process (left) and after the HIP run and subsequent machining performed with the objective of matching the required sizes in order to house the blocks in the target inner tank (right).}
\end{figure}

Once the machining was performed, the target parts were sent to an external contractor for the prototype blocks production via HIPing. The manufacturing process of the prototype blocks is identical to the one of the final target blocks described in Section~\ref{Sec:TGT:MechDesign}. Figure~\ref{fig:TGT:proto_assy_block} shows the target parts of one target block mounted before HIPing, and a finished target block after HIPing and final machining.

\subsubsection{Prototype testing and assessment}

Prior to the full assembly of the target prototype, several tests on the target prototype housing were carried out:

\begin{itemize}
    \item The target outer tank was tested at a pressure of 31 bar in order to assess the leak tightness and pressure compliance of the tank with a safety factor of 1.43 with respect to the operation pressure of 22 bar~\cite{EN-13445};
    \item The fully metallic plug-in for the water and electrical connections was connected to the target outer tank and tested at 31 bar static pressure and 22 bar with circulating water flow. Figure~\ref{fig:TGT:proto_staubli} presents the plug-in connection of the female and male plates before installation and the plug-in installed in the outer tank with the connection performed for testing; 
\begin{figure}[htbp]
\centering %
\includegraphics[width=1\linewidth]{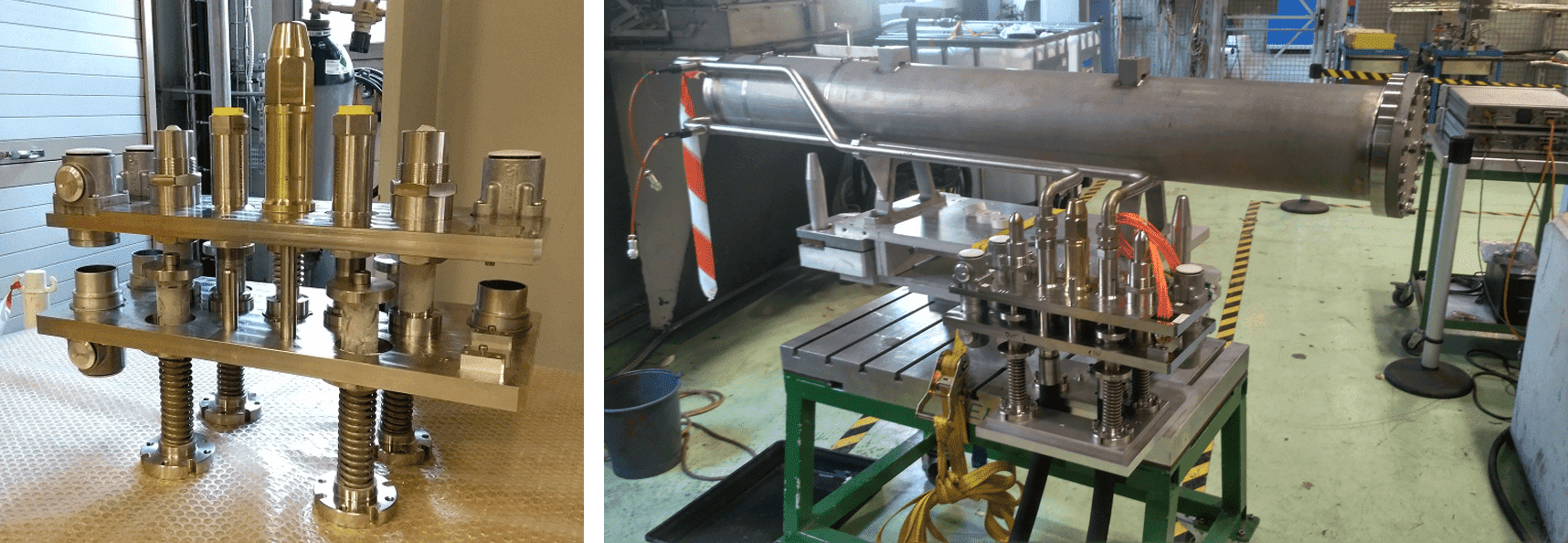}
\caption{\label{fig:TGT:proto_staubli} Fully metallic connection for the water and electrical interfaces. Coupled male and female connection before installation (left); installed plug-in with coupled connection for testing (right).}
\end{figure}
    
    \item Several "dummy" cooper blocks were mounted on the inner tank lower shell in order to perform and consolidate the assembly of the inner and outer tank, as can be seen in Figure~\ref{fig:TGT:proto_assy_copper}. Some of the copper blocks were equipped with instrumentation cabling in order to simulate the prototype blocks instrumentation.
    
\begin{figure}[htbp]
\centering %
\includegraphics[width=1\linewidth]{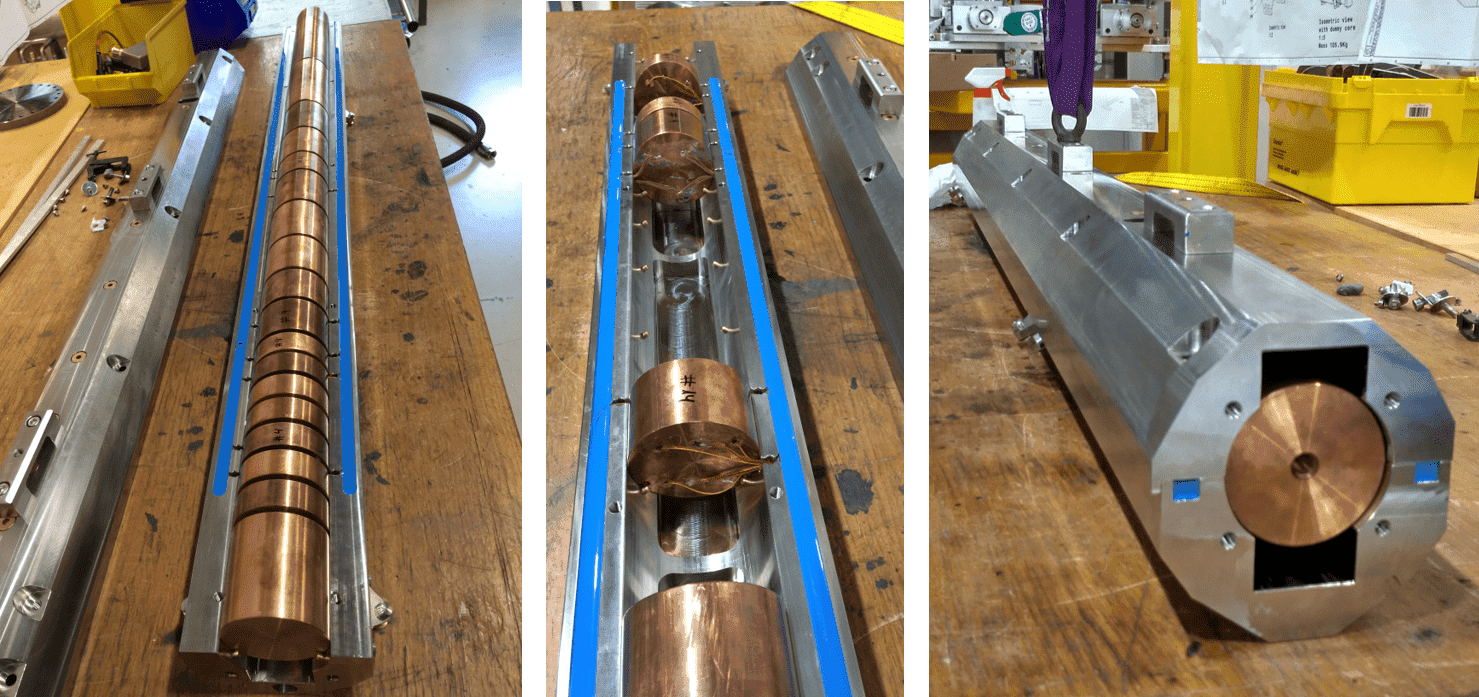}
\caption{\label{fig:TGT:proto_assy_copper} Copper dummy blocks installed in the target prototype inner tank to validate the prototype assembly.}
\end{figure}
    
    \item Cooling tests were executed with the whole target assembly (with dummy copper blocks) to confirm the good functioning of the target cooling system. The measured value of pressure drop was $\approx$ 3 bar, very close to the calculated value of 2.5 bar via CFD simulations. Figure~\ref{fig:TGT:proto_test_cooling} presents the experimental setup used for the cooling system preliminary tests. 
    
\begin{figure}[htbp]
\centering %
\includegraphics[width=1\linewidth]{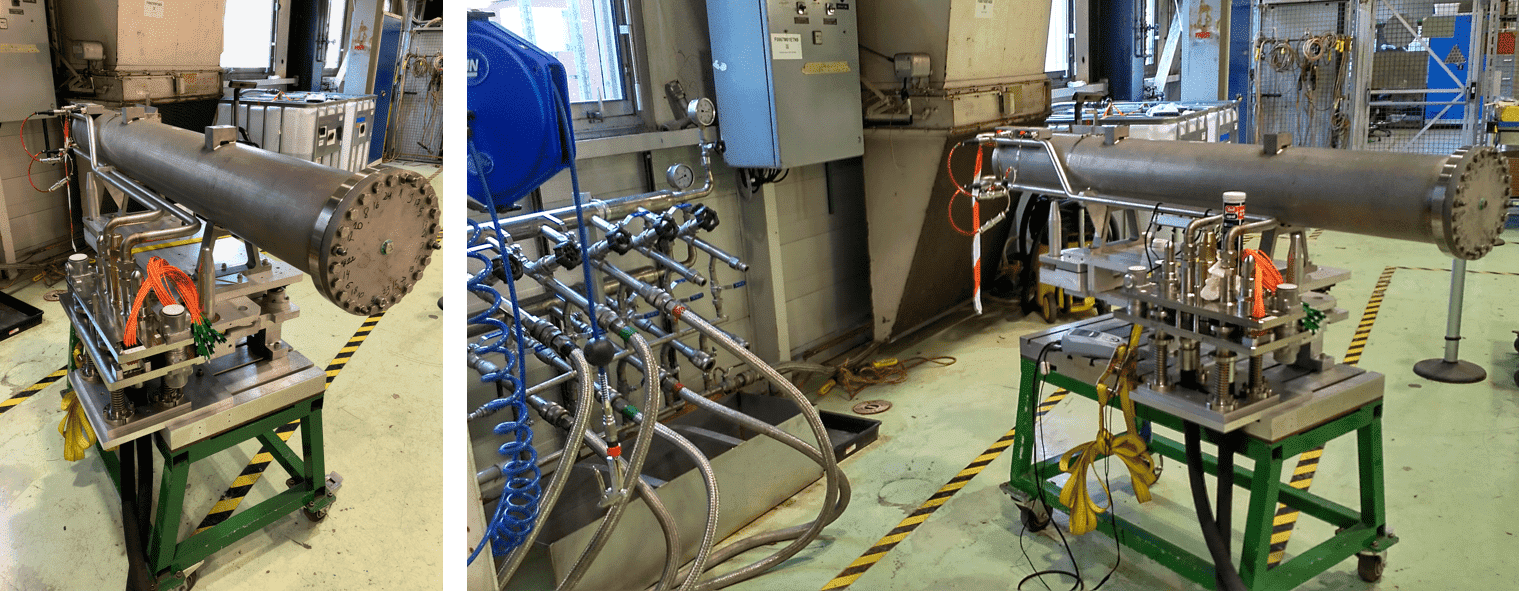}
\caption{\label{fig:TGT:proto_test_cooling} Target prototype preliminary cooling tests experimental setup. The target prototype was tested under static pressure and under water circulation with and without "fake" instrumentation in the copper blocks.}
\end{figure}

    \item Different tests were carried out with the remote manipulation and heavy handling teams in order to simulate several operations of remote installation and dismantling (lifting, unscrewing, extracting the inner tank). Figure~\ref{fig:TGT:proto_robot} illustrates some of the robotics and handling tests performed.
    
\begin{figure}[htbp]
\centering %
\includegraphics[width=1\linewidth]{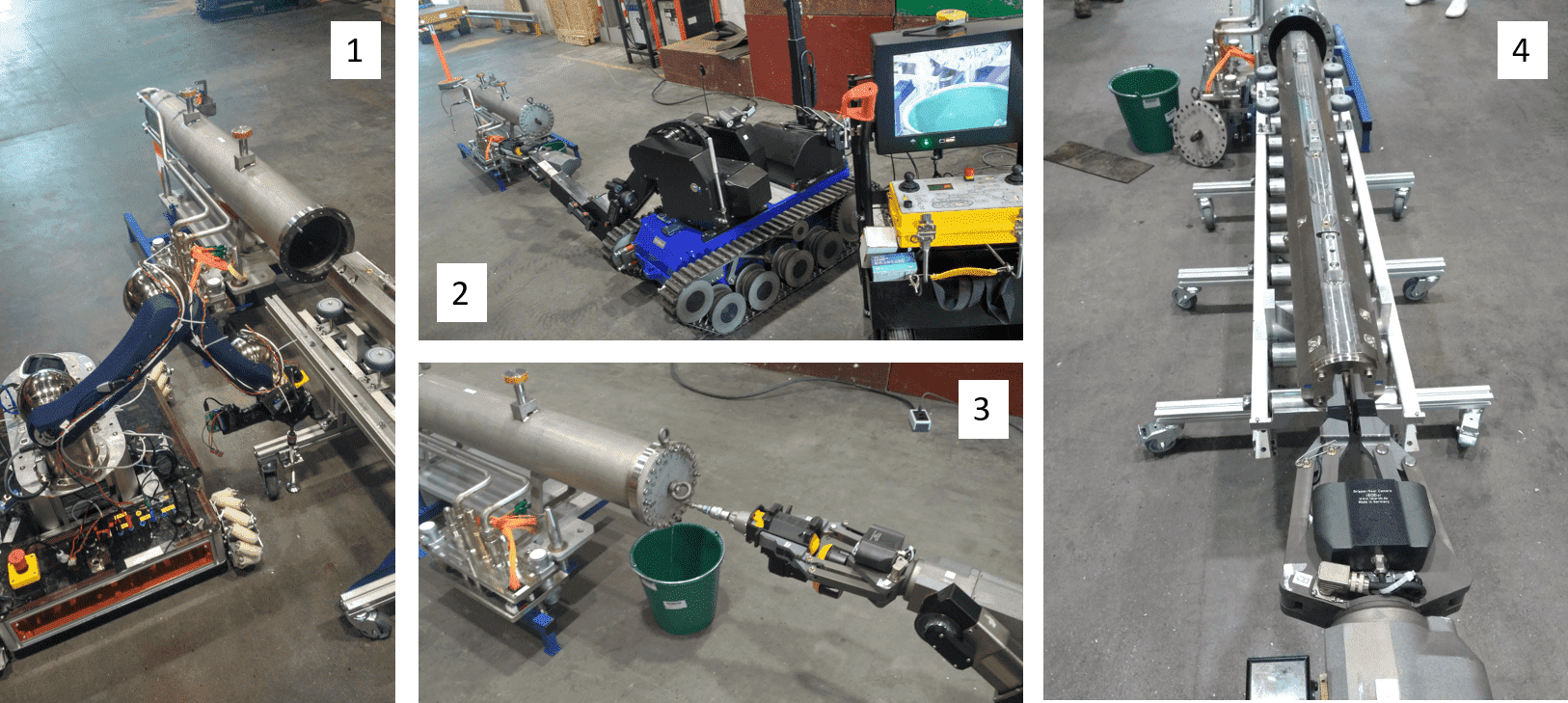}
\caption{\label{fig:TGT:proto_robot} Robot tests executed on the target prototype assembly with CERNBot (1) and Teodor (2). Simulation of dismantling operations such as: target outer tank draining (3), where a screw is unbolted by the robot in order to remove all the water remaining in the outer tank; and inner tank extraction (4), where the robot pulls the inner tank out of the outer tank with help of an additional conveyor support.}
\end{figure}

\end{itemize}

\subsubsection{Target prototype assembly}

\begin{figure}[htbp]
\centering %
\includegraphics[width=1\linewidth]{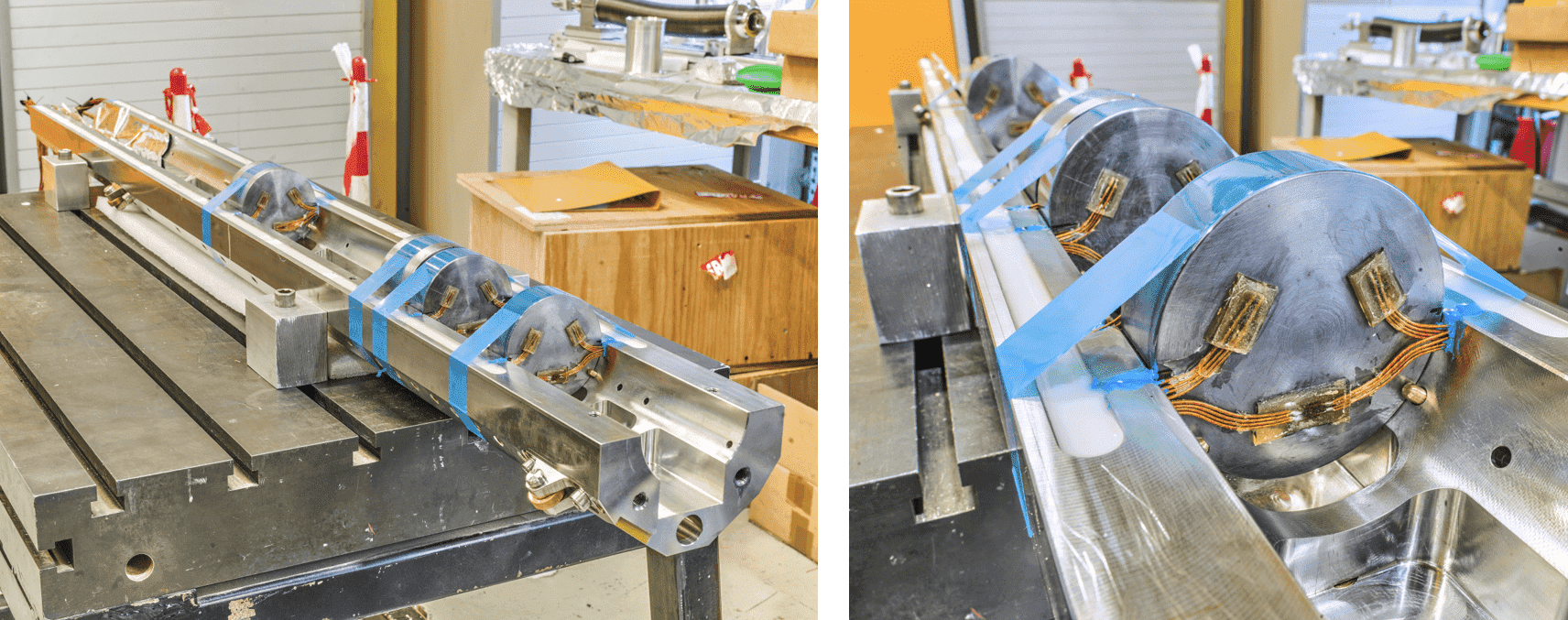}
\caption{\label{fig:TGT:proto_assy_instru} Instrumented prototype blocks installed on the inner tank lower half shell. Photograph: J. M. Ordan~\cite{CDS_pictures1}.}
\end{figure}

The final assembly of the target prototype was carried out once the production of the final blocks was achieved. Four of the target blocks were instrumented and installed on the inner tank lower half-shell. More details on the target blocks instrumentation will be given in Section~\ref{Sec:TGT:Proto:Instru}. Figure~\ref{fig:TGT:proto_assy_instru} presents the instrumented target blocks laying on the lower half-shell. 

\begin{figure}[htbp]
\centering %
\includegraphics[width=1\linewidth]{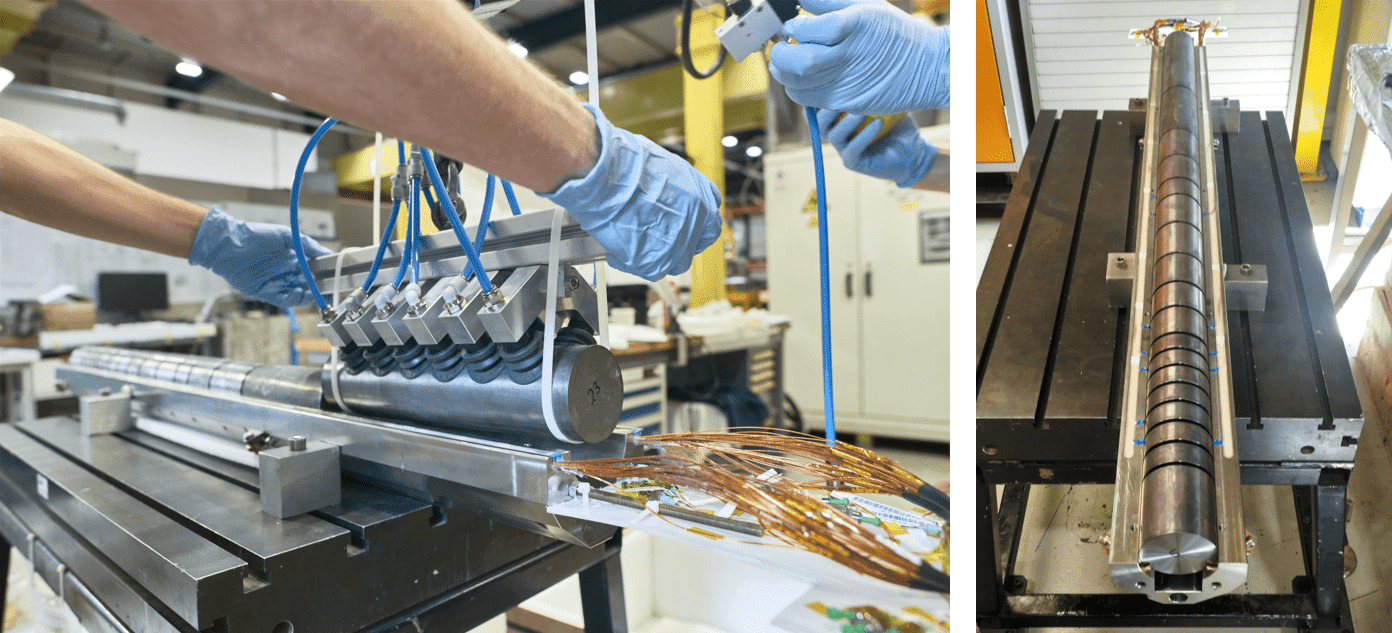}
\caption{\label{fig:TGT:proto_blocks_installation} (Left) Installation of the last tungsten block with dedicated handling equipment (Photograph: J. M. Ordan)~\cite{CDS_pictures1}. (Right) Full target core assembly laying on the inner tank lower shell.}
\end{figure}

Finally, the remaining target blocks were installed on the inner tank lower shell. Some of the target cylinders required dedicated handling tools due to their heavy weight (see Table~\ref{tab:TGT:proto_dimensions}). Figure~\ref{fig:TGT:proto_blocks_installation} illustrates the installation procedure as well as the full target core installed in the inner tank.

\begin{figure}[htbp]
\centering %
\includegraphics[width=1\linewidth]{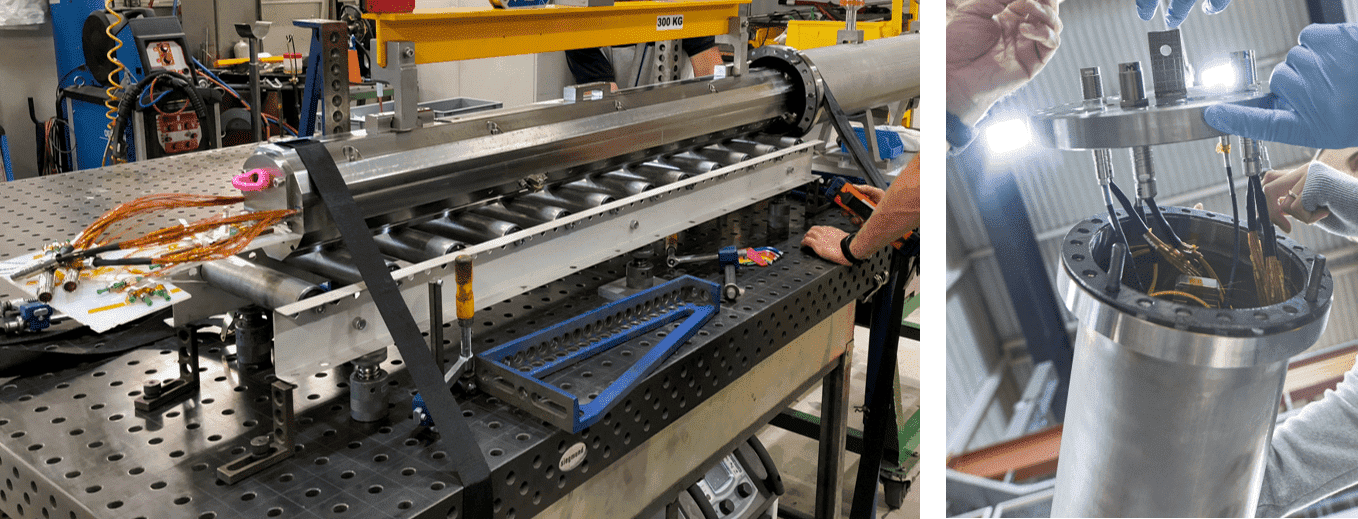}
\caption{\label{fig:TGT:proto_assy_closing}Left: insertion of the inner tank into the outer tank with dedicated tooling. Right: connection of the instrumentation electrical wiring to the feedthroughs installed in the outer tank downstream flange before the tank closing (Photograph: J. M. Ordan).}
\end{figure}

Once the inner tank assembly was completed, it was inserted into the outer tank. Once again, dedicated tooling was necessary to safely insert the inner tank in the outer tank: a dedicated support with a conveyor mechanism aligned with the outer tank was manufactured in order to facilitate the inner tank insertion. The electrical connections from the inner tank to the feedthroughs in the outer tank downstream flange were performed, and the outer tank was finally closed to be ready for installation. Figure~\ref{fig:TGT:proto_assy_closing} illustrates the described process. Before the target installation in TCC2, cooling water tests with the final target prototype assembly were carried out in order to assess the leak-tightness of the whole assembly.
 
Detailed pictures of the target assembly and installation can be found in the CERN Document Server \cite{CDS_pictures1, CDS_pictures2}.

\FloatBarrier

\subsection{Design and development of the BDF target prototype instrumentation}
\label{Sec:TGT:Proto:Instru}

\subsubsection{Introduction}
This section summarises the design and selection of the instrumentation and electronics for the measurements of strain and temperature of four disk-shaped blocks of the BDF target prototype installed upstream the T6 target in the TCC2 area. These sensors and services will be subjected to an underwater and pressurised environment, high-speed water stream, high temperatures and radiation. 

\begin{figure}[htbp]
\centering %
\includegraphics[width=1\linewidth]{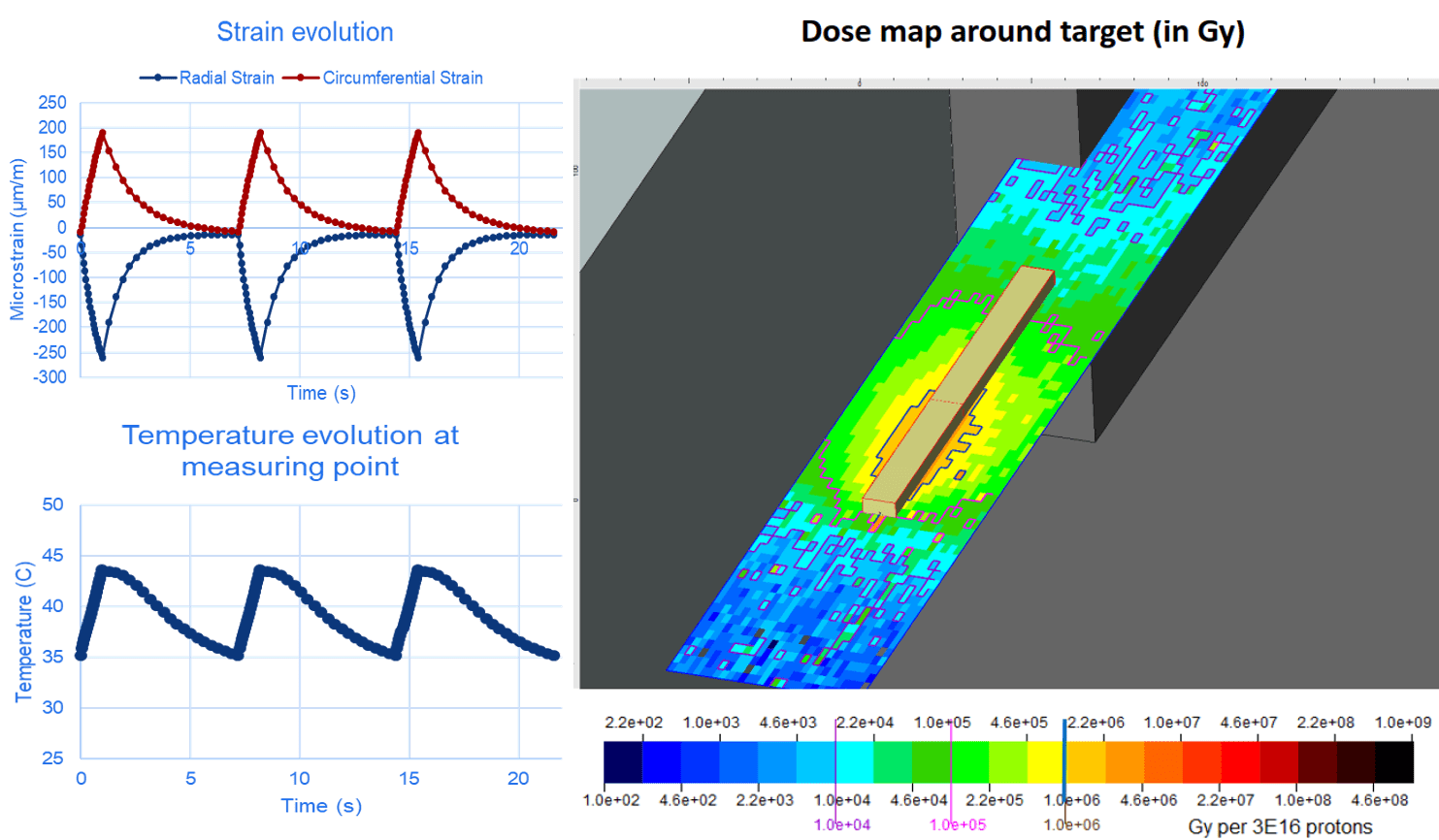}
\caption{\label{fig:TGT:Proto_simulations} (Left) Simulation of the temperature, radial and circumferential strain vs time at the measuring point. (Right) Total dose (in Gy) around the target at the end of the irradiation.}
\end{figure}

The design of the instrumentation and services has been done according to preliminary simulations results: some of these guidelines, such as expected temperature rise, strain response and radiation dose, are shown in Figure~\ref{fig:TGT:Proto_simulations}. The aim of the instrumentation is to validate the mechanical response and thermal models assumed for the materials and to compare the measured behaviour with the FEA simulations carried out. 

\begin{figure}[htbp]
\centering %
\includegraphics[width=1\linewidth]{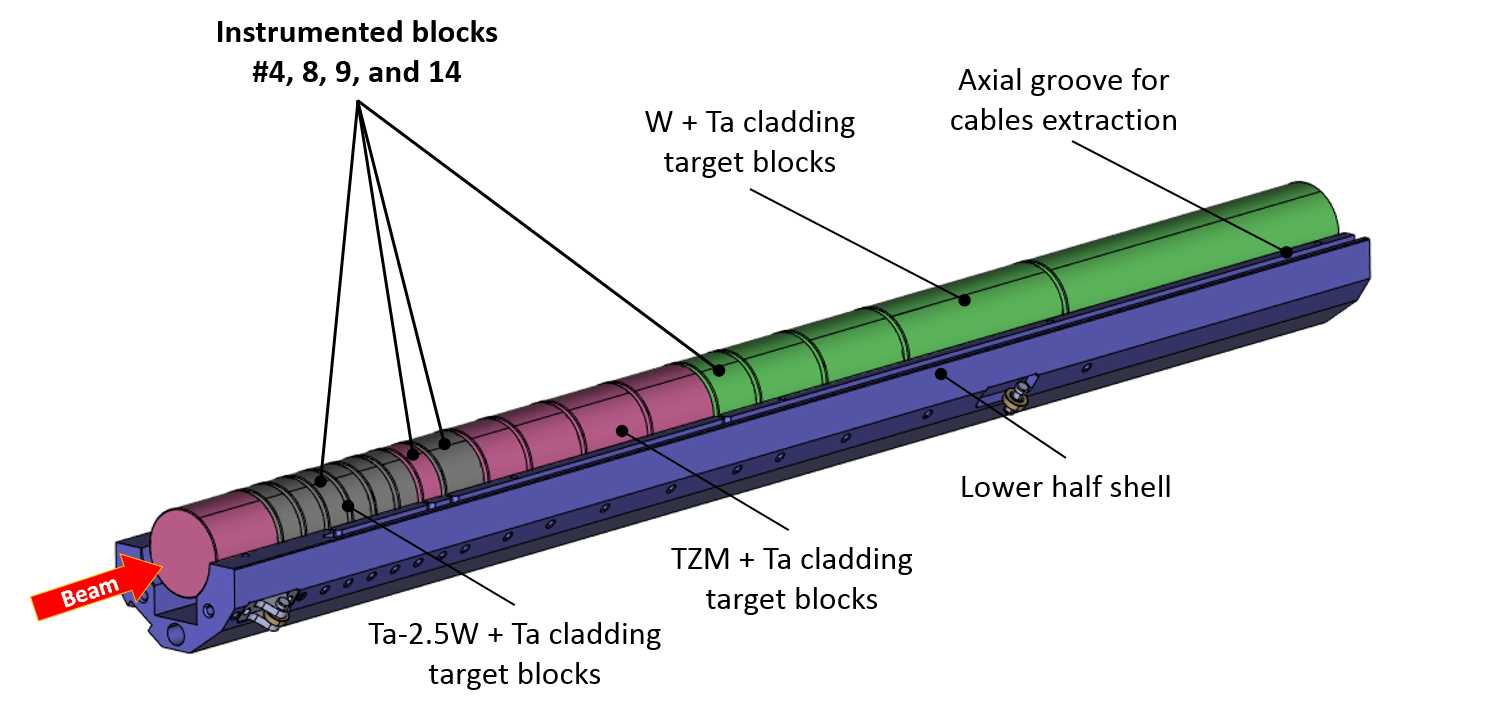}
\caption{\label{fig:TGT:Instru_block_labels_v2} Schematic view of the BDF target prototype instrumented blocks.}
\end{figure}

The instrumented blocks locations and materials, as well as the position of the groove for the extraction of the cabling, are indicated in Figure~\ref{fig:TGT:Instru_block_labels_v2}.

\subsubsubsection{Instrumentation requirements}

It is requested to measure temperature, radial and circumferential strain of several points of the flat surfaces (upstream and downstream ones) of the blocks. The very first step in the choice of the has been to consider the harsh working environment of the experiment, which imposes the minimum requirements of the sensors for their functioning, as shown in Table \ref{tab:TGT:environment_conditions}.

\begin{table}[htbp]
\centering
\caption{BDF target prototype instrumental requirements imposed by the environmental conditions of the experiment.}
\label{tab:TGT:environment_conditions}
\begin{tabular}{cc}
\toprule
\textbf{Environment feature} & \textbf{Requirement/max value} \\ \hline
Underwater & Waterproofness \\
Pressure {[}bar{]} & 22 \\
Water flow {[}m/s{]} & 4.2 \\
Max temperature ${[}^{\circ}\mathrm{C}{]}$ & 200 \\
Total dose {[}MGy{]} & 100 \\
\bottomrule
\end{tabular}
\end{table}

These requisites impose the choice of waterproof/watertight, radiation hard and pressure resistant equipment. The second step is represented by the capability to accurately measure the physical quantities of interest, meaning to acquire the related signal amplitude and frequency, as required in Table~\ref{tab:TGT:Sensors_requirements}.

\begin{table}[htbp]
\centering
\caption{Requirements of the measurements.}
\label{tab:TGT:Sensors_requirements}
\begin{tabular}{cc}
\toprule
\textbf{Physical quantity} & \textbf{Requirement/max value} \\ \hline
Strain {[}\textmu m/m{]} & 500 \\
Strain time response {[}ms{]} & 100 \\
Temperature ${[}^{\circ}\mathrm{C}{]}$ & 200 \\
Temperature time response {[}s{]} & 1\\
\bottomrule
\end{tabular}
\end{table}

Other restrictions are imposed by the design of the BDF target prototype and should be considered to correctly perform the application of the sensors and their cabling (max number of cables and length). These are summarised in Table~\ref{tab:TGT:proto_design_parameters}.

\begin{table}[htbp]
\centering
\caption{BDF target prototype design requirements and impact on the instrumentation.}
\label{tab:TGT:proto_design_parameters}
\begin{tabular}{cc}
\toprule
\textbf{Geometrical parameter} & \textbf{Value} \\ \hline
Gap between neighbouring blocks [mm] & 5 \\
Max distance Sensor - Feedthrough flange [mm] & 1560 \\
Cable passage section per block [mm$^2$] & 10x2 \\
Axial cables passage per section [mm$^2$] & 10x12\\
\bottomrule
\end{tabular}
\end{table}

\subsubsection{Proposed instrumentation and services}
\begin{figure}[htbp]
\centering %
\includegraphics[width=0.5\linewidth]{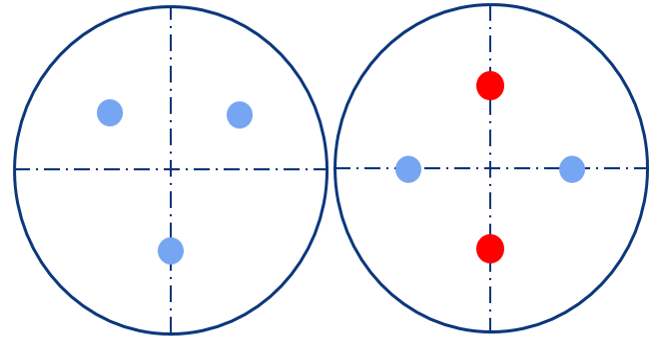}
\caption{\label{fig:TGT:sensors_position} Strain (blue) and temperature (red) measuring points. Design for blocks 4, 9 and 14 blocks' upstream (left) and downstream (right) faces.}
\end{figure}

Taking into account requests and constraints of the beam experiment, it was proposed and accepted to instrument the following measuring points of the blocks as follows (see Figure \ref{fig:TGT:sensors_position}):

\underline{Blocks \#4, 8 and 14}: three measurement points at 120\textdegree\:of the upstream faces for radial and circumferential strain, one of the points coincides with the vertical (resistive bi-axial strain gauges). Two measurement points at 180\textdegree\:of the downstream faces for radial and circumferential strain, coinciding with the horizontal (optical fiber sensors based on FBG). Two measurement points at 180\textdegree\:on the vertical of the downstream faces for temperature sensing (Pt100).

\underline{Block \#9}: same design of the other blocks with the instrumentation of the upstream face placed on the downstream one and vice versa. 
For information about radiation hardness of the different materials employed see Ref.~\cite{radiation_damage}.
The following paragraphs explain the working principle of the sensitive elements selected.

\subsubsubsection{Strain measurements}

\underline{Electrical Strain Gauges}
\smallskip

The working principle of the strain gauges relies on the relationship between the change of electrical resistance and deformation: it is well known that all electrical conductors change their electrical resistance along with mechanical strain. The change of resistance during the strain measurement is rather small, therefore an electronic amplifier is necessary to perform the measurement.
The sensitivity of the strain gauges is determined by a gauge factor \textit{k} (approximately 2) defined in equation~\ref{eqn:TGT:gage_factor}, stated by the manufacturer by testing a sample from each production lot, according to the standard procedure.

\begin{equation}
    \label{eqn:TGT:gage_factor}
    \textit{k}=(\Delta R/R_0)/(\Delta l/l)=(\Delta R/R_0)/\epsilon  [(\Omega/\Omega)/(m/m)]
\end{equation}

The strain gauges for BDF target prototype application are the 1-XY91-1,5/120 by HBM\textsuperscript{\textregistered}, where two measuring grids are placed in a rosette to measure strain in perpendicular direction, whose characteristics are listed in Table \ref{tab:TGT:strain_gages}. The gauges feature a polyimide carrier and cover while the measuring grids are embedded in a constantan foil. 

\begin{table}[htbp]
\centering
\caption{1-Y91-1.5/120 resistive strain gauge characteristics.}
\label{tab:TGT:strain_gages}

\begin{tabular}{cc}
\toprule
\textbf{Property} & \textbf{Value} \\ \midrule
Nominal resistance ${[}\Omega{]}$ & 120 +/- 0.5 \\
\textit{a}, lenght of measuring grid (circumferential strain) {[}mm{]} & 1.5 \\
\textit{b}, width of the measuring grid (radial strain) {[}mm{]} & 1.2 \\
\textit{k} factor, \textit{a} & 1.94 +/- 1.5 \% \\
\textit{k} factor, \textit{b} & \multicolumn{1}{l}{1.97 +/- 1.5 \%}\\
\bottomrule
\end{tabular}
\end{table}

\underline{Bonding \& protection of the strain gauges}
\smallskip

A good bonding is required between the specimen surface and the strain gauge, in order to minimize the strain dissipation; curing adhesives, ceramic putty, flame-deposited ceramics and spot welded joints are some possible methods. The selected strain gauges are not waterproof but they can be employed in combination with a protective cover agent. The glue and protective agent are listed in Table~\ref{tab:TGT:glue_properties}.

\begin{table}[htbp]
\centering
\caption{Selected bonding and cover agents.}
\label{tab:TGT:glue_properties}
\begin{tabular}{ccc}
\toprule
\textbf{Product} & \textbf{Label} & \textbf{Characteristics} \\ \midrule
Bonding material & Vishay\textregistered M-Bond 610 & \begin{tabular}[c]{@{}c@{}}2-component epoxy phenolic adhesive, hot curing\\{[}-269 \textdegree C; +260 \textdegree C{]} \end{tabular} \\
Cover agent & HBM\textregistered X280 & \begin{tabular}[c]{@{}c@{}}Epoxy resin, cold-curing\\{[}-200 \textdegree C; +200 \textdegree C{]} \end{tabular}\\
\bottomrule
\end{tabular}
\end{table}

\underline{Wheatstone bridge circuit}
\smallskip

A Wheatstone bridge circuit is used to determine the absolute value of an electrical resistance or the relative change in resistance (10\textsuperscript{-4} - 10\textsuperscript{-2} $\Omega/\Omega$).

\begin{figure}[htbp]
\centering %
\includegraphics[width=0.4\linewidth]{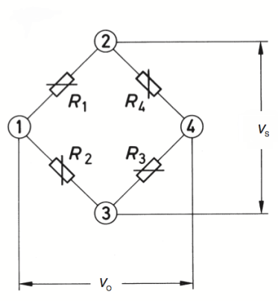}
\caption{\label{fig:TGT:wheatstone_bridge} Wheatstone bridge circuit.}
\end{figure}

With reference to Figure~\ref{fig:TGT:wheatstone_bridge}, resistances R\textsubscript{1} ... R\textsubscript{4} represent the four arms of the bridge. V\textsubscript{S} is the bridge excitation voltage, which is connected to the points 2 and 3. V\textsubscript{0} is the output voltage of the bridge (the measuring signal), it is connected to the point 1 and 4. A direct or alternative voltage is usually applied as bridge excitation. In a voltage-fed bridge configuration, the deviations from linearity of the strain gauges are automatically corrected.

The Wheatstone bridge is a voltage divider. Voltage V\textsubscript{S} is applied to point 2 and 3 and it is divided up in the two arms of the bridge R\textsubscript{1}, R\textsubscript{2} and R\textsubscript{4}, R\textsubscript{3}. If the bridge is unbalanced, different voltages are measured in the two arms. This is calculated by equation \ref{eqn:TGT:bridge_voltages}.

\begin{equation}
    \label{eqn:TGT:bridge_voltages}
    V_0/V_S =1/4 ((\Delta R_1)/R_1 -(\Delta R_2)/R_2 +(\Delta R_3)/R_3 -(\Delta R_4)/R_4 )
\end{equation}

Or, in terms of strain:

\begin{gather}
    \label{eqn:TGT:bridge_voltages_strain}
    V_0/V_S =\textit{k}/4 (\epsilon_1-\epsilon_2+\epsilon_3-\epsilon_4)
\end{gather}

\begin{figure}[htbp]
\centering %
\includegraphics[width=0.4\linewidth]{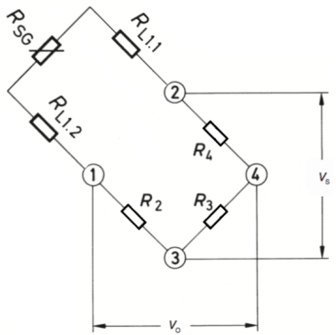}
\caption{\label{fig:TGT:quarter_bridge} Wheatstone bridge circuit.}
\end{figure}

Equation \ref{eqn:TGT:bridge_voltages_strain} is true if all the arms of the bridge are active. If only one gauge is active for the measurement, the circuit is called a quarter bridge circuit. The strain gauge and the connecting leads form one arm of the bridge and R\textsubscript{2}, R\textsubscript{3} and R\textsubscript{4} are fixed resistances to complete the bridge circuit (Figure \ref{fig:TGT:quarter_bridge}). R\textsubscript{SG} is the measuring strain gauge, R\textsubscript{L1.1} and R\textsubscript{L1.2} are the resistances of the connecting leads between the strain gauge and the rest of the bridge, R\textsubscript{2}...R\textsubscript{4} are the completion resistors. It is assumed that the strain measurement is due to the the bridge unbalance causes by the strain gauge [eq. \ref{eqn:TGT:quarter_bridge}].

\begin{gather}
    \label{eqn:TGT:quarter_bridge}
    V_0/V_S =1/4 (\Delta R/R)_{SG} \Rightarrow V_0/V_S =k/4 \epsilon
\end{gather}

All the resistive strain gauges in the experiment are installed in quarter bridge, 8 of them in a 4-wire configuration and the others in a 2-wire configuration.

\underline{Optical Fiber Sensors for strain measurements}
\smallskip

\begin{figure}[htbp]
\centering %
\includegraphics[width=1\linewidth]{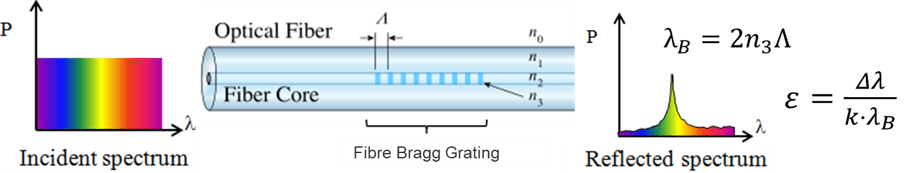}
\caption{\label{fig:TGT:FBG_principle} FBG working principle.}
\end{figure}

Optical fibers can be used to measure strain by means of the Fiber Bragg Gratings (FBG) technology, which exploits the photo-sensitivity of the fiber core material. The main features of this kind of sensors are: robustness, extremely small dimensions, waterproofness, radiation hardness, immunity to electromagnetic field, power supply not necessary. Typically, FBG are obtained from standard single-mode telecom fibers with an external diameter of 200 \textmu m and a core diameter of 9 \textmu m. The working principle is based on the periodic modulation of refractive indexes between the core and the grating, obtained by UV laser impression and a phase mask grating. The grating is able to reflect light with a wavelength \textlambda\:given by equation~\ref{eqn:TGT:lambda_bragg_mask} (see Figure~\ref{fig:TGT:FBG_principle}).

\begin{gather}
    \label{eqn:TGT:lambda_bragg_mask}
    \lambda = 2nD
\end{gather}

Where \textit{n} is the refractive index of the core, \textit{D} the distance in between the gratings. The FBG is glued on the surface of the piece subjected to deformation, hence as soon as deformation occurs the fiber stretches/shrinks and \textit{D} changes. By monitoring the corresponding wavelength change of the reflected light, it is possible to determine the deformation with the linear relationship expressed by equation~\ref{eqn:TGT:lambda_bragg_mask2}:

\begin{gather}
    \label{eqn:TGT:lambda_bragg_mask2}
    \epsilon = \Delta \lambda / k \lambda_B
\end{gather}

Where the \textit{k}-factor of the sensor is 0.78 +/- 2\%. 

\begin{figure}[htbp]
\centering %
\includegraphics[width=0.9\linewidth]{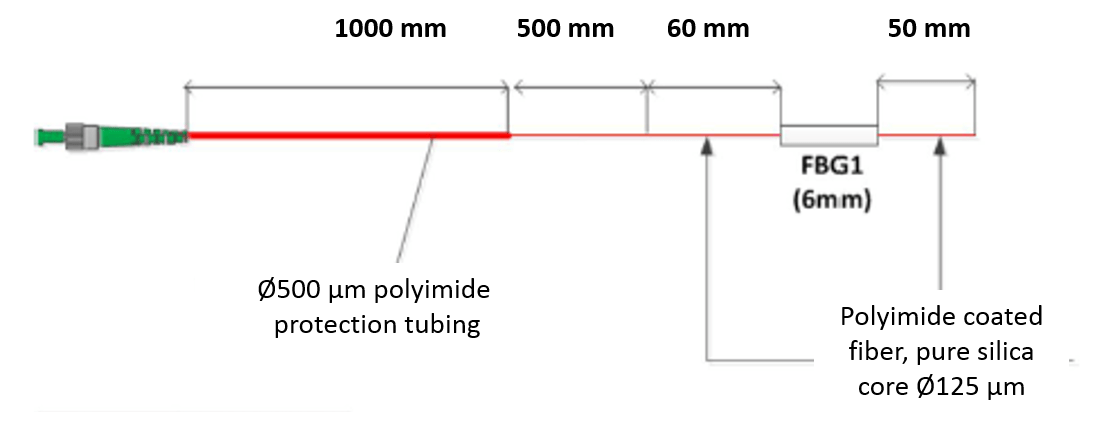}
\caption{\label{fig:TGT:FBG_array} Optical array design.}
\end{figure}

The sensors chosen are provided by HBM\textsuperscript{\textregistered} FiberSensing and their accuracy has been determined in the Mechanical Measurement Laboratory of the Engineering department of CERN \cite{EDMS-FBG}. The selected sensors have FBG inscribed on the fly with a new technique based on femtosecond long laser pulses. Their design is shown in Figure \ref{fig:TGT:FBG_array}, while some of their characteristics are stated in Table \ref{tab:TGT:fbg_charact}.

\begin{table}[htbp]
\centering
\caption{Selected bonding and cover agents.}
\label{tab:TGT:fbg_charact}
\begin{tabular}{cc}
\toprule
\textbf{Property} & \textbf{Value} \\ \hline
Core (pure silica) diameter {[}\textmu m{]} & 9 \\
Coating (polyimide) diameter {[}\textmu m{]} & 125 \\
Wavelegths {[}nm{]} & From 1515 to 1560 \\
Strain measurements accuracy & \textless{}5\% \\
Max strain {[}\textmu m/m{]} & 5000 \\
Grating length {[}mm{]} & 6 \\
Temperature influence {[}pm/\textdegree C{]} & 10 \\
Maximum temperature {[}\textdegree C{]} & 350\\
\bottomrule
\end{tabular}
\end{table}

\underline{Bonding of the optical fiber sensors}
\smallskip

The fibers are bonded to the surfaces following a procedure similar to the one employed for resistive strain gauges which especially guarantees the reduction of the quantity of glue. The adhesive used is Araldite\textsuperscript{\textregistered}  Standard (Ultra Strong) which is a cold curing two components epoxy glue. Since the fibers are waterproof, no protective agent has been used for these sensors. 

\subsubsubsection{Temperature measurements}

The electrical resistance of a metal changes with the temperature. The relationship between the two is non-linear and can be described by a high order polynomial (eq. \ref{eqn:TGT:pt100_r}):

\begin{gather}
    \label{eqn:TGT:pt100_r}
    R(t)=R_0 (1+A\cdot t+B\cdot t^2+C\cdot t^3+ \dots)
\end{gather}

Where R\textsubscript{0} is the nominal resistance at 0 \textdegree C. The number of high order terms is chosen depending on the required accuracy of the measurement.
Platinum is the most commonly used material, thanks to its chemical stability, its simple manufacturing, the availability of wire in a highly pure state and the excellent reproducibility of its electrical characteristic. The temperature measurement range varies from -200 \textdegree C to 961.78 \textdegree C. For measurements between 0 and 100 \textdegree C, the increase of resistance is linear with \textalpha, defined by eq. \ref{eqn:TGT:pt100_alpha}.

\begin{gather}
    \label{eqn:TGT:pt100_alpha}
    \alpha=(R_{100}-R_0)/(R_0\cdot100^{\circ}C)
\end{gather}

For platinum, \textalpha\:is 3.85055 10 \textdegree C \textsuperscript{-1}. The temperature can be determined from equation \ref{eqn:TGT:pt100_T} (R\textsubscript{T} is the Pt100 resistance at the temperature \textit{T}, R\textsubscript{0}=100 \textOmega).

\begin{gather}
    \label{eqn:TGT:pt100_T}
    T = (R_T/R_0 -1)/\alpha	
\end{gather}

The Pt100 selected for the experiment are provided by Radiospares (6117801), they are suitable for submerged environment and the temperature range is from -50 \textdegree C to +500 \textdegree C. The accuracy class is B. 

\underline{Bonding of Pt100}
\smallskip

A key factor to consider when selecting the glue for a temperature probe is that it must have a good thermal conductivity to minimise the error on the precision of the measurement. The selected glue is a cold curing two components epoxy adhesive called Stycast\textsuperscript{\textregistered} 2850 FT. To protect the electrical connection from the water stream these probes were embedded with the same protective agent used for resistive strain gauges (see Table \ref{tab:TGT:glue_properties}).

\subsubsection{DAQ (Data Acquistion) and software}
A description of the DAQ and Software used for data acquisition has already been provided in paragraph 4 and 5 of the EDMS document \cite{EDMS-T6-instru}, to which we refer. This paragraph aims to add information about the optical strain sensors acquisition device (optical interrogator) and the combination of modules used.

\underline{Optical Interrogator}
\smallskip

Initially, a dynamic optical interrogator was intended to use, in order to sample at 1000 Hz. Due to reasons explained in the following chapters, eventually it was decided to employ 2 static interrogators by HBM\textsuperscript{\textregistered} FS22SI which have a lower sampling frequency (1 Hz). 

\underline{Final Acquisition System}
\smallskip

\begin{figure}[htbp]
\centering %
\includegraphics[width=1\linewidth]{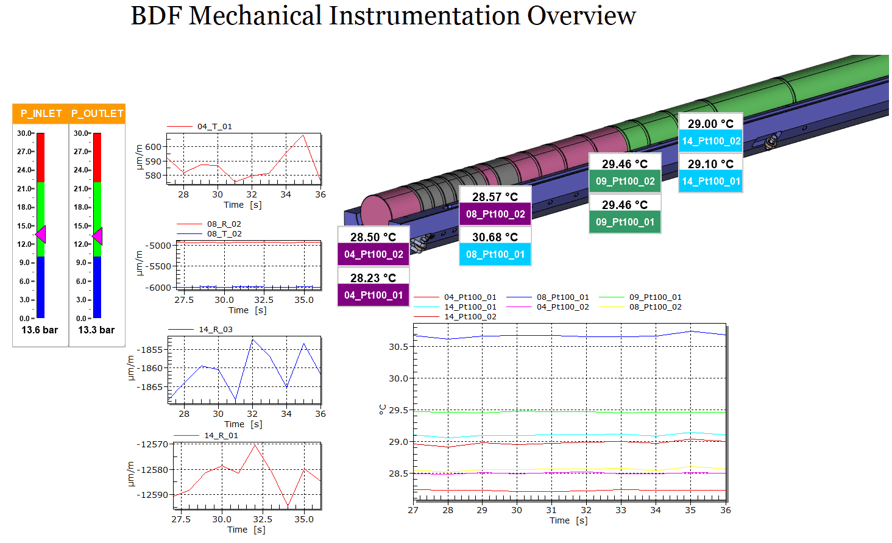}
\caption{\label{fig:TGT:catman} Catman\textsuperscript{\textregistered} front panel for online measurement overview.}
\end{figure}

The acquisition system installed in BA80 is composed by the following equipment:
\begin{itemize}
    \item 2 HBM\textsuperscript{\textregistered} Quantum X MX1615B, each one with 16 channels to read resistive strain gages and Pt100 signals
    \item 1 HBM\textsuperscript{\textregistered}  Quantum X M640 dedicated to the pressure sensors installed by EN/SMM on the inlet and outlet water pipes of the tank
    \item 1 HBM\textsuperscript{\textregistered}  CX27 for the synchronisation in between these 3 modules
    \item 2 optical interrogators HBM\textsuperscript{\textregistered}  FS22SI
    \item An Ethernet hub for the connection to the PC of all the devices
\end{itemize}
The software allows the synchronisation in between all the signals sampled at different frequencies and with different devices thanks to the Network Time Protocol (NTP) and to monitor online the measured quantities of interested through a tailored front panel, shown in Figure \ref{fig:TGT:catman}. The resistive strain gauges and pressure sensors are sampled at 1000 Hz while the temperature probes at 100 Hz. 

\subsubsection{Integration}

\subsubsubsection{Connectors/Feedthroughs}

On the connection flange of the tank, there are five electrical feedthroughs and four optical ones. These components, together with the plugs must be watertight and resistant against the pressure and radiation damage. These components allow the extraction of the signal from electrical strain gauges, optical fibres and temperature probes from the inside of the tank and they also allow the connection to the Staubli table. On this interface there are 3 electrical connectors 48 pins and 1 connector (Combitac) which allows 16 optical connections.  
Both electrical and optical feedthroughs are provided by Fischer Connectors and had been tested in the same test set-up employed for the sensors (see \ref{instrBDF:validation}). 
The electrical feedthroughs and plugs have 27 pins and belong to the Stainless Steel Fischer core series. The regular Viton O-rings have been substituted with EPDM ones, due to their better radiation hardness. These components are watertight and guaranteed for pressures around 30 bar. To increase their effectiveness against the pressure the feedthroughs are mounted with the female side towards the inner of the tank. 
The optical feedthroughs are composed by a gas tight receptacle and each one can receive four optical fibres with 8\textdegree FC/APC connections. The manufacturer does not provide any specification about the maximum pressure that the connectors can withstand. 

\subsubsubsection{Cabling and Installation}

The following equipment are installed to extract the signals from the sensors on the blocks to the upper Staubli plate:
\begin{itemize}
    \item SCEM 04.01.63.C \textendash  Polyimide cables 0,25 mm\textsuperscript{2}
    \item Radiation hard optical fibers
    \item Fisher Optical Feedthroughs: FO4 P01P8 S9-001.4-00.3 CFCN8 AAA
    \item Fisher Electrical Feedthroughs: ST-WDE 105 AZ102-130E
    \item Staubli connection system 
\end{itemize}
The following equipment are used from the fixe table up to the measurement devices:
\begin{itemize}
    \item Staubli connection system 
    \item 3x NE48 Cables for electrical instrumentation
    \item 16 Optical fibers DRAKKA Radiation Hard: 
    \begin{itemize}
        \item 10x XPSS031 PCORD SIMPLEX SM-7A2 E2000/APC C1|FC/APC C1 1m
        \item 6x XPSS032 PCORD SIMPLEX SM-7A2 E2000/APC C1|FC/APC C1 2m
        \item 15x XPSS080 PCORD SIMPLEX SM-7A2 E2000/APC C1|SC/APC C1 5m
        \item 1x XPSS081 PCORD SIMPLEX SM-7A2 E2000/APC C1|SC/APC C1 7.5m
    \end{itemize}
\end{itemize}
DAQ systems will be installed in TCC2. DIC are in attachments of the EDMS document \cite{EDMS-T6-instru}.

\subsubsection{Validation tests}
\label{instrBDF:validation}
A validation campaign was necessary to establish the behaviour and suitability of the equipment, due to the lack of technical information available at the BDF Target prototype operational conditions. None of the manufacturers could provide specifications about the flow rate effects and sometimes about the water tightness and maximum admissible pressure. Therefore, a test-rig had been designed and manufactured, to test strain gauges, optical fibre sensors, plugs and feedthroughs against pressurised water and high flow rate.

\begin{figure}[htbp]
\centering %
\includegraphics[width=1\linewidth]{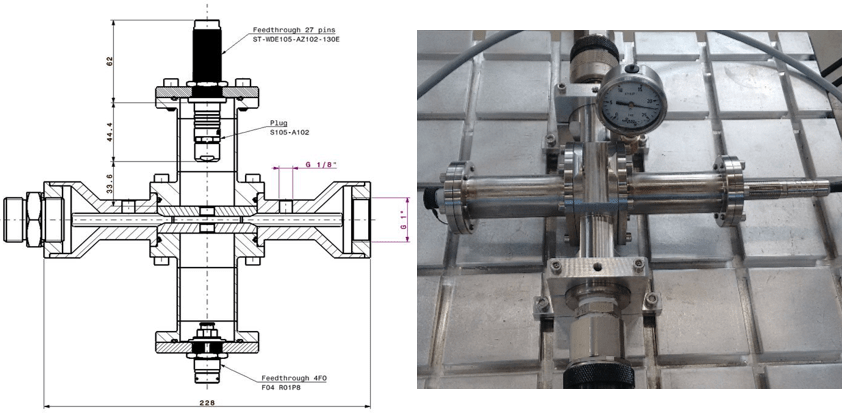}
\caption{\label{fig:TGT:testrig} Technical drawing and installation at building 181 of a dedicated test rig aimed at testing the instrumentation for the BDF Target prototype.}
\end{figure}

As shown in Figure~\ref{fig:TGT:testrig}, the set-up hosts two inserts, which are instrumented with the strain sensors. The candidates' measurements were compared with a reference electrical strain gauge (LE11-3/350), that is encapsulated and waterproof but not radiation hard, also glued on the same insert. Due to changes in the requirements of the instrumentation (different measuring points and directions), the tests were run over different electrical strain gauges than the Y series ones. The second insert was instrumented with an optical fibre FBG. The system run for several days with a circulating water flow of 4.2 m/s at 22 bar.

\begin{figure}[htbp]
\centering %
\includegraphics[width=0.5\linewidth]{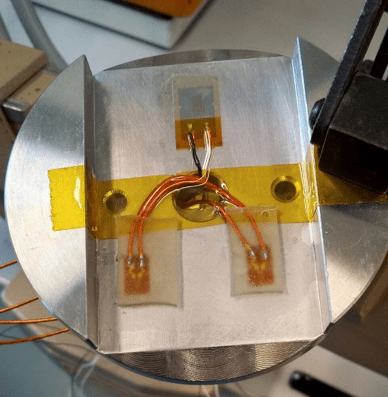}
\caption{\label{fig:TGT:instru_insert} Test-rig insert instrumented with the electrical strain gauges. The LE11 is on the top, it is encapsulated and waterproof and two LY61 on the bottom covered with protective agent.}
\end{figure}

\begin{figure}[htbp]
\centering %
\includegraphics[width=0.5\linewidth]{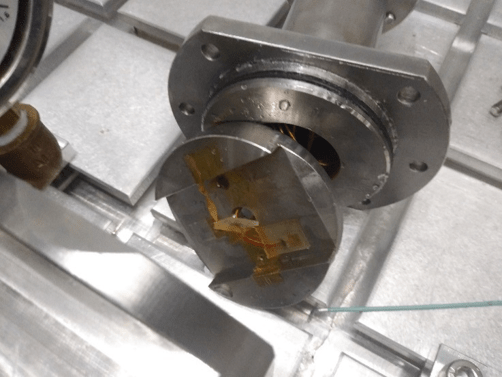}
\caption{\label{fig:TGT:tested_insert} One of the two LY61 strain gauges is detached after the tests. This insert was tested underwater and then exposed to air during CERN annual closure (2 weeks).}
\end{figure}

The main results of the tests were:
\begin{itemize}
    \item The electrical and optical feedthroughs can withstand the operational pressure without leakages.
    \item There is agreement in between the strain measurements carried out with the two types of resistive strain gauges. In particular, it is possible to observe thermal expansion and contraction of the inserts due to the change of temperature during the days/nights (see \cite{EDMS-T6} for examples of these measurements).
    \item The cover agent successfully protected the Y series strain gauges, which are not guaranteed against underwater environment and do not influence the measurements. 
    \item It was observed that the cover agent becomes stiff and easy to detach if alternately exposed to air and water (see Figure \ref{fig:TGT:tested_insert}). Therefore, once it is submerged it should stay wet until the end of the experiment. 
    \item Tests done with and without protecting the OF with the cover agent shows that its protection is not necessary. 
\end{itemize}

\subsubsection{Instrumentation steps}

\subsubsubsection{Instrumentation of the blocks}

The first step was to instrument the faces of the blocks with the electrical strain gauges (marking of the position, gluing of the gauges, application and curing of the cover agents). Only at the end, the other faces were instrumented first with Pt100s, while the last ones to be placed on the blocks were the optical fiber sensors (to avoid possible damaging manipulation in their surroundings).

\begin{figure}[htbp]
\centering %
\includegraphics[width=1\linewidth]{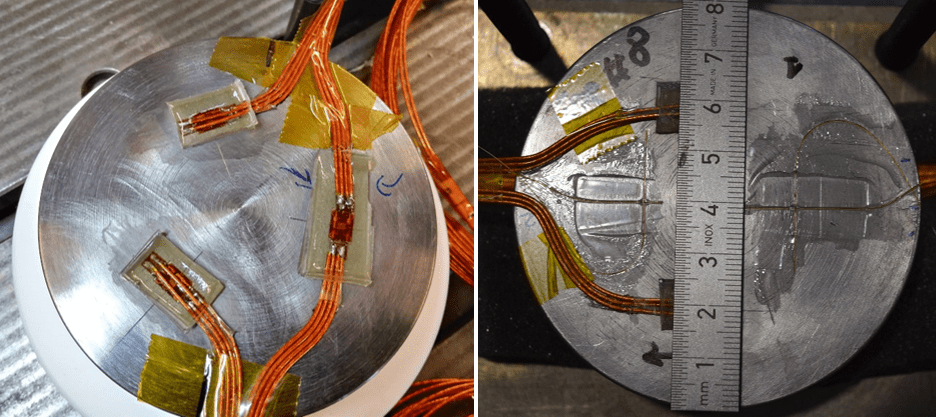}
\caption{Target Block\textquotesingle s surface instrumented with biaxial resistive strain gauges, positioned at 120\textdegree and covered with protective agent to guarantee waterproofness (left); surface instrumented with 2 Pt100s and 4 FBGs (right).}
\end{figure}

\subsubsubsection{Installation of the blocks}

\begin{figure}[htbp]
\centering %
\includegraphics[width=1\linewidth]{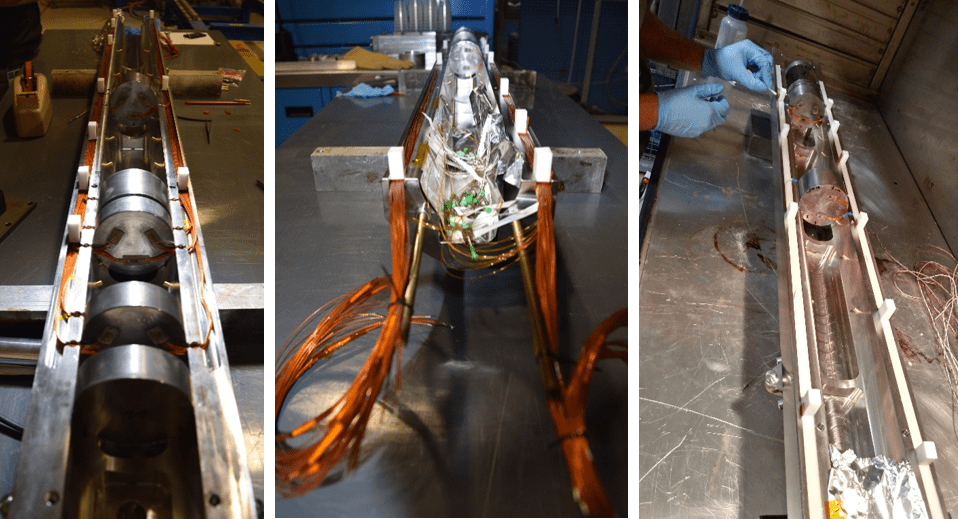}
\caption{\label{fig:TGT:blocks_installation} Installation of blocks and wires (left-centre); application of the resin in the axial grooves (right).}
\end{figure}

After application of the sensors the blocks were positioned in the lower half shell of the tank and the cabling extracted towards the connection flange thanks to radial grooves machined next to every instrumented block and two bigger axial grooves machined all along the shell thickness. 
Once in place, all the radial and axial grooves hosting the wires and optical fibers were covered with resin to avoid water leaks, and hot cured (Figure~\ref{fig:TGT:blocks_installation}). 

\begin{figure}[htbp]
\centering %
\includegraphics[width=0.8\linewidth]{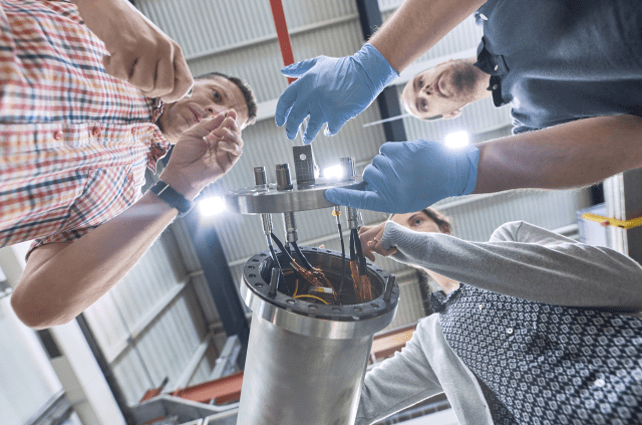}
\caption{\label{fig:TGT:flange_closing} Adjustment of the wires inside the BDF target prototype tank and closing with the connectors flange.}
\end{figure}

After these operations, the placement of the other targets\textquotesingle blocks and the installation of the optical and electrical feedthroughs, it was possible to place the upper half shell and place the closing flange (Figure \ref{fig:TGT:flange_closing}).

\subsubsection{Test of the acquisition chain and signals}
\begin{figure}[htbp]
\centering %
\includegraphics[width=0.5\linewidth]{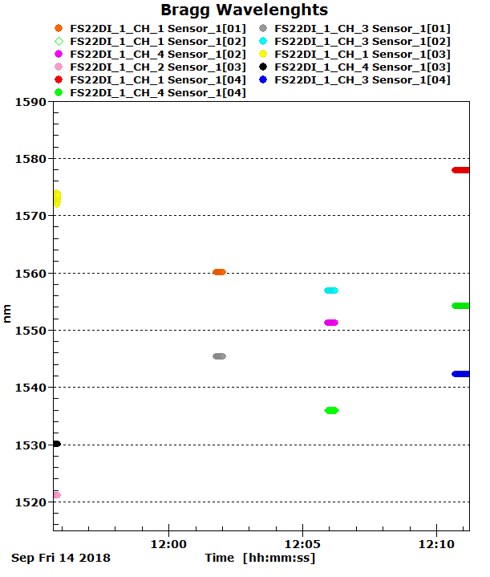}
\caption{\label{fig:TGT:blocks_FBG_check} Measurements of the Bragg wavelength of the optical sensors installed on the blocks during a water pressure test.}
\end{figure}

During the aforementioned operation the electrical continuity of resistive strain gauges and Pt100 was often checked, always giving positive outcomes, and the optical fibers where checked after the curing of the resin in the grooves, also giving 100\% positive results (one fiber over 16 was broken during installation). During the pressure tests, 4 more fibers were lost. Figure \ref{fig:TGT:blocks_FBG_check} depicts the measured Bragg wavelength of the 11 working sensors during the last pressure tests.

The optical connection line from the electronic rack of BA80 up to the mobile Staubli plate connector was checked in March 2018 by means of a test specimen instrumented with 4 FBGs. The line included the facility cables from the rack to the Combitac of the fixed Staubli plate, the Staubli Combitac connection, and the sensing elements connected to the Combitac of the mobile Staubli plate with FC/ACP-FC/APC adapters.

\begin{figure}[htbp]
\centering %
\includegraphics[width=1\linewidth]{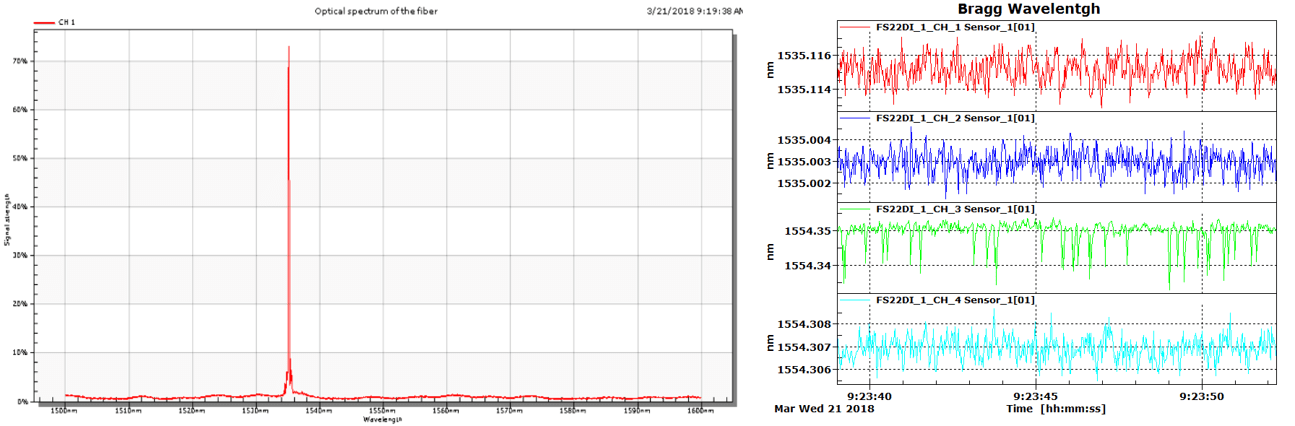}
\caption{\label{fig:TGT:specimen_lambda0} Spectrum of one FBG on the test specimen (left) and measured wavelength for all of the sensors (right).}
\end{figure}

The connections resulted to be well done as the signal back from the test specimen was strong, allowing a good tracking of the peaks, with an example shown in Figure~\ref{fig:TGT:specimen_lambda0}. 

Despite the positive outcomes of the tests performed on the different sectors of the connection chain (experiment tank and BA80 installations) when the BDF target prototype was lowered underground and connected to the fixed Staubli plate several problems raised in detecting the Bragg peaks. The signal to noise ratio was extremely low making it impossible to distinguish the position of the Bragg peak in the full spectrum, and therefore to measure the strain. On the other hand, 100\% of the signals from resistive strain gauges, Pt100s and pressure sensors were detectable.

\subsubsection{Operation}
\begin{figure}[htbp]
\centering %
\includegraphics[width=1\linewidth]{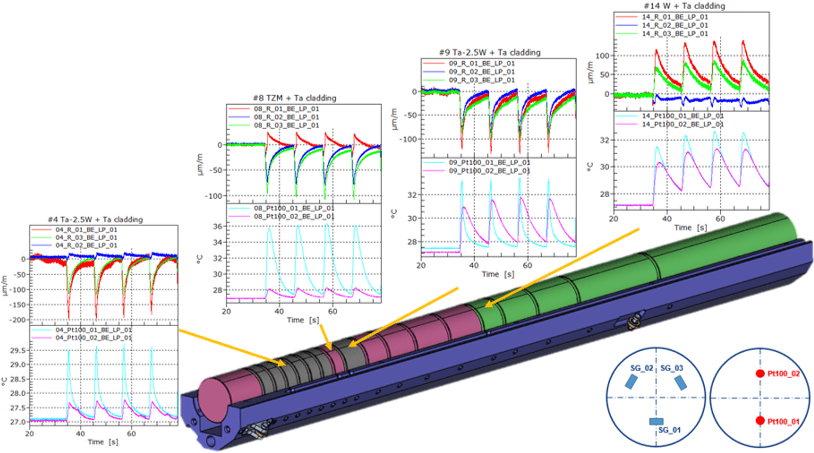}
\caption{\label{fig:TGT:online_meas} Examples of strain and temperature measurements visible online during tests.}
\end{figure}

During the 3 days of experiment, the sensors were monitored online by remotely access the PC installed in BA80 from the CCC. The online measurements allowed the observation real time of the target moving, the beam steering and the beam impacts, as shown in Figure \ref{fig:TGT:online_meas}. Only a few number of sensors were lost since the beginning of the experiment. After more than 1 month since the installation in T6 area, under a continuous water flow, pressurised environment and a total POT of \num{2.4e16} cumulated during the experiments, 23/24 resistive strain gauges and 6/8 Pt100s were still working at the moment of the disconnection of the acquisition system. 

\subsection{Beam tests area preparation and execution}
\subsubsection{Overall layout of the area}

\begin{figure}[!htb]
\centering
\includegraphics[width=0.9\textwidth]{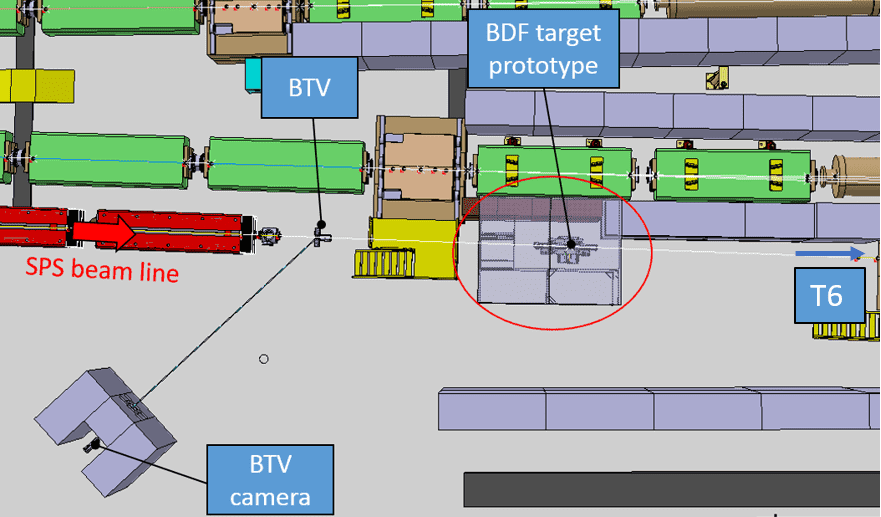}
\captionsetup{width=0.85\textwidth} \caption{Layout of TCC2 area, with a view of the target prototype bunker, the target beam screens (BTV), and the upstream BTV camera and camera shielding.}
\label{fig:T6layout}
\end{figure}

The target prototype experimental setup of the BDF prototype was located in the North Area target zone (TCC2), upstream the T6 target assembly (see Figure~\ref{fig:T6layout}). A new concrete bunker was realized directly downstream the T6 target in order to house the experiment.

As the test was expected to receive a significant number of protons, the target was placed in a dedicated shielding to keep the dose to personnel low in case of interventions in the area. The main target shielding is made out of standard concrete blocks, 2400x1600x800 mm. The first level is a base of 100 mm spacers to allow the passage of cables and pipes on the floor below the shielding. The construction was planned to be executed by stapling the different blocks by the overhead crane by EN/HE.

On the upstream side of the shielding, a 100 x 200 mm opening for the beam passage was foreseen. On the downstream side, a passage to enter the bunker was established. If personnel access to T6 was required, an additional concrete block was placed downstream the target to reduce the residual dose rate coming from the BDF target. If required, the small hole upstream the shielding could be closed as well in order to protect people intervening on the T4 target shielding assembly. The BDF target, once available, was inserted and removed by the crane from the top. During beam operation, the bunker was also closed on top by two concrete blocks.

\begin{figure}[!htb]
\centering
\includegraphics[width=0.8\textwidth]{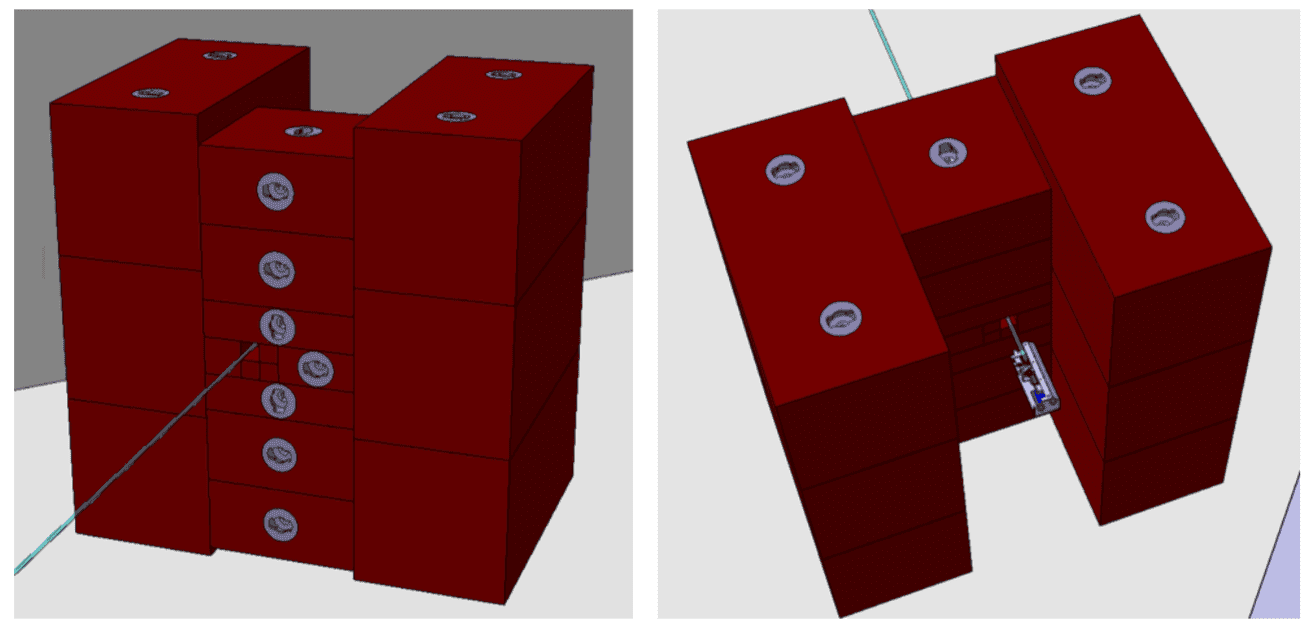}
\captionsetup{width=0.85\textwidth} \caption{Cast iron shielding for the BTV camera bunker.}
\label{fig:T6camera}
\end{figure}

Two dedicated BTVs (beam screens) were installed, one upstream and one downstream the experimental setup in order to measure the beam profile impinging on the target and to help aligning the beam on target. The digital camera aiming at the upstream BTV had to be protected from radiation damage: to that purpose it was located inside a U-shape shielding wall, made out of standard cast iron blocks (Figure~\ref{fig:T6camera}). 

\begin{figure}[!htb]
\centering
\includegraphics[width=0.8\textwidth]{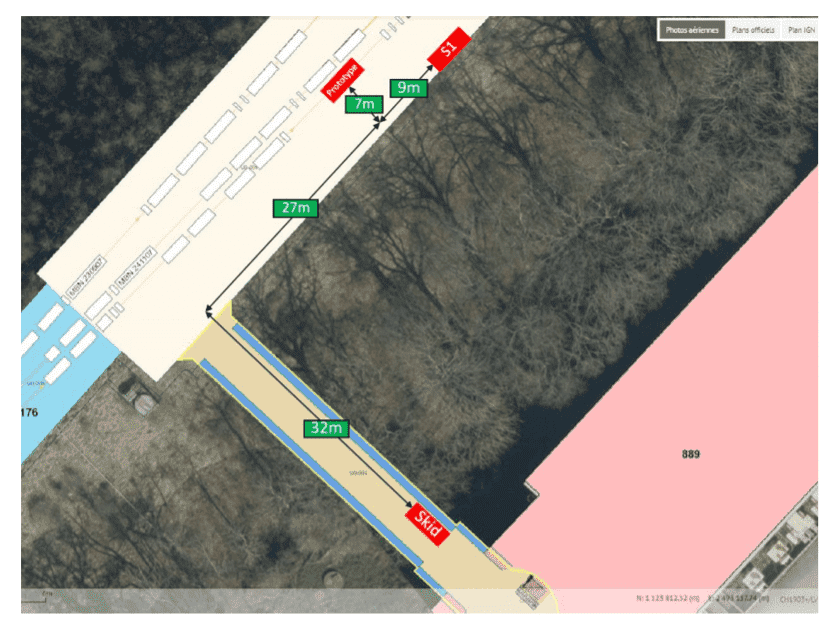}
\captionsetup{width=0.85\textwidth} \caption{Layout of TCC2 and TA801 access tunnel where the position of the target prototype and the dedicated cooling skid can be seen.}
\label{fig:T6skid_map}
\end{figure}

A dedicated cooling system skid was foreseen, in order to avoid any possible contamination to the North Area primary cooling circuit in case of target cladding failure. The cooling system skid was installed in the access tunnel of TCC2, TA801 (see Figure~\ref{fig:T6skid_map}). 

\subsubsection{Preparation of the area during YETS 2017-2018}
\begin{figure}[!htb]
\centering
\includegraphics[width=0.8\textwidth]{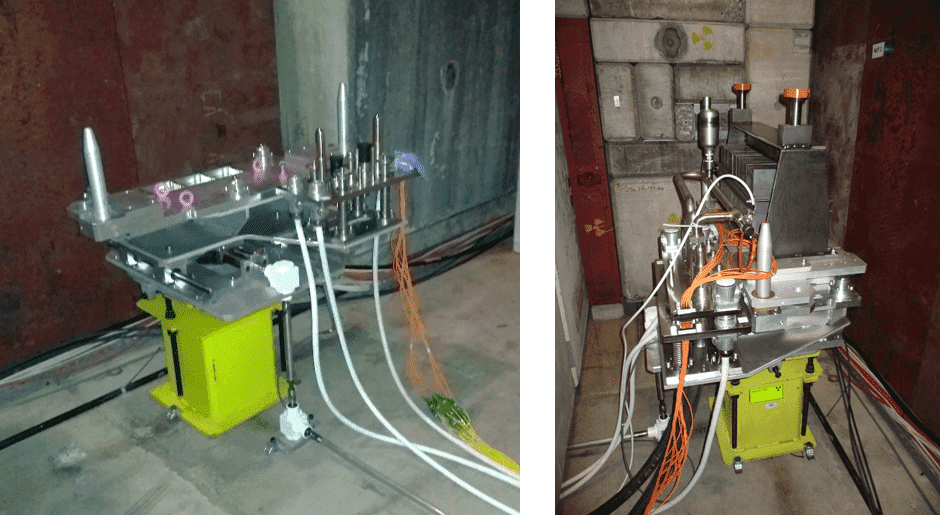}
\captionsetup{width=0.85\textwidth} \caption{Target prototype support installed in the TCC2 area (left); and target prototype mock-up of robot, cooling and aligment tests (right).}
\label{fig:TGT:Proto:YETS}
\end{figure}

The prototype test area was prepared during the Year-End Technical Stop (YETS) 2017-2018, in order to profit from the lower radiation levels in TCC2. The target support and its motorization system were installed, as well as the surrounding shielding blocks. A mock-up of the target was installed on the target support for alignment, remote handling and cooling testing (Figure~\ref{fig:TGT:Proto:YETS}.

\begin{figure}[!htb]
\centering
\includegraphics[width=0.8\textwidth]{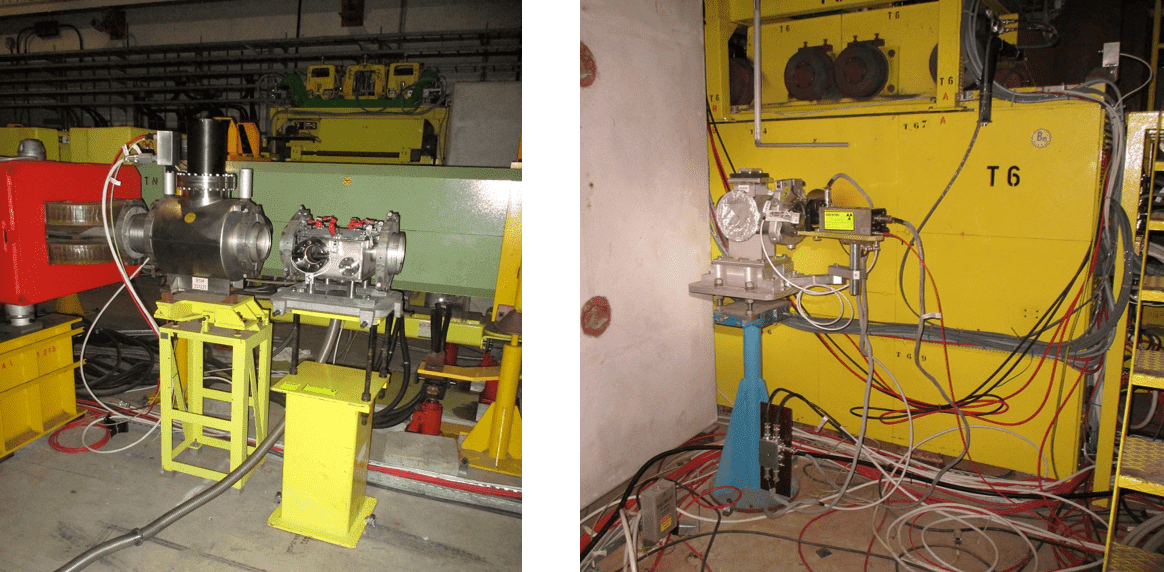}
\captionsetup{width=0.85\textwidth} \caption{Upstream (left) and downstream (right) dedicated BTVs installed in the TCC2 area in order to provide beam alignment capabilities on the BDF target.}
\label{fig:TGT:Proto:BTV}
\end{figure}

\begin{figure}[!htb]
\centering
\includegraphics[width=0.5\textwidth]{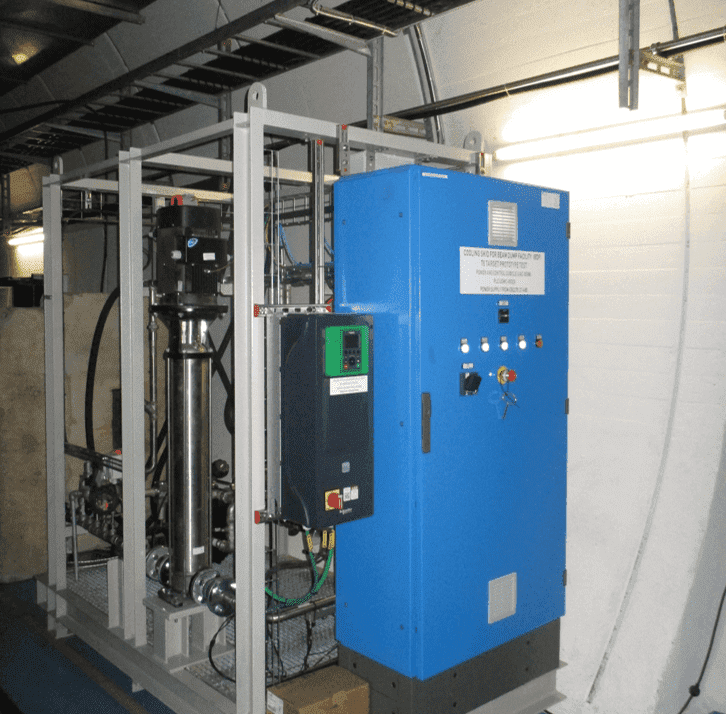}
\captionsetup{width=0.85\textwidth} \caption{Target prototype cooling system skid installed in the access tunnel to TCC2 (TA801).}
\label{fig:T6skid}
\end{figure}

The upstream and downstream BTVs, as well as the shielding for the upstream BTV camera were also installed (Figure~\ref{fig:TGT:Proto:BTV}). Most of the cabling necessary for the target blocks instrumentation and motor operation was put in place. The cooling skid was installed in the access tunnel, and most of the pipe connections required were performed, awaiting for the target prototype installation at a later stage (Figure~\ref{fig:T6skid}).

\subsubsection{Prototype target installation}

The installation of the prototype target took place in September 2018. The installation required the combined effort of several teams, in order to cope with the radiation protection, remote handling, transport, cooling and mechanical aspects of such an operation during a Technical Stop of the North Area. 

\begin{figure}[!htb]
\centering
\includegraphics[width=0.9\textwidth]{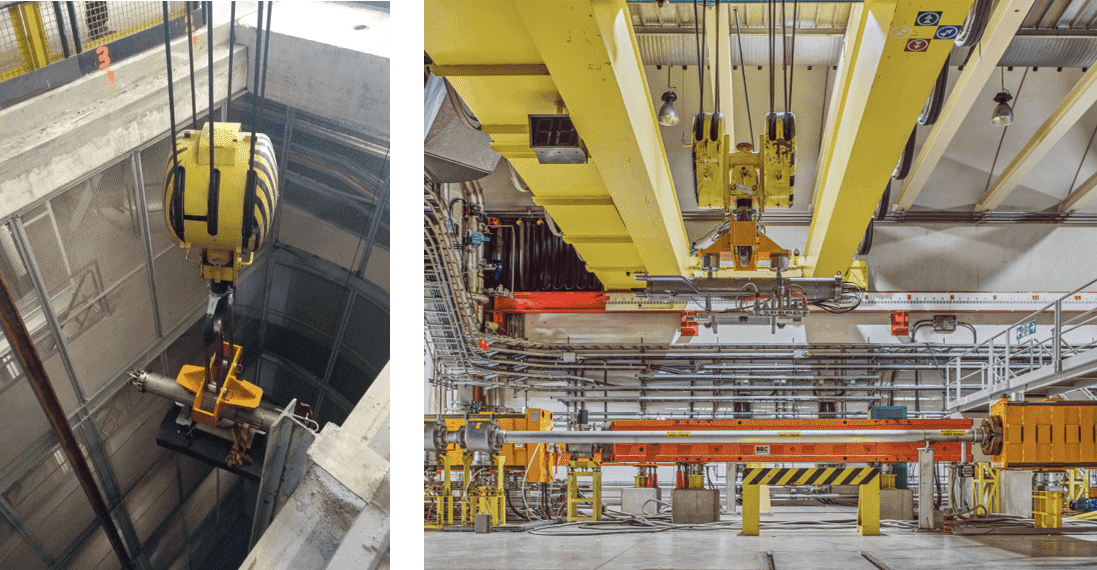}
\captionsetup{width=0.85\textwidth} \caption{BDF target prototype handling with overhead crane towards the support location (Photograph: J. M. Ordan)~\cite{CDS_pictures2}.}
\label{fig:fig:TGT:Proto:install}
\end{figure}

The installation was performed fully remotely with the main crane of TCC2 and with telemanipulation tools; the target was transported and placed on the support with the TCC2 area overhead crane, with the additional help of the robot cameras. The water and electrical connections were performed by the robot, and a brief one-minuted personnel access was granted for visual inspection of possible leaks (Figure~\ref{fig:fig:TGT:Proto:install}).

Additionally, concrete shielding was installed around the cooling system skid in order to mitigate the potential Radiation-to-Electronics (R2E) events on the electronics of the cooling system. 

\FloatBarrier
\label{Sec:TGT:Proto:preparation}

\subsection{Target prototype radiation protection considerations}
\label{sec:TGT:Radprot}
\begin{figure}[!htb]
\centering
\includegraphics[width=0.9\textwidth]{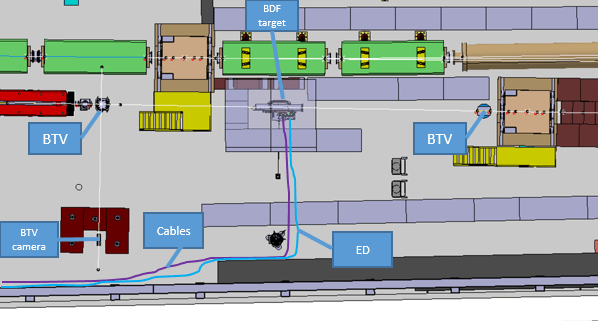}
\captionsetup{width=0.85\textwidth} \caption{Layout of TCC2 with the T6 bunker, where BTV's are beam screens and ED (Eau Demineralis\'ee) the connection to the cooling circuit.}
\label{fig:T6position}
\end{figure}

This section summarizes the radiological assessment of the BDF target prototype beam test executed in September-October 2018. The target has been irradiated with representative BDF/SHiP beam scenarios in order to have an insight of the material response. The evaluation of radiation protection risks was crucial to perform this experiment. The test beam required a slow extraction process at high intensity, currently available in the North Area target zone (TCC2). The experiment took place upstream of the T6 primary target (see Figure \ref{fig:T6position}). Quantifying the impact of this experiment on the water and air activation was mandatory to ensure that the radiation protection requirements are fulfilled in the area.  The studies are based on extensive simulations with the FLUKA Monte Carlo particle transport code~\cite{Ferrari2005,FLUKA_Code} and ActiWiz3~\cite{Actiwiz}. Studies prior to the test assumed an integrated total of $3\times10^{16}$ protons on target (POT), while for the test two beam periods an integrated total of $0.9\times10^{16}$ and $1.6\times10^{16}$ POT were collected, respectively.

\begin{figure}[!htb]
 \centering
\includegraphics[width=0.9\textwidth]{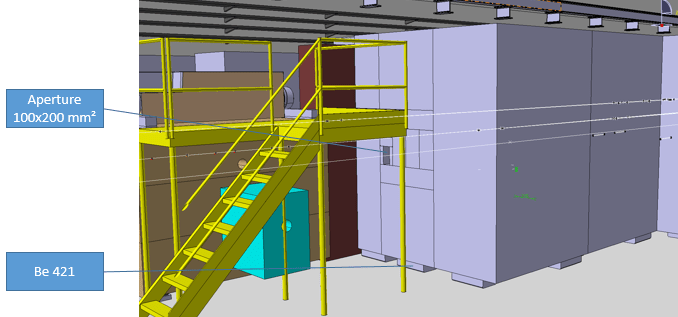}
\captionsetup{width=0.85\textwidth} \caption{\small View of the BDF T6 target test bunker.}
\label{fig:T6shielding}
\end{figure}

As a significant number of protons were received during the test, the target was placed in a dedicated shielding as shown in Figure~\ref{fig:T6shielding}. This allowed to reduce the dose to personnel in case of interventions in the area. A new concrete bunker was placed between the T4 target passerelle and the T6 target. The main target shielding was constructed with standard concrete blocks of $2400\times1600\times800$~mm$^3$. 

\begin{figure}[!htb]
 \centering
\includegraphics[width=0.9\textwidth]{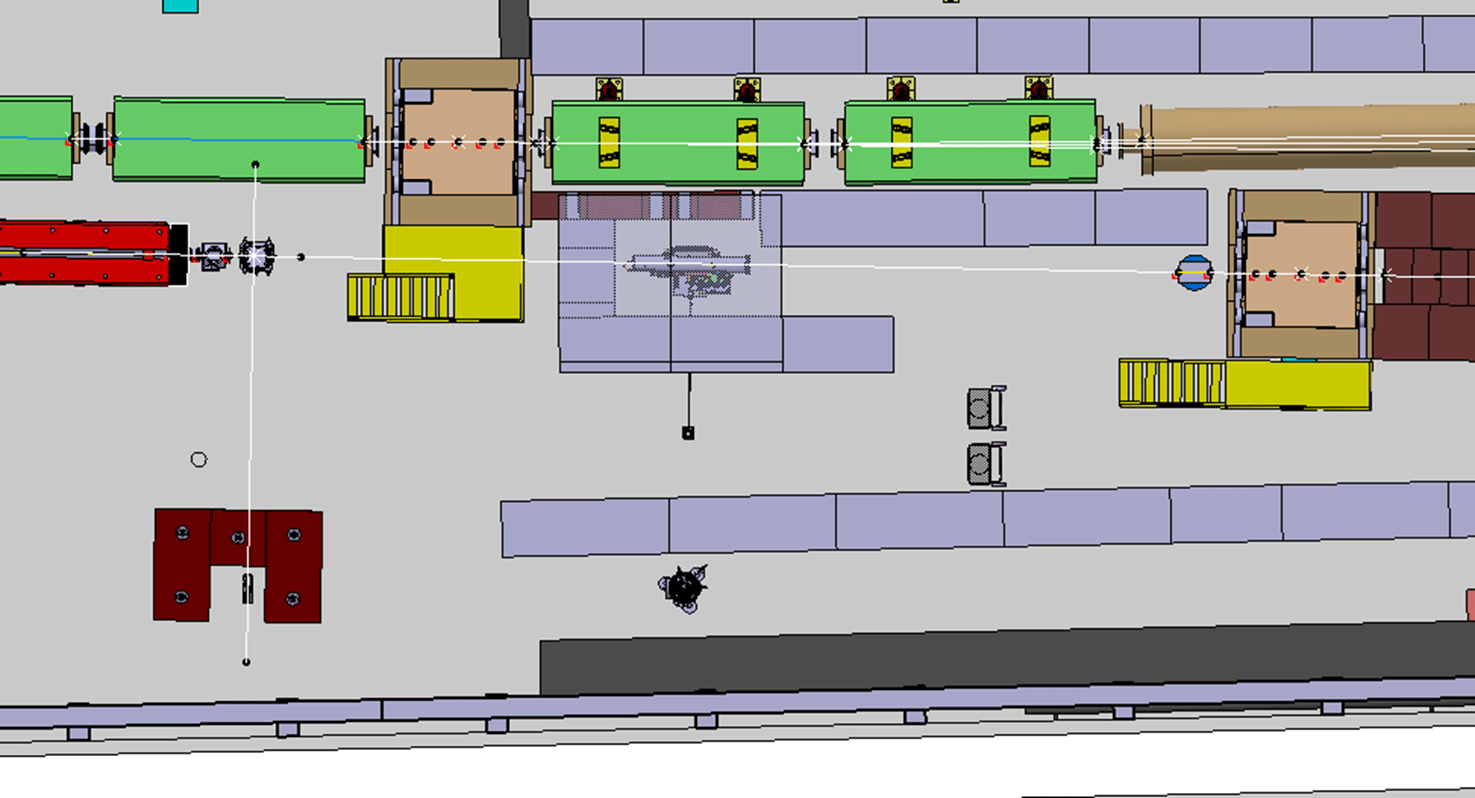}
\captionsetup{width=0.85\textwidth} \caption{\small Shielding configuration in the ``open'' position.}
\label{fig:T6open}
\end{figure}

\begin{figure}[!htb]
 \centering
\includegraphics[width=0.9\textwidth]{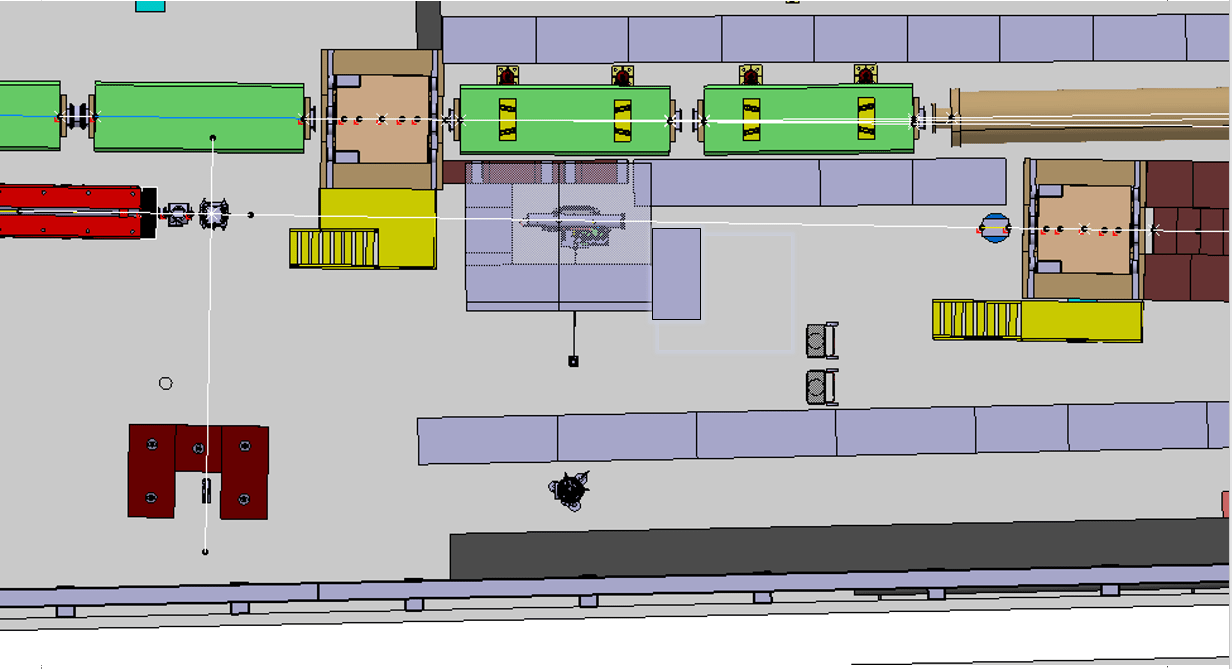}
\captionsetup{width=0.85\textwidth} \caption{\small Shielding configuration in the ``closed” position.}
\label{fig:T6closed}
\end{figure}

\begin{figure}[!htb]

    
     \begin{subfigure}[b]{0.5\linewidth}
            \centering
            \includegraphics[width=\linewidth]{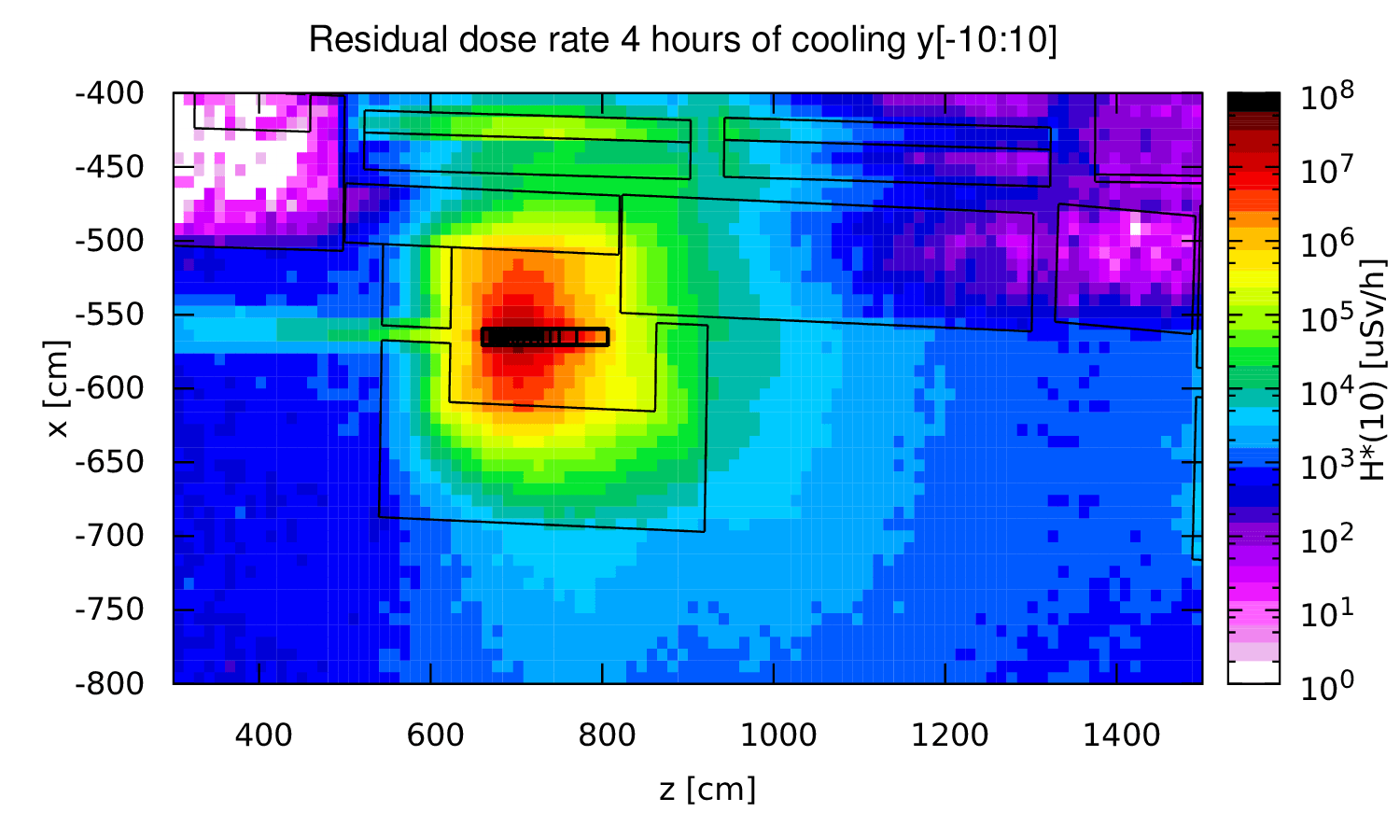}
    \label{fig:bert1}
    \end{subfigure}   
     \begin{subfigure}[b]{0.5\linewidth}
            \centering
            \includegraphics[width=\linewidth]{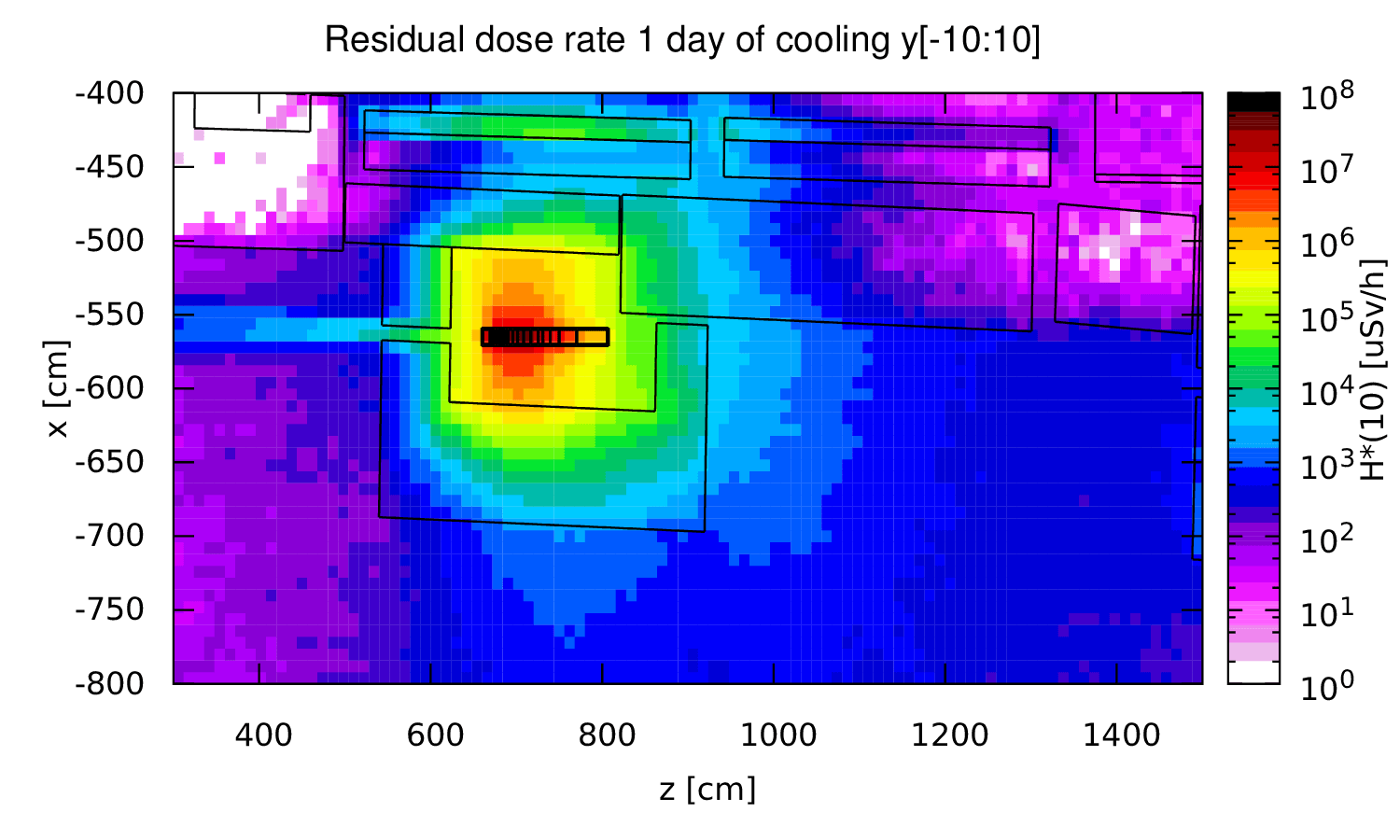}
    \label{fig:bert2}
    \end{subfigure}       
     \begin{subfigure}[b]{0.5\linewidth}
            \centering
            \includegraphics[width=\linewidth]{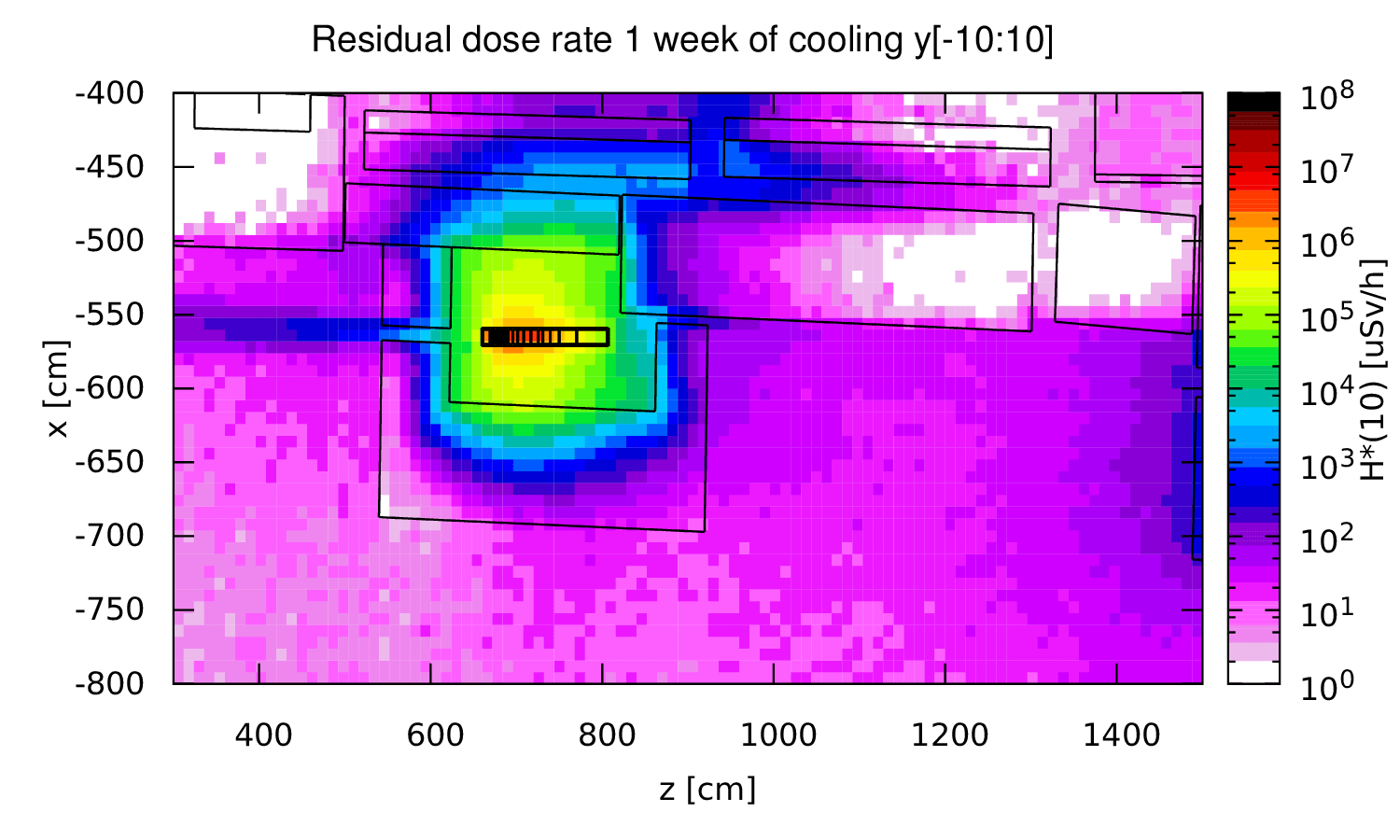}
    \label{fig:bert3}
    \end{subfigure}       
     \begin{subfigure}[b]{0.5\linewidth}
            \centering
            \includegraphics[width=\linewidth]{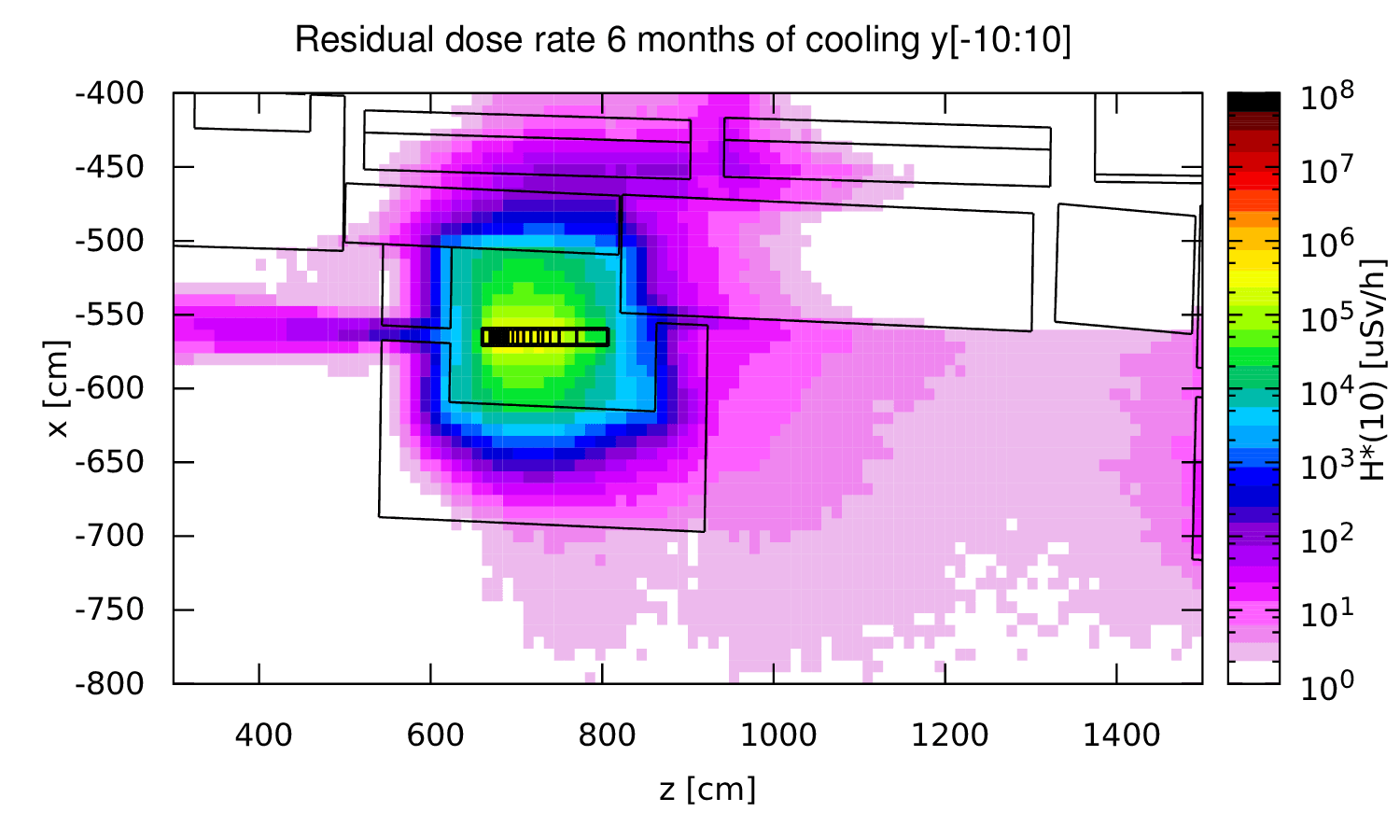}
    \label{fig:bert4}
    \end{subfigure}

\captionsetup{width=0.85\textwidth} \caption{\small Residual dose rates  in $\mu$Sv/h after (a) 4 hours, (b) 1 day, (c) 1 week and (d) 6 months cooling.}
\label{fig:T6RD}
\end{figure}

On the downstream side, the shielding blocks were put in the “open” position (see Figure \ref{fig:T6open}) to have enough space in case of remote handling by a robot. If personnel access to T6 was required, the concrete block closest to T6 were moved to the “closed” position (see Figure \ref{fig:T6closed}) in order to reduce the residual dose rate coming from the prototype target. The residual dose rates in the “closed” configuration, evaluated at different cooling times (4h, 1 day, 1 week and 6 months), are shown in Figure~\ref{fig:T6RD}. The residual dose rates around the bunker for short cooling times ($\leq$1week), which are interesting in case of interventions, range between hundreds of $\mu$Sv/h to few mSv/h. For at least 6 months after the irradiation, the target assembly remains in the bunker in the “closed” configuration before its removal during LS2. The removal of the target can be performed only by using remote handling due to the high residual dose rates (O(20 mSv/h) at 40 cm) even after 6 months of cooling.

After the first beam period, the dose rates at a distance of 40 cm from the target were measured remotely after 1 week of cooling reaching a maximum of 43 mSv/h. The simulations result in 30 mSv/h for the same number of POT and therefore show relatively good agreement with the measurements, considering the uncertainty on the positioning of the detector held by the robot, which performed the measurements. 

The impact of the test on the activation of the air in TCC2 was negligible. As a matter of fact, the test required only $3\times10^{16}$ POT that is equivalent to $\approx1\%$ of the total POTs received by the T2, T4 and T6 targets during the year. Furthermore those targets, in contrast to the BDF prototype target, are air cooled leading to higher activation due to the proximity of the air to the targets itself. Moreover, the BDF target provides a significant shielding power, reducing the radiation field prior to reaching air regions. During beam operation in TCC2, the ventilation system is set in recirculation mode and the air is flushed only in case of an intervention. Flushing the air requires 2 hours of cool-down and 2 hours of flushing, thus 4 hours is the minimum amount of time needed before an intervention can take place.

High water speed (5 m/s) was required in the channels between the prototype target blocks to remove the power deposited on the target. The circuit was supplied by demineralized water and equipped with mechanical filters and an ion exchanger to catch impurities and activated ions. The water activation was estimated using the ActiWiz3 code. It was expected that some components of the cooling circuit would become contaminated with activated ions. For example, assuming that all the isotopes are caught in the cartridge of the ion exchanger, the residual dose rate after 1 h of cooling at 40 cm from the cartridge is expected to be around 14.5 mSv/h. For the first beam period of the test, the residual dose rate after 1 h of cooling at 40 cm from the cartridge was around 10.3 mSv/h. When comparing it to the measurements by a PMI, which was placed 40 cm away from the cartridge, the dose rate with 15.2 mSv/h after 1 hour of cooling showed good agreement. In case of an intervention a longer cooling time should be considered ($>$4 h). In that case the residual dose rate at 40~cm from the cartridge due to water activation are around 75~$\mu$Sv/h with Be-7 as main contributor. The main purpose of this test was to study the material response therefore the possibility of contaminating the water circuit with 1~g of material from the most radioactive part of the target was considered as an accident case(conservative assumption, reasonable estimate in case of accident O(1~$\mu$g) ). These debris, assuming they are collected in the filters or in the ion exchanger,  results in a dose rate of $\approx$300 $\mu$Sv/h after 1 h cooling at 40 cm. The irradiation of the high-Z material of the target produces also alpha emitters, which are considered as radiologically problematic. The main contributor among the ions produced is Gd-148 and in the aforementioned accident case $\approx$100 Bq of Gd-148 would be released in the cooling circuit. To avoid the contamination of the existing cooling circuit, a dedicated cooling circuit for this experiment was put in place, as detailed in Section~\ref{Sec:TGT:Proto:design}.  
The analysis of the gamma radiation emitted from water through a sampling valve showed the presence of spallation products in the water. Ongoing studies will characterize the dust found in the water.

\begin{figure}[!htb]
 \centering
\includegraphics[width=0.55\textwidth]{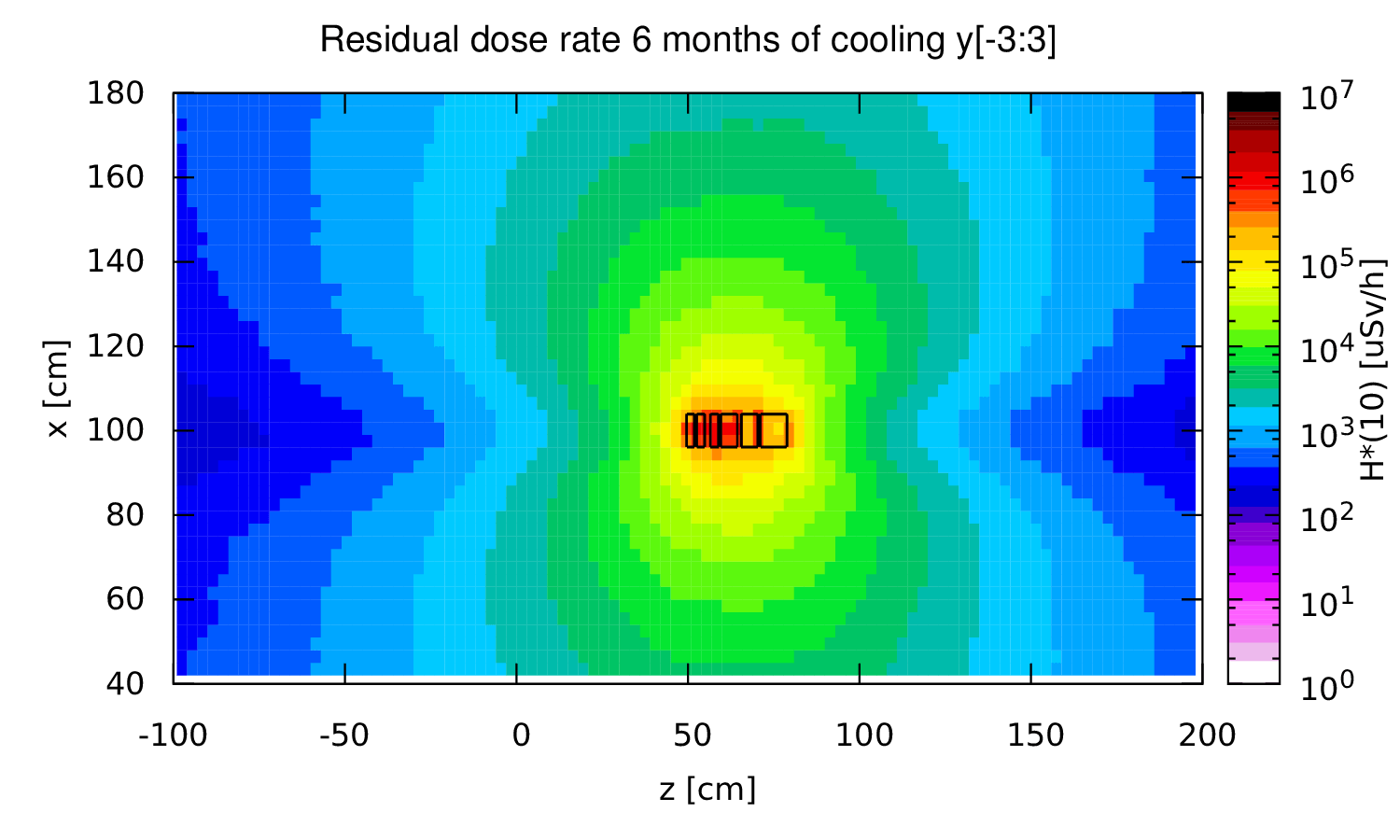}
\captionsetup{width=0.85\textwidth} \caption{Expected Residual dose rate in $\mu$Sv/h after 6 months cooling for the blocks which will be shipped.}
\label{fig:T6shippingblock}
\end{figure}

To investigate the response of the materials a PIE analysis is foreseen. Therefore, six blocks of the prototype BDF target will be sent out of CERN to a specialized company for analysis. The residual dose rates for these blocks after 6 months of cool-down are shown in Figure \ref{fig:T6shippingblock}, preliminary calculations with ActiWiz 3 interfaced with e-SHIP in Nucleonica~\cite{Nucleonica} suggest that a Type A package is needed for the shipment. After the cool-down period of at least 6 months, the BDF prototype target will be transferred out of the bunker into a transport container for storage before disposal. For transportation of radioactive materials, as indicated in Ref.~\cite{Transport}, the maximum dose rate level at any point on the external surface of a package shall not exceed 2 mSv/h. Therefore, the container will be made out of iron and with the thickness of roughly 18 cm for the lateral sides and of 15 cm for the upstream and downstream sides.

\begin{figure}[!htb]
 \centering
\includegraphics[width=0.7\textwidth]{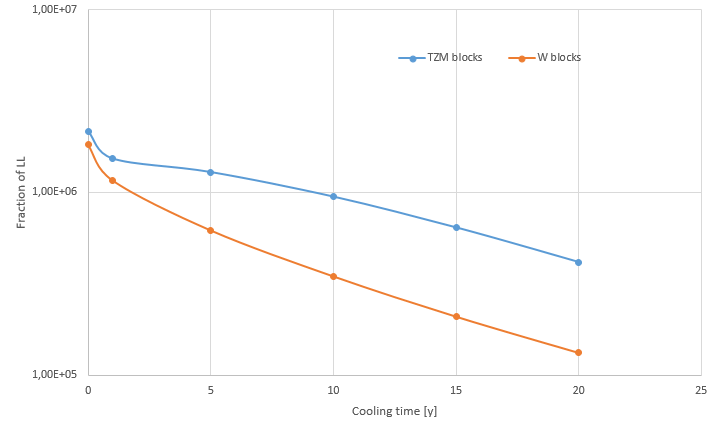}
\captionsetup{width=0.85\textwidth} \caption{\small Time evolution of the LL for TZM and W blocks.}
\label{fig:T6LL}
\end{figure}

 Some preliminary studies were performed for the disposal of the BDF prototype target after irradiation and the PIE tests. The results of ActiWiz3 calculations are reported in terms of the Swiss Liberation Limits (see Ref.~\cite{ORAP}) (LL) for the two most radioactive blocks (one of TZM and one of W). In the calculation the same irradiation profile was assumed, but with 1 year cooling time. The respective total LL is $1.53\times10^{6}$ and $1.17\times10^{6}$ for the TZM and W block with Y-88 ($\approx50\% $LL) and W-185 ($\approx40\%$ LL) as top contributors for TZM and W, respectively. The time evolution of the LL value is shown in Figure~\ref{fig:T6LL}, even after 20 years cool down period the LL is significantly above the exemption limit. The target should be treated as radioactive waste in case of disposal. A proper assessment of the disposal will be performed in the future.
 
\FloatBarrier

\subsection{Target prototype beam tests}
\label{Sec:TGT:Proto:MD}

\subsubsection{Beam tests day 1}

On the 3$^{\text{rd}}$ October 2018, the BDF target prototype received the first shots on target. The beam steering and tuning to reach the target center with the required beam spot dimensions was performed profiting from the beam instrumentation installed (one BTV upstream the target, another one downstream, see Section~\ref{Sec:TGT:Proto:preparation}). 

\begin{figure}[htbp]
\centering %
\includegraphics[width=0.8\linewidth]{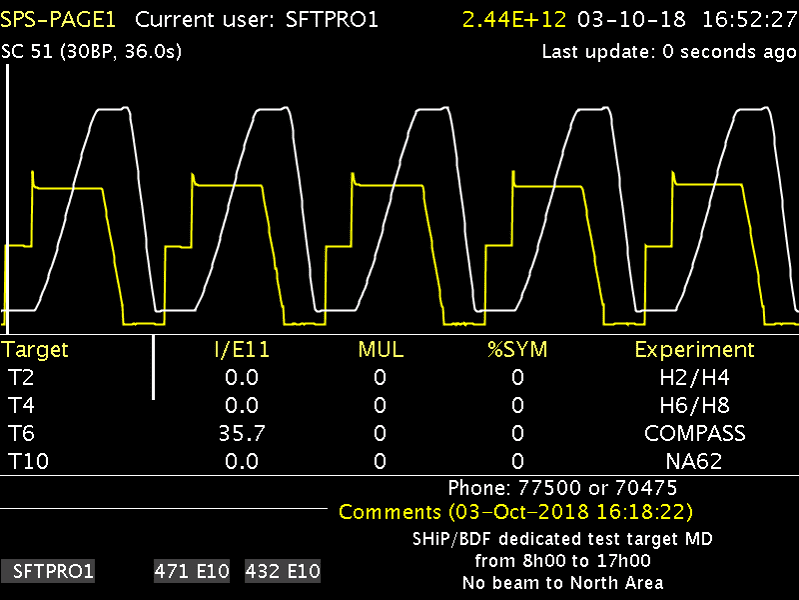}
\caption{\label{fig:TGT:Proto_MD_Ship}SPS Page 1 screen during the execution of the BDF target prototype beam tests; dedicated SHiP super-cycle: 1 second spill, 7.2 seconds cycle, corresponding to a beam power of 35 kW.}
\end{figure}

The target prototype operated under the SHiP super-cycle (i.e. 1 second spill, 7.2 seconds pulse) for several hours, and a total of 6 hours of beam on target were achieved. Figure~\ref{fig:TGT:Proto_MD_Ship} presents a capture of the SPS Page 1 screen with the dedicated SHiP supercycle. The accumulated Protons-On-Target (POT) during the first day of beam tests was around \num{1e16}. One of the goals of the beam tests was to achieve \num{3e16} protons on target, in order to have a representative number of cycles and a meaningful target irradiation, even if far from the expected POT of the final BDF target. 

The target instrumentation operated successfully during the whole test duration, despite the challenging experimental conditions. An example of the online monitoring recorded by the instrumentation sensors can be seen in Figure~\ref{fig:TGT:online_meas}. The response of the temperature and strain sensors to the beam impacts on target can be appreciated, in this case four beam pulses and the subsequent cooling times are recorded.

\subsubsection{Beam tests day 2}

The second day scheduled for the prototype tests was the 24$^{\text{th}}$ October 2018. Before the first shots on target, the upstream camera failed, most probably due to long-time radiation exposure. As a consequence, the beam tuning (alignment on target) was performed using the downstream BTV, much less precise in terms of beam positioning and beam size. A total of 5.5 hours of dedicated beam were accomplished during the day, and the total POT after the test was \num{2.4e16}.

\begin{figure}[htbp]
\centering %
\includegraphics[width=1\linewidth]{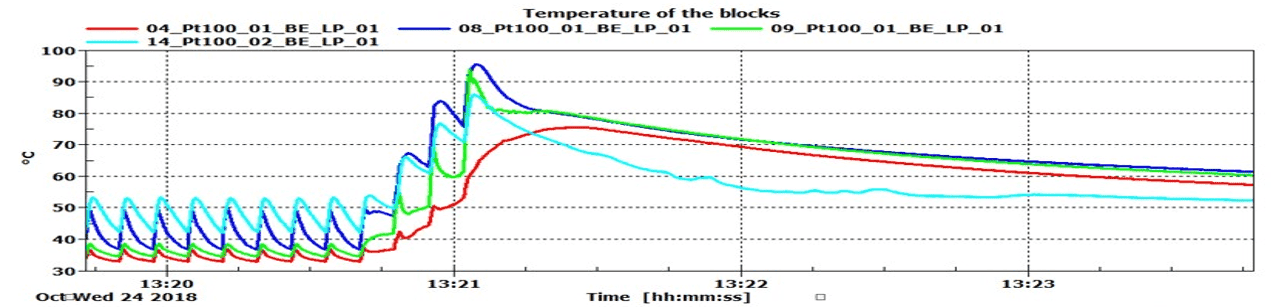}
\caption{\label{fig:TGT:Proto_MD_pump} Temperature sensors recording during the second day of beam tests on the BDF target prototype. The pump failure occurred around 13:20:45, and the temperature increased drastically for the next four pulses until the beam trip.}
\end{figure}

During the execution of the test, the pump of the dedicated cooling skid for the target prototype failed. This event had been already registered after the cooling skid installation, and is thought to be related to Radiation to Electronics (R2E) stochastic events on the control system of the skid. In order to avoid the target failure in case of such an event, a beam interlock was implemented to trip the beam after the pump failure. The beam trip occurred after 4 pulses on target without cooling, which was clearly visible in the strain and temperature sensors online measurements. Figure~\ref{fig:TGT:Proto_MD_pump} displays a view of the temperature measurements recorded following the cooling skid pump failure. The accident was not thought to have damaged the target, hence, the system was successfully restarted and the beam tests were resumed.

\subsubsection{Beam tests day 3}

On the 7$^{\text{th}}$ November 2018, the target prototype received two hours of non-dedicated beam, i.e. shared with other accelerators and experiments. The exchange of the upstream BTV camera was performed one week before, in order to ensure a good accuracy in the beam position and dimensions. 

The major aim of the test during this period was to crosscheck the instrumentation measurements carried out in the previous tests. Several iterations were performed with several beam axis positions on target in order to record the instrumentation response to the different beam impact coordinates.

\begin{figure}[htbp]
\centering %
\includegraphics[width=1\linewidth]{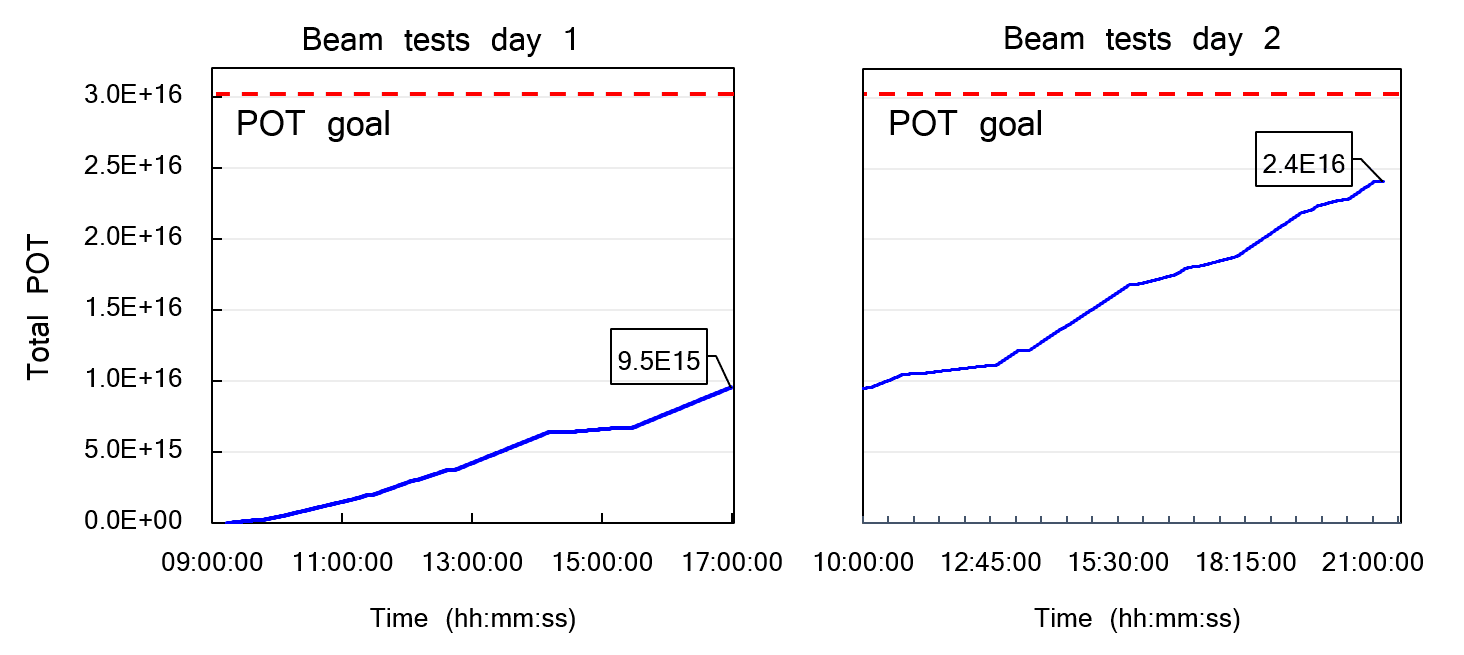}
\caption{\label{fig:TGT:Proto_MD_POT} Accumulation of Protons-On-Target (POT) during the first two days of beam tests. The total POT achieved is around \num{2.4e16}, close to the goal of \num{3e16}}
\end{figure}

In summary, at the end of the three beam test days, more than 14 hours of beam were allocated to the target prototype, and a total of around \num{2.4e16} POT was reached. Figure~\ref{fig:TGT:Proto_MD_POT} presents the POT evolution during the two first days of beam tests (the contribution of the third day to the total POT is negligible). 

It can be concluded that the beam tests have been very successful, given that the beam instrumentation, the target blocks instrumentation, and all the related equipment worked proficiently during the tests. Additionally, many useful information could be extracted from the tests and lots of data have been gathered from the instrumentation measurements.

\FloatBarrier

\subsection{Post Irradiation Experiment (PIE) plans}
\label{Sec:TGT:Proto:PIE}
\subsubsection{Objectives of the PIE}
A Post Irradiation Examination (PIE) of the BDF target prototype after the beam test is envisaged to reach several objectives, aiming at understanding the effect of the proton beam on the target materials:

\begin{itemize}
    \item Study the surface state and integrity of the cladding to validate the cladding material resistance to cooling conditions. 
    \item Validate the Ta2.5W as new cladding material and compare its performance with unalloyed Ta.
    \item Study the state of the interface cladding-target to assess its reliability under thermal cycling due to beam impacts. Identify any degradation of the interface properties (strength, thermal conductivity) or apparition of defects (detachments, segregations) for each different cladding-target couple (TZM-Ta2.5W, TZM-Ta and W-Ta)
    \item Study the state of the target and cladding materials (W, TZM, Ta and Ta2.5W) after thermal cycling due to beam impacts. Identify any degradation of the material properties (strength, thermal conductivity) or apparition of defects (cracks, voids, microstructure changes, segregations)
    \item Foresee any other potential issue related to target operation (target blocks movements or deformation, cooling channels blocking...) and provide feedback to the actual target design
\end{itemize}
To this aim, several target blocks will be extracted at CERN and shipped to an external contractor with the required PIE capabilities. 

\subsubsection{Samples description}
\begin{figure}[htbp]
    \centering
    \includegraphics[scale=0.8]{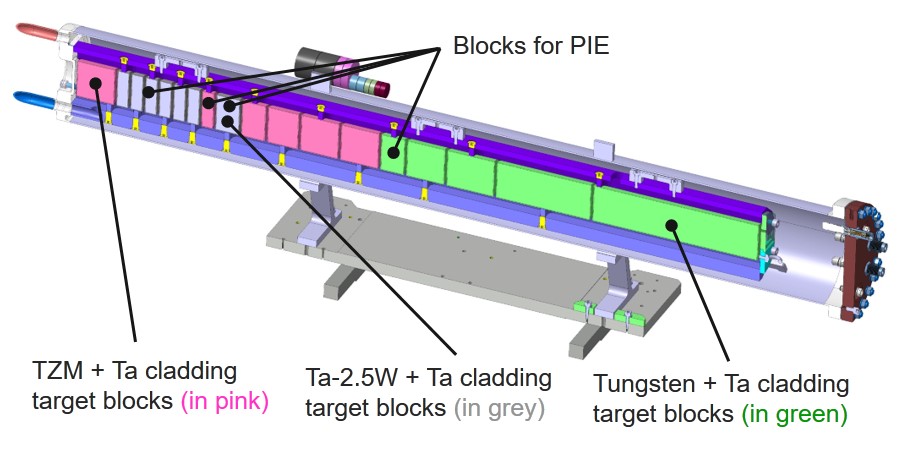}
    \caption{BDF target prototype longitudinal cross-section, with the three different target blocks highlighted in different colours}
    \label{Fig:TGT:PIE:targetblocks}
\end{figure}

\begin{figure}[htbp]
    \centering
    \includegraphics[scale=0.6]{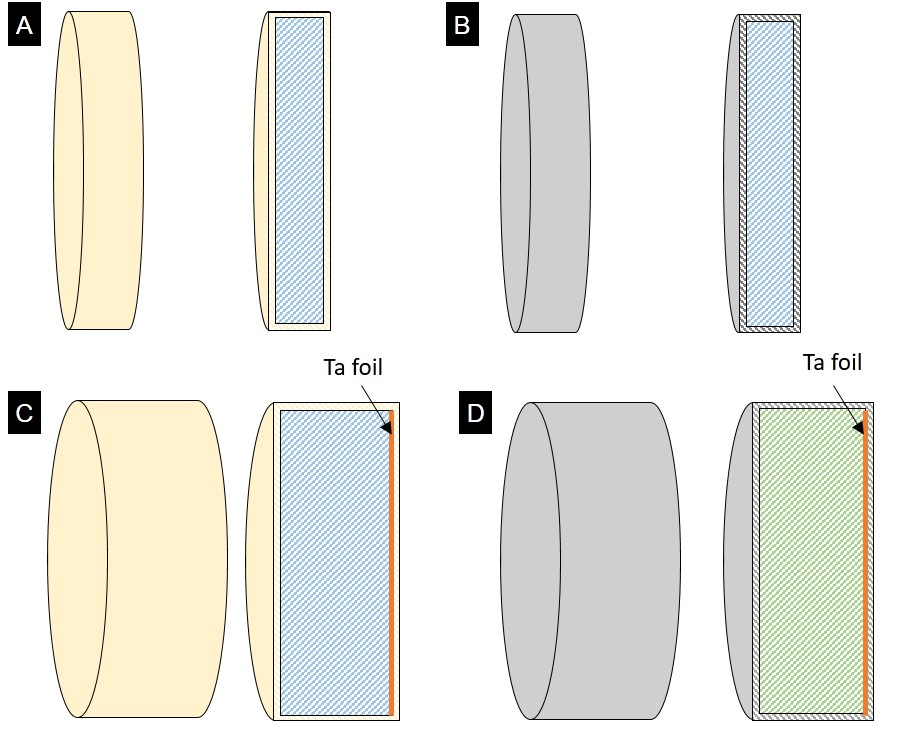}
    \caption{Sketch of the samples a) 4, b)  8, c) 9 and d) 14. 3D view (left) and axial cross-sectional view (right) given for each sample. Colours are indicative of Ta2.5W (yellow), TZM (blue) and W (green). Blocks 9 and 14 contain a 50 µmTa foil in the downstream side, between the target and the cladding material}
    \label{fig:TGT:PIE:PIEblocks}
\end{figure}

The PIE is foreseen to be carried out on several target blocks, representative of the different block dimensions, all the target and cladding materials combinations, and several levels of cyclic stress. The four blocks foreseen for the PIE are highlighted in Figure~\ref{Fig:TGT:PIE:targetblocks}. The characteristics of these blocks are listed in Table~\ref{tab:TGT:PIE:blocks} and summarized in Figure~\ref{fig:TGT:PIE:PIEblocks}. The target blocks will present residual dose rates in the order of several hundreds of mSv per hour and therefore special considerations will be taken for handling and shipping of the blocks to the PIE contractor, as described in section~\ref{sec:TGT:Radprot}.

\begin{table}[htbp]
\caption{Target prototype blocks foreseen for PIE with the respective dimensions and materials}
\label{tab:TGT:PIE:blocks}
\small
\begin{tabular}{cccccccc}
\hline
\multirow{2}{*}{Sample name} & \multirow{2}{*}{Shape} & \multicolumn{3}{l}{Dimensions} & \multicolumn{2}{l}{Materials} & \multirow{2}{*}{Comments} \\  
 &  & \makecell{Diameter\\ (mm)} & \makecell{Length\\(mm)} & \makecell{Cladding\\ thickness (mm)} & Core & Cladding &  \\ \hline
4 & \multirow{4}{*}{Cylindrical} & \multirow{4}{*}{80} & 25 & \multirow{4}{*}{1.5} & TZM & Ta2.5W &  \\
8 &  &  & 25 &  & TZM & Ta &  \\
9 &  &  & 50 &  & TZM & Ta2.5W & \multirow{2}{*}{\makecell{50 µm thick Ta\\ foil downstream}}  \\ 
14 &  &  & 50 &  & W & Ta &   \\ \hline
\end{tabular}
\end{table}

\subsubsection{Description of the foreseen tests}
In the following section, a first proposal of the envisaged tests is depicted. Final PIE will be subject to discussions with external contracts to match the desired testing, the available contractor capabilities and the dedicated financial resources. Each one of the following described PIE steps will be performed for each of the studied blocks.

\subsubsubsection{Surface inspection}

\begin{figure}[htbp]
    \centering
    \includegraphics[scale=0.7]{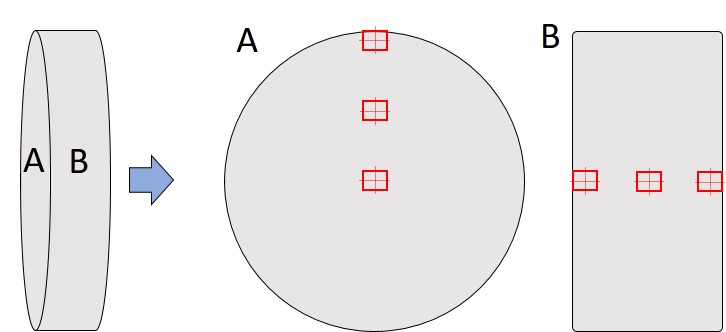}
    \caption{Target block 3D view and 2D view in flat surface (A) and curved surface (B). Areas for OM outlined in red.}
    \label{fig:TGT:PIE:OM}
\end{figure}

Surface inspection is planned to be carried out on the samples before any destructive tests.
\begin{itemize}
    \item Visual inspection: as a first approach, visual inspection shall be carried out over entire the block surfaces. Any change in the surface colour, texture and the presence of any type of heterogeneity such cracks, holes, marks shall be reported and properly localized;
    \item Optical microscopy (OM): any reported features following visual inspection shall be imaged with optical microscopy and micrographs shall be saved at different representative magnifications (e.g. 20x, 100x, 500x...). Moreover, optical micrographs shall be acquired on each flat surface (at center, mid radius and outer radius) and curved surfaces (next to the cylinder edges and in the middle length) according to Figure~\ref{fig:TGT:PIE:OM} also at several representative magnifications;
\end{itemize}

\subsubsubsection{Metrology requirements}

\begin{figure}[htbp]
    \centering
    \includegraphics[scale=0.7]{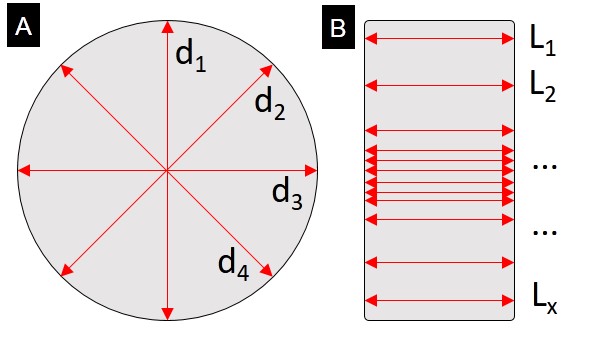}
    \caption{Schema of a target block with the a) diameter measurements and b) length measurements.}
    \label{fig:TGT:PIE:metro}
\end{figure}

\begin{itemize}
    \item Dimensional measurements: cylinders diameter shall be measured in several diameters, according to Figure \ref{fig:TGT:PIE:metro}A. Cylinders length shall be measured at different distances to the center (or continuously) along two different perpendicular diameters, according to Figure~\ref{fig:TGT:PIE:metro}B and to detect any cladding or target swelling;
    \item Surface roughness shall be measured in several areas (same areas than for optical microscopy in Figure~\ref{fig:TGT:PIE:OM}), to detect any surface morphology change due to cooling water;
\end{itemize}

\subsubsubsection{Non-destructive testing}

\begin{itemize}
     \item Penetrant testing: the entire block external surface shall be inspected with dye penetrant testing in order to reveal any weld imperfections, surface cracking and other surface defects on the cladding which might not be perceptible with visual inspection;
     \item Ultrasonic testing of interfaces: the totality of interface cladding-target shall be examined with ultrasonic testing in order to detect any loss of contact cladding-target. Sensitivity shall be adjusted with artificial defects performed on un-irradiated samples already available (see an example of interfacial defect performed on the cylinder curved surface in Figure \ref{fig:TGT:PIE:UTref});
    \item Ultrasonic testing of bulk material: the entire blocks volume shall be examined with ultrasonic testing in order to detect any defect in the bulk material. Sensitivity shall also be adjusted with artificial defects performed on un-irradiated samples;

\begin{figure}[htbp]
    \centering
    \includegraphics[scale=0.6]{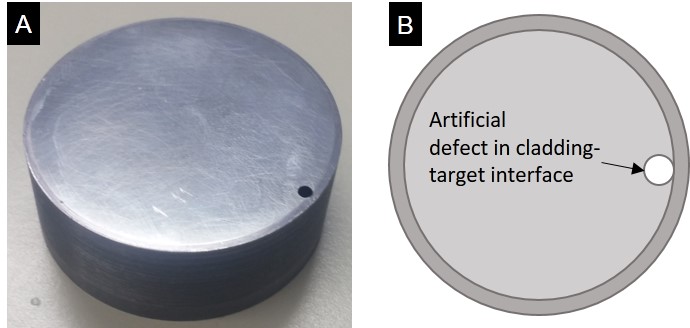}
    \caption{a) Image of an artificial interface defect prepared for internal previous studies and b) schema of the defect positioning with the hole wall tangent to the cladding target interface}
    \label{fig:TGT:PIE:UTref}
\end{figure}

\end{itemize}

\subsubsubsection{Specimen extraction}

Once all the surface inspections, metrology and non-destructive testing are performed, specimen extraction shall be carried out to allow further inspection. Careful cutting techniques such as EDM or low speed sawing shall be employed for specimen extraction in order to minimize artefact introduction due to cutting. 

\subsubsubsection{Metallographic inspection}

\begin{figure}[htbp]
    \centering
    \includegraphics[scale=0.6]{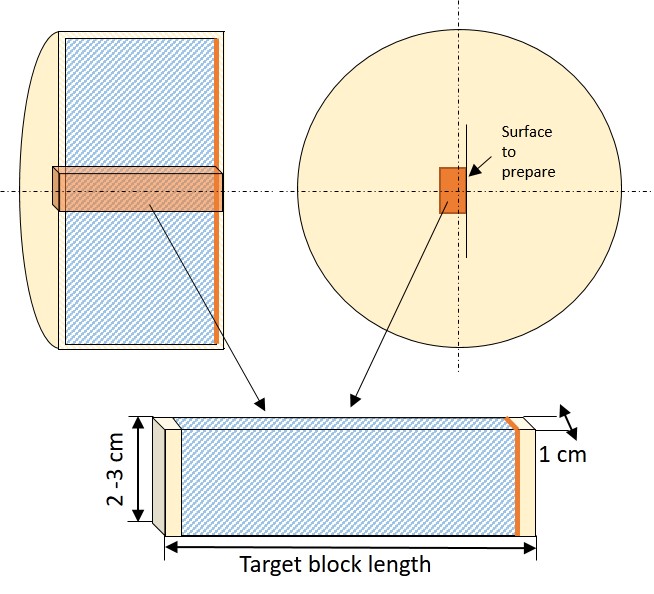}
    \caption{Schema of one target block with the placemenet of the specimen for metallographic preparation}
    \label{fig:TGT:PIE:trench}
\end{figure}

Metallographic inspection shall be performed on a slice of block length x 20-30 mm x 10 mm located at the center of each target block, such as outlined in Figure~\ref{fig:TGT:PIE:trench}. Slice shall be extracted in a way that the surface to be prepared is slightly offset from the center (0.5 mm - 2 mm). After the material removal consequence of metallographic preparation (grinding and polishing) the surface to examine will be coincident with the cylinder axis.

The specimen will be mounted in electrically conductive resin and the surface prepared with conventional metallographic preparation. Optical micrographs at representative magnifications shall be acquired at the cladding surface level and at cladding -target interface level, upstream and downstream sides in both cases. Additionally optical micrographs shall be acquired along the target material bulk. Any detected feature or heterogeneity shall be additionally imaged and reported.

Scanning electron micrographs shall be acquired at the same places than the optical micrographs with secondary electron imaging to observe the surface morphology with additional resolution (e.g. 50x, 1000x and 5000x). In parallel, backscattered electron imaging shall be employed to image any cladding surface layer (Ta oxides, pollution) and the diffusion layer in the cladding target interfaces at representative magnifications.
In case that any chemical composition heterogeneity is detected with backscattered electron contrast (external cladding layer, diffusion layer, bulk material heterogeneity...) microanalysis techniques shall be employed to identify the chemical nature of the feature. 

\subsubsubsection{Hardness testing}

\begin{figure}[htbp]
    \centering
    \includegraphics[scale=0.7]{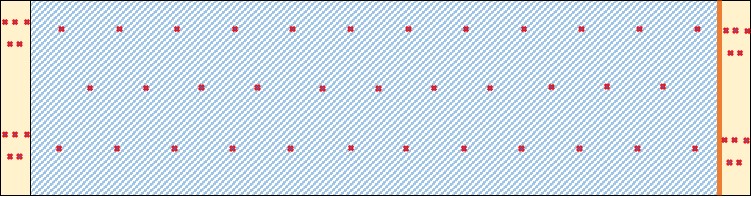}
    \caption{Schema of the specimen used for metallographic inspection with the micro-hardness indentations indicate in red.}
    \label{fig:TGT:PIE:harndess}
\end{figure}

Micro-hardness testing shall be carried out along several profiles in the same specimen used for metallographic inspection, following an indentation pattern as given in Figure \ref{fig:TGT:PIE:harndess}. Any phenomena affecting the mechanical properties of the material, such plastification, recrystallisation, grain growth, appearance of voids or precipitates would be measurable.  

\subsubsubsection{Thermophysical properties}

Thermal conductivity across the interface cladding-target shall be measured in order to detect any change in the thermal contact resistance. Specimens shall be extracted form both sides of the target blocks (upstream and downstream), the closest possible to the cylinder axis and with the dimensions imposed by the measuring system. 

\subsubsubsection{Cladding-target interface mechanical properties}

\begin{figure}[htbp]
    \centering
    \includegraphics[scale=0.6]{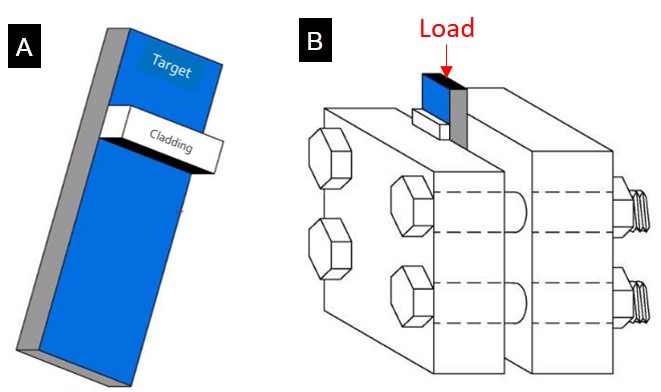}
    \caption{a) Schema of the shear specimen and b) schema of the setup for the shear testing}
    \label{fig:TGT:PIE:shear}
\end{figure}

The small thickness of the cladding materials in the target blocks does not allow the extraction of tensile specimens such as those performed during the R\&D activities on the cladding (see section~\ref{Sec:TGT:mat:RandD}). Therefore, the strength of the interface cladding-target materials can only be measured by shear testing. Sample specimens and setup necessary for this measurements are depicted in Figure~\ref{fig:TGT:PIE:shear}. Test will be carried out with miniaturised specimens due to the little amount of material available. Specimens shall be extracted in a way that the cladding lip is obtained from as close as possible to the cylinder axis. 

\FloatBarrier


\printbibliography[heading=subbibliography]

 \chapter{Target complex design and development}
\label{Chap:TargetComplex}

\section{Introduction to the Target Complex}
\label{Sec:TC:Intro}
The target will be at the heart of the new facility. High levels of radiation (both prompt and residual) will be produced by the SPS beam hitting the target; a total cumulated dose near the target of around 500 MGy/year is expected. The target will be located in an underground area (to contain radiation as much as possible) located at about 15 metres below ground level. The depth of the infrastructure is determined by the location of the extraction line (TT20) from where the SPS beam will be deflected.

The target will be surrounded by approximately 5500~tonnes of cast iron and steel shielding with outer dimensions of around 6.8~m x 9.5~m x 8~m high (the so-called hadron absorber) to reduce the prompt dose rate during operation and the residual dose rate around the target during shutdown. The target and its surrounding shielding will be housed in a vessel containing gaseous helium slightly above atmospheric pressure in order to reduce air activation and reduce the radiation accelerated corrosion of the target and surrounding equipment.

The SPS beam will enter the helium vessel through a beam window, then pass through a collimator which serves to protect the target and adjacent equipment from misalignment of the incident SPS beam and to protect the equipment in the extraction tunnel from particles (essentially neutrons generated by the target) travelling backwards relative to the incident beam. Downstream of the target a magnetic coil and US1010 steel yoke are used to produce a magnetic field of 1.5-1.6 T in order to sweep high energy muons produced in the target to reduce experimental backgrounds. The secondary beam will pass through the downstream helium vessel wall and leave the target complex via a 5 cm thick steel plate "window" at the upstream end of the detector hall.

\begin{figure}[ht]
\centering
\includegraphics[width=0.7\linewidth]{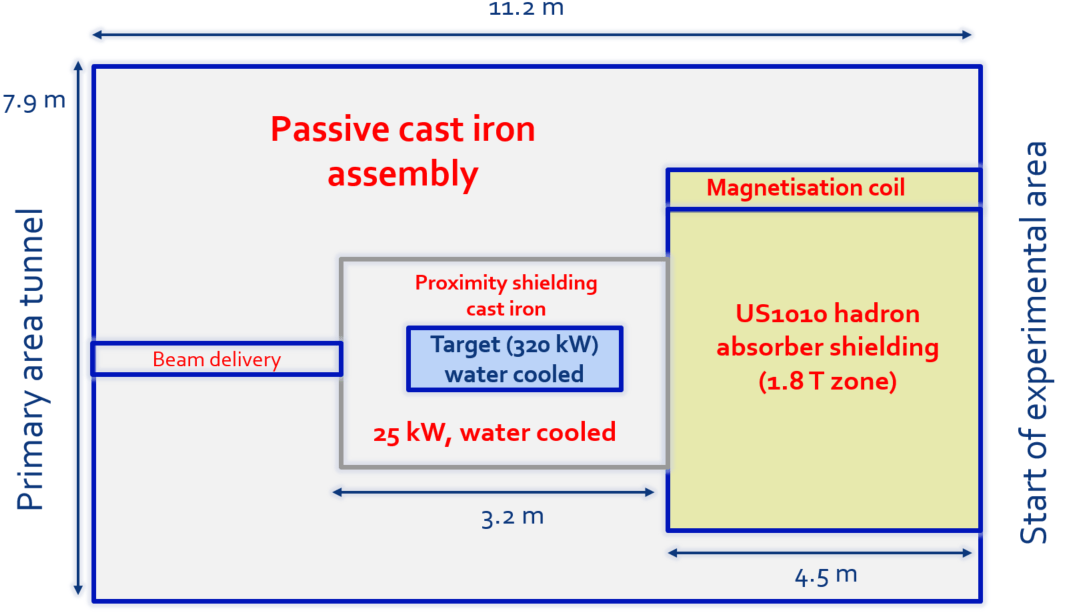}
\caption{Schematic view of the BDF Target Complex.}
\label{Fig:TC:intro-1}
\end{figure}

\begin{figure}[ht]
\centering
\includegraphics[width=0.75\linewidth]{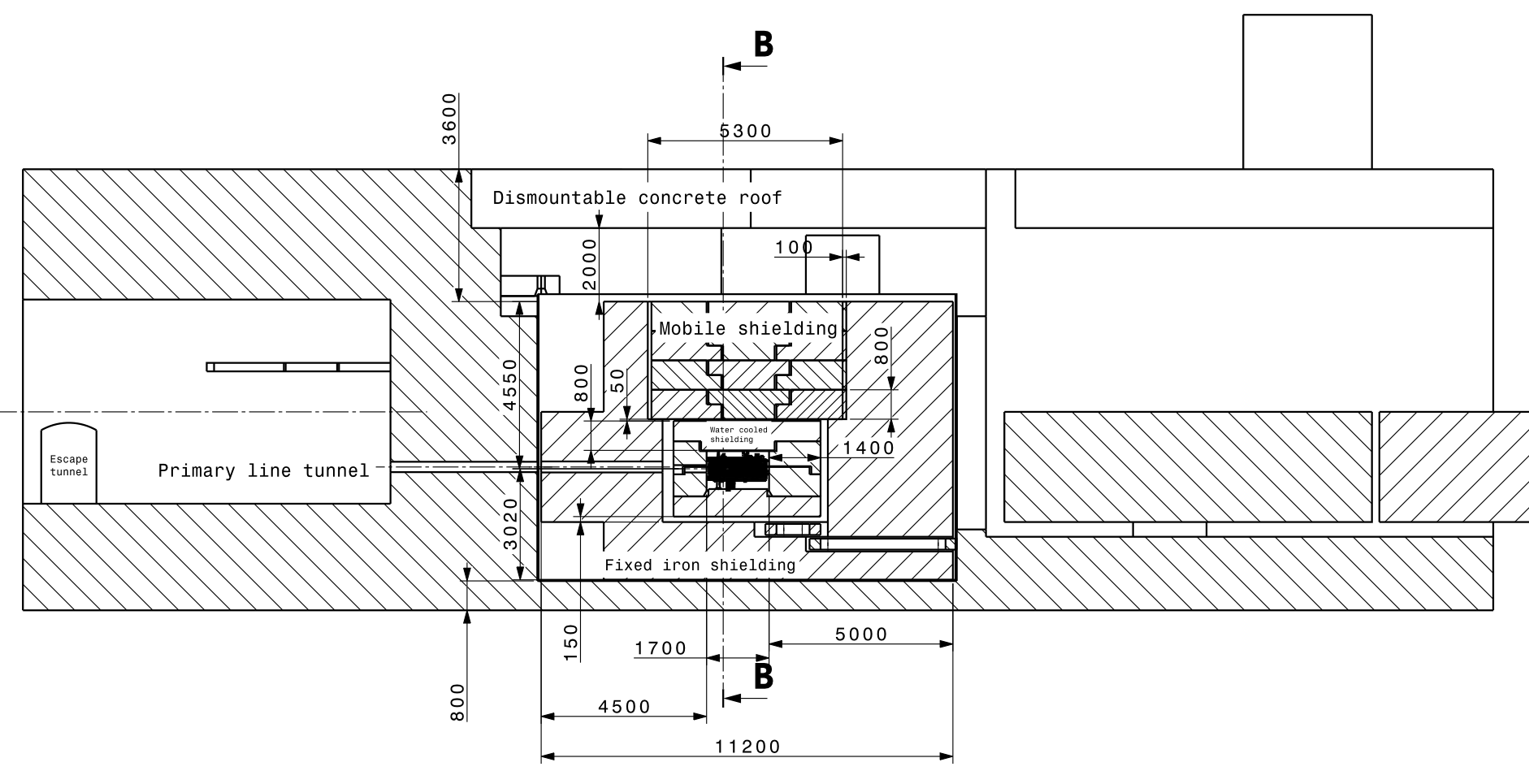}
\caption{Cross-sectional view of the BDF Target Complex.}
\label{Fig:TC:intro-2}
\end{figure}

 A schematic view is presented in Figure~\ref{Fig:TC:intro-1} and \ref{Fig:TC:intro-2}.

The target and the shielding immediately around it will be water cooled. All the shielding in the helium vessel will be built up of blocks; the layout and geometries of which are designed to avoid direct radiation shine paths and to minimise the number of block movements needed to allow exchange of failed equipment. A helium purification system is foreseen to allow flushing of the air to ensure a 99.9\% pure helium atmosphere in the vessel after initial installation and after maintenance interventions. Ventilation of the target complex will use a cascade of pressure between different zones to provide containment of any radioactive contamination that could potentially be released according to ISO standard 17873.

\begin{figure}[ht]
\centering
\includegraphics[width=0.8\linewidth]{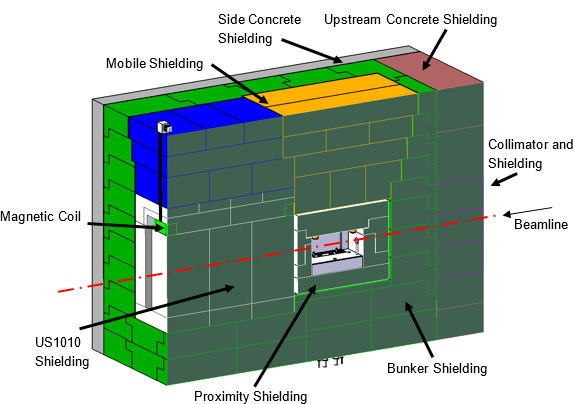}
\caption{Isometric view of the BDF Target Complex.}
\label{Fig:TC:intro-3}
\end{figure}

Remote handling and manipulation of the target and surrounding shielding will be mandatory due to the high residual dose rates. The target complex has been designed to house the target and its shielding in the helium vessel along with the cooling, ventilation and helium purification services below ground level. The target complex design allows for removal and temporary storage of the target and shielding blocks in the cool-down area below ground level and includes dedicated shielded pits for storage of the highest dose rate equipment.

A 40-tonne capacity overhead travelling crane in the target complex building will be used for initial installation and will carry out the handling and remote handling of the shielding blocks and other equipment as needed for assembly and maintenance of the facility. All the shielding blocks and other heavy equipment are designed to be compatible with the crane's lifting capacity. Target and beam elements are to be aligned within +/-10 mm with respect to the impinging proton beam.

\clearpage

\section{Design study of handling and integration of the BDF target complex}
\label{Sec:TC:Design}

\subsection{Introduction to the handling and integration study}
\label{Sec:TC:Intro-Handling-Integration-study}

\subsubsection{Aim of the study}
\label{Sec:TC:StudyAim}

After the initial work to determine the main requirements and basic layout of the target complex as explained in the introduction, the target complex design was further developed by going into more detail on the handling and remote handling operations required throughout the life of the facility. This work aimed to demonstrate the feasibility of the construction, operation, maintenance of the BDF target complex along with decommissioning of the key elements and to provide an integrated design of the target complex. The output from the study was fed into the design work on the target, the magnetised muon shield system, the helium vessel as well as cooling and ventilation systems.

\subsubsection{Study deliverable}
The study deliverables were in the form of reports, 2-D drawings, 3-D models of the target complex and animations of the handling sequences. They are stored in CERN's EDMS System: EDMS 1977049 v1 ``BDF Target Complex Design Final Results May 2018'' \cite{HePurifTCdesign}. The study results were published in Ref.~\cite{BDFcomplex}.

\subsubsection{Scope of the study}
\label{Sec:TC:StudyScope}

The remote handling of highly activated radioactive objects, such as target, collimator, beam window, shielding blocks and magnetic coil, along with their connection and disconnection within the target complex building was studied by CERN in collaboration with Oxford Technologies Ltd in order to arrive at the integrated designs presented here. In addition to  ``foreseen'' remote handling operations, such as target exchange, the study considered ``unforeseen'' remote handling operations needed to recover from failures or damage to equipment. The study included the conceptual design of lifting, handling and remote handling equipment for the highly activated objects along with the necessary water, helium and electrical connections compatible with the radiation environment and remote handling constraints. These designs were then integrated in to the target complex as a whole. For both concepts, the handling and integration study concentrated on the target hall and its underground area - the integration of an adjacent service building housing the ventilation plant, access systems and controls etc. was dealt with separately (Chapter~\ref{Chap:Integration}).

\subsubsection{Key design considerations}
\label{Sec:TC:KeyDesignConsiderations}

\subsubsubsection{Shielding design}
\label{Sec:TC:ShieldingDesign}
A large amount of the design is driven by radiation protection and safety considerations. For the shielding blocks in the helium vessel this leads to requirements such as: 

\begin{itemize}
\item Minimise and stagger gaps to avoid direct radiation shine paths;
\item Provide gaps to allow purging by the helium purification system;
\item Equipped with lifting features to allow remote handling;
\item Proportioned to be stable during transport and installation;
\item Compatibility with building crane capacity / road transport;
\end{itemize}

\subsubsubsection{Water and electrical connections}
\label{Sec:TC:WaterConnections}
Experience at CERN has shown that connections in radiation environments are a source of problems such as water leaks -- for this reason particular attention was addressed to connection and disconnection methods compatible with the radiation environment along with repair in the event of failures and damage.

\subsubsubsection{Whole life cycle}
\label{Sec:TC:LifeCycle}
The study took into account the whole life cycle of the facility -- in particular:

\begin{itemize}
\item First installation and alignment, 
\item Potential failures and damage during operation and maintenance along with how to recover from them (using remote techniques where needed)
\item Reconfiguration of the facility for a different experiment in the future
\item Decommissioning.  
\end{itemize}

\subsubsection{The two handling concepts that were studied (crane and trolley concepts)}
\label{Sec:TC:TwoHandlingConcepts}

Target complex designs based on two different handling concepts have been developed: the ``crane concept'' and the  ``trolley concept''.

\subsubsubsection{Crane concept - using remotely controlled overhead travelling crane}
\label{Sec:TC:CraneConcept}

The crane concept relies on the overhead travelling crane in the target complex building for remotely controlled handling and transfers of the target, shielding and magnetic coil etc. during the life of the facility. 

\begin{figure}[!htb]
\centering
\includegraphics[width=0.65\linewidth]{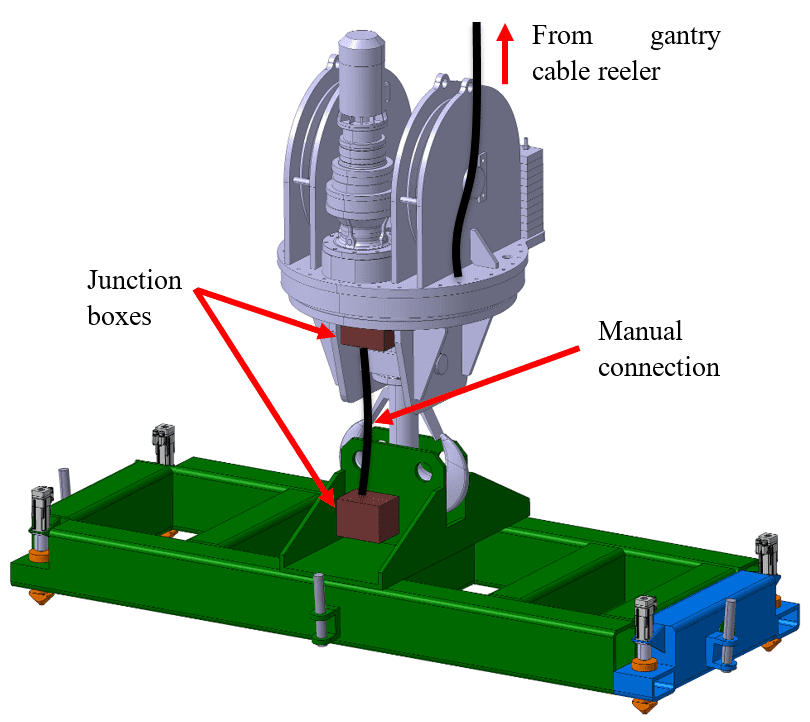}
\caption{Overhead travelling crane pulley block and hook (in grey) with remotely operated spreader beam for handling of shielding. The spreader shown has remotely operated ISO load attachments at each corner and is equipped with cameras.}
\label{Fig:TC:1-overheadCrane}
\end{figure}

The crane hook has motorised rotation to allow remote handling of radioactive equipment. Power and signal connection services are supplied to the pulley block of the hook to operate the spreader attachment features used to remotely pick up loads and for the cameras on the spreaders (Fig.~\ref{Fig:TC:1-overheadCrane}). 
More details of the building crane are given in section \ref{Sec:TC:CraneConcept}.

Remotely operated spreader beams are needed to enable the crane to pick up and handle the different radioactive elements of the facility. More details of the spreader beam designs are given in the sections dealing with the handling of the key components of the facility.

\subsubsubsection{Trolley concept}
\label{Sec:TC:TrolleyConcept}

\begin{figure}[!htb]
\centering
\includegraphics[width=0.9\linewidth]{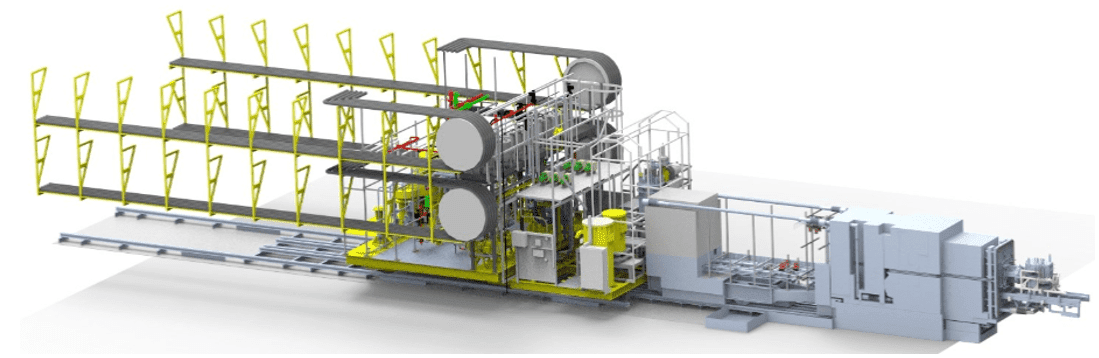}
\caption{ISIS target trolley (courtesy of RAL).}
\label{Fig:TC:2-ISIS-target-trolley}
\end{figure}

The trolley concept has the target and its main services installed on a mobile trolley running on rails. When the trolley retracts the target and shielding from the helium vessel, the target enters a remote handling  ``hot-cell'' equipped with master-slave manipulators which allow the target to be disconnected from its services remotely and then exchanged without disturbing other equipment in the helium vessel. This concept is already used for ISIS at the Rutherford Appleton Laboratory in the UK (Fig.~\ref{Fig:TC:2-ISIS-target-trolley}). For the handling operations on other equipment, the building crane is used in the same way as in the crane concept.

\subsubsubsection{Common features and main differences between the crane and trolley concepts}
\label{Sec:TC:features-differencens-crane-trolley}

For both concepts the main elements (target core, beam window, helium vessel, collimator, magnetic coil and US1010 shielding) are essentially the same and are common to the target complex designs produced for both concepts. The designs for both concepts include cool-down areas, remote handling areas, cooling and ventilation areas and sump rooms.

The main differences between the crane and trolley concepts are in the way the target and water-cooled proximity shielding are supported, installed and removed from the helium vessel and how their services are connected and disconnected. These differences can potentially have a big impact on the reliability and operation of the facility.

\subsubsection{The crane concept -- overview of target complex}
\label{Sec:TC:CraneConcept-Overview}

\begin{figure}[!htb]
\centering
\includegraphics[width=0.7\linewidth]{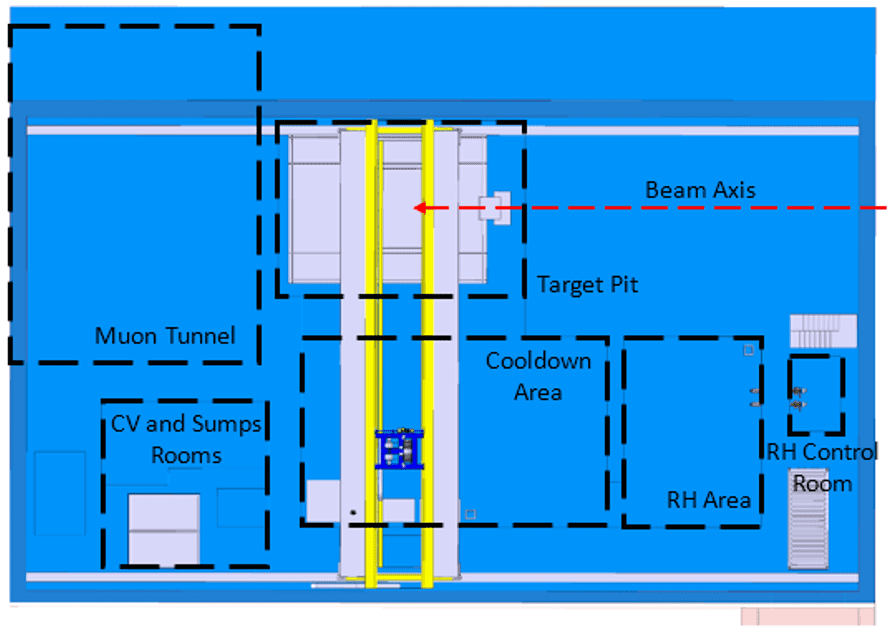}
\caption{Plan view showing the main areas of the crane concept target complex. The proton beam from the SPS enters from the right in the figure.}
\label{Fig:TC:3-craneConcept}
\end{figure}

\begin{figure}[!htb]
\centering
\includegraphics[width=0.8\linewidth]{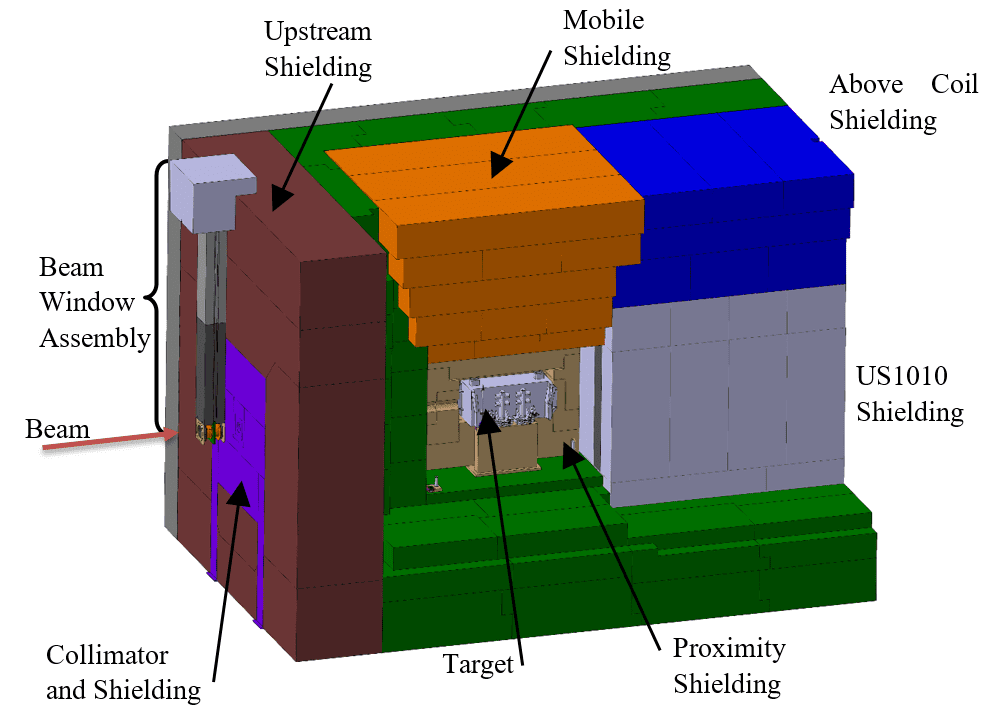}
\caption{Isometric cut-away view of crane concept equipment in the helium vessel.}
\label{Fig:TC:4-cut-away-Crane}
\end{figure}

\begin{figure}[!htb]
\centering
\includegraphics[width=0.7\linewidth]{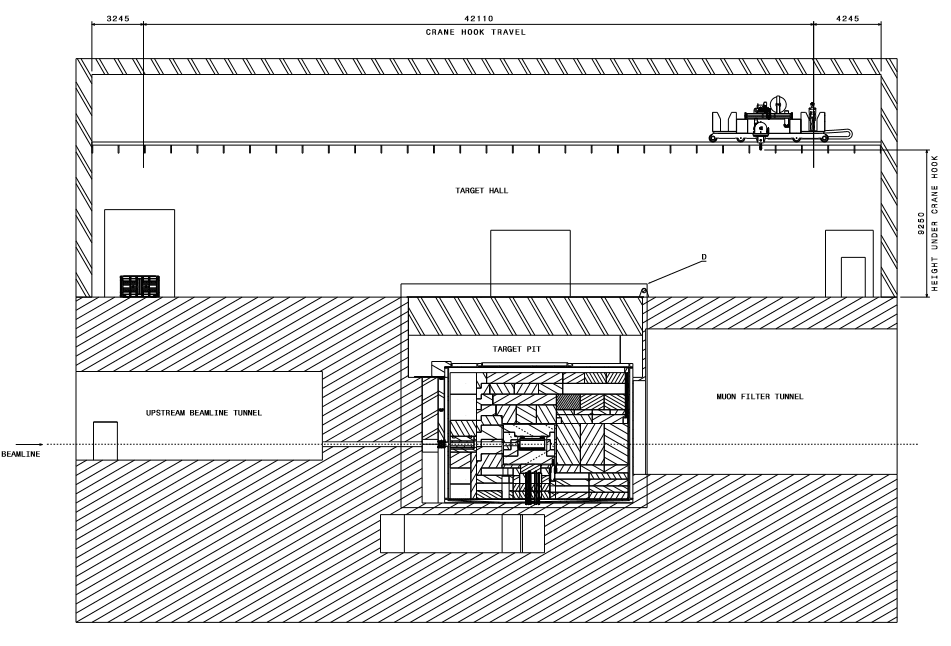}
\caption{Section view of crane concept target complex showing underground area and surface building - showing proton beam entering from left side of figure.}
\label{Fig:TC:Extra-1-section-view-crane}
\end{figure}

\begin{figure}[!htb]
\centering
\includegraphics[width=0.6\linewidth]{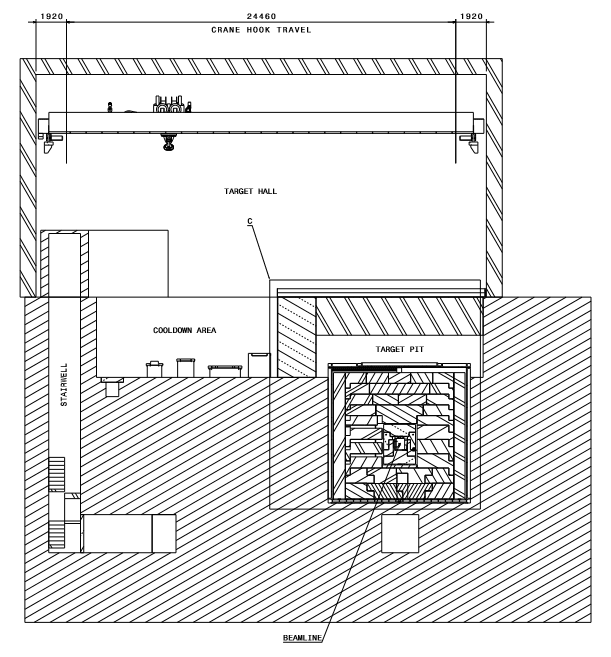}
\caption{Section view of crane concept target complex showing underground area and surface building - view along beam axis.}
\label{Fig:TC:Extra-1b-section-view-crane-2}
\end{figure}

\begin{figure}[!htb]
\centering
\includegraphics[width=0.8\linewidth]{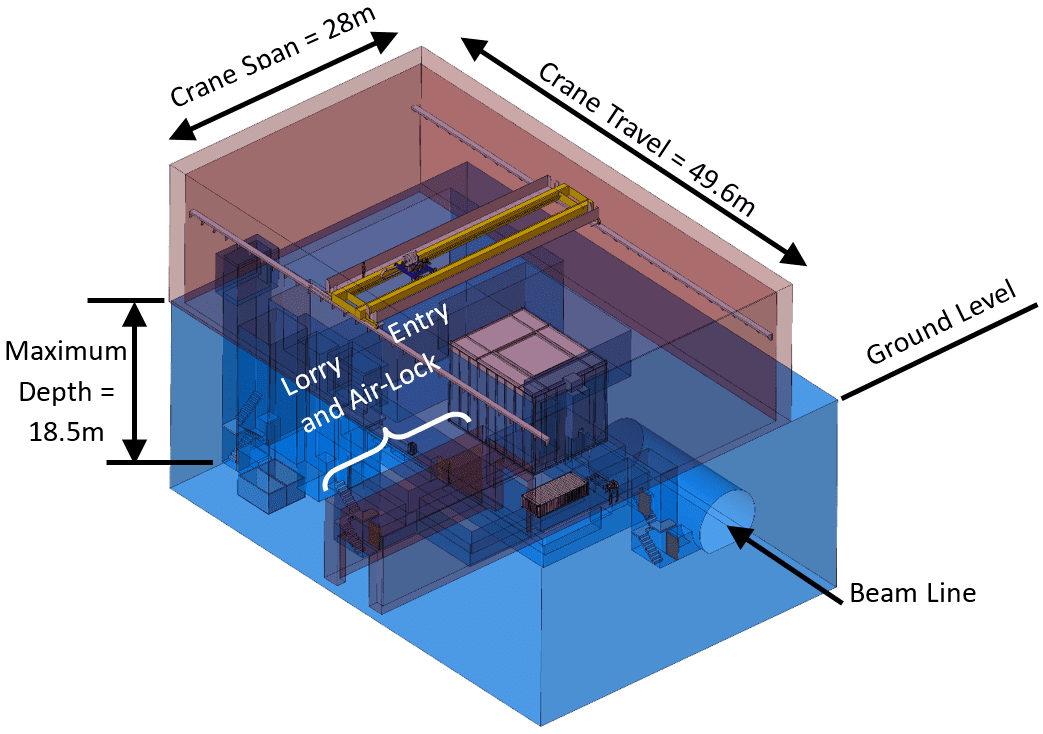}
\caption{Isometric view of crane concept target complex building.}
\label{Fig:TC:5-isometric-crane}
\end{figure}

The crane concept target complex design is illustrated in Figs.~\ref{Fig:TC:3-craneConcept} to~\ref{Fig:TC:5-isometric-crane}. The target, and its surrounding shielding are housed in a helium vessel in the target pit. An underground cool-down area including a remote handling area is used for storage and dismantling of activated components. A surface building equipped with an overhead travelling crane covers the underground areas. Transfers of radioactive equipment removed from the helium vessel in the target pit are carried out using the building crane whilst keeping the load below ground level to ensure compliance with radiation dose rate limits outside the building. 

Cooling plant (for cooling of the equipment in the helium vessel and for helium purification) is also installed underground. Separate sump rooms are used to collect any water leaking from the equipment in the helium vessel. A surface building equipped with an overhead travelling crane covers all the underground areas. The complex is equipped with a ventilation system to ensure confinement for radiological safety reasons. A vehicle airlock is attached to the surface building to allow vehicle movements without disrupting the ventilation system. More details of cooling and ventilation systems are given in section~\ref{Sec:TC:CV}.

\subsubsection{The trolley concept - overview of target complex}
\label{Sec:TC:TrolleyConcept-Overview}

The target complex design for the trolley concept is illustrated in Figs.~\ref{Fig:TC:7-Trolley-layout} to~\ref{Fig:TC:11-cut-away-withdrawn-trolley}. As for the crane concept, the target and its surrounding shielding are housed within a helium vessel in the target pit. The main difference is that the target is supported on a trolley running on rails and enters the helium vessel from the side. 

\begin{figure}[!htb]
\centering
\includegraphics[width=0.7\linewidth]{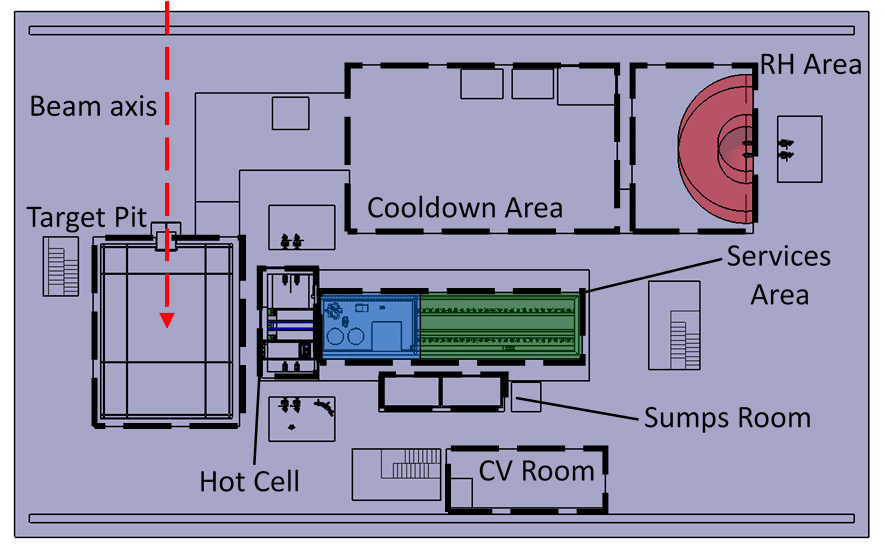}
\caption{Plan view showing main areas of trolley concept target complex}
\label{Fig:TC:6-area-trolley}
\end{figure}

\begin{figure}[!htb]
\centering
\includegraphics[width=0.8\linewidth]{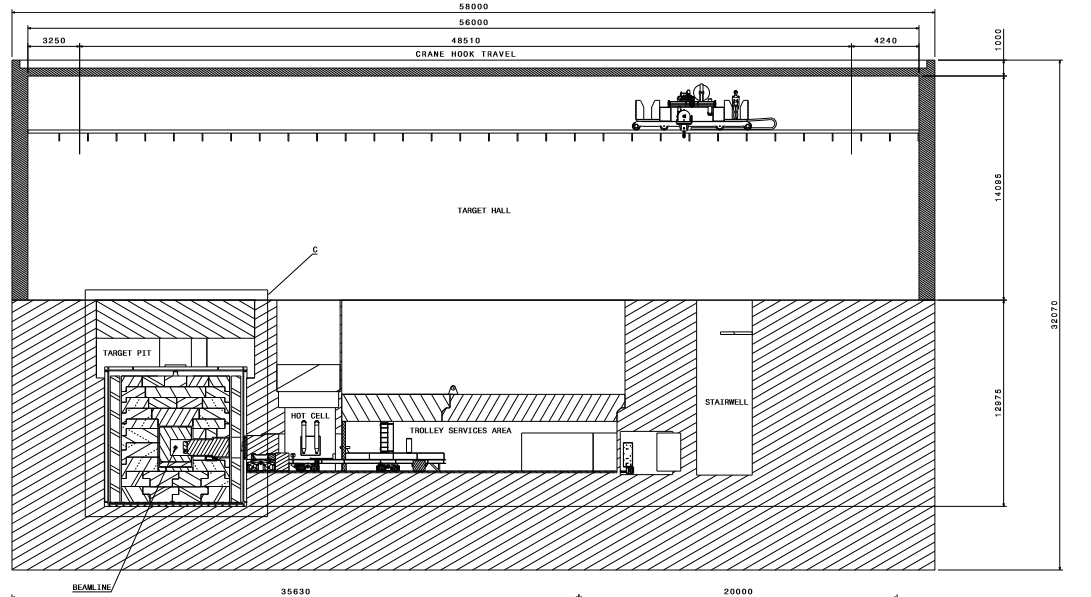}
\caption{Section of trolley concept target complex showing underground area and surface building - view along beam axis.}
\label{Fig:TC:Extra-2-section-view-trolley}
\end{figure}

The target can be withdrawn from the helium vessel, without the need to remove the lid of the helium vessel or remove the shielding above the target, by rolling back the trolley. When the target is withdrawn from the helium vessel it enters a hot cell equipped with a 3 tonne overhead travelling crane and two pairs of master-slave manipulators which are primarily used to operate the target securing clamp and make and break water and electrical connections to the trolley in the event of target exchange. More details of the hot cell crane are given in Section~\ref{Sec:TC:hot-cell-trolley}.

\begin{figure}[!htb]
\centering
\includegraphics[width=0.75\linewidth]{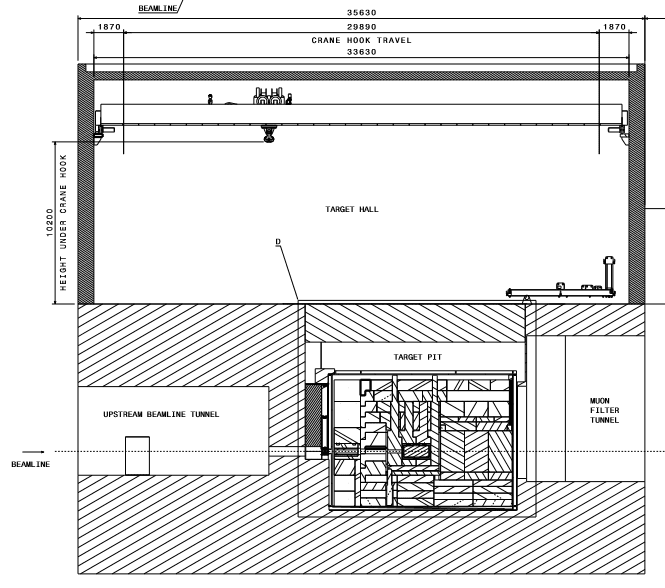}
\caption{Section view of trolley concept target complex showing underground area and surface building - proton beam entering from left side of figure.}
\label{Fig:TC:Extra-2b-section-view-trolley-2}
\end{figure}

The trolley supports the target on cantilever beams which also support a  ``plug'' of cast iron shielding that moves into the helium vessel along with the target. All the services to the target, and the trolley-mounted water-cooled proximity shielding, are incorporated into the trolley.

\begin{figure}[!htb]
\centering
\includegraphics[width=\linewidth]{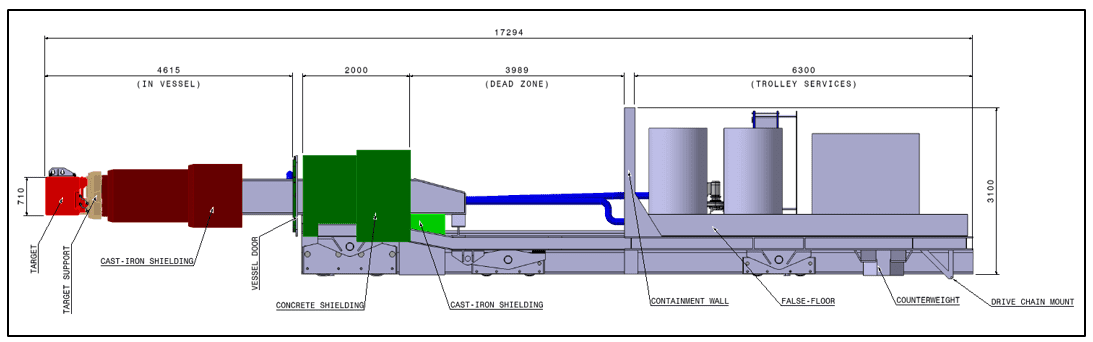}
\caption{Trolley layout with main elements and dimensions.}
\label{Fig:TC:7-Trolley-layout}
\end{figure}

\begin{figure}[!htb]
\centering
\includegraphics[width=0.8\linewidth]{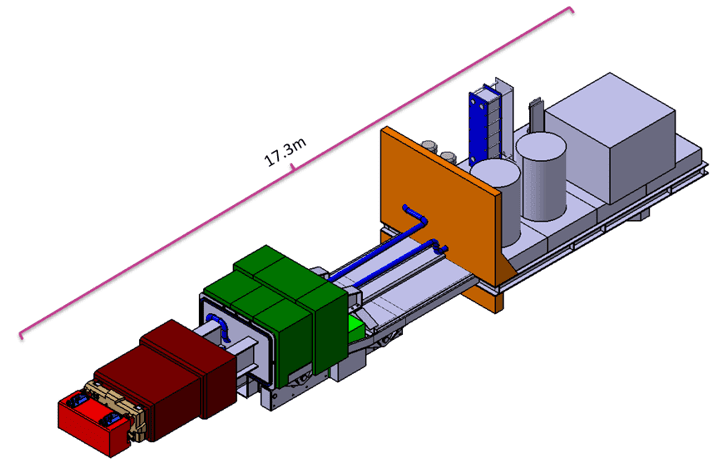}
\caption{Isometric view of the BDF trolley with the main components and the size.}
\label{Fig:TC:8-isometric-trolley}
\end{figure}

\begin{figure}[!htb]
\centering
\includegraphics[width=0.8\linewidth]{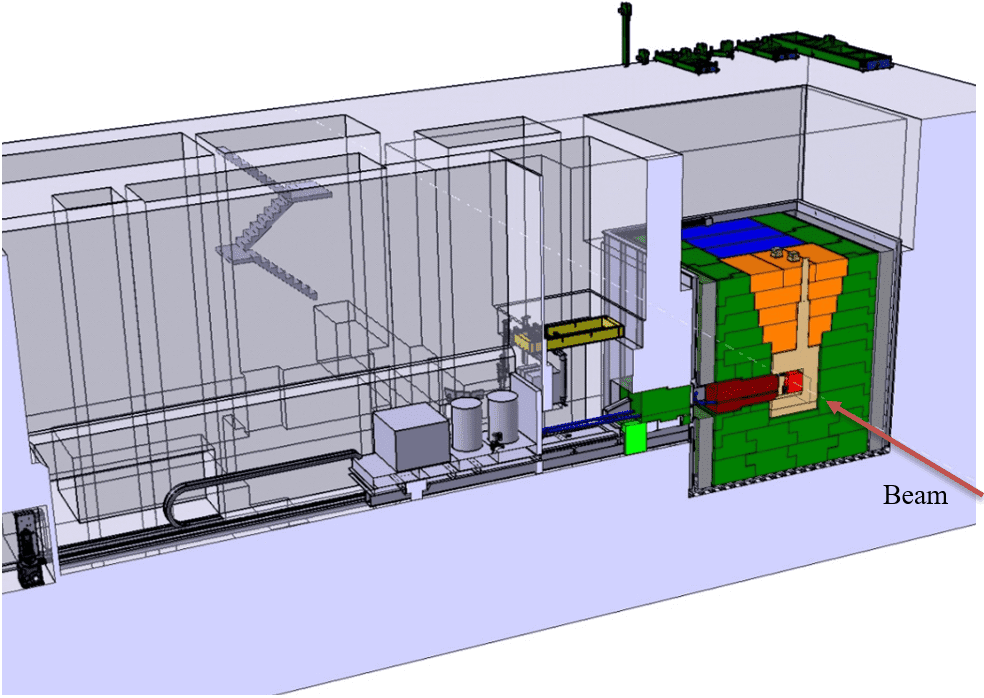}
\caption{Cut away view of trolley concept target complex underground area: target in operating position inside the helium vessel. Hot cell crane in yellow.}
\label{Fig:TC:9-cut-away-trolley}
\end{figure}

\begin{figure}[!hbt]
\centering
\includegraphics[width=0.8\linewidth]{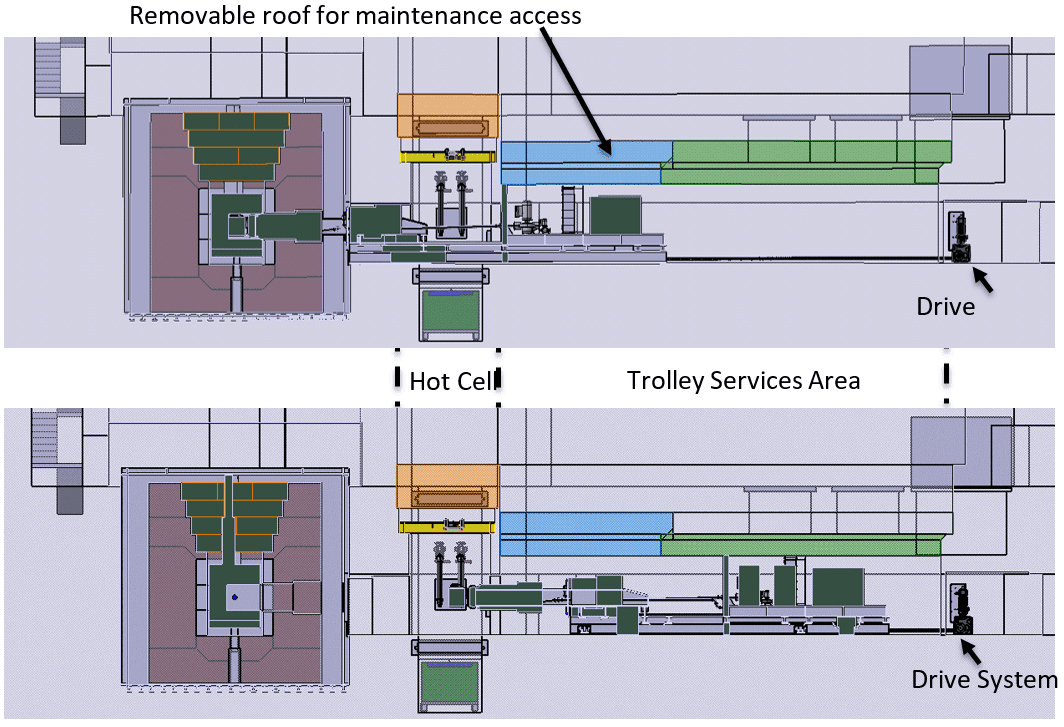}
\caption{Section views showing the trolley with the target in its operating position in the helium vessel (upper image) and withdrawn into the maintenance hot-cell (lower image).}
\label{Fig:TC:10-section-view-trolley}
\end{figure}

\begin{figure}[!htb]
\centering
\includegraphics[width=0.8\linewidth]{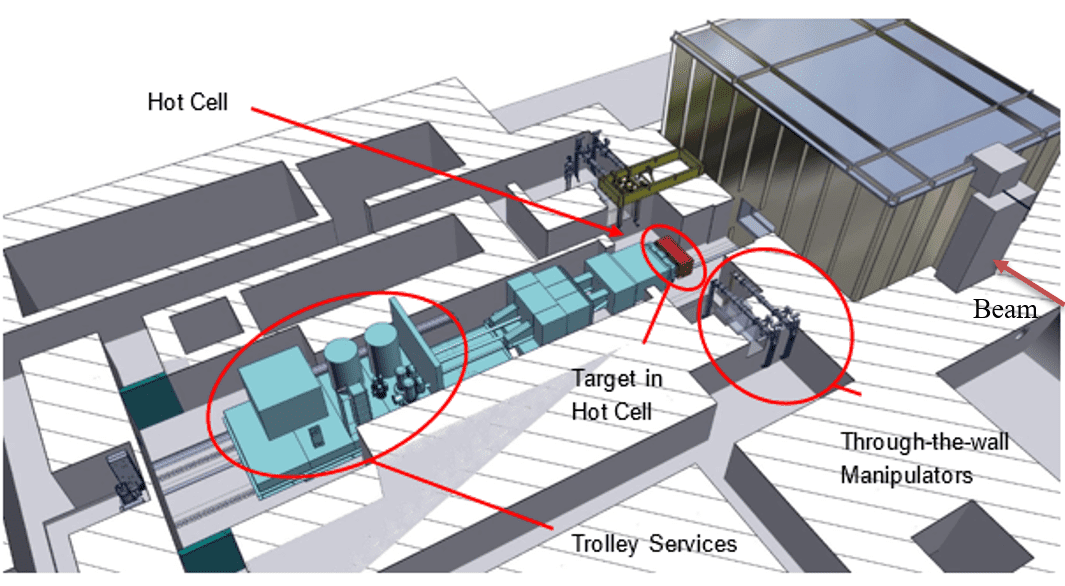}
\caption{Cut-away view showing trolley withdrawn from helium vessel with target in hot cell.}
\label{Fig:TC:11-cut-away-withdrawn-trolley}
\end{figure}

\begin{figure}[!htb]
\centering
\includegraphics[width=0.75\linewidth]{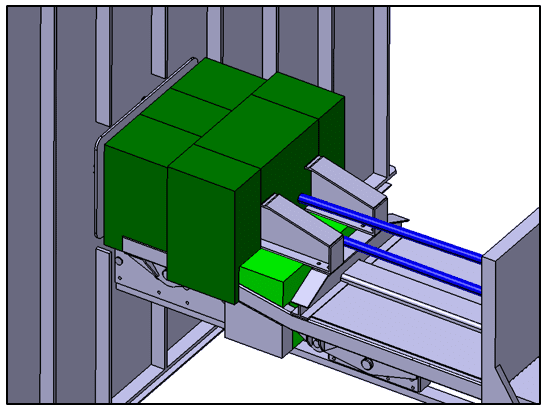}
\caption{Trolley moved to fully forward position -- helium vessel door closed and concrete shielding plug (in green) fills the opening in the concrete shielding wall between the helium vessel and the hot cell (Shielding wall not shown for clarity).}
\label{Fig:TC:12-forward-position-trolley}
\end{figure}

\begin{figure}[!htb]
\centering
\includegraphics[width=0.6\linewidth]{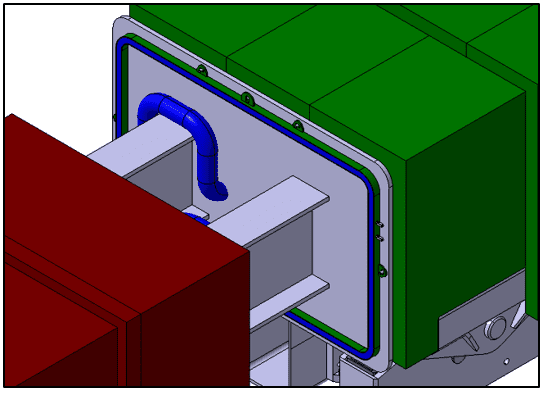}
\caption{Helium vessel door on trolley: the door is equipped with an inflatable EPDM seal which seals against the helium vessel wall when the trolley is in its fully forward position.}
\label{Fig:TC:13-helium-door-trolley}
\end{figure}

A door equipped with inflatable EPDM seals is fitted to the trolley to close the helium vessel when the target is in its operating position (Figs.~\ref{Fig:TC:12-forward-position-trolley} and \ref{Fig:TC:13-helium-door-trolley}). The EPDM seal can be replaced using the manipulators in the hot-cell. 

\begin{figure}[!htb]
\centering
\includegraphics[width=0.8\linewidth]{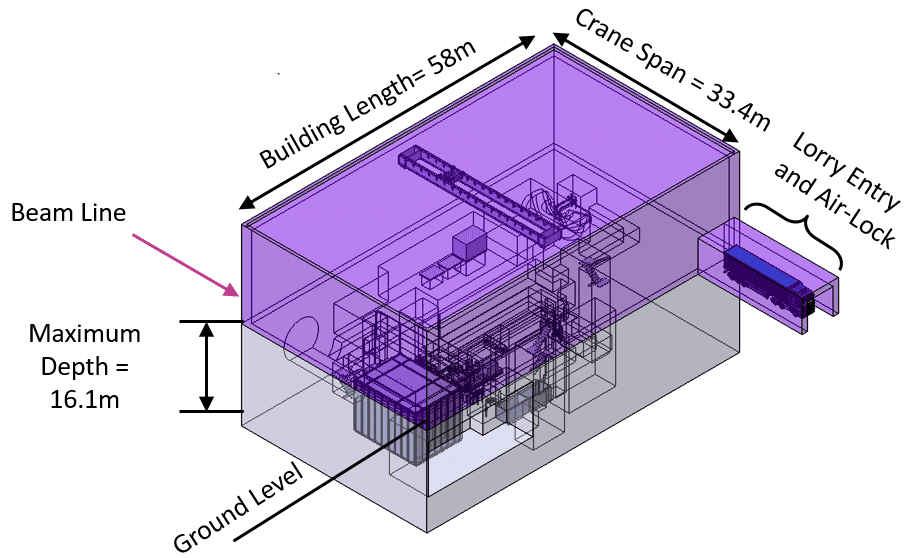}
\caption{Isometric view of trolley concept target complex building.}
\label{Fig:TC:14-isometric-trolley-concept}
\end{figure}

\subsection{Target support and handling}
\label{Sec:TC:Target-support-handling}

\subsubsection{The target housing, supports, services and connections -- crane concept}
\label{Sec:TC:Target-housing-supports-crane-concept}

\begin{figure}[!htb]
\centering
\includegraphics[width=0.7\linewidth]{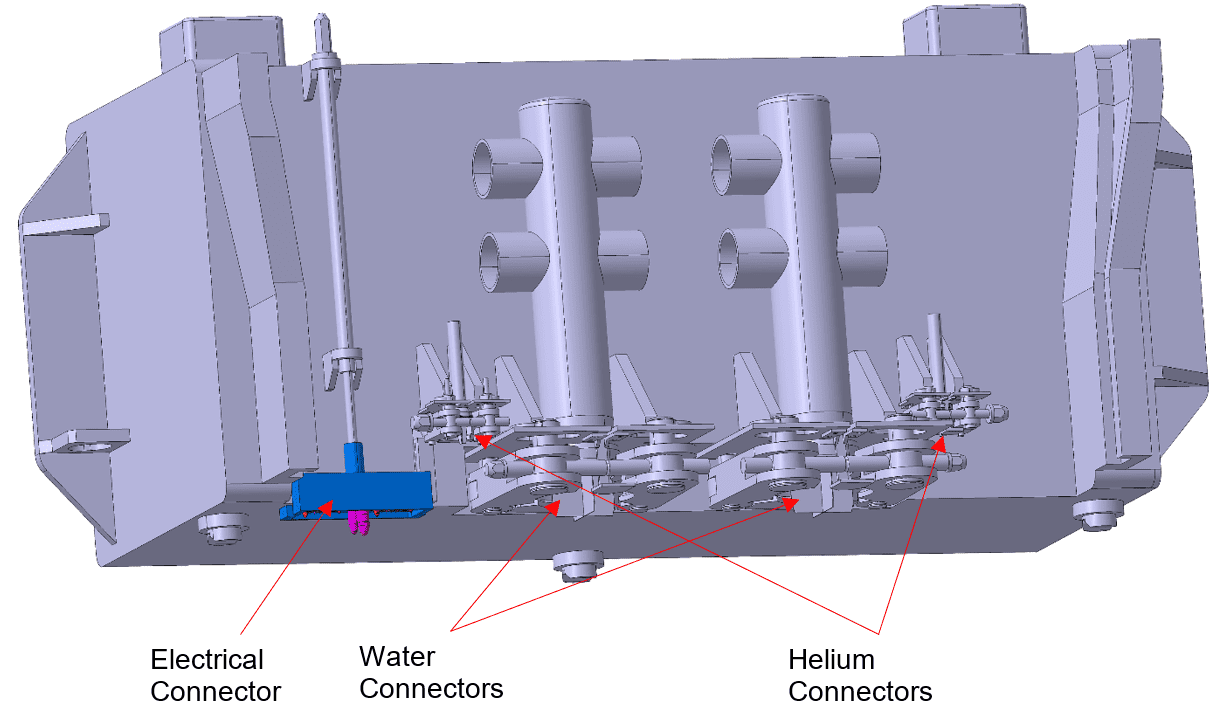}
\caption{Target container showing the service connections (crane concept)}
\label{Fig:TC:15-target-container}
\end{figure}

The crane concept target is housed in a rectangular stainless-steel container which incorporates support and alignment features, lifting points, handling guides, water and electrical connectors (Fig.~\ref{Fig:TC:15-target-container}). Internal pipework connects the target cooling channels to the remotely operated connector clamps near the base of the target container. The details of the internal pipework are covered by a separate study (Chapter~\ref{Chap:Target}).

\begin{figure}[!htb]
\centering
\includegraphics[width=0.7\linewidth]{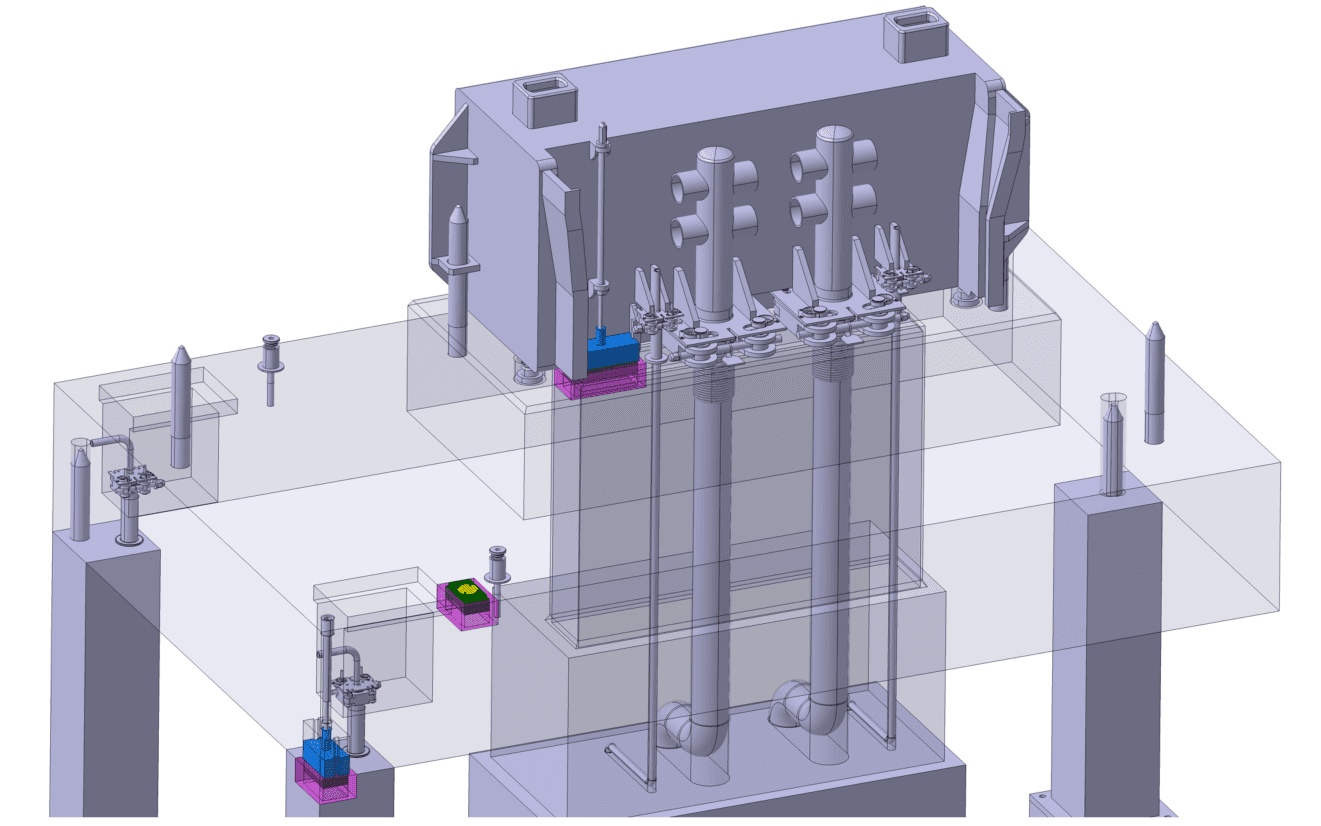}
\caption{Target sitting on bottom layer of proximity shielding (crane concept). Helium, water cooling and electrical connections for the target pass though the central support pillar. Services for the proximity shielding pass through the proximity shielding support pillars. Pipework connections are made by remotely operated screw-clamp connections.}
\label{Fig:TC:16-target-sitting}
\end{figure}

\begin{figure}[!htb]
\centering
\includegraphics[width=0.7\linewidth]{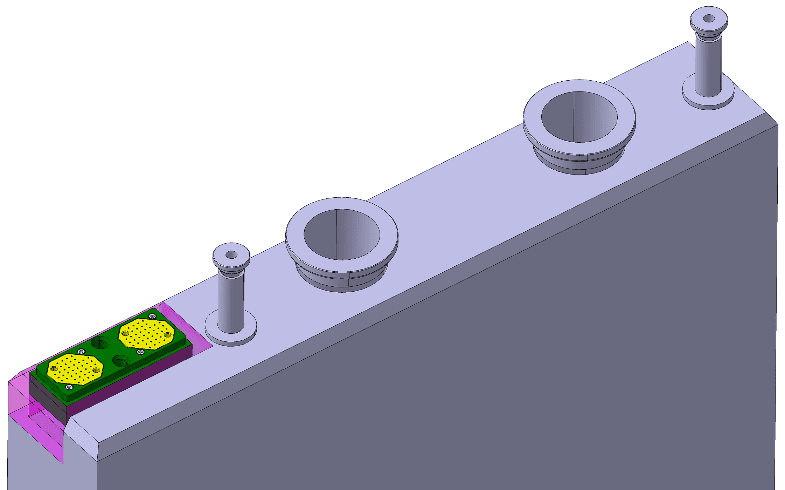}
\caption{Upper part of crane concept target service pillar showing helium, water and electrical connections}
\label{Fig:TC:17-upper-crane}
\end{figure}

The target is supported on three V and ball supports on the bottom layer of the proximity shielding, which in turn sits on top of three pre-aligned pillars. Cooling water pipework and temperature sensor cables to the proximity shielding pass through the pillars. A separate service pillar supplies the target with its water, helium and electrical connections (Figs. \ref{Fig:TC:16-target-sitting} and \ref{Fig:TC:17-upper-crane}). Service galleries are therefore needed underneath the helium vessel to allow the passage of the services and provide personnel access for installation and repair of the services. 

\begin{figure}[!htb]
\centering
\includegraphics[width=0.7\linewidth]{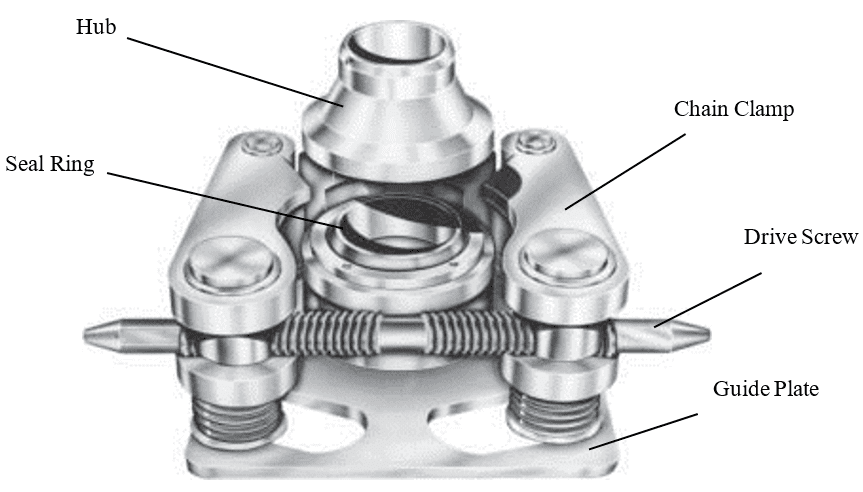}
\caption{Grayloc{\textregistered} connector clamping system elements: rotation of the drive screw clamps the hub welded onto one pipe to the other pipe hub, compressing the captive seal between the hubs}
\label{Fig:TC:18-grayloc-connector}
\end{figure}

\begin{figure}[!htb]
\centering
\includegraphics[width=0.7\linewidth]{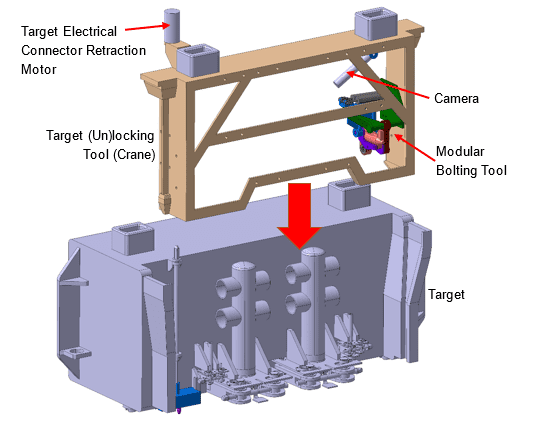}
\caption{(Un)locking tool for the target remotely operated screw-clamp water and helium pipework connections (crane concept). Electrical connections are also connected using a motor drive on the (un)locking tool. The tool is lowered into position by the crane; guidance rails on the target ensure the correct position of the tool.}
\label{Fig:TC:19-unlocking-tool}
\end{figure}

\begin{figure}[!htb]
\centering
\includegraphics[width=0.6\linewidth]{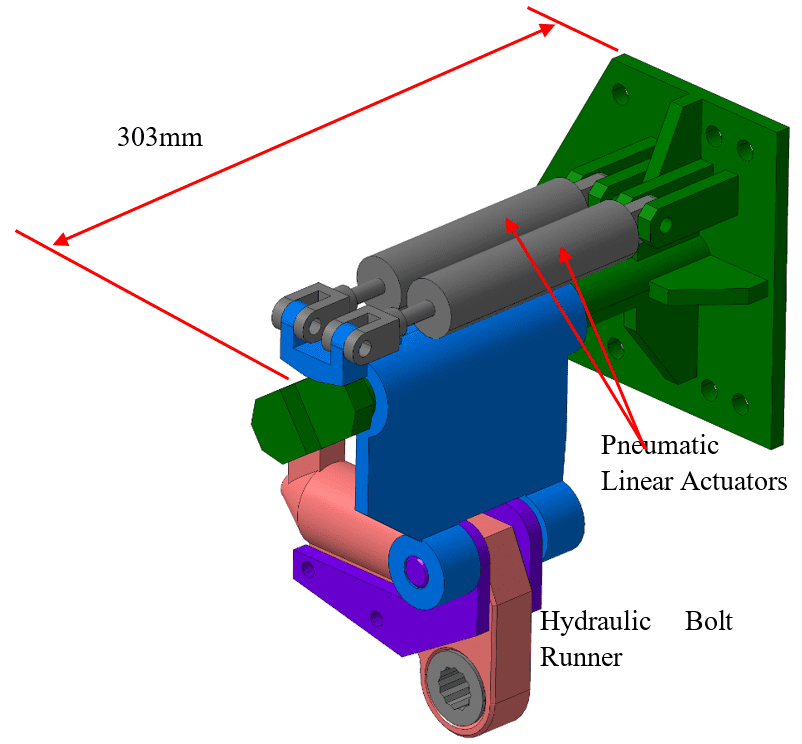}
\caption{Bolting module fitted to (un)locking tool -- it is used to remotely tighten and loosen screw-driven pipe clamps on the crane concept target.} 
\label{Fig:TC:20-bolting-module}
\end{figure}

\begin{figure}[!htb]
\centering
\includegraphics[width=0.7\linewidth]{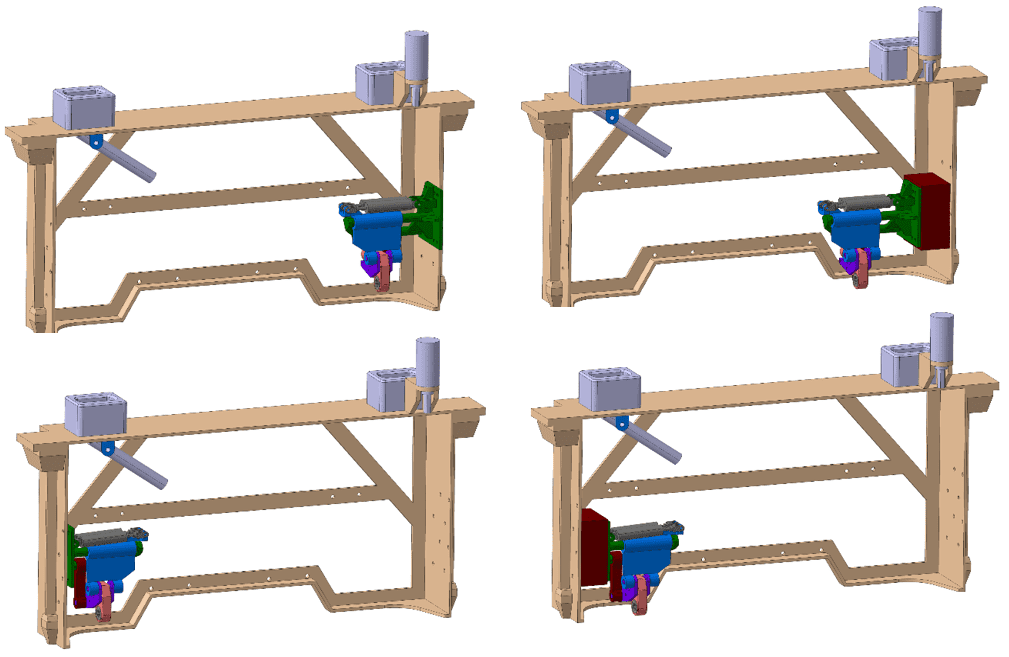}
\caption{(Un)locking tool with bolting module fitted in different positions in order to loosen or tighten the different pipe clamps.}
\label{Fig:TC:21-unlocking-tool-positions}
\end{figure}

Water connections to the proximity shielding, and water and helium connections to the target, are made using remotely operated clamp connections based on Grayloc{\textregistered} connectors (Fig.~\ref{Fig:TC:18-grayloc-connector}). Electrical connections are made remotely by radiation-tolerant, custom designed connectors once the proximity shielding and target are lowered onto their supports. Clamping and unclamping for the water and helium connections to the target are made by an  ``(un)locking tool'' lowered into position by the crane (Fig.~\ref{Fig:TC:19-unlocking-tool} to \ref{Fig:TC:21-unlocking-tool-positions}).

\begin{figure}[!htb]
\centering
\includegraphics[width=0.7\linewidth]{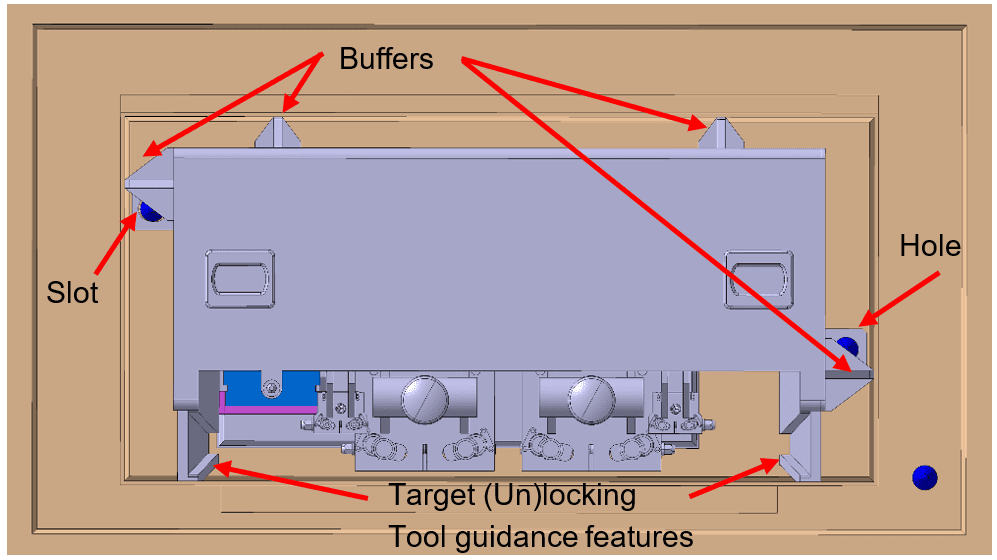}
\caption{The crane concept target container showing buffers for first stage (rough) guidance during installation, the hole and slot that engage with the guide pins for more precise installation guidance and the (un)locking tool guidance features}
\label{Fig:TC:22-crane-buffers}
\end{figure}

\begin{figure}[!htb]
\centering
\includegraphics[width=0.8\linewidth]{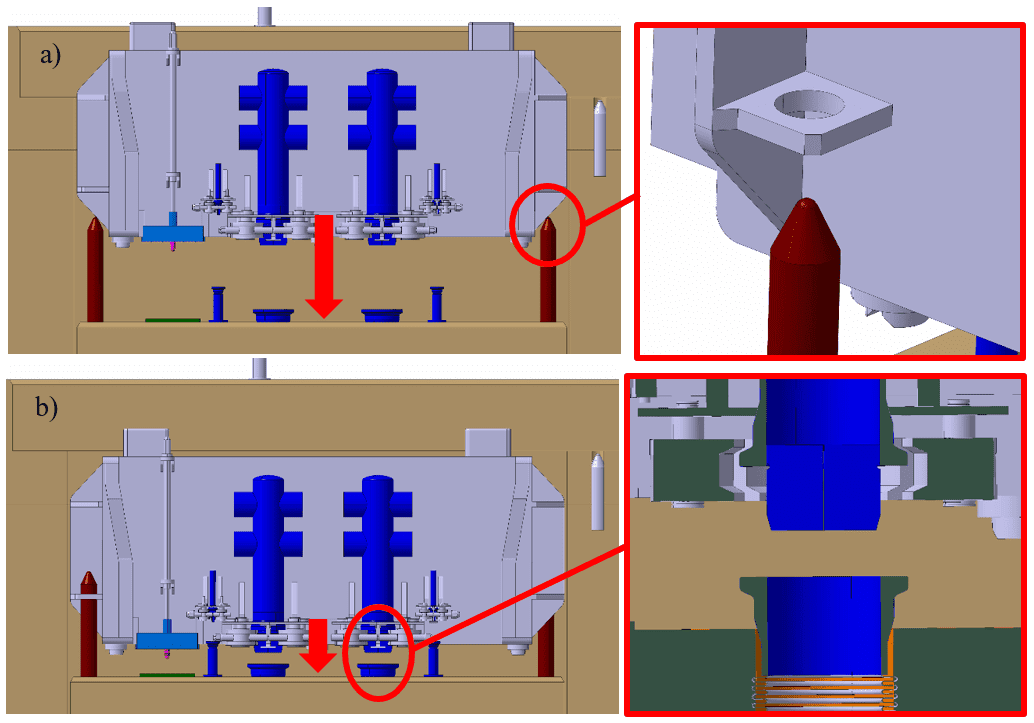}
\caption{Illustration of guidance stages during installation of target (crane concept). Upper images: As the target is lowered into position guide pins guide the target to align it with the water and electrical connections. Lower images: Guide fins in the target water connections align the fixed water connections (which are mounted on flexible bellows) as the target is lowered.}
\label{Fig:TC:23-guidance-stages}
\end{figure}

A two -- stage guidance arrangement is used to guide the target into position as it is lowered by the crane onto its supports. In the first stage, buffers on the target container (Fig.~\ref{Fig:TC:22-crane-buffers}) provide an initial rough alignment to allow the target to engage on the conical upper sections of the second stage guide pins. The guide pins ensure that the target lowers accurately in order to allow helium and water connections to be made. Water and helium pipe connections include bellows and internal guides to ensure correct alignment and compliance during installation and clamping operations (Fig.~\ref{Fig:TC:23-guidance-stages}).

\begin{figure}[!htb]
\centering
\includegraphics[width=0.7\linewidth]{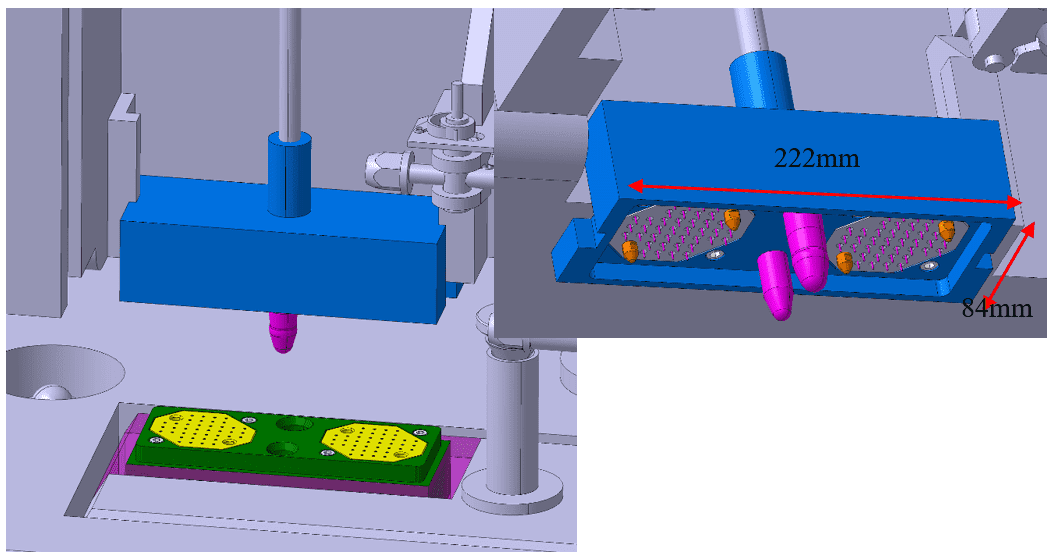}
\caption{Electrical connectors on the target and service pillar (crane concept).They are driven into their connected position using a leadscrew which is turned by a motor on the (un)locking tool.}
\label{Fig:TC:24-electrical-connectors}
\end{figure}

Electrical connections are not made as the target is lowered; once the target is in position the electrical connections are driven into contact using leadscrew drives mounted on the target container and remotely operated by the (un)locking tool (Figure \ref{Fig:TC:24-electrical-connectors}). The electrical connections are disconnected by reversing the leadscrew drive; however the target can still be extracted if the lead screw drive fails.

\begin{figure}[!htb]
\centering
\includegraphics[width=0.7\linewidth]{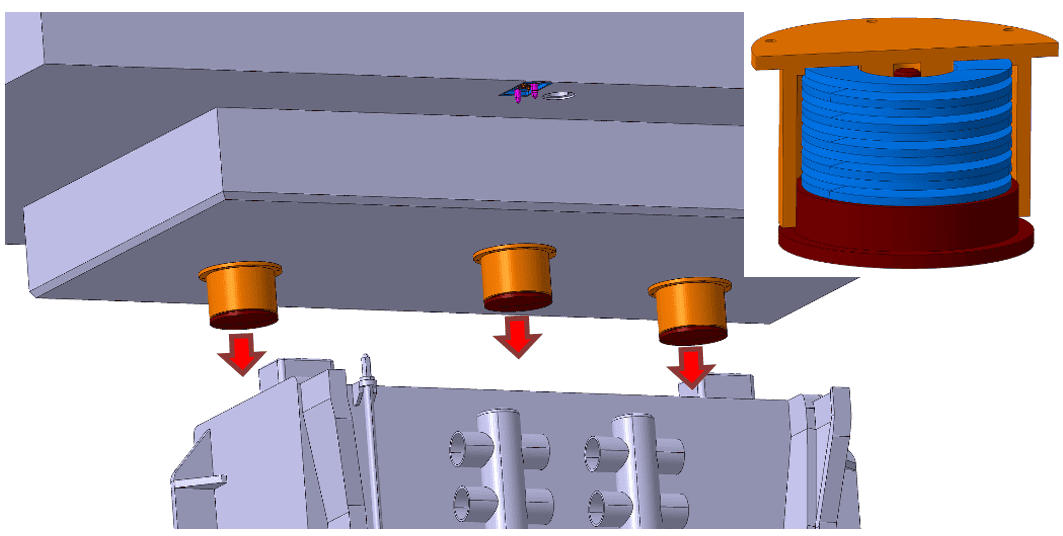}
\caption{Springs made up of Belleville washers attached to the underside of the upper layer of crane concept proximity shielding hold the target down on its supports}
\label{Fig:TC:25-belleville-springs}
\end{figure}

The crane concept target is held down onto its supports by springs (made up of Belleville washers) attached to the underside of the upper layer of proximity shielding (Fig. \ref{Fig:TC:25-belleville-springs})

\subsubsection{Target exchange sequence -- crane concept}
\label{Sec:TC:Target-exchange-crane}
\begin{figure}[!htb]
\centering
\includegraphics[width=0.7\linewidth]{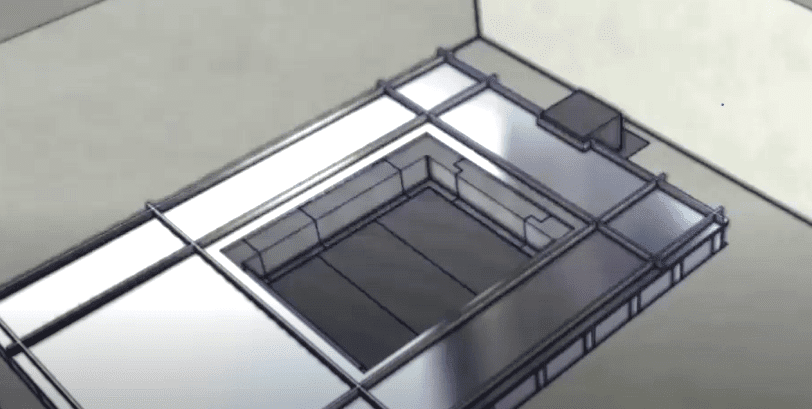}
\caption{Target exchange (crane concept): helium vessel with small lid and first layer of mobile shielding removed}
\label{Fig:TC:26-target-exchange-1}
\end{figure}

\begin{figure}[!htb]
\centering
\includegraphics[width=0.6\linewidth]{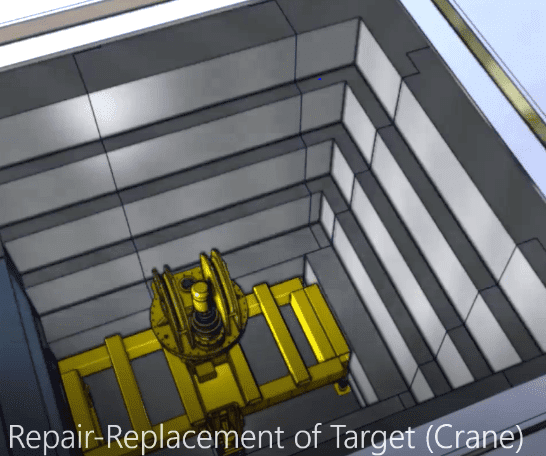}
\caption{Target exchange (crane concept): remotely operated spreader removing proximity shielding. Spreader equipped with (un)locking tools to disconnect electrical and water connections.}
\label{Fig:TC:27-target-exchange-2}
\end{figure}

In the crane concept all installation and removal operations are carried out vertically using the building's overhead travelling crane. In order to exchange the target, it is necessary to first remove the helium vessel small lid, the  ``mobile shielding'' and the top layer of proximity shielding (Figs.~\ref{Fig:TC:26-target-exchange-1} and \ref{Fig:TC:27-target-exchange-2}). 

A series of remotely operated spreader beams are used to lift the mobile and proximity shielding above the target and then the target itself. Disconnection of services must be carried out for the proximity shielding and the target before they can be lifted out.  For the target, disconnection is carried out remotely using the target (un)locking tool described in section~\ref{Sec:TC:Target-housing-supports-crane-concept}. For the proximity shielding disconnection, the (un)locking tools are incorporated into the remotely operated lifting spreaders; details of this are given in section~\ref{Sec:TC:proximity-shielding-crane}. The steps involved in target removal are given in Table~\ref{Tab:TC:target-removal-steps-crane}. The installation of a new target follows the same procedure but in reverse.

\begin{table}[!htb]
\begin{tabular}{|p{0.1\textwidth}|p{0.4\textwidth}|p{0.4\textwidth}|} \hline 
\centering
\textbf{Step} & \textbf{Task} & \textbf{Tooling} \\ \hline 
a & Open lid of helium vessel & Hands-on operation \\ \hline 
b & Remove mobile shielding above target and transfer to cool down area & Crane and remotely operated spreaders \\ \hline 
c & Disconnect water connections to top layer of proximity shielding & Crane and spreader with (un)locking tool \\ \hline 
d & Remove top layer of proximity shielding and transfer to cool-down area & Crane and spreader with (un)locking tool \\ \hline 
e & Disconnect water connections to target & Crane and (Un)locking tool \\ \hline 
f & Lift out target with shielded spreader and transfer to cool down area & Crane and target transfer shielded spreader  \\ \hline 
\end{tabular}
\caption{Steps for removal of a target from the helium vessel and transfer to the cool down area (crane concept).}
\label{Tab:TC:target-removal-steps-crane}
\end{table}

\begin{figure}[!htb]
\centering
\includegraphics[width=0.6\linewidth]{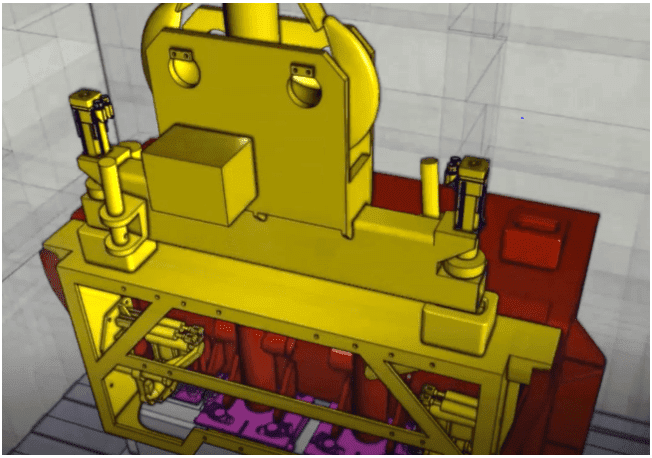}
\caption{Target exchange (crane concept): (un)locking tool lowered into place by crane to open target connections.}
\label{Fig:TC:28-target-exchange-3}
\end{figure}

\begin{figure}[!htb]
\centering
\includegraphics[width=0.6\linewidth]{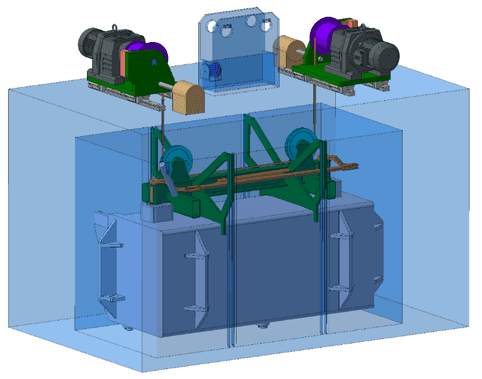}
\caption{Target exchange (crane concept): transfer spreader for shielded transfer of target from helium vessel to cool down area. Two hoists are used for redundancy so that operations can continue if one hoist fails. Target shown inside shielding.}
\label{Fig:TC:29-target-exchange-4}
\end{figure}

A shielded spreader is used for the target lifting and transfers, it is equipped with a hoist system (with remotely operated twist-lock attachments) to lift the target inside the shielding before the target is transferred to the cool down area. This transfer is carried out with the spreader kept as close as possible to the floor to minimise external radiation shine from the bottom of the target. Additional concrete shielding blocks may be installed along the transfer path to further reduce radiation levels in the target hall during transfers. The crane concept target transfer spreader is shown in Fig. \ref{Fig:TC:29-target-exchange-4}.

\subsubsection{The target housing, supports, services and connections -- trolley concept}
\label{Sec:TC:Target-housing-trolley}

\begin{figure}[!htb]
\centering
\includegraphics[width=0.7\linewidth]{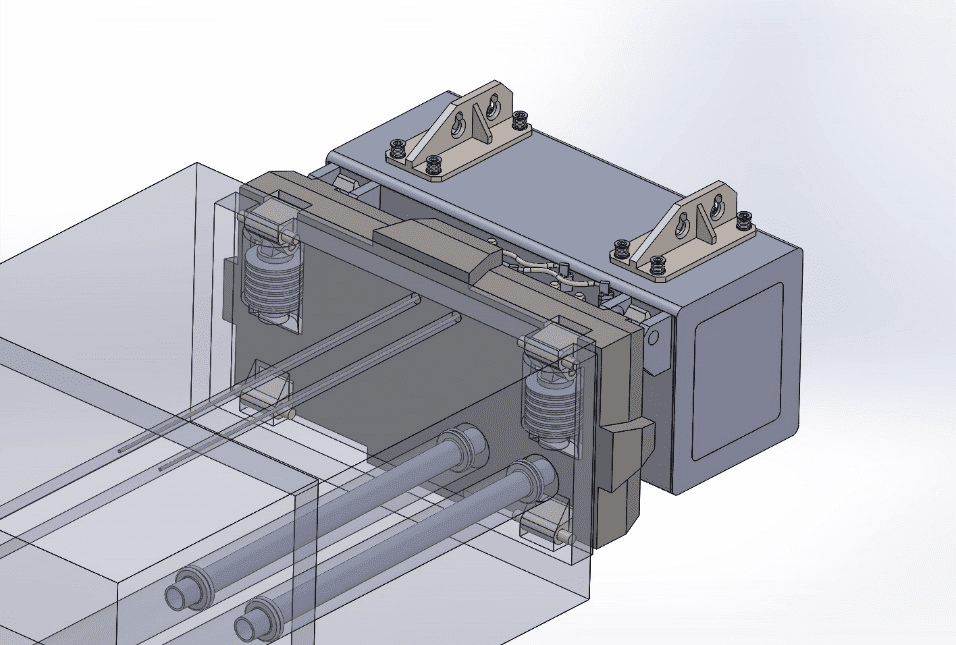}
\caption{Target supported on nose of trolley. Coolant and helium pipes pass through the shielding to the connectors on the side of the target housing.}
\label{Fig:TC:30-target-supported-nose}
\end{figure}

The trolley concept target is housed in a rectangular stainless-steel container which incorporates the support interfaces used to attach it to the front of the trolley, lift attachment points, water, helium and electrical connectors (Fig. \ref{Fig:TC:30-target-supported-nose}). Internal pipework connects the target cooling channels to the remotely operated connector clamps on the side of the target container. The design of the internal pipework were outside the scope of the handling study but was covered in the target design work (~\ref{Chap:Target}).

\begin{figure}[!htb]
\centering
\includegraphics[width=0.9\linewidth]{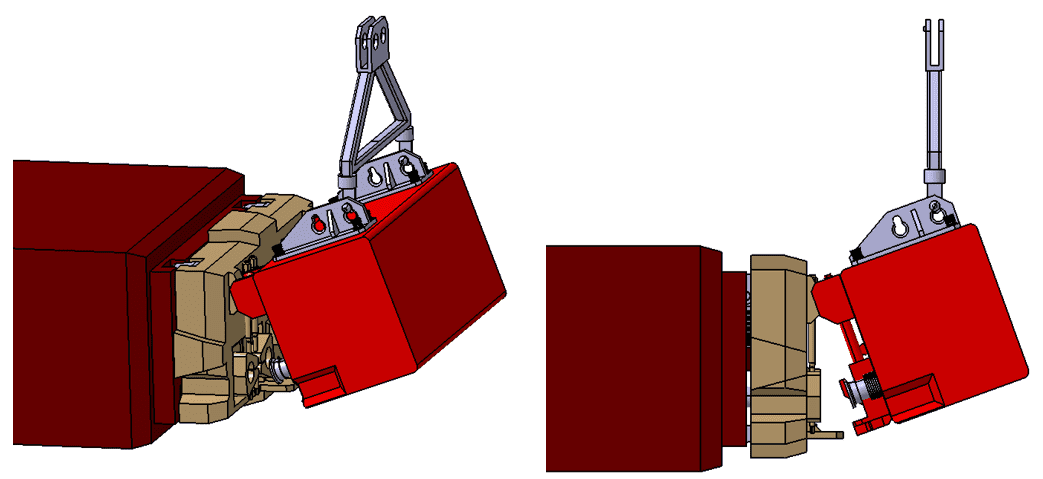}
\caption{Target being lowered onto support hooks at the end of the trolley. Once the target is engaged on the hooks, further lowering rotates the target so that the water connections on the side of the target engage with the screw operated connection clamps on the nose of the trolley. The screw clamps are operated by the hot-cell master-slave manipulator using appropriate tooling.}
\label{Fig:TC:31-target-hooked}
\end{figure}

\begin{figure}[!htb]
\centering
\includegraphics[width=0.9\linewidth]{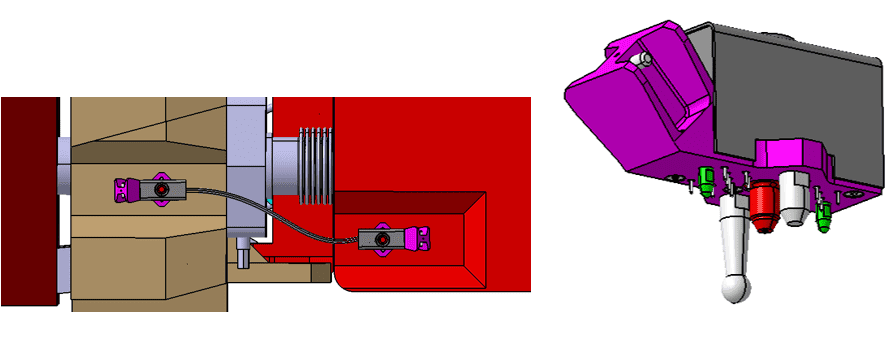}
\caption{Electrical connections to trolley concept target: Left -- leads with connectors at each end are used for the connections. Right -- electrical connectors are designed to be compatible with remote handling by the master-slave manipulators in the hot cell.}
\label{Fig:TC:32-electrical-connections}
\end{figure}

\begin{figure}[!htb]
\centering
\includegraphics[width=0.7\linewidth]{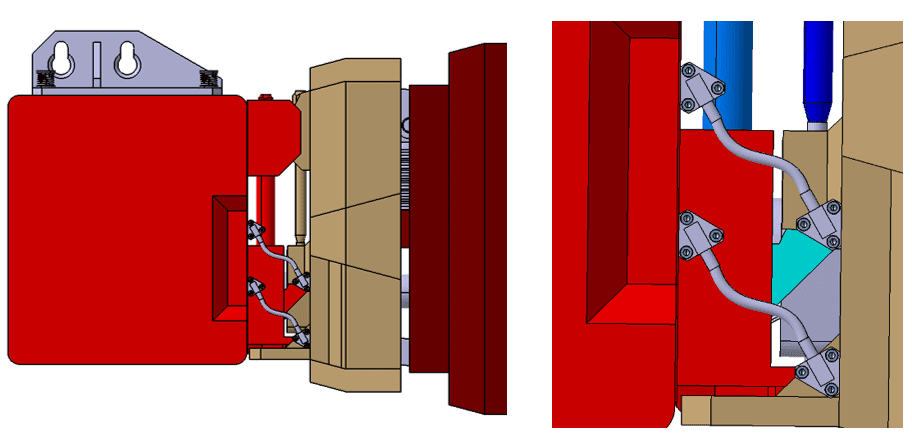}
\caption{Helium connections to the trolley concept target; the connections are screwed into position using the hot-cell master-slave manipulators}
\label{Fig:TC:33-helium-connection}
\end{figure}

The target is supported on hooks at the  ``nose'' end of the trolley. It is lowered onto the hooks by the hot cell crane and guided into position by master-slave manipulators (Fig. \ref{Fig:TC:31-target-hooked}). It is secured by a screw-driven clamp operated by a power tool that is positioned using the hot-cell master-slave manipulators. All the service connections to the target arrive at the nose of the trolley. Once the target is fully lowered and the water connection pipes are engaged, the screw clamp water connections are tightened using the hot-cell manipulators. Electrical and helium connections are also made using the hot-cell manipulators (Figs.~\ref{Fig:TC:32-electrical-connections} and \ref{Fig:TC:33-helium-connection}). 

\subsubsection{Target exchange -- trolley concept}
\label{Sec:TC:target-exchange-trolley}

\begin{figure}[!htb]
\centering
\includegraphics[width=0.7\linewidth]{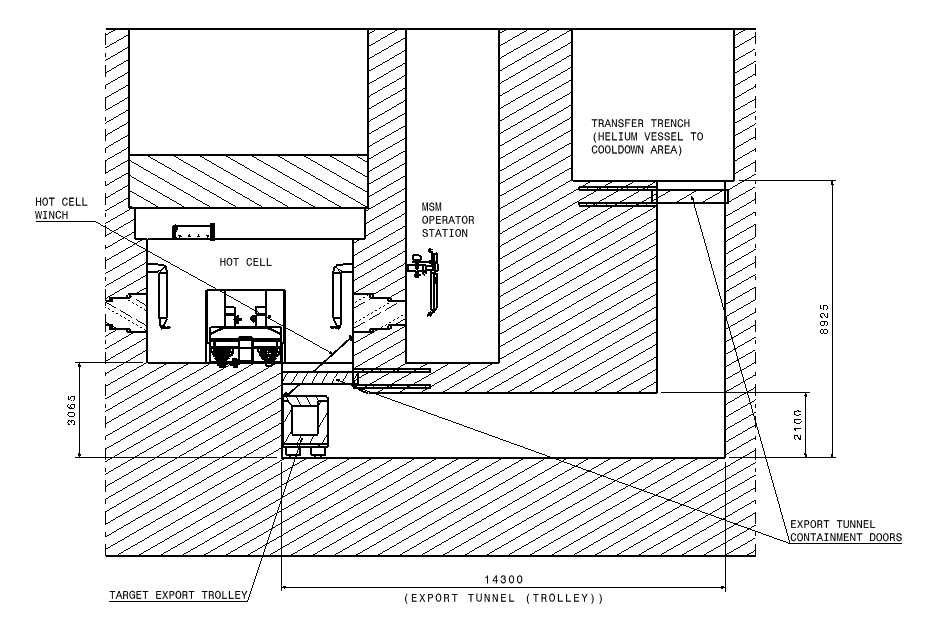}
\caption{Section through the trolley concept underground area showing the hot cell and target export trolley and the export tunnel linking the hot cell to the cool-down area.}
\label{Fig:TC:34-section-trolley-concept}
\end{figure}

\begin{figure}[!htb]
\centering
\includegraphics[width=0.7\linewidth]{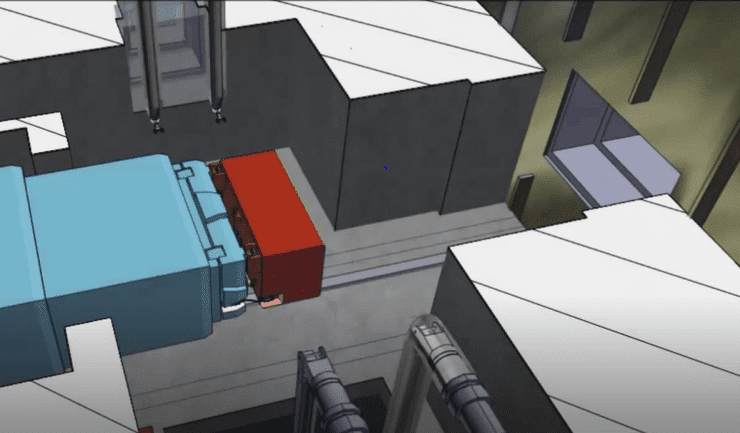}
\caption{Target exchange (trolley concept): withdrawal of target on end of trolley from helium vessel into hot cell.}
\label{Fig:TC:35-target-exchange-5b}
\end{figure}

\begin{figure}[!htb]
\centering
\includegraphics[width=0.6\linewidth]{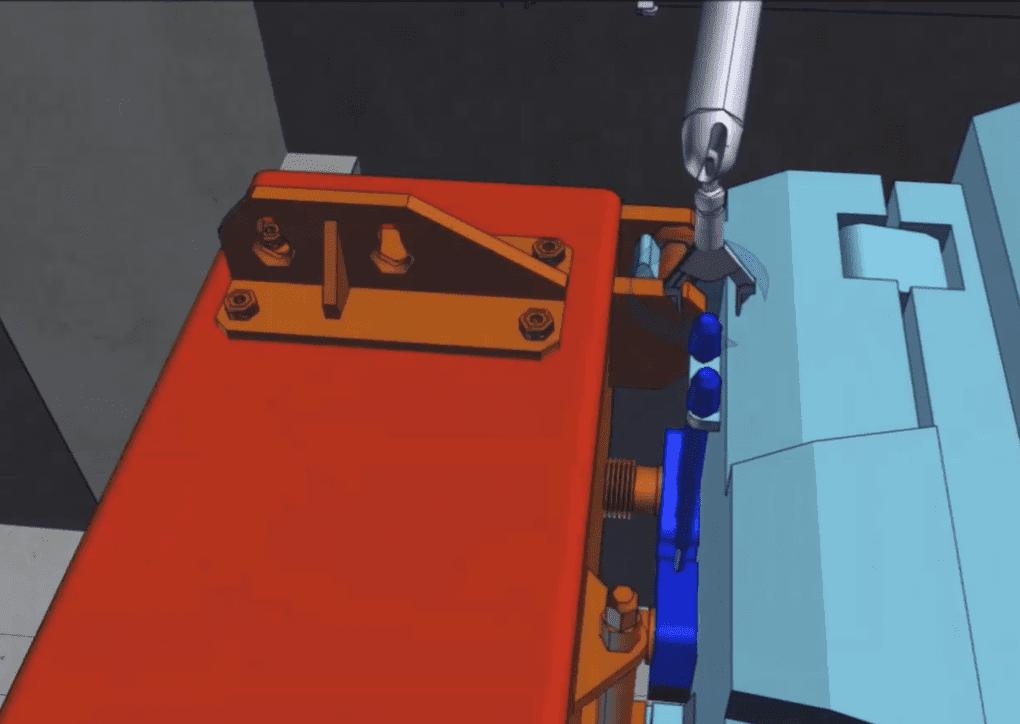}
\caption{Target Exchange (trolley concept): screw driven pipe clamp disconnection using Master-slave manipulator - drive shafts used to provide an accessible rotation drive point. The master slave manipulator would be used to handle a power tool (not shown) to undo the pipe clamps.}
\label{Fig:TC:36-target-exchange-6b}
\end{figure}

\begin{figure}[!htb]
\centering
\includegraphics[width=0.55\linewidth]{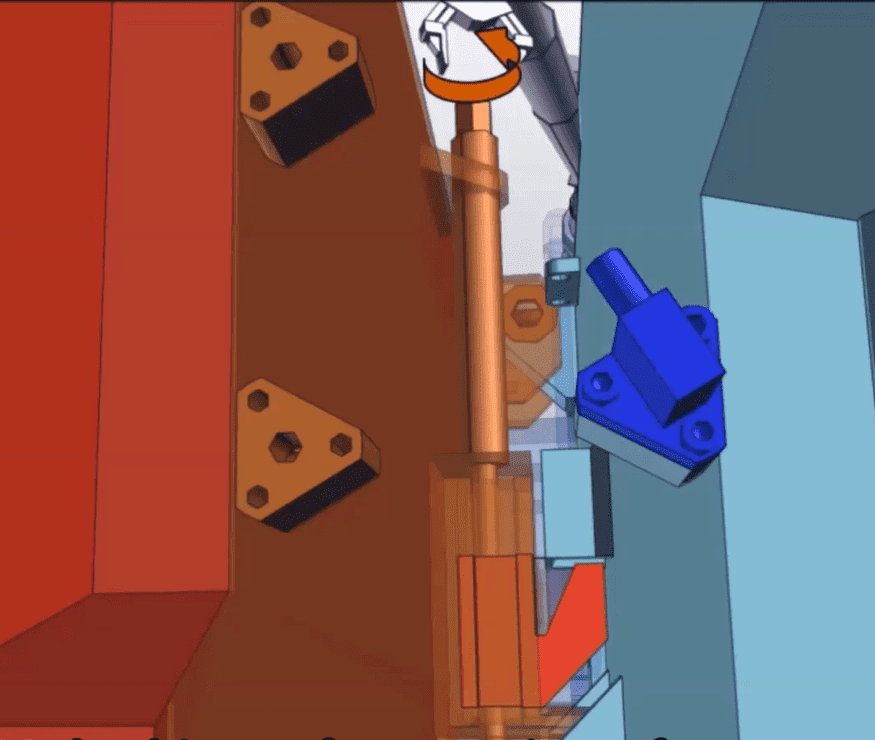}
\caption{Target Exchange (trolley concept): release of screw driven target securing clamp (downward movement to release) using Master-slave manipulator}
\label{Fig:TC:37-target-exchange-7b}
\end{figure}

\begin{figure}[!htb]
\centering
\includegraphics[width=0.55\linewidth]{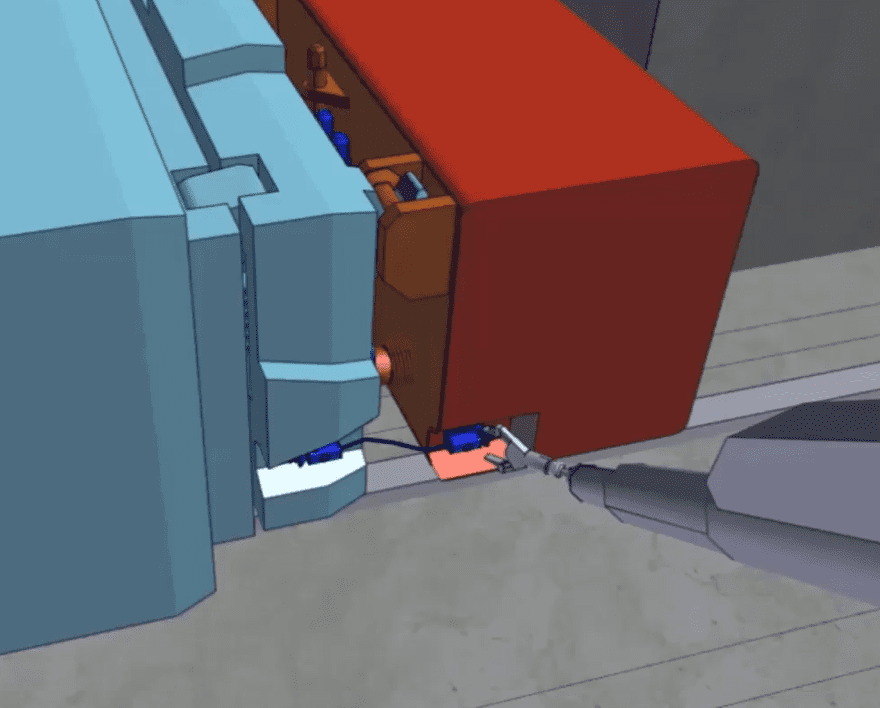}
\caption{Target exchange (trolley concept) Disconnection of target electrical connections using hot-cell master slave manipulators}
\label{Fig:TC:38-target-exchange-8b}
\end{figure}

\begin{figure}[!htb]
\centering
\includegraphics[width=0.55\linewidth]{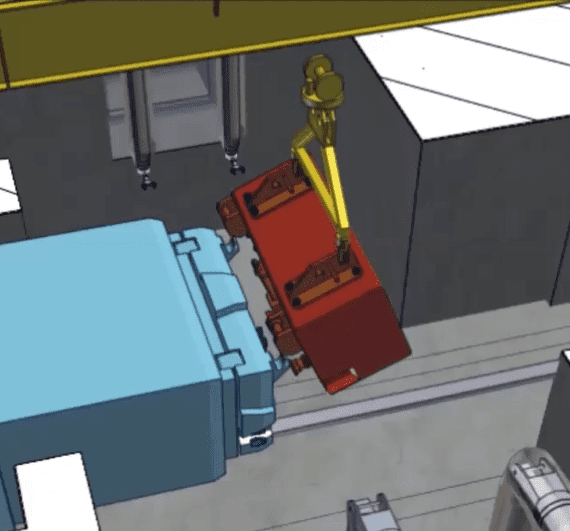}
\caption{Target exchange (trolley concept): Lifting target off trolley support hooks using hot cell crane and spreader beam}
\label{Fig:TC:39-target-exchange-9b}
\end{figure}

\begin{figure}[!htb]
\centering
\includegraphics[width=0.55\linewidth]{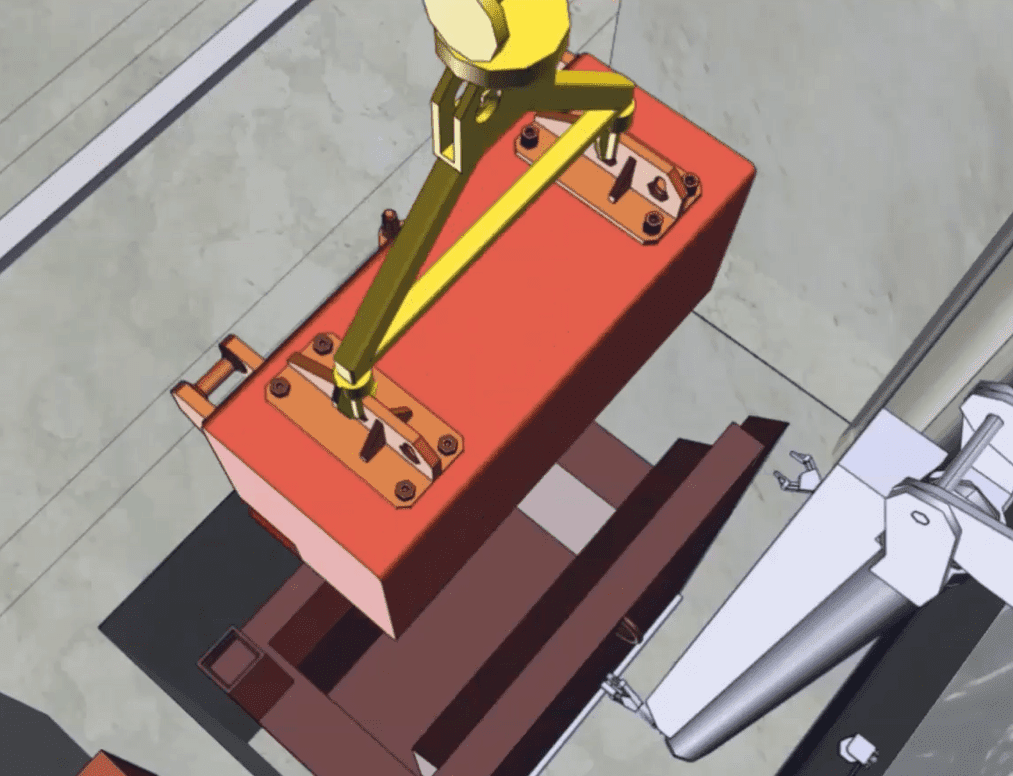}
\caption{Target exchange (trolley concept): Lowering target into export trolley cask using hot cell crane and manipulators}
\label{Fig:TC:40-target-exchange-10b}
\end{figure}

\begin{figure}[!htb]
\centering
\includegraphics[width=0.55\linewidth]{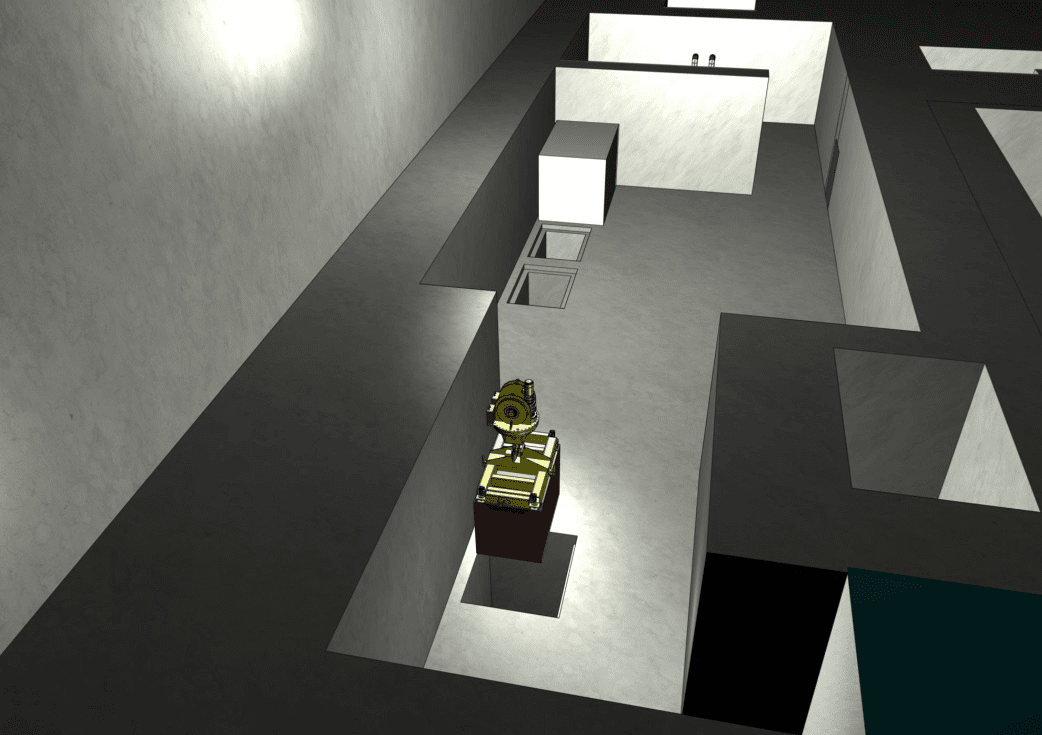}
\caption{Target exchange (trolley concept): Moving target inside export trolley cask inside cooldown area using building crane and remotely operated spreader}
\label{Fig:TC:41-target-exchange-11b}
\end{figure}

\begin{figure}[!htb]
\centering
\includegraphics[width=0.55\linewidth]{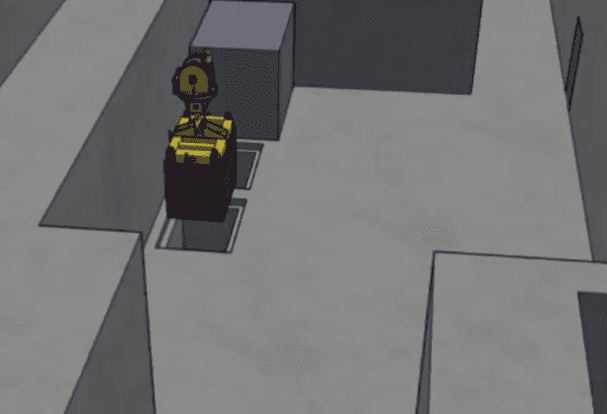}
\caption{Target exchange (trolley concept): Moving target inside export trolley cask inside cooldown area using building crane and remotely operated spreader}
\label{Fig:TC:42-target-exchange-12b}
\end{figure}

Target exchange is carried out by driving the trolley backwards to withdraw it from the helium vessel and into the hot cell. After disconnection of services in the hot cell and release of the securing clamp using the master-slave manipulators, the target is lifted off the nose of the trolley and placed in a shielded target export trolley using the hot cell crane. The target is then transferred to the cool-down area in the target export trolley that runs along the export tunnel under the hot cell (Fig. \ref{Fig:TC:34-section-trolley-concept}). Installation of a new target is a reversal of the removal procedure. The steps involved are summarised in Table \ref{Tab:TC:target-removal-steps-trolley} and illustrated in Figs \ref{Fig:TC:35-target-exchange-5b} to \ref{Fig:TC:42-target-exchange-12b}.

\begin{table}[!htb]
\begin{tabular}{|p{0.1\textwidth}|p{0.4\textwidth}|p{0.4\textwidth}|} \hline 
\centering
\textbf{Step} & \textbf{Task} & \textbf{Tooling} \\ \hline 
a & Withdraw target on end of trolley from helium vessel into hot cell & Trolley \\ \hline 
b & Disconnect helium, water and electrical connections, release target securing clamp & Hot-cell manipulators and hot-cell crane. \\ \hline 
c & Attach lift attachments to target, remove target from trolley nose & Hot-cell manipulators and hot-cell crane. \\ \hline 
d & Lower target into export trolley and close lid & Hot-cell manipulators, in-cell winch, hot-cell crane. \\ \hline 
e & Transfer target to cool down area and put into storage pit & Export trolley and cask, building crane \\ \hline 
\end{tabular}
\caption{Main steps for removal of a target from the helium vessel and transfer to the cool down area (trolley concept).}
\label{Tab:TC:target-removal-steps-trolley}
\end{table}

\subsection{Design of the target proximity shielding}
\label{Sec:TC:design-target-proximity-shielding}

\subsubsection{Thermo-mechanical performance}
\label{Sec:TC:thermo-mechanical-performance}

The thermo-mechanical performance results presented here are based on the trolley concept proximity shielding design. The trolley concept proximity shielding is in two main parts: the fixed proximity shielding - made up of blocks that are permanently installed in the helium vessel - and the trolley-mounted proximity shielding block.

\subsubsubsection{Fixed proximity shielding}
\label{Sec:TC:fixed-proximity-shielding}

The primary beam interaction with the BDF target produces a shower of secondary particles that interact with the proximity shielding. The energy deposited on the proximity shielding is not negligible, and will be dissipated via active cooling. Fig. 1 shows the energy deposition map in the proximity shielding for the trolley concept obtained via FLUKA simulations. The total power deposited in the shielding blocks is around 12 kW.

\begin{figure}[!htb]
\centering
\includegraphics[width=0.75\linewidth]{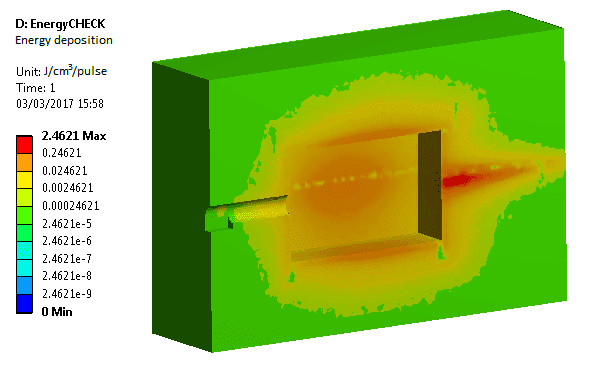}
\caption{Energy deposition distribution in a transverse cut of the BDF proximity shielding blocks (logarithmic scale)}
\label{Fig:TC:43-energy-deposition-BDF}
\end{figure}

The heat dissipation in the proximity shielding is performed by the forced circulation of cooling water. The water flows through stainless steel pipes that are embedded in the different cast iron blocks. In the trolley concept design, the water is brought to the shielding blocks through the service chimneys. The cooling system design is based on the SPS Internal Dump (TIDVG) shielding cooling principle, where 10 mm thick stainless steel pipes are embedded in the cast iron shielding of the dump~\cite{TIDVG4}. The preliminary water circuit design consists of long pipes, with internal diameter of 16 mm and outside diameter of 36 mm, that enter and exit the shielding blocks from the top. The internal and external diameter dimensions are equivalent to those of the SPS TIDVG shielding cooling, given the proven compatibility of the 10 mm thick pipe walls with the casting process.

The mass flow rate foreseen for each block water circuit is 0.25 kg/s, leading to a total mass flow rate of 1 kg/s at the water supply level. The water inlet temperature considered is 28$\mathrm{{}^\circ}$C, and the maximum temperature increase expected is 5$\mathrm{{}^\circ}$C. The calculated water velocity in the cooling pipes is around 1.2 m/s - well below the design limit of 2 m/s, applied to avoid damage to the pipes or fittings. 

\begin{figure}[!htb]
\centering
\includegraphics[width=0.6\linewidth]{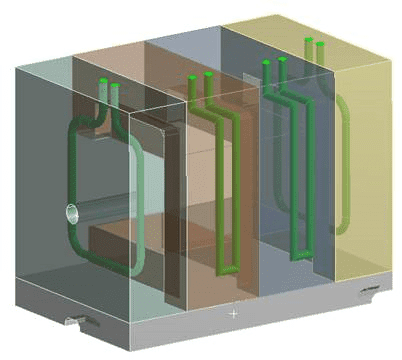}
\caption{3D Model of the proximity shielding blocks with embedded stainless steel pipes used for FEM calculations.}
\label{Fig:TC:44-3D-model-FEM}
\end{figure}

Thermal calculations have been performed to evaluate the maximum temperatures reached in the proximity shielding. The heat loads are given by the beam energy deposition in the shielding and the heat is mainly dissipated by the water cooling system. The heat removal by the cooling system is evaluated in two steps: first, the heat transfer coefficient from the water to the stainless steel pipes is obtained. Then, it is necessary to calculate the conduction through the stainless steel pipes that are in contact with the cast iron blocks. It is assumed that 50\% of the outer surface of the stainless steel pipes is in perfect contact with the surrounding cast iron, as measured in a similar experimental setup via ultrasound testing~\cite{TIDVG4_UT}. The 3D model used for the thermal calculations is presented in Fig.~\ref{Fig:TC:44-3D-model-FEM}.

\begin{figure}[!htb]
\centering
\includegraphics[width=0.7\linewidth]{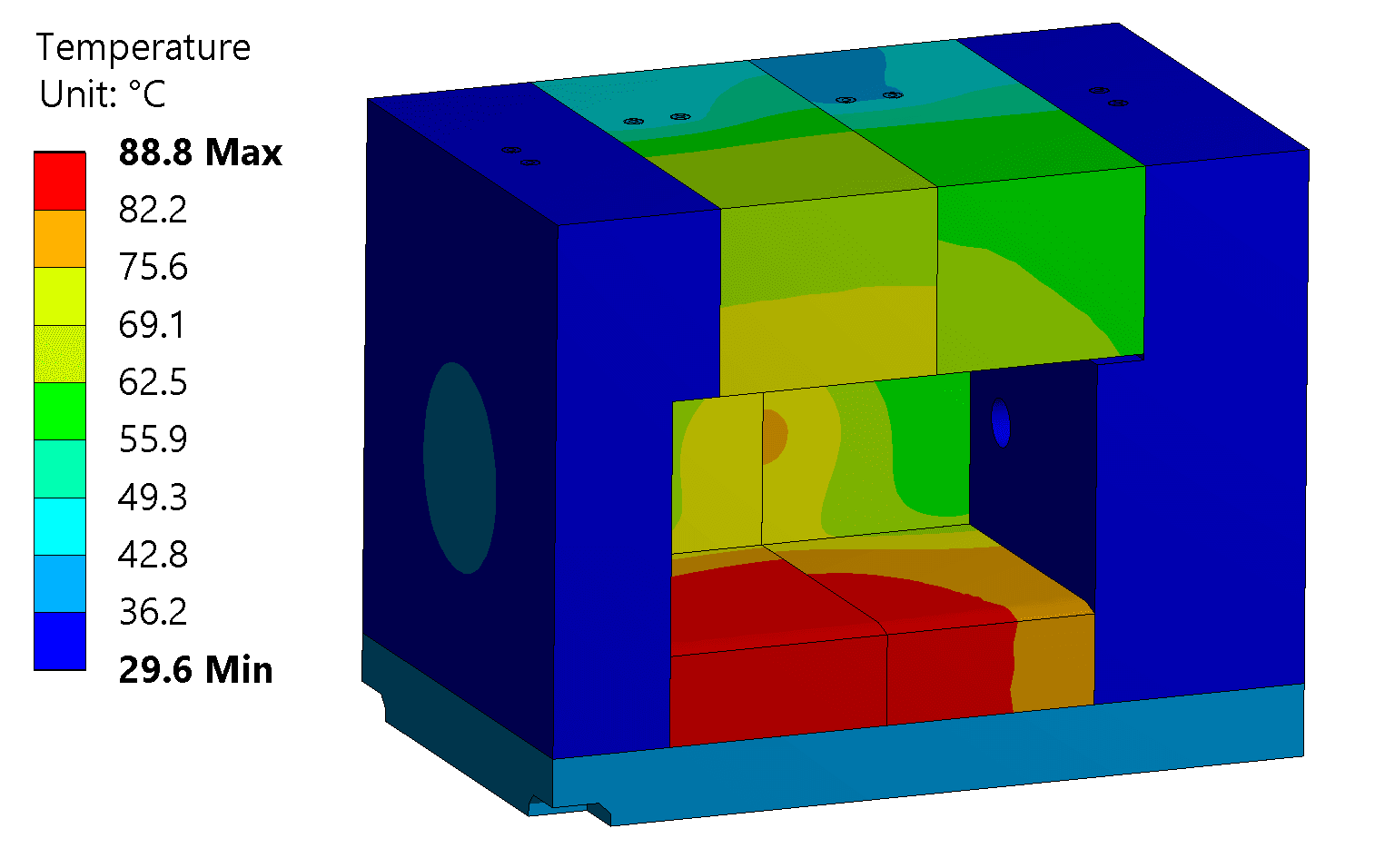}
\caption{Steady-state temperature distribution in the proximity shielding blocks. The maximum temperature reached is around 90$\mathrm{{}^\circ}$C}
\label{Fig:TC:45-steady-state}
\end{figure}

Fig.~\ref{Fig:TC:45-steady-state} shows the steady-state temperature distribution obtained in the proximity shielding blocks. As a conservative approach, it has been assumed that no heat conduction takes place from one block to the other. The maximum temperature reached is around 90$\mathrm{{}^\circ}$C, well below the maximum service temperature of cast iron which is between 230 and 300$\mathrm{{}^\circ}$C.

\begin{figure}[!htb]
\centering
\includegraphics[width=0.8\linewidth]{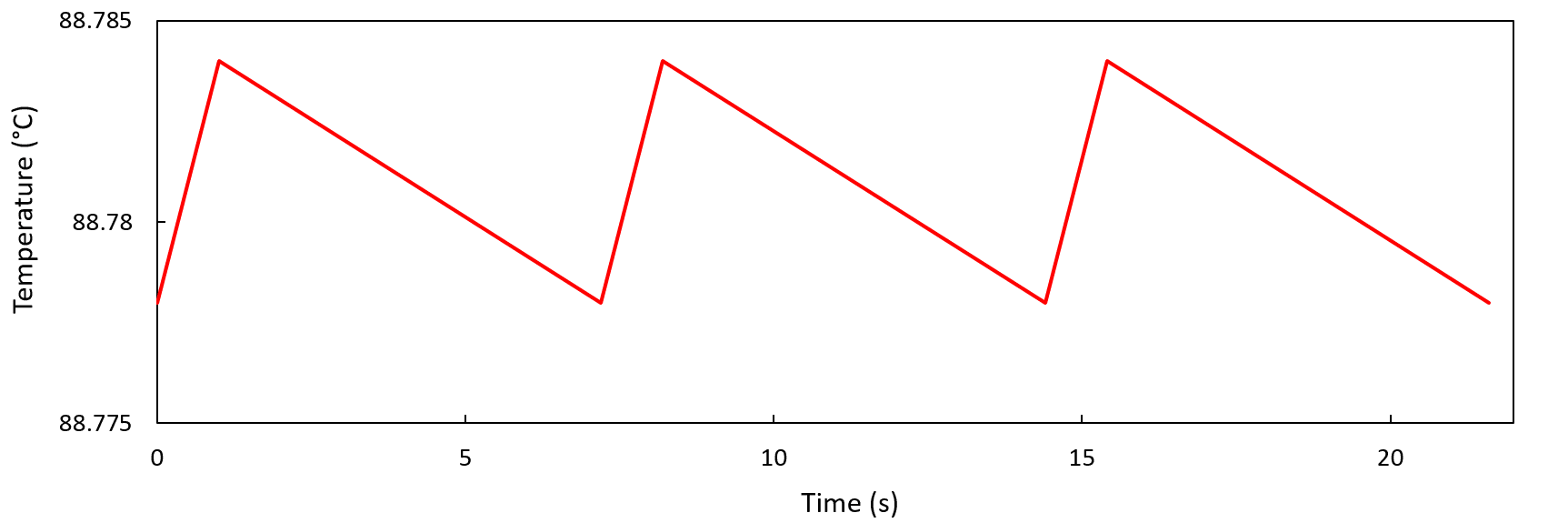}
\caption{Transient evolution of the temperature in the location of maximum temperature during 3 pulses after a long period of operation. The temperature increase is below 0.1$\mathrm{{}^\circ}$C.}
\label{Fig:TC:46-steady-state-curve}
\end{figure}

Additionally, transient thermal calculations have been carried out, in order to evaluate the temperature increase during the beam pulse impact on target. Fig.~\ref{Fig:TC:46-steady-state-curve} shows the maximum temperature evolution in the most thermally loaded shielding block. It can be observed that the maximum temperature increase with every pulse is below 0.1$\mathrm{{}^\circ}$C, and is therefore negligible.

\subsubsubsection{Trolley-mounted proximity shielding block}
\label{Sec:TC:trolley-mounted-shielding}

\begin{figure}[!htb]
\centering
\includegraphics[width=0.7\linewidth]{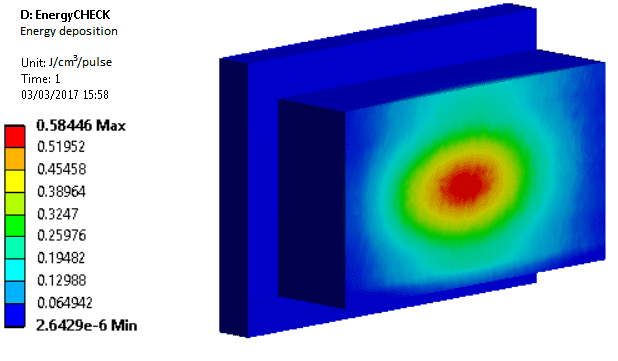}
\caption{Energy deposition distribution in the trolley shielding block}
\label{Fig:TC:47-energy-deposition-trolley-shielding}
\end{figure}

The shielding block mounted on the target trolley also requires active water cooling, given the level of energy deposition in the block. The total heat deposited reaches 6 kW during beam operation. Fig. \ref{Fig:TC:47-energy-deposition-trolley-shielding} displays the energy deposition map in the trolley shielding block.

The cooling system design is based on the same principle as for the rest of the proximity shielding blocks, with the exception that the water pipes are routed to the side, in the direction of the services that lay on the trolley.

\begin{figure}[!htb]
\centering
\includegraphics[width=0.5\linewidth]{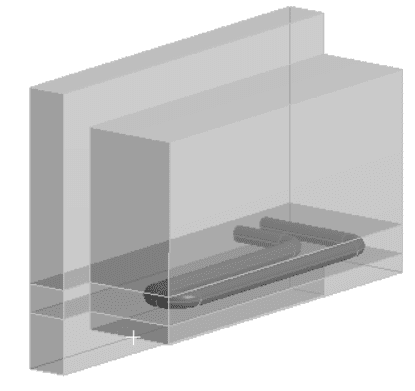}
\caption{3D Model of the trolley shielding block with embedded stainless steel pipe used for FEM calculations}
\label{Fig:TC:47b-3D-model-trolley-shielding}
\end{figure}

The mass flow rate foreseen for the trolley block water circuit is 0.3 kg/s. The water inlet temperature considered is 28$\mathrm{{}^\circ}$C, with an expected temperature increase of 5$\mathrm{{}^\circ}$C from inlet to outlet. The calculated water velocity in the cooling pipes is around 1.4 m/s, also below the design limit of 2 m/s. The 3D model used for the FEM simulations along with the pipeline design is shown in Fig. \ref{Fig:TC:47b-3D-model-trolley-shielding}.

The boundary conditions considered for the thermal analysis are equivalent to the rest of shielding blocks. Fig. \ref{Fig:TC:49-crane-concept-shielding} shows the steady-state temperature distribution in the trolley shielding block. The maximum temperature reached is 73$\mathrm{{}^\circ}$C, well within the safety limits of the material. The transient thermal simulation performed have shown that the maximum increase of temperature during every beam pulse is negligible.

\begin{figure}[!htb]
\centering
\includegraphics[width=0.7\linewidth]{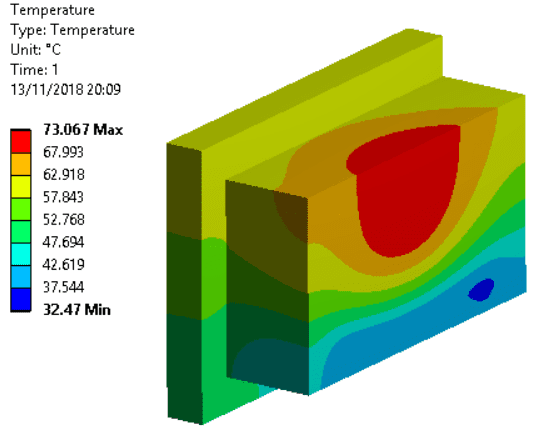}
\caption{Steady-state temperature distribution in the trolley shielding block. The maximum temperature reached is around 70$\mathrm{{}^\circ}$C.}
\label{Fig:TC:48-steady-state-trolley-shielding}
\end{figure}

\subsubsection{Proximity shielding handling -- crane concept}
\label{Sec:TC:proximity-shielding-crane}

\begin{figure}[!htb]
\centering
\includegraphics[width=0.6\linewidth]{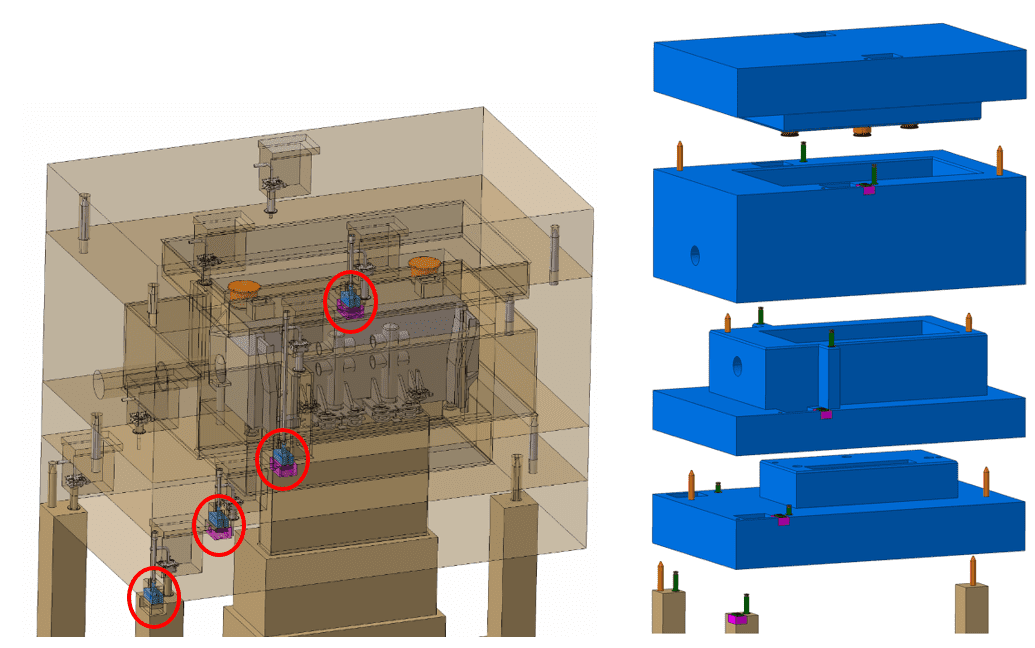}
\caption{Crane concept proximity shielding. On the left - the layers of proximity shielding (crane concept) when assembled with the target in the middle -- electrical connectors are circled in red. On the right - the proximity shielding layers are separated showing alignment pins and cooling pipework stubs.}
\label{Fig:TC:49-crane-concept-shielding}
\end{figure}

The crane concept proximity shielding is built up of layers of cast iron with internal cast-in stainless steel thick-wall pipes to provide cooling water passages (Fig.~\ref{Fig:TC:49-crane-concept-shielding}). The proximity shielding layers are installed on top of support pillars which provide the cooling and electrical services to the proximity shielding. The coolant pipes are connected using remotely operated screw clamps as used for the target; the connections are made once each layer has been installed on top of the previous layer. Two (un)locking tools, similar to the ones used for the target, are mounted on the lifting spreader and used to operate the clamps (Figs.~\ref{Fig:TC:51-crance-shileding-unlocking} and~\ref{Fig:TC:52-shielding-bolting-tool}).

\begin{figure}[!htb]
\centering
\includegraphics[width=0.6\linewidth]{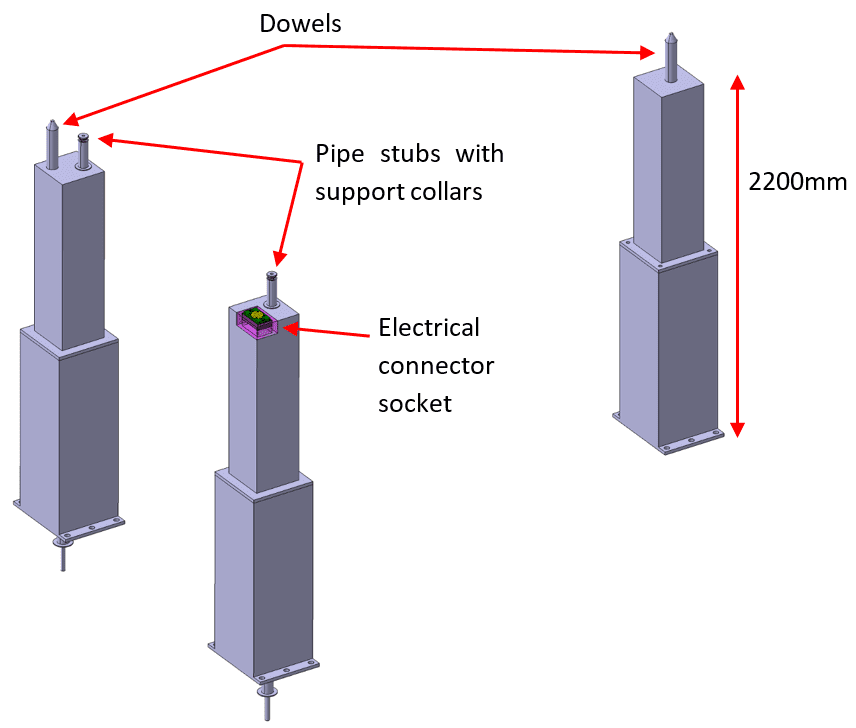}
\caption{Crane concept proximity shielding support pillars with water cooling and electrical connections}
\label{Fig:TC:50-crance-shileding-pillars}
\end{figure}

\begin{figure}[!htb]
\centering
\includegraphics[width=0.6\linewidth]{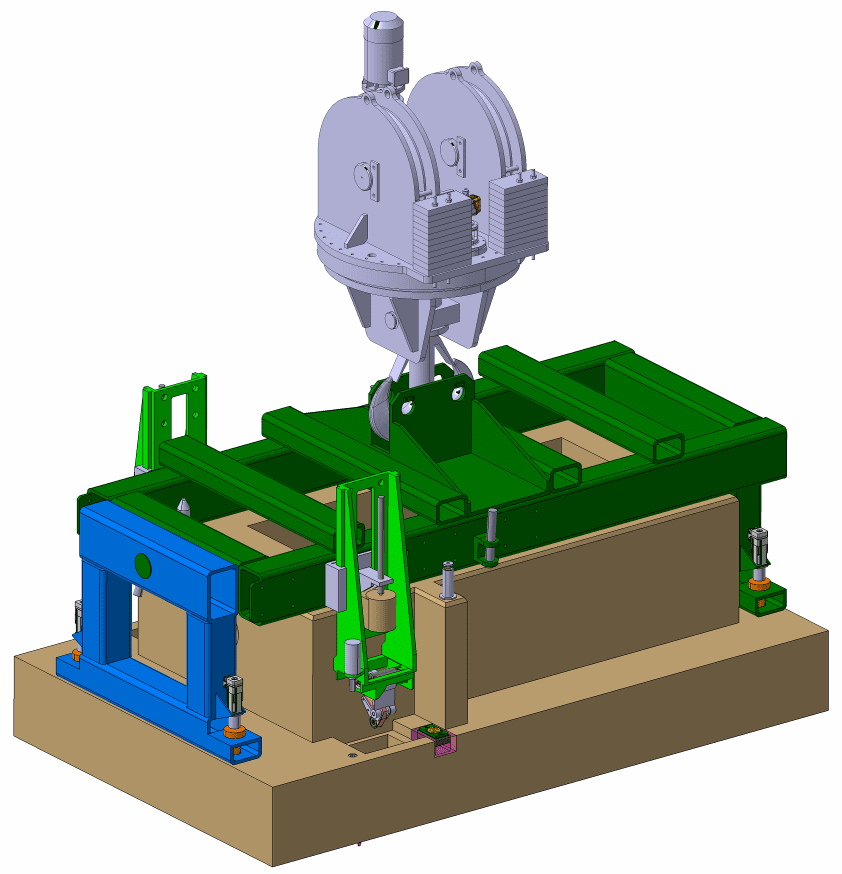}
\caption{Crane concept proximity shielding spreader beam equipped with two (un)locking tools -- shown lifting proximity shield block 3. The unlocking tools can be re-positioned on the spreader to suit the connection positions on the different layers of the proximity shielding}
\label{Fig:TC:51-crance-shileding-unlocking}
\end{figure}

\begin{figure}[!htb]
\centering
\includegraphics[width=0.5\linewidth]{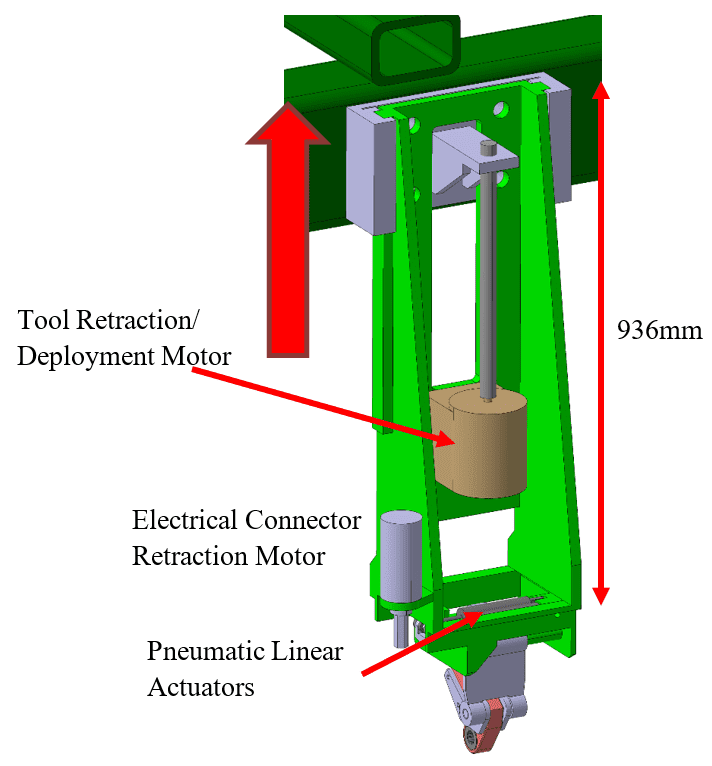}
\caption{Crane concept proximity shielding bolting tool used to remotely loosen or tighten water cooling pipe clamps}
\label{Fig:TC:52-shielding-bolting-tool}
\end{figure}

\subsubsection{Proximity shielding handling -- trolley concept}
\label{Sec:TC:proximity-shielding-trolley}

\begin{figure}[!htb]
\centering
\includegraphics[width=0.7\linewidth]{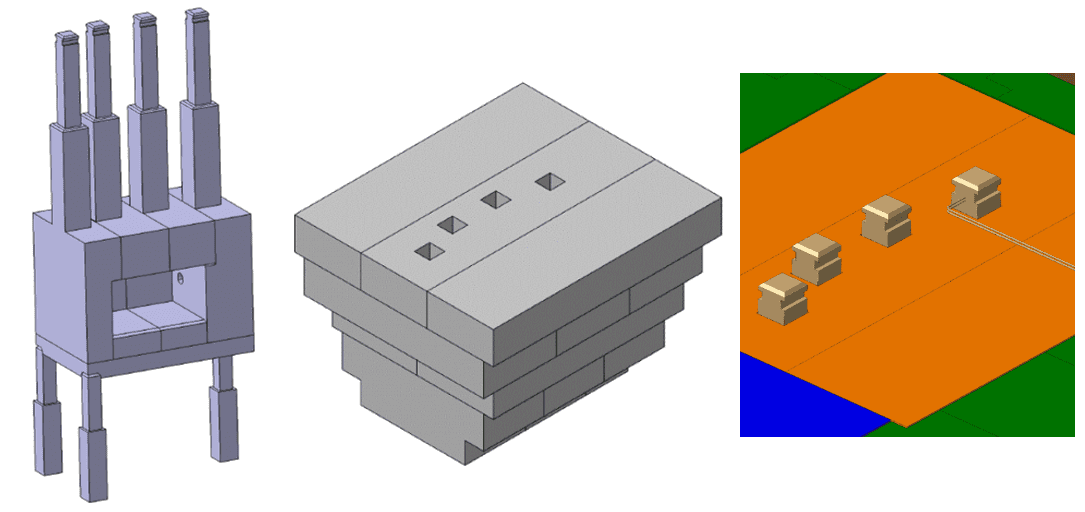}
\caption{Proximity shielding for trolley concept. On the left - the four blocks with their service  ``chimneys'' installed on a flat plate supported by three pillars. Note that the service chimneys incorporate steps to avoid direct radiation shine paths. In the middle -- the mobile shielding above the proximity shielding with passages for the service chimneys. On the right -- the tops of the proximity shielding service chimneys protruding above the mobile shielding with service connections shown on the right-hand chimney.}
\label{Fig:TC:53-proximity-shielding-crane}
\end{figure}

\begin{figure}[!htb]
\centering
\includegraphics[width=0.9\linewidth]{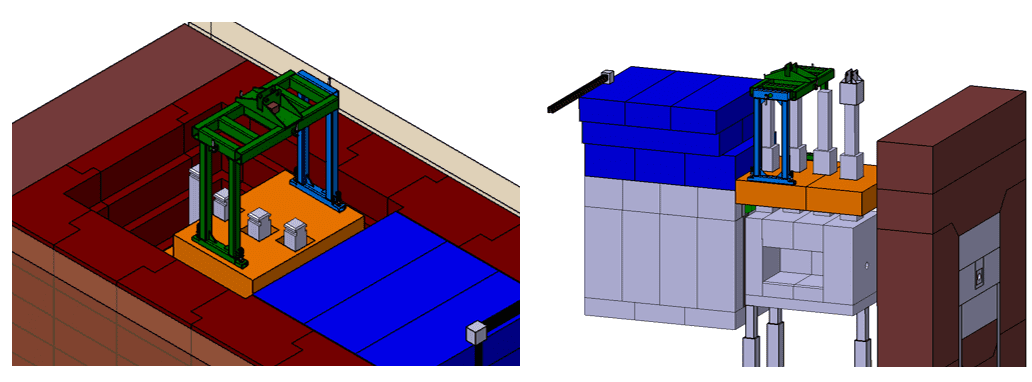}
\caption{Removing mobile shielding from above the trolley concept proximity shielding: the. remotely operated spreader has  ``legs'' in order to be able to reach down to the lower levels of mobile shielding while the main structure of the spreader is above the top of the service chimneys}
\label{Fig:TC:54-removing-mobile-shielding}
\end{figure}

\begin{figure}[!htb]
\centering
\includegraphics[width=0.7\linewidth]{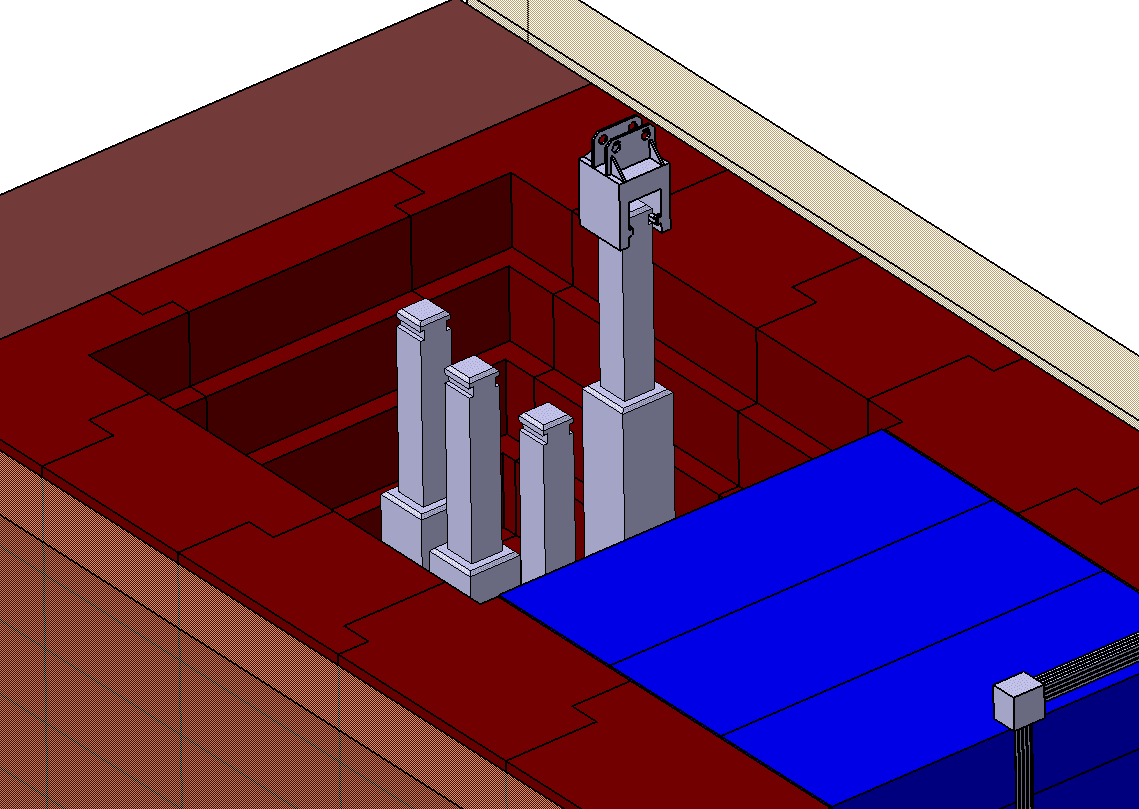}
\caption{Lifting the trolley concept proximity shielding: a lifting attachment fitted to the crane hook engages with lifting features at the top of the service chimneys. Lifting attachment shown engaged on right hand service chimney}
\label{Fig:TC:55-lifting-trolley}
\end{figure}

\begin{figure}[!htb]
\centering
\includegraphics[width=0.8\linewidth]{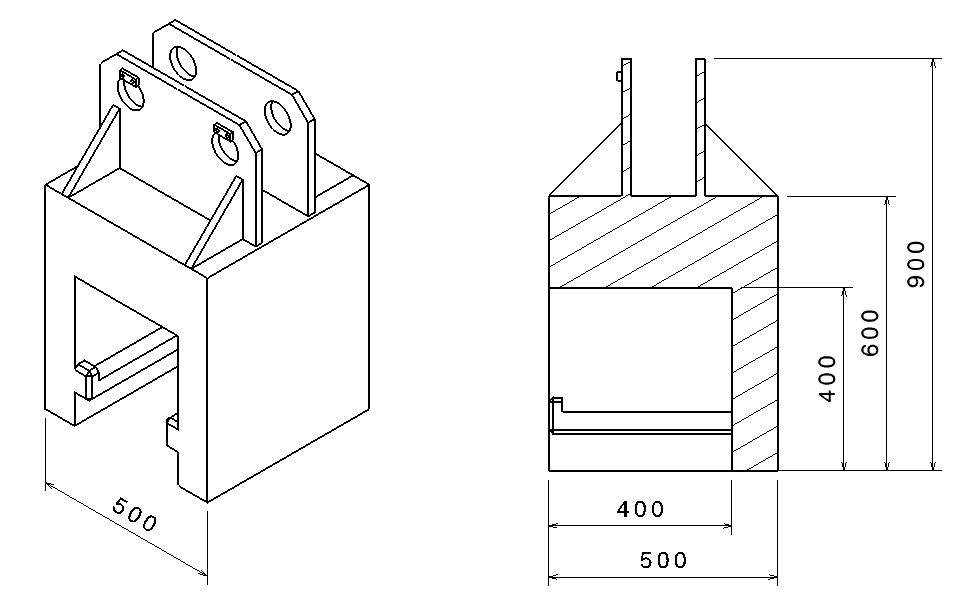}
\caption{Trolley concept proximity shielding lifting attachment. The attachment is fitted manually to the crane hook and permits the remote lifting of the proximity shielding}
\label{Fig:TC:56-lifting-attachement}
\end{figure}

 The cast iron proximity shielding for the trolley concept is designed to allow the target to be installed and removed from the side. It is made up of four elements which include service  ``chimneys'' which are used to provide the water cooling and temperature sensor services and radiation protection shielding. The water and electrical connections to the proximity shielding are all made at the top of the helium vessel shielding where residual radiation levels are sufficiently low to allow hands-on work. For alignment reasons the proximity shielding is supported on a flat plate which is mounted on support pillars (Fig.~ \ref{Fig:TC:53-proximity-shielding-crane}). 
 The mobile shielding above the trolley concept proximity shielding is handled using spreader beams equipped with legs to be able to work around the proximity shielding service chimneys (Fig.~\ref{Fig:TC:54-removing-mobile-shielding}). The service chimneys include a lifting feature at the top which interfaces with a lifting attachment fitted to the crane hook (Figs.~\ref{Fig:TC:55-lifting-trolley} and~\ref{Fig:TC:56-lifting-attachement}).

\begin{figure}[!htb]
\centering
\includegraphics[width=0.7\linewidth]{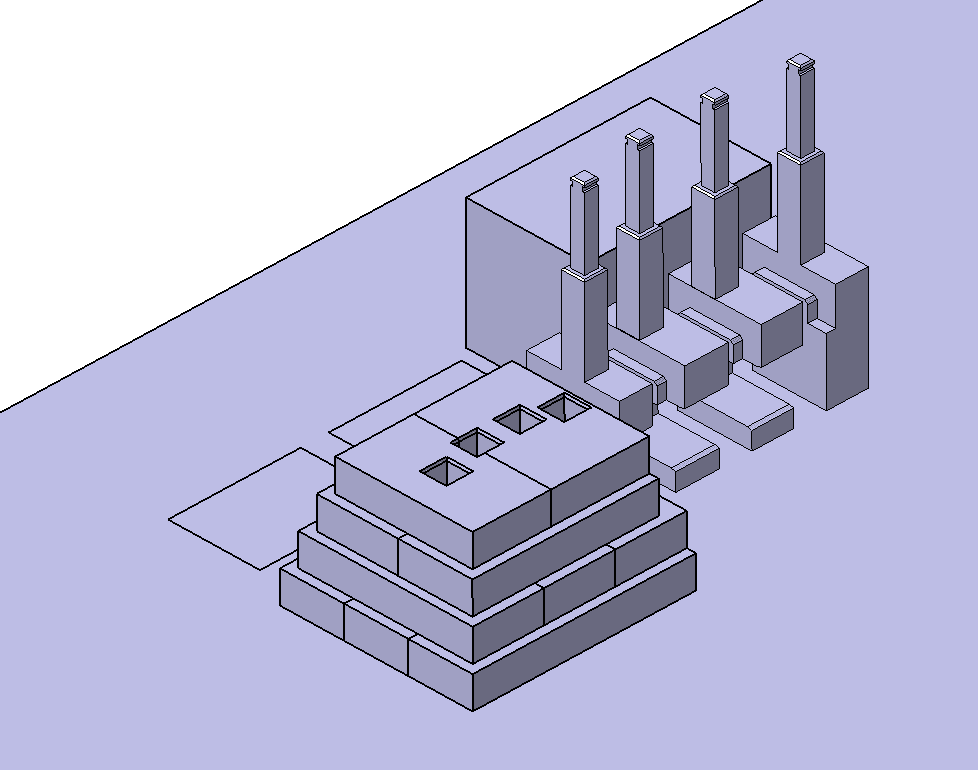}
\caption{Trolley concept fixed proximity shielding elements and mobile shielding stacked in the cooldown area}
\label{Fig:TC:57-trolley-mobile-shielding}
\end{figure}

\subsection{Handling of other beam line equipment in the target complex}
\label{Sec:TC:handling-beamline-equipment}

The handling of beam line equipment (other than the target and proximity shielding) is essentially the same for both the crane and trolley concepts.

\subsubsection{Proton beam window design and handling}
\label{Sec:TC:proton-beam-window}

\begin{figure}[!htb]
\centering
\includegraphics[width=0.5\linewidth]{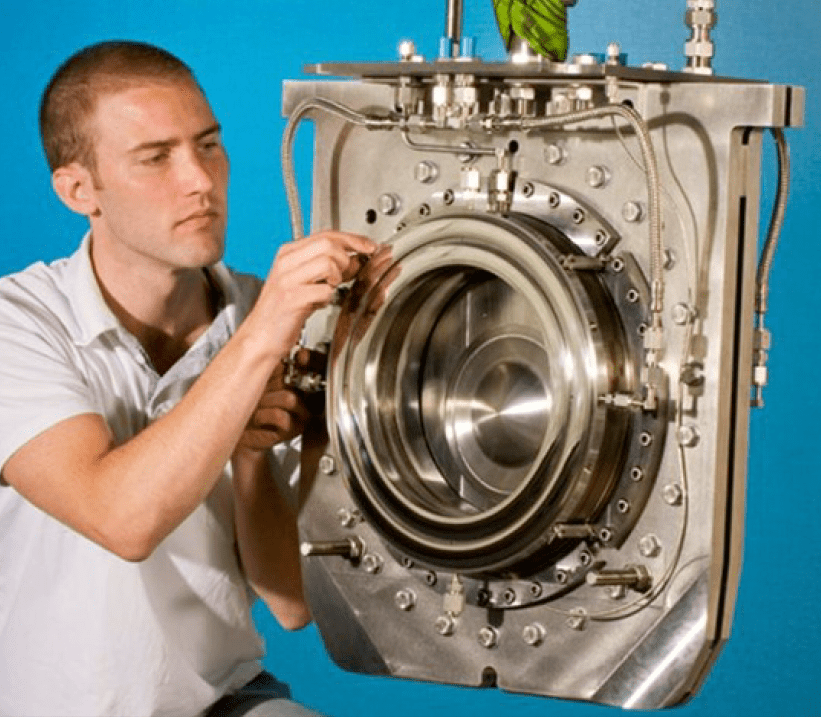}
\caption{T2K beam window with inflatable stainless steel pillow seals on both sides (Image courtesy T2K (Japan) and STFC (UK)).}
\label{Fig:TC:58-T2K-beam}
\end{figure}

A beam window is required to isolate the the helium vessel (which is at atmospheric pressure or above) from the vacuum in the beam pipe delivering protons from the SPS. It is necessary to design the beam window and its supports to allow replacement during the life of the facility. Due to high levels of induced radioactivity it is necessary to design for replacement using remote handling techniques. The beam window design produced as part of the handling study and the remote handling operations needed to exchange it are described below.  The beam window itself is similar in concept to the one used at T2K shown in Fig.~\ref{Fig:TC:58-T2K-beam}.

\begin{figure}[!htb]
\centering
\includegraphics[width=0.9\linewidth]{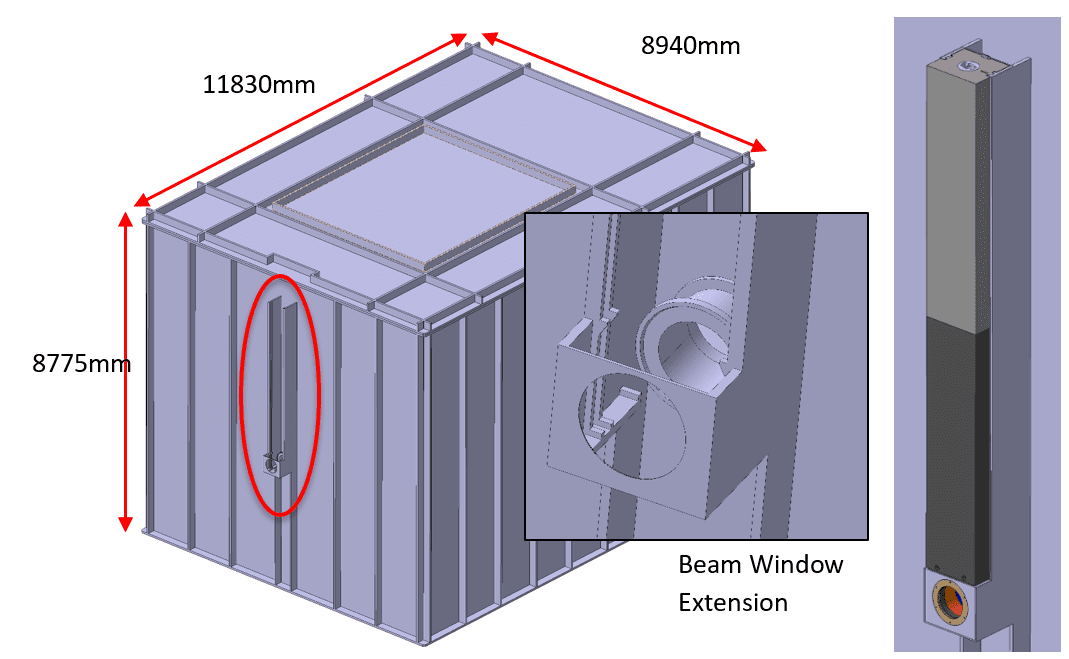}
\caption{Proton beam window: On the left -- BDF helium vessel with support features for the beam window and its shielding (ringed). Inset -- beam window extension with helium vessel sealing face for the beam window and the support for the end of the upstream beam pipe. On the right -- beam window and its shielding installed in the support on the helium vessel.}
\label{Fig:TC:59-proton-beam}
\end{figure}

\begin{figure}[!htb]
\centering
\includegraphics[width=0.9\linewidth]{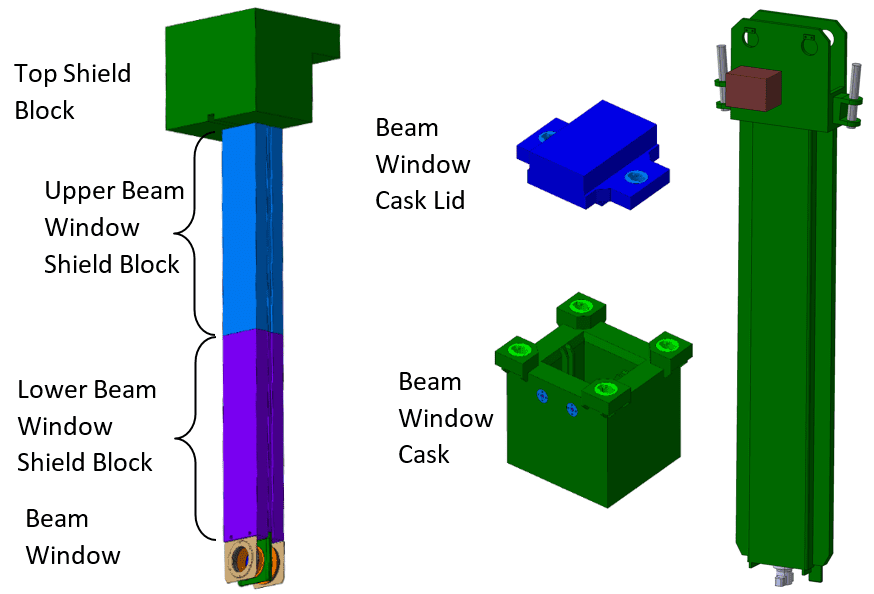}
\caption{Proton beam window; On the left -- main elements of the beam window and its shielding. In the middle -- shielded storage cask and lid for the beam window itself. On the right -- extended remote lift attachment for lifting lower shield block / beam window module.}
\label{Fig:TC:60-proton-beam-window-2}
\end{figure}

\begin{figure}[!htb]
\centering
\includegraphics[width=0.9\linewidth]{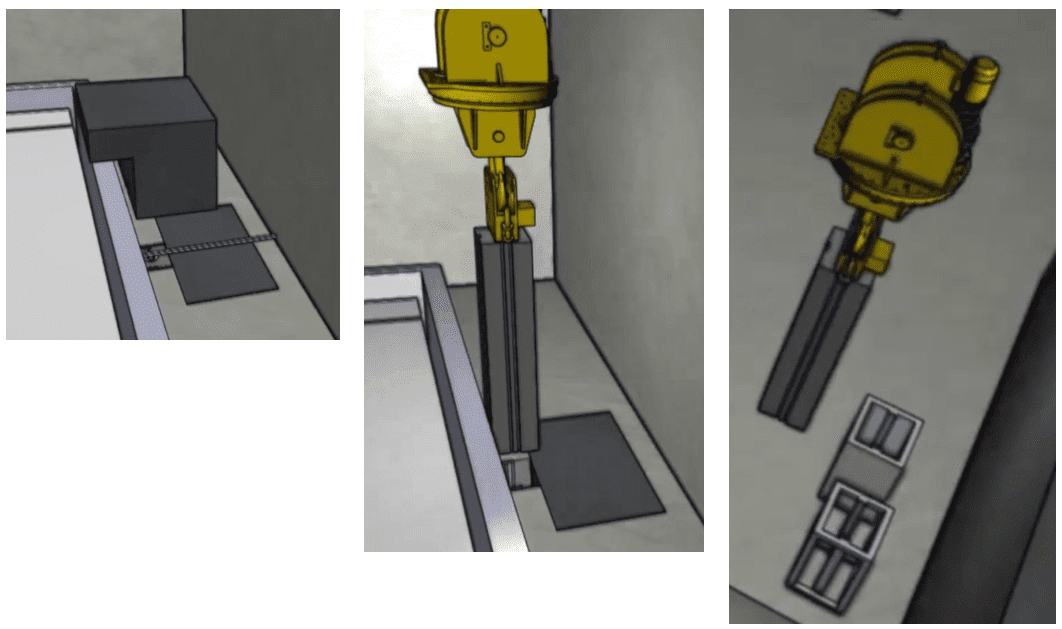}
\caption{Proton beam window removal (1): Left -- removal of top shield block Centre -- withdrawal of upper shield block using building crane. Right -- transfer of upper shield block to storage support in cooldown area. Note: the dark rectangle next to the beam window shaft represents shielding blocks which can be removed to access the beam window extension.}
\label{Fig:TC:61-proton-beam-window-3.}
\end{figure}

\begin{figure}[!htb]
\centering
\includegraphics[width=0.6\linewidth]{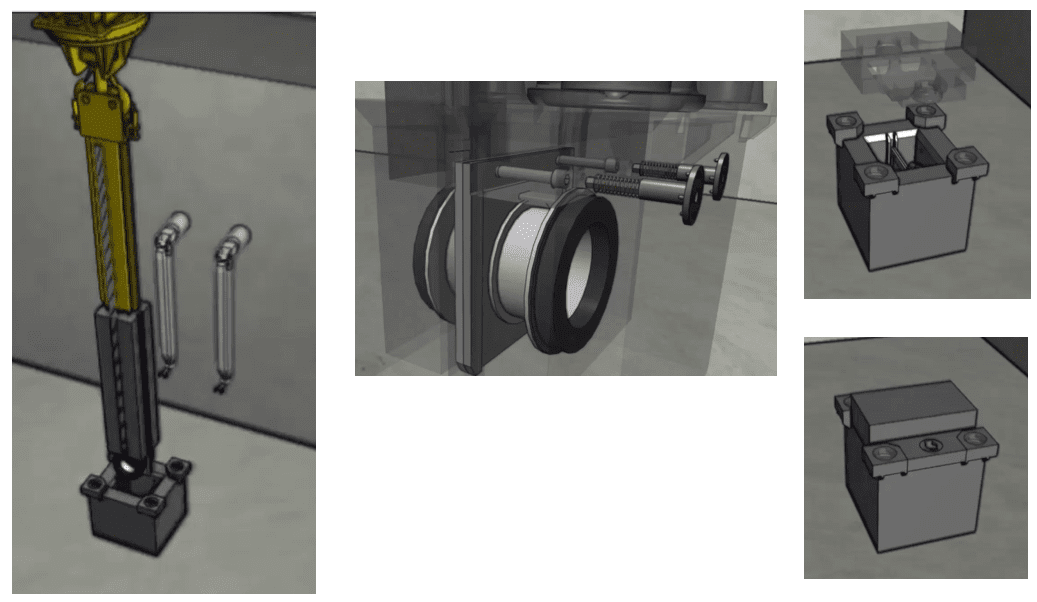}
\caption{Proton beam window removal (2): Left -- withdrawal of lower shielding block and beam window assembly using extended lifting attachment and transfer to shielded cask in remote handling area. Centre -- Disconnection of beam window from lower shielding block using manipulators (via captive drivers in cask wall). Right -- placing of lid on cask. (after shearing beam window service pipes)}
\label{Fig:TC:62-proton-beam-window-4}
\end{figure}

The beam window is mounted on the outside of the helium vessel (Fig.~\ref{Fig:TC:59-proton-beam}). It uses two inflatable pillow seals to seal it to the upstream beam pipe on one side, and to the helium vessel on the other. The beam window assembly is designed to allow remote replacement using the target hall overhead travelling crane. The main elements of the beam window,  its shield blocks, shielded cask and extended remote lift attachment are shown in Fig.~\ref{Fig:TC:60-proton-beam-window-2}.   To remove the beam window, first the crane lifts out the upper shielding block and then the lower shielding block with the beam window attached to it. The shielding blocks are transferred to the remote handling section of the cool-down area where the beam window is then disconnected from the lower block using a pair of through-the-wall master-slave manipulators. The main steps in the removal of the beam window are listed in table \ref{Tab:TC:window-removal} and illustrated in Figs.~\ref{Fig:TC:61-proton-beam-window-3.} and~\ref{Fig:TC:62-proton-beam-window-4}.

\begin{table}[!htb]
\begin{tabular}{|p{0.1\textwidth}|p{0.4\textwidth}|p{0.4\textwidth}|} \hline 
\centering
\textbf{Step} & \textbf{Task} & \textbf{Tooling} \\ \hline 
a & Remove top shield block & Building crane (hands-on operation) \\ \hline 
b & Disconnect services & By hand. \\ \hline 
c & Lift out upper shield block, transfer to storage support in cooldown area & Building crane and remote lift attachment \\ \hline 
d & Lift out lower shield block and beam window assembly then transfer to beam window shielded cask in remote handling area & Building crane and extended remote lift attachment. \\ \hline 
e & Disconnect beam window from lower shielding block (screw attachment designed for remote disconnection). Put lower shielding block in storage support & Manipulators in remote handling area and building crane with remote lift attachment \\ \hline 
f & Cut off service connections to beam window & Manipulators + shear \\ \hline 
g & Put lid on beam window cask & Building crane + manipulators \\ \hline 
\end{tabular}
\caption{Main steps and tooling for removal of proton beam window from the helium vessel and transfer to the cool down area for storage}
\label{Tab:TC:window-removal}
\end{table}

Replacement is essentially a reverse of the removal procedure and can be carried out using a new beam window and new lower shield block or by fitting a new beam window onto the used lower shielding block. In front of the beam window a second set of shielding blocks are used to fill-in an access shaft -- these blocks can be removed to provide access to the beam window extension on the vacuum vessel (Fig. \ref{Fig:TC:59-proton-beam}) This access will be used for initial installation and connection of the vacuum pipe and (remote) repair of the sealing faces on the helium vessel or vacuum pipe.

\subsubsection{Collimator handling}
\label{Sec:TC:collimator-handling}

\begin{figure}[!htb]
\centering
\includegraphics[width=0.6\linewidth]{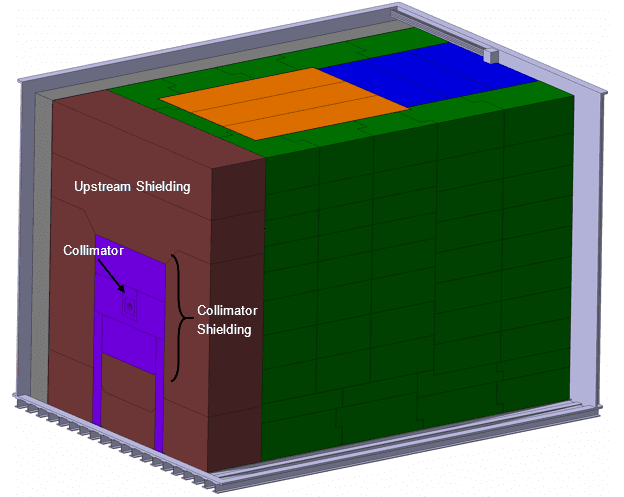}
\caption{The collimator and surrounding shielding in cut-away helium vessel.}
\label{Fig:TC:63-collimator-shielding}
\end{figure}

\begin{figure}[!htb]
\centering
\includegraphics[width=0.9\linewidth]{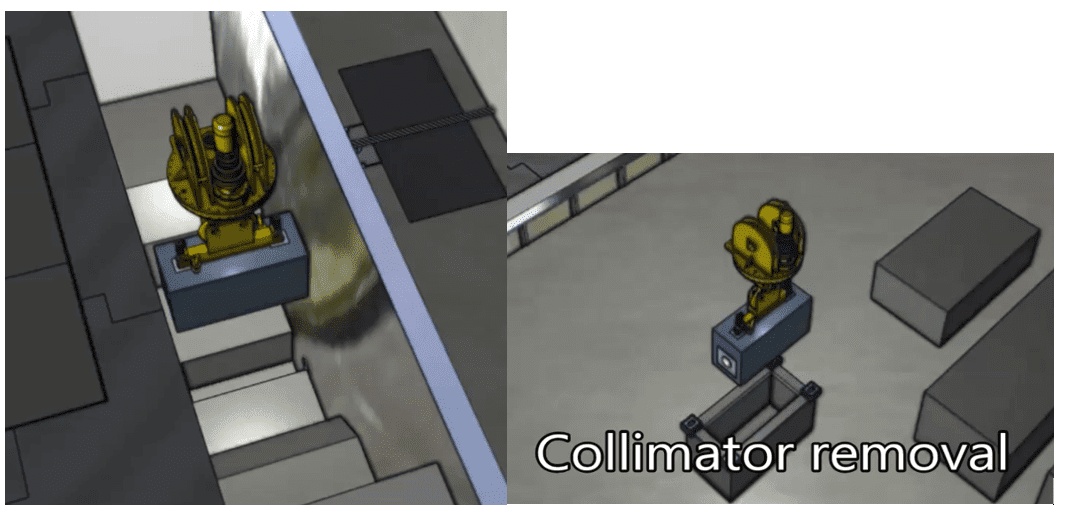}
\caption{Collimator removal. Left -- Crane lifting collimator out of helium vessel after removal of upstream shielding. Right -- Placing collimator in shielded cask.}
\label{Fig:TC:64-collimator-removal}
\end{figure}

The collimator is formed as part of a removable block within the upstream shielding inside the helium vessel. It consists of a 150 cm long, 20 cm diameter graphite mask employed to protect the downstream target and shielding assembly against beam misalignment. The graphite mask is incorporated into a shielding block with lift points so that it can be remotely installed and removed by the building crane. The shielding immediately around the collimator is supported on pillars to allow precise alignment of the collimator during initial installation (Fig.~\ref{Fig:TC:63-collimator-shielding}). The shielding above the collimator is designed to minimise handling operations needed in order to exchange the collimator. The main steps and tooling required for collimator removal are listed in table \ref{Tab:TC:collimator-removal} and illustrated in Fig. \ref{Fig:TC:64-collimator-removal}.

\begin{table}[!htb]
\begin{tabular}{|p{0.1\textwidth}|p{0.4\textwidth}|p{0.4\textwidth}|} \hline 
\centering
\textbf{Step} & \textbf{Task} & \textbf{Tooling} \\ \hline 
a & Remove helium vessel lid & Building crane (hands-on operation) \\ \hline 
b & Disconnect electrical connections & By hand. \\ \hline 
c & Lift out two upstream shielding blocks above collimator and transfer to cooldown area & Building crane and remotely operated spreaders \\ \hline 
d & Lift out collimator then transfer to shielded cask in remote handling area & Building crane and remotely operated spreader. \\ \hline 
e & Cut off service connections & Manipulators + shear \\ \hline 
f & Put lid on cask & Building crane + manipulators \\ \hline 
\end{tabular}
\caption{The table shows the main steps for removal of collimator from the helium vessel and transfer to the cool down area.}
\label{Tab:TC:collimator-removal}
\end{table}

\subsubsection{Magnetic coil and US1010 shielding -- handling aspects}
\label{Sec:TC:magnetic-coil-US1010-shielding}

\begin{figure}[!htb]
\centering
\includegraphics[width=0.9\linewidth]{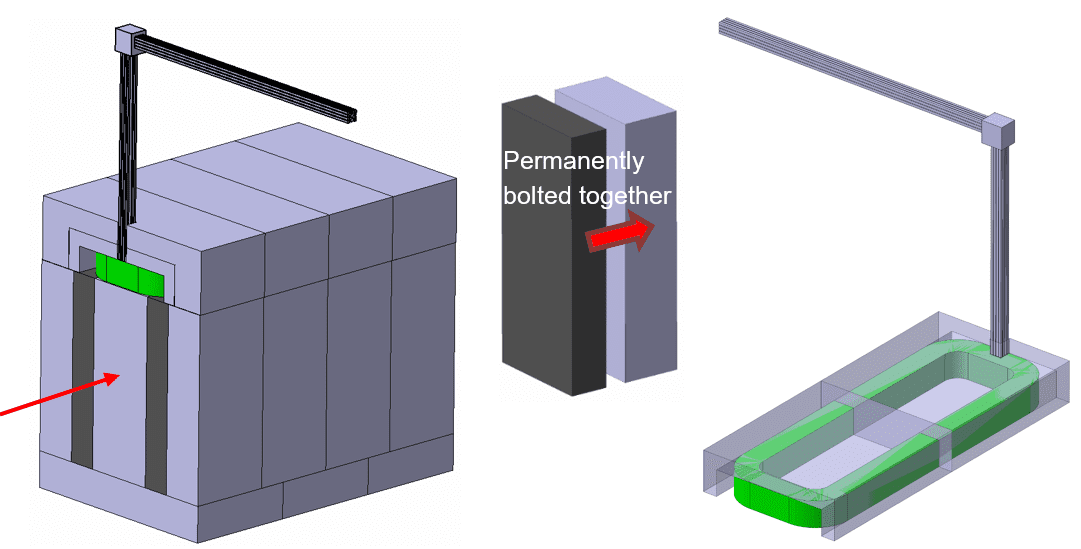}
\caption{Magnetic coil and US1010 shielding. On the left -- the US1010 shielding that is downstream of the target and the magnetic coil.  On the right - the magnetic coil with its surrounding US1010 support structure which restrains the coil during operation and provides lifting points for the crane. The junction box between the horizontal and vertical portion of the services connections represents the connections which are made and disconnected by hands-on interventions.}
\label{Fig:TC:65-magnetic-coil-US1010}
\end{figure}

\begin{figure}[!htb]
\centering
\includegraphics[width=0.9\linewidth]{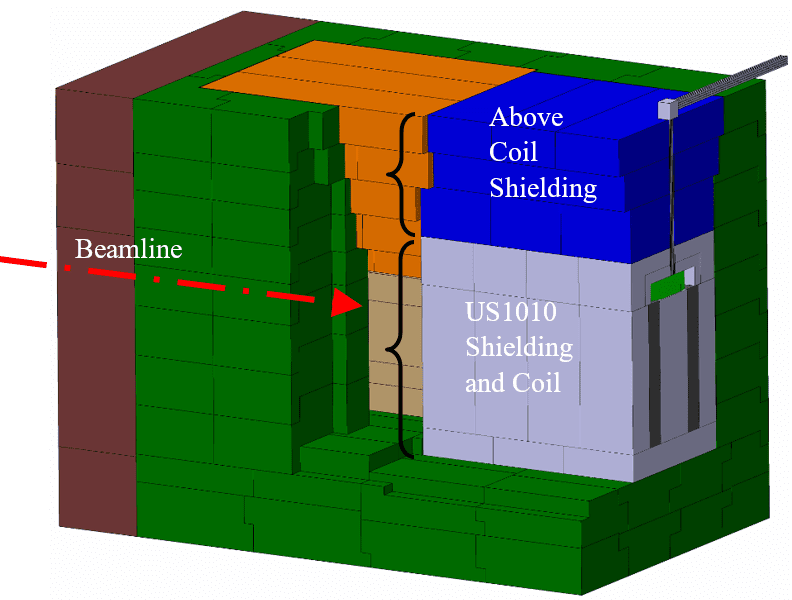}
\caption{Cut-away view of shielding in helium vessel showing the US1010 shielding and coil along with the above coil shielding. Note staggering of joints between blocks to avoid direct radiation shine paths}
\label{Fig:TC:66-cut-away-shielding-helium}
\end{figure}

\begin{figure}[!htb]
\centering
\includegraphics[width=0.9\linewidth]{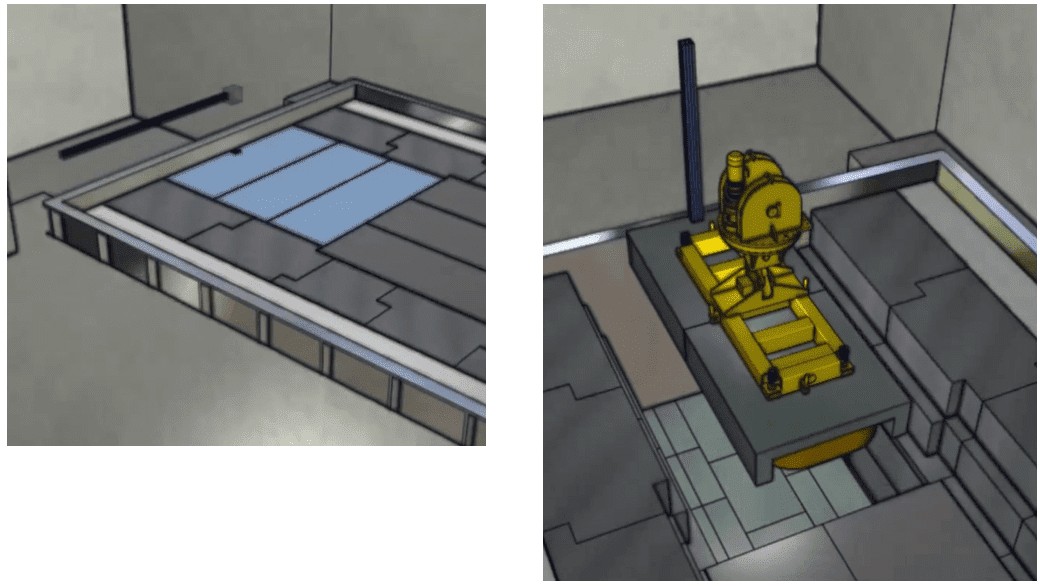}
\caption{Magnetic coil removal: On the left -- disconnection of services above the shielding in the helium vessel. On the right -- lifting of the coil and its supporting shielding out of the helium vessel using the building crane and a remotely operated spreader.}
\label{Fig:TC:67-magnetic-coil-removal}
\end{figure}

\begin{figure}[!htb]
\centering
\includegraphics[width=0.9\linewidth]{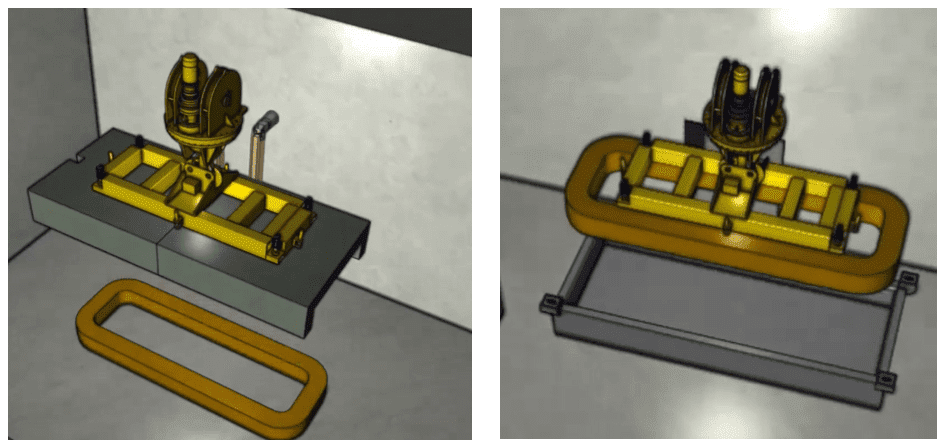}
\caption{Magnetic coil separation and placing in storage cask: On the left -- separation of the coil from the shielding in the remote handling area. On the right -- placing the coil in a storage cask.}
\label{Fig:TC:68-magnetic-coil-separation}
\end{figure}

The magnetic coil and US1010 shielding are designed to work together to provide a magnetic field downstream of the target of 1.5-1.6 T in order to sweep the muons produced in the target away from the detector acceptance to reduce experimental backgrounds. More details are given in section 3. The shielding includes two sections of non-magnetic stainless-steel blocks (shown in black in Fig.~\ref{Fig:TC:65-magnetic-coil-US1010}) to ensure that the magnetic field is correctly guided through the US1010 steel yoke. The coil is lifted in and out of the helium vessel along with its surrounding US1010 shielding and its service connections which protrude above the top layer of the shielding in the helium vessel (Fig.~\ref{Fig:TC:66-cut-away-shielding-helium}). The location of this service connection allows for hands-on connection and disconnection above the bunker shielding. The design and layout of the US1010 blocks have been optimised to give the minimum number of gaps seen by the magnetic field generated by the coil whilst remaining compatible with achievable manufacturing and handling precision. To ensure that the individual components of the shielding can be safely left in a standing position during storage and initial installation, the US1010 and stainless steel parts are bolted together (Fig.~\ref{Fig:TC:65-magnetic-coil-US1010}) The coil is assembled into a module with a surrounding support structure made of US1010 steel and is handled by the crane during installation and removal from the helium vessel using lift points in the support structure (Fig.~\ref{Fig:TC:65-magnetic-coil-US1010}). The main handling steps for removal of the coil in the event of failure are listed in Table~\ref{Tab:TC:coil-removal} and illustrated in Figs.~\ref{Fig:TC:67-magnetic-coil-removal} and~\ref{Fig:TC:68-magnetic-coil-separation}.

\begin{table}[!htb]
\begin{tabular}{|p{0.1\textwidth}|p{0.4\textwidth}|p{0.4\textwidth}|} \hline 
\centering
\textbf{Step} & \textbf{Task} & \textbf{Tooling} \\ \hline 
a & Remove helium vessel lid & Building crane (hands-on operation) \\ \hline 
b & Disconnect electrical connections & By hand. \\ \hline 
c & Lift out the above-coil shielding and mobile shielding blocks and transfer to cooldown area & Building crane and remotely operated spreaders \\ \hline 
d & Lift out coil then transfer to remote handling area & Building crane and remotely operated spreader. \\ \hline 
e & Cut off service connections and separate coil from its support shielding & Manipulators + shear, building crane and remotely operated spreader \\ \hline 
f & Place in cask and put lid on cask & Building crane and remotely operated spreader \\ \hline 
\end{tabular}
\caption{Main steps and tooling for removal of coil from the helium vessel and transfer to the remote handling area}
\label{Tab:TC:coil-removal}
\end{table}

\subsection{Design of the fixed target bunker shielding -- handling aspects}
\label{Sec:TC:design-target-bunker-handling}

\begin{figure}[!htb]
\centering
\includegraphics[width=0.9\linewidth]{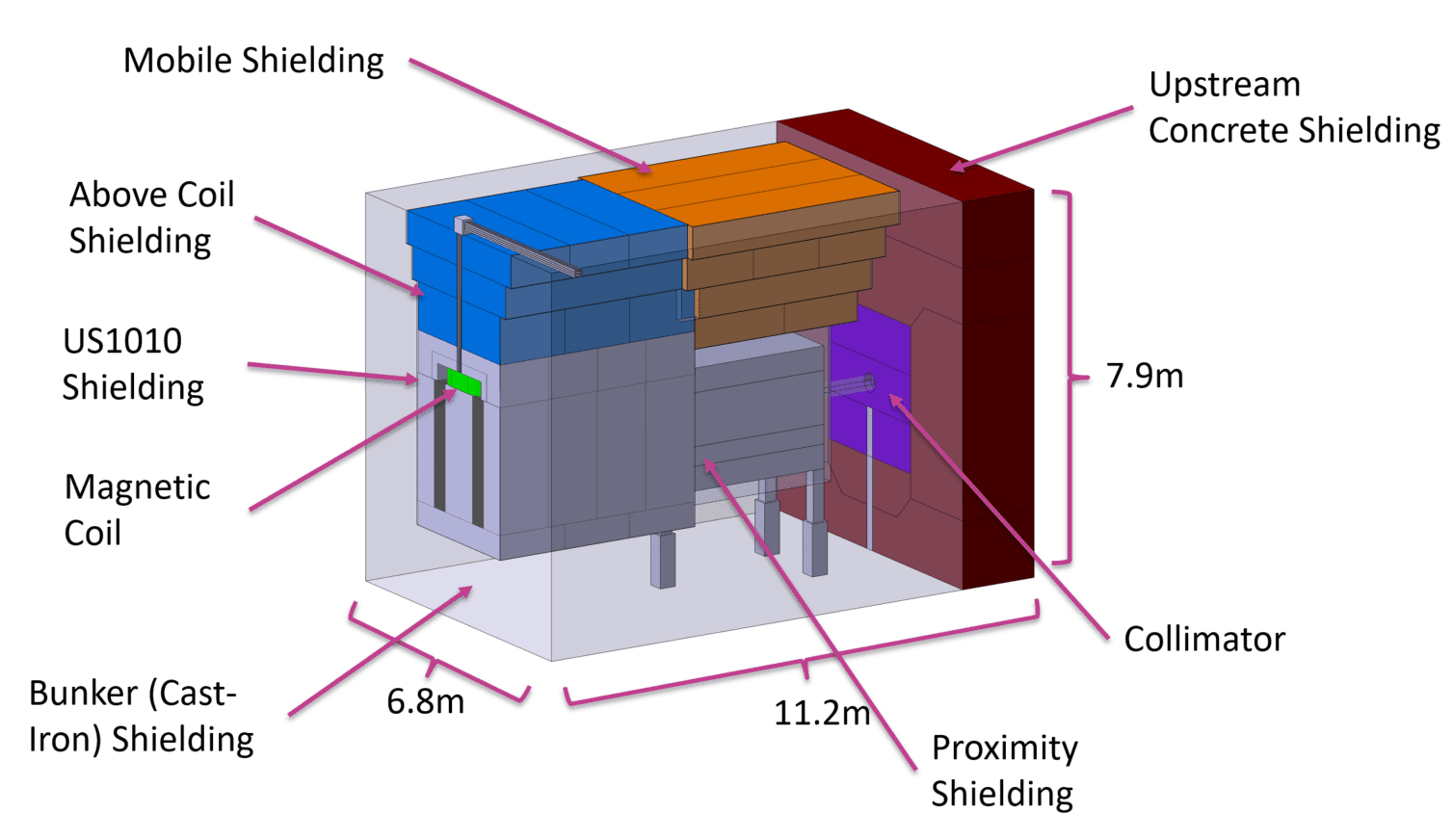}
\caption{Different sections of shielding in the helium vessel. The fixed target cast iron bunker shielding is shown as semi-transparent.}
\label{Fig:TC:69-shielding-helium-vessel}
\end{figure}

The fixed target bunker shielding is the shielding that fills in the space in the helium vessel around the sections of shielding that may need to be moved for maintenance reasons during the life of the facility. Fig. \ref{Fig:TC:69-shielding-helium-vessel} shows the different sections of iron and steel shielding in the helium vessel; The fixed bunker shielding is shown as transparent in the figure. The design of the bunker shielding was carried out once all the other shielding had been designed. The requirements (for avoiding shine paths etc) are, in general, the same as for the rest of the shielding as set out in section \ref{Sec:TC:KeyDesignConsiderations}.

\subsection{Global handling, integration and operational aspects}
\label{Sec:TC:global-handling}

\subsubsection{Overhead travelling crane in target complex surface building (crane and trolley concept)}
Both the trolley and crane concepts have a remotely operated overhead travelling crane in the surface building to carry out the necessary handling operations for the installation and operation of the target complex. The crane runs on rails that run the length of the building, supported on the building walls.

The building crane capacity was set as 40 tonnes; giving a useful working load of around 36 tonnes if 10\% is reserved for spreader beams; this capacity is in line with road transport limitations which is an important consideration given the quantities of shielding blocks to be delivered and installed.

The crane hook rotation is motorised so that radioactive loads can be oriented without the need for the crane operator to be next to the load to rotate it manually – as is the case for normal crane operations. The crane hook is a size 20 ramshorn hook (according to DIN 15402). Electrical power and signal services are supplied down to the crane hook to operate spreader beams and their cameras. The crane is equipped with cameras and lighting on the crane bridge, trolley and hook pulley block to illuminate the area around the loads being handled.

Spreader beams with remotely operated load attachment mechanisms are used in conjunction with the building crane. The spreaders are attached manually to the crane hook in an area away from radioactive equipment; electrical power and signal services are connected to the spreader beams at the same time to operate the remote load attachments and the spreader cameras.

Remote operation of the building crane is carried out from the control room in the experimental area service building. In addition to the controls for the crane, the control room has TV monitors for the different camera views and controls for the pan-tilt-zoom drives of the cameras mounted on the crane, on the spreader beams and around the building. 

The crane has no on-board electronics; electrical control cubicles are not installed on the crane in order to avoid damage to the electronic components due to radiation and also to allow repair of the control electronics without the need to access the crane in the event of a failure. Position feedback of mechanisms on the crane is by means of resolvers rather than encoders to avoid the presence of electronics on the crane. A cable festoon running the length of the building is used to transfer power and control signals between the control cubicles and the crane.

In the event of a crane breakdown during handling of a radioactive load it is necessary to be able to move to load to a safe, shielded area. Recovery in the event of breakdown of the crane is ensured by siting the control electronics off the crane and designing-in features such as redundancy of the hoisting, long travel, cross travel and hook rotation drives along with their cabling. Repair of the crane can be carried out once the load is safely stored and the crane moved to an area at the end of the building where radiation levels permit access for the repairs.

\subsubsection{Hot cell crane for trolley concept}
\label{Sec:TC:hot-cell-trolley}

The trolley concept hot cell is equipped with a 3 tonne remotely operated overhead travelling crane to carry out target handling operations inside the hot cell when the trolley is in its withdrawn position. The crane is fitted with a ramshorn hook with motorized rotation.  

The crane is controlled from the area outside the hot cell next to the master-slave manipulator master arms; viewing of operations is via the hot cell shielded windows and radiation tolerant pan-tilt-zoom cameras installed in the hot cell.

The spreader beam used to lift the target is attached to the crane hook with the aid of the master-slave manipulators and the manipulators can also guide the spreader beam to engage it with the target. 

The crane has no on-board electronics; resolvers are used for position feedback of the crane motions.  Recovery in the event of breakdown of the crane is ensured by siting the control electronics off the crane and designing-in features such as redundancy of the hoisting, long travel, cross travel and hook rotation drives along with their cabling.  Repairs to the crane can be carried out by accessing the hot cell once the target is removed – either by moving it on the trolley into the helium vessel or by lowering it into the transfer cask and moving it into the storage area.

\subsubsection{Cameras and viewing}
\label{Sec:TC:cameras-viewing}

\subsubsubsection{Cameras and viewing for remote handling operations with the target hall crane}
\label{Sec:TC:cameras-viewing-remote-handling-crane}

\begin{figure}[!htb]
\centering
\includegraphics[width=0.6\linewidth]{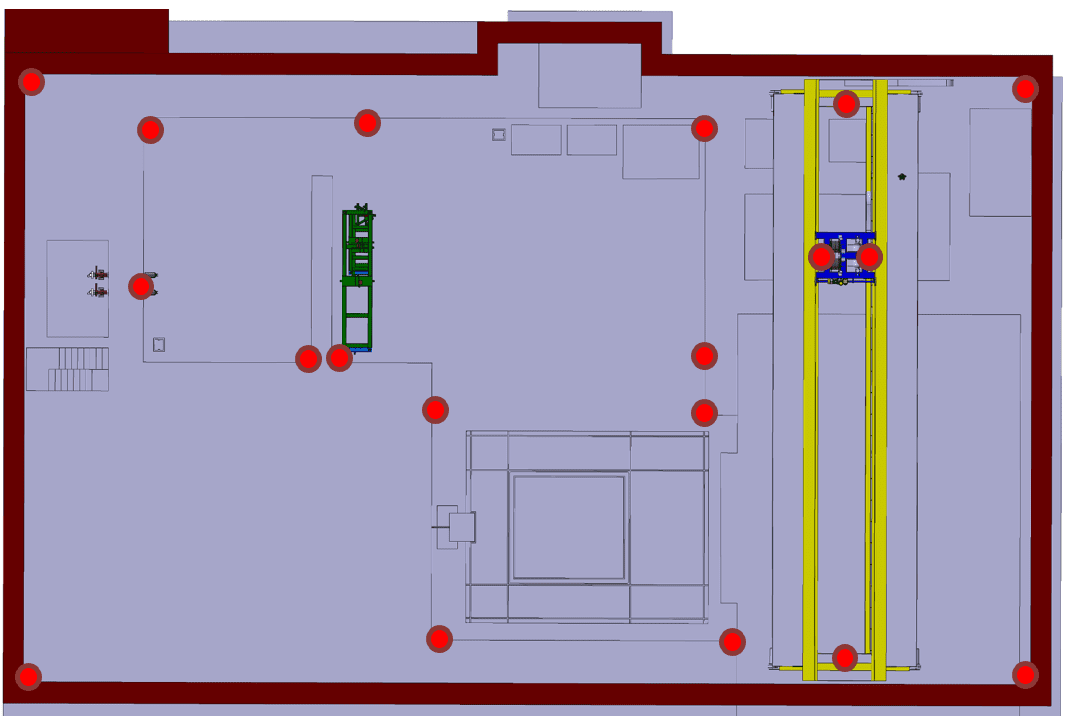}
\caption{Crane concept camera positions on the crane and in the target hall building (around the cooldown and remote handling areas and around the target pit)}
\label{Fig:TC:75-crane-concept-camera}
\end{figure}

Cameras are needed for remote handling operations carried out by the crane. Pan-tilt--zoom Cameras are mounted on the crane itself and around the target hall building (figure \ref{Fig:TC:75-crane-concept-camera}). Fixed cameras are mounted on the lifting spreaders (see Figs. \ref{Fig:TC:70-collimator-lifting} and \ref{Fig:TC:71-upper-beam-window}).

\subsubsubsection{Cameras and viewing for the remote handling area and trolley cell}
\label{Sec:TC:cameras-viewing-remote-handling-trolley}

The trolley concept hot cell and the remote handling areas for both concepts are fitted with lead glass shielded viewing windows and also pan-tilt-zoom cameras. In addition cameras may be fitted to the manipulator arms in order to get a close-up view of the grippers. 

\begin{figure}[!htb]
\centering
\includegraphics[width=0.6\linewidth]{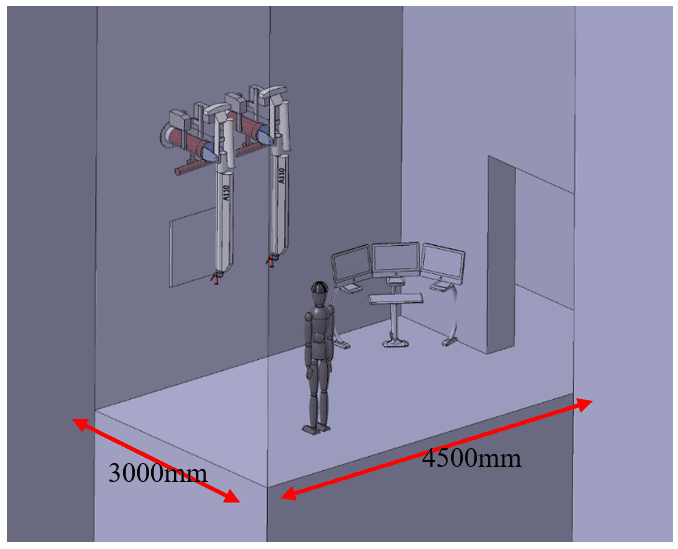}
\caption{Operating area for the trolley concept hot cell with manipulator master arms, shielded viewing window in the hot cell wall and camera monitors}
\label{Fig:TC:76-operating-area}
\end{figure}

\subsubsection{Spreader beams and lifting interfaces}
\label{Sec:TC:spreader-beams}

The study included the conceptual design of spreader beams and lift attachments used to lift the equipment that will become radioactive. Designs of some of the spreader beams and lift attachments are explained and illustrated in the different sections of this report describing handling methods and sequences for the exchange of key components. Most of the spreader beams have motorised lifting interfaces to allow fully remote connection and disconnection of the loads.

\begin{figure}[!htb]
\centering
\includegraphics[width=0.8\linewidth]{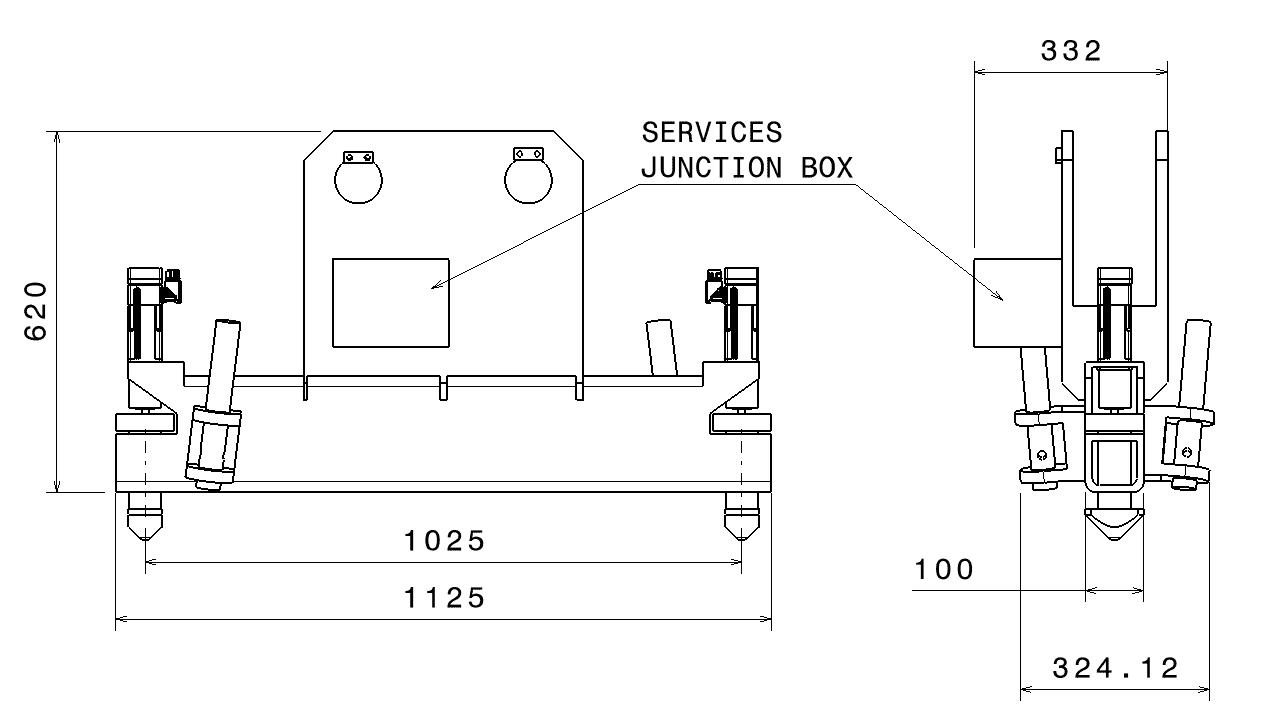}
\caption{Collimator lifting spreader equipped with four motorised ISO twist locks. Inclined cameras are installed on the spreader to allow the crane operator to view the engagement of the twist locks during remote operations.}
\label{Fig:TC:70-collimator-lifting}
\end{figure}

\begin{figure}[!htb]
\centering
\includegraphics[width=0.7\linewidth]{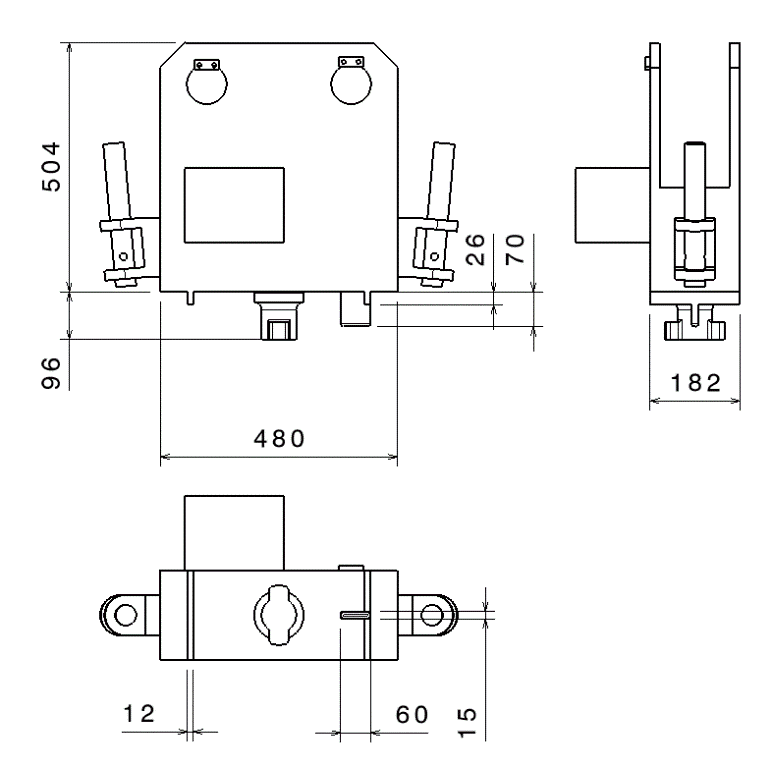}
\caption{Upper beam window shield block spreader equipped with a CERN twist lock interface}
\label{Fig:TC:71-upper-beam-window}
\end{figure}

Two different lifting interfaces are used in these spreader beams:

\begin{itemize}
\item ISO  ``twist lock'' lifting interfaces (as used on standardised transport containers) -- these have conical ends to allow easier remote operations and are rotated through 90 degrees to lock on to the load. They have a load capacity of approximately 15 tonnes each (see Fig. \ref{Fig:TC:70-collimator-lifting}).

\item CERN lift attachments -- these also are rotated through 90 degrees to lock onto the load but do not have conical ends to facilitate remote use -- They have a load capacity of 7.5 tonnes (see Fig. \ref{Fig:TC:71-upper-beam-window}).
\end{itemize}

\subsubsection{Helium vessel and services -- handling and integration}
\label{Sec:TC:helium-vessel-handling}

\begin{figure}[!htb]
\centering
\includegraphics[width=0.95\linewidth]{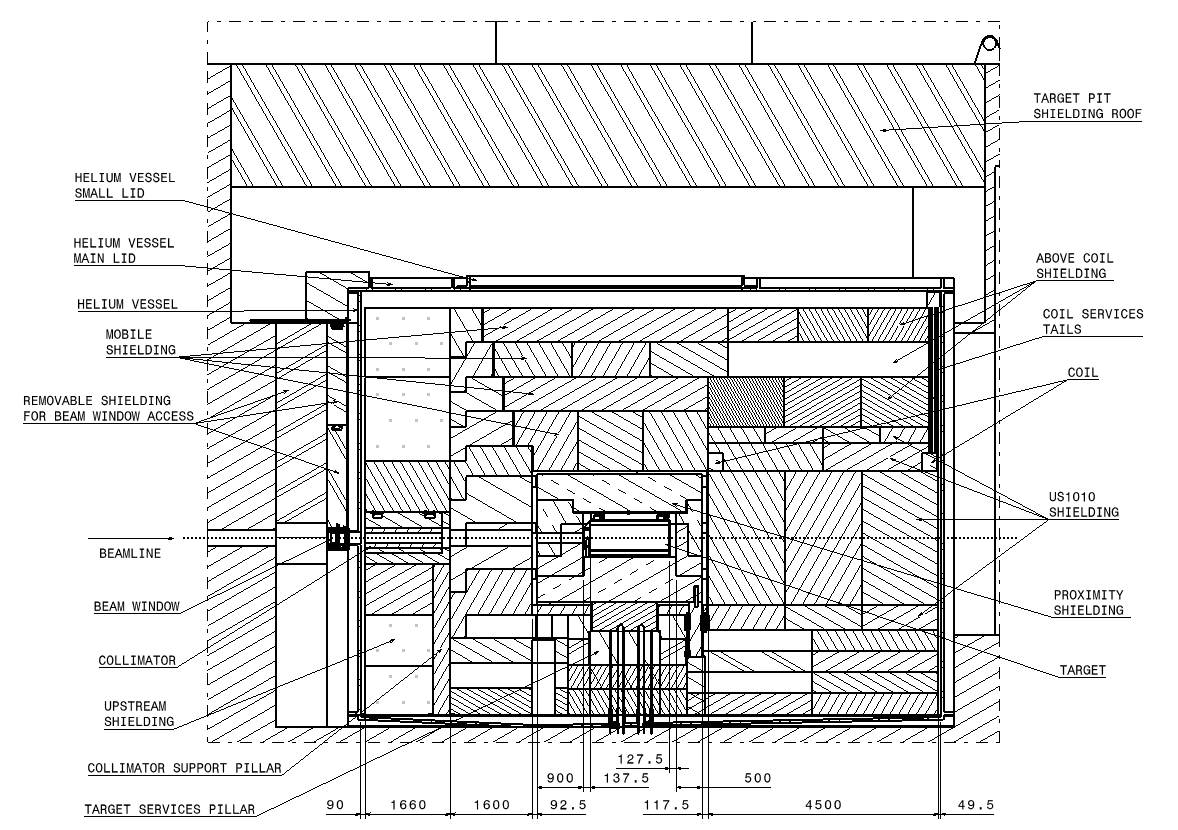}
\caption{Section view of helium vessel in the target pit showing the main elements and different shielding areas (crane concept shown - which has a small lid as part of the main lid; only the small lid needs to be removed for target exchange)}
\label{Fig:TC:72-helium-vessel-target-pit}
\end{figure}

The helium vessel (Fig.~\ref{Fig:TC:72-helium-vessel-target-pit}) has to support the loads due to the shielding inside it, ensure good helium leak tightness, allow the passages of services, permit flushing of air with helium and also allow draining of any water leaks. 

To allow for flushing of helium and draining of any water leaks, the floor of the vessel slopes down to a drain point. This drain point is connected to a sump room so that any water leaking into the vessel drains even in the event of pump failure. Service galleries are included in the civil engineering structure to allow connections between the CV and sump rooms. 

The helium vessel designs for the two handling concepts have some differences in the position of service passages and access openings for target and shielding movements. In the crane concept design, the services (water, helium, electrical) for the target and proximity shielding enter the helium vessel from below, whilst the services for the magnetic coil enter from the vessel side close to the top (Fig. \ref{Fig:TC:72-helium-vessel-target-pit}). For the trolley concept design no services enter the helium vessel from below; target services pass via the trolley while proximity shielding and coil services enter the helium vessel from the side above the upper surface of the shielding inside the vessel. 

The structural design of the helium vessel is covered separately in Section~\ref{Sec:TC:HeV}. 

\subsubsection{Cool-down and remote handling areas}
\label{Sec:TC:cool-down-remote-handling}

\begin{figure}[!htb]
\centering
\includegraphics[width=0.6\linewidth]{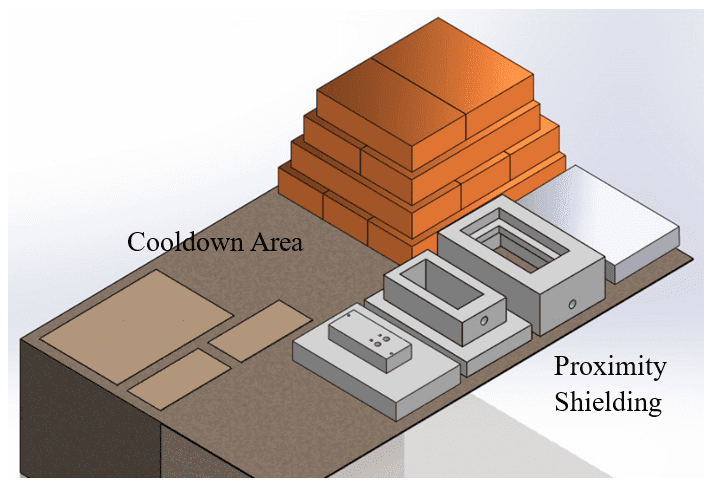}
\caption{Example of storage space checks used to determine the required dimensions of the cooldown area (in this case the crane concept proximity shielding and mobile shielding are shown).}
\label{Fig:TC:72b-example-storage}
\end{figure}

\begin{figure}[!htb]
\centering
\includegraphics[width=0.7\linewidth]{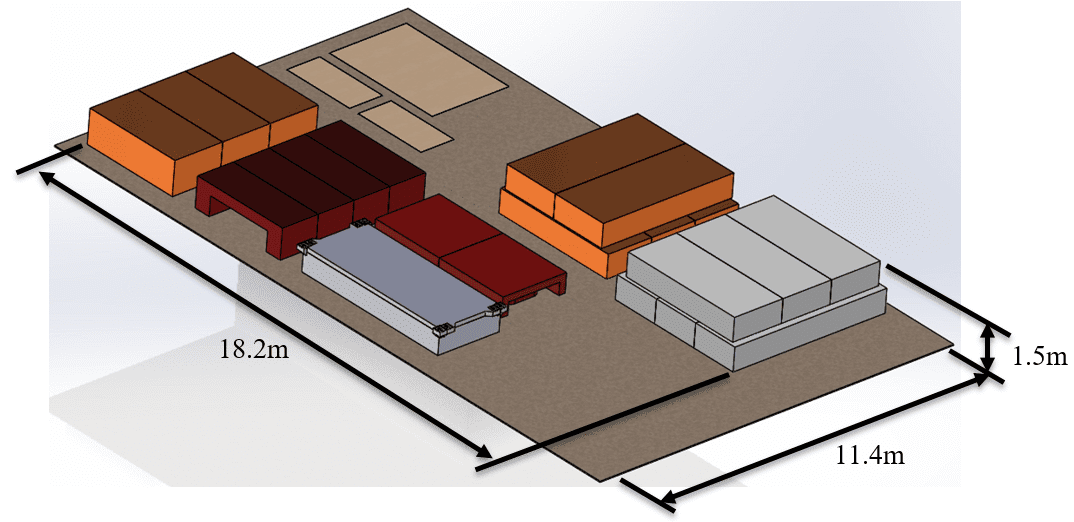}
\caption{Dimensions of cooldown area required for coil removal.}
\label{Fig:TC:72c-dimensions-cooldown}
\end{figure}

A cool-down area below ground level is provided for temporary storage of the target, shielding, magnetic coil, beam window etc. The area has been designed to allow temporary storage and transfer as required, for all foreseen maintenance operations during the life of the facility. As part of the study, the space requirements in the cooldown area for each maintenance operation were determined in order to define the required dimensions of the cooldown area.  Figs. \ref{Fig:TC:72b-example-storage} and \ref{Fig:TC:72c-dimensions-cooldown} illustrate typical storage space checks carried out as part of the design process. 

In addition, a remote handling area, equipped with a pair of through-the-wall master-slave manipulators, has been included as part of the cool-down area to carry out operations such as the disconnection of the beam window from its shielding block, removal of the magnetic coil from its surrounding shielding and unforeseen repair work if needed (see Figs. \ref{Fig:TC:3-craneConcept} and \ref{Fig:TC:6-area-trolley}).

\subsubsection{Cooling and ventilation services and equipment -- handling and integration}
\label{Sec:TC:cooling-ventilation-handling}
Cooling and Ventilation (CV) systems for the target complex consist of: water cooling systems for the target, proximity shielding and magnetic coil, a helium purge \& purification system for the helium vessel and target, and a pressure cascade ventilation system for the target complex to ensure containment of any radioactive contamination. ANSYS{\textregistered} simulations have been used extensively to validate the cooling of the target and of the proximity shielding; the results of this work have been fed into the design of the CV systems. More details of the CV systems are given in Section~\ref{Sec:TC:CV}. 

The equipment for water cooling and helium purification are housed underground in the target complex along with the sump room to collect any water leaks in the helium vessel. The pressure cascade ventilation ducts are integrated into the target complex building whilst the rest of the ventilation system equipment and cooling towers are housed in an auxiliary building to be constructed next to the target complex building.

For the crane concept design, all the cooling and helium system equipment for the target and proximity shielding is housed in the underground area (see Fig.~\ref{Fig:TC:3-craneConcept}); for the trolley concept the cooling and helium systems for the target are installed separately on the rear of the trolley (see Figs.~\ref{Fig:TC:7-Trolley-layout} and \ref{Fig:TC:11-cut-away-withdrawn-trolley}). Integration of the CV equipment in the target complex is explained in Section~\ref{Sec:TC:CV}. 

\subsubsection{Recovery from beam-line equipment failures -- remote handling capabilities needed}
\label{Sec:TC:failure-recovery}

Designing for recovery from failures was an important aspect of the handling and integration study as the very high residual radiation dose rates for the target and the proximity shielding necessitate the use of remote handling techniques. In the event of failure of the beam window, collimator or  magnetic coil, the failed item would be replaced using the techniques and procedures described in section~\ref{Sec:TC:handling-beamline-equipment} which are common to the crane and trolley concepts. However, for the connections to the target and proximity shielding the two concepts use different remote handling approaches. 

\subsubsubsection{Trolley concept connector repairs}
\label{Sec:TC:trolley-repairs}
For the trolley concept, failures of connections to the target would be dealt with using the remote handling manipulators and custom-designed tools (adapted for remote use) within the hot-cell. The trolley concept proximity shielding connections are outside the shielding and so can be repaired hands-on.

\subsubsubsection{Crane concept connector repairs}
\label{Sec:TC:crane-repairs}

\begin{figure}[!htb]
\centering
\includegraphics[width=0.9\linewidth]{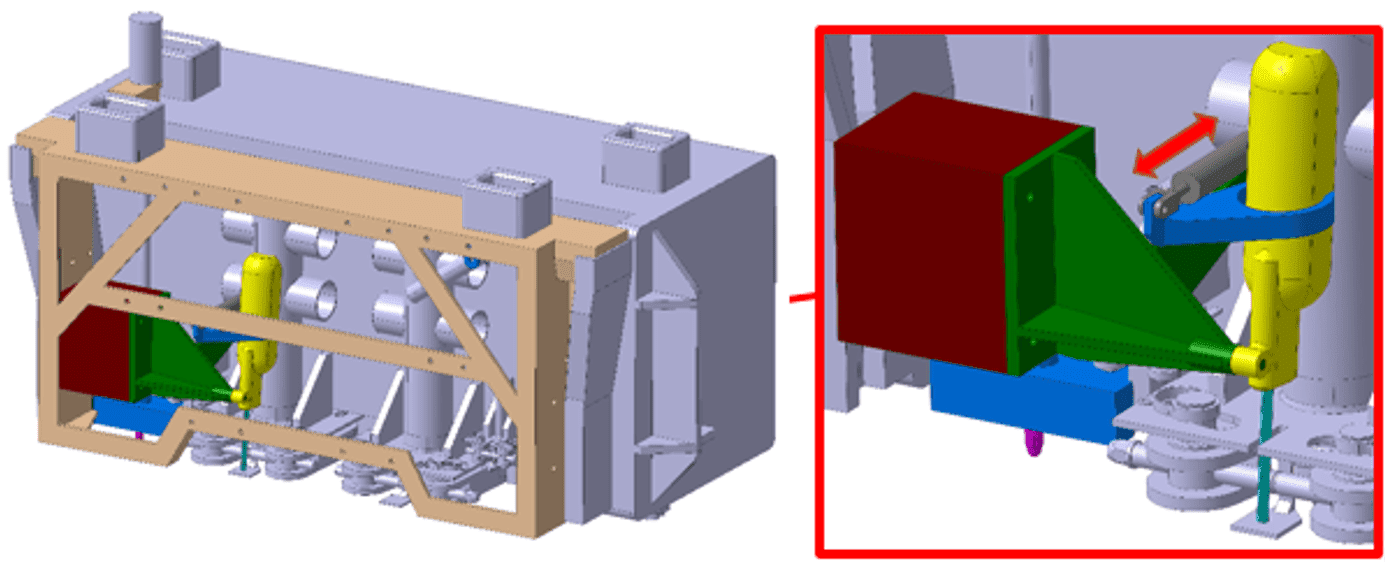}
\caption{Cutting a seized clamping screw on the target water connection using a saw module installed in the (un)locking tool frame. Saw module shown in zoom view on right. A second device then prises open the failed connector so that the target can be lifted.}
\label{Fig:TC:73-cutting-seized-clamping-screw}
\end{figure}

\begin{figure}[!htb]
\centering
\includegraphics[width=0.9\linewidth]{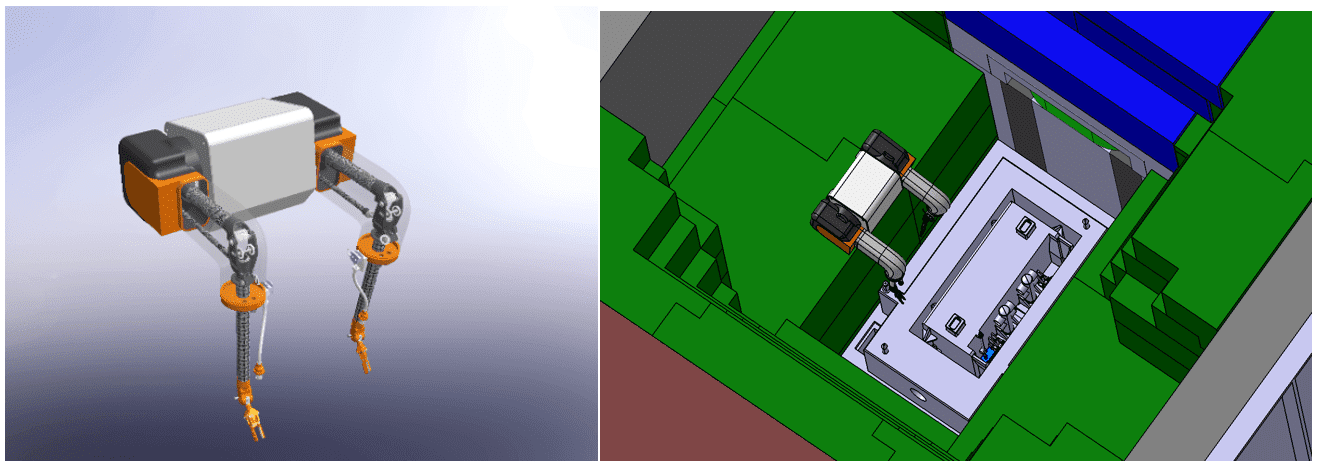}
\caption{Additional remote handling system for repair of damage to service pillar connections in crane concept: On the left -- twin slave arm master-slave servo-manipulator  ``Dexter'' produced by Oxford Technologies. On the right -- an illustration showing the slave arms in the helium vessel.}
\label{Fig:TC:74-remote-handling-system}
\end{figure}

For the crane concept there is a need for special tooling to deal with connector failures; a solution based on the use of cutting tools incorporated in the (un)locking tool frame is used to deal with failure of the connector clamp (Fig.~\ref{Fig:TC:73-cutting-seized-clamping-screw}). However, if the fixed connections on the service pillars in the crane concept helium vessel are damaged, a mobile remotely operated manipulator system capable of reaching into the helium vessel and working at the level of the service pillar connections will be needed for repair. One potential solution is the use of a twin arm force reflecting servo manipulator mounted on a support platform which can be lifted into position by the building crane in order to carry out repairs using adapted tools. The master arms would be in the facility control room with the power and control signals to the slave arm via a mixture of Ethernet and electrical cables (Fig.~\ref{Fig:TC:74-remote-handling-system}).

\subsubsection{Reconfiguration and Decommissioning}
\label{Sec:TC:reconfiguration-decommissioning}
Reconfiguration of the facility could be required in the future in order to install a different experiment. For the purposes of the handling and integration study reconfiguration was taken to mean replacement of all the equipment in the helium vessel. Reconfiguration will therefore necessitate removal of the equipment in the helium vessel using the remote handling equipment and procedures already described. In addition, the service pillars for the crane concept could need to be removed or the nose of the trolley could need to be replaced.  

Decommissioning of the facility would involve removal of all the equipment in the helium vessel and of all the other radioactive or contaminated equipment in the building. It is similar in scope to the equipment removal needed for reconfiguration but with extra work to dismantle and dispose of the helium vessel and, in the case of the trolley concept, the activated / contaminated parts of the trolley itself and the hot cell.

\subsection{Comparison of crane and trolley concepts -- conclusions}
\label{Sec:TC:comparison-crane-trolley}
Analysing the handling and remote handling operations needed during the lifetime of the facility (including those to recover from failures and damage) as part of the integration design work has led to a clearer understanding of the design requirements for the target complex; the design work forms a sound basis for further work as the BDF design study advances. 

 The remote handling methods for many elements of both concepts are essentially the same so they do not influence any decision on the choice of concept -- for example the beam window, collimator, magnetic coil and US1010 shielding. To compare the two concepts, the following elements related to operation and construction of the target complex have been considered:

\begin{enumerate}[label=\alph*]
\item Target exchange
\item Water and electrical connections
\item Shielding
\item Helium vessel
\item Handling equipment design and development risks 
\item Civil engineering
\item Reconfiguration 
\item Decommissioning
\end{enumerate}

\subsubsection{Target exchange}
\label{Sec:TC:conclusion-target-exchange}

The trolley concept design offers a simpler and faster target exchange than the crane concept as the target is withdrawn from the helium vessel without the need to remove the lid, remove the mobile shielding then disconnect and remove the proximity shielding. The pipework design in both concepts allows draining of the target before disconnection.

\subsubsection{Water and electrical connections}
\label{Sec:TC:conclusion-connections}

Both concepts have remotely operated clamp water connections and remotely operated electrical connections to the target and proximity shielding that are exposed to high levels of radiation -- this is a potential source of major problems because conventional industrial sealing solutions and materials cannot be used.

The trolley concept water and electrical connections offer some major advantages over the crane concept connections:

\begin{itemize}
\item The connections to the trolley concept fixed proximity shielding are above the helium vessel shielding -- this allows hands-on access to connect, disconnect, diagnose faults and repair once the helium vessel lid has been removed.

\item There are no fixed (permanently installed) connectors in the trolley concept helium vessel. The crane concept has fixed connections in the service pillars which will require an additional (relatively complex) mobile remote handling system along with custom-designed remotely operated tooling in order to repair them if they are damaged.

\item The trolley concept design offers good remote handling access (and viewing via the hot cell shielding windows) of the target connections on the trolley. This will allow the remote repair of failed or damaged connectors on the trolley by means of custom -- designed tooling put in place by the hot cell master-slave manipulators. 

\item The connections to the crane concept fixed proximity shielding are built up in layers -- this is a major disadvantage in the event of a leak. First of all it will be hard to leak test the circuits until all the layers are installed. Secondly it will be difficult to diagnose the source of a leak -- especially once the facility is in operation and it is not possible to have personnel access the area of the proximity shielding. Thirdly to repair a leak will require removal and replacement of several layers of shielding and the use of relatively complicated custom -- made remotely operated (un)locking and recovery tooling put in place by the building crane.  
\end{itemize}

\subsubsection{Helium vessel}
\label{Sec:TC:conclusion-helium-vessel}

The trolley concept helium vessel has the added complexity of the side opening for the trolley and the resulting risk of leaks in the event of door seal damage or damage to the sealing face. The seal would need to be replaced remotely using the hot cell master slave manipulators and specially developed tooling.  Damage to the seal face would be very difficult to repair with remote techniques.

The crane concept helium vessel has added complexity due to the provision of services to the target and proximity shielding through the vessel floor.

\subsubsection{Shielding}
\label{Sec:TC:conclusion-shielding}

The key differences between the crane and trolley concept shielding arise from the differences in the proximity shielding layouts. The proximity shielding blocks for both the crane and trolley concepts are of similar levels of complexity - and CERN has already experience of design and manufacture of large iron castings with embedded water cooling pipes. The trolley concept service chimneys and surrounding mobile shielding will need to be particularly carefully designed, built and installed to avoid shine paths above the target.

\subsubsection{Design and development risks}
\label{Sec:TC:conclusion-risks}

A disadvantage of the trolley concept is the additional design and development work needed for the trolley itself and the additional space required in the building underground area for it to operate. 

The crane concept offers the advantage of a simpler facility/building design than the trolley concept. However, the tooling required to operate and recover from failure of the water connections in the helium vessel will be relatively complex and require extensive development and testing. For the crane concept, repair of damaged water or electrical connections in the helium vessel will also require an additional mobile remote handling manipulator system.

\subsubsection{Civil engineering}
\label{Sec:TC:conclusion-civil-engineering}

The crane concept has the advantage of a smaller underground area than the trolley concept, however the need to provide service galleries directly underneath the crane concept's heavily loaded helium vessel will add complexity and risk of settlement movement over the life of the facility.

\subsubsection{Reconfiguration}
\label{Sec:TC:conclusion-reconfiguration}

The BDF shielding thicknesses in the helium vessel have been determined in order to have minimal activation of the helium vessel. In order to allow the maximum flexibility for reconfiguration of the target complex, the design should ideally allow the all the equipment in the helium vessel to be easily removed so that a new kind of target (possibly of completely different dimensions) and shielding etc can then be installed.

Both the crane and trolley concepts place some restrictions on reconfiguration of the target complex. 
The service pillars in the crane concept (which will be radioactive) would need to be cut out to provide complete flexibility for a new design. 
The trolley concept is more restrictive because, in order to keep the cantilevered weight of shielding on the trolley as low as possible, the size of the opening in the shielding in the helium vessel has been kept to a minimum compatible with the existing target design. A bigger target would therefore not be possible without major modification of the trolley and the helium vessel. Further restrictions on the size of a new target could be caused by the size of the opening in the concrete wall between the helium vessel and the hot cell and the dimensions of the transfer passage between the hot cell and the cooldown area.

\subsubsection{Decommissioning}
\label{Sec:TC:conclusion-decommissioning}

Decommissioning of the trolley concept facility would produce a slightly higher volume of waste due to the trolley itself if this became contaminated. Decommissioning of the crane concept facility would be complicated by the need to cut the service pillars out of the helium vessel and remove potentially contaminated pipework from the service galleries below the helium vessel.

\subsubsection{Choice between crane concept and trolley concept}
\label{Sec:TC:conclusion-choice}

At the end of the study, after analysis and comparison of the two designs - mainly in view of the concerns explained above relating to the water and electrical connections inside the helium vessel -- it was decided to consider the trolley concept as the baseline for the ongoing work being carried out on the integration of the whole facility, cooling and ventilation studies, helium vessel studies etc.

\subsection{The  ``Crane++'' concept - a third concept combining crane and trolley concept benefits}
\label{Sec:TC:conclusion-crane++-concept}

\subsubsection{Explanation of Crane++ concept}
\label{Sec:TC:conclusion-crane++-concept-explanation}
After analysing and comparing the designs produced during the handling and integration study based on the crane and trolley concepts it was realised that a third concept could combine the best features and advantages of both concepts. This third concept uses the crane for all handling but has service chimneys (as used for the trolley concept proximity shielding) for both the target and the proximity shielding. This third concept has been named the  ``Crane++'' concept.

\begin{figure}[!htb]
\centering
\includegraphics[width=0.95\linewidth]{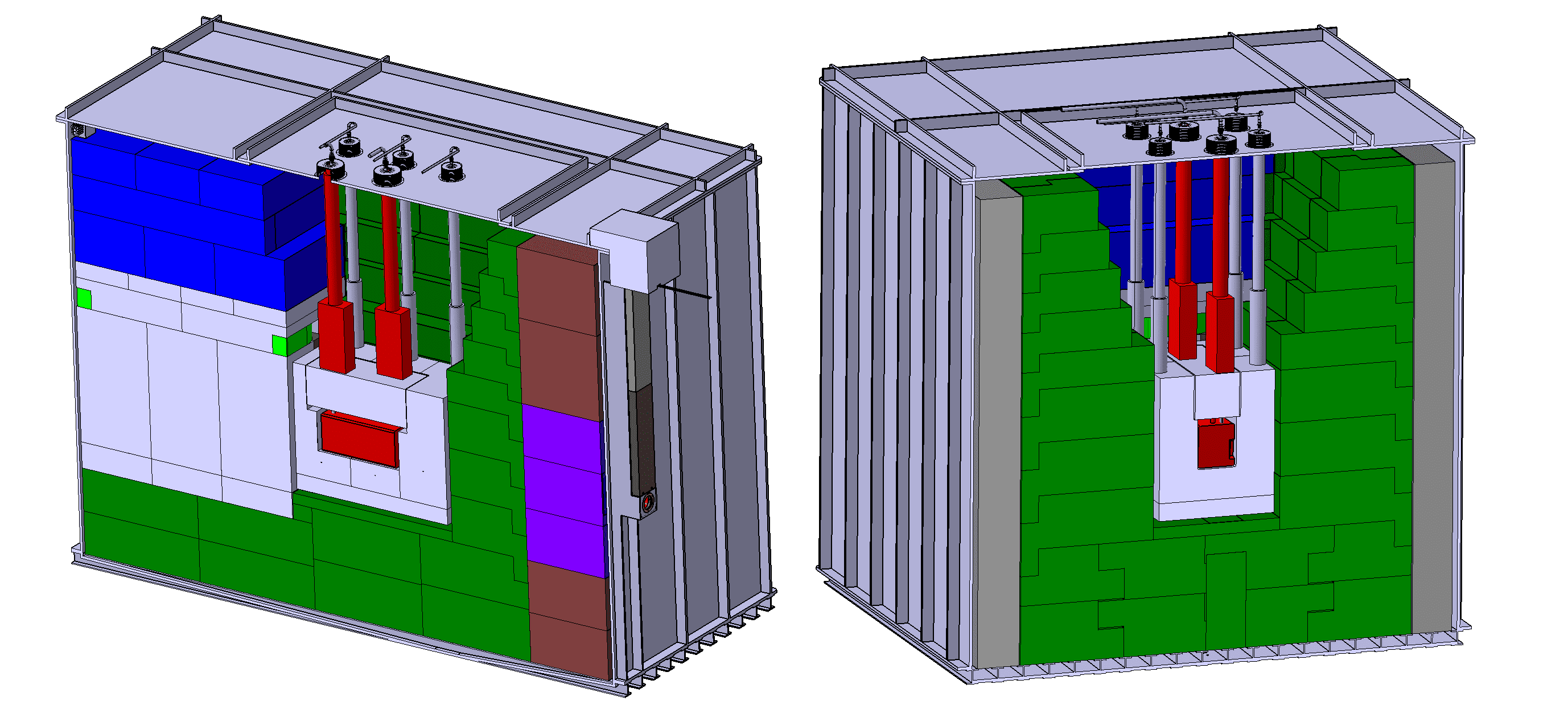}
\caption{Crane++ concept; On the left -- vertical section through the helium vessel parallel to the beam line showing the target (in red) in the centre surrounded by the proximity shielding (in grey).  The target is suspended from an upper plug in the proximity shielding. The mobile shielding above the proximity shielding has been removed to show the service chimneys for the target (in red) and for the proximity shielding (in grey). On the right -- section through the helium vessel perpendicular to the beam line showing U shaped proximity shielding.}
\label{Fig:TC:77-crane++}
\end{figure}

The Crane++ proximity shielding is similar to the trolley concept proximity shielding but the fixed portion is in a U shape rather than a C shape to allow the target to be installed and removed vertically by the building crane (Fig.~\ref{Fig:TC:77-crane++}) . The target is suspended from an upper plug of water cooled shielding that is supported by the fixed portion U shaped proximity shielding castings. The method of suspension will allow adjustment of the target alignment if necessary. Water cooling of the proximity shielding uses cast-in thick walled stainless steel pipes as used for the trolley concept. Pipework for water cooling of the target and electrical services passes through the two service chimneys attached to the upper plug of proximity shielding. 

The crane++ concept includes service passages in the helium vessel lid to provide fully accessible water and electrical service connections. Any connections in the pipework below the helium vessel lid are welded to avoid the risk of leaks from mechanically clamped seal connectors. Water and electrical connections at the top of the service chimneys can be made, disconnected and repaired hands-on rather than remotely; this reduces the complexity of the connectors, the risk of damage during connection work and greatly simplifies fault diagnosis and repair. 

\begin{figure}[!htb]
\centering
\includegraphics[width=0.9\linewidth]{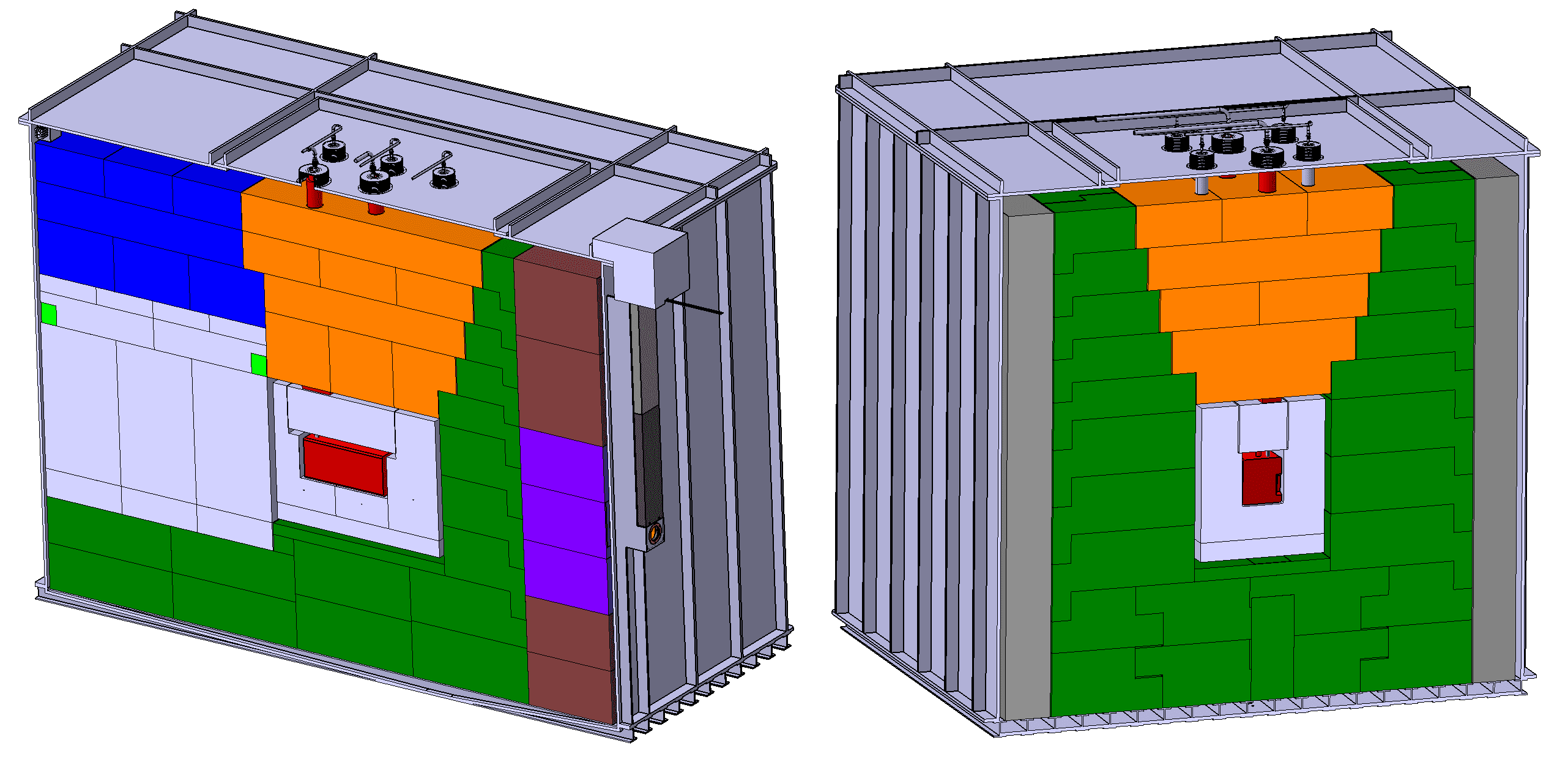}
\caption{Crane++ concept : vertical sections through the helium vessel with the mobile shielding (orange) installed above the proximity shielding.}
\label{Fig:TC:after77-crane++}
\end{figure}

\begin{figure}[!htb]
\centering
\includegraphics[width=0.7\linewidth]{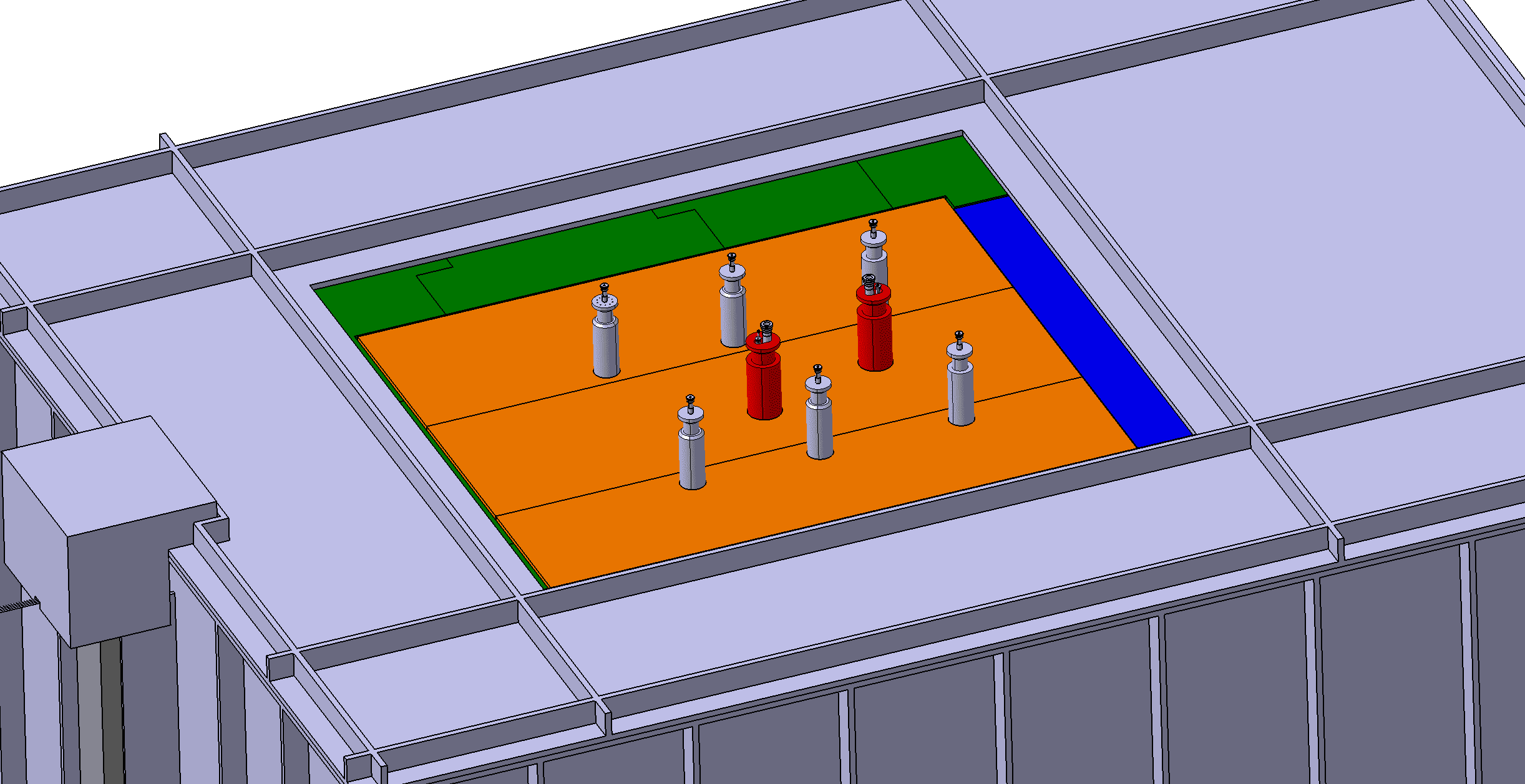}
\caption{Crane++ concept: view of the upper surface of the shielding in the helium vessel with the small lid removed, showing the tops of the service chimneys for the proximity shielding (in grey) and for the target (in red). Note that the services for the proximity shielding plug above the target also pass through the target service chimneys.}
\label{Fig:TC:Extra-3-crane++-upper-surface}
\end{figure}

\begin{figure}[!htb]
\centering
\includegraphics[width=0.8\linewidth]{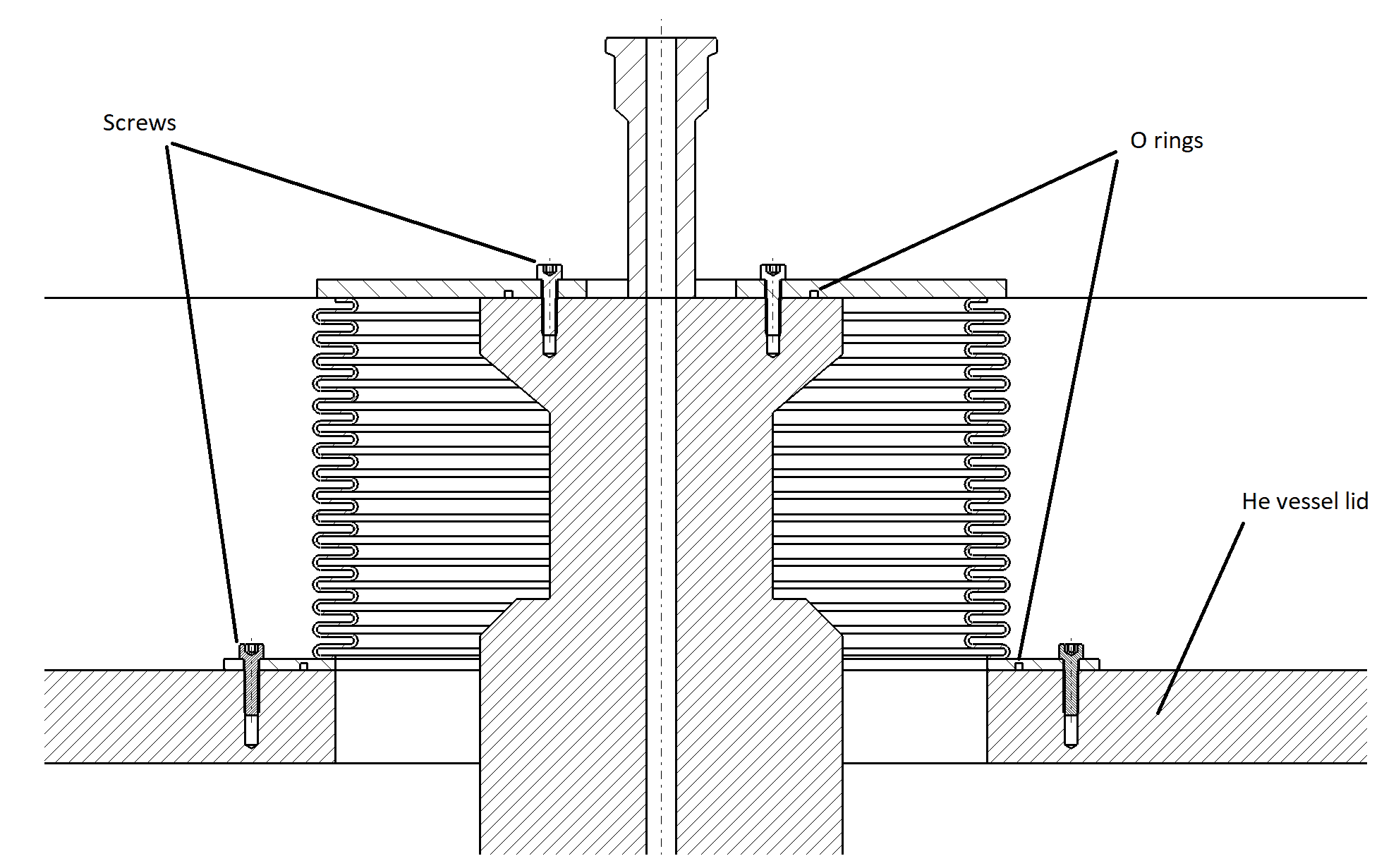}
\caption{Crane++ concept: service passages in the helium vessel lid. Section view illustrating outline concept for target water cooling connections at the top of service chimneys.  A bellows unit is used to provide flexibility to adapt to misalignments (in XYZ+ tilt) of the service chimneys with respect to the He vessel lid and also to any movements of the lid due to pressure changes etc.}
\label{Fig:TC:Extra-4-crane++-service-passage}
\end{figure}

\begin{figure}[!htb]
\centering
\includegraphics[width=0.55\linewidth]{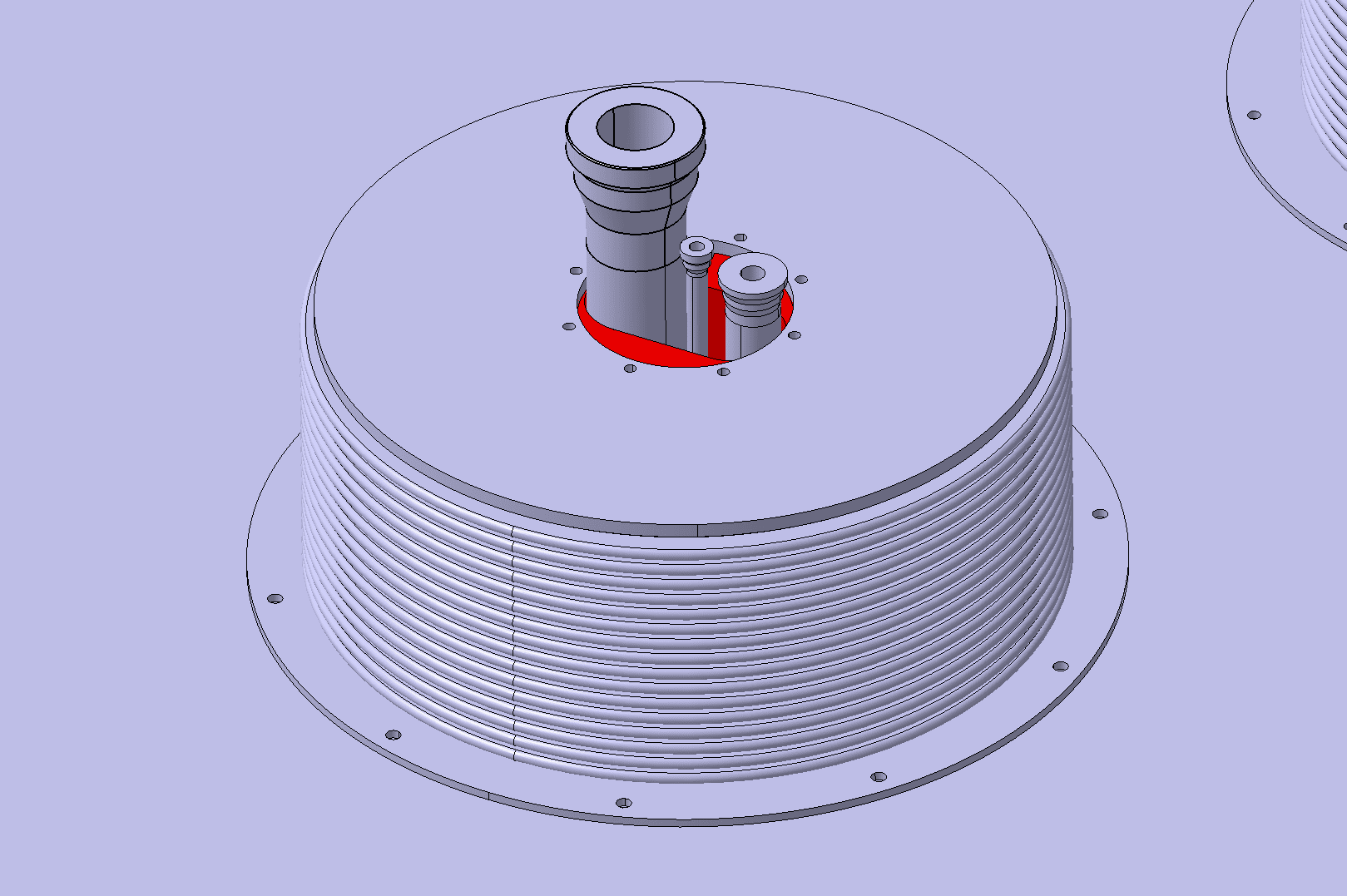}
\caption{Crane++ concept: service passages in the helium vessel lid.  Close-up view of pipes coming out of target service chimney viewed from above the service vessel lid.  Pipe clamps not shown.  The large diameter pipe is for target cooling, smaller diameter pipes are for cooling of the proximity shielding top plug and helium supply to the target.}
\label{Fig:TC:Extra-5-crane++-service-passage-2}
\end{figure}

\begin{figure}[!htb]
\centering
\includegraphics[width=0.5\linewidth]{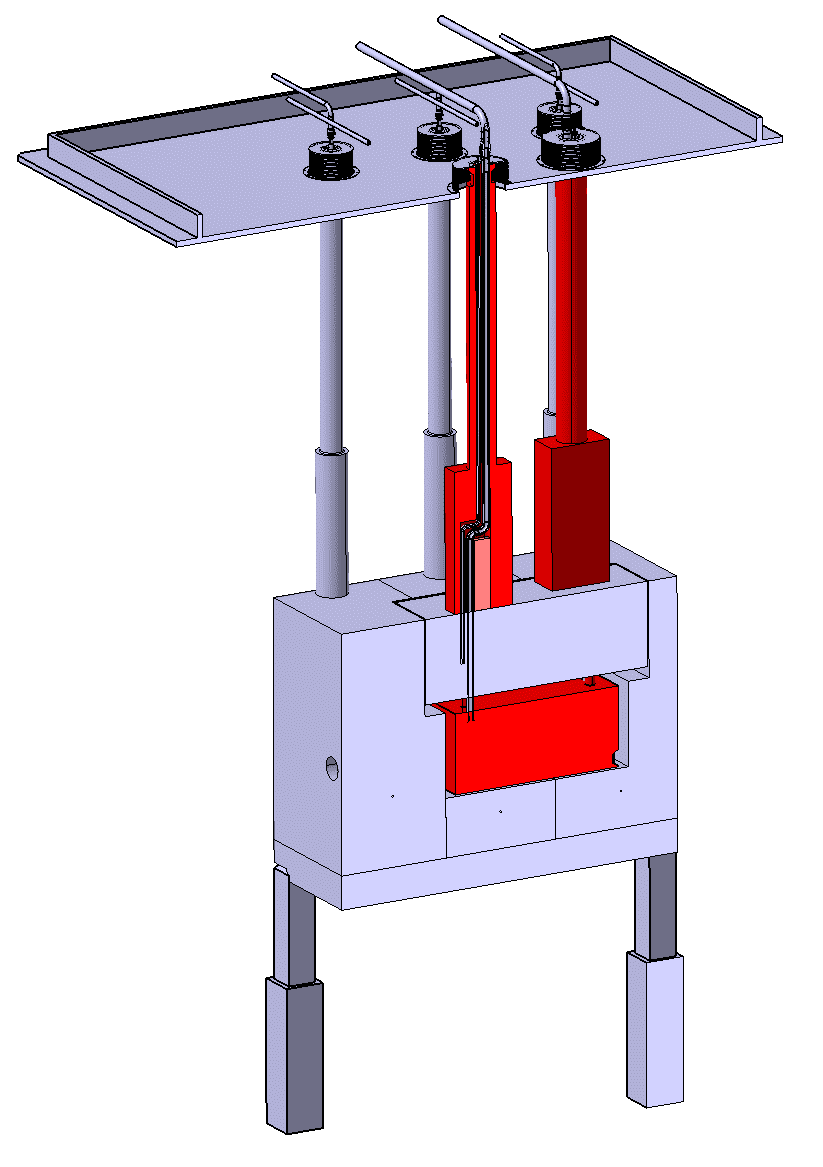}
\caption{Crane++ service passages in the small helium vessel lid.  Cut-away view of proximity shielding and target with service chimneys and small helium vessel lid.}
\label{Fig:TC:Extra-6-crane++-service-passage-3}
\end{figure}

\begin{figure}[!htb]
\centering
\includegraphics[width=0.7\linewidth]{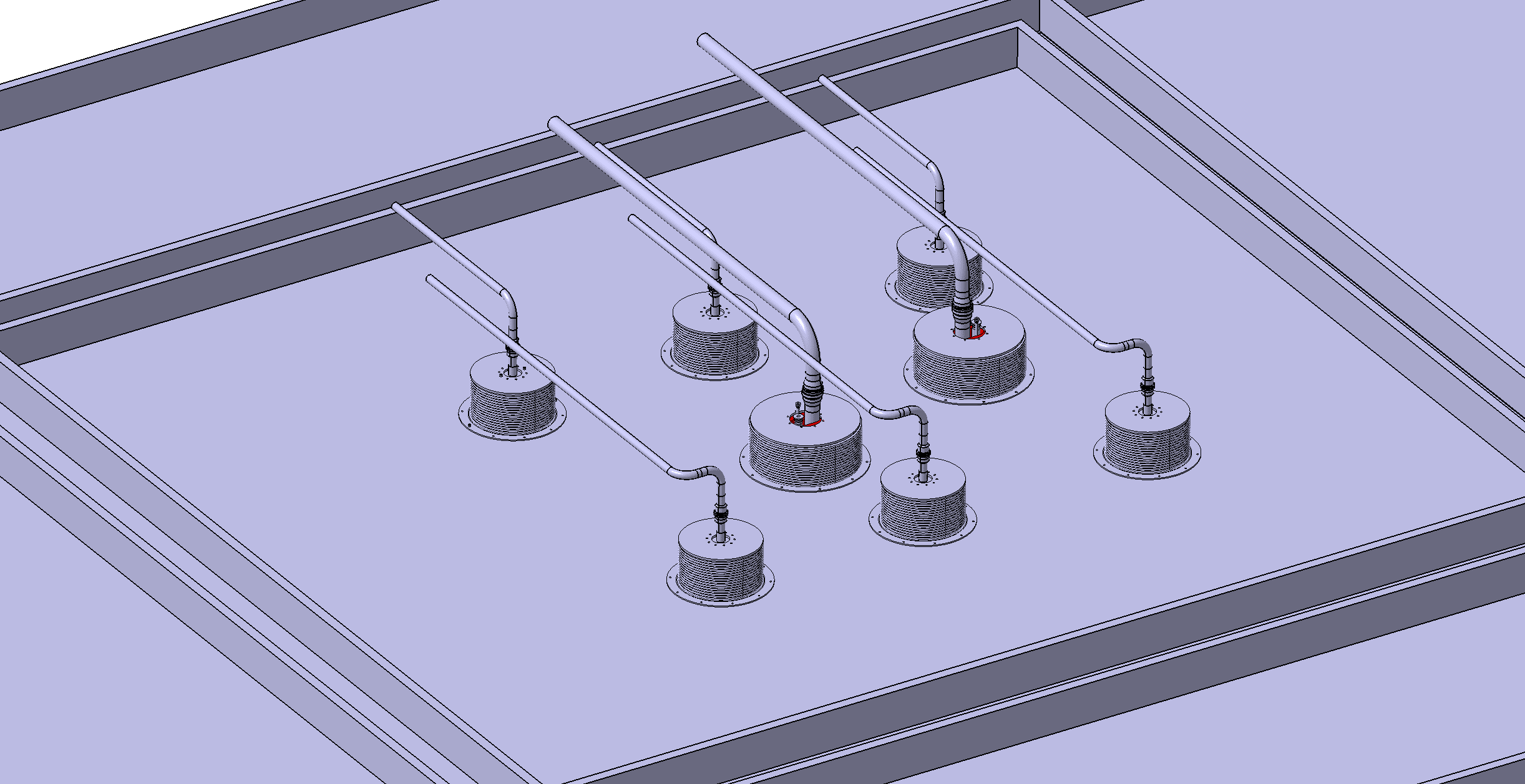}
\caption{Crane++ concept: service passages in the helium vessel lid.  Isometric view of the main cooling pipes coming out of target and proximity shielding service chimneys above the small helium vessel lid.  Smaller cooling pipes for the proximity shielding top plug and and helium pipes for the target are not shown.}
\label{Fig:TC:Extra-7-crane++-service-passage-4}
\end{figure}

The main features of the crane++ concept are illustrated in Figs.~\ref{Fig:TC:77-crane++} to~\ref{Fig:TC:Extra-7-crane++-service-passage-4}; it is proposed to further develop the concept in the next phase of the project.

\subsubsection{Crane++ target exchange steps}
\label{Sec:TC:conclusion-crane++-target-exchange}

For the Crane++ concept (as in the crane concept) all installation and removal operations are carried out vertically using the building's overhead travelling crane. As for the crane concept, it is proposed to have a small lid as part of the helium vessel lid so that it is not necessary to remove the complete lid in order to exchange the target.

Before starting lifting operations, all the connections to the target and proximity shielding service chimneys are manually disconnected from outside the lid of the helium vessel. In order to exchange the target, it is necessary to remove the helium vessel small lid and the  ``mobile shielding'' (Figs. \ref{Fig:TC:80-crane++-4} and \ref{Fig:TC:81-crane++-5}). 

Once the mobile shielding is removed, a shielded cask is lowered into position above the target. The target and proximity shielding upper plug are then lifted -- via a lift attachment that interfaces with the top of the target service chimney in a similar way to the one used for the trolley concept (Figs. \ref{Fig:TC:51-crance-shileding-unlocking} and \ref{Fig:TC:52-shielding-bolting-tool}) As the target and proximity shielding are lifted, the target enters the shielded transfer cask (Fig.~\ref{Fig:TC:82-crane++-6}). Further lifting of the target also lifts the shielded cask so that the target is shielded during transfers to the storage area and lowering of the target into a cooldown area storage pit (Figs.~\ref{Fig:TC:83-crane++-7} and~\ref{Fig:TC:84-crane++-8}).

A series of remotely operated spreader beams (as used for the trolley concept -- see Fig. \ref{Fig:TC:54-removing-mobile-shielding}) are used to lift the mobile shielding above the target and then the target itself. The steps involved in target removal are given in table \ref{Tab:TC:crane++-removal}. The installation of a new target follows the same procedure but in reverse.

\begin{table}[!htb]
\begin{tabular}{|p{0.1\textwidth}|p{0.4\textwidth}|p{0.4\textwidth}|} \hline 
\centering
\textbf{Step} & \textbf{Task} & \textbf{Tooling} \\ \hline 
 &  Disconnect water and electrical connections to target and proximity shielding upper plug &  Hands-on operation -- working on top of helium vessel \\ \hline 
 a &  Open lid of helium vessel &  Hands-on operation \\ \hline 
 b &  Remove mobile shielding above target and transfer to cool down area & Crane and remotely operated spreaders \\ \hline 
 c &  Lower shielded transfer cask into position above target &  Crane and remotely operated spreader \\ \hline 
 d &  Lift out target with shielded transfer cask and transfer to cool down area -- placing target (and proximity shielding upper plug) into underground storage pit &  Crane and remote lift attachment and shielded transfer cask\\ \hline
\end{tabular}
\caption{Steps for removal of a target from the helium vessel and transfer to the cool down area (crane++ concept)}
\label{Tab:TC:crane++-removal}
\end{table}

The main steps in target removal for the crane++ concept are illustrated below.

\begin{figure}[!htb]
\centering
\includegraphics[width=0.6\linewidth]{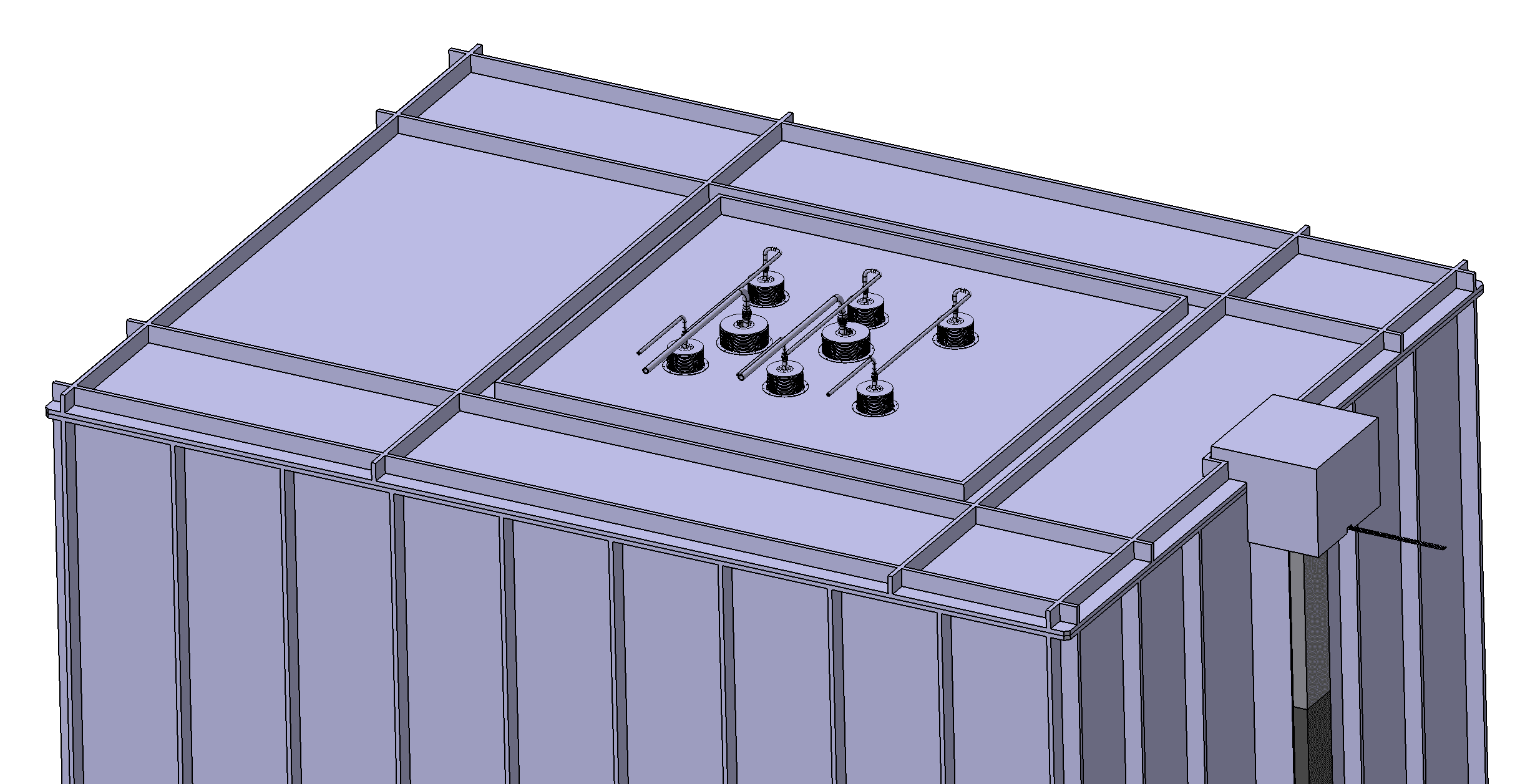}
\caption{Crane++ concept target exchange: isometric view of top of the helium vessel at the start of operations.}
\label{Fig:TC:2before80-crane++}
\end{figure}

\begin{figure}[!htb]
\centering
\includegraphics[width=0.6\linewidth]{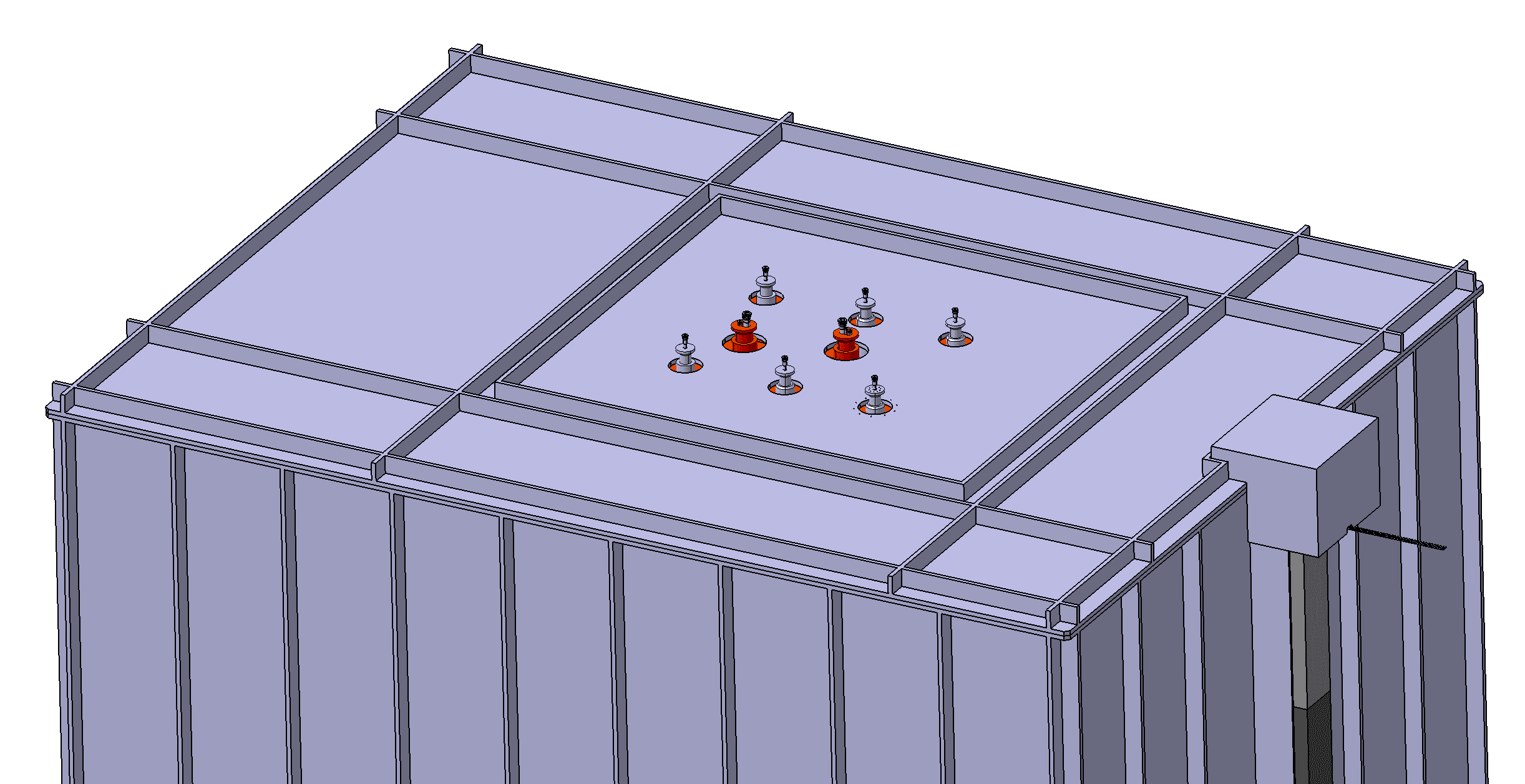}
\caption{Crane++ concept target exchange: pipework disconnected and removed.  Note that this work is carried out "hands-on" without the need for remote handling. }
\label{Fig:TC:before80-crane++}
\end{figure}

\begin{figure}[!htb]
\centering
\includegraphics[width=0.6\linewidth]{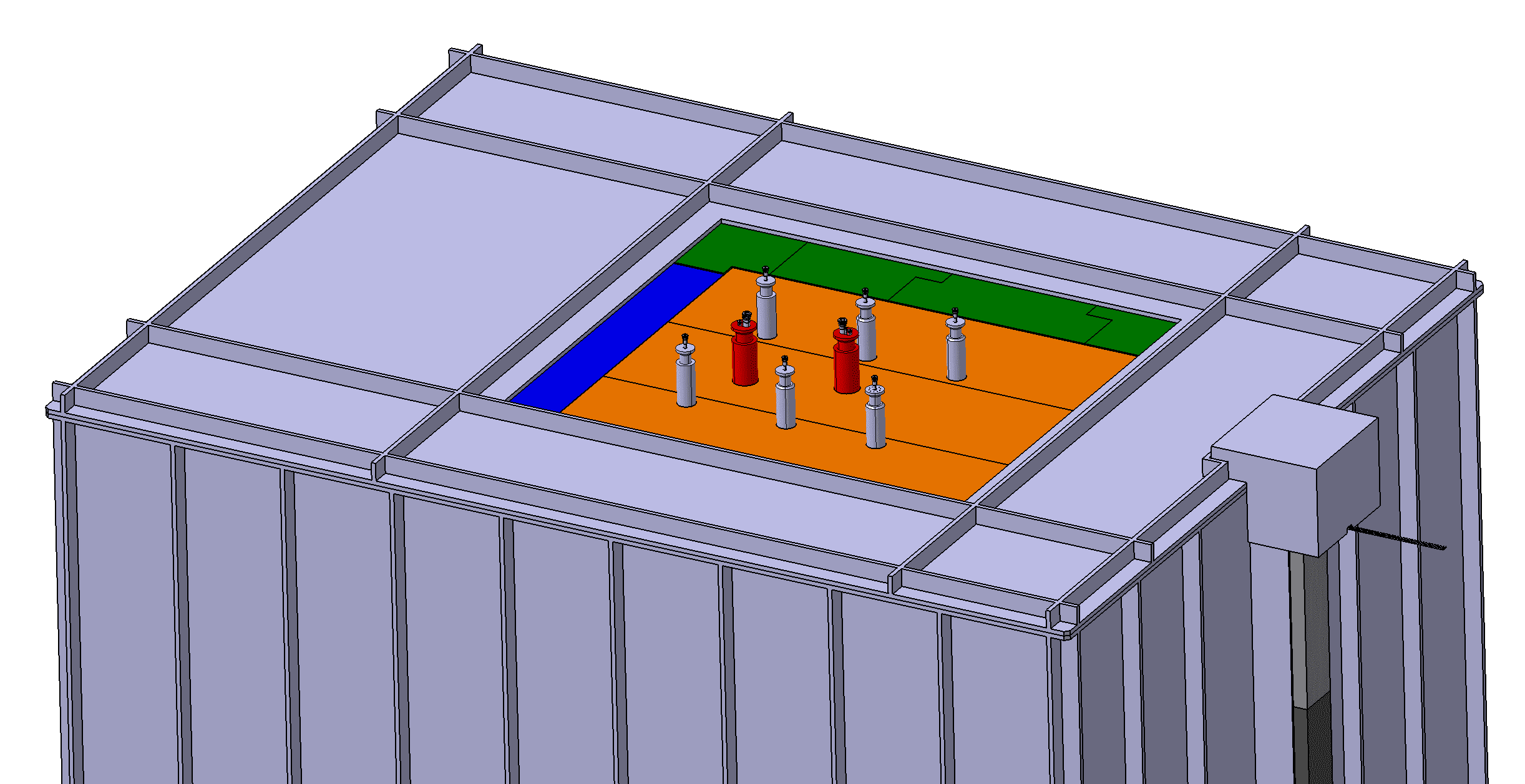}
\caption{Crane++ concept target exchange: small lid removed from helium vessel }
\label{Fig:TC:80-crane++-4}
\end{figure}

\begin{figure}[!htb]
\centering
\includegraphics[width=0.7\linewidth]{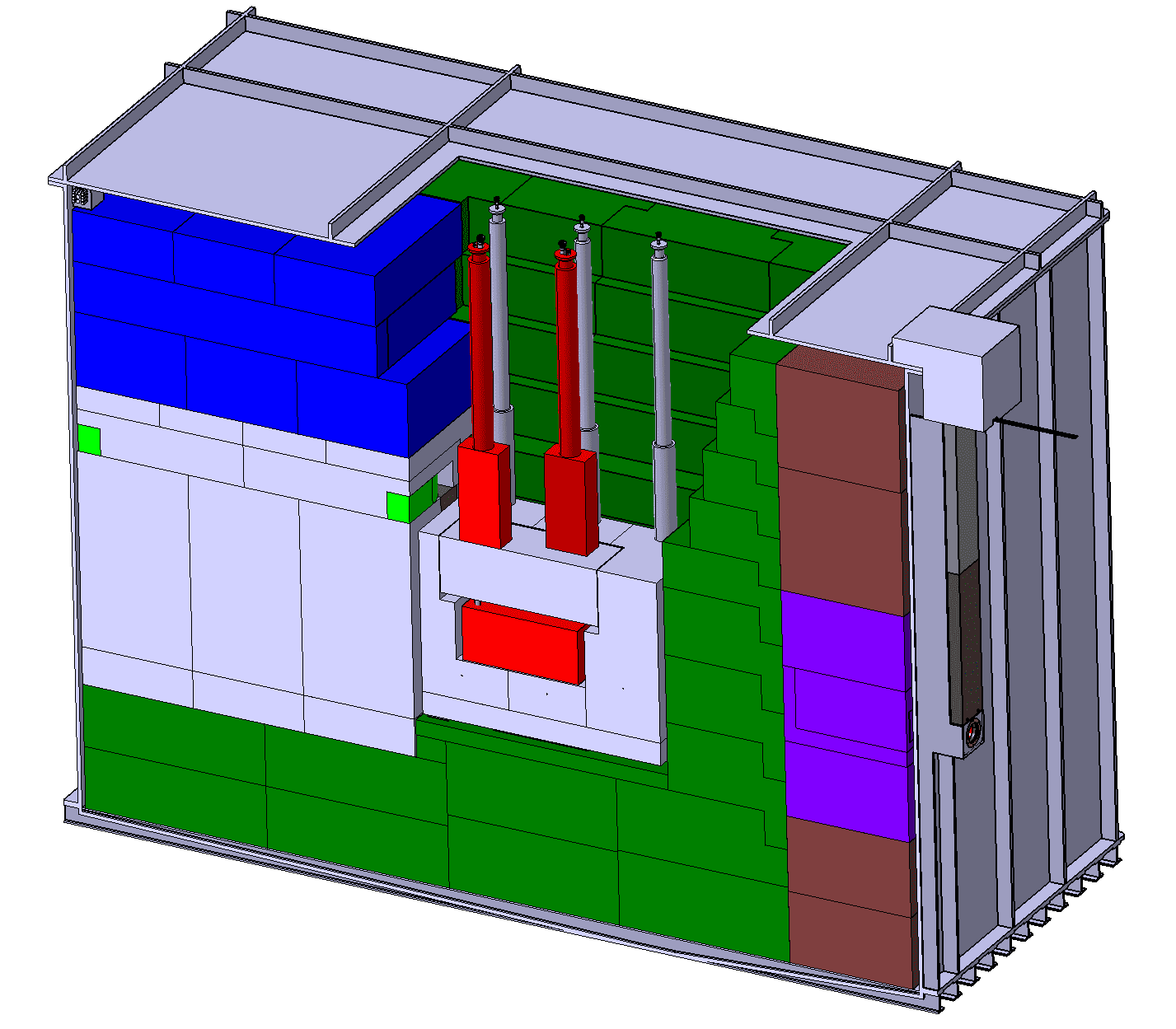}
\caption{Crane++ concept target exchange: mobile shielding above target removed. Note that this is carried out remotely.}
\label{Fig:TC:81-crane++-5}
\end{figure}

\begin{figure}[!htb]
\centering
\includegraphics[width=\linewidth]{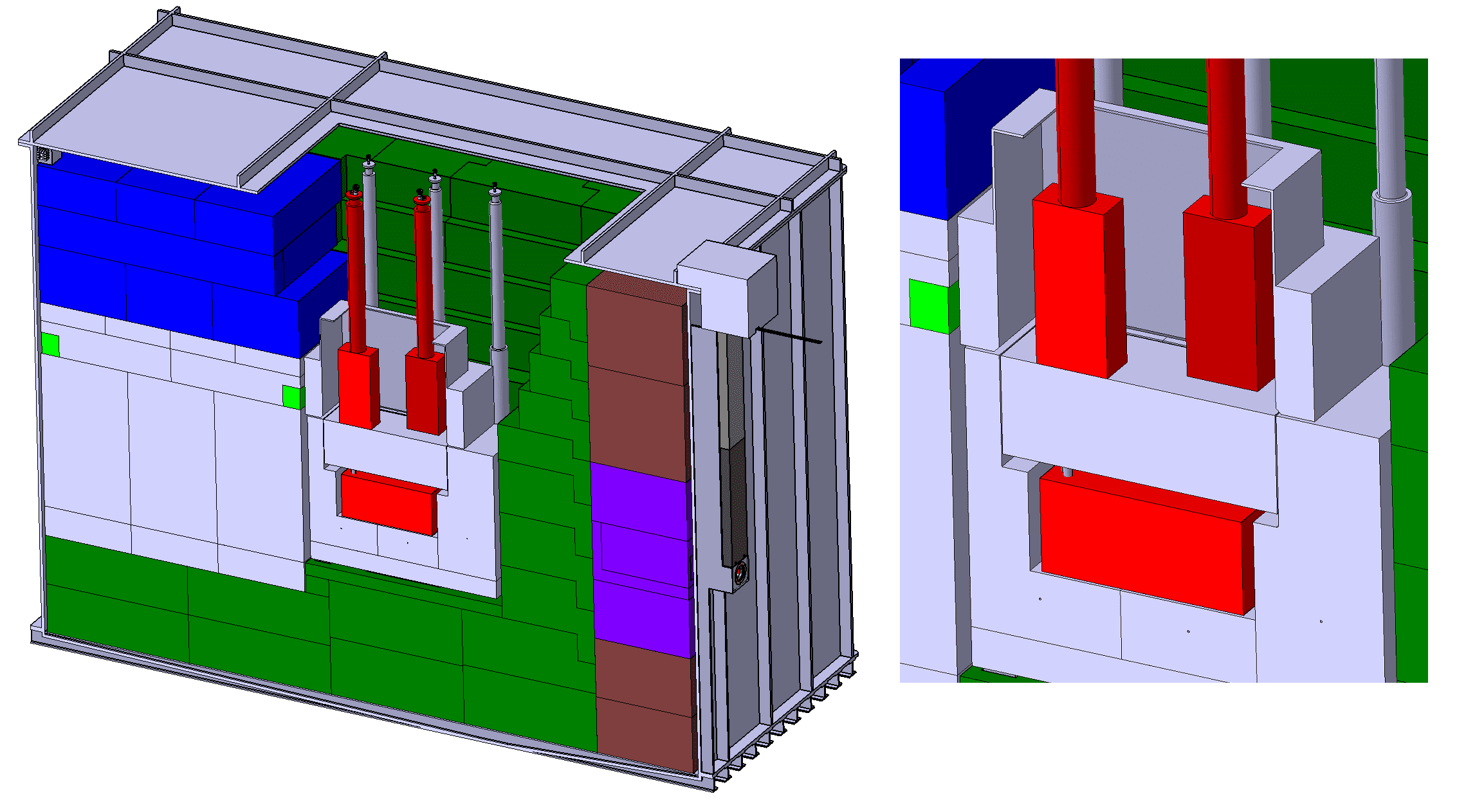}
\caption{Crane++ concept target exchange: shielded transfer cask lowered into position above target. On the left -- overview. On the right -- close up of shielded transfer cask after being lowered into position over target, before the target is raised.} 
\label{Fig:TC:82-crane++-6}
\end{figure}

\begin{figure}[!htb]
\centering
\includegraphics[width=0.6\linewidth]{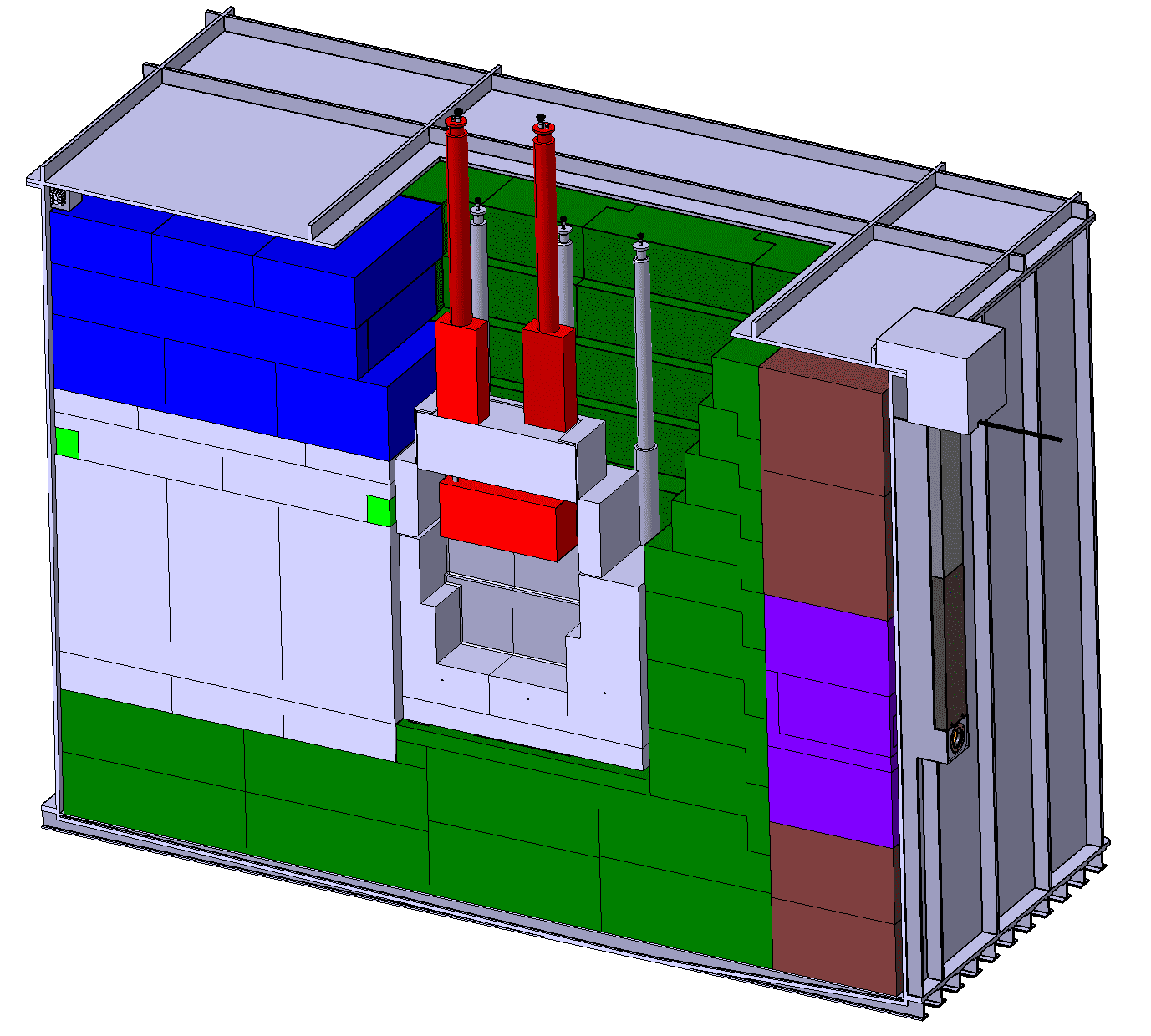}
\caption{Crane++ concept target exchange: target and proximity shielding plug lifted by target complex building crane into shielded transfer cask.}
\label{Fig:TC:83-crane++-7}
\end{figure}

\begin{figure}[!htb]
\centering
\includegraphics[width=0.6\linewidth]{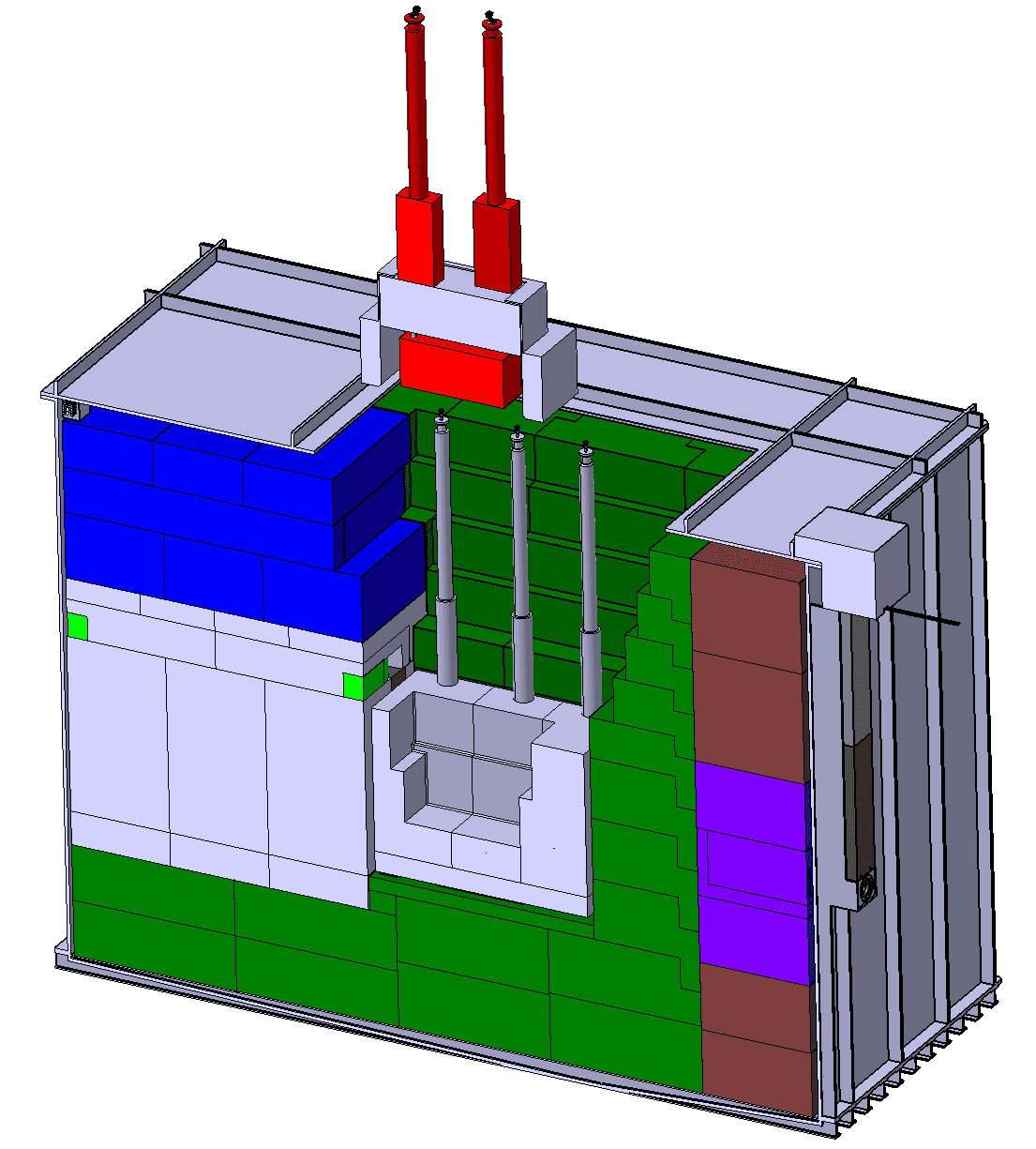}
\caption{Crane++ concept target exchange: further lifting of the target and shielding plug lifts the shielded transfer cask. The target is lifted out of the helium vessel inside the shielded transfer cask before transfer to cooldown area storage pit (keeping close to the floor during transfer)}
\label{Fig:TC:84-crane++-8}
\end{figure}

\subsection{Comparison of crane++ concept with crane and trolley concepts}
\label{Sec:TC:comparison-crane++-trolley}
Many elements of all three concepts are essentially the same so they do not influence any decision on the choice of concept -- for example the beam window, collimator, magnetic coil and US1010 shielding. However it should be noted that the crane++ concept service chimney design approach with connections accessible from outside the helium vessel could also be applied to the magnetic coil and collimator; this could improve availability due to shorter fault diagnosis and repair times.

To compare the crane++ concept with the other two concepts, the following elements related to operation and construction of the target complex have been considered:

\begin{enumerate}[label=\alph*]
\item Target exchange \label{Item:TC:exchange}
\item Water and electrical connections \label{Item:TC:connections}
\item Helium vessel \label{Item:TC:helium}
\item Shielding \label{Item:TC:shielding}
\item Handling equipment design and development risks \label{Item:TC:handling-equip}
\item Civil engineering \label{Item:TC:engineering}
\item Reconfiguration \label{Item:TC:reconfiguration}
\item Decommissioning \label{Item:TC:decommissioning}
\end{enumerate}

\subsubsection{Target exchange}

The Crane ++ target exchange is similar to the crane concept but with some key advantages:

\begin{itemize}
\item Water and electrical connections to the target and proximity shielding upper plug are disconnected by hand.
\item The shielded transfer cask does not need its own, high integrity, hoist system.
\end{itemize}

\subsubsection{Water and electrical connections}
The major advantage of the crane++ concept is that the water and electrical connections are accessible for hands-on connection, disconnection, fault diagnosis and repair. This can be expected to lead to higher reliability and availability of the facility along with lower construction and operation costs.

\subsubsection{Helium vessel}
As the water and electrical connections for the crane ++ concept are accessible outside the helium vessel, the risk of leaks and electrical problems requiring removal of the lid to repair them is minimised; there will therefore be less need to remove the whole helium vessel lid. This will reduce the number of times the helium has to be purified and reduce the risk of problems with the main and small lid seals.

Not having the side door opening in the helium vessel wall that is needed for the trolley concept avoids the risk of leaks or damage to the inflatable seal or sealing faces which would need to be repaired via the trolley hot-cell using remote handling techniques.

\subsubsection{Shielding}
The proximity shielding is very similar to the trolley concept with the same sort of service chimneys. The mobile shielding above the proximity shielding is also similar to the trolley concept with stepped openings for the service chimneys.

\subsubsection{Handling equipment design and development risks}
The handling equipment is much simpler than for the crane concept as there are no remotely operated (un)locking tools, remotely operated recovery tools and no mobile remote telemanipulator system to design and develop. The handling system is much simpler than for the trolley concept as there is no trolley and trolley hot cell to design and develop. The remote handling area at the end of the cooldown area will be the same as for the crane and trolley concepts and will be used to disconnect beam window and coil services as well as to disconnect the service chimneys from the crane ++ target and proximity shielding before disposal.

\subsubsection{Civil engineering}
The civil engineering will be simpler than for the trolley concept as there will be no need for the trolley operating area and hot cell. The civil engineering will be simpler than for the crane concept as there will be no need for service galleries under the helium vessel.

\subsubsection{Reconfiguration}
The crane++ concept is more flexible than the trolley concept as it does not have the restrictions on target size, weight and cooling that are inherent in the trolley concept. The crane++ concept reconfiguration would be more flexible than the crane concept which has fixed service pillar positions and extensive remote tooling for connection operation which is specific to the crane concept target vessel and proximity shielding designs.

\subsubsection{Decommissioning}
The crane++ offers the easiest decommissioning of all the concepts as all radioactive parts inside the helium vessel can be easily lifted out (no need to cut out radioactive service pillars from the helium vessel and no need to dismantle the radioactive parts at the front of the trolley). The dismantling of the service chimneys would be carried out in the cooldown remote handling area.

\subsection{Further work -- Choice between crane, trolley and crane++ concepts}
\label{Sec:TC:choice-crane=trolley-crane++}
At the current stage of development the crane++ concepts appears to offer a simpler, cheaper and potentially more reliable (hence offering higher availability) target complex design. It is therefore proposed to develop the design to a similar level of detail as already done for the other two concepts. If this does not show up any major weak points then the crane++ concept will be used as the basis for the further work listed below.

\subsection{Further work -- integration of results of other studies}
\label{Sec:TC:further-work}
As increasingly detailed results from design studies on aspects of the target complex become available, the integration of the target complex will need to be updated. Examples of the aspects concerned are:

\begin{itemize}
\item Radiation protection
\item Target and target container
\item Helium vessel
\item Magnetic coil and shielding
\item Cooling and ventilation (including helium purification)
\item Access systems
\item General safety
\item Electrical services 
\item Overhead travelling cranes
\end{itemize}

\begin{figure}[!htb]
\centering
\includegraphics[width=0.9\linewidth]{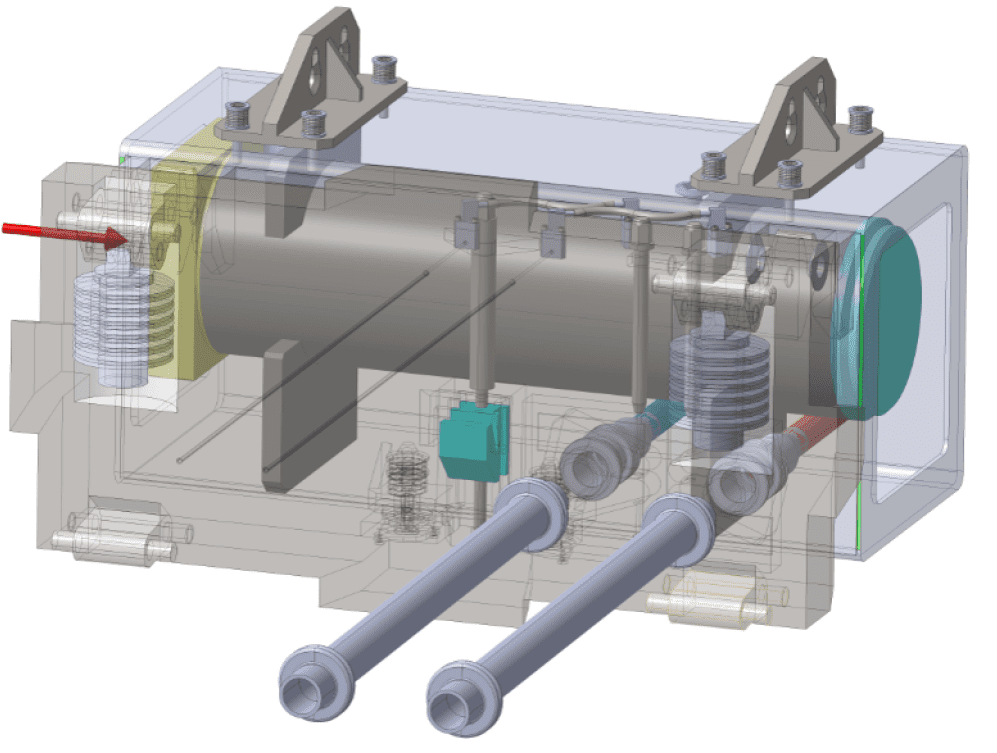}
\caption{Example of design work integrating the results of different studies - integration of BDF target core in target container (trolley concept container shown).}
\label{Fig:TC:85-integration-target-core}
\end{figure}

As explained above, it is expected that this next phase of the target complex integration work will be based on the crane++ concept. 

\clearpage

\section{Hadron absorber and muon shield}
\label{Sec:TC:hadron_absorber}

One of the primary requirements of the SHiP experiment is to operate in an environment of
extremely low background from ordinary physics processes.  To meet this objective, the 
target system is immediately followed downstream by an important mass of shielding with several 
functionalities. Firstly, the shielding should absorb the electromagnetic radiation and 
hadrons emerging from the proton target. Secondly, SHiP employs a magnetic system referred
to as {\it muon shield}~\cite{Akmete:2017bpl} to deflect the large flux of muons emerging 
from the target away from the fiducial volume of the detector. The muon shield is designed 
to reduce the flux of muons by six orders of magnitude in the detector acceptance. 

The shield consists of a first set of magnets that deflect the positively and negatively 
charged muons on either side of the beam axis, irrespective of their initial 
direction. This allows creating a region around the beam line beyond this first set of 
magnets in which there are no charged particles. A second set of magnets which have 
their return field in this unoccupied region then provides further deflection. This 
configuration prevents muons deflected in the first section from being deflected back 
towards the detector by the return field in the second section.  

Despite the aim to search for 
particles with relatively long lifetimes, the sensitive volume of the SHiP experiment should 
be situated as close as possible to the proton target due to the relatively large production 
angles. The length and the aperture of the magnetic system is minimised by applying the 
highest possible magnetic field as close as possible to where the muons are produced.  
Technically it is very difficult to introduce a field across the target system and the proximity 
shielding, which makes up the first 0.8~m of the hadron absorber, and the field would 
interfere with the instrumentation of the target system. However, the rest of the absorber may be 
magnetised with the help of a magnetic coil integrated into the shielding. The subsequent magnetic 
system downstream of the hadron absorber is composed of a set of 
free-standing magnets located inside the underground experimental hall. Their field configuration 
is optimized using machine learning techniques. The separation between the magnetised volumes 
in the hadron absorber and in the first free-standing magnet is minimised by having a roughly 30~mm thick
stainless-steelwindow in the wall separating the target bunker from the 
experimental hall.  

\begin{figure}[htbp]
\centering
\includegraphics[width=0.7\columnwidth]{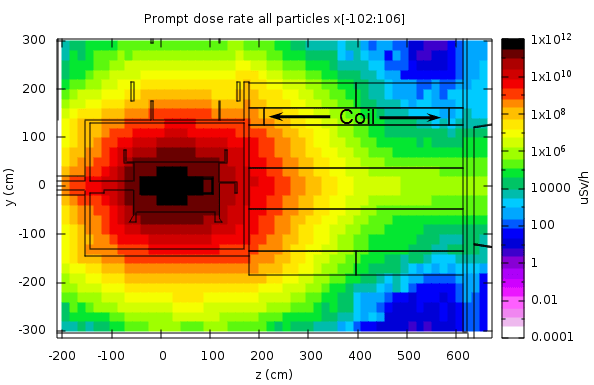}
\caption{Average annual radiation dose at coil height. Note that the coil body will see an annual radiation dose of around 10$^5$ Gray at its closest point to the target, i.e. 0.1 MGy per year. At the location where the services arrive, the radiation dose is approximately 10$^{-3}$ Gy/year.}
\label{Fig:TC:CoilRadiationMaps}
\end{figure}

\begin{figure}[htbp]
\centering
\includegraphics[width=0.9\columnwidth]{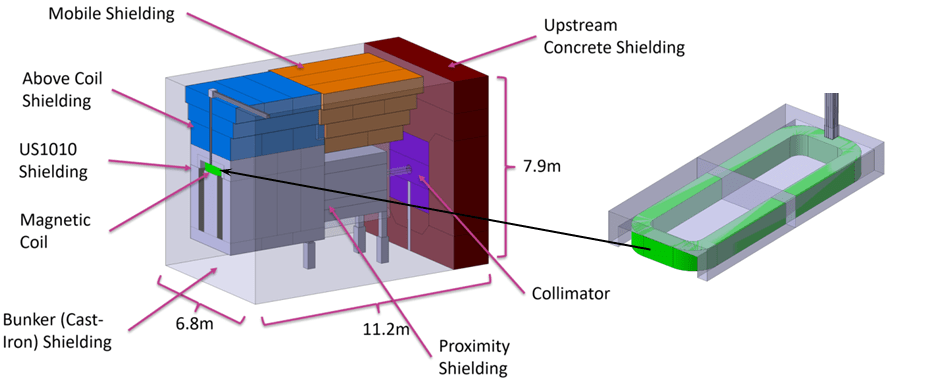}
\caption{View of the magnetized section of the target shielding with the yoke configuration and the coil. The coil is embedded in specially designed shielding blocks which restrains the coil during operation and which provides crane lifting points should an intervention be necessary (Credits: V. Bayliss, J. Boehm, RAL (UK)). }
\label{Fig:TC:HadronStopperMagnetization}
\end{figure}

\begin{figure}[htbp]
\centering
\includegraphics[width=0.6\columnwidth]{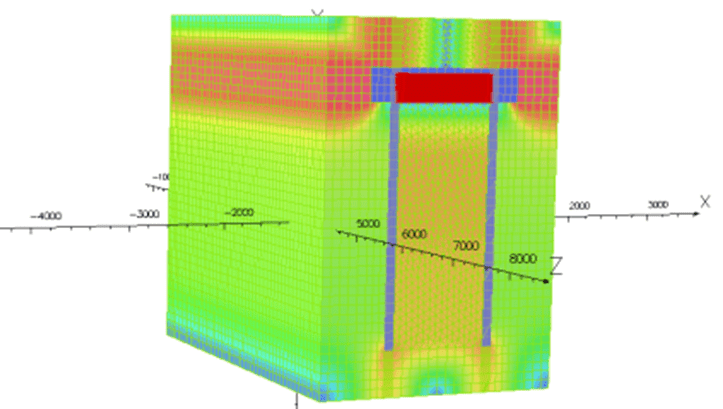}
\caption{The magnetic field in the magnetised volume of the hadron stopper. The coil is shown in red near the top of the shielding (Credits: V. Bayliss, J. Boehm, RAL (UK)).}
\label{Fig:TC:HadronStopperMagneticField}
\end{figure}

The overall physical dimensions of the hadron absorber are primarily driven by the radiological 
considerations as described in Section~\ref{Sec:TC:RP}. Studies show that it is possible to almost
entirely prevent radio-activation of the experimental area and its components with 5~m of iron 
shielding in total. However, the coil and the integration of the coil in the target shielding 
is subject to several severe constraints related to the radiation exposure, 
(see Figure~\ref{Fig:TC:CoilRadiationMaps}), powering, heat extraction, and handling. As shown in Figure~\ref{Fig:TC:HadronStopperMagnetization} the preliminary design is based on a single 
set of coils which magnetises the last 4~m of the hadron absorber after the proximity shielding. In order reduce the radiation to 
the coil it is located at 1.30~m above the beam axis, which then also drives the requirement 
on the height of the volume that is magnetised. Two vertical non-magnetic stainless-steel plates 
(shown in black in Figure~\ref{Fig:TC:HadronStopperMagnetization}) ensure that the dipole field 
is correctly guided through the US1010 steel blocks making up the yoke. The design and layout of the 
US1010 blocks have been optimised to ensure the minimum number of gaps seen by the 
magnetic field, whilst a the same time remaining compatible with the achievable precision 
in manufacturing and handling. The magnetic field modelling of the whole assembly in Opera Vector 
Fields shows that a field of >1.6 T can be achieved in the critical volume, as shown in 
Figure~\ref{Fig:TC:HadronStopperMagneticField}. In order to deflect a sufficient fraction of the 
angular spread of the muon flux that would otherwise reach the detector acceptance, the width 
of the magnetised region must be around 1.60~m.

\begin{figure}[htbp]
\centering
\includegraphics[width=0.8\columnwidth]{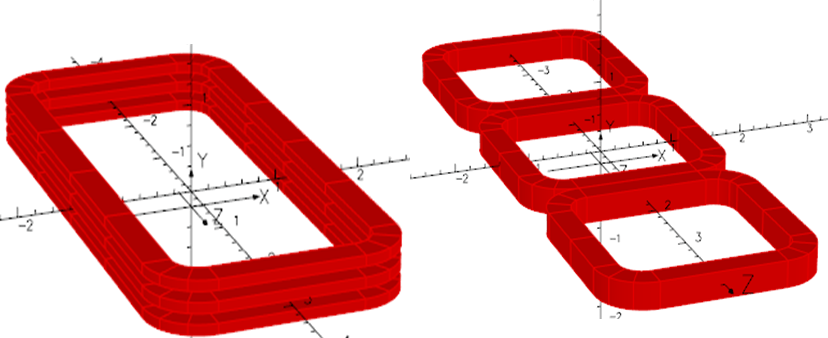}
\caption{The modular coil assemblies being considered, with either coils stacked vertically (left) or aligned horizontally (right).}
\label{Fig:TC:CoilStudies}
\end{figure}

Initial calculations indicate that the thermal and the electrical engineering of a coil of this 
size are manageable. An aluminium tape conductor is preferred for mass, radiation and thermal reasons. 
The preliminary design considerations for the coil assembly pursues two different configurations,
both of which are based on multiple coils powered and cooled independently to ensure redundancy 
(Figure~\ref{Fig:TC:CoilStudies}). The baseline consists of a single 
coil assembly comprised of several layers of coils stacked vertically and stretching the entire length of 
the volume to magnetise. The second option comprises several coils located in a chain horizontally.
The latter option have the advantage of reducing the mass of the individual 
coils and reducing the manufacturing complexity, but at the penalty of slightly reducing the integrated 
field. The impact on the routing of the services through the shielding and the helium vessel surrounding 
the entire shielding assembly is also under investigation. 

The remote handling of the magnetic yoke blocks and the coil assembly, along with their 
connection and disconnection within the target complex has been considered in the design study 
target complex handling and integration (Section~\ref{Sec:TC:magnetic-coil-US1010-shielding}).

\clearpage

\section{He-vessel containment}
\label{Sec:TC:HeV}

\subsection{Introduction and requirements}
The He-vessel is an essential part of the BDF Target Complex, as it contains all beam intercepting devices and shielding elements. Its role is to guarantee an inert atmosphere in the high radiation area surrounding the spallation target, aiming at reducing formation of short lived air activation products and reducing NOx formation that could attack materials and produced radiation-accelerated corrosion.

The He-vessel will be designed to be slightly over-pressure with respect to the neighboring areas in order to avoid air entry into the vessel during operation. The following Sections will detail the design of the He-vessel as well as the purification and circulation system.

\subsection{Design and preliminary structural assessment of the He-vessel}

\subsubsection{Design requirements}
\label{Sec:TC:design_requirements}

\subsubsubsection{Dimensions and surrounding space}
The minimum internal dimensions of the He-vessel (minimum available space for the target and its surrounding radio-protection shielding) shall be 11330 mm x 8440 mm and 8275 mm high (approximately 790 m$^3$). The minimum space reserved for the mechanical structure shall comply with the load resistance criteria described in Section~\ref{Sec:TC:HV:load_case}. Adequate space shall be reserved inside and outside the structure to allow assembly, testing, maintenance and dismantling operations. A man access route (4 persons maximum at the same time) shall be reserved on the top surface of the structure to allow time-limited manual interventions outside the vessel during the operational phase when the beam is off.

\subsubsubsection{Interface with civil engineering works}
The He-vessel shall be installed in a rectangular pit of approximately 12000 mm x 9000 mm and 9000 mm deep. The mechanical structure shall lay on the concrete floor without any other connections to civil engineering works. The interface between the mechanical structure and the concrete floor shall be designed in compliance with civil engineering tolerances and permissible floor load.

\subsubsubsection{Material constraints}
The choice of materials for the design of the He-vessel shall consider the following criteria: compliance with the European construction code (Eurocode), resistance to corrosion, water and helium leak tightness, respect of radiation protection rules, cost.

\subsubsubsection{Load case}
\label{Sec:TC:HV:load_case}
The He-vessel structure shall be designed to withstand its own weight as well as an internal helium gas pressure of 0.1 barg (test pressure up to 0.5 barg). The floor of the He-vessel shall be designed to withstand the weight of the target assembly as well as its surrounding shielding (total weight: ~ 5500 t, contact pressure: ~ 60 t/m$^2$). The weight of four people working at the same time on the top surface of the structure shall also be considered. The structure and what it contains shall be designed to withstand a seismic action having a peak ground acceleration of 1.1 m/s$^2$. The He-vessel shall not be designed for internal vacuum.

\subsubsubsection{Lids and side window}
\begin{itemize}
\item Top lid (large):
    
The entire top surface of the He-vessel (11830 mm x 8940 mm) shall be removable in one piece to allow installation, maintenance and dismantling of internal components (lid to be opened once a year at most along the 5-10 years lifetime of the experiment). The interface with the rest of the structure (full perimeter of the lid) shall be sealed according to the helium leak lightness requirements described in section \ref{Sec:TC:HV:helium_leak_tightness}. The seal shall be put in place or removed manually, since the top of the He-vessel is man-accessible when the beam is off. Materials like EPDM can be used as a sealing solution, as long as they are replaced at regular periods to avoid their invalidity due to long term radiation and loading exposure.

\item Small lid (crane version only):

In the crane version, a smaller lid shall be fitted in the top lid, in order to perform the maintenance of the target assembly and its proximity shielding only. It shall be opened approximately three times a year along the 5-10 years lifetime of the experiment. A similar solution to the one used for the top lid shall ensure helium leak tightness.

\item Side window (trolley version only):

In the trolley version, the maintenance of the target assembly shall be done using an opening made in the side wall of the He vessel. Maintenance operations shall occur a few times a year along the 5-10 years lifetime of the experiment.
\end{itemize}

\subsubsubsection{Beam window}
The beam window shall be mounted on the outside of the He-vessel. The interface with the mechanical structure shall be compatible with the remote handling installation and maintenance process, as well as a leak tight connection of the window.

\subsubsubsection{Service feedthroughs (water, gas and electrical)}
Since the top part of the He-vessel will be man-accessible once the shielding is in place, the service connections shall be preferably routed to ports located in the upper part of the He-vessel side walls, where feedthroughs can be permanently installed and where maintenance can be done without any consequence on the opening and closing of the lids and without removing the shielding blocks. The routing of the service connections for the target and its proximity shielding depends on the option which will be chosen for handling (trolley or crane).
\begin{itemize}
\item He vessel services:
    
Feedthroughs shall be provided in the side wall of the tank for the connection to the helium purification system. They shall be installed between the top level of the internal shielding and the He-vessel main lid, to allow manual intervention.
\item Target services (crane version only):

Water cooling, helium gas and electrical connections for the target shall be supplied through pillars towards the floor, so the shielding can be removed without disconnecting the circuits.
\item Proximity shielding services:

For the trolley version, water cooling feedthroughs shall be provided in the side wall of the tank. They shall also be installed between the top level of the internal shielding and the He-vessel main lid. For the crane version, water cooling shall be supplied through pillars towards the floor.
\item Coil services:

Feedthroughs shall be provided in the side wall of the tank for the supply and water cooling of the coil. They shall also be installed above the top level of shielding and below the He-vessel main lid.
\item Collimator services:

Electrical feedthroughs shall be provided in the side wall of the tank for the connection of temperature sensors (PT100).
\end{itemize}

\subsubsubsection{Draining of water and helium}
If a leak occurs in any of the cooling circuits contained within the He-vessel, the coolant shall be contained by the vessel itself and be drained by gravity through a dedicated circuit without having to remove any embedded component. If leakage occurs outside the He-vessel, the coolant will be collected by the drainage system designed by civil engineering.

\subsubsubsection{Helium leak tightness}
\label{Sec:TC:HV:helium_leak_tightness}
All joints and components of the He-vessel related to helium leak tightness shall be designed in order to comply with a maximum room temperature leak rate (both localized and global) of 0.1 mbar l s$^{-1}$.

\subsubsubsection{Temperature level}
Any part of the He-vessel shall be designed to withstand a maximum temperature of \SI{40}{\celsius}  without any permanent damages or impact on the functional behaviour of the assembly.

\subsubsubsection{Installation, maintenance and dismantling}
The design shall allow the mounting of the He-vessel inside the area within the constraints of space and access defined by integration work. The design shall be compatible with transport of He-vessel parts within handling capabilities (maximum weight and range). The design shall be compliant with the CERN safety regulation as well as radiation protection rules (ALARA) during installation, operation, maintenance and dismantling.

\subsubsection{Applicable norms and standards}
\label{Sec:TC:norms_standards}

\subsubsubsection{Design according to the Eurocode}
The BDF He-vessel, having a relatively low pressure of 0.1 barg, is not considered a pressure vessel under the scope of PED directive 2014/68/EU~\cite{directive_2014}. It shall therefore be assessed as a steel structure for which the Eurocode 3 applies. The structure has been classified as level 2 in the:
\begin{itemize}
\item class of consequences (EN 1990 \cite{standard_en1990}) which is used for the purpose of reliability differentiation by considering the consequences of failure or malfunction of the structure. Class of consequences CC2 is applied when there are medium consequences for loss of human life, economic, social and environmental;
\item reliability class (EN 1990) is associated to the class of consequences and has a direct impact on the reliability index $\beta$, on the multiplication factor K$_{\text{FI}}$  = 1.0 to the partial factors on actions, on design supervision and inspections during execution;
\item execution class (EN 1090 \cite{standard_en1090}) is related to quality of production in general, to welding, bolting, riveting and assembling of the structural members;
\end{itemize}

The Eurocode 3 shall be applied in the design of the structure with the relevant documents:
\begin{itemize}
\item EN 1993-1-1 for General rules and rules for buildings~\cite{standard_en1993_1_1};
\item EN 1993-1-5 for Plated structural elements~\cite{standard_en1993_1_5};
\item EN 1993-1-6 for Strength and stability of shell structures~\cite{standard_en1993_1_6} and
\item EN 1993-1-8 for Design of joints~\cite{standard_en1993_1_8}.
\end{itemize}

While there are good practices for choosing columns, beams and joints, it is generally assumed that unconventional designs are accepted by the Eurocode as long as it is proved that it complies with the Ultimate Limit State (ULS) and Serviceability Limit State (SLS), see Section~\ref{Sec:TC:assessment_eurocode}. 
All materials used shall be considered in accordance with EN 1993 and EN 10025~\cite{standard_en10025}. At the time of writing this document, the material of choice is the steel S235 for all columns and beams which properties are in agreement with \ref{Tab:TC:material_s235}.

\begin{table}[htbp]
\centering
\caption{\label{Tab:TC:material_s235} Material properties considered for S235 in accordance with EN 10025.}
\smallskip
\begin{tabular}{r|c|c}
\hline
 &  $f_y [\text{N/mm}^2]$  &  $f_u [\text{N/mm}^2]$ \\
\hline
S235 & 235 & 360 \\
\hline
\end{tabular}
\end{table}

\subsubsubsection{Assessment according to the Eurocode}
\label{Sec:TC:assessment_eurocode}

The structure shall be assessed against the Ultimate Limit State (ULS), which will guarantee a structural integrity of the He vessel against all permanent, variable and accidental actions. On the other hand, the assessment against the Serviceability Limit State (SLS) will guarantee that the structure is fit for purpose. This last condition will limit the maximum displacements and the natural frequencies.
Considering the STR limit state, which concerns the strength of the structure, the combination of actions may be expressed as equation 6.10 of EN 1990, in which $''+''$ means combination:
\begin{equation}
\sum_{j\geq1} \gamma_{G,J} {G_{k,j}} ''+'' {\gamma_{Q,1}}{{Q_{k,1}} ''+'' \sum_{i\geq1} \gamma_{Q,i}\psi_{0,i}Q_{k,i}}
\label{eurocode_actions_1}
\end{equation}

in which:
\begin{itemize}
\item $\gamma_{G,j}$is taken from the EN 1990 French national annex for design of structural members not involving geotechnical actions (STR) and corresponds to 1.35 for unfavorable actions and 1.0 for favorable actions;
\item $G_{k,j}$ are the permanent actions of self-weight (including the weight of the iron blocks inside the structure) and the inner pressure due to helium;
\item $\gamma_{Q,1} $ is also taken from the same national annex and corresponds to 1.5 in an unfavorable action or 0 if favorable;
\item $Q_{k,1}$ is the main variable action;
\item $\gamma_{Q,i}$ is a factor for the remaining variable actions being 1.5 or 0 in an unfavorable or favorable case, respectively;
\item $\psi_{0,i}$ is a factor for buildings (table A1.1 of the EN 1990);
\item $ Q_{k,i} $ are the remaining variable actions.
\end{itemize}

Other variations of this equation, suggested in the Eurocode, will also be taken into account to find the least favorable combination.
In the Beam Dump Facility He-vessel, permanent actions are the self-weight of structural and non-structural members and the internal pressure while variable actions are the weight of four persons on the top of the roof. The only accidental action is the over-pressure (1.5x) which shall be applied as indicated in EN 1991-1-7:

\begin{equation}
\sum_{j\geq1} {G_{k,j}} ''+'' {A_d} ''+'' {\Psi_{1,1}}{Q_{k,i}} ''+'' \sum_{i\geq1} \Psi_{2,i}Q_{k,i}
\label{eurocode_actions_2}
\end{equation}

The seismic assessment shall be performed in accordance with EN 1998.
As part of the ULS, a non-linear buckling analysis will be performed by applying the nominal boundary conditions in the structure with a deformation equivalent to the fabrication tolerances. Such deformation shall take a scaled shape of the first mode obtained from a linear buckling analysis. The Eurocode does not consider such analysis and therefore the procedure will be followed according to the direct route guidelines defined in the European code for pressure vessels EN 13445 - 3~\cite{standard_en13445_3}.

\subsubsubsection{Load cases}
\label{Sec:TC:eurocode_load_cases}

The actions imposed on the structure shall be defined in accordance to the Eurocode 1, in specific EN 1991-1-1~\cite{standard_en1991_1_1}: Densities, self-weight, imposed loads for buildings. Table~\ref{Tab:TC:eurocode_load_cases} presents the different load cases and the actions that shall be considered.

\begin{table}[htbp]
\centering
\caption{\label{Tab:TC:eurocode_load_cases} Multipliers on actions to be considered at each load case.}
\smallskip
\begin{tabular}{r|c|c|c|c|c|c|c}
\hline
\textbf{ } & \multicolumn{2}{c|}{\textbf{Permanent}} & \textbf{ Variable } & \textbf{ Accidental } & \multicolumn{3}{c}{\textbf{ Analysis }}\\
\textbf{ } & \multicolumn{2}{c|}{\textbf{Actions}} & \textbf{ actions } & \textbf{ actions } & \multicolumn{3}{c}{\textbf{ }}\\
\hline
Load Cases / & Self	& Pressure & Weight of & Over-pressure & Static & Modal & Buckling \\
Actions & weight & $P$ &  4 people &  &  &  &  \\
\hline
\multicolumn{8}{c}{Serviceability Limit State (SLS)} \\
\hline
Nominal & 1 & 1 & 1 & - & Yes & Yes & - \\
conditions & & & & & & & \\
\hline
\multicolumn{8}{c}{Ultimate Limit State (ULS)} \\
\hline
Self weight  & 1.35 & 1.35 & 1.5 & - & Yes & No & Yes \\
with roof & & & & & & & \\
\hline
Self weight  & 1.35 & - & - & - & Yes & No & Yes \\
without roof & & & & & & & \\
\hline
Leak test & 1.35 & $1.35 \times P_{test}$ & 1 & - & Yes & No & Yes \\
\hline
Assembly & 1.35 & - & - & - & Yes & No & Yes \\
\hline
Accidental & 1 & - & 1.5 & $1.5 \times P$ & Yes & No & Yes \\
\hline
Seismic & 1 & 1 & - & - & \multicolumn{3}{c}{Response spectrum analysis} \\
\hline
\end{tabular}
\end{table}

In the assembly load case, different joints between columns and beams shall be assessed under cantilever load cases. These cases are expected to appear during assembly and are likely to be the most demanding load scenario for the joints.
A fatigue assessment does not need to be performed as the expected number of cycles is low. The roof lid will be open 3x per year so a maximum of 4 pressure cycles per year can be expected for a time frame of 15 years, which amounts to a total of 60 cycles.
While fire will not be assessed as an action, a risk analysis can be performed and prevention measures should be put in place as part of the installation of the equipment. Creep does not therefore need to be taken into account as the considering temperature is only $40^\circ$C.

\subsubsection{Design principles}
\label{Sec:TC:design_principles}

\subsubsubsection{State of the art}
The design work will be carried out in compliance with the acknowledged state of the art in the field of mechanical design. The rules issued by CERN for this type of activity as well as the applicable norms and standards (described in section \ref{Sec:TC:norms_standards}) will be respected.

\subsubsubsection{Self-supporting structure}
Since the He vessel cannot be in contact with civil engineering works except the pit ground, the principle of a self-supporting structure will be applied in this case. This structure will be designed and built in compliance with the applicable norms and standards as described in section \ref{Sec:TC:norms_standards}.

\subsubsubsection{Dis-mountable structure}
Since the design will have to comply with the CERN radiation protection rules, especially during the decommissioning phase, the principle of a dis-mountable structure will be applied in this case. For components to be assembled on site, bolted connections will be preferred to welded joints. Welding operations to be made on site (and therefore to be cut on site during dismantling operations) will be minimized and the welded joints will be designed in order to be easily removed.

\subsubsubsection{Transportable parts}
In order to comply with transport and handling capabilities as well as space constraints in the Beam Dump Facility, any prefabricated component of the He vessel will have a maximum size of 6000 mm x 2000 mm x 500 mm and a maximum mass of 35000 kg (lifting capacity of the BDF crane). Some components will be assembled inside the facility near the pit (in a dedicated area) and therefore the size of the assembled elements may be greater than the value specified above, but the total mass of the assembly will not exceed 35000 kg.

\subsubsubsection{Separated functions}
The two main functions of this mechanical structure are to guarantee a structural resistance to the load case described in section~\ref{Sec:TC:HV:load_case} and to ensure helium leak tightness as described in section~\ref{Sec:TC:HV:helium_leak_tightness}. The design principle used in this case will be to separate these two functions, so that no structural element will be used to fulfill the leak tightness requirement and vice versa. This will minimize, for example, the mechanical stresses on components dedicated to the leak tightness.

\subsubsubsection{Narrow or confined spaces}
Since most of the structure will be assembled inside a pit, the design will be done in such a way that manual assembly, maintenance or dismantling operations in narrow or confined spaces are minimized and can be performed in accordance with CERN safety standards. 

\subsubsubsection{Verification or testing}
The design will be made in such a way that all quality, performance or reliability checks (inspections or tests) can be carried out in accordance with CERN safety standards.

\subsubsubsection{Cleaning}
The design will be compatible with the application of a proper cleaning procedure to rid the tank (and in particular its internal part) of metallic and non-metallic particles that could be activated during beam operation.

\subsubsection{Conceptual design}
\label{Sec:TC:conceptual_design}

\subsubsubsection{Concept}
The basic concept will be to build an helium tight envelope (skin) surrounded by a support structure (skeleton).

\subsubsubsection{Self-supporting structure}

\begin{itemize}
\item Geometry and structural distribution:

The mechanical structure of the He-vessel, as shown on Figure~\ref{fig:TC:HV:Support_structure}, will consist of a floor whose main function will be to support the 5500 tons of shielding of the target. On this floor will be fixed walls that will have the main mechanical functions to support their own weight, the weight of the top cover, the weight of equipment or people who could be placed on it and an internal pressure of 0.1 barg. The upper part of the He vessel will consist of a removable structure (lid) whose main mechanical functions will be to support its own weight, the weight of the equipment and people who could be placed on it, as well as an internal pressure of 0.1 barg.

\begin{figure}[htbp]
\centering %
\includegraphics[width=0.6\linewidth]{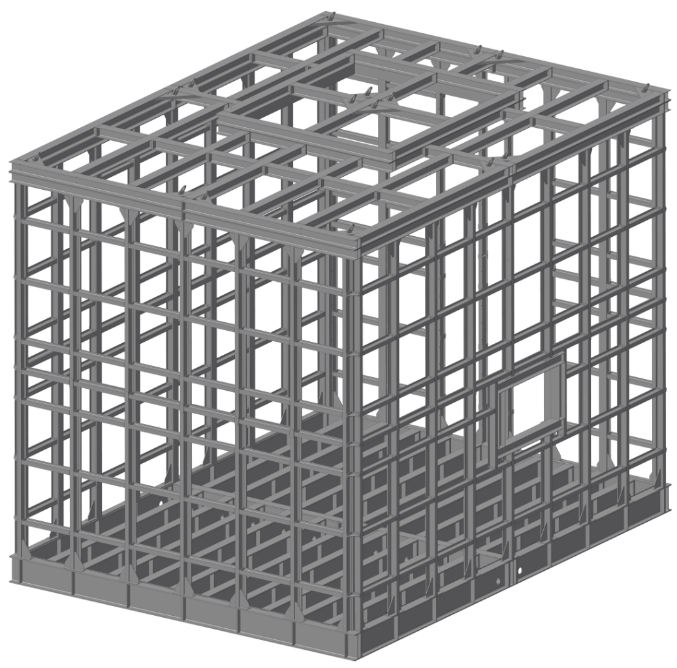}
\caption{He vessel support structure (skeleton).}
\label{fig:TC:HV:Support_structure}
\end{figure}

\item Material:

The support structure will be built mostly of structural steel (S235 type) standard profiles such as IPE400 and HEA240, as shown on Figure~\ref{fig:TC:HV:Standard_steel_profiles}. Welded structural parts will be made using EE, MAG or TIG processes using the respective filler metals E7018, S35 and ER7056. The material used for bolts and nuts will be at least Steel 8.8. An adequate long life surface treatment will be applied to profiles, welds and bolts in order to minimize erosion-corrosion effects. 

\begin{figure}[htbp]
\centering %
\includegraphics[width=0.8\linewidth]{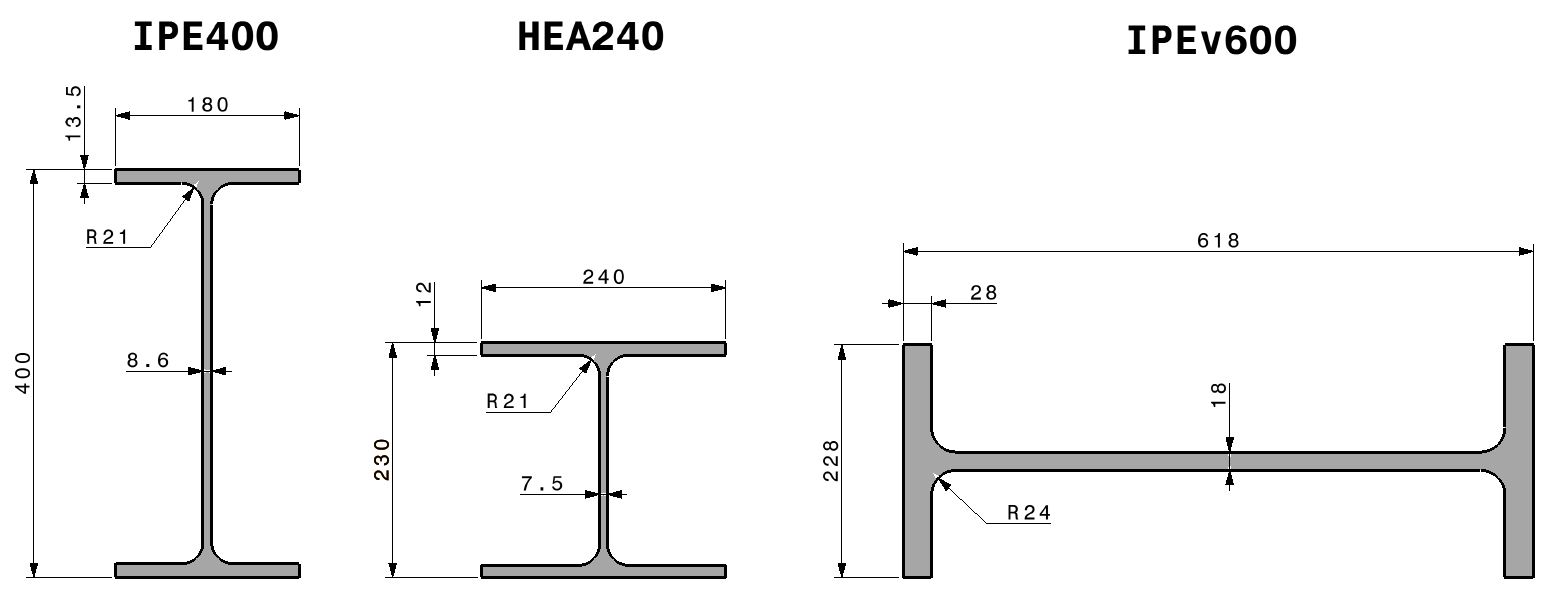}
\caption{Standard steel profiles used in the self-supporting structure.}
\label{fig:TC:HV:Standard_steel_profiles}
\end{figure}

\item Total weight:

The total weight of the He vessel mechanical structure will be approximately 130 t, which is less than 3\% of the total mass of the target assembly (target/internal shielding/He vessel).
\end{itemize}

\subsubsubsection{Floor}
\begin{itemize}
\item Geometry and structural distribution:

The floor structure will be made of reinforced IPEv600 profiles bolted together, as shown on Figure~\ref{fig:TC:HV:HV_Floor}. These components will be pre-assembled on site (in a dedicated area) and then transported to the pit. The differences in flatness with the concrete floor will be compensated by shims. Since the shims used will have a thickness of 1 mm and the flatness monitoring system is expected to have a resolution of 1 mm, the flatness of the He-vessel floor over its entire surface will be less than 10 mm, with local flatness differences of the order of a few millimeters. The use of IPEv600 profiles provides the necessary space for the installation of supply service (such as the water drainage system) under the floor.

\begin{figure}[htbp]
\centering %
\includegraphics[width=0.7\linewidth]{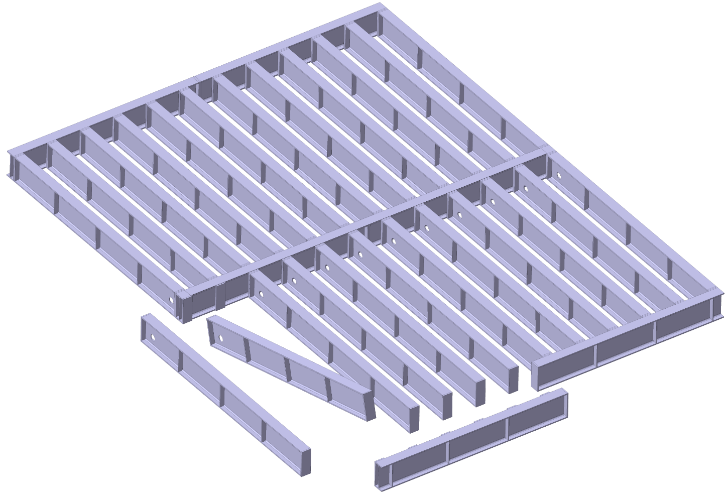}
\caption{He-vessel floor structure.}
\label{fig:TC:HV:HV_Floor}
\end{figure}

\item Water/air draining:

As shown on Figure~\ref{fig:TC:HV:HV_draining_system}, the water drainage system will consist of stainless steel helium-tight retention containers (sloped) that will be attached to the mechanical structure of the floor. The tightness between the containers will be ensured by stainless steel welded profiles that will cover the joints. The floor surface will then be covered with perforated stainless steel plates. The containers will be connected to the water drainage circuit that can also be used for pushing air out with injection of helium (from the top of the vessel).
\end{itemize}

\begin{figure}[htbp]
\centering %
\includegraphics[width=0.8\linewidth]{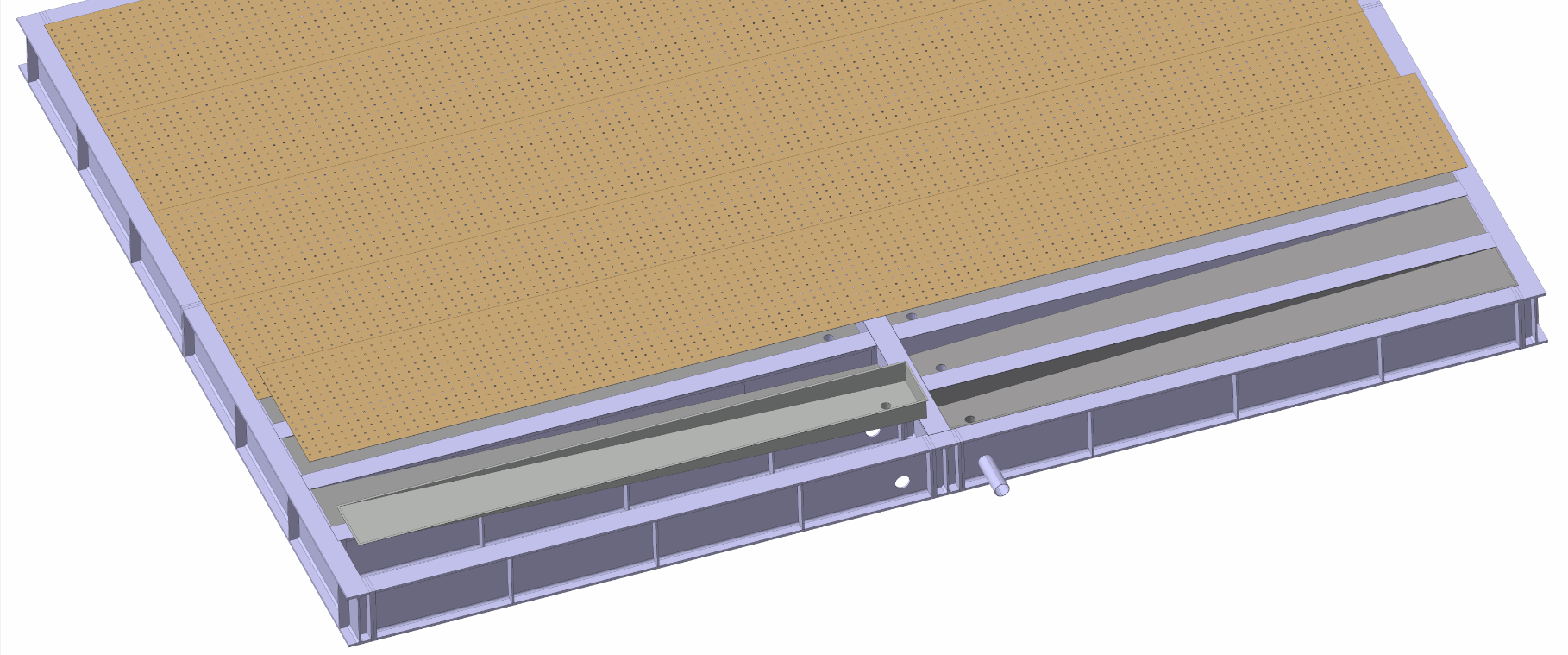}
\caption{He-vessel Water and air draining system.}
\label{fig:TC:HV:HV_draining_system}
\end{figure}

\subsubsubsection{Helium leak tightness}
\begin{itemize}
\item Insulation:

Figures~\ref{fig:TC:HV:Helium_tight_walls} and \ref{fig:TC:HV:Helium_tight_joints} show that the helium tight wall will be built using thin stainless steel panels bolted on the inside of the support structure. These panels will cover the inner surface of the He-vessel. The joints and fasteners will be covered by stainless steel profiles that will be welded to the panels, to ensure the required leak tightness. These profiles can be easily cut out to provide access to bolts during dismantling phase.

\begin{figure}[htbp]
\centering %
\includegraphics[width=0.7\linewidth]{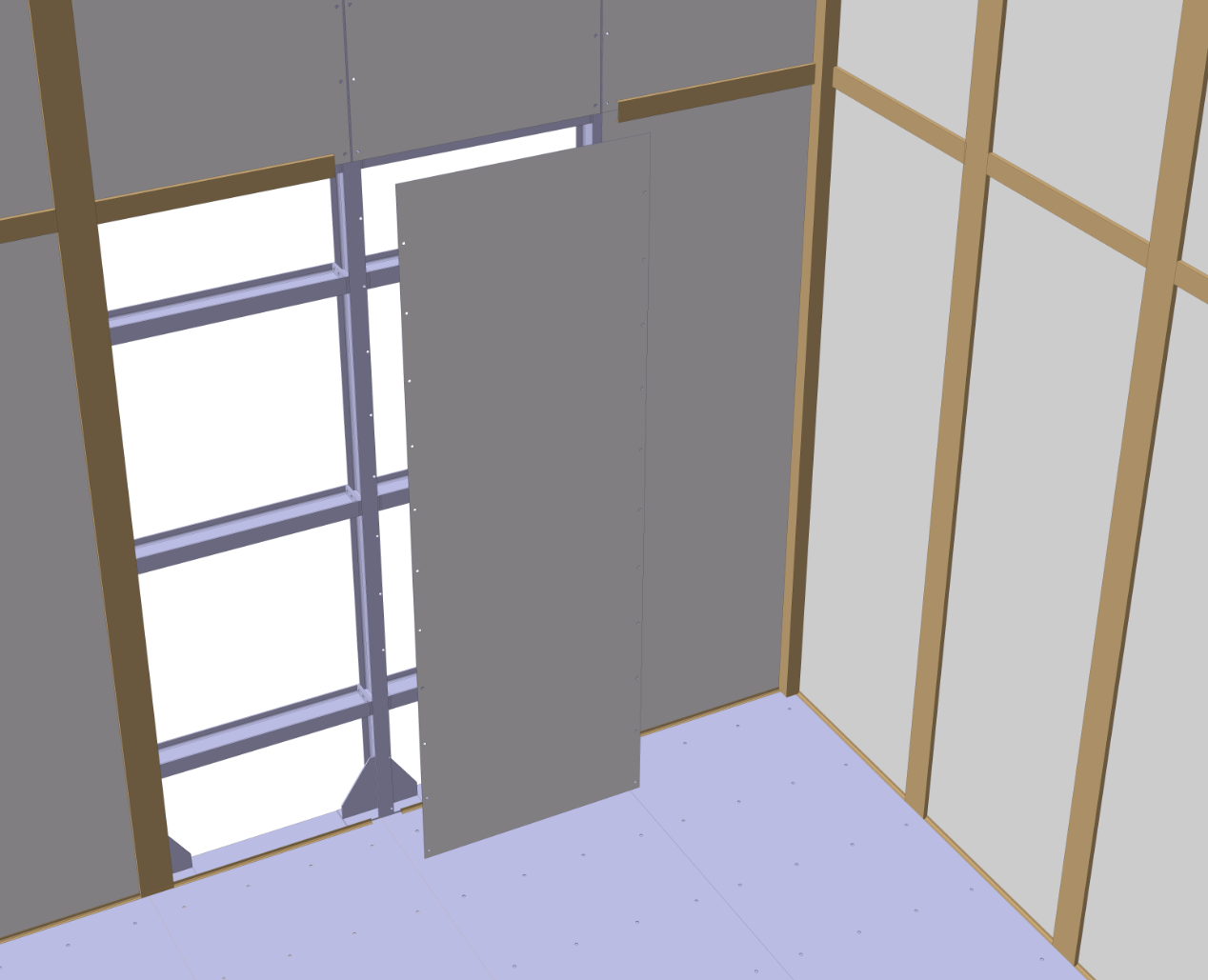}
\caption{Helium tight walls (skin).}
\label{fig:TC:HV:Helium_tight_walls}
\end{figure}

\begin{figure}[htbp]
\centering %
\includegraphics[width=0.7\linewidth]{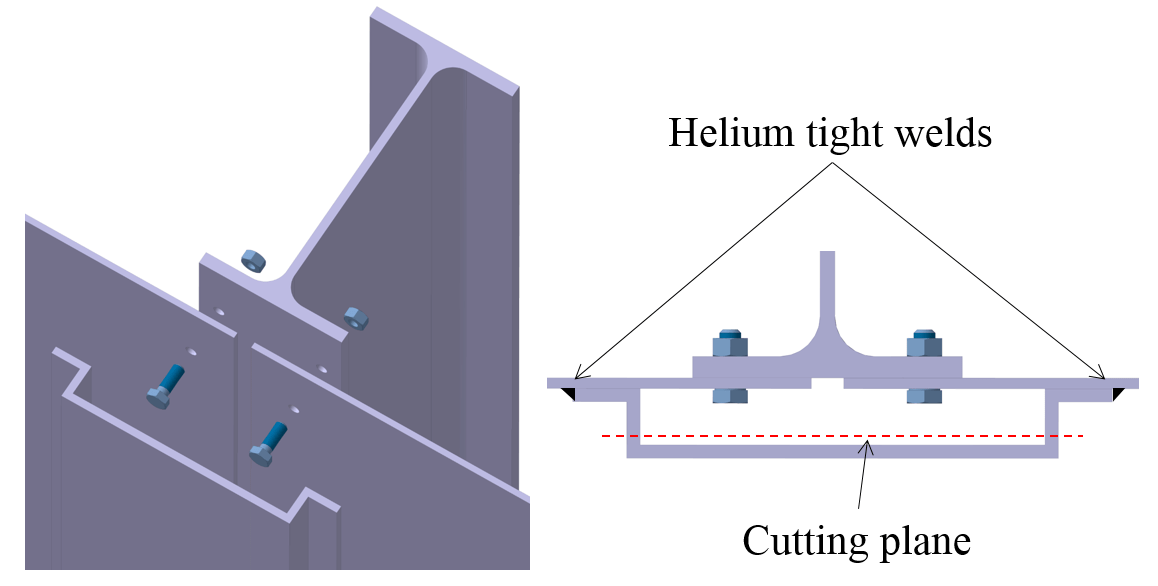}
\caption{Helium leak tightness at joints.}
\label{fig:TC:HV:Helium_tight_joints}
\end{figure}

\item Material:

The leak tight envelope will be made out of thin plates of stainless steel (EN 1.4306 type, thickness 2 to 4 mm). The material used for bolts and nuts will be at least Steel 8.8. Filler used for insulation TIG welds will be stainless steel 317LN (no filler in case of plasma welding).
\item Weld leak detection during construction phase:

Each weld procedure will be qualified on samples. A strict follow-up on the production welds (operator, equipment, procedure, independent visual inspection, x-ray where possible) will be implemented. Samples will be produced again during the installation phase to justify that there is no deviation with respect to agreed quality. Multipass welds will be specified where possible to reduce the probability of leakage. Depending on geometry, intermediate leak tests will be specified.
\end{itemize}

\begin{figure}[htbp]
\centering %
\includegraphics[width=0.7\linewidth]{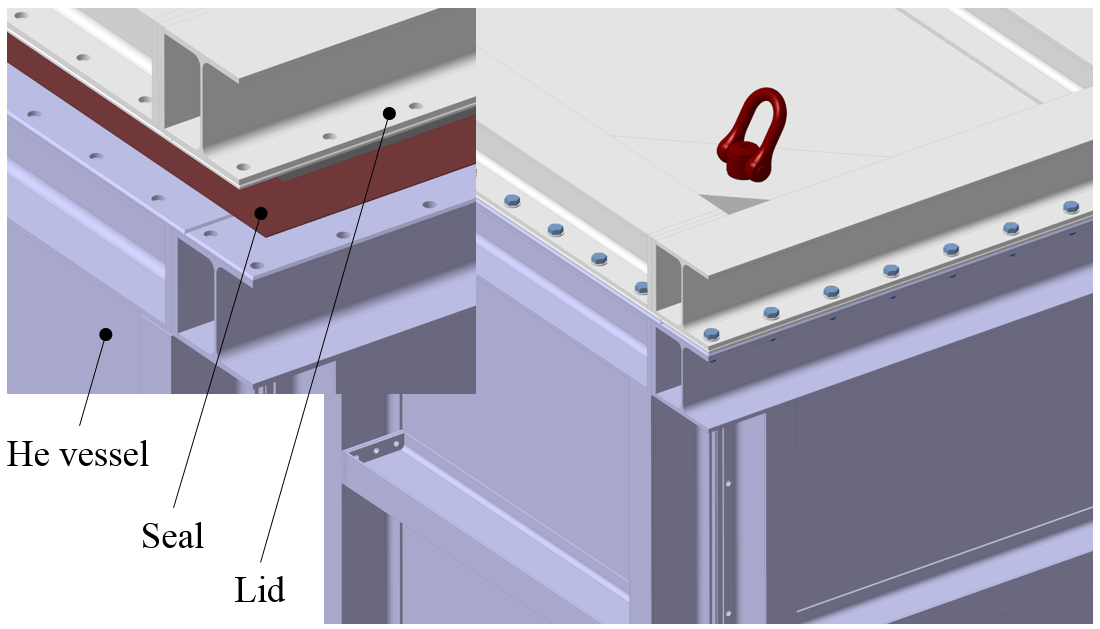}
\caption{Lid in open and closed positions using bolted connections and EPDM seal.}
\label{fig:TC:HV:HV_Lid}
\end{figure}

\subsubsubsection{Lids and side window}
In order to comply with the CERN radiation protection rules, the lids allowing access to the internal components of the helium tank will be held in position by bolted connections (welded connections will be excluded), Figure \ref{fig:TC:HV:HV_Lid}. The leak tightness of the joint will be ensured by replaceable EPDM flat seals.

\begin{figure}[htbp]
\centering %
\includegraphics[width=0.9\linewidth]{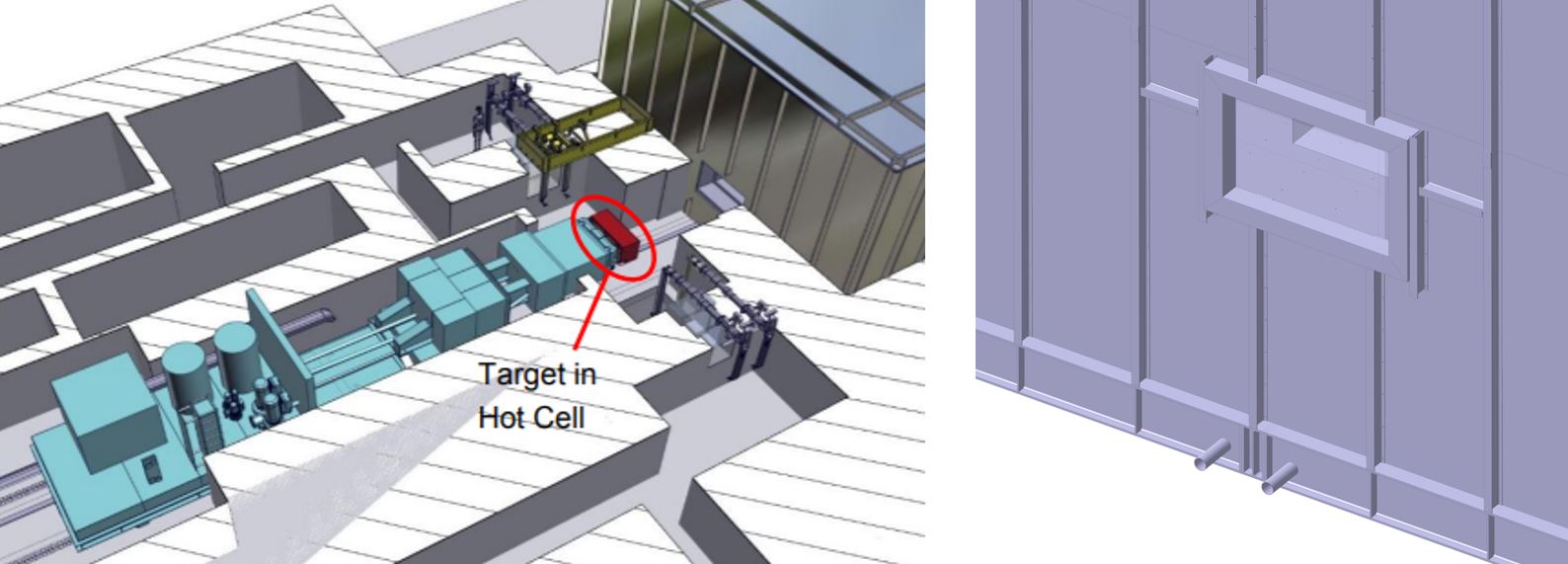}
\caption{Trolley side window opening-closing principle.}
\label{fig:TC:HV:HV_Trolley_side_window}
\end{figure}

In the trolley version only, an opening (side window) will be built in the side wall of the He vessel, Figure \ref{fig:TC:HV:HV_Trolley_side_window}. This opening will be tightly closed using a moveable door, which will be fixed to the trolley and closed by movement of the trolley.

\begin{figure}[htbp]
\centering %
\includegraphics[width=0.9\linewidth]{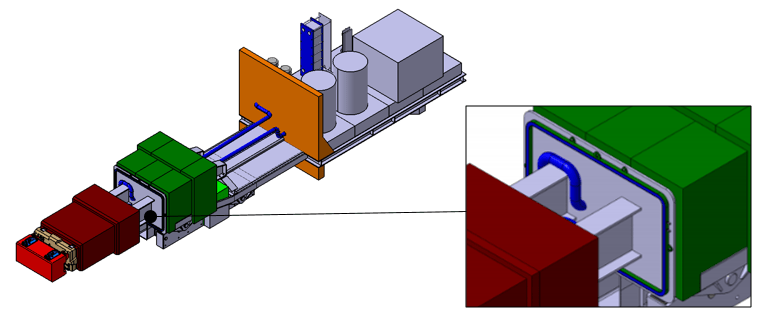}
\caption{Moveable door and sealing solution for the BDF target trolley version.}
\label{fig:TC:HV:HV_Moveable_door}
\end{figure}

This opening/closing system will be compatible with the misalignment tolerance of the target (+/- 10 mm in all directions) and with the maximum deformation in this area of the structure. EPDM type seals will be used due to a relatively low level of radiations in this location. Figure \ref{fig:TC:HV:HV_Moveable_door} illustrates the sealing solution in the trolley.

\begin{figure}[htbp]
\centering %
\includegraphics[width=0.5\linewidth]{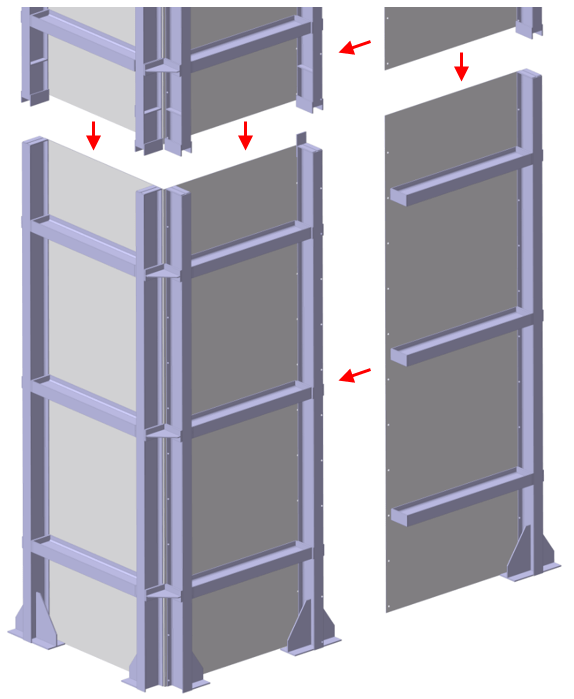}
\caption{Sketch of the assembly of the He-vessel prefabricated components.}
\label{fig:TC:HV:HV_Prefabricated_components}
\end{figure}

\subsubsubsection{Prefabricated components}
The structure will be constructed from prefabricated elements (welded or bolted) composed of beams and panels that can be manufactured off-site and easily transported, with the objective of minimizing on-site assembly operations. See Figure \ref{fig:TC:HV:HV_Prefabricated_components}.

\begin{figure}[htbp]
\centering %
\includegraphics[width=0.8\linewidth]{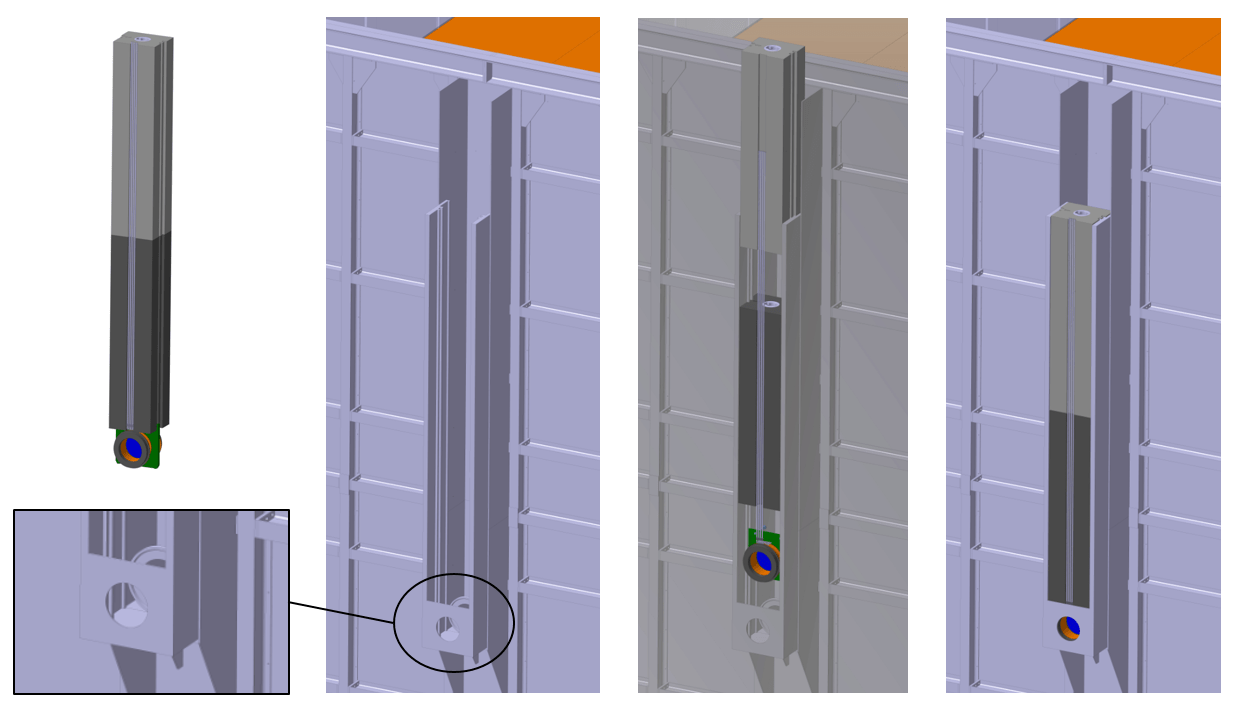}
\caption{BDF He-vessel beam window assembly.}
\label{fig:TC:HV:HV_Beam_window}
\end{figure}

\subsubsubsection{Beam window}
A sliding system is located at the front of the structure, as shown on Figure \ref{fig:TC:HV:HV_Beam_window}. It allows precise positioning of the beam window and its fixation on the structure (by remote handling). The slide is also used as a guide for the precise positioning of the shielding elements.

\begin{figure}[htbp]
\centering %
\includegraphics[width=0.8\linewidth]{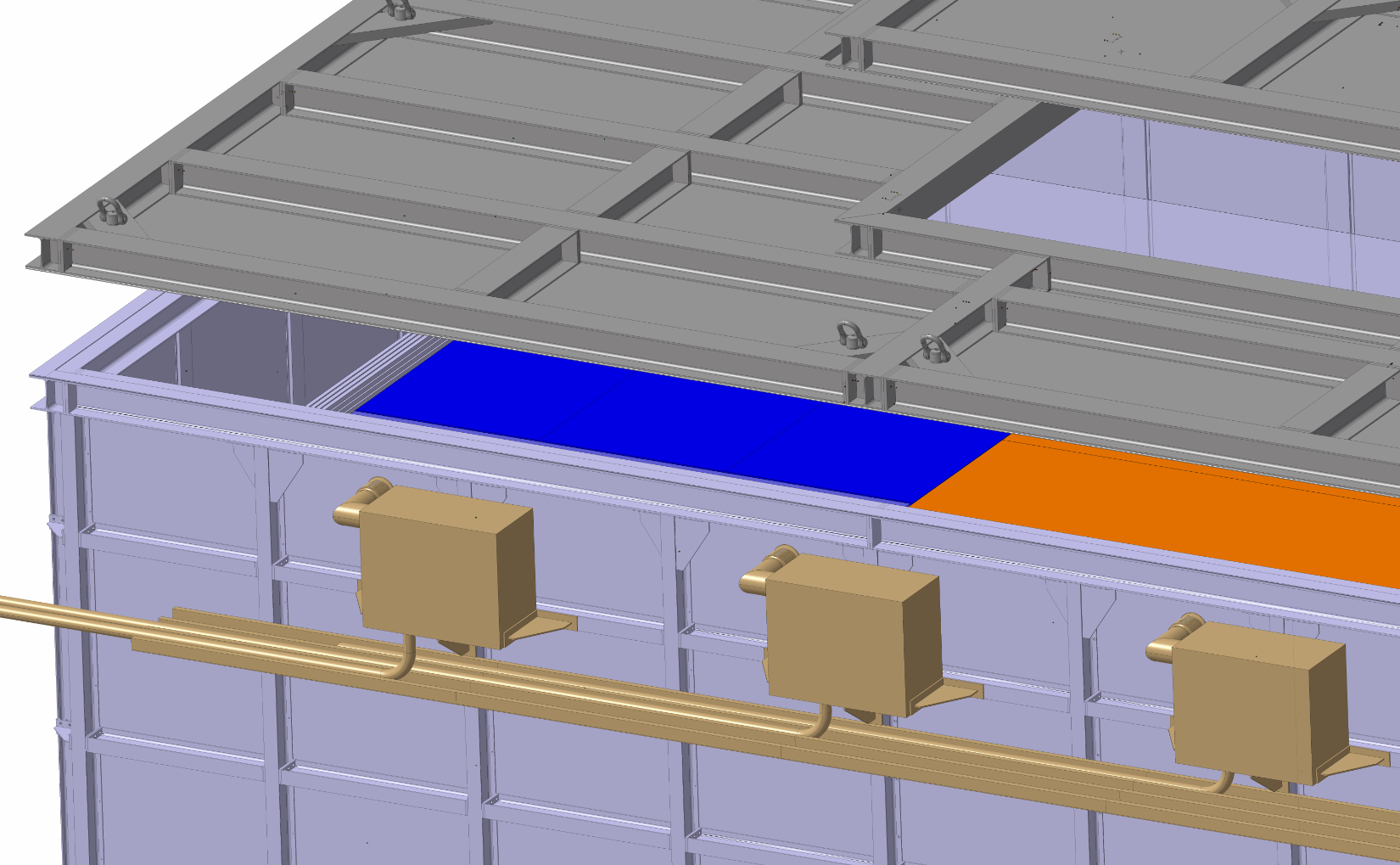}
\caption{Standard interface modules and feedthroughs.}
\label{fig:TC:HV:HV_Interfaces}
\end{figure}

\subsubsubsection{Interfaces}
Standard interface modules will be bolted to the He-vessel support structure, Figure~\ref{fig:TC:HV:HV_Interfaces}. These modules will contain fluid and electrical distribution systems (connectors, pumps, valves, etc...). Openings will be made in the helium-tight panels to allow services to be supplied inside the vessel. The use of flexible feedthroughs will minimize the transmission of mechanical stresses in the walls.

\subsubsection{Structural assessment}
\label{Sec:TC:preliminary_design}

The structural assessment of the He-vessel will be performed using finite element methods (FEM). To model the structure, shell elements will be used for the beams webs and horizontal flanges as well as for the helium tight plates.
The joints will be assessed for their worst-case scenario (many of the joints will be similar and there is no need to assess all cases) using a sub-model discretized with solid elements and detailing the welds and bolt interfaces. The displacement fields ($u$, $v$, $w$, $r_x$, $r_y$, $r_z$) will be introduced in the boundaries of the sub-models.
While this approach is reliable for assessing the structural integrity of the He-vessel, it is also computationally intensive and therefore will only be deployed once the geometry of the skeleton is fully iterated. 
For preliminary iterations, much simpler models will be used which discretize the geometries into shells and beam elements. Such modeling, even though simplified, allows a fast comparison between two models and can be useful again in the future to quickly check the resistance and deformations of the structure under different load cases.
In the following sub-chapters, preliminary results of the He-vessel and in particular the roof and the floor are presented under SLS (nominal working conditions).

\begin{figure}[htbp]
\centering %
\includegraphics[width=0.8\linewidth]{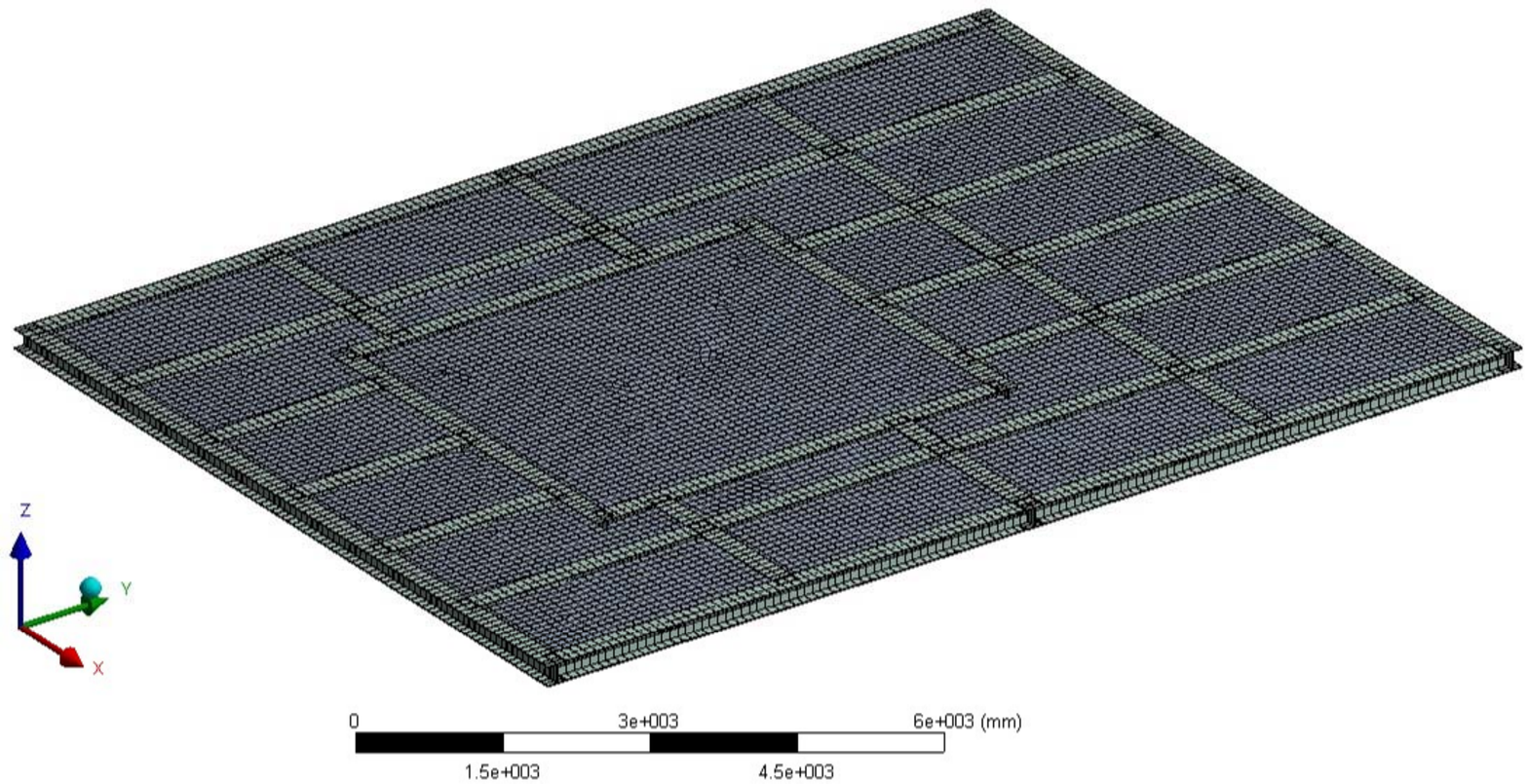}
\caption{Mesh representation of the He-vessel roof. The model has 14000 nodes.}
\label{fig:he-vessel_roof_mesh}
\end{figure}

\begin{figure}[htbp]
\centering %
\includegraphics[width=0.8\linewidth]{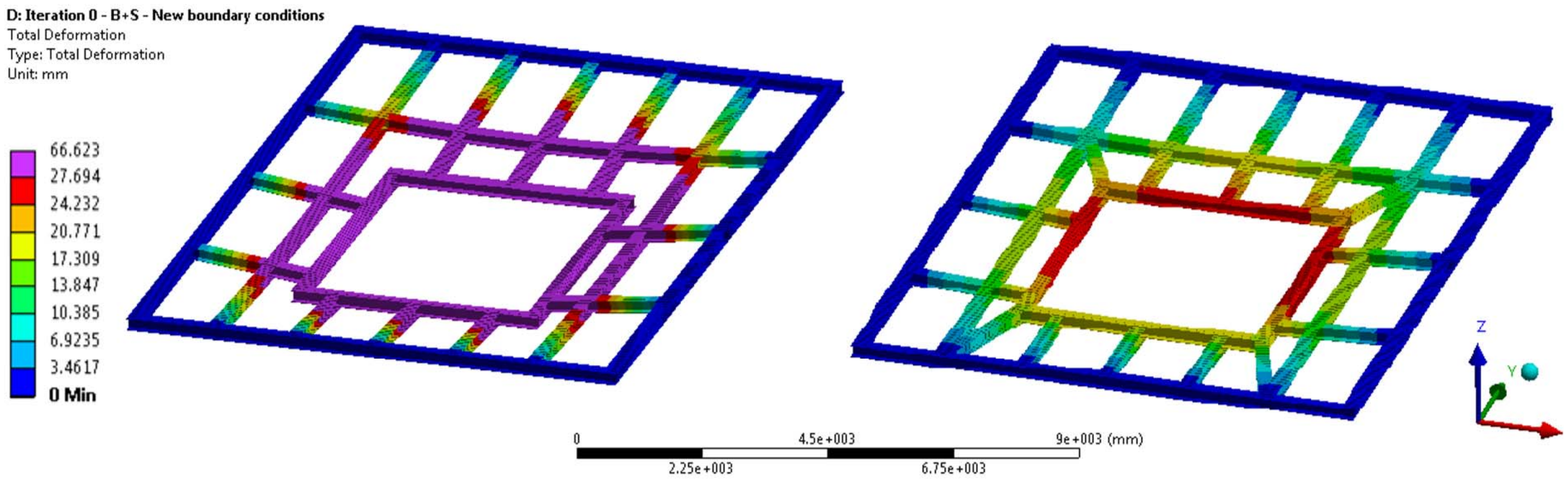}
\caption{Expected deflection of the roof when placed on the He-vessel with no internal pressure. Shells were hidden from the image for best visualization. Left image represents the first model and the right one corresponds to the second iteration. There is a clear evolution between both models.}
\label{fig:he-vessel_roof_iteration}
\end{figure}

\subsubsubsection{Self-weight analysis of the roof}
As the roof is joined by one perimeter edge of bolted connections, it can be considered as a pinned connection in a conservative fashion. The roof of the vessel is to be removed once per year and it contains a lid which is to be removed three times per year which justifies the bolted connections at both interfaces.
Using a simplified model made of beam elements and shell elements, as shown in Fig.~\ref{fig:he-vessel_roof_mesh}, containing only 14000 nodes, the roof beams distribution was quickly optimized. Figure~\ref{fig:he-vessel_roof_iteration} shows an evolution of different configurations. 

\begin{figure}[!htb]
\centering %
\includegraphics[width=0.8\linewidth]{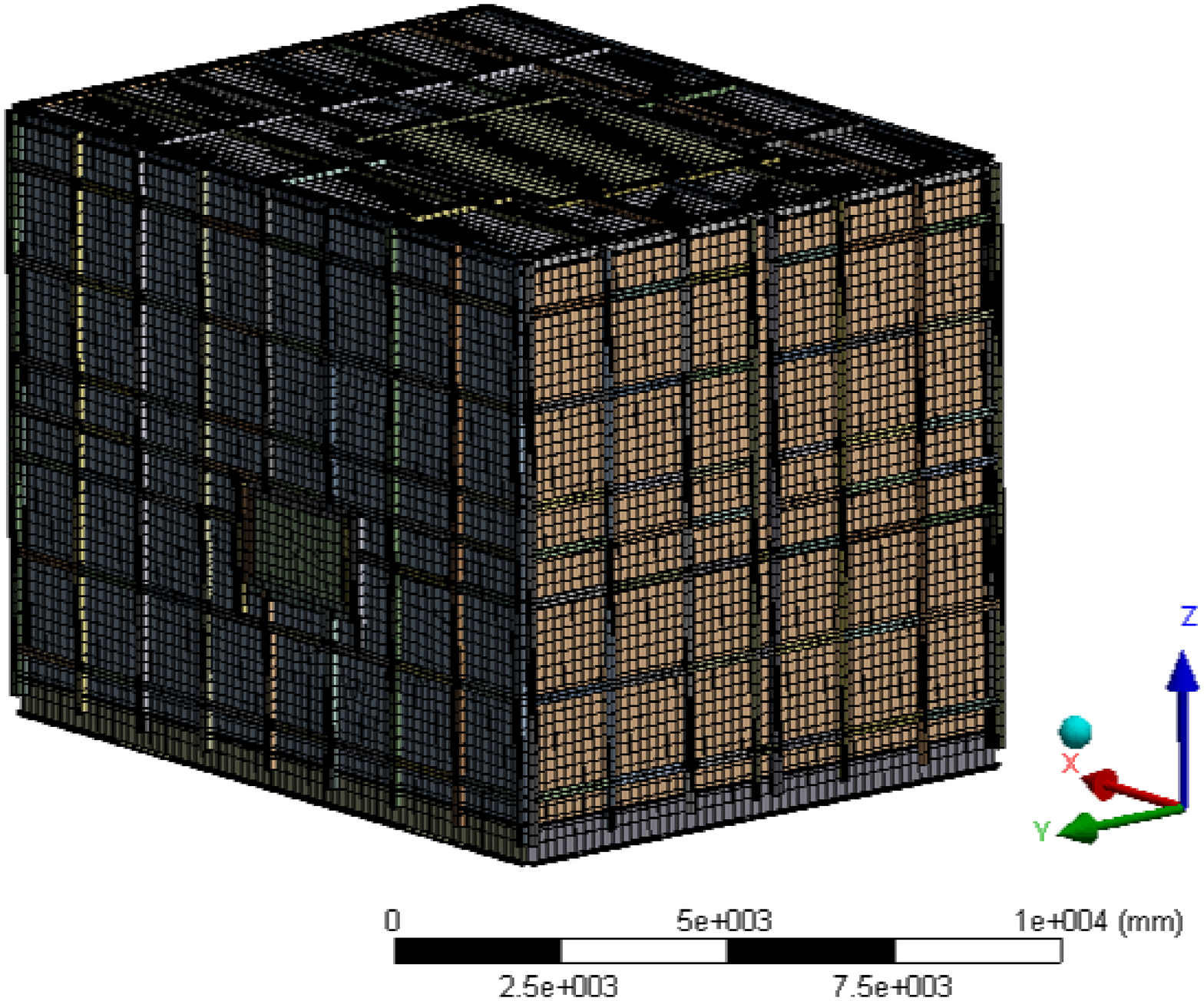}
\caption{Model of the He vessel using shell and beam elements. The model amounts to 25000 nodes.}
\label{fig:HV_mesh}
\end{figure}

\begin{figure}[!htb]
\centering %
\includegraphics[width=0.9\linewidth]{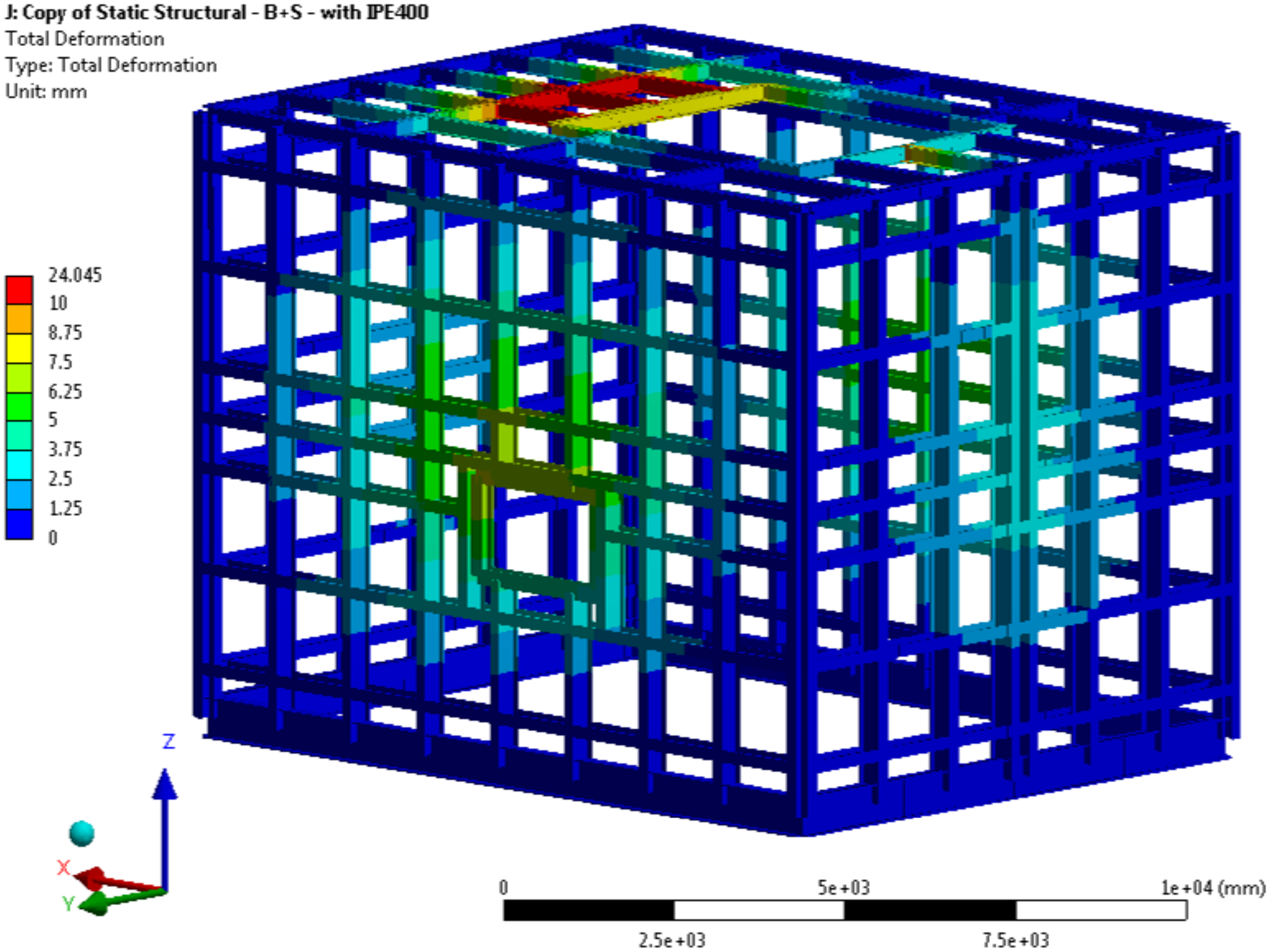}
\caption{Deflections on the He vessel due to internal pressure. Shell elements representing the leak tight envelope have been hidden for best visualization.}
\label{fig:HV_displacements}
\end{figure}

\subsubsubsection{He-vessel under internal pressure}
The He-vessel has been modeled against internal pressure and self-weight. Using a simplified discretization, with only shell and beam elements, the model amounts to a total of 25000 nodes as shown in Fig. \ref{fig:HV_mesh}. With this model it is possible to quickly evaluate the reaction forces in the joints, the impact of self-weight in the beams (compression/ buckling) and the impact of internal pressure in the deflection of the beams and columns. Fig. \ref{fig:HV_displacements} shows an estimation of the deflection due to these loads on the beams.

\begin{figure}[!htb]
\centering %
\includegraphics[width=0.9\linewidth]{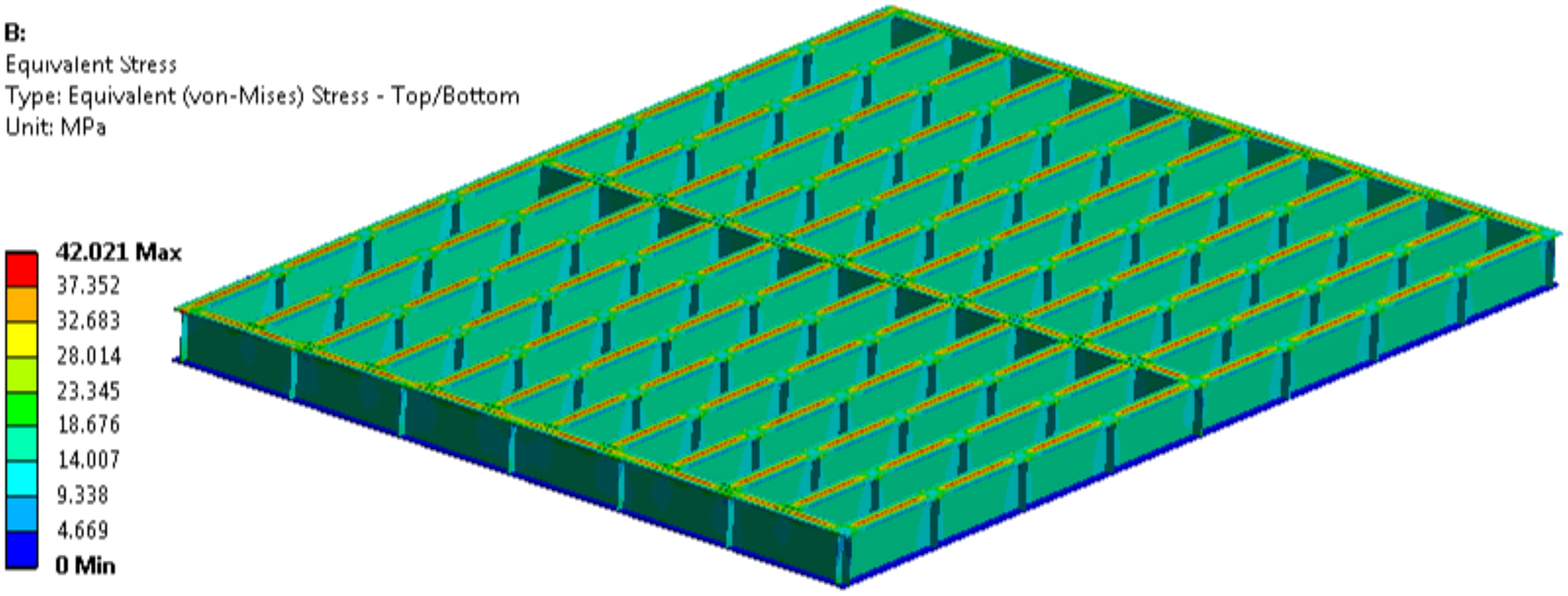}
\caption{Stress distribution in the lower beams}
\label{fig:HV_floor_stress}
\end{figure}

\subsubsubsection{Weight acting on the lower beams}
The equivalent weight of the iron blocks inside the vessel amounts to 5500 tons acting on the lower beams. Using shell elements to model the flanges and webs of the beams, an estimation of the stress involved in supporting this weight was obtained. A maximum stress of 42MPa is expected, which is quite low when compared to the yield strength of steel S235, Fig. \ref{fig:HV_floor_stress}.

\clearpage

\subsection{CFD analysis of the He-vessel system}
\label{Sec:TC:HeV:CFD}
\subsubsection{Purpose}
\label{Sec:TC:CFDHeVess_Purpose}
CFD simulations for the helium vessel have been run in support of the helium vessel design effort. The purpose of these simulations is to confirm that helium circulates properly in the helium vessel, and that the following general constraints are met:
\renewcommand{\labelenumi}{\Alph{enumi}.}
\begin{enumerate}
\item During start-up of the helium passivation system, the formation of air pockets shall be prevented so that the air-to-helium replacement procedure happens smoothly. This constraint will help to minimize the amount of pure helium lost to the atmosphere during the air-helium mixture flushing; it will also minimize the overhead on the helium purification system during the purification phase of the start-up procedure;
\item During operation of the helium passivation system, the helium flow stagnation areas shall be minimized. This constraint will help to minimize the amount of impurities remaining in the helium vessel (not treated by the helium purification system), thus reducing the potential activation of the gas mixture.
\end{enumerate}
Due to the size of the problem being solved and the characteristics of the mesh, the transient version of the problem results to be more challenging than the steady-state version. For this reason, the steady-state circulation of helium has been treated first, and it is presented in this report before the time-dependent version, in opposition to what would be the chronological order of the operations (flushing first, then circulation).

\subsubsection{Scope}
The simulation of the helium circulation in the helium vessel is one aspect of the design of the helium passivation system~\cite{HePurifPredesignRep,HePurifITdocs} that will be operated in the BDF target complex. This helium passivation system will supply purified helium to the helium vessel that contains the BDF target and its shielding and its primary purpose is to remove impurities from the gas mixture in the helium vessel.
Helium provides a low level of activation caused by radiation from the primary beam; combined with the purification capability of the passivation system, this results in minimal activation of the gas mixture contained in the helium vessel. Moreover, a pure helium atmosphere will help protecting the materials from oxidation, thereby increasing the lifetime of the components contained in the helium vessel.
The helium is supplied at the top of the helium vessel, and it is collected from a location at the bottom; further details on layout and components are provided in~\ref{Sec:TC:CFDHeVess_3DMod}. During the start-up mode, pure helium from cylinders is injected and helium-air mixture is extracted from the system and vented to the atmosphere, until helium purity reaches about 85\%vol. At this point, the flushing is interrupted and the helium passivation system begins to purify the mixture by recirculating it through the vessel. During operation, whenever the purity decreases below a given threshold, the purification system would turn on and purify the mixture to the desired level. In this context, optimal flow distribution as described in~\ref{Sec:TC:CFDHeVess_Purpose} is desired. 
Section~\ref{Sec:TC:CFDHeVess_3DMod} presents the 3D model used for the simulation and a summary of the simulation setup, while Sections~\ref{Sec:TC:CFDHeVess_Steady} and \ref{Sec:TC:CFDHeVess_Transient} describe the steady-state and transient simulations, respectively.

\subsubsection{3D Model} 
\label{Sec:TC:CFDHeVess_3DMod}

\begin{figure}[htbp]
\centering %
\includegraphics[width=0.65\linewidth]{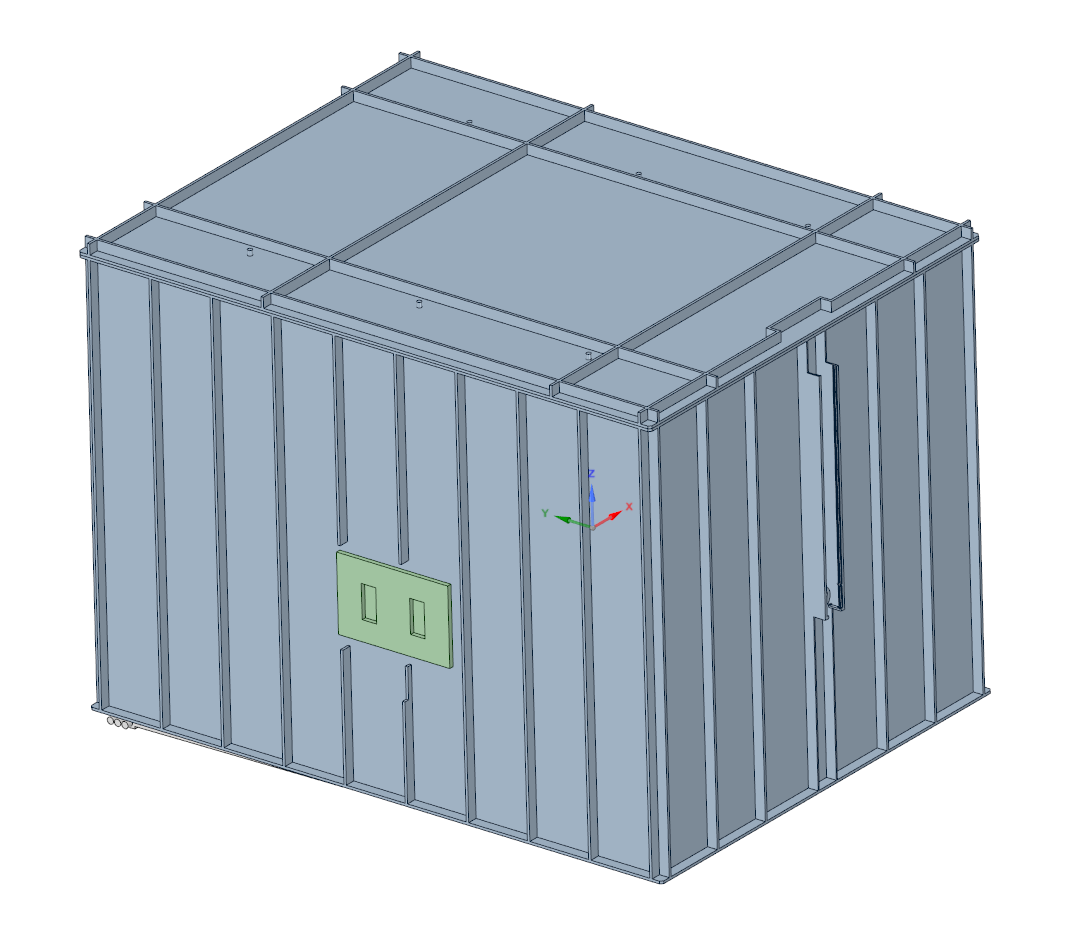}
\caption{\label{Fig:TC:BDF_HeVe_3Dout} 3D model of the helium vessel employed for the CFD studies.}
\end{figure}

The 3D model that has been used for the CFD simulations has been generated in SpaceClaim~\cite{HePurifSpaceClaim}, starting from a model developed in Catia V5 by Oxford Technologies LTD for CERN~\cite{HePurifTCdesign}.
The initial model has been partially cleaned up from components that are not relevant for the CFD simulations; also, some of the remaining components have been simplified in order to ease the solution process.
Figure \ref{Fig:TC:BDF_HeVe_3Dout} shows a view of the helium vessel after model clean-up. A series of ribs supports the steel containment on each face. For the purpose of this simulation, the lid is provided with six cylindrical penetrations (70 mm each) that represent the helium supply and distribution system. Similarly, the outlet is represented by three cylindrical penetrations at the bottom-left corner in Figure \ref{Fig:TC:BDF_HeVe_3Dout}. Figure \ref{Fig:TC:BDF_HeVe_3Dout} also shows the penetration and the door for the trolley that supports the target and allows its insertion and extraction from the helium vessel \cite{HePurifTCdesign}.

The final design of the helium vessel will likely have one single inlet penetration for the helium flow, positioned on one of the side walls of the structure, in order to minimize penetrations through the helium vessel. This is expected not to affect the conclusions of the CFD analysis, since the flow is mainly determined by the layout of the shielding blocks and the gaps between them.

\begin{figure}[!htbp]
\centering %
\includegraphics[width=0.7\linewidth]{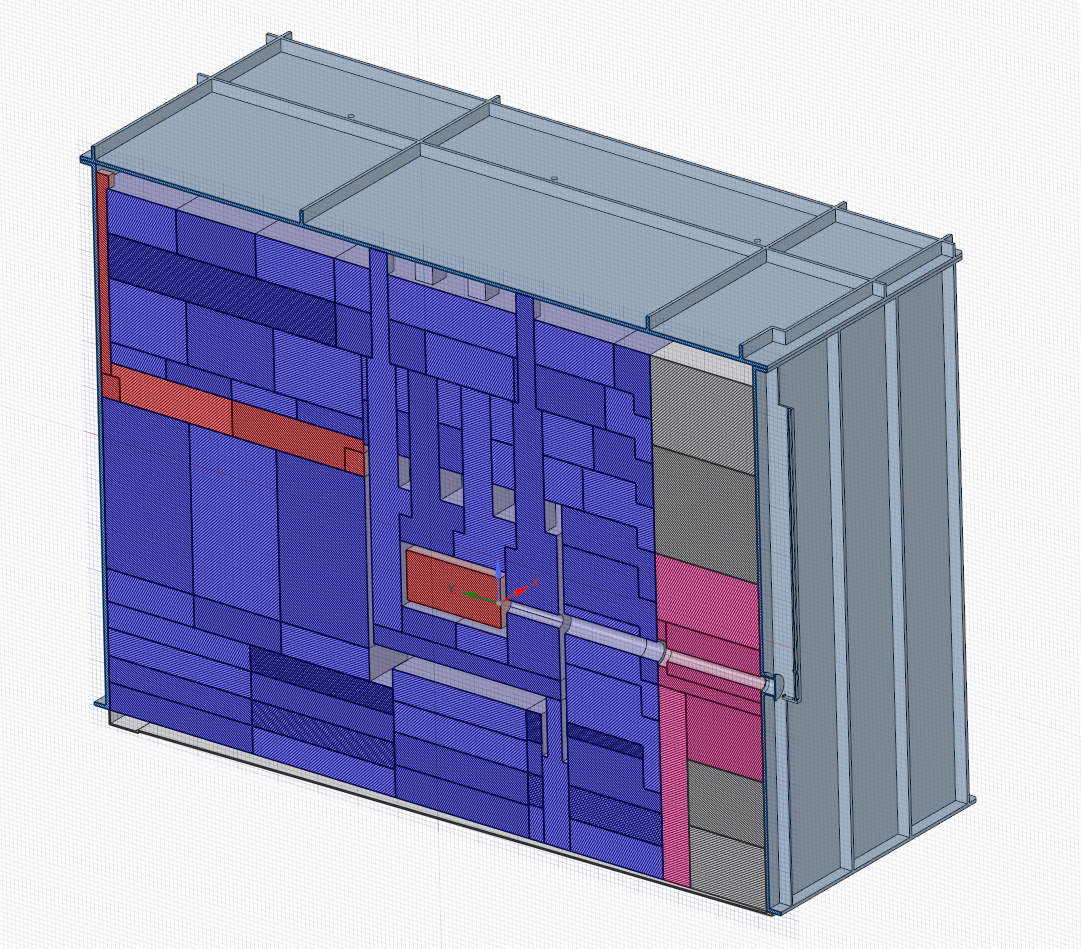}
\caption{\label{Fig:TC:BDF_HeVe_3Dslice} Vertical slice of the helium vessel model on a plane normal to the trolley insertion/extraction line.}
\end{figure}

Figure~\ref{Fig:TC:BDF_HeVe_3Dslice} shows a slice of the helium vessel on a vertical plane passing through the beam axis; the target is visible in red at the center of the figure. Figure~\ref{Fig:TC:BDF_HeVe_3Dslice} also shows the following components:
\begin{itemize}
\item Collimator assembly components (pink);
\item Collimator concrete shielding (gray);
\item Proximity shielding (blue);
\item Coil assembly blocks (blue);
\item Bunker blocks (blue);
\item Magnetic coil (red);
\end{itemize}

\begin{figure}[!htbp]
\centering %
\includegraphics[width=0.7\linewidth]{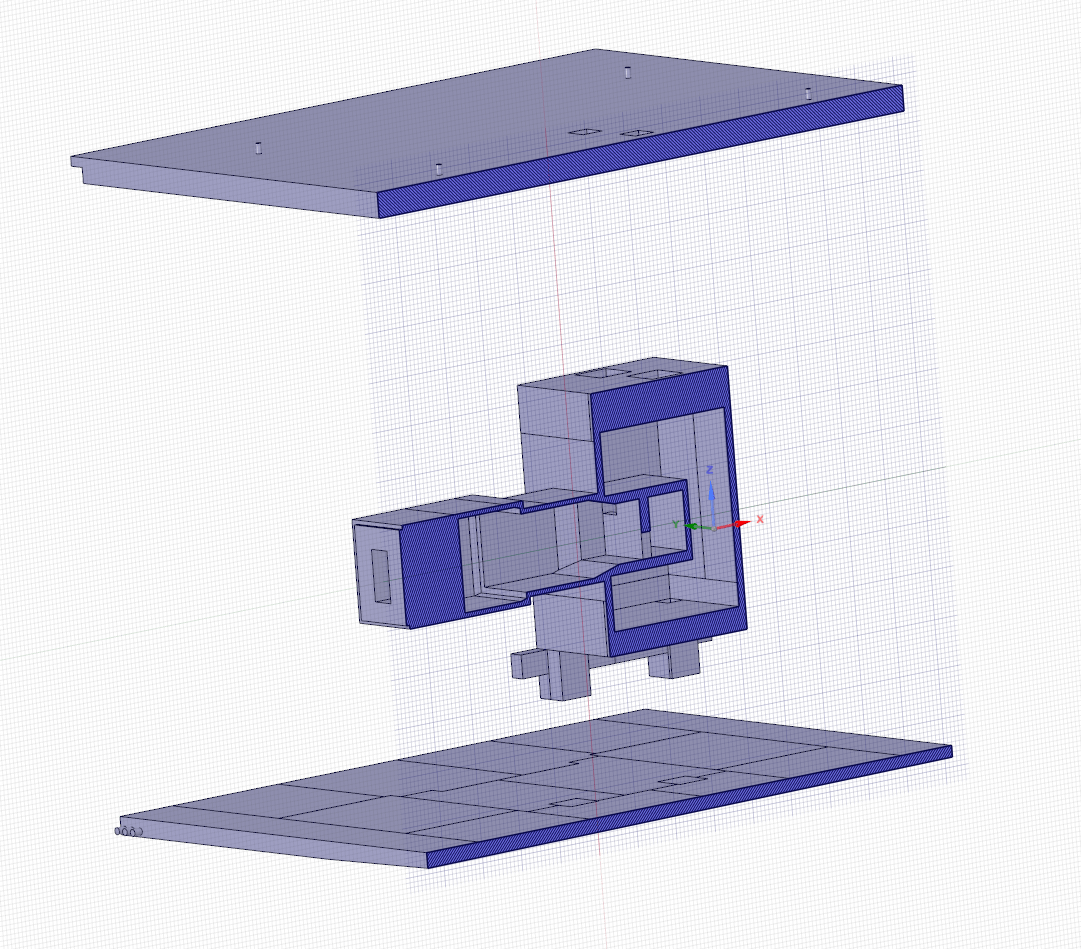}
\caption{\label{Fig:TC:BDF_HeVe_3Dfrslice} Slice of the free volume on a plane normal to the beam axis and crossing the center of the BDF target.}
\end{figure}

Figure~\ref{Fig:TC:BDF_HeVe_3Dslice} also gives an idea of the free volume available in the helium vessel. In order to facilitate the modelling, all gaps between blocks (except for those thicker than 10 mm) have been eliminated; they will be generated again during the inflation phase of the meshing process. The resulting remaining free volume has been extracted and it is shown in Figure \ref{Fig:TC:BDF_HeVe_3Dfrslice}.

\subsubsection{Meshing}
\label{Sec:TC:CFDHeVess_Mesh}

The size of the model and the presence of a high number of thin volumes make the meshing process relatively challenging. The mesh has been developed using Fluent Meshing, using the following approach:
\begin{itemize}
\item Step 1. The SpaceClaim model (\ref{Sec:TC:CFDHeVess_3DMod}) is imported into Fluent Meshing using the CFD surface mesh option and the parameters listed in Table \ref{Tab:TC:BDF_HeVe_SurfMesh}. This option produces a fully connected and conformal mesh based on an automatic size function;
\item Step 2. Volumetric regions are computed and a first scoped prism layer is defined according to parameters in Table \ref{Tab:TC:BDF_HeVe_Prisms}. With this prism growth configuration, each face grows two layers on both sides, for a total of four layers and 10 mm gap thickness. Note that this mesh is based on the assumption that the gap between blocks is 10 mm for all gaps in the model;  
\item Step 3. A preliminary volume mesh with prisms is computed using the auto-fill-volume option. This step is performed with the aid of a journal file that, for each volume, grows the prism layer according to the definition of step 2 and fills with tetrahedrons. This process requires several hours to fill all 197 volumes in the model. At the end of the process, unutilized tetrahedral cell zones (corresponding to solid blocks) are deleted and the remaining volumes are merged to create a single fluid cell zone. Note that this is the step in which the gaps between blocks are created;
\item Step 4. Two prism layers are grown on each face by morphing the existing mesh, thus resulting in a total of eight transversal cells per gap. The thickness of the two layers is 1 mm for the first and 1.5 mm for the second. The purpose of these two more layers is to improve the velocity profile near the walls and enhance convergence. The mesh is now ready for final clean-up, transfer to solution mode and conversion to polyhedra.
\end{itemize}

\begin{table}[!htbp]
\centering
\caption{\label{Tab:TC:BDF_HeVe_SurfMesh} He-vessel CFD surface mesh parameters.}
\smallskip
\begin{tabular}{l|r|l}
\hline
\textbf{Parameter} & \textbf{Value} & \textbf{Unit}\\
\hline
Min. size & 25 & mm \\
Max. size & 80 & mm \\
Growth rate & 1.2 & - \\
\hline
\end{tabular}
\end{table}

\begin{table}[htbp]
\centering
\caption{\label{Tab:TC:BDF_HeVe_Prisms} He-vessel CFD prism definition parameters.}
\smallskip
\begin{tabular}{l|r|l}
\hline
\textbf{Parameter} & \textbf{Value} & \textbf{Unit}\\
\hline
Type & Uniform & - \\
First layer thickness & 2.5 & mm \\
Number of layers & 2 & - \\
Max. thickness & 5 & mm \\
Growth rate & 1 & - \\
\hline
\end{tabular}
\end{table}

\begin{figure}[htbp]
\centering %
\includegraphics[width=0.6\linewidth]{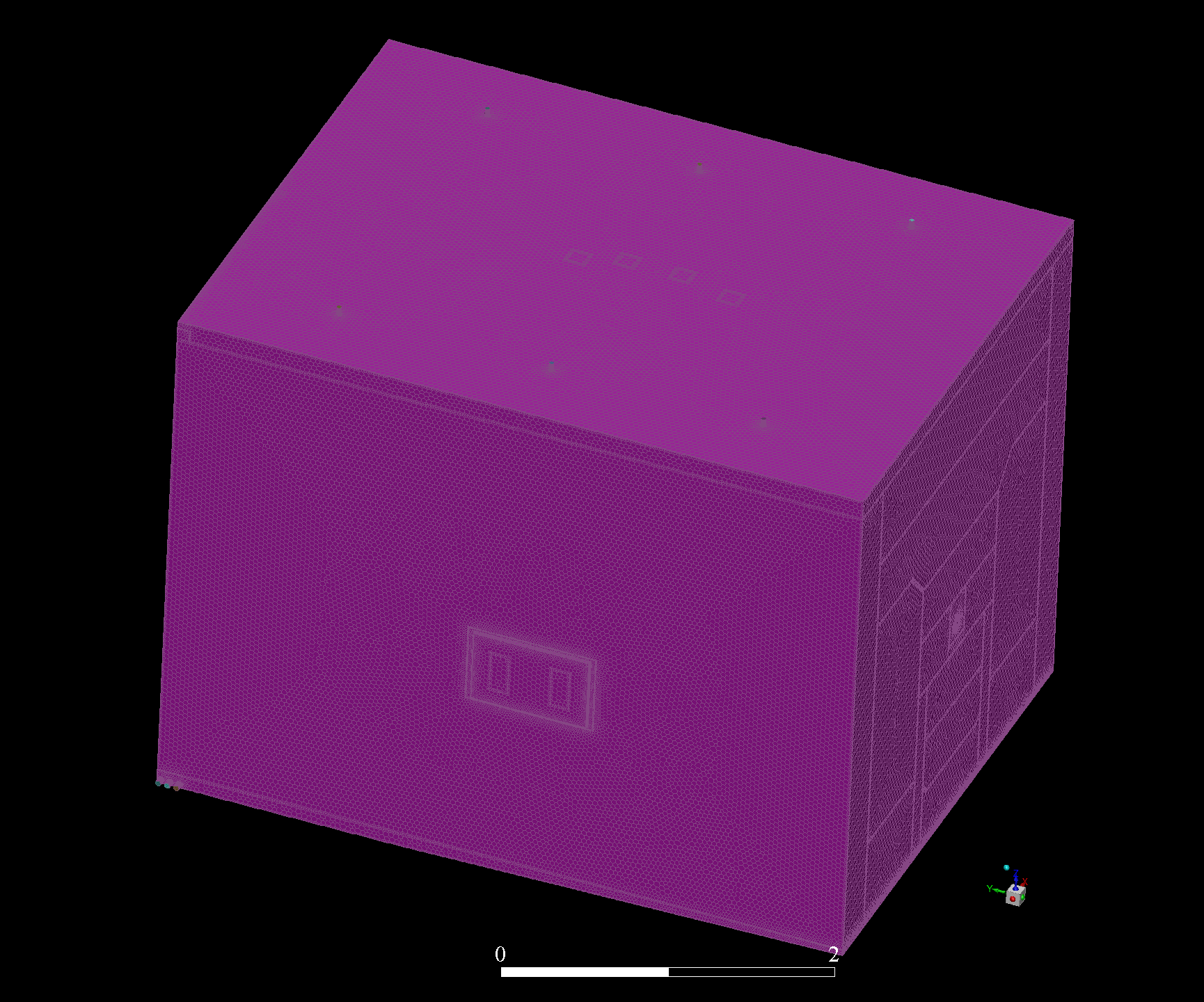}
\caption{\label{Fig:TC:BDF_HeVe_Mesh2} Overview of the polyhedral mesh.}
\end{figure}

\begin{figure}[htbp]
\centering %
\includegraphics[width=0.7\linewidth]{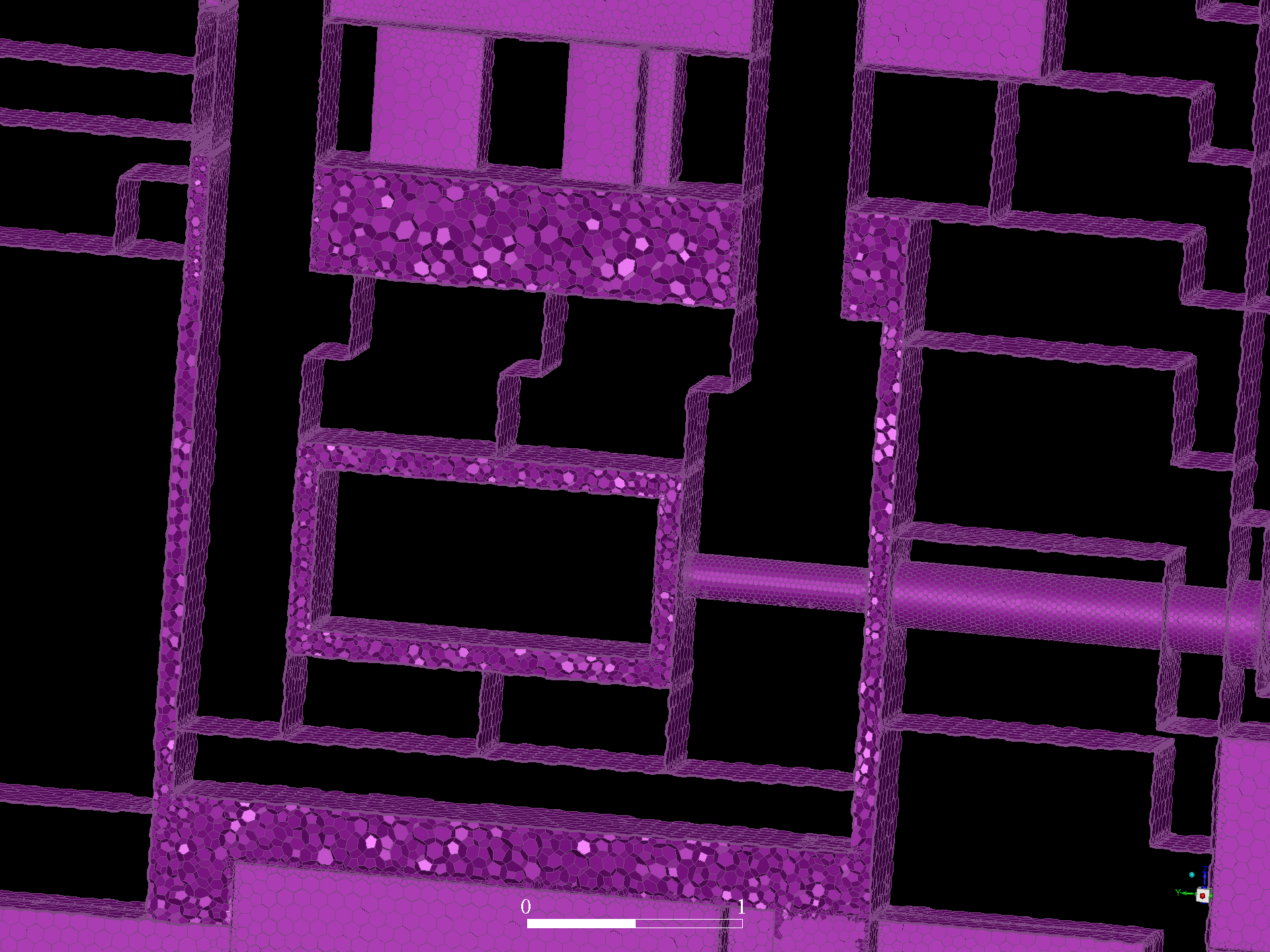}
\caption{\label{Fig:TC:BDF_HeVe_Mesh1} Mesh slice in the collimator-target region of the He-vessel.}
\end{figure}

Figure~\ref{Fig:TC:BDF_HeVe_Mesh2} shows the polyhedral mesh resulting from the meshing process. Table~\ref{Tab:TC:BDF_HeVe_MeshStat} presents the main parameters, statistics and size of the mesh. Figure~\ref{Fig:TC:BDF_HeVe_Mesh1} shows a slice of the mesh in the area between the collimator and the target.

\begin{table}[htbp]
\centering
\caption{\label{Tab:TC:BDF_HeVe_MeshStat} He-vessel CFD mesh statistics.}
\smallskip
\begin{tabular}{l|r|l}
\hline
\textbf{Parameter} & \textbf{Value} & \textbf{Unit}\\
\hline
Size & 6782628 & cells \\
Allocated memory & 10982 & Mbytes \\
Total volume & 74.5 & m$^3$ \\
Min. orthogonal quality & 3.34 & \%\\
Max. aspect-ratio & 328 & -\\
Mesh check & Passed & -\\
\hline
\end{tabular}
\end{table}

\begin{figure}[htbp]
\centering %
\includegraphics[width=0.7\linewidth]{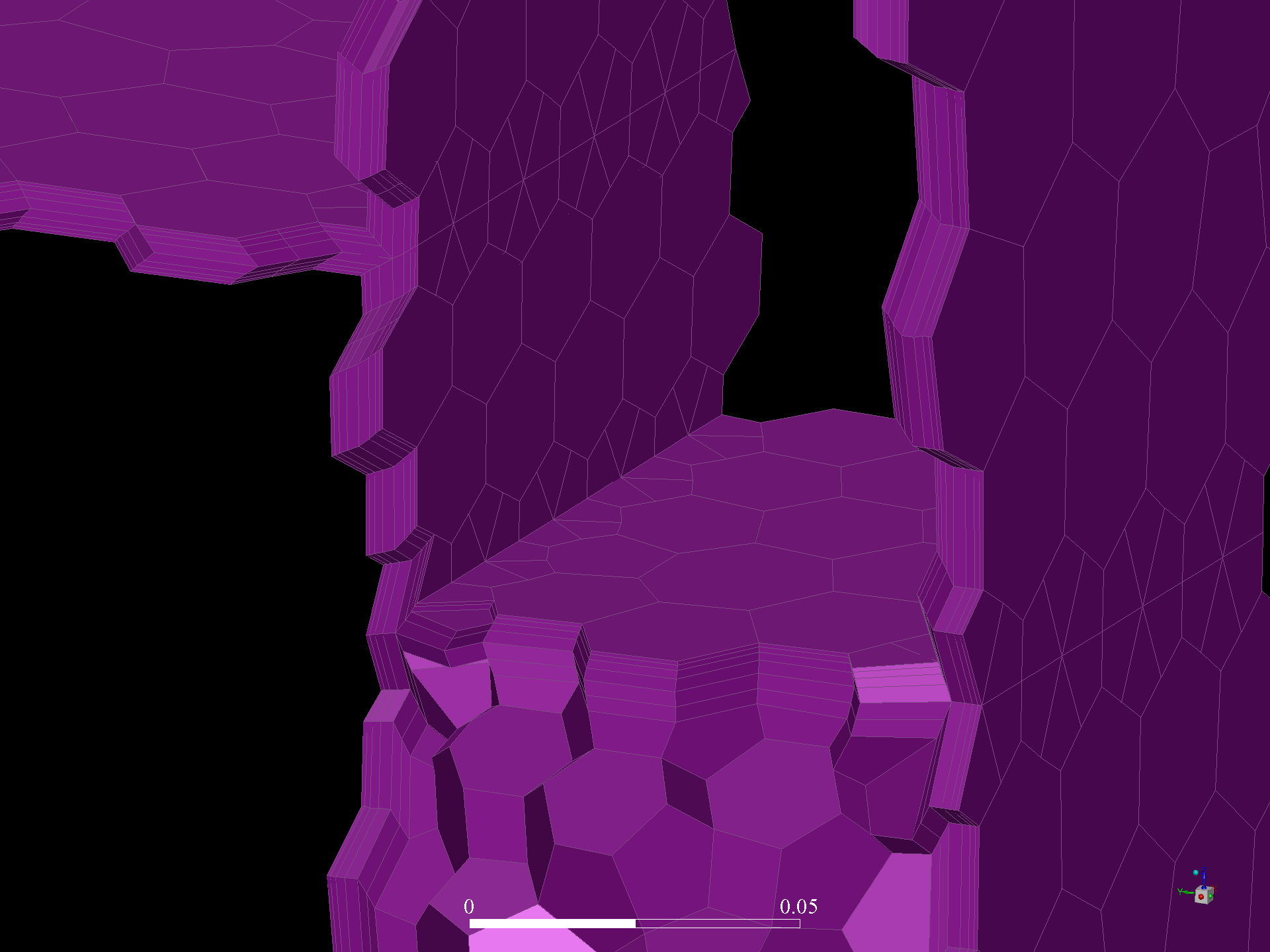}
\caption{\label{Fig:TC:BDF_HeVe_Mesh5} Detail of mesh in gaps between blocks.}
\end{figure}

The main contribution to the mesh size is given by the thin cells in the gaps between blocks. Figure~\ref{Fig:TC:BDF_HeVe_Mesh1} shows the gaps distribution on a vertical slice passing through the beam axis and gives an idea of the number of gaps present in the model. 
Figure \ref{Fig:TC:BDF_HeVe_Mesh5} shows a detail of the mesh in some of these gaps. The cells in the gaps have a high aspect-ratio, which becomes particularly critical in proximity of intersections of different gaps, where the flow direction changes sharply. At these locations, convergence of the solution is more challenging; however a reasonably converging solution has been obtained. On the other side, a finer mesh could not be used due to limitations in computational power.

\subsubsection{Steady-State Simulation: Helium Circulation}
\label{Sec:TC:CFDHeVess_Steady}

A steady-state simulation of helium circulation has been run to analyze the pressure and velocity distributions and to identify locations where helium velocity is particularly small.

\subsubsubsection{Simulation setup}

The setup of the simulation has been done according to the following points:
\begin{itemize}
\item Solver type: pressure-based, for incompressible flow (maximum velocity of helium is less than 30\% of the speed of sound);
\item Turbulence modeling: realizable k- with enhanced wall treatment;
\item Materials: the fluid has been assumed as pure helium gas with properties listed in Table \ref{Tab:TC:BDF_HeVe_HeMatProp};
\item Boundary conditions:
\renewcommand{\labelitemii}{$\bullet$}
\begin{itemize}
\item Six velocity inlets; inlet velocity equal to 10 m/s, normal to boundary; total flow for six inlets equal to 764.45 m\textsuperscript{3}/h;
\item Three pressure outlets;
\item Remaining surfaces modeled as walls.
\end{itemize}
\end{itemize}

\begin{table}[htbp]
\centering
\caption{\label{Tab:TC:BDF_HeVe_HeMatProp} Thermophysical properties of helium as implemented in the CFD simulation.}
\smallskip
\begin{tabular}{l|r|l}
\hline
\textbf{Parameter} & \textbf{Value} & \textbf{Unit}\\
\hline
Density & 0.1625 & kg/m\textsuperscript{3} \\
Molecular weight & kg/kmol & 4.0026 \\
Specific heat & 5193 & J/kg*K \\
Thermal conductivity & 0.152 & W/m*K \\
Dynamic viscosity & 1.99e-5 & kg/m*s\\
\hline
\end{tabular}
\end{table}

The flow rate selected for the simulation, 764.45 m\textsuperscript{3}/h, represents the flow provided by the cooling compressor of the helium passivation system; note that the design flow of the purification part of the helium passivation system is much smaller, 75 m\textsuperscript{3}/h \cite{HePurifITdocs}.

\subsubsubsection{Solution setup}

The setup of the solution has been done according to the following points:
\begin{itemize}
\item Solution method: SIMPLE scheme with second order discretization for all equations and default under-relaxation factors;
\item Default hybrid initialization;
\item Calculation run for several hundred iterations; steady-state solution is reached after about 525 iterations.
\end{itemize}

\subsubsubsection{Steady-state results}
Due to the large number of gaps and the relatively low number of cells across these gaps (which reduce the effectiveness of pathlines plots), it is difficult to provide an overall 3D representation of the resulting velocity distribution.

\begin{figure}[htbp]
\centering %
\includegraphics[width=0.7\linewidth]{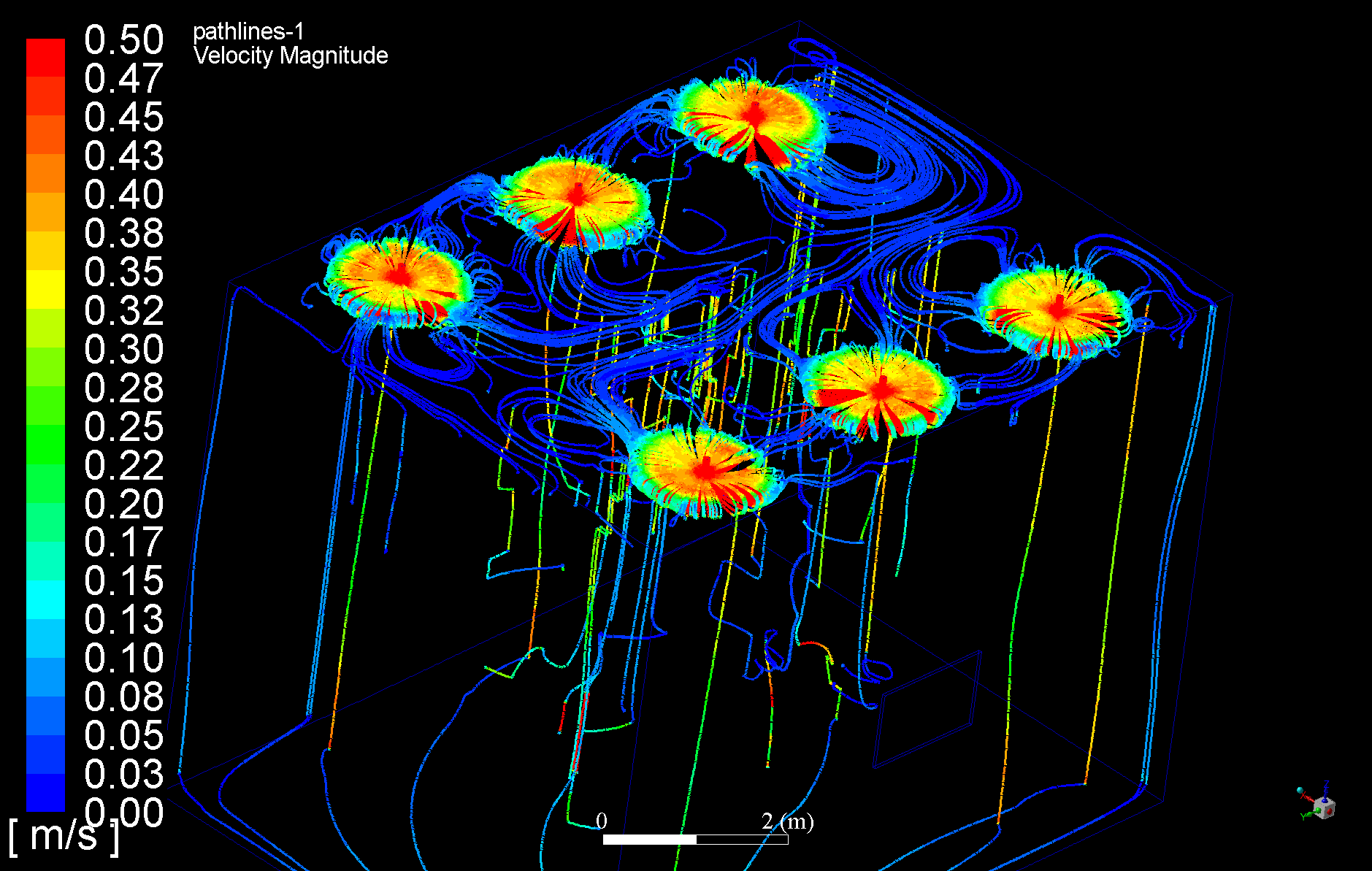}
\caption{\label{Fig:TC:BDF_HeVe_St1} Pathlines colored by velocity magnitude in the upper plenum of the He-vessel.}
\end{figure}

\begin{figure}[htbp]
\centering %
\includegraphics[width=0.7\linewidth]{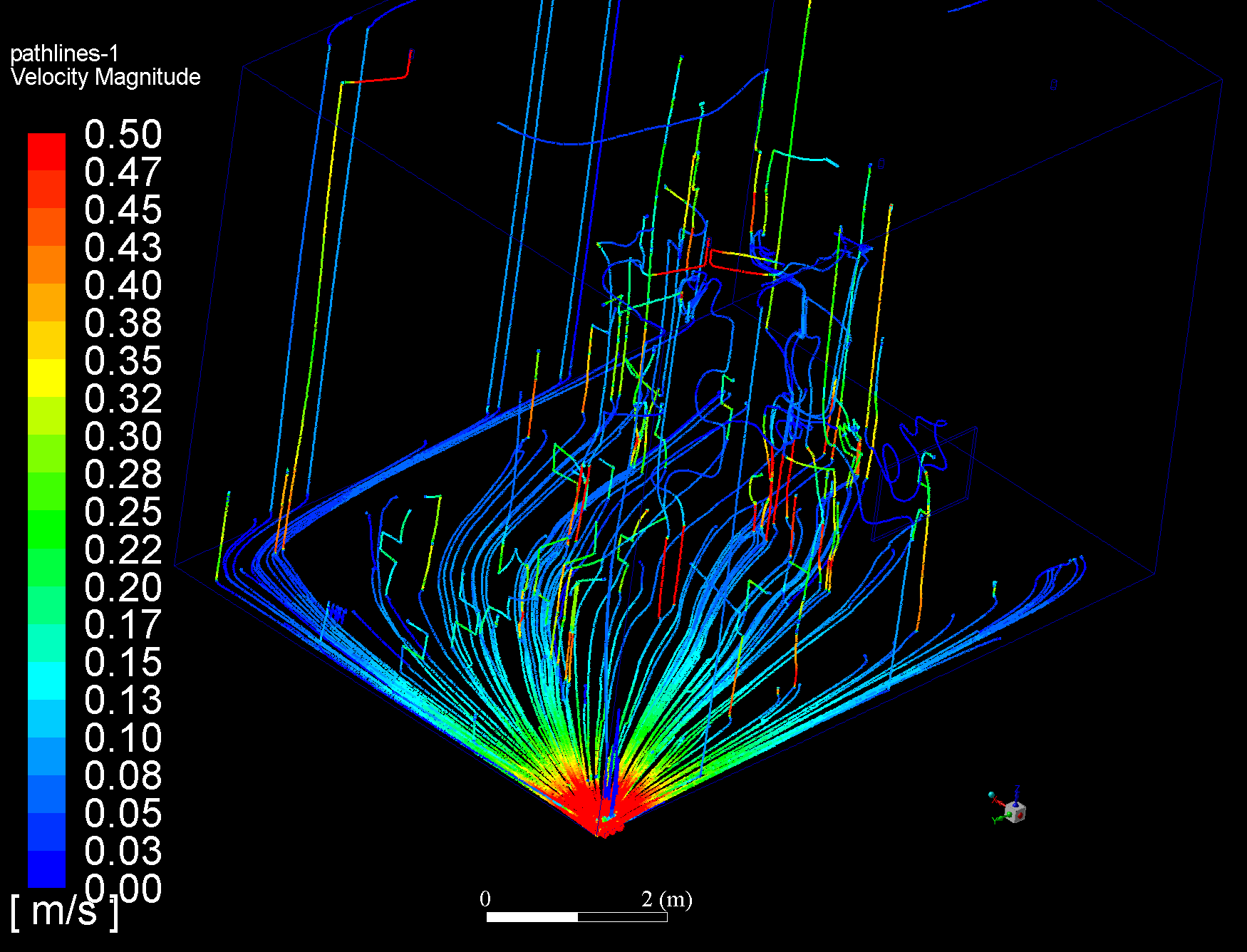}
\caption{\label{Fig:TC:BDF_HeVe_St2} Pathlines colored by velocity magnitude in the lower plenum of the He-vessel.}
\end{figure}

Figure~\ref{Fig:TC:BDF_HeVe_St1} and Figure \ref{Fig:TC:BDF_HeVe_St2} show the pathlines colored by velocity magnitude in the upper and lower plenum in the He-vessel, respectively. The plots are constrained to a maximum velocity of 0.5 m/s.

In the upper plenum, the flow distribution is dominated by recirculation zones in proximity of the six inlet locations. Everywhere else the velocity is smaller and the pathlines are less well-defined and less dense. Figure \ref{Fig:TC:BDF_HeVe_St1} also gives an idea of the complicated path of the flow through the upper blocks of the shielding.

The helium flow through the shielding is also visible in Figure \ref{Fig:TC:BDF_HeVe_St2}. Sudden acceleration and deceleration of flow at the gap intersection areas make the flow distribution relatively irregular. The flow distribution however becomes more regular when helium approaches the outlets.

\begin{figure}[htbp]
\centering %
\includegraphics[width=0.7\linewidth]{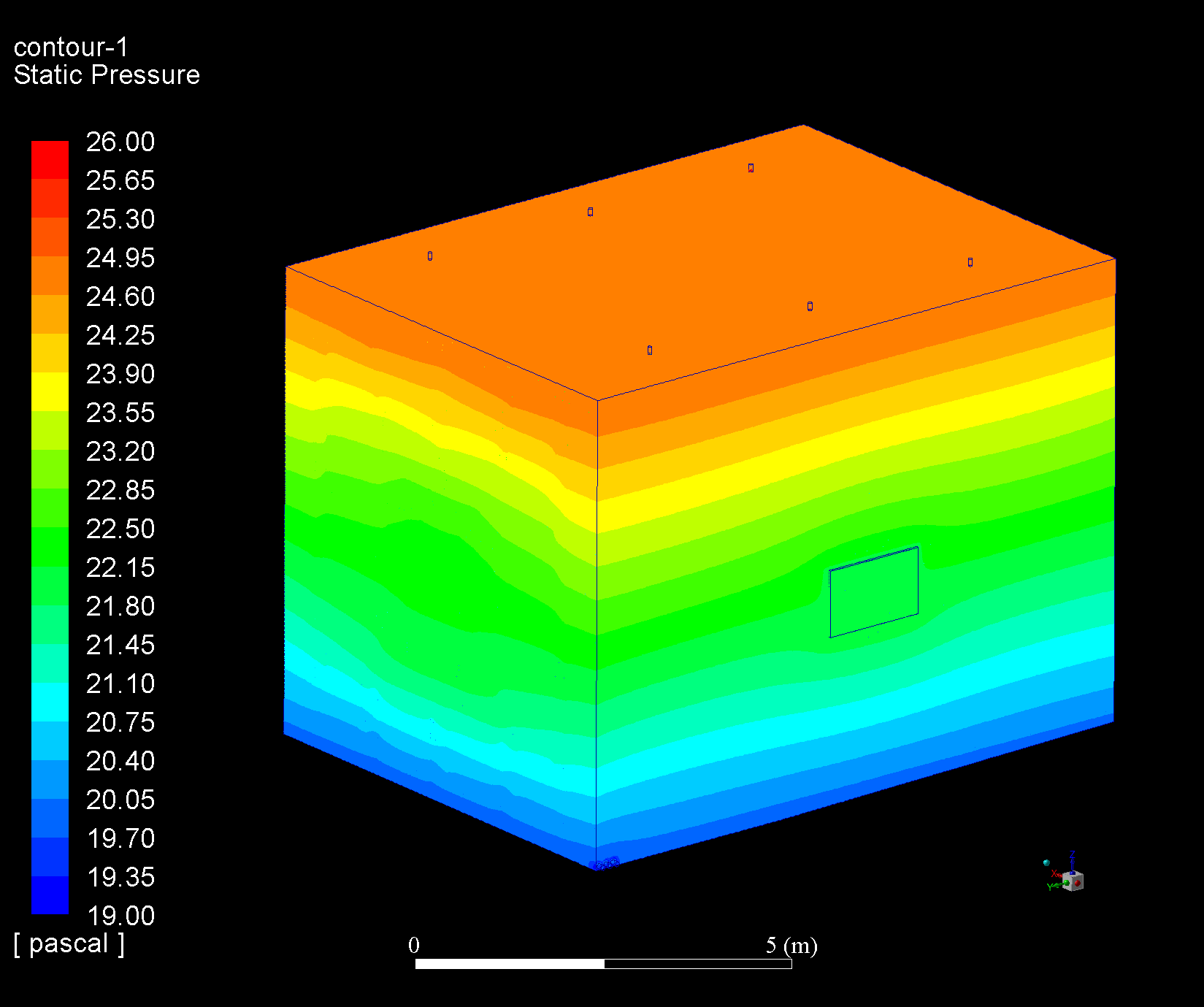}
\caption{\label{Fig:TC:BDF_HeVe_St3} Static pressure contours on the boundaries of the model}
\end{figure}

\begin{figure}[htbp]
\centering %
\includegraphics[width=0.7\linewidth]{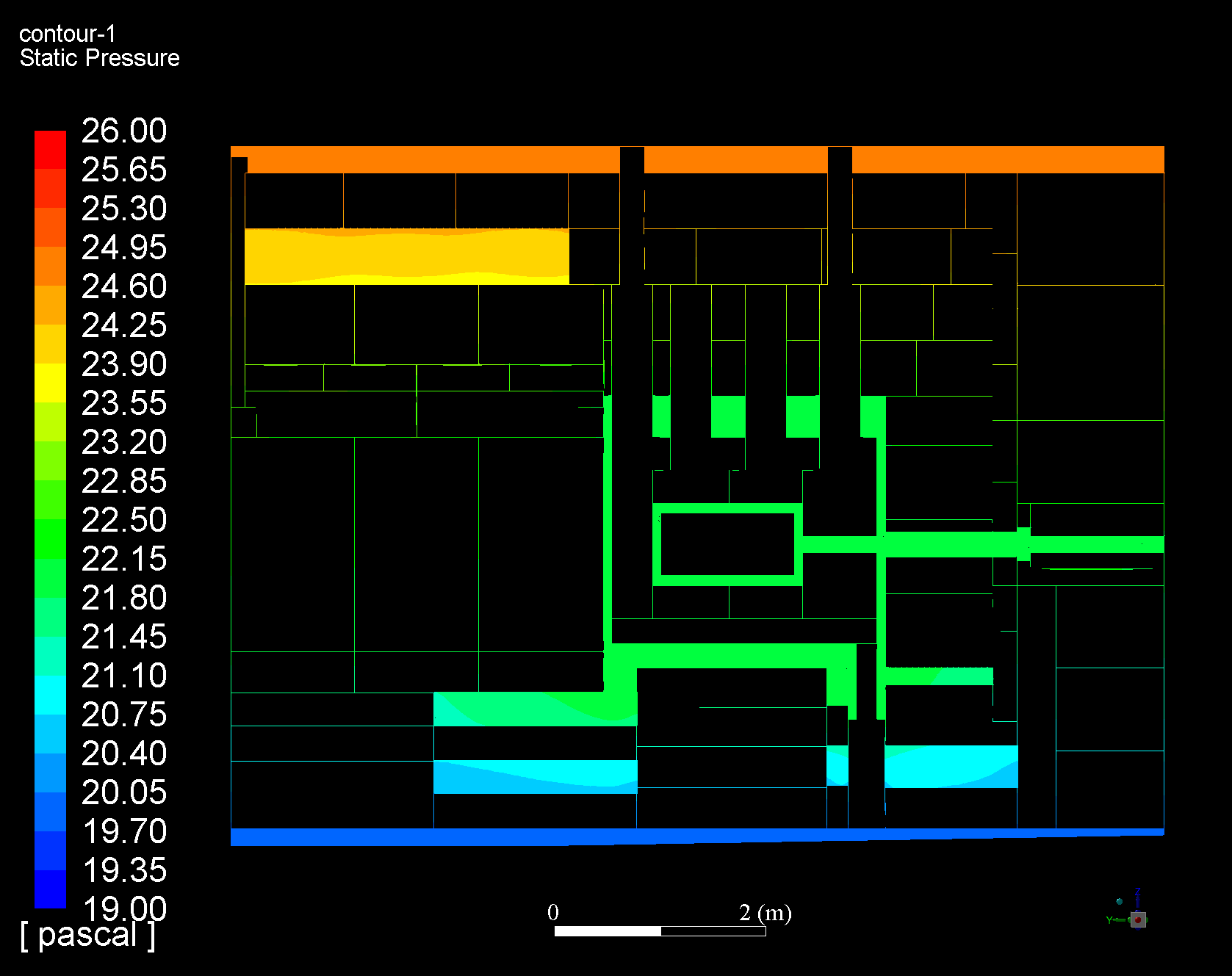}
\caption{\label{Fig:TC:BDF_HeVe_St4} Static pressure contours on a vertical plane containing the beam axis}
\end{figure}

Looking at pressure distribution, Figure~\ref{Fig:TC:BDF_HeVe_St3} shows the static pressure contours on the outer boundary faces of the model, those in contact with the helium vessel structure. Figure~\ref{Fig:TC:BDF_HeVe_St4} shows the same contours on a vertical plane containing the beam axis. 

Both figures show that the pressure distribution grows uniformly from upper plenum to lower plenum, in such a way that the pressure drop across the helium vessel is on the order of 25 Pa. 

Figure~\ref{Fig:TC:BDF_HeVe_St4} shows that the presence of gaps does not significantly alter the pressure distribution in the internal channels of the helium vessel, in particular the fact that pressure uniformly increases from top to bottom; this indicates that the flow distribution is acceptable.

\begin{figure}[htbp]
\centering %
\includegraphics[width=0.7\linewidth]{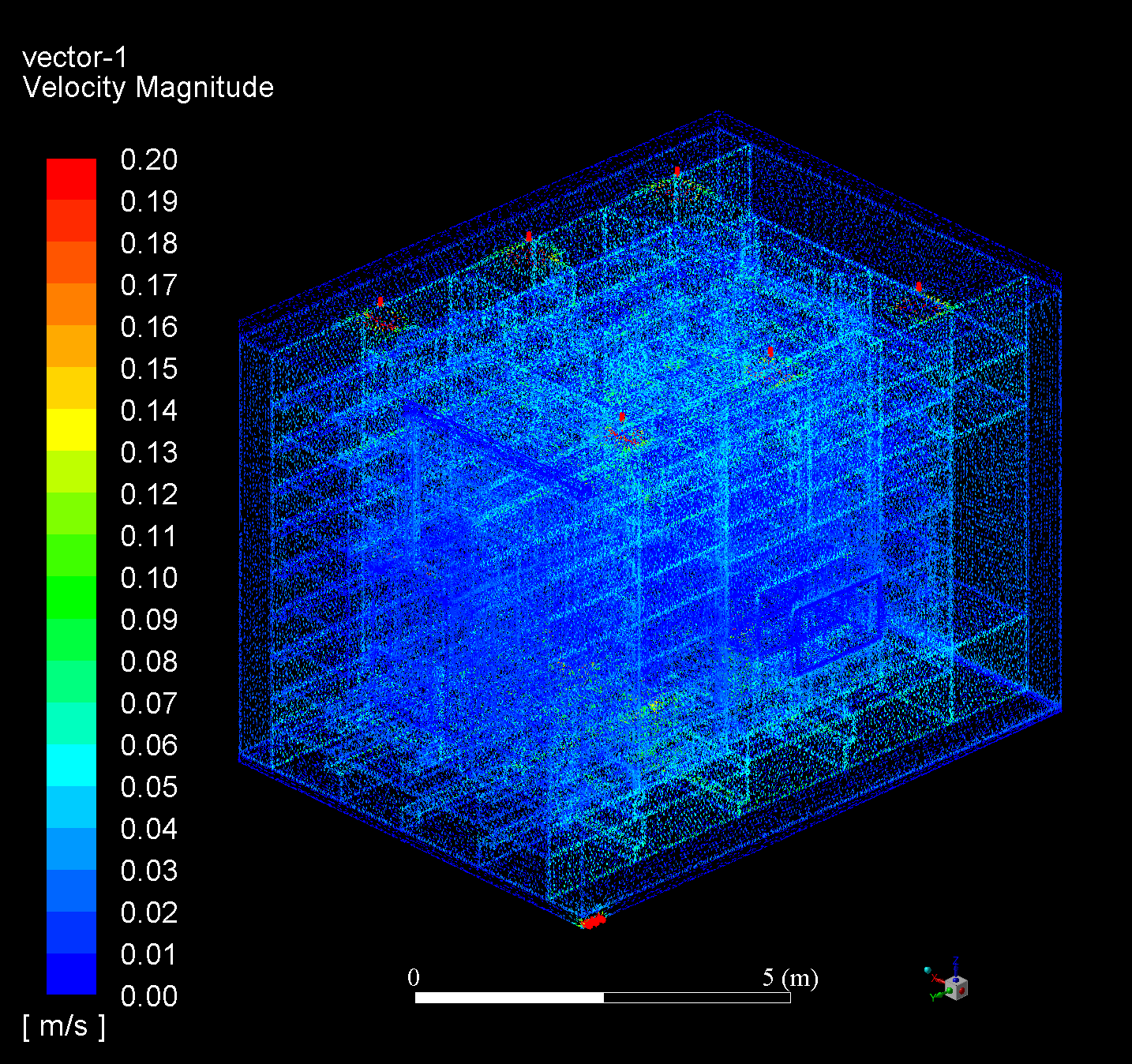}
\caption{\label{Fig:TC:BDF_HeVe_St5} Vector plot colored by velocity magnitude and limited to 0.2 m/s.}
\end{figure}

The primary intent of the steady simulation is the identification of areas where the flow velocity is close to 0 m/s. Figure \ref{Fig:TC:BDF_HeVe_St5} shows a vector plot colored by velocity magnitude and limited to 0.2 m/s; note that all vectors are scaled to the same length and, due to the size of the model, vectors resemble points.
Excluding the inlet and outlet locations (where velocity is well beyond 20 cm/s), the velocity distribution generally looks uniform and not higher than 10 cm/s. However the plot does not allow to identify any critical locations.

\begin{figure}[htbp]
\centering %
\includegraphics[width=0.7\linewidth]{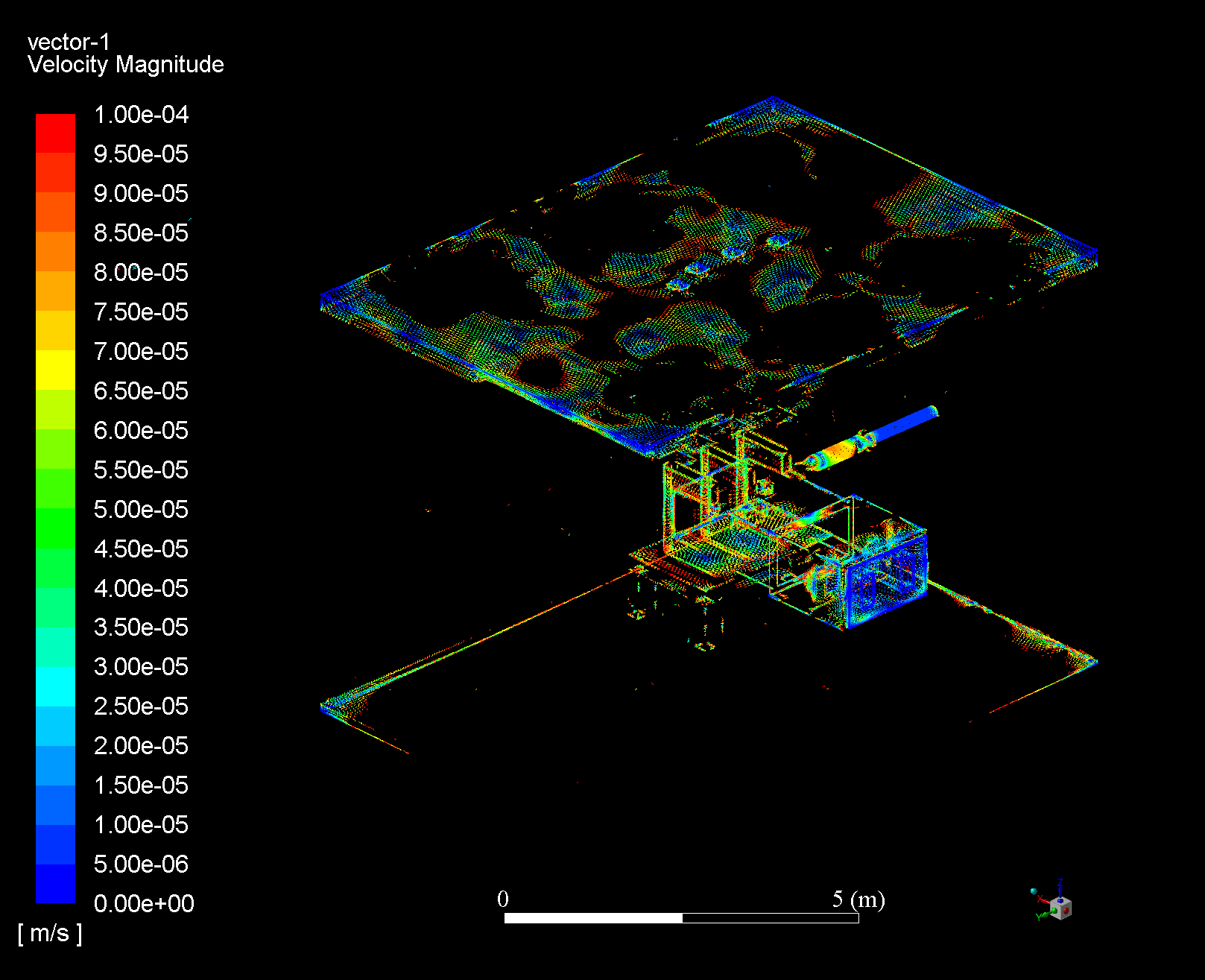}
\caption{\label{Fig:TC:BDF_HeVe_St6} Vector plot colored by velocity magnitude and clipped to 0.1 mm/s.}
\end{figure}

In order to show areas of low velocity, the scale has been reduced to a maximum velocity of 0.1 mm/s, corresponding to 6 mm/min. Figure \ref{Fig:TC:BDF_HeVe_St6} shows a vector plot colored by velocity magnitude and clipped to 0.1 mm/s. The figure shows that the main critical areas are the trolley, the target the proximity shielding and the collimator assembly. Note that the areas identified in the upper and lower plenum are not as critical because they are located in open volumes, for which recirculation and gravity would help providing some mixing effect.

\begin{figure}[htbp]
\centering %
\includegraphics[width=0.7\linewidth]{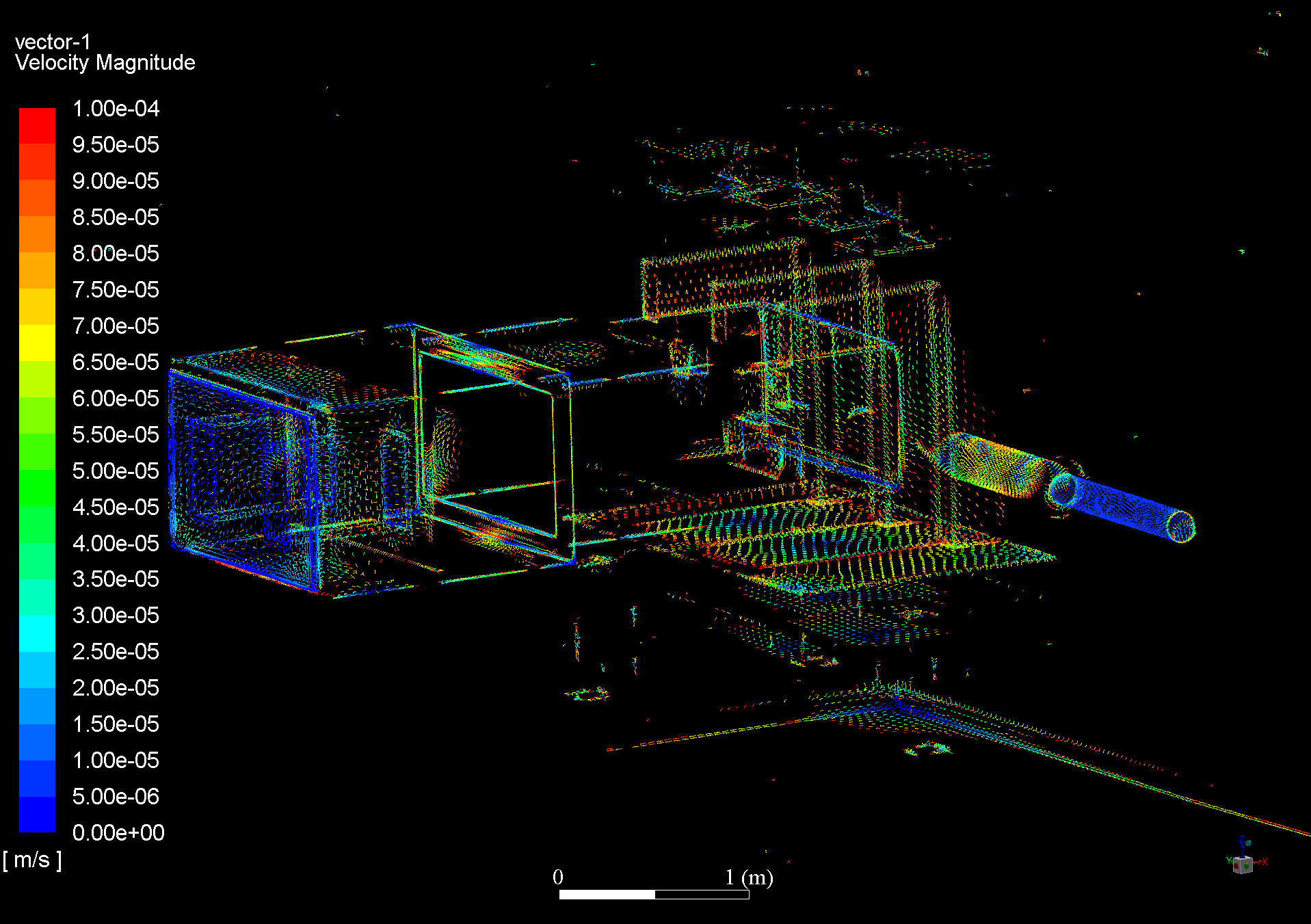}
\caption{\label{Fig:TC:BDF_HeVe_St7} Vector plot colored by velocity magnitude and clipped to 0.1 mm/s.}
\end{figure}

Figure \ref{Fig:TC:BDF_HeVe_St7} represents the same plot of Figure~\ref{Fig:TC:BDF_HeVe_St6} (vector plot colored by velocity magnitude clipped to 0.1 mm/s) zoomed on the target area. The following low-velocity areas can be identified:
\begin{itemize}
\item Collimator and beam pipe; 
\item Gaps between proximity shielding blocks;
\item Gaps below proximity shielding;
\item Trolley structure.
\end{itemize}
Collimator and Trolley areas are less critical, due to the fact that the gaps are not excessively thin; the transient simulation (\ref{Sec:TC:CFDHeVess_Transient}) provides deeper insight on the impurities left in these areas.
On the other side, the gaps between and below the proximity shielding blocks are more critical: since these blocks are open on one side, the helium prefers to flow around them than through them.

\subsubsection{Transient Simulation: Helium Filling Process}
\label{Sec:TC:CFDHeVess_Transient}

A transient simulation of the helium filling process has been run to identify air pockets that remain in the helium vessel during the flushing.

\subsubsubsection{Simulation setup}

The setup of the simulation is analogous to Section~\ref{Sec:TC:CFDHeVess_Steady} (Steady-state simulation), except for the following points:
\begin{itemize}
\item Energy equation: on;
\item Species model: species transport;
\item Materials: mixture of two fluids, helium and air. Properties of helium are listed in Table \ref{Tab:TC:BDF_HeVe_HeMatProp}; properties of air are listed in Table \ref{Tab:TC:BDF_HeVe_AirMatProp}; properties of the mixture are listed in Table \ref{Tab:TC:BDF_HeVe_MixMatProp}; 
\item Boundary conditions are identical to those for the steady-state simulation; helium concentration at inlets is 100\%;
\end{itemize}

\begin{table}[htbp]
\centering
\caption{\label{Tab:TC:BDF_HeVe_AirMatProp} Thermophysical properties of air as implemented in the CFD simulation.}
\smallskip
\begin{tabular}{l|r|l}
\hline
\textbf{Parameter} & \textbf{Value} & \textbf{Unit}\\
\hline
Molecular weight & 28.966 & kg/kmol \\
Specific heat & 1006.43 & J/kg*K \\
Thermal conductivity & 0.0242 & W/m*K \\
Dynamic viscosity & 1.7894e-5 & kg/m*s\\
\hline
\end{tabular}
\end{table}

\begin{table}[htbp]
\centering
\caption{\label{Tab:TC:BDF_HeVe_MixMatProp} Selected models for calculation of thermophysical properties of he-air mixture.}
\smallskip
\begin{tabular}{l|r}
\hline
\textbf{Parameter} & \textbf{Model} \\
\hline
Density & Ideal gas \\
Specific heat & Mixing law \\
Thermal conductivity & Mass-weighted mixing law \\
Dynamic viscosity & Mass-weighted mixing law \\
Mass diffusivity & Constant dilute approximation \\
\hline
\end{tabular}
\end{table}

\subsubsubsection{Solution setup}

The setup of the solution is analogous to Section~\ref{Sec:TC:CFDHeVess_Steady} (Steady-state simulation), except for the following points:
\begin{itemize}
\item Solution method: PISO scheme with 4 skewness correction iterations, 0 neighbor correction iterations;
\item Spatial discretization: see Table \ref{Tab:TC:BDF_HeVe_SpDiscr};
\item Transient formulation: bounded second order implicit;
\item Under-relaxation factors: see Table \ref{Tab:TC:BDF_HeVe_URF};
\item Hybrid initialization: air concentration equal to 100\% at start of simulation in all cells;
\item Transient calculation parameters: fixed time-step (10 max iterations per time-step), 0.1 s up to 10 s, then 1 s up to 784 s (final time). Total simulation time is equivalent to 166.5 m\textsuperscript{3} of pure helium injected at 764.45 m\textsuperscript{3}/h. Since the helium vessel free volume is 74.5 m\textsuperscript{3}, the mentioned pure helium amount is equivalent to 2.23 gas changes. 
\end{itemize}
Note that one auto-save file is about 2.29 GB, so auto-saving every 2 seconds results in an output files folder of about 898 GB on the hard-drive. 

\begin{table}[htbp]
\centering
\caption{\label{Tab:TC:BDF_HeVe_SpDiscr} Selected models for spatial discretization of the Navier–Stokes equations. Higher-order models are used to reduce numerical diffusion.}
\smallskip
\begin{tabular}{l|r}
\hline
\textbf{Parameter} & \textbf{Model} \\
\hline
Gradient & Least squares cell based \\
Pressure & PRESTO! \\
Momentum & Third order MUSCL \\
Turbulent kinetic energy & Third order MUSCL \\
Turbulent dissipation rate & Third order MUSCL \\
Air & Second order upwind \\
Energy & Second order upwind \\
\hline
\end{tabular}
\end{table}

\begin{table}[htbp]
\centering
\caption{\label{Tab:TC:BDF_HeVe_URF} Under-relaxation factors. Values lower than default have been implemented in order to increase the numerical stability of the solution.}
\smallskip
\begin{tabular}{l|c}
\hline
\textbf{Parameter} & \textbf{Under-relaxation factor} \\
\hline
Pressure & 0.7 \\
Density & 0.9 \\
Body forces & 0.9 \\
Momentum & 0.3 \\
Turbulent kinetic energy & 0.7 \\
Turbulent dissipation rate & 0.7 \\
Turbulent viscosity & 0.7 \\
Air & 1.0 \\
Energy & 1.0 \\
\hline
\end{tabular}
\end{table}

\subsubsubsection{Transient results}

The transient simulation has been performed to characterize the helium distribution inside the helium vessel during the flushing process.

\begin{figure}[htbp]
\centering %
\includegraphics[width=0.7\linewidth]{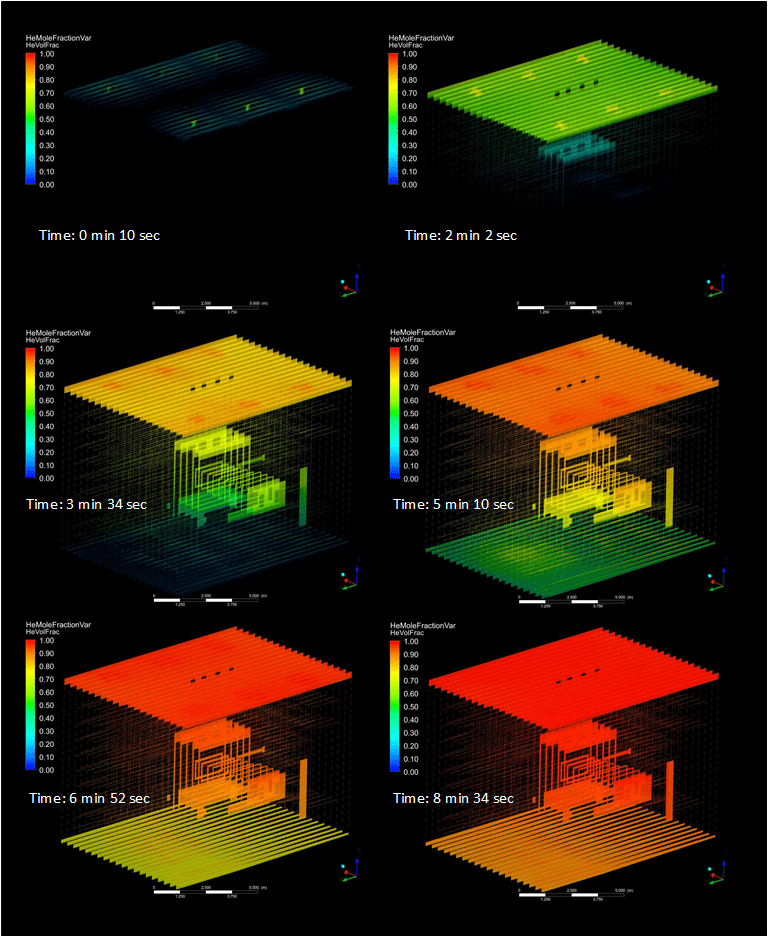}
\caption{\label{Fig:TC:BDF_HeVe_Tr1} Volume rendering of helium mole fraction during filling}
\end{figure}

\begin{figure}[htbp]
\centering %
\includegraphics[width=0.7\linewidth]{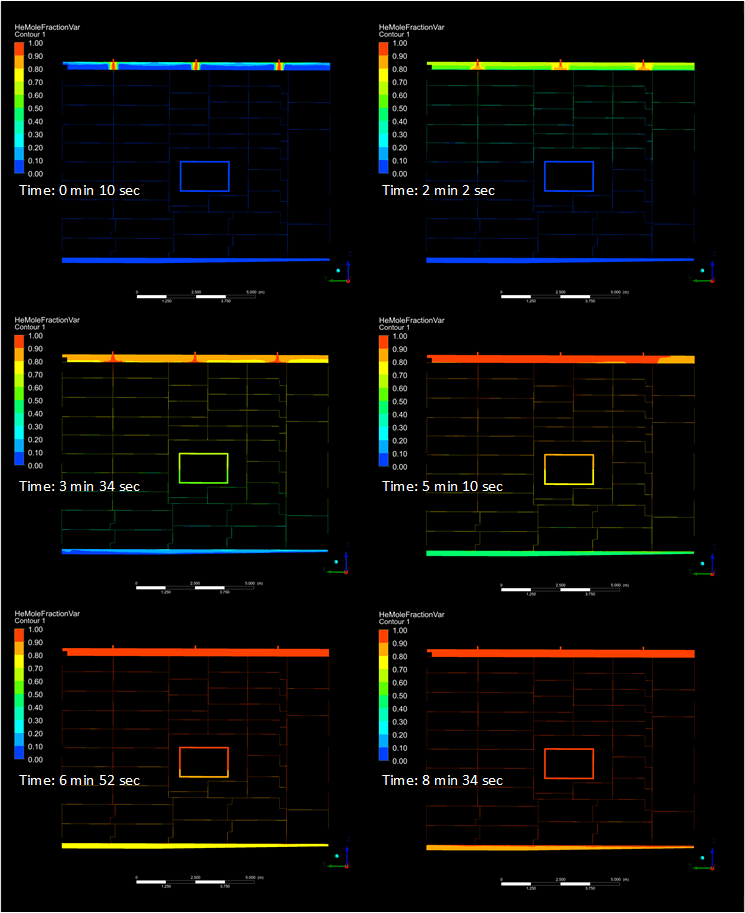}
\caption{\label{Fig:TC:BDF_HeVe_Tr2} Contours of helium mole fraction on a vertical plane crossing three inlets}
\end{figure}

Figure \ref{Fig:TC:BDF_HeVe_Tr1} shows a volume rendering of the helium mole fraction (helium purity) distribution in the helium vessel at six instants of time during the transient. As a reference, the last snapshot of the series (8 minutes and 34 seconds) corresponds to about 1.46 volumes of pure helium injected in the system. The volume rendering is a series of semi-transparent contour plots on about 20 vertical parallel planes normal to the trolley translational direction. Figure \ref{Fig:TC:BDF_HeVe_Tr2} shows the same purity contours on one of these planes, specifically the one that intersects the three inlets on the side of the trolley door. 

The series of plots shows that the helium flow gradually and uniformly fills the helium vessel, including its gaps, from top to bottom. This suggests that the helium purity level that is reachable via flushing is high; however the plots do not allow the identification of critical areas, where the purity is lower than the average in the vessel.

The inlet flow rate, as mentioned in \ref{Sec:TC:CFDHeVess_Steady}, is equal to 764.45 m\textsuperscript{3}/h, resulting in 10 m/s inlet flow velocity. One volume change (74.5 m\textsuperscript{3}) is achieved in 5 minutes and 51 seconds. Note that the specified flow rate is:
\begin{itemize}
\item Equal to the flow rate required for cooling of the helium vessel internals \cite{HePurifITdocs};
\item Higher than the purification flow rate, which is 75 m\textsuperscript{3}/h \cite{HePurifITdocs}.
\end{itemize}

The pure helium for the flushing process would be provided by 200 bar compressed helium cylinders. The 764.45 m\textsuperscript{3}/h flow rate has been selected because it is expected to be more critical than a lower flow rate (such as the purification flow rate) in terms of helium-air mixing: high velocity and high turbulence should enhance mixing between the two gases, thus increasing the time needed for flushing and the amount of pure helium lost. However the results of the simulation seem to contradict this expectation; factors that may influence this behavior are the complicated geometry, which enhances separation between helium and air, and mixing by diffusion, which has a smaller impact on small time scales. Deeper investigation could be done on this aspect, however the optimization of the helium flow rate is not the main purpose of these simulations, as long as it can be shown that the pure helium required for reaching 85\% average purity is less than two vessel volumes.

\begin{figure}[htbp]
\centering %
\includegraphics[width=0.8\linewidth]{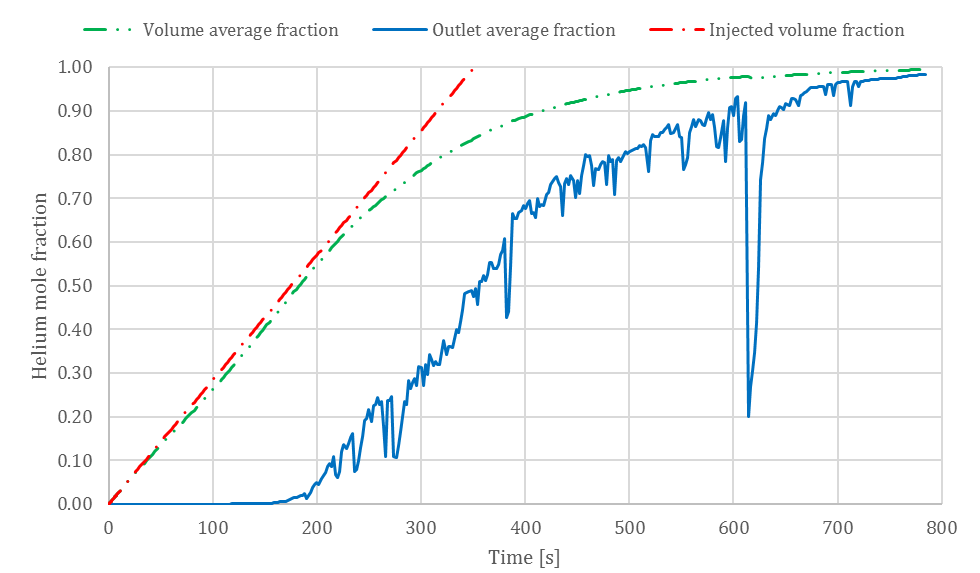}
\caption{\label{Fig:TC:BDF_HeVe_Tr3} Helium mole fraction as function of time.}
\end{figure}

Figure \ref{Fig:TC:BDF_HeVe_Tr3} shows the helium purity evolution with time. The three datasets represent the following quantities:
\begin{itemize}
\item Injected volume fraction: volume of pure helium injected in the vessel, normalized by the vessel free volume (74.5 m\textsuperscript{3}); this parameter is useful to identify the number of gas changes: it reaches 1.0 when one volume change has been achieved. Since 1.0 is achieved at about 350 s, 2 volume changes (threshold for pure helium loss) are achieved at about 700 s; 
\item Outlet average fraction: surface average helium purity (weighted by mass) at the outlets of the helium vessel. This is the purity that would be measured by a purity meter at the inlet of the purification system. Note that the peak at ~615 s is not physical (generated by local convergence issues during the simulation) and should be disregarded;  
\item Volume average fraction: average helium purity in the helium vessel as function of time.
\end {itemize}

The parameter of interest is the volume average fraction or purity. The flushing process allows to reach a high purity without losing a large amount of helium before being able to start the purification process: the volume average fraction reaches 85\% (threshold for purification system [1]) at about 350 s, equivalent to about 1 volume change (Injected volume fraction equal to 1).

The behavior of the volume average fraction is asymptotically increasing to 1; the average purity is 85\% at 1 volume change and >98\% at 2 volume changes (~700 s).

The outlet average fraction is delayed (due to the inertia of the system) and more oscillating; it consistently exceeds 90\% only towards the end of the second volume change.

\subsubsubsection{Identification of low-purity areas}

\begin{figure}[htbp]
\centering %
\includegraphics[width=0.7\linewidth]{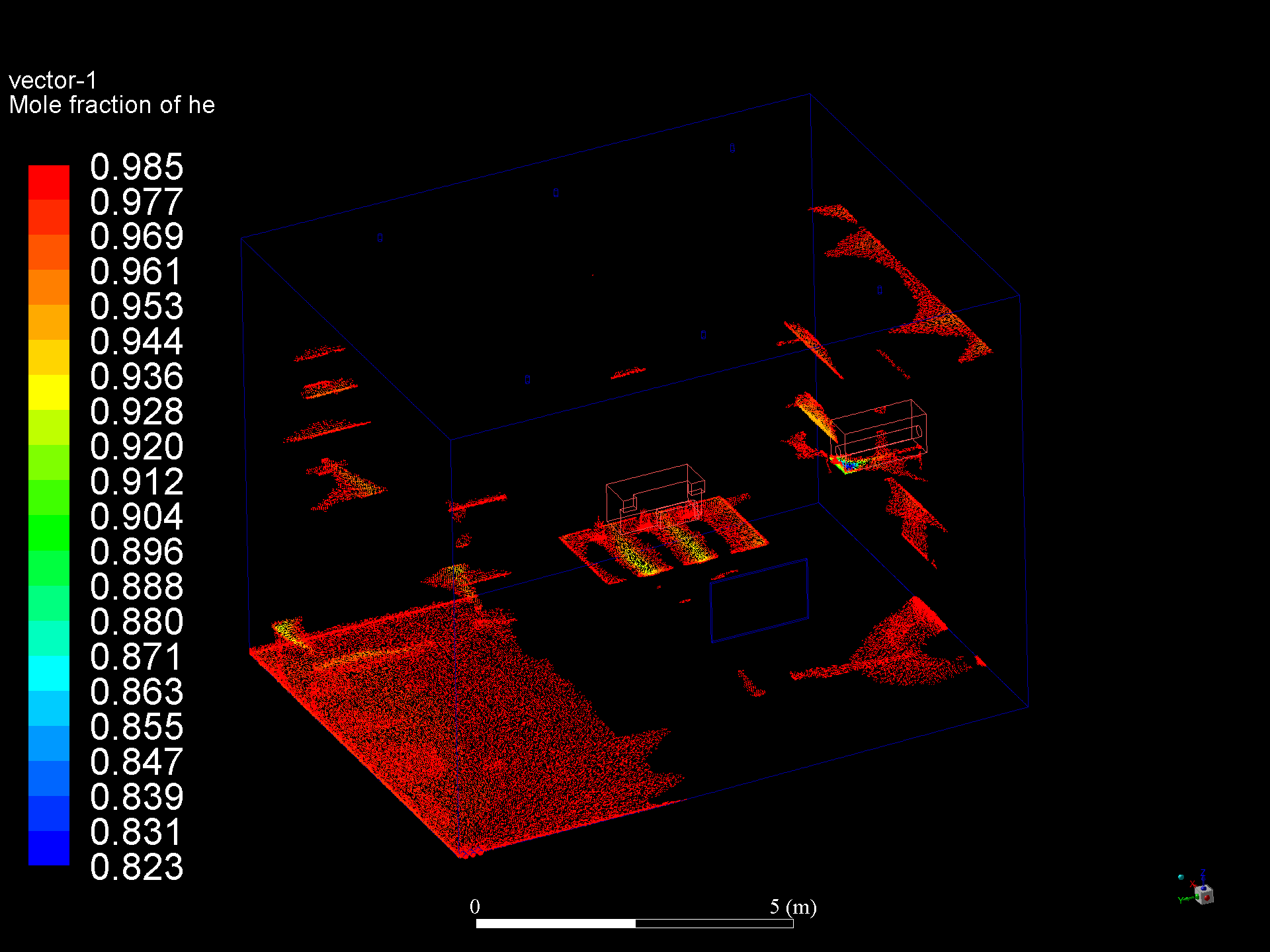}
\caption{\label{Fig:TC:BDF_HeVe_Tr61} Vector plot of helium mole fraction clipped at 98.5\% (2.23 volume changes).}
\end{figure}

\begin{figure}[htbp]
\centering %
\includegraphics[width=0.7\linewidth]{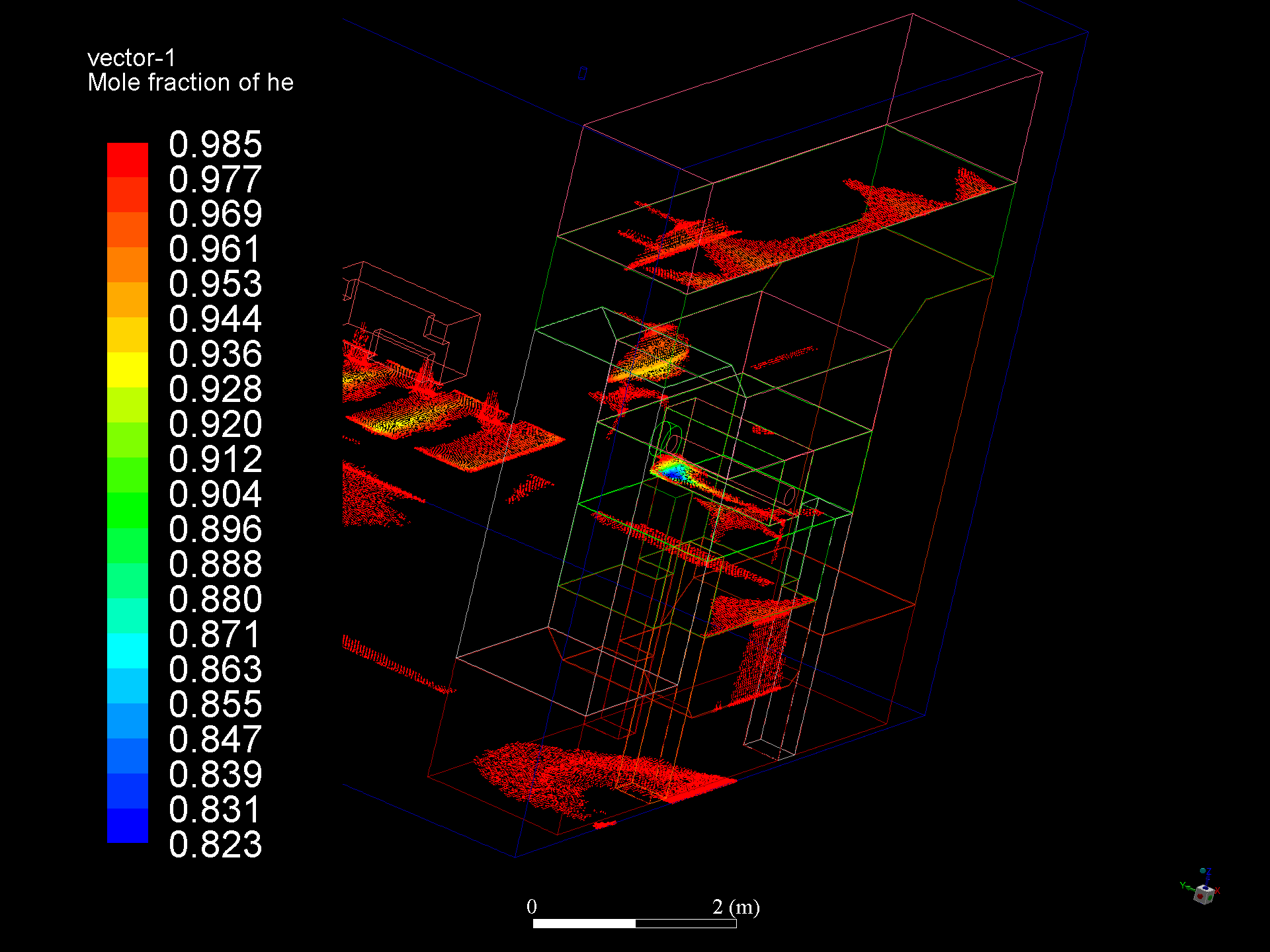}
\caption{\label{Fig:TC:BDF_HeVe_Tr62} Vector plot of helium mole fraction clipped at 98.5\% (2.23 volume changes) - front view, collimator area.}
\end{figure}

\begin{figure}[htbp]
\centering %
\includegraphics[width=0.7\linewidth]{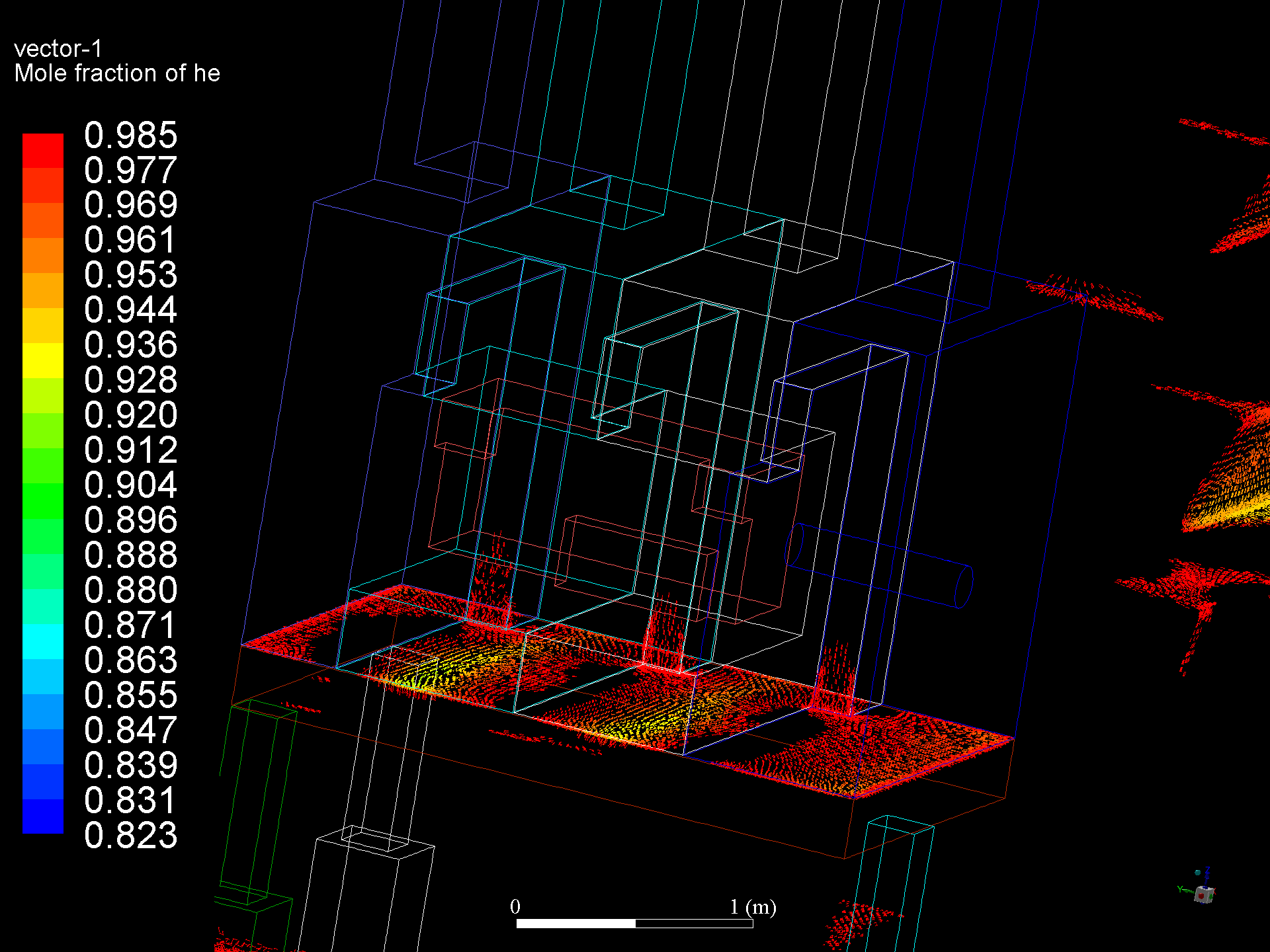}
\caption{\label{Fig:TC:BDF_HeVe_Tr63} Vector plot of helium mole fraction clipped at 98.5\% (2.23 volume changes) - proximity shielding.}
\end{figure}

\begin{figure}[htbp]
\centering %
\includegraphics[width=0.7\linewidth]{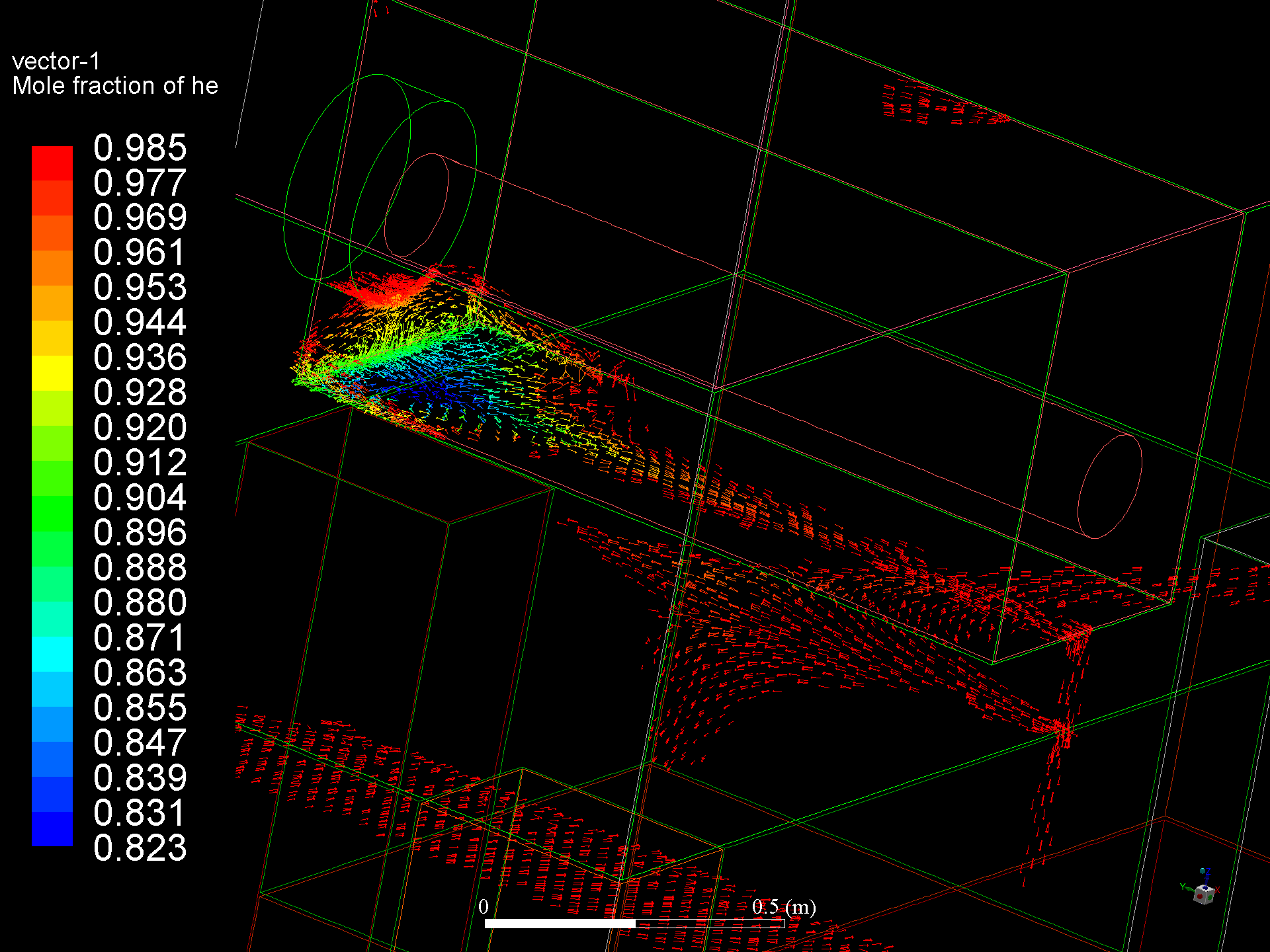}
\caption{\label{Fig:TC:BDF_HeVe_Tr64} Vector plot of helium mole fraction clipped at 98.5\% (2.23 volume changes) - collimator.}
\end{figure}

\begin{figure}[htbp]
\centering %
\includegraphics[width=0.7\linewidth]{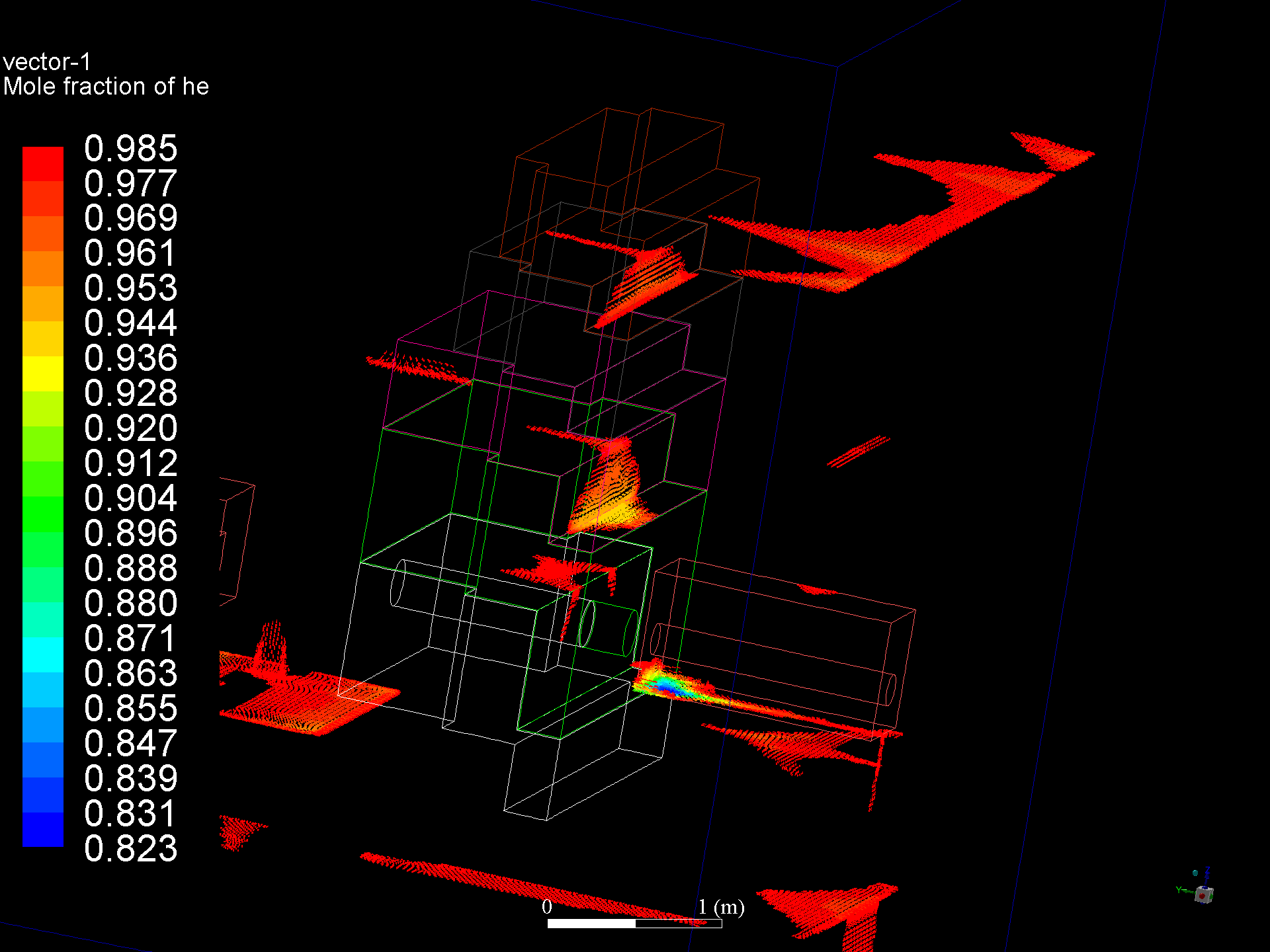}
\caption{\label{Fig:TC:BDF_HeVe_Tr65} Vector plot of helium mole fraction clipped at 98.5\% (2.23 volume changes) - front bunker blocks.}
\end{figure}

\begin{figure}[htbp]
\centering %
\includegraphics[width=0.7\linewidth]{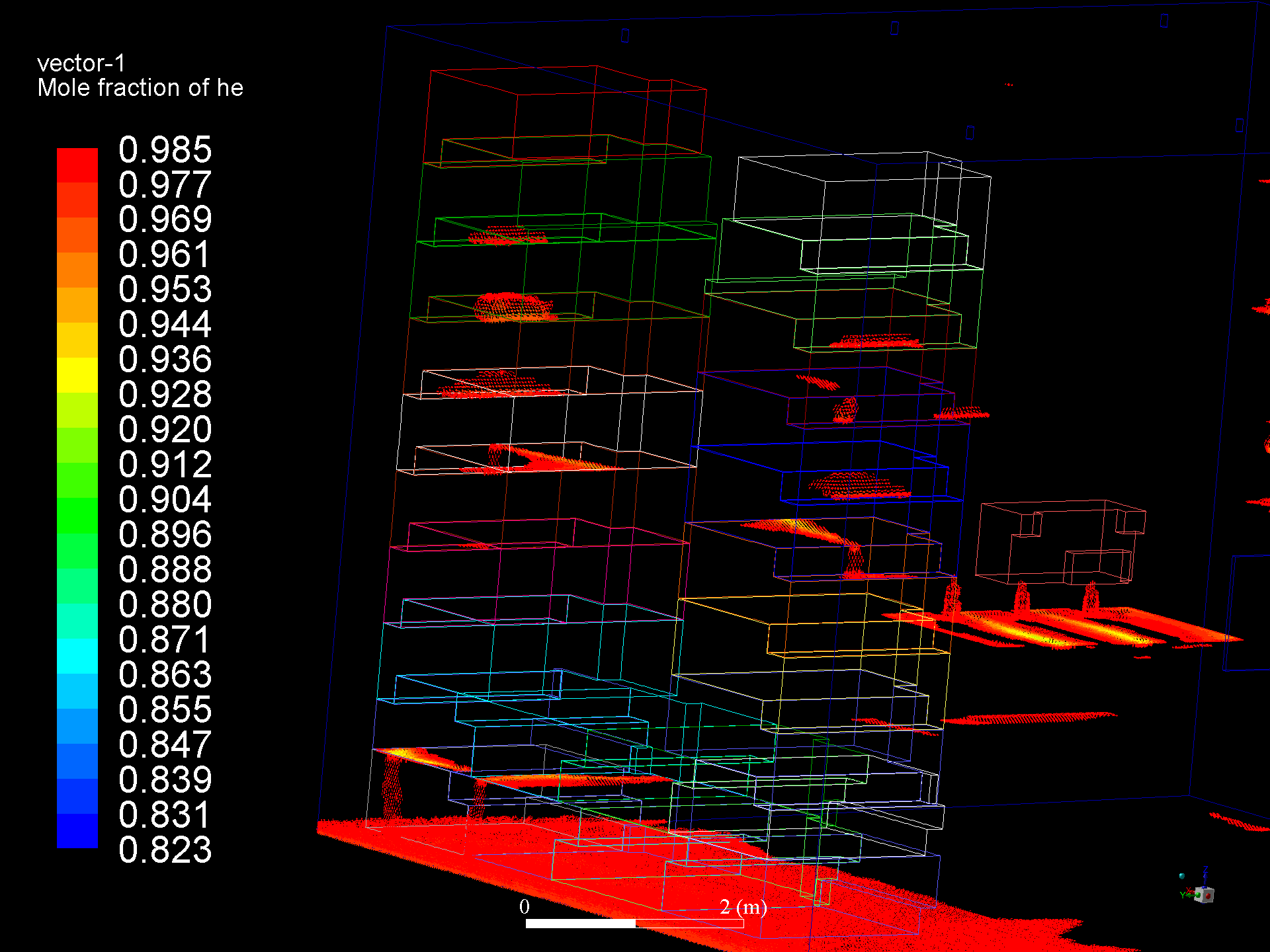}
\caption{\label{Fig:TC:BDF_HeVe_Tr66} Vector plot of helium mole fraction clipped at 98.5\% (2.23 volume changes) - back bunker blocks.}
\end{figure}

Figure \ref{Fig:TC:BDF_HeVe_Tr61} to Figure \ref{Fig:TC:BDF_HeVe_Tr66} represent helium purity at different locations in the helium vessel at the end of the simulated transient, corresponding to 13 min and 4 sec, or 2.23 volume changes; the scale is clipped to a maximum purity of 98.5\% to show areas of low purity. 
Figure \ref{Fig:TC:BDF_HeVe_Tr61} shows an overview of the low-purity areas. The minimum purity is about 82\% and it is likely located in the collimator area. Not many low-purity areas are identified, proving that the current design provides good circulation patterns for helium. Note that the large impurity volume located in the lower plenum should be neglected since it would naturally disappear with a longer simulation time. 
Figure \ref{Fig:TC:BDF_HeVe_Tr62} shows the same results from a different point of view, looking at the frontal collimator blocks.

A first low-purity area can be identified right below the C-shaped blocks of the proximity shielding (Figure \ref{Fig:TC:BDF_HeVe_Tr63}), in the horizontal gap between these blocks and the flat horizontal block. This area had already been identified in the steady simulation. The reason for the presence of this stagnation area is the fact that the helium prefers flowing around the proximity shielding blocks than through the gaps. Also note that vertical gaps still allow gravity to drive a flow induced by different gas molecular masses, so that impurities are removed, but horizontal gaps have no means to let these impurities flow out. 

A second important stagnation area is located between the collimator and the block surrounding it (Figure \ref{Fig:TC:BDF_HeVe_Tr64}). This is the location where the purity is lowest. The stagnation is generated by the presence of multiple corners in series combined with a horizontal gap along the flow path, so that the pressure differential is not able to generate a flow through the gaps.

A series of stagnation areas can also be identified in the bunker shielding blocks. Figure \ref{Fig:TC:BDF_HeVe_Tr65} shows the front bunker blocks; stagnation areas can be found again where the already slow flow of horizontal gaps is forced to stop at corners. However these corners are required to prevent stray radiation from reaching the outside rooms and buildings without shielding.

Analogous considerations hold for the back bunker blocks represented in Figure \ref{Fig:TC:BDF_HeVe_Tr66}.

\subsubsubsection{He-vessel design recommendations in view of the CFD studies}

The following conclusions and recommendations can be drawn based on the simulation analysis:
\begin{itemize}
\item The presence of a 10 mm gap between blocks and vessel internals provides an acceptable level of helium purity after flushing, even at high inlet flow rate;
\item The current design allows to achieve 85\% average purity (minimum required to start the purification system) in the helium vessel well before reaching 2.0 pure helium volume changes;
\item 2.0 volume changes are still recommended before starting the helium purification system, so that not only the volumetric average purity, but also the outlet surface average purity is beyond 85\% before starting the purification; 
\item The control over the accuracy of gap thicknesses is important to prevent the formation of unexpected stagnation areas or flow imbalances. This is particularly true for the lateral concrete shielding blocks, or any location where a bypass flow could be generated by unplanned variation of gap sizes, leaving some areas unventilated;
\item The presence of corners combined with horizontal gaps generally makes stagnation areas more likely to appear;
\item Horizontal gaps with large surface should be avoided as much as possible; if this is not possible, the dimensions and geometry of the gap shall be defined and a CFD analysis shall prove that stagnation is limited as much as possible.
\end{itemize}

\clearpage

\subsection{Preliminary technical considerations on the He purification and circulation system}
\label{Sec:TC:HeV:HePass}
\subsubsection{General Description}
\label{Sec:TC:HePassSys_GenDescr}
The helium passivation system supplies purified helium to the helium vessel that contains the BDF target and its shielding; its primary purpose is to remove impurities from the gas mixture so that it does not activate and remains inert from a chemical standpoint.

Helium provides a low level of activation caused by radiation from the primary beam; combined with the purification capability of the passivation system, this results in minimal activation of the gas mixture contained in the helium vessel. Moreover, a pure helium atmosphere will help to protect the materials from oxidation, thereby increasing the lifetime of the components contained in the helium vessel.

\begin{figure}[htbp]
\centering %
\includegraphics[width=1\linewidth]{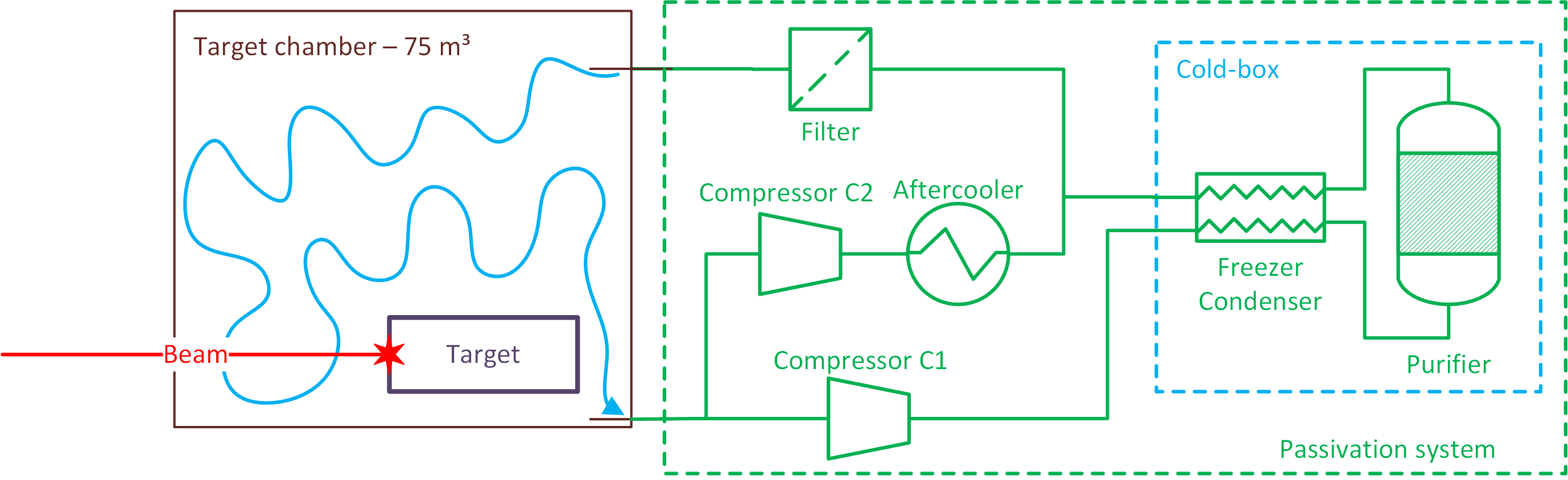}
\caption{\label{Fig:TC:HePassSys_Schematic} Schematic representation of the helium purification system, with the different components of the system.}
\end{figure}

A simplified representation of the overall system is presented in Figure~\ref{Fig:TC:HePassSys_Schematic}; in particular, the passivation system consists of the green part of Figure~\ref{Fig:TC:HePassSys_Schematic}. The system is divided in two parallel units:

\begin{itemize}
\renewcommand{\labelitemi}{$\bullet$}
\item A cooling unit equipped with the compressor C2 (C2 unit); 
\item A purification unit equipped with compressor C1 (C1 line), which includes a cold-box unit (light blue box).
\end{itemize}

The two main units (cooling and purification) are designed to comply with the following main requirements:
\begin{enumerate}
\item Purify helium so that helium purity at the outlet of the system is at least 99.99\% (design purity);
\item Remove 3 kW from the helium in the chamber, with a maximum helium temperature increase in the chamber of 20 K. The mentioned value for the cooling power will need to be revised at a more advanced design stage to take into account the actual thermal dissipation to the helium.
\end{enumerate}

The helium is supplied from a distribution system at the top of the helium vessel, and it is collected from a location at the bottom. During the start-up mode, pure helium from cylinders is injected and helium-air mixture is extracted from the system and vented to the atmosphere, until helium purity reaches about 85\%vol. At this point, the flushing is interrupted and the helium passivation system begins to purify the mixture by recirculating it in the vessel. During operation, whenever the purity decreases below a given threshold, the purification system would turn on and purify the mixture to the desired level. 

The system has three operational modes:
\begin{enumerate}
\item Preparation mode: this mode is operated during the initial start-up phase, and prepares the cold-box unit, the target chamber and the other components of the system for nominal mode operation;
\item Nominal mode: this mode is operated during the main experimental phase, when the beam is hitting the target; the purpose of this mode is to treat the helium so that its conditions are compliant with the two main requirements listed in the previous paragraph;
\item Purge mode: this mode is operated at the end of the experimental phase; this mode purges the helium from the target chamber and shuts down the system in safe conditions.
\end{enumerate}

With respect to helium purity, the following two definitions are assumed:
\begin{itemize}
\renewcommand{\labelitemi}{$\bullet$}
\item Design purity: 99.99\%. This purity is the purity at which the helium shall be provided at the outlet of the system after the purification process. The system shall be capable of producing helium with purity at least equal to the specified design purity;
\item Nominal purity: 99.9\%. This purity is the desired minimum purity in the target chamber during the operation of the system. In case of construction, the compliance with this requirement would not be under the responsibility of the contractor, because it is affected by the target chamber leak tightness and internal configuration.
\end{itemize}

\subsubsection{P\&ID and Process Description}
\label{Sec:TC:HePassSys_ProcDescr}
The low-temperature adsorption (LTA) method is used for the helium purification unit. The implementation of this purification technology allows to achieve the design purity. The basic principle of LTA-based purification is to remove impurities (moisture, air, nitrogen, etc.) by condensation and adsorption on an adsorbent bed that is maintained at cryogenic temperature (about 77 K). For reaching the design purity, the process works at a pressure level of at least 15 bar.

The adsorbent bed allows to capture the gas impurities from the main helium stream. For simultaneous adsorption of several gas components, a mixed bed of sorbents is used, typically a multi-sorbent based on coconut activated carbon (washed) and CMS-T3A.

\begin{figure}[htbp]
\centering %
\includegraphics[width=0.8\linewidth]{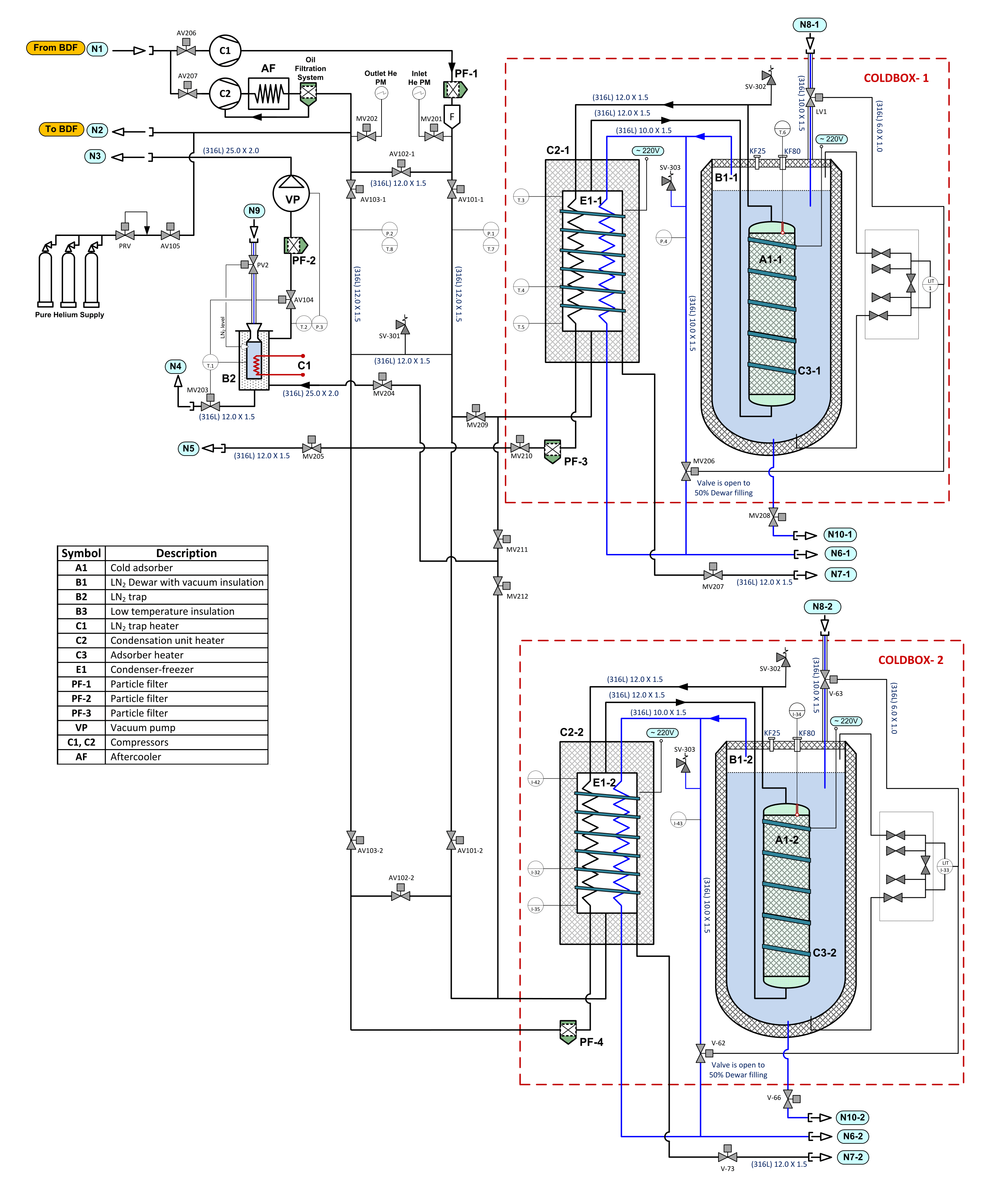}
\caption{\label{Fig:TC:HePassSys_PFD} Preliminary process flow diagram for the helium passivation system, showing two redundant coldbox units for continuous operation and a cooling and circulation unit.}
\end{figure}

A preliminary simplified process flow diagram of the helium purification unit is shown in Figure~\ref{Fig:TC:HePassSys_PFD} for the redundant cold-box version.The process of helium purification performs the following steps: 

\begin{enumerate}
\item The impure helium from the target chamber with pressure slightly above the atmospheric one and temperature below 46\textsuperscript{\degree}C is supplied to the C1 compressor suction (see Figure~\ref{Fig:TC:HePassSys_PFD}); 
\item In compressor C1 (compressor for purification), helium is compressed to a pressure of at least 15 bar; 
\item Helium enters the shell space of the condenser-freezer E1 (see Figure \ref{Fig:TC:HePassSys_PFD}). In the condenser-freezer, helium is cooled to a temperature of about 88 K by the flow in the return line and the nitrogen vapors from the adsorber dewar. All impurities like moisture, CO2 and residual oil vapors from the compressor freeze out in the condenser-freezer; 
\item After the condenser-freezer, the helium enters the purification block. The flow of helium is further cooled down in the helium purification unit to a temperature of 77 K, and then enters the adsorbent bed. The adsorber is located in a tank, which is filled with liquid nitrogen; 
\item Pure helium after the adsorber is directed back to the condenser-freezer E1 for precooling of the main flow of impure helium;
\item After the condenser-freezer, pure helium is filtered (cleaning from the adsorbent particles, at PF3) and returns to the target chamber at the design purity.
The system is fully automated, because the target complex area is not accessible during the experimental phase and all operations need to be performed remotely.
\end{enumerate}
The system is fully automated, because the target complex area is not accessible during the experimental phase and all operations need to be performed remotely.

\subsubsection{Layout and Components}
\label{Sec:TC:HePassSys_Layout}

The system is divided in two pieces: the cooling unit (C2 line: compressor C2 and aftercooler) and the purification unit (compressor C1 + cold-box). The units are compliant with the following requirements: 
\begin{itemize}
\item The purification unit is supplied on a skid, whose maximum dimensions are 2.5 m [L], 1.2 m [W], 2 m [H]; the skid is assembled and transported on site, for connection and testing; 
\item The cooling unit, whose maximum dimensions are 2.5 m [L], 1.5 m [W], 2 m [H], is transportable and provided with lifting points on top of the structure;
\item Both units are equipped with their own electrical and control cubicles;
\item Both units are provided with connection flanges for required services (Water connection, helium cylinders connection, LN connection, second cold-box connections);
\end{itemize}

\begin{figure}[htbp]
\centering %
\includegraphics[width=0.7\linewidth]{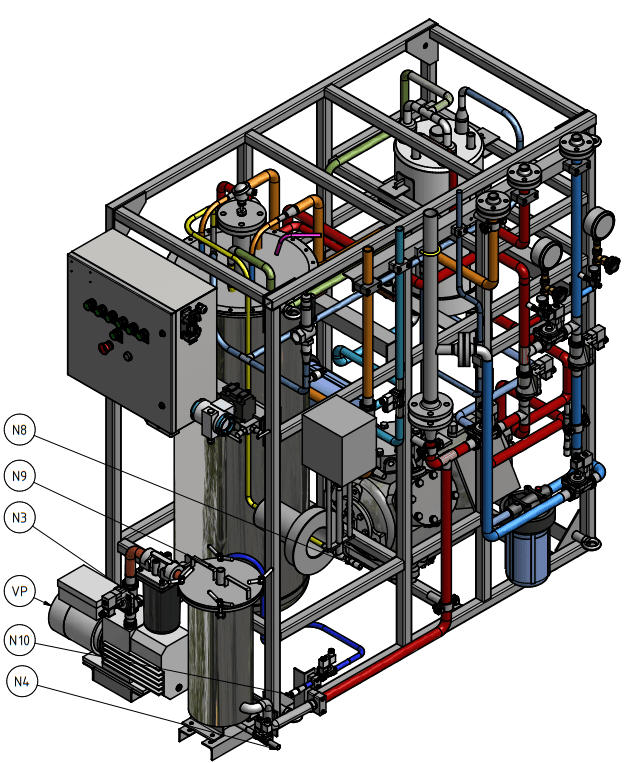}
\caption{\label{Fig:TC:HePassSys_3DMod} Preliminary 3D model of the purification unit, including the coldbox, the regeneration system and the electrical cubicle.}
\end{figure}

\begin{figure}[htbp]
\centering %
\includegraphics[width=0.8\linewidth]{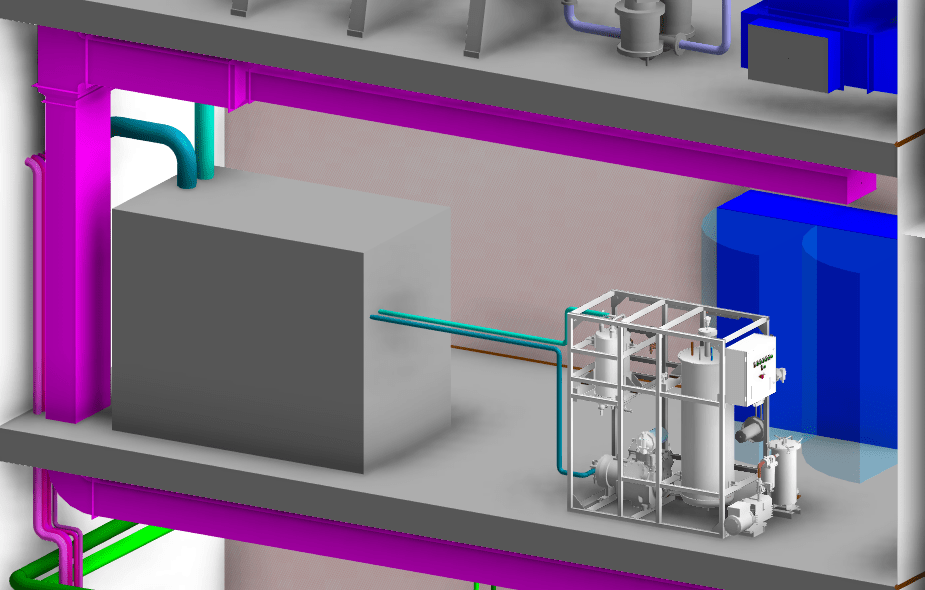}
\caption{\label{Fig:TC:HePassSys_Integr} Preliminary integration of the helium passivation system in the CV room of the BDF target complex. Both the purification and circulation units are illustrated.}
\end{figure}

The preliminary design of the system features two compressors, one for purification (C1) and another for circulation/cooling (C2).
Compressor C1 is provided with an adjustable flow-rate and with automatic control unit. The flow-rate is determined so that the preparation process can be completed in less than 24 hours. The compressor C1 has the following characteristics:

\begin{itemize}
\item In order to limit as much as possible and potentially eliminate oil vapor/aerosol content in the outlet helium flow, compressor C1 is oil-free;
\item Compressor C1 will be operated in a radiation-controlled area. Selection of compressor will assure need of preventive maintenance to no more than once per year;
\item Compress helium to at least 15 bar;
\item Feature an aftercooler.
\end{itemize}

Similar considerations hold for the compressor for circulation (C2) and its aftercooler; the compressor C2 and aftercooler have the following characteristics: 
\begin{itemize}
\item The combination compressor-aftercooler is capable of cooling down helium from 46\textsuperscript{\degree}C to 26\textsuperscript{\degree}C for a maximum of 3 kW thermal power (corresponding to roughly 700 m\textsuperscript{3}/h);
\item The pressure head is at least 1000 Pa;
\end{itemize}
The purification system consists of two main blocks: the condenser-freezer and the adsorber.

The adsorber is a vessel made of stainless steel and filled with sorption material. For increasing the sorption capacity, the absorber is submerged into a liquid nitrogen dewar. The main technical characteristics of the adsorber are listed in Table \ref{Tab:TC:HePassSys_Adsorber}.

\begin{table}[htbp]
\centering
\caption{\label{Tab:TC:HePassSys_Adsorber} Technical requirements for the purification unit and the adsorber.}
\smallskip
\begin{tabular}{m{8cm}|r|l}
\hline
\textbf{Parameter} & \textbf{Unit} & \textbf{Value}\\
\hline
Design purity & \% & 99.99\\
\hline
Minimum operating pressure & bar & 15.0\\
\hline
Minimum allowable pressure & bar & 0.0\\
\hline
Maximum allowable pressure & bar & 30.0\\
\hline
Inlet temperature & K & 93.0 - 95.0\\
\hline
Operating temperature & K & 77.0\\
\hline
Purification capacity (minimum working time before saturation) at 85\% constant inlet gas purity & hour & 6\\
\hline
\end{tabular}
\end{table}
The adsorber is connected in series with the condenser-freezer and they work simultaneously. The restoration of the sorption capacity (regeneration) of the adsorbent is done by heating and pumping out the cold-box cavities. A cold trap (B2 in Figure \ref{Fig:TC:HePassSys_PFD}) is installed to collect the impurities during the regeneration procedure.

The helium purification unit is designed in such a way that a single pass through the condenser-freezer and the adsorber is needed to achieve the design purity. The adsorber is dimensioned to provide a purification capacity of six hours at 85\% constant inlet gas purity and nominal flow rate (as mentioned in Table~\ref{Tab:TC:HePassSys_Adsorber}). Note that in this context the purification capacity is defined as the operational time (at the specified constant flow and inlet purity) after which the adsorber saturates and requires regeneration. For the BDF nominal operating conditions, assuming an average impurity level in the helium vessel of 0.05\%vol (coherent with 1 m\textsuperscript{3}/day impurity ingress), the unit can work continuously for more than a month before regeneration is needed. However, regeneration might be needed much more frequently if high level of moisture is present inside the vessel during operation, causing blockage of the condenser-freezer unit due to excessive icing.

The condenser-freezer is designed to freeze moisture, vapour and CO\textsubscript{2}, thus significantly reducing the load on the absorber and increasing its lifetime. The purpose of the condenser-freezer is be two-fold:
\begin{enumerate}
\item to precool the impure helium via heat exchange with reverse streams of pure helium and cold nitrogen vapors;
\item to freeze and trap the high boiling components, like moisture, carbon dioxide, hydrocarbons.
\end{enumerate}
The condenser-freezer is designed as a shell-and-tube heat exchanger. The tubes are double-walled, so that one fluid can flow in the annular gaps and another can flow in the internal, cylindrical pipes. The fluids distribution and flow direction in the condenser-freezer volumes are organized as follows:
\begin{itemize}
\item Shell volume: direct flow of impure helium (coming from compressor C1);
\item Annular gaps: reverse flow of nitrogen gas (coming from dewar B1, see Figure \ref{Fig:TC:HePassSys_PFD});
\item Cylindrical pipes: reverse flow of pure helium (coming from adsorber in dewar B1);
\end{itemize}
As illustrated in Figure \ref{Fig:TC:HePassSys_PFD}, the purification unit is connected in bypass mode on the cooling unit. The entire system is finally connected to the helium vessel via piping going from the CV room to the helium vessel area.

\subsubsection{Purification System Upgrades}
\label{Sec:TC:HePassSys_Upgrades}

The following optional upgrades are foreseen for increasing the functionality of the purification system:
\begin{enumerate}
\item An additional hydrogen removal system; two options are possible:
\begin{itemize}
\item An additional catalytic recombination unit before entering the cold-box;
\item An additional specific cryogenic unit for hydrogen adsorption.
\end{itemize}
\item An additional gas chromatography unit, including: gas chromatograph, helium ionization detector, thermal conductivity detector, control software;
\item An upgrade of the compressor for circulation and its aftercooler unit to remove 10 kW thermal power from the helium flow.
\end{enumerate}

\subsubsection{Pure Helium Supply}
\label{Sec:TC:HePassSys_HeSupply}

\begin{figure}[htbp]
\centering %
\includegraphics[width=0.7\linewidth]{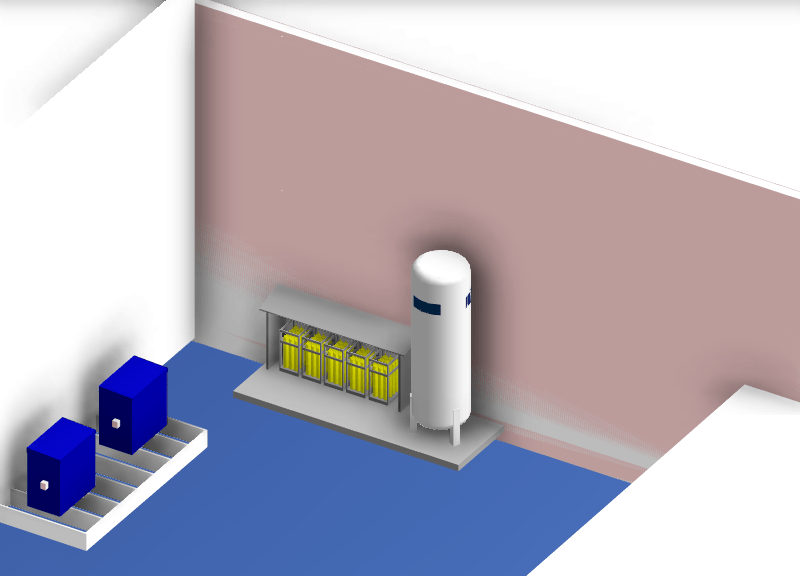}
\caption{\label{Fig:TC:HePassSys_HeLNSupply} Layout of pure helium and liquid nitrogen supply next to the target complex wall.}
\end{figure}

Pure helium is supplied from pressurized cylinders located in a designated area on the surface, next to the target complex (Figure~\ref{Fig:TC:HePassSys_HeLNSupply}). One cylinder holds 50 liters of pure helium at 200 bar, corresponding to roughly 10 Nm\textsuperscript{3}; cylinders are supplied in groups of 12 on dedicated racks: each rack has a capacity of 120 Nm\textsuperscript{3} of pure helium. Accounting for two startups per year, plus the equivalent for 1-year leaks, the amount of pure helium needed for the operation of the helium purification system for one year is roughly 600 Nm\textsuperscript{3}. In order to provide this storage capacity, five racks (60 cylinders) are placed in the helium storage area (Figure \ref{Fig:TC:HePassSys_HeLNSupply}). Each rack is provided with a pressure indicator to allow timely replacement of exhausted cylinders; the helium storage reserve needs to be periodically checked (at least every six months and in any case after every helium flushing operation) to make sure that the system does not run out of helium during operation. The helium storage area is provided with supply modules that include pressure regulators/reducers, relief valves, drain valves and shut-off valves. If necessary, the supply pressure can be monitored on the module via pressure transmitter, which is linked to alarms/warnings on the supervision system that activate whenever the pressures in the racks are below a specified threshold.

\subsubsection{Liquid Nitrogen Supply}
\label{Sec:TC:HePassSys_LNSupply}

Liquid nitrogen is needed to maintain the adsorber at cryogenic temperature; the amount of LN needed for the operation of the passivation system is roughly 120 l for every start-up and a maximum of 10 l/h for nominal conditions. Assuming 4 start-ups per year, the overall need is roughly 90000 l/year or 7500 l/month. A reasonable tank refilling interval at CERN is on the order of 1 month.

LN suppliers provide tanks of standard sizes; the sizes of interest for BDF are 6000 l, 11000 l and 19000 l, corresponding to an autonomy of 0.7, 1.3 and 2.1 months, respectively (accounting for losses). In order to avoid issues in case one monthly refill is skipped, the 19000 l tank is selected; this tank has a diameter of 2.4 m and height of 8 m (Figure \ref{Fig:TC:HePassSys_HeLNSupply}). Its pressure-building coil needs to be sized for a LN supply of 10 l/h.
The tank can be either rented or purchased. The pay-off time for such tanks is on the order of 6-7 years. The BDF target operation is planned to be 5 years; in this context, the rental option is likely to be preferable. Instead, if the operation of the helium purification system is extended beyond 5 years, the purchase option is recommended.

\subsubsection{Chilled Water Cooling}
\label{Sec:TC:HePassSys_CoolDesign}

The helium purification system is designed to remove a maximum of 10 kW thermal load from the helium vessel. In order to provide this cooling, chilled water is supplied to the target complex CV room. Table~\ref{Tab:TC:HePassSys_CoolingDesign} lists the main parameters of the aforementioned chilled water cooling system.

\begin{table}[htbp]
\centering
\caption{\label{Tab:TC:HePassSys_CoolingDesign} Chilled water cooling parameters for the helium purification system }
\smallskip
\begin{tabular}{l|r|l}
\hline
\textbf{Parameter} & \textbf{Unit} & \textbf{Value}\\
\hline
Location & - & CV room \\
T in & \textsuperscript{\degree}C & 6 \\
T out & \textsuperscript{\degree}C & 8.5 \\
T & \textsuperscript{\degree}C & 2.5 \\
Flow rate & m\textsuperscript{3}/h & 3.6 \\
Piping size & - & DN25 \\
Thermal load & kW & 10 \\
Type & - & Chilled water \\
Activation & - & No \\
\hline
\end{tabular}
\end{table}


\section{Considerations on the cooling and ventilation (CV) aspects}
\label{Sec:TC:CV}

\subsection{Introduction}
\label{Sec:TC:CV_Intro}
This section presents the preliminary design of the cooling and ventilation systems for the target and target complex of the Beam Dump Facility; a detailed description of the mentioned design is provided in Ref.~\cite{BDF_CVWorkPackage}. 

\begin{figure}[!htbp]
\centering %
\includegraphics[width=0.8\linewidth]{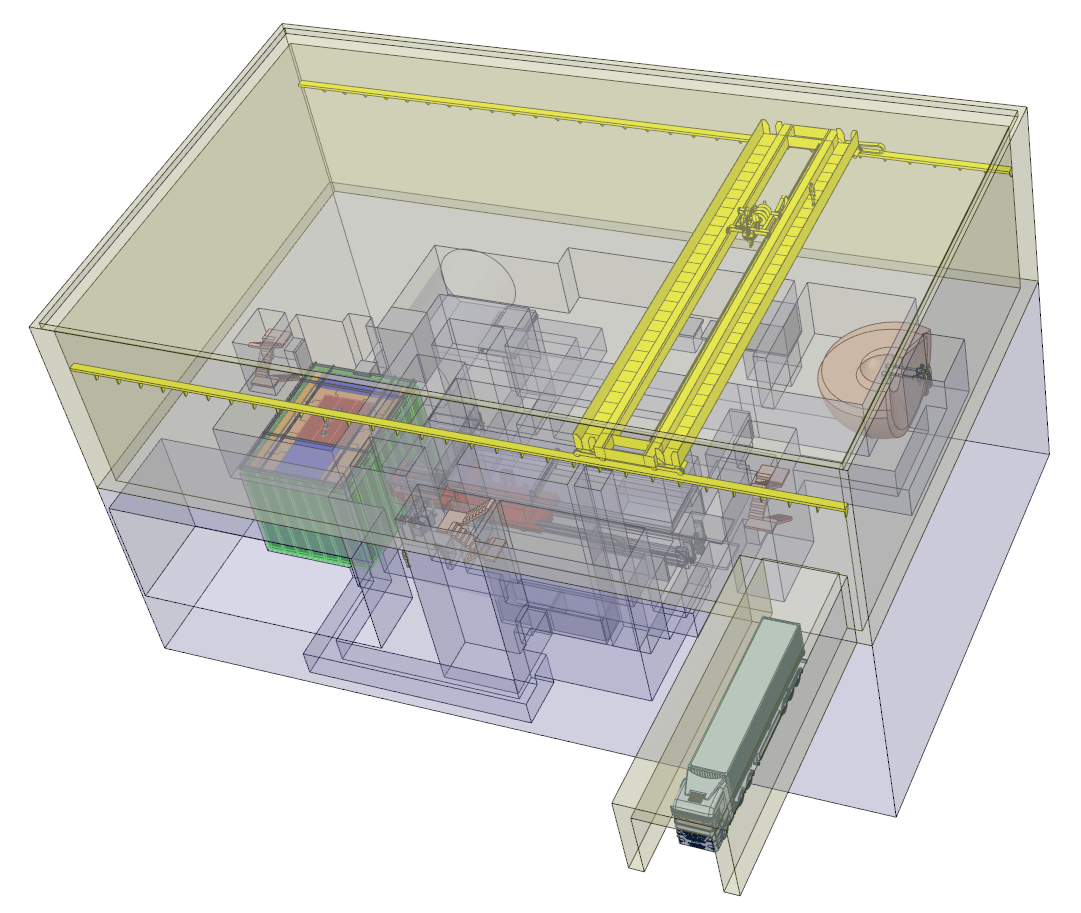}
\caption{\label{Fig:TC:CV_Intro_010} View of preliminary 3D model for BDF target complex (Section~\ref{Sec:TC:Design}).}
\end{figure}

The BDF target complex (Figure~\ref{Fig:TC:CV_Intro_010}) houses the production target and its shielding, as well as services, handling facilities, cooling and ventilation systems, cool down areas and general equipment for supporting the reliable and long-term operation of the facility. The area surrounding the target is passivated by means of a helium cooling, circulation and purification system, which preserves the integrity of materials and enhance safety from a radiation protection standpoint. Parts of the cooling and ventilation systems are also located in the BDF auxiliary building, located upstream the target complex building.

\subsubsection{Building Layout}
\label{Sec:TC:CV_BldgLayout}

\begin{figure}[!htbp]
\centering %
\includegraphics[width=0.8\linewidth]{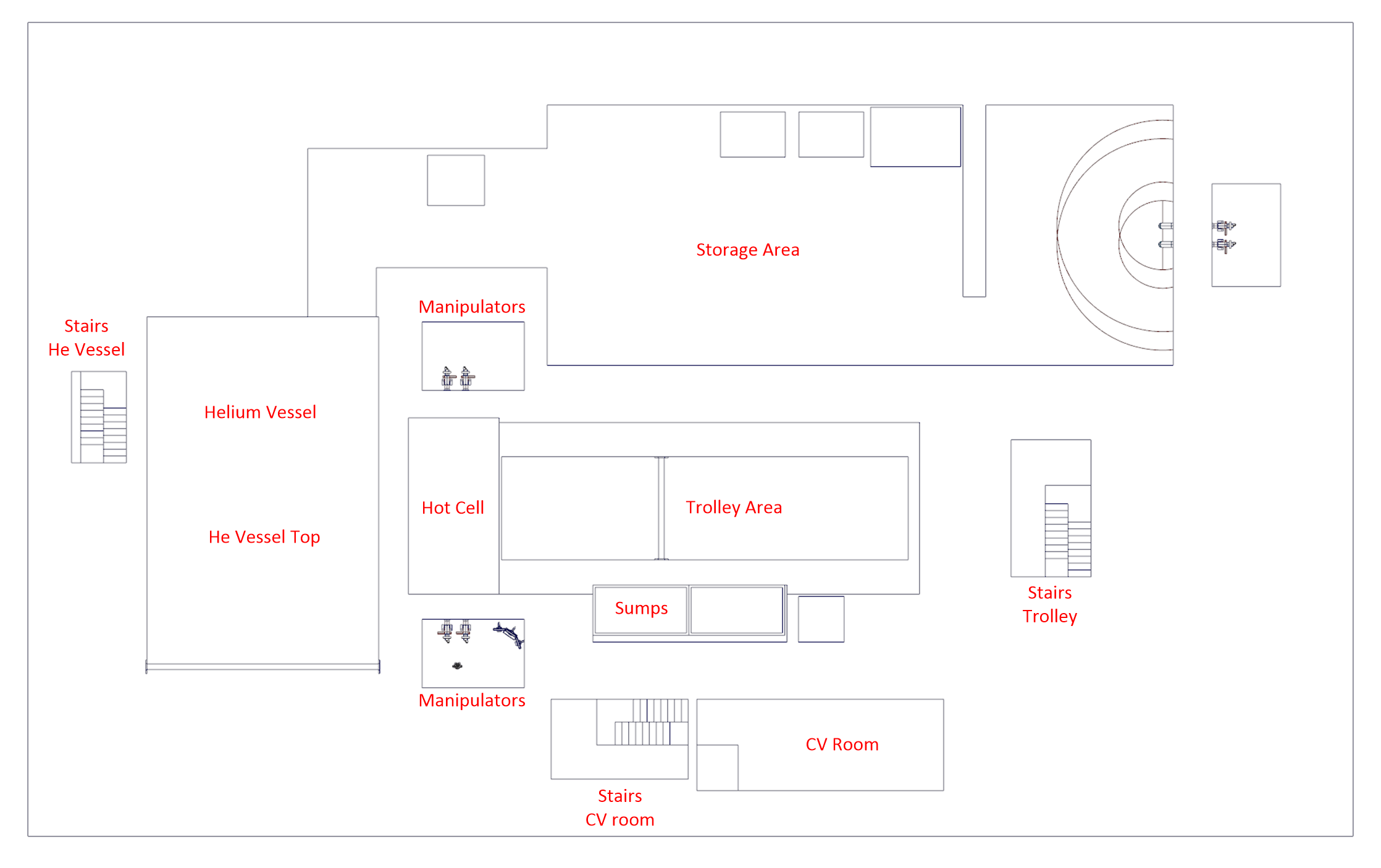}
\caption{\label{Fig:TC:CV_Intro_020} Layout of rooms in the BDF target complex (top view from surface hall)}
\end{figure}

Figure~\ref{Fig:TC:CV_Intro_010} shows the preliminary 3D model of the BDF target complex and Figure \ref{Fig:TC:CV_Intro_020} shows a top view of the rooms inside the target complex.
The building is composed of the rooms listed in Table \ref{Tab:TC:CV_IntroRooms}.

\begin{table}[htbp]
\centering
\caption{\label{Tab:TC:CV_IntroRooms} Size of rooms in the BDF target complex}
\smallskip
\begin{tabular}{l|c}
\hline
\textbf{Room} & \textbf{Volume [m\textsuperscript{3}]} \\
\hline
Surface hall & 28786 \\
Stairs (He vessel) & 47 \\
Stairs (CV room) & 340 \\
Stairs (Trolley) & 231 \\
Manipulators rooms & 292 \\
CV room & 562 \\
Sumps shaft & 340 \\
He vessel top & 333 \\
Hot cell & 173 \\
Trolley area & 240 \\
\hline
\end{tabular}
\end{table}

The main purpose of the rooms listed in Table~\ref{Tab:TC:CV_IntroRooms} is as follows:
\begin{itemize}
\item Surface hall: allow access and manipulation of components during maintenance;
\item Stairs: allow access to the underground areas;
\item Manipulators rooms: manipulate target components during maintenance;
\item CV room: host activated CV equipment underground;
\item Sumps shaft: allow collection of water leaks at the lowest point;
\item He vessel top: provide separation layer between He vessel and surface hall;
\item Hot cell: host target during maintenance;
\item Trolley area: host target CV systems;
\item Service room: host non-activated CV equipment and other services (this area is located in auxiliary building).
\end{itemize}

The storage area serves as storage space during maintenance operations; from a ventilation standpoint, it is not separated from the surface hall, despite being underground. 
The service room is located in the auxiliary building and it is shared with the services needed for the extraction tunnel (cooling and ventilation, power converters, etc.).

\subsubsection{Systems Description}
\label{Sec:TC:CV_IntroSysDesc}
The BDF target complex is designed around a trolley which supports the target and allows its insertion and extraction from the helium vessel for maintenance.
The helium vessel contains the following elements (Figure~\ref{Fig:TC:BDF_HeVe_3Dfrslice}):
\begin{itemize}
\item Target;
\item Shielding;
\item Collimator;
\item Beam window;
\item Magnetic coil.
\end{itemize}
All the services that are directly related to the target are located on the trolley, outside of the helium vessel, which is also equipped with a door that seals the helium vessel when the target is inside the helium vessel.

The target can be transported to the storage area and vice-versa without having to remove the helium vessel lid and shielding; here, hydraulic and electrical connections can be maintained via manipulators.
The proximity shielding is made of cast iron and it allows the target to enter from the side; its cooling connections come from the top of the helium vessel. 
The entire facility is conceived to allow recovery from failures via remote handling capabilities.

\subsection{Cooling and Ventilation Requirements}
\label{Sec:TC:CV_UR}

\subsubsection{Cooling Requirements}
\label{Sec:TC:CV_UR_Cool}

The cooling requirements for the BDF water cooling systems are listed in Table~\ref{Tab:TC:CV_URCool}.
The non-activated water cooling systems (such as the primary system) are located in a service area in the auxiliary building (separated from the target complex), whereas the equipment that will likely be activated is located underground in the CV room, inside the target complex. All secondary cooling circuits are demineralized, in order to minimize the activation of the water. 

The target cooling circuit is pressurized, in order to increase the water boiling temperature and avoid undesired increase of the blocks' surface temperature that could damage the target. The power deposition is done over a 1 s pulse, every 7.2 s; the power level is roughly 2.5 MW during the first second, and 0 MW during the remaining 6.2 s. The overall thermal load on the target circuit, averaged over 7.2 s, is equal to 350 kW, with the peak power being 7.2 times larger than the average power (see Section~\ref{Sec:TGT:Design:optimisation}).
The proximity shielding cooling system provides cooling to the shielding blocks immediately surrounding the target; the thermal load for all blocks is roughly 20 kW (averaged over the 7.2 s cycle).
The magnetic coil cooling system provides cooling for the magnetic coil located inside the helium vessel, immediately downstream the target. The thermal load generated by the current is 150 kW.

\begin{table}[htbp]
\centering
\caption{\label{Tab:TC:CV_URCool} Cooling system requirements for the BDF target complex.}
\smallskip
\begin{tabular}{l|c|c|c|c}
\hline
\textbf{Parameter} & \textbf{Unit} & \textbf{Target} & \textbf{Proximity Shielding} & \textbf{Magnetic Coil} \\
\hline
Location & - & Trolley area & CV room & CV room \\
T supply & \textsuperscript{\degree}C & 28.0 & 28.0 & 28.0 \\
Flow rate & m\textsuperscript{3}/h & 45 & 6 & 15 \\
Thermal load & kW & 350 & 20 & 150 \\
P supply & bar & 22 & - & - \\
$\Delta$P & bar & 3.5 & - & - \\
Type & - & Demineralized & Demineralized & Demineralized \\
\hline
\end{tabular}
\end{table}

\subsubsection{Helium Systems Requirements}
\label{Sec:TC:CV_UR_Helium}

The BDF requires two helium systems for the following purposes:
\begin{itemize}
\item Helium target circulation (\ref{Sec:TC:CV_UR_HeCirc}): detect potential leaks from the primary containment of the target;
\item Helium passivation system (\ref{Sec:TC:CV_UR_HePass}): purify, circulate and cool helium in the helium vessel, to prevent activation and degradation of materials.
\end{itemize}
The requirements for the two systems are presented in the following sections.

\subsubsubsection{Target Helium Circulation Requirements}
\label{Sec:TC:CV_UR_HeCirc}

The target helium circulation system has the primary purpose of circulating pure helium around the target primary containment, to detect potential leaks from it and prevent undesired target failure or contamination.

The requirements for the target helium circulation system are presented in Table \ref{Tab:TC:CV_URHeCirc}.
The helium for the target circulation system is not continuously purified.

\begin{table}[htbp]
\centering
\caption{\label{Tab:TC:CV_URHeCirc} Design requirements for the helium circulation system for the target containment. The system is positioned on the trolley and allows leak detection from the target assembly.}
\smallskip
\begin{tabular}{l|c|c}
\hline
\textbf{Parameter} & \textbf{Unit} & \textbf{Value} \\
\hline
Location & - & Trolley area \\
T supply & \textsuperscript{\degree}C & 28.0 \\
Flow rate & g/s & 1.5 \\
Thermal load & kW & 0.2 \\
P return & bar & 1.0 \\
P max & bar & 0.1 \\
Activation & - & Yes \\
\hline
\end{tabular}
\end{table}

\subsubsubsection{Helium Passivation System Requirements}
\label{Sec:TC:CV_UR_HePass}
The helium passivation system has the purpose of purifying the helium that circulates through the helium vessel and to cool it.
A preliminary design has been developed by ILK (DE) in collaboration with CERN \cite{HePurif_Predesign}; ILK has also performed a detailed design of the system~\cite{HePurif_CollAgree,HePurif_DetDesign}.

The cooling requirements for the helium passivation system are presented in Table \ref{Tab:TC:CV_URHePassC}.
The system is capable of circulating 780 m\textsuperscript{3}/h of helium and cooling a thermal load of 3 kW. Preliminary CFD simulations (Section~\ref{Sec:TC:HeV:CFD}) have been run on the helium vessel to demonstrate that a 0.2 bar maximum pressure head requirement is sufficient to compensate for the system pressure drop and that impurities do not stagnate inside the helium vessel.

\begin{table}[htbp]
\centering
\caption{\label{Tab:TC:CV_URHePassC} Cooling requirements for the helium passivation system.}
\smallskip
\begin{tabular}{l|c|c}
\hline
\textbf{Parameter} & \textbf{Unit} & \textbf{Value} \\
\hline
Location & - & CV room \\
T supply & \textsuperscript{\degree}C & 28.0 \\
Flow rate & m\textsuperscript{3}/h & 780 \\
Thermal load & kW & 3 \\
P return & bar & 1.0 \\
$\Delta$P & bar & 0.2 \\
Activation & - & Yes \\
\hline
\end{tabular}
\end{table}

The purification requirements for the helium passivation system are presented in Table \ref{Tab:TC:CV_URHePassP}. The system is designed to provide a minimum helium purity of 99.9~vol\% (averaged over the vessel volume).

\begin{table}[htbp]
\centering
\caption{\label{Tab:TC:CV_URHePassP} Purification requirements for the helium passivation system.}
\smallskip
\begin{tabular}{l|c|c}
\hline
\textbf{Parameter} & \textbf{Unit} & \textbf{Value} \\
\hline
Location & - & CV room \\
Flow rate & m\textsuperscript{3}/h & 75 \\
Design helium purity & \% & 99.9 \\
P purifier & bar & 15.0 \\
Activation & - & Yes \\
\hline
\end{tabular}
\end{table}

\subsubsection{Ventilation Requirements}
\label{Sec:TC:CV_UR_Vent}
The BDF target complex requires a ventilation system that provides dynamic confinement, in order to avoid dispersion of contaminants. The design of the ventilation system is done according to the guidelines provided in~\cite{CVBDFTTC_ISOnorm}; Section~\ref{Sec:TC:CV_UR_VentISO} illustrates the application of the norm to the BDF target complex case. The radiation protection group at CERN has classified the ventilation areas in the target complex according to these guidelines, and a series of requirements on the ventilation system has been produced (Section~\ref{Sec:TC:CV_UR_VentRP}). These requirements, together with the thermal requirements (Section~\ref{Sec:TC:CV_UR_VentTH} and \ref{Sec:TC:CV_UR_VentLoads}), constitute the basis on which the ventilation system is designed. All ventilation units are located in a CV service area located in the auxiliary building, next to the target complex.

\subsubsubsection{Nuclear Facility Ventilation Design Guidelines}
\label{Sec:TC:CV_UR_VentISO}

The mentioned ISO norm \cite{CVBDFTTC_ISOnorm} provides criteria for the design and operation of ventilation systems for nuclear installations other than nuclear reactors.
The guidelines described in the norm are applied according to the following procedure:
\begin{enumerate}
\item Each ventilation area is classified on a scale one to four, on the basis of permanent (surface and airborne) and accidental contamination;
\item The underpressure value is defined for each area, in order to provide dynamic confinement;
\item An estimated air change rate is determined;
\item Recommendations (such as filtration level, etc.) for the design of the specific ventilation system are given, based on the respective classification.
\end{enumerate}
This procedure has been applied to the BDF target complex and the resulting requirements are listed in Section~\ref{Sec:TC:CV_UR_VentRP}.

\subsubsubsection{Radiation protection requirements}
\label{Sec:TC:CV_UR_VentRP}

The application of the ISO norm to BDF is essentially based on the potential accidental conditions that have been postulated by RP; these accidental conditions result to be more stringent than the normal operating conditions in terms of classification, since expected permanent contamination levels for BDF are low.

Table \ref{Tab:TC:CV_URRP1} shows the RP requirements for each ventilated room, in terms of class, underpressure and minimum air change rate.
The pressure cascade produces a dynamic confinement that prevents the contaminants from being transported from highly-critical to non-critical areas in the target complex. The air change rate is also determined in order to replace activated air in passage areas (such as stairs) and most critical rooms.

\begin{table}[htbp]
\centering
\caption{\label{Tab:TC:CV_URRP1} Radiation protection requirements for BDF target complex and nuclear classification of the different areas of the complex.}
\smallskip
\begin{tabular}{l|c|c|c}
\hline
\textbf{Room} & \textbf{RP class} & \textbf{Underpressure [Pa]} & \textbf{Air changes [ac/h]} \\
\hline
Surface hall & C1 & -20 & 1 \\
Stairs (He vessel) & C1 & -40 & 2 \\
Stairs (CV room) & C1 & -40 & 2 \\
Stairs (Trolley) & C1 & -40 & 2 \\
Manipulators rooms & C1 & -60 & 4 \\
CV room & C2 & -80 & 2 \\
Sumps shaft & C2 & -80 & 2 \\
He vessel top & C2 & -100 & 5 \\
Hot cell & C2 & -100 & 5 \\
Trolley area & C2 & -100 & 5 \\
\hline
\end{tabular}
\end{table}

As shown in Table~\ref{Tab:TC:CV_URRP1}, different underpressure values can be associated to the same class; for this reason the ventilation rooms have been reclassified according to the subclassification shown in Table \ref{Tab:TC:CV_URRP2}, in which each subclass is associated to a specific combination of class and underpressure.

\begin{table}[htbp]
\centering
\caption{\label{Tab:TC:CV_URRP2} Radiation protection classes for subclassification}
\smallskip
\begin{tabular}{l|c|c|l}
\hline
\textbf{RP subclass} & \textbf{RP class} & \textbf{Underpressure [Pa]} & \textbf{Zones}\\
\hline
C1A & C1 & -20 & Surface hall \\
C1B & C1 & -40 & Staircases \\
C1C & C1 & -60 & Manipulators rooms \\
C2A & C2 & -80 & CV room, sumps \\
C2B & C2 & -100 & He vessel top, hot cell, trolley area \\
\hline
\end{tabular}
\end{table}

\begin{figure}[htbp]
\centering %
\includegraphics[width=0.8\linewidth]{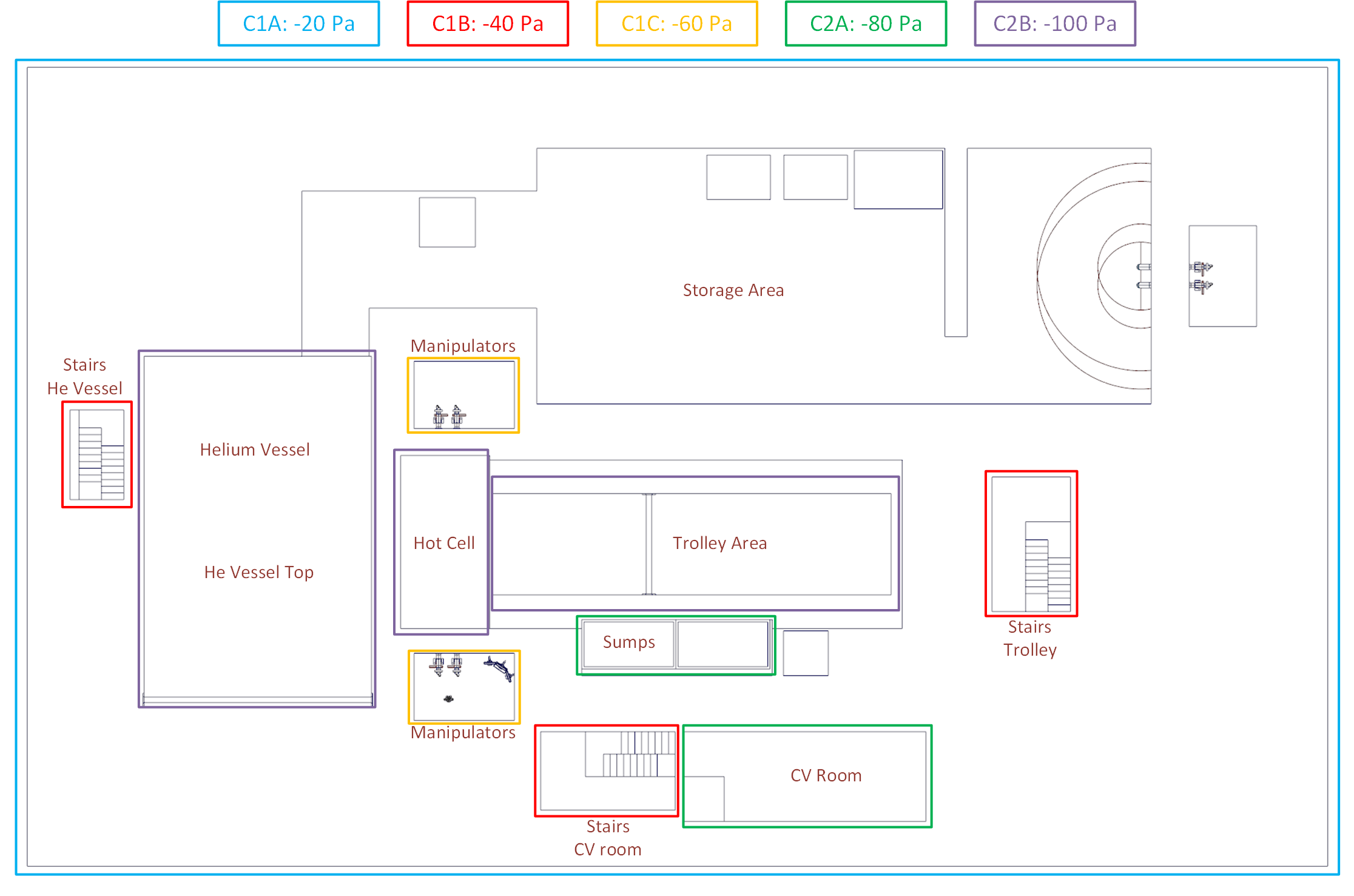}
\caption{\label{Fig:TC:CV_UR_010} Pressure cascade layout for dynamic confinement.}
\end{figure}

The resulting RP requirements following the subclassification are shown in Table \ref{Tab:TC:CV_URRP3}. Figure \ref{Fig:TC:CV_UR_010} shows the layout of the pressure cascade for the dynamic confinement. The minimum supply flow in Table \ref{Tab:TC:CV_URRP3} only accounts for the radio protection requirements on the air change rate listed in Table \ref{Tab:TC:CV_URRP1}; the actual design values presented in Section~\ref{Sec:TC:CV_Design_Vent} also account for requirements other than RP (e.g. thermal load). The supply air may be recirculated air for subclass C1A; the supply air shall be fresh air for all other subclasses.

\begin{table}[htbp]
\centering
\caption{\label{Tab:TC:CV_URRP3} Summary of radiation protection requirements according to subclassification.}
\smallskip
\begin{tabular}{l|c|c|c}
\hline
\textbf{RP subclass} & \textbf{Total volume [m\textsuperscript{3}]} & \textbf{Underpressure [Pa]} & \textbf{Minimum RP flow [m\textsuperscript{3}/h]} \\
\hline
C1A & 28786 & -20 & 28786 \\
C1B & 618 & -40 & 1236 \\
C1C & 292 & -60 & 1166 \\
C2A & 902 & -80 & 1803 \\
C2B & 746 & -100 & 3731 \\
\hline
\end{tabular}
\end{table}

\subsubsubsection{Temperature and Humidity Requirements}
\label{Sec:TC:CV_UR_VentTH}

Table~\ref{Tab:TC:CV_URTH1} presents requirements for temperature and humidity range in each room.

\begin{table}[htbp]
\centering
\caption{\label{Tab:TC:CV_URTH1} Temperature and humidity requirements for BDF target complex.}
\smallskip
\begin{tabular}{l|c|c}
\hline
\textbf{Room} & \textbf{T range [\textsuperscript{\degree}C]} & \textbf{RH range [\%]} \\
\hline
Surface hall & 18-26 & Not controlled \\
Stairs & 18-26 & 40-70 \\
Manipulators rooms & 18-26 & 40-70 \\
CV room & 18-26 & 40-70 \\
Sumps shaft & 18-26 & 40-70 \\
He vessel top & 18-24 & 40-70 \\
Hot cell & 18-24 & 40-70 \\
Trolley area & 18-24 & 40-70 \\
\hline
\end{tabular}
\end{table}

\subsubsubsection{Internal Heat Dissipation}
\label{Sec:TC:CV_UR_VentLoads}

The internal dissipations, due to equipment and occupational activities inside the conditioned spaces are listed in Table \ref{Tab:TC:CV_URLoads}. At the current design level, most of the spaces in the target complex do not present any significant thermal load other than people and lighting; further investigation will have to be done once more accurate information becomes available (EL loads, eventual power converter requirements, etc.).

\begin{table}[htbp]
\centering
\caption{\label{Tab:TC:CV_URLoads} Internal heat load and occupancy for each room in the target complex. Relevant loads are expected only for technical rooms.}
\smallskip
\begin{tabular}{l|c|c}
\hline
\textbf{Room} & \textbf{Thermal load [kW]} & \textbf{Max. occupancy} \\
\hline
Surface hall & - & 10 \\
Stairs & - & 2 \\
Manipulators rooms & - & 3 \\
CV room & 10 & 2 \\
Sumps shaft & 2 & 2 \\
He vessel top & - & 2 \\
Hot cell & - & 2 \\
Trolley area & 5 & 2 \\
\hline
\end{tabular}
\end{table}

\subsubsection{Other Requirements}
\label{Sec:TC:CV_UR_Other}

\subsubsubsection{Compressed Air}

Compressed air supply is required inside the BDF target complex. Other than general use in the surface hall, compressed air is mainly required in the trolley area and CV room, for purging the water circuits and for pneumatic actuators. Table~\ref{Tab:TC:CV_URCompAir} shows the compressed air requirements; as the compressed air will be supplied from a technical gallery, the standard dew-point and filtration available on the Pr\'evessin site are acceptable.

\begin{table}[htbp]
\centering
\caption{\label{Tab:TC:CV_URCompAir} Compressed air requirements}
\smallskip
\begin{tabular}{l|c|c}
\hline
\textbf{Parameter} & \textbf{Value} & \textbf{Unit} \\
\hline
Flow rate & 200 & Nm\textsuperscript{3}/h \\
Supply pressure & 7 & bara \\
Dew-point & NS & \textsuperscript{\degree}C \\
Filtration & NS & - \\
\hline
\end{tabular}
\end{table}

\subsubsubsection{Rising System}
The BDF target complex needs to be provided with sumps to collect the activated water of the underground services and leakage through the building walls. The sumps area is equipped with pumps to rise the water to surface storage tanks, located on the top floor of the CV room and in the auxiliary building. The sump room is provided with a steel liner to prevent contamination of concrete. 

Table~\ref{Tab:TC:CV_URRisingVols} lists the volumes of water to be discharged to the sump system every time the circuits are drained and refilled. Due to the high level of activation of the water, the target circuit is the driving factor in sizing the sump system.

\begin{table}[htbp]
\centering
\caption{\label{Tab:TC:CV_URRisingVols} Estimated water volume of cooling circuits}
\smallskip
\begin{tabular}{l|c}
\hline
\textbf{Circuit} & \textbf{Volume [m\textsuperscript{3}]} \\
\hline
Target & 1.0 \\
Proximity Shielding & 0.3 \\
Magnetic Coil & 0.6 \\
\hline
\end{tabular}
\end{table}

The system is provided with two sump tanks, each sized at 10 m\textsuperscript{3}, allowing to store 20 times the volume of the target circuit and eventually accommodating for water infiltration. Each sump tank is provided with two redundant sump pumps.
The target circuit is drained based on RP instructions; the drain water is left in the sump for two weeks to allow the decay of the short-lived products. After the decay time, the water is pumped to the dedicated tank at the ground level.

\subsubsubsection{Evaporation System}
The auxiliary building is equipped with a room specifically designed for hosting the evaporation system. The evaporation system consists of three evaporation units of the type Encon Drum Dryer (or similar), allowing a maximum evaporation rate of 90~m\textsuperscript{3}/year. 
The room includes two redundant 5~m\textsuperscript{3} tanks that hold the water before evaporation. The transfer of the water from the tank to the evaporator is automatic.

\subsubsubsection{Smoke Extraction System}
A smoke extraction system is needed for the target complex. The design of the smoke extraction system needs to be compatible with the confinement of radioactive products provided by the ventilation system. During a fire event the pressure cascade shall be kept in place as much as possible to avoid unforeseen release of potentially contaminated smoke.
A 36 m\textsuperscript{3}/h/m\textsuperscript{2} extraction flow rate is provided by the surface hall smoke extraction system. Regarding the underground areas, the smoke is extracted using the standard ventilation system, after the smoke cool-down.
The rooms in the underground area are considered as separated fire compartments; each fire compartment is provided with fire walls and doors according to EN13501, and the required fire dampers for smoke confinement.

\subsection{Design of Cooling and Ventilation Systems}
\label{Sec:TC:CV_Design}

\subsubsection{Cooling Systems}
\label{Sec:TC:CV_Design_Cool}

\subsubsubsection{General Description}
\label{Sec:TC:CV_Design_Cool_Intro}

\begin{figure}[htbp]
\centering %
\includegraphics[width=0.8\linewidth]{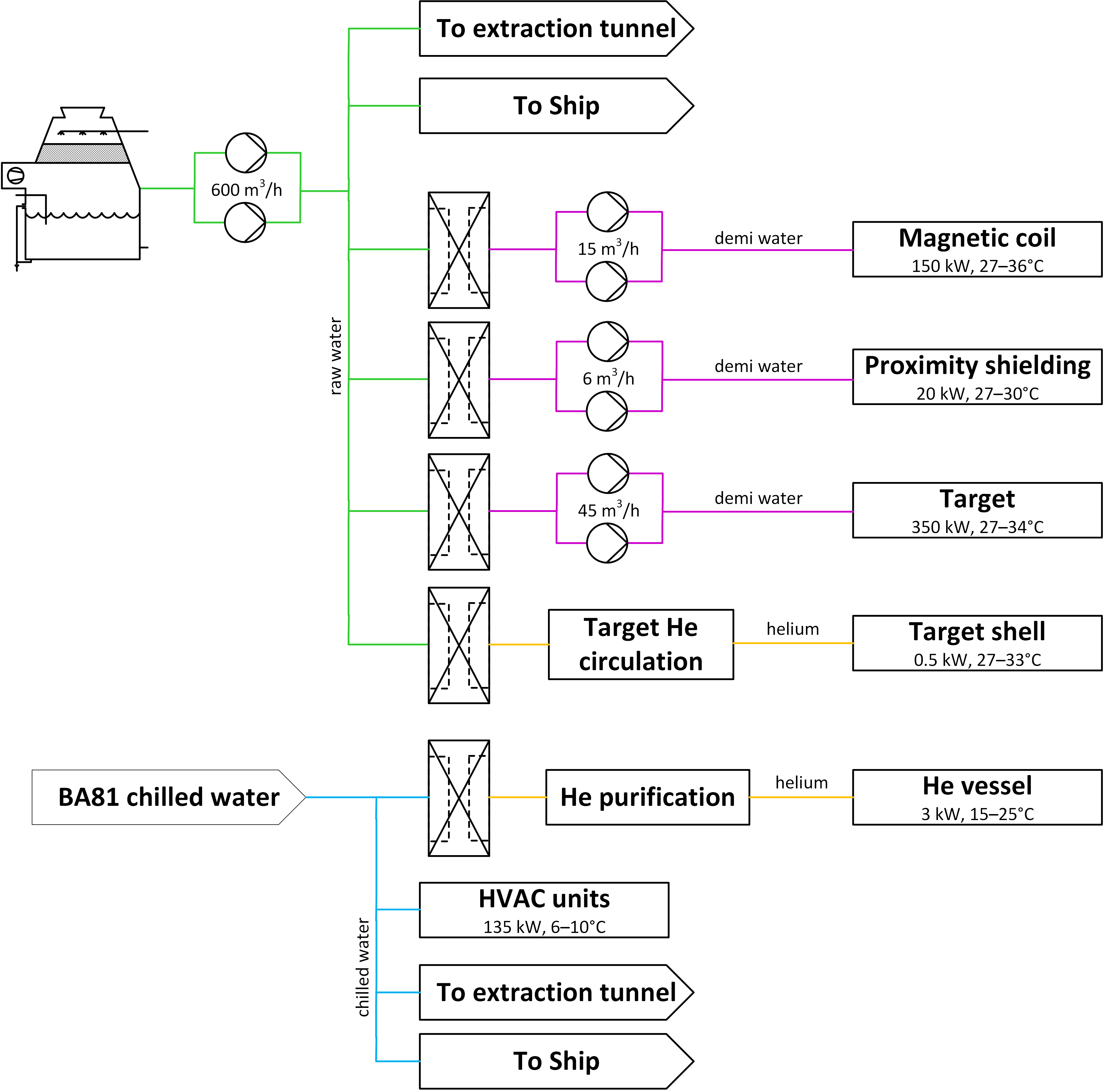}
\caption{\label{Fig:TC:CV_DesCool_010} Overview of the cooling system structure for the BDF target complex. Primary (green) and chilled (blue) water circuits are shared with extraction and experimental area on the BDF complex.}
\end{figure}

The BDF cooling system is based on a raw water primary cooling system supplying water at 25\textsuperscript{\degree}C via cooling towers. Local cooling is provided via demineralized water secondary circuits; chilled water provides cooling for the AHUs and the helium purification system. Figure \ref{Fig:TC:CV_DesCool_010} shows a schematic view of the cooling systems for the target complex, which are structured as follows:
\begin{enumerate}
\item Primary raw water (\S\ref{Sec:TC:CV_Design_Cool_Prim}):
\begin{enumerate}
\item Target helium circulation system (\S\ref{Sec:TC:CV_Design_Cool_HeCirc}): located in the trolley area;
\item Secondary demineralized water cooling systems:
\begin{enumerate}
\item Target (\S\ref{Sec:TC:CV_Design_Cool_Targ}): located in the helium vessel;
\item Proximity shielding (\S\ref{Sec:TC:CV_Design_Cool_Prox}): located in the helium vessel;
\item Magnetic coil (\S\ref{Sec:TC:CV_Design_Cool_Mag}): located in the helium vessel;
\end{enumerate}
\end{enumerate}
\item Chilled water (\S\ref{Sec:TC:CV_Design_Cool_Chilled}):
\begin{enumerate}
\item Helium passivation system: located in the CV room;
\item AHUs: located in the auxiliary building.
\end{enumerate}
\end{enumerate}
Table \ref{Tab:TC:CV_DesCool_Site} presents a preliminary evaluation of the cooling requirements for the target complex compared to the cooling requirements of the BDF complex, including the extraction tunnel and the Ship experimental area. A preliminary estimate of the total cooling power needed by the target complex is roughly 600 kW on the demineralized water circuits and 135 kW on the chilled water system, whereas the total cooling power for the BDF complex is roughly 5500 kW for the demineralized water and 750 kW for the chilled water. 

\begin{table}[htbp]
\centering
\caption{\label{Tab:TC:CV_DesCool_Site} Cooling requirements of the BDF target complex compared to the overall BDF complex needs.}
\smallskip
\begin{tabular}{l|c|c}
\hline
\textbf{Cooling type} & \textbf{TC only [kW]} & \textbf{BDF complex [kW]} \\
\hline
Demineralized water & 600 & 5500 \\
Chilled water & 135 & 750 \\
\hline
\end{tabular}
\end{table}

\subsubsubsection{Cooling System P\&ID}
\label{Sec:TC:CV_Design_Cool_PID}

The following documentation has been prepared during the design process:
\begin{itemize}
\item Preliminary cooling synoptic \cite{CVBDFTTC_CoolSyn};
\item Preliminary cooling PID \cite{CVBDFTTC_CoolPID};
\item Preliminary CV 3D integration \cite{CVBDFTTC_CVIntegration};
\end{itemize}

\subsubsubsection{Primary Cooling System}
\label{Sec:TC:CV_Design_Cool_Prim}

The BDF cooling system is based on a primary circuit that supplies raw water at 25\textsuperscript{\degree}C. Two options have been considered for the supply of the primary water:
\begin{enumerate}
\item Installation of local independent cooling towers;
\item Connection to CT2 cooling tower system in the Prevessin area;
\end{enumerate}
In both cases the primary system will have to be sized to cope with the cooling load of the entire experimental facility (5500 kW), which includes the target complex needs (600 kW max).
Table \ref{Tab:TC:CV_DesCoolPrim} presents a high-level comparison between the two options, showing the advantages and disadvantages of one over the other.

\begin{table}[htbp]
\centering
\caption{\label{Tab:TC:CV_DesCoolPrim} Comparison of primary cooling design options.}
\smallskip
\begin{tabular}{>{\centering\arraybackslash}m{7cm}|>{\centering\arraybackslash}m{7cm}}
\hline
\textbf{Option 1 \linebreak Local cooling towers} & \textbf{Option 2 \linebreak CT2 cooling towers} \\
\hline
Operation independent from other cooling systems in the Prevessin area & Operation dependent on CT2 towers availability \\ \hline
Operation and maintenance more complex (water treatment, etc.) & No added complexity from operation and maintenance standpoint \\ \hline
Potential activation confined to new experimental area & Potential activation spreading over CT2 cooling system \\ \hline
Requires installation of cooling towers and pumping station on the new site & Requires the upgrade of the CT2 pumps and a new pipeline (technical gallery) \\
\hline
\end{tabular}
\end{table}

Considering the historical operational data, the experience of the operators, the recent upgrades of the CT2 towers and the future increase in load demand, the following considerations can be made:
\begin{itemize}
\item The piping and pumping capacity is not sufficient to supply raw water for the installations in the new experimental area;
\item The CT2 towers will likely not have enough capacity to accommodate the addition of 5.5 MW cooling load.
\end{itemize}
Being option 1 the more constraining one, it has been selected for the preliminary design phase; option 2 is maintained as backup option, in case the resulting total cooling load on the CT2 towers is lower than expected. 

\begin{figure}[htbp]
\centering %
\includegraphics[width=0.7\linewidth]{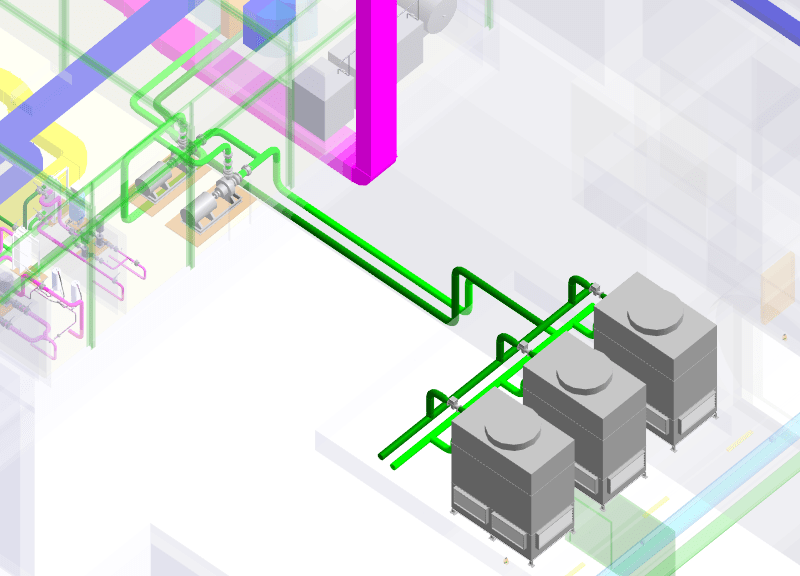}
\caption{\label{Fig:TC:CV_DesCool_020} Primary system pumping station and cooling towers.}
\end{figure}

The primary cooling circuit supplies 600 m\textsuperscript{3}/h at 25\textsuperscript{\degree}C for a total cooling load of 5.5 MW for the entire BDF facility; the expected temperature increase is 8\textsuperscript{\degree}C. The estimated main piping diameter is DN400; the pressure drop of the circuit is estimated around 3 bar and the required pumping power is roughly 88 kW. The pumping power is provided by 2 redundant pumps regulated via VFD, each one sized for 600 m\textsuperscript{3}/h; the cooling will be provided by 3 cooling towers (n+1 redundancy), 2.75 MW each. The pumps, the sand filter, the filling circuit and the other main components of the primary circuit are located in the auxiliary building service area, which is shared with the extraction tunnel services. Figure \ref{Fig:TC:CV_DesCool_020} shows the layout of the primary pumping station in the auxiliary building.

The BDF target complex uses a fraction of the 600 m\textsuperscript{3}/h flow rate, corresponding to 72 m\textsuperscript{3}/h and 550 kW, for cooling the demineralized water circuits and the target helium circulation system. The reference size for the primary piping going to the target complex circuits is DN125; within the BDF target complex, primary water reaches the CV room and it is distributed locally and in the trolley area.

\subsubsubsection{Target Cooling System}
\label{Sec:TC:CV_Design_Cool_Targ}

\begin{figure}[htbp]
\centering %
\includegraphics[width=0.8\linewidth]{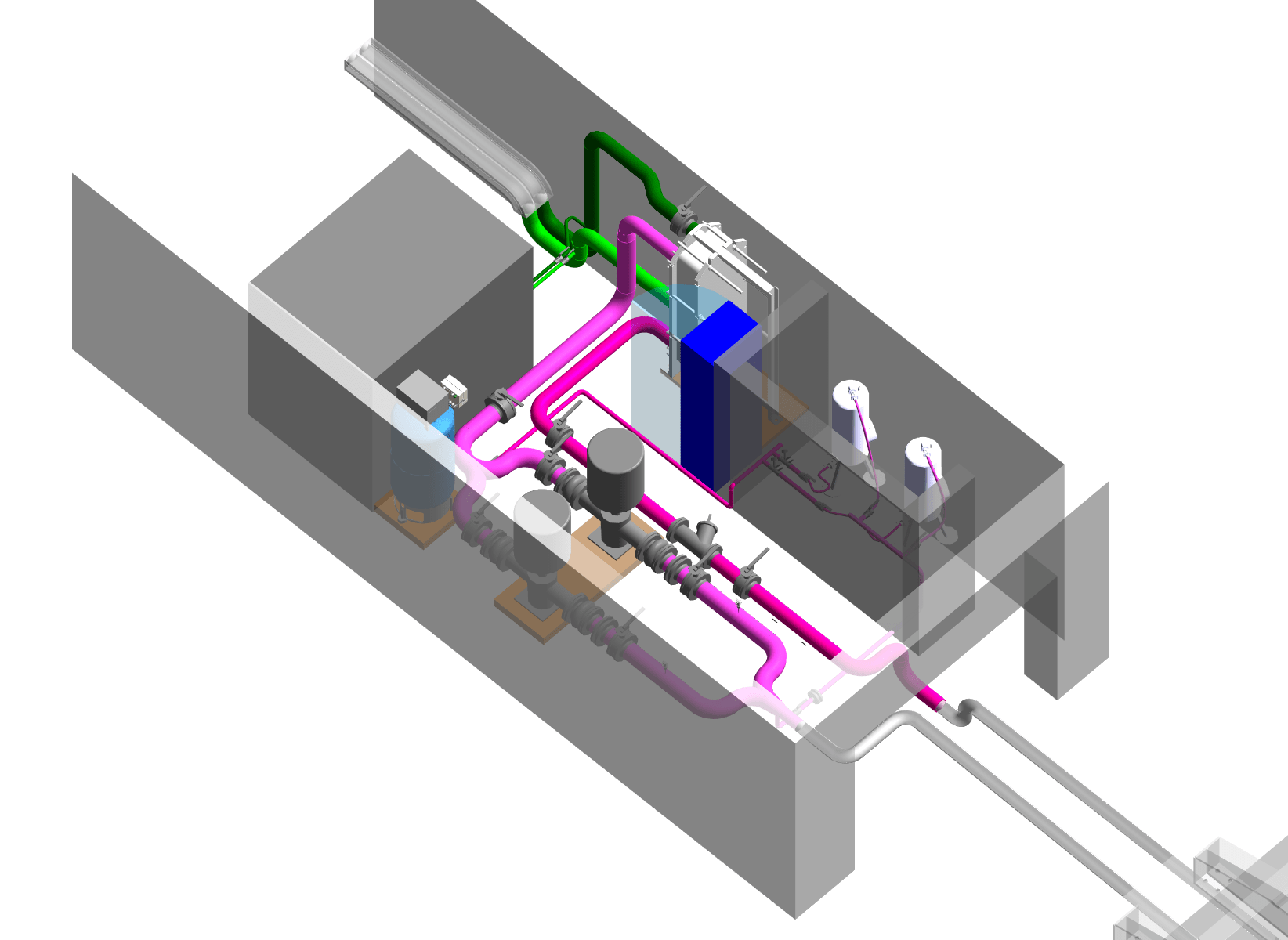}
\caption{\label{Fig:TC:CV_DesCool_030} Integration of the target cooling system and the helium circulation system on the trolley. The model also shows the heat exchanger for heat rejection to the primary system.}
\end{figure}

The target cooling system provides demineralized water cooling for the target and it is located on the trolley in the underground area (Figure~\ref{Fig:TC:CV_DesCool_030}). The system supplies 45 m\textsuperscript{3}/h at 28\textsuperscript{\degree}C for a thermal load of about 350 kW on average; the estimated return temperature is 35\textsuperscript{\degree}C. The system is pressurized at 19 bar in order to avoid local water boiling on the surface of the target blocks when the beam impacts the target. This pressurization is provided by an automatic expansion tank connected to the circuit before the pumps. Two redundant pumps provide the additional 3.5 bar head to compensate the pressure losses of the circuit, resulting in a supply pressure at the outlet of the pump of 22.5 bar. The size of the piping for the target circuit is DN100 and the estimated pumping power is 10 kW.

All operations on the circuit, such as filling, pressurization or draining, are automated and do not require physical human intervention to open valves. Draining of the circuit is performed indicatively once a year and it is planned based on activation assessments performed by RP; the drain water is discharged to the sump; details on draining operations and volumes of drain water are provided in \S\ref{Sec:TC:CV_Design_Rising}. Draining is followed by a complete demineralized water refill of the circuit. The actuation of valves for performing this operation is done via pneumatic actuators, which need to be functioning under irradiation; this aspect will be analyzed more in depth during the detailed design phase.

The equipment on the skid is provided with a retention basin to collect potential leaks from the system and directly drain them to the sump.

\begin{figure}[htbp]
\centering %
\includegraphics[width=0.8\linewidth]{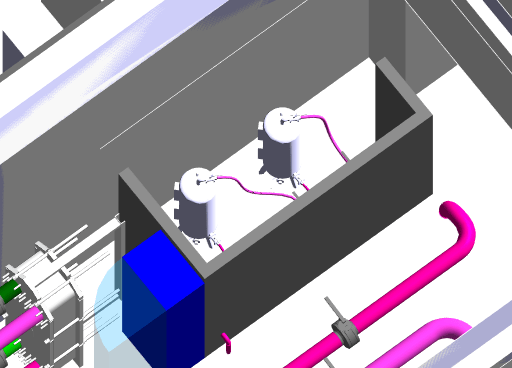}
\caption{\label{Fig:TC:CV_DesCool_031} Shielding for primary ion-exchanger cartridges.}
\end{figure}

The system exchanges the thermal power with the primary circuit via a plate heat exchanger located on the trolley. In order to prevent cross-contamination, the heat exchanger plates are double-walled, with EPDM double gaskets. The primary water is delivered to the primary side of the heat exchanger on the trolley via flexible piping. 
The system operates in radioactive area: the selection of the equipment, in particular electronics and plastic/elastomeric materials for hydraulic components, needs to account for this aspect. From preliminary evaluations, EPDM is acceptable for gaskets, piping and other components; the correct behavior of components under irradiation shall be confirmed in the detailed design phase, once expected dose estimated via simulations become available.
The water in the target cooling system is demineralized by a demineralization circuit in bypass mode, provided with two redundant cartridges; the maximum conductivity of the supplied water is 0.5 \textmu S/cm. During operation the cartridges activate, because of absorption of radioactive ions from the cooling water; in order to allow safe maintenance on the trolley during system shutdown, the cartridges are surrounded by iron shielding (Figure \ref{Fig:TC:CV_DesCool_031}) 20 cm thick, which reduces the radiation levels generated by the mixed-bed resin.

\subsubsubsection{Proximity Shielding Cooling System}
\label{Sec:TC:CV_Design_Cool_Prox}

\begin{figure}[htbp]
\centering %
\includegraphics[width=0.8\linewidth]{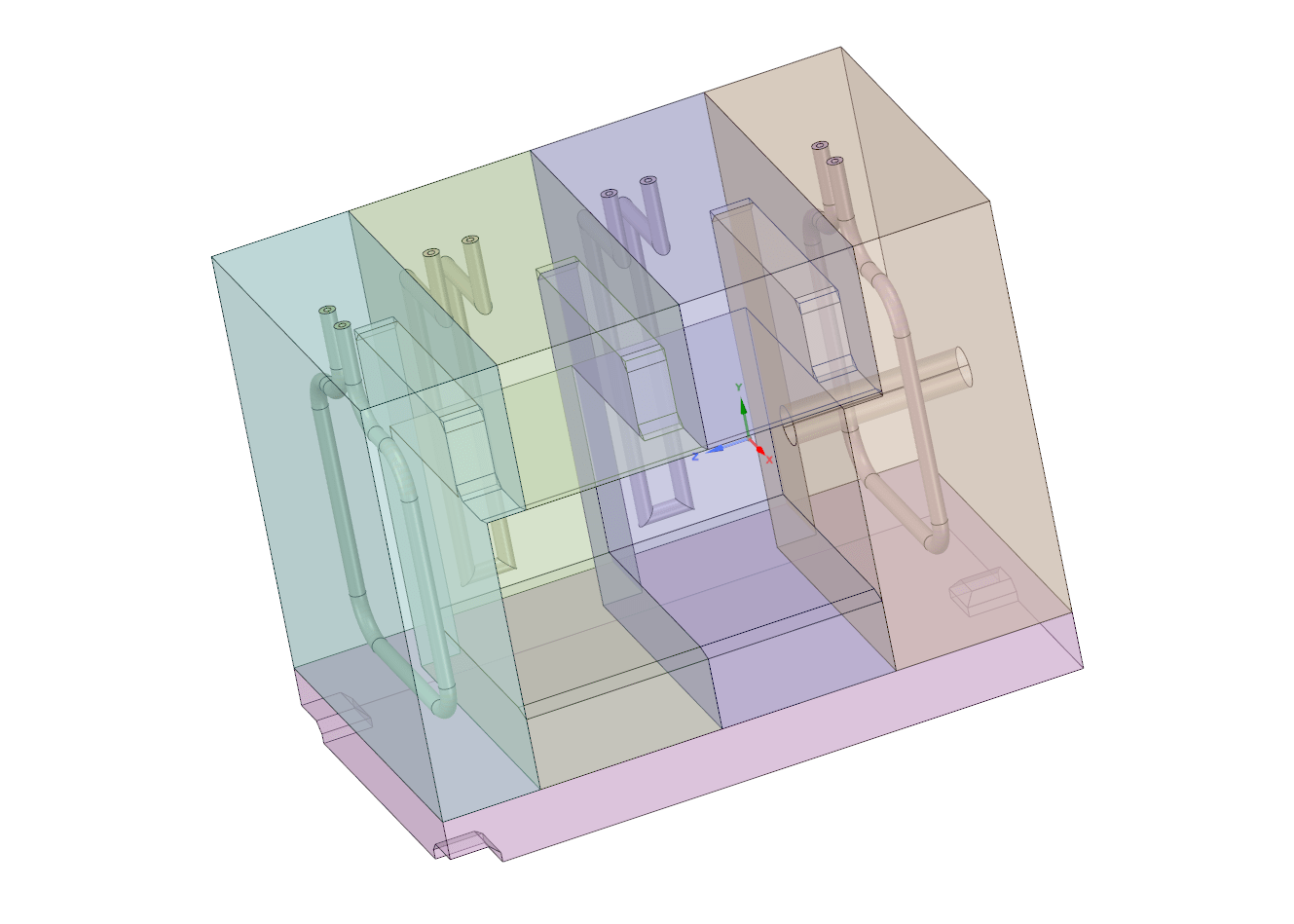}
\caption{\label{Fig:TC:CV_DesCool_040} Cooling pipes layout inside proximity shielding blocks.}
\end{figure}

The proximity shielding is a set of 5 blocks located around the BDF target. The beam delivers a thermal load of roughly 20 kW on average on these shielding blocks. This thermal load is cooled by cooling pipes internal to the shielding blocks; the preliminary layout of the cooling pipes is shown in Figure \ref{Fig:TC:CV_DesCool_040}.

\begin{figure}[htbp]
\centering %
\includegraphics[width=0.7\linewidth]{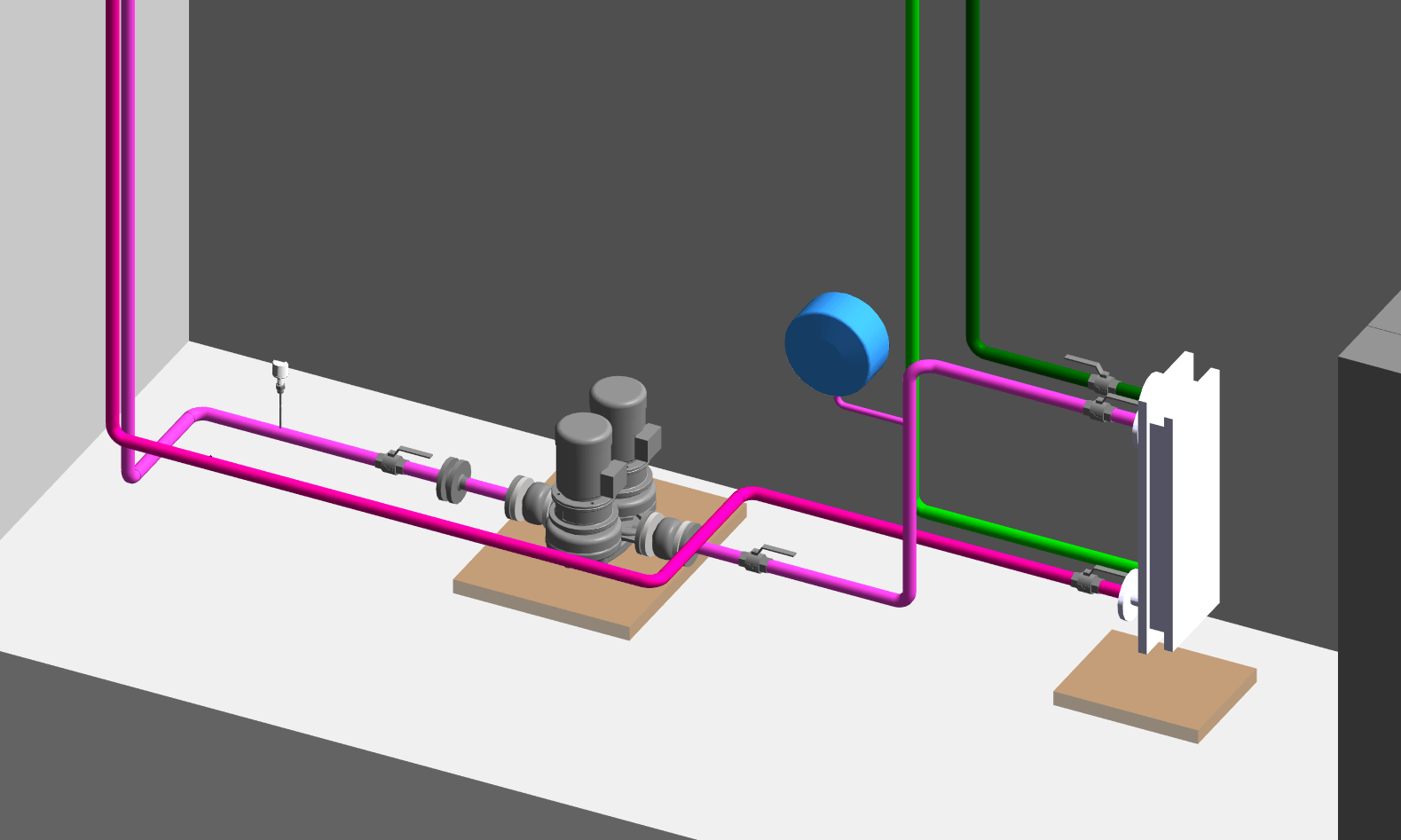}
\caption{\label{Fig:TC:CV_DesCool_041} Proximity shielding cooling station in the CV room.}
\end{figure}

The cooling is provided by a cooling station located in the CV room (Figure~\ref{Fig:TC:CV_DesCool_041}). This pumping station supplies 5.75 m\textsuperscript{3}/h of demineralized water at 28\textsuperscript{\degree}C at a supply pressure of 5 bar; the expected pressure drop in the proximity shielding is expected to be less than 1 bar. The return temperature is 30\textsuperscript{\degree}C and the reference size for the circuit piping is DN32. The pumping is provided by two redundant pumps; the power of each pump is 0.5 kW.
The water is demineralized by two redundant cartridges installed in the CV room; the maximum conductivity of the supplied water is 0.5 \textmu S/cm. During operation the cartridges activate, similarly to the target demineralization circuit; however, the activation level is lower than for the target circuit. A 40 cm concrete shielding is placed around the cartridge to reduce the dose rate in the CV room during access for maintenance.

\begin{figure}[htbp]
\centering %
\includegraphics[width=0.7\linewidth]{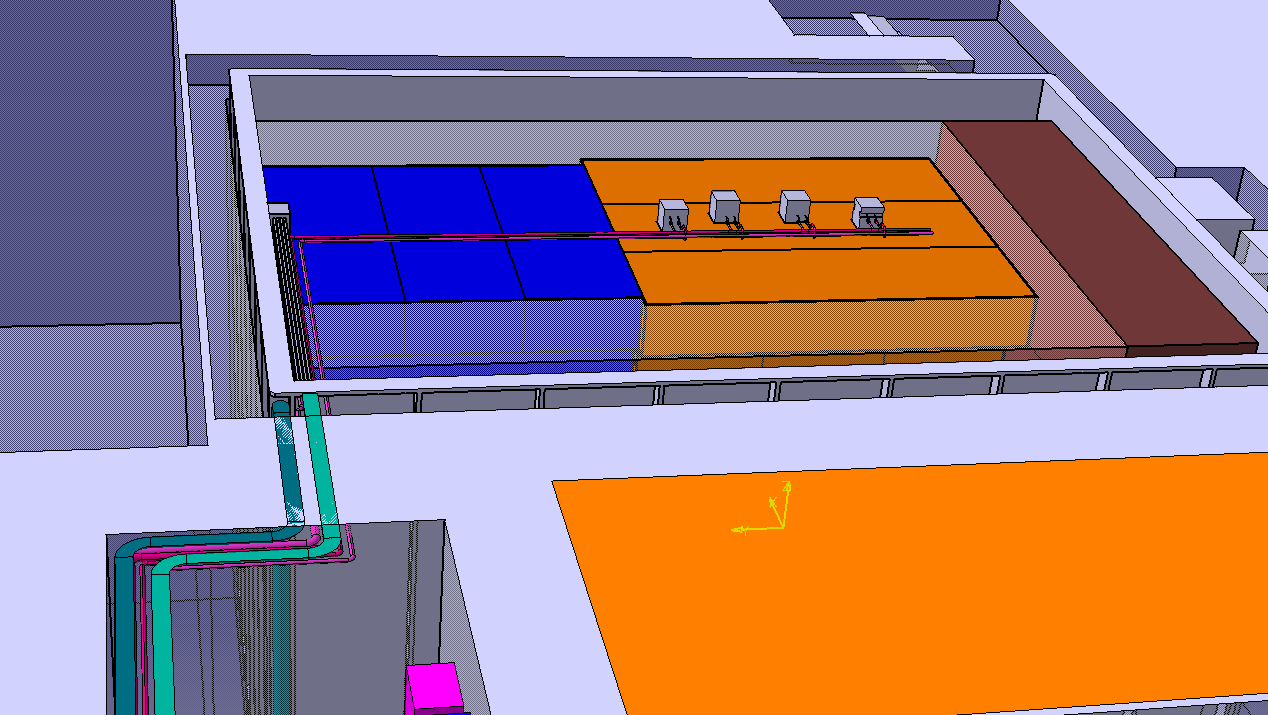}
\caption{\label{Fig:TC:CV_DesCool_042} Proximity shielding cooling connections on top of the bunker shielding.}
\end{figure}

The cooling system is connected to the proximity shielding blocks from the top of the helium vessel (Figure \ref{Fig:TC:CV_DesCool_042}); the portion of the piping contained inside the helium vessel is removable, to allow the positioning and removal of the bunker blocks.

\subsubsubsection{Magnetic Coil Cooling System}
\label{Sec:TC:CV_Design_Cool_Mag}

\begin{figure}[htbp]
\centering %
\includegraphics[width=0.5\linewidth]{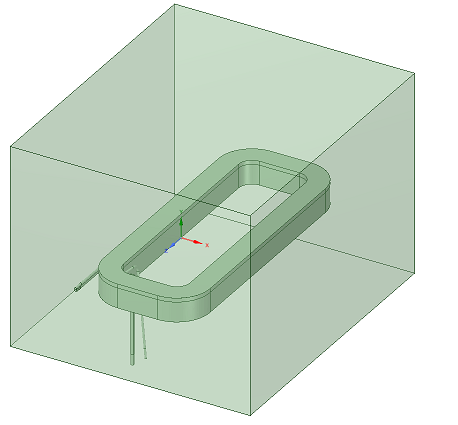}
\caption{\label{Fig:TC:CV_DesCool_051} 3D model of the muon shield magnet downstream the target.}
\end{figure}

\begin{figure}[htbp]
\centering %
\includegraphics[width=0.7\linewidth]{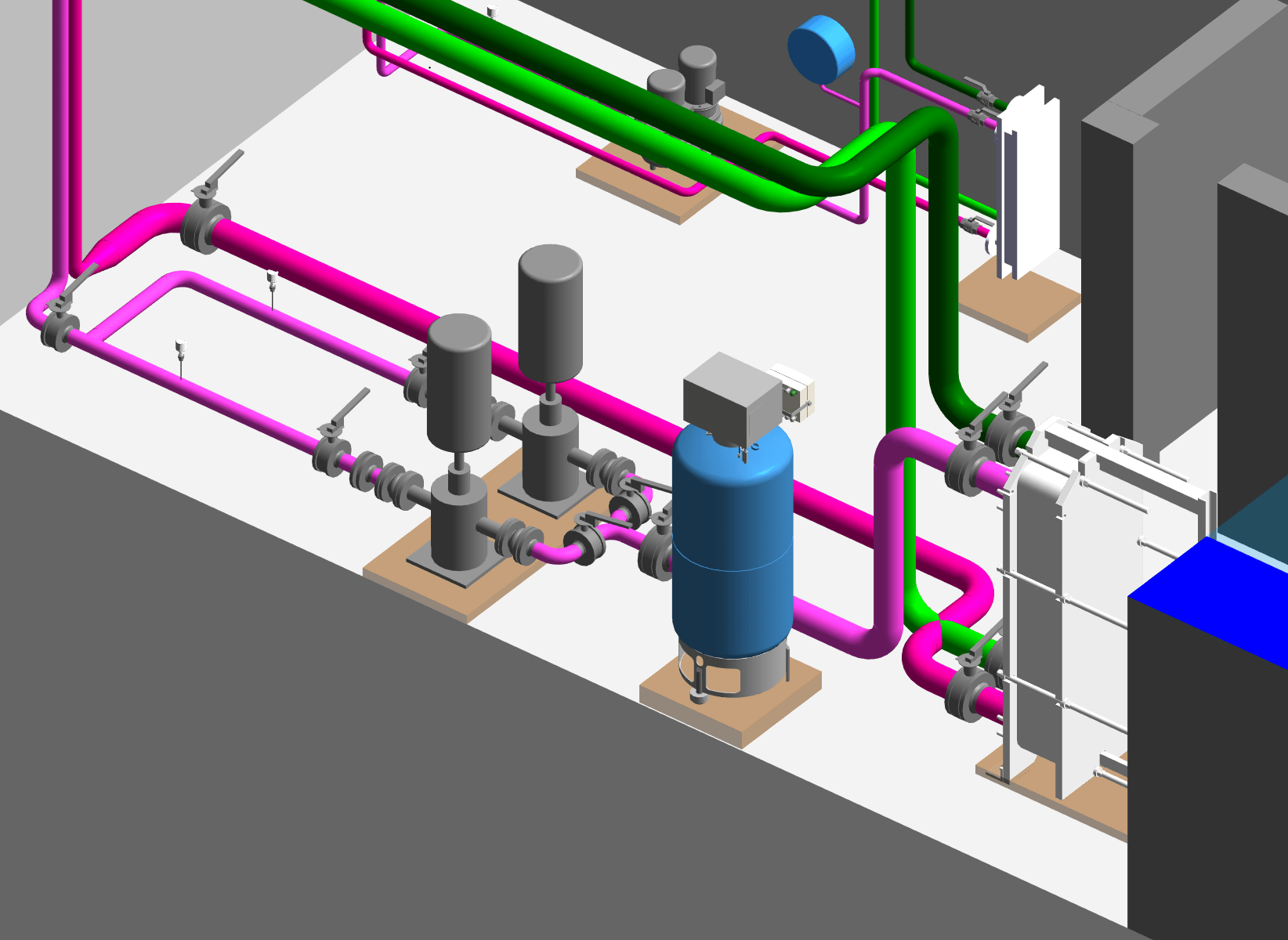}
\caption{\label{Fig:TC:CV_DesCool_052} Magnetic coil cooling station in CV room.}
\end{figure}

A preliminary design of the magnetic coil has been performed (Figure \ref{Fig:TC:CV_DesCool_051}), featuring a conductive plate that cools the heat generated by the current flowing through the coil windings. The magnetic coil is located inside the helium vessel, immediately downstream the target. The cooling for the magnetic coil is provided by a cooling station located in the CV room (Figure \ref{Fig:TC:CV_DesCool_052}), supplying demineralized water at 28\textsuperscript{\degree}C and 15 m\textsuperscript{3}/h. The cooling system is rated to absorb 150 kW for a return temperature of 37\textsuperscript{\degree}C. The pumping is provided by two redundant pumps, 15 kW each, supplying water at a maximum pressure of 20 bar; the reference size for the piping is DN65. The demineralization of the water is achieved via two redundant cartridges located in the CV room and shielded by a 40 cm concrete wall, to allow maintenance of the rest of the equipment after their activation; the maximum conductivity of the supplied water is 0.5 \textmu S/cm.

\subsubsubsection{Cooling for Target Helium Circulation System}
\label{Sec:TC:CV_Design_Cool_HeCirc}

The primary cooling system supplies cooling to the target helium circulation system. The power that needs to be removed from the helium circulation system is expected to be less than 0.5 kW. Details on the circulation system and its cooling are provided in the target helium circulation section (\ref{Sec:TC:CV_Design_HeCirc}).

\subsubsubsection{Chilled Water}
\label{Sec:TC:CV_Design_Cool_Chilled}

Chilled water is needed in the BDF target complex building for the cooling of the helium purification system and the ventilation units in the auxiliary building. The overall requirements in terms of chilled water for the BDF experimental area amount to roughly 750 kW max (Table \ref{Tab:TC:CV_DesCool_Site}); this chilled water is provided by the chilled water production plant at building BA81, which currently has enough margin to accept this load. If in the future this margin will be reduced, a new chilled water plant should be foreseen in the BDF auxiliary building, supplying chilled water for the entire experimental area.

Of the total amount of chilled water cooling (750 kW), about 135 kW (corresponding to 29 m\textsuperscript{3}/h of chilled water at 6-10\textsuperscript{\degree}C) are needed for the cooling coils of the three AHUs for the BDF target complex (located in the auxiliary building) and the helium purification system (located in the CV room).

\subsubsection{Helium Systems}
\label{Sec:TC:CV_Design_He}

The BDF target complex uses two helium systems for cooling, passivation and monitoring purposes:
\begin{enumerate}
\item Target helium circulation system (\S\ref{Sec:TC:CV_Design_HeCirc}): this system, located on the trolley, is used to flow helium around the target (through the target external shell) to detect any leak or activated gas release from the target containment;
\item Helium passivation system (\S\ref{Sec:TC:CV_Design_HePass}): this systems, located in the CV room, is used to cool and purify the helium flowing through the helium vessel.
\end{enumerate}

\subsubsubsection{Target Helium Circulation}
\label{Sec:TC:CV_Design_HeCirc}

\begin{figure}[htbp]
\centering %
\includegraphics[width=0.9\linewidth]{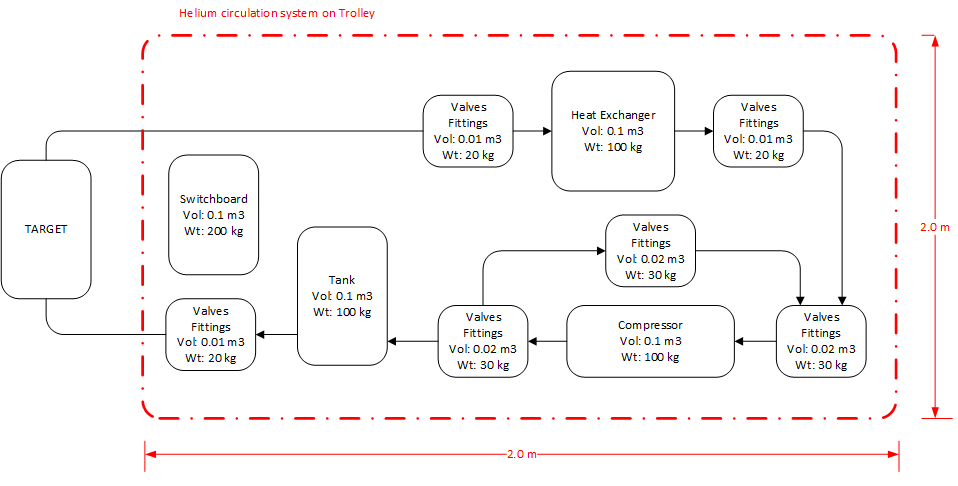}
\caption{\label{Fig:TC:CV_DesCool_060} Layout of the helium target circulation system.}
\end{figure}

The helium target circulation system blows helium around the target containment in order to detect any potential leak from its containment.
Figure \ref{Fig:TC:CV_DesCool_060} shows the layout of the main components of the circulation system skid, positioned on the trolley; a preliminary PID for the system is provided in \cite{CVBDFTTC_HeCircPID}. Table \ref{Tab:TC:CV_DesCool_HeCircDesign} shows the main design parameters for the system.

\begin{table}[htbp]
\centering
\caption{\label{Tab:TC:CV_DesCool_HeCircDesign} Design parameters for the helium circulation system. The system is designed for a maximum flow rate of 4.1 g/s and thermal power of 500 W.}
\smallskip
\begin{tabular}{l|c|c}
\hline
\textbf{Parameter} & \textbf{Unit} & \textbf{Value} \\
\hline
Location & - & Trolley \\
Flow rate & g/s & 4.1 \\
Changes per hour & h\textsuperscript{-1} & 1000 \\
T in & \textsuperscript{\degree}C & 28 \\
T out max & \textsuperscript{\degree}C & 50 \\
$\Delta$T max & \textsuperscript{\degree}C & 22 \\
Thermal load & kW & 0.5 \\
Piping size & - & DN25 \\
Pressure supply & bar & 1.1 \\
Pressure drop & Pa & 6000 \\
\hline
\end{tabular}
\end{table}

The main components of the system are a compressor, a helium storage and pressurization tank (positioned immediately after the compressor) and a heat exchanger on the return line. This heat exchanger transfers the heat to the primary circuit.

The target helium circulation system supplies 4.1 g/s of helium at nearly atmospheric pressure; the compressor is sized to overcome a pressure drop of 6000 Pa. The maximum power that can be removed by the system is about 500 W, assuming a maximum temperature for the target containment of 50\textsuperscript{\degree}C; note that no simulations have been run yet on the helium circulation through the shell, so this assumption will have to be more deeply analyzed at a detailed design stage. In any case, the purpose of the system is primarily helium circulation, and not cooling, so a higher or lower equilibrium temperature for the helium in the target shell is acceptable.
The size of the skid is expected to be about 2 m [L] x 2 m [W] x 1.5 m [H], resulting in 6 m\textsuperscript{3} volume and 1300 kg maximum weight.

The helium circulation system is cooled via primary raw water at 25\textsuperscript{\degree}C; the water is supplied by the same primary piping that is connected to the primary side of the heat exchanger of the target secondary circuit (Figure \ref{Fig:TC:CV_DesCool_030}); a regulation valve allows to set the desired flow on this bypass line. Table \ref{Tab:TC:CV_DesCool_HeCircCOolPar} shows the main parameters of the water cooling circuit.

\begin{table}[htbp]
\centering
\caption{\label{Tab:TC:CV_DesCool_HeCircCOolPar} Water cooling specifications for the helium circulation system}
\smallskip
\begin{tabular}{l|c|c}
\hline
\textbf{Parameter} & \textbf{Unit} & \textbf{Value} \\
\hline
Location & - & Trolley \\
T in & \textsuperscript{\degree}C & 25 \\
T out & \textsuperscript{\degree}C & 25.4 \\
T & \textsuperscript{\degree}C & 0.4 \\
Flow rate & m\textsuperscript{3}/h & 1.1 \\
Piping size & - & DN15 \\
Thermal load & kW & 0.5 \\
Type & - & Raw water \\
Activation & - & No \\
\hline
\end{tabular}
\end{table}

\subsubsubsection{Helium Passivation System}
\label{Sec:TC:CV_Design_HePass}

The helium passivation system supplies purified helium to the helium vessel that contains the BDF target and its shielding; its primary purpose is to remove impurities from the gas mixture so that it does not activate and remains inert from a chemical standpoint. A complete description of the system is provided in Section~\ref{Sec:TC:HeV:HePass}.
Its cooling layout is shown on the preliminary cooling synoptic \cite{CVBDFTTC_CoolSyn};
Section~\ref{Sec:TC:HeV:HePass} provides a preliminary P\&ID for the system and a preliminary integration model is available in \cite{CVBDFTTC_CVIntegration}.

\subsubsection{Ventilation Systems}
\label{Sec:TC:CV_Design_Vent}

\subsubsubsection{General Description}
\label{Sec:TC:CV_Design_VentDescr}

\begin{figure}[htbp]
\centering %
\includegraphics[width=0.8\linewidth]{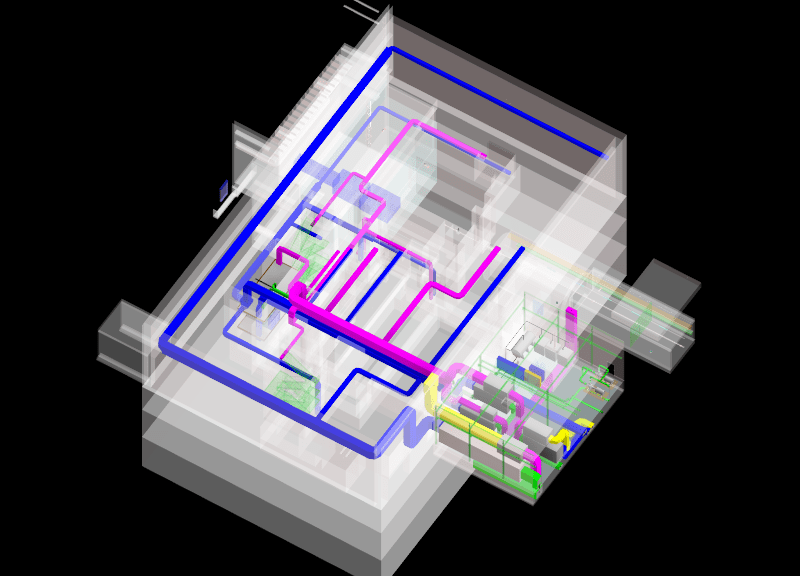}
\caption{\label{Fig:TC:CV_DesVent_010} Overview of the ventilation ducts in the target complex building. Blue ductwork represents supply lines, purple ductwork represents extraction lines, yellow ductwork represents recirculation lines.}
\end{figure}

\begin{figure}[htbp]
\centering %
\includegraphics[width=0.8\linewidth]{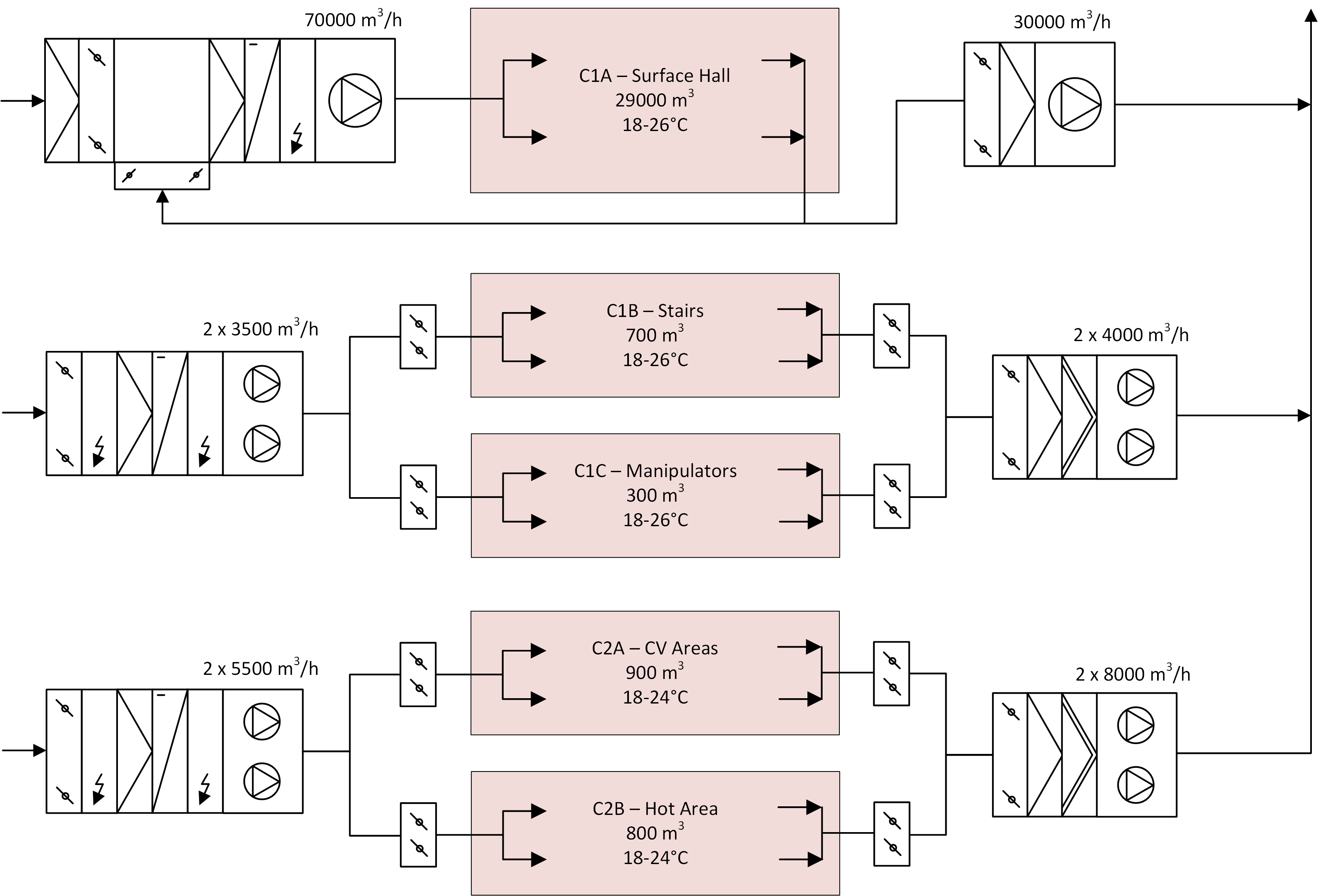}
\caption{\label{Fig:TC:CV_DesVent_011} Ventilation system schematic for the target complex.}
\end{figure}

\begin{figure}[htbp]
\centering %
\includegraphics[width=0.8\linewidth]{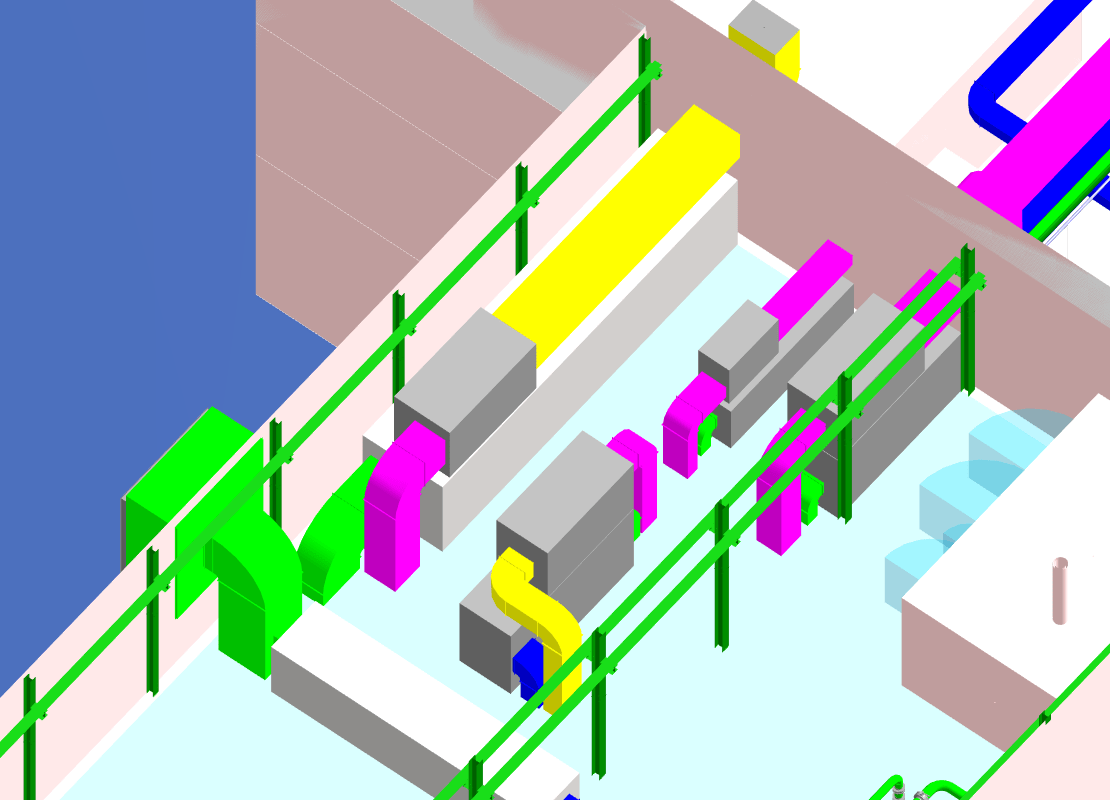}
\caption{\label{Fig:TC:CV_DesVent_012} Layout of ventilation units in a dedicated area of the auxiliary building, including all ventilation units for the target complex, the auxiliary building and the extraction tunnel.}
\end{figure}

The design approach for the ventilation system is based on the radiation protection and thermal requirements, presented in Section~\ref{Sec:TC:CV_UR_Vent}, and the sizing of the ventilation flow rates is performed in such a way that adequate dynamic confinement and heat removal can be provided. The number of ventilation units is defined based on the number of radiation protection classes involved in the building design (C1/C2) and the position of the room to be ventilated (surface or underground); the following three units are needed for the target complex ventilation:
\begin{itemize}
\item Surface hall ventilation (C1, on surface): ventilation of surface hall;
\item Access area ventilation (C1, underground): ventilation of staircases and manipulators rooms;
\item Hot area ventilation (C2, underground): ventilation of CV room, trolley and helium vessel areas.
\end{itemize}
Figure \ref{Fig:TC:CV_DesVent_011} provides a schematic of the ventilation units and their characteristics; Figure \ref{Fig:TC:CV_DesVent_010} shows a preliminary integration of the ducts within the target complex. All AHUs for the BDF target complex are located in the auxiliary building, as shown in Figure \ref{Fig:TC:CV_DesVent_012}.

A pressure cascade is defined according to the radiation protection requirements listed in Section~\ref{Sec:TC:CV_UR_Vent}. In order to maintain the underpressure in each room, it is necessary to have an extraction flow rate that compensates the air leaks into the room coming from higher pressure levels and from the outside environment. For sizing purposes, it has been assumed that the permeability (at 50 Pa) through the walls (both internal and external) of the target complex is equal to 5 m\textsuperscript{3}/h/m\textsuperscript{2}; the actual value of permeability is determined by several factors, mainly the civil engineering design, and will be estimated with more precision in a detailed design phase. In order to be conservative, outgoing leakage (on interfaces where outside pressure is lower than inside pressure) has been neglected. The incoming leakage flow rate has been calculated for each room in the target complex from the permeability, the pressure distribution determined by the pressure cascade and the surface of the walls; the resulting incoming leakage flow rates are listed in Table \ref{Tab:TC:CV_DesVent_MinRPflow}. The leakage flow is the minimum air flow rate that needs to be extracted from each volume in order to maintain the underpressure.

\begin{table}[htbp]
\centering
\caption{\label{Tab:TC:CV_DesVent_MinRPflow} Minimum RP flow and leakage flow for each RP subclass.}
\smallskip
\begin{tabular}{l|c|>{\centering\arraybackslash}m{2.5cm}|>{\centering\arraybackslash}m{2.3cm}|>{\centering\arraybackslash}m{2.3cm}|>{\centering\arraybackslash}m{2.5cm}}
\hline
\textbf{RP subclass} & \textbf{AHU} & \textbf{Underpressure [Pa]} & \textbf{Min RP flow [m\textsuperscript{3}/h]} & \textbf{Leakage flow [m\textsuperscript{3}/h]} & \textbf{Leakage / Min RP flow [\%]} \\
\hline
C1A	& Surface hall & -20 & 28786 & 13884 & 48 \\
C1B	& Access area & -40	& 1236 & 163 & 13 \\
C1C	& Access area & -60	& 1166 & 206 & 18 \\
C2A	& Hot area	& -80 & 1803 & 644 & 36 \\
C2B	& Hot area	& -100 & 3731 & 1770 & 47 \\
\hline
\end{tabular}
\end{table}

The underpressure set-point for the surface hall is not large (-20 Pa), but, being the surface of the walls large, the leakage flow rate is consistent with respect to the supply flow. The hot area also has a consistent leakage flow rate, caused instead by the stronger depressurization value (-80/-100 Pa). 

Table~\ref{Tab:TC:CV_DesVent_MinRPflow2} summarizes the minimum flow rates for each ventilation system deriving from RP requirement and dynamic confinement considerations.

\begin{table}[htbp]
\centering
\caption{\label{Tab:TC:CV_DesVent_MinRPflow2} Minimum supply and extraction flow based on RP requirements.}
\smallskip
\begin{tabular}{l|>{\centering\arraybackslash}m{4cm}|>{\centering\arraybackslash}m{5cm}}
\hline
\textbf{AHU} & \textbf{Min supply flow [m\textsuperscript{3}/h]} & \textbf{Min extraction flow [m\textsuperscript{3}/h]} \\
\hline
\textbf{Surface hall} & 28786 & 13884 \\
\textbf{Access area} & 2403 & 369 \\
\textbf{Hot area} & 5534 & 2413 \\
\hline
\end{tabular}
\end{table}

RP preliminary requirements are used as starting point for the thermal calculations needed to size the ventilation system. The outdoor design conditions for the Geneva region, defined by the standards and used for the sizing of the air handling units are presented in Table \ref{Tab:TC:CV_DesVent_Outdoor}.

\begin{table}[htbp]
\centering
\caption{\label{Tab:TC:CV_DesVent_Outdoor} Reference outdoor conditions for design of ventilation systems in winter and summer in the Geneva area. Actual conditions may be out of this range for a limited time during the year.}
\smallskip
\begin{tabular}{l|c|c}
\hline
\textbf{Season} & \textbf{Winter} & \textbf{Summer} \\
\hline
\textbf{Dry Bulb Temperature} & -11\textsuperscript{\degree}C & 32\textsuperscript{\degree}C \\
\textbf{Relative Humidity} & 90\% & 40\% \\
\textbf{Soil Temperature} & 13\textsuperscript{\degree}C & 17\textsuperscript{\degree}C \\
\hline
\end{tabular}
\end{table}

These values are to be understood as mean value maxima (95\% percentile) for calculation and sizing purpose only. Values outside the range can be observed for short periods of time.

The ventilation units are sized based on the worst cooling and heating conditions in winter and summer; the heat loads for the two conditions are calculated from the temperature requirements (Table \ref{Tab:TC:CV_URTH1}), the internal heat dissipation (Table \ref{Tab:TC:CV_URLoads}) and the outdoor conditions (Table \ref{Tab:TC:CV_DesVent_Outdoor}). Table \ref{Tab:TC:CV_DesVent_Loads} shows the total heat loads in winter and summer for each ventilation area.

The ventilation units sizing based on these heat loads is described in Section~\ref{Sec:TC:CV_Design_VentSurfHall} (Surface hall), Section~\ref{Sec:TC:CV_Design_VentAccessArea} (Access area) and Section~\ref{Sec:TC:CV_Design_VentHotArea} (Hot area).

\begin{table}[htbp]
\centering
\caption{\label{Tab:TC:CV_DesVent_Loads} Overall (internal+external) heat loads for ventilation areas.}
\smallskip
\begin{tabular}{l|>{\centering\arraybackslash}m{2.5cm}|>{\centering\arraybackslash}m{2.7cm}}
\hline
\textbf{Ventilation area} & \textbf{Min heat load (Winter) [kW]} & \textbf{Max heat load (Summer) [kW]} \\
\hline
\textbf{Surface hall} & -158 & 62 \\
\textbf{Access area} & -10 & -8 \\
\textbf{Hot area} & -10 & 7 \\
\hline
\end{tabular}
\end{table}

\subsubsubsection{Ventilation P\&ID and Synoptic}
\label{Sec:TC:CV_Design_VentPID}

Figure \ref{Fig:TC:CV_DesVent_011} shows a schematic of the ventilation system for the target complex. The ventilation of the target complex is provided by three ventilation units. These units share a common fresh air inlet and discharge exhaust air to a shared stack, located in proximity of the auxiliary building.

The surface ventilation unit takes advantage of recirculation in order to minimize the energy loss. The underground units instead use 100\% fresh air, which helps to remove activated components from the ventilation volumes. 
The two ventilation units for the underground areas have two pressure levels each; automatic dampers will be used to regulate underpressure of different rooms within each ventilation line.

The following documentation on the BDF ventilation has been developed:
\begin{itemize}
\item Preliminary ventilation synoptic \cite{CVBDFTTC_VentSyn};
\item Preliminary P\&ID for each ventilation unit \cite{CVBDFTTC_VentPID};
\item Preliminary 3D integration \cite{CVBDFTTC_CVIntegration};
\end{itemize}

\subsubsubsection{Surface Hall Ventilation}
\label{Sec:TC:CV_Design_VentSurfHall}

The preliminary P\&ID for the surface hall ventilation unit is provided in \cite{CVBDFTTC_VentPID}.
Supply and extraction units are located in the BDF auxiliary building and supply/extract air from the surface hall via ductwork. Air is supplied to the surface hall via distribution ducts, whereas the extraction is localized at a single extraction grille in proximity of the wall that separates the hall from the auxiliary building.
Since the volume and the heat loads are large, recirculation is used to minimize energy loss; a minimum amount of fresh air is supplied to maintain optimal conditions.
In worst winter conditions, 70000 m\textsuperscript{3}/h of air at 29\textsuperscript{\degree}C are supplied in order to maintain at least 18\textsuperscript{\degree}C in the hall; this compensates for both the heat loss (-158 kW) and the infiltrations due to depressurization (14000 m\textsuperscript{3}/h at -11\textsuperscript{\degree}C). The extraction unit extracts at least 14000 m\textsuperscript{3}/h in order to maintain -20 Pa inside the hall. The supply unit needs a 256 kW heating coil to treat the recirculated air.
In worst summer conditions, 40000 m\textsuperscript{3}/h of air at 17\textsuperscript{\degree}C maintain the average temperature of in the surface hall below 26\textsuperscript{\degree}C (accounting for 14000 m\textsuperscript{3}/h infiltration at 32\textsuperscript{\degree}C). The cooling coil needs to be sized for 98 kW.

\begin{figure}[htbp]
\centering %
\includegraphics[width=0.8\linewidth]{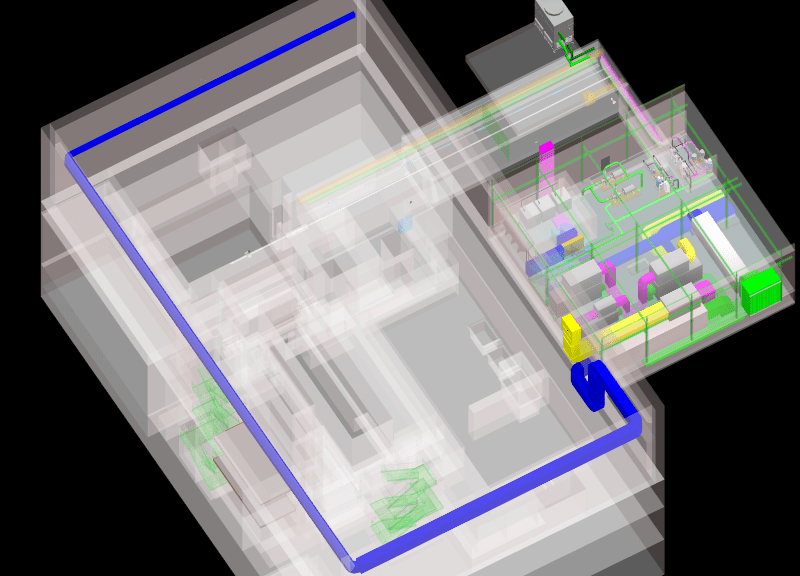}
\caption{\label{Fig:TC:CV_DesVent_020} Preliminary integration of surface hall ventilation. Supply duct is shown in blue; air is extracted at a grille in proximity of the separation wall with the auxiliary building.}
\end{figure}

The layout of the ducts is shown in Figure~\ref{Fig:TC:CV_DesVent_020} and the main parameters of the supply and extraction units are shown in Table~\ref{Tab:TC:CV_DesVent_AHUsurf}. The extraction flow rate is sized as two times the expected infiltration flow rate in order to allow for some margin for pressure regulation; this margin on the extraction flow can also be used to accelerate the air change rate in the hall and reduce the wait time before being able to access the area because of air activation.

\begin{table}[htbp]
\centering
\caption{\label{Tab:TC:CV_DesVent_AHUsurf} Surface hall ventilation unit parameters.}
\smallskip
\begin{tabular}{l|c|c}
\hline
\textbf{Parameter} & \textbf{Unit} & \textbf{Value} \\
\hline
Supply AHU flow & m\textsuperscript{3}/h & 70000 \\
Extraction AHU flow & m\textsuperscript{3}/h & 30000 \\
Supply fan power & kW & 27 \\
Extraction fan power & kW & 12 \\
Reference duct diameter & m & 1.76 \\
Preheat (30\% flow) power & kW & 100 \\
Heating coil & kW & 250 \\
Cooling coil & kW & -100 \\
\hline
\end{tabular}
\end{table}

\subsubsubsection{Access Area Ventilation}
\label{Sec:TC:CV_Design_VentAccessArea}

\begin{figure}[htbp]
\centering %
\includegraphics[width=0.8\linewidth]{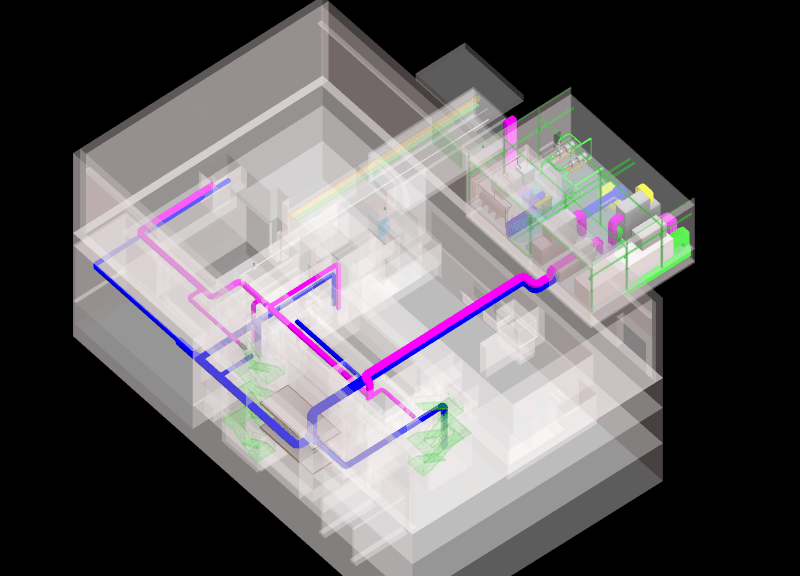}
\caption{\label{Fig:TC:CV_DesVent_030} Preliminary integration of access area ventilation. Supply lines to staircases and manipulators areas are shown in blue, whereas extraction lines are shown in purple.}
\end{figure}

The access area includes three staircases and two manipulators' areas. The preliminary P\&ID for the access area ventilation unit is provided in \cite{CVBDFTTC_VentPID}. The layout of the ducts is shown in Figure \ref{Fig:TC:CV_DesVent_030} and the main parameters of the supply and extraction units are shown in Table \ref{Tab:TC:CV_DesVent_AHUaccess}.

The supply and extraction units for this ventilation area are located in the auxiliary building. The supply and extraction flow are 3500 m\textsuperscript{3}/h and 4000 m\textsuperscript{3}/h, respectively; the infiltration leakage due to depressurization amounts to about 500 m\textsuperscript{3}/h in nominal conditions. The ventilation system does not feature recirculation, so that activated air can be immediately extracted and exhausted to the atmosphere.
In worst winter conditions, air is supplied at 27\textsuperscript{\degree}C and extracted at 19\textsuperscript{\degree}C. In worst summer conditions, since these rooms are located underground and do not have any major heat load, about 10 kW cooling capacity is needed to maintain the temperature in the rooms below 24°C. Table \ref{Tab:TC:CV_DesVent_AHUaccess} lists the heating and cooling power for the ventilation unit.

\begin{table}[htbp]
\centering
\caption{\label{Tab:TC:CV_DesVent_AHUaccess} Access area ventilation unit parameters.}
\smallskip
\begin{tabular}{l|c|c}
\hline
\textbf{Parameter} & \textbf{Unit} & \textbf{Value} \\
\hline
Supply AHU flow & m\textsuperscript{3}/h & 3500 \\
Extraction AHU flow & m\textsuperscript{3}/h & 4000 \\
Supply fan power & kW & 1.5 \\
Extraction fan power & kW & 2.0 \\
Reference duct diameter & m & 0.4 \\
Preheat (100\% flow) power & kW & 20 \\
Heating coil & kW & 30 \\
Cooling coil & kW & -10 \\
\hline
\end{tabular}
\end{table}

\subsubsubsection{Hot Area Ventilation}
\label{Sec:TC:CV_Design_VentHotArea}

\begin{figure}[htbp]
\centering %
\includegraphics[width=0.8\linewidth]{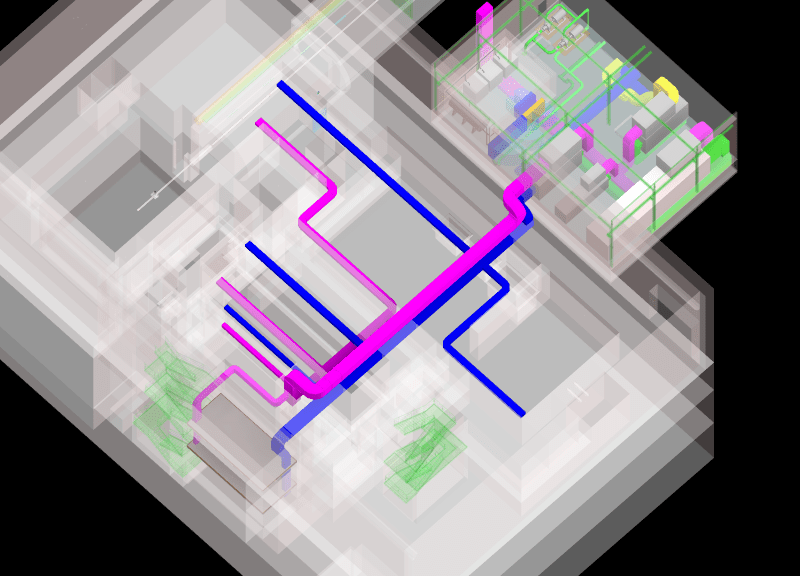}
\caption{\label{Fig:TC:CV_DesVent_040} Preliminary integration of hot area ventilation. Blue ducts are supply lines, whereas purple ducts are extraction lines.}
\end{figure}

The hot area ventilation unit supplies ventilation to the CV areas, the trolley, the helium vessel and part of the storage area. The preliminary P\&ID for the hot area ventilation unit is provided in \cite{CVBDFTTC_VentPID}. The layout of the ducts is shown in Figure \ref{Fig:TC:CV_DesVent_040} and the main parameters of the supply and extraction units are shown in Table \ref{Tab:TC:CV_DesVent_AHUhot}.

Its structure is similar to the access area ventilation. Since the depressurization is higher, the leakage rate is expected to be about 2500 m\textsuperscript{3}/h, whereas the supply are extraction flow rates are 5500 m\textsuperscript{3}/h and 8000 m\textsuperscript{3}/h, respectively. The air is not recirculated through the unit, so that activated air is directly discharged to the atmosphere.
In worst winter conditions, air is supplied at 26\textsuperscript{\degree}C and extracted at 20\textsuperscript{\degree}C, in worst summer conditions, air is supplied at 20\textsuperscript{\degree}C and extracted at 24\textsuperscript{\degree}C.

\begin{table}[htbp]
\centering
\caption{\label{Tab:TC:CV_DesVent_AHUhot} Design parameters for the hot area ventilation unit.}
\smallskip
\begin{tabular}{l|c|c}
\hline
\textbf{Parameter} & \textbf{Unit} & \textbf{Value} \\
\hline
Supply AHU flow & m\textsuperscript{3}/h & 5500 \\
Extraction AHU flow & m\textsuperscript{3}/h & 8000 \\
Supply fan power & kW & 2 \\
Extraction fan power & kW & 3 \\
Reference duct diameter & m & 0.5 \\
Preheat (100\% flow) power & kW & 30 \\
Heating coil & kW & 40 \\
Cooling coil & kW & -25 \\
\hline
\end{tabular}
\end{table}

\subsubsubsection{Smoke Extraction System}
\label{Sec:TC:CV_Design_SmokeExtr}

The smoke extraction system design for the BDF target complex needs to take into account the fact that smoke can potentially contain activated gases that shall not be released to the environment without proper treatment. Moreover, the operation of the smoke extraction system shall not compromise the pressure cascade determined by the ventilation system, so that flow of activated smoke from contaminated areas to not contaminated areas is prevented. In a fire scenario the pressure levels can be changed but the ranking of the areas within the pressure cascade needs to be maintained.

In this context, two different approaches are used for the underground areas and the surface hall.

The approach for the underground areas is based on shutting off the air supply and letting the fire burn the oxygen in the room as much as possible, potentially until the fire is chocked. In case of fire in one of the underground areas, the supply fire damper closes and the respective extraction damper and fan are regulated to maintain the under-pressure in the room. If the fire is extinguished, after cool-down and RP activation check, the smoke can be extracted by the ventilation unit. Instead, if the temperature at the top of the room exceeds 300\textsuperscript{\degree}C or the exhaust air filtration capability is lost, the extraction fire damper needs to be closed; in this situation, no air is extracted from the room and confinement is lost. In order to implement this method, all supply and extraction ducts for underground areas are provided with fire dampers, as well as flow regulation dampers. 

Regarding the surface hall, the method implemented for the underground areas cannot be considered, due to the large amount of air and oxygen in the room. Moreover, since the area is classified and its air requires filtration before being released, natural circulation via skydomes is not applicable. Forced ventilation is needed via smoke extractors; the surface of the hall is approximately 19000 m\textsuperscript{2}, requiring a nominal flow rate for the smoke extractors of 70000 m\textsuperscript{3}/h.

\subsubsection{Rising System}
\label{Sec:TC:CV_Design_Rising}

\subsubsubsection{Overview and Layout}
\label{Sec:TC:CV_Design_RisingOverview}

\begin{figure}[htbp]
\centering %
\includegraphics[width=0.7\linewidth]{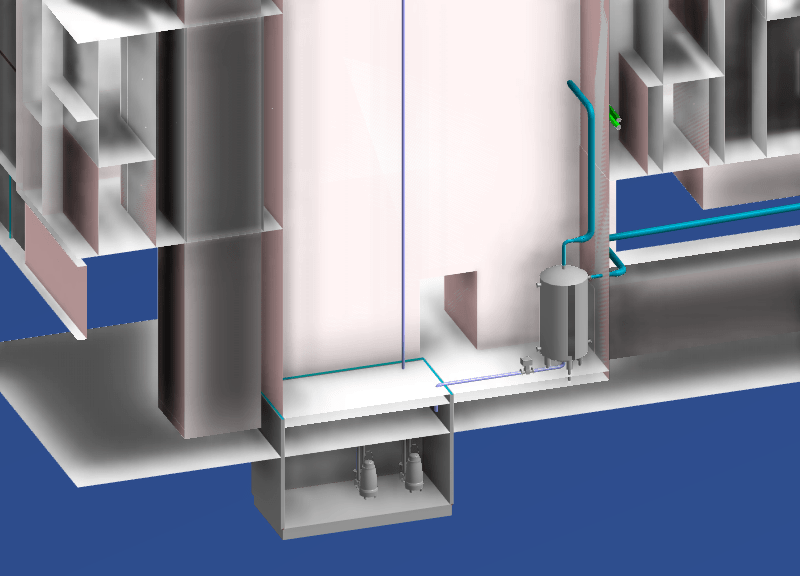}
\caption{\label{Fig:TC:CV_DesRis_010} Preliminary integration of the rising system in the sump area.}
\end{figure}

\begin{figure}[htbp]
\centering %
\includegraphics[width=0.7\linewidth]{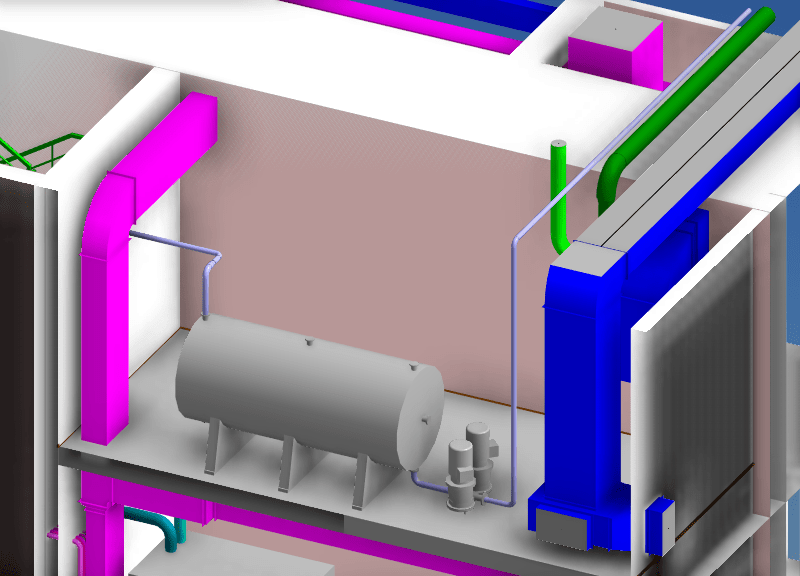}
\caption{\label{Fig:TC:CV_DesRis_011} Preliminary integration of the intermediate storage tank in the CV room.}
\end{figure}

\begin{figure}[htbp]
\centering %
\includegraphics[width=0.7\linewidth]{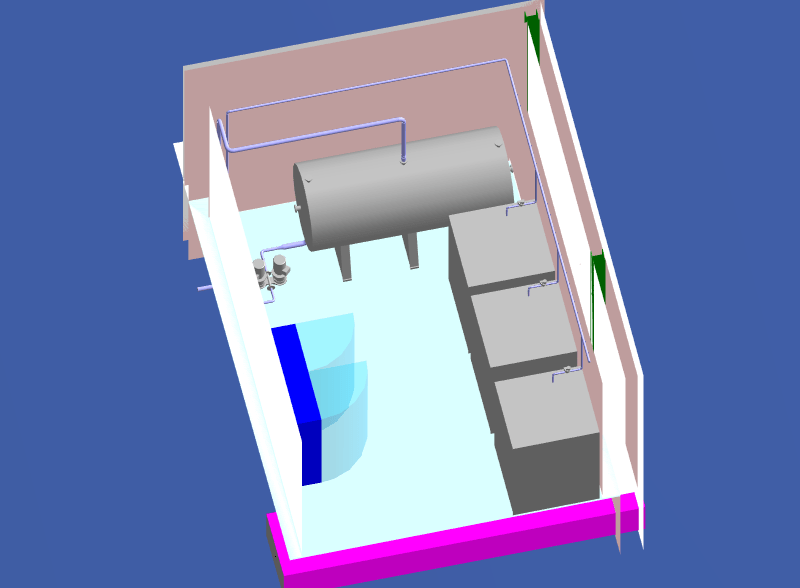}
\caption{\label{Fig:TC:CV_DesRis_012} Preliminary integration of the evaporators room, including three evaporation units.}
\end{figure}

The purpose of the rising system is to pump the drain and waste water from the sump area to the surface, to allow its disposal after its activation has decreased below the safety limits. Two sumps are foreseen to be included in the target complex, one for the trolley and helium vessel area, and another for the CV room and the rest of the building. Figure \ref{Fig:TC:CV_DesRis_010}, Figure \ref{Fig:TC:CV_DesRis_011} and Figure \ref{Fig:TC:CV_DesRis_012} show a preliminary integration for the sump system.

\subsubsubsection{Rising System Description}
\label{Sec:TC:CV_Design_RisingDescr}

The preliminary cooling synoptic \cite{CVBDFTTC_CoolSyn} and preliminary P\&ID \cite{CVBDFTTC_CoolPID} include the layout and components of the rising system.
This system is designed to collect the water leaking from the three secondary circuits and potential infiltrations.

The drainage of the leaks inside the helium vessel is done by gravity via pneumatic valve activated by a water leak detector. The detector and the valve system is located in a dedicated separation tank next to the sump tanks. The pneumatic valve opens whenever a leak is detected and discharges the water to the sump; during normal operation the valve is closed and maintains a separation between the helium and the air in the sump.  

The sump tanks are provided with a flood detector and a level detector and two redundant sump pumps that pump the water to the surface. The sump pumps shall provide a flow rate of 15 m$^3$/h and a minimum pressure head of 3 bar; this flow rate allows moving 2.5 m$^3$ of water from the sump tank to the storage tank in the evaporators room in roughly 10 minutes.

In order to allow pump motor maintenance without having to physically enter into the sump tanks, where expected dose is high, the sump pumps are provided with auto-coupling mechanism.

After a decay time that is determined by RP, the water is pumped to the surface, where it is stored in two redundant storage tanks, 5 m$^3$ each; subsequently, water is treated according to RP indications. 

Two evaporation units are foreseen in the evaporators room, of the type ENCON Drum Dryer; these units are provided with mixer, mist eliminator, power and control cubicle, and can produce a volumetric air flow of 200 m$^3$/h. In order to minimize re-condensation of water at the evaporator exhaust, the installation of a ventilation unit will be investigated, specifically dedicated to the evaporator air flow.

Since the evaporators room contains activated components and water, it is provided with shielding to reduce the dose to the outside environment.

\subsubsection{Electrical and Control System}
\label{Sec:TC:CV_Design_EL}

\subsubsubsection{General Description}
\label{Sec:TC:CV_Design_ELdescr}

The CV electrical and control system involves the installation of power and control cubicles in the target complex and auxiliary building. A preliminary sizing has been performed for the target complex components. The installation of sensors and control components in the target complex rooms needs to account for expected dose in order to prevent component failure due to radiation. The dose is expected to be low everywhere in the target complex, except in the helium vessel; the most critical area for sensors and control components, in terms of radiation dose, is the trolley area, followed by the CV room. For this reason (and for maintenance/access reasons) control components should be placed in the service area in the auxiliary building, and, only if strictly needed, in the target complex CV room.

\subsubsubsection{Electrical Cubicles Integration}
\label{Sec:TC:CV_Design_ELintegr}

\begin{figure}[htbp]
\centering %
\includegraphics[width=0.7\linewidth]{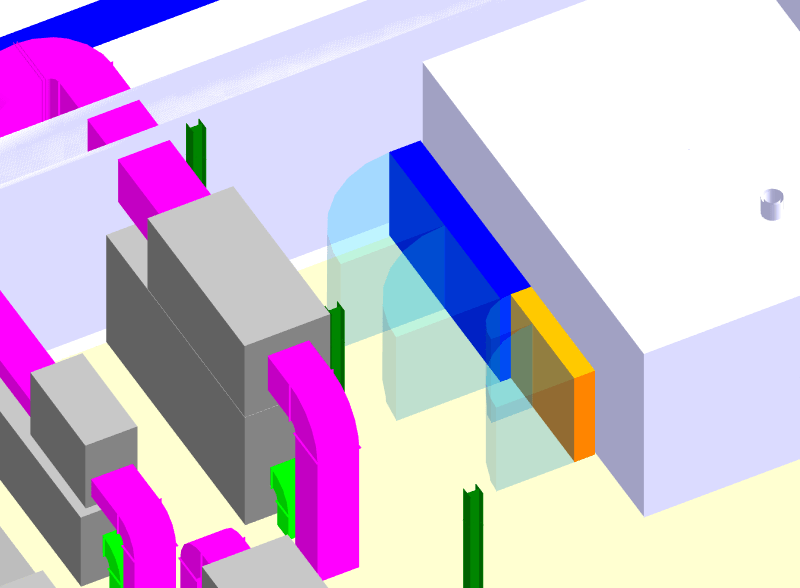}
\caption{\label{Fig:TC:CV_DesEL_010} Preliminary integration of electrical cubicles for the target complex services}
\end{figure}

Figure \ref{Fig:TC:CV_DesEL_010} illustrates a preliminary integration for the target complex CV cubicles. The installation is based on an ABB switchboard (5.8 m [W] x 2.37 m [H] x 0.6 m [D]) located in the service room; the following cubicles are also foreseen in the predesign:
\begin{itemize}
\item Cooling systems: all pumps are provided with variable frequency drive and relative cubicle (0.3 m [W] x 0.9 m [H] x 0.4 m [D]); a control cubicle is also foreseen (3.2 m [W] x 2.0 m [H] x 0.4 m [D]);
\item Ventilation systems: all supply and extraction units are provided with variable frequency drive cubicles (0.3 m [W] x 0.7 m [H] x 0.3 m [D]); all preheating and postheating coils are provided with a thyristor cubicle (0.6 m [W] x 0.6 m [H] x 0.4 m [D]); all supply-extraction unit couples are provided with a control cubicle (1.6 m [W] x 2.0 m [H] x 3.4 m [D]).
\end{itemize}

\clearpage

\section{Safety considerations for the BDF target complex}
\label{Sec:TC:Safety}

\subsection{Identification of main hazards in the target complex and risk assessment}
\subsubsection{Introduction}
\label{Sec:TC:Safety_Intro}
The objective of the hazard identification study is to identify the principal possible failure scenarios related to process control within the Target-Target Complex, and to suggest possible mitigation measures to the BDF project management. The decisions taken on the suggested mitigation measures are considered as an input for the team in charge of safety system engineering, on both underground and surface parts. This hazard identification study is focused on areas accessible to personnel.

\vfill
\pagebreak

\small
\begin{adjustbox}{angle=90}
\noindent\begin{tabu} to 1\textheight { | X[l] | X[l] | X[l] | X[l] | X[l] | X[l] | X[l] | }
\hline
    \multicolumn{7}{| c |}{\textbf{Hazard Identification Study for the Target \& Target Complex 1}} \\
\hline
    \centering\textbf{Location} & \centering\textbf{Equipment / Operations} & \centering\textbf{Hazardous Situation} & \centering\textbf{Causes} & \centering\textbf{Consequences} & \centering\textbf{Existing Barriers} & \centering\textbf{Recommendations} \\
\hline
    Surface building / Building crane SB\_BC\_Hz1 & For the crane concept, concerning the repair/replacement of target using the building crane, the sequence order of component removal from the helium vessel is as follows: 
    
        \noindent 1) He vessel lid   
    
        \noindent 2) Mobile shielding
    
        \noindent 3) Proximity shielding
    
        \noindent 4) Target
        
 & Building crane (40T) gets stuck once the floor shielding has been removed & - Failure of electrical supply / motor
 
\noindent- Mechanical obstruction

 & Radiation (a few microSv/h) in the area due to activated materials present underground \cite{BDFcomplex}
 
 & - Fence the area

\noindent- Definition of operational safety procedures

\noindent- Shielded transfer cask (iron) used for the transfer of the target from the helium vessel to cool down area 

 & - Full redundancy to be foreseen : double electrical circuit
 
\noindent- Access control system: fence and restrict the access to the building.

\noindent- Interlock of access to the area with the target exchange procedure. 

\noindent- Local evacuation system.

\noindent- CCC remote supervision and accesses management to the area. \\\hline

    Surface building / Building crane SB\_BC\_Hz2 & For the crane concept, concerning the repair/replacement of target using the building crane, the sequence order of component removal from the helium vessel is as follows: 
    
        \noindent 1) He vessel lid   
    
        \noindent 2) Mobile shielding
    
        \noindent 3) Proximity shielding
    
        \noindent 4) Target
        
 & Building crane (40T) gets stuck during operation of target transfer from the He vessel sited underground to target storage area sited on surface building 
 
 & - Failure of electrical supply / motor
 
\noindent- Mechanical obstruction

 & Radiation (a few microSv/h) in the area due to activated materials present underground \cite{BDFcomplex}
 
 & - Fence the area

\noindent- Organisational measure: RP procedure

\noindent- Target iron shielding cask foreseen and walls around

 & - Full redundancy to be foreseen : double electrical circuit

\noindent- Interlock of access to the area with the target exchange procedure. 

\noindent- Local evacuation system.

\noindent- CCC remote supervision and accesses management to the area. 

\noindent- Forbid access to the building roof \\ \hline
\end{tabu}
\end{adjustbox}

\small
\begin{adjustbox}{angle=90}
\noindent\begin{tabu} to 1\textheight { | X[l] | X[l] | X[l] | X[l] | X[l] | X[l] | X[l] | }
\hline
    \multicolumn{7}{| c |}{\textbf{Hazard Identification Study for the Target \& Target Complex 2}} \\
\hline
    \centering\textbf{Location} & \centering\textbf{Equipment / Operations} & \centering\textbf{Hazardous Situation} & \centering\textbf{Causes} & \centering\textbf{Consequences} & \centering\textbf{Existing Barriers} & \centering\textbf{Recommendations} \\
\hline
    Surface building / Floor Shielding SB\_FS\_Hz1 
    
 & Proton beam extracted from SPS and sent to the target. The shielding floor is closed. 
        
 & Radiation from target pit
 
 & Floor shielding blocks not correctly positioned. Faulty / inadequate shielding

 & Elevated radiation levels. Some 5 cm gaps already foreseen in RP simulations, with acceptable radiation levels still maintained. Much higher levels would be seen if a piece of shielding were to be missing. \cite{BDFcomplex}
 
 &- Definition of shielding verification procedure

\noindent- DSO tests before after any intervention on the floor shielding 

 & - RP monitoring system activates local alarms if threshold exceeded.
 
\noindent- Interlock RP monitors with Beam operations.

\noindent- Visual check to confirm all parts of the shielding are present shall be part of the patrol before the beam can be turned on

\\\hline

    Surface building / Electrical room SB\_ER\_Hz1 
    
 & A technical room is dedicated to the installation of all cabinets needed for the power distribution. High tension and high current devices are located in this area.  Maintenance procedures can imply the removal of several protection parts.
        
 & Exposure to high current / voltage of untrained people present in the area during co-activities. The magnetic coil is expected to be the only high voltage circuit in the area.
 
 & Electrical cabinets with door open (maintenance operation: removal of IP3X protection)

 &- Electrical contact (Skin burns)
 
\noindent- Serious Injuries

\noindent- Death

 & - IP3X protection

\noindent- French regulation/ standard 

\noindent- CERN Code C1

& - Training and certification of people who has to intervene in such electrical cabinets. Only trained and certified people has the authorisation to open an electrical cabinet door. Otherwise, keep the door closed.

\noindent- Interlock of access to electric room to the power to the magnetic coil during hazardous maintenance operation. 

\\ \hline

\end{tabu}
\end{adjustbox}

\small
\begin{adjustbox}{angle=90}
\noindent\begin{tabu} to 1\textheight { | X[l] | X[l] | X[l] | X[l] | X[l] | X[l] | X[l] | }
\hline
    \multicolumn{7}{| c |}{\textbf{Hazard Identification Study for the Target \& Target Complex 3}} \\
\hline
    \centering\textbf{Location} & \centering\textbf{Equipment / Operations} & \centering\textbf{Hazardous Situation} & \centering\textbf{Causes} & \centering\textbf{Consequences} & \centering\textbf{Existing Barriers} & \centering\textbf{Recommendations} \\
\hline
    Surface building / Roof SB\_Ro\_Hz1 
    
 & Maintenance operation on target. Opening of floor shielding.
        
 & Someone present on the roof during target exchange.
 
 &- Work planed on the roof when floor shielding is removed or during target exchange
 
 \noindent- Intrusion

 & Exposure to radiation coming from the remote handling of activated materials.
 
 & Due to the distance (15m), even if the target is exchanged the shielding is sufficient to have a supervised area.

 & Padlock ladders, controlled by IMPACT in conjunction with RP.

\\\hline

    Hot Cell area / Trolley HC\_Tro\_Hz1 
    
 & For the trolley concept, the initial operation is to remove the trolley from the He vessel by withdrawing it along rails into the hot cell and trolley services area. The Target and services are located on the trolley on rails.

 &- Trolley immovable 
 
\noindent- Trolley fails to stop

 &- Rails issue
 
\noindent- Broken wheel

\noindent- Motor failure

 & Damage to the trolley should it fail to stop

\noindent- Trolley stuck without remote handling access

 & - Easy access for visual inspection (containment between services which are personnel-accessible areas and Hot Cell area)
 
\noindent- Motor/drive chain designed to overcome the friction of seized wheel condition. Target is removed and then the wheel is replaced

 & - Compliant buffers at each end to cushion movement. Speed kept to a minimum. 

\noindent- Dual brakes can be considered / located in man accessible area

\noindent- Define clear recovery procedure.

\noindent- The trolley drive mechanism should capable of manual override, so that it can be driven into the hot cell in the event of a malfunction.

\\ \hline
\end{tabu}
\end{adjustbox}

\small
\begin{adjustbox}{angle=90}
\noindent\begin{tabu} to 1\textheight { | X[l] | X[l] | X[l] | X[l] | X[l] | X[l] | X[l] | }
\hline
    \multicolumn{7}{| c |}{\textbf{Hazard Identification Study for the Target \& Target Complex 4}} \\
\hline
    \centering\textbf{Location} & \centering\textbf{Equipment / Operations} & \centering\textbf{Hazardous Situation} & \centering\textbf{Causes} & \centering\textbf{Consequences} & \centering\textbf{Existing Barriers} & \centering\textbf{Recommendations} \\
\hline
    Hot Cell area / MSMs arms HC\_MSM\_Hz1 
    
 & The 2 MSM slave arms can be disconnected from the Hot Cell side in case of failure or required maintenance. The arms can be exported from the Hot Cell using the target export route (used to import and export equipment into Hot Cell) and a new one can be imported in a similar way and installed using the Hot Cell crane.

 & The 2 MSM arms fail at the same time (highly unlikely). MSMs become blocked. To repair them, MSMs have to be removed through the shielding wall into the operator area.
 
 &- Failure of electrical supply / motor
 
 \noindent- Mechanical obstruction

 & - Radiation in operator area coming from MSM’s arm 
 
 \noindent- Contamination of the operator zone
 
 & The MSM system has been designed to allow the extraction of the MSM without having people entering the Hot Cell. The MSM have detachable slave arms which allow them to be lifted off through-tubes that pass the mechanical drives from the master arms through the shielding wall. This lifting will be done via Hot Cell crane.

 & - Implement special access modes during this procedure.
 
 \noindent- RP procedure with decontamination before allowing access (ensuring distance between the activated components and personnel)

\\ \hline
   
    Hot Cell area / Hot Cell crane HC\_Cr\_Hz1 
    
 & The Hot Cell includes a 3 Tonne crane, which is capable of lifting the target from the trolley as well as assisting the MSMs in lifting and moving heavier items around the Hot Cell. The crane will be recoverable from failure but maintained during manned access to the Hot Cell.

 & Hot Cell crane gets blocked during the repair / replacement of the target or before starting the repair / replacement of the target
 
 & - Failure of electrical supply / motor
 
\noindent- Mechanical obstruction
 
&- During remote handling operation: increase of radiation level due to remote handling operation stuck while radioactive object is lifted (highly unlikely).

 &

 & Full redundancy to be implemented: double electrical circuit.\\ \hline

\end{tabu}
\end{adjustbox}

\small
\begin{adjustbox}{angle=90}
\noindent\begin{tabu} to 1\textheight { | X[l] | X[l] | X[l] | X[l] | X[l] | X[l] | X[l] | }
\hline
    \multicolumn{7}{| c |}{\textbf{Hazard Identification Study for the Target \& Target Complex 5}} \\
\hline
    \centering\textbf{Location} & \centering\textbf{Equipment / Operations} & \centering\textbf{Hazardous Situation} & \centering\textbf{Causes} & \centering\textbf{Consequences} & \centering\textbf{Existing Barriers} & \centering\textbf{Recommendations} \\
\hline
    Hot Cell area / Helium connections HC\_HeC\_Hz1 
    
 & Helium connections removed by MSM slave’s arms.

 & Helium leak
 
 &- Human error / bad manipulation during disconnection operation
 
 \noindent- Obsolescence of the connections

 & - Helium release (asphyxiation)
 
 &
 
 & - ODH alarm to be foreseen (ODH risk assessment)
 
  \noindent- Self rescue mask needed for people who has to intervene
  
  \noindent- Check the He connections (integrity management program)

\\\hline

   Hot Cell area / Water connections HC\_WC\_Hz1 
    
 & Water connections are un-clamped from the Grayloc connectors using the extension screw and remote handling unbolting tool (Bolt Runner Tool).

 & Leak of activated water
 
 & Bad manipulation to unclamp

 & Activated water into the sump

 & - Grayloc connector (fully metal seals, operating in the elastic domain,  allowing seal to be reused
 
 \noindent- Double walled sumps
 
 \noindent- Facility to pump fluids from the sumps via a filter to a containment area in the Target Hall, and then to an evaporator in the Service Area.

 & - During target exchange, the operator should stay inside the operator area for unclamping water connections.\\ \hline

\end{tabu}
\end{adjustbox}

\small
\begin{adjustbox}{angle=90}
\noindent\begin{tabu} to 1\textheight { | X[l] | X[l] | X[l] | X[l] | X[l] | X[l] | X[l] | }
\hline
    \multicolumn{7}{| c |}{\textbf{Hazard Identification Study for the Target \& Target Complex 6}} \\
\hline
    \centering\textbf{Location} & \centering\textbf{Equipment / Operations} & \centering\textbf{Hazardous Situation} & \centering\textbf{Causes} & \centering\textbf{Consequences} & \centering\textbf{Existing Barriers} & \centering\textbf{Recommendations} \\
\hline
    Hot Cell area / Hot Cell roof HC\_HCroof\_Hz1 
    
 & Above the crane is the Hot Cell roof which is lined to provide a containment barrier. The roof will be placed on installation and will not be removed until decommissioning, unless there are unforeseen circumstances such as replacement of the Hot Cell crane.
Once the Hot Cell has been repaired / replaced, the Hot Cell roof is put in the same position as before the repair / replacement of the Hot Cell crane. Then the beam is on. 

 & Radiation from the Hot Cell roof
 
 & Hot Cell roof not put correctly, not put exactly in the same position as before the repair / replacement of the Hot Cell crane

 & Radiation in the area
 
 & Procedural verification / DSO tests before making the beam on
 
 & - RP monitoring to detect if radioactive threshold is exceeded 
 
  \noindent- Interlock the hot cell roof with Beam Operations
  
  \noindent- Need to fence and restrict/forbid the access + DIMR\textsuperscript{1}. \\

\hline

    \multicolumn{7}{ c }{\begin{minipage}{\textheight}
    \hfill \break
    
    \textsuperscript{1}The DIMR procedure is linked to the access request (IMPACT) and concerns work in activated areas: the interventions are classified into one of three levels of concern with the aim of limiting received doses. For the higher level interventions a detailed work and dose planning is made and approved before working. This procedure will be required for interventions during technical stops and in shutdown following operation, it will also cover the times during shutdowns when shielding blocks are removed exposing activated materials.\end{minipage}} \\

\end{tabu}
\end{adjustbox}

\small
\begin{adjustbox}{angle=90}
\noindent\begin{tabu} to 1\textheight { | X[l] | X[l] | X[l] | X[l] | X[l] | X[l] | X[l] | }
\hline
    \multicolumn{7}{| c |}{\textbf{Hazard Identification Study for the Target \& Target Complex 7}} \\
\hline
    \centering\textbf{Location} & \centering\textbf{Equipment / Operations} & \centering\textbf{Hazardous Situation} & \centering\textbf{Causes} & \centering\textbf{Consequences} & \centering\textbf{Existing Barriers} & \centering\textbf{Recommendations} \\
\hline
    Hot Cell area / Electrical winch HC\_EW\_Hz1 
    
 & Electric winch

 & Winch not operational
 
 &- Failure of the electrical system
 
 \noindent- Pulleys failure

 & - Cask not open
  
 \noindent- Target remained lifted and radiation level increases (due to time of lifting longer than the one foreseen)
 
 &
 
 & - Full redundancy of the electric winch
 
  \noindent- Maintenance of the pulleys
  
  \noindent- Fire extinguisher to be foreseen in the operator area
    
  \noindent- Sprinklers in the storage area of the Lower Target Building
  
\\ \hline
   
   Trolley services area / Cartridge TS\_Car\_Hz1 
    
 & Cartridge with resin (currently foreseen to be Purolite NRW3240 as employed in n\_TOF target).Both the NFPA fire hazard rating and the HMIS Flammability rating are 1 on a scale 0 to 4 for this material.

 & Cartridge using flammable resin in contact with ignition source
 
 & External ignition source (electrical cables beneath the trolley in the false floor)

 & Fire (Highly unlikely)

 & Concrete shielding around the cartridge  
 
 \noindent- The resin is immersed in water, inside a stainless steel tank

 & Suggested extinguishing media: CO2, Dry chemical, Water fog \\ 
 
  \hline
    Underground area / Access HC\_Ac\_Hz1 
    
 & Short access periods after beam operations

 & Activation of material, water and air due to high level of radiations during beam
 
 & Long phases of beam operations

 & Minimum 4h to access the underground part only if it was possible to flush all the air
 
 & - CCC supervision
 
  \noindent- RP access authorisation procedure

 & Interlock accesses to underground area with RP monitors \\ \hline

\end{tabu}
\end{adjustbox}

\small
\begin{adjustbox}{angle=90}
\noindent\begin{tabu} to 1\textheight { | X[l] | X[l] | X[l] | X[l] | X[l] | X[l] | X[l] | }
\hline
    \multicolumn{7}{| c |}{\textbf{Hazard Identification Study for the Target \& Target Complex 8}} \\
\hline
    \centering\textbf{Location} & \centering\textbf{Equipment / Operations} & \centering\textbf{Hazardous Situation} & \centering\textbf{Causes} & \centering\textbf{Consequences} & \centering\textbf{Existing Barriers} & \centering\textbf{Recommendations} \\
\hline
    Target export tunnel / “Little” trolley TET\_Tro\_Hz1 
    
 & “Little” trolley: The target export tunnel is accessed directly from the Hot Cell and is located beneath a containment door. As for the far end of the export tunnel, there is a second containment door which provides access for the building crane to remove the target cask (in the trolley concept) from the export tunnel and move it to the cool-down area. A “little” trolley and a static drive system allow the target cask to be exported via the target export tunnel.

 & “Little” trolley gets blocked during the target export operations
 
 &- Electrical failure
 
 \noindent- Static drive system failed

 \noindent- Rails issue

 & - Impossible to export the target cask into the cool-down area

 &
 
 & - Foresee a redundant system
 
  \noindent- Integrity management program for the rails
  
  \noindent- Cask should be such that the dose rate outside is reasonably low

\\ \hline
\end{tabu}
\end{adjustbox}

\small
\begin{adjustbox}{angle=90}
\noindent\begin{tabu} to 1\textheight { | X[l] | X[l] | X[l] | X[l] | X[l] | X[l] | X[l] | }
\hline
    \multicolumn{7}{| c |}{\textbf{Hazard Identification Study for the Target \& Target Complex 9}} \\
\hline
    \centering\textbf{Location} & \centering\textbf{Equipment / Operations} & \centering\textbf{Hazardous Situation} & \centering\textbf{Causes} & \centering\textbf{Consequences} & \centering\textbf{Existing Barriers} & \centering\textbf{Recommendations} \\
\hline
   Sump pumps SB\_Pu\_Hz1 
    
 & Activated water has to be pumped out in a dedicated tank + evaporator on surface

 & Stagnation of activated water into sumps + radiation
 
 &- Deficiency of the pump 
 
 \noindent- Failure of power supply
 
 \noindent- Pump not well designed
 
 \noindent- Evaporator not well dimensioned
 
 & Radiation in the surface building hall due to activated water in case of leak or evaporator not operational. Current estimation of 1000 litres of cooling water in the Target circuit, 300 litres in the Proximity shielding circuit and 600 litres in the Magnetic coil circuit. 

 & The pumps will be located outside of the sump, so that they can be accessed in case of a malfunction.

 & Alarm level 3

 \noindent- Redundancy of the system
 
 \noindent- Monitoring with interlocks to the pumps
 
 \noindent- Fence evaporator area on surface and some shielding

 \\ \hline
    CV room / Compressor CV\_Com\_Hz1 
     
 & Compressors
   
 & Fluid leakage

 & Worn, misaligned or damaged sealing system
 
 & - Leakage of compressor fluid in the CV room could pollute the water in the drainage system
 
 \noindent - Leakage from the compressors in the exchangers to the cooling water could potentially pollute the whole cooling circuit
 
 & 

 & - Consider the implementation of an intermediate closed circuit between the compressor fluid and the water to the cooling towers
  
  \noindent - Consider the implementation of a dual-skin system, to prevent leakage, even in the event of damage
  
   \noindent - Compressors shall be oil-free 
   
  \\ \hline
\end{tabu}
\end{adjustbox}

\small
\begin{adjustbox}{angle=90}
\noindent\begin{tabu} to 1\textheight { | X[l] | X[l] | X[l] | X[l] | X[l] | X[l] | X[l] | }
\hline
    \multicolumn{7}{| c |}{\textbf{Hazard Identification Study for the Target \& Target Complex 10}} \\
\hline
    \centering\textbf{Location} & \centering\textbf{Equipment / Operations} & \centering\textbf{Hazardous Situation} & \centering\textbf{Causes} & \centering\textbf{Consequences} & \centering\textbf{Existing Barriers} & \centering\textbf{Recommendations} \\
\hline
    CV room / Compressor CV\_Com\_Hz2 
     
 & Compressors
   
 & Noise

 & Compressors in operation
 
 & Pain for the ears / Loss of hearing
 
 & The compressors' noise levels shall be below the regulation limits by design

 & Ear plugs / system to protect ears; these, along with the appropriate signage shall be readily available for use. 
 
  \\ \hline

    CV room / Motors CV\_Mot\_Hz1 
    
 & Motors

 & Hot surface, which can reach more than 80°C.
 
 & Ineffective or missing thermal insulation

 & Possible burns through skin contact

 & The pump motor has thermal protection, does not have thermal insulation to prevent contact with surface
 
 & - Cage-off the hazard to prevent accidental contact
 
  \noindent- Appropriate signage indicating the “Hot Surface” hazard must be in place
  
  \noindent- Training and certification of people who need to intervene

\\ \hline

  CV room / Cooling pipes CV\_Pip\_Hz1 
    
 & The cooling pipes for the Target, Proximity Shielding and Magnetic Coil circuits

 & Leakage of cooling water
 
 & Loss of integrity of the cooling pipes
  
 & Flooding
 
 \noindent- Activated water outside of the cooling circuits' piping
 
 & Ventilation system already foreseen, should there be any vapour from leaked water

 & Alarm /detector in case of water cooling leak

 \noindent- Automatic shut-off of the beam on leak detection
 
 \noindent- Retention tank for activated water
 
 \noindent- Integrity management program: 100\% radiographic welding tests, thickness measures of pipe, pressure tests 

 \\ \hline
\end{tabu}
\end{adjustbox}

\small
\begin{adjustbox}{angle=90}
\noindent\begin{tabu} to 1\textheight { | X[l] | X[l] | X[l] | X[l] | X[l] | X[l] | X[l] | }
\hline
    \multicolumn{7}{| c |}{\textbf{Hazard Identification Study for the Target \& Target Complex 11}} \\
\hline
    \centering\textbf{Location} & \centering\textbf{Equipment / Operations} & \centering\textbf{Hazardous Situation} & \centering\textbf{Causes} & \centering\textbf{Consequences} & \centering\textbf{Existing Barriers} & \centering\textbf{Recommendations} \\
\hline
    CV room / Cryogenic CV\_Cryo\_Hz1 
    
 & Cryogenic equipment for the purification system

 & Cold surfaces
 
 & Ineffective or missing thermal insulation

 & Possible cold burns through skin contact

 & 
 
 & - Training and certification of people who need to intervene
 
  \noindent- Appropriate signage indicating the hazard must be in place
  
  \noindent- Self rescue masks to be foreseen

\\ \hline
  
  CV room / Filters CV\_Fil\_Hz1 
    
 & Filters: 2 filters at the exhaust, 1 medium and 1 HEPA (according to EN 799:2012). Ventilation rate: 5000 m3/h

 & Release of contaminated air into the atmosphere
 
 & Filters not well dimensioned
  
 & Air contaminated

 &

 &- Maintenance plan for the filters has to be defined  

 \noindent- Environmental monitoring must be put in place
 
 \\ \hline
 
    CV room / Electrical Cabinets CV\_EC\_Hz1 
    
 & Electrical cabinets will be located in the CV room

 & High current and voltage exposure
 
 & Electrical cabinets with door open (maintenance operation: removal of IP3X protection)

 & Electrical shocks and burns
 
 & - Design of the cabinets to the French standards, and CERN code C1
 
  \noindent- IP3X protection

 & Restricted access to the CV room to CV operators and authorised workers

\\ \hline
   
   CV Room / Cartridge CV\_Resi\_Hz1 
    
 & Cartridge with resin (currently foreseen to be Purolite NRW3240).Both the NFPA fire hazard rating and the HMIS Flammability rating are 1 on a scale 0 to 4 for this material.

 & Cartridge using flammable resin in contact with ignition source
 
 & External ignition source (electrical cables)

 & Fire (Highly unlikely)

 & 

 & Suggested extinguishing media: CO2, Dry chemical, Water fog \\ 
 
  \hline
\end{tabu}
\end{adjustbox}

\clearpage
\FloatBarrier


\section{Summary of target complex radiation protection considerations}

\label{Sec:TC:RP}

This section summarizes the radiological assessment for the design of the BDF target complex. The high intensity beam power deposited on the target poses challenges to the radiation protection in several locations. In order to reduce the effect and mitigate the impact, the radiological aspects have been carefully addressed at the design stage. The studies include expected prompt and residual dose rates in the various areas of the BDF target complex. The risk due to activated air and helium and the consequence of its release into the environment were also evaluated. Finally, studies on radioactive waste zoning were conducted.

\begin{figure}[!htb]
\begin{subfigure}[b]{0.8\textwidth}
\centering
\includegraphics[width=\textwidth]{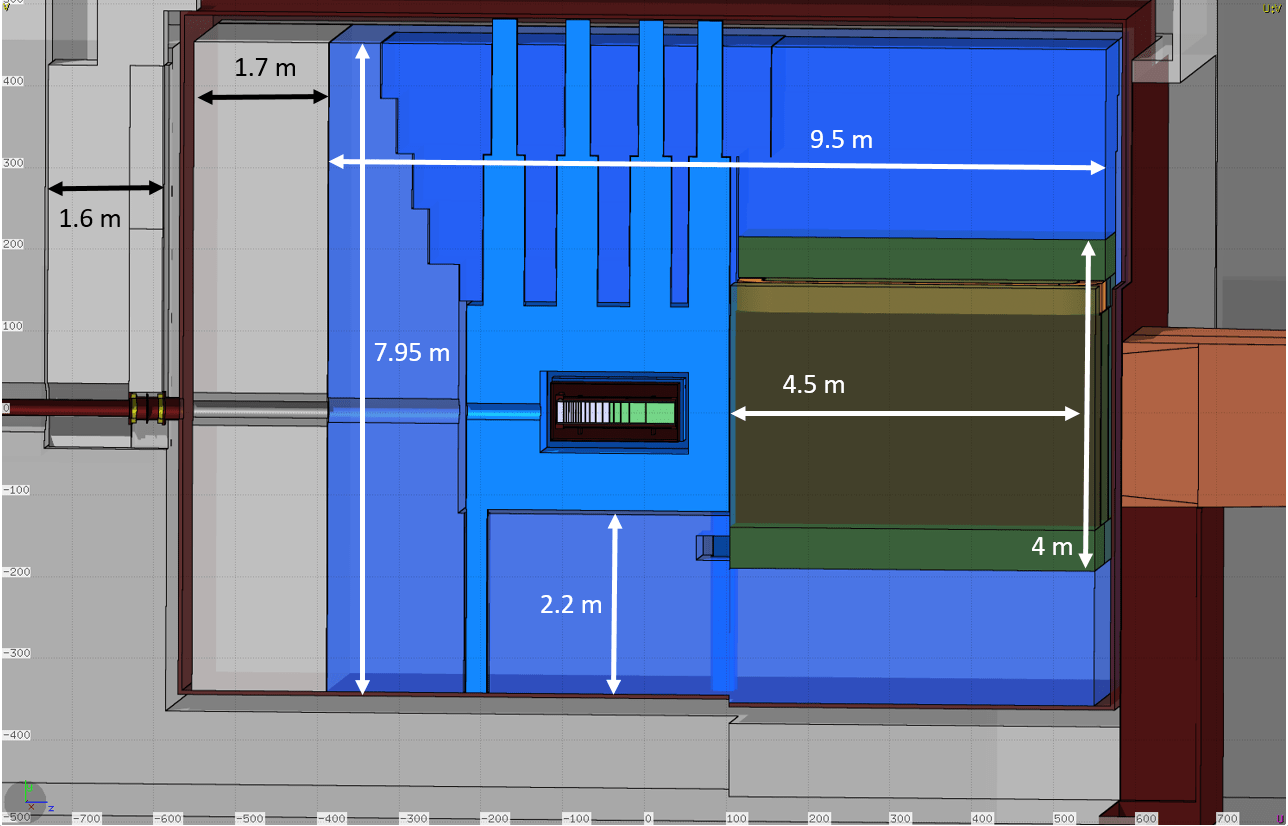}
\end{subfigure}
\begin{subfigure}[b]{0.8\textwidth}
\includegraphics[width=\textwidth]{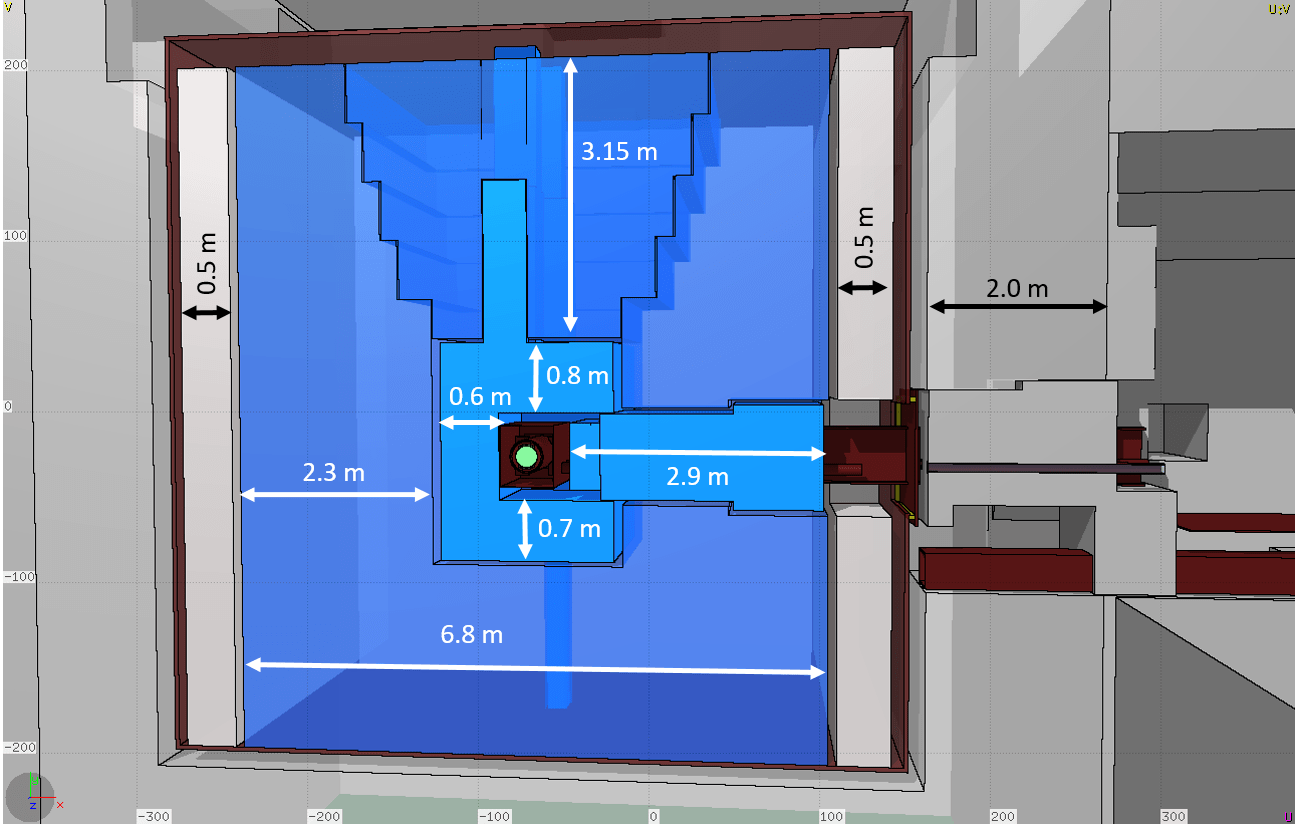}
\end{subfigure}
\caption{\small Lateral (a) and perpendicular (b) view of the FLUKA-modelled BDF target complex for the respective radiation protection calculations.}
\label{fig:ShM2}
\end{figure}

The studies are based on extensive simulations with the FLUKA Monte Carlo particle transport code~\cite{BOHLEN2014211,fluka1} and Actiwiz3~\cite{Actiwiz}. Figure \ref{fig:ShM2} shows the layout of the facility as implemented in FLUKA. All studies assume $4\times10^{13}$ protons on target per spill (duration of 7.2 s) and an integrated total of $2\times10^{20}$ protons on target over five years operation, each with 83 days of operation followed by 272 days of shutdown.

The target complex was designed under the condition that the target hall can be accessed during beam operation and classified as a Supervised Radiation Area (< 3 $\mu$Sv/h)~\cite{zonage}. On the contrary, no access during beam operation will be permitted to the underground target bunker or the experimental hall. 

In addition to personnel protection regarding prompt dose rates, considerable shielding is indispensable to reduce the residual dose rates and the environmental impact from activated air and soil as well as to relax radiation levels on electronics equipment. The shielding was consequently designed with the objective to keep the various radiological hazards originating from the operation of the BDF/SHiP facility as low as reasonably possible, while taking the constraints from the different stages of the experiment, that is the construction, operation, maintenance and dismantling, into account. The envisaged configuration is such as to avoid activation of the fixed concrete civil engineering structures simplifying not only the dismantling, but also possible changes of scope of the installation. The shielding blocks were specially designed and optimized for remote handling, since they, as well as the target, will become highly activated. The air volumes of the facility were minimized to reduce the production of airborne radioactivity. In the most critical area, that is the central region around the target and hadron absorber, the air was further replaced by a helium environment. This is motivated by the fact that pure helium gives only rise to the formation of tritium, which has a significantly lower radiological impact than the radionuclides arising from air. A pressure cascade between the various compartments of the BDF is therefore foreseen, at least during beam operation, in order to compensate for the defects of the static confinement.

\begin{figure}[!htb]
\begin{subfigure}[b]{0.5\textwidth}
\centering
\includegraphics[width=\textwidth]{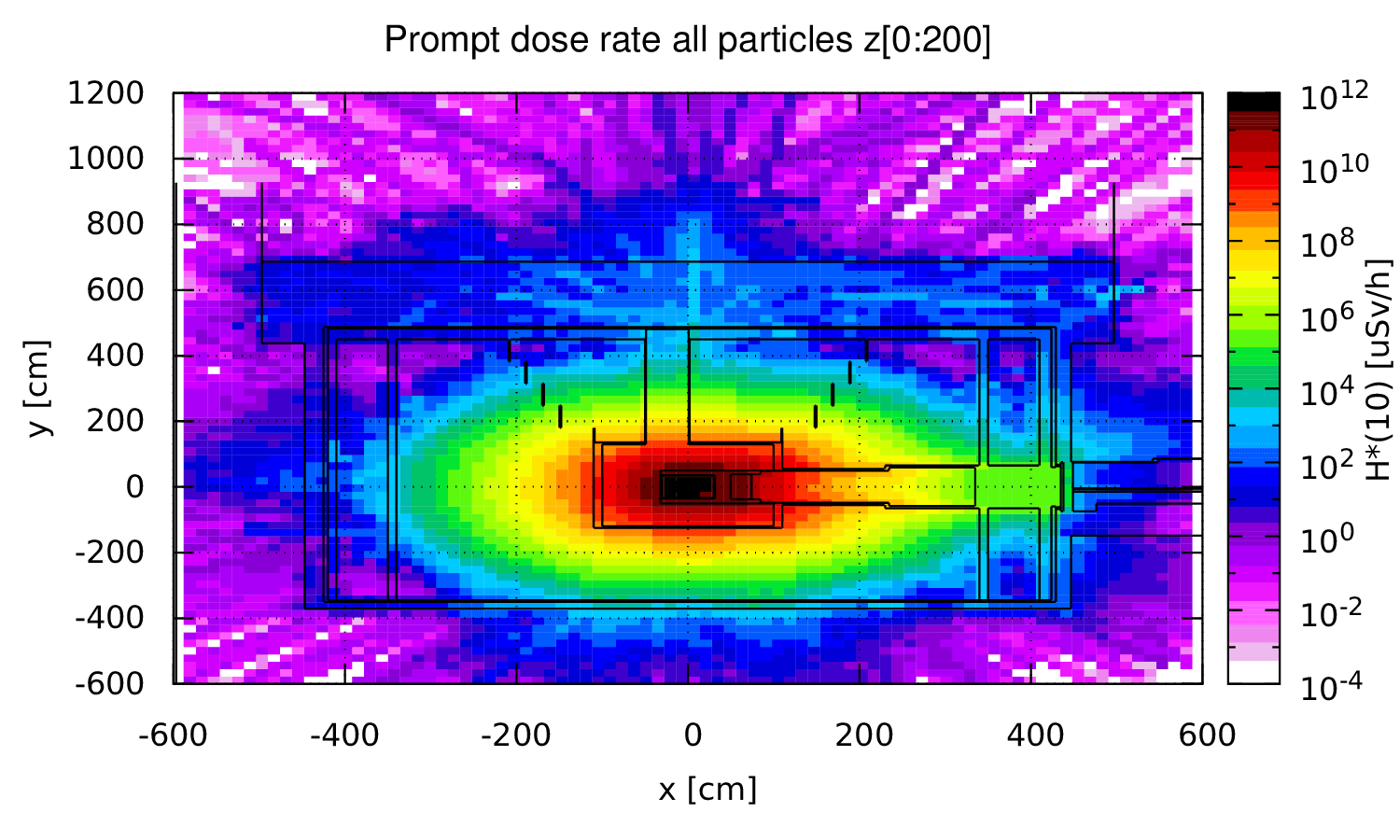}
\end{subfigure}
\begin{subfigure}[b]{0.5\textwidth}
\includegraphics[width=\textwidth]{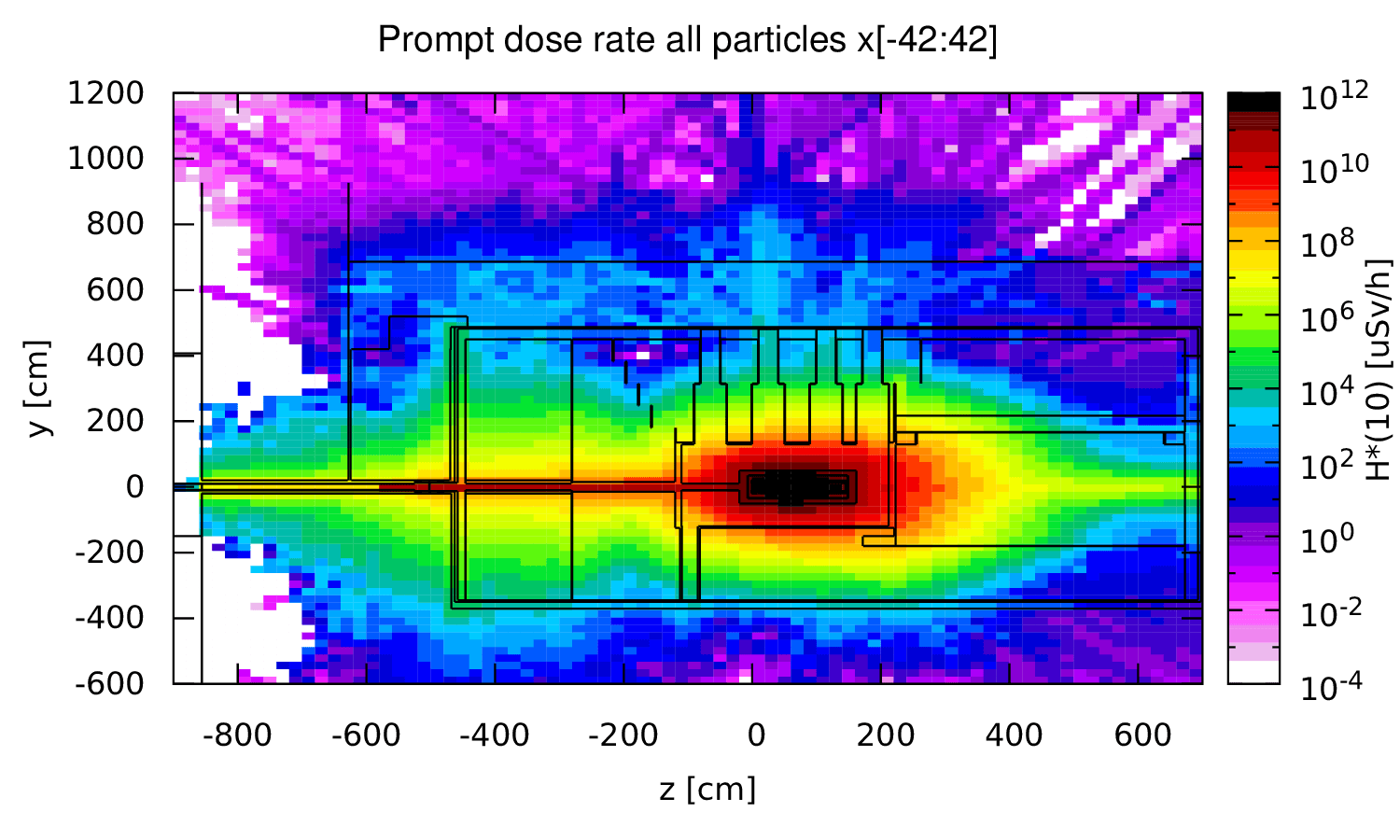}
\end{subfigure}
\caption{Prompt dose rates in $\mu$Sv/h in the BDF target complex for all particles. Left plot shows the perpendicular view at target level and the right plot shows the side view cut through the beam line.}
\label{fig:RPpDR1}
\end{figure}

The prompt dose rates in the BDF target complex are depicted in Figure~\ref{fig:RPpDR1}. As expected, the highest dose rates can be found in the region of the target reaching a few $10^{12}$ $\mu$Sv/h. They are reduced by a few orders of magnitude in the surrounding iron shielding. Above the helium vessel enclosing the shielding, the prompt dose rates amount up to 3~mSv/h. The prompt dose rates are further reduced by the above concrete shielding, such that they drop down to below a 1 $\mu$Sv/h in the target hall. 

\begin{figure}[!htb]
\begin{subfigure}[b]{0.5\textwidth}
\centering
 \includegraphics[width=\textwidth]{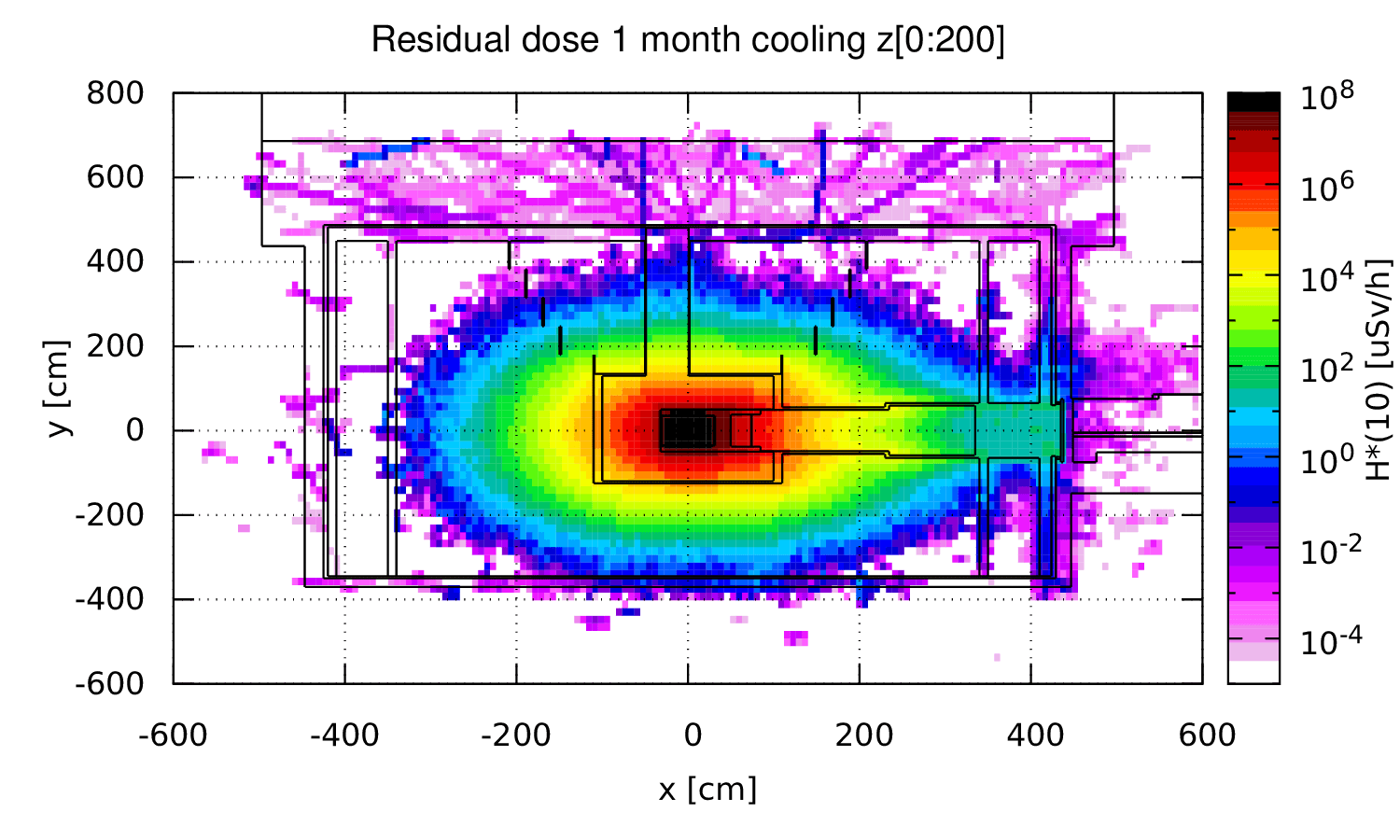}
\end{subfigure}
\begin{subfigure}[b]{0.5\textwidth}
\includegraphics[width=\textwidth]{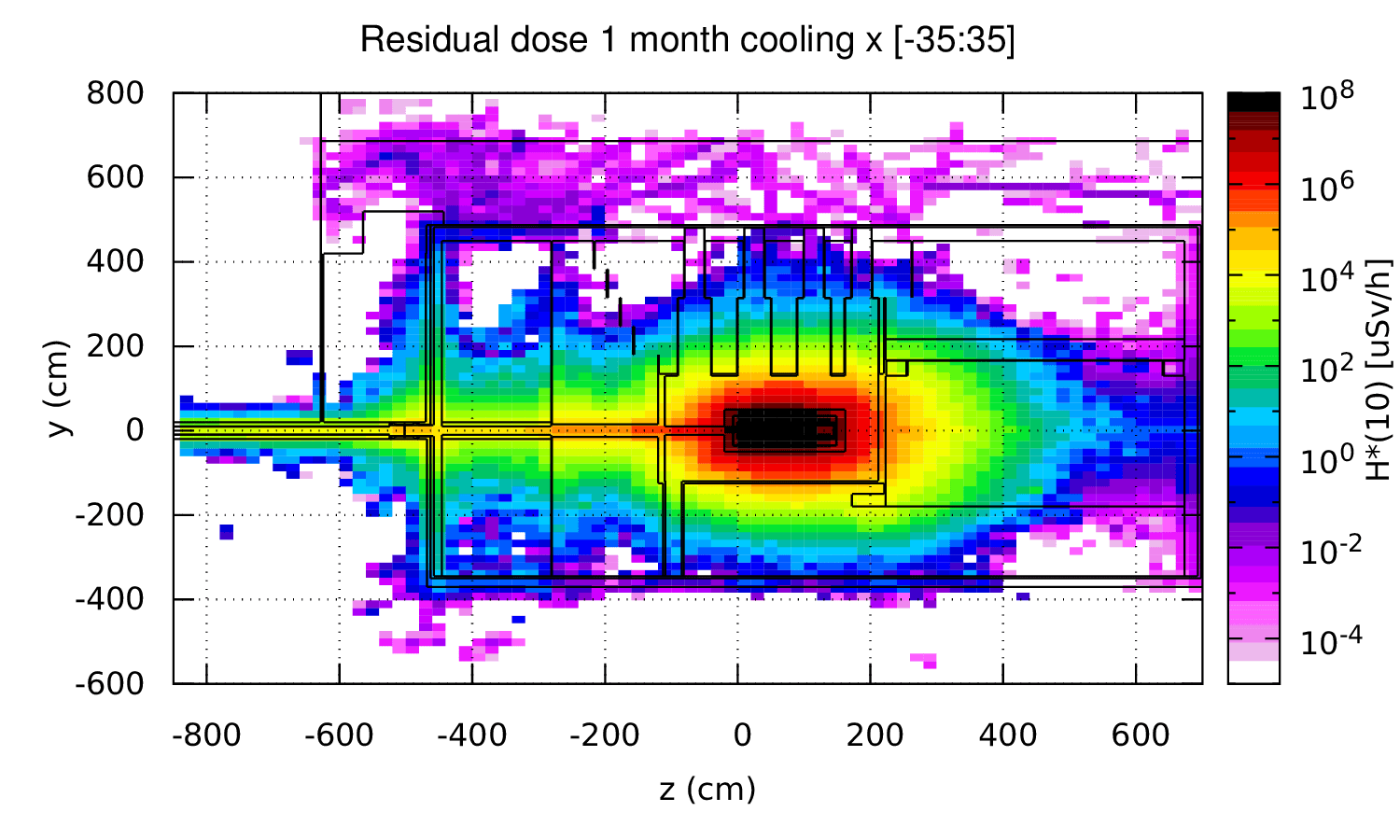}
\end{subfigure}
\caption{Perpendicular (a) and side (b) view cut through the beam line of residual dose rates in $\mu$Sv/h in the BDF target complex for 1 month cooling.}
\label{fig:RPrDR2}
\end{figure}

Figure \ref{fig:RPrDR2} show the expected residual dose rates in the BDF target complex for 1 month cooling time. The highest dose rates can be found in the  region of the target and they are in the order of a few $10^{8}$ $\mu$Sv/h after 1 month of cooling.

The air and helium activation in the target complex was evaluated assuming the maximum beam power for five operational years. When considering the accident of a helium vessel breakdown with a complete mixing of the activated air and helium, this would result in 2.7 CA\footnote{Person working 40 hours per week, 50 weeks per year with standard breathing rate in activated air with CA = 1 receives 20 mSv.} and a committed effective dose per hour of stay of 8 $\mu$Sv on top of the He vessel.  This was used to define the classification of the BDF ventilation system, for which the ISO norm 17873:200 for nuclear installations was taken as guideline for BDF. Another important aspect, which has to be taken into account from an environmental point of view, is the activation of the water of the cooling circuits.

The environmental impact of air and water releases was studied in detail. One can conclude that the maximum effective dose for airborne release is about 3.3 nSv/y. The maximum effective dose for airborne release to agriculture is only about 0.2 nSv/y. Such doses are sufficiently low so that their contribution to other sources of exposure on the site (the transfer tunnels TT20, TT26, EHN1, NA62 etc.) would be low enough to fulfill the dose objective of < 10 $\mu$Sv/y for the whole site. The annual release tritiated water would result in an effective dose of less than 50 nSv/y. This value is far below the dose constraint of 10 $\mu$Sv/y fixed for the Organization for new facilities.

\begin{figure}[!htb]
\centering
\includegraphics[width=0.9\textwidth]{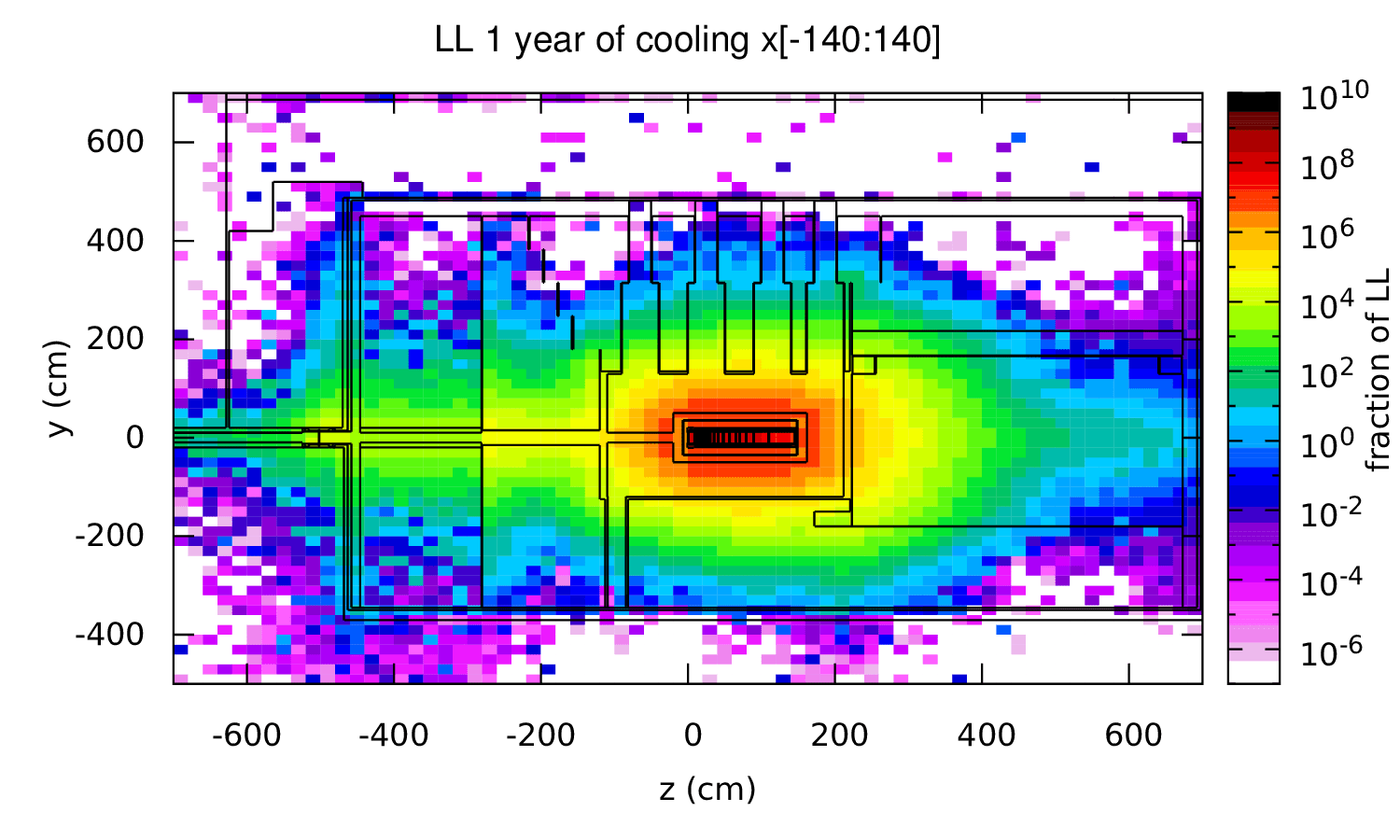}
\caption{Side view cut through the beam line of residual activity represented as fraction of Swiss LL.}
\label{fig:BDFLL}
\end{figure}

A waste study was performed to predict the amount and the characteristics of the radioactive waste that will be produced during BDF operation (see Figure~\ref{fig:BDFLL}). The objectives of such a study are to improve the management of radioactive waste and to eventually reduce the overall radioactive waste production. To distinguish areas of radioactive waste from conventional ones the liberation limits LL from Swiss legislation~\cite{ORAP} were used. Materials with a fraction of LL larger than one are radioactive according to the Swiss legislation. The zoning plots show that the most activated parts of BDF are the target and the cast iron and steel shielding elements. The radioactive waste production was minimized by making activated parts of the shielding removable such that they can easily be separated from the non-radioactive parts.

\clearpage


\printbibliography[heading=subbibliography]

 \chapter{Experimental Hall}
\label{Chap:ExpHall}

\section{Introduction}

The underground Experimental Hall (ECN4) is located immediately downstream of the target complex, centred on the beam axis, as shown in Fig.~\ref{fig:ExperimentalAreaOverview}. 
The design of the Experimental Area is mainly dictated by the requirements of the proposed SHIP experiment~\cite{2015arXiv150404956S, SHiP:2018yqc, Ahdida:2654870V2} as the first user of the Beam Dump Facility, but also taken into account are possible future extensions and reuse. 
All phases of the experiment, including assembly, construction and installation, the operational phase as well as the dismantling phase, have been taken into consideration. 
Experience from existing CERN facilities (LHCb, NA62) and other studies of future facilities (CENF, HL-LHC) have provided important input and guidelines.

\section{ General requirements }

\begin{figure}[htbp]
  \centering
  \includegraphics*[width=0.95\linewidth]{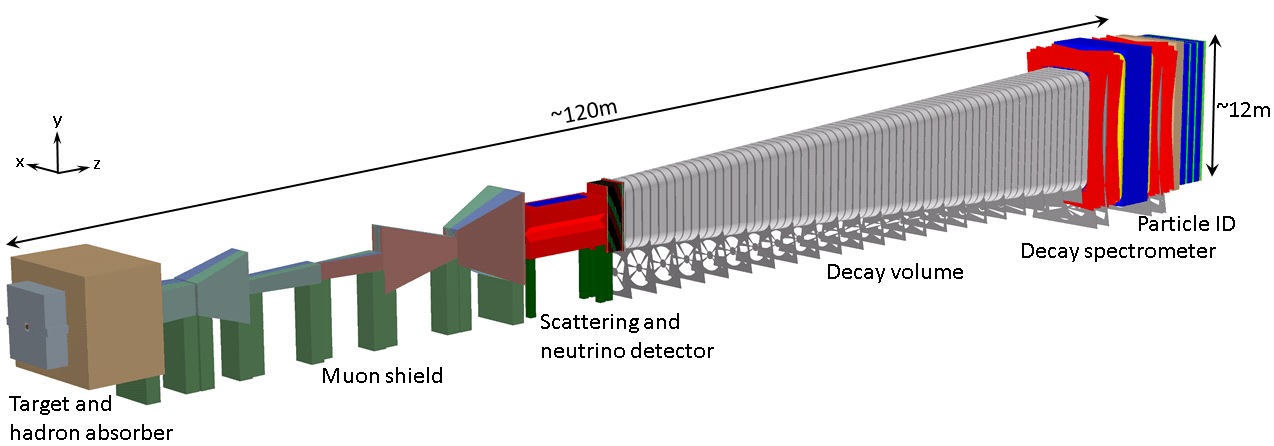}
  \caption{SHiP experimental setup as implemented in the physics simulation.}
  \label{fig:SHiP}
 \end{figure}

The Beam Dump Facility is designed to host a multi-purpose large-scale experimental program using a single beam line and a single main target station. The SHiP experimental setup (Fig.~\ref{fig:SHiP}) consists of three main components, a magnetic deflector composed of six free-standing magnets, each $\sim$\SI{5}{\meter} in length, followed by two complementary detector systems. The first detector system is the Scattering and Neutrino Detector (SND) capable of detecting light dark matter through recoil signatures of electrons or nuclei, and performing measurements on neutrino interactions, in particular on tau neutrinos.  The second detector system is designed to measure decays of Hidden Sector particles, and consists of a large decay volume followed by a spectrometer and particle identification detectors with an aperture of \SI{5x10}{m}. The decay volume is kept under a vacuum of \SI{1}{mbar} by means of a vacuum chamber, \SI{50}{\meter} in length, with transverse dimensions upstream of \SI{2.2}{\meter}$\times$\SI{5}{\meter} and downstream of \SI{6}{\meter}$\times$\SI{12}{\meter}.

The total length of the SHiP experimental setup is \SI{113}{\meter}, defining the minimum length of the underground Experimental Hall. In order to reduce background from particle scattering in the cavern walls, the width of the cavern should be minimum \SI{20}{\meter} along the entire length with the detector located in the centre. No structures or other components should be located between the cavern walls and the detector setup.

\begin{figure}[htbp]
  \centering
  \includegraphics[width=\linewidth]{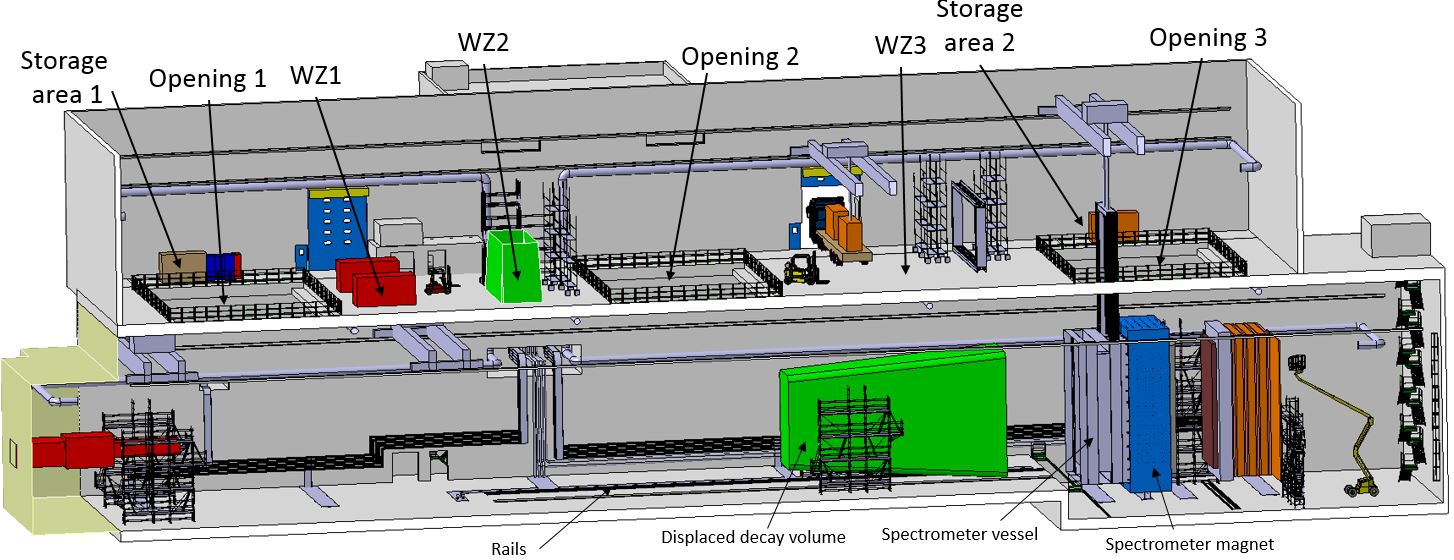}
  \caption{Preliminary organisation of the assembly work zones (WZ) in the Surface Hall for the SHIP subsystems.}
  \label{fig:SHiPassembly}
 \end{figure}

The preliminary dimensions and the loads of the different detector components are summarised in Table~\ref{tab:SHiPsubsystems}. The assembly and the installation of the muon shield, the two spectrometer magnets and the vacuum vessel are major undertakings which largely determines the layout of the Surface Hall, its access doors and the crane configuration, and drives the installation schedule. The current strategy foresees parallel construction of the muon shield magnets and the main spectrometer magnet together with its associated vacuum chamber. A large part of the assembly of both systems will take place in the underground Experimental Hall. Two overhead bridge cranes are preferred in the underground hall for these and subsequent activities, one with a single-hoist \SI{40}{\tonne} capacity and the other with a double-hoist \SI{40}{\tonne} capacity that can be combined into a joint \SI{80}{\tonne} capacity. They may be located on common rails. This phase is followed by the the construction of the decay volume starting from the larger end and the structural elements of the downstream particle identification detectors. The decay volume will arrive from the factory as pre-manufactured panels. The plan foresees pre-assembly of \SI{7.5}{\meter} long sections in the Surface Hall, with the rectangular vessel sections welded "laying down", as shown in Work Zone 2 (WZ2) in Fig.~\ref{fig:SHiPassembly}. Each ring is then rotated to its final orientation while being lowered onto the supports in the underground area. This limits the number of welding operations underground, and allows other installation activities to continue in the underground hall. Before shipping the sections underground, high-pressure water-cleaning will be required to remove dust and oil from the black steel production and welding. A similar type of cleaning will also be required after the final welding underground. Suspended protective curtains and recuperation of the waste water will be required during the cleaning process. The preliminary installation scheme considers installing the decay volume underground on a set of chariots on rails to allow longitudinal displacement of the decay volume during the final assembly and leak tests. As shown in Fig.~\ref{fig:SHiPassembly}, a gap between the decay volume (in light green) and the spectrometer section will allow works and leak tests to continue on both systems. 

The upstream spectrometer magnet and the upstream muon filters will be installed after the completion of the muon shield, the decay volume, and the spectrometer section.

\begin{table}[tpb]
\small
 \centering
 \caption{Preliminary space reservation and loads for the SHiP detector components. The "start" position of each component is relative to the centre of the proton target.}
 \label{tab:SHiPsubsystems}
 \begin{tabular}{lccccc}
  \hline 
 \multicolumn{1}{c}{\textbf{System}} & \textbf{Start [m]} & \textbf{Length [m]} & \textbf{Width [m]} & \textbf{Height [m]} & \textbf{Weight [t]} \\ \hline
 Target & -0.72 & - & - & - & - \\
 Absorber magnet & 0.82 & 5.32 & 7.90 & 6.80 & - \\
 Upstream wall & 6.10 & 0.20 & 20.00 & 0.00 & - \\
 Muon shield magnet 1 & 6.30 & 4.20 & 3.08 & 2.02 & 300.00 \\
 Muon shield magnet 2 & 10.60 & 4.10 & 2.46 & 4.60 & 330.00 \\
 Muon shield magnet 3 & 14.80 & 5.60 & 1.46 & 1.24 & 115.00 \\
 Muon shield magnet 4 & 20.60 & 5.00 & 1.44 & 1.14 & 100.00 \\
 Muon shield magnet 5 & 25.60 & 6.10 & 1.54 & 4.62 & 200.00 \\
 Muon shield magnet 6 & 31.80 & 5.00 & 3.60 & 6.36 & 550.00 \\
 Emulsion spectrometer magnet & 37.90 & 7.00 & 2.20 & 3.40 & 330.00 \\
 Upstream muon system & 45.00 & 1.60 & 2.20 & 4.90 & 85.00 \\
 Decay volume front cap & 46.60 & 0.20 & 2.20 & 5.00 & 0.40 \\
 Decay volume & 46.80 & 50.00 & 2.2-5.4 & 5.0-10.7 & 650.00 \\
 Straw Tracker tandem box 1 & 96.80 & 3.40 & 8.50 & 13.00 & 120.00 \\
 Main spectrometer magnet & 100.20 & 4.00 & 7.50 & 13.00 & 1100.00 \\
 Straw Tracker tandem box 2 & 104.20 & 3.10 & 8.50 & 13.00 & 120.00 \\
 End cap & 107.30 & 0.20 & 5.80 & 10.70 & 3.00 \\
 Timing detector & 107.55 & 0.10 & 7.00 & 12.00 & 10.00 \\
 Split ECAL & 107.70 & 0.40 & 5.30 & 10.60 & 400.00 \\
 Muon pre-filter & 108.20 & 1.10 & 6.30 & 12.60 & 700.00 \\
 Muon detector 0 & 109.40 & 0.20 & 6.00 & 12.00 & 10.00 \\
 Muon filter 0 & 109.70 & 0.60 & 6.00 & 12.00 & 400.00 \\
 Muon detector 1 & 110.40 & 0.20 & 6.00 & 12.00 & 10.00 \\
 Muon filter 1 & 110.70 & 0.60 & 6.00 & 12.00 & 400.00 \\
 Muon detector 2 & 111.40 & 0.20 & 6.00 & 12.00 & 10.00 \\
 Muon filter 2 & 111.70 & 0.60 & 6.00 & 12.00 & 400.00 \\
 Muon detector 3 & 112.40 & 0.20 & 6.00 & 12.00 & 10.00 \\
 Muon filter 3 & 112.70 & 0.10 & 6.00 & 12.00 & 60.00 \\ 
 \hline
 \end{tabular}
 \label{tab:space_reservation}
 \end{table}

With the exception of the spectrometer straw tracker stations, which are housed in the spectrometer section vacuum chamber, all other detector elements will be assembled directly in-situ in the final location. To allow access, the large detectors downstream of the spectrometer straw tracker are either installed on trolleys on lateral floor rails, or are suspended with frames on girders, in a similar manner to the LHCb experiment.

\begin{figure}[htbp]
  \centering
  \includegraphics[width=0.9\linewidth]{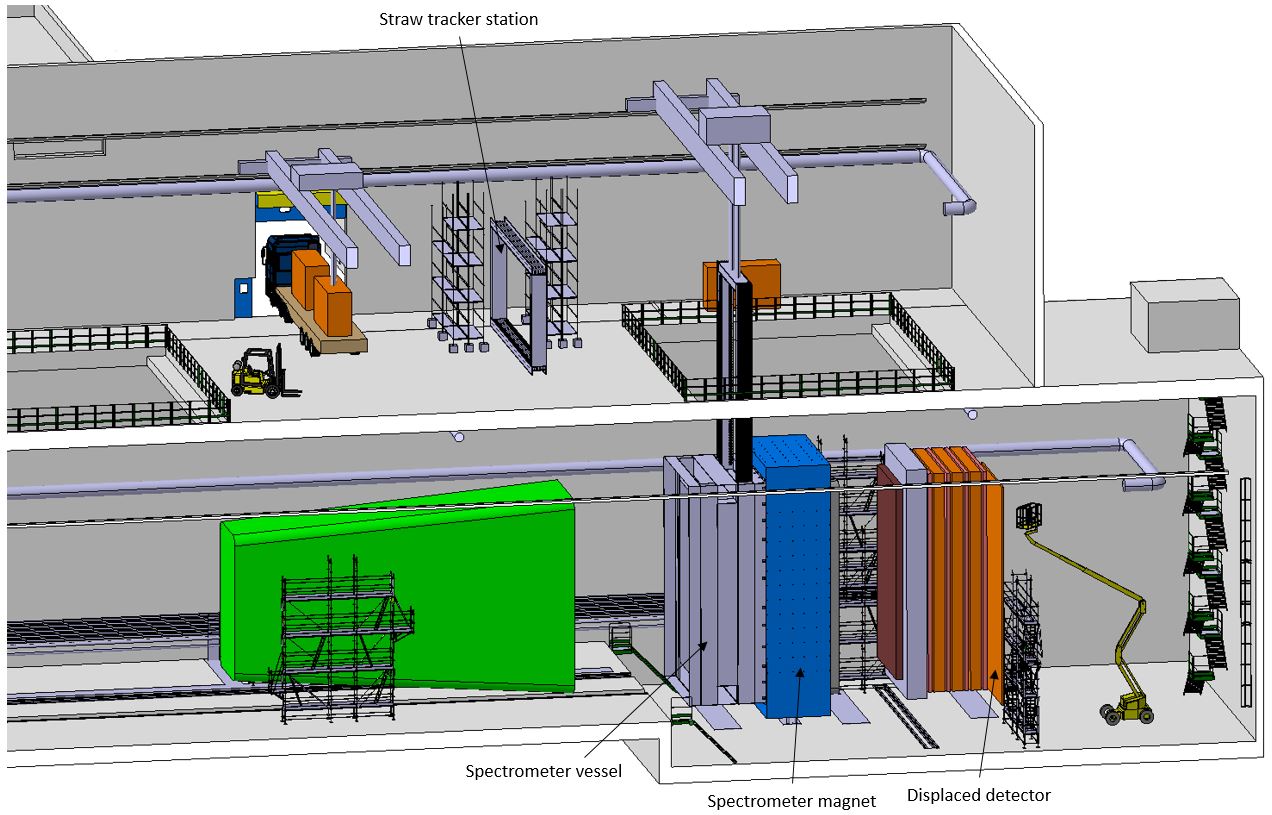}
  \caption{Illustration of the SHiP straw tracker installation.}
  \label{fig:SHiPassemblydwn}
 \end{figure}

A clean room will be installed in the Surface Hall, once the large components are installed, in order to assemble the straw tracker stations before lowering them directly into position in the spectrometer vacuum chamber. The insertion of the \SI{5}{\meter} straws will take place in the Surface Hall, and will require several rotations of the entire frames. Fig.~\ref{fig:SHiPassemblydwn} shows the installation of a straw tracker station inside the spectrometer vacuum vessel.

A preliminary zone arrangement has been made according to the SHiP assembly procedure.  Fig.~\ref{fig:SHiPassembly} shows the proposed layout of the Surface Hall with three work zones (WZ) where the SHiP components will be kept for intermediate storage and pre-assembly, with two reception areas and two storage areas. The work zones are associated with three strategically located openings down to the underground hall. Equipment and components will enter through the two doors of the Surface Building, shown in dark blue in Fig.~\ref{fig:SHiPassembly}, by a truck or semi-trailer and then moved along the hall by a movable platform, overhead cranes or forklift. The SHiP assembly activities are  distributed  over  the  three  different  work  zones  allowing  parallel operations  in  order  to  reduce  the duration of the construction and installation. Two overhead bridge cranes on separate rails, with a capacity of \SI{40}{\tonne} and \SI{10}{\tonne}, respectively, are required to support these activities. 

The muon shield and the Scattering and Neutrino Detector magnet will use Zone 1. The decay volume will alternate between Zone 2 and 3 while the main spectrometer magnet will use Zone 3.  The third zone will also be used in the second phase for intermediate storage and pre-assembly of the downstream detector systems, and finally for a clean room for the straw tracker stations.

Opening 3 is strategically located directly on top of the main spectrometer section (see Fig.~\ref{fig:SHiPassemblydwn}), allowing construction of the magnet yoke and insertion of the coils. The spectrometer vacuum chamber, consisting of two identical twin boxes, will be constructed in Zone 3 and inserted into the aperture of the magnet from both sides. The opening will then allow direct insertion of the straw tracker stations, as well as future access, with the help of the overhead bridge crane in the Surface Hall. 

Table~\ref{tab:SHiPservices} summarises the use of the work zones for the different detector components. The entire installation phase is expected to take 2~-~2.5 years.

During the operation of the facility, the three openings must be covered by concrete slabs, and detector tooling and "detector access” equipment (scaffolding, cherry pickers, etc.) should be stored in the Surface Hall.

\begin{table}[tp]
\centering
\caption{Preliminary service infrastructure requirements for SHiP, and summary of the use of the work zones for storage and assembly.}
 \label{tab:SHiPservices}
 \begin{scriptsize}
 \begin{tabular}{p{2cm}p{1cm}p{1.2cm}p{1.2cm}p{0.5cm}p{1cm}p{1.5cm}p{1.5cm}}
 \multicolumn{1}{l}{} & \multicolumn{5}{c}{Preliminary infrastructure requirements and services} & \multicolumn{2}{c}{Installation} \\ 
  \multicolumn{1}{l|}{\textbf{Subsystem}}  & \textbf{\begin{tabular}{c} Floor \\ footprint\\ (LxW) \\ \SI{}{[cm]} \end{tabular}} & \textbf{\begin{tabular}{c}Supply \\ current \\ \\ \SI{}{[A]} \end{tabular}} & \textbf{\begin{tabular}{c}Power \\ cons. \\ detector \\ \SI{}{[kW]}\end{tabular}} & \multicolumn{1}{c}{\textbf{Cooling}} & \multicolumn{1}{c|}{\textbf{\begin{tabular}{c}Power \\ dissipation \\ into air \\ \SI{}{[kW]}\end{tabular}}} & \textbf{\begin{tabular}{l} Storage/ \\ pre-assembly\end{tabular}} & \textbf{\begin{tabular}{l}Main \\ assembly\end{tabular}}  \\ 
 \hline
 \multicolumn{1}{l|}{Muon shield magnet 1} & \multicolumn{1}{c}{416x400} & \multicolumn{1}{c}{50} & \multicolumn{1}{c}{1}   & \multicolumn{1}{c}{air} & \multicolumn{1}{c|}{1} & \multicolumn{1}{c}{WZ1/ECN4} & \multicolumn{1}{c}{ECN4}   \\
 \multicolumn{1}{l|}{Muon shield magnet 2} & \multicolumn{1}{c}{414x400} & \multicolumn{1}{c}{50} & \multicolumn{1}{c}{1.5} & \multicolumn{1}{c}{air} & \multicolumn{1}{c|}{1.5} & \multicolumn{1}{c}{WZ1/ECN4} & \multicolumn{1}{c}{ECN4}  \\
 \multicolumn{1}{l|}{Muon shield magnet 3} & \multicolumn{1}{c}{562x300} & \multicolumn{1}{c}{50} & \multicolumn{1}{c}{0.9} & \multicolumn{1}{c}{air} & \multicolumn{1}{c|}{0.9} & \multicolumn{1}{c}{WZ1/ECN4} & \multicolumn{1}{c}{ECN4}  \\
 \multicolumn{1}{l|}{Muon shield magnet 4} & \multicolumn{1}{c}{496x300} & \multicolumn{1}{c}{50} & \multicolumn{1}{c}{1.4} & \multicolumn{1}{c}{air} & \multicolumn{1}{c|}{1.4} & \multicolumn{1}{c}{WZ1/ECN4} & \multicolumn{1}{c}{ECN4}   \\
 \multicolumn{1}{l|}{Muon shield magnet 5} & \multicolumn{1}{c}{610x300} & \multicolumn{1}{c}{50} & \multicolumn{1}{c}{3.5} & \multicolumn{1}{c}{air} & \multicolumn{1}{c|}{6.5} & \multicolumn{1}{c}{WZ1/ECN4} & \multicolumn{1}{c}{ECN4}   \\
 \multicolumn{1}{l|}{Muon shield magnet 6} & \multicolumn{1}{c}{484x460} & \multicolumn{1}{c}{50} & \multicolumn{1}{c}{5}   & \multicolumn{1}{c}{air} & \multicolumn{1}{c|}{5} & \multicolumn{1}{c}{WZ1/ECN4} & \multicolumn{1}{c}{ECN4}  \\
 \multicolumn{1}{l|}{SND detector magnet}  & \multicolumn{1}{c}{700x220} & \multicolumn{1}{c}{14600}& \multicolumn{1}{c}{1900} & \multicolumn{1}{c}{water}& \multicolumn{1}{c|}{100} & \multicolumn{1}{c}{WZ1/ECN4} & \multicolumn{1}{c}{ECN4}  \\
 \multicolumn{1}{l|}{}                     & \multicolumn{1}{c}{}        & \multicolumn{1}{c}{}   & \multicolumn{1}{c}{}    & \multicolumn{1}{c}{(60-80\,\SI{}{m^3/h})} & \multicolumn{1}{c|}{} & &  \\
 \multicolumn{1}{l|}{Emulsion target}      & \multicolumn{1}{c}{incl.}   & \multicolumn{1}{c}{0}  & \multicolumn{1}{c}{0}   & \multicolumn{1}{c}{air} & \multicolumn{1}{c|}{0} & \multicolumn{1}{c}{WZ1} & \multicolumn{1}{c}{ECN4}  \\
 \multicolumn{1}{l|}{Target tracker}       & \multicolumn{1}{c}{incl.}   &                        & \multicolumn{1}{c}{8}   & \multicolumn{1}{c}{air} & \multicolumn{1}{c|}{8} & \multicolumn{1}{c}{WZ2} & \multicolumn{1}{c}{ECN4}   \\
 \multicolumn{1}{l|}{Upstream muon ID}     & \multicolumn{1}{c}{160x220} &                        & \multicolumn{1}{c}{10}  & \multicolumn{1}{c}{air} & \multicolumn{1}{c|}{10} & \multicolumn{1}{c}{WZ2} & \multicolumn{1}{c}{ECN4}   \\
 \multicolumn{1}{l|}{Decay volume/vacuum}  & \multicolumn{1}{c}{5000x600}&                        & \multicolumn{1}{c}{19}  & \multicolumn{1}{c}{water}& \multicolumn{1}{c|}{2} & \multicolumn{1}{c}{outside} & \multicolumn{1}{c}{WZ2/WZ3}  \\
 \multicolumn{1}{l|}{}                     & \multicolumn{1}{c}{}        &                        & \multicolumn{1}{c}{}    & \multicolumn{1}{c}{(\SI{65}{m^3/h})} & \multicolumn{1}{c|}{} &  &    \\
  \multicolumn{1}{l|}{Surround bkg tagger}  & \multicolumn{1}{c}{incl.}   &                        & \multicolumn{1}{c}{10}  & \multicolumn{1}{c}{air} & \multicolumn{1}{c|}{10} & \multicolumn{1}{c}{WZ2} & \multicolumn{1}{c}{ECN4}  \\
 \multicolumn{1}{l|}{Main spectr.  magnet} & \multicolumn{1}{c}{550x763} & \multicolumn{1}{c}{3000}& \multicolumn{1}{c}{1083} & \multicolumn{1}{c}{n/a} & \multicolumn{1}{c|}{100} & \multicolumn{1}{c}{WZ3} & \multicolumn{1}{c}{ECN4}  \\
 \multicolumn{1}{l|}{Straw tracker}      & \multicolumn{1}{c}{300x650} &                        & \multicolumn{1}{c}{10}  & \multicolumn{1}{c}{air} & \multicolumn{1}{c|}{10} & \multicolumn{1}{c}{WZ3} & \multicolumn{1}{c}{WZ3}   \\
 \multicolumn{1}{l|}{Timing detector}      & \multicolumn{1}{c}{100x700} &                        & \multicolumn{1}{c}{10}  & \multicolumn{1}{c}{air} & \multicolumn{1}{c|}{10} & \multicolumn{1}{c}{WZ3} & \multicolumn{1}{c}{ECN4}  \\
 \multicolumn{1}{l|}{Split calorimeter}    & \multicolumn{1}{c}{250x700} &                        & \multicolumn{1}{c}{10}  & \multicolumn{1}{c}{air} & \multicolumn{1}{c|}{10} & \multicolumn{1}{c}{WZ3} & \multicolumn{1}{c}{ECN4}   \\
 \multicolumn{1}{l|}{Muon system}          & \multicolumn{1}{c}{550x800} &                        & \multicolumn{1}{c}{10}  & \multicolumn{1}{c}{air} & \multicolumn{1}{c|}{10} & \multicolumn{1}{c}{WZ3} & \multicolumn{1}{c}{ECN4}   \\  \hline
 & \multicolumn{1}{l}{} & \multicolumn{1}{l}{}  & \multicolumn{1}{c}{3093.3}   & \multicolumn{1}{c}{293.3} & \multicolumn{1}{l}{} & \multicolumn{1}{l}{} & \multicolumn{1}{l}{}
 \end{tabular}
 \end{scriptsize}
\end{table}

As the Surface Hall is not accessible during beam operation, control racks, safety and power distribution equipment, computing infrastructure, offices and a control room will be housed in an adjacent building Service Building (BAE91). Table~\ref{tab:SHiPservices} lists the service infrastructure requirements for the SHiP detector, and Table~\ref{tab:SHiPspaces} summarises the space requirements in the Service Building.  As listed in the second column, control, computing and network racks with cooling, ventilation and power distribution equipment will share the ground floor and the first floor, while labs, offices, cafeteria, control room and a conference room are housed on the second and third floors. In addition, a workshop to support the SHiP construction is foreseen. For the gas supply and storage, a dedicated Gas Building (BG91) is required. The liquid scintillator option for the Surrounding Background Tagger of SHiP will also require a platform outside for the storage tanks.

\begin{table}[htpb]
\small
\centering
 \caption{Space reservation for SHiP’s services and operational needs. SB stands for Service Building, and SH for Surface Hall.}
 \label{tab:SHiPspaces}
 \begin{tabular}{lcccc}
 \hline 
 \textbf{Description} & \textbf{Location} & 
 \textbf{\begin{tabular}[c]{@{}c@{}}Space \\ {[\SI{}{m^2}]}\end{tabular}} & 
 \textbf{\begin{tabular}[c]{@{}c@{}}Height \\ {[\SI{}{cm}]}\end{tabular}} & 
 \textbf{\begin{tabular}[c]{@{}c@{}}Load \\ {[\SI{}{kg/m^2}]}\end{tabular}} \\ \hline
 Services: Voltage transformers & Outside, adjacent to SB & 24 &  &  \\ Gas building & Separated building & 150 & 300 &  \\
 Liquid scintillator tanks (\SI{270}{\cubic\meter}) & Outside, adjacent to SB & 80 &  &  \\ 
 Services: underground ventilation & Upstream inside SH & 100 &  &  \\ 
 Services: safety & SB, ground floor, adjacent to SH & 25 & 250 & 400 \\
 Services: cooling & SB, ground floor, adjacent to SH & 100 & 250 &  \\ 
 Services: power distribution & SB, ground floor, adjacent to SH & 50 & 250 &  \\
 Services: liquid scintillator & SB, ground floor, adjacent to SH & 50 & 250 &  \\ 
 Electronics room & SB, ground floor, adjacent to SH & 80 & 250 & 400 \\
 Computing infrastructure & SB, ground floor & 80 & 250 & 400 \\
 Control room & SB, first floor & 150 & 250 & 300 \\
 Lab rooms (\SI{4x5}{\meter}) & SB, first floor & 150 & 250 & 300 \\
 Common lab & SB, first floor & 100 & 250 & 300 \\
 Workshop & SB, ground floor & 150 & 400 &  \\
 Clean room & Temporary in SH & 100 & 400 &  \\
 System management room & SB, ground floor & 50 & 250 & 300 \\
 Office space & SB, second floor & 60 & 250 &  \\
 Conference room & SB, second floor & 150 & 250 &  \\
 Kitchen, coffee area & SB, second floor & 50 & 250 &  \\
 Elevator/staircase & SB & 100 & 250 &  \\ 
 \hline 
 \end{tabular}
\end{table}

\FloatBarrier
\section{ Radiation protection  }

\begin{figure}[!htb]
  \centering
  \includegraphics[width=0.75\textwidth]{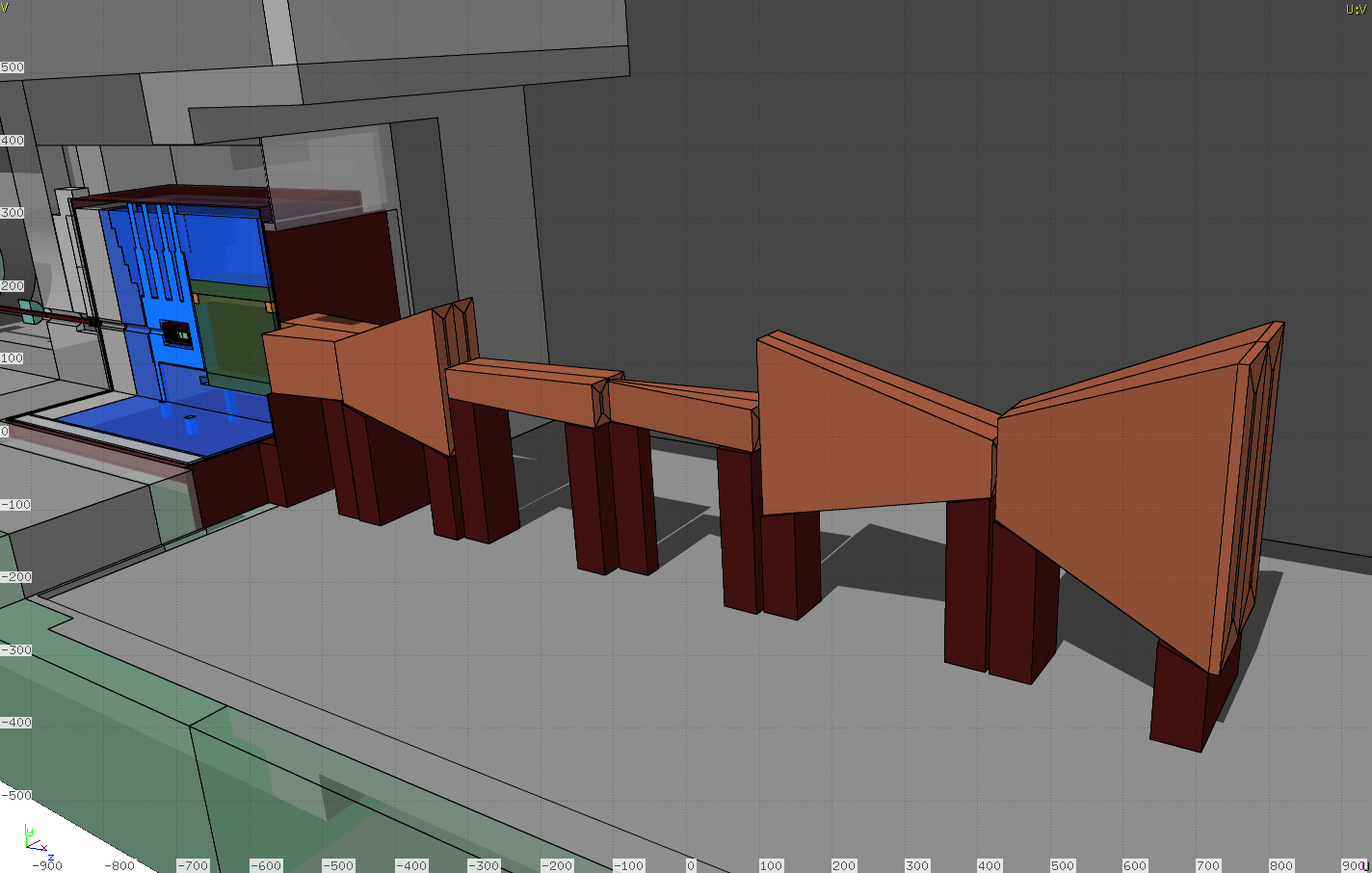}
\captionsetup{width=0.85\textwidth} \caption{\small Partial layout of the facility (target assembly and muon shield) as implemented in FLUKA.}
\label{fig:exphall-layout}
\end{figure}

The high intensity beam power deposited on the target poses challenges to the radiation protection in several locations. In order to reduce the effect and mitigate the impact, the radiological aspects of the experimental area have been carefully addressed at the design stage.
The studies include expected prompt and residual dose rates in the various areas of the SHiP experimental area and public areas. The facility has been designed under the condition that there is no access to the Experimental Hall, nor the Surface Hall on top, during beam operation.
The studies are based on past measurements and extensive simulations with the FLUKA Monte Carlo particle transport code \cite{FLUKA_Code}\cite{fluka1} and Actiwiz3 \cite{Actiwiz}. Fig.~\ref{fig:exphall-layout} shows the partial layout of the facility (target assembly and muon shield) as implemented in FLUKA.

\noindent All studies assume $4\times10^{13}$ protons on target per spill (duration of 1\,s every 7.2\,s) and an integrated total of $2\times10^{20}$ protons on target over five years operation, each with $\sim$280 days of nominal operation followed by $\sim$80 days of shutdown.

\begin{figure}[!htb]
\begin{subfigure}{0.5\textwidth}
  \centering
  \includegraphics[width=\textwidth]{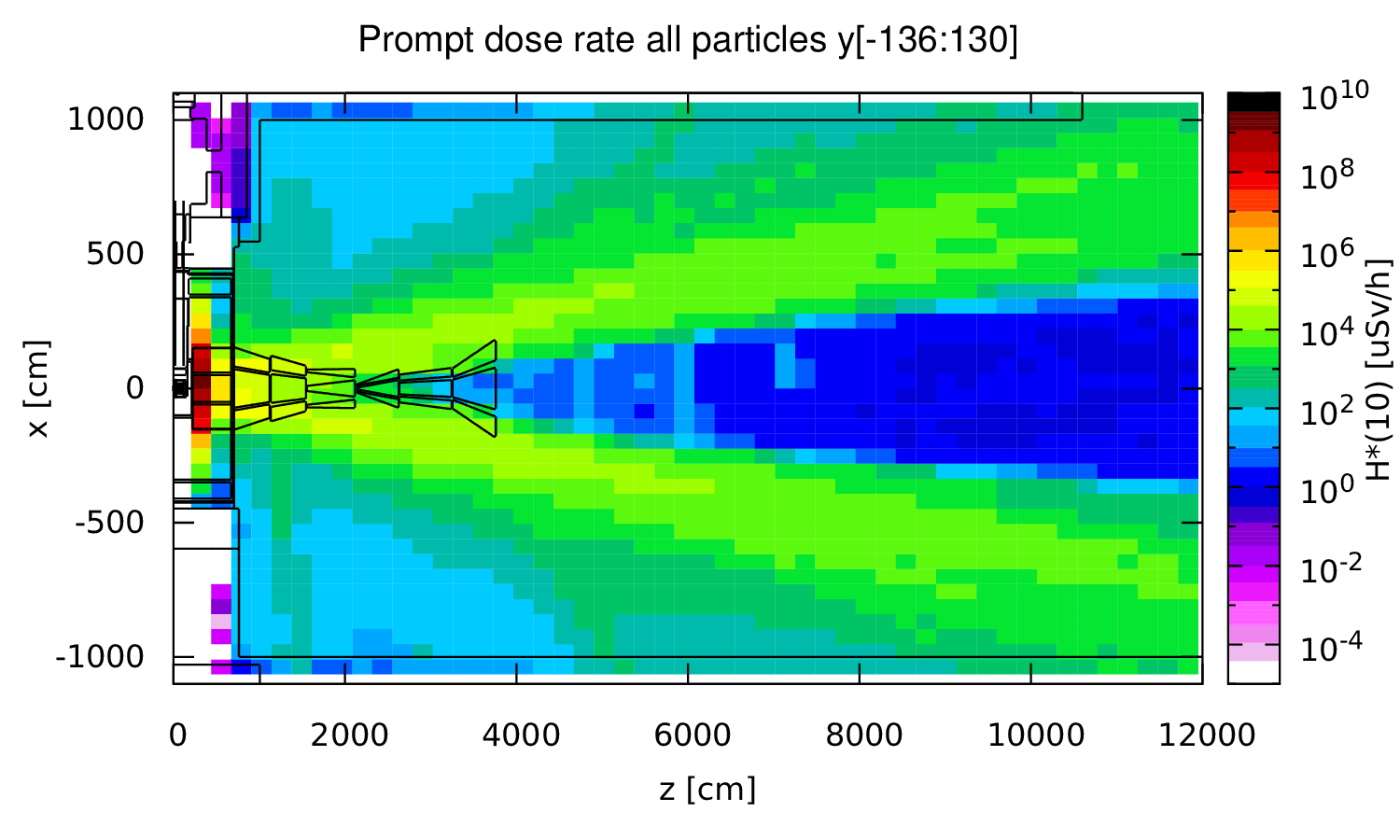}
  \caption{top view}
  \label{fig:prompt1}
\end{subfigure}
\begin{subfigure}{0.5\textwidth}
  \centering
  \includegraphics[width=\textwidth]{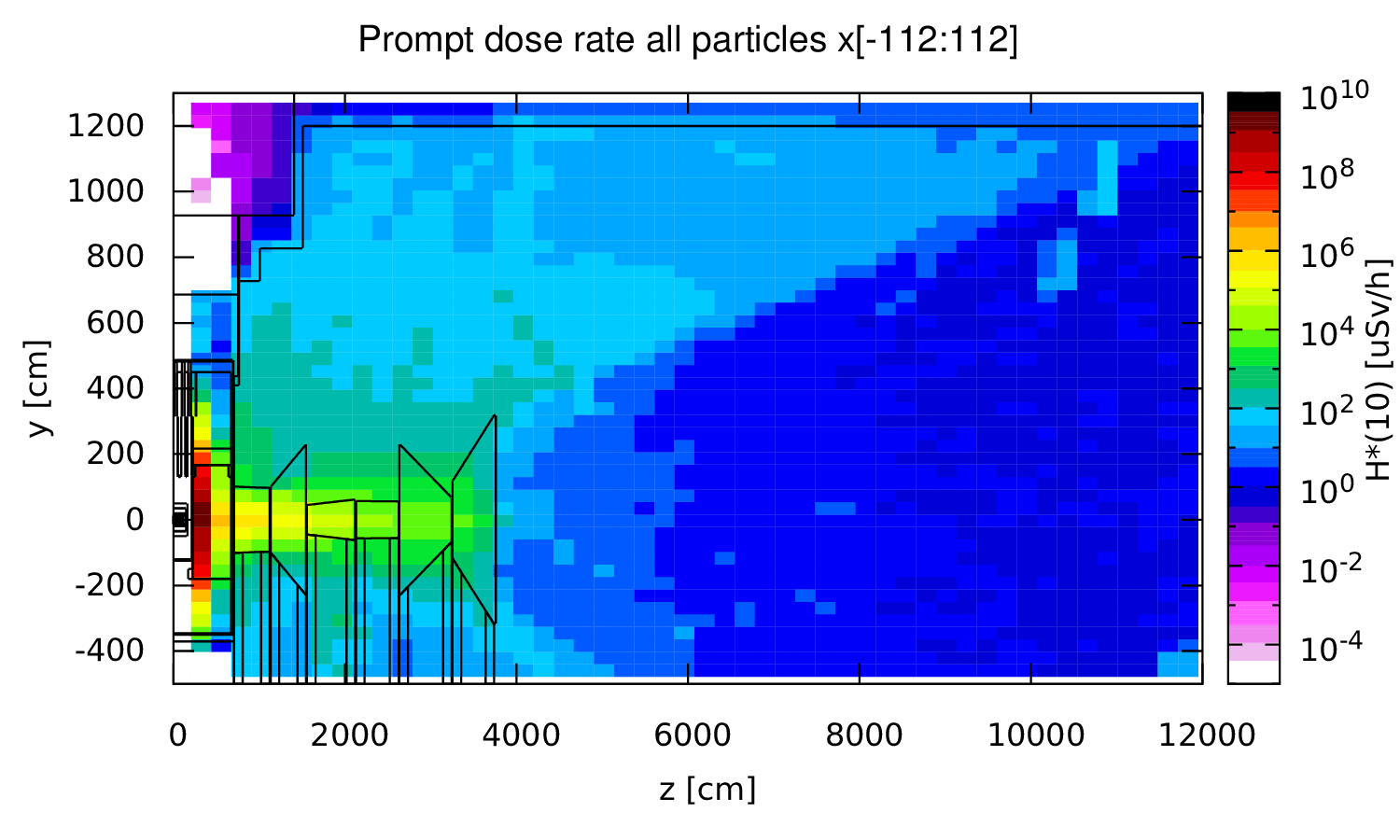}
  \caption{side view}
  \label{fig:prompt2}
\end{subfigure}

\captionsetup{width=0.85\textwidth} \caption{\small Prompt dose rates in $\mu$Sv/h in the SHiP Experimental Hall for all particles. The left plot shows the top view and the right plot shows the side view cut at the level of the beam line.}
\label{fig:pDR1}
\end{figure}

Fig.~\ref{fig:pDR1} shows the prompt dose rates from all particles along the muon shield, in the Experimental Hall, and in the Surface Hall in the horizontal plane (left) and vertical plane (right) cut through the beam-line.

During operation the dose rates, which are mainly due to muons, reach a few \SI{}{mSv/h} along the walls of the Experimental Hall behind the muon shield and drop to below \SI{1}{mSv/h} in the surrounding soil such that levels of soil activation are considered acceptable.

The side view of the Experimental Hall shows that muons are also bent towards the ceiling of the Experimental Hall. The dose reaches a few \SI{}{\micro Sv/h} in the Surface Hall on top of the underground hall, which is not designed to be accessible during beam operation. Further simulations demonstrate that the existing beam lines TT81, TT82 and TT83 are not affected by the prompt dose rates originating from the BDF facility with SHiP operation. The SHiP operation neither influences the present area classification of the EHN1 experimental hall, which corresponds to a Supervised Radiation Area ($<3$\,$\mu$Sv/h). It should be noted that these results on long-range effects are conservative due to the assumption of a moraine density which is 20$\%$ lower than measured in ground samples. Thus, the operation of the SHiP experiment has no impact on the neighbouring experimental areas.

\begin{figure}[!htb]
\begin{subfigure}{0.5\textwidth}
  \centering
  \includegraphics[width=\textwidth]{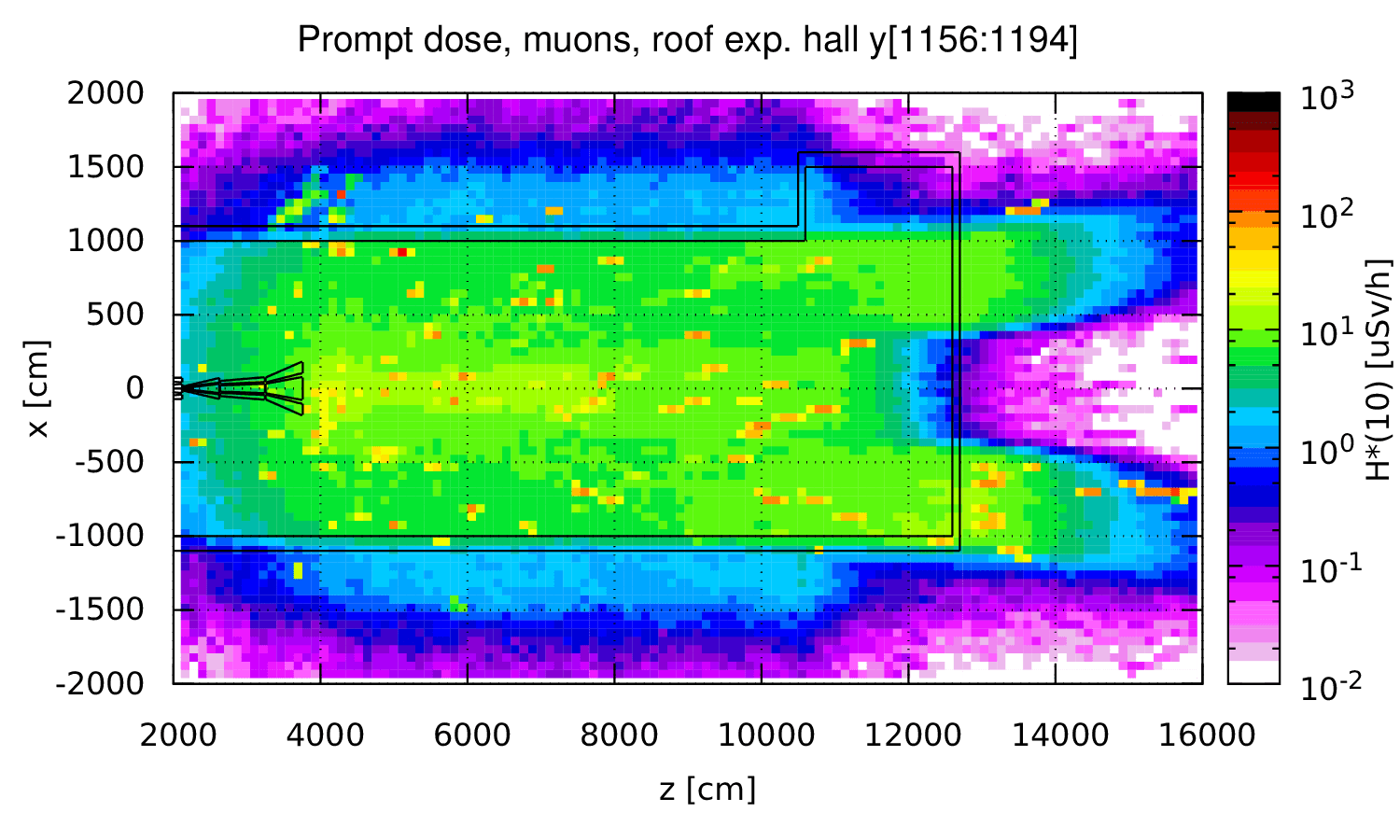}
  \caption{}
  \label{fig:prompt3}
\end{subfigure}
\begin{subfigure}{0.5\textwidth}
  \centering
  \includegraphics[width=\textwidth]{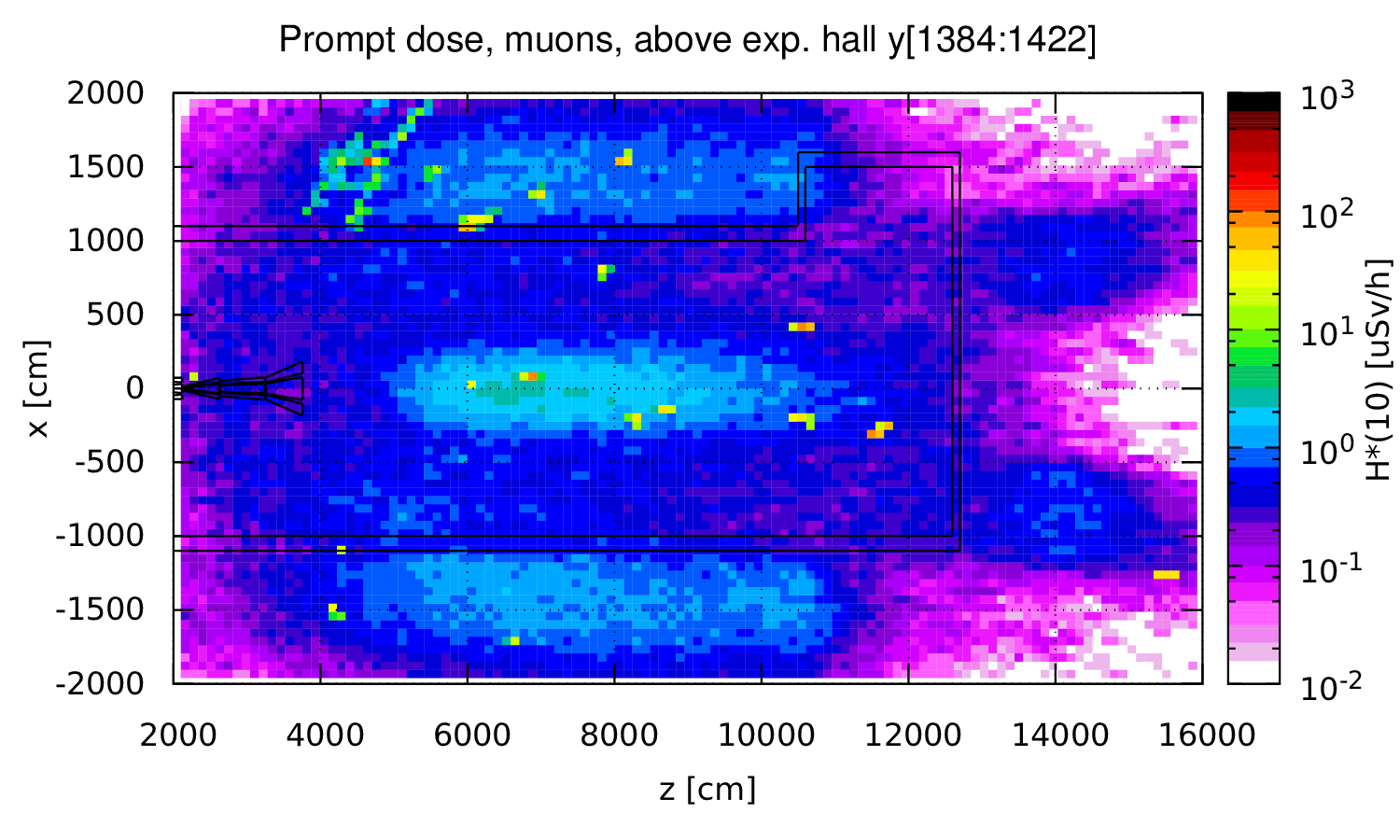}
  \caption{}
  \label{fig:prompt4}
\end{subfigure}

\captionsetup{width=0.85\textwidth} \caption{\small Prompt dose rates in $\mu$Sv/h inside the Experimental Hall at level of the roof (a) and at the ground level of Surface Hall (b).}
\label{fig:pDR2}
\end{figure}

Fig.~\ref{fig:pDR2} shows the prompt dose rates from muons  at roof level of the underground Experimental Hall and at ground level of the above-lying Surface Hall to show the effectiveness of the 1 m of concrete ceiling. 

It shows that a few $\mu$Sv/h are expected on top of the ceiling of the underground Experimental Hall behind the detector Surface Hall. It is planned to fence and cover that area with 6 m of the remaining soil from the excavations, such that the dose rate level is brought down to below 0.5\,$\mu$Sv/h allowing for a Non-Designated Area \cite{zonage}. A standard stray radiation monitor for photons, muons and neutrons is envisaged at the CERN fence closest to the most exposed area. 

\begin{figure}[!htb]
  \centering
  \includegraphics[width=0.8\textwidth]{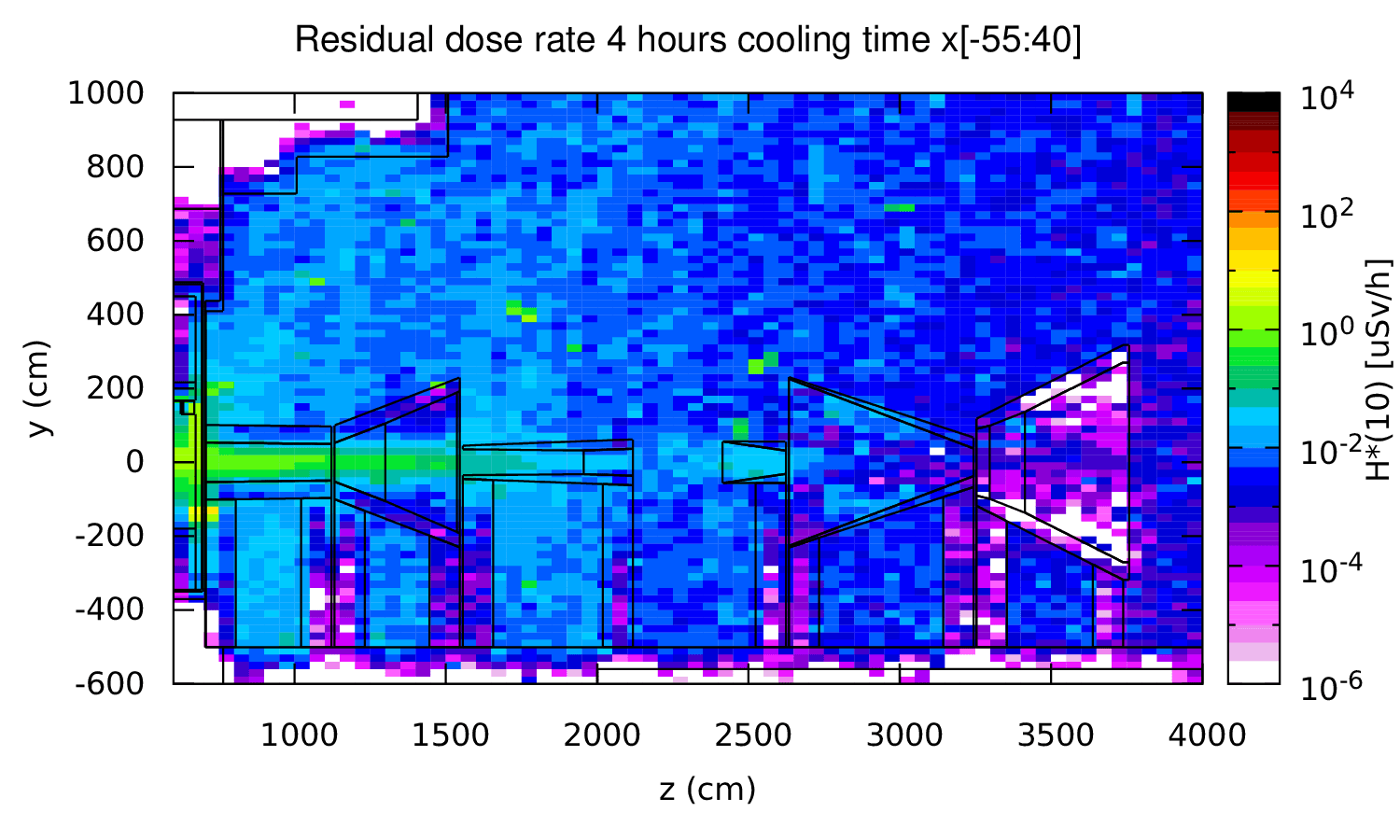}

\captionsetup{width=0.80\textwidth} \caption{\small Residual ambient dose equivalent rate in $\mu$Sv/h for the active muon shield after five years operation and four hours of cool-down time.}
\label{fig:rDR1}
\end{figure}

Assuming five years of operation the residual dose rates in the SHiP Experimental Hall are expected to be at most a few \,$\mu$Sv/h at contact with the first part of the active muon shield after four hours of cool-down (see Fig.~\ref{fig:rDR1}), thus allowing for access to this area. The Experimental Hall up to the end of the active muon shield will be classified as Simple Controlled, while the rest as Supervised Radiation Area.

\FloatBarrier 
\printbibliography[heading=subbibliography]

 \chapter{Integration}
\label{Chap:Integration}

\section{Introduction}

\begin{figure}[!h]
  \centering
  \includegraphics[width=0.74\textwidth]{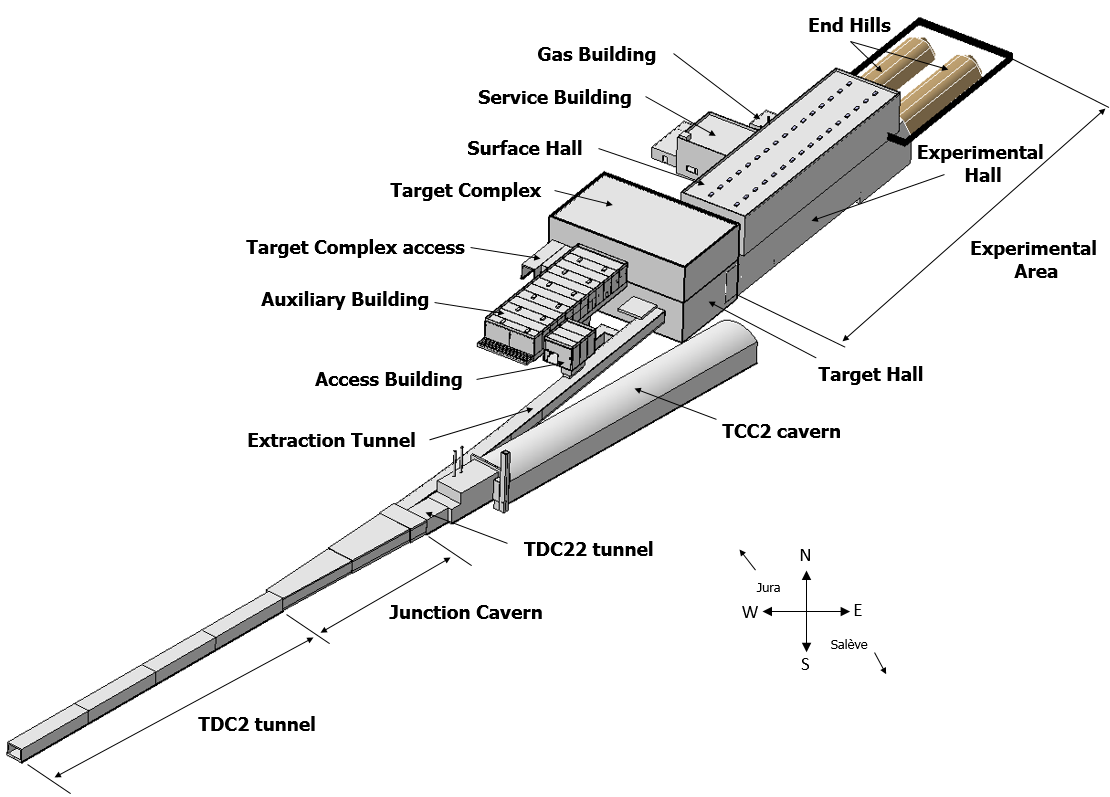}
  \caption{BDF integration overview}
  \label{fig:BDF integration overview}
\end{figure}

Detailed integration studies have been performed to evaluate the feasibility of siting the Beam Dump Facility within CERN's Pr\'evessin site.  
All the infrastructure requirements have been defined and integrated within the Civil Engineering layout and the locations and number of structures and the services have been optimized in terms of radiation protection, general safety, accessibility, and practicality. Fig.~\ref{fig:BDF integration overview} shows the integration layout of the overall facility. 

The integration of the facility has been divided into the following four areas:

\begin{enumerate}

\item \textbf{Transfer Tunnel}: houses the BDF beamline equipment and services; it consists of the existing TDC2 tunnel, the Junction Cavern, and the Extraction Tunnel.
\item \textbf{Access Building and Auxiliary Building}: the Access Building allows the installation of large beamline equipment directly into the transfer tunnel. 
The Auxiliary Building houses the cooling, ventilation and electrical infrastructure for the transfer tunnel, the Target Complex and the Experimental Area. 
In addition, it includes the transfer tunnel’s personnel access. 
\item \textbf{The Target Complex access}: the personnel access and a pressurized vehicle access for the Target Complex.
\item \textbf{The Experimental Area}: defined and sized to house and allow the assembly of the SHiP experiment inside its Experimental Hall while providing the services and infrastructure required for the operation of the detector.

\end{enumerate}

A preliminary proposal for the technical galleries layout has been undertaken assuming that the BDF electrical connection to the Pr\'evessin network is made through CERN’s existing service building, BA80 (see Chapter~\ref{Chap:CivEng}, Fig.~\ref{fig:Layout}). The connection from the existing technical gallery to the Beam Dump Facility first reaches the Auxiliary Building, where the primary cooling pump and main electrical distribution equipment are located. From here, the water and electrical power are distributed throughout the facility. 
Further study should be undertaken to determine the optimised layout.

The following sections describe in detail the integration studies
performed for each of the four areas listed above.
In the appendices to this chapter, Table \ref{tab:IntegrationModelsReferences} contains the  
SmarTeam numbers of the integration models
(ENOVIA SmarTeam is a Product Data Management tool that enables organisations to manage and collaborate on component information.)
Table \ref{tab:EDMSdocuments} lists the structural specification documents that have been developed for the facilities listed above.

\section{Transfer tunnel}
\label{TransferTunnel}

The transfer tunnel for the BDF beamline is located upstream (South-West) of the Target Complex as shown in Fig.~\ref{fig:BDF integration overview}. 
It is composed of the existing TDC2 tunnel, the Junction Cavern and the Extraction Tunnel. The transfer tunnel starts at the existing TDC2 tunnel, from which the BDF bending magnets bend the beam away from the existing North Area beamlines into a new Junction Cavern. The beamline then branches off the Junction Cavern into the Extraction tunnel and runs alongside the existing TCC2 cavern up to the Target Complex.

The purpose of the structure is to house the beamline equipment and services that deliver the beam to the target. 
The equipment includes magnets, beam instrumentation, vacuum, safety, and alignment equipment. 
The services include the general electrical distribution, cooling, and ventilation. 

\subsection{TDC2 Tunnel}
\subsubsection{Summary}

The TDC2 tunnel (shown in Fig.~\ref{fig:BDF integration overview}) is an existing underground tunnel that houses North Area beamlines (TT22, TT23, TT24 and TT25) and their services.
The access to the beamlines is from the TCC2 cavern via building BA80, in which the transport access path is located on the Salève (South-East) side.

\begin{figure}[ht!]
\centering
    \begin{subfigure}[b]{.39\linewidth}
        \centering
        \includegraphics[width=\linewidth]{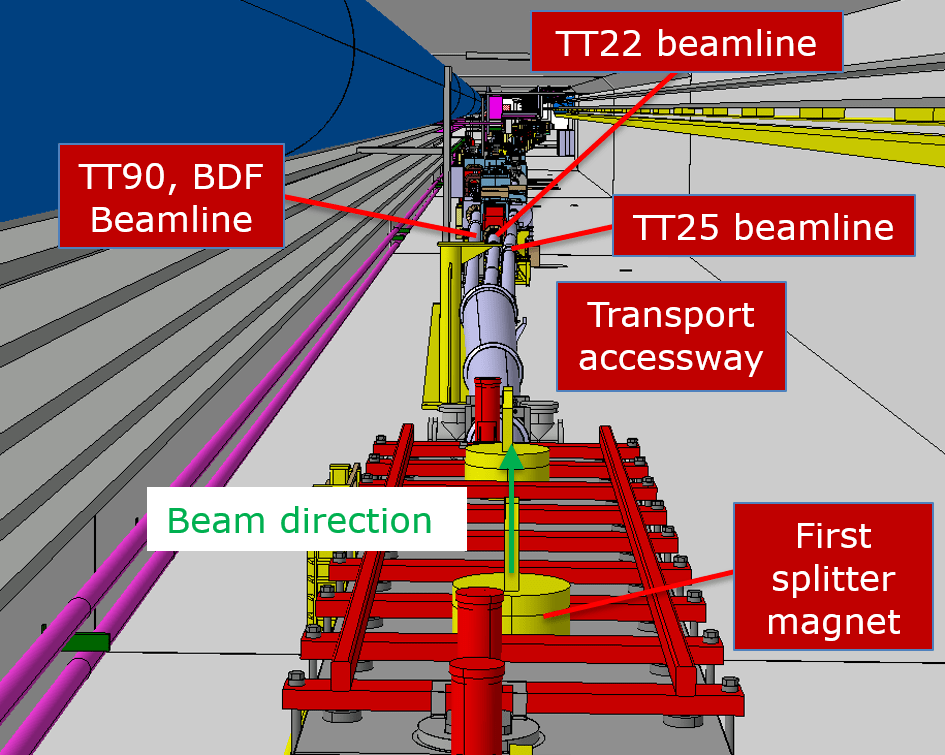}
        \caption{} \label{fig:4TT}
    \end{subfigure}
    \begin{subfigure}[b]{.39\linewidth}
        \centering
        \includegraphics[width=\linewidth]{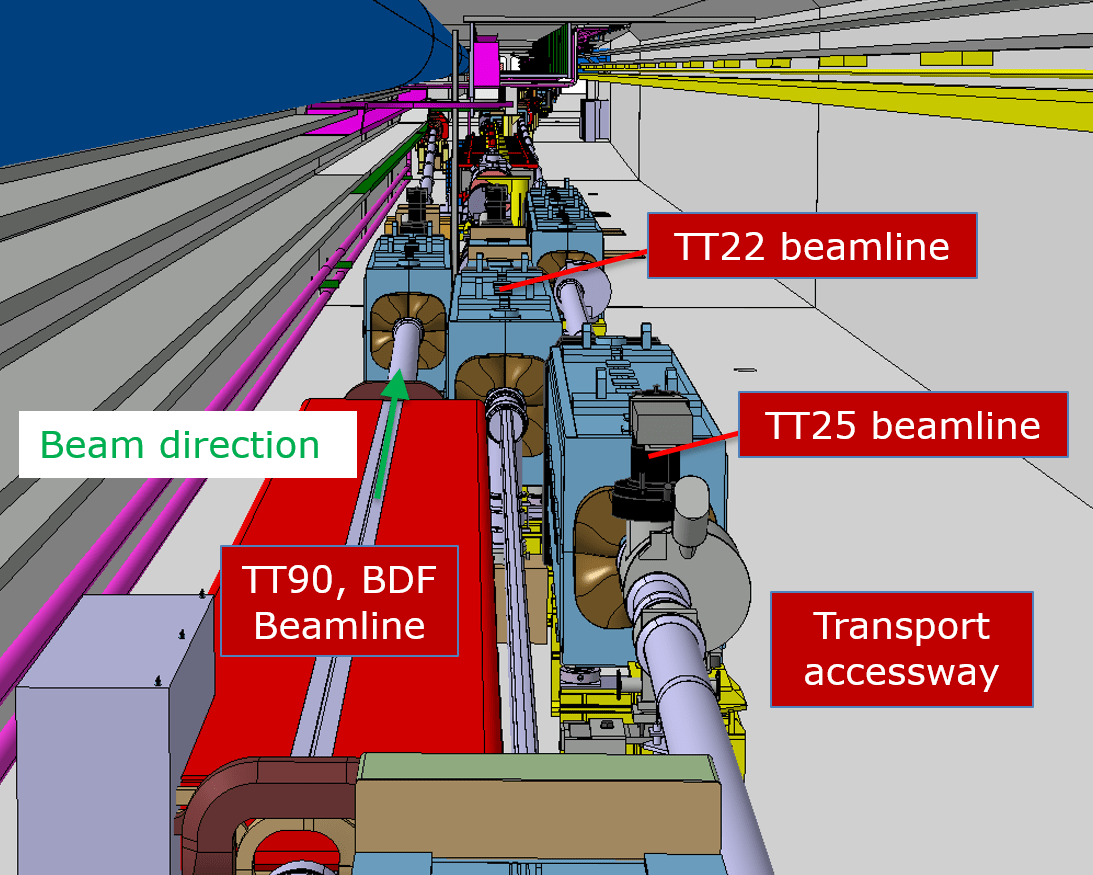}
        \caption{} \label{fig:5TT}
    \end{subfigure}
    \caption{Start of BDF beamline (TT90) at the first MSSB splitter magnet \subref{fig:4TT}. BDF beamline (TT90) integrated with the existing North Area beamlines \subref{fig:5TT}.}
    \label{fig:4TTand5TT}
\end{figure}

The new BDF beamline (TT90) starts downstream of the new splitter magnet in the TDC2 tunnel (shown in Fig.~\ref{fig:4TT}) and runs alongside the existing North Area beamlines for approx.~\SI{110}{\meter} as shown in Fig.~\ref{fig:5TT}. 
As there are only five BDF magnets and one beam monitor on the BDF beamline in the TDC2 tunnel, and as the BDF beamline does not significantly diverge away from the existing North Area beamlines until reaching the Junction Cavern, the existing tunnel’s geometry is sufficient to house this part of the BDF beamline.

\newpage

\subsubsection{TDC2 dismantling}

\begin{figure}[h]
  \centering
  \includegraphics[width=\linewidth]{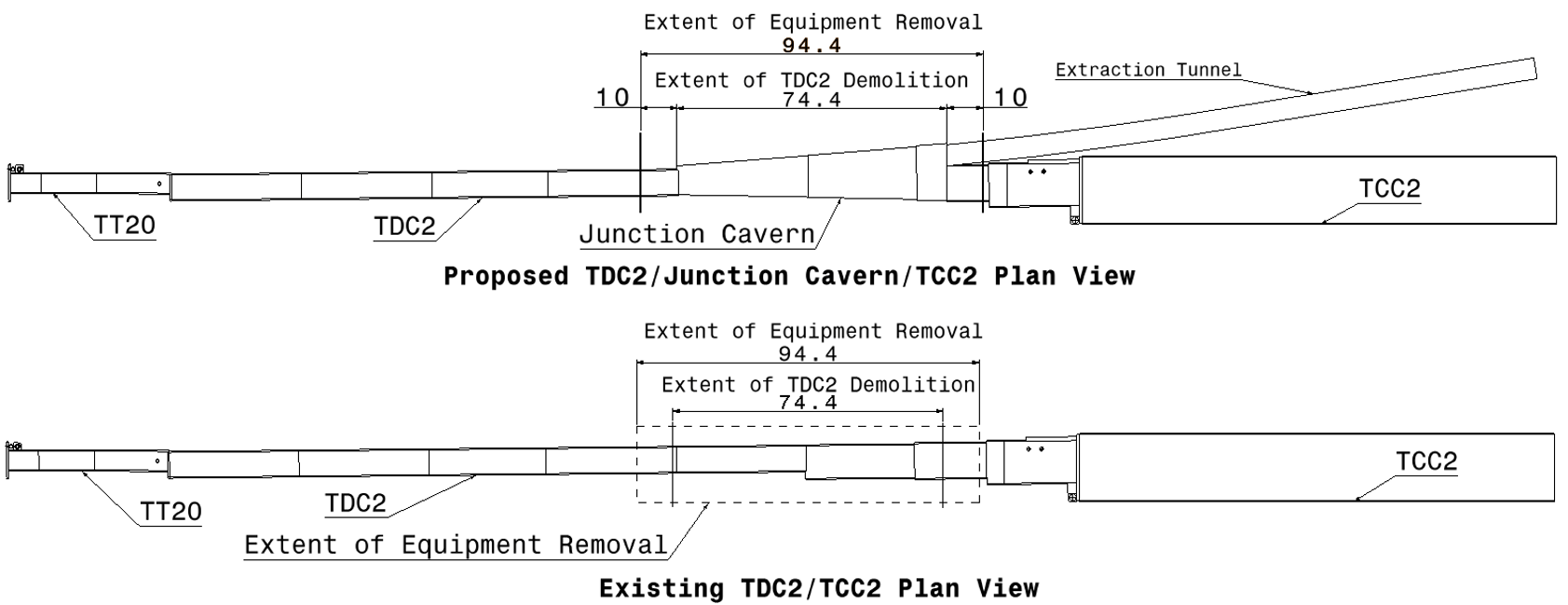}
  \caption{Extent of tunnel demolition and equipment removal in the new Junction Cavern area~\cite{BDFequipmentremoval}}
  \label{fig:RemovalTDC2}
\end{figure}

\begin{figure}[htbp]
  \centering
  \includegraphics[width=6.1in]{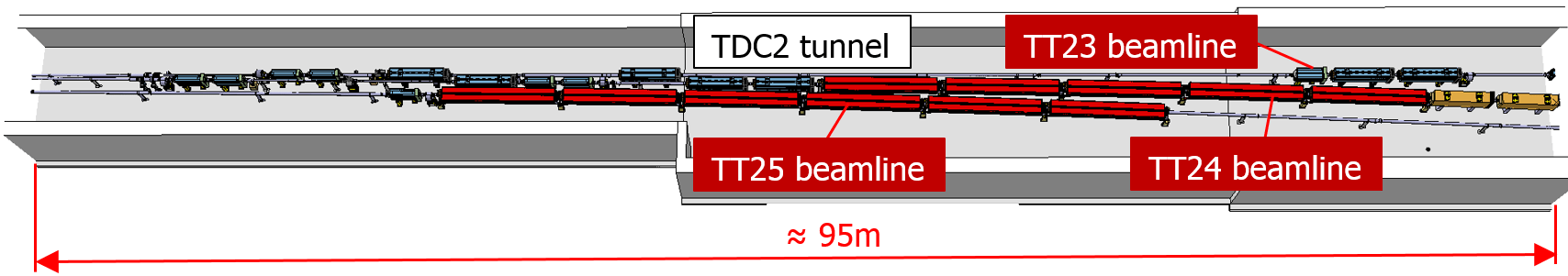}
  \caption{Detail of the demolition region, with lines TT23, TT24 and TT25}
  \label{fig:EquipmentTDC2}
\end{figure}

As described in Chapter~\ref{Chap:CivEng}, a section of the TDC2 tunnel will be demolished for the construction of the new Junction Cavern, and all the beamline elements and associated auxiliary infrastructure will be dismantled, starting and finishing 10 m away from the demolition region (see Fig.~\ref{fig:RemovalTDC2}). The total extent of the TDC2 demolition will be approximately 75 m, while equipment (including cabling, trays, ventilation ducts, lightning, fire detection and extinguishing systems, etc.) will be removed in a 95 m long section. Some of this equipment, including magnets, vacuum equipment and beam instrumentation, will be stored and may be reused (see Table~\ref{fig:ELEMENTSREMOVED} and Fig.~\ref{fig:EquipmentTDC2}) while other equipment and associated infrastructures will be replaced with new equipment after the construction. 

The approach assumed the construction of two shielding walls on both sides of the future Junction Cavern to reduce the dose from the splitters and the Target Attenuator eXperimental areas (XTAX) and protect the regions from dust and debris, and the use of a portable booster ventilation system before removing the ventilation ducts (this provisional unit is supposed to run continuously to guarantee fresh air in the TDC2).
The beamline equipment of TT22, TT23, TT24 and TT25 within this region will have to be removed with minimum disassembly wherever possible, while destructive works will be acceptable for the associated infrastructure (cutting of ducts and cabling). 

\begin{table}[htpb]

  \centering
  \caption{Main elements to be removed before demolition of TDC2 at the new junction cavern }
  \includegraphics[width=6.2in]{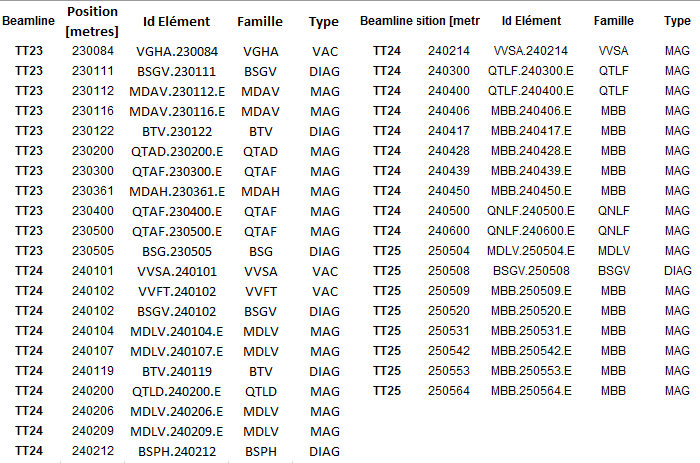}
    \label{fig:ELEMENTSREMOVED}
\end{table}

For evaluating the necessary cool-down time, a detailed Work and Dose Planning (WDP) for the given works will be produced. 
At this stage, only a rough estimate of the WDP can be done as it largely depends on the detailed methodology of the works (remote handling, shielding, work optimization etc.). 

The estimation of the dose was undertaken by reviewing similar past activities (e.g. removal of magnets) and the associated data available for magnets and beamline elements with different level of activation.
The main assumption for the dose estimation during removal of beamline elements was to spend a continuous time in a environment dominated by the background radiation scenario 
(which assumes that the magnets have been removed). 

When looking at the expected dose rates in TDC2 for various cool-down times, the most significant decrease occurs, as expected,  in the very first weeks of the cool-down
(e.g. a factor 13 over 4 weeks -- typically given by an end of year lead ion run). 
Further cool-down, for example, between 9 weeks and 5 months after the proton run, only amounts to a further gain of factor 0.3. 
At the downstream end of the given area the dose rates at a distance of 40 cm from the equipment are relatively low (30 $\mu$Sv/h, 9 weeks after proton (p+) run), and thus the cool-down time becomes less critical.
For the most activated elements to be removed (e.g. beam monitor BSPH.240212), the dose rates are of the order of 590 $\mu$Sv/h after 9 weeks after p+ run. 
These are still much lower than the dose rates of the splitters (11 mSv/h at 40 cm, 9 weeks after p+ run), which fortunately do not have to be removed at the same time.

It has to be noticed that due to the position of some magnets (surrounded by elements from other beamlines), the classical approach of removing the most activated objects (namely, the MDAVs, QTADs and MBBs upstream near the splitters) needs further detailed analysis.

\begin{figure}[htbp]
  \centering
  \includegraphics[width=4in]{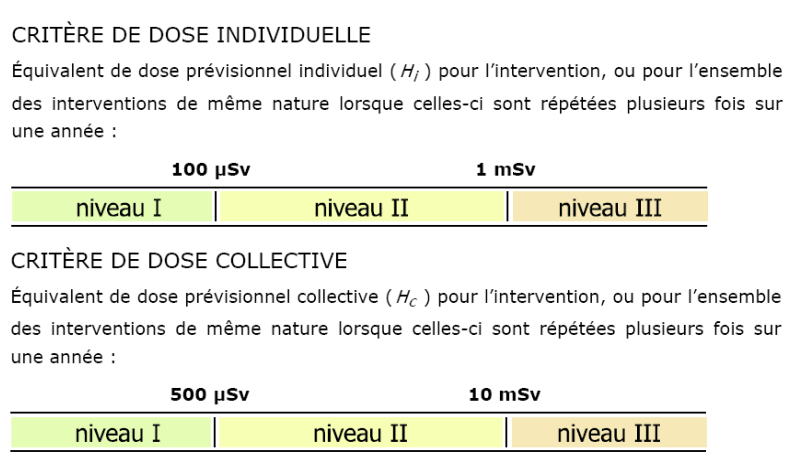}
  \caption{Criteria for ALARA classification at CERN concerning individual and collective doses}
  \label{fig:ALARA}
\end{figure}

The first estimation (based on previous intervention of similar experience and assuming dose rates per equipment based on RP measurements and estimates on equipment position along the line), results in a total collective dose of the order of 10-20 mSv if the equipment removal starts after 13 weeks (3 months) of cool-down (including a 4 weeks ion run). 
This means, that the works will clearly be an ALARA 3 level (see Fig.~\ref{fig:ALARA}) from a collective dose view, but does not seem infeasible. 
Reducing the cool-down time to 9 weeks (including 4 weeks ion run) would increase the collective dose by a factor 0.09 (1-2 mSv). 

At this stage of the project and in view of ALARA, CERN-HSE-RP representatives therefore recommend to use the maximum cooling time that is acceptable for the project
and which is at least 9 weeks after the end of the proton run (i.e. 5 weeks in addition to 4 weeks of an ion run). 
A further recommendation is to use if possible 13 weeks (3 months) of cooling in the planning, such that even without an ion run the planning could be met. 
Furthermore, the dismantling should start at the downstream end of TDC2, giving more cooling time for the hotter elements further upstream. 
Further deinstallation details are discussed in Sections \ref{ENELTT} and \ref{Coolingandventilation}.

\subsection{Junction Cavern (TDC21)}
\subsubsection{Summary}

\begin{figure}[htbp]
  \centering
  \includegraphics[width=4.8in]{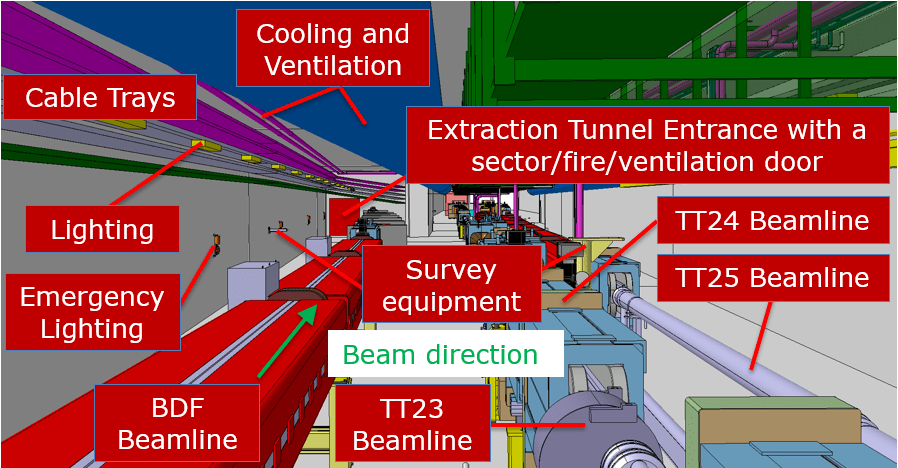}
  \caption{The Junction Cavern’s integration layout}
  \label{fig:6TT}
\end{figure}

The Junction Cavern (TDC21) (shown in Fig.~\ref{fig:BDF integration overview}) is an underground cavern that will house four beamlines (TT91, TT23, TT24 and TT25) and their corresponding services. 
The beamlines enter the Junction Cavern upstream (South-West) at the TDC2 tunnel, the existing North Area beamlines continue into the TDC22 tunnel, while the BDF beamline (TT90) deviates away towards the Jura (North-West) into the Extraction Tunnel. 
The beamline services include cooling and ventilation, cabling/cable trays and lighting among other services as shown in Fig.~\ref{fig:6TT}. 
For the alignment of the beamline equipment in the cavern, there are survey brackets secured to the Jura (North-West) side wall, survey pillars secured to the floor and survey floor points installed in the cavern floor.
During technical stops and long shutdowns of the beamline, personnel and transport vehicles may access the Junction Cavern for maintenance purposes. 
The access to the existing North Area beamlines is from the TCC2 cavern via building BA80, in which the transport accessway is located on the Salève (South-East) side. 
The access to the BDF beamline is from the Extraction Tunnel in which the transport accessway is located on the Jura (North-West) side.

\subsubsection{Geometry}

\begin{figure}[htbp]
  \centering
  \includegraphics[width=5.91in]{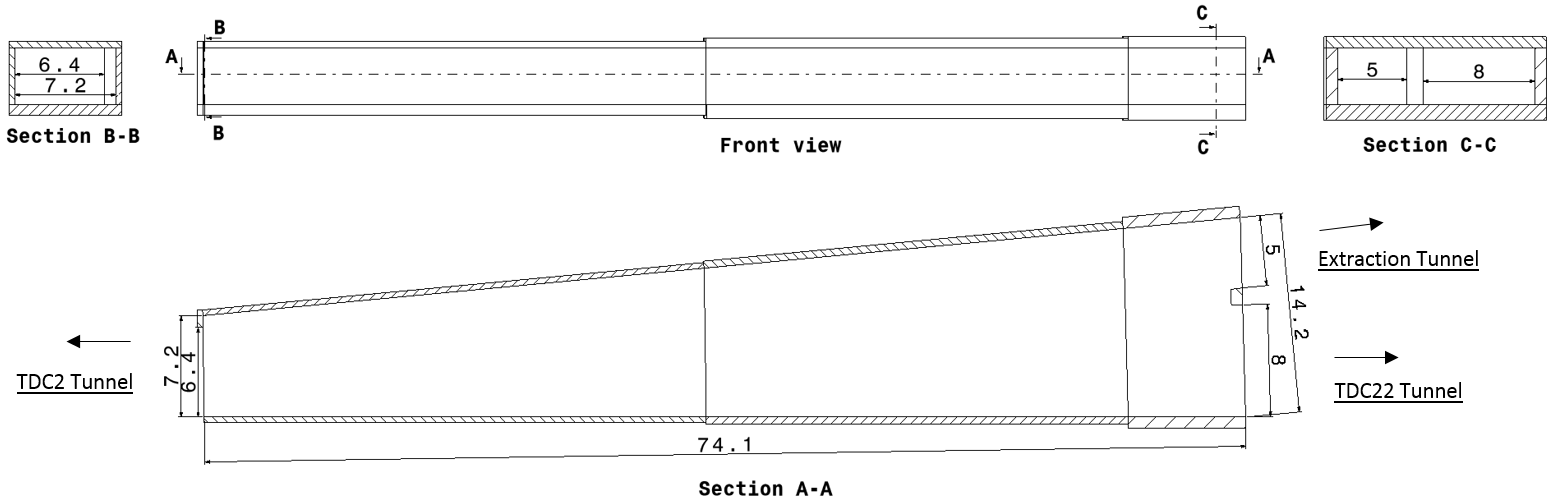}
  \caption{Plan and section views of the Junction Cavern}
  \label{fig:8TT}
\end{figure}

\begin{figure}[htbp]
  \centering
  \includegraphics[width=5.1in]{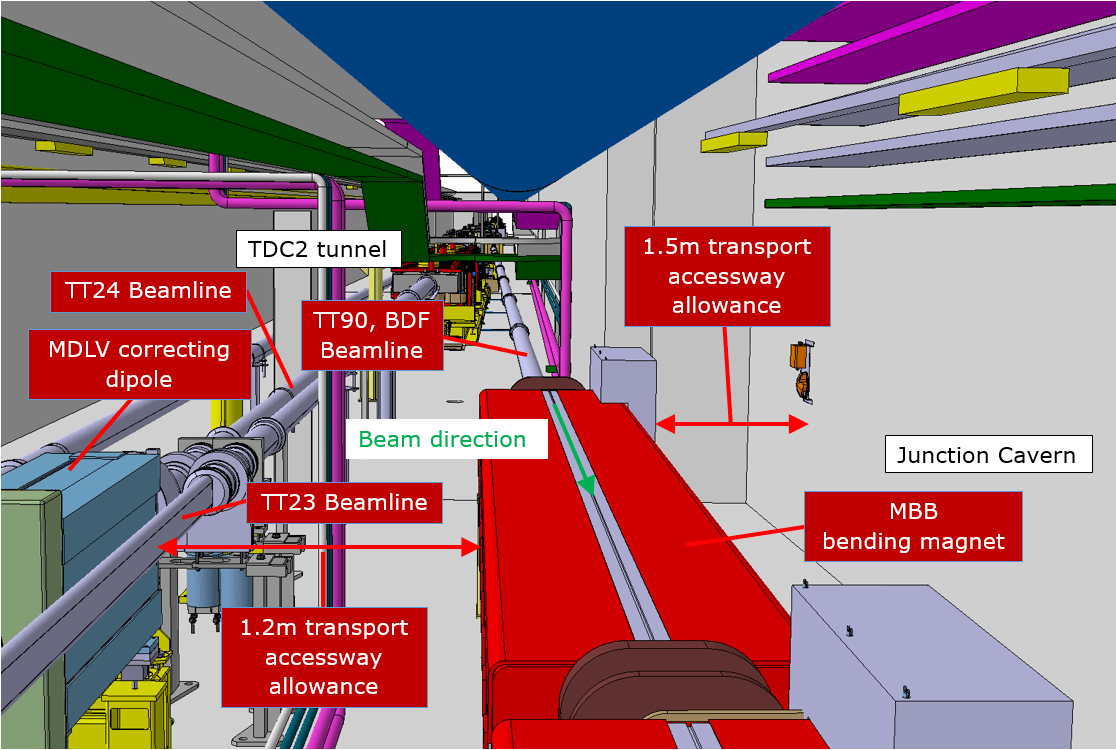}
  \caption{Transport vehicle allowance in the Junction Cavern}
  \label{fig:9TT}
\end{figure}

Fig.~\ref{fig:8TT} shows the plan view and sections of the Junction Cavern. The cavern has a length of \SI{74.1}{\meter}, the width of the upstream TDC2 tunnel (South-West) end is \SI{7.2}{\meter} and the width of the downstream (North-East) end is \SI{14.2}{\meter}. 
The upstream end is based on an internal opening of \SI{6.4}{\meter} that ties into the existing TDC2 tunnel and an allowance of \SI{1.5}{\meter} between the BDF beamline and the Junction Cavern’s Jura (North-West) side wall. 
This allowance is required for a transport vehicle to manoeuvre the MBB bending magnets into position as shown in Fig.~\ref{fig:9TT}. 
There is also a clearance of approx.~\SI{1.2}{\meter} between the vacuum chamber on the existing beamline TT23 and the new BDF beamline.  
This allows a smaller handling vehicle to reach the MDLV correction dipoles on the existing beamlines TT23 and TT24 for which a minimum of \SI{1.0}{\meter} is required. 
The downstream end is based on two openings, one \SI{5}{\meter} wide that ties into the Extraction Tunnel and the other opening \SI{8}{\meter} wide that ties into the existing TDC22 tunnel. 
The cavern has a height of \SI{4}{\meter}, which is the same as the existing TDC2 tunnel. 

\newpage

\subsection{Extraction Tunnel (TT90)}
\subsubsection{Summary}

The Extraction Tunnel (TT90) (shown in Fig.~\ref{fig:BDF integration overview}) is a new underground tunnel that houses the BDF beamline and its corresponding services. 
It allows the beamline to deviate away from the existing North Area beamlines to the Target Complex. 
The beamline services include cooling and ventilation, cable trays and lighting among other things. 
For the alignment of the beamline equipment in the tunnel, there are survey brackets secured to the Jura side wall and survey floor points installed in the tunnel floor.

\begin{figure}[htbp]
  \centering
  \includegraphics[width=5in]{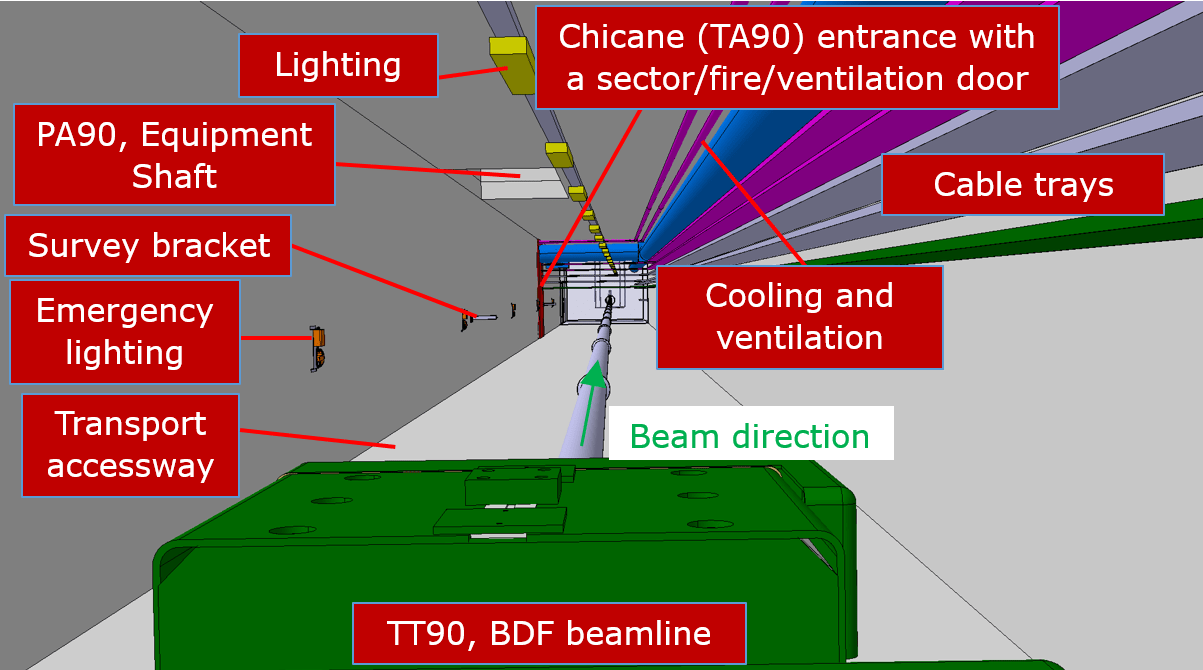}
  \caption{Extraction Tunnel’s integration layout}
  \label{fig:11TT}
\end{figure}

During technical stops and long shutdowns of the beamline, personnel and transport vehicles may access the Extraction Tunnel for maintenance purposes, with the transport accessway located on the Jura side of the tunnel. Personnel access into the Extraction Tunnel is via the chicane (TA90) connected to the side wall of the Extraction Tunnel, whilst transport vehicles and heavy equipment enter the tunnel via the Equipment Shaft (PA90), as shown in Fig.~\ref{fig:11TT} (see Ref.~\cite{SBstructuralspec} for further details). 
The equipment shaft is located on the Jura side of the tunnel roof, such that the vehicles and equipment lowered in the tunnel do not clash with the services on the Salève side.

\begin{figure}[htbp]
\centering
\includegraphics[width=4.6in,height=3in]{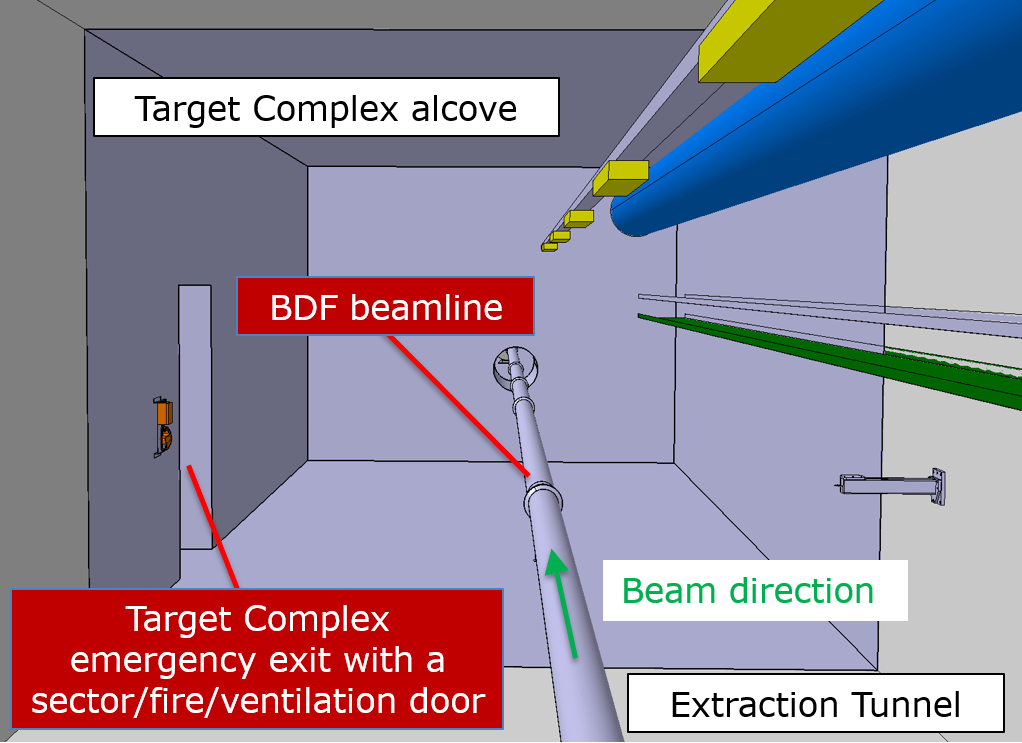}
\caption{Extraction Tunnel’s interface with the Target Complex}
\label{fig:12TT}
\end{figure}

The downstream end of the Extraction Tunnel is connected to the Target Complex. 
For emergency purposes, and for ventilation and fire compartmentalisation, a sector/fire/ventilation door is located in the Target Complex alcove, see Fig.~\ref{fig:12TT} for further details.

\subsubsection{Geometry}

\begin{figure}[htbp]
  \centering
  \includegraphics[width=5.5in]{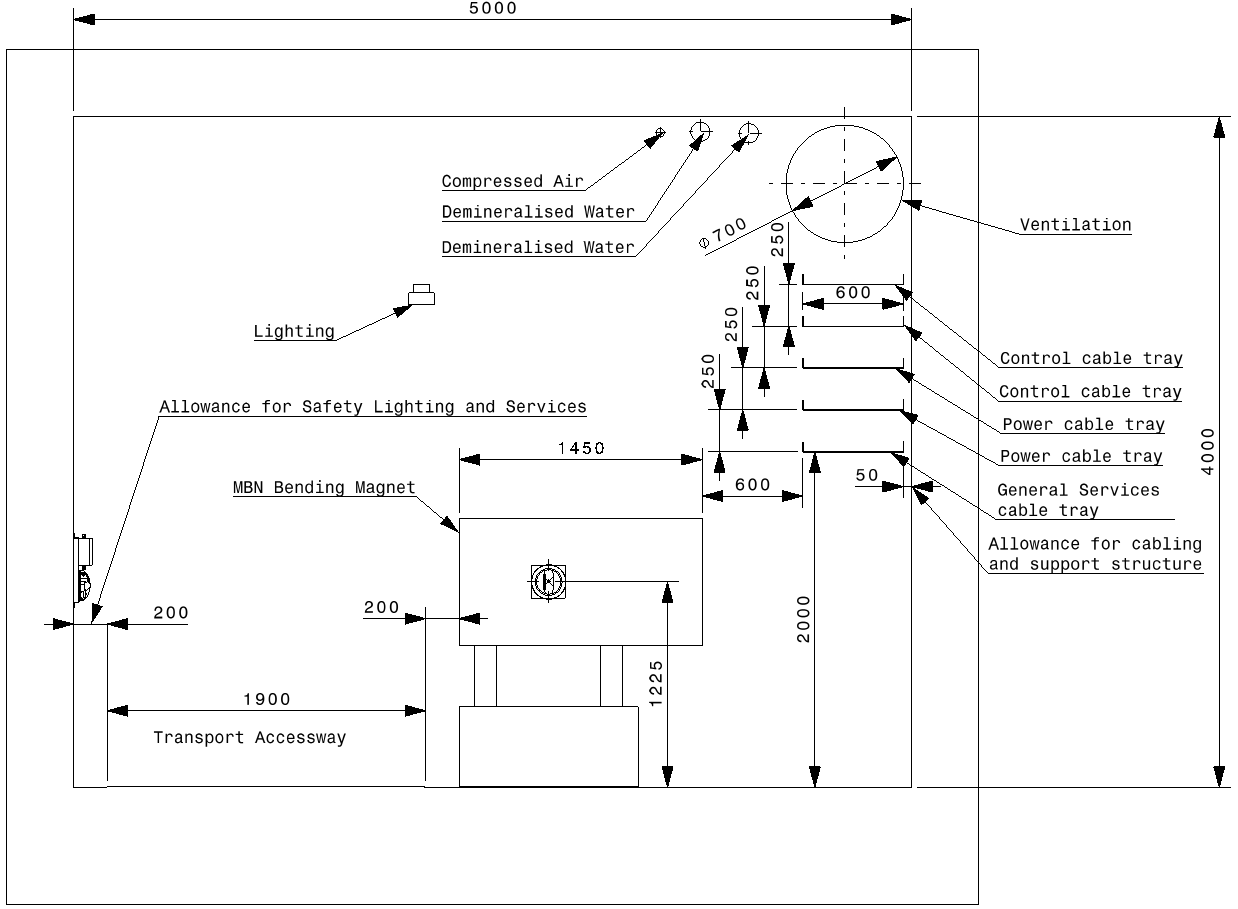}
  \caption{Extraction Tunnel cross-section}
  \label{fig:13TT}
\end{figure}

The structure has a length of approx.~\SI{165}{\meter} with an internal height of \SI{4}{\meter} and a width of \SI{5}{\meter}. 
The height of the tunnel is the same as the Junction Cavern and the width of the tunnel is based on the required dimensions for the equipment and personnel/transport access as shown in Fig.~\ref{fig:13TT}:

\begin{itemize}

\item \SI{200}{\milli\meter} allowance for safety lighting and services;
\item \SI{1900}{\milli\meter} allowance for the transport vehicles;
\item \SI{200}{\milli\meter} allowance for the sector door between the Junction Cavern and the Extraction Tunnel;
\item \SI{1450}{\milli\meter} width of the MBN bending magnet;
\item \SI{600}{\milli\meter} allowance for personnel access and maintenance of the cable trays;
\item \SI{600}{\milli\meter} allowance for cable trays;
\item \SI{50}{\milli\meter} allowance for cabling and the cable tray support structure.

\end{itemize}

The transport accessway in the Extraction Tunnel is larger than the \SI{1.5}{\meter} allowance in the Junction Cavern because the MBN bending magnets require a larger transport vehicle to manoeuvre the magnets than that of the MBB bending magnets.

\begin{figure}[htbp]
  \centering
  \includegraphics[width=\linewidth]{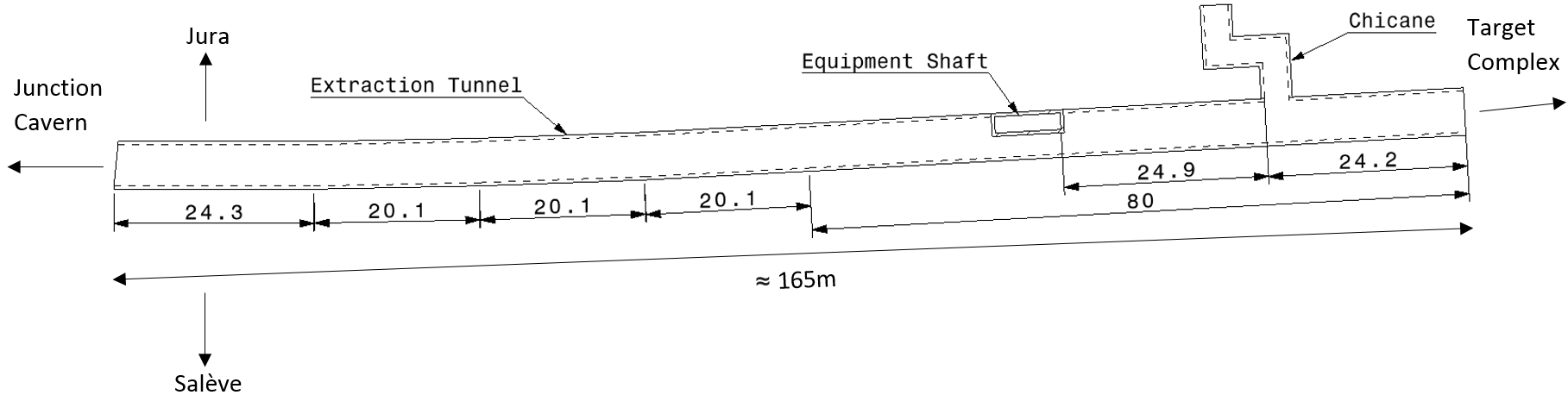}
  \caption{Plan view of the Extraction Tunnel}
  \label{fig:14TT}
\end{figure}

The chicane is connected to the Jura side of the Extraction Tunnel, approx.~\SI{24}{\meter} upstream of the Target Complex. Approx. \SI{25}{\meter} upstream of the chicane entrance is the equipment shaft, connected to the top of the Extraction Tunnel, as shown in Fig.~\ref{fig:14TT}.

\subsection{Beamline Equipment}

The new BDF beamline (TT90) is approx.~\SI{380}{\meter} in length, and it is housed in the following structures:

\begin{itemize}

\item \SI{120}{\meter} in the existing TDC2 tunnel;
\item \SI{75}{\meter} in the Junction Cavern, TDC21;
\item \SI{165}{\meter} in the Extraction Tunnel, TT90 tunnel;
\item \SI{20}{\meter} in the Target Complex.

\end{itemize}

The BDF beamline starts at the first MSSB splitter magnet in the existing TDC2 tunnel (shown in Fig.~\ref{fig:4TT}) and runs alongside the existing North Area beamlines TT22 and TT25, as shown in Fig.~\ref{fig:5TT} (refer to Section~\ref{transf:sec:splitter} for further information on the MSSB splitter magnet). 
The maintenance of the beamline equipment for both the BDF and existing North Area beamlines was considered to set the layout of the beamline. 
To avoid any clashes between the new BDF beamline and beamline TT22, there were a few modifications made to beamline TT22. 
To determine the precise location of the existing beamline equipment, 3D survey scans were undertaken in the TDC2 tunnel during a technical stop in March 2018. 
For information on the cooling and ventilation equipment and the transport and handling requirements of the beamline see Sections \ref{Transportandhandling} and \ref{Coolingandventilation}.

\subsubsection{Magnets} \label{integ:sec:magnets}

For the layout of the magnets in the BDF beamline see Ref.~\cite{BDFbeamlineIS}. The type of magnets used in the BDF beamline and their geometry is shown in Table \ref{Table2TT}.

\begin{table}[ht!]
 \center
 \caption{The BDF beamline’s magnet characteristics}
 \label{Table2TT}
\begin{tabular}{cccccc}
\hline 
\textbf{Magnet}  & \textbf{ Qty.} & \textbf{ Length [m]} & \textbf{ Width [m]} & \textbf{ Height [m]} & \textbf{ Weight [t]}  \\
\hline
MBB & 5 & 6.7 & 0.844 & 0.685 & 18 \\
MBN & 18 & 5.51 & 1.448 & 0.676 & 23  \\
MDX & 10 & 0.655 & 0.73 & 0.68 & 1.1 \\
QTG & 1 & 2.2 & 0.473 & 0.473 & 2.65 \\
QTL & 5 & 3.3 & 0.6 & 1.057 & 9.9 \\
\hline 
\end{tabular} 
\end{table}

\newpage

\noindent \textbf{Changes to existing beamline TT22:} 

\begin{figure}[htbp]
  \centering
  \includegraphics[width=5.5in]{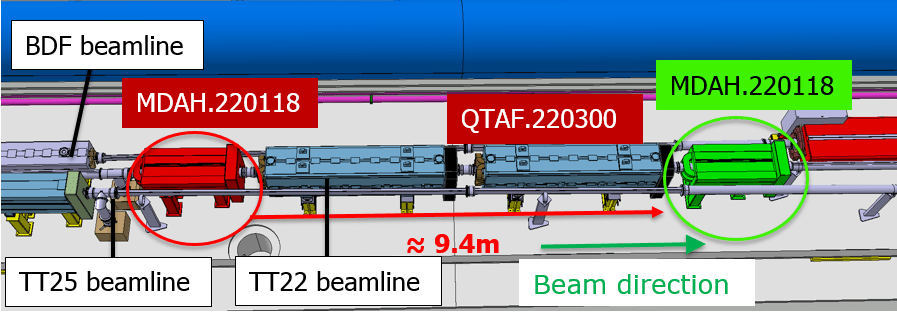}
  \caption{Dipole MDAH.220118 moved \SI{9.4}{\meter} downstream}
  \label{fig:16TT}
\end{figure}

The existing dipole MDAH.220118 on beamline TT22 clashes with the vacuum chamber on the BDF beamline. 
The proposed solution is to move dipole MDAH.220118 downstream approx.~\SI{9.4}{\meter} next to quadrupole QTAF.220300 as shown in Fig.~\ref{fig:16TT}.

\begin{figure}[ht!]
  \centering
  \includegraphics[width=4.9in,height=2.8in]{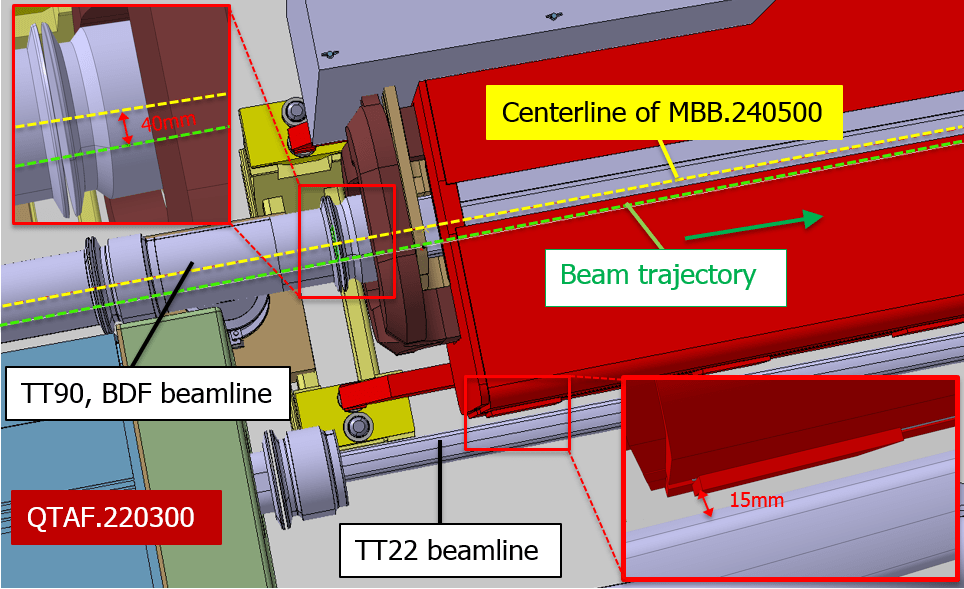}
  \caption{MBB bending magnet orientated with respect to the beamline trajectory}
  \label{fig:17TT}
\end{figure}

The bending magnet MBB.240500 on the BDF beamline clashes with the vacuum chamber on the TT22 beamline. 
The magnet is orientated by \ang{0.368} about the centre of the magnet core’s downstream end with respect to the beamline trajectory. 
The orientation creates a \SI{40}{\milli\meter} offset between the centreline of the magnet and the beam trajectory, it also creates a \SI{15}{\milli\meter} clearance between the magnet and the vacuum chamber as shown in Fig.~\ref{fig:17TT}.

\subsubsection{Beam Instrumentation}  \label{integ:sec:instrumentation}

For the layout of the beam instrumentation in the BDF beamline see Ref.~\cite{BDFbeamlineIS}. 
The beam instrumentation’s control system is composed of two racks that are to be housed in the Auxiliary Building.

\noindent \textbf{Changes to existing beamline TT22:} 

\begin{figure}[htbp]
  \centering
  \includegraphics[width=5.1in]{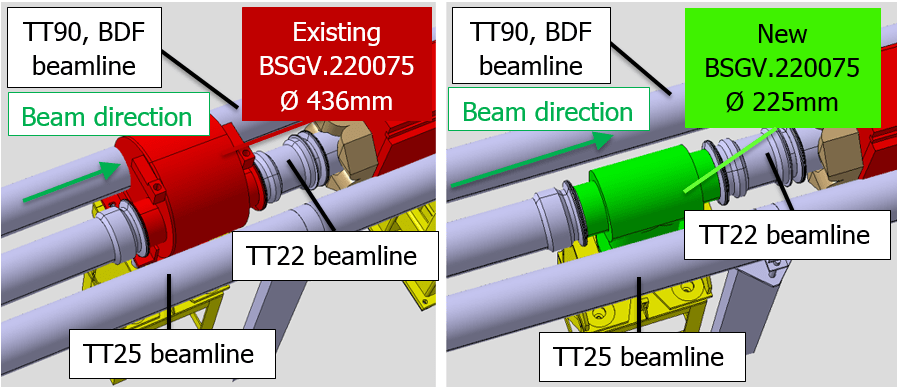}
  \caption{Secondary emission monitor BSGV.220075 clash with the BDF beamline}
  \label{fig:18TT}
\end{figure}

The existing BSGV.220075\footnote{BSGV -- Secondary Emission Monitor (SEM)} monitor’s exterior casing (Ø~\SI{436}{\milli\meter}) on beamline TT22 clashes with the vacuum chamber on the BDF beamline. 
A preliminary design was undertaken by the beam instrumentation group of a new beam monitor with a small exterior casing (Ø~\SI{225}{\milli\meter}), of which fit between the new BDF beamline and the existing TT22 beamline as shown in Fig.~\ref{fig:18TT}. 

\subsubsection{Vacuum} \label{integ:sec:vac}

For the vacuum layout of the BDF beamline see Ref.~\cite{BDFvacuumIS}. 

\noindent \textbf{New BDF beamline vacuum equipment:} 

Approximately~\SI{200}{\meter} of new vacuum chamber is required for the BDF beamline (each type of magnet has a specific vacuum chamber of which is not included within the \SI{200}{\meter} length). 
The new vacuum chamber consists of two types of varying lengths, which are consistent with the existing vacuum chambers in the TDC2 tunnel. 
A standard circular vacuum chamber with an outer diameter of \SI{159}{\milli\meter} and a rectangular vacuum chamber with an external envelope of 156$\times$\SI{64}{\milli\meter}. 
Each length of vacuum chamber is supported by at least two supports, one of which is fixed and the other free in the longitudinal direction. 
The number of bellows has been chosen such that between any two magnets there is a bellow. 
The number and type of vacuum components for the BDF beamline were qualitatively chosen, as shown in Table \ref{Table1TT}.

\begin{table}[htpb]
 \center
 \caption{ New vacuum equipment for the BDF beamline }
 \label{Table1TT}
\begin{tabular}{cc}
\hline 
\textbf{Component}  & \textbf{ Quantity}  \\
\hline
Recombination chamber & 1 \\
 Vacuum pump, VPIA & 7  \\
Vacuum pump, VPIB & 5 \\
Roughing valve, VVRA & 6 \\
 Sector valve, VVS & 2 \\
 Vacuum gauge, VG & 2 \\
 Venting valve, VVV & 2 \\
Vacuum window & 1 \\
\hline
\end{tabular} 
\end{table}

The vacuum equipment includes: 

\begin{itemize}

\item A recombination chamber, for which a new design is required to be undertaken. This vacuum chamber creates a three way split for the BDF, TT22 and the TT25 beamlines (shown in Fig.~\ref{fig:4TT}). For the integration purposes a preliminary model has be used in the beamline model.
\item Two types of vacuum pumps (VPIAs and VPIBs).
\item Roughing valves (VVRA) equally spaced along the beamline, the first one located at the recombination chamber.
\item  Two sector valves (VVS), one after the recombination chamber and one after the MDX magnets dilution system; therefore, creating two sectors.
\item  One vacuum gauge assembly (VG) per sector (a combination of two types pirani penning). For interlock requirements, a vacuum gauge is required to be located before the sector valve.
\item One venting valve (VVV) per sector.
\item   A vacuum window is required at end of the beamline, for the preliminary integration study, a titanium window was used (see Ref.~\cite{BDFvacuumwindow} for further details).

\end{itemize}

A compressed airline is required for the sector valves and the roughing valves. The sector valves requires a constant compressed air connection; however, it only requires the supply during opening and closing. The VVR only requires the compressed air connection during interventions. It is noted that the vacuum specifications are to be the same as the existing compressed airline in the TDC2 tunnel. The pressure required is between \SIrange{6}{7}{\bar}; however, a pressure of \SI{10}{\bar} is required to account for losses along the compressed airline of which will be regulated down to the required pressure. 

A \SI{220}{\volt} connection is required for vacuum maintenance along the beamline. The vacuum pressure of the BDF beamline should be that of the existing TDC2 tunnel vacuum system. There are to be three 45U racks housed in the Auxiliary Building for the vacuum control system of which have three power outlets \SI{16}{\ampere} – 10 sockets (General network) and three power outlets \SI{10}{\ampere} – 5 sockets (uninterruptible power supply).

\noindent \textbf{Changes to existing beamline TT22:} 

\begin{figure}[htbp]
  \centering
  \includegraphics[width=5.2in]{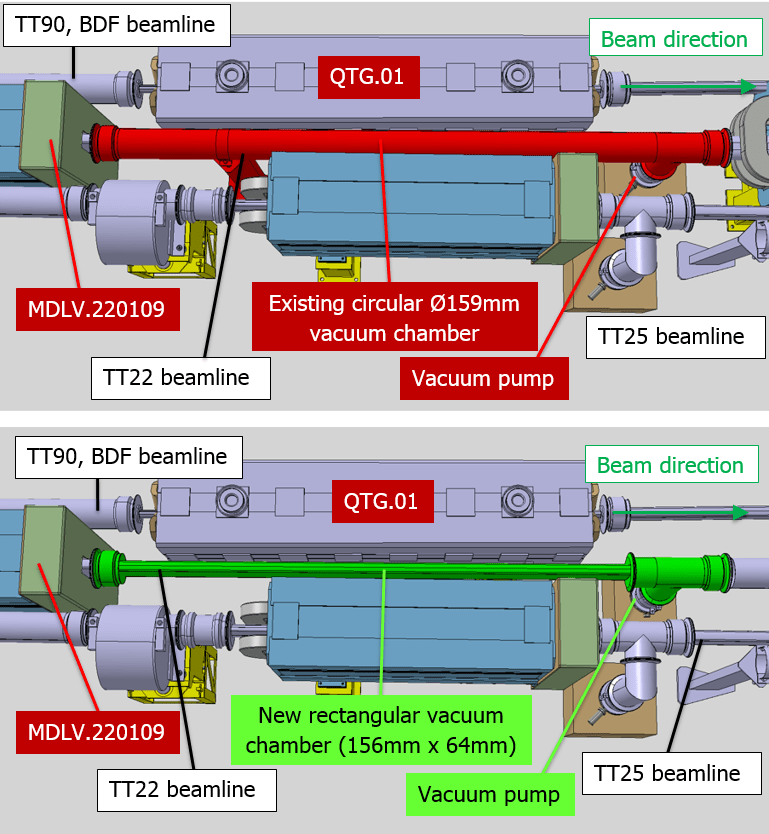}
  \caption{Quadrupole QTG.01 clash with a vacuum chamber on the TT22 beamline}
  \label{fig:20TT}
\end{figure}

The vacuum chamber downstream of dipole MDLV.220109 on the TT22 beamline clashes with quadrupole QTG.01 on the BDF beamline. 
The existing circular vacuum chamber (Ø~\SI{159}{\milli\meter}) with a port for a VPIB vacuum pump was changed to rectangular vacuum chamber (156$\times$\SI{64}{\milli\meter}) with a short  circular vacuum chamber (Ø~\SI{159}{\milli\meter}) with a port for a VPIB vacuum pump as shown in Fig.~\ref{fig:20TT}. 

\begin{figure}[htbp]
  \centering
  \includegraphics[width=4.9in]{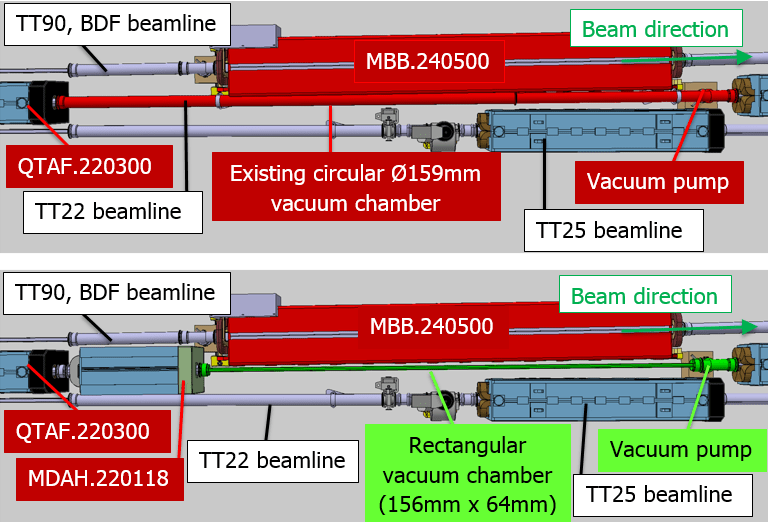}
  \caption{BDF magnet MBB.240500 clash with a vacuum chamber on the TT22 beamline}
  \label{fig:21TT}
\end{figure}

The vacuum chamber downstream of quadrupole QTAF.220300 on the TT22 beamline clashes with bending magnet MBB.240500 on the BDF beamline. The existing circular vacuum chamber (Ø~\SI{159}{\milli\meter}) with a port for a VPIB vacuum pump was changed to rectangular vacuum chamber (156$\times$\SI{64}{\milli\meter}) with a short circular vacuum chamber (Ø~\SI{159}{\milli\meter}) with a port for a VPIB vacuum pump as shown in Fig.~\ref{fig:21TT}.

\subsection{Electrical Network System}
\label{ENELTT}

For details on the main electrical engineering infrastructure required for the beamline, see Section \ref{ENEL}.

To construct the Junction Cavern, part of the existing TDC2 tunnel will be demolished. 
During LS3 long shutdown, all of the equipment in the Junction cavern will be removed prior to the demolition (see Ref.~\cite{BDFequipmentremoval} for further information on the extent of the equipment removal).
The existing cabling in the TDC2 tunnel should be checked during the LS2 long shutdown such that the demolition and construction of the Junction Cavern can be undertaken during the LS3 long shutdown. 
This exercise requires full identification of all electrical services (including cables and fibres pulled for the equipment groups) running within the tunnel from the upstream TT20 tunnel that connects with the TDC2 tunnel, to the downstream TCC2 cavern as potentially all these services will be cut for the duration of the works. 
For example, if critical services (Level 3 alarms, Network, fixed or mobile telephone connectivity) are compromised, a significant period of time will be required to put a temporary solution in place prior to the start of the demolition.

It is foreseen that all the existing cabling in the Junction Cavern area will be reused after the equipment deinstallation as the infrastructure is relatively new.

During the construction of the Junction Cavern and demolition of the TDC2 tunnel section, lighting is required in the upstream area between the Junction Cavern and the TDC2 sector door. 
Therefore, a temporary electrical feed will be required from the upstream TT20 tunnel, for which cabling can be brought from a sump located close to the TDC2 sector door.

The existing TCC2 cavern, downstream of the Junction Cavern, may be used as a marshalling area during the construction of the cavern. 
CERN’s electrical engineering group’s recommendation is that the LED lighting in the TCC2 cavern should be upgraded to increase the power by \SIrange{30}{50}{\percent}.

Another recommendation is that HALFEN rails (Ref.~\cite{HALFEN}) or equivalent are installed in the Junction Cavern and the Extraction Tunnel, as this would prevent drilling into the tunnel and allow some flexibility during installation and maintenance of the electrical infrastructure. 

It is foreseen that the BDF beamline in the existing TDC2 tunnel will not require new cable trays as the cabling requirements can be integrated into the existing cable trays currently in place.

In the Junction Cavern, new cable trays are to be installed on the Jura side wall.  It is foreseen that the number of cable trays will be similar than that of the TDC2 tunnel, two for control, two for power and one for general services. The cable trays and cabling should be installed before the beamline. As the Junction Cavern is a new structure, wider than the existing structure, conventional and emergency lighting are to be installed on the Jura side, as shown in Fig.~\ref{fig:6TT}. The cable trays will continue from the Junction Cavern into the Extraction Tunnel, however, they will be located on the Salève side wall of the Extraction Tunnel. Conventional and emergency lighting are to be installed on the Jura side, as shown in Fig.~\ref{fig:11TT}. A typical cross-section of the Extraction Tunnel, showing the different service and geometry requirements is shown in Fig.~\ref{fig:13TT}.

\subsection{Survey}

During the Junction Cavern equipment removal process, one of the existing survey pillars will be removed in the Junction Cavern area, of which will be replaced with a new survey pillar after construction. During the construction, survey floor points are to be installed at \SI{25}{\meter} intervals; either in the middle of transport side or the personnel side. Cast-iron covers are to be used to enable transport vehicles to drive over the survey floor points. The floor points are to be installed by civil engineering and the survey group will provide the steel cup and screws, see Ref.~\cite{GGPSOprocedure} for further details.

The reference survey points (fiducials), which are supplied by the equipment group, will be installed permanently on the equipment before the equipment is installed inside the Transfer Tunnel. The equipment owner may request the survey group’s help to align the survey fiducial (fiducialisation). 

To align the equipment, the reference survey points include at least: one reference point at the beam entry, one reference point at the beam exit and one reference surface or cylinder for the roll angle measurement around the beam axis. Each equipment must also include an alignment system that can move the equipment vertically, radially and longitudinally. This alignment system and survey reference must be validated by the survey section. 

After the Junction Cavern construction and equipment installation, the survey group will align all the equipment (new and existing).  As there is currently some structural movement in the TT20 and the TDC2 tunnels, survey has foreseen four different structural movements in the TDC2 tunnel, the TCC2 cavern, the Extraction Tunnel and the SHiP Experimental Hall. Therefore, the beam equipment has to be monitored every 4 to 6 months to check for movements.

The new survey reference network consists of new survey pillars and brackets. From at least one pillar position, each equipment must be measurable. The positions of the brackets are located within the \SI{200}{\milli\meter} allowance for emergency lighting and services (see Fig.~\ref{fig:13TT}). The pillars are located between the existing magnets and vacuum chambers. A clear line of sight is required between the brackets and pillars with the maximum distance between any pillar and bracket between \SIrange{25}{30}{\meter}. Therefore, the new survey system is composed of:

\begin{itemize}

\item 1 survey pillar in the TDC2 tunnel;
\item 2 survey pillars in the Junction Cavern;
\item 1 survey pillar in the TCC2 cavern;
\item 1 survey bracket in the Junction Cavern;
\item 1 survey bracket in TDC22 tunnel;
\item 7 survey brackets in the Extraction Tunnel.

\end{itemize}

\begin{figure}[htbp]
  \centering
  \includegraphics[width=5.4in]{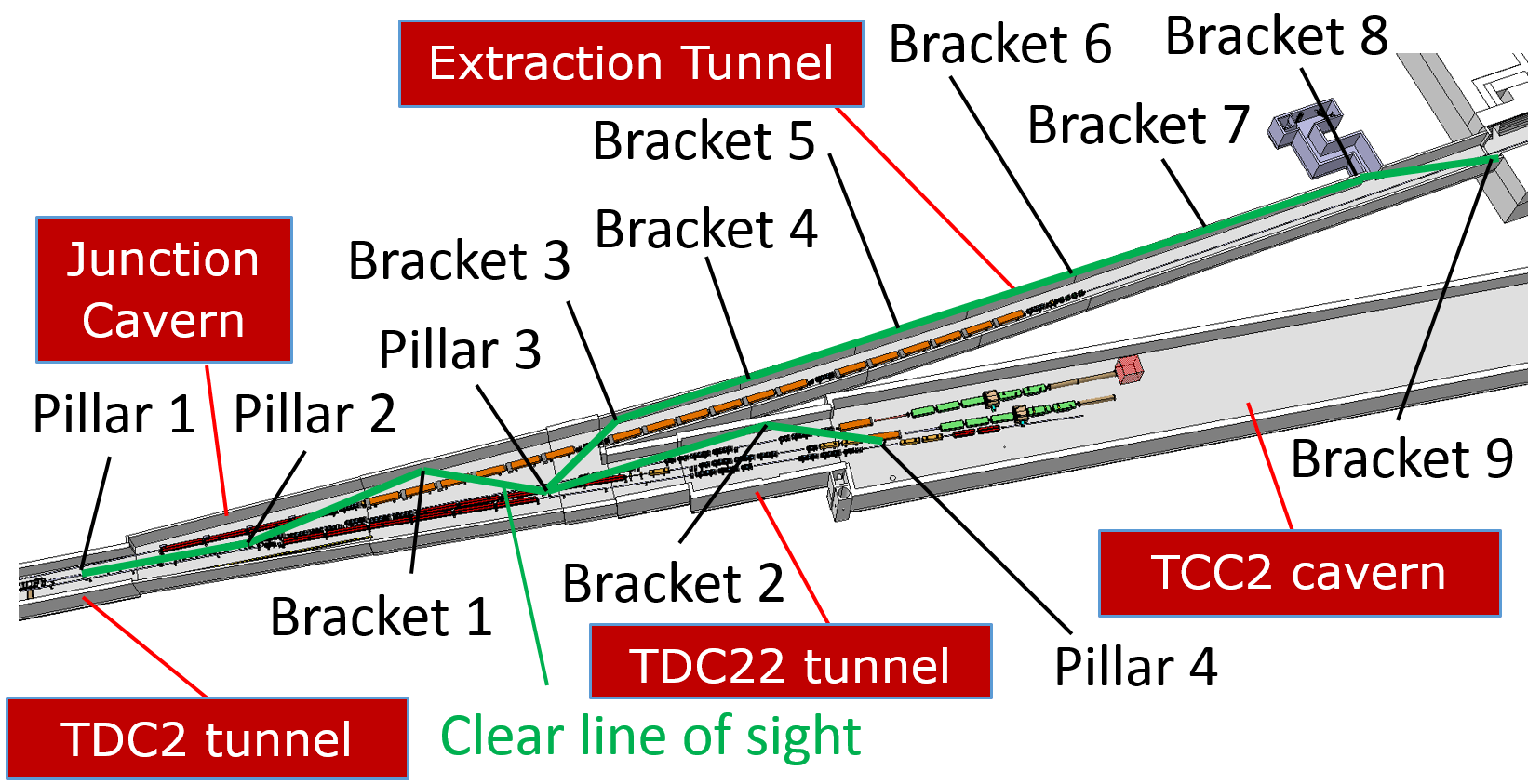}
  \caption{Survey pillar and wall bracket locations in the transfer tunnel}
  \label{fig:22TT}
\end{figure}

The position of the pillars and brackets are shown in Fig.~\ref{fig:22TT}. Some additional small references may be added later to add redundancy in the survey network. The beamline survey network will be connected to the SHiP alignment network via a clear line of sight through the Target Complex (see Section \ref{SHiPsurvey} for further details).

\FloatBarrier

\subsection{Radiation Protection}

The BDF beamline has required important radiation protection constraints to be incorporated into the overall facility layout. This has included specific thicknesses of soil above the tunnel for shielding purposes (see Ref.~\cite{TTstructuralspec}) and suitable fencing requirements at ground level to prevent personnel access (see Section \ref{Auxperchic}). Refer to Chapter~\ref{Chap:RP} for details on the RP specifications for the facility.

\subsection{Safety}

\subsubsection{Fire and Alarms}

The safety requirements for the transfer tunnel has been a key aspect in the overall layout of the facility. This has included fire protection, detection, alarms and evacuation. Refer to Chapter~\ref{Chap:Safety} for further details.

\subsubsection{Access Control}

\begin{figure}[htbp]
  \centering
  \includegraphics[width=\linewidth]{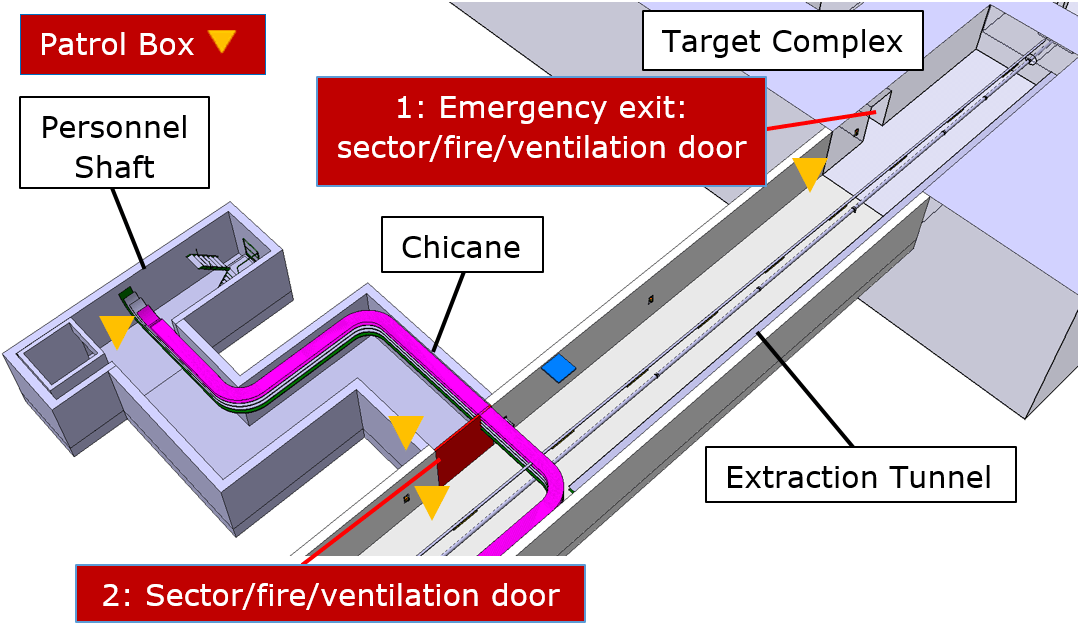}
  \caption{Access control at the downstream end of the Extraction Tunnel }
  \label{fig:25TT}
\end{figure}

\begin{figure}[htbp]
  \centering
  \includegraphics[width=\linewidth]{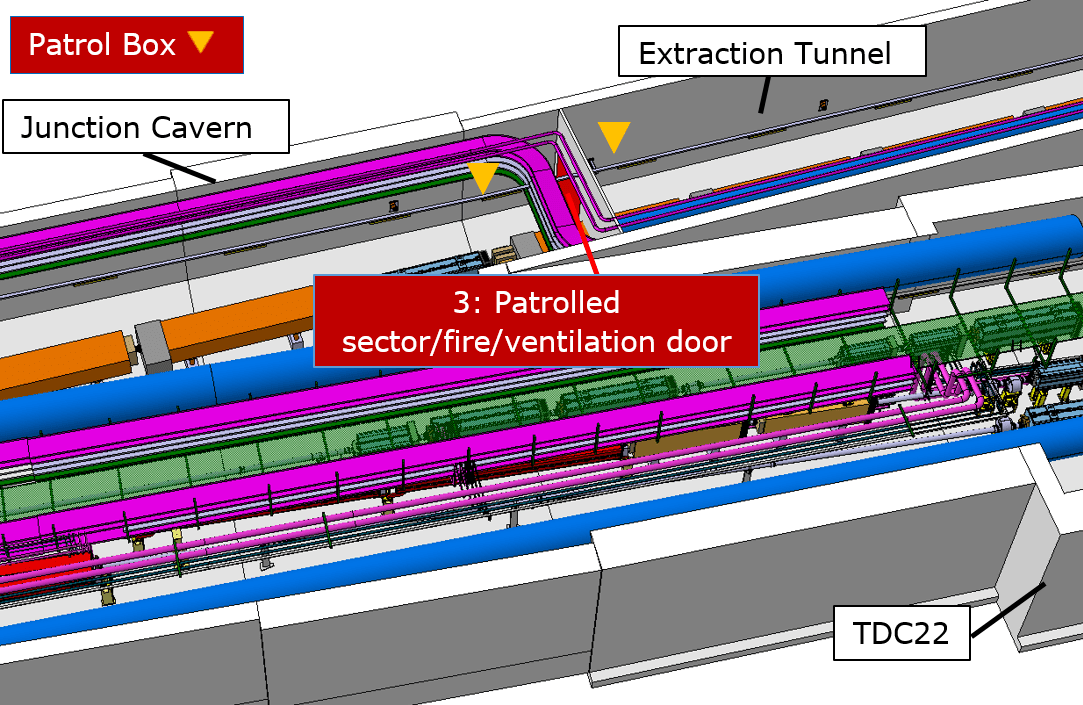}
  \caption{Access control at the upstream end of the Extraction Tunnel}
  \label{fig:26TT}
\end{figure}

The BDF access control will be controlled from the CERN Control Center (CCC) located in building 874 in the CERN Pr\'evessin site. As the Target Complex, Access Building and Auxiliary Building (including the personnel access to the transfer tunnel) are located next to one another, they will share the same badge control station. The access control system for the transfer tunnel is shown in Fig.~\ref{fig:25TT} and Fig.~\ref{fig:26TT}, it includes:

\begin{enumerate}

\item  Emergency exit with a sector/fire/ventilation door between the Extraction Tunnel and the Target Complex;
\item  Sector/fire/ventilation door between the Chicane and the Extraction Tunnel;
\item  Patrolled sector/fire/ventilation door between the Junction Cavern and the Extraction Tunnel;
\item  Six patrol boxes located throughout the transfer tunnel in accordance with Access Control’s recommendations.

\end{enumerate}

\newpage
\section{Access Building and Auxiliary Building}
\label{AuxiliaryB}

The service building for the transfer tunnel and the Target Complex and the equipment access building for the transfer tunnel are located upstream (South-West) of the Target Complex, on the Jura (North-West) side of the Extraction Tunnel as shown in Fig.~\ref{fig:BDF integration overview}.

\begin{figure}[htbp]
  \centering
  \includegraphics[width=\linewidth]{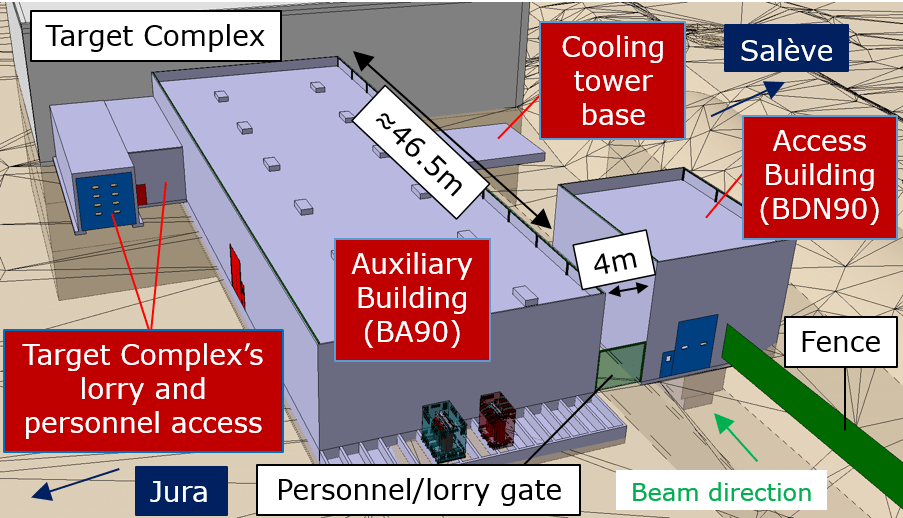}
  \caption{Access and Auxiliary Buildings}
  \label{fig:2SB}
\end{figure}

\begin{figure}[htbp]
  \centering
  \includegraphics[width=5.8in]{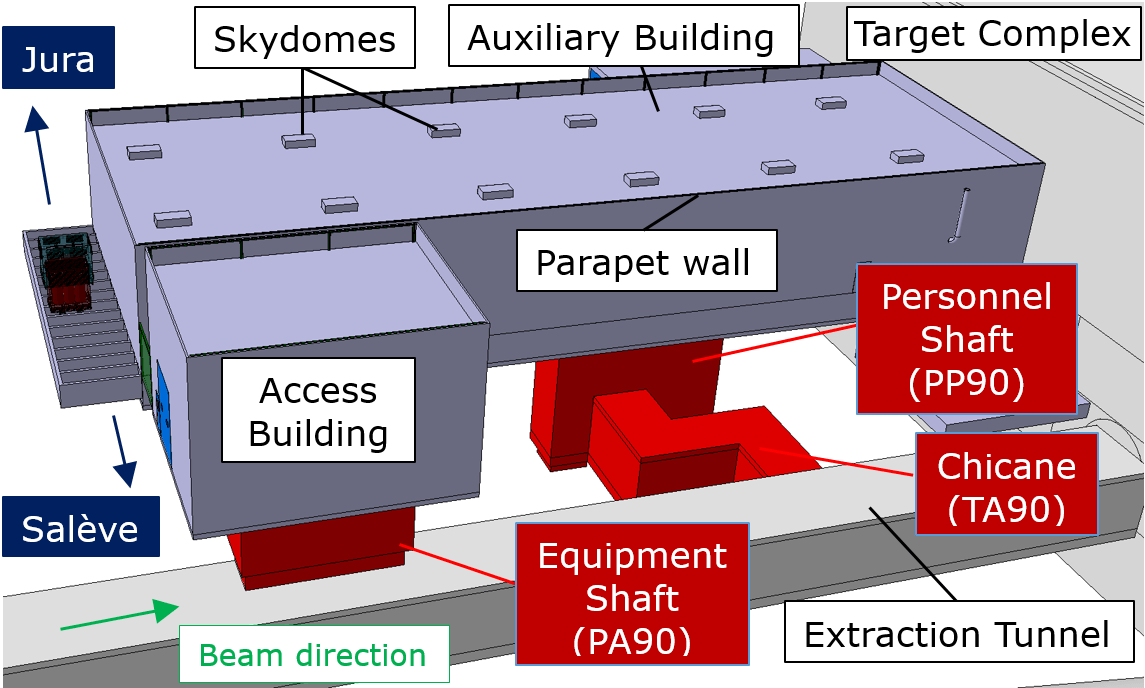}
  \caption{Equipment Shaft, Personnel Shaft and Chicane}
  \label{fig:3SB}
\end{figure}

The structures consists of two surface buildings (Access Building and Auxiliary Building) of which have a Finished Floor Level (FFL) of \SI{451.1}{\meter} above sea level. The Auxiliary Building's downstream (North-East) wall is directly next to the Target Complex and its Jura (North-West) side wall is next to the Target Complex's lorry and personnel access. The Access Building is approx.~\SI{46.5}{\meter} from the Target Complex and \SI{4}{\meter} from the Auxiliary Building. Located next to the Auxiliary Building is a cooling tower base as shown in Fig.~\ref{fig:2SB}. Underneath the Auxiliary Building is a Personnel Shaft and Chicane and inside the Access Building is an Equipment Shaft, both of which are connected to the Extraction Tunnel as shown in Fig.~\ref{fig:3SB}.

The purpose of these structures includes the following: 

\begin{itemize}

\item Beamline equipment access to the transfer tunnel;
\item Personnel access to the transfer tunnel;
\item Service storage and distribution for the transfer tunnel, the Target Complex and the Experimental Area.

\end{itemize}

The equipment includes magnets, beam instrumentation, vacuum equipment, general electrical and safety equipment. The services include the electrical services, cooling and ventilation. 

\subsection{ Access Building (BDN90) and Equipment Shaft (PA90)}
\subsubsection{Summary}

\begin{figure}[htbp]
  \centering
  \includegraphics[width=5in]{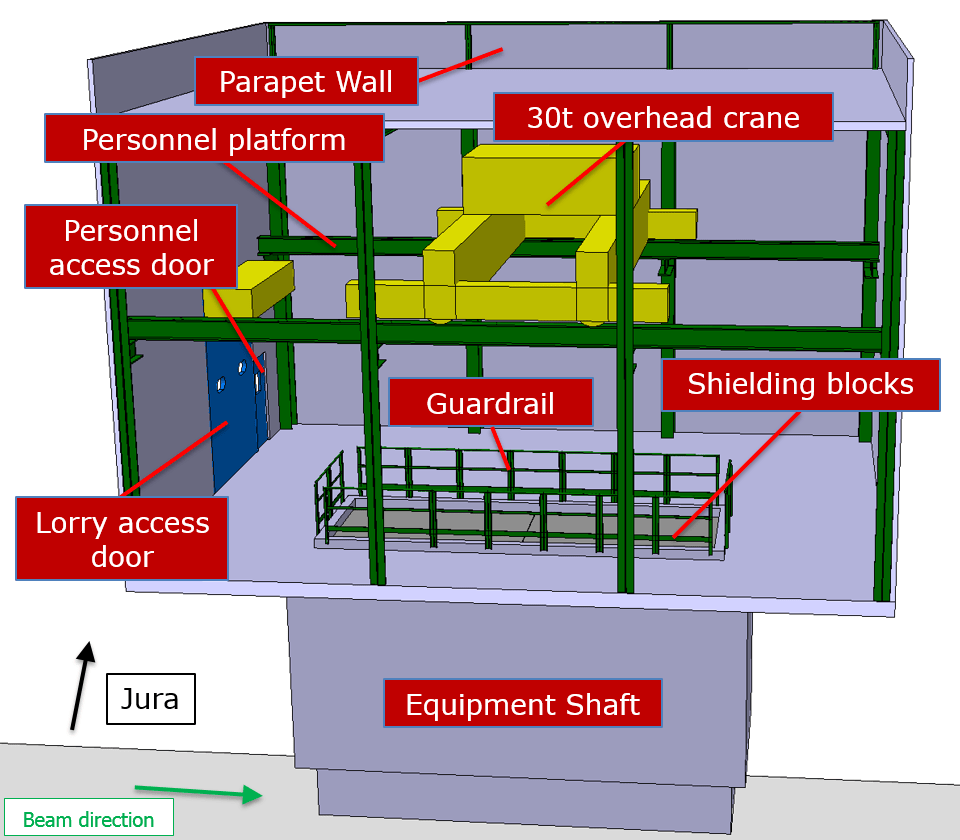}
  \caption{Access Building’s integration layout}
  \label{fig:4SB}
\end{figure}

The Access Building (BDN90) is a single storey industrial building located directly over the Extraction Tunnel. Located inside the building is an Equipment Shaft (PA90) that extends from the floor level of the Access Building to the top of the Extraction Tunnel. For the integration studies, a simple single storey steel framed building with an exterior cladding and a concrete floor and equipment shaft has been modelled. Fig.~\ref{fig:4SB} shows the proposed integration layout of the Access Building and the Equipment Shaft.

The Equipment Shaft is located directly over the Extraction Tunnel as this allows the equipment to be lowered directly onto the Extraction Tunnel floor. The alternative method would have been a layout similar to that of the Personnel Shaft and required the tunnel to be locally widened to allow the transport vehicle sufficient space to turn, for which the design was foreseen to be less effective in terms of cost and feasibility.

During the equipment installation, transport vehicles including a \SI{40}{\tonne} semi-trailer, a \SI{19}{\tonne} truck and a forklift will move the equipment into the Access Building, from which a \SI{30}{\tonne} overhead crane will lift and lower the equipment into the Equipment Shaft. A transport handling vehicle inside the Extraction Tunnel will collect the equipment for positioning along the beamline. 

The Equipment Shaft is filled with nine concrete shielding blocks during beam operation for RP considerations and surrounding the Equipment Shaft is a guard rail for personnel safety. During installation and deinstallation of equipment into the tunnel, the shielding blocks are removed from the shaft and stored outside the building. The specific storing location for the shielding blocks outside the building has not been defined; however, it is foreseen that the surface outside the building should be of sufficient capacity to temporarily store the shielding blocks during the installation and deinstallation procedure.

Inside the building, there is a personnel platform at the height of the crane rail on the Jura (North-West) side to allow personnel to access the crane. To access the platform, there is a ladder attached to the side wall. For ventilation requirements, there will be an Air Handling Unit supported by the wall/columns inside the building. 

\subsubsection{ Geometry}

\begin{figure}[htbp]
  \centering
  \includegraphics[width=5.8in]{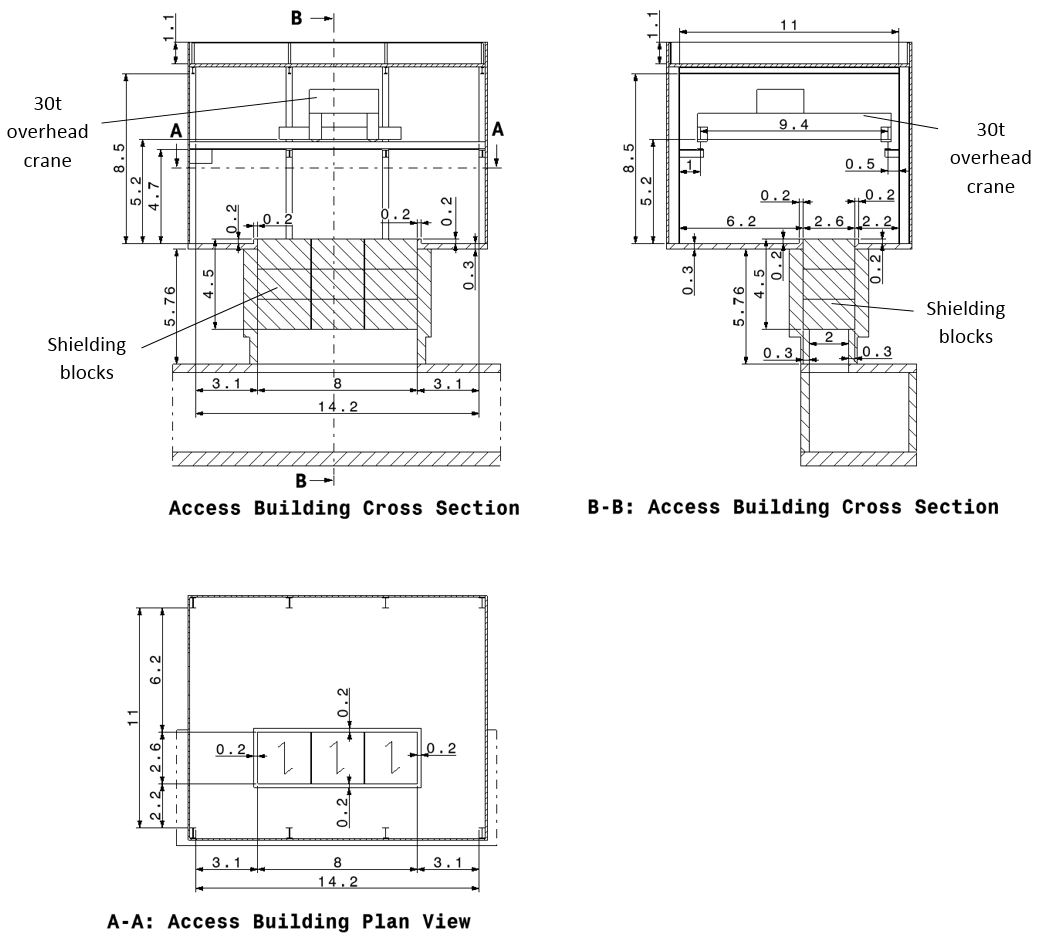}
  \caption{Plan and section views of the Access Building}
  \label{fig:5SB}
\end{figure}

Fig.~\ref{fig:5SB} shows the plan view and sections of the Access Building and the Equipment Shaft. For a detailed layout see Ref.~\cite{Accessbuildinglayout}.

The building has an internal geometry of \SI{14.2}{\meter} in length, \SI{11}{\meter} in width and \SI{8.5}{\meter} in height. The geometry of the structure is based on the required distances for the \SI{30}{\tonne} overhead crane, transport vehicles and personnel inside the building. 

The width of the structure (\SI{11}{\meter}) is based on the equipment shaft opening (\SI{2.6}{\meter}), a personnel accessway (\SI{1.5}{\meter}), a semi-trailer (\SI{2.5}{\meter}) and the clearance of the overhead crane at both sides (2 x \SI{2.2}{\meter}). The length of the structure (\SI{14.2}{\meter}) is based on the equipment shaft (\SI{8}{\meter}) and the clearance of the overhead crane at both sides (2 x \SI{3.1}{\meter}).

The shaft length is \SI{8}{\meter} long and \SI{2}{\meter} wide at the bottom, for transporting heavy equipment in and out of the Extraction Tunnel (the largest magnets are approx.~\SI{7}{\meter} in length and approx.~\SI{1.5}{\meter} in width). The top of the shaft is \SI{2.6}{\meter} wide as there is a \SI{300}{\milli\meter} wide support on either side of the shaft for the shielding blocks. There is a \SI{200}{\milli\meter} upstand all around the Equipment Shaft to prevent water entering into the Extraction Tunnel. A preliminary one way spanning shielding block support arrangement has been outlined, as shown in Fig.~\ref{fig:5SB}. 

The height of the structure (\SI{8.5}{\meter}) is based on the height of the lorry access door (\SI{4.7}{\meter}), clearance between the top of the door and the underside of the crane (\SI{0.5}{\meter}), the height of the crane (\SI{2.5}{\meter}), clearance between the top of the crane and the underside of the services (\SI{0.5}{\meter}) and an allowance for lighting and services (\SI{0.3}{\meter}).

To access the overhead crane for maintenance there is a personnel platform, for which the crane is offset from the wall/column by \SI{1}{\meter}. On the other side, the crane is offset by \SI{0.5}{\meter} for the minimum crane clearance. On the roof there is a \SI{1.1}{\meter} parapet wall for personnel access safety. 

For further details on the minimum required thickness of concrete shielding for the structures see Ref.~\cite{SBstructuralspec}. Refer to Chapter~\ref{Chap:RP} for details on the RP specifications for the structures. 

\subsection{ Auxiliary Building (BA90), Personnel Shaft (PP90) and Chicane (TA90)}
\label{Auxperchic}

\subsubsection{Summary}

\begin{figure}[htbp]
  \centering
  \includegraphics[width=5.5in]{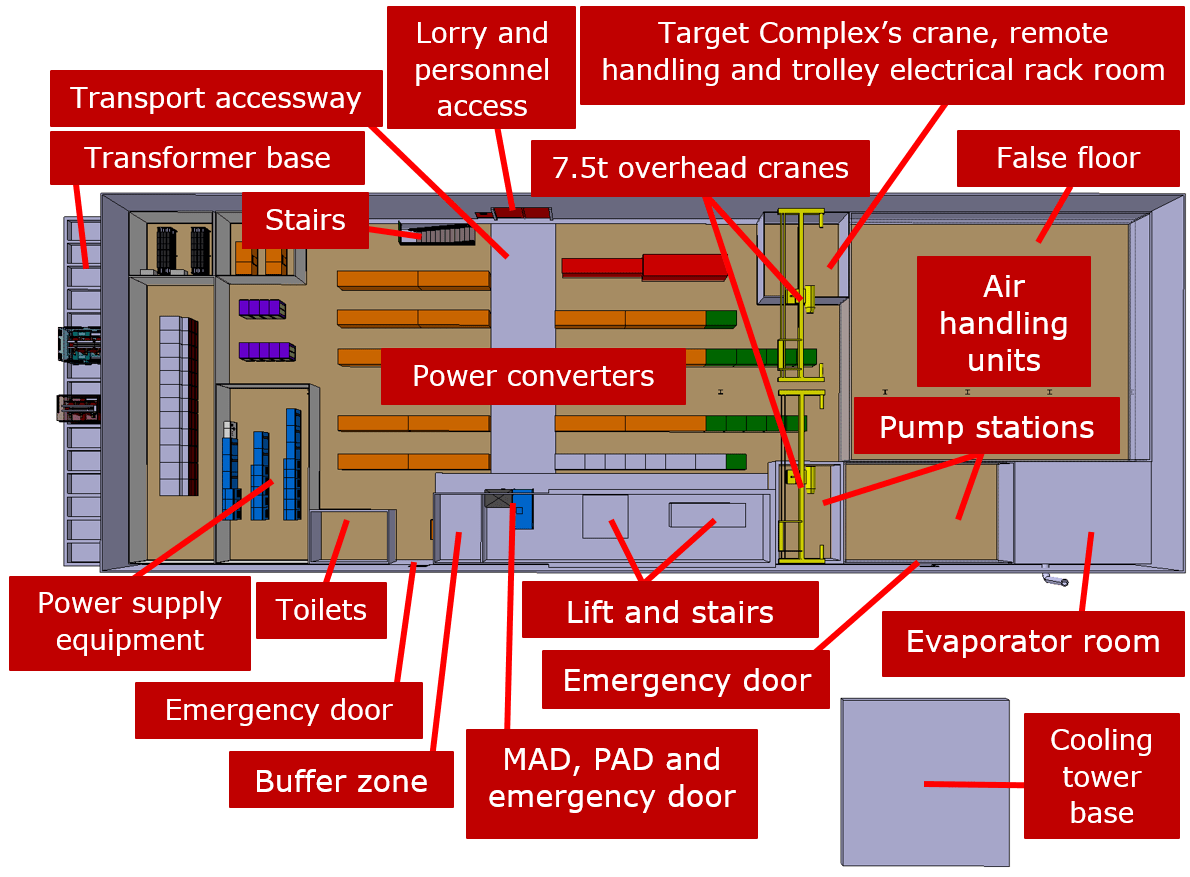}
  \caption{Auxiliary Building’s integration layout}
  \label{fig:6SB}
\end{figure}

The Auxiliary Building (BA90) houses the power supply, control racks and the cooling and ventilation equipment for the Extraction Tunnel and the Target Complex (Target Complex's service requirements provided in Ref.~\cite{Targetservices}). The proposed integration layout is shown in Fig.~\ref{fig:6SB}, it follows that of a typical service building at CERN.

\begin{figure}[htbp]
  \centering
  \includegraphics[width=5.1in]{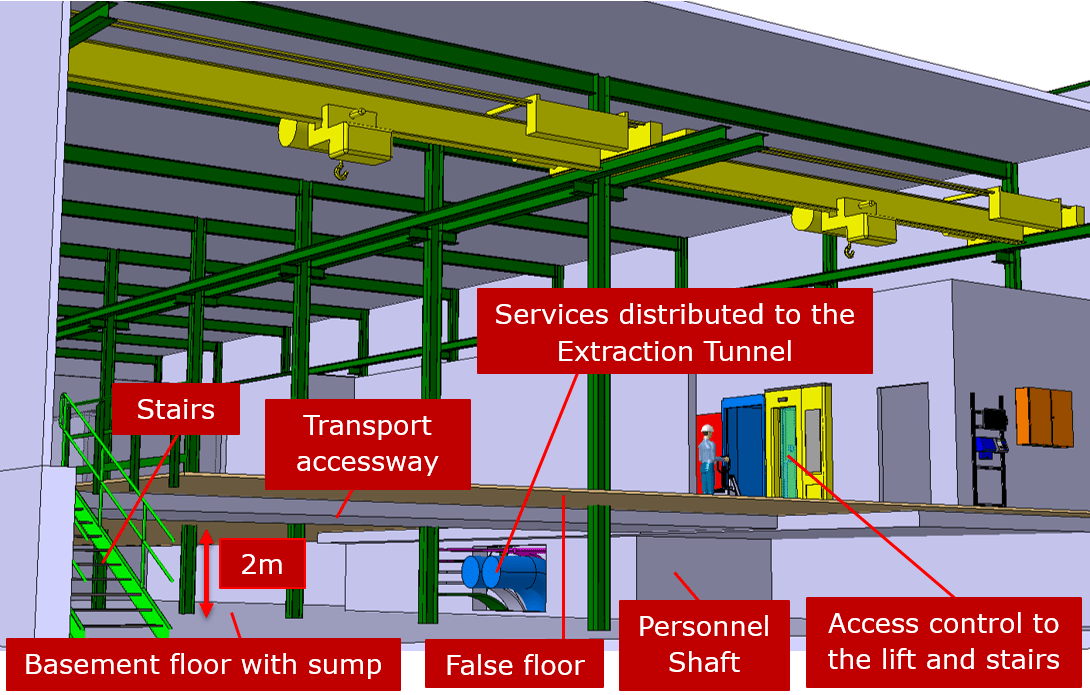}
  \caption{Auxiliary Building’s false floor, transport accessway and Personnel Shaft}
  \label{fig:7SB}
\end{figure}

The cable trays, cooling piping and ventilation ducts are distributed to the Extraction Tunnel via the Personnel Shaft, from which services pass underneath the personnel accessible false floor and into the Personnel Shaft via a hole as shown in Fig.~\ref{fig:7SB}. Services for the Target Complex housed in the Auxiliary Building, also pass through a hole in the walls between the two structures. Both holes are to be sufficiently waterproofed and airtight.

The access control for the Extraction Tunnel is housed inside the Auxiliary Building, of which is composed of a material access door (MAD), a personnel access door (PAD) and an emergency door. Through the access control there is a personnel lift and stairs of which brings personnel and light equipment down to the Chicane. The lift and stairs are fire protected and not connected to the general electrical circuit for safety. A buffer zone for storing material brought up from the tunnel during maintenance works is located next to the access control.

Transport vehicles including a \SI{19}{\tonne} truck and a standard forklift drive the equipment into the building on the transport accessway. The transport accessway is suspended to allow the services to pass beneath the accessway. At end of the transport accessway, next to the access control there are toilets. 

Two \SI{7.5}{\tonne} overhead cranes move the equipment into position throughout the building. There is a ladder attached to the side wall for personnel to access and maintain the overhead crane.

On the roof of the Auxiliary Building there are natural smoke extraction units (skydomes) distributed throughout the roof, as shown in Fig.~\ref{fig:3SB}. Outside the building on the upstream side, is a transformer base to support the transformers. The cabling for the transformers will run from the base to underneath the false floor to be distributed to the various electrical equipment. Also outside, on the Salève side of the building, is a cooling tower base. For RP considerations, to access the cooling tower there is a personnel/lorry gate between the Access and Auxiliary Building and a fence running alongside the Extraction Tunnel, as shown in Fig.~\ref{fig:2SB}. 

\begin{figure}[htbp]
  \centering
  \includegraphics[width=5.2in]{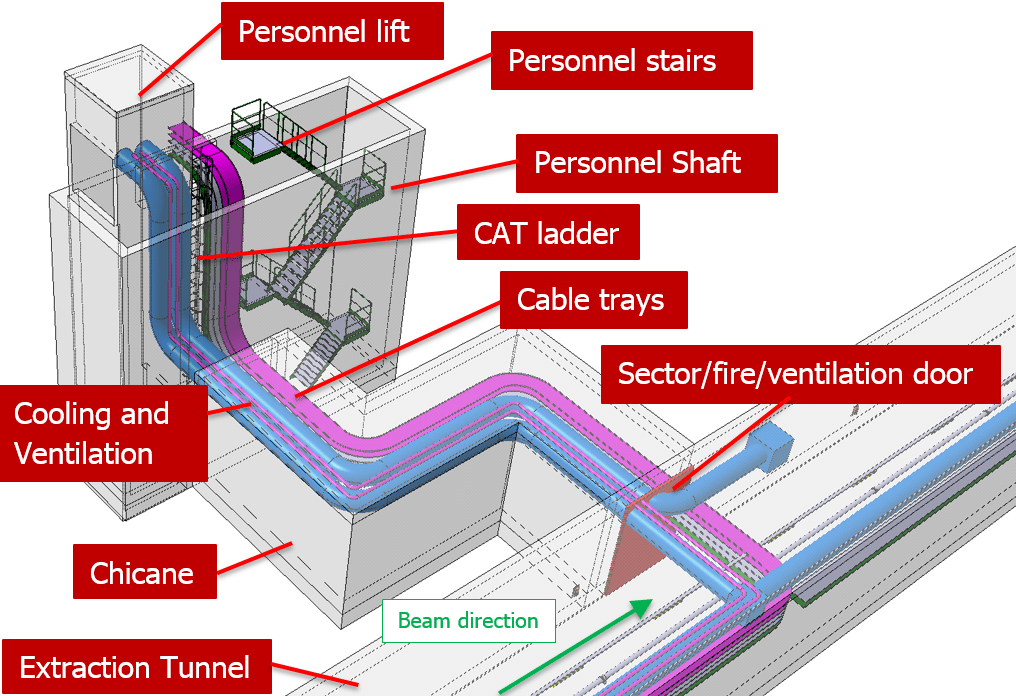}
  \caption{Personnel Shaft’s and Chicane’s integration layout}
  \label{fig:9SB}
\end{figure}

The Personnel Shaft (PP90) and Chicane (TA90) are located underneath the Auxiliary Building. The shaft houses a personnel lift, a CAT ladder, a metallic personnel stairs and the services (ventilation, cooling and cable trays) from the Auxiliary Building. The CAT ladder is composed of two sections (maximum height \SI{6}{\meter}) with an intermediate platform and a platform at the top. The shaft brings personnel from the FFL of the Access Building to the FFL of the Extraction Tunnel. The services are brought into the Personnel Shaft from underneath the false floor in the Auxiliary Building to the ceiling level of the Chicane (see Fig.~\ref{fig:7SB} and Fig.~\ref{fig:9SB}). Personnel and services enter into the Extraction Tunnel via the Chicane. At the end of the Chicane at the entrance to the Extraction Tunnel there is a sector/fire/ventilation door.

\subsubsection{ Geometry}

\begin{figure}[htbp]
  \centering
  \includegraphics[width=\linewidth]{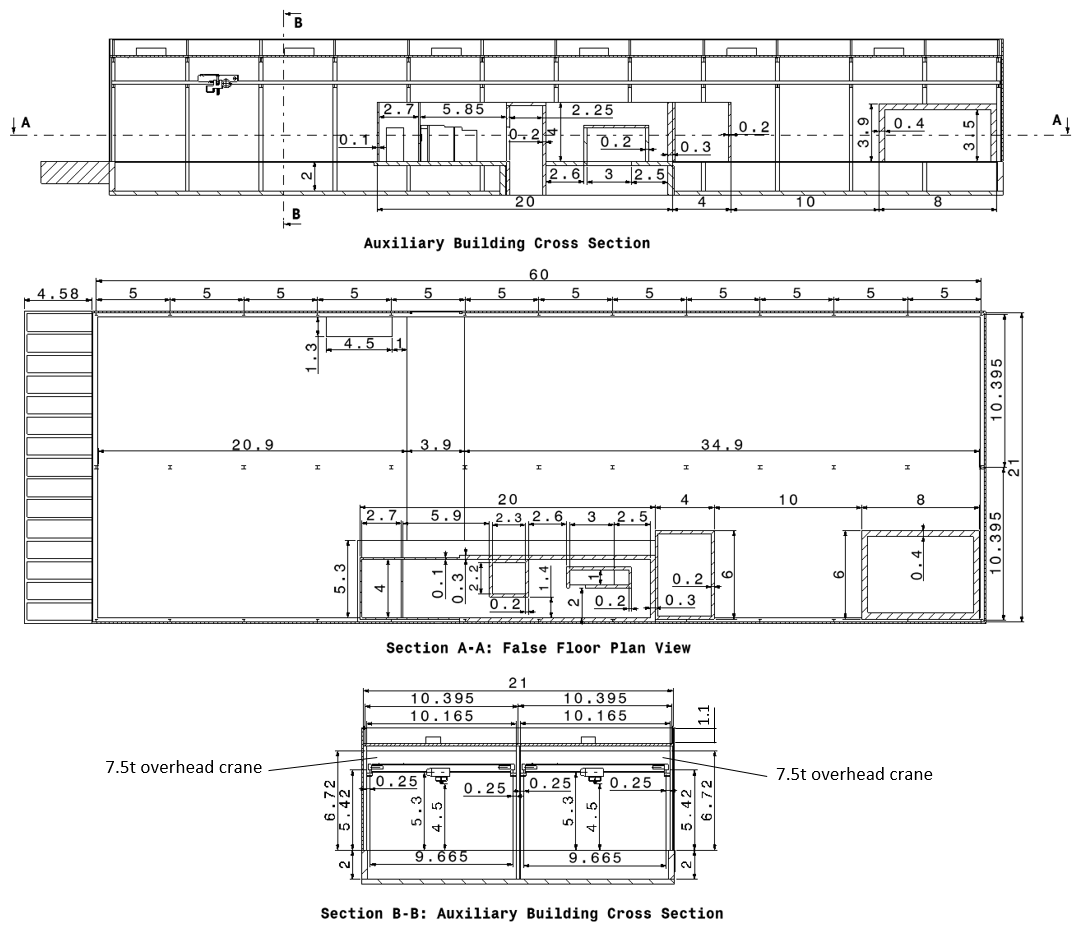}
  \caption{Auxiliary Building’s structural geometry}
  \label{fig:10SB}
\end{figure}

The Auxiliary Building is a single storey steel framed industrial building, \SI{21}{\meter} in width between the outer edges of the external columns of which is composed of 3 columns with a transversal c/c spacing of approx.~\SI{10.4}{\meter}. The building is approx.~\SI{60}{\meter} in length with a longitudinal c/c column spacing of \SI{5}{\meter}. The ceiling to floor height is approx.~\SI{6.7}{\meter} and there is a \SI{2}{\meter} deep false floor to allow personnel to access below the false floor. On the roof, there is a \SI{1.1}{\meter} parapet wall for personnel access safety, (see Fig.~\ref{fig:10SB}). For a detailed layout of the Auxiliary Building see Ref.~\cite{Auxiliarybuildinglayout}.

The height of the structure (\SI{6.7}{\meter}) is based on the height of the lorry access door (\SI{4}{\meter}), clearance between the top of the door and the underside of the crane (\SI{0.5}{\meter}), the height of the crane (\SI{1.4}{\meter}), clearance between the top of the crane and the underside of the services (\SI{0.5}{\meter}) and an allowance for lighting and services (\SI{0.3}{\meter}).

The \SI{2}{\meter} deep false floor allows the services to be distributed throughout the building and into the Extraction Tunnel and the Target Complex.

There is an evaporator room with \SI{0.4}{\meter} thick concrete walls and ceiling, with an internal height of \SI{3.5}{\meter}. It is foreseen that equipment housed inside the room may need to be removed for maintenance purposes. Therefore, a proposed structural arrangement would be to use removal concrete panels or shielding blocks for the roof and walls, to avoid demolition of the concrete structure during the removal of the equipment. 

There is also a secondary pumping station room with \SI{0.2}{\meter} thick concrete walls. Surrounding the access control is a \SI{0.3}{\meter} thick concrete wall decreasing to a \SI{0.1}{\meter} thick concrete wall at the location of the MAD, PAD and emergency door. Surrounding the buffer zone is a \SI{0.1}{\meter} thick concrete wall.  There is also a \SI{1}{\meter} wide concrete accessway surrounding the access control connected to the transport accessway.

As shown in Fig.~\ref{fig:6SB}, there is a \SI{4.58}{\meter} wide transformer base outside the building that the transformers will be supported on. Also outside the building is a \SI{10}{\meter} x \SI{10}{\meter} cooling tower base. The base is located \SI{8}{\meter} from both the Target Complex and Auxiliary Building in accordance with cooling and ventilation requirements. 

It is noted, that the geometry of the chicane is based on RP requirements, such that there is no direct line of sight between the Extraction Tunnel and the Personnel Shaft and its length is such that the Personnel Shaft is located at a required distance away from the Extraction Tunnel. The geometry was checked to ensure that CERN’s tunnel fire vehicle (PEFRA \cite{TractorPEFRA}) could safely manoeuvre through the chicane to and from the transfer tunnel.

\begin{figure}[htbp]
  \centering
  \includegraphics[width=\linewidth]{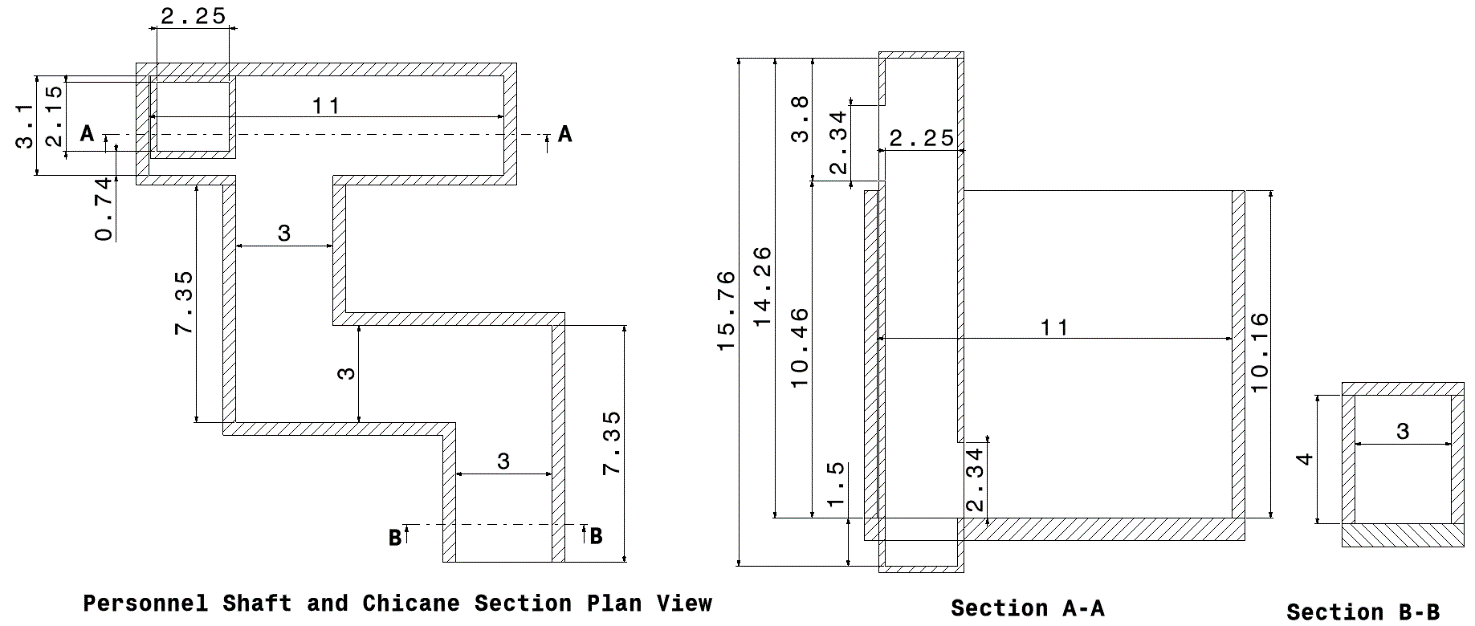}
  \caption{Personnel Shaft’s and Chicane’s structural geometry}
  \label{fig:11SB}
\end{figure}

The proposed dimensions of the Personnel Shaft and Chicane are shown in Fig.~\ref{fig:11SB}. For a detailed layout of the Personnel Shaft and Chicane see Ref.~\cite{Chicanelayout}.  The personnel lift basic geometry is based on that of a typical lift used at CERN (see Ref.~\cite{Schindlerlift}). 

For further details on the minimum required thickness of concrete shielding for the structures see Ref.~\cite{SBstructuralspec}. 

\FloatBarrier

\subsection{ Transport and Handling}
\label{Transportandhandling}

The possible transport and handling vehicles foreseen to be used in the Access Building include a \SI{40}{\tonne} semi-trailer, a forklift and a \SI{19}{\tonne} truck.  The transport and handling vehicles foreseen to be used inside the transfer tunnel are typical CERN magnet handling vehicles, as used in the existing TDC2 tunnel.

During technical stops and long shutdowns of the beamline, transport vehicles and beamline equipment may be required in the transfer tunnel. Personnel with light equipment pass through the access control in the Auxiliary Building (see Fig.~\ref{fig:6SB}) and down to the floor level of the transfer tunnel via the lift in the Personnel Shaft (see Fig.~\ref{fig:9SB}).

\begin{figure}[htbp]
  \centering
  \includegraphics[width=6in]{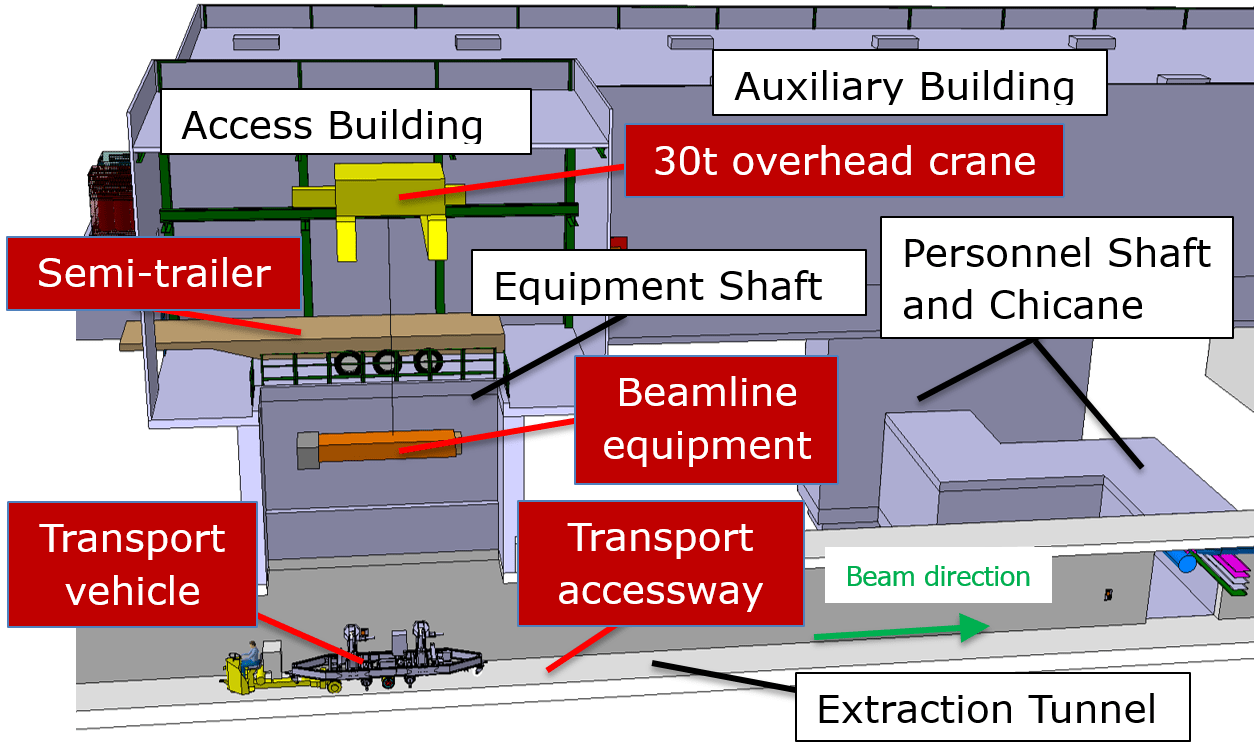}
  \caption{Equipment transported into the transfer tunnel}
  \label{fig:14SB}
\end{figure}

For the larger equipment, the beam interlocked lorry door of the Access Building is opened and a semi-trailer/truck is driven in. The \SI{30}{\tonne} overhead crane lifts the shielding blocks onto the semi-trailer/truck. The semi-trailer/truck moves the shielding blocks outside the building for storage and the semi-trailer/truck is driven into the Access Building loaded with the handling vehicle. The overhead crane lifts the transport vehicle off the semi-trailer/truck and into the Equipment Shaft to the Extraction Tunnel floor. The semi-trailer/truck is driven into the Access Building loaded with the beamline equipment (see Table \ref{Table2TT} for the BDF beamline magnet geometry). The overhead crane lifts the beamline equipment off the semi-trailer/truck and into the Equipment Shaft to the Extraction Tunnel floor (see Fig.~\ref{fig:14SB}). 

The transport vehicle moves along the Extraction Tunnel and into the Junction Cavern on the transport accessway. After the equipment has been positioned inside the transfer tunnel, the overhead crane lifts the transport vehicle from the Extraction Tunnel floor, through the Equipment Shaft and onto the semi-trailer. The semi-trailer/truck drives the shielding blocks inside the Access Building and the overhead lifts the shielding blocks inside the Equipment Shaft.

The possible Transport and Handling vehicles foreseen to access the Auxiliary Building include a standard forklift and a \SI{19}{\tonne} truck. During installation of equipment in the Auxiliary Building, a truck/forklift drives onto the transport accessway with the equipment (for typical equipment geometry see Table \ref{Table2SB}  and Table \ref{Table3SB}). The \SI{7.5}{\tonne} overhead cranes lift and move the equipment throughout the building.

\subsection{ Cooling and Ventilation}
\label{Coolingandventilation}

The CV installations for the Target Complex and the Extraction Tunnel are located in the Auxiliary Building. A description of the CV systems for the Extraction Tunnel and the surface buildings is provided in \cite{BDFCVworkpackage}.

The Extraction Tunnel and the surface buildings require the following CV systems: 

\begin{itemize}

\item Demineralized water cooling for magnets; 
\item Demineralized water cooling for power converters;
\item Compressed air for the vacuum equipment in the tunnel and general use in the Auxiliary Building;
\item Ventilation for the Auxiliary Building, the Access Building and the Extraction Tunnel.

\end{itemize}

Compressed air for the Junction Cavern and the Extraction Tunnel will be supplied from the existing compressed air infrastructure in TDC2. The existing TDC2 ventilation infrastructure will be used for the TDC2 tunnel and the new Junction Cavern. The existing TDC2 cooling infrastructure will used for the equipment in the TDC2 tunnel, whereas cooling for the new magnets installed in the Junction Cavern will be provided by the Extraction Tunnel’s demineralized water cooling. The piping and ductwork layout in TDC2 will remain unchanged, except for minor adaptions in the Junction Cavern section and the addition of piping (demi water and compressed air) for the new BDF beamline magnets on the Jura side of TDC2 and the Junction Cavern.

The construction of the Junction Cavern during LS3 will require the demolition of a 75 m long section of the TDC2 tunnel (see Chapter~\ref{Chap:CivEng} for further details). The CV equipment located in this part of the tunnel is to be dismantled before the work starts. The CV equipment located in the area includes the following:

\begin{itemize}

\item Supply ventilation duct;
\item Demineralized water cooling for magnets;
\item Raw water piping (currently not in use);
\item Compressed air for vacuum equipment.

\end{itemize}

The listed systems all reach the upstream end of the TDC2 tunnel (leading to the TT20 tunnel).
Before the demolition, the parts of the CV systems (mainly piping and ductwork) that are located in the portion of TDC2 that will be demolished are to be dismantled. Components that can be reused, are stored according to RP instructions; parts that cannot be reused will be replaced with new equipment. After the construction of the Junction Cavern is finished, piping and ductwork is to be  reinstalled according to the new layout of the Junction Cavern; since the layout of the Junction Cavern simply expands the width of the tunnel on one side, the piping and ductwork can be reinstalled at the same locations as they were before the demolition; minor adaptions to the layout of the piping will need to be performed in a detailed design phase.

\subsubsection{ Cooling Systems}

\begin{figure}[htbp]
  \centering
  \includegraphics[width=5.2in]{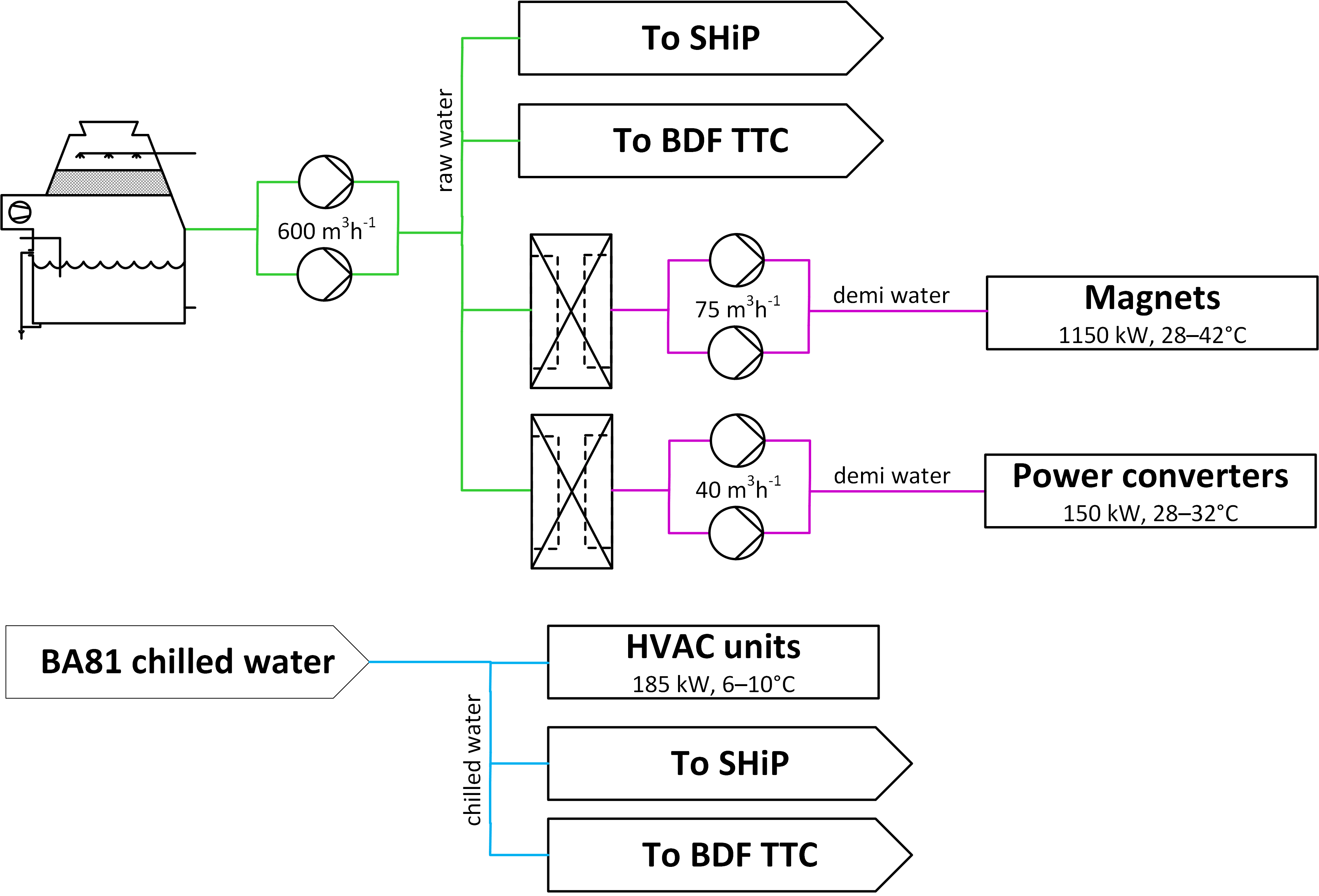}
  \caption{Overview of the cooling systems for the facility}
  \label{fig:17SB}
\end{figure}

The BDF cooling system is based on a raw water primary cooling system supplying water at \SI{25}{\degreeCelsius} via cooling towers. Local cooling is provided via demineralised water secondary circuits; chilled water provides cooling for the air handling units (AHUs). The primary system is located in the BDF Auxiliary Building and provides primary cooling for the entire BDF facility, including the Extraction Tunnel, the Target Complex and the Experimental Area.
Fig.~\ref{fig:17SB} shows a synoptic view \cite{BDFcooling} of the cooling systems for the facility.

The magnets cooling system provides demineralised water cooling for the magnets in the Extraction Tunnel; its pumping station is located in the Auxiliary Building. The system supplies \SI{75}{\cubic\metre\per\hour} at \SI{28}{\degreeCelsius} and \SI{20}{\bar} for a thermal load of \SI{1200}{\kW} on average; the estimated return temperature is \SI{42}{\degreeCelsius} and pressure reducers are used to locally reduce the pressure to the needs of the specific magnet. Since the water is likely to activate, the pumping station located on the surface in the Auxiliary Building is separated from the rest of the CV equipment.

The power converters cooling system provides demineralised water cooling for the converters in the Auxiliary Building. Its pumping station is located in the CV area of the Auxiliary Building; since the level of activation of this water is expected to be low, the circuit does not need to be separated from the rest of the equipment. The system supplies \SI{40}{\cubic\metre\per\hour} at \SI{28}{\degreeCelsius} for a maximum thermal load of \SI{150}{\kW}.

Chilled water is needed in the facility for the ventilation units in the Auxiliary Building and it is provided by the chilled water production plant in one of CERN’s existing service buildings (BA81).

\subsubsection{ Ventilation Systems}

Ventilation units are required for the following structures:

\begin{itemize}

\item Extraction Tunnel;
\item Access Building;
\item Auxiliary Building.

\end{itemize}

\begin{figure}[htbp]
  \centering
  \includegraphics[width=5.28in,height=3.37in]{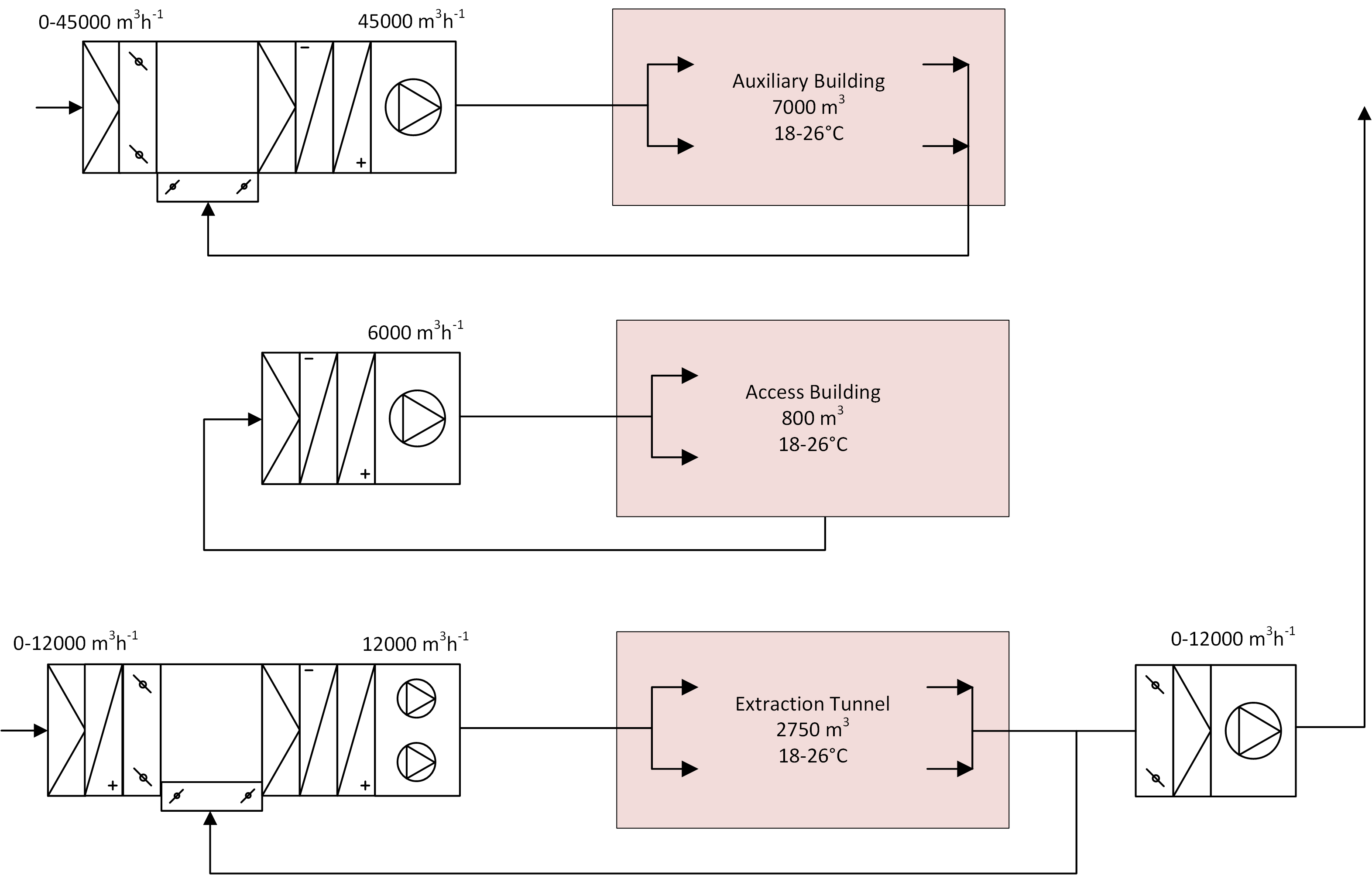}
  \caption{Ventilation system synoptic for the Extraction Tunnel, Access and Auxiliary Buildings}
  \label{fig:18SB}
\end{figure}

\begin{figure}[htbp]
  \centering
  \includegraphics[width=6in]{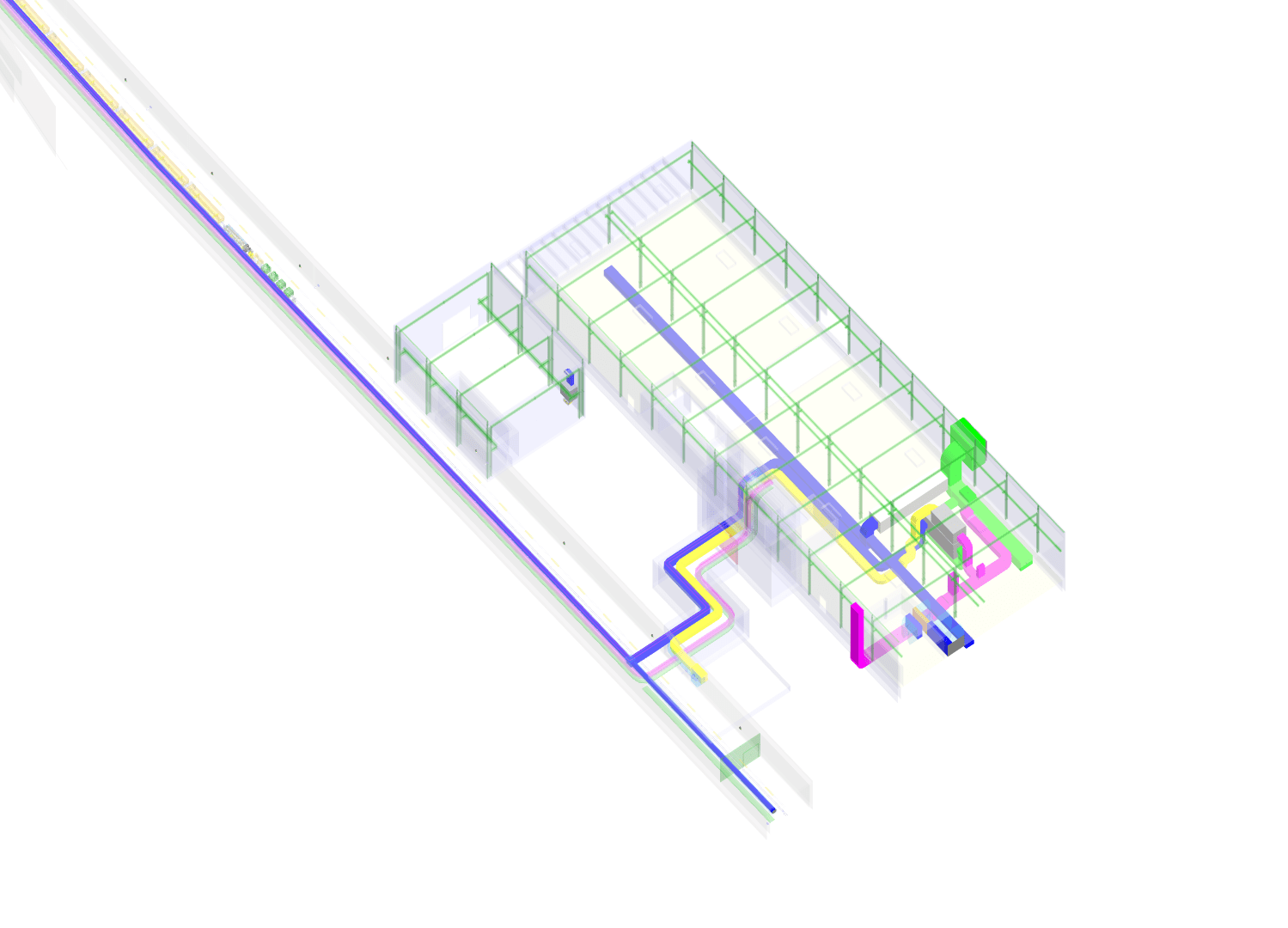}
  \caption{Ventilation duct layout for the Extraction Tunnel, Access and Auxiliary Buildings}
  \label{fig:19SB}
\end{figure}

Fig.~\ref{fig:18SB} provides a synoptic view \cite{BDFventilation} of the ventilation units and their characteristics; Fig.~\ref{fig:19SB} shows a preliminary integration of the ducts within the facility.

Supply and extraction units for the tunnel ventilation are located in the Auxiliary Building. Ducts reach the Extraction Tunnel via the false floor in the Auxiliary Building and through the Chicane. Inside the Extraction Tunnel, two supply ducts reach the ends of the tunnel and one extraction grill in the proximity of the Chicane extracts the entire flow. Recirculation is normally used during operation; full extraction is activated during shutdowns. The design flow rate for the ventilation units is \SI{12000}{\cubic\metre\per\hour}.

The ventilation of the Access Building is done via recirculation; the ventilation unit is located inside the Access Building and it is designed for a flow rate of \SI{6000}{\cubic\metre\per\hour}.

The ventilation supply unit for the Auxiliary Building is based on recirculation; air is supplied to the room via the false floor and collected directly in the unit. The ventilation approach takes advantage of free cooling whenever possible, in order to minimize energy loss. The design flow rate is \SI{45000}{\cubic\metre\per\hour}.

With respect to the smoke extraction system, smoke extraction for the tunnel is provided via portable extractor units and flexible ducts. The smoke extraction method for the auxiliary building is based on natural ventilation and skydomes located on the ceiling or grilles located on the side walls for improved leak-tightness.

\FloatBarrier

\subsection{ Electrical Network System}
\label{ENEL}

The BDF electrical feed is foreseen to be supplied from building BA80 for which a minimum of a \SI{1}{\meter} wide trench is required from BA80 to the Auxiliary Building. There may possibly be a technical gallery with cooling and ventilation infrastructure (compressed air, demineralised water, raw water and chilled water) included, see Chapter~\ref{Chap:CivEng}, Fig.~\ref{fig:Layout} for the proposed layout of the technical gallery. 

\begin{figure}[htbp]
  \centering
  \includegraphics[width=\linewidth]{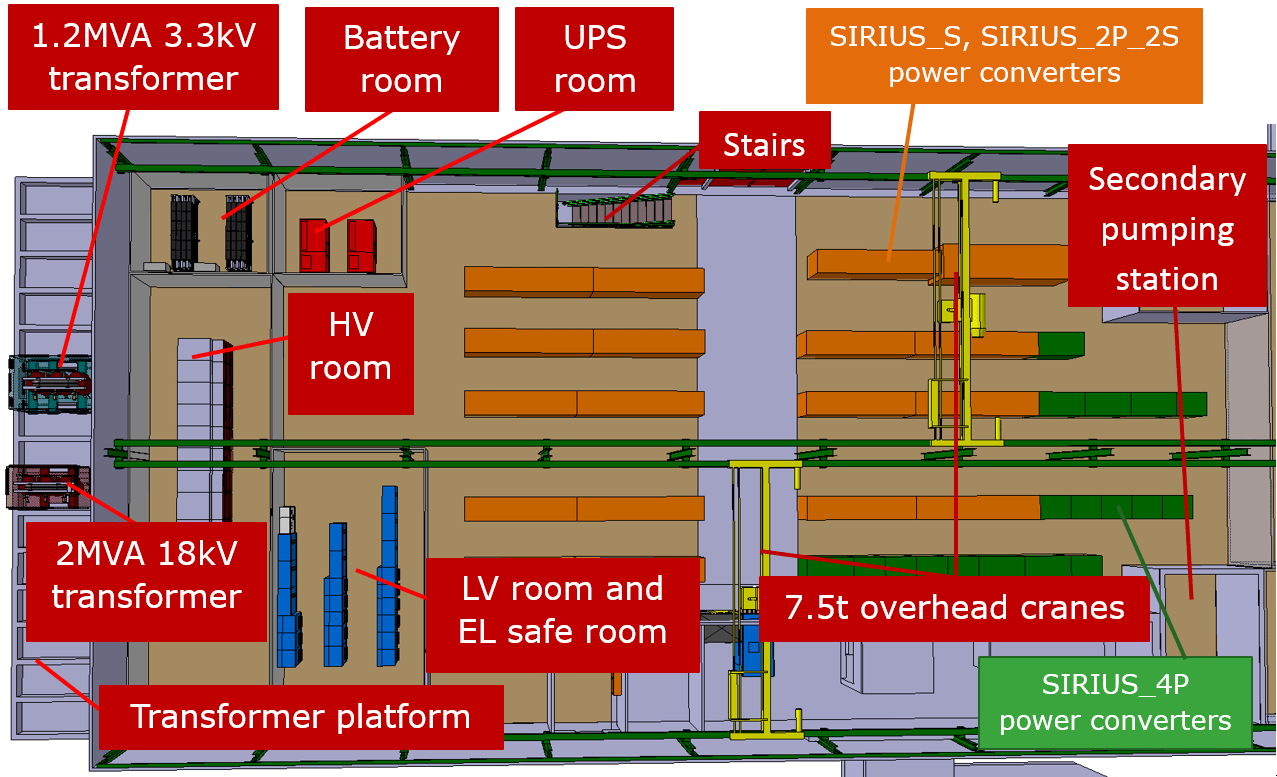}
  \caption{Auxiliary Building’s electrical infrastructure integration layout}
  \label{fig:20SB}
\end{figure}

Preliminary space reservations in the Auxiliary Building for the BDF electrical engineering infrastructure have been outlined, as shown in Fig.~\ref{fig:20SB}.

The infrastructure includes:

\begin{itemize}


\item An Uninterruptible Power Supply (UPS) room for the 800 kVA high heat load, of which can be in shared space (the cooling for this area must be supplied by UPS). Two 400 kVA systems provide the requirements for the transfer tunnel, Access Building, Auxiliary Building and the Target Complex with a margin for future expansion. 
\item A dedicated battery room with smoke extraction in case of fire. It is to have an autonomy of 10 minutes at 800 kVA, and which must be cooled, at an ideal temperature of \SI{18}{\degreeCelsius} to maximise battery life. 
\item An electrical engineering EL safe room, a small rack room with the space for 6 racks to contain electrical protection and network supervision hardware. This safe room is in a shared space with the low voltage room. The low voltage room, sized for the transfer tunnel, Access Building, Auxiliary Building and the Target Complex. 2 MVA normal, 1.2 MVA ESD and 800 kVA UPS with no non-electrical engineering racks allowed in front of switchboards. 
\item The HV stable switch-room (Stable and Pulsed), \SI{3}{\meter} in height with full height double door is accessible to electrical engineers only. The doors shall face onto the external perimeter of the building. One switchboard for stable and one switchboard for pulsed. There shall be a \SI{1.2}{\meter} space between the two switchboards and a space reservation in the corner of the switch-room for a spare HV circuit breaker (\SI{1}{\meter} x \SI{1}{\meter}) for each switchboard. 
\item Two transformers, a 2 MVA \SI{18}{\kilo\volt} transformer for distribution and a 1.2 MVA \SI{3.3}{\kilo\volt} transformer for diesel backup.

\end{itemize}

A false floor, \SI{2}{\meter} deep, of which is personnel accessible via a stairs has been included for service distribution throughout the Auxiliary Building and into the transfer tunnel. 

In the Access and Auxiliary Building, an \SI{800}{\milli\meter} space reservation has been allowed for between the top of the crane and the underside of the ceiling, of which the first \SI{300}{\milli\meter} below the ceiling has been reserved for lighting. It is noted that the lighting will be installed and maintained from the overhead crane.

CERN's electrical engineering group's recommendation is that the cable trays have a minimum of \SI{1.2}{\meter} bending radius horizontally and vertically. The control trays will need to be accessed at least once per year for the first 5 years, therefore there is a CAT ladder and a metallic stairs in the Personnel Shaft to inspect and maintain the cable trays as shown in Fig.~\ref{fig:9SB}. 

\subsection{ Electrical Power Converters}

Power converters are required for the magnets in the BDF beamline.
They are housed in the Auxiliary Building as shown in Fig.~\ref{fig:20SB}.

A preliminary design of the power converters was provided by CERN's power converter group. Table \ref{Table1SB} lists the magnet features taken into account in the design. 
Note that 4 quadrant (4Q) operation mode implies bipolar current and voltage capabilities.

\begin{table}[htpb]
 \center
 \begin{scriptsize}
 \caption{Design parameters of beamline power converters}
 \label{Table1SB}
\begin{tabular}{ccccccc}
\hline 
\vspace*{0.1cm}
\begin{minipage}{0.5in}\centering \vspace{0.1cm} \textbf{Circuit name}\end{minipage} &
\begin{minipage}{1in}\centering \vspace{0.1cm} \textbf{No. of magnets in series}\end{minipage} &
\begin{minipage}{0.5in}\centering \vspace{0.1cm} \textbf{$R_{mag}$ [m$\Omega$]}\end{minipage} &
\begin{minipage}{0.5in}\centering \vspace{0.1cm} \textbf{$L_{mag}$ [mH]}\end{minipage} &
\begin{minipage}{1in}\centering \vspace{0.1cm} \textbf{$I_{max}$ requested [A]}\end{minipage} &
\begin{minipage}{0.5in}\centering \vspace{0.1cm} \textbf{Qty}\end{minipage} &
\begin{minipage}{0.5in}\centering \vspace{0.1cm} \textbf{Operation mode}\end{minipage} \\

\hline
 MSSB2117 & 3 & 66 & 140 & 1100 & 1 & 4Q \\
 MBB & 5 & 4.42 & 9.9 & 5750 & 1 & 4Q \\
 MBN & 3 & 51.5 & 170 & 1180 & 6 & 4Q \\
 QTG.01 & 1 & 57 & 31 & 385 & 1 & 4Q \\
 QTG.02 & 1 & 276 & 390 & 345 & 1 & 4Q \\
 QTG.03 & 1 & 276 & 390 & 304 & 1 & 4Q \\
 QTG.04 & 1 & 276 & 390 & 192 & 1 & 4Q \\
 QTG.05 & 1 & 276 & 390 & 354 & 1 & 4Q \\
 QTG.06 & 1 & 276 & 390 & 444 & 1 & 4Q \\
 MDX1 corrector & 1 & 320 & 221 & 120 & 6 & 4Q \\
 MDX1 bend & 1 & 320 & 221 & 240 & 1 & 4Q \\
 MDX1 dilution & 1 & 320 & 221 & 170 & 4 & 4Q \\
 \hline 
\end{tabular} 
\end{scriptsize} 
\end{table}

\begin{figure}[htbp]
  \centering
  \includegraphics[width=4.17in,height=2.29in]{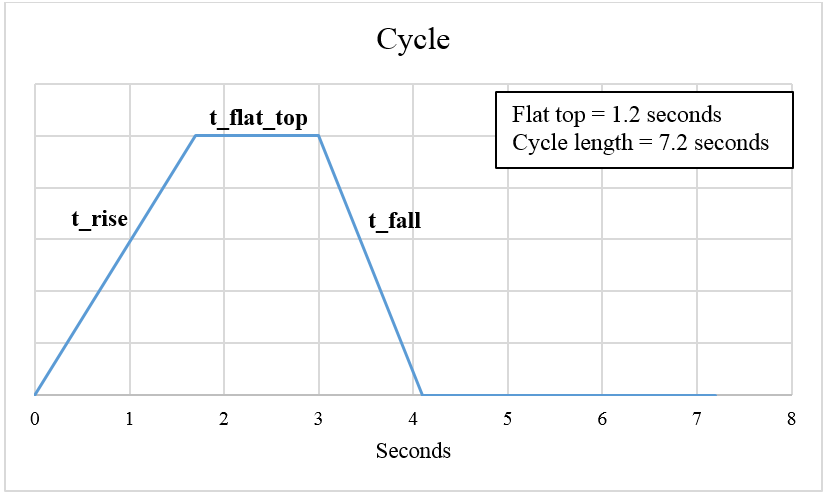}
  \caption{Magnet current reference}
  \label{fig:22SB}
\end{figure}

The initial request was for 1 string of 18 magnets with a voltage of \SI{2.7}{\kilo\volt}. The implemented solution consists of 6 strings of 3 magnets, to guarantee the same current in each magnet. For standardisation, the SIRIUS family of power converters are used. The magnet current reference (per unit) for the transfer line is shown in Fig.~\ref{fig:22SB}.

The performance of the current sources requested is “accuracy class 4” which establishes a long term stability (1 year) of 100 ppm, 20 ppm during 12 hours and in the short term (20 min) a stability of 5 ppm. 

Table \ref{Table2SB} and Table \ref{Table3SB} show the geometrical specifications of the power converters and the energy storage racks.

The layout of the power converters were arranged in two groups (SIRIUS S, SIRIUS 2P 2S) and (SIRIUS 4P), shown in green and orange in Fig.~\ref{fig:20SB}. The power converters are positioned in the middle of the building, next to both the electrical engineering infrastructure and the transfer tunnel to optimise the length of cabling. The secondary pumping station is located next to the power converters to optimise the length of cooling piping required.

The distance to the front and the rear of any two converters/storage racks and the distance between a converter/storage rack and a wall was kept to a minimum of \SI{1.4}{\meter} in accordance with the SIRIUS installation guide (see Ref.~\cite{BDFSirius} ). This distance was increased to \SI{2}{\meter} for the MBB power converter. The power converters and racks were arranged such that the two overhead cranes could manoeuvre them in and out of position for installation and de-installation purposes. The distance between the edge of the wall and the centre of the crane hook was also considered in the layout of the converters.

The existing MSSB2117 splitter magnets will be replaced with new laminated splitter magnets. The existing power converter for these magnets are housed in building BA80. However, the new magnets require the SIRIUS 4P+ power converter, which is larger than the existing; for which, the new power converter is too large to be integrated into building BA80.  In addition, the reserve SIRIUS 4P+ is to be housed in the Auxiliary Building; therefore, the MSSB2117 power converter is to be housed in the Auxiliary Building.

\begin{table}[htpb]
 \center
 \caption{Power converters' geometry}
 \label{Table2SB}
\begin{tabular}{cccccccc}
\hline
\\
\textbf{Circuit}  & \textbf{Converter} & \textbf{}  & \multicolumn{5}{c}{\textbf{ Converter Dimensions (per unit)}}  \\
\textbf{Name}  & \textbf{Type} & \textbf{Qty.}  & \textbf{Length} & \textbf{Width} & \textbf{Height} & \textbf{Surface} & \textbf{Weight/unit}  \\
\textbf{}  & \textbf{} & \textbf{} & \textbf{[m]} & \textbf{[m]} & \textbf{[m]} & \textbf{[m${}^{2}$]} & \textbf{[kg]}  \\
\hline
MSSB2117 & SIRIUS 4P+ & 1 & 4.2 & 0.9 & 2.1 & 3.8 & 2600  \\
 MBB & NEW 4P+ & 1 & 5 & 1.3 & 2.4 & 6.5 & 5500  \\
 MBN & SIRIUS 4P+ & 6 & 4.2 & 0.9 & 2.1 & 22.7 & 2600  \\
 QTG.01 & SIRIUS S & 1 & 1.2 & 0.9 & 2.1 & 1.1 & 700  \\
 QTG.02 & SIRIUS 2P & 1 & 1.8 & 0.9 & 2.1 & 1.6 & 1100  \\
 QTG.03 & SIRIUS S & 1 & 1.2 & 0.9 & 2.1 & 1.1 & 700  \\
 QTG.04 & SIRIUS S & 1 & 1.2 & 0.9 & 2.1 & 1.1 & 700  \\
 QTG.05 & SIRIUS S & 1 & 1.2 & 0.9 & 2.1 & 1.1 & 700  \\
 QTG.06 & SIRIUS 2P & 1 & 1.8 & 0.9 & 2.1 & 1.6 & 1100  \\
 MDX1 corrector & SIRIUS S & 6 & 1.2 & 0.9 & 2.1 & 6.5 & 700  \\
 MDX1 bend & SIRIUS S & 1 & 1.2 & 0.9 & 2.1 & 1.1 & 700  \\
 MDX1 dilution & SIRIUS 2S & 4 & 1.8 & 0.9 & 2.1 & 6.5 & 1100  \\
Reserve & SIRIUS 4P+ & 1 & 4.2 & 0.9 & 2.1 & 3.8 & 2600  \\
Reserve & SIRIUS 2P & 1 & 1.8 & 0.9 & 2.1 & 1.6 & 1100  \\
 \hline
\end{tabular} 
\end{table}

\begin{table}[htpb]
 \center
 \caption{Energy storage racks' geometry}
 \label{Table3SB}
\begin{tabular}{ccccccc}
\hline
\\
\textbf{Circuit}  & \textbf{} & \multicolumn{5}{c}{\textbf{Energy storage dimensions per unit}}  \\
\textbf{Name}  & \textbf{Qty.}  & \textbf{Length} & \textbf{Width} & \textbf{Height} & \textbf{Surface} & \textbf{Weight/unit}  \\
\textbf{}  & \textbf{}  & \textbf{[m]} & \textbf{[m]} & \textbf{[m]} & \textbf{[m${}^{2}$]} & \textbf{[kg]}  \\
\hline
 MSSB2117 Splitter & 1 & 4.8 & 0.9 & 2.1 & 4.3 & 2400 \\
 MBB & 1 & 4.8 & 0.9 & 2.1 & 4.3 & 2400 \\
 MBN & 6 & 4.8 & 0.9 & 2.1 & 25.9 & 2400 \\
 Reserve & 1 & 4.8 & 0.9 & 2.1 & 4.3 & 2400 \\
 \hline
\end{tabular} 
\end{table}

The total estimated power dissipated in the Auxiliary Building air from the power converters is \SI{80}{\kW} and the total cooling power requested is \SI{150}{\kW}. The cooling is from demineralised water with the following features:

\begin{itemize}
\item  Maximum water temperature in the circuit \SI{27}{\degreeCelsius} with a $\Delta$$\theta$ of approx.~\SI{10}{\K}.

\item  Total flow rate: \SI{660}{\litre\per\minute}.

\item  $\Delta$P of \SI{3}{\bar} with a nominal pressure less than \SI{6}{\bar}.
\end{itemize}

It is noted that the AC and DC cabling is to be covered by CERN’s electrical engineering group. Water cooled cabling is required for the link between the MSSB2117 splitter, the MBB magnets and their respective power converters. Table \ref{Table4SB} shows the AC and DC distribution considered in the study. The third column indicates the cables cross-section as a recommendation from CERN’s power converter group in order to keep the voltage drop at a reasonable level. In addition, thermal aspects have to be evaluated by CERN’s electrical engineering group. The cable length equals to two times the distance between the converter and the magnet. The total power to be installed is 1548 kVA.

\begin{table}[htpb]
\begin{scriptsize} 
 \center
 \caption{Power converters' AC and DC distribution}
 \label{Table4SB}
\begin{tabular}{ccccccccc}

\hline

\vspace*{0.1cm}
\begin{minipage}{0.5in}\centering \vspace{0.1cm} \textbf{Circuit name}\end{minipage} &
\begin{minipage}{0.5in}\centering \vspace{0.1cm} \textbf{Qty}\end{minipage} &
\begin{minipage}{0.5in}\centering \vspace{0.1cm} \textbf{Total cable length [m]}\end{minipage} &
\begin{minipage}{0.5in}\centering \vspace{0.1cm} \textbf{$I_{rms}$ [A]}\end{minipage} &
\begin{minipage}{0.5in}\centering \vspace{0.1cm} \textbf{DC cable sec. Cu/Al {[}mm${}^{2}${]}}\end{minipage} &
\begin{minipage}{0.5in}\centering \vspace{0.1cm} \textbf{Converter type}\end{minipage} &
\begin{minipage}{0.5in}\centering \vspace{0.1cm} \textbf{Electrical feeder [V]}\end{minipage} &
\begin{minipage}{0.5in}\centering \vspace{0.1cm} \textbf{$I_{rms}$ for $I_{rms}$ \mbox{converters}}\end{minipage} &
\begin{minipage}{0.5in}\centering \vspace{0.1cm} \textbf{Total installed RMS power [kVA]}\end{minipage} \\

\hline
MSSB2117 & 1 & 1000 & 800 & 480 & SIRIUS 4P+ & 400 & 125 & 87  \\
 MBB & 1 & 435 & 3250 & 1000 & NEW 4P+ & 400 & 630 & 436  \\
 MBN & 6 & 230 & 800 & 480 & SIRIUS 4P+ & 400 & 125 & 520  \\
 QTG.01 & 1 & 278 & 200 & 120 & SIRIUS S & 400 & 32 & 22  \\
 QTG.02 & 1 & 250 & 400 & 240 & SIRIUS 2P & 400 & 63 & 44  \\
 QTG.03 & 1 & 168 & 200 & 120 & SIRIUS S & 400 & 32 & 22  \\
 QTG.04 & 1 & 129 & 200 & 120 & SIRIUS S & 400 & 32 & 22  \\
 QTG.05 & 1 & 90 & 200 & 240 & SIRIUS S & 400 & 32 & 22  \\
 QTG.06 & 1 & 52 & 400 & 240 & SIRIUS 2P & 400 & 63 & 44  \\
 MDX1 corrector & 6 & 500 & 200 & 120 & SIRIUS S & 400 & 32 & 133  \\
 MDX1 bend & 1 & 500 & 200 & 120 & SIRIUS S & 400 & 32 & 22  \\
 MDX1 dilution & 4 & 50 & 200 & 120 & SIRIUS 2S & 400 & 62 & 175  \\
\hline
\end{tabular} 
\end{scriptsize} 
\end{table}

\FloatBarrier

\subsection{ Radiation Protection}

Radiation protection has been one of the key aspects in the overall layout of the Access and Auxiliary Buildings. This has included specific concrete and soil thicknesses as well and specific distances and positions of the buildings away from the beamline.  For further details on the minimum required thicknesses of shielding for the structures, see Ref.~\cite{SBstructuralspec}. Refer to Chapter~\ref{Chap:RP} for details on the RP specifications for the facility. 

\subsection{ Safety}

\subsubsection{ Fire and Alarms}

The safety requirements for the Access and Auxiliary building has been a key aspect in the overall layout of the buildings. This has included natural smoke extraction and safe emergency access throughout the buildings in accordance with safety regulations. Refer to Chapter~\ref{Chap:Safety} for further details.

\subsubsection{ Access Control}

\begin{figure}[htbp]
  \centering
  \includegraphics[width=4.6in]{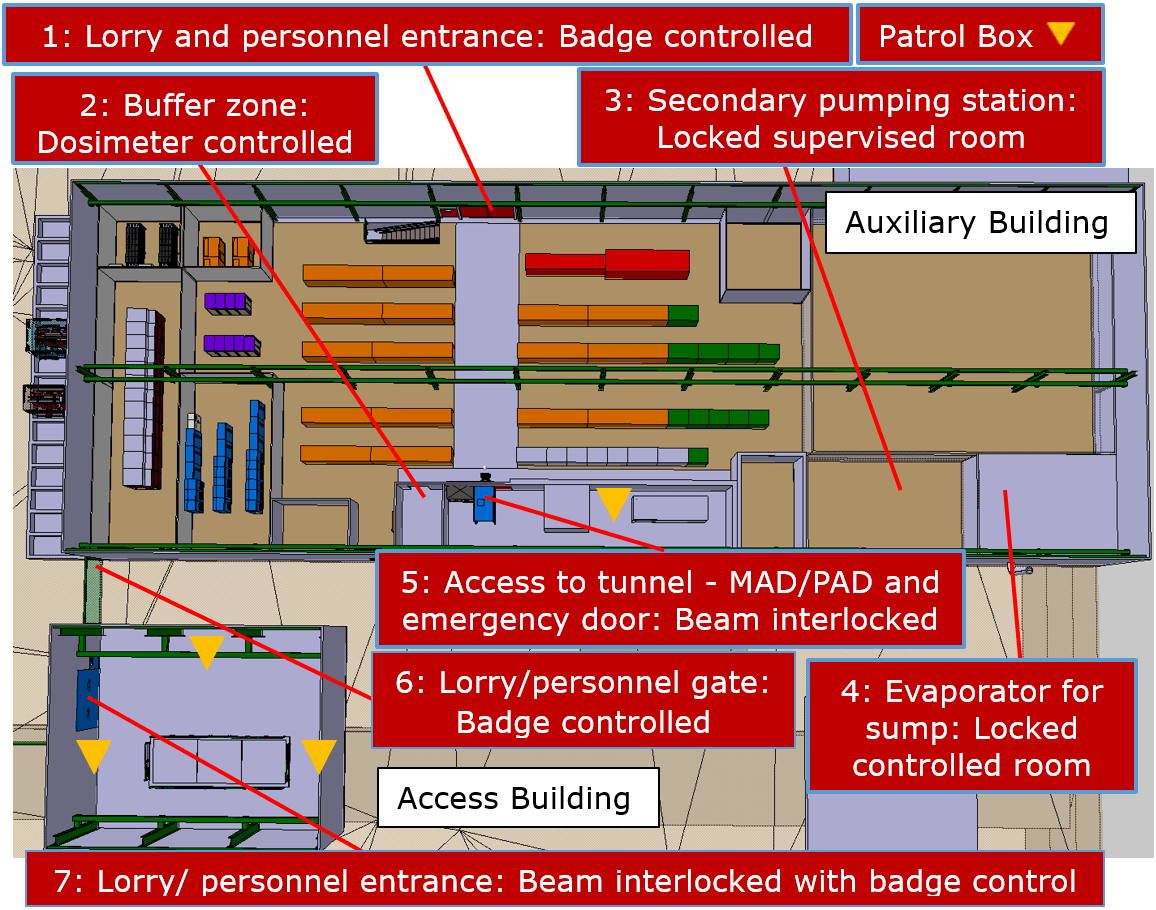}
  \caption{Access control for the Access and Auxiliary Buildings}
  \label{fig:24SB}
\end{figure}

The access control system for the Access and Auxiliary Building is shown in Fig.~\ref{fig:24SB}, it includes:

\begin{enumerate}

\item  Lorry and personnel entrance, badge controlled to allow access to the Auxiliary Building;
\item  Buffer zone with a grilled dosimeter controlled door;
\item  Secondary pumping station is a locked supervised room;
\item  Evaporator for sump is a locked controlled room;
\item  MAD/PAD and emergency door, beam interlocked to allow access to the transfer tunnel;
\item  Lorry/personnel gate, badge controlled to allow access to the cooling tower;
\item  Lorry/personnel entrance, beam interlocked with badge control for access to the Access Building;
\item  Four patrol boxes located throughout both buildings in accordance with Access Control’s recommendations.
\end{enumerate}

\newpage
\section{Target Complex Access}
\label{TransferT}

The access for the Target Complex is located on the upstream side of the Target Complex on the Jura side of the Auxiliary Building as shown in Fig.~\ref{fig:2SB}.

\noindent\textbf{Personnel Access}

\begin{figure}[htbp]
  \centering
  \includegraphics[width=4.9in]{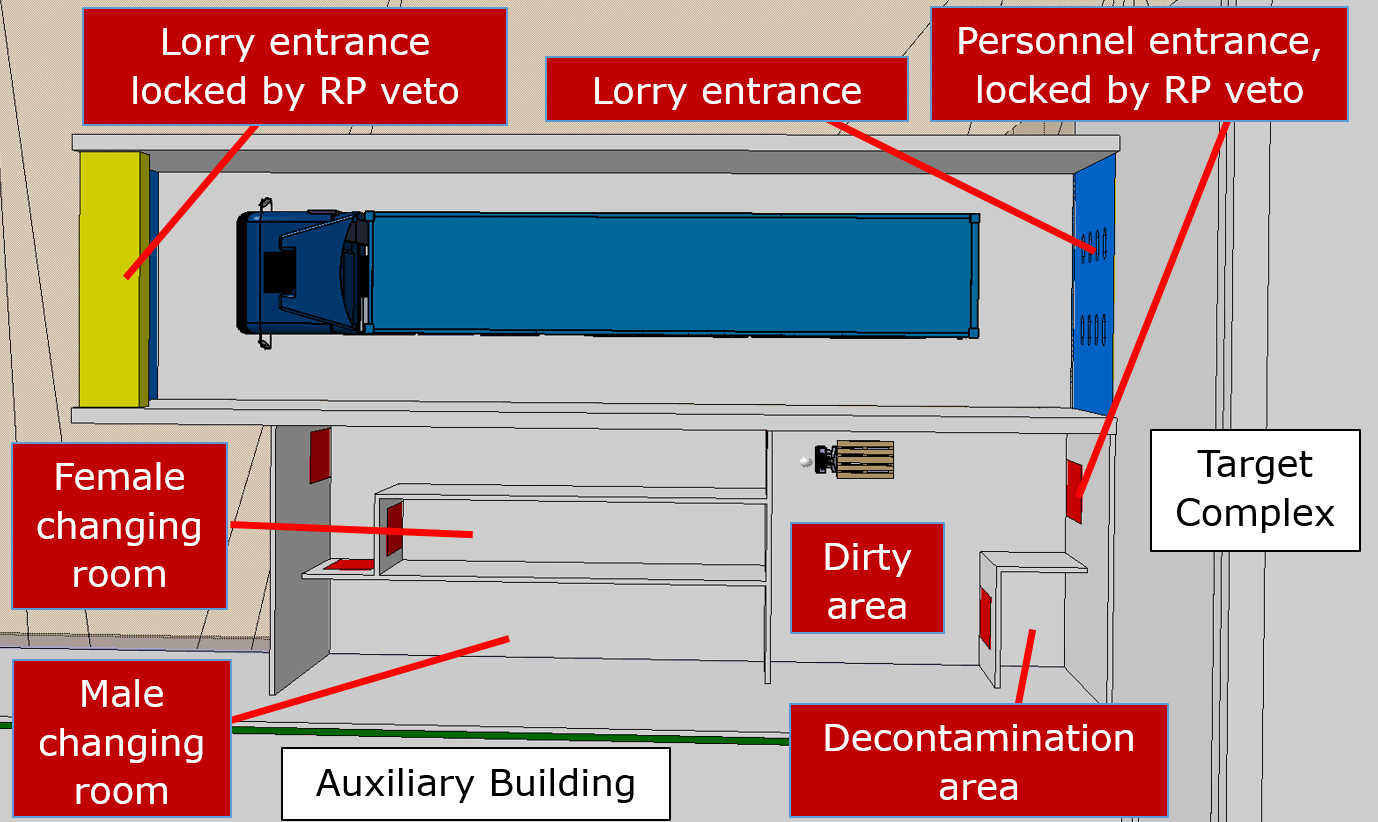}
  \caption{Target Complex's personnel and vehicle access}
  \label{fig:2TC}
\end{figure}

The Target Complex’s personnel access is composed of a male and female changing room, a dirty area, and a decontamination area, as shown in Fig.~\ref{fig:2TC}. 
The layout is in accordance with RP’s requirements of which follows a similar arrangement as that of changing rooms at two of CERN’s other target related facilities, MEDICIS and building 867. 
The facility has a plan area of approx.~\SI{80}{\meter\per\squared} and the height of the structure is the same as the lorry entrance, as this allows a practical structural design and constructability as well as allowing sufficient space for ventilation at the ceiling level. To access the Target Complex, there is door locked by an RP veto.

\noindent\textbf{Lorry Access}

The Target Complex’s lorry access (shown in Fig.~\ref{fig:2TC}) is an air-locked compartment to keep the pressure cascade in the Target Complex, with a large lorry access door locked with an RP veto.
The pressure inside the lorry access has a negative pressure of \SI{-20}{\pascal}, consistent with that of the Target Complex. 
This ensures that no contaminated air is released to the external environment during the transport and handling process. The geometry of the facility is in accordance with CERN’s Transport and Handling group’s requirements. Its geometry is such that the full length of an articulated lorry can fit inside the structure and there is sufficient room for the lorry doors to open on either side allowing personnel to enter and exit the vehicle. The structure is \SI{20}{\meter} in length, \SI{5}{\meter} in width and \SI{5}{\meter} in height. To access the Target Complex with a vehicle, the external lorry access door is opened and the vehicle is driven in. The pressure inside is decreased to the same negative pressure as the Target Complex and the internal lorry door is opened to allow the lorry to access the facility.

\noindent\textbf{Access to the Underground Area}

Inside the Target Complex there are two staircases to access the controlled underground area. In front of one staircase there is an end of zone door and in front of the other there is an interlocked PAD. A fence connects the PAD and end of zone door, ensuring personnel must pass through the interlocked access control to access the underground area. There are three patrol boxes foreseen in the Target Complex.

\newpage
\section{Experimental Area}
\label{ExperimentalArea}

\begin{figure}[ht]
  \centering
  \includegraphics[width=\linewidth]{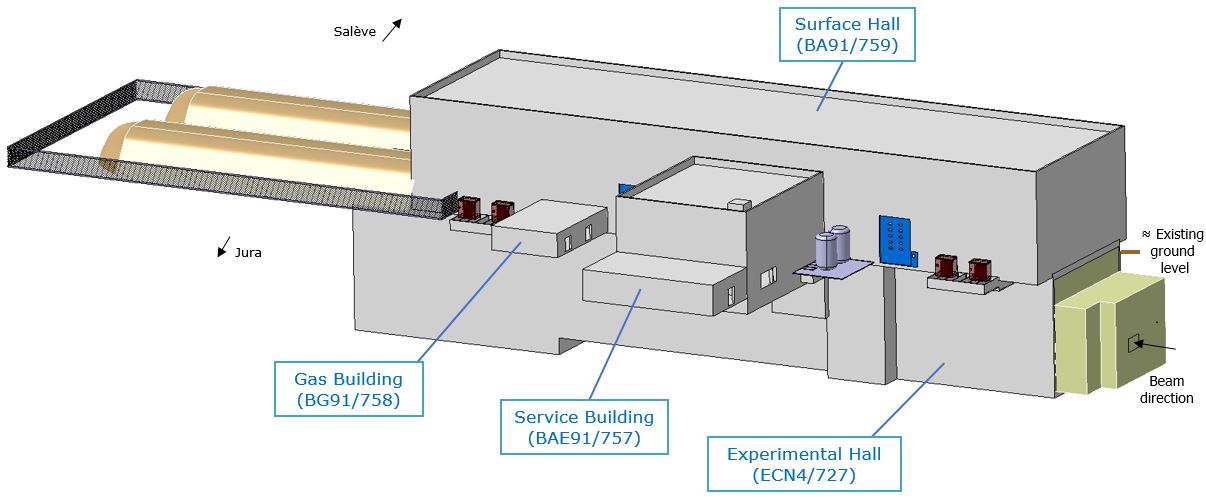}
  \caption{Experimental Area layout}
  \label{fig:Experimental area layout}
 \end{figure}

Particular effort has been placed on services integration studies to guarantee a cost efficient conceptual design. The Experimental Area is located immediately downstream of the Target Complex as shown in Fig.~\ref{fig:BDF integration overview}. Its overall footprint has a maximum length of \SI{180}{\meter}, a maximum width of approx. \SI{60}{\meter}, \SI{20}{\meter} of depth below the ground level and \SI{16.3}{\meter} of height above the ground level. The structure consists of an underground Experimental Hall and three surface buildings (Service and Gas buildings and Surface Hall) 
which have Finished Floor Levels (FFL) of approximately \SI{2.1}{\meter}  (between \SI{1.6}{\meter} and \SI{2.6}{\meter}) above the existing ground level ($\approx$ 452.4 masl). The current approach is to fill the surrounding area  with structural fill to reach the requested FFL. The underground Experimental Hall is situated underneath the Surface Hall as shown in Fig.~\ref{fig:Experimental area layout}. Adjacent to the Surface Hall on the Jura side, the experimental Service Building and the Gas Building are located and, at the back of the Surface Hall and in contact with its downstream wall, there are two \SI{6}{\meter} high fenced land hills as required for Radiation Protection (RP) considerations. The layout has been defined to house and assemble the SHiP experiment whilst guaranteeing the operation of the facility from RP and safety point of view. The naming of the buildings are shown in Fig.~\ref{fig:Experimental area layout}.

 \noindent The purpose of these structures includes the following:
 \begin{itemize}
  \item Service supply and distribution to the Search for Hidden Particles (SHiP) experiment \cite{SHiP2018} located in the Experimental Hall (requirements are described in Chapter \ref{Chap:ExpHall}). 
  \item Assembly and installation of SHiP’s components and equipment inside the Experimental Hall. 
  \item Personnel access to the Experimental Hall.
  \item House the operational complex that includes a workshop, labs, a control room, offices and a conference room.
 \end{itemize}

The services include electrical power supply, cooling, ventilation, gas supply, a smoke extraction system, detector supporting systems, gas distribution, electronics, readout, network and computing systems. In addition, the layout provides the required handling equipment, an alignment connection with the beam line and access control for the facility. The location of the buildings has been optimized in accordance to the muon flux during beam operation (see Chapter \ref{Chap:RP})  and thus, the Service Building and Gas Building are the only two buildings accessible during beam operation due to radiation protection constraints. To protect personnel from the muon flux, the Service Building is located approx. \SI{33}{\meter} away from the target interface and approx. \SI{25}{\meter} offset with respect to the “beam axis”. 

Preliminary integration studies to house the SHiP detector inside the Experimental Area have been performed to evaluate the feasibility. The following sections describes the buildings and halls geometries located inside the Experimental Area. In addition, they contain an overview of the preliminary design solutions adopted for the key infrastructures, such as ventilation, water-cooling, electrical supply, transport, survey, radioprotection, safety and the access system.

\subsection{Experimental hall (ECN4)}

\begin{figure}[ht!]
  \centering
  \includegraphics[width=\linewidth]{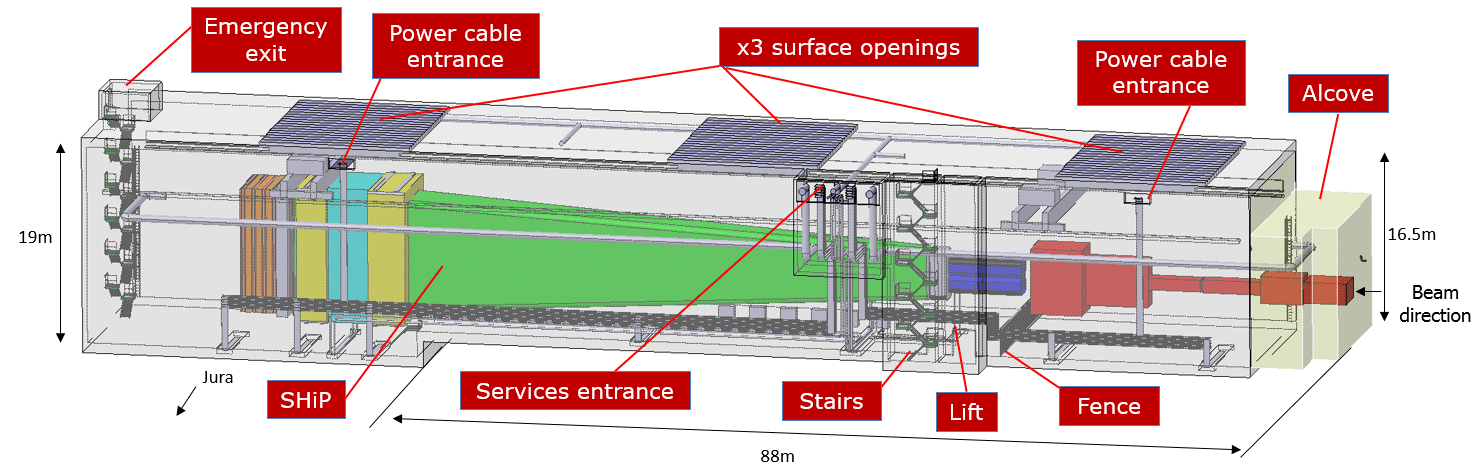}
  \caption{Underground Experimental Hall layout}
  \label{fig:Underground Experimental Hall layout}
 \end{figure}

The underground Experimental Hall has been defined to house the SHiP detector taking into account the assembly of the experiment and its services supply. The layout of the Experimental Hall is detailed in Fig.~\ref{fig:Underground Experimental Hall layout} and its geometry is defined in Ref.~\cite{Exphalldra}.

The SHiP experiment will be located along the \SI{120}{\meter} hall and centred inside its \SI{20}{\meter} width. The internal height of the structure (\SI{19}{\meter}) is driven by the space reservation of the spectrometer magnet centred with respect to the beam axis (\SI{13}{\meter}), clearance between the top of the magnet and the underside of the crane (\SI{0.5}{\meter}), the space reservation of the \SI{80}{\tonne} crane (\SI{3.2}{\meter}), clearance between the top of the crane and the underside of the services (\SI{0.3}{\meter}) and an allowance for lighting and services (\SI{0.5}{\meter}). In order to optimize the height structure in accordance to the SHiP detector geometry, the cavern floor has two levels. The floor along the first \SI{88}{\meter} is located \SI{5.5}{\meter} below the beam axis and a \SI{2.5}{\meter} step brings the floor level down to 8 m below the beam axis to accommodate the large spectrometer section.

There are \textbf{three openings for equipment access} between the Surface Hall and the underground Experimental Hall (free space: \SI{12.5}{\meter}$\times$\SI{18}{\meter}) through which the large components of the experiment can be lowered. The openings will be left fully open during the assembly phase.

\begin{figure}[h]
  \centering
  \includegraphics[width=4.22in,height=2.5in,keepaspectratio=false]{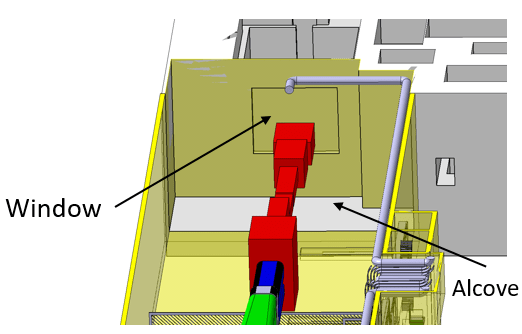}
    \caption{Interface between the Experimental Hall and the Target Complex}
  \label{fig:alcoveh}
 \end{figure}

\begin{figure}[ht]
  \centering
  \includegraphics[width=3in,height=3in]{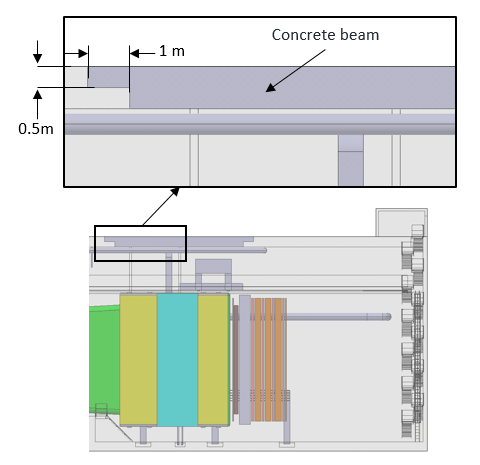}
  \caption{Preliminary support system for the concrete shielding beams}
  \label{fig:Preliminary support system}
 \end{figure}

For shielding purposes, each opening will be covered by 18 concrete beams during beam operation. Due to crane availability for assembly and disassembling, each beam has an approximate weight of \SI{37}{\tonne} and approximate volume of 14.5$\times$1$\times$1:\SI{14.5}{\cubic\meter}. The concrete beams will only need temporary storage in the Surface Hall in case of an intervention or exchange of a large SHiP component. The proposed configuration for the support system for the concrete beams is detailed in Fig.~\ref{fig:Preliminary support system}. In order to avoid the ‘diagonal’ slip of the concrete along the opening, a dowel joint shall be considered in their design. The \textbf{alcove} (highlighted in Fig.~\ref{fig:Underground Experimental Hall layout}) is the volume of the underground hall which is “inserted” into the Target Complex structure and it is the interface between the Target Hall and the Experimental Hall (see Fig.~\ref{fig:alcoveh}). The target bunker and the Experimental Hall are separated by a concrete wall with a window made of a thin (approx. \SI{30}{\milli\meter} in thickness) steel plate centred on the beam axis allowing the first muon shield magnet to be as close as possible to the magnetized volume in the target shielding (\textit{i.e.} the hadron stopper).
The lateral dimensions of the window, which are driven by the lateral size of the first muon shield magnet, are not yet precisely defined. 
The window should provide complete air-tightness from the Target Complex.

There are \textbf{two main access to the underground hall}: at approximately \SI{33}{\meter} from the the Target Complex interface, via an elevator and stairs  and at the end of the hall, and via a second stairs for emergency purposes (shown in Fig.~\ref{fig:Underground Experimental Hall layout} as emergency exit) of which ends in a small-dedicated access shed located at ground level and behind the Surface Hall. 

\begin{figure} [ht!]
     \centering
     \includegraphics[width=\linewidth,height=3in]{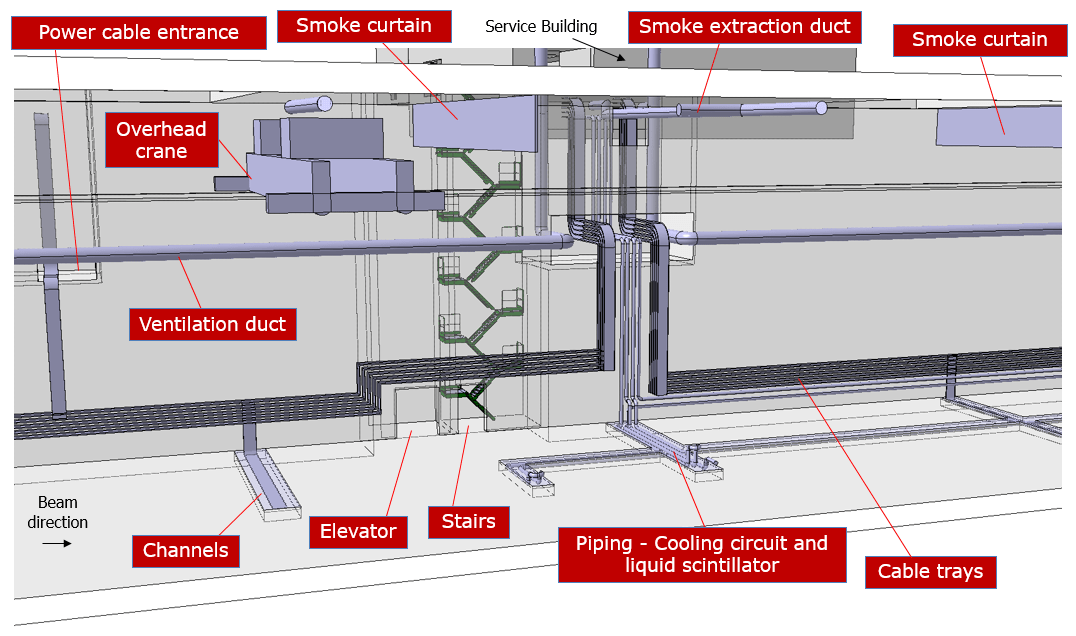}
     \caption{Services integration of the underground Experimental Hall. SHiP detector is hidden for better illustration}
     \label{fig:Exphallservices}
 \end{figure}

There is an opening (indicated in Fig.~\ref{fig:Underground Experimental Hall layout} as \textbf{services entrance}) between the Service Building's basement and the Experimental Hall through which the infrastructure and the detector services are routed \textit{i.e.} ventilation ducts, extraction smoke system, cooling, cables trays, etc. In addition, there are two more openings upstream and downstream of the personnel access for the routing of the power cables of the magnets. The \textbf{services are routed along the walls and the ceiling} (see Fig.~\ref{fig:Exphallservices}). Channels, approximately \SI{0.5}{\meter} deep and \SI{1}{\meter} wide, covered by steel plates and fitted with trays for pipes and cabling in the cavern floor allow routing the services from the wall to the SHiP detector. 

\begin{figure} [ht!]
     \centering
     \includegraphics[width=\linewidth,height=1.8in]{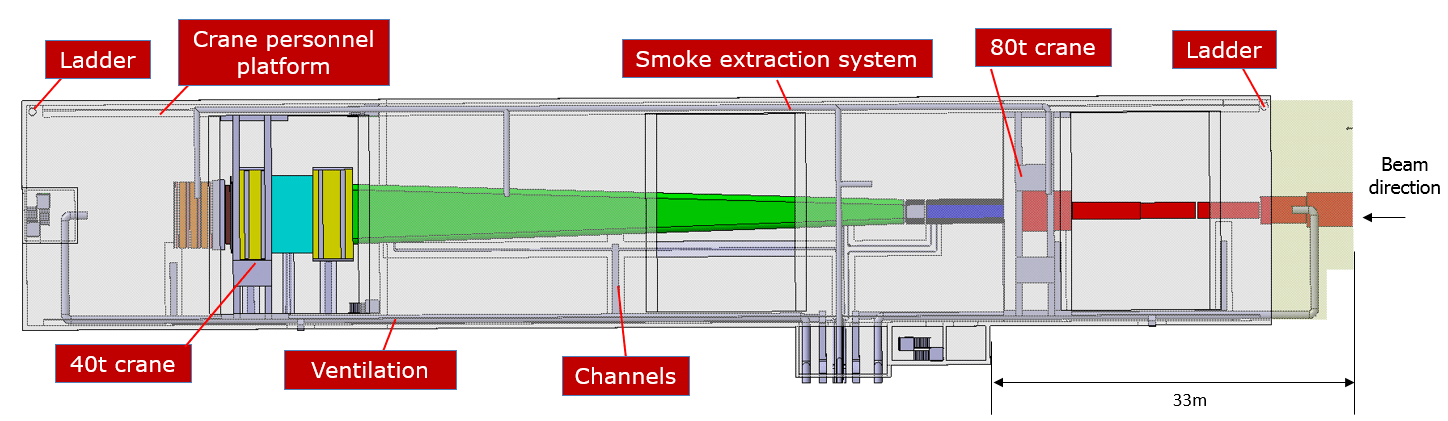}
     \caption{Top view of the underground Experimental Hall}
     \label{fig:Exphalltop}
 \end{figure}

For construction and handling purposes, a \SI{40}{\tonne} overhead crane and an overhead crane with a dual \SI{40}{\tonne}-hoist (\SI{80}{\tonne} capacity) are foreseen in the underground hall. There is a personnel platform for maintenance purposes on the right wall (taking as reference the beam direction) of which is accessible by two ladders (see Fig.~\ref{fig:Exphalltop} and Fig.~\ref{fig:Crossexp}). Smoke extraction system, sprinklers, lighting etc. will be located above the cranes. Ventilation, cable trays and cooling pipes will be routed along the walls below the cranes as shown in Fig.~\ref{fig:Exphallservices}. The smoke extraction curtains can be raised manually for the cranes displacement along the hall.
  
\subsection{Surface hall (BA91)}

The Surface Hall will be mainly dedicated to the assembly of the SHiP experiment. The building has an internal length of \SI{100}{\meter}, a internal  width of \SI{26.5}{\meter}, and a FFL of approximately 452.4 masl. Its maximum external height is \SI{16.3}{\meter} that does not exceed the height of the EHN1 extension (the highest building of the CERN Prevessin site). The internal height of the structure (\SI{14.7}{\meter}) is driven by the following requirements: 
 
\begin{itemize}

 \item height of the lorry access door (\SI{7}{\meter});
 \item clearance between the top of the door and the underside of the \SI{10}{\tonne} crane for services installation (\SI{1.2}{\meter});
 \item height of the \SI{10}{\tonne} crane (\SI{2}{\meter});
 \item clearance between the top of the \SI{10}{\tonne} crane and the underside of the \SI{40}{\tonne} crane (\SI{0.5}{\meter});
 \item height of the \SI{40}{\tonne} crane (\SI{3}{\meter});
 \item clearance between the top of the \SI{40}{\tonne} crane and the underside of the services (\SI{0.5}{\meter});
 \item allowance for lighting and services (\SI{0.5}{\meter}). 
  \end{itemize}

\begin{figure}[ht!]
    \centering
    \includegraphics[width=6.2in,height=2in]{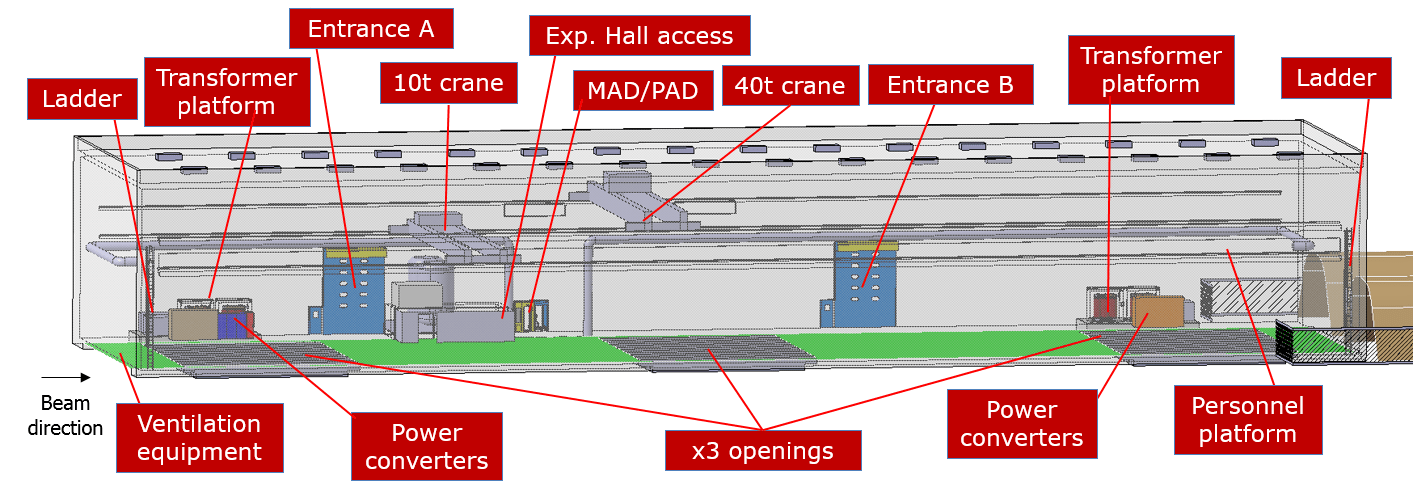}
    \caption{Surface Hall layout. Surface highlighted in green corresponds to the underground Experimental Hall ceiling}
    \label{fig:Surfacehall}
\end{figure}

The roof has a \SI{1.1}{\meter} parapet wall for personnel access safety. The layout of the Surface Hall is shown in Fig.~\ref{fig:Surfacehall}, where the surfaces highlighted in green correspond to the underground Experimental Hall ceiling. The plan view and sections can be found in Ref.~\cite{Surfhalldra}. 
 
The Surface Hall building is \SI{6.5}{\meter} wider than the underground Experimental Hall due to RP requirements and handling purposes (see Fig.~\ref{fig:Experimental area layout}). It has \textbf{two exterior doors} (\SI{7}{\meter} high and \SI{5}{\meter} wide), shown as entrance A and B, and \textbf{one access to the Service Building}.  The large SHiP components will be lowered into the underground Experimental Hall through the \textbf{three openings} that will be fully opened during the SHiP assembly period. For this reason, they will be fenced for personnel safety. Concrete beams will cover them when the SHiP installation is completed.
 
During operation, a \textbf{MAD/PAD} (Material Access Door/Personnel Access Door) controls the access of personnel and light material to the Surface Hall via the Service Building.  The underground hall is accessed by the \textbf{elevator and stairs} located approximately 25m from the upstream wall of the Surface Hall.
 
For handling purposes, the Surface Hall is equipped with \textbf{two cranes}, a \SI{40}{\tonne} crane and, below, a \SI{10}{\tonne} crane that run along the full length of the \SI{100}{\meter} long building. In order for the two cranes to be used independently, each crane is on its own dedicated rails. A personnel platform on the right wall (taking as reference the beam direction) provides access for maintenance purposes. The platform is reachable by two ladders (see Fig.~\ref{fig:Surfacehall}).

Two \textbf{transformer platforms} are located outside the Surface Hall on the Jura side to connect the Experimental Area with the CERN's electrical network. The base shall be designed such that water does not pass from the transformer base into the Surface Hall. Cabling will pass underground and through a false floor inside the Surface Hall to where the power converters are located.
 
The \textbf{ventilation equipment} for the Surface Hall and the Experimental Hall is installed in the upstream end of the building between the first access opening and the front wall. Its allocated space reservation is indicated in Fig.~\ref{fig:Surfacehall} as ventilation equipment.

\begin{figure}[ht!]
    \centering
    \includegraphics[width=6in,height=3in]{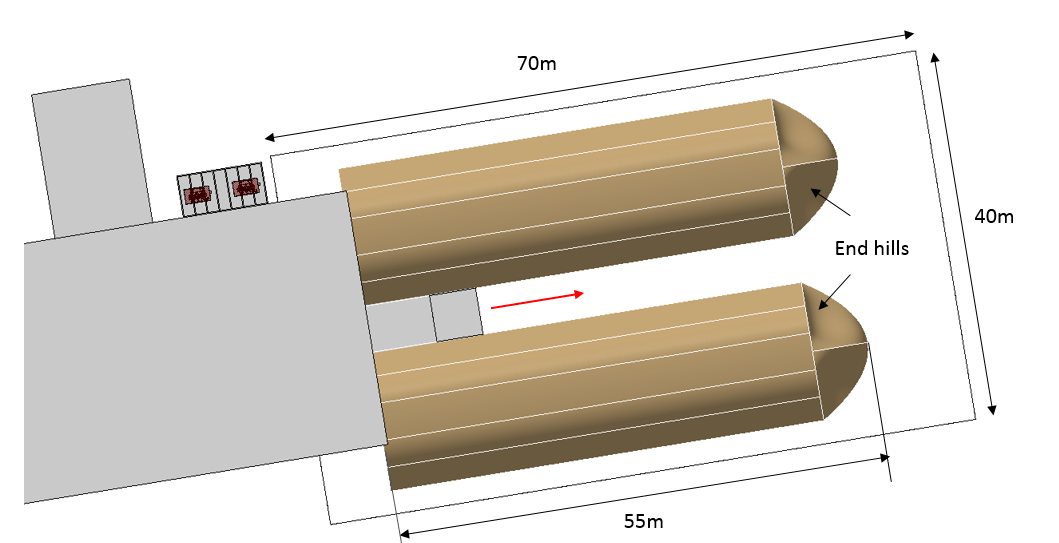}
    \caption{Fenced area and land hills dimensions. Unobstructed emergency escape route is indicated by the red arrow}
    \label{fig:endhills}
\end{figure}

As shown in Fig.~\ref{fig:endhills}, there are two \SI{6}{\meter} high \textbf{land hills} downstream in contact with the Surface Hall for shielding. The hills leave space for a free escape route from the Experimental Hall's downstream emergency exit. The hills are \textbf{fenced} to avoid personnel accessing the area during beam operation.
 
\subsection{Service Building (BAE91)}

\begin{figure}[ht!]
  \centering
  \includegraphics[width=\linewidth]{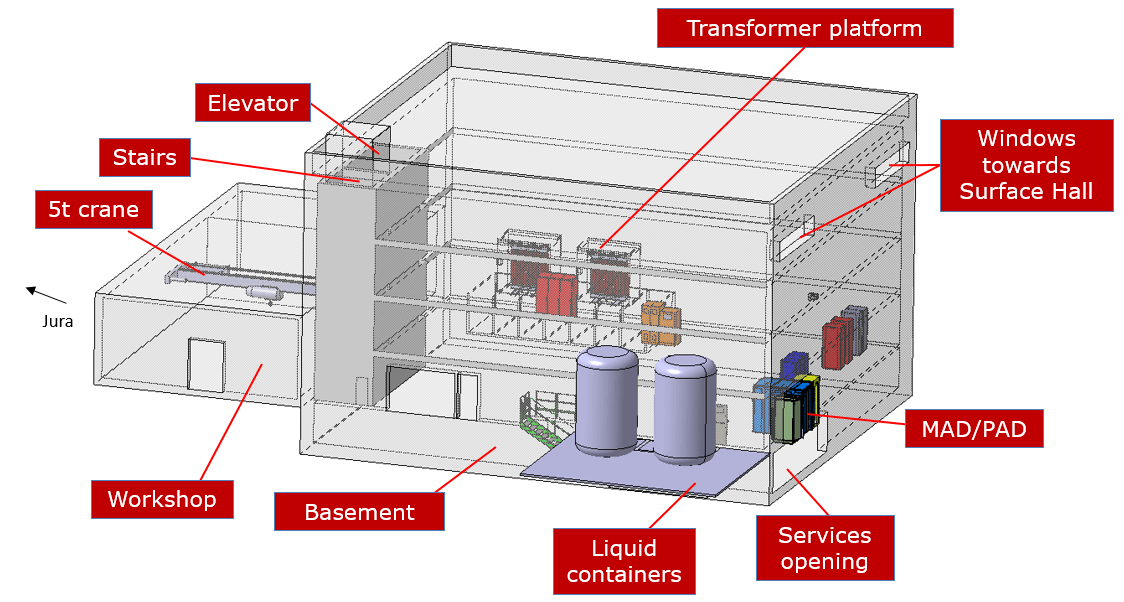}
  \caption{Service Building layout (floor layouts are hidden for clarification)}
  \label{fig:Service Building layout}
 \end{figure}

The Service Building is located next to the Surface Hall. Its purpose is to house all the power supplies, control and electronics racks, network, computing, cooling and ventilation equipment and the area required for the detector operation. With its \SI{20}{\meter} length and \SI{34}{\meter} width, its three floors also include space for offices, laboratories, a control room (for the SHiP detector and the Target remote handling \cite{Targetservices}), a conference room and a workshop as an adjacent independent building. Electrical and electronics racks, ventilation equipment, pumps etc. will be located on the ground floor. The roof has a \SI{1.1}{\meter} parapet wall for personnel safety when accessing the roof (\textit{i.e.} for elevator maintenance). The layout of the Surface Building is shown in Fig.~\ref{fig:Service Building layout}.

The three floors of the Service Building are accessed by stairs and an elevator with a capacity of 13 people. The installation and de-installation of electrical racks, AHUs, pumps, etc. along the ground floor is foreseen to be performed with a pallet truck. As the system management room and the computing infrastructure are located on the first floor, the racks (loaded on a pallet truck) are to reach them through the elevator. There is a \textbf{personnel and light material access through a MAD/PAD} to the Surface Hall on the ground floor. A \textbf{\SI{5}{\tonne} overhead crane} is foreseen to be installed in the workshop.

\textbf{Storage tanks} (\SI{300}{\cubic\meter} volume in a \SI{50}{\meter\squared} space reservation) for the liquid scintillator required by one of the detectors of SHiP are located on one side of the building. The \textbf{transformers platform} is located on the opposite side. Their bases shall be designed such that water does not pass from the base into the Service Building.

\begin{figure}[ht!]
  \centering
  \includegraphics[width=5.01in,height=3.3in,keepaspectratio=false]{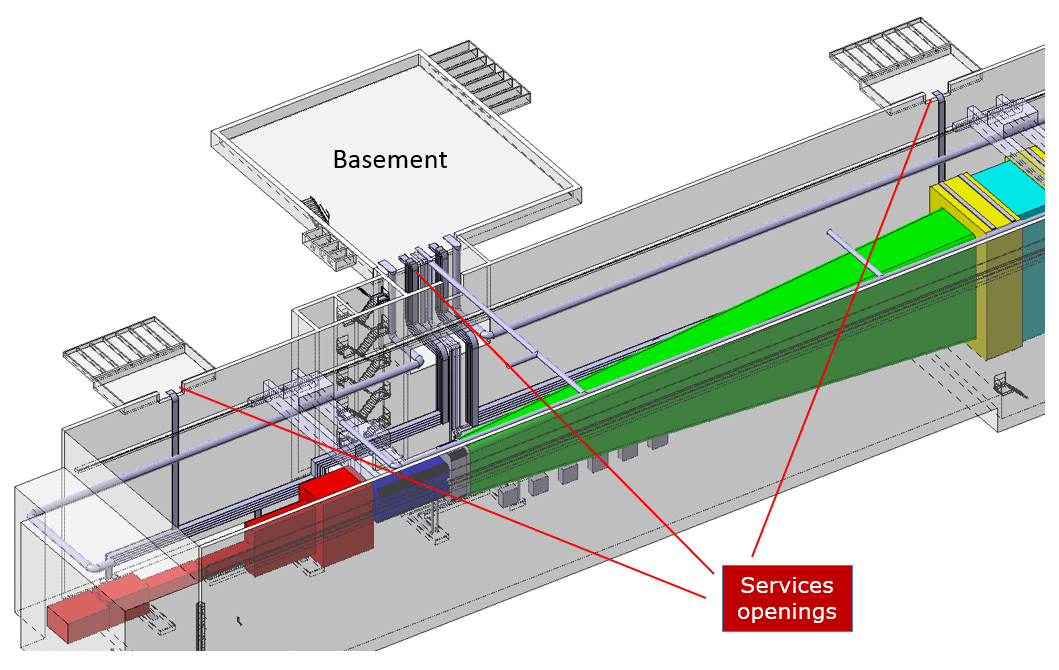}
  \caption{Overview of the distribution of the services along the Experimental Area}
  \label{fig:Overview of the distribution of the services}
 \end{figure}

The Service Building is constructed with a basement dedicated for the distribution of all the services throughout the building. The cables and services will pass through the basement wall, which is the interface between the Service Building and the Surface Hall, to the underground Experimental Hall (see Fig.~\ref{fig:Overview of the distribution of the services}). The detailed cable layout has yet to be defined.

\begin{figure}[ht!]
  \centering
  \includegraphics[width=6in,height=2.8in,keepaspectratio=false]{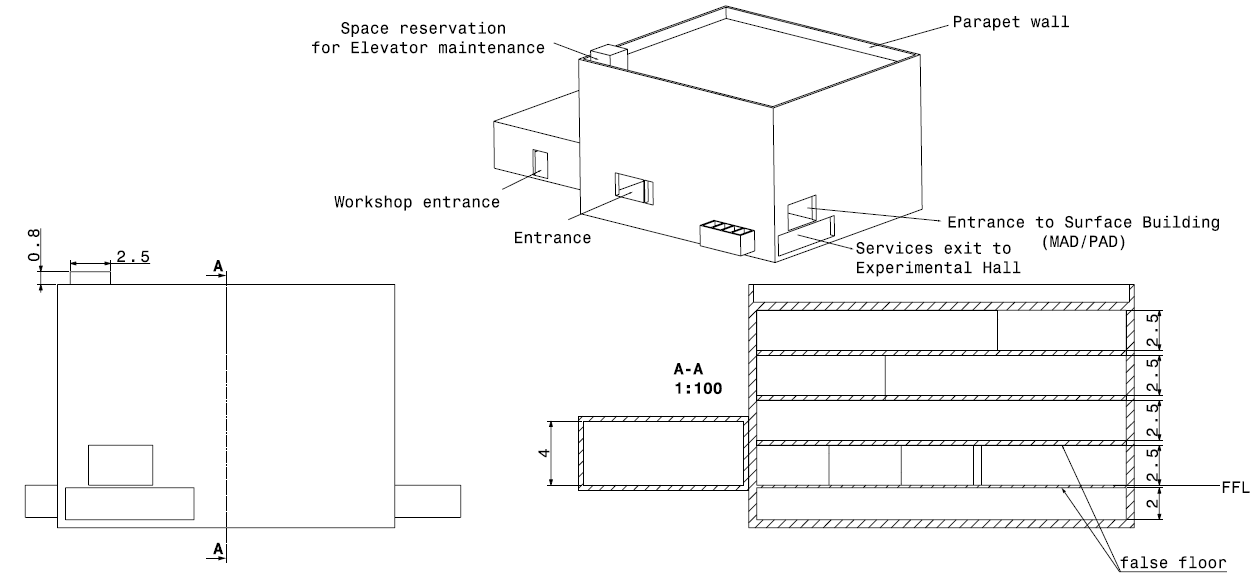}
  \caption{Plan view and sections of the Service Building}
  \label{fig:Plan view Service Building}
 \end{figure}

\begin{figure}[ht!]
    \begin{subfigure}[b]{.45\linewidth}
        \centering
        \includegraphics[width=\linewidth]{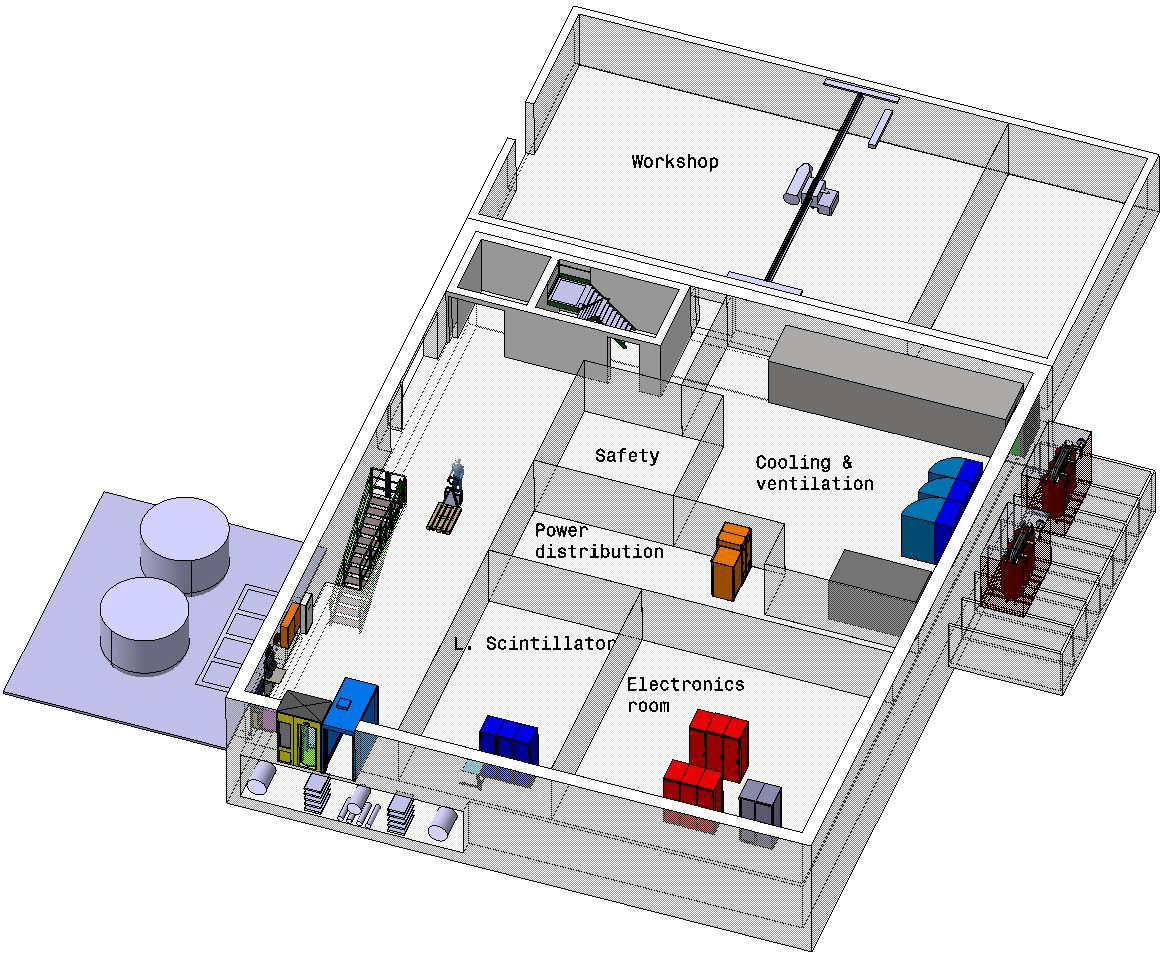}
        \caption{Ground floor} 
        \label{fig:Serbground}
    \end{subfigure}
    \begin{subfigure}[b]{.45\linewidth}
        \centering
        \includegraphics[width=\linewidth]{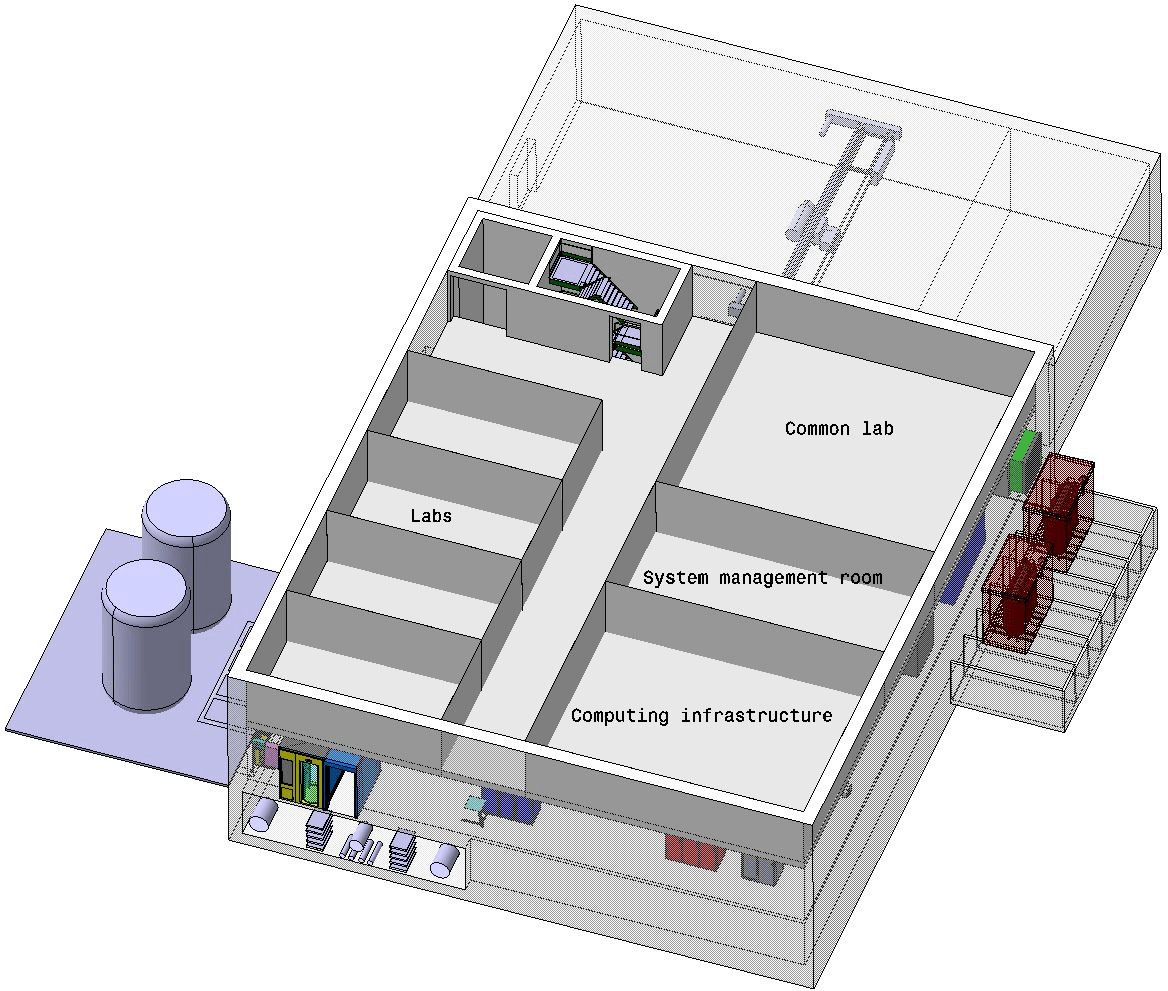}
        \caption{First floor} 
        \label{fig:Serbfirst}
    \end{subfigure}
    \begin{subfigure}[b]{.45\linewidth}
        \centering
        \includegraphics[width=\linewidth]{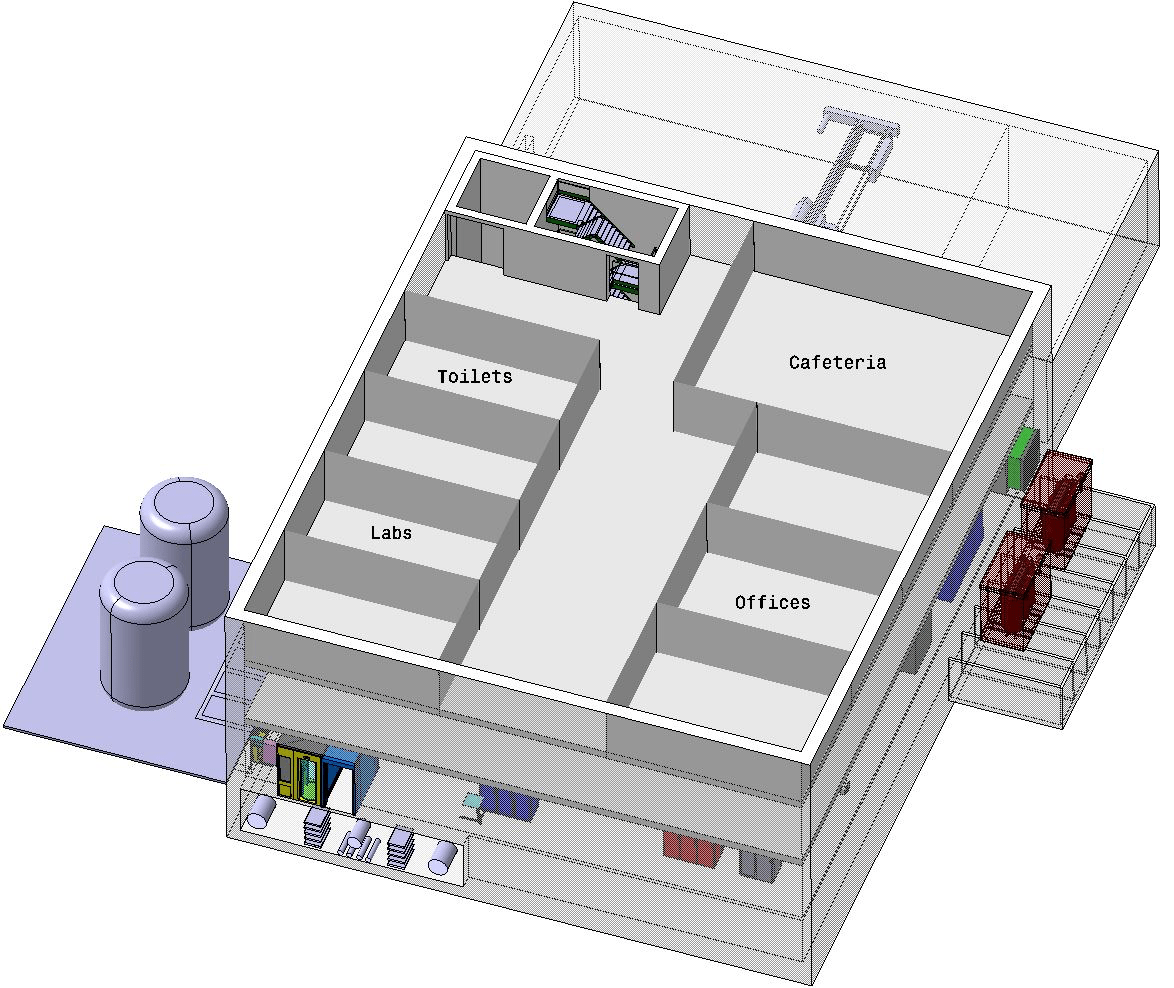}
        \caption{Second floor}
        \label{fig:Serbsecond}
    \end{subfigure}
    \begin{subfigure}[b]{.45\linewidth}
        \centering
        \includegraphics[width=\linewidth]{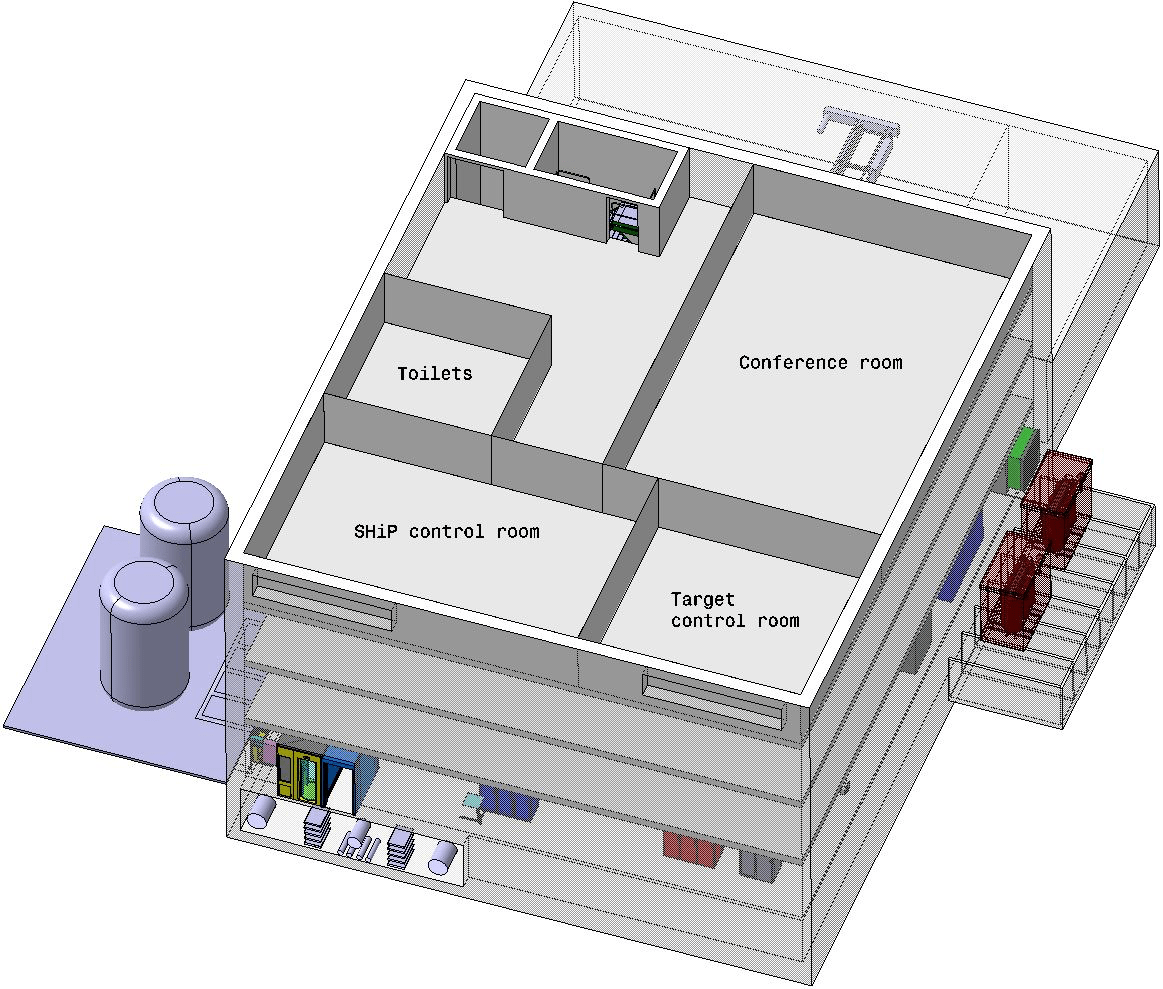}
        \caption{Third floor}
        \label{fig:Serbthird}
    \end{subfigure}
    \caption{Internal layout of the Service Building}
    \label{fig:SerB1}
\end{figure}

Fig.~\ref{fig:Plan view Service Building} and  Ref.~\cite{Servdra} show the plan view, layout and sections of the Service Building. Its internal layout, which has been distributed based on the SHiP requirements (see Table \ref{tab:SHiPservices} and \ref{tab:SHiPspaces}), is shown in Fig.~\ref{fig:SerB1}).

 \subsection{Gas building (BG91)}

Its purpose is to store and supply all the gases foreseen to be used in the Experimental Area. For instance, the straws tracker of the SHiP detector will need a gas mixture of CO${}_{2}$/Ar.  A gas line coming from the Pr\'evessin gas distribution network is not foreseen as the required quantity estimated is only a few m${}^{3}$ at present. The installation will be designed  to be compatible with the storage of standard gas bottles.
 
Its internal layout is not yet specified; however, a reception and a storage area are required. Its estimated ground surface is \SI{15}{\meter}$\times$\SI{10}{\meter} and a \SI{3}{\meter} high ceiling is assumed. The building shall comply with the Eurocode guidelines and the relevant National Annexes for a gas installation. 

\subsection{Surface Hall and Experimental Hall cross-sections}
\label{crosssections}

\begin{figure}[h]
  \centering
  \includegraphics*[width=4in,height=1.5in,keepaspectratio=false]{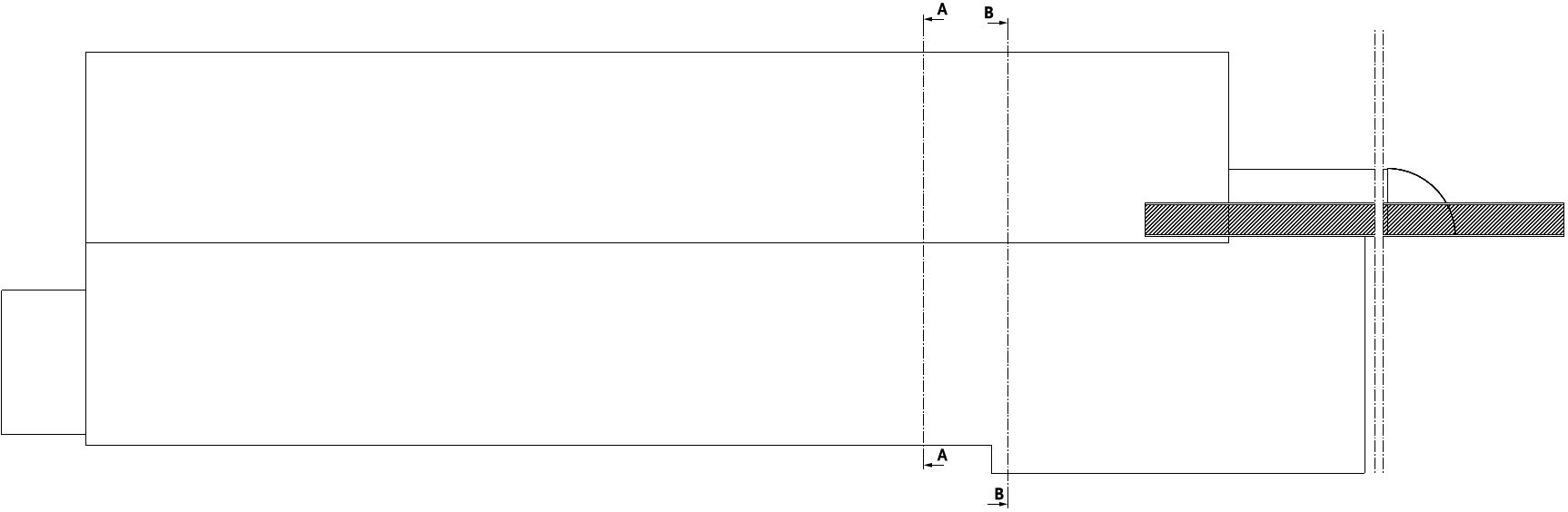}
  \caption{Underground Experimental Hall section cut}
  \label{fig:Cross2}
 \end{figure}

\begin{figure}[ht!]
    \begin{subfigure}[b]{.49\linewidth}
        \centering
        \includegraphics[width=\linewidth]{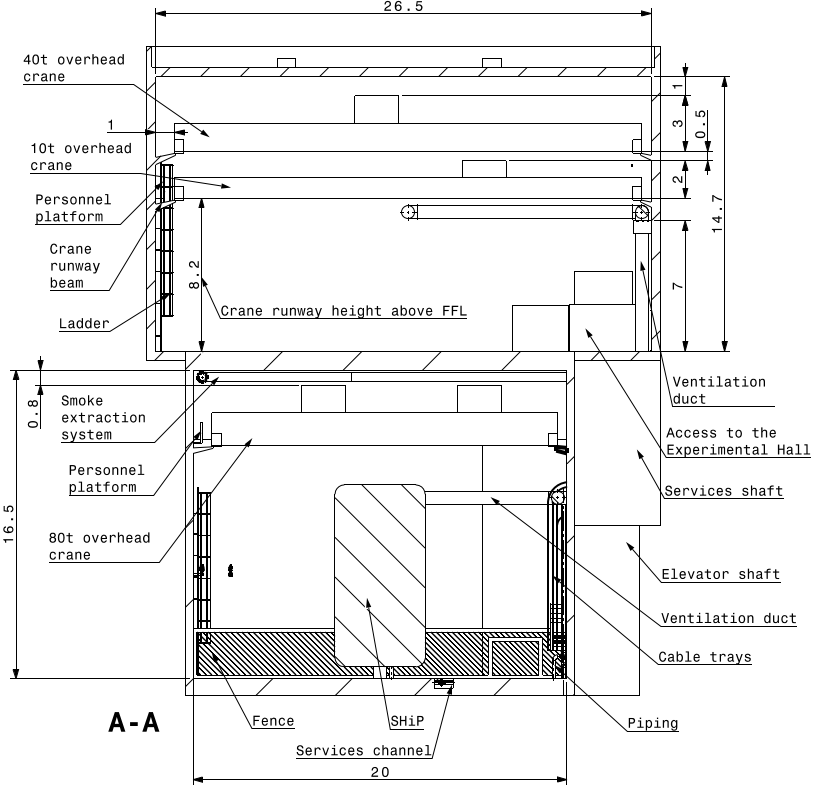}
        \caption{Surface Hall height breakdown} 
        \label{fig:Crosssur}
    \end{subfigure}
    \begin{subfigure}[b]{.49\linewidth}
        \centering
        \includegraphics[width=\linewidth]{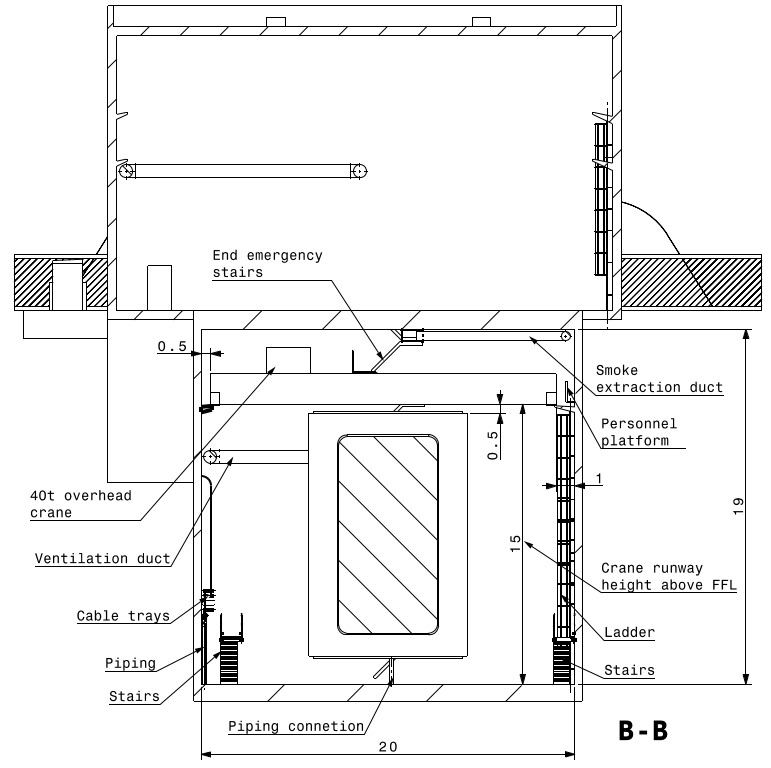}
        \caption{Experimental Hall height breakdown} 
        \label{fig:Crossexp}
    \end{subfigure}
    \caption{Experimental Area cross-sections}
    \label{fig:Cross1}
\end{figure}

Fig.~\ref{fig:Cross2} indicates the location of the section cuts shown by Fig.~\ref{fig:Crosssur} and \ref{fig:Crossexp} along the Experimental Area.

Fig.~\ref{fig:Crosssur} describes the section cut of the Surface Hall and the upper floor of the Experimental Hall. Cable trays and piping run along the walls, below and above the \SI{80}{\tonne} crane while the SHiP detector is centred in the cavern. In relation with the Surface Hall, the proposed height breakdown allows the two overhead cranes to run along the entire hall independently. 

Fig.~\ref{fig:Crossexp} shows in detail the integration performed in the most restrictive section of the cavern, the lower floor of the Experimental Hall. At this location, the \SI{13}{\meter}  spectrometer magnet has to be centred with respect to the beam axis and, in addition, the crane must pass above it for installation purposes. As well as in the Experimental Hall's upper floor, services run along the walls, below and above the crane.

\newpage

\subsection{Civil engineering}
 
The civil engineering studies carried out were based on the foreseen activities and the equipment/SHiP components to be housed inside the buildings.

The study of the estimated service loads foreseen in the Service Building, Gas Building, Surface Hall and Experimental Hall based on the infrastructures proposed is detailed in Ref.~\cite{Expareastructural}. In particular, the SHiP experiment service loads are described in Ref.~\cite{SHiPload}. 
The heavier SHiP components which require to be displaced for maintenance or equipment replacement (\textit{i.e.} the particle ID sub-components) will be installed on trolleys on floor rails or suspended with frames or girders. Air-pads \cite{airpads} could be used for their displacement along the rails.

The key features of the proposed layout are:
 \begin{itemize}
  \item Underground Experimental Hall located behind the Target Complex to house the SHiP detector. 
  \item Surface Hall above the Experimental Hall for the assembly of the SHiP experiment. 
  \item Service Building on the Jura side of the Surface Hall to provide electrical and CV (among others) supply to the SHiP Experiment.
  \item Gas Building for gas supply to the BDF facility.
 \end{itemize}

As the facility is foreseen to be located in the CERN site at Prévessin, it is assumed that the existing facilities inside this campus such as; restaurants, main access, and road network are sufficient. However, for the parts which surround the BDF facility, the following items will have to be included:
 \begin{itemize}
  \item Roads and car parks. 
  \item Drainage networks. 
  \item Lighting.
  \item Fire water hoses.
  \item Landscaping and planting.
 \end{itemize}
 
 Refer to Chapter \ref{Chap:CivEng} for further details of the CE studies.

\subsection{Transport and handling}

The quantity and variety of equipment to be installed in the Experimental Area is defined by the SHiP experiment. As defined in Chapter \ref{Chap:ExpHall}, the detector will be assembled in the Surface Hall with a maximum crane lifting load of \SI{40}{\tonne} and the pre-assembled SHiP components will be lowered into the underground Experimental Hall.  In addition, supports, cooling and ventilation ducts, electrical cables and cable trays are to be lowered and installed to the underground area. Detailed transport and handling solutions will need to be defined for all of the industrial equipment and the SHiP pre-assembled  and detector components, developing special tooling wherever necessary.
 
For cooling and ventilation equipment, electrical cables and cable trays, industrial standard handling equipment are foreseen. 

 The transport and installation operations include: 
 \begin{itemize}
  \item Unloading material and equipment. 
  \item Transfer within the surface buildings and from the Surface Hall to the Experimental Hall for the purposes of assembly, test and storage.
  \item Transport along the Experimental Hall and final installation.
\end{itemize}

The SHiP parts and materials will arrive to the Surface Hall by a \SI{40}{\tonne} semi-trailer or a normal \SI{19}{\tonne} truck. Then, the overhead cranes or a forklift (\SI{30}{\tonne}/axle) will unload the trucks and move the equipment along the Surface Hall. A \SI{40}{\tonne} overhead crane and a 10t crane will lift and lower the pieces from the Surface Hall to the Experimental Hall. It is noted that the maximum distance from the FFL to the \SI{10}{\tonne} crane's hook is about \SI{8}{\meter} and, in the case of the \SI{40}{\tonne} crane, $\sim$\SI{11}{\meter}. 

A \SI{40}{\tonne} overhead crane and an overhead crane with a dual \SI{40}{\tonne}-hoist (\SI{80}{\tonne} capacity), which can access the full length of the hall, are foreseen in the Experimental Hall. In addition, a few small temporary cranes will support the construction and installation of the detector.

Table~\ref{tab:Crane specifications} lists the crane specifications for the Experimental Area. The cross section of the height of the Surface Hall and Experimental Hall are shown in Fig.~\ref{fig:Cross1}.
 
 \begin{table} [htpb]
 \centering
 \caption{Experimental Area crane specifications}
 \label{tab:Crane specifications}
 \begin{scriptsize}
 \begin{tabular}{ccccccccc}
   \hline 
   \multicolumn{1}{c}{\textbf{Location}} & \textbf{\begin{tabular}[c]{@{}c@{}}Lifting \\ {capacity [}t{]}\end{tabular}} & \textbf{\begin{tabular}[c]{@{}c@{}}Distance from \\ the FFL to hook {[}m{]}\end{tabular}} & \textbf{\begin{tabular}[c]{@{}c@{}}Span \\ {[}m{]}\end{tabular}} & \textbf{\begin{tabular}[c]{@{}c@{}}Travel \\ {[}m{]}\end{tabular}} & \textbf{\begin{tabular}[c]{@{}c@{}}Length \\ {[}m{]}\end{tabular}} & \textbf{\begin{tabular}[c]{@{}c@{}}Width \\ {[}m{]}\end{tabular}} & \textbf{\begin{tabular}[c]{@{}c@{}}Height \\ {[}m{]}\end{tabular}} & \textbf{Quantity} \\ \hline
   Surface Hall & 40 & 11.2/28.7/31.2 & 25 & 97 & 6 & 20 & 3 & 1 \\ 
   Surface Hall & 10 & 8.3/25.8/28.3 & 25 & 97 & 5 & 20 & 2 & 1 \\ 
   Experimental Hall & 40 & 13.2/15.7 & 18.5 & 109 & 6 & 20 & 3 & 1 \\ 
   Experimental Hall & 2x40 & 13.2/15.7 & 18.5 & 109 & 6 & 20 & 3.2 & 1 \\ 
   Workshop & 5 & 2.5 & 9 & 20 & 3 & 10 & 1.5 & 1 \\ 
   \hline 
  \end{tabular}
  \end{scriptsize}
 \end{table}

Two lifts are required in the Experimental Area. The first one is located in the Service Building to allow personnel and light material access the three floors. As the system management and computer rooms are housed on the first floor (see Ref.~\cite{Servdra}), the lift should have the capacity to raise electronics racks and laboratory equipment. The second lift provides personnel and material access (e.g. pallets, chariots, light equipment etc.) from the Surface Hall to the Experimental Hall. Technical details of the elevators are described in Ref.~\cite{Exparealift}. Table~\ref{tab:Lifts specifications} shows the main lifts specifications of the Experimental Area.

 \begin{table}[htpb]
 \centering
 \caption{Experimental Area lifts specifications}
 \label{tab:Lifts specifications}
\begin{tabular}{cccc} 
\hline
\textbf{Location} & \textbf{Number of levels} & \textbf{Depth [m]} & \textbf{Capacity} \\ \hline 
Access to experimental cavern & 2 & 17.5 & \SI{3}{\tonne} \\  
Service building & 4 & 9 & \SI{1}{\tonne} \\ 
\hline
\end{tabular}

 \end{table}
 
\subsection{Electrical network system}

The SHiP detector is foreseen to consume approx. \SI{3.1}{\mega\watt} (see Table \ref{tab:SHiPservices}) with a total of six magnets for the muon shielding, one emulsion spectrometer magnet and a spectrometer magnet with a weight of more than \SI{1000}{\tonne}. 
The electrical power system is divided in two sub-systems:
  \begin{itemize}
  \item Conventional power (magnet power supplies, electronic racks, cooling and ventilation systems, and infrastructure components); 
  \item Emergency power provided by back-up generators (emergency lighting, sump pumps and ventilation systems for sub-surface enclosures). 
 \end{itemize}

Any critical system that cannot accept any power interruption should be provided with an Uninterruptible Power Supply (UPS) system. The emergency supply system is based on batteries of which are located in the Auxiliary Building (see Fig.~\ref{fig:6SB}); therefore, its supply will be routed through the technical galleries to reach the Experimental Area. A proposal for the technical galleries layout has been performed assuming that the BDF electrical connection to the Pr\'evessin network is made through the CERN's service building BA80 (see Fig.~\ref{fig:Layout}).
 
The electrical power supply description has been estimated from the CENF study \cite{CENFENEL}. The proposal considers only the electrical power infrastructure, thus, cabling and fibre optical infrastructure are yet to be studied. The number of cable trays shown in Fig.~\ref{fig:Overview of the distribution of the services} and Fig.~\ref{fig:Cross1} are represented only for illustration. 

The general infrastructure required for an electrical and personnel safe operation is the following:
   \begin{itemize}
    \item Installation of EL SCADA monitoring equipment to supervise the installation and the status of the electrical services from the CERN Control Center.
    \item Electrical containment, normal lighting, emergency lighting and power distribution in all the buildings and technical galleries.  
   \end{itemize}

\begin{figure}[ht!]
  \centering
  \includegraphics[width=5.47in,height=2.3in]{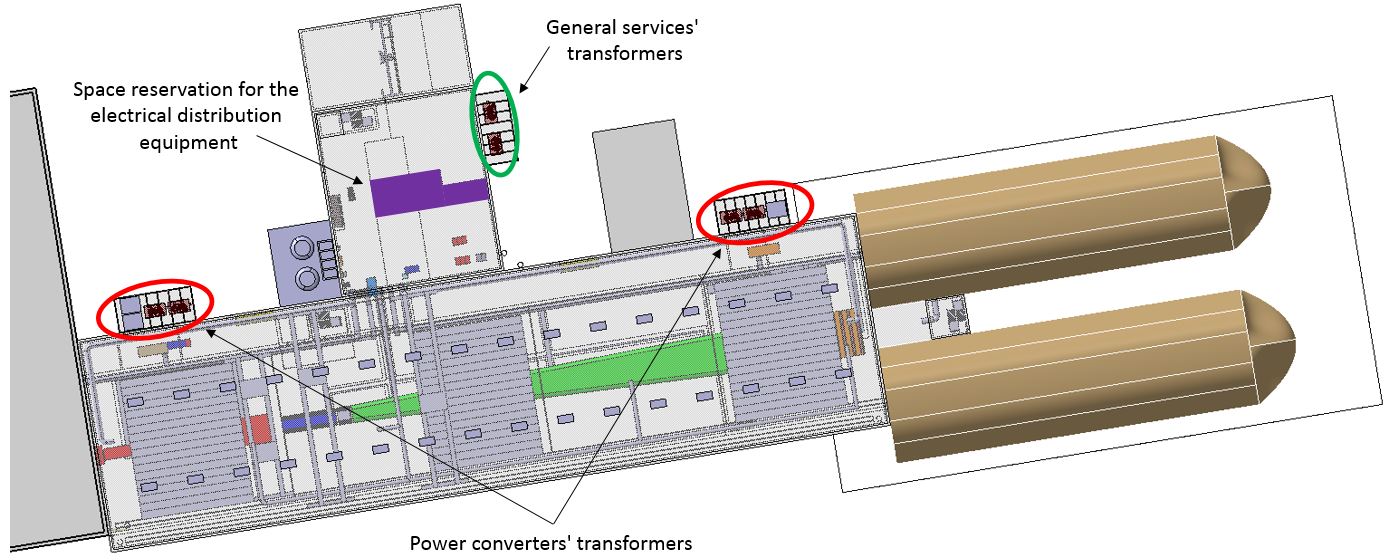}
  \caption{Transformers location along the Experimental Area}
  \label{fig:Transformers}
 \end{figure}

The transformers location and the space reserved for the electrical distribution equipment inside the Service Building are shown in Fig.~\ref{fig:Transformers}. There are in total three slabs foreseen for the transformers, of which two of them are positioned in such a way that the distance between them and the power converters is optimized. The number of transformers has yet to be defined. The transformers located next to the Service Building (highlighted in green in Fig.~\ref{fig:Transformers}) will supply the general services distribution of the Experimental Area, of which includes:
   \begin{itemize}
    \item Overhead cranes;
    \item Cooling and ventilation supplies;  
    \item General lighting and small power;
    \item Racks and electronics; 
    \item Offices;
    \item Workshop.
   \end{itemize}
   
The loads for experiment magnets will be supplied through the transformers pointed out in red in Fig.~\ref{fig:Transformers} from a dedicated Type 1 and Type 2 distribution boards located in the Surface Hall. Table~\ref{tab:ENELrequi} lists the estimated power of the Experimental Area loads.
   
\begin{table} [htpb]
  \centering
  \caption{Experimental Area electrical power loads}
  \label{tab:ENELrequi}
  \begin{tabular}{cccccc}
  \hline 
   \multicolumn{2}{c}{\textbf{Pulsed loads}} \\ \hline
   Power converters (experiments magnets) & \SI{3.1}{\mega\watt} RMS \\ 
   TOTAL & \SI{3.1}{\mega\watt} RMS \\ \hline
   \multicolumn{2}{c}{\textbf{Stable loads}} \\ \hline 
   Ventilation and cooling & \SI{320}{\kilo\watt} \\
   General services & \SI{280}{\kilo\watt} \\
   Racks and electronics & \SI{200}{\kilo\watt} \\ 
   TOTAL & \SI{800}{\kilo\watt} \\ \hline
   Secure networks loads & Located in the Auxiliary Building \\
   \hline 
  \end{tabular}
\end{table}

\FloatBarrier  
\subsection{Electrical power converters}

\begin{figure}[ht]
  \centering
  \includegraphics[width=\linewidth]{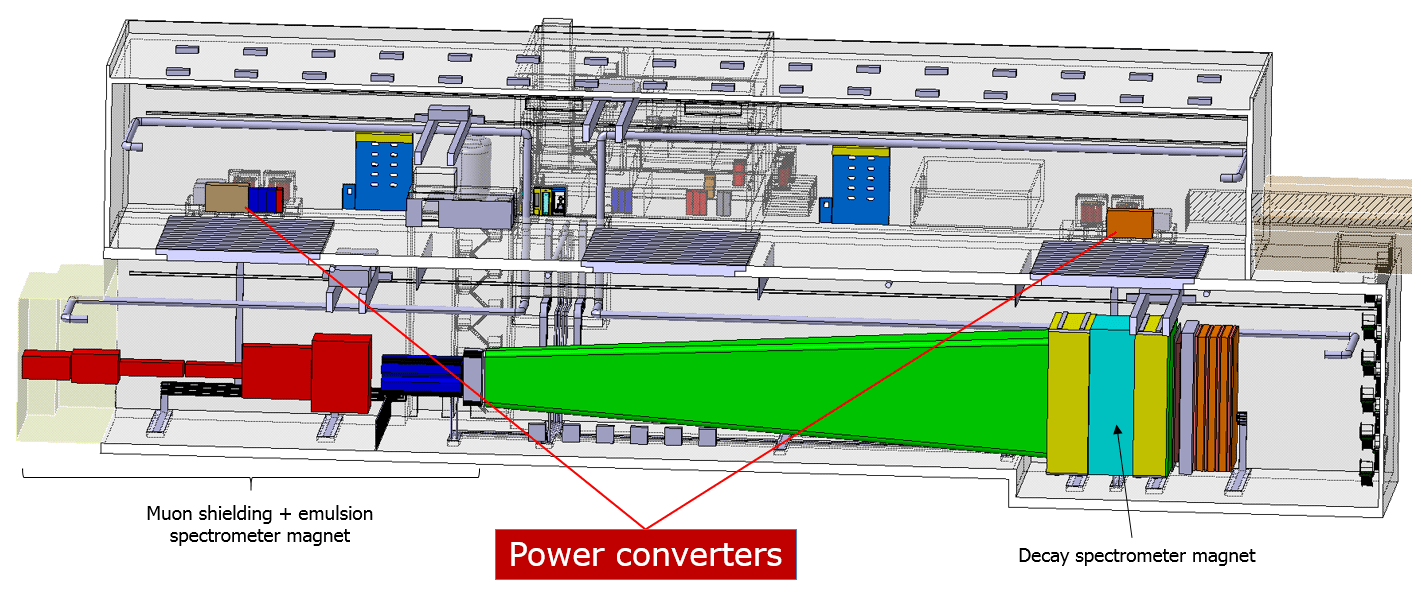}
  \caption{Power converters layout}
  \label{fig:Power converters layout}
 \end{figure}

Power converters are required for the magnets of the muon shielding, the emulsion spectrometer magnet and the decay spectrometer magnet. The criteria is to locate the power converters in such a way that the cable length between the magnets and them is minimized.  Therefore, as the decay spectrometer magnet is about \SI{60}{\meter} away from the emulsion spectrometer magnet, two locations for the power converters have been established (see Fig.~\ref{fig:Power converters layout}).

A preliminary design of the power converters has been provided based on the current and power inputs given in Chapter \ref{Chap:ExpHall}, excluding the resistance of the DC cables. 
Table~\ref{tab:Power converters design parameters} lists the magnets characteristics taken into account in the design. 
The performance of the current sources requested is ``accuracy class 4" which establishes a long-term stability (1 year) of 100~ppm, 20~ppm during 12 hours and in the short term (20 min.) a stability of 5~ppm. 
The modules will be controlled by a FGC3 platform taken in consideration that in DC operation mode, there is no control of the current during ramp up and ramp down.

 \begin{table} [htpb]
  \centering
  \caption{Power converters design parameters}
  \label{tab:Power converters design parameters}
  \begin{tabular}{cccccc}
  \hline 
  \textbf{Circuit name} & 
  \textbf{\begin{tabular}[c]{@{}c@{}} No. of magnet \\ {in series}\end{tabular}} &
  \textbf{\begin{tabular}[c]{@{}c@{}} $R_{mag}$ \\ {at 20ºC}\end{tabular}} & 
  \textbf{\begin{tabular}[c]{@{}c@{}} $I_{max}$ \\ {requested {[}A{]}}\end{tabular}} &
  \textbf{\begin{tabular}[c]{@{}c@{}} No. of \\ {converters}\end{tabular}} & 
  \textbf{\begin{tabular}[c]{@{}c@{}} Operation \\ {mode}\end{tabular}} \\ \hline
  \begin{tabular}[c]{@{}c@{}} Emulsion Spectrometer \\ {magnet}\end{tabular} & 1 & 8.9 & 14600* & 1 & \begin{tabular}[c]{@{}c@{}} DC – \\ {bipolarity}\end{tabular} \\ 
  \begin{tabular}[c]{@{}c@{}} Decay Spectrometer \\ {magnet}\end{tabular} & 1 & 111 & 3000 & 1 & DC \\ 
  \begin{tabular}[c]{@{}c@{}} Muon Shielding \\ {magnet 1}\end{tabular} & 1 & 400 & 50 & 1 & \begin{tabular}[c]{@{}c@{}} DC – \\ {unipolar}\end{tabular} \\ 
  \begin{tabular}[c]{@{}c@{}} Muon Shielding \\ {magnet 2}\end{tabular} & 1 & 600 & 50 & 1 & \begin{tabular}[c]{@{}c@{}} DC – \\ {unipolar}\end{tabular} \\ 
  \begin{tabular}[c]{@{}c@{}} Muon Shielding \\ {magnet 3}\end{tabular} & 1 & 360 & 50 & 1 & \begin{tabular}[c]{@{}c@{}} DC – \\ {unipolar}\end{tabular} \\ 
  \begin{tabular}[c]{@{}c@{}} Muon Shielding \\ {magnet 4}\end{tabular} & 1 & 560 & 50 & 1 & \begin{tabular}[c]{@{}c@{}} DC – \\ {unipolar}\end{tabular} \\ 
  \begin{tabular}[c]{@{}c@{}} Muon Shielding \\ {magnet 5}\end{tabular} & 1 & 1400 & 50 & 1 & \begin{tabular}[c]{@{}c@{}} DC – \\ {unipolar}\end{tabular} \\ 
  \begin{tabular}[c]{@{}c@{}} Muon Shielding \\ {magnet 6}\end{tabular} & 1 & 2000 & 50 & 1 & \begin{tabular}[c]{@{}c@{}} DC – \\ {unipolar}\end{tabular} \\ \hline 
  \multicolumn{6}{l}{*Aluminium pancake design} \\
  \end{tabular}
 \end{table}

The design criteria is to use standard CERN power converters or the combination of them in order to reduce the cost and optimize the design timing (proposed design is shown in Table \ref{tab:SHiP power converters preliminary design}). The Muon Shield Magnets design is based on three racks, each one of them being populated by 2 converters and a spare. The total numbers of power converters is eight and its running cost considering \SI{4000}{\hour} of operation per year is about 690~kCHF per year. 

   \newpage
   
   \begin{table}[htpb]
      \centering
      \caption{SHiP power converters preliminary design}
      \label{tab:SHiP power converters preliminary design}
      \begin{tabular}{cccc}
      \hline 
      \textbf{Circuit name} & \textbf{Converter type} & \textbf{\begin{tabular}[c]{@{}c@{}}No. of \\ converters\end{tabular}} & \textbf{Operation mode} \\ \hline
      Emulsion Spectrometer magnet & 
      \begin{tabular}[c]{@{}c@{}}New ALICE\_4P – \\ 2 x{[}150V/8kA{]}\end{tabular} & 1 & \begin{tabular}[c]{@{}c@{}}DC – bipolar\\ (4 quadrants operation)\end{tabular} \\ 
      Decay Spectrometer magnet & \begin{tabular}[c]{@{}c@{}}New design – \\ 4P\_B\end{tabular} & 1 & \begin{tabular}[c]{@{}c@{}}DC – unipolar \\ (1 quadrant operation)\end{tabular} \\
      Muon Shielding magnet 1 & Combo & 1 & \begin{tabular}[c]{@{}c@{}}DC – unipolar\\ (1 quadrant operation)\end{tabular} \\
      Muon Shielding magnet 2 & Combo & 1 & \begin{tabular}[c]{@{}c@{}}DC – unipolar \\ (1 quadrant operation)\end{tabular} \\ 
      Muon Shielding magnet 3 & Combo & 1 & \begin{tabular}[c]{@{}c@{}}DC – unipolar\\ (1 quadrant operation)\end{tabular} \\ 
      Muon Shielding magnet 4 & Combo & 1 & \begin{tabular}[c]{@{}c@{}}DC – unipolar \\ (1 quadrant operation)\end{tabular} \\ 
      Muon Shielding magnet 5 & Combo & 1 & \begin{tabular}[c]{@{}c@{}}DC – unipolar\\ (1 quadrant operation)\end{tabular} \\ 
      Muon Shielding magnet 6 & Combo & 1 & \begin{tabular}[c]{@{}c@{}}DC – unipolar \\ (1 quadrant operation)\end{tabular} \\ 
      \hline 
     \end{tabular}
   \end{table}
   
Regarding the cooling requirements and the ventilation contribution, the total estimated power dissipated to the Surface Hall environment is \SI{56}{\kilo\watt} and the total power cooling requested is \SI{126}{\kilo\watt}. Demineralised water is foreseen for the cooling circuit in addition to the following features:

   \begin{itemize}
    \item Maximum water temperature in the circuit \SI{27}{\celsius} with a $\Delta\theta$ of approximately \SI{10}{\kelvin}.
    \item Total flow rate equals to \SI{480}{\litre\per\minute}.  
    \item $\Delta$P of \SI{3}{\bar} with a nominal pressure less than \SI{6}{\bar}.
   \end{itemize}

The integration specifications of the power converters are listed in Table \ref{tab:Integration and cooling specifications of the power converters}. The requirements include:
      
      \begin{itemize}
       \item Each of the power converters shall be accessible from the front and the rear.
       \item Modules ALICE4P for the emulsion spectrometer magnet require an outdoor platform for \SI{18}{\kilo\volt} transformers and an additional \SI{15}{\meter\squared} space reservation for two electrical boxes.  
       \item New4PB modules for the decay spectrometer magnet requires an outdoor platform for \SI{18}{\kilo\volt} transformers and an additional \SI{8}{\meter\squared} space reservation for one electrical box.
       \item Additional control rack for Ethernet (dimensions \SI{0.6}{\meter}$\times$\SI{0.9}{\meter}$\times$\SI{2}{\meter}).
      \end{itemize} 
   
   \begin{table}[htpb]
     \caption{Integration and cooling specifications of the power converters}
     \label{tab:Integration and cooling specifications of the power converters}
     \begin{scriptsize}
     \begin{tabular}{cccccc}
    \hline 
    \textbf{Circuit name} & \textbf{Converter type} & \textbf{\begin{tabular}[c]{@{}c@{}}Dimensions \\ (length/width/height) {[}m{]}\end{tabular}} & \textbf{\begin{tabular}[c]{@{}c@{}}Weight/unit \\ {[}t{]}\end{tabular}} & \textbf{\begin{tabular}[c]{@{}c@{}}Power dissipated \\ to the air {[}kW{]}\end{tabular}} & \textbf{\begin{tabular}[c]{@{}c@{}}Water cooling \\ {[}kW{]}\end{tabular}} \\ \hline
    Emulsion Spectrometer magnet & \begin{tabular}[c]{@{}c@{}}New ALICE\_4P –\\ 2x{[}150V/8kA{]}\end{tabular} & 7.6/1.3/2.4 & 6 & 36 & 84 \\ 
    Decay Spectrometer magnet & \begin{tabular}[c]{@{}c@{}}New design \\ - (4P\_B)\end{tabular} & 4.0/1.3/2.4 & 4 & 18 & 42 \\ 
    Muon Shielding magnet 1 & Combo & 0.8/0.9/2.1 & \multirow{2}{*}{0.5} & \multirow{2}{*}{0.84} & 0 \\  
    Muon Shielding magnet 2 & Combo & 0.84 &  &  & 0 \\ 
    Muon Shielding magnet 3 & Combo & 0.8/0.9/2.1 & \multirow{2}{*}{0.5} & \multirow{2}{*}{0.84} & 0 \\  
    Muon Shielding magnet 4 & Combo & 0.84 &  &  & 0 \\ 
    Muon Shielding magnet 5 & Combo & 0.8/0.9/2.1 & \multirow{2}{*}{0.5} & \multirow{2}{*}{0.84} & 0 \\  
    Muon Shielding magnet 6 & Combo & 0.84 &  &  & 0 \\ 
    \hline 
    \end{tabular}
    \end{scriptsize}
   \end{table}
   
\begin{figure}[htbp]
    \begin{subfigure}[b]{.49\linewidth}
        \centering
        \includegraphics[width=\linewidth]{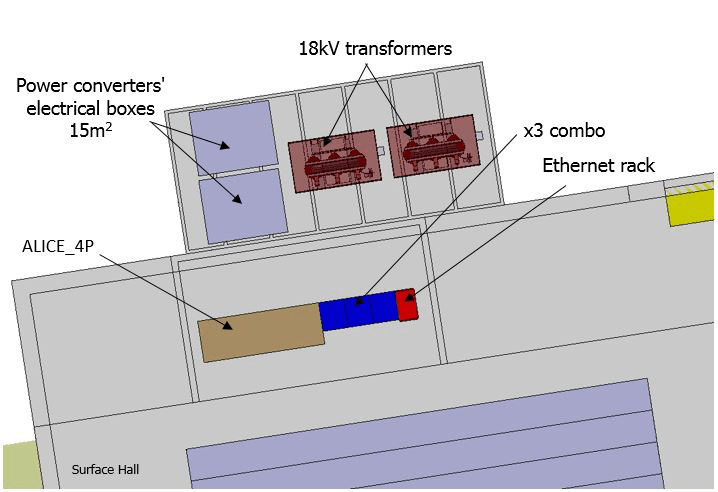}
        \caption{Emulsion spectrometer magnet and muon shield PC} 
        \label{fig:intpcemu}
    \end{subfigure}
    \begin{subfigure}[b]{.49\linewidth}
        \centering
        \includegraphics[width=\linewidth]{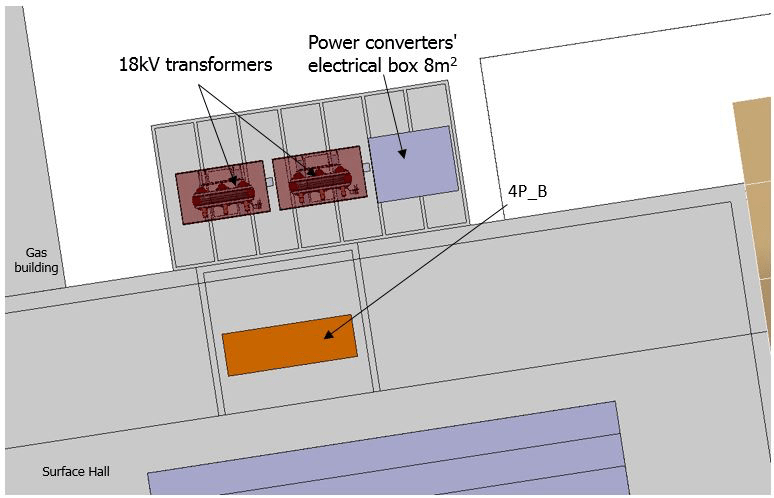}
        \caption{Decay spectrometer magnet PC} 
        \label{fig:intpcdec}
    \end{subfigure}
    \caption{Integration of the power converters (PC)}
    \label{fig:Integration of the power converters}
\end{figure}

Fig.~\ref{fig:Integration of the power converters} shows the integration of the power converters inside the Surface Hall that complies with the requirements specified. As the Surface Hall is not accessible during beam operation due to radiation protection requirements, maintenance when the beam is running is not allowed.  

\newpage
   
\FloatBarrier
Due to the requested high current, water-cooled cables could be considered for the link between emulsion spectrometer and the decay spectrometer magnets and their respective power converters. Table \ref{tab:Power converters AC and DC distribution} shows the AC and DC distribution considered in the study. The second column indicates the cables cross-section, as a recommendation from the CERN's power converters group, in order to keep the voltage drop at a reasonable level. In addition, thermal aspects have to be evaluated. The total cable length has been estimated from the integration model and it equals to two times the distance between the converter and the magnet. The total power to be installed is about 3823 kVA.
 
   \begin{table}[htpb]
       \centering
       \caption{Power converters AC and DC distribution }
       \label{tab:Power converters AC and DC distribution}
       \begin{scriptsize}
       \begin{tabular}{cccccccc}
       \hline 
       \textbf{Circuit name} & 
       \textbf{\begin{tabular}[c]{@{}c@{}}Cable total \\ length {[}m{]}\end{tabular}} & \textbf{\begin{tabular}[c]{@{}c@{}}$I_{rms}$ \\ {[}A{]}\end{tabular}} & \textbf{\begin{tabular}[c]{@{}c@{}}DC cable sect.\\Cu/Al {[}mm${}^{2}${]}\end{tabular}} & \textbf{\begin{tabular}[c]{@{}c@{}}Converter \\ type\end{tabular}} & \textbf{\begin{tabular}[c]{@{}c@{}}Electrical \\ feeder {[}V{]}\end{tabular}} & \textbf{\begin{tabular}[c]{@{}c@{}}$I_{rms}$ \textbf{for} $I_{rms}$ \\ converters\end{tabular}} & \textbf{\begin{tabular}[c]{@{}c@{}}Total installed\\   RMS power {[}kVA{]}\end{tabular}} \\ \hline
       \begin{tabular}[c]{@{}c@{}} Emulsion Spectrometer \\ {magnet}\end{tabular} & 80 & 16000 & 2000 & ALICE\_4P & 18000 & 0 & 2526 \\ 
       \begin{tabular}[c]{@{}c@{}} Decay Spectrometer \\ {magnet}\end{tabular} & 100 & 3000 & 1440 & New 4P\_B & 18000 & 0 & 1263 \\ 
       \begin{tabular}[c]{@{}c@{}} Muon Shielding \\ {magnet 1}\end{tabular} & 110 & 50 & 25 & Combo & 400 & 16 & 11 \\ 
       \begin{tabular}[c]{@{}c@{}} Muon Shielding \\ {magnet 2}\end{tabular} & 100 & 50 & 25 & Combo & 400 & 0 & 0 \\ 
       \begin{tabular}[c]{@{}c@{}} Muon Shielding \\ {magnet 3}\end{tabular} & 90 & 50 & 25 & Combo & 400 & 16 & 11 \\ 
       \begin{tabular}[c]{@{}c@{}} Muon Shielding \\ {magnet 4}\end{tabular} & 80 & 50 & 25 & Combo & 400 & 0 & 0 \\ 
       \begin{tabular}[c]{@{}c@{}} Muon Shielding \\ {magnet 5}\end{tabular} & 80 & 50 & 25 & Combo & 400 & 16 & 11 \\ 
       \begin{tabular}[c]{@{}c@{}} Muon Shielding \\ {magnet 6}\end{tabular} & 90 & 50 & 25 & Combo & 400 & 0 & 0 \\ 
       \hline 
       \end{tabular}
       \end{scriptsize}
   \end{table}
   
\subsection{Cooling and ventilation}
\label{CVsection}
The CV installations for the SHiP Experimental Area comprises the following systems:
 
 \begin{itemize}
     \item Raw water system;
     \item Demineralised water system for the cooling of magnets and power converters;
     \item Chilled water system for the cooling of ventilation units;
     \item Ventilation units and duct work;
     \item Smoke extraction system.
 \end{itemize}

Specific details on the design of the cooling and ventilation systems are provided in Ref.~\cite{ExpareaCV}. Inside the Experimental Area, two areas are reserved for the installation of CV equipment: 
 \begin{itemize}
    \item \SI{100}{\meter\squared} on the ground floor of the Service Building \cite{Servdra} which will house the secondary pumping system, AHU units for the Service Building and smoke extraction system. This area is accessible at any time.
    \item \SI{100}{\meter\squared} area next to the separation wall between the Target Complex and the Surface Hall (see Fig.~\ref{fig:Surfacehall}) for the Surface Hall and Experimental Hall AHU units. These units will be accessible only during technical shutdowns of the machine and in case of a major failure, an intervention of 1 day is allowed.
 \end{itemize}

\noindent\textbf{Cooling}

\noindent Since little activation of cooling water is expected in the Underground Hall, pumping stations and the main cooling equipment can be placed in the Service Building. Preliminary user requirements for the cooling of SHiP components is presented in Ref.~\cite{ExpareaCV}.

\begin{figure}[htbp]
  \centering
  \includegraphics{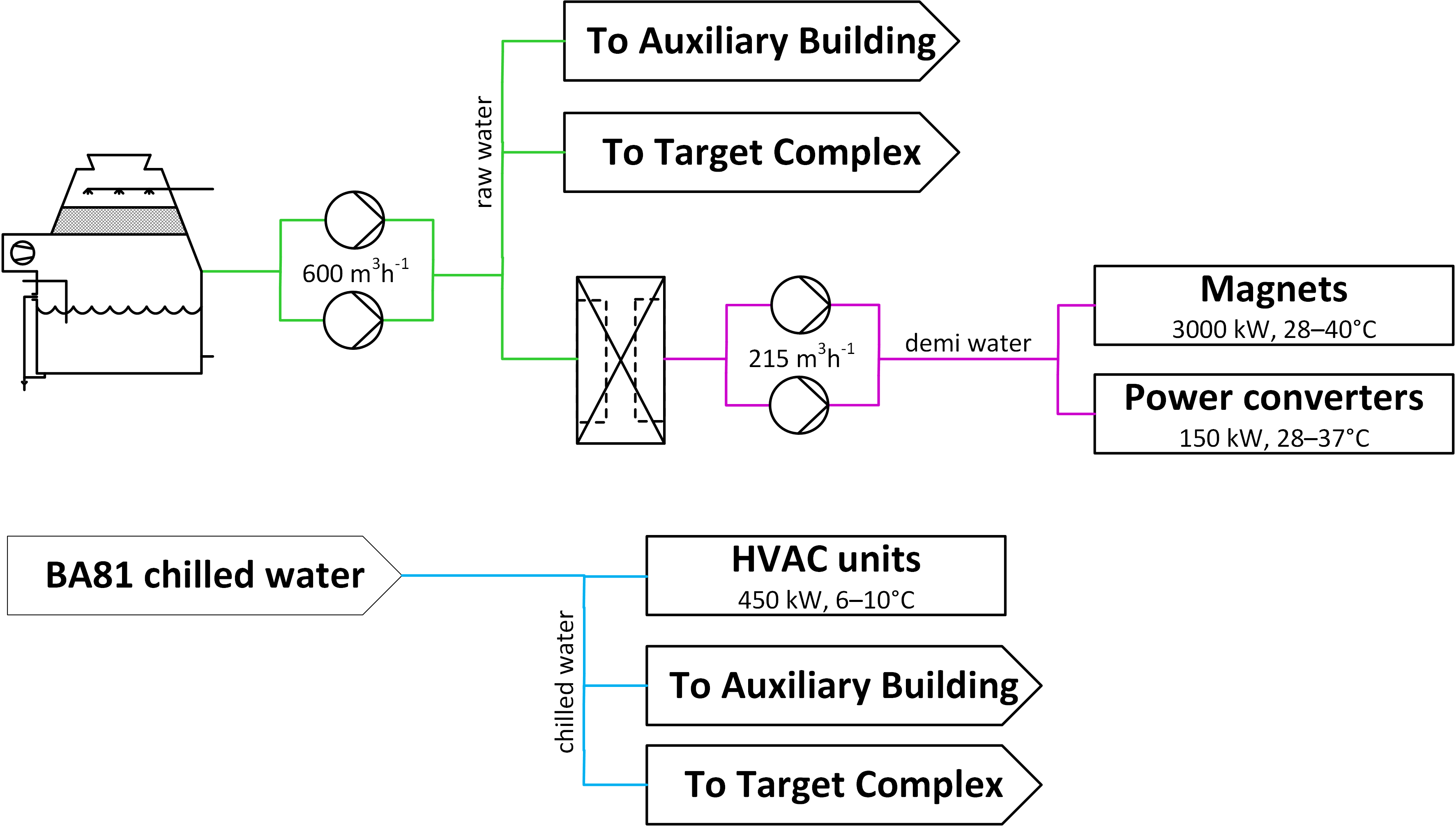}
  \caption{Schematic of the Experimental Area cooling system}
  \label{fig:EAcooling}
 \end{figure}

The SHiP cooling system is based on a raw water primary cooling system supplying water at \SI{25}{\celsius} via cooling towers. Local cooling is provided via demineralised water secondary circuits (\textit{e.g.} for the power converters cooling circuit); chilled water provides cooling for the AHUs.  Fig.~\ref{fig:EAcooling} illustrates the cooling system structure.

SHiP cooling represents about half of the cooling requirements of the entire BDF facility primary circuit. The primary flow rate to the Experimental Area amounts to \SI{270}{\cubic\meter\per\hour} at \SI{25}{\celsius} for a total cooling load of \SI{3.1}{\mega\watt}; the expected temperature increase is \SI{10}{\celsius} and the estimated main piping diameter is DN250. The pressure drop of the circuit is assumed to be \SI{3}{\bar} and the required pumping power is roughly \SI{40}{\kilo\watt}.

A secondary cooling system provides demineralised water-cooling to the magnets and detectors in the Experimental Hall. Located at ground level and inside the Surface Building, it supplies \SI{215}{\cubic\meter\per\hour} at \SI{28}{\celsius} for a thermal load of about \SI{3100}{\kilo\watt} on average; the estimated return temperature is \SI{40.5}{\celsius}. The system is equipped with an expansion tank and two redundant pumps which supply demineralised water at \SI{20}{\bar}; pressure reducers are used to lower the pressure for equipment that needs a lower supply pressure. The size of the piping circuit is DN250 and the estimated pumping power equals to \SI{180}{\kilo\watt}.

Chilled water is needed in the Experimental Area for the ventilation units in the Service Building and the Surface Hall. The overall requirements in terms of chilled water amount to roughly \SI{750}{\kilo\watt} max; this cooling capacity is provided by the BA81 chilled water production plant. Of this amount, about \SI{450}{\kilo\watt} (corresponding to \SI{80}{\cubic\meter\per\hour} of chilled water at \SIrange{6}{10}{\celsius}) is needed for the cooling coils of the three AHUs.

Chilled water, demineralised water filling and compressed air are supplied from a new technical gallery. Moreover, since the primary water production is located next to the Auxiliary Building, a trench from the Auxiliary Building to the Service Building is required. The pipes are routed through the trench up to the Service Building ground floor. A proposal of the trenches layout is detailed in Fig.~\ref{fig:Layout}. 

\noindent\textbf{Ventilation}

\begin{figure}[htbp]
  \centering
  \includegraphics[width=6in]{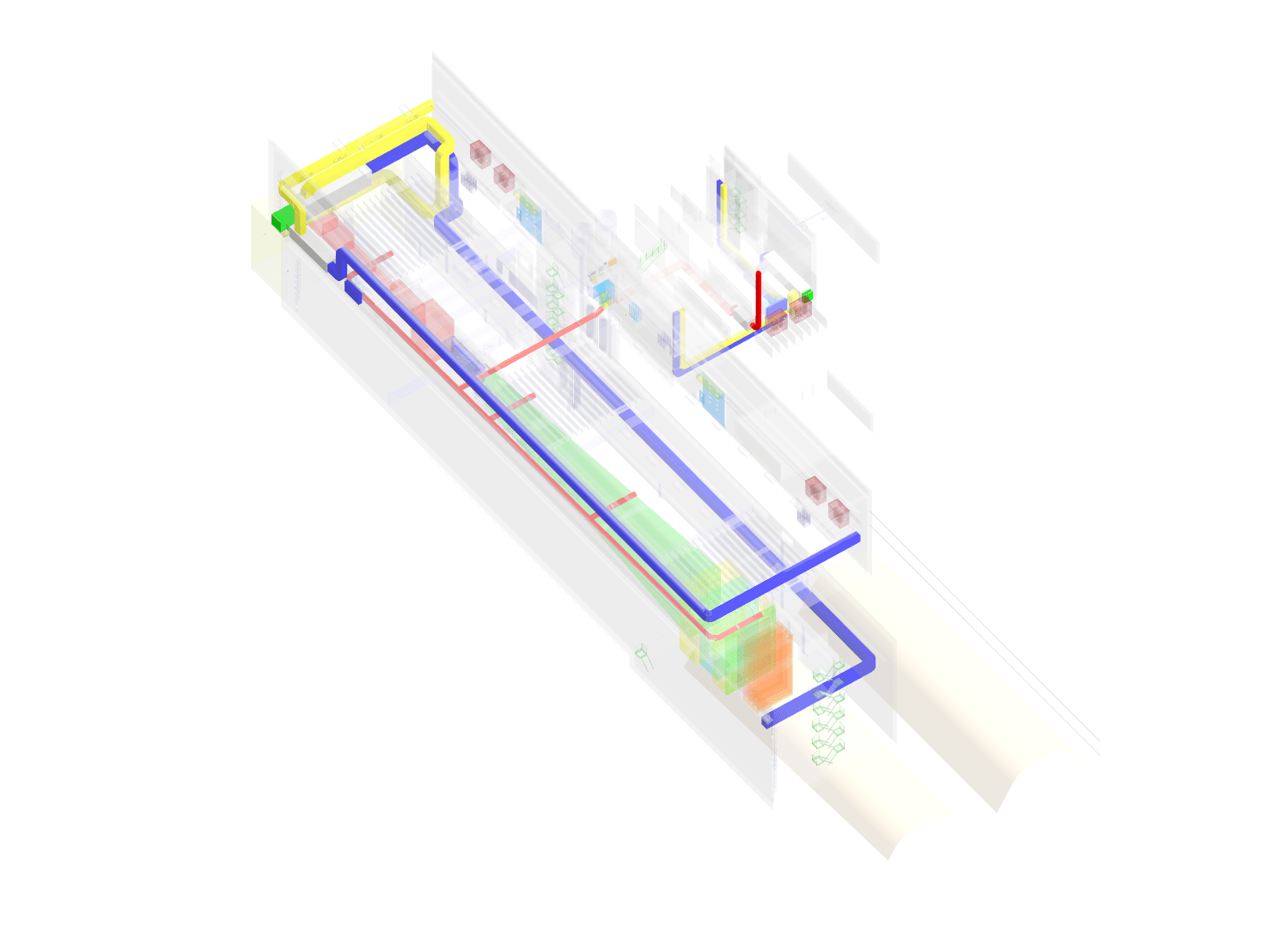}
  \caption{Preliminary integration of the ventilation ducts in the Experimental Area}
  \label{fig:EAventilation}
\end{figure}

\noindent Preliminary user requirements for the ventilation of SHiP equipment are presented in Ref.~\cite{ExpareaCV}. As mentioned in the cooling section, due to the low expected activation of the area, dynamic confinement is not required. The ventilation system is designed to supply fresh air and provide adequate temperature and humidity conditions in the buildings. Fig.~\ref{fig:EAventilation} shows a preliminary integration of the ventilation ducts inside the buildings.
     
The following ventilation units are needed in the context of the Experimental Area's ventilation system:
     \begin{itemize}
         \item Experimental Hall ventilation (plus smoke extraction);
         \item Surface Hall ventilation; 
         \item Service Building ventilation.
     \end{itemize}
     
Regarding the Experimental Hall system, the ventilation unit is located in the Surface Hall, next to the upstream wall that separates the Surface Hall from the Target Complex; at this location, activation of equipment is minimized. Air is supplied to the Experimental Hall via distribution ducts and extracted from grills, whose optimal position will be determined via CFD analysis. Due to the volume of the hall and the heat loads, re-circulation is used to minimize the energy requirements; a minimum amount of fresh air is supplied to maintain optimal conditions. The nominal flow rate of the unit is \SI{85000}{\cubic\meter\per\hour} for a reference duct diameter of \SI{1.94}{\meter}. The emulsion spectrometer magnet will require a dedicated local temperature and humidity control due to its special requirements. 

The Surface Hall is a large building located on top of the Experimental Hall. The supply unit for this ventilation area is located inside the Surface Hall, so that duct work is minimized. The maximum supply flow is \SI{55000}{\cubic\meter\per\hour} for a reference duct diameter of \SI{1.56}{\meter}; the ventilation system takes advantage of re-circulation.

In relation to the Service Building, ventilation needs to be supplied to the service equipment area at the first floor, and offices, control rooms and laboratories at higher floors. The preliminary design foresees a single ventilation unit for the entire building. A more detailed design will be performed when the internal rooms layout is established and the heat load distribution computed. The nominal flow rate considered of the unit is \SI{45000}{\cubic\meter\per\hour}, for a reference duct diameter of \SI{1.4}{\meter}.

With respect to the smoke extraction system (design parameters detailed in Section \ref{firesystem}), the Experimental Hall has been divided into 4 smoke extraction sectors (see Fig.~\ref{fig:Smoke extraction}); the same extraction unit is used for all sectors, and only one sector at a time can be activated. The maximum extraction flow per sector is \SI{24000}{\cubic\meter\per\hour}. In case of fire in the Surface Hall, the smoke will be ventilated via natural ventilation and sky domes located on the ceiling (see Fig.~\ref{fig:skydomes}).

 \subsection{Survey}
 \label{SHiPsurvey}

In the SHiP experiment, the large-scale metrology survey work is performed at different steps: design, construction, assembly and alignment of the detector elements. That means:
\begin{itemize}
    \item Participating in the project at an early stage, \textit{e.g.} collecting the geometrical parameters and alignment needs or discussing the integration of the survey needs in the design of the infrastructure, tools or detector elements.
    \item Providing the necessary geometrical data for adjustment and control of the assembly infrastructures.
    \item Providing the position and orientation information related to the SHiP detector assembly and tests, \textit{i.e.} the geometrical information:
    \begin{itemize}
        \item for the detector assembly tooling alignment;
        \item for the geometrical follow-up and adjustment of the detector elements during the assembly;
        \item for the positioning of the detector elements for tests.
    \end{itemize}
    \item Establishing, measuring, computing and maintaining geodetic networks or coordinate systems and defining the parameters linking different ones when needed.
    \item Providing surveyed position and orientation information to locate the detector in the CERN Coordinate System (CCS) and to link it to the upstream beam line.
    \item Providing when necessary metrology measurements for the fiducialisation of the detector elements as well as of module assemblies.
    \item Providing the control of the position and the alignment of standalone elements and module assemblies in the Experimental Area.
    \item Participating to any upgrades at early stage to ensure high quality geometrical information during the lifetime of the detector.
\end{itemize}

This covers both the theoretical and the practical aspects of the large scale metrology work for the SHiP Project. Knowing that the theoretical position of the beamline elements are provided by the accelerator optics team.

In field, measurements should be performed using suitable survey instrumentation and methods such as total stations, optical levelling, photogrammetry, 3D laser scanners or laser trackers.

\noindent\textbf{Specifications}

The positioning and stability tolerances and requirements of each sub-system of the SHiP experiment need to be identified in order to have an estimation of the necessary means of adjustment. The base for a more detailed analysis of the survey tasks would need an input provided by the different persons responsible for the detector inside the experiment collaboration:
\begin{itemize}
    \item the expected positioning tolerances after installation of all sub-systems,
    \item the further displacements that may be observed during the assembly process of the experiments,
    \item the future displacements induced by the operation of the detector as vacuum forces or deformations in time at the civil engineering level (floor stability).
\end{itemize}

The positioning tolerances should be expressed in a physicist reference frame where Z is along the beam and X lays in a horizontal plane.

\noindent\textbf{Support and alignment reference system}

A geodetic network has to be established that includes the upstream beam line up to the SHiP Experimental Hall. Additional geodetic points have to be installed in the SHiP experimental hall. Those points are materialised by either wall brackets, permanent tripods or ground inserts equipped with the CERN standard survey (CSS) reference sockets. The complete network is determined in the CSS and measured by total stations, laser trackers and direct levelling. The geodetic network of the area needs to be updated periodically.

All the elements to be aligned have to be equipped with survey reference points and/or a tilt reference surface. Typical detector elements are equipped with a minimum of three (preferably more) fiducials that are distributed in space to be able to determine its position and rotation in space. 

The supports of the elements should dissociate the movements in the horizontal and vertical planes as well as shifts and rotations. Each element needs a corresponding adjustment system with the sufficient adjustment range. 

The survey team should help with the definition and placement of these fiducials/supports and should be included in the approval of the final designs.

\noindent\textbf{Link of machine geometry to Experimental Area}

\begin{figure}[ht!]
  \centering
  \includegraphics[width=\linewidth,height=2.73in]{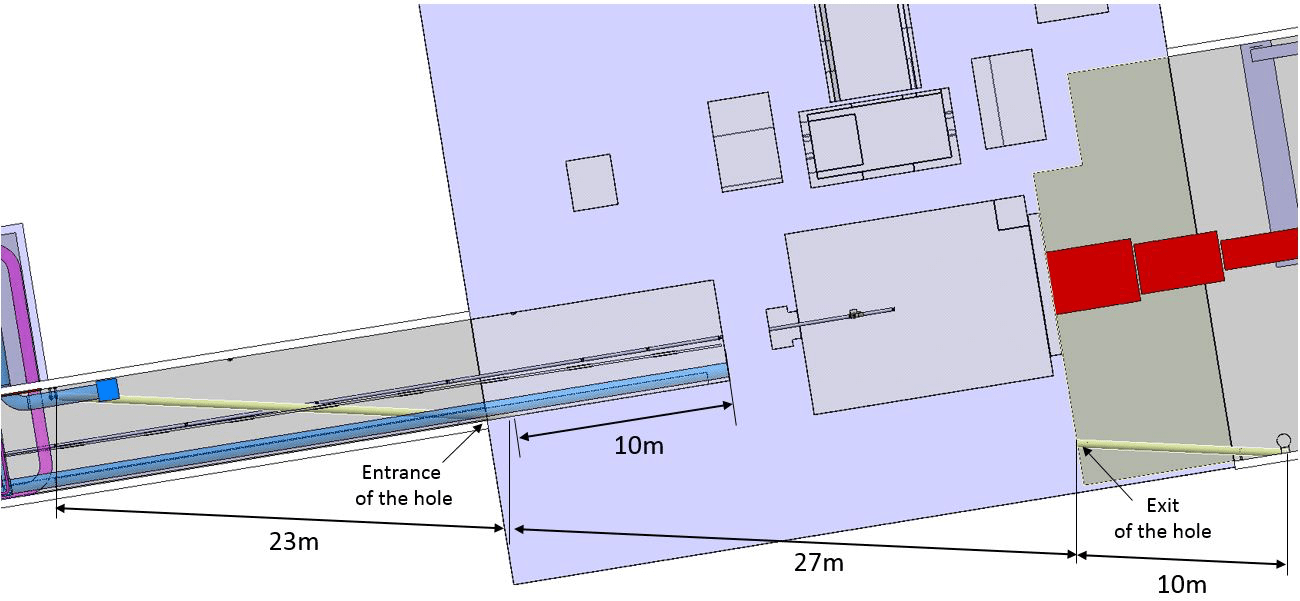}
  \caption{Preliminary integration of the connection of the beamline and SHiP alignment networks}
  \label{fig:alignemnt}
 \end{figure}

The very first geometrical link between the machine geometry and the Experimental Hall is expected to pass by the Target Complex while the line of sight is still available. In parallel, a dedicated 0.4 m diameter pipe that traverses the Target Complex should be used for the geometrical link. At least a minimum of four survey reference points on dedicated brackets (at minimum two on the machine side and at minimum two in the Experimental Hall) should be the considered as the main geometrical references for the transfer of the machine geometry to the Experimental Hall after completion of the Target Complex. The length of the line of sight passing by the machine ($\sim$\SI{23}{\meter}), the Target Complex ($\sim$\SI{27}{\meter}) and the Experimental Hall ($\sim$\SI{10}{\meter}) is in total nearly \SI{60}{\meter} (see Fig.~\ref{fig:alignemnt}).  

\noindent\textbf{Geometrical quality control measurement}

The survey team should provide on demand some of the geometrical and dimensional control measurements for a prototype and production elements.

\noindent\textbf{Fiducialisation measurements}

The determination of the survey reference target positions with respect to the reference system of the element or module on which they are placed is called a fiducialisation. If the fiducialisation of an element cannot be guaranteed by construction or achieved by the Metrology Laboratory due to the object size or specificity, it could be carried out by the survey team after a dedicated discussion with the technical coordination and the survey team.

The parameters of the fiducialisation are stored in the survey reports and, depending on the object, can be extracted from metrology reports, geometrical quality control measurement or from a specific fiducialisation operation. The external fiducials are supposed to stay visible and accessible for the different measurement operations.

\noindent\textbf{Theoretical data}

The spatial position and orientation data for the beamline elements including the SHiP detector needs to be extracted from the beam optics calculations such as BEATCH/MADX beamline definition files. Additional parameters necessary for the large-scale metrology can be derived from the layout drawings of the detector. The MADX file and the layout plans have to be provided to the survey team prior to any survey or alignment work. 
If possible, all measurements and results from metrology, survey and alignment activities are documented in measurement reports and stored on EDMS.

\noindent\textbf{Marking out}

With respect to the geodetic network, reference marks representing the projected beam line and the elements to be aligned can be painted on the floor and walls by the survey team. These marks help the installation of the services and beamline elements. Everybody working in their vicinity shall ensure that these marks remain visible. The required marks and annotations have to be defined in collaboration with the technical coordination of the experiment.

\noindent\textbf{Positioning}

Prior to the installation of the detector elements, their supports are installed by others and pre-adjusted to their nominal position by the survey team. These survey interventions can happen also before the support installation if shims are required to overcome local floor deformations.
Once the detector elements are installed, their initial positioning is carried out with respect to the geodetic network of the area. Precise survey methods and instrumentation have to be used, such as laser trackers, total stations and direct levelling. 

\noindent\textbf{As-Built Measurements}

In order to save time during the infrastructure installation, and to provide 3D documentation of the Experimental Area, a number of as-built measurements of the civil engineering structures could be done, followed by as-built measurements of the installed infrastructure and finally the experiment. 

It’s recommended to make 3D scans in the experimental hall at the end of the civil engineering work and eventually after the installation work. These measurements should be pre-processed by the survey team to give the point clouds and their 3D coordinates in the CCS or a defined experimental coordinate system to the SHiP integration team.

 \subsection{Radiation protection}

The high intensity beam power deposited on the target poses challenges to the radiation protection in several locations. In order to reduce the effect and mitigate the impact, the radiological aspects have been carefully addressed at the design stage. The studies include expected prompt and residual dose rates in the various areas of the SHiP Experimental Area and public areas. The studies are based on past measurements and extensive simulations with the FLUKA Monte Carlo particle transport code and Actiwiz3. The results of the study are shown in Chapter \ref{Chap:RP}.
 
The RP requirements integrated inside the Experimental Area layout are the following:
\begin{itemize}
    \item \SI{1}{\meter} concrete thickness for the underground Experimental Hall ceiling.
    \item \SI{0.4}{\meter} concrete thickness for the Experimental Hall walls, floor and elevator/stairs shaft.
    \item \SI{6}{\meter} high and \SI{40}{\meter} long land hills behind the Surface Building (see Fig.~\ref{fig:endhills}).
    \item A surface fence must be behind the building to avoid personnel accessing the area during operation (see Fig.~\ref{fig:endhills}).
    \item Underground Experimental Hall and Surface Hall are not accessible during beam operation.
\end{itemize}

 \subsection{Safety}

The proposed Experimental Area's layout complies with the CERN safety rules and has being analysed from the safety point of view. The main safety requirements taken into account are:
\begin{itemize}
    \item The lift and stairs are protected against fire and their lighting is not connected to the general electrical circuit (they will be connected to the UPS network); therefore, they can be used at any time.
    \item Fire equipment and Fire Brigade vehicle (e.g. PEFRA \cite{TractorPEFRA}) can descend to the underground Experimental Hall through the lift, especially for rescue operation.
    \item Two escapes routes from the Experimental Hall are foreseen (see Fig.~\ref{fig:Underground Experimental Hall layout}). One is located \SI{33}{\meter} away from the Target Hall/alcove interface and the other at the end of the underground hall.
    \item \SI{1.1}{\meter} height parapet around on the Surface Hall and Service Building roof (see Fig.~\ref{fig:skydomes}).
    \item Ladders to access the roofs and the crane platforms foreseen inside the Surface Hall and the underground Experimental Hall. Maximum step height and exact location will be determined in further studies.
    \item As the concrete beams will close the openings inside the Surface Hall at the end of the SHiP installation phase, barriers must be installed around the three openings.
    \item Implementation of a safe zone against fire in the underground area is required. An option could be a room prepared against fire at the entrance of the lift allocated out of the \SI{20}{\meter} wide hall in order to reduce the detector background. The safe area must be over pressurized and linked to vertical egress path. Location and configuration will be studied in more detail.
    \item Regarding the fire/sector door locations and sizing, the underground Experimental Hall is considered as one fire compartment in itself completely independent from the Target Complex.
    \item Installation of a fire detection system to ensure early detection. Early detection is such that it allows evacuation (last occupant out) before untenable conditions are reached.
    \item Install a system capable of transmitting an alarm, along with a message containing safety instructions, to occupants anywhere in the North Area. This alarm shall be triggered upon fire detection, action on evacuation push buttons, CERN FB action out of CERN FB SCR/CCC or BIW (Beam Imminent Warning) situations. Evacuation push buttons shall cover all premises.
    \item Integration of fire detection and evacuation push buttons with safety actions such as compartmentalization, ventilation stop and other machine functions according to a predefined fire protection logic.
    \item Respect the fire resistance of ducts and fire compartments with the proper fire dampers as they go across fire resistance partitions.
\end{itemize}

Refer to Chapter \ref{Chap:Safety} for further information on the safety considerations.

 \subsubsection{Fire system and alarms}
 \label{firesystem}
\begin{figure}[ht]
  \centering
  \includegraphics[width=\linewidth]{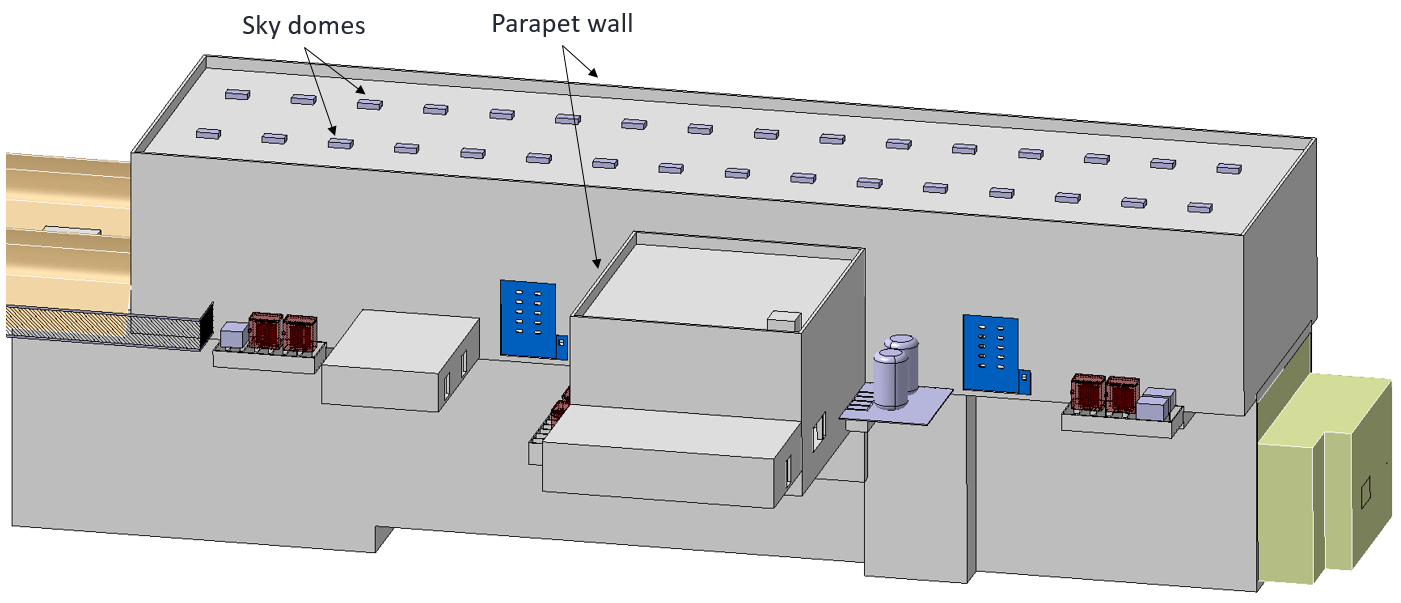}
  \caption{Fire evacuation system of the Surface Hall}
  \label{fig:skydomes}
 \end{figure}

For the Service Building, standard fire safety requirements on building codes are very likely to be applicable wherever fire induced radiological risk is not an issue. Concerning the Surface Hall, natural ventilation through sky domes (see Fig.~\ref{fig:skydomes}) is an easy implementation/integration technical solution.

The smoke extraction system of the Experimental Hall falls outside of the applicable building codes. Therefore, it should be fine-tuned by means of CFD simulations on a Performance-Based design effort. The HSE-OHS-XP Fire Safety Engineering Team has provided preliminary estimates based on regulations and previous similar experiences. Further studies have to be performed by CFD simulations. These rough estimates cannot be taken as definitive functional requirements, however, at this stage of the project, they help size the elements for integration purposes. 

\begin{figure}[ht]
  \centering
  \includegraphics[width=2.3in,height=2in]{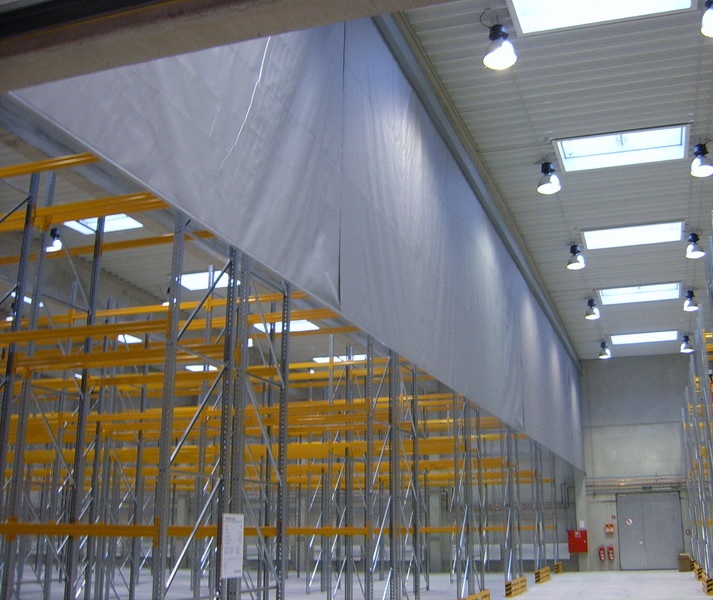}
  \caption{Smoke curtain \cite{scurtain}}
  \label{fig:smoke curtain}
 \end{figure}

The features of the proposed smoke extraction system for the Experimental Hall are:

\begin{itemize}
    \item The \SI{2400}{\meter\squared} (\SI{120}{\meter}$\times$\SI{20}{\meter}) Experimental Hall is split in 4 smoke extraction areas of approximately \SI{600}{\meter\squared}.
    \item Smoke extraction flowrates shall be in the order of \SI{1}{\cubic\meter\per\second} for every \SI{100}{\meter\squared} surface.
    \item The smoke extraction system shares the main duct and extraction equipment for the four areas.
    \item Only one smoke extraction area is triggered at a time (the one where the origin of the fire is located).
    \item A target velocity on \SI{10}{\meter\per\second} on the ducts.
    \item \SI{0.5}{\meter} diameter shaft.
    \item Four extraction sectors.
    \item Four smoke extraction points.
    \item Three “smoke curtains” (\SI{2}{\meter}$\times$\SI{20}{\meter}). They are screens directly connected to the ceiling with the aim of containing the smoke in case of fire (see Fig.~\ref{fig:smoke curtain}). They should remain extended but they can be raised punctually and manually.
\end{itemize}

\begin{figure} [ht]
  \centering
  \includegraphics[width=\linewidth]{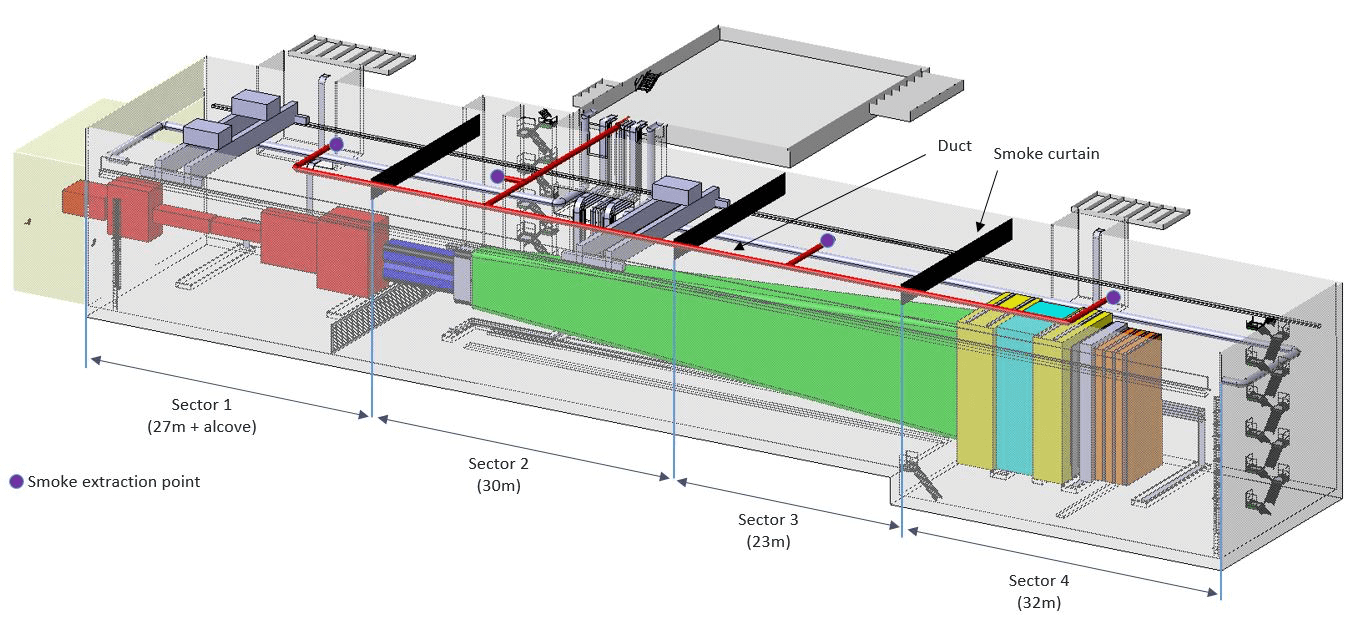}
  \caption{Smoke extraction system of the Experimental Hall}
  \label{fig:Smoke extraction}
 \end{figure}

Therefore, the smoke extraction flow rate should be in the order of \SI{6.6}{\cubic\meter\per\second} (~\SI{24000}{\cubic\meter\per\hour}) and a DN1000 extraction duct (\SI{0.785}{\meter\squared} cross section) allows a velocity of \SI{8.4}{\meter\per\second}, below the \SI{10}{\meter\per\second} target. The smoke extraction equipment is foreseen to be allocated inside the Service Building where it can be accessible during beam operation. It will be integrated with the cooling and ventilation equipment (see Section \ref{CVsection}). Fig.~\ref{fig:Smoke extraction} shows the smoke extraction ducts and the smoke curtains highlighted in red and black, respectively. The ducts run above the overhead cranes and on the right wall of the underground hall (taking as reference the beam direction).  

\FloatBarrier
\subsection{Access control}

The BDF access control will be controlled from the CERN Control Center (CCC) located in building 874 in the CERN Pr\'evessin site. Therefore, a connection between the both is required.   

The access control system of the Experimental Area includes:
\begin{itemize}
    \item Badge access to enter in the Service Building, workshop and control room.
    \item No access to the Surface Hall and underground Experimental Hall during beam operation due to RP constraints.
    \item Material Access Door/Personnel Access Door (MAD/PAD) to access from the Service Building to the Surface Hall. A maintenance fence inside the Surface Hall could be added to allow the MAD/PAD maintenance during beam operation.
    \item Surface Hall exterior doors (personnel and material) beam interlocked. Emergency exit of the Experimental Hall is also beam interlocked.
    \item Two patrolled installations, one in the Surface Hall and the second one in the Experimental Hall.
    \item Sector door to access from the Surface Hall to the Experimental Hall (dosimeter requested).
 \end{itemize}
 
\begin{figure} [ht]
  \centering
  \includegraphics[width=\linewidth]{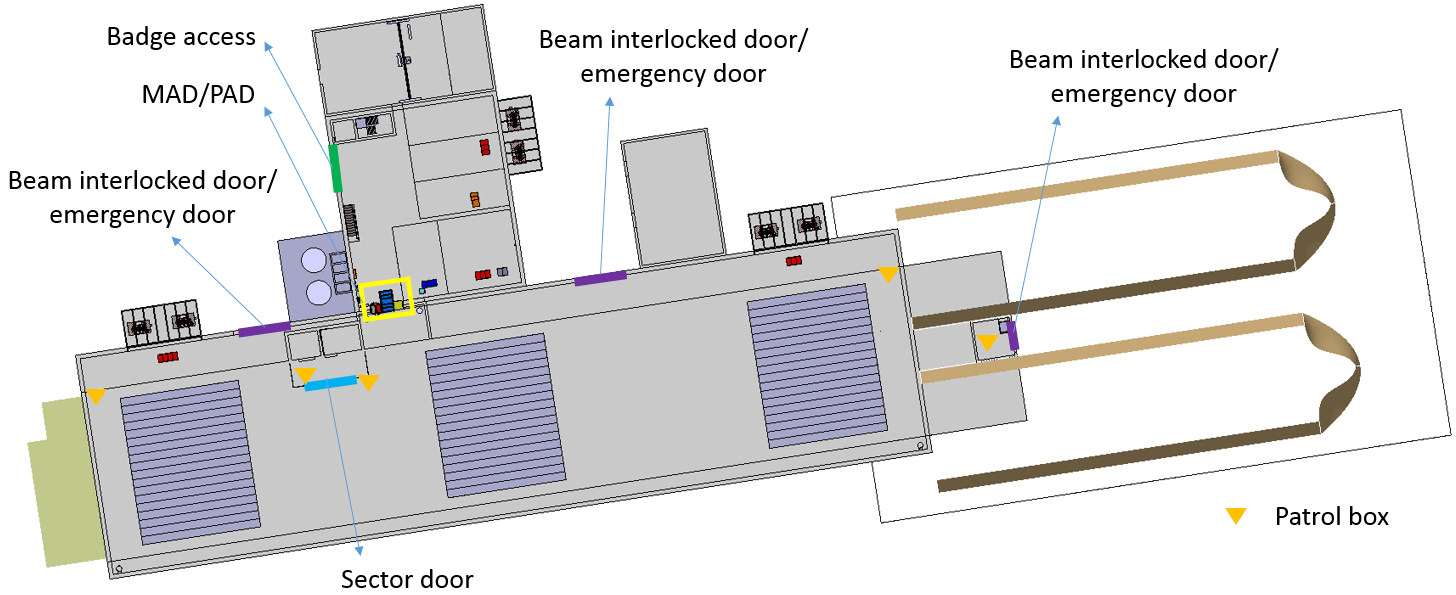}
  \caption{Access control scheme of the Experimental Area}
  \label{fig:Access control}
 \end{figure}

Fig.~\ref{fig:Access control} represents the scheme of the access control system at ground level. During construction and installation, large pieces, pre-assembled SHiP components and personnel will enter to the Surface Hall through the doors located on the Jura side (as shown in Fig.~\ref{fig:Access control} as beam interlock door/emergency door). During beam operation and technical stops, these doors will be interlocked and the personnel access to the Surface Hall must be through the MAD/PAD to avoid losing the patrol of the Surface Hall. To access to the underground Experimental Hall, personnel, tooling and material have to pass through the sector door. If a large detector part has to be replaced, it will enter through the large Surface Hall’s doors and lowered underground through the openings. In consequence, the patrol of the Surface Hall and Experimental Hall will be lost. A presence detection system on the concrete beams may be required to ensure that the openings are completely closed.

\begin{figure} [ht]
  \centering
  \includegraphics[width=\linewidth]{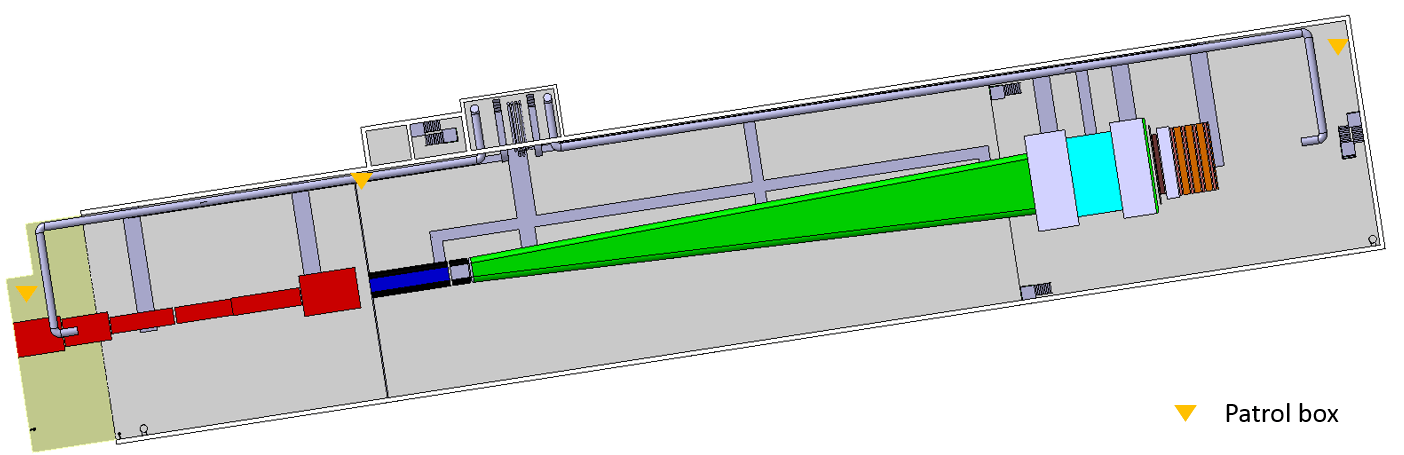}
  \caption{Access control equipment location inside the Experimental Hall}
  \label{fig:Access control equipment location}
 \end{figure}

Fig.~\ref{fig:Access control equipment location} shows the patrol boxes approx. location in the underground Experimental Hall. There are two at the hall extremes and one more next to the elevator.

\newpage 
Finally, the estimated list of the hardware accounted in the access control system is shown in Table \ref{tab:Access system hardware}. Beam interlock logic modifications, cabling and PLCs are not included.

\begin{table} [ht]
    \centering
    \caption{Access system hardware}
    \label{tab:Access system hardware}
    \begin{tabular}{cc} 
    \hline
      \textbf{Item} & \textbf{Quantity} \\ \hline
      MAD/PAD & 1 \\  
      Beam interlocked doors & 3 \\  
      Badge control station & 1 \\ 
      Badge control readers & $\mathrm{\approx}$5 \\  
      Patrol boxes & 8 \\ 
      \hline 
    \end{tabular}
\end{table}


\FloatBarrier
\printbibliography[heading=subbibliography]

 \newpage
 \begin{subappendices}

\begin{flushleft}
\mbox{}\\[1mm]
\bfseries\LARGE Integration: appendices\\[1cm]
\end{flushleft}

\label{App:Integration}
{\centering}

\section{SmarTeam numbers}

The top-level assemblies' SmarTeam references for the 3-D models of the Beam Dump Facility are listed in Table \ref{tab:IntegrationModelsReferences}.

 \begin{table}[h]
 \centering
 \caption{SmarTeam references of the integration 3-D models}
 \label{tab:IntegrationModelsReferences}
 \begin{tabular}{cc}
 \hline 
 \textbf{Description} & \textbf{SmarTeam number} \\ \hline
 BDF Transfer Line Integration Study & ST0890663 \\ 
 BDF Access and Auxiliary Buildings and Target Complex Access & ST1054816 \\
 BDF Experimental Area Layout & ST0964387 \\ 
 BDF - SHiP Assembly Mock-up & ST1095050 \\ 
 SHiP Conceptual Layout & ST0947198 \\
 \hline 
 \end{tabular}
 \end{table}
 
 \section{Structural specification documents}
The structural specification documents for the transfer tunnel, surface buildings and the Experimental Area stored in CERN's Engineering Data Management Service (EDMS) are listed in Table \ref{tab:EDMSdocuments}.

 \begin{table}[h]
 \centering
 \caption{Structural specification documents}
 \label{tab:EDMSdocuments}
 \begin{tabular}{cc}
 \hline 
 \textbf{Title} & \textbf{EDMS number} \\ \hline
Transfer Tunnel  & 2037312 \\ 
Access Building, Auxiliary Building, Personnel Shaft and Chicane  & 2000961 \\
Experimental Area  & 2027772 \\ 
 \hline 
 \end{tabular}
 \end{table}

 \end{subappendices}

 \chapter{Radiation Protection}
\label{Chap:RP}

\section{Introduction}

The main Radiation Protection (RP) challenges for BDF arise from the high beam power, the proximity to the surface, other experimental facilities and the CERN fence, but also from keeping the flexibility for future installations. In order to respect the applicable CERN radiation protection legislation regarding doses to personnel as well as the environmental impact, a full radiological assessment was carried out for the design of the BDF facility. The facility was optimised based on general radiation protection guidelines and specific studies on prompt and residual dose rates, air and helium activation, ground and water activation as well as radioactive waste production as they heavily influence the design. To assess the above-mentioned radiation protection aspects, extensive simulations were performed with the FLUKA Monte Carlo particle transport code \cite{fluka1}\cite{fluka2}.


Next to the radiological assessment of the facility itself also the evaluation of its primary beam extraction from the SPS is a crucial factor. 
Since the dedicated BDF beam line branches off at the top of the existing TT20, in the TDC2 cavern, studies on ground activation around TT20 and TDC2 were conducted. These are particularly relevant for the civil engineering works (see chapter \ref{Chap:CivEng}). Also the necessary cool-down times for the existing tunnels before any works can start were evaluated. In addition, beam losses in the dedicated BDF transfer line were studied with the help of FLUKA.

 Furthermore, tritium out-diffusion experiments for all materials relevant for BDF are performed. The contribution of tritium out-diffusion is particularly relevant for the activation of the cooling water of the BDF target. The report is divided into four sections, which focus on the primary beam extraction for BDF, the BDF facility, the SHiP experimental area and the tritium out-diffusion measurement.

\section{BDF beamline}

The SHiP beam extraction from the SPS will be a slow resonant extraction from  the SPS LSS2 using existing extraction equipment and transfer of the beam along TT20 up to a switch in the TDC2 tunnel that leads into a dedicated BDF transfer line \ref{Sec:ExtractionTunnel}. The TDC2 tunnel is an existing underground tunnel that houses the North Area beamlines (TT22, TT23, TT24 and TT25) and their services. The new BDF beamline (TT90) starts downstream of the new splitter magnet in the TDC2 tunnel and runs alongside the existing North Area beamlines for approximately 110 m as shown in Figure \ref{fig:BDFintegrationoverview}. 

With the SHIP extraction in addition to the North Area requirement the total number of protons extracted per year will increase by a factor of approximately 4. Thus, to keep the activation of the SPS extraction region at a comparable level as of today a reduction of the beam losses during extraction by a similar factor is envisaged by several techniques as described in Chapter \ref{Chap:Extraction}.

\subsection{Junction cavern}

\begin{figure}[!htb]
  \includegraphics[width=0.9\textwidth]{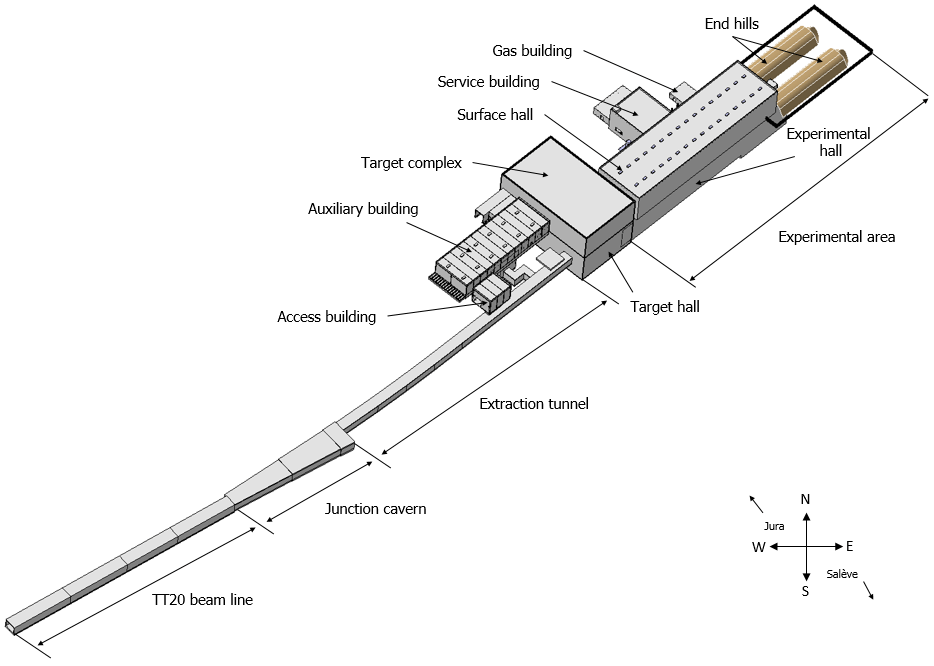}
  \caption{BDF integration overview}
  \label{fig:BDFintegrationoverview}
\end{figure}

 The Junction Cavern (TDC21) is an underground cavern that houses four beamlines (TT90, TT23, TT24 and TT25) and their corresponding services (see Figure \ref{fig:BDFintegrationoverview}). To enable the construction of the junction cavern,  a part of the existing TDC2 tunnel must first be demolished. To assess the radioactivity levels that would be encountered during the civil engineering works 
concrete and soil samples have been taken and analysed.

\begin{figure}[!htb]
 \centering
\includegraphics[width=0.9\textwidth]{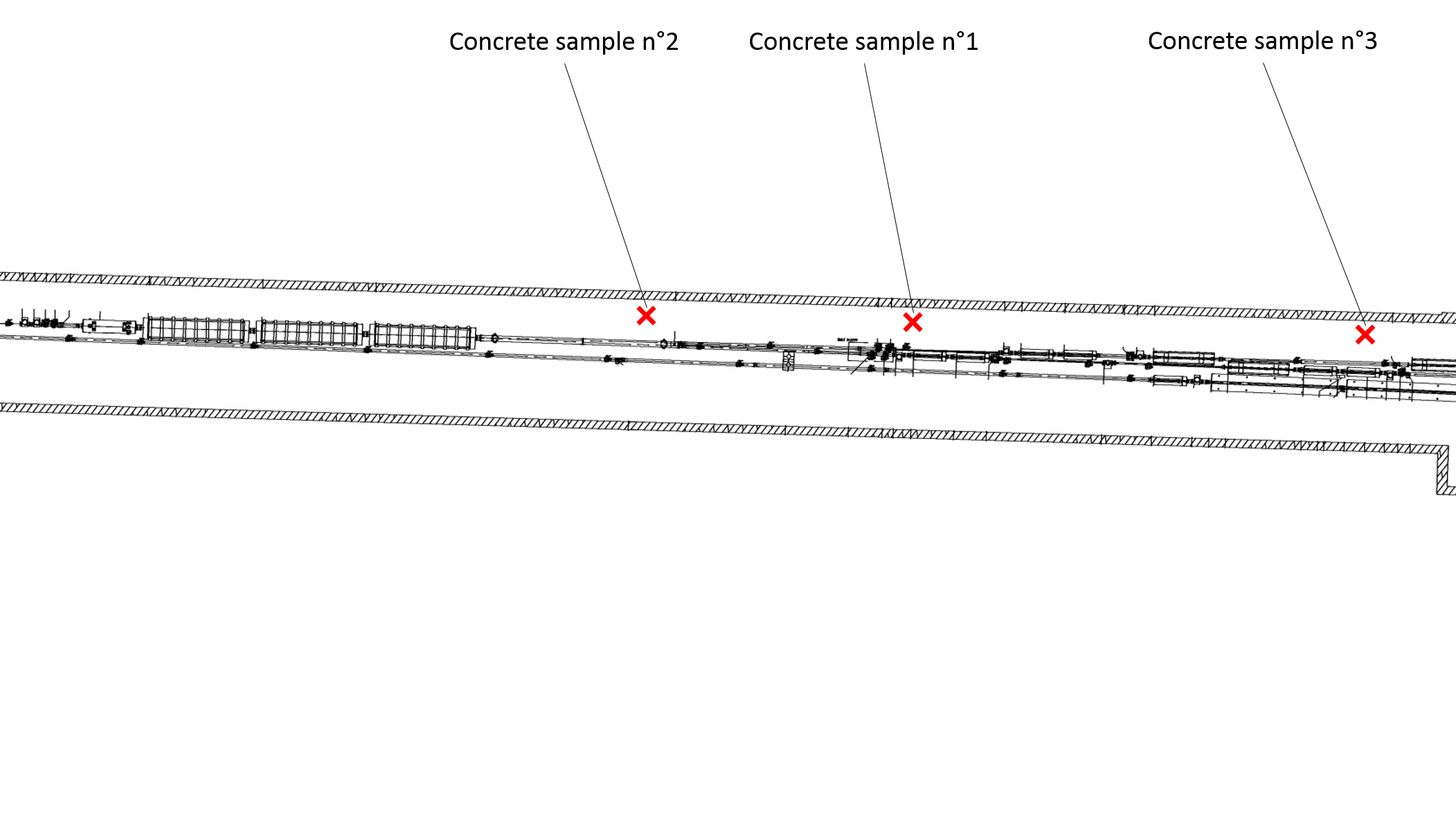}
\captionsetup{width=0.85\textwidth} \caption{\small Locations of the concrete sampling.}
\label{fig:concretesamples}
\end{figure}

Concrete samples have been taken at three different locations from the slab of the existing TDC2 tunnel (see Figure \ref{fig:concretesamples}) during the 2016–2017 
year-end technical stop (beginning of April) after a cool-down period 
of 4.5 months (stop of proton run in mid-November, stop of ion run in mid-December). The results of the gamma spectrometry of the concrete samples showed for all a radioactivity above the Swiss Liberation Limit (LL) \cite{ORAP} with 145 to 404 times the LL. This means that the concrete shall be considered as radioactive in every location, and at any depth, of the TDC2 tunnel.

\begin{figure}[!htb]
 \centering
\includegraphics[width=0.9\textwidth]{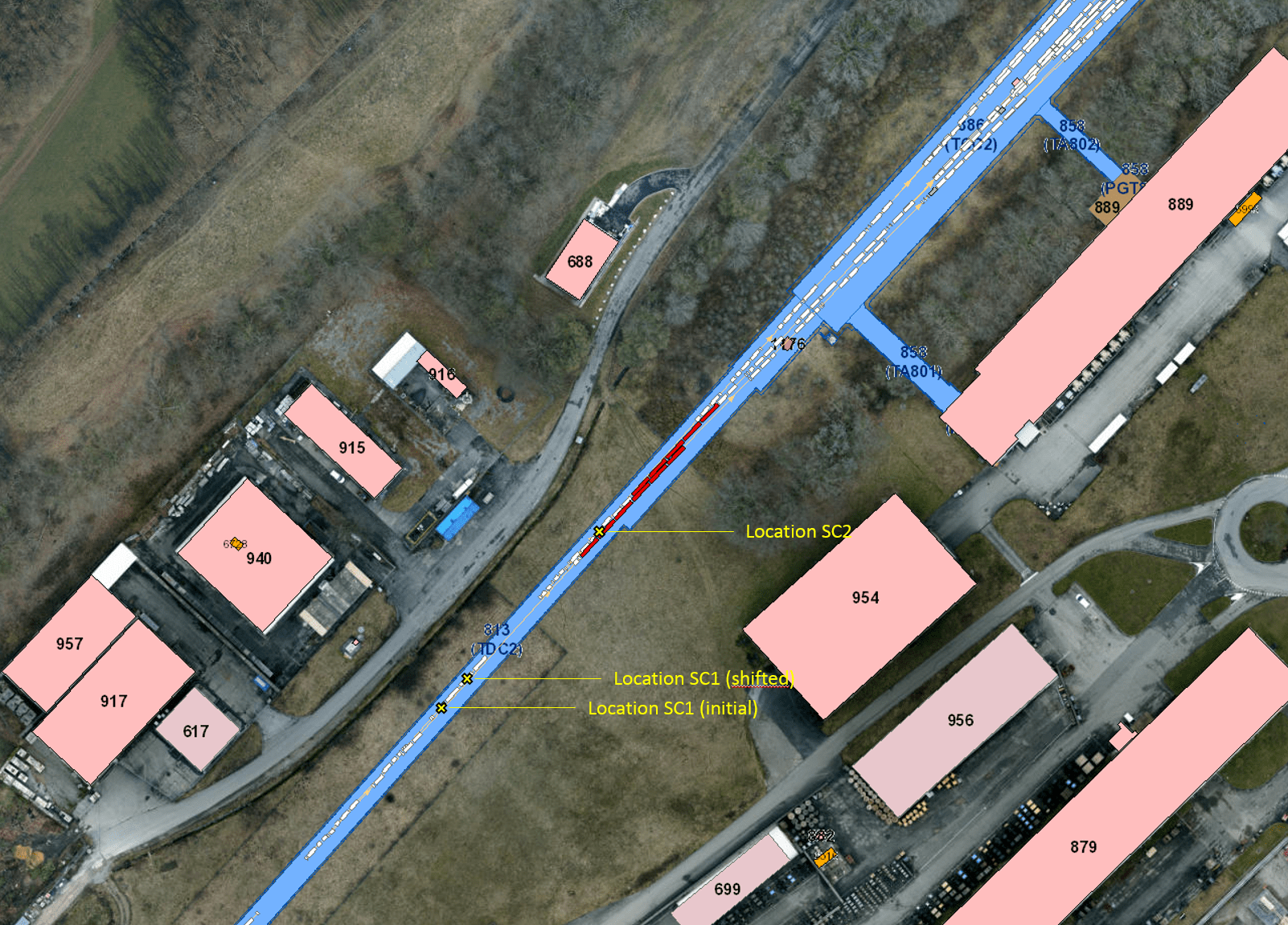}
\captionsetup{width=0.85\textwidth} \caption{\small Soil sampling locations.}
\label{fig:soilsamples}
\end{figure}

Surface soil sampling was carried out during the 2017 – 2018 Year End Technical Stop (YETS) (on the 21st of March 2018) after a cool down of approximately 5 months (stop of proton run at the end of October, stop of ion run at the beginning of December). The locations, shown in Figure \ref{fig:soilsamples}, were chosen where the activation of the equipment in the tunnel is the highest. The drilling stopped at 1.5 meters from the roof of the tunnel, which corresponds to approximately 5.5 meters depth. Na-22 was found in the results of the gamma spectrometry with values of up to 5$\%$ of the LL. Based on the results, the soil up to 5.5 depth from the surface can be considered non-radioactive. However, the soil even closer to the cavern could show a different gradient of activation, wherefore special measures shall be taken if activation levels are proven to be significant. Thus, the soil excavations, demolishing and construction works for the new Junction cavern require Radiation Workers.

For any civil engineering works in the given area, it further has to be taken into account that a minimum soil thickness of 8 m must be kept around the walls of the TDC2 and TCC2 tunnels during beam operation in the North Area \cite{Anelli:2015pba}. For the start of the soil excavations and demolishing works it should furthermore be noted, that from an ALARA point of view the cooling time should be as long as possible. However, it is clear that from a project point of view this is the contrary. Therefore, a compromise has to be found, which is acceptable for the project. 

When looking at the expected dose rates in TDC2 for various cooling times the most significant decrease occurs in the very first weeks of cool-down (factor of approximately 13 during a 4 weeks Pb run). In between for example 9 weeks and 5 months after the proton run, the further gain  amounts to about 30$\%$. At the downstream end of the TDC2 area, in which the equipment will have to be removed, the dose rates at 40 cm distance from the equipment are relatively low (30 $\mu$Sv/h, 9 weeks after the proton run), wherefore the cooling time becomes less critical. For the hottest elements to be removed (e.g. BSPH.240212), the dose rates are of the order of 590 $\mu$Sv/h after 9 weeks after the proton run. These are still much lower than the dose rates of the splitters (11 mSv/h at 40 cm, 9 weeks after the proton run) that fortunately do not have to be removed for the civil engineering works.

For evaluating the necessary cool-down time, a Work and Dose Planning (WDP) for the given works should usually also be looked at. At this stage, only a preliminary estimate of the WDP can be done as it largely depends on the detailed methodology of the works (remote handling, shielding, work optimisation, etc.). However, an estimate can suggest an order of magnitude of the cool-down time. A collective dose of about 10 - 20 mSv for equipment removal after 13 weeks (3 months) of cooling (including 4 weeks of Pb run) was estimated. Reducing the cooling time to 9 weeks (including 4 weeks of Pb run) would increase the collective dose by approximately 9$\%$ (1-2 mSv).

At this stage of the project and in view of ALARA, it is recommended to use the maximum cooling time, which is acceptable for the project and which is at least 9 weeks after the end of the proton run (i.e. 5 weeks in addition to 4 weeks of Pb run). From a radiation protection point of view it is recommended to use if possible 13 weeks (3 months) of cooling in the planning, such that even without an ion run the planning could be met. Note, that for both construction scenarios, there is a contingency of more than 15 weeks, of which eventually part can be used for additional cool-down. Furthermore, the dismantling should start at the downstream end of TDC2, giving more cooling time for the hot elements further upstream.

\subsection{Extraction Tunnel}

\begin{figure}[htbp]
  \centering
  \includegraphics[width=\linewidth]{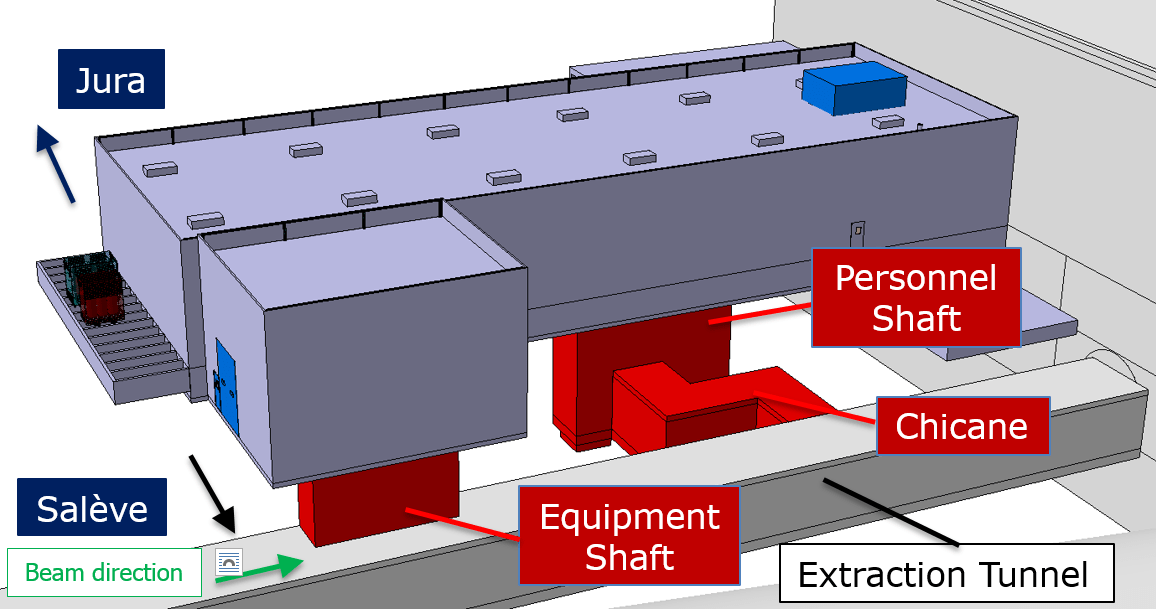}
  \caption{Equipment Shaft, Personnel Shaft and Chicane}
  \label{fig:3SB-v2}
\end{figure}

\begin{figure}[!htb]
 \centering
\includegraphics[width=0.9\textwidth]{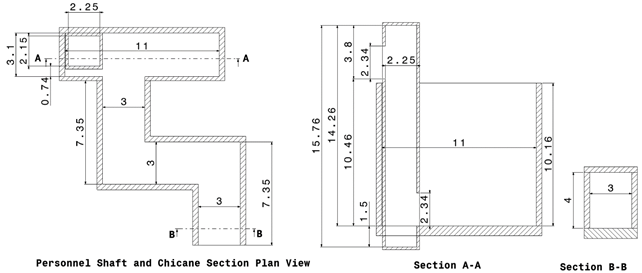}
\captionsetup{width=0.85\textwidth} \caption{\small Chicane Section Plan view.}
\label{fig:chicane1}
\end{figure}

The Extraction Tunnel is an underground tunnel that houses the BDF beamline and its corresponding services. A personnel shaft from the auxiliary building serves as entrance to the extraction tunnel (see Figure \ref{fig:3SB-v2}). A chicane was designed in order to reduce the direct shining from beam losses happening in the extraction tunnel into the auxiliary building (see Figure \ref{fig:chicane1}). 

\begin{figure}[!htb]
 \centering
\includegraphics[width=0.7\textwidth]{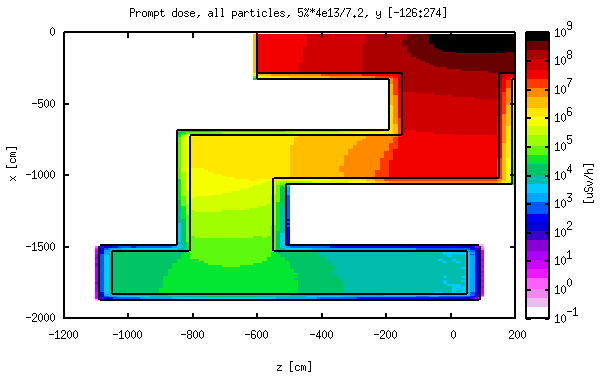}
\captionsetup{width=0.85\textwidth} \caption{\small Prompt dose rate in $\mu$Sv/h for the access chicane to the SHiP transfer line.}
\label{fig:chicane2}
\end{figure}

In principle, no major beam losses are expected in the BDF beamline. In order to assess the radiological aspects, a conservative beam loss of $5\%$ of the BDF spill ($4\times10^{13}$ protons on target in 7.2~s) on a cylindrical copper target (loss on a magnet) was assumed. However, such kind of continuous loss will not be acceptable for the operation of the facility. In case of such losses further measures (e.g. optimisation of beam transfer, shielding around hot spots) shall be put in place in order to minimise the activation of the beam line, the soil and the radioactive waste production.
The prompt dose rate is depicted in Figure \ref{fig:chicane2}. The chicane reduces the prompt dose rate by 5 orders of magnitude. They are further reduced in the personnel shaft and the shielding provided by the above-ground building structures (see Section \ref{AuxiliaryB}) 
such that the radiation reaches acceptable levels in the accessible areas of the auxiliary building. Furthermore, the access building and the personnel access in the auxiliary building are not accessible during beam. 
 
\section{The BDF FLUKA model}
\label{RP:Flumodel_comp}

\begin{figure}[!htb]
\begin{subfigure}{0.9\textwidth}
  \centering
  \includegraphics[width=0.9\textwidth]{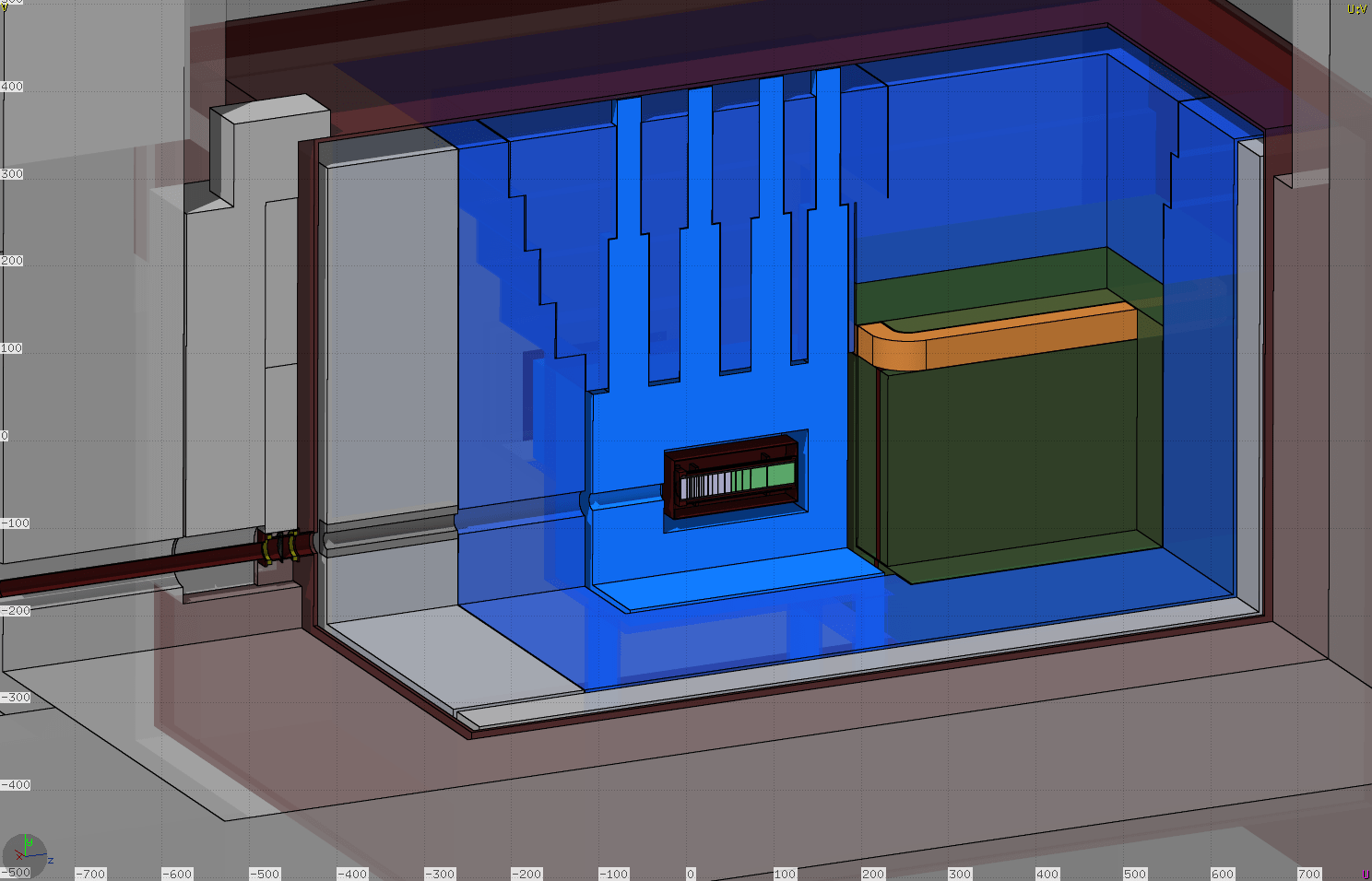}
  \caption{}
  \label{fig:TC}
\end{subfigure}
\begin{subfigure}{0.9\textwidth}
  \centering
  \includegraphics[width=0.9\textwidth]{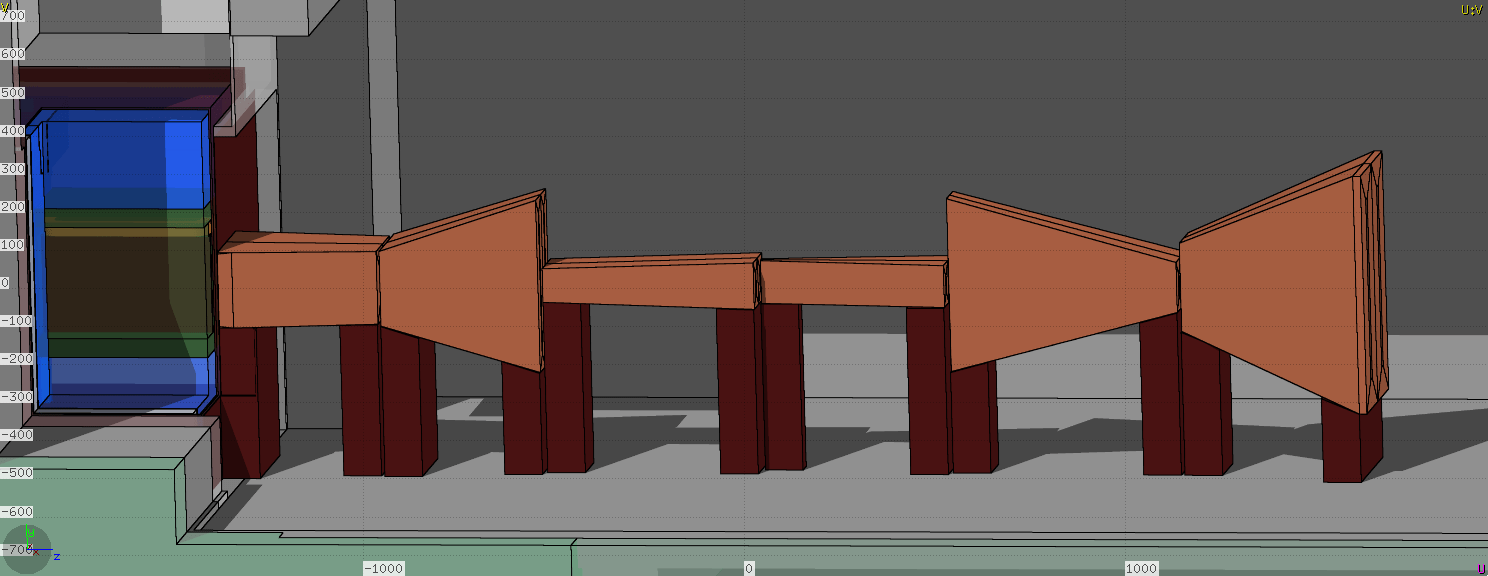}
  \caption{}
  \label{fig:AMS}
\end{subfigure}

\captionsetup{width=0.85\textwidth} \caption{\small View of the target complex (\ref{fig:TC}) and of the active muon shield (\ref{fig:AMS}).}
\label{fig:Views}
\end{figure}

The Monte Carlo particle code FLUKA was used to evaluate the radiation protection requirements for the entire BDF/SHiP facility. The FLUKA model of the facility was developed in collaboration with EN-STI. Figure \ref{fig:Views} depict, from a radiation protection point of view, the most critical areas of the facility: the target complex and the active muon shield. 

The coordinate system used in the model is a right-handed Cartesian coordinate system with origin at the target. The orientation of the coordinate system is defined by the width (x) and height (y) of the target complex and the beam direction (z). Due to the proximity of SHiP to the ground level ($\approx$ 10~m), other experimental facilities ($\approx$ 20~m) and public areas ($\approx$ 70 m), massive shielding is required to keep the prompt radiation in the various accessible areas of the facility and the surrounding reasonably low. Next to personnel protection regarding prompt dose rates, considerable shielding is indispensable to reduce the residual dose rates and the environmental impact from activated air and soil as well as to relax radiation levels on electronics equipment (see also [6]). The shielding was consequently designed with the objective to keep the various radiological hazards originating from the operation of the facility as low as reasonably possible, while taking the constraints from the different stages of the experiment, that is the construction, operation, maintenance and dismantling, into account. The envisaged configuration is such as to avoid activation of the fixed concrete civil engineering structures simplifying not only the dismantling, but also possible changes of the scope of the installation.

\begin{figure}[!htb]
\begin{subfigure}{0.9\textwidth}
  \centering
  \includegraphics[width=0.9\textwidth]{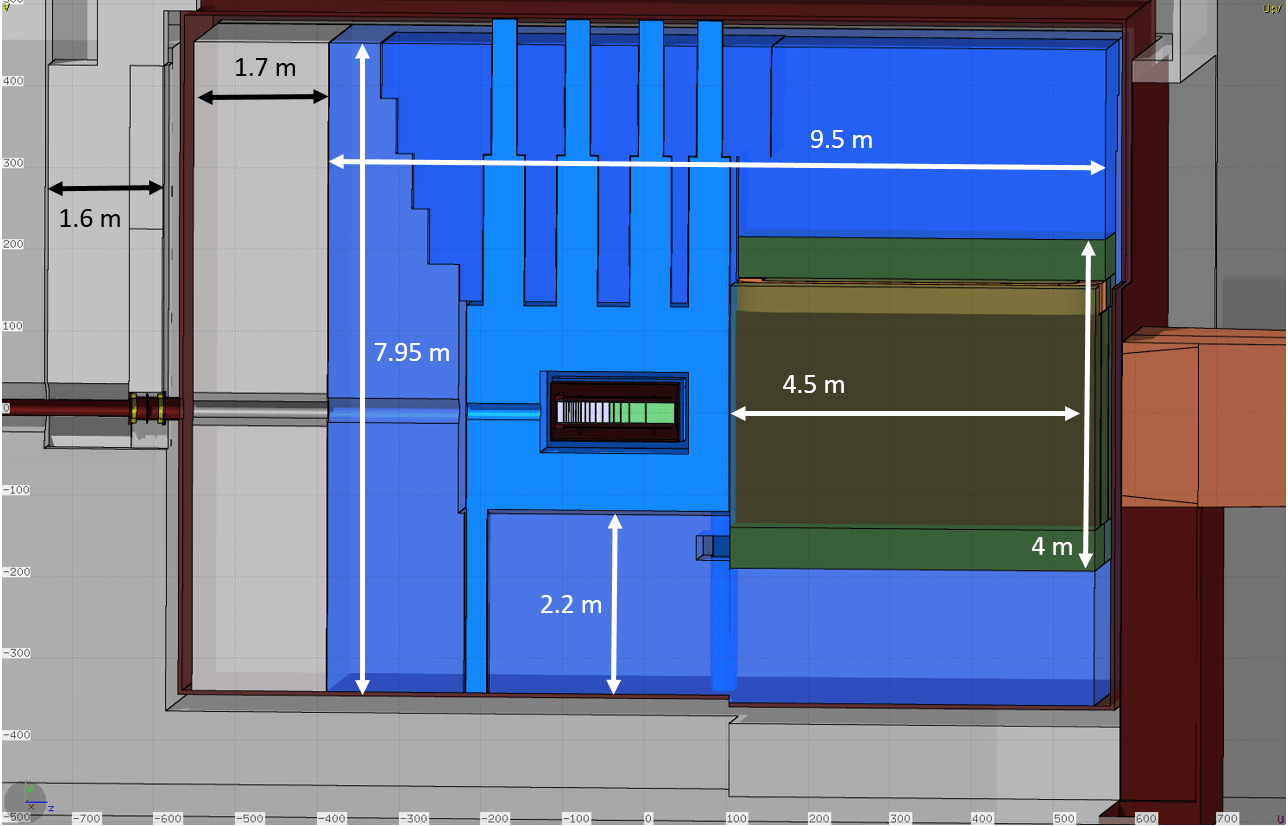}
  \caption{}
  \label{fig:TCShM1}
\end{subfigure}
\begin{subfigure}{0.9\textwidth}
  \centering
  \includegraphics[width=0.9\textwidth]{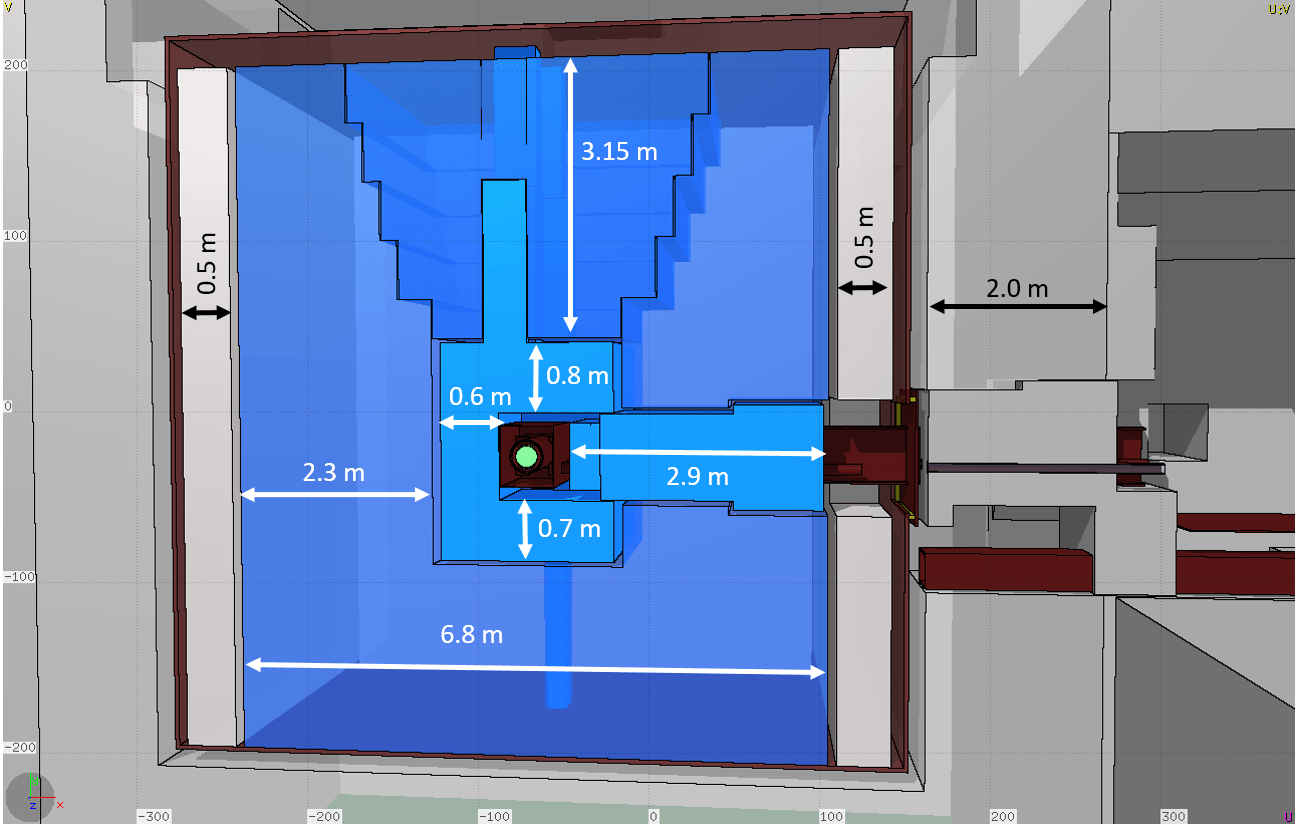}
  \caption{}
  \label{fig:TCShM2}
\end{subfigure}

\captionsetup{width=0.85\textwidth} \caption{\small Lateral (\ref{fig:TCShM1}) and perpendicular (\ref{fig:TCShM2}) view of the target complex.}
\label{fig:ShM}
\end{figure}

The shielding in the target area was therefore modelled with massive iron and concrete blocks of thicknesses as specified in Figure \ref{fig:ShM}. The iron blocks are specially designed for remote handling as they will become highly activated. Several gaps were included in between the blocks in order to account for imperfect alignment, ducts for cooling, electronics, etc. The innermost shielding blocks will include stainless steel water cooling pipes for heat removal. The water cooling circuits for these elements as well as for the target will be closed and separated from others. The downstream shielding, which has a thickness of 4.8 m, is magnetised and also acts as a hadron stopper with the objectives of absorbing the secondary hadrons and the residual non-interacting protons emerging from the target, to significantly reduce the exposure of the active muon shield to radiation and to start bending the muons to increase the fiducial volume of the SHiP experiment. The iron shielding is embedded in a helium vessel to reduce the corrosion. The remaining gaps between the iron shielding and the helium vessel structure are filled with removable concrete shielding blocks. The helium vessel is further surrounded by the fixed concrete civil engineering structures. On the sides and on the bottom of the vessel the main requirement for the concrete shielding thickness comes from civil engineering. The minimum concrete shielding thickness of 2.4 m towards the target hall was further estimated based on the required prompt dose rate reduction. The iron shielding around the beam window has to minimise the air volume and to be dismountable in order to minimise prompt dose rates and air activation in the backward region. Upstream of the window the shielding has an aperture for the primary beam of 20 cm in radius. This passage  towards the primary beam line will be filled with removable iron bricks passage to reduce the “back splash” of particles into the primary beam area, which leads to activation of the upstream beam-line components and the surrounding air. To further reduce the ``back splash'', two absorbers (preferably made out of polyethylene of 1 m and 1.5 m length, respectively) will be placed in the extraction tunnel. For further information about the conceptual design of the target area station please refer to Chapter \ref{Chap:TargetComplex}.

The material properties employed for the shielding components were chosen such that they result in rather conservative prompt and residual dose rate estimates. The composition of the shielding materials is given in Table 1. Note that for cast iron and US1010 iron the self-shielded low energy neutron cross-sections were utilised in order to correct for self-shielding effects. A pessimistic cobalt concentration of 0.035$\%$ and 0.04$\%$ was furthermore assumed respectively for the cast iron and the US1010 iron. A density of 7.85 g/cm$^3$, 7.87 g/cm$^3$ and 2.34 g/cm$^3$ was utilised for the cast iron, the US1010 iron and concrete components, respectively. The soil surrounding the whole facility was modelled with a density of 1.9 g/cm$^3$, which is lower than the measurement performed nearby for CENF, which resulted in 2.3 g/cm$^3$. In this way, location-dependent density differences and a local decrease due to civil engineering works are conservatively taken into account. The chemical composition of the soil as specified in Table 1 was determined from core samples taken for CENF. A water content of 7.5$\%$ as measured from the samples was furthermore assumed. Results from the soil sample analysis can be found in [9]. The soil around the SHiP facility was modelled according to the current ground level in that area with the least distance to the SHiP beam line of 10.3 m.

\begin{table}[!htb]
\centering
\begin{small}
\begin{tabular}{ccc}
\toprule
\textbf{Material} & \textbf{Element} &    \textbf{Weight percentage ($\%$)}\\

  \midrule
   Cast Iron & Iron (self-shielded) & 94.267\\
   &Carbon&3.399\\
   &Silicon&1.799\\
   &Manganese&0.5\\
   &Cobalt&0.035\\
   
      \midrule
   US1010 & Iron (self-shielded) & 99.14\\
   &Carbon&0.105\\
   &Sulphur&0.05\\
   &Manganese&0.45\\
   &Phosphorus&0.04\\
   &Cobalt&0.04\\

\bottomrule
\end{tabular}
\captionsetup{width=0.85\textwidth} \caption{\small Chemical composition of cast iron and US1010 iron as used in the FLUKA studies.}\label{tab:tab_Iron}
\end{small}
\vspace{1cm}
\end{table}

\begin{table}[!ht]
\centering
\begin{footnotesize}
\begin{tabular}{ccc}
\toprule
\textbf{Material} & \textbf{Element} &    \textbf{Weight percentage ($\%$)}\\
\midrule
Concrete& Hydrogen        &   0.6\\
&Carbon&5.62\\
 &  Silicon &  18.867\\ 
&Magnesium        & 0.663\\  
    & Sulphur   &  0.012 \\ 
&Oxygen        &   49.287\\
    &  Potassium  &  0.656\\

& Sodium        &   0.453  \\
 & Calcium  & 20.091 \\ 
  
&Aluminium         &  2.063\\
   & Iron &  1.118 \\
   &Phosphorus&0.048\\
   &Titanium&0.347\\
   &Manganese&0.0387\\
   &Zinc&0.0241\\
   &Zirconium&7.4e-3\\
   &Barium&0.0179\\
   &Lead& 0.0464\\
   &Strontium&0.399\\
   &Europium&4.2e-5\\

\midrule
Moraine& Oxygen& 3.90E-01\\
 &Calcium &2.41E-01\\
 &Silicon &1.83E-01\\
 &Carbon& 5.03E-02\\
 &Iron &4.88E-02\\
 &Aluminium& 4.35E-02\\
 &Potassium& 2.16E-02\\
 &Magnesium& 8.07E-03\\
 &Titanium& 4.46E-03\\
 &Sodium& 3.34E-03\\
 &Manganese& 1.47E-03\\
 &Barium& 9.41E-04\\
 &Strontium& 8.68E-04\\
 &Phosphorus& 6.20E-04\\
 &Chromium& 5.28E-04\\
 &Zinc& 2.92E-04\\
 &Zirconium& 2.57E-04\\
 &Sulphur &2.32E-04\\
 &Nickel &1.72E-04\\
 &Vanadium& 1.40E-04\\
 &Cerium& 1.32E-04\\
 &Chlorine& 1.25E-04\\
 &Lanthanum& 1.10E-04\\
 &Tungsten& 1.05E-04\\
 &Copper& 6.37E-05\\
 &Neodymium& 5.48E-05\\
 &Cobalt& 4.27E-05\\
 &Yttrium& 4.07E-05\\
 &Lead &3.78E-05\\
 &Gold& 3.48E-05\\
 &Gallium& 2.54E-05\\
 &Lithium &5.73E-06\\
 &Europium& 6.87E-08\\

\bottomrule
\end{tabular}
\captionsetup{width=0.85\textwidth} \caption{\small Chemical composition of concrete and soil~\cite{CENF} as used in the FLUKA studies.}\label{tab:tab_Con}
\end{footnotesize}
\vspace{1cm}
\end{table}

\clearpage
\section{BDF target area}
\subsection{Prompt and residual dose rates}
\begin{figure}[!htb]
\centering
\begin{subfigure}{0.8\textwidth}
  \centering
  \includegraphics[width=0.8\textwidth]{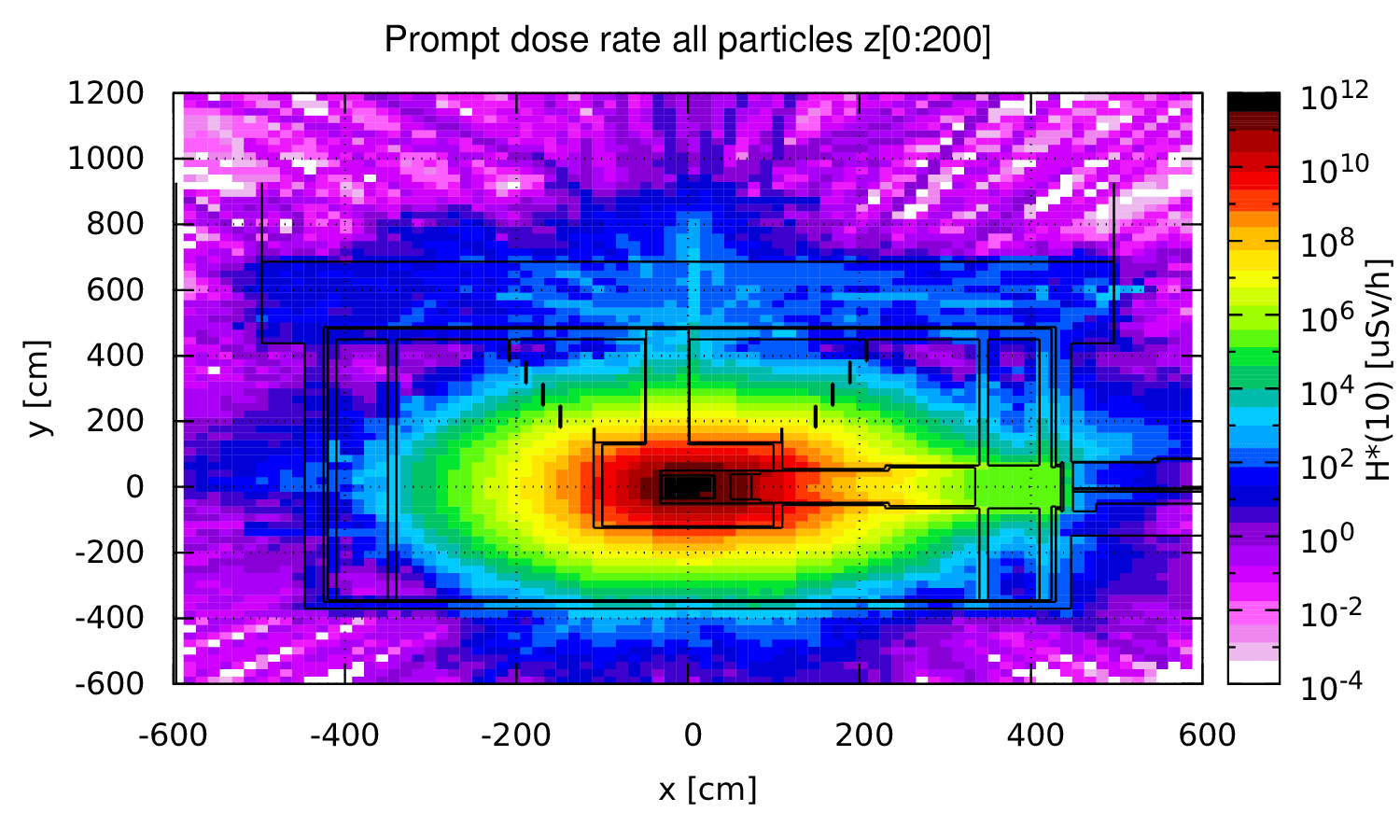}
  \caption{}
  \label{fig:RPprompt1}
\end{subfigure}
\begin{subfigure}{0.8\textwidth}
  \centering
  \includegraphics[width=0.8\textwidth]{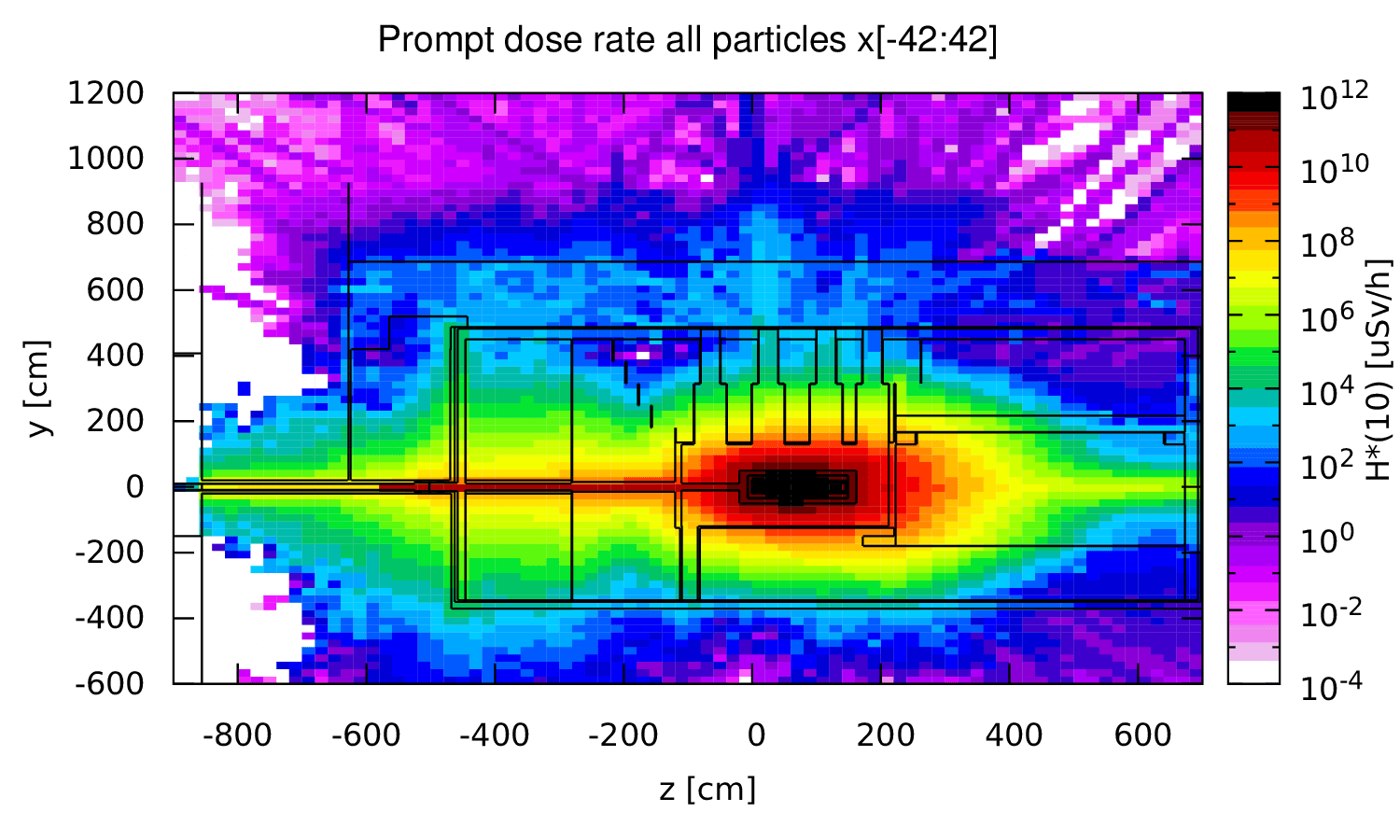}
  \caption{}
  \label{fig:RPprompt2}
\end{subfigure}

\captionsetup{width=0.85\textwidth} \caption{\small Prompt dose rates in $\mu$Sv/h in the BDF target complex for all particles. Left plot shows the perpendicular view at target level and the right plot shows the side view cut through the beam line.}
\label{fig:RPpDR1-v2}
\end{figure}

The BDF target complex was designed under the condition that the target hall can be accessed during beam operation and classified as a Supervised Radiation Area (< 3 $\mu$Sv/h). On the contrary no access during beam operation will be permitted to the target bunker. The prompt dose rates in the BDF target complex are depicted in Figure \ref{fig:RPpDR1-v2}. As expected, the highest dose rates can be found in the region of the target reaching a few $10^{12}$ $\mu$Sv/h. They are reduced by a few orders of magnitude in the surrounding iron shielding. Above the helium vessel enclosing the shielding, the prompt dose rates amount up to 3~mSv/h. The prompt dose rates are further reduced by the above concrete shielding, such that they drop down to below a 1 $\mu$Sv/h in the target hall.

\begin{figure}[!htb]
  \centering
\begin{subfigure}{0.7\textwidth}
  \centering
  \includegraphics[width=0.9\textwidth]{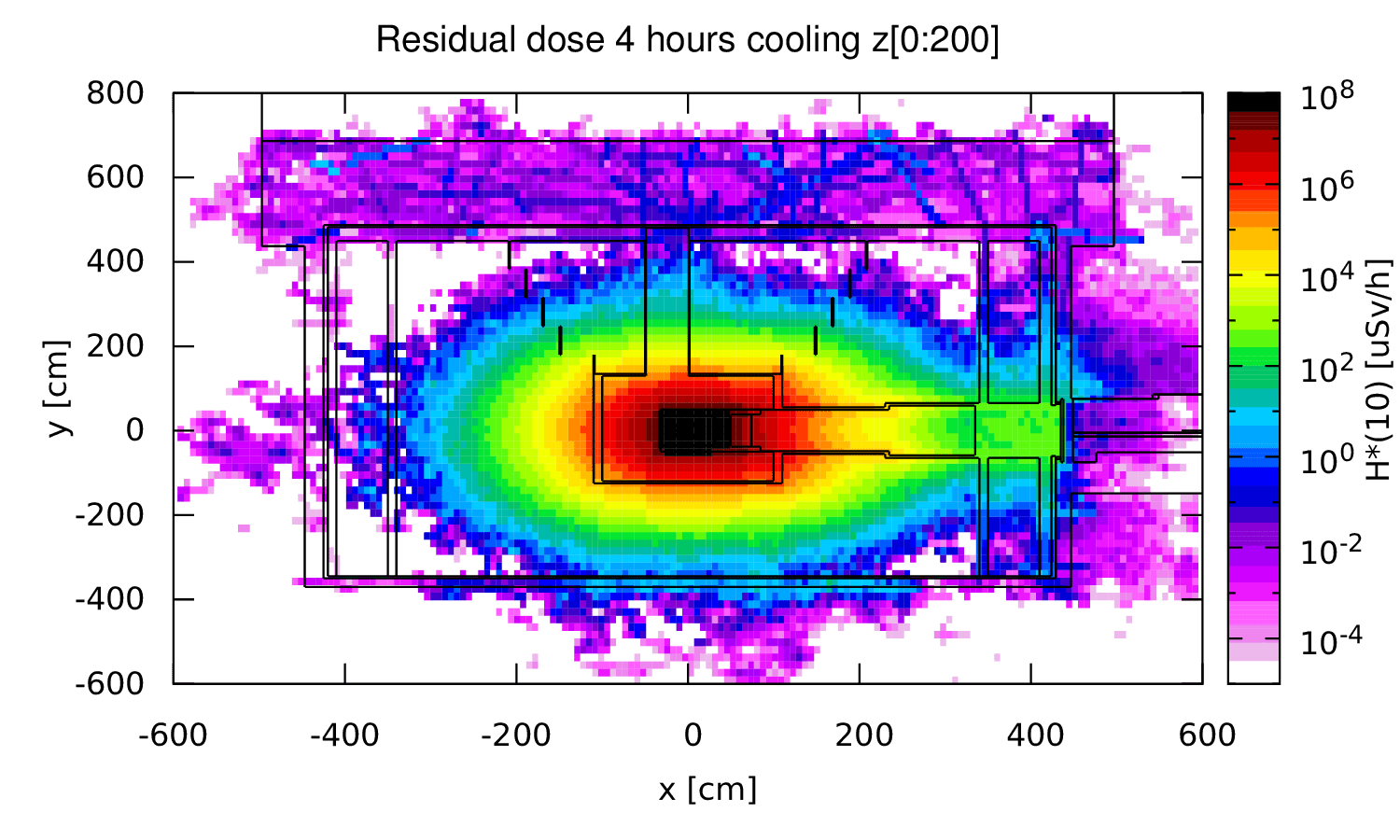}
  \caption{}
  \label{fig:RPresidual1}
\end{subfigure}
\begin{subfigure}{0.7\textwidth}
  \centering
  \includegraphics[width=0.9\textwidth]{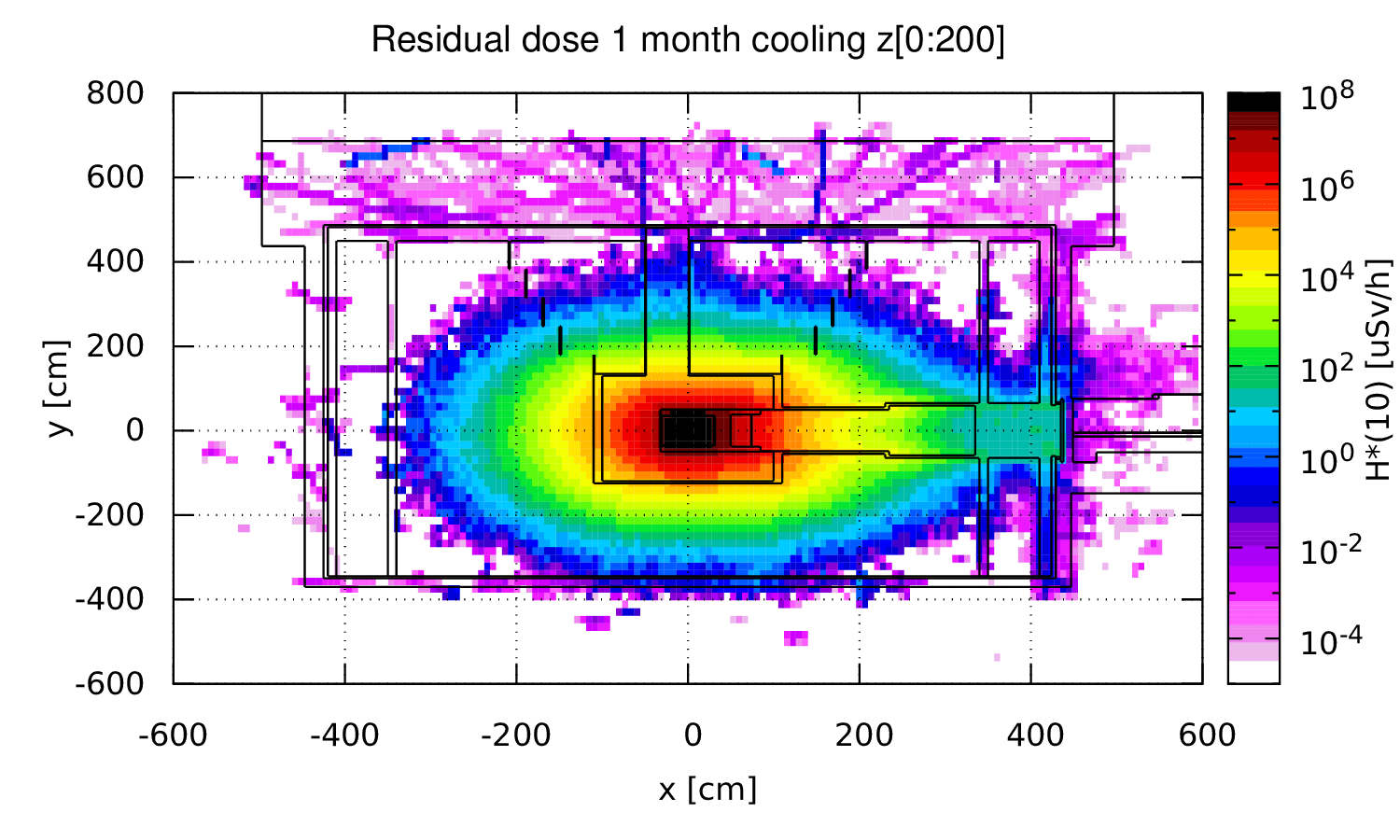}
  \caption{}
  \label{fig:RPresidual2}
  \end{subfigure}
  \begin{center}
  \begin{subfigure}{0.7\textwidth}
  \centering
  \includegraphics[width=0.9\textwidth]{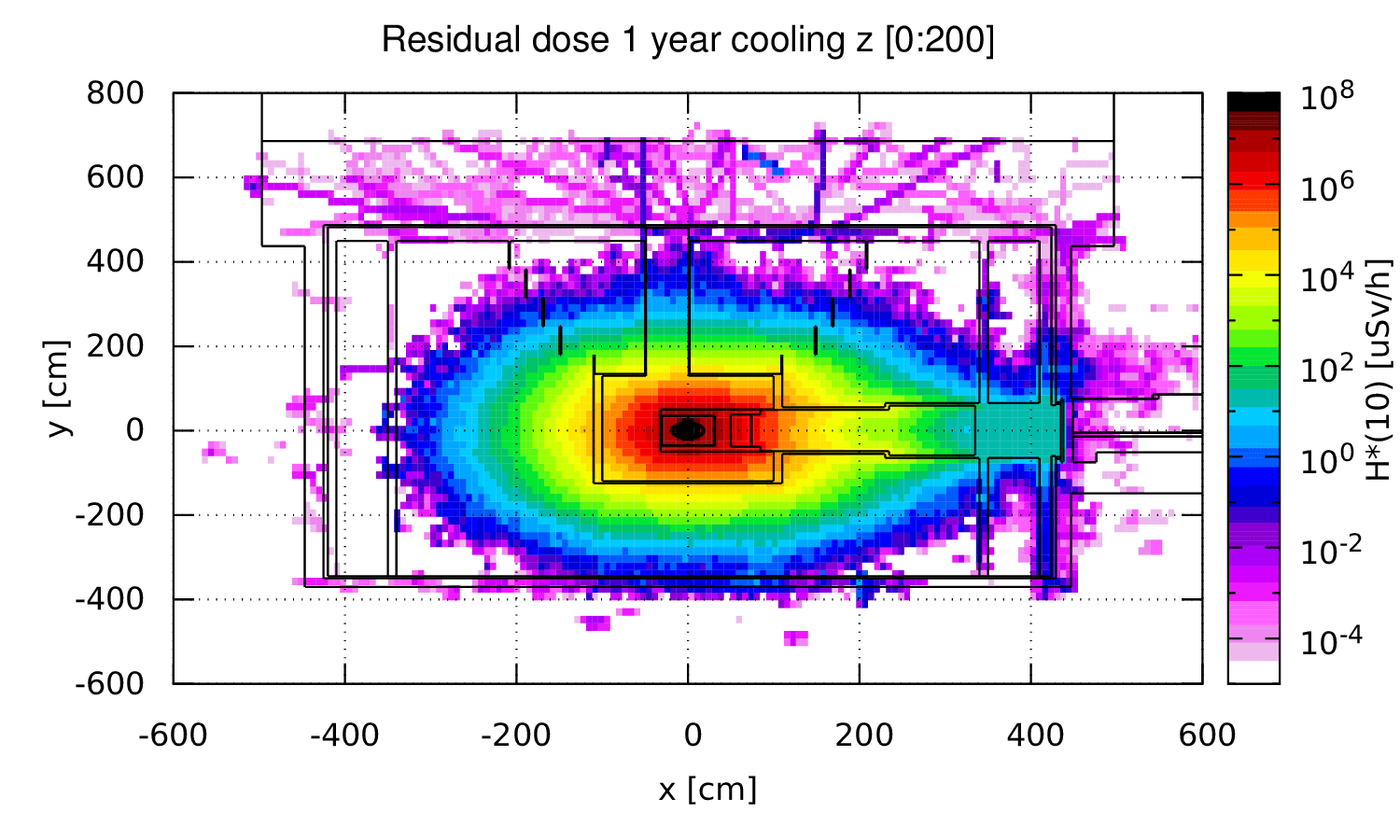}
  \caption{}
  \label{fig:RPresidual3}
\end{subfigure}
\end{center}

\captionsetup{width=0.85\textwidth} \caption{\small  Perpendicular view at target level of residual dose rates in $\mu$Sv/h in the BDF target complex for 4 hours (a), 1 month (b) and 1 year (c) cooling.}
\label{fig:RPrDR1}
\end{figure}

\begin{figure}[!htb]
\centering
\begin{subfigure}{0.7\textwidth}
  \centering
  \includegraphics[width=0.9\textwidth]{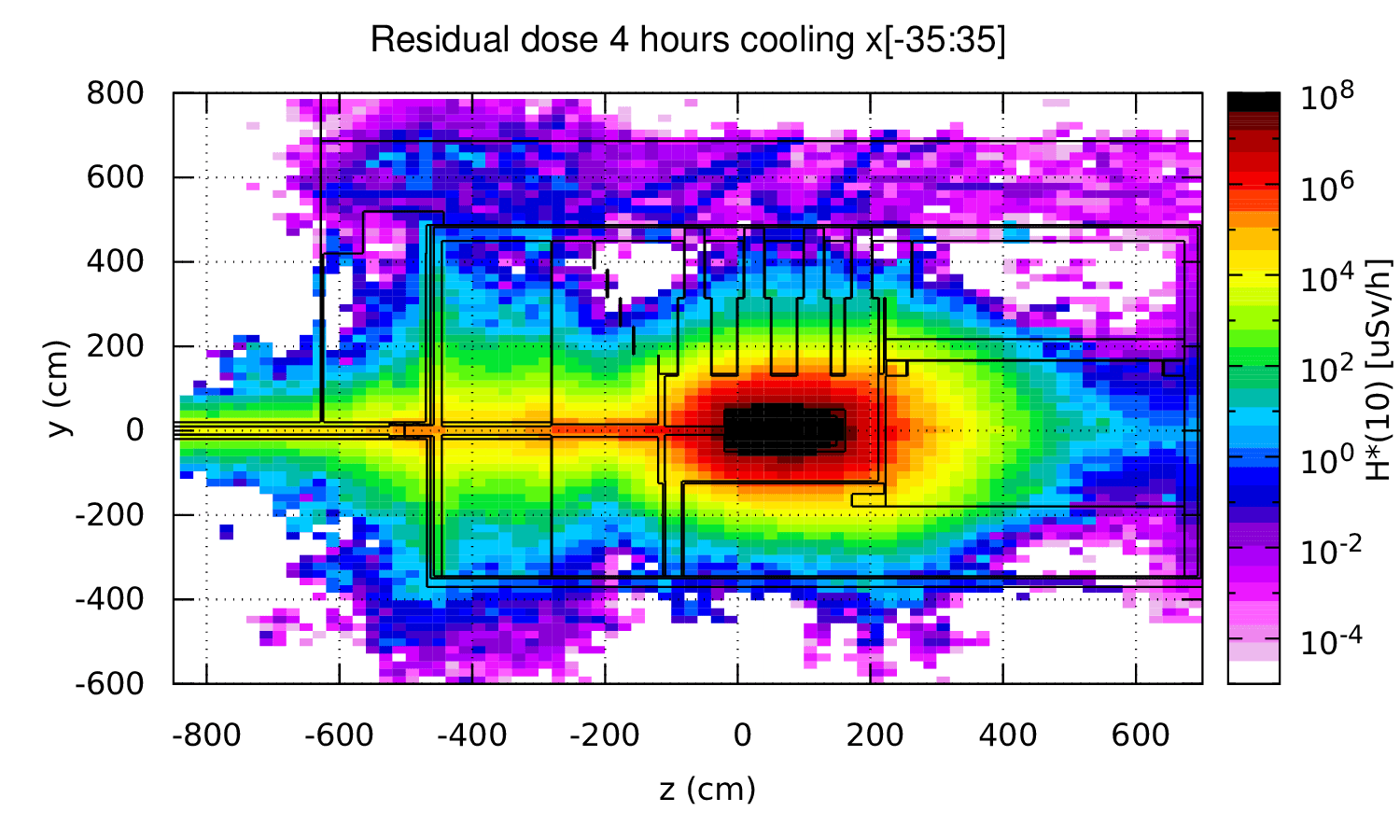}
  \caption{}
  \label{fig:RPresidual4}
\end{subfigure}
\begin{subfigure}{0.7\textwidth}
  \centering
  \includegraphics[width=0.9\textwidth]{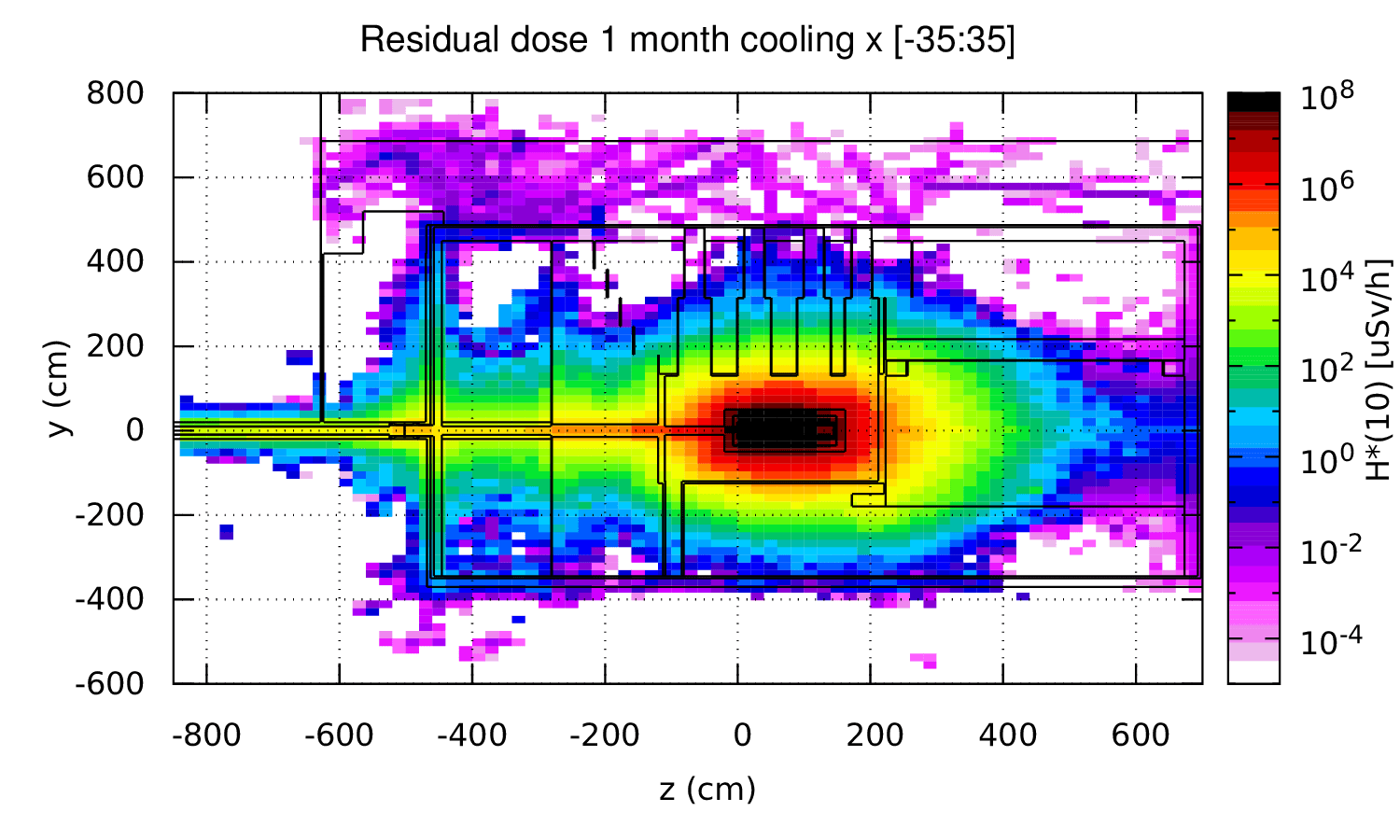}
  \caption{}
  \label{fig:RPresidual5}
\end{subfigure}
  \begin{center}
\begin{subfigure}{0.7\textwidth}
  \centering
  \includegraphics[width=0.9\textwidth]{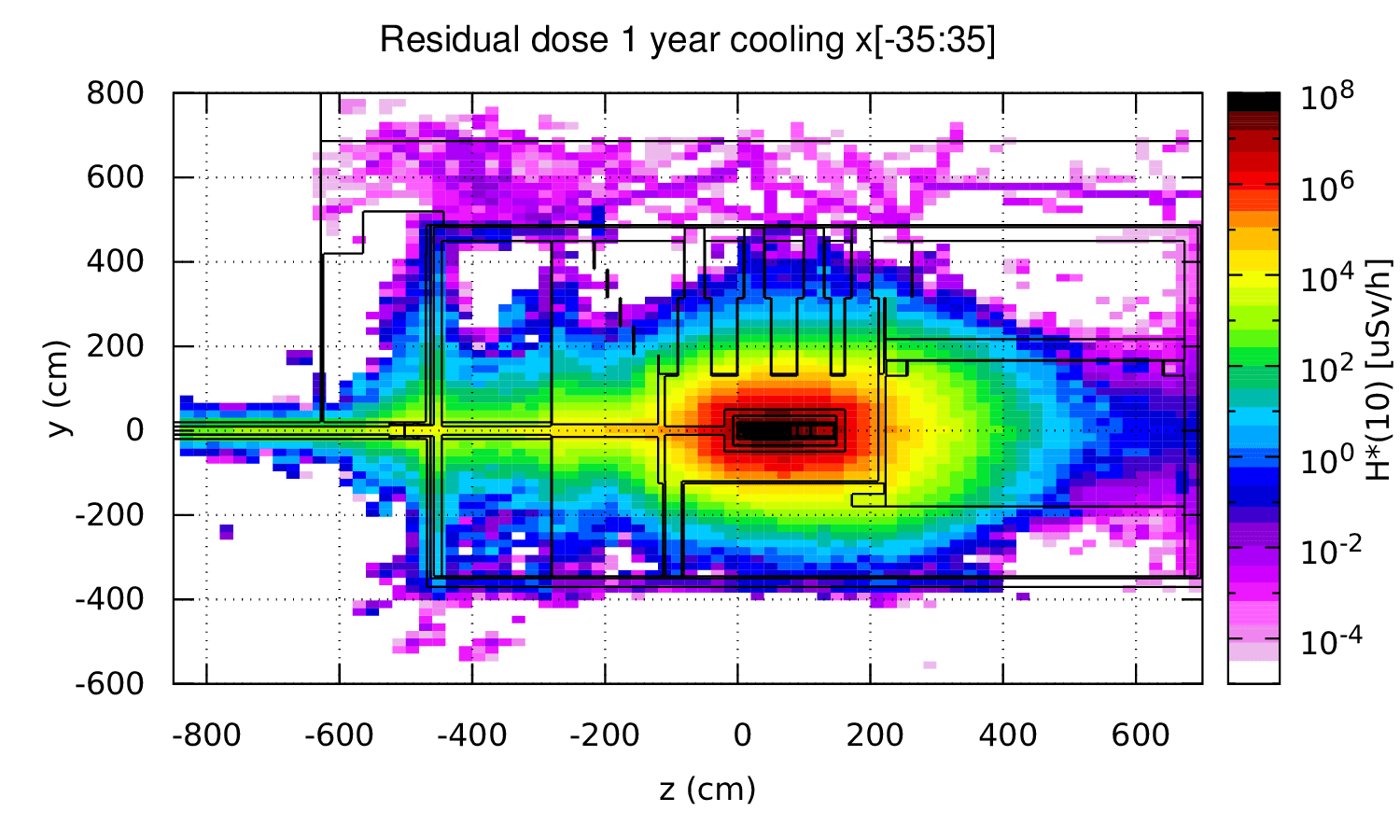}
  \caption{}
  \label{fig:RPresidual6}
\end{subfigure}
\end{center}

\captionsetup{width=0.85\textwidth} \caption{\small Side view cut through the beam line of residual dose rates in $\mu$Sv/h in the BDF target complex for 4 hours (a), 1 month (b) and 1 year (c) cooling.}
\label{fig:RPrDR2-v2}
\end{figure}

Figure \ref{fig:RPrDR1} and Figure \ref{fig:RPrDR2-v2} show the expected residual dose rates in the SHiP target complex for different cooling times. The highest dose rates can be found in the  region of the target and they are in the order of a few $10^{8}$ $\mu$Sv/h after 1 month of cooling.

For this reason, the facility design was such that all the interventions planned in the target area will be executed remotely. The closest man-accessible area is above and next to the helium vessel enclosing the shielding. Here, maximum residual dose rates of a few $\mu$Sv/h after 1 week of cooling are reached considering that the helium vessel is closed and all shielding elements are in place. The residual dose rates in the target hall can further be considered negligible.

\subsection{Helium and air activation}

The air and helium activation in the target complex was evaluated assuming the maximum beam power for five operational years. To evaluate the production of radionuclides in air a total of 39 isotopes were considered, including the radiologically most relevant short-lived isotopes  $^{11}C$, $^{13}N$, $^{14}O$, $^{15}O$ and $^{41}Ar$ as well as $^{3}H$, $^{7}Be$, $^{14}C$, $^{32}P$, $^{33}P$ and $^{35}S$ among those with long half-lives. The highest air activation resulted in the air surrounding the helium vessel with $1.7\times10^{7}$ Bq after 60 s of cooling. When comparing it to the CA\footnote{Person working 40 hours per week, 50 weeks per year with standard breathing rate in activated air with CA = 1 receives 20 mSv.} values of the Swiss legislation \cite{ORAP} it results in 0.7 CA. 

For all helium-filled regions, a realistic purity of 99.9$\%$ helium and 0.1$\%$ air contamination was considered. In the most critical helium region, that is the innermost region of the helium vessel surrounding the target, a total activity of $2.8\times10^{9}$ Bq for helium and $6.1\times10^{7}$ Bq for air after 60 s of cooling was evaluated. It results in 0.4 CA for helium and $7.5\times10^{5}$ CA for air. This demonstrates the effectiveness of helium, which only gives rise to the formation of tritium that has a significantly lower radiological impact than the radionuclides arising from air.

Note that these results do not yet take into account the tritium out-diffusion from the iron and concrete shielding into the air and helium environment due to the deficient availability of diffusion constants for tritium at the moment. However, tritium out-diffusion experiments for all of the material relevant for BDF are currently performed (see Section~\ref{sec:tritium-out-diff}).

When considering the accident of a helium vessel breakdown with a complete mixing of the activated air and helium, it would result in 2.7 CA and a committed effective dose per hour of stay of 8 $\mu$Sv.  This was used to define the classification of the BDF ventilation system, for which the ISO norm 17873:2004 for nuclear installations, which was taken as guideline for BDF. The ventilation system guarantees a pressure cascade from low to high contaminated areas (see Section \ref{Sec:TC:CV}) to sufficiently compensate the defects of the static confinement. The air around the helium vessel will be injected with a flow rate of 5500 m$^{3}$/h and extracted with a flow rate of 8000 m$^{3}$/h leading to an average irradiation time of $\approx$7 minutes. The activation in the air surrounding the helium vessel with 7 minutes irradiation and no cooling time results in a total activity of $2.61\times10^{6}$ Bq, leading to 0.11 CA. 

The ventilation circuits should be equipped with high-efficiency particle and aerosol (HEPA) filters to remove activated dust particles and aerosol-bound radionuclides from the air. Also the air exhaust should be foreseen with such filters and the airborne radioactivity released into the environment should be monitored. 

\subsection{Water activation}
\label{sub:wateractivation}
Another important aspect, which has to be taken into account from an environmental point of view, is the activation of the water from the cooling circuits. The expected radioactivity in the water was evaluated assuming the expected beam power in the five operational years and concentrating all the water statically in the target. The results give a conservative estimate due to the fact that the water speed in the circuits is 5 m/s and the cooling circuits of the proximity shielding as well as the magnetic coil are far away from the target area. The production of radionuclides in the water and the resulting activities were used to define the shielding around the demineralisation cartridges, where most of the radionuclides will be stopped.
For the cartridge of the target water cooling circuit, 50 cm cylindrical concrete shielding is foreseen and for the roof of the trolley services area 165 cm concrete. For the cartridge of the shielding/coil water cooling circuit, which is located in the CV room, a 40 cm cylindrical shielding will be envisaged like for the roof of the room itself.
The water after passing through the cartridges will mostly contain tritium. The tritium concentration coming from direct activation was estimated to be 0.5 GBq/l. Due to the high tritium production in the target (approximately $18~\mbox{TBq}$ during 5 years operation), a significant contribution to the tritium concentration in the water can come from out-diffusion of gaseous tritium from the blocks of the target and subsequent trapping in the cooling water. Tritium out-diffusion experiments at CERN are being performed to overcome the lack of experimental data (see Section~\ref{sec:tritium-out-diff}). Assuming 1$\%$ of out-diffusion every 2 months and 100$\%$ trapping, the tritium concentration from out-diffusion will be around 60 MBq/l every 2 months. The exchange of cooling water (1 m$^3$) in one year would result in 280~GBq of tritium activity.

\subsection{Radioactive waste zoning}
\begin{figure}[!htb]
    \centering
    \includegraphics[width=0.9\textwidth]{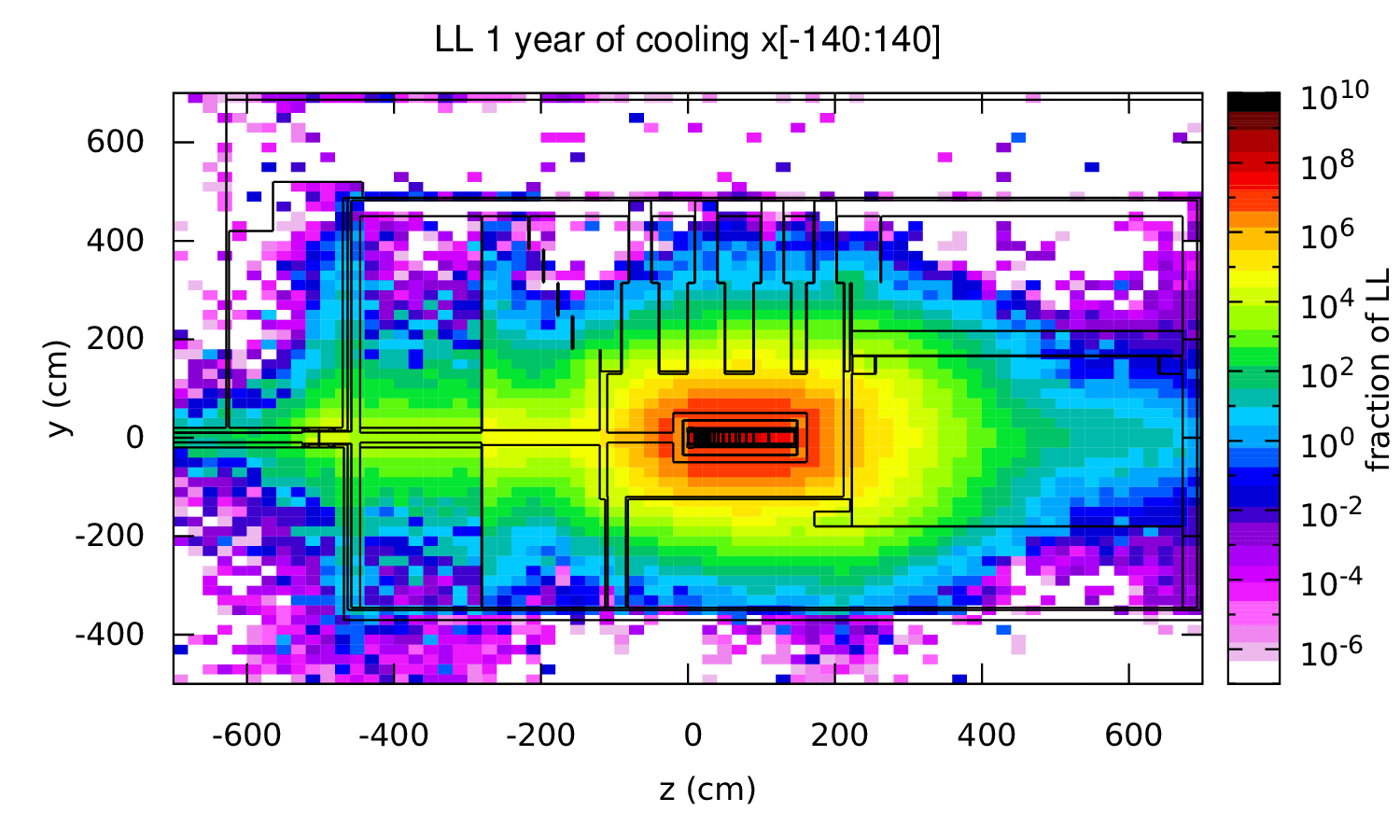}

    \captionsetup{width=0.85\textwidth} \caption{\small Side view cut through the beam line of residual activity represented as fraction of Swiss LL.}
    \label{fig:BDFLL-v2}
\end{figure}

A waste study was performed to predict the amount and the characteristics of the radioactive waste that will be produced during BDF operation. The objectives of such a study are to improve the management of radioactive waste and to eventually reduce the overall radioactive waste production. To distinguish areas of radioactive waste from conventional ones the liberation limits from Swiss legislation~\cite{ORAP} were used. To exempt a material containing a mixture of radionuclides of artificial origin from any further regulatory control, the following sum rule should be respected: 
\begin{equation}
\sum_{i=1}^{n} \frac{a_{i}}{LL_{i}}<1
\end{equation}
where $a_i$ is the specific activity (Bq/kg) or the total activity (Bq) of the $i^{th}$ radionuclide of artificial origin in the material, $LL_i$ is the respective Swiss liberation limit for the radionuclide $i$ in the material and $n$ is the number of radionuclides present. If the sum rule is not fulfilled, the material is radioactive according to the Swiss legislation. Twelve different cooling times ranging from 15 minutes to 30 years were altogether assessed. Figure~\ref{fig:BDFLL-v2} shows the waste zoning of the BDF target area after one year of cooling time. Note that the values in air- and helium-filled regions are insofar not representative, as the air and helium will be diluted by leakage and/or extraction into the environment. 
The zoning plots show that the most activated parts of BDF are the target and the iron shielding elements. It can be seen that the target and part of the iron shielding remains radioactive. In the design of the shielding, the minimisation of radioactive waste was taken into account having a modular iron bunker such, that activated parts might easily be separated from the ones that are below the liberation limits. 
\subsection{Soil activation}
The production of radioactivity in the soil and water surrounding the SHiP facility is a  significant  environmental  concern.  Particularly  soluble  radionuclides  likely  to  pass  through  the  karstic  system  are  critical  for  the  protection  of  groundwater  resources.  To  minimise  related  radiological  risks,  the  specific  activities  of  the leachable radionuclides H-3 and Na-22 should lie below the following design goals \cite{CENF}:  
\begin{itemize}
    \item H-3 < 10 Bq/kg, 
    \item Na-22 < 2 Bq/kg. 
\end{itemize}

The  leachable  radionuclide  Na-24  was  neglected  due  to  the  fact  that it  is  too  short-lived  to  survive  the  way  from  its  place  of  creation  to  its  place  of  consumption.  When  studying  the  relation  between  the  prompt  radiation  and  the  soil  activation  for  the  CENF  facility,  it  
was  estimated  that  the  above-given  limits are not exceeded with prompt dose rates of 1 mSv/h or below.
As it could be seen in the previous subsection the design of the facility is such that avoid activation of the fixed concrete civil engineering structures and by consequence the soil, in fact the LL of the soil is below 1 and the prompt dose rate at the soil level was kept below 1 mSv/h.

\section{Environmental impact}
The environmental impact from releases of radioactive air and helium was studied in detail. The annual releases from the the target pit around the He vessel and from the He vessel are listed in Table \ref{tab:releases}.

\begin{table}[!ht]
\centering
\begin{footnotesize}
\begin{tabular}{ccc}
\toprule
\textbf{Radioisotope} & \textbf{Flushed activity form target pit [Bq/y]} & \textbf{Flushed activity form helium vessel [Bq/y]} \\
\midrule
H-3& 5.5$\times10^{4}$&1.44$\times10^{9}$\\
Be-7& 9.0$\times10^{5}$&1.46$\times10^{6}$\\
Be-10& 1.5$\times10^{-1}$&3.57$\times10^{-1}$\\
C-11&3.9$\times10^{9}$&2.77$\times10^{6}$\\
C-14&9.4$\times10^{3}$&2.66$\times10^{4}$\\
N-13&1.8$\times10^{10}$&7.81$\times10^{6}$\\
O-14&7.5$\times10^{8}$&1.29$\times10^{5}$\\
O-15&2.0$\times10^{10}$&3.60$\times10^{6}$\\
O-19&2.0$\times10^{6}$&1.02$\times10^{3}$\\
F-18&2.0$\times10^{5}$&1.39$\times10^{3}$\\
Ne-23&3.2$\times10^{6}$&1.06$\times10^{3}$\\
Ne-24&7.9$\times10^{5}$&4.60$\times10^{2}$\\
Na-22&2.4$\times10^{1}$&9.44$\times10^{1}$\\
Na-24&3.3$\times10^{4}$&1.51$\times10^{3}$\\
Na-25&5.9$\times10^{6}$&1.73$\times10^{3}$\\
Mg-27&5.0$\times10^{6}$&2.70$\times10^{3}$\\
Mg-28&1.2$\times10^{4}$&8.41$\times10^{2}$\\
Al-26&6.6$\times10^{-5}$&2.10$\times10^{-4}$\\
Al-28&4.0$\times10^{7}$&7.89$\times10^{3}$\\
Al-29&1.4$\times10^{7}$&5.01$\times10^{3}$\\
Si-31&1.7$\times10^{6}$&8.91$\times10^{3}$\\
Si-32&5.3$\times10^{-1}$&1.57\\
P-30&1.6$\times10^{7}$&2.86$\times10^{3}$\\
P-32&3.7$\times10^{4}$&2.17$\times10^{4}$\\
P-33&2.2$\times10^{4}$&2.13$\times10^{4}$\\
P-35&2.6$\times10^{7}$&4.15$\times10^{3}$\\
S-35&1.3$\times10^{4}$&2.22$\times10^{4}$\\
S-37&5.7$\times10^{7}$&1.35$\times10^{4}$\\
S-38&6.2$\times10^{5}$&4.01$\times10^{3}$\\
Cl-34&1.2$\times10^{6}$&1.19$\times10^{3}$\\
Cl-36&1.7$\times10^{-2}$&3.83$\times10^{-2}$\\
Cl-38&3.6$\times10^{7}$&3.94$\times10^{4}$\\
Cl-39&6.8$\times10^{7}$&1.12$\times10^{5}$\\
Cl-40&1.4$\times10^{8}$&1.82$\times10^{4}$\\
Ar-37&6.2$\times10^{4}$&6.67$\times10^{4}$\\
Ar-39&1.4$\times10^{2}$&3.08$\times10^{2}$\\
Ar-41&5.4$\times10^{8}$&1.70$\times10^{6}$\\
K-38&3.1$\times10^{4}$&8.54\\
K-40&4.3$\times10^{-9}$&1.19$\times10^{-8}$\\
\bottomrule
\end{tabular}
\captionsetup{width=0.85\textwidth} \caption{\small Annual releases from the the target pit around the He vessel.}\label{tab:releases}
\end{footnotesize}
\vspace{1cm}
\end{table}

\begin{figure}[!htb]
    \centering
    \includegraphics[width=0.9\textwidth]{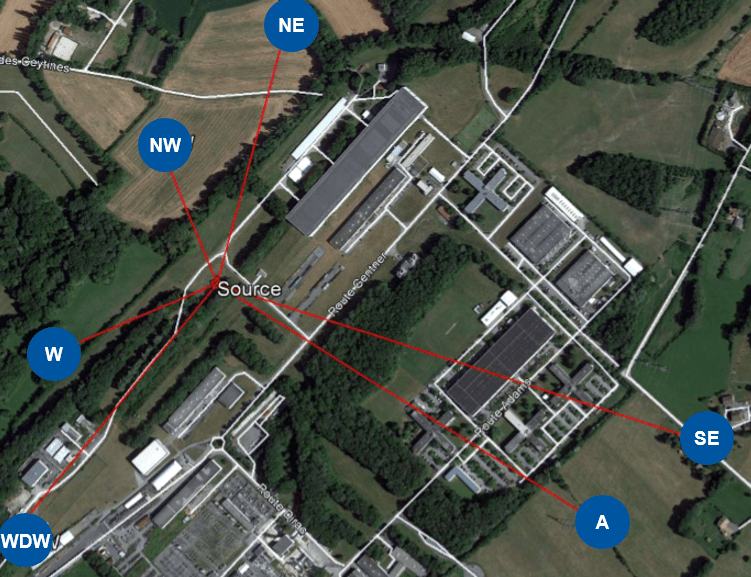}

    \captionsetup{width=0.85\textwidth} \caption{\small Locations of the reference groups around the CERN site.}
    \label{fig:RefGroup}
\end{figure}

In total 44 GBq and 16 MBq of the short-lived gases ($^{11}C$, $^{13}N$, $^{14}O$, $^{15}O$ and $^{41}Ar$) will be released annually from the target pit and the helium vessel respectively. It shall be noted that the BDF target complex will be equipped with standard ventilation monitoring stations measuring the short-lived radioactive gases on-line, taking samples of aerosol-bound radioactivity and enabling tritium sampling (the two latter categories are analysed in a laboratory).
Different residential groups and a farmer group were assumed as representative for airborne exposure.
The locations of the reference groups relative to the source are shown in Figure \ref{fig:RefGroup}.

The dose coefficients [Sv/Bq], which have to be multiplied with the annual long-term releases [Bq/y] for each radionuclide to obtain the annual effective dose [Sv/y], are listed in Table \ref{tab:coefficientsgroups} for each reference population group and each radionuclide. The release from the helium vessel is assumed to happen once at the end of the yearly run thus the dose coefficients in table \ref{tab:coefficientsgroups} have been multiplied by a factor 20 to take into account short term effects.
The greatest effective doses to the residential reference groups were reached for the group NW for all sources. To simplify the presentation only effective doses for this group and for the agricultural group will be presented. They are listed for each radionuclide and each group in Table \ref{tab:effectivedose}.
\begin{table}[!ht]
\centering
\begin{footnotesize}
\begin{tabular}{c|cccccc}
\toprule
 & \multicolumn{6}{c}{\textbf{Dose coefficients [Sv/Bq]}}\\
\midrule
Radioisotopes	&	WDW	&	NE	&	NW	&	W	&	SE	&	A	\\
H-3	&	2.54$\times10^{-20}$	&	9.66$\times10^{-20}$	&	1.64$\times10^{-19}$	&	1.55$\times10^{-19}$	&	3.31$\times10^{-20}$	&	8.40$\times10^{-20}$	\\
Be-7	&	1.04$\times10^{-17}$	&	3.21$\times10^{-17}$	&	5.46$\times10^{-17}$	&	5.18$\times10^{-17}$	&	1.10$\times10^{-17}$	&	2.23$\times10^{-19}$	\\
B-10	&	5.84$\times10^{-17}$	&	2.35$\times10^{-16}$	&	3.99$\times10^{-16}$	&	3.78$\times10^{-16}$	&	8.05$\times10^{-17}$	&	3.16$\times10^{-17}$	\\
C-11	&	2.14$\times10^{-20}$	&	8.74$\times10^{-20}$	&	1.53$\times10^{-19}$	&	1.46$\times10^{-19}$	&	2.59$\times10^{-20}$	&	0.00	\\
C-14	&	2.83$\times10^{-18}$	&	1.39$\times10^{-17}$	&	2.35$\times10^{-17}$	&	2.23$\times10^{-17}$	&	4.75$\times10^{-18}$	&	5.83$\times10^{-17}$	\\
N-13	&	1.18$\times10^{-20}$	&	5.33$\times10^{-20}$	&	9.59$\times10^{-20}$	&	9.12$\times10^{-20}$	&	1.28$\times10^{-20}$	&	0.00	\\
O-14	&	2.56$\times10^{-21}$	&	1.33$\times10^{-20}$	&	3.65$\times10^{-20}$	&	2.90$\times10^{-20}$	&	3.54$\times10^{-22}$	&	0.00	\\
O-15	&	1.55$\times10^{-21}$	&	1.10$\times10^{-20}$	&	2.18$\times10^{-20}$	&	1.90$\times10^{-20}$	&	5.20$\times10^{-22}$	&	0.00	\\
O-19	&	5.50$\times10^{-23}$	&	6.37$\times10^{-22}$	&	3.59$\times10^{-21}$	&	1.77$\times10^{-21}$	&	5.79$\times10^{-24}$	&	0.00	\\
F-18	&	3.67$\times10^{-19}$	&	1.26$\times10^{-18}$	&	2.17$\times10^{-18}$	&	2.07$\times10^{-18}$	&	4.08$\times10^{-19}$	&	6.13$\times10^{-24}$	\\
Ne-23	&	2.48$\times10^{-23}$	&	2.89$\times10^{-22}$	&	8.58$\times10^{-22}$	&	7.10$\times10^{-22}$	&	1.39$\times10^{-24}$	&	0.00	\\
Ne-24	&	1.78$\times10^{-21}$	&	1.08$\times10^{-20}$	&	2.07$\times10^{-20}$	&	1.94$\times10^{-20}$	&	1.10$\times10^{-21}$	&	0.00	\\
Na-22	&	4.38$\times10^{-15}$	&	1.35$\times10^{-14}$	&	2.30$\times10^{-14}$	&	2.18$\times10^{-14}$	&	4.64$\times10^{-15}$	&	3.14$\times10^{-14}$	\\
Na-24	&	9.49$\times10^{-18}$	&	3.03$\times10^{-17}$	&	5.16$\times10^{-17}$	&	4.89$\times10^{-17}$	&	1.03$\times10^{-17}$	&	2.26$\times10^{-17}$	\\
Na-25	&	1.94$\times10^{-22}$	&	1.51$\times10^{-21}$	&	4.25$\times10^{-21}$	&	3.61$\times10^{-21}$	&	2.60$\times10^{-23}$	&	0.00	\\
Mg-27	&	2.98$\times10^{-19}$	&	1.31$\times10^{-18}$	&	2.06$\times10^{-18}$	&	2.20$\times10^{-18}$	&	2.63$\times10^{-19}$	&	0.00	\\
Mg-28	&	7.53$\times10^{-18}$	&	3.10$\times10^{-17}$	&	5.27$\times10^{-17}$	&	5.00$\times10^{-17}$	&	1.05$\times10^{-17}$	&	1.30$\times10^{-16}$	\\
Al-26	&	5.32$\times10^{-14}$	&	1.61$\times10^{-13}$	&	2.73$\times10^{-13}$	&	2.59$\times10^{-13}$	&	5.51$\times10^{-14}$	&	1.68$\times10^{-16}$	\\
Al-28	&	6.13$\times10^{-20}$	&	7.41$\times10^{-19}$	&	8.11$\times10^{-19}$	&	9.42$\times10^{-19}$	&	3.24$\times10^{-20}$	&	0.00	\\
Al-29	&	1.72$\times10^{-20}$	&	8.88$\times10^{-20}$	&	1.44$\times10^{-19}$	&	1.47$\times10^{-19}$	&	1.59$\times10^{-20}$	&	0.00	\\
Si-31	&	1.01$\times10^{-19}$	&	5.28$\times10^{-19}$	&	9.08$\times10^{-19}$	&	8.66$\times10^{-19}$	&	1.71$\times10^{-19}$	&	2.30$\times10^{-23}$	\\
Si-32	&	1.56$\times10^{-16}$	&	4.99$\times10^{-16}$	&	8.46$\times10^{-16}$	&	8.02$\times10^{-16}$	&	1.71$\times10^{-16}$	&	1.47$\times10^{-17}$	\\
P-30	&	6.03$\times10^{-21}$	&	3.85$\times10^{-20}$	&	5.65$\times10^{-20}$	&	5.51$\times10^{-20}$	&	2.14$\times10^{-21}$	&	0.00	\\
P-32	&	5.02$\times10^{-18}$	&	8.10$\times10^{-17}$	&	1.38$\times10^{-16}$	&	1.30$\times10^{-16}$	&	2.77$\times10^{-17}$	&	4.96$\times10^{-15}$	\\
P-33	&	2.12$\times10^{-18}$	&	1.31$\times10^{-17}$	&	2.22$\times10^{-17}$	&	2.10$\times10^{-17}$	&	4.47$\times10^{-18}$	&	6.32$\times10^{-16}$	\\
P-35	&	3.26$\times10^{-20}$	&	3.05$\times10^{-19}$	&	3.27$\times10^{-19}$	&	3.55$\times10^{-19}$	&	1.78$\times10^{-21}$	&	0.00	\\
S-35	&	1.99$\times10^{-18}$	&	1.16$\times10^{-17}$	&	1.97$\times10^{-17}$	&	1.87$\times10^{-17}$	&	3.97$\times10^{-18}$	&	3.65$\times10^{-16}$	\\
S-37	&	2.19$\times10^{-20}$	&	1.10$\times10^{-19}$	&	1.80$\times10^{-19}$	&	1.84$\times10^{-19}$	&	1.75$\times10^{-20}$	&	0.00	\\
S-38	&	1.10$\times10^{-18}$	&	4.22$\times10^{-18}$	&	7.23$\times10^{-18}$	&	6.89$\times10^{-18}$	&	1.39$\times10^{-18}$	&	1.55$\times10^{-20}$	\\
Cl-34	&	2.06$\times10^{-19}$	&	8.13$\times10^{-19}$	&	1.41$\times10^{-18}$	&	1.38$\times10^{-18}$	&	2.38$\times10^{-19}$	&	9.63$\times10^{-33}$	\\
Cl-36	&	2.48$\times10^{-17}$	&	1.08$\times10^{-14}$	&	1.84$\times10^{-14}$	&	1.74$\times10^{-14}$	&	3.70$\times10^{-15}$	&	5.00$\times10^{-13}$	\\
Cl-38	&	1.74$\times10^{-19}$	&	6.99$\times10^{-19}$	&	1.22$\times10^{-18}$	&	1.18$\times10^{-18}$	&	2.08$\times10^{-19}$	&	1.04$\times10^{-30}$	\\
Cl-39	&	2.49$\times10^{-19}$	&	9.22$\times10^{-19}$	&	1.60$\times10^{-18}$	&	1.53$\times10^{-18}$	&	2.86$\times10^{-19}$	&	7.85$\times10^{-27}$	\\
Cl-40	&	3.47$\times10^{-21}$	&	2.31$\times10^{-20}$	&	5.12$\times10^{-20}$	&	4.27$\times10^{-20}$	&	9.10$\times10^{-22}$	&	0.00	\\
Ar-37	&	6.77$\times10^{-26}$	&	2.04$\times10^{-25}$	&	3.47$\times10^{-25}$	&	3.29$\times10^{-25}$	&	7.00$\times10^{-26}$	&	0.00	\\
Ar-39	&	2.57$\times10^{-22}$	&	7.75$\times10^{-22}$	&	1.32$\times10^{-21}$	&	1.25$\times10^{-21}$	&	2.66$\times10^{-22}$	&	0.00	\\
Ar-41	&	4.47$\times10^{-20}$	&	1.51$\times10^{-19}$	&	2.40$\times10^{-19}$	&	2.27$\times10^{-19}$	&	5.68$\times10^{-20}$	&	0.00	\\
K-38	&	1.10$\times10^{-19}$	&	4.92$\times10^{-19}$	&	7.64$\times10^{-19}$	&	8.13$\times10^{-19}$	&	9.09$\times10^{-20}$	&	0.00 \\
K-40	&	3.18$\times10^{-15}$	&	2.44$\times10^{-14}$	&	4.14$\times10^{-14}$	&	3.92$\times10^{-14}$	&	8.34$\times10^{-15}$	&	2.06$\times10^{-13}$	\\

\bottomrule
\end{tabular}
\captionsetup{width=0.85\textwidth} \caption{\small Effective doses for the NW group and for the agricultural group.}\label{tab:coefficientsgroups}
\end{footnotesize}
\vspace{1cm}
\end{table}

\begin{table}[!ht]
\centering
\begin{footnotesize}
\begin{tabular}{c|cc}
\toprule
 & \multicolumn{2}{c}{\textbf{Effective dose [Sv/y]}}\\
\midrule
Radioisotope&NW&A\\
H-3	&	4,73379$\times10^{-9}$	&	2,42462$\times10^{-9}$	\\
Be-7	&	1,64454$\times10^{-9}$	&	6,71669$\times10^{-12}$	\\
Be-10	&	2,9091$\times10^{-15}$	&	2,30395$\times10^{-16}$	\\
C-11	&	6,10442$\times10^{-10}$	&	0	\\
C-14	&	1,27261$\times10^{-11}$	&	3,15714$\times10^{-11}$	\\
N-13	&	1,79319$\times10^{-9}$	&	0	\\
O-14	&	2,73733$\times10^{-11}$	&	0	\\
O-15	&	4,48658$\times10^{-10}$	&	0	\\
O-19	&	7,182$\times10^{-15}$	&	0	\\
F-18	&	4,96779$\times10^{-13}$	&	1,40334$\times10^{-18}$	\\
Ne-23	&	2,76003$\times10^{-15}$	&	0	\\
Ne-24	&	1,65388$\times10^{-14}$	&	0	\\
Na-22	&	4,39769$\times10^{-11}$	&	6,0038$\times10^{-11}$	\\
Na-24	&	3,25718$\times10^{-12}$	&	1,42659$\times10^{-12}$	\\
Na-25	&	2,53092$\times10^{-14}$	&	0	\\
Mg-27	&	1,03207$\times10^{-11}$	&	0	\\
Mg-28	&	1,4961$\times10^{-12}$	&	3,69057$\times10^{-12}$	\\
Al-26	&	1,16683$\times10^{-15}$	&	7,18052$\times10^{-19}$	\\
Al-28	&	3,22696$\times10^{-11}$	&	0	\\
Al-29	&	2,02626$\times10^{-12}$	&	0	\\
Si-31	&	1,74293$\times10^{-12}$	&	4,41491$\times10^{-17}$	\\
Si-32	&	2,70759$\times10^{-14}$	&	4,70469$\times10^{-16}$	\\
P-30	&	9,3639$\times10^{-13}$	&	0	\\
P-32	&	6,49838$\times10^{-11}$	&	2,33565$\times10^{-9}$	\\
P-33	&	9,95586$\times10^{-12}$	&	2,83428$\times10^{-10}$	\\
P-35	&	8,69143$\times10^{-12}$	&	0	\\
S-35	&	9,02045$\times10^{-12}$	&	1,6713$\times10^{-10}$	\\
S-37	&	1,02981$\times10^{-11}$	&	0	\\
S-38	&	5,07684$\times10^{-12}$	&	1,0884$\times10^{-14}$	\\
Cl-34	&	1,78834$\times10^{-12}$	&	1,2214$\times10^{-26}$	\\
Cl-36	&	1,44043$\times10^{-14}$	&	3,91422$\times10^{-13}$	\\
Cl-38	&	4,53519$\times10^{-11}$	&	3,86606$\times10^{-23}$	\\
Cl-39	&	1,12714$\times10^{-10}$	&	5,53001$\times10^{-19}$	\\
Cl-40	&	7,20983$\times10^{-12}$	&	0	\\
Ar-37	&	4,84585$\times10^{-19}$	&	0	\\
Ar-39	&	8,3041$\times10^{-18}$	&	0	\\
Ar-41	&	1,37001$\times10^{-10}$	&	0	\\
K-38	&	2,39231$\times10^{-14}$	&	0	\\
K-40	&	1,00467$\times10^{-20}$	&	4,9991$\times10^{-20}$	\\

\midrule
Total	&	9.78$\times10^{-9}$	&	5.31$\times10^{-9}$	\\

\bottomrule
\end{tabular}
\captionsetup{width=0.85\textwidth} \caption{\small The effective dose for the reference groups NW and A.}\label{tab:effectivedose}
\end{footnotesize}
\vspace{1cm}
\end{table}

One can conclude that the maximum effective dose for the group NW is about 9.8 nSv/y. The maximum effective dose due to agriculture is only about 5.3 nSv/y. 
Such doses are sufficiently low so that they will not contribute significantly to the total dose received by any member of the public due to operation of all facilities on the Pr\'evessin site (TT20, TT26, EHN1, NA62 etc.), the new included.  CERN aims at keeping this dose below the dose objective of 10 $\mu$Sv/y. Considerably exceeding the latter would require optimisation of the facilities. With BDF in operation, this will not be the case. In line with our experience, the radionuclides, which stick to aerosols, are very efficiently removed by HEPA filters. Hence, the radionuclides like $^{7}$Be, $^{22,24}$Na, $^{32,33}$P and $^{35}$S could be released with activities a few orders of magnitude lower than those calculated.

The environmental impact of releases of the water was studied in detail (see Section~\ref{sub:wateractivation}). The tritium activity concentration of 0.3 GBq/l at the end of each operational year is too high for its rapid discharge from a sump into the receiving river Le Lion, because the immission limit for water accessible to the public could be exceeded. A new evaporator was therefore included in the design of the BDF facility to evaporate the water slowly into the atmosphere profiting from the long term variations of the wind direction. The dosimetric impact of a similar facility was studied in the past \cite{CENF}. Several hypothetical population groups were examined and the maximum effective dose per Bq of released activity of 1.64×10$^{-19}$~Sv/Bq was obtained \cite{CENF}. The annual release of 280 GBq of tritium would result in an effective dose of less than 50 nSv/y. This value is far below the dose constraint of 10 $\mu$Sv/y fixed by the Organization for new facilities.
\clearpage 
\section{Experimental area}
\subsection{Prompt radiation and shielding requirements}
\begin{figure}[!htb]
  \centering
\begin{subfigure}{0.8\textwidth}
  \centering
  \includegraphics[width=0.9\textwidth]{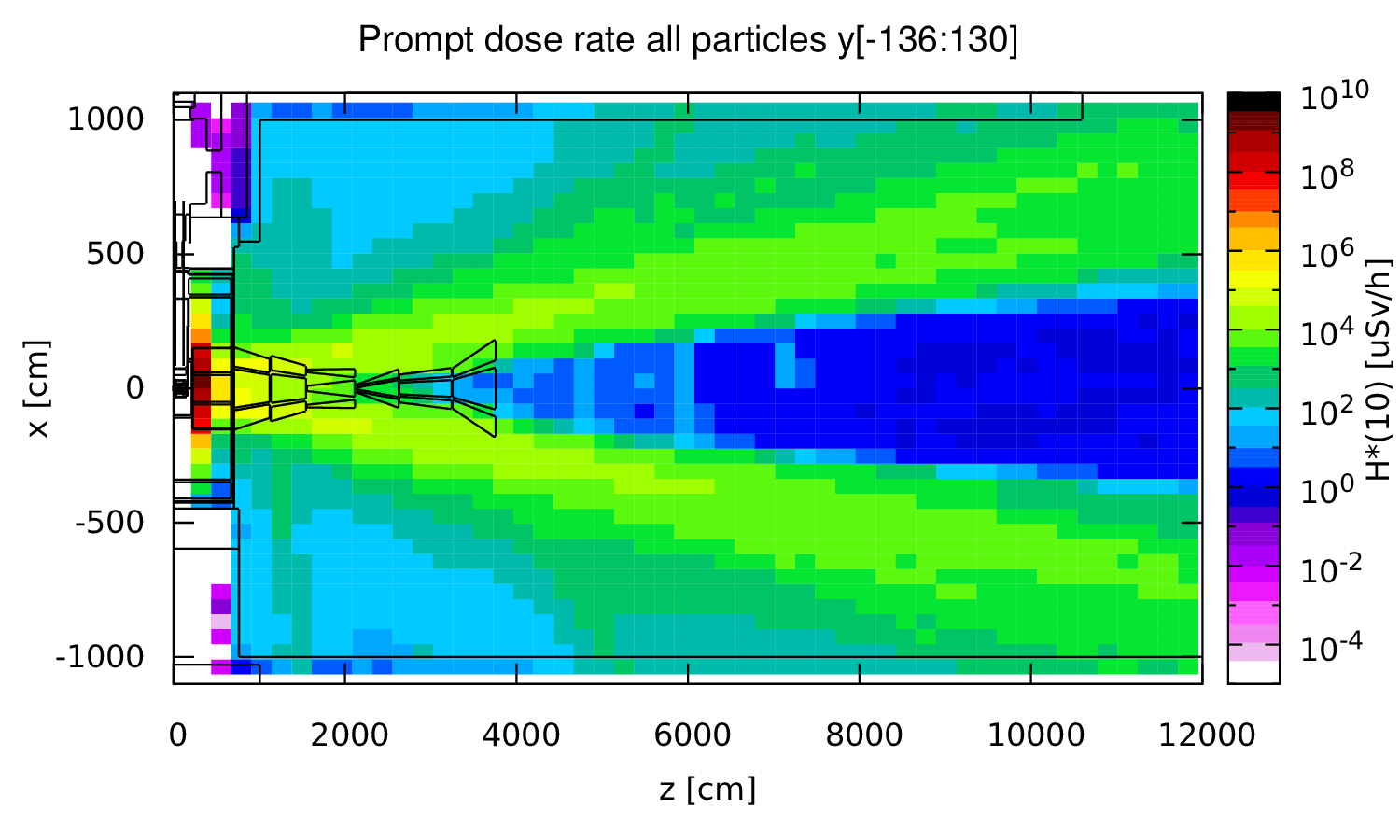}
  \caption{}
  \label{fig:MF1}
\end{subfigure}
\begin{subfigure}{0.8\textwidth}
  \centering
  \includegraphics[width=0.9\textwidth]{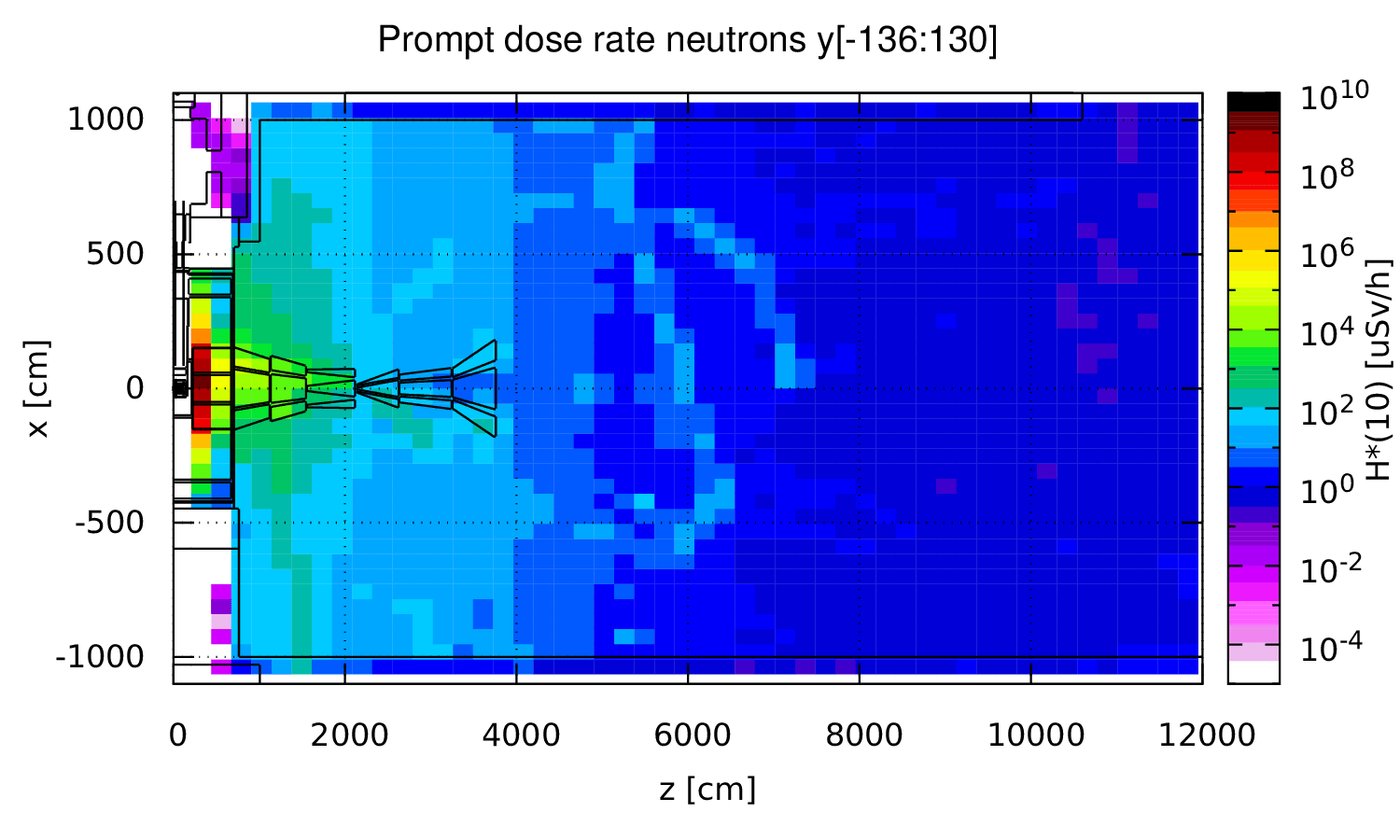}
  \caption{}
  \label{fig:MF1n}
\end{subfigure}
  \begin{center}
\begin{subfigure}{0.8\textwidth}
  \centering
  \includegraphics[width=0.9\textwidth]{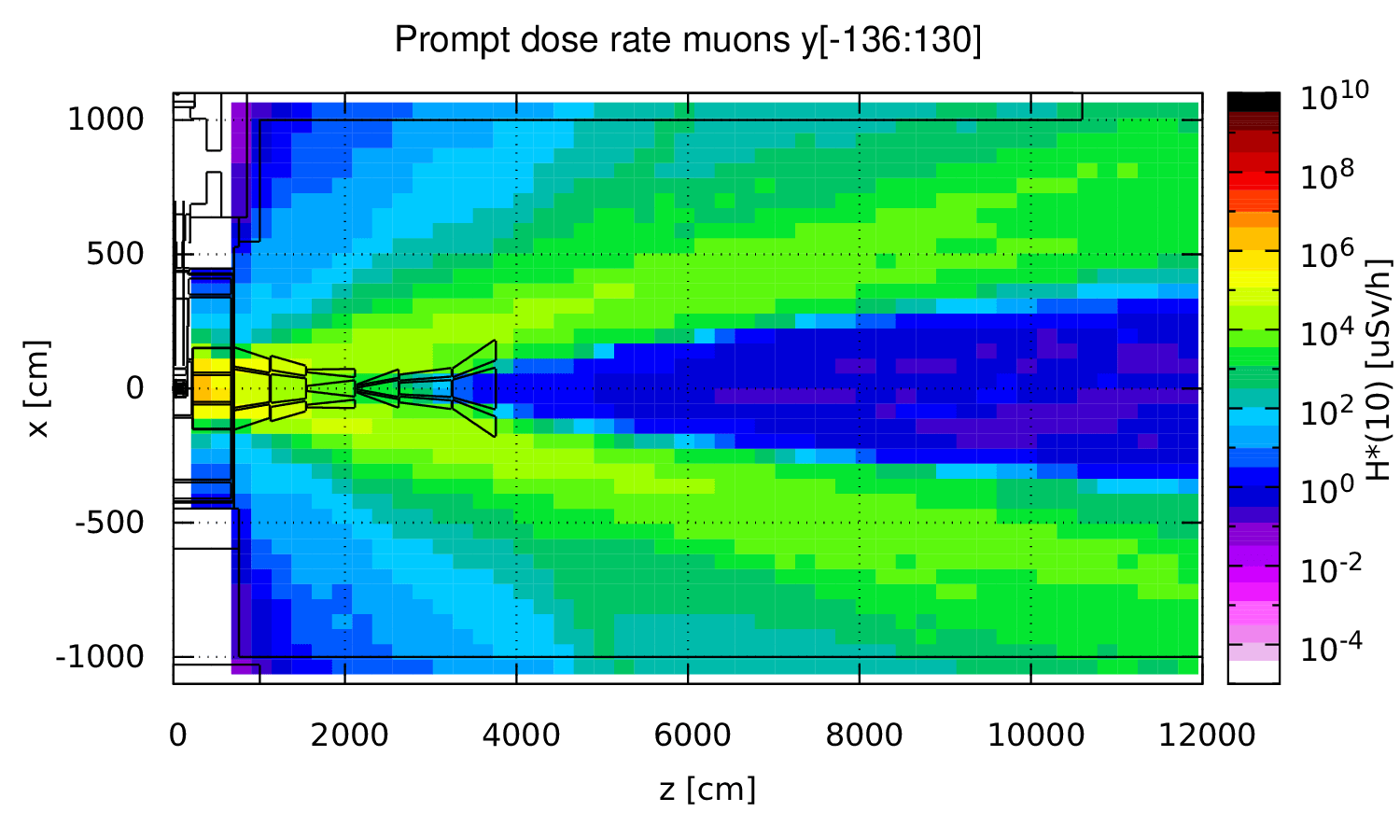}
  \caption{}
  \label{fig:MF1mu}
\end{subfigure}
\end{center}

\captionsetup{width=0.85\textwidth} \caption{\small Top view of prompt dose rates in $\mu$Sv/h in the experimental cavern for all particles (a), only neutrons (b) and only muons (c) .}
\label{fig:MFTop}
\end{figure}

\begin{figure}[!htb]
  \centering
\begin{subfigure}{0.75\textwidth}
  \centering
  \includegraphics[width=0.9\textwidth]{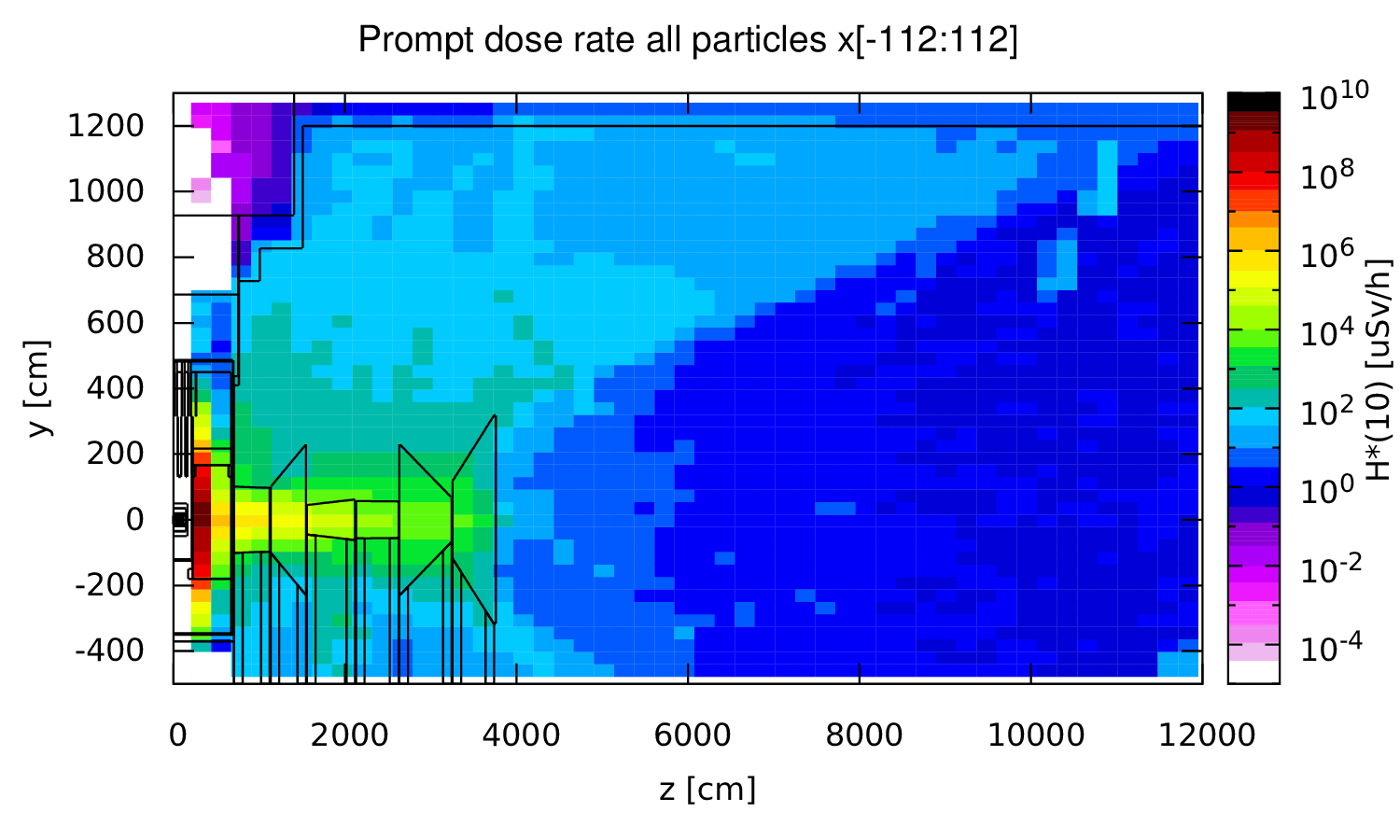}
  \caption{}
  \label{fig:MF2}
\end{subfigure}
\begin{subfigure}{0.7\textwidth}
  \centering
  \includegraphics[width=0.9\textwidth]{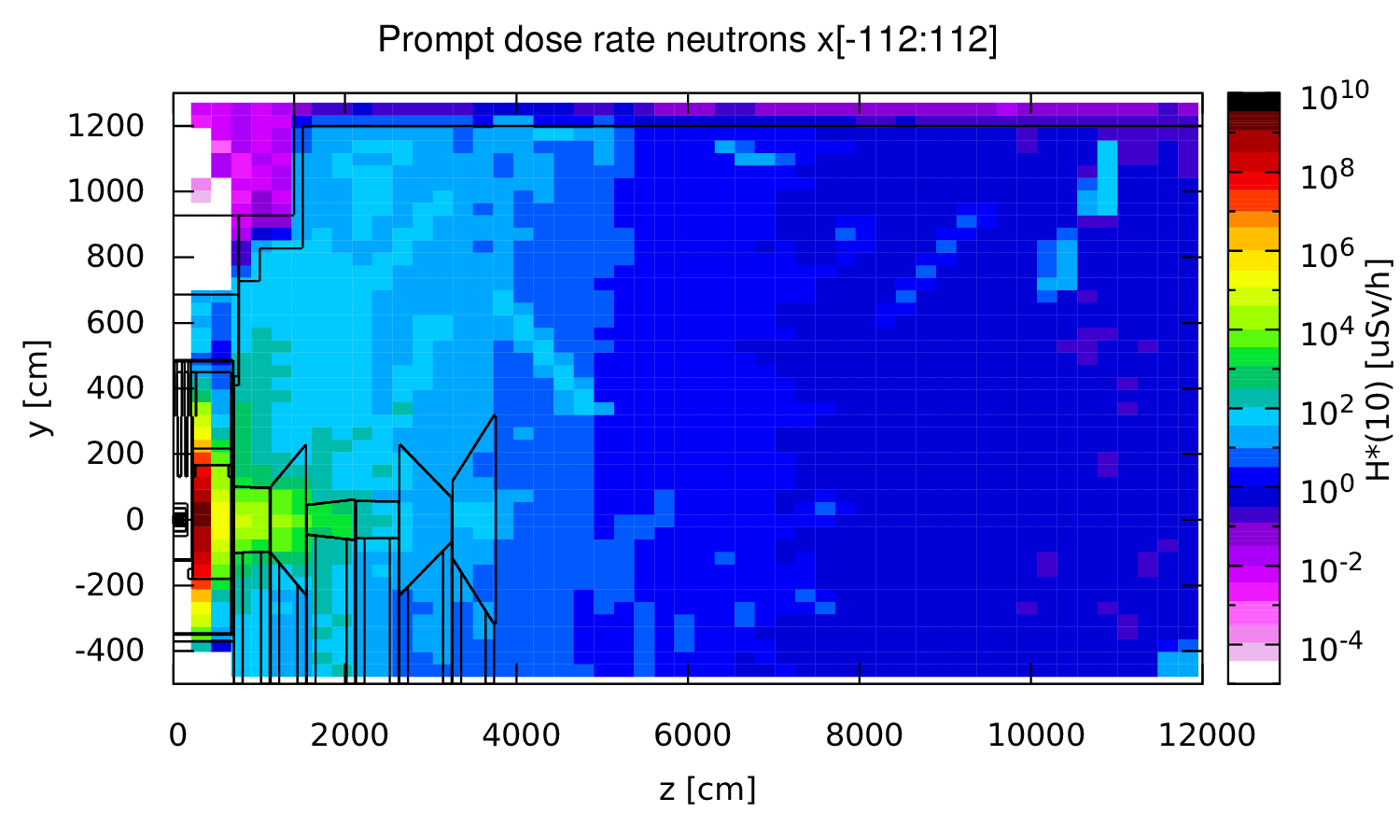}
  \caption{}
  \label{fig:MF2n}
\end{subfigure}
  \begin{center}
\begin{subfigure}{0.7\textwidth}
  \centering
  \includegraphics[width=0.9\textwidth]{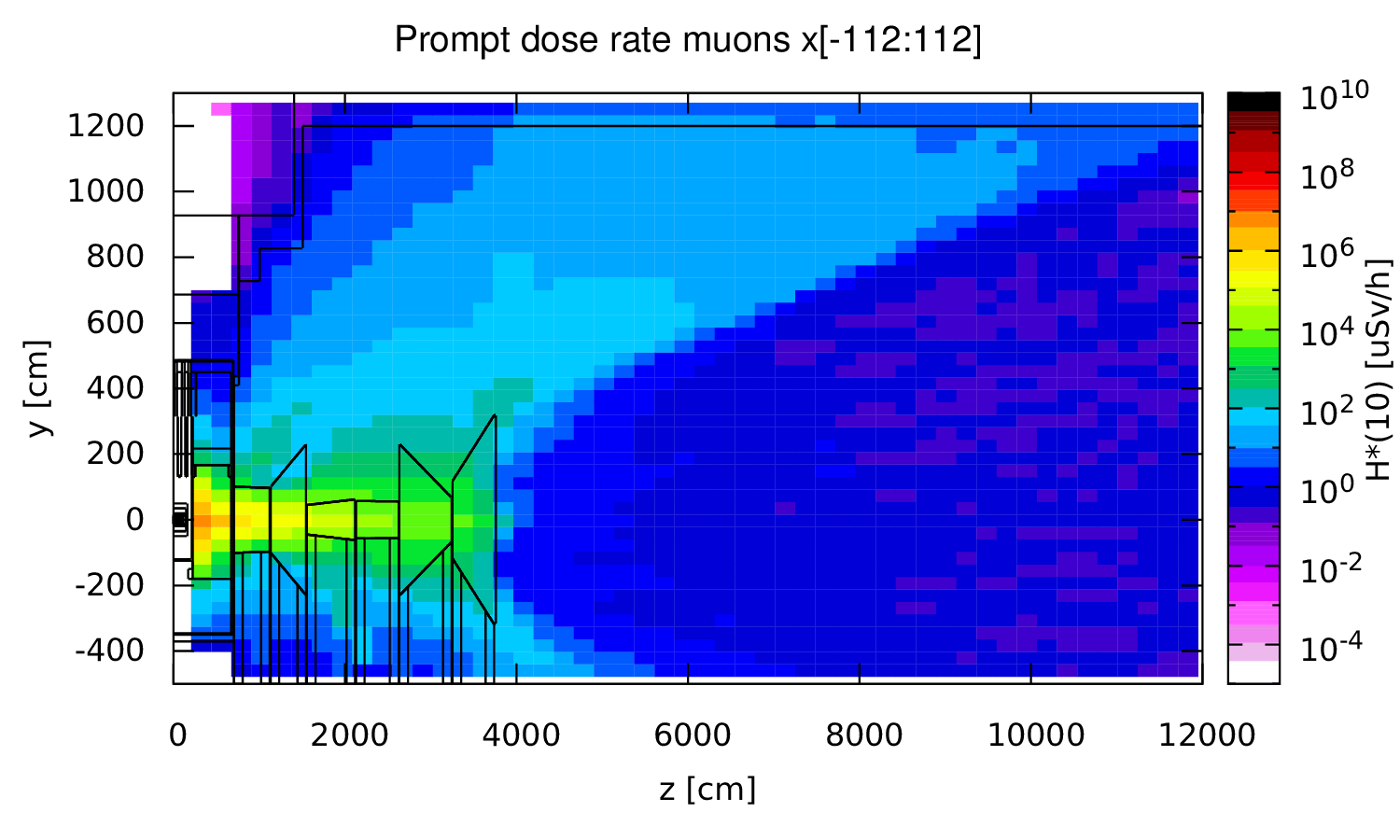}
  \caption{}
  \label{fig:MF2mu}
\end{subfigure}
\end{center}

\captionsetup{width=0.85\textwidth} \caption{\small Lateral view of prompt dose rates in $\mu$Sv/h in the experimental cavern for all particles (a), only neutrons (b) and only muons (c).}
\label{fig:MFLat}
\end{figure}

Similar to the target complex, no access during beam operation will be permitted to the underground experimental hall. Furthermore, no access during operation is required to the above-ground access building to the experimental hall (surface hall building). The prompt dose rates were investigated to illustrate the effectiveness of the active muon shield and to provide information for a further risk analysis.
Figure \ref{fig:MFTop} and Figure \ref{fig:MFLat} present different views of the prompt dose rate distributions in the underground experimental hall from all particles, muons and neutrons. They demonstrate that the muons are swept away from the beam line by the active muon shield and keep their direction due to their small large-angle scattering behind the muon shield, while the neutrons show a relatively direction-independent shape. The dose rates reach a few mSv/h on the side of the experimental hall behind the muon shield and drop below 1 mSv/h in the surrounding soil. The level of soil activation is considered acceptable, particularly due to the fact that the dose rates are dominated by muons. The side view of the experimental hall illustrates that the muons are also bent towards the top of the experimental hall. Thanks to the 1 m concrete shielding in between the experimental and the surface hall, only a few $\mu$Sv/h are reached in the surface hall. The latter could principally be classified as Supervised Radiation Area, however since there is no need for personnel in this area during beam operation, no permanent radiation monitoring is foreseen for this area.

\subsection{Muon flux in surrounding areas}

\begin{figure}[!htb]
    \centering
    \includegraphics[width=0.9\textwidth]{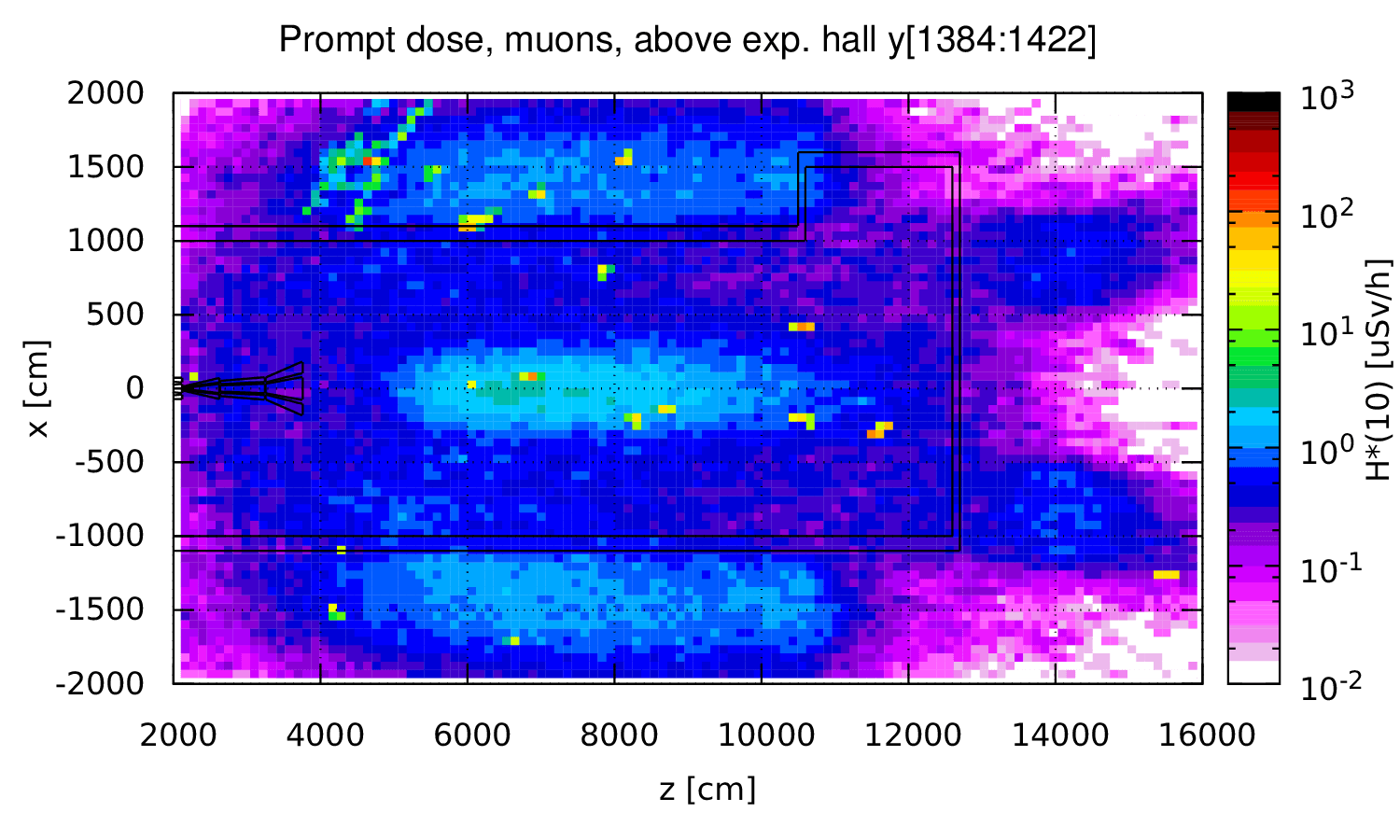}

    \captionsetup{width=0.85\textwidth} \caption{\small Muon flux in $\mu$Sv/h at ground level of the surface hall.}
    \label{fig:MFsurf}
\end{figure}

The prompt radiation at the ground level above the underground experimental hall was analysed in order to define the dose rates next to the surface hall building, which covers only the first 100 m of the underground experimental hall. Figure \ref{fig:MFsurf} presents the above-ground prompt dose rates in the area of the experimental hall. It shows that the highest dose rates are reached behind the surface hall building amounting to a few $\mu$Sv/h until approximately 30 m behind and 5 m next to the experimental hall. This area will be fenced off and covered with 6 m of soil from the excavations such that the dose rates would be further reduced down to a level allowing for a non-designated area (<0.5 $\mu$Sv/h).

\begin{figure}[!htb]
\begin{subfigure}{0.5\textwidth}
  \centering
  \includegraphics[width=0.9\textwidth]{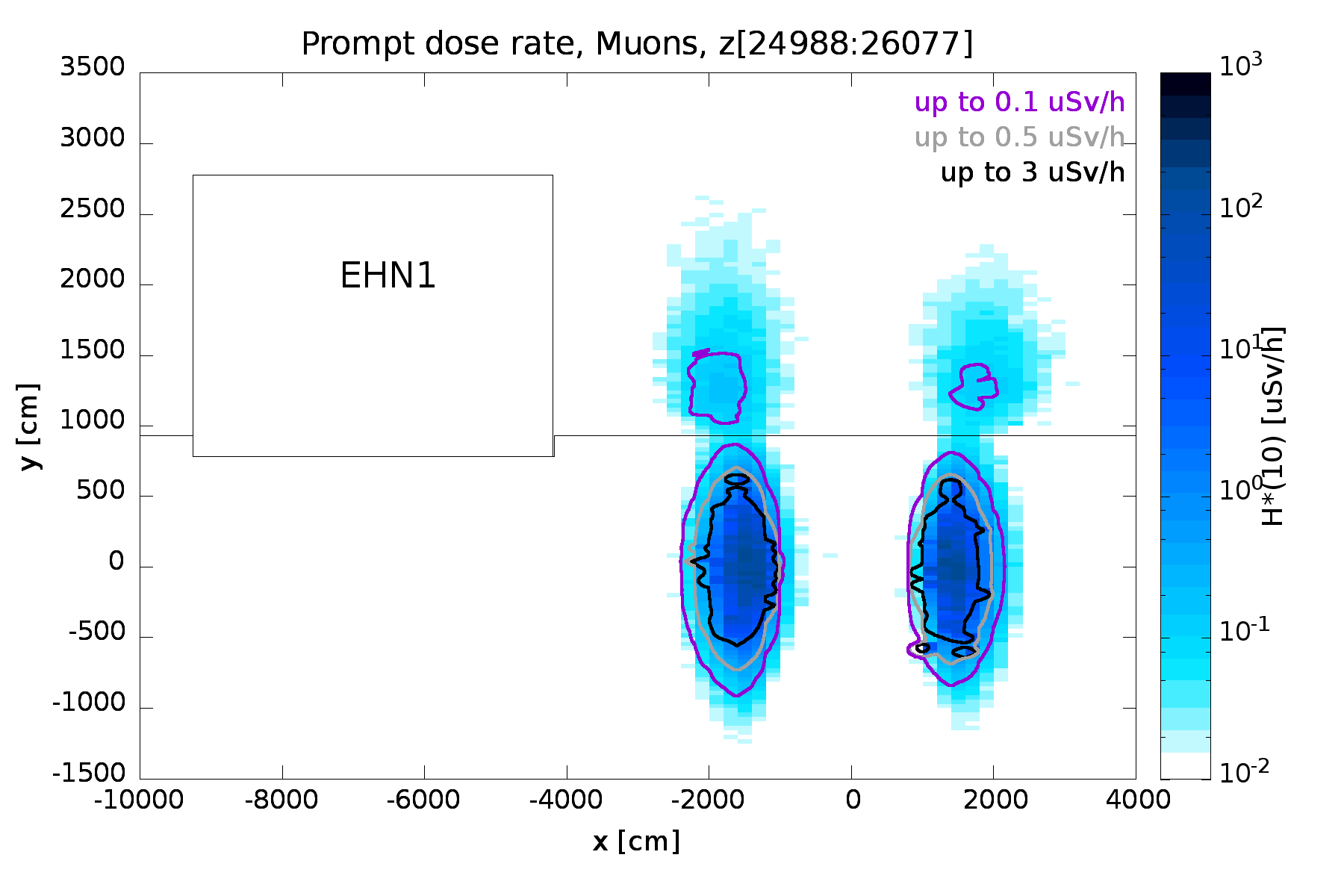}
  \caption{}
  \label{fig:MF4}
\end{subfigure}
\begin{subfigure}{0.5\textwidth}
  \centering
  \includegraphics[width=0.9\textwidth]{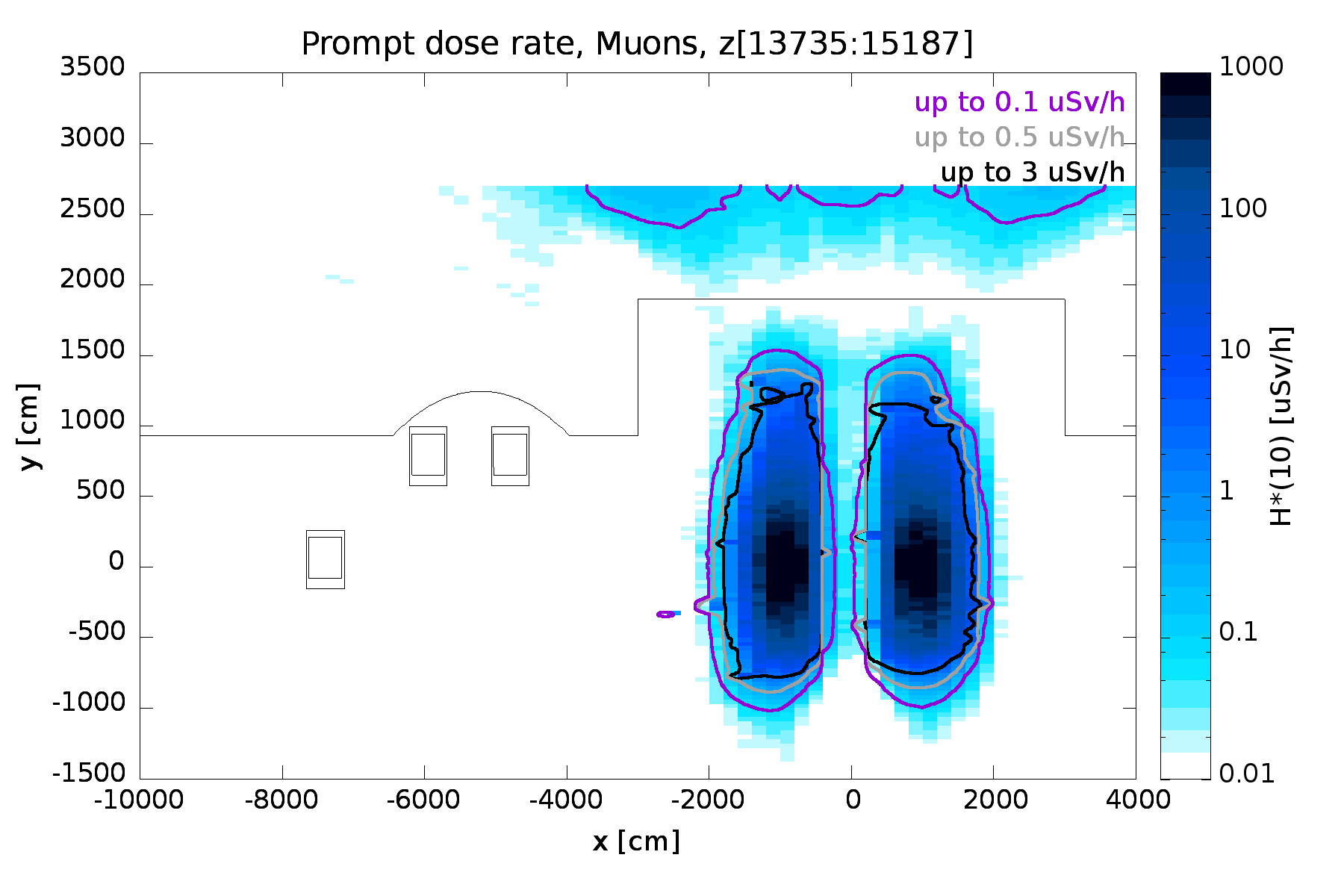}
  \caption{}
  \label{fig:MF4t}
\end{subfigure}

\captionsetup{width=0.85\textwidth} \caption{\small Muon flux in $\mu$Sv/h at the proximity of EHN1 (a) and TT81, TT82 and TT83 tunnels (b).}
\label{fig:MFTun}
\end{figure}

Figure \ref{fig:MFTun} shows the expected prompt dose rates in the ground and experimental facilities surrounding the SHiP facility. It demonstrates that the existing beam lines TT81, TT82 and TT83 are not affected by the prompt dose rates originating from the SHiP facility. To be noted, that any excavation of soil downstream of the underground experimental hall is forbidden.

\begin{figure}[!htb]
    \centering
    \includegraphics[width=0.9\textwidth]{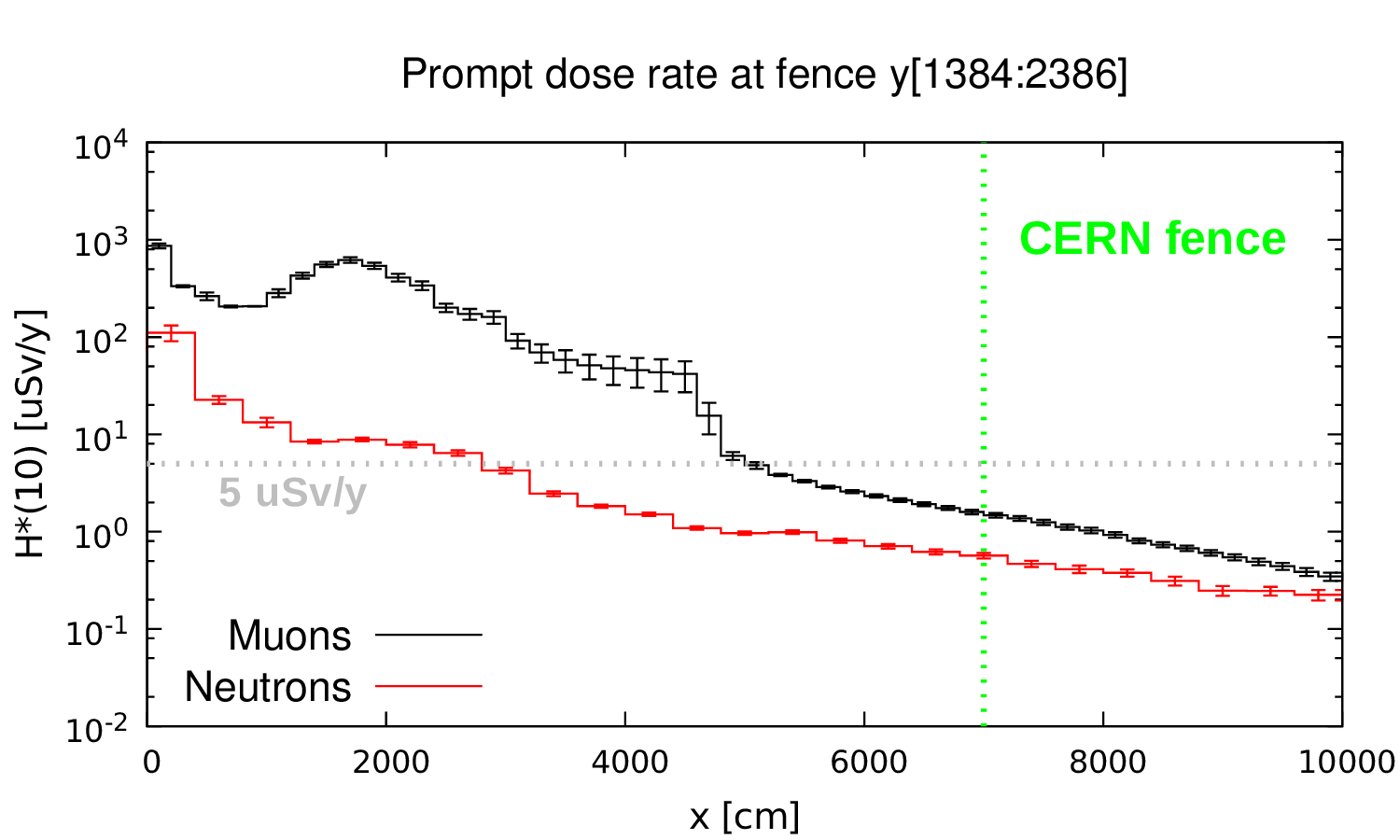}

    \captionsetup{width=0.85\textwidth} \caption{\small Muons and neutrons prompt dose rates in $\mu$Sv/y at fence. The histograms were averaged on different ranges in z ( z[-59:24625] for the muons and z[-59:11557] for the neutrons) in order to catch the respective maximum.}
    \label{fig:fence}
\end{figure}

The SHiP operation also does not influence the present area classification of the EHN1 experimental hall, which corresponds to a permanently occupied Supervised Radiation Area (< 3 $\mu$Sv/h). One should bear in mind that the given results are conservative estimates due to the fact that a moraine density 20$\%$ lower than the measured one was assumed. The operation of the SHiP facility – as designed – should therefore not have any impact on its surrounding experimental areas. According to CERN’s radiation protection code F \cite{SCF}, if the total annual effective dose from all CERN facilities to any member of the public remains below 10 $\mu$Sv/y the exposure does not require any justification and facilities are considered as optimised. In SHiP, the effective dose to members of the public is expected to be dominated by stray radiation from muons (see Figure \ref{fig:fence}), which means prompt radiation that still penetrates outside the shielded zones to the environment and beyond the fenced areas of CERN. The prompt radiation outside of the fenced CERN site, thus the publicly accessible area, was envisaged to stay below 5 $\mu$Sv/y. The latter is fulfilled for SHiP as can be seen from Figure \ref{fig:fence}. A standard stray-radiation monitor for photons, muons and neutrons shall be installed at the fence closest to the most exposed area.

\subsection{Residual dose rate}
\begin{figure}[!htb]
  \centering
  \includegraphics[width=0.8\textwidth]{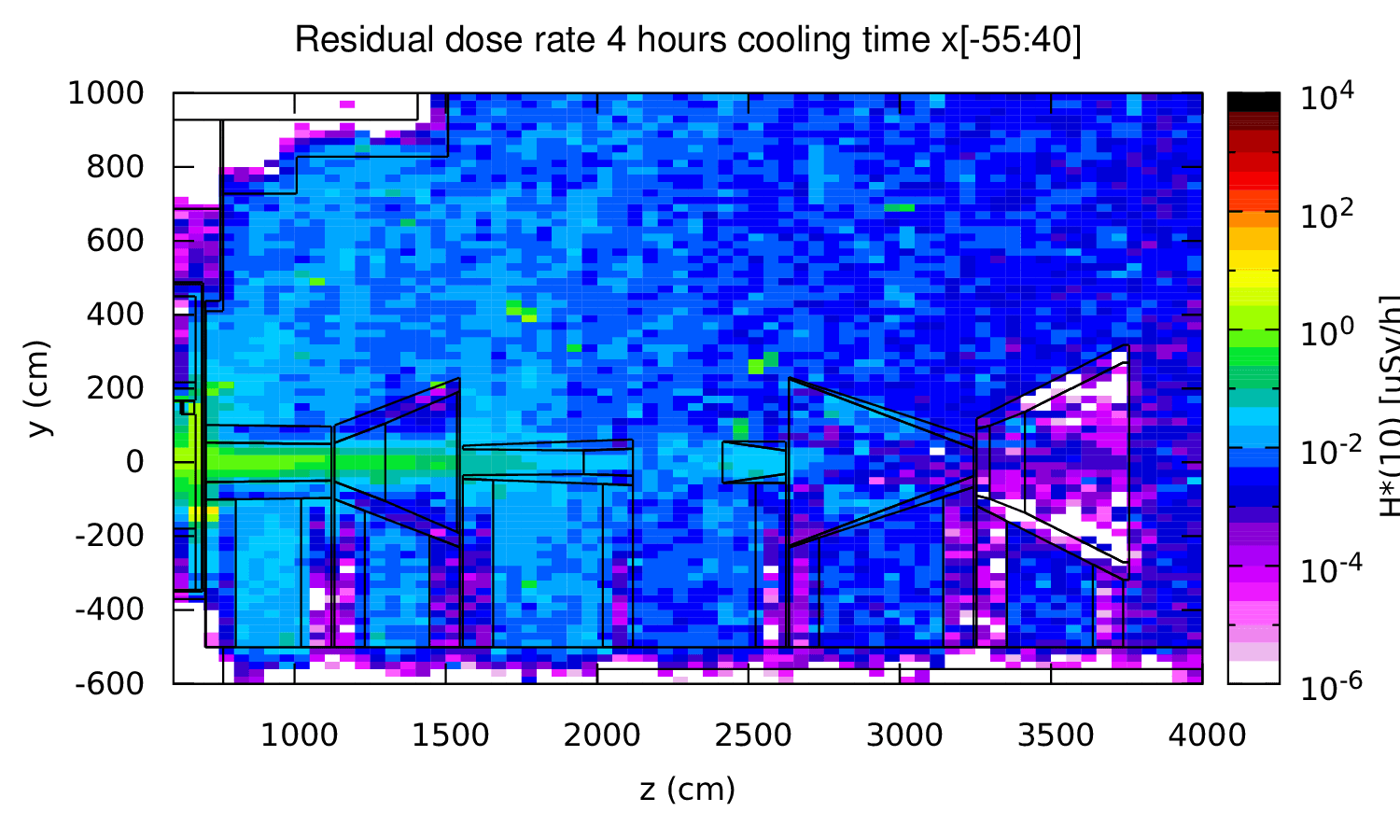}

\captionsetup{width=0.85\textwidth} \caption{\small Residual ambient dose equivalent rate in $\mu$Sv/h for the active muon shield after 5years operation and 4 hours of cooling time.}
\label{fig:rDR1-v2}
\end{figure}

 Assuming five years of operation the residual dose rates in the SHiP target bunker are expected to be at the order of a few $\mu$Sv/h at contact with the first part of the active muon shield after four hours of cooling (see Figure \ref{fig:rDR1-v2}), thus allowing for access to this area. The experimental hall up to the active muon shield will be classified as Simple Controlled while the rest as Supervised Radiation area. 

\section{Tritium out-diffusion}
\label{sec:tritium-out-diff}
 
\begin{figure}[!htb]
 \centering
\includegraphics[width=0.7\textwidth]{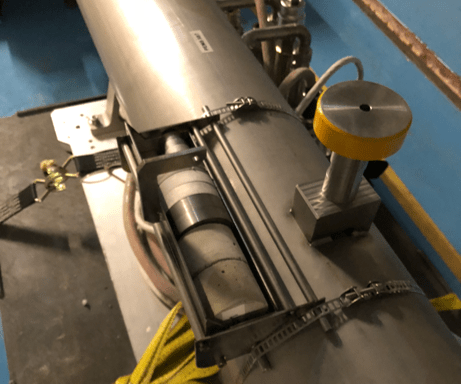}
\captionsetup{width=0.85\textwidth} \caption{\small Location of the samples around the BDF prototype target.}
\label{fig:RPsamples}
\end{figure}

As discussed in Section~\ref{sub:wateractivation} most of the tritium ($\approx 95\%$) produced at BDF will be contained in the target ($\approx 18$~TBq). Due to high volatility of tritium, releases of tritium from the target are expected (out-diffusion) into the target cooling water during operation and into air during shutdowns or after target decommissioning. Tritium can also outgas from the iron and concrete shielding into the helium and air environment. Even if tritium releases have a negligible dosimetric impact they may have negative public-relation impact. Based on available literature, mostly from fusion facilities and nuclear power plants, tritium diffusion coefficients for the BDF materials in the operational temperature range are available with considerable uncertainty. The released fraction of tritium to the water circuit and out-diffusion in air during storage would need an experimental evaluation to assess all safety requirements for the new facility without a consistent over-sizing. For this reason during the BDF prototype target beam test several samples of the materials, that compose the target (TZM, W and Ta) and that will be used in the facility (concrete, iron, stainless steel, aluminium), were placed on the side of the prototype target (see Figure \ref{fig:RPsamples}) to be irradiated under conditions similar to the ones of the future BDF facility.

The dose rate for those samples are shown in Table \ref{tab:tab_RPSamples}. 

\begin{table}[!htb]
\centering
\begin{small}
\begin{tabular}{cc}
\toprule
\textbf{Material} & \textbf{Dose rate @ 40 cm ($\mu$Sv/h)} \\

  \midrule
   Concrete & 2.5\\
  Aluminium & 6.5\\
  Cast Iron & 35\\
  SS316 & 18\\
  TZM & 25\\
  W & 18\\
  Ta & 200\\

\bottomrule
\end{tabular}
\captionsetup{width=0.85\textwidth} \caption{\small Dose rate for the samples used for the tritium out-diffusion measurement. Cooling time was 2 months except for Concrete and Aluminium which was 1 week.}\label{tab:tab_RPSamples}
\end{small}
\vspace{1cm}
\end{table}

\begin{figure}[!htb]
 \centering
\includegraphics[width=0.9\textwidth]{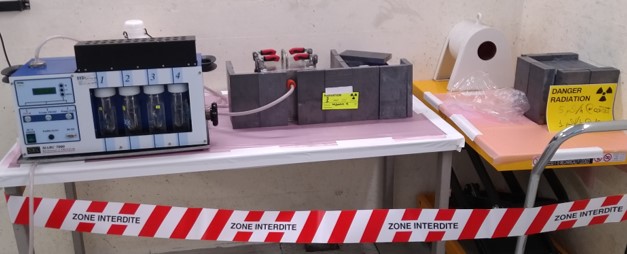}
\captionsetup{width=0.85\textwidth} \caption{\small Bubbler (left), box hosting the sample where the air is recirculated (centre), lead shielding for samples (right).}
\label{fig:bubbler}
\end{figure}

For the measurement of the tritium out-diffusion, the samples are placed in a container connected to a bubbler (MARC 7000 see Figure \ref{fig:bubbler} ), which circulates air in the container. 

The air flow passes through water flasks to capture the tritium releases. The flasks are analysed through liquid scintillation to quantify the release. Another measurement method, which has been employed, is the immersion in water to measure the out-diffusion in a different environment. These measurements will benchmark an out-diffusion plugin for FLUKA and furnish for the first time tritium out-diffusion rates, which can be used for CERN applications.

\FloatBarrier
\printbibliography[heading=subbibliography]

 \chapter{Safety Engineering}
\label{Chap:Safety}

\section{Introduction} \label{Sec:Introduction}

In all of the technical considerations for a facility such as BDF, it is essential to consider safety and environmental protection as an integral part of the comprehensive design, taking conscious decisions early-on to eliminate or significantly reduce potential hazards, in a cost-effective, pragmatic manner.

\section{Legal context of CERN} \label{Sec:Legal}

By virtue of its intergovernmental status, CERN is entitled to adopt its own internal organisational rules, which prevail over national laws to facilitate the execution of its mission. In the absence of specific CERN regulations, the laws and regulations of the Host States generally prevail.

In response to its unique geographical situation (straddling without discontinuity across the Swiss-French border) and its highly specific technical needs, the Organization stipulates its safety policy, in the frame of which it establishes and updates rules aimed to ensure uniform safety conditions across its sites. CERN’s safety rules apply to the Organization's activities, as well as to persons participating in CERN’s activities or present on its site. 

When establishing its own safety rules, CERN takes into account the laws and regulations of the Host States, EU regulations and directives as well as international regulations, standards and directives and, as a general principle, CERN aligns with these as much as reasonably possible. 

Where such compliance is not possible or desirable due to technical or organisational constraints, such as for equipment and facilities not covered by normal standards, specific clearance from CERN’s Occupational Health \& Safety and Environmental Protection Unit (HSE Unit)
based on a risk assessment and compensatory measures is required.

\section{Occupational Health and Safety} \label{Sec:OH&S}

CERN’s Safety Policy, in order of priority, sets out to protect all persons affected by its activities, to limit the impact of the Organization’s activities on the environment, and to protect its equipment and ensure continuity of operations. 
The agreed safety objectives are shown in Table \ref{Tab:Safety:AgreedObjectives}:

\begin{table}[htbp]
\centering
\caption{\label{Tab:Safety:AgreedObjectives} Agreed safety objectives for the BDF project}
\smallskip
\begin{tabularx}{400pt}{|c|X|X|X|X|} 
\hline
\textbf{} & \textbf{A: Life Safety} & \textbf{B: Environmental Protection} & \textbf{C: Property \newline Protection} & \textbf{D: Continuity of Operation}\\
\hline
 1 & Safe evacuation of valid occupants & Limited release of pollutants to air & Continuity of \newline essential services & Limit downtime \\ \hline
 2 & Safe evacuation or staging of injured occupants & Limited release of pollutants to water & Incident shall not cause further \newline incidents &  \\ 
\hline
 3 & Safe intervention of rescue teams &  & Limit property loss &  \\ 
\hline
\end{tabularx}
\end{table}

\section{Hazard identification study for BDF TTC} \label{Sec:HazID}

A specific hazard identification study for the BDF Target and Target Complex is shown in Chapter 7, Section 8.1. The objective of the hazard identification study was to identify as early as possible the principal failure scenarios leading to critical accident situations as well as the consequences and define essential mitigation measures or redesign. 

\section{Fire safety} \label{Sec:Fire}

The goal of fire safety is to protect occupants, rescuers, the external population, the environment, the facility itself, and continuity of operation. To this end, all buildings, experimental facilities, equipment and experiments installed at CERN shall comply with CERN Safety Code E. In view of the special nature of the use of certain areas, in particular underground, with increased fire risk, the HSE Unit is to be considered the authority of approving and stipulating special provisions.  

As the project moves to the Technical Design Report stage, finalising layouts and interconnecting ventilation systems, detailed fire risk assessments will have to be made for all areas of the BDF complex, i.e., Extraction Tunnel, Target Complex and Experimental Hall. At this stage, a general fire safety strategy has been produced, based upon the location and current level of design, along with the latest fire safety strategies employed at CERN. The complex will be considered as an extension of the fire concept developed for the whole of the North Area \cite{NorthArea_FireRA}. 

The most efficient protection strategy is one that uses multi-level “safety barriers”, with a bottom-up structure, to limit fires at the earliest stages with the lowest consequences, thus considerably limiting the probability and impact of the largest events. 

In order to ensure that large adverse events are possible only in very unlikely cases of failure of many barriers, measures at every possible level of functional design need to be implemented: 

\begin{itemize}
\item in the conception of every piece of equipment (e.g. materials used in electrical components, circuit breakers, etc.); 
\item in the grouping of equipment in racks or boxes (e.g. generous cooling of racks, use of fire-retardant cables, and fire detection with power cut-off within each rack, etc.); 
\item in the creation and organisation of internal rooms (e.g., fire detection, power cut-off and fire suppression inside a room with equipment); 
\item in the definition of fire compartments; 
\item in the definition of firefighting measures. 
\end{itemize}

The key fire safety strategy concepts can be split into compartmentalisation, fire detection, smoke extraction and fire suppression, as set out below.

\subsection{Compartmentalisation}

Compartmentalisation impedes the propagation of fire and potentially activated smoke through a facility, allowing occupants to escape to a comparatively safe area much more quickly than otherwise, as well as facilitating the effective fighting of the fire, and evacuation of victims by the Fire Brigade. In the North Area fire concept, the following requirements have been set:

\begin{itemize}
\item All ventilation doors must be fire doors EI90;
\item Isolate communicating galleries with fire doors EI90; 
\item Isolate neighbouring surface facilities with fire doors EI120;
\item Avoid compartments longer than 450m;
\item Normally opened fire doors to be equipped with remote action release mechanism, monitoring position and self-action thermal fuse. 
\end{itemize}

\subsubsection{Extraction Tunnel}

\begin{figure}[htbp]
\centering %
\includegraphics[width=.87\linewidth]{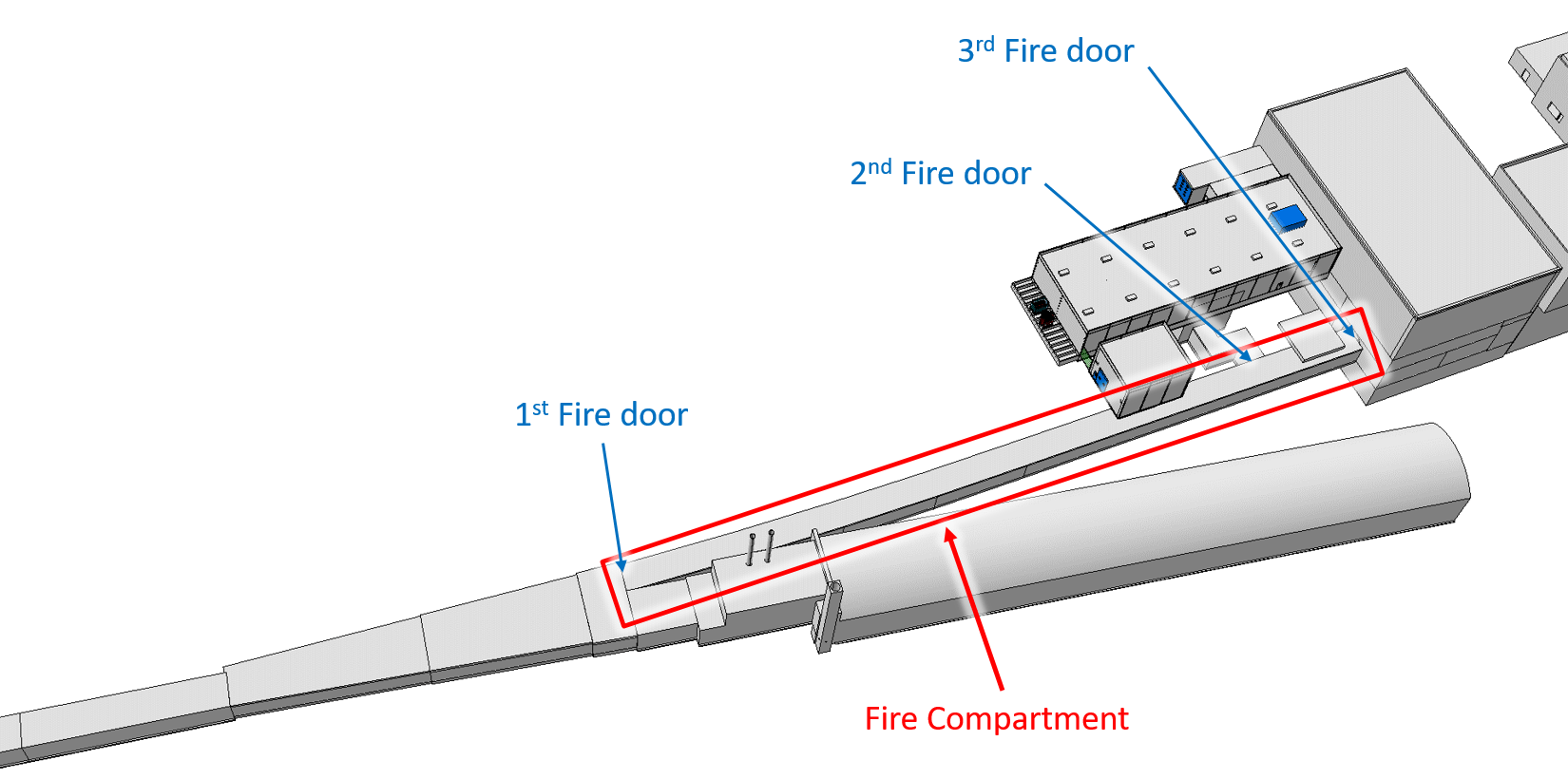}
\caption{Compartmentalisation in the BDF Extraction Tunnel}
\label{fig:extraction_tunnel}
\end{figure}

The new Extraction Tunnel shall be considered as one single fire compartment, requiring 3 fire doors (as shown in Figure \ref{fig:extraction_tunnel}). 

The fire doors will follow the standard requirements for the North Area Fire Concept: 0.9 m minimum width, with fire doors EI90 or greater. They will be set to a normally-closed configuration, due to the ventilation specification and needs of the Extraction Tunnel. The door separating the Extraction Tunnel from the SPS shall be EI120.

\subsubsection{Extraction Tunnel}
\begin{figure}[htbp]
\centering %
\includegraphics[width=.87\linewidth]{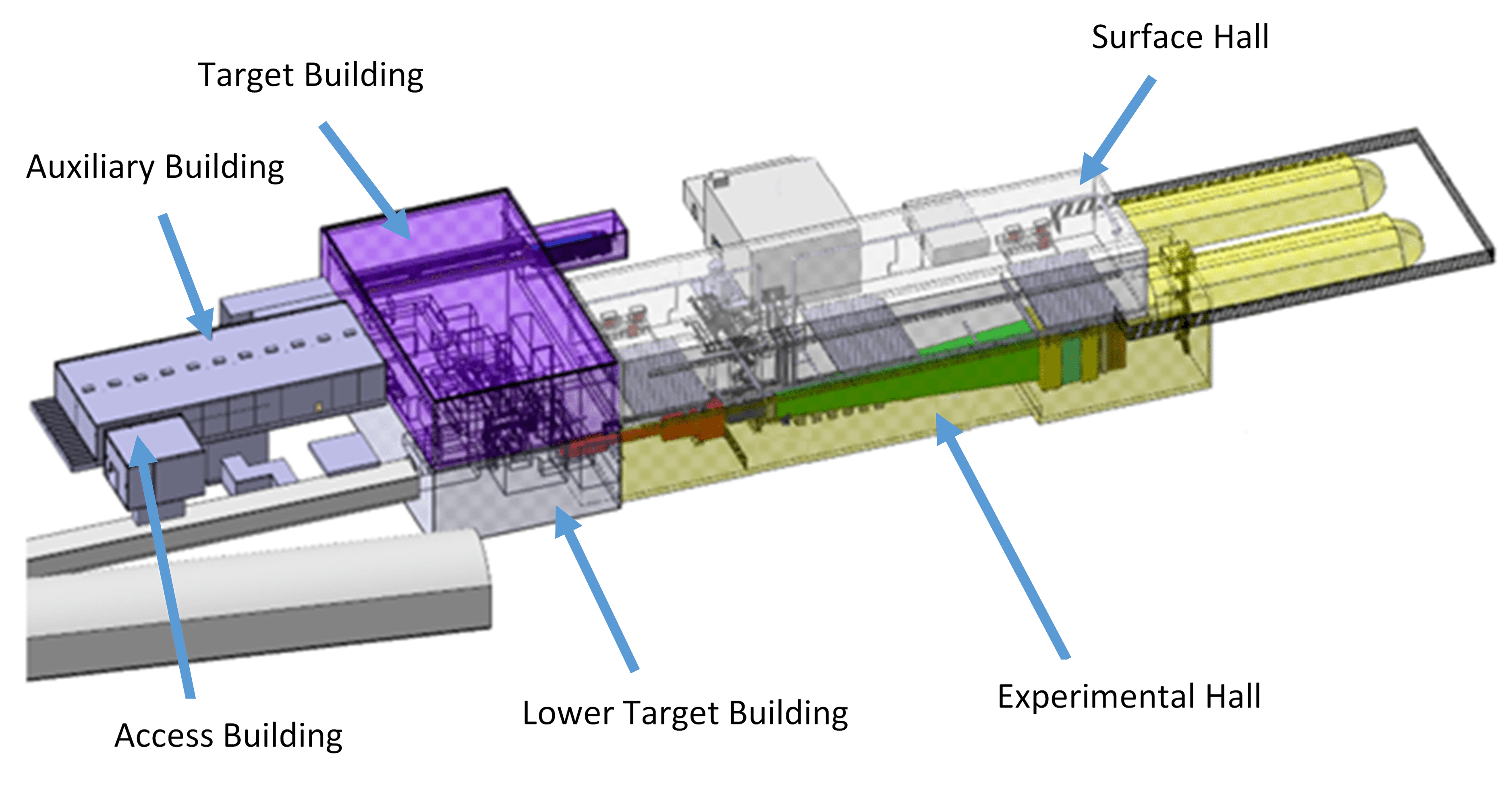}
\caption{Compartmentalisation areas of the BDF complex}
\label{fig:BDF_complex}
\end{figure}

The following areas shall be divided into separate fire compartments (shown in Figure \ref{fig:BDF_complex}):

\begin{itemize}
\item Auxiliary Building;
\item Access Building;
\item Target Building;
\item Lower Target Building;
\item Surface Hall;
\item Experimental Hall.
\end{itemize}

It is to be noted that both the Experimental Hall and the Lower Target Building shall be separate compartments from the corresponding Surface Hall and Target Building, accounting for the differing uses of the spaces. The Lower Target Building shall also be divided into several separate fire compartments, corresponding to the pressure cascades implemented for radiation containment.

\subsection{Fire Detection}

An early fire detection system, integrated into the safety action system is a crucial component of the North Area fire strategy. Early detection is such that allows evacuation (last occupant out) before untenable conditions are reached; the CERN HSE Fire Safety team shall be consulted for this design. The system must be capable of transmitting an alarm, along with a message containing safety instructions, to occupants anywhere in the North Area. This alarm shall be triggered upon fire detection, action on evacuation push buttons, CERN Fire Brigade action out of CERN FB SCR/CCC or BIW (Beam Imminent Warning) situations. Evacuation push buttons shall cover all premises. The fire detection and evacuation push buttons must also be integrated with safety actions such as compartmentalisation, ventilation stop and other machine functions according to a predefined fire protection logic.
Fire detection encompasses the following actions: 

\begin{itemize}
\item Close all fire doors in the fire compartment of origin and the adjacent fire compartments;
\item Trigger an evacuation alarm in the fire compartment of origin and the adjacent fire compartments;
\item Stop free-cooling ventilation; 
\item Broadcast an information message through the public address system; 
\item Trigger a level 3 alarm in the CSAM system, which alerts the CERN Fire Brigade control room, and results in crews being dispatched immediately.
\end{itemize}

\subsection{Smoke Extraction}

One of the key findings from the North Area Fire Risk Assessment was the need for careful risk assessment of the effect of smoke in the event of a fire in the underground and surface areas, taking into account both the safe evacuation of occupants, and the effective intervention of the fire brigade to locate victims and prevent the further spread of a fire. A buildup of smoke can also result in lasting damage to the sensitive and valuable equipment present, an effect that can be limited through extraction. 

In accordance with CERN Safety Guideline SG-FS-0-0-2, premises with a surface area exceeding 2000 m² or with a length exceeding 60 m shall be divided in smoke zones with a maximum surface area of 1600 m² and a maximum length of 60 m. Smoke curtains between smoke zones shall be made out of a noncombustible material (B-s3, d0) with a fire resistance of at least ¼ hour. The height of roof screens shall be 25\% of the average height building when this height is less than 8 m, or 2 m when the height building is more than 8 m. Depending on the specific requirements of the building design, these can be retractable, with automatic release on fire detection.

An additional consideration for fires in accelerator tunnels is the danger of potentially activated smoke, and the need to handle this in a controlled manner to limit the release of polluting agents to the environment. A fire assessment methodology that entails the radiological hazard induced by a fire event is currently under development by the FIRIA Project led by HSE. Many of the concepts featured in the FIRIA methodology have been incorporated in the safety support provided by HSE to the BDF project. However, once the FIRIA Methodology is fully available in July 2021, it is recommended to consider carrying out a FIRIA exercise as part of the Technical Design Report phase.
A description of the FIRIA project is available in \cite{FIRIA_Kickoff}.

\subsubsection{Extraction Tunnel}

The extraction tunnel shall follow the strategy set out for the North Area, namely through containment via both the automatic closing of any “Normally Open” fire doors, and the stopping of ventilation on fire detection. Potentially activated smoke shall therefore remain in the fire compartment for assessment by the Radiation Protection (HSE-RP) group, who can then make a decision on the release delay, extraction flowrate, and smoke filtering / scrubbing options to minimise the public dose, using a portable system of ducts and smoke extraction devices to remove the smoke as needed.

\subsubsection{Auxiliary Building}
The Auxiliary Building is generally unclassified, with just two smaller radiation classified zones, and controlled access to the Extraction Tunnel. The surface footprint is less than 2000 m², and no side is longer than 60 m. Natural smoke extraction is therefore appropriate, through the use of Sky Domes.

\subsubsection{Lower Target Building}
The lower target building will contain radioactive areas, with cascaded pressures to prevent the escape of activated material to areas with lower radiation classifications. In the absence of prescriptive codes for the design of the smoke extraction strategies for such areas the detailed design shall be \textit{performance based}, and made in conjunction with the HSE Fire Safety team. Mechanical extraction is an intrinsic part of the design for this system, and the fire safety strategy will govern the logic behind the automatic actions on detection of a fire in a particular area. Where the air temperature in the compartment of origin of the fire is below 300°C (and so will not damage the extraction ducts), the strategy will be to immediately turn off the air supply to the compartment, while maintaining the air extraction. The strategy is to starve the fire of oxygen, choking it before significant quantities of smoke can be produced. In this way, the negative impacts of the fire can be minimised. Where the air temperature has already exceeded 300°C, both the air supply and air extraction shall be turned off, and the containment actions shall be extended to the surrounding fire compartments. For the Technical Design Report stage, further studies will be required in this area, in line with the FIRIA methodology. 

\subsubsection{Target Building}
The Target Building shall be a separate fire compartment from the Lower Target Building, but will not be sealed against smoke. A mechanical smoke extraction strategy shall therefore be employed, in tandem with a central Smoke Curtain, which will divide the hall into two areas of less than 1600 m², ensuring compliance with the CERN Safety Guideline. 

\subsubsection{Experimental Hall}
The Experimental Hall shall use mechanical smoke extraction, with four Smoke Curtains dividing the hall into five sectors of less than 1600 m². As the Experimental Hall will be a Radiologically Classified Area, the strategy will be to contain any potentially activated smoke for assessment by the HSE-RP group, who can then make a decision on the release delay, extraction flowrate, and smoke filtering / scrubbing options to minimise the public dose. The initial design details of this system are discussed in Chapter 9, and shall be fine-tuned using Computational Fluid Dynamics simulations and performance based design methodology as the project moves to the Technical Design Report stage.

\subsubsection{Surface Hall}
The Surface Hall shall be a separate fire compartment from the Experimental Hall below, and will have a lower radiological classification. Natural smoke extraction strategy is therefore currently foreseen, using Sky Domes, in tandem with Smoke Curtains. The arrangement of Smoke Curtains shall be the same as that employed for the Experimental Hall, dividing the hall into five sectors of less than 1600 m². 

\subsection{Fire Suppression}
 The CERN Fire Brigade need adequate means of fighting a fire on arrival, including a surface hydrant network, which shall be foreseen as the project moves to the Technical Design Report stage, in tandem with the HSE Fire Safety specialists.
 
 \subsubsection{Extraction Tunnel, Lower Target Building, Experimental Hall}
Dry Risers shall be installed throughout the underground areas, allowing the CERN Fire Brigade to choose to supply water to them for firefighting as required. The additional fire loads and radiation risks present in the Lower Target Building require the addition of a suppression system, to be specified in detail as the project moves into the Technical Design Report phase. A number of fire suppression solutions for such challenging environments are being considered at CERN, such as the delivery of fire suppression by remotely operated drones, which may be applied in this area.

\subsection{Access Safety}
For the underground access, it should be ensured that:
\begin{itemize}
\item The lift and stairs are protected against fire, not connected to the general electrical circuit (i.e., can be used at any time);
\item A safe area, with an overpressure relative to the surroundings is available at the base of the lift (or other vertical egress path). The size of this area shall be commensurate with the number of occupants, in addition to the time taken for evacuation, and shall be determined as the project moves to the Technical Design Report stage.
\end{itemize}

\begin{figure}[htbp]
\centering %
\includegraphics[width=.87\linewidth]{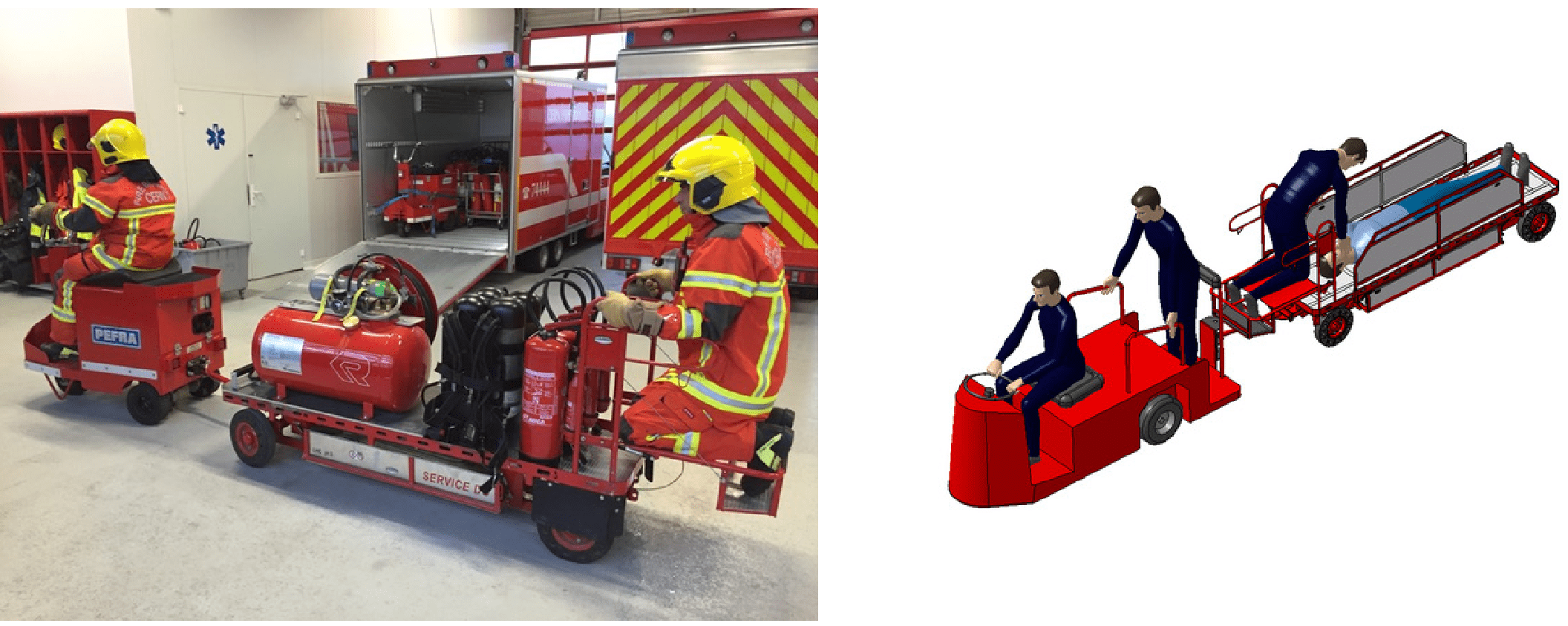}
\caption{The CERN Fire Brigade "PEFRA" vehicle}
\label{fig:PEFRA}
\end{figure}

For the extraction tunnel, it is additionally important that fire equipment and Fire Brigade vehicle (for example, the “PEFRA”, shown in Figure \ref{fig:PEFRA}), can freely move in and out the lift and pass through the "chicane" immediately before the lift and stairs without any problem, especially for a rescue operation.

\section{Safety of Civil Structures} \label{Sec:Civil}
The installation of the BDF complex will involve a significant amount of civil works, including tunnel modifications, numerous new buildings and a challenging target complex. At CERN, all structures are to be designed and manufactured according to the Eurocodes, especially accounting for the local seismic action. Due to the estimated levels of radiation foreseen across the facility, along with environmental considerations, the control of water ingress and egress is of particular interest. It is important to note that this is not an exhaustive list of requirements, and that further analyses will be required as the project moves into the Technical Design Report stage.

\subsection{Drainage}
The objective of the design of underground drainage systems in areas subject to HSE-RP control or chemical contamination is to ensure any water at risk of activation or contamination is collected and controlled before it can pass into the external drainage networks.  

A preliminary study  of the ground water of the Pr\'evessin site was performed during the comprehensive design study phase in order to tackle the risk of radionuclides transfer in the groundwater \cite{GADZ}. The aim of this study was to provide information on the present hydrological situation by means of collecting historical data and by performing new water level measurements in the concerned area, since the most recent previous data was taken in 2014.

\begin{figure}[htbp]
\centering %
\includegraphics[width=.87\linewidth]{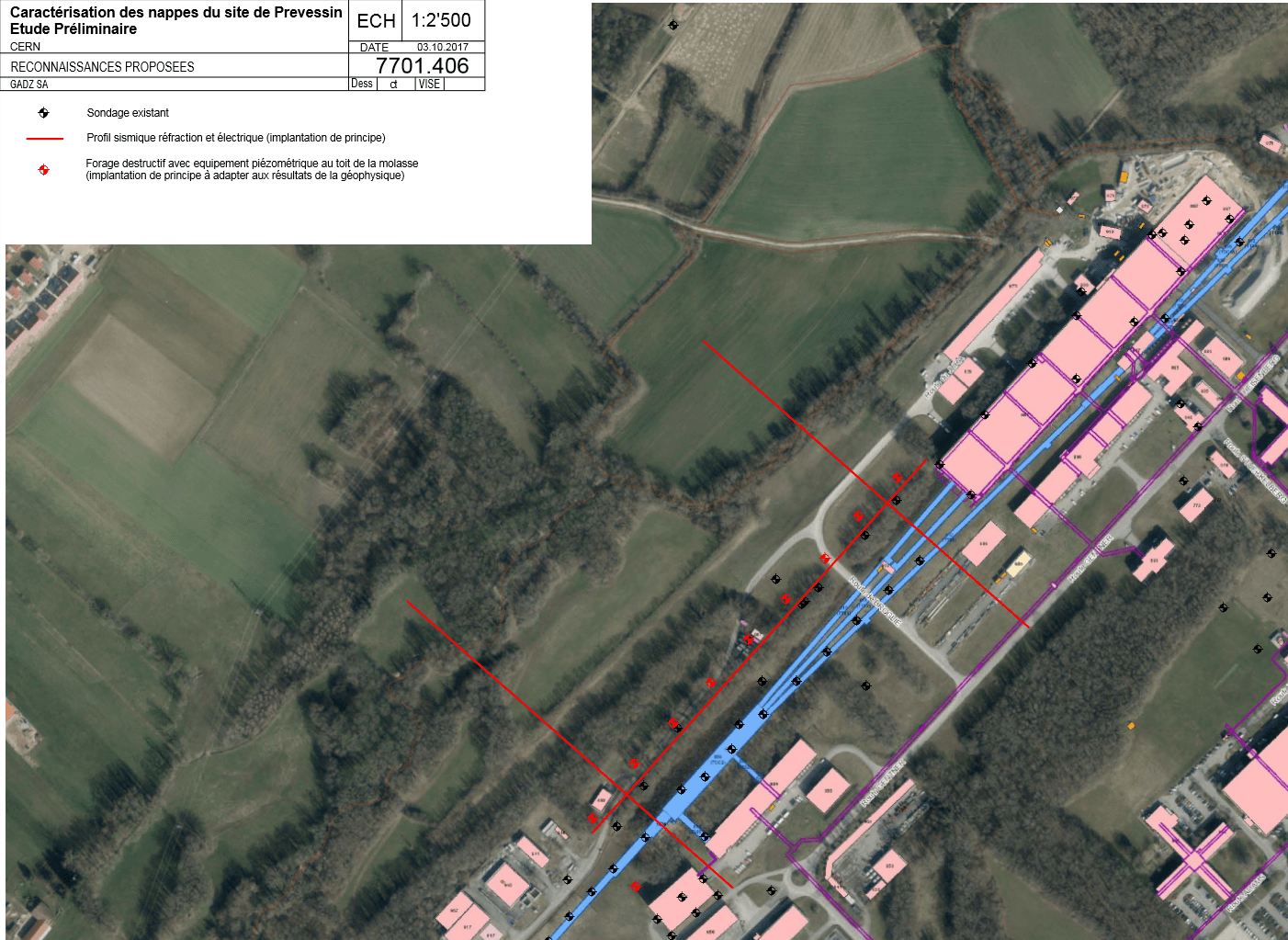}
\caption{Key area of interest identified in the hydrological and geotechnical studies - North Area of Pr\'evessin site}
\label{fig:HydrogeologicalStudy}
\end{figure}

The outcome of the study is that further investigations by CERN should focus on an area 400 m to the south west of EHN1, as shown in Figure \ref{fig:HydrogeologicalStudy}. Whilst the shallow water lenses identified are not thought to represent a risk from an environmental point of view, a ~30 m deep aquifer may require further consideration. Water from the surface could potentially reach the deep aquifer if the water infiltrated the ground locally or if the existing drainage was not perfect (i.e. tight). Most probably, this deep aquifer flows in a north westerly direction towards the \textit{Sillon de Saint-Genis} and joins the \textit{Allondon} river. The study has therefore highlighted an existing area of consideration for CERN, separate to the BDF project.  It can be assumed that the lower aquifer exists and that the current TT20 tunnel links the level of the BDF to the aquifer. The integrity of the drainage in the TT20 is potentially compromised by construction quality issues and ground movement. The tunnel could therefore in principle transmit water from the upper level to the lower aquifer either via the central drain or via the peripheral drainage of the tunnel itself. The BDF project shall therefore ensure that no water from the facility drains into the TT20 tunnel, employing the strategies outlined below.

\noindent The source of water entering the drainage network can be considered to originate from:

\begin{itemize}
    \item Ground water – subsurface water occupying the saturation zone, from which wells, springs, or streams are fed;
    \item Machine originated water – water from chilled water supply leaks, machine cooling water leaks or water escaping from other pipework such as dry risers;
    \item Infiltration water (ground water, plus rain infiltration) – usually enters the underground area through failed or failing waterproofing membranes and cracks in the lining at construction joints or shrinkage cracks. In some areas with a high water table and hydrostatic head, water may enter though the floor or surcharging of the main drains;
    \item Fire suppression water – water originating from the fire extinguishing system in the Lower Target Building.
\end{itemize}

It is assumed that ground water is unlikely to be activated, whereas water originating from the underground building or tunnel has a higher risk of activation and water contained in closed circuit cooling is highly likely to be activated.
The design approach for each source of water shall be as follows:

Ground water should not be allowed to enter the underground area directly. A free-flowing drainage layer should be provided to allow water to drain away from the underground area to a central drain pipe which carries the water to a sump and pump station for evacuation to the surface and appropriate handling. The central drain should be sealed from the underground area to prevent water ingress into it. Water tight covers every 50 m must be provided to allow access for inspection and cleaning of the network.  Any collector drains (usually slotted land drains) must also have rodding eyes accessible from the inside of the underground area. Any change in section size or direction must also have an access chamber to allow for inspection and cleaning. The waterproofing and drainage systems should always be detailed to minimise the risk of water leaving the drainage network and finding its way back into the surrounding water table. Syphons and areas of standing water should be avoided and alternative details found.

Machine originated water is mainly from closed circuit cooling systems, and has the potential to be highly activated (including the danger of contamination with Tritiated water). This water must be collected at its source and a drip tray created to collect any leaking water from the installation. A drain tap must be provided to channel the water into the storage containers. The potentially contaminated water must then be stored locally in containers for verification by HSE-RP prior to being transported to the surface for disposal. 

Both infiltration water and fire suppression water can be considered as at risk of activation as they may collect near radiation sources and stand for periods of time leading to activation. For this reason, a second drain dedicated to water from the interior of the underground area must be provided, from which it should be pumped to the surface for testing. The capacity of this network should be sized to cope with these loads, based on analysis of the flows involved, as the project moves to the Technical Design Report stage. Water from the network must not reach the TT20 tunnels, in which leaks paths are known to exist. There should additionally be no connection between this network and the drain taking the ground water. Constriction joints should be detailed to provide either a self-sealing gasket or a re-sealable injection tube to keep the joint watertight. In areas of known high water table pressures both measures may be needed. The joint should also be designed to allow easy channelling of any water ingress by the installation of a simple gutter to guide the water to the floor and away from the machine. This same approach is equally valid for shrinkage crack inducers.

Water from a tunnel vault must be collected in a channel at the interface of the tunnel vault and floor. This channel should then be connected to the central drain (internal water) via pipes buried in the concrete invert. The design of the channels and interceptor drains must be done so as to minimise the risk of water on the transport path of the tunnel.

\subsection{General Building Structural Safety Aspects}
\subsubsection{Fire Resistance}
New structures and infrastructure shall be designed and executed to guarantee a mechanical resistance for 120 minutes of exposure to the design fire. Eventual passive protection systems, e.g. intumescent paintings and plasters, will be foreseen only for those elements that are unable to respect such a requirement. The structural assessment will need to be carried out in accordance with EN 1991-1-2, EN 1992-1-2 and EN 1993-1-2.

\subsubsection{False Floors}
A false or raised floor is a floor that provides a void or space for the technical installations and maintenance thereof in a building or part of the infrastructure. Such floors have been the cause of a number of incidents or near misses at CERN in recent years, and so incorporating safe design at an early stage will help to avoid potential hazards. 

The objective of the design of false floors is therefore to provide a suitable environment for the foreseen services, allowing access for the installation maintenance and eventual removal of the installed equipment. The services may include, but are not limited to:

\begin{itemize}
    \item Electrical cabling;
    \item Signal cabling;
    \item Heating and ventilation;
    \item Water supply and distribution;
    \item Gas supply and distribution;
    \item Fire detection;
    \item Fire main.
\end{itemize}

A number of considerations must be taken into account when deciding the type of floor to install and the space required to meet the above criteria. Each phase of the floor’s life must also be designed for and must allow for safe installation, utilisation and finally decommissioning of the floor. Key design criteria include:

\begin{itemize}
    \item Minimise the area of floor that is freely removable (this reduces the risks of damage and deterioration of the floor from multiple interventions);
    \item Where possible provide a full height access to the cable trays and pipework;
    \item Identify cable pulling routes for the current layout and where possible future modifications and try to fix smaller access points for pulling the cables;
    \item Identify points of access and egress from the floor for maintenance purposes that have the minimum impact on the normal walkways through the building;
    \item Detail the pulling points and access points such that there is a rigid barrier integrated into the access trap (access via a trap door which in the open position is supported by barriers that protect the opening);
    \item Provide a suitable ladder or steps to access under the floor;
    \item Consult all groups who may need to pull cables to ensure the layout chosen is suitable and if necessary pre-equip the cable trays with pulleys and pulling wires;
    \item Consult with transport to identify the method for transport of foreseen equipment into the building and verify the floor loadings;
    \item Develop a plan of the building showing all access and egress points for daily use, access under the floor for works, fire escape routes and transport routes. This can be used to quickly validate the loadings and mark the floor areas with the permissible loads;
    \item Once the basic layout of the floor and structures is made, the removable parts can be either designed in-house or bought in as a system. In either case it is highly recommended to have a system that fixes the position of the tiles with a solid frame to prevent creep of the tiles when repeatedly lifted and replaced. It is also worth considering a numbering system or pattern to help ensure tiles are replaced with the correct orientation and position.
\end{itemize}

\subsection{Target Complex}
The target of BDF Target Complex is embedded in shielding blocks and a study of the impact of such heavy shielding blocks on the floor level with respect to the beam alignment has been taken into account. The structural safety of the BDF helium vessel was considered in accordance to the following parameters:

\begin{itemize}
    \item Material: Structural steel;
    \item Self-supporting structure: 12 m x 9 m x 9 m;
    \item Operational helium pressure 0.05 to 0.1 barg (test pressure 0.5 barg);
    \item Maximum load on the vessel’s floor 60 tons/m²;
    \item Compatible with the building’s flat concrete floor (civil engineering tolerances);
    \item One drain point on the bottom to evacuate water leaks, also for flushing air out with helium. 
\end{itemize}

\noindent The requirements are the following :

\begin{itemize}
    \item Partial safety factor as per EN 1990;
    \item Seismic action having a peak ground acceleration of 1.1 m/m² (soil type A) and EN 1998-1 for the definition of the response spectrum;
    \item Structural assessment in accordance with EN 1993-1-1, EN 1993-1-5 and EN 1993-1-6;
    \item Anchors to the surrounding concrete elements to be assessed according to the EN 1992-4;
    \item Executional aspects to be treated in accordance with EN 1090.
\end{itemize}

\section{Chemical Safety} \label{Sec:Chemical}
The chemicals currently foreseen for the BDF project represent standard risks seen in many other facilities at CERN. As these are subject to change as the project moves to the Technical Design Report stage, and as the exact quantities and storage conditions are not yet known, the installations will require proper risk assessment to CERN Safety Form C-0-0-1 when these details become fixed. As for all such facilities at CERN, activities involving chemical agents shall comply with the following CERN Safety rules:

\begin{itemize}
    \item Safety Regulation on chemical agents (SR-C);
    \item General Safety Instruction (GSI-C-1) on prevention and protection measures;
    \item General Safety Instruction (GSI-C-3) on monitoring of exposure to hazardous chemical agents in workplace atmospheres (where required).
\end{itemize}

\noindent Activities involving asphyxiant chemical agents shall comply with the following CERN Safety rules:

\begin{itemize}
    \item Safety Regulation on chemical agents (SR-C);
    \item General Safety Instruction (GSI-C-1) on prevention and protection measures.
\end{itemize}

\noindent Should any additional chemicals be proposed for use in the facility, the Chemical Specialists within the HSE group must be consulted. 

\subsection{Liquid Scintillator used in the BDF Experimental Area}
\begin{figure}[htbp]
\centering %
\includegraphics[width=.87\linewidth]{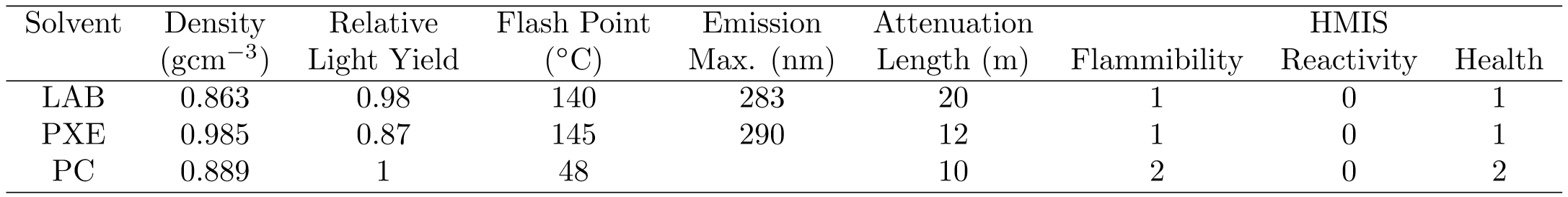}
\caption{Summary Table of the Properties of Possible Solvents for the Liquid Scintillator Candidates \cite{Anelli:2015pba}}
\label{fig:LiquidScintillator}
\end{figure}

Within the integration scope of the BDF facility, certain chemicals are already considered. The liquid scintillator chemicals foreseen to be used in the initial detectors of the proposed SHiP experiment consist of two organic compounds: a solvent and a solute powdered fluor. Such mixtures were evaluated by the SHiP collaboration \cite{Anelli:2015pba} , taking into account each chemical’s flammability and flashpoint, as well as other health, safety and environmental risks. Figure \ref{fig:LiquidScintillator} summarises the properties of these solvents and fluors.

The currently selected mixture is Linear Alkyl Benzene (LAB) with 2,5-Diphenyloxazole (PPO). The solvent, LAB, has a relatively high flash point of 140°C, with PPO added in small quantities (1.5-3 g/l). The chemicals are foreseen to be stored in two liquid containers of total volume 270 m$^3$, located immediately outside the Service Building. This quantity falls below the thresholds classified by the applicable regulations in force at the time of authoring this document. However, should the quantities being stored change significantly this must be revaluated and, in addition, respect the legislation in force at the time of installation.

From a safety perspective, both LAB and PPO are classified as hazardous chemicals, and must be handled with suitable precautions and protection measures. LAB represents an aspiration hazard, and must therefore be stored and disposed of appropriately. PPO represents acute toxicity (oral), eye irritation, and chronic aquatic toxicity hazards. A chemical risk assessment must be carried out at the Technical Design Report stage, based on a safety data sheet prepared for the mixture in accordance with the REACH directive (see CERN Safety Guideline C-0-0-4). All the appropriate preventive measures shall be taken against environmental pollution; the regulatory requirements and the best available techniques shall be applied, including putting in place a suitable retention basin to retain spilt chemicals, create a loading/ unloading area, and installation of all necessary means to detect a chemical leak. 

In addition to the risk assessment, the following Safety forms shall be completed, when required, for the use of hazardous chemical agents:

\begin{itemize}
    \item Safety Form C-1-0-2 – Chemical Inventory (example);
    \item Safety Form C-1-0-3 – Tests of safety showers/eye washes;
    \item Safety Form C-1-0-4 – Respirator use (example);
    \item Safety Form C-3-0-1 – Exposure Form for hazardous chemicals and CMR;
\end{itemize}

\subsection{Helium}
The Target Bunker will be housed within a helium vessel to prevent air activation, with a nominal volume of approximately 82 m$^3$ of helium in the system, split between the vessel, the helium passivation system, and the helium circulation system on the trolley. This will be supplied from 60 cylinders located immediately outside of the Target Complex. The current level of design detail is insufficient to assess the Oxygen Deficiency Hazard (ODH) level represented by the helium in this area. This will depend on what (if any) rate of leakage is found from the system, and if such leakages will reach confined spaces within the facility. Moving into the Technical Design Report stage, a chemical risk assessment shall be carried out to determine the level of risk present, and the mitigation strategies required. Due to its low density, helium will predominantly collect at the ceilings of buildings; in the case of the Target Complex, this could include rising to the roof of the surface Target Building. It will therefore be necessary to include this area in the risk assessment, in particular due to the potential hazard it could represent to personnel operating cranes at this height in the building.

\subsection{Liquid Nitrogen}
Liquid nitrogen is foreseen for the helium passivation system; this is currently expected to be provided by a cryogenic Dewar located in the CV room, containing approximately 120 litres of LN$_2$. At 10 litres per hour, the design flow rate in the system is relatively low, and will all be extracted and vented to atmosphere. As for the helium in the system, a chemical risk assessment shall be conducted for the area, to determine if the nitrogen presents an ODH risk.

\subsection{Gas Building}
The supply and facilities design includes a gas building, intended to store and supply all the gases foreseen to be used in the Experimental Area. As the design of the Experimental Area is still developing, a full inventory of the gases that will be contained in this building is not available. However, it is currently expected to house CO$_2$ and Ar for the Straw Tracker of the SHiP detector. As for the helium and nitrogen for the Target Complex, once the details of the gas inventory in this area have been finalised, a chemical risk assessment will be required to determine the risks present. Should flammable gases be added to the inventory, an explosion risk assessment shall be carried out, and the hazardous areas shall be classified as required. The following Safety forms shall be completed for the use of flammable gas:

\begin{itemize}
    \item Safety Form C-2-0-1 - Explosion risk assessment;
    \item Safety Form C-2-0-2 - Classification of hazardous areas (when required);
    \item Safety Form C-2-0-3 - Declaration/Cancellation of the use of flammable gas in an experiment area (when required).
\end{itemize}

\noindent The following Guideline documents shall be consulted, when completing the Safety forms:

\begin{itemize}
    \item Safety Guideline C-2-0-1 - Explosion protection measures;
    \item Safety Guideline C-2-0-2 - Identification and prevention of explosion hazards;
    \item Safety Guideline C-2-0-3 - Practical Guide for Classification of Hazardous Areas
    \item Safety Guideline C-1-0-1 - Storage of hazardous chemical agents.
\end{itemize}

Any purchase of flammable gases shall be authorised by the FGSO (Flammable Gas Safety Officer) of the relevant Department, and the use of such areas shall be subject to authorisation by the HSE Unit.

\subsection{Lead}
Lead is not currently part of the shielding design for the Target Area. However, as the design is still at a preliminary stage, and that in the Experimental Hall is still under development, it is important to note that lead can present significant hazards. Care must be taken that the necessary procedures are followed for purchasing, shipping, storing and handling of the blocks to limit the dangers of lead poisoning or exposure to activated materials. In particular, blocks should arrive at CERN pre-painted or adequately protected by equivalent means, to ensure that risks from dust are contained. Should lead be required, a chemical risk assessment shall be carried out, and the following safety form shall be completed:

\begin{itemize}
    \item Safety Guideline C-0-0-3 – Lead.
\end{itemize}

\section{Electrical Safety} \label{Sec:Electrical}
The electrical infrastructure design for the BDF facility is currently at a general level, but will incorporate subsystems that either produce, or use, high voltage or current, both of which represent electrical hazards to personnel. Dedicated electrical rooms will be used to contain all the electrical cabinets required for the power distribution. The hazards are expected to be standard for such an installation, and shall be mitigated through sound design practice and execution. The CERN Electrical Safety rules, alongside NF C 18-510, shall be followed throughout the design process; where exceptions are required, this shall be subject to an appropriate level of risk assessment to evaluate the residual risk, and determine the mitigation strategies required. NF C 18-510 compliant covers, interlocks preventing access to high voltage equipment, and restriction of access to the electrical rooms to those with the appropriate level of CERN electrical \textit{habilitation} training shall be used to protect personnel from any electrical hazards present.

\subsection{Electromagnets}
Of particular interest from an electrical safety perspective will be the magnetic coil and steel yoke, located downstream of the Target. This is likely to be the highest voltage component in the system. As for all electromagnets, appropriate grounding measures shall be implemented for the magnet yokes, and all live parts protected to a minimum of IPXXB for Low Voltage and IPXXC for High Voltage circuits or locked out for any intervention in their vicinity. Interventions may only be carried out by personnel with the necessary training, after following the work organisation procedures and authorisation of the facility coordinator (VICs, IMPACT etc.).

\section{Mechanical safety and design of the Target Complex} \label{Sec:Mechanical}
The cooling requirements of the target due to the high energy levels being deposited by the beam, along with the high radiological activation levels of the materials surrounding the target have led to the need for a number of different pressurised fluid circuits within the Target Complex.

\noindent All pressure equipment shall comply with the following CERN Safety rules:

\begin{itemize}
    \item CERN Safety Regulation SR-M - Mechanical equipment;
    \item CERN General Safety Instruction GSI-M-2 - Standard Pressure Equipment.
\end{itemize}

Moreover, there are specific sets of rules applicable only to certain types of standard pressure equipment. Those rules are defined in the following Specific Safety Instructions, of which the following may be applicable to the BDF project:

\begin{itemize}
    \item CERN Specific Safety Instruction on pressure vessels (SSI-M-2-1);
    \item CERN Specific Safety Instruction on safety accessories for standard pressure equipment (SSI-M-2-3);
    \item CERN Specific Safety Instruction on metallic pressurised piping (SSI-M-2-4);
    \item CERN Specific Safety Instruction on vacuum chambers and beam pipes (SSI-M-2-5);
    \item CERN Specific Safety Instruction on transportable pressure equipment (SSI-M-2-6).
\end{itemize}

According to CERN Safety rules, pressure equipment shall meet the essential requirements set by the following applicable European Directives:

\begin{itemize}
    \item Directive 2014/68/EU on pressure equipment - Pressure Equipment Directive (PED);
    \item Directive 2010/35/EU on transportable pressure equipment – Transportable Pressure Equipment Directive (TPED).
\end{itemize}

Pressure equipment designed and manufactured according to harmonised European standards benefit from presumption of conformity with the essential requirements laid down in the above mentioned European Directives. In accordance with the CERN Safety rules, the use of harmonised European standards is compulsory for pressure equipment designed and/or manufactured at CERN. The use of other design codes or national standards shall be reviewed and approved by the HSE Unit.

\subsection{Cooling Water Circuit}
The highest pressure seen in the pressurised systems foreseen within the facility is in the pressurised cooling water circuit, pumping demineralised water around the Target and Proximity Shielding. The system is currently foreseen to operate at approximately 25 barg, and due to the high radiation levels within the target area, will be inaccessible to personnel during operation. However, limiting the spillages of the highly activated (and tritiated) water will be a key part of the risk assessment of the area. It is therefore essential that the system should be able to withstand the internal pressure loads without critical failure in all conceivable modes of operation. Standard mitigation strategies must be put in place, in line with the requirements from the CERN rules and Pressure Equipment Directive, but a number of complicating factors make the Target relatively unique as a pressure vessel. A key failure mode for the system will be the eventuality of a cooling water pump malfunctioning, causing a “Natural Convection Scenario”. The design must show (through a combination of simulation and testing) that the high energy deposition at the end of the target will not cause localised boiling (and high pressures) sufficient to cause critical failure of the pressure envelope. Complicating factors in this will include the significant pressure drop across the Target, and the high localised stresses in the material due to heating. In light of the criticality of the facility, a higher Quality Assurance programme shall also be implemented for this system during fabrication and installation, requiring 100\% volumetric inspection of all welds.

In addition to the dangers of overheating in operation, consideration must be given to the effects of afterheat after the beam is turned off. As a mitigation measure, a jockey pump should be considered, to provide additional redundancy and to keep pressure and some flow in the Target should the main pump fail, or be turned off outside of the beam time.

A key mitigation measure (and an essential feature for the Target Complex) will be the ability of the system to deal with water leaks, and prevent contamination of other water sources. To this end, there must be capacity within the system to detect leaks, retain the entire inventory of cooling water in the event of a leak, and to be able to effectively drain the whole circuit if required (with no significant areas inaccessible to drainage of the activated water).

\subsection{Helium Purification Circuit}
In order to ensure that the helium delivered to the helium vessel remains at 99.9\% purity, the helium purification circuit will operate at 15 barg in the condenser-freezer and adsorber components of the system, with a maximum allowable pressure of 30 barg. As for the Target Cooling Water circuit, the criticality of this system requires a higher degree of Quality Assurance during fabrication and installation, in addition to fulfilling the design and QA requirements of the Pressure Equipment Directive and the CERN Safety rules, including 100\% volumetric inspection of all of the welds within the system.

\subsection{Cryogenics}
The helium purification system will also require cooling at the condenser-freezer and adsorber components. To this end, the adsorber shall be submerged into a liquid nitrogen dewar, whilst the condenser-freezer shall pre-cool the impure helium via heat exchange with cold nitrogen vapour supplied from the dewar at 1.5 barg. This equipment shall comply with the following CERN Safety rules:

\begin{itemize}
    \item CERN Safety Regulation SR-M - Mechanical equipment;
    \item CERN General Safety Instruction GSI-M-4 - Cryogenic Equipment.
\end{itemize}

\noindent In accordance with the rules above, cryogenic pressure equipment shall comply with the European directives:

\begin{itemize}
    \item Directive 2014/68/EU on pressure equipment - Pressure Equipment Directive (PED);
\end{itemize}

\noindent For the pressurised cryogenic piping, it is required that:

\begin{itemize}
    \item the equipment complies with the Directive 2014/68/EU on pressure equipment;
    \item the equipment is delivered with EC declaration of conformity.
\end{itemize}

Cryogenic pressure vessels shall be equipped with pressure relief devices to ensure the safe release of the working fluid in case of overpressure. Piping sections that may become isolated with cryogenic liquid or cold gas shall be also equipped with pressure relief devices to ensure the safe release of the working fluid in case of overpressure.

\noindent The pressure relief devices shall:

\begin{itemize}
    \item be installed, commissioned, and periodically tested according to General Safety Instruction on cryogenic equipment (GSI-M-4);
    \item be sized in accordance with:
    \begin{itemize}
        \item EN 13648-3 : Cryogenic vessels – Safety devices for protection against excessive pressure – Part 3: Determination of required discharge; Capacity and sizing; and the applicable part of
        \item ISO 4126 : Safety devices for protection against excessive pressure;
    \end{itemize}
    \item be CE marked and delivered with an EC declaration of conformity with respect to Directive 2014/68/EU. Pressure relief devices must be classified in category IV and their conformity assessed accordingly. However, by way of exception and in agreement with the HSE Unit, pressure relief devices manufactured for specific equipment may be classified in the same category as the equipment they protect. They shall be additionally Pi-marked whenever installed on transportable cryogenic storage vessels.
    \item The exhausts of all cryogenic safety valves shall vent to a safe location.
\end{itemize}

\subsection{HVAC}
\subsubsection{Pressurised Areas}
Pressurised areas and controlled ventilation are key parts of both the radiation protection and fire safety strategies for this facility, helping to prevent the spread of airborne particles or smoke beyond the areas of origin. Typically at CERN pressurised areas meet the fire safety requirements set by:

\begin{itemize}
    \item French “instruction technique 246” – overpressure in stairwells of buildings with public access of 20 Pa to 80 Pa;
\end{itemize}

\noindent and the safety requirements for the airborne radioactive particles set by:

\begin{itemize}
    \item Ventilation Working Group Report EDMS N. 1226988 – for particle accelerator type environments, where the overpressure between accelerators and “safe area” is 20 Pa to 30 Pa.
\end{itemize}

For areas with higher radiation levels, a case-specific risk assessment together with the BDF project and the HSE-RP Group shall be made to define the correct values of pressure cascade. The Target Complex, due to the significant containment needs driven by the high radiation levels in this area, may require values well above the ones mentioned above. In any case, the maximum force to manually open emergency exit doors is 100 N.

\subsection{HVAC Equipment}
Equipment purchased on the market (e.g.: Air Handling Units, chillers, boilers, fan coils) shall comply with the applicable European Directives and shall bear the CE marking.

Ductwork (supply or exhaust air) and piping systems incorporated in a permanent manner in a building, shall comply with the following European Regulations:

\begin{itemize}
    \item European Regulation 305/2011 - Construction Products Regulation;
    \item EN 1505 - Ventilation for buildings. Sheet metal air ducts and fittings with rectangular cross-section; 
    \item EN 1506 - Ventilation for buildings. Sheet metal air ducts and fittings with circular cross-section; 
    \item EN 12097 - Ventilation for buildings. Requirements for ductwork components to facilitate maintenance of ductwork systems;
    \item EN 13480 – Metallic Industrial piping. 
\end{itemize}

These standards provide presumption of conformity to the Safety requirements regarding the design laid down in the applicable European Regulations. Electrical parts related to HVAC installations shall respect the general Safety requirements as indicated in Section \ref{Sec:Electrical}.

\subsection{Trolley Mechanics}
A key mechanical consideration for the Target Trolley itself will be the danger of a malfunction causing the trolley to be stuck at a point along its travel position. A recovery procedure will therefore be needed for each of the potential failure modes, identified through Failure Mode and Effects Analysis (FMEA) risk assessment. The most likely of these would be a failure of drive motor; this could be overcome by ensuring that the Trolley drive system can be overridden, either manually using a crank handle, or with a supplementary backup motor. Additionally, care shall be taken that the drive motor has sufficient capacity to ensure that the Trolley can still be driven in the event of a seized wheel or similar obstruction.

\subsection{End of Life Considerations for Mechanical Components}
An important consideration for a facility such as the BDF Target Complex is the design for end of operational life. Thought should be given to facilitating disassembly and disposal of the highly activated materials at the decommissioning of the facility. This may include separating out, as far as possible, different materials within the helium vessel, and making disassembly of dissimilar materials as easy as possible. Keeping witness samples of all the used materials inside the vessel (and exposed to the highest levels of radiation) will allow easy examination of the effects of the actual radiation dose on the mechanical properties of the exact materials used. This could be invaluable for disassembly, for example revealing the radiation hardening that has occurred in the metallic components should cutting operations be required. Maintaining modularity, and designing in significant temperature measurement capacity in components in the helium vessel will also be important for extending the life of the facility, should this be required towards the end of the predicted initial operational time.

\section{Lifting and Handling Equipment} \label{Sec:Lifting}
The BDF facility is expected to use a significant quantity of lifting equipment, including hoists, cranes and personnel lifts. Most notably, all three of the proposed concepts for the Target Complex require a 40 tonne crane in the ground level target hall to perform maintenance activities, and lift equipment (including highly activated material) in and out of the helium vessel. All lifting and handing equipment installed at CERN shall comply with the CERN Safety rules:

\begin{itemize}
    \item Safety Regulation on Mechanical Equipment (SR-M); 
    \item General Safety Instruction on Lifting Equipment and Accessories (GSI-M-1).
\end{itemize}

\subsection{Target Complex}
Due to the radiation levels foreseen in the Target Complex, remote handling equipment has been designed to deal with the need for regular maintenance to the technical equipment in the area. All of the concepts, but most particularly the Crane Concept, will require remote handling operations to be carried out using the cranes in the Target Area. Due to the activation of the material, it is essential that there is sufficient redundancy in this system to deal with crane malfunctions. Mitigation will include redundancy for lifting capacity, as well as planned strategies for handling malfunctions using detailed Failure Mode and Effects Analysis (FMEA). 
All Cranes, Bridge Cranes, Gantry Cranes and Power-driven Hoists must conform to the following:

\begin{itemize}
    \item CERN Specific Safety Instruction (SSI-M-1-2) for design, installation and use; 
    \item CERN Specific Safety Instruction (SSI-M-1-4) – Manually powered lifting equipment.
\end{itemize}

\noindent All Remote Handling Equipment and Tooling must additionally follow the following:
\begin{itemize}
    \item CERN Specific Safety Instruction (SSI-M-1-3) – Non-fixed load-lifting accessories;
\end{itemize}

\noindent Overhead travelling cranes shall also respect the following general standards: 

\begin{itemize}
    \item EC Machinery directive (European Directive 98/37/EC) 
    \item FEM 1.001 3rd edition revised 1998.10.01 
\end{itemize}

\noindent As well as CERN safety documents:

\begin{itemize}
    \item Safety Regulation on Mechanical Equipment (SR-M) 
    \item General Safety Instruction on Lifting Equipment and Accessories (GSI-M-1).
    \item Electrical Safety Code C1 
    \item Safety Instruction IS23 Rev. 2. Criteria and standard test methods for the selection of electrical cables, wires and insulated parts with respect to fire safety and radiation resistance 
    \item Safety instruction IS24. Regulations applicable to electrical installations 
    \item Safety Instruction IS41. The use of plastic and other non-metallic materials at CERN with respect to fire safety and radiation resistance 
    \item Safety code A3 Rev. Safety colours and safety signs 
\end{itemize}

\noindent The following specific safety aspects for cranes shall also be respected: 

\begin{itemize}
    \item Drainage apertures in all places where water and oil may collect;  
    \item Catwalk to access to the rails;
    \item Catwalk on the main beam and on crab for a safe maintenance on the crane. Headroom above the catwalks of at least 1.8 m. Mobile ladder to reach the crane catwalk. Safe access to the crab catwalk from the crane one;
    \item Maintenance access platform;
    \item Noise level index (defined by ISO standard No. R.1996): maximum 65 dB; 
    \item Protection against overturning;
    \item Overhanging system for equipment mounted beyond the crane steelworks;
    \item Travelling, traversing and hoisting end stops; hoisting upper over travel end stop; 
    \item Structural steelworks and crab earthing;
    \item Load limiter;
    \item Floodlights to illuminate the area under the crane;
    \item Overspeed detection.
\end{itemize}

\subsubsection{Hot Cell Winch}
\noindent As for all hoists in the facility, the winch shall respect the following general standards: 

\begin{itemize}
    \item EC Machinery directive (European Directive 2006/42/EC); 
    \item FEM 1.001 3rd edition revised 1998.10.01; 
\end{itemize}

\noindent As well as CERN safety documents: 

\begin{itemize}
    \item Safety Regulation on Mechanical Equipment (SR-M) 
    \item General Safety Instruction on Lifting Equipment and Accessories (GSI-M-1).
    \item Electrical Safety Code C1; 
    \item Safety Instruction IS23 Rev. 2. Criteria and standard test methods for the selection of electrical cables, wires and insulated parts with respect to fire safety and radiation resistance; 
    \item Safety instruction IS24. Regulations applicable to electrical installations; 
    \item Safety Instruction IS41. The use of plastic and other non-metallic materials at CERN with respect to fire safety and radiation resistance; 
    \item Safety code A3 Rev. Safety colours and safety signs. 
\end{itemize}

\noindent The following specific safety aspects for winches and hoists shall also be respected:

\begin{itemize}
    \item Noise level index (defined by ISO standard No. R.1996): maximum 65 dB; 
    \item Protection against overturning;  
    \item Overhanging system for equipment mounted beyond the hoist steelworks; 
    \item Travelling and hoisting end stops. Hoisting upper over travel end stop; 
    \item Structural steelworks and crab earthing; 
    \item Load limiter; 
    \item Over speed detection.
\end{itemize}

\subsection{Personnel Lifts}
\noindent Personnel lifts in the facility shall respect the following general standards: 

\begin{itemize}
    \item European standard EN 81.1; 
    \item EN 294:1992 and EN 811:1997; 
    \item EC Directive 95-16 (EC conformity certification).
\end{itemize}
 
\noindent As well as safety documents: 

\begin{itemize}
    \item European Directive 95/16C; 
    \item EN81-72 Firefighter lifts; 
    \item EN81-73 Behaviour of Lifts in case of fire;
    \item Electrical Safety Code C1; 
    \item Safety Code E Rev. Fire protection; 
    \item Safety Instruction IS23 Rev. 2. Criteria and standard test methods for the selection of electrical cables, wires and insulated parts with respect to fire safety and radiation resistance; 
    \item Safety instruction IS24. Regulations applicable to electrical installations; 
    \item Safety instruction IS37 Rev. 2. Alarms and alarm systems; 
    \item Safety Instruction IS41. The use of plastic and other non-metallic materials at CERN with respect to fire safety and radiation resistance; 
    \item Safety code A3 Rev. Safety colours and safety signs. 
\end{itemize}

\noindent The following specific safety aspects on personnel lifts shall also be respected: 

\begin{itemize}
    \item Automatic lighting, with economiser, provided by recessed ceiling fitting with emergency lighting; 
    \item Hands-free telephone integrated in car control panel; 
    \item Overload/full load indicator via electronic sensors with notification in cab (full load = 100\%, overload = 110\%). Load-sensor precision: 5\% rated load; 
    \item Car door with variable speed drive and preliminary unlocking in levelling zone; 
    \item Electrical manoeuvring system; 
    \item Emergency escape hatch onto the car roof.
\end{itemize}

\section{Non-Ionising Radiation}
The high magnetic fields from the proposed electromagnets in the current BDF design represent a hazard similar to that found in many of the facilities at CERN, and shall be handled with standard mitigation strategies. The facility shall follow the Directive 2013/35/EU on the occupational exposure of workers, alongside CERN Safety Instruction IS 36 and its Amendment. Any activity inside the static magnetic field shall be subject to risk assessment and ALARA. Personnel shall be informed about the hazards and appropriately trained. Areas with magnetic flux densities exceeding 0.5 mT shall be delimited (use pacemaker warning signs), while areas with magnetic flux densities exceeding 10 mT shall be rendered inaccessible to the public. 

\subsection{Extraction Tunnel}
Electromagnets are foreseen in the BDF extraction beam line, with modified splitter magnets, quadrupoles, dilution kickers and dipole magnets present in the optics design, all of which are to be largely identical to those already in use in the North Area. All magnets foreseen for the facility are warm magnets.

\subsection{Target Area}
The major novelty will be the Active Muon Shield to be installed immediately downstream of the Target and Hadron Stopper. A warm 1.4 T magnetic coil will be installed within the larger shielding surrounding the Target. The restrictions around this magnet are expected to be enforced as a natural result of the radiation levels around the magnet when it is in operation. The expected risk is therefore only foreseen for commissioning, testing, and maintenance of the magnet, which shall be mitigated by ensuring that these operations are only carried out by appropriately qualified personnel, in line with the rules and processes outlined at the start of this section.

\section{Noise}
The BDF facilities are located in the North area of the Prévessin site, close to the fence line of CERN property. In order to ensure occupational health and safety to people exposed to noise, the BDF project shall be compliant with the following rules and Regulations:

\begin{itemize}
    \item CERN Safety Code A8 - Protection against noise
    \item Directive 2003/10/EC of the European Parliament and of the Council of 6 February 2003 on the minimum health and safety requirements regarding the exposure of workers to the risks arising from physical agents (noise)
    \item French Code du travail
\end{itemize}
    
\noindent Emissions of environmental noise related to neighbourhoods shall respect the thresholds indicated in:
    
\begin{itemize}    
    \item \textit{French Arrêté du 23 janvier 1997 relatif à la limitation des bruits émis dans l'environnement par les installations classées pour la protection de l'environnement.}
\end{itemize}

\section{Asbestos}
As the extraction tunnel was constructed before 1997, an asbestos survey was carried out by HSE Unit. The findings of the investigation are summarised in EDMS 2054048.
In case the presence of asbestos is ascertained, necessary protective measures have to be put in place following the applicable CERN Safety rule:

\begin{itemize}    
    \item Safety Instruction IS 43 – Asbestos : Dangers and precautions
\end{itemize}

\section{Protection of the environment}

\noindent With regard to protection of the environment, CERN Safety Policy states that the Organization is committed to ensuring the best possible protection of the environment. This can be achieved by ensuring that the given regulations are followed for the different activities and experiments. 

As the project moves to the Technical Design Report stage, a review with the Environmental Protection specialists within HSE shall be held to determine whether the relevant technical provisions of the following regulation shall apply for the BDF project:

\begin{itemize}    
    \item \textit{Arrêté du 29/05/00 relatif aux prescriptions générales applicables aux installations classées pour la protection de l'environnement soumises à déclaration sous la rubrique n° 2925;}
    \item \textit{Arrêté du 14/12/13 relatif aux prescriptions générales applicables aux installations relevant du régime de l’enregistrement au titre de la rubrique n° 2921 de la nomenclature des installations classées pour la protection de l'environnement.}
\end{itemize}

\noindent The technical environmental mitigation measures shall be determined more precisely once technical detail information are available such as the type and the quantity of chemicals or the solution chosen for the cooling.

\subsection{Air}
Atmospheric emissions shall be limited at the source and shall comply with the relevant technical provisions of the following regulations:

\begin{itemize}    
    \item \textit{Arrêté du 02 février 1998 relatif aux prélèvements et à la consommation d'eau ainsi qu'aux émissions de toute nature des installations classées pour la protection de l'environnement soumises à autorisation, Articles 26, 27, 28, 29, 30.}
\end{itemize}

\noindent The design of exhaust air discharge points shall comply with the requirements of the section 5.1.3 of the CERN Safety Guideline C-1-0-3 - Practical guide for users of Local exhaust ventilation (LEV) systems. 

\noindent Whenever greenhouse gases are used, relevant technical provisions contained in the following regulations apply:

\begin{itemize}    
    \item Regulation (EU) No 517/2014 of the European Parliament and of the Council of 16 April 2014 on fluorinated greenhouse gases and repealing Regulation (EC) No 842/2006;                       
    \item \textit{Code de l’environnement Livre V: Titre II (Art. R521-54 to R521-68) and Titre IV (Art. R543-75 to R543-123).}
\end{itemize}

\noindent All the appropriate preventive measures shall be taken against the release of greenhouse gases into the atmosphere. Working procedures shall be established and implemented for activities involving the use of those gases including the storage, handling, transport, recovery and disposal. Additionally such activities shall be performed by trained personnel. The emissions of fluorinated gases shall be registered during the entire life-cycle of the equipment or Experiment.

In accordance with the siting of the facility on French territory, the design, operation and maintenance of cooling tower water circuits shall comply with the relevant technical provisions of the following regulations and standards in order to limit the risk of legionella bacteria and its dispersion in the atmosphere:

\begin{itemize}    
    \item \textit{NF E38-424 Aéroréfrigérants humides : terminologie et exigences de conception vis-à-vis du risque légionellose;}                       
    \item \textit{Arrêté du 14 décembre 2013 relatif aux prescriptions générales applicables aux installations relevant du régime de l’enregistrement au titre de la rubrique n° 2921 de la nomenclature des installations classées pour la protection de l'environnement;}
    \item \textit{Guide des bonnes pratiques Legionella et tours aéroréfrigérantes.}
\end{itemize}

\subsection{Water}
The BDF project shall ensure the rational use of water. The discharge of effluent water into the CERN clean and sewage water networks shall comply with the relevant technical provisions contained in the following regulations:

\begin{itemize}    
    \item \textit{Loi n° 2006-1772 du 30 décembre 2006 sur l'eau et les milieux aquatiques;}                       
    \item \textit{Arrêté du 02 février 1998 relatif aux prélèvements et à la consommation d'eau ainsi qu'aux émissions de toute nature des installations classées pour la protection de l'environnement soumises à autorisation.}
\end{itemize}

\noindent The direct or indirect introduction of potentially polluting substances into water, including their infiltration into ground is prohibited. Applicable emission limit values for effluent water discharged in the Host States territory are defined in the following regulations:

\begin{itemize}    
    \item \textit{Arrêté du 02 février 1998 relatif aux prélèvements et à la consommation d'eau ainsi qu'aux émissions de toute nature des installations classées pour la protection de l'environnement soumises à autorisation Art. 31 and art.32.}
\end{itemize}

\noindent In complement, in case independent cooling towers are installed, the cooling circuit shall be equipped with a recycling process, and the effluent resulting from the recycling process shall be discharged into the sanitary network.

Retention measures for fire extinguishing water are required for any CERN project in which large quantities of hazardous, or potentially polluting substances are used or stored. As the project moves to the Technical Design Report stage, through discussions with the Environmental Protection specialists from HSE, it will be determined whether the following guidance document shall be applied (in accordance with the French \textit{Code de l’Environnement}:

\begin{itemize}    
    \item \textit{Référentiel APSAD D9 : Dimensionnement des besoins en eau pour la défense contre d'incendie and Référentiel APSAD D9A : Dimensionnement des rétentions des eaux d'extinction available in the Centre National de Prévention et de Protection (CNNP) (http://www.cnpp.com/).}
\end{itemize}

\subsection{Energy}
The use of energy shall be done as efficiently as possible. For the entire facility, adequate measures shall be taken to comply with the relevant technical provisions contained in the following regulations:

\begin{itemize}    
    \item \textit{Loi n° 2010-788 du 12 juillet 2010 portant engagement national pour l’environnement (Grenelle II).}
\end{itemize}

\noindent In addition, construction of new buildings sited in France shall comply with the relevant technical provisions relating to thermal efficiency contained in the following regulation:

\begin{itemize}
    \item \textit{Décret n° 2012-1530 du 28 décembre 2012 relatif aux caractéristiques thermiques et à la performance énergétique des constructions de bâtiments;}
    \item \textit{Arrêté du 26 octobre 2010 relatif aux caractéristiques thermiques et aux exigences de performance énergétique des bâtiments nouveaux et des parties nouvelles de bâtiments and the French Réglementation Thermique 2012 (RT 2012);} 
    \item \textit{NF EN 15232 Performance énergétique des bâtiments - Impact de l'automatisation, de la régulation et de la gestion technique.} 
\end{itemize}

\subsection{Soil}

The natural physical and chemical properties of the soil must be preserved. All the relevant technical provisions related to the usage and/or storage of hazardous substances to the environment shall be fulfilled to avoid any chemical damage on soil. Furthermore, the excavated material shall be handled adequately and prevent further site contamination. All excavated material must be disposed of appropriately in accordance with the associated waste regulations. 

\subsection{Waste}

The selection of construction materials, design and fabrication methods shall be such that the generation of waste is both minimised and limited at the source. Waste shall be handled from its collection to its recovery or disposal according to: 

\begin{itemize}
    \item \textit{Code de l’environnement,  Livre V: Titre IV -Déchets ;}
    \item \textit{LOI n° 2009-967 du 3 août 2009 de programmation relative à la mise en œuvre du Grenelle de l'environnement (1), Art. 46;} 
\end{itemize}

\noindent The traceability of the waste shall be guaranteed at any time.

\subsection{Preservation of the natural environment}

The BDF project shall ensure the preservation of the natural environment (e.g. landscaping, fauna, floral reserve, etc.) according to the relevant technical provisions contained in the following regulations:

\begin{itemize}    
    \item \textit{Code de l’environnement, Art. L411-1.}
\end{itemize}

\noindent Vegetal species listed in this regulation shall be protected, restored or adequately replaced (e.g. orchids). Whenever CERN natural areas are affected by a project, the Civil Engineering and Buildings (SMB-SE-CEB) section of the SMB Department shall be contacted for authorisation and definition of appropriate measures.

\section{Bibliography}
\printbibliography

\chapter{Civil Engineering}

\label{Chap:CivEng}

\section{Overview}

Civil Engineering (CE) costs for projects such as BDF typically represent a significant proportion of the overall implementation budget. For this reason, particular emphasis has been placed on CE studies to ensure a cost efficient conceptual design and construction methodology. This chapter provides an overview of the designs adopted for CE along with the key considerations. The design developed has been used to provide an estimated cost and schedule for CE elements.  

\begin{figure}[ht]
\centering\includegraphics[width=.55\linewidth]{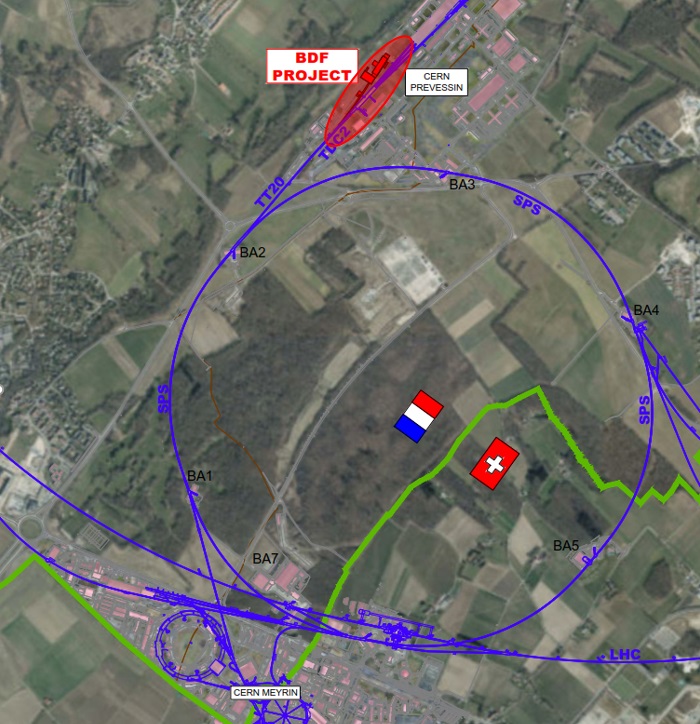}
\caption{Location of the BDF Project on the French CERN Site in Pr\'{e}vessin }
\label{fig:Location}
\end{figure}

The CE studies have been based on the assumption that the BDF facility will be sited at the CERN Pr\'evessin laboratory in France. A junction with the existing tunnel TDC2 will be required to enable a new machine extraction tunnel in the North Area, leading to a new target complex and experimental facility. All CE works for the project are fully located within existing CERN land. 
The area foreseen for the development of the new BDF facility is highlighted in Fig.~\ref{fig:Location}.

\begin{figure}[ht]
\centering\includegraphics[width=.87\linewidth]{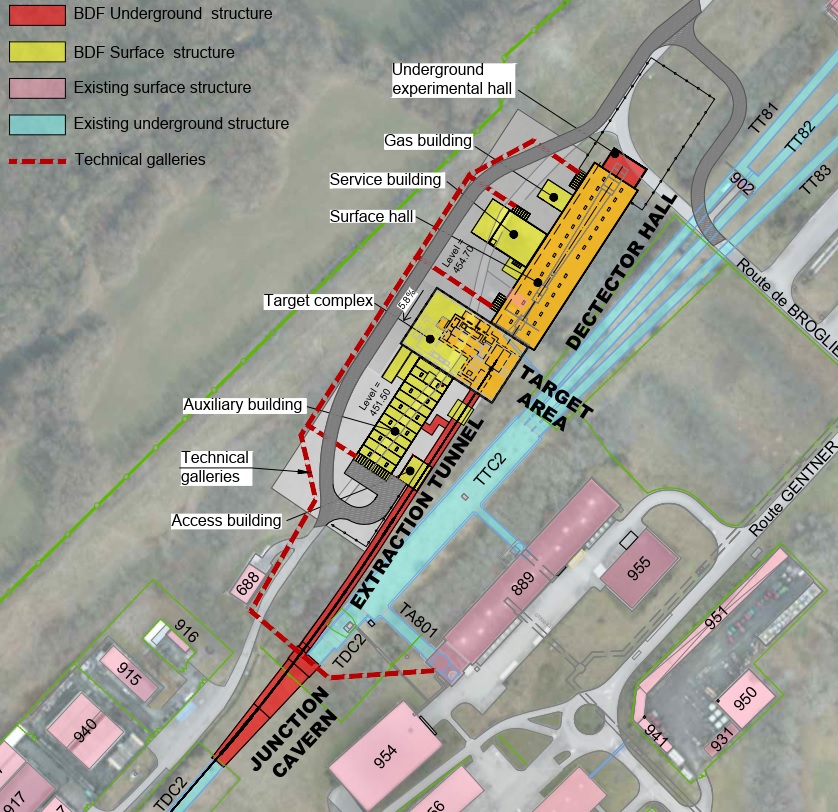}
\caption{Existing CERN infrastructure and proposed BDF facilities indicating above and below ground installations}
\label{fig:Layout}  
\end{figure}

A plan overlay of the proposed BDF facilities on the existing site is shown in Fig.~\ref{fig:Layout}. 
The key features of this layout are: 

\begin{itemize}
\item  Demolition of a section of the existing TDC2 tunnel to form a \SI{75}{\metre} long Junction Cavern
\item \SI{165}{\metre} long extraction tunnel - \SI{5}{\metre} wide by \SI{4}{\metre} high (internal dimensions) 
\item \SI{15}{\metre} long by \SI{12}{\metre} wide access building including a heavy equipment access shaft  
\item \SI{60}{\metre} long by \SI{20}{\metre} wide auxiliary building servicing  the extraction tunnel and target complex
\item \SI{36}{\metre} long by \SI{58}{\metre} wide target complex along with associated vehicle airlock and personnel access
\item \SI{120}{\metre} long by \SI{20}{\metre} wide experimental hall 
\item \SI{100}{\metre} long by \SI{27.5}{\metre} wide surface building above the experimental hall
\item \SI{21}{\metre} long by \SI{35}{\metre} wide service building and workshop and \SI{10}{\metre} long by \SI{15}{\metre} wide gas building
\end{itemize}
 
The civil engineering studies presented in this chapter have been performed by the SMB-SE Future Accelerator Studies (FAS) section with various input from technical experts across the BDF project team. Engineering consultant Arup have also worked on the study along with specialist radiation sub-consultant Studsvik. A more detailed report on the CE study produced in collaboration with Arup is also available 
\cite{ARUP2018-2}.

\section{Location and existing infrastructure}

\subsection{Location}
The proposed site for the BDF project is entirely located within existing CERN land on the Pr\'{e}vessin campus. The existing site is composed of green areas and existing roads. There is  also a small SMB depot/storage area designated `Zone 9103'. This comprises building 687, de-mountable buildings 6357 and 6575 and a steel container 6361.

This location is well suited to housing the BDF project from a CE perspective, primarily due to the relatively stable and well understood ground conditions. Quite detailed geological records exist in this area and have been utilised for this study to minimise the costs and risk to the project. The underground works will be constructed in the stable Moraine glacial deposits at depths of up to \SI{20}{\metre} below ground level. 

\subsection{Geology and Hydrology}

\begin{figure}[ht]
\centering\includegraphics[width=.87\linewidth]{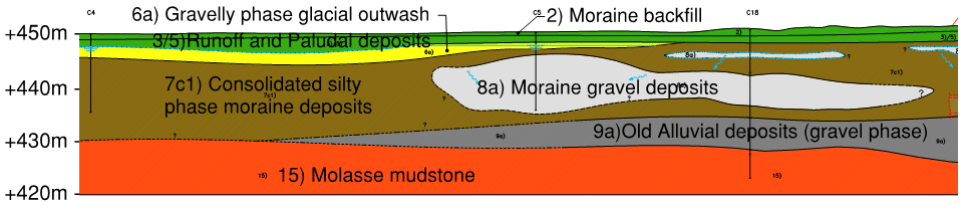}
\caption{Typical geological long section for the area (based on GADZ, 2014 \cite{GADZ}).}
\label{fig:Geosection}
\end{figure}

The proposed location of BDF is situated within the Geneva Basin, a sub-basin of the large North Alpine Foreland (or Molasse) Basin. This is a large basin which extends along the entire Alpine Front from South-Eastern France to Bavaria, and is infilled by clastic `Molasse' deposits of the Oligocene and Miocene ages. The basin is underlain by crystalline basement rocks and formations of Triassic, Jurassic and Cretaceous age. The Molasse, comprising an alternating sequence of marls and sandstones (and formations of intermediate compositions), is overlain by Quaternary glacial moraines related to the Wurmien and Rissien glaciations.  
Fig.~\ref{fig:Geosection} shows a typical geological cross-section for the Moraines in the vicinity of BDF. 

A ground investigation has previously been carried out in this area. The findings were summarised by geotechnical consultant GADZ \cite{GADZ}. Although the ground investigation was carried out for an earlier stage of the CERN Neutrino Platform (CENF) project, the results are both applicable to BDF and sufficiently detailed to be used for feasibility review of foundations and other CE infrastructure. 

According to the GADZ report, several groundwater tables have been identified in the area, independent of one another. Careful consideration will be required during detailed design and construction both to optimise the design and also to prevent any contamination of groundwater. Future project specific ground investigation will provide an opportunity for more detailed examination. 

\subsection{Existing infrastructure}

\begin{figure}[ht]
\centering\includegraphics[width=.6\linewidth]{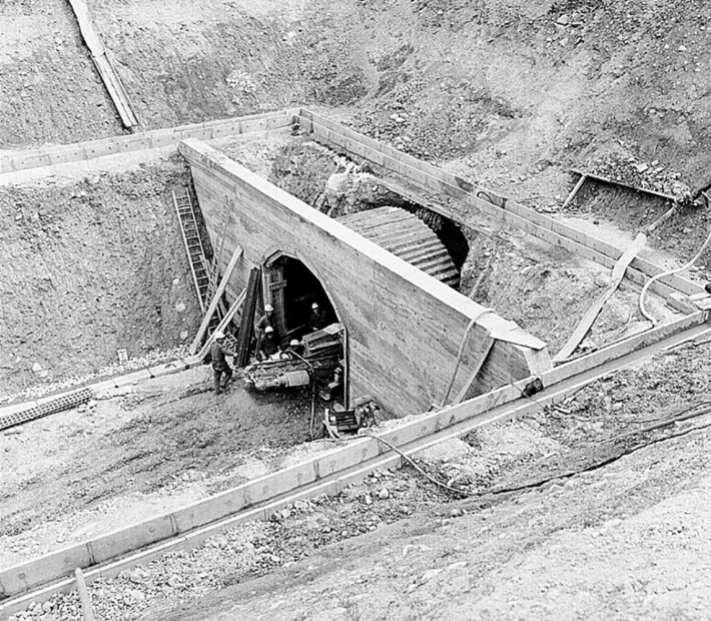}
\caption{Photograph from May 1972 taken at the start of tunnelling for TT20 looking towards the SPS.}
\label{fig:Cutandcover}
\end{figure}
The proposed works connect with the existing TDC2 tunnel which was excavated in 1972 using the`cut and cover' technique as shown in Fig.~\ref{fig:Cutandcover}.

The proposed BDF facilities will be close to the existing TCC2 cavern as well as transfer tunnels TT81, 82 and 83. Existing tunnels are likely to be sensitive to vibration and excavation carried out as part of the construction works. The CE study proposals aim to avoid any impact on existing infrastructure, however tunnel monitoring will be required before, during and after construction works. It will be particularly important to monitor and manage construction induced vibration during `beam on' to avoid disruption to operations.

The age and use of existing infrastructure also has an impact on the works planning due to the radiological activation of the tunnels and surrounding earth. Existing concrete in TDC2 has been tested and was found to be up to 400 times the liberation limit (LL), meaning it must be classified and treated as radioactive. Simulations also indicated it was likely the soil surrounding TDC2 would be highly irradiated. As part of this study, soil samples were taken and testing carried out to a depth \SI{1.5}{\metre} above the tunnel roof slab. No levels above LL were found. A conservative assumption that all soil within \SI{1.5}{\metre} of the tunnel walls is activated has been proposed by CERN's radiation protection (RP) team until further testing can confirm or otherwise. The full findings and conclusions are detailed in a RP report \cite{Aberle}.

RP studies \cite{Calviani} have shown that work cannot be carried out within \SI{8}{\metre} of the existing beam line. Again, this is an important factor as to how CE works can be planned and scheduled. This has been taken into account in this study.

\section{Civil engineering design description}

This section details the proposals for each part of the planned infrastructure including requirements and considerations in relation to CE.

\subsection{TDC2 junction cavern}

\begin{figure}[ht]
\centering\includegraphics[width=.80\linewidth]{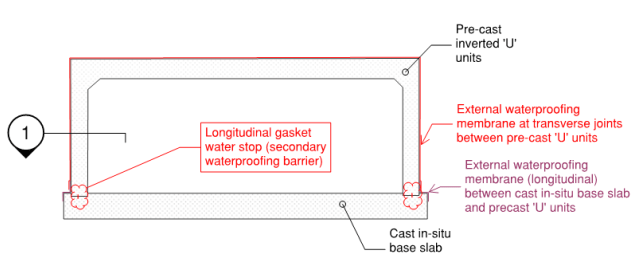}
\caption{Typical cross-section through the junction cavern showing the proposed form of construction.}
\label{fig:JCXsec}
\end{figure}

The junction cavern consists of a new tunnel measuring \SI{75}{\metre} long, \SI{5.3}{\metre} high and varying between 8 and \SI{16}{\metre} in width (external dimensions). The proposed design for the tunnel involves a cast \textit{in situ} reinforced concrete (RC) base slab and pre-cast RC `n' sections on top. Pre-cast sections were chosen for the increase in speed of construction and reduced time operatives will need to be next to activated soil. Sections will be between 1 and \SI{2}{\metre} in length. The cast \textit{in situ} base will reduce differential settlement. 

Waterproofing will be provided via a multi-layer passive system as shown in Fig.~\ref{fig:JCXsec}, with an external membrane and gaskets between sections. The junction cavern will provide a connection with the existing TDC2 tunnel, allowing sufficient space for extraction beam line equipment, services and future maintenance access. The dimensions of the cavern have been sized accordingly.

\begin{figure}[ht]
\centering\includegraphics[width=.87\linewidth]{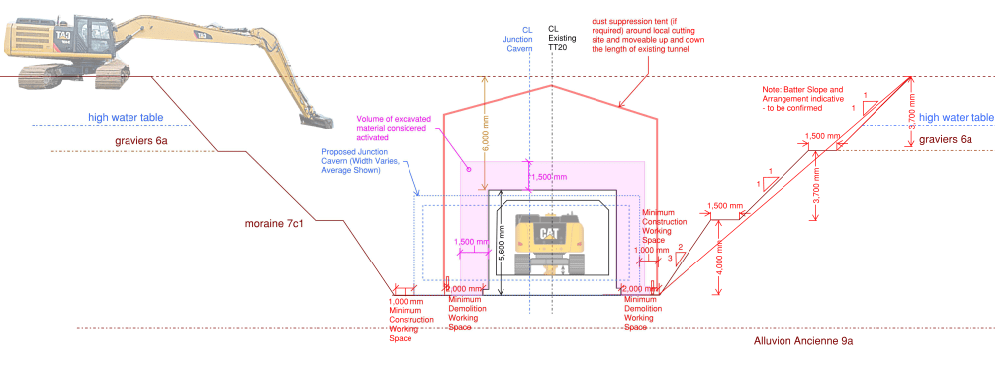}
\caption{Illustration of cross-section during works for traditional demolition of the existing TDC2 tunnel.}
\label{fig:JCXsecworks}
\end{figure}

To enable construction of the junction cavern, first the relevant part of the existing TDC2 tunnel must be demolished. A detailed options analysis was undertaken to determine the best demolition approach, cognisant of both civil engineering and radiation protection constraints \cite{ARUP2018-1}.

The outcome of the process was a decision to demolish TDC2 in full over the required length via traditional methods using crushing and hammering in an open cut excavation as shown in Fig.~\ref{fig:JCXsecworks}. This has significant advantages in reducing the duration of construction works in comparison with more specialist methods such as dry diamond rope saw cutting (DDRWC) into sections. The shorter duration led to a reduction in cost and radiation dose received for the construction workforce. DDRWC will however be used at the points where TDC2 will be retained to ensure a neat joint can be achieved.

It  should  be  noted  that  Arup’s  study concluded that a tent should not be needed for this operation. Standard methods of dust suppression and control of inhalation will be sufficient to prevent any internal dose issues with construction operatives. Spread of dust to areas outside CERN must be evaluated in more detail to ensure any dose received by the public is within acceptable levels. This will need to be evaluated further at later stages of project to ensure the measures in place will prevent any release of contaminated material. A tent is shown in Fig.~\ref{fig:JCXsecworks} for illustrative purposes only. 

Prior to demolition, approximately 100\,m length of machine and services will have to be removed from TDC2 to allow demolition. To prevent the spread of dust and debris from construction into the retained sections of TDC2, sealed walls will be built at each end. 

The demolished concrete and activated earth will be reused as backfill above the new junction cavern and extraction tunnel to avoid disposal off-site and to avoid producing additional activated soil in the future around the BDF facility. The soil shall be backfilled such that the most activated soil will be placed closest to the new junction cavern or extraction tunnel structure.

Construction sequencing will also be optimised to avoid double-handing i.e. storage of material prior to deposition will be minimised.

\subsection{Extraction tunnel}
\label{Sec:ExtractionTunnel}

\begin{figure}[ht]
\centering\includegraphics[width=.75\linewidth]{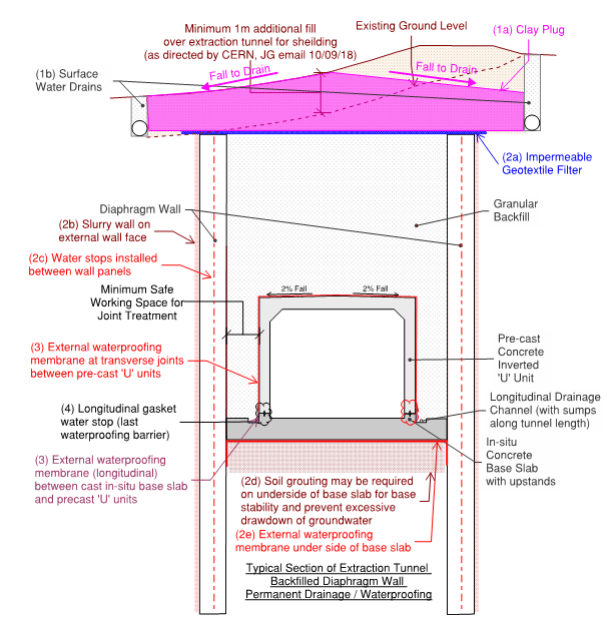}
\caption{Typical cross-section through the extraction tunnel showing diaphragm walls and multi-layer passive waterproofing systems.}
\label{fig:ETsec}
\end{figure}

Following a detailed integration study, the internal dimensions required for the extraction tunnel have been set. The resulting dimensions of the \SI{165}{\metre} long extraction tunnel will be \SI{5.8}{\metre} wide by \SI{5.1}{\metre} in height. 

The form of construction will be similar to that used for the junction cavern ie \SIrange{1}{2}{\metre} long pre-cast RC `n' units on an \textit{in situ} cast RC base slab. For the majority of its length, however, diaphragm walls will also be required for two reasons: to allow construction close to TCC2 and to significantly reduce the extent of excavation required in comparison to an open cut solution. Some propping within TCC2 may still be required for stability during works depending on the extents of asymmetric loading applied.

The junction cavern approach will also be replicated here for waterproofing, but with an additional clay `plug' and geomembrane used above the diaphragm walls and formalised drainage provided to collect surface water.

\begin{figure}[ht]
\centering\includegraphics[width=.87\linewidth]{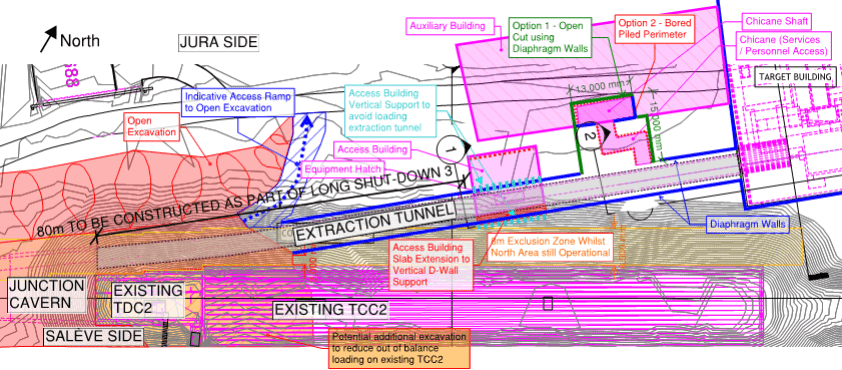}
\caption{Plan view of extraction tunnel showing extents of open cut (shaded red) and diaphragm walls (in blue).}
\label{fig:Worksplan}
\end{figure}

As noted earlier in the chapter, the works to construct the junction cavern and the first approximately \SI{80}{\metre} of the extraction tunnel (within \SI{8}{\metre} offset of the beam line) will need to be executed outside of operational beam runs. Despite selecting options to reduce the duration of these works as far as practical, this will still mean aligning the programme with a long shutdown. 

\subsection{Access and auxiliary buildings}

Above ground, the auxiliary and access buildings will be  basic steel portal frame structures with cladding to provide water-tightness and insulation.

The \SI{60}{\metre} long by \SI{20}{\metre} wide auxiliary building has a false floor to allow maintenance access for the large quantity of services contained within. It will be supported on RC shallow strip foundations since it does not directly bear on the extraction tunnel. 

 \begin{figure}[ht]
\centering\includegraphics[width=.87\linewidth]{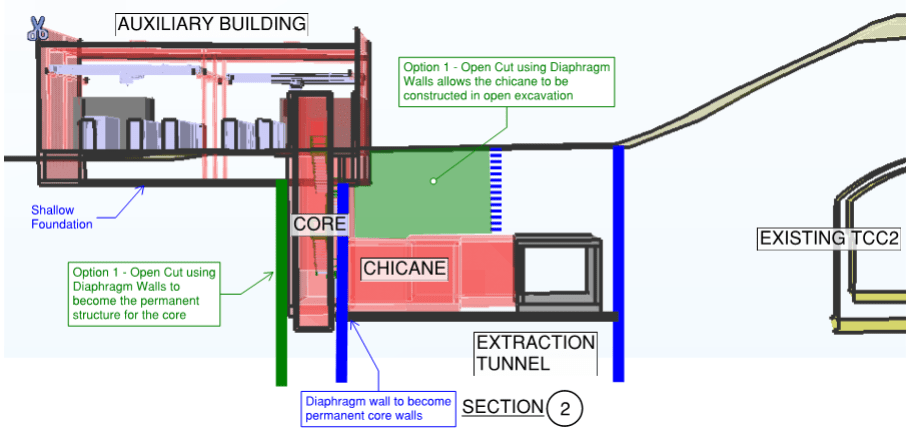}
\caption{Sectional schematic view showing the access core and chicane within diaphragm wall `box'}
\label{fig:Core}
\end{figure}

 At the auxiliary building, the extraction tunnel diaphragm walls will extend to form a `box' to enable construction within. This will maintain a watertight perimeter. The walls will continue to provide a permanent shell to an underground core providing personnel and service accesses to the extraction tunnel.
 
 The \SI{15}{\metre} long by \SI{12}{\metre} wide access building is located above the extraction tunnel. It will require a deep foundation to avoid uneven loading on the extraction tunnel. A transfer structure will be required to support the shaft and concrete shielding, potentially utilising the diaphragm walls via a shear connection if required.
 
\begin{figure}[ht]
\centering\includegraphics[width=.85\linewidth]{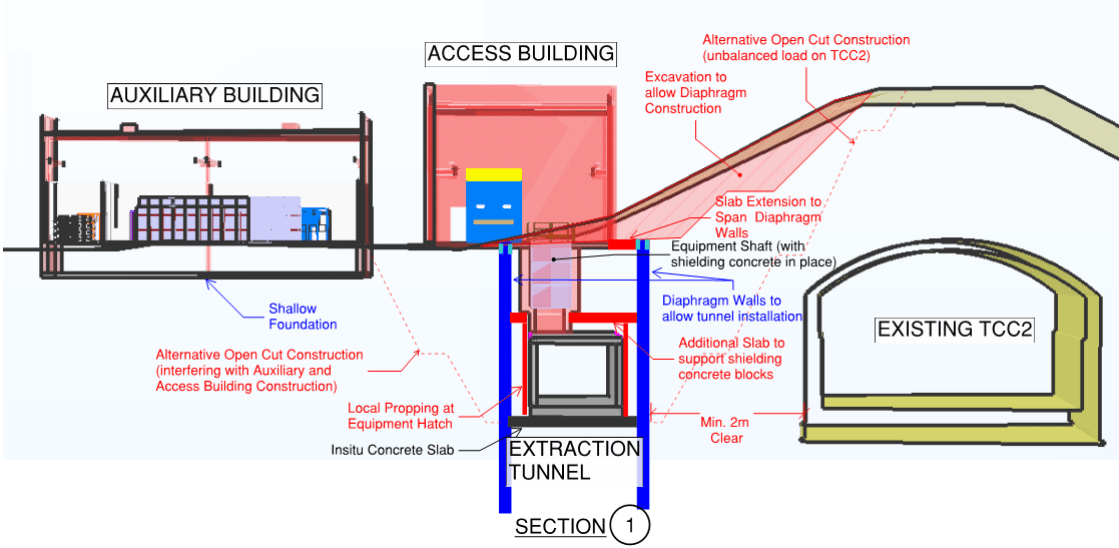}
\caption{Sectional schematic view showing the proposed arrangement around the access building heavy equipment shaft and the relationship with TCC2.}
\label{fig:Access}
\end{figure}

\subsection{Target complex}

The target complex will be a \SI{58}{\metre} by \SI{36}{\metre} building with \SI{15}{\metre} above ground and \SI{17}{\metre} below. The structure will be characterised at surface by a steel frame main hall equipped with a \SI{40}{\tonne} crane. The wall to the north-west, will however be part-formed from a retaining wall to accommodate a ramp. The basement structure will be formed from diaphragm walls with RC internal structures.

\begin{figure}[ht]
\centering\includegraphics[width=.58\linewidth]{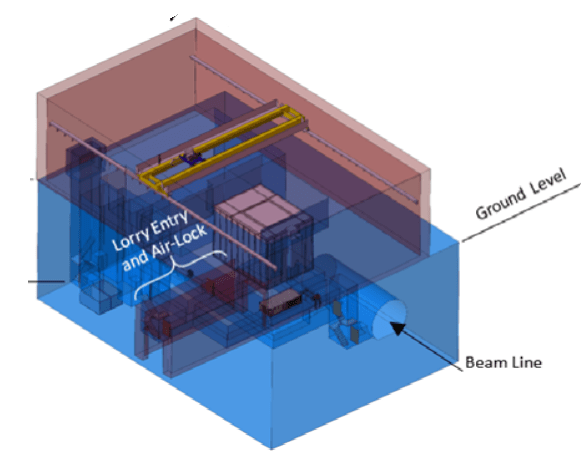}
\caption{Isometric view of `trolley concept' target building showing above ground surface hall in red with underground areas blue.}
\label{fig:Target}
\end{figure}

Several target handling options remained in discussion at the time the CE study was undertaken. The `trolley concept' option as shown in Fig.~\ref{fig:Target} was agreed on as a baseline for CE work. The principles of the CE study could be equally applied to other options nonetheless.

Due to potential radiation contamination, special measures need to be taken in the target complex to minimise the amount of groundwater that is able to seep into the underground facility or alternatively permeate from within to the surrounding soil and groundwater. This has dictated the use of diaphragm walls which provide the best solution in terms of permeability to groundwater. Further detail on the necessary measures is noted within the full civil engineering feasibility review \cite{ARUP2018-2}. 

The diaphragm walls will be lined to provide an acceptable interior to the basement of the complex. 

Tolerances and settlement were looked at as part of the study since they are key to the operation and installation of the target and helium vessel (being as they are very sensitive to movement). 

Slab roughness and evenness in the area supporting the helium vessel are critical to evenly distributing the very significant point loading generated by the cast iron shielding. An evenness of \SI{3}{\milli\metre} per \SI{3}{\metre} straight edge could be achieved by use of a combination of the following measures: 
\begin{itemize}

\item Enhanced preparation of sub-grade (if ground bearing) or specialist formwork and falsework; 

\item Pouring concrete in thin strips (would add to works duration); 

\item Specialist concrete mixes to reduce shrinkage;

\item Trial slabs; and 

\item Self-levelling toppings (may be placed at final fit-out stage to allow construction traffic). 

\end{itemize}
 
Steel lamina reinforced elastomeric strips or pads, similar to those used for bridge bearings are also suggested to reduce the sensitivity to any discontinuities. Elsewhere, tolerances are achievable within industry norms for a very flat/even specification.
 
 An overarching strategy has been adopted to minimise settlement between the extraction tunnel, target complex and experimental area. The following measures are proposed:
 
 \begin{itemize}
     \item The target complex is massive and will act as an anchor point for other structures which will be stabilised by it. 
     \item Shared diaphragm walls between the target complex and both the experimental area and extraction tunnel will help to limit differential movement between areas.
     \item Specific connection details at the joints between areas.
     \item This arrangement will allow continuity of base slab between target and experimental areas.
    
 \end{itemize}

\begin{figure} [ht]

\centering

\begin{subfigure}[b]{0.4\textwidth}
            \includegraphics[width=\textwidth]{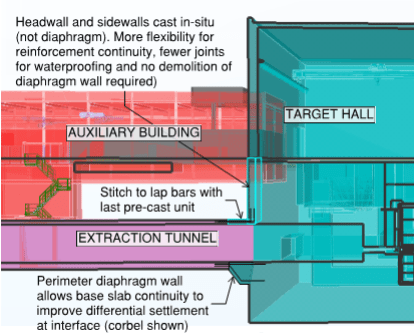}
    \label{fig:d1}
    \end{subfigure}
\begin{subfigure}[b]{0.4\textwidth}
            \includegraphics[width=\textwidth]{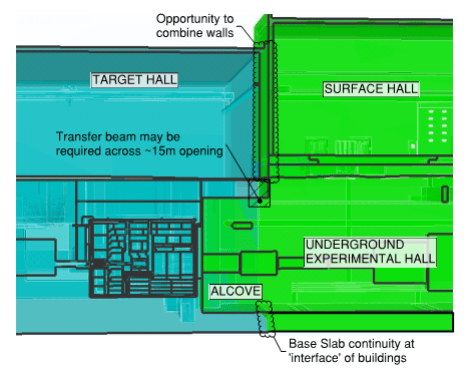}
    \label{fig:dd1}
    \end{subfigure}

    
   \caption{Sectional schematic views showing the specific connection details to be used at interfaces between target complex and extraction tunnel (left) and experimental area (right).}
   \label{joints}
   \end{figure}

Since the target is the most sensitive element for settlement, differential settlement will be controlled relative to it. The above measures should be sufficient to limit differential settlement to levels which can be dealt with through adjustment of supports along the beam line and detector equipment. Elsewhere, however ground movements are hard to predict and detailed numerical analysis will be required as part of design development. 

The target complex is also characterised by several underground areas with different size, depth and internal dimensions (i.e. target bunker, muon shield tunnel, cooling and ventilation rooms and a storage area) as well as a trolley and crane for remote handling and operation. The technical aspects of these areas have been fully described within the target chapter. 

\subsection{Experimental area}

\begin{figure}[ht]
\centering\includegraphics[width=.87\linewidth]{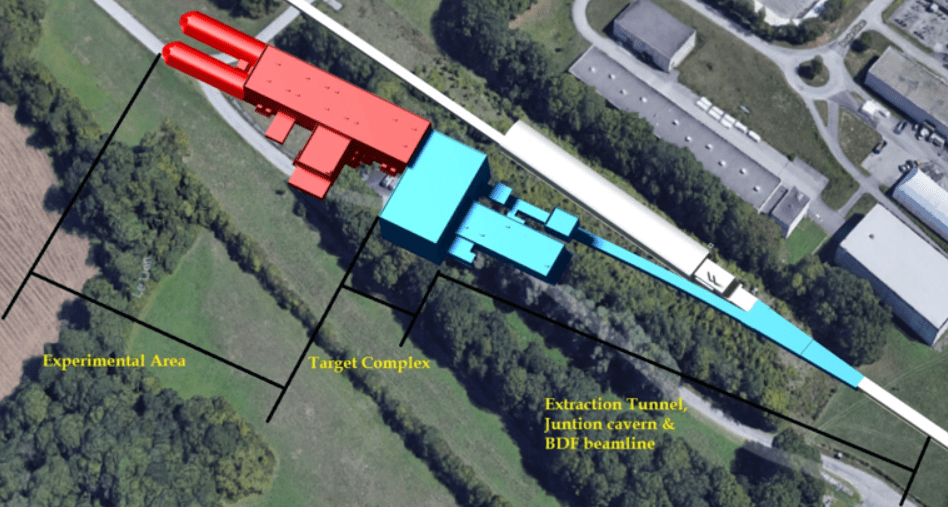}
\caption{3D Aerial view of experimental area in relation to other infrastructure}
\label{fig:Experimentalarea}
\end{figure}

The experimental area's main structures are the underground experimental hall and surface hall. 
The experimental hall measures \SI{120}{\metre} in length by \SI{20}{\metre} width with two sections of 88 and \SI{32}{\metre} length at floor levels of 16.5 and \SI{19}{\metre} depth below ground level respectively. The building will be composed of a diaphragm wall perimeter with a large RC base slab. At the top of this `box', another RC floor slab spans between walls with three openings of 14.5 by \SI{18}{\metre} to allow vertical access for assembled components from above. The openings will be filled with eighteen \SI{1}{\metre} deep removable precast, pre-stressed pre-tensioned RC beams. The underground area will be equipped with two cranes of 40 and \SI{80}{\tonne} capacity running at the same height. Access will also be provided for services and personnel via an underground core with stairs and lift, offset from the main hall. An alcove projects into the target hall at the joint between structures as shown in Fig.~\ref{joints}. The foundation for the hall will be thickened beneath main detector components where necessary to accommodate the large punching shear and allow distribution of loads. 

The surface hall will be a more conventional clad steel frame structure, \SI{100}{\metre} long, \SI{26.5}{\metre} wide with a maximum height of \SI{16.5}{\metre}. The structure will be just slightly wider than the hall below. This building will be served by independent \SI{40}{\tonne} and \SI{10}{\tonne} cranes supported by separate rails. 

It is envisaged that the experimental hall could be extended a further \SI{100}{\metre} on the same alignment if required in the future for further developments, assuming similar extents of shielding and fenced supervised area beyond.

\subsection{Service and gas buildings}

A three storey service building with basement structure will be provided within the experimental area, although also providing facilities for the target complex. This building measures \SI{20}{\metre} by \SI{34}{\metre} in plan, extending \SI{12}{\metre} above ground and \SI{2.4}{\metre} below. The building extends on the Jura side at ground level to house a workshop with \SI{5}{\tonne} capacity crane. The building foundation will also form the basement structure with a perimeter of RC cantilever retaining walls sharing a common footing constituting a floor slab. The reminder of the structure will be a simple steel frame building with lift and stairs in an RC core providing stability. This building is at a preliminary design stage. 

A basic steel frame gas building measuring \SI{15}{\metre} by \SI{10}{\metre}, \SI{3}{\metre} tall on a simple RC raft foundation is to be located close-by.

\subsection{Ancillary infrastructure and general considerations}

\subsubsubsection{Loading}

Detailed specifications have been produced for loading in each area by the teams responsible in collaboration with the FAS team \cite{Dougherty1, Dougherty2, Kershaw, Santos-Diaz}.
These have been used in the development of CE proposals.

\subsubsubsection{Site levels and hardstanding}

The existing site levels vary significantly and the design has looked to tie into these, while ensuring the finished floor levels of buildings are suitable to avoid surface water flooding issues \cite{Arcadis}. Approximate finished levels for areas of hard standing have been proposed along with a ramp of suitable gradient between differing site levels outside the experimental area and for the remainder of the site as noted in Fig.~\ref{fig:Layout}. A requirement for a maximum gradient of 6\% has been respected in line with the requirements of CERN's transportation service. 

Vehicle swept path analysis has been carried out to ensure the operation of the access roads and access to facilities is suitable for all necessary vehicles.

Assumptions have been made on areas of parking to be provided for each facility which will need to be refined at a later stage when building occupancy is fully defined. Estimated parking needs can be accommodated within the area of hardstanding provided. Specific parking areas are not shown on drawings.

\subsubsubsection{Access roads and technical galleries}

An access road will be provided linking to existing roads at Route de Broglie and an unnamed road to the south-west, allowing access to each part of the BDF facility. 

Technical galleries have been assumed alongside access roads to allow connection of proposed facilities to existing service networks. \SI{2}{\metre} wide, \SI{1}{\metre} deep culvert-type galleries with  RC removable concrete lids at regular intervals are envisaged for access. An indicative layout is shown in Fig.~\ref{fig:Layout} but this will need to be refined when service supplies are studied in more detail.

\subsubsubsection{Earthworks}

The volume of earthworks arising out of the project is significant, predominantly because of the substantial underground structures required. Excavated material will be reused as far as possible for backfill above structures and to create the required levels across the site. A large amount of fill material will be required to bring the access road to the north and hardstanding levels at various points across the site up to the required levels. Imported structural fill will be used to provide a solid foundation for the roads and hardstanding, while lower levels, verges and the surrounding area will be infilled with excavated material. Careful selection and control of materials during placement and re-compaction will be needed close to roads. 

There is still a large volume of excess `cut' material beyond this and it is assumed these can be stockpiled permanently on site to minimise costs. Areas have been identified suitable for stockpiling. Weaker materials will be permanently stockpiled in mounds.

\subsubsubsection{Drainage}

Drainage needs have been looked at along with the impact on existing site drainage. Outline design assumptions have been made in order to enable costing although a full design cannot be completed until the capacity of existing drainage systems is better understood.

\subsubsubsection{Seismic design}

According to the \textit{`D\'{e}cret no 2010-1255 du 22 Octobre 2010 portant d\'{e}limitation des zones de sismicit\'{e} du 
territoire fran\c{c}ais'} CERN is classified as seismic zone 3, \textit{`sismicit\'{e} mod\'{e}r\'{e}e'}. The requirements for the project for dealing with seismic response should be achievable through careful design to Eurocode 8. The seismic response is highly dependant on ground conditions, so more detailed work will be needed on this following ground investigation at the next stage of development. 

\subsubsubsection{Existing infrastructure}

`Zone 9103' buildings and an area of hardstanding will need to be re-located or re-provided as part of the project.

\section{Recommendations for work at the next stage of project development and opportunities} \label{Recommendations}

The CE study carried out has necessarily worked to an appropriate level of detail for this stage of project development. At the next stage, more detailed studies will be required to inform the detailed design. The following further studies and investigation work is identified:

\begin{itemize}

\item A detailed ground investigation will be required to confirm or disprove existing assumptions and in particular obtain more detailed boreholes extending to bedrock in this area. The opportunity should be taken to gain detailed information on groundwater movement and perched water tables. 

\item Activation of soil around the existing TDC2 should be confirmed as this would help to enable planning and tendering of works.

\item Further study will be required to confirm service supplies from existing networks and then optimise the location of technical galleries needed.

\item Consideration will be required to avoid or minimise the impact on adjacent beam line operations and experiments during construction work. Vibration could be an issue and mitigation measures may be required for work carried out during ‘beam-on’. Tunnel monitoring is also likely to be required during works.

\item Flood risk assessment and basic hydrological modelling to ensure the risks and requirements are fully understood to be able to confirm design levels for parking areas, thresholds and waterproofing details.

\item A ground penetrating radar survey should be carried out to confirm the location of services.

\item A survey and study of existing drainage systems should be made to assess condition and capacity in order to design connections and replacement drainage systems where affected.

\item An optimisation exercise for reduction or substitution of concrete fills in the target complex should be carried out.

\end{itemize}

It is recommended that, if the project receives approval, the additional work be carried out as soon as possible.

\printbibliography[heading=subbibliography]

\begin{appendices}
\renewcommand{\appendixname}{Appendix}
\fancyhead[RE]{\textsc{\appendixname\ \thechapter}}
\fancyhead[RO]{}

 \chapter{Road Map}
\section{Status}

In 2016 the BDF team were charged by CERN management to complete key technical feasibility studies of the proposed facility in time for the European Strategy for Particle Physics (ESPP) update process.
This was in conjunction with a recommendation by the CERN SPS and PS Experiments Committee (SPSC) to the SHiP experiment to prepare a Comprehensive Design Study as input to the ESPP update. 
Material and personnel resources were made available to perform in-depth studies and prototyping.
The key areas addressed were:  extraction and beam transfer; the target and target complex; radiation protection; safety engineering; integration; and civil engineering. 

Since inception, there has been sustained effort with good progress on all fronts as described in detail in this report. 
In brief, the study has addressed all pertinent technological challenges,
and in-depth studies and prototyping have been performed, or are already well underway, for all critical components and systems.
Through a mixture of novel hardware development, beam physics and technology, the study and prototype validation
have shown that the SPS can deliver the beam with the required characteristics, with acceptable losses, to a robust target housed in a functional and secure target complex. The radiological implications appear manageable, 
the technologies and their deployment,  although challenging, appear to be within CERN's established competencies,
and the project, given the resources, is ready to move towards the detailed design and execution phase.


\section{Feasibility study phase}

The feasibility study phase will come to an end with the delivery of this Comprehensive Design Study which is timed to coincide with the ESPP update process.

With the go ahead from CERN management, under the continuing auspices of Physics Beyond Colliders,
it is foreseen to continue these studies in 2019 and 2020.
Specific areas include:
\begin{itemize}
    \item Post Irradiation Examination (PIE) of the prototype target tested in 2018;
    \item helium vessel design and preparation for prototyping;
    \item construction and testing of the prototype laminated switch/splitter magnet;
    \item intermediate iteration of system integration;
    \item development and possible prototyping of the hadron absorber magnetisation system;
    \item continued studies of loss reduction techniques during slow extraction in the SPS and the deployment of these techniques onto operational beams (planned for the SPS run starting in 2021); 
    \item site investigation required for a definitive civil engineering study. 
\end{itemize}

\section{Studies during the TDR phase}

By 2020, the BDF team will to be in a position to seek approval to go ahead with the
preparation of a technical design report (TDR).
While working towards the delivery of a TDR, it is envisioned to commence:
\begin{itemize}
    \item detailed engineering studies and specification of the deliverables for
    both standard and novel systems;
    \item  detailed integration plans;
    \item civil engineering pre-construction activities i.e.
    environmental impact study; build permit submission/approval; tender process and detailed civil engineering  design;
    \item definitive radiation protection studies.
\end{itemize}
Appropriate progress with these activities -- some would follow after approval -- could lead to the delivery of the TDR in 2022.
The aim would be to seek project approval in 2023, allowing subsequent project execution to take advantage of the North Area stop during LS3. 

A summary of the possible evolution of the BDF project is shown in Table \ref{tab:time-line}.

\begin{table}[h]
\begin{center}
\caption{Outline of a possible BDF time-line}
\label{tab:time-line}
\begin{tabular}{lp{10cm}}
\hline
2020 & Continued design studies and prototyping \\
\hline
End 2020 & Approval to go ahead with TDR \\
\hline
2021 - 2022  & Engineering design studies towards TDR \\
 & Detailed integration studies \\
 & Specification towards production  \\
 & Begin CE pre-construction activities: environmental impact study;
 detailed CE design and pre-tender process. \\
 \hline
2022  & TDR delivery \\
\hline
 2023 & Seek approval \\
 2023+ &  Tender, component production, CE contracts  \\
\hline
\end{tabular}
\end{center}
\end{table}

Some further investigation work as detailed in the Civil Engineering chapter will need to be carried out before detailed civil engineering design can begin with specialist external consultancies. Once these designs are complete, tendering for the civil construction contracts can start. 
In parallel, an environmental impact study must be prepared and approved by the local authorities, prior to the timely submission of the building permit application for the project, to allow construction works to commence. 

\subsection{Civil Engineering Schedule and Resource Considerations} \label{programme}

\begin{figure}[ht]
\centering\includegraphics[width=.87\linewidth]{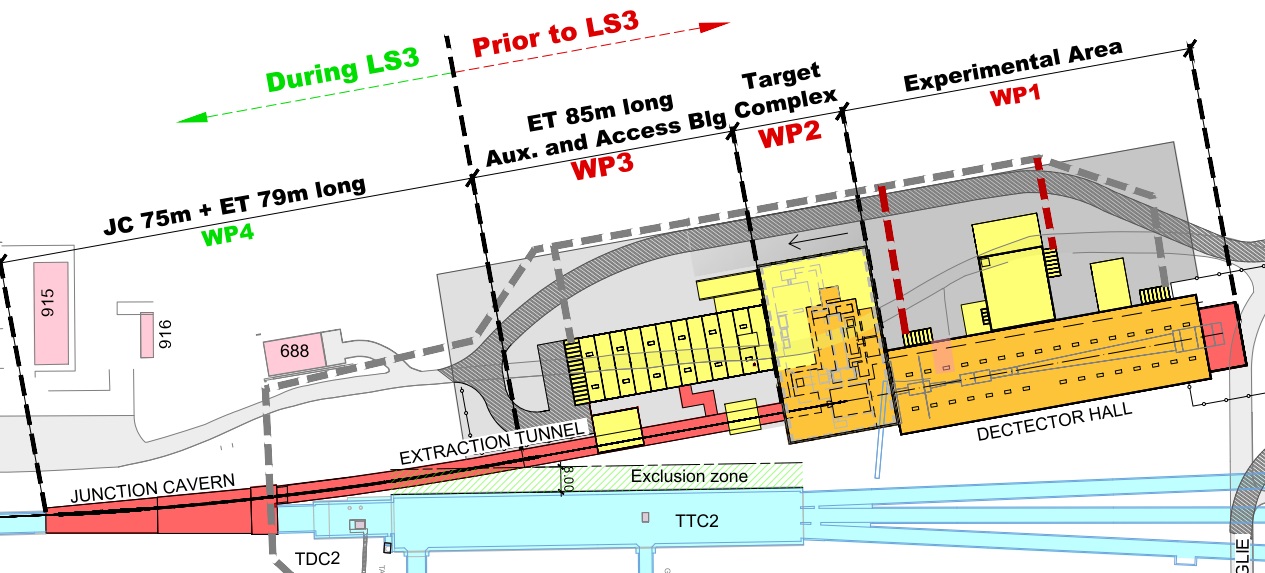}
\caption{Indicative division and sequence of work packages for delivery of BDF project}
\label{fig:Workpackages}
\end{figure}

The Civil Engineering work is split into four main packages, as indicated in Fig. \ref{fig:Workpackages}.  WP1, WP2 and WP3 activities  can be performed during normal beam operation in the North Area, where the only constraint is a requirement to maintain \SI{8}{\metre} of earth shielding to TCC2 given by Radiation Protection considerations. This is discussed in more detail in the civil engineering chapter (\ref{Chap:CivEng}) of this report.

WP4 includes all the activities that must be carried out during beam stop which is likely to be during the injectors' Long Shutdown 3 (LS3). This will include the junction cavern demolition and reconstruction along with the construction of the extraction tunnel up to a point where the distance between proposed extraction tunnel and TCC2 increases to \SI{8}{\metre}.

The current beam operation schedule includes approximately 12 months of shut-down without beam during the injectors' LS3. This is  unlikely to be sufficient to carry out the required civil engineering works. Other activities such as the machine and tunnel cool-down, infrastructure dismantling/re-installation and testing/commissioning must also be considered in the works schedule in order to calculate the required period of North Area shutdown.
An extension of the North Area shutdown to, say, 24 months would also be beneficial to address important consolidation requirements of the overall SPS NA complex. Any extension of an injectors' long shutdown would be in parallel with an ongoing LHC stop.

The working assumption for the moment is that the re-construction of the junction cavern and construction of the first part of the new transfer line will take around 1.25 years. The required cool-down period after the stop of North Area operation could be partially covered by an ion run at the end of the preceding operating period. As this work package is not on the critical path, a generous cool down period can be allowed increasing the duration of the work package to 1.75 years including 1.5 months for removal of services and equipment.

In principle, work packages can be partially executed in parallel, although this needs to be studied in more detail and carefully set out in contract documents to avoid cost increases from contractors working in close proximity to one another. Various activities, including access roads, technical galleries and other non-area specific tasks, have been included as miscellaneous works and are not assigned specifically to any work package. It is likely these activities will need to be done as advanced enabling works or as part of the first work package (WP4).

The key drivers of the construction phase are outlined below.
Figures shown do not include contingency. 
Radiological considerations have been taken into account.

\begin{itemize}
    \item  { \bf Civil engineering underground works} are estimated to take 3 years:
    
 \begin{itemize}
     \item 1.75 years for the re-construction of junction cavern and the construction of the first part  of the new transfer line including cooldown etc. (WP4)
     \item 1.25 years for the final part of the new transfer line and access shafts (WP3)
     \item 1.5 years for the target complex (WP2)
     \item 1.75 years for the experimental hall  (WP1)
    \item 0.75 years for the service buildings' foundations
 \end{itemize}  

\vspace{4 mm}
    
    \item {\bf Civil engineering surface works}, following on from the associated underground works, are estimated to take 2.5 years:
    
 \begin{itemize}
     \item 0.75 year for access building and auxiliary building
     \item 1.0 year for service building and target hall
     \item 1.0 year for the experimental hall
     \item 1.0 year for experimental area service buildings
 \end{itemize}   
    These work packages can be executed in parallel with underground work where appropriate.
    
\end{itemize}

\subsection{Component production and installation}

Component production and preparation can continue during the execution of the civil engineering phase.
This would include production/preparation of the standard beam line elements (magnet refurbishment, splitters, power converters, beam instrumentation, vacuum etc.) and delivery of the target assembly, helium vessel, target handling, shielding and other components for the target complex.

Installation can be staged as the given structures become available. 
Approximate time estimates for this phase are given below.
    \begin{itemize}
    \item Junction cavern hardware and services: 6 months re-installation including testing and commissioning;          
    \item Installation of hardware and services for the extraction beamline: 6 months;
    \item Installation of hardware and services in the target complex: 2.5 years;
    \item Access and auxiliary building installation of hardware, services, access and safety systems, metallic structure, electrical and cabling activities IT network: 15 months;
    \item Experimental hall: the entire installation phase, including detector, around 2.5 years.
    \end{itemize}

\subsection{Schedule}

The medium-term schedule (circa 2019) of the injector complex and the LHC, foresees  long shutdown for the injectors in 2025 (LS3). 
LS3 for the LHC stretches from 2024 to mid-way through 2026.

A indicative schedule for the preparation and execution of the BDF project is shown below in Table \ref{tab:execution-time-line}.
Uncertainties are significant at the study phase of the project. 
Time estimates on the civil engineering front are based  on  the knowledge acquired from the construction of previous similar projects at CERN, 
and the analysis performed in conjunction with an external company as described in chapter \ref{Chap:CivEng}.
The production and preparation of beamline components is well-versed.
The truly novel exercise will be the installation, testing, and commissioning of the target complex, and this will be an exploratory process for CERN.

\begin{table}[h]
\begin{center}
\caption{Outline of an indicative BDF project execution time-line}
\label{tab:execution-time-line}
\begin{tabular}{lcl}
\hline
Indicative dates & Years & Activity  \\ 
\hline 

2023 - 2024 & 2.0 & \textbf{CE pre-construction}  \\
            & & Environmental impact study  \\
            & & Building permit submission/approval  \\
            & & Tender and CE detailed design  \\
\hline

2023 - 2025 & 3.0  & \textbf{Component production} \\
            & & Beamline systems and components  \\
            & & Tender technical services production \\
            & & Target assembly  \\
            & & Target complex handling systems etc. \\
\hline

2025-2027 & 3.0  & \textbf{Underground CE}  \\
            & 1.25 & Junction Cavern/Beamline-part-1 \\
            & 1.25 & Beamline-part-2/Access building  \\
            & 1.5  & Target complex  \\
            & 1.75 & Experimental hall  \\

\hline

2026 - 2028 & 2.5 &  \textbf{Surface CE}  \\
            & 0.75 & Access and auxiliary buildings  \\
            & 1.0 & Service building/Target Hall  \\
            & 1.0  & Experimental hall   \\
\hline
2026 - 2028 & 2.5 & \textbf{Installation} \\
            & 0.5 & Junction Cavern/Beamline-part-1  \\
            & 0.5 & Beamline-part-2  \\
            & 1.25 & Access and auxiliary buildings  \\
            & 2.5 & Service building/Target Hall  \\           
            & 2.0 & Experimental Hall  \\

\hline
\end{tabular}
\end{center}
\end{table}

\FloatBarrier
Note:
\begin{itemize}
    \item Component production and preparation can continue in parallel with civil engineering.
    \item Non junction cavern work packages can take place in parallel with beam operation. They would be staged and start as soon as feasible.
    \item Surface work can begin when associated underground structures are complete or in some cases before to provide protection from the weather.
    \item The overall goal is to commence data taking as soon as early as reasonably possible in Run 4.
\end{itemize}



\chapter{Cost Estimates}

\section{Introduction}

The cost estimates presented here essentially map on to the work package breakdown of the study:
\begin{itemize}
    \item Extraction targeting SPS losses, activation and mitigation measures
    \item TT20, switch, and new transfer line design
    \item Target and target complex
    \item Radiation protection
    \item Safety engineering 
    \item Integration (junction cavern, beam-line, target complex, detector hall) 
    \item Civil engineering 
\end{itemize}

Overall a class 4 (intermediate -- concept study or feasibility) cost estimate has been performed.
An outline justification for this categorization follows.



\section{Material}

\subsection{Junction cavern and dump line}

In term of relative expense, the slow extraction system of the SPS and the transfer line TT20 are in place, 
and, although critical, future developments will not represent a major capital investment.
Once constructed the new beam line is relatively standard CERN design and the associated cost estimates for 
services and integration can be taken as solid.
The hardware for the switch and new transfer line are well defined and certainly well within CERN's expertise
and, again, the figures quoted below can be taken with a reasonable degree of confidence.

The hardware cost for the junction cavern and the new transfer line to the target complex are shown in Table \ref{tab:cost-summary}.
The costs for global systems (cooling and ventilation, electricity distribution,
survey, access system, radiation monitoring, and handling engineering)
are covered in subsection \ref{ss-GS} below.

\subsection{Target and Target Complex}

The target itself is a challenging development. Good anticipatory understanding of the costs has been gained
in the production of the prototype and it will certainly be a relatively expensive component.

 The Target Complex design team have worked with an external company that specialises in high radiation complex design and handling. 
In collaboration with this company two handling concepts, trolley and overhead access, were developed and costed
(see chapter \ref{Chap:TargetComplex}).

In the summary shown in table \ref{tab:cost-summary}, the trolley concept is taken as baseline.
Another proposal incorporating features of both concepts is under study with 
anticipated costs potentially lower the pure trolley concept.

\subsection{Global Systems}
\label{ss-GS}

The standard global systems have benefited by working with the respective CERN groups 
(cooling and ventilation,  access,  radiation protection, safety, survey).
The groups are generally well versed in the production and deployment of the foreseen solutions and the estimates may be considered sound. Due to lack of resources, the cost of electrical distribution are based on a previous estimate and must be regarded as approximate.
The cost estimates for global systems and services are summarized in table \ref{tab:cost-summary}

\subsection{Civil engineering}

The largest work package in terms of cost, civil engineering, has benefited from collaboration with 
specialised external companies with a history of working with CERN.
The companies  have scrutinised and endorsed the CE cost estimate presented below.
Here uncertainties have been reduced with respect to the 2015 exercise \cite{2015arXiv150404956S} and the quoted total
can be regarded with some degree of confidence. 
However, until the required ground investigations are performed, the estimate remains class 4.

\subsubsection{Basis of estimate}

The cost estimate for the BDF Project has been based on the layouts presented in the civil engineering chapter. The estimate produced includes all aspects of construction, detailed engineering design work and construction management except where stated otherwise. Many of the rates used to formulate this estimate have been based on real construction costs from the large hadron collider experience (1998-2005), from consultant (ILF's) future circular collider cost studies \cite{ILF} and following recent tendering for similar projects at CERN such as CENF and the HL-LHC. In addition to this process, Arup have independently reviewed the cost estimate as part of their scope of work and the results have been fed back into the costing process to ensure a robust estimate. 

The civil engineering activities have been split into four primary work packages as shown in the chapter discussing the BDF Roadmap. 
The provisional cost for the main tasks identified and included within each package is shown in the summary table \ref{tab:cost-summary}.

\begin{table}[htb]
\begin{center}
\caption{Summary cost estimates for the CE work packages}
\label{tab:cost-summary}
\begin{tabular}{lc}
\hline
Work package & Cost [MCHF]      \\
\hline	
Work Package 1  & 	31.7 \\
Work Package 2 & 	16.5 \\
Work Package 3 & 	10.9 \\
Work Package 4  & 	7.3 \\
Work Package 5  & 	1.3 \\
Site investigation  & 	0.6 \\
\hline
Facility total & 	68.3 \\
\hline
\end{tabular}
\end{center}
\end{table}

The accuracy of the estimate is considered Class 4 - Study or Feasibility which could be 15 to 30\% lower or 20 to 50\% higher (in line with AACE international's best practice recommendations \cite{christensen} as has been used for previous CERN projects). Until the project requirements are further developed, it is suggested that a suitable band to adopt would be -20\% to +40\% for CE costs.

\subsubsection{Costing assumptions and exclusions}

The cost estimate is based on the following assumptions:

\begin{itemize}
\item Costs have been based on retaining spoil on CERN land with no tipping and disposal costs. If this were to change, the cost increase would be significant.
\item The proposed drainage can be connected into existing drainage without significant capacity enhancement of the existing.
\item Build up of hardstanding areas and access road can be carried out using site won material with only 1.5\,m depth road construction.
\item Radiation protection measures required for the works and assumptions agreed are as stated in \cite{ARUP2018-1,ARUP2018-2}.
\item The construction works programme will be as stated in the schedule as noted in the BDF roadmap chapter. This programme is outline and will need to be reviewed to optimise activities to allow multiple work packages to progress in parallel. 
\item Ground and groundwater conditions do not vary significantly from those previously found in the area.
\item No improvements are required to the wider Pr\'{e}vessin site for office space, restaurant, main access, road network etc
\item Building 687 will be demolished and a replacement constructed elsewhere while de-mountable buildings 6357 and 6575 and a container - 6361, will be relocated. An area of hardstanding will be re-provided to replace the existing.
\item Assumptions have been made on the required depth of diaphragm walling, however the cost is very sensitive to changes in depth due to the large proportion of total scheme costs relating to diaphragm wall construction (approx. 27\%).
\item Allowances have been made for temporary propping of TCC2 and excavations during diaphragm wall construction, however this will depend on ground conditions meaning they are subject to change.
\item An allowance has been made for four tensile fabric structures to temporarily house activated soil before re-use with two thirds of material re-used directly without intermediate storage. 
\end{itemize}

All temporary facilities needed for the civil engineering works have been included in the cost estimate, but nothing for any temporary areas/buildings needed for machine or detector assembly/installation. 

For clarity, the overall cost estimate does not include:

\begin{itemize}
\item SMB-SE Resources 
\item Special foundation support for facilities (e.g. Detector, etc) 
\item Shielding precast concrete blocks 
\item Tenting over excavation and demolition of TDC2
\end{itemize}

\FloatBarrier
\subsection{Material Costs: Overall Summary}

A breakdown of the material cost estimate is shown in table \ref{tab:cost-summary-material}.
In this table, a breakdown of the items in each of the main work packages is shown.

\begin{table}[htb]
\begin{center}
\caption{Summary of material costs for BDF construction}
\label{tab:cost-summary-material}
\noindent
\begin{tabular}{@{}lc@{}}
\hline
Work package & Estimate  [MCHF] \\
\hline

{\bf Civil engineering} & \bf{68.2}  \\
\hline	
\hspace{3mm} Work Package 1  & 	31.7 \\
\hspace{3mm} Work Package 2 & 	16.5 \\
\hspace{3mm} Work Package 3 & 	10.9 \\
\hspace{3mm} Work Package 4  & 	7.3 \\
\hspace{3mm} Work Package 5  & 	1.3 \\
\hspace{3mm} Site investigation  & 	0.6 \\

\hline
{\bf Extraction and beamline }  &    \bf{9.6}   \\
\hspace{3mm} Magnet refurbishment & 0.7 \\
\hspace{3mm} Switch/Splitters    & 2.1 \\
\hspace{3mm} Dilution system  (MDX)  &   0.52   \\
\hspace{3mm} Power converters   & 3.0 \\
\hspace{3mm} Beam instrumentation &  0.8   \\
\hspace{3mm} Vacuum system      &  0.6 \\
\hspace{3mm} Interlocks         &  0.2 \\ 
\hspace{3mm} DC cabling & 1.62 \\

\hline
{\bf Target and target complex }  & \bf{45.5} \\
\hspace{3mm} 	Target assembly \& annex equipment &		5.6	    \\
\hspace{3mm} 	Instrumentation package &	 	0.9	 \\
\hspace{3mm} 	Bunker iron shielding \& US1010 &		14.0  \\
\hspace{3mm} 	Concrete shielding &		0.85	  \\
\hspace{3mm} 	Magnetized coil for US1010 shield &		1.0	 \\
\hspace{3mm} 	Target \& coil handling system &		19.1 \\
\hspace{3mm} 	Target and coil shielding casks	 &	0.2 \\
\hspace{3mm} 	Helium vessel (HW) &		2.3  \\
\hspace{3mm} 	Integration	 &	0.6	  \\
\hspace{3mm} 	Beam window(s)	 &	0.6	  \\
\hspace{3mm} 	Collimator(s) &		0.5	  \\

\hline
{\bf Infrastructure and services } &  \bf{31.5} \\
\hspace{3mm} Cooling & 7.4 \\
\hspace{3mm} Ventilation & 6.3 \\
\hspace{3mm} Electrical distribution  &   5.6  \\
\hspace{3mm} Survey and alignment  & 1.1   \\ 
\hspace{3mm} Access system &   1.4  \\
\hspace{3mm} Fire safety system   &  3.5  \\
\hspace{3mm} Radiation monitoring &  0.4   \\
\hspace{3mm} Transport (inc. cranes and lifts)  &  5.1 \\ 
\hspace{3mm} Controls infrastructure & 0.7 \\

{ \bf Integration and installation} & \bf{3.6}  \\
\hspace{3mm} Installation & 2.6  \\ 
\hspace{3mm} Installation, design support & 0.6 \\
\hspace{3mm} Preparation, dismantling  &  0.4 \\
\hline
{ \bf Total } &     \bf{158.4}  \\
\hline
\end{tabular}
\end{center}
\end{table}

\FloatBarrier

A high-level summary of estimated material costs for construction is shown in table \ref{tab:cost-summary-material-overview}.

\begin{table}[htb]
\begin{center}
\caption{Overview summary of the material cost estimates for the main BDF work packages.}
\label{tab:cost-summary-material-overview}
\begin{tabular}{lc}
\hline
\textbf{Work Package/System} & \textbf{Estimate [MCHF]}      \\
\hline
Beam-line and junction cavern hardware       & 9.6  \\
Target and Target Complex infrastructure     &  45.5  \\
Civil Engineering        &  68.2  \\
Cooling and ventilation  &  13.7  \\  
Electrical distribution  &   5.6 \\
Survey and alignment  & 1.1   \\
Access, safety, RP, controls &   6.0 \\
Transport (inc. cranes and lifts)  &  5.1 \\
(De)Installation & 3.6 \\
\hline
Total &  158.4 \\
\hline 
\end{tabular}
\end{center}
\end{table}

\subsection{Personnel}

The breakdown shown below addresses foreseen requirements for staff and fellows, both classes of which are identified as members of CERN personnel (MPE).
The installation of hardware, cabling etc. will generally be performed by field support units (FSU).
This is covered in the material budget summarized above.

The personnel estimate is based largely on the experience of the groups concerned.
For well-established activities such as power converters and magnets, the estimates can be 
taken with reasonable confidence; for more novel endeavours such as the target complex, the estimates should be regarded as preliminary at this stage.



\begin{table}[htb]
\begin{center}
\caption{Summary of personnel requirements for BDF construction}
\label{tab:cost-summary-personnel}
\noindent
\begin{tabular}{@{}lcc@{}}
\hline
\bf{Work Package} & Staff & Fellows \\
 &  [FTE.y] & [FTE.y] \\
\hline

{\bf Civil engineering} & \bf{23.5} & \bf{7}\\
\hspace{3mm} Site investigation & & \\
\hspace{3mm} Junction cavern & & \\
\hspace{3mm} Extraction tunnel & & \\
\hspace{3mm} Target complex  &  &  \\
\hspace{3mm} Experimental area & & \\
\hline
{\bf Infrastructure and services } & \bf{18.9}  & \bf{10} \\
\hspace{3mm} Cooling  &  3 & -\\
\hspace{3mm} Ventilation  & 3 & -\\
\hspace{3mm} Electrical infrastructure  & 1.3  & 1 \\
\hspace{3mm} Access \& safety systems  & 5 & -\\
\hspace{3mm} Radiation protection  & 4.3  & 7 \\
\hspace{3mm} Survey & 1 & 2 \\
\hspace{3mm} Transport: cranes, lifts, tooling  & 1.3 & -\\

\hline
{\bf Extraction and beamline }  & \bf{28.0}  & \bf{11.5} \\

\hspace{3mm} Power converters & 12 & 3 \\
\hspace{3mm} Magnets & 4.5  & -\\
\hspace{3mm} Vacuum  & 2 & -\\
\hspace{3mm} Beam Instrumentation & 4 & 4  \\
\hspace{3mm} Interlocks & 1 & 0.5  \\
\hspace{3mm} Optics design and commissioning  & 2 & 2   \\
\hspace{3mm} Software, layout, integration  & 2.5 & 2  \\

\hline
{\bf Target and target complex }  & \bf{77.0} & \bf{5.4} \\
\hspace{3mm} 	Target assembly \& annex equipment	 & 	15 & 	0.9  \\
\hspace{3mm}  	Instrumentation package  & 		5 & 	0.45  \\
\hspace{3mm}  	Bunker iron shielding \& US1010	 & 	3 & 	0.6  \\
\hspace{3mm}  Concrete shielding & 		3 & 	-  \\
\hspace{3mm} Magnetizing coil for US1010 shielding & 		3 & 	0.3  \\
\hspace{3mm} 	Target \& coil handling system & 		25 & 	1.8  \\
\hspace{3mm}  	Target and coil shielding casks & 		0.5 & 	-  \\
\hspace{3mm} Helium vessel (HW)	 & 	11 & 	0.6  \\
\hspace{3mm} 	Integration	 & 	6 & 	-  \\
\hspace{3mm}  	Beam window(s) & 		2 & 	0.3  \\
\hspace{3mm}  	Survey \& alignment & 		3 & 	0.3  \\
\hspace{3mm} 	Collimator(s) & 		0.5 & 	0.15  \\
\hline			

{ \bf Integration and installation} & \bf{10.2} & \bf{6} \\
\hspace{3mm} Integration and coordination  & 3 & -\\
\hspace{3mm} Installation, design support & 4.2 & 6 \\
\hspace{3mm} Removal/installation TDC2  & 3 & -\\
\hline
\bf{Total}  & \bf{158} & \bf{40} \\
\hline
\end{tabular}
\end{center}
\end{table}

 \chapter{Potential siting of a Lepton Flavour Violation experiment}
\label{App:taufv}

\section{Introduction and experimental requirements}
\label{sec:taufv_intro}

The BDF beam line offers a potential opportunity to host and operate in parallel an experiment ("TauFV"~\cite{TauFV}) to search for Lepton Flavour Violation and rare decays by taking advantage of the possibility to produce a very large sample of tau leptons and D mesons. Using a thin in-line target to intercept about 2\% of the intensity delivered to the SHiP target, the experiment would have access to close to \num{8e13} tau lepton and \num{5e15} D$^0$ meson decays. The detector under study consists of a 7.5\, long and 3\,m wide spectrometer including a vertex detector, tracker, potentially a fast timing detector, electromagnetic calorimeter and muon system. The experimental dipole produces a field of 2.5\,Tm over a length of 2\,m. It is desirable to be able to swap the polarity to manage systematic errors in the physics measurements.

The reconstruction of secondary vertices and suppression of background can be significantly enhanced by 
using a target made up of several very narrow blades, orthogonal to the beam and spaced by a few centimetres, in conjunction with a highly elliptic transverse beam profile with a beam sigma of >4\,mm in the plane parallel to the target blade, and preferably <1\,mm orthogonal to the target blade. This allows discerning the secondary vertices in two dimensions, and diluting the interactions along the blade to reduce pile-up.
As discussed in detail below, from the beam optics point of view, several locations can provide the required beam conditions 
and the beam drift space to accommodate the detector along the BDF transfer tunnel between 
the TDC2 switch yard cavern and the BDF target station without affecting the location of the BDF 
experimental area and without significant changes to the configuration of the beam line. The 
choice is instead driven by considerations related to the civil engineering in the vicinity of 
the existing installations, radiological protection, and to access and transport requirements 
above ground and underground.  Lateral space is required on both sides for shielding in order 
to limit the radiation exposure of the surrounding underground area to levels typical for the 
rest of the beam line.

Further simulation studies must be made to understand the impact on the SHiP experiment 
in terms of background. The flux from intercepting 2\% of the beam is small but the residual
muon flux penetrating the target bunker enters into the SHiP muon shield at a location and
angular distribution for which it was not designed. This could potentially lead to capture 
of a fraction of the muons increasing significantly the flux in the SHiP acceptance.
This will be studied as the TauFV experimental configuration reaches more maturity.

\section{Implementation on BDF beam-line}
\label{sec:taufv_line}


\subsection{Lattice modifications}

\begin{figure}[htbp]
\centering
\includegraphics[width=16cm]{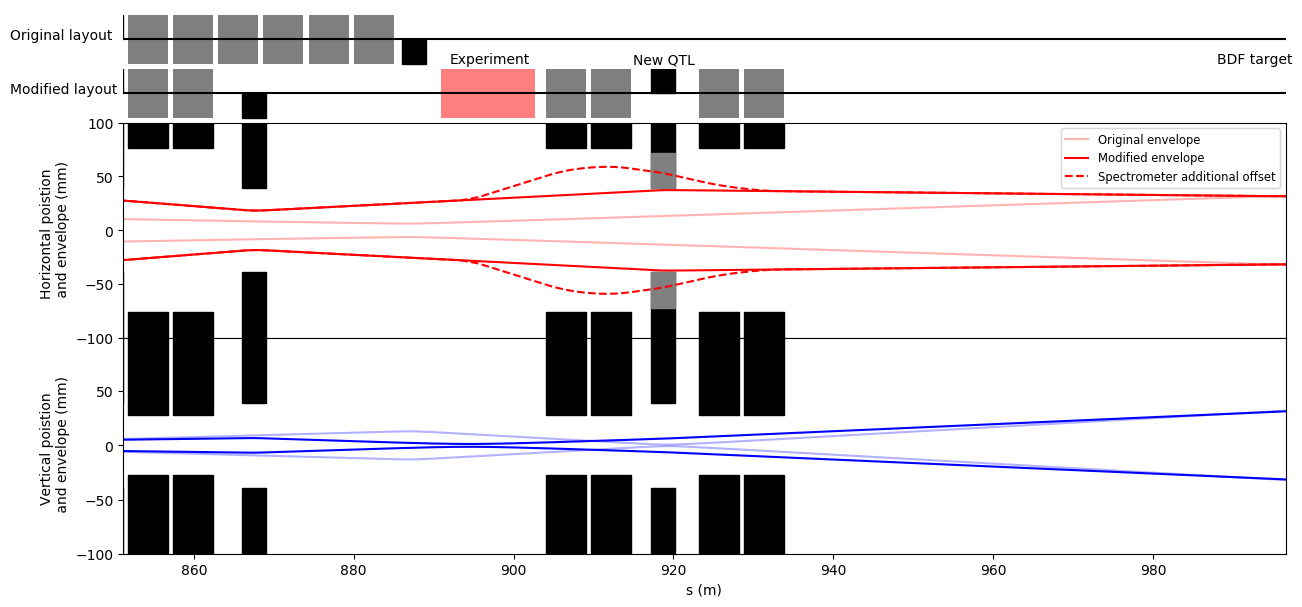}
\caption{Synoptic, apertures and beam sizes in the later part of the new BDF beam line without and with the TauFV experiment. In the synoptic, dipoles are represented as grey bars, while quadrupoles are represented as black squares above the centre line for focusing and below for defocusing magnets.}
\label{tau:fig:zoom_apers}
\end{figure}

The most suitable location to implement the TauFV experiment is about $\SI{100}{m}$ upstream of the BDF target. Fig.~\ref{tau:fig:zoom_apers} summarises all the changes required to accommodate the TauFV requirements and the associated shielding. The detector is represented by the plain red rectangle. 

The last four main dipoles must be moved downstream by around \SI{40}{m}. This allows separating the beam axis of the TauFV experiment and the SHiP experiment by an angle of $\SI{21.9}{mrad}$. An additional quadrupole, referred as \emph{New QTL}, needs to be added after the TauFV detector to control the evolution of the beam size. 

Beam sizes are represented by solid lines in Fig.~\ref{tau:fig:zoom_apers}, and the apertures of the elements in the modified layout are shown as black rectangles. The beam size is plotted as $\pm 4\sigma$, using $N=4$ in Eq. \ref{transf:eq:beamsize} with the beam parameters discussed Sec. \ref{transf:sec:beamparms}. Comparing to the original beam size in the nominal layout, shown as light red lines, the figure also demonstrates that the beam dimensions on the BDF target are preserved. The new beam-line design allows producing a beam spot of $\sigma_x = \SI{6.8}{mm}$ and $\sigma_y = \SI{0.34}{mm}$, respectively, in the horizontal and vertical plane at the TauFV target. 

The beam size and the magnet currents along the first part of the transfer line are very similar to the nominal case. Very small adjustments are required to satisfy the combined requirements of the TauFV target and the BDF target. The complete trajectory correction scheme has not been simulated for the new configuration but with only small differences in the strength of the quadrupoles, the correction scheme can remain the same. Studies of specific locations, such as around and immediately downstream of the splitter (see Fig.~\ref{transf:fig:zoom_H_apers}) confirm the modified optics to be compatible with the nominal design. 

\subsection{Spectrometer compensation}

The experiment will use a \SI{2.5}{T m} spectrometer that will cause a deflection of the primary beam of up to \SI{1.87}{mrad}. It is foreseen to be able to switch polarity of the spectrometer magnet. In order to correct for the angle produced by the experimental magnet, the four downstream main dipoles are employed in pairs. The first pair of dipoles generates a kick with an opposite direction to the spectrometer while the second pair provides a kick in the same direction as the spectrometer. At most the first pair will be powered to around \SI{1400}{A} to provide up to \SI{9.45}{T m} per magnet. This is slightly beyond the maximal nominal strength of the magnet but remains achievable since those will only be pulsed to follow the BDF cycle. The second pair will be powered up to \SI{1240}{A} or \SI{8.32}{T m}, which is below the maximal nominal strength for this type of magnet. This scheme allows compensating the effect of the spectrometer without an additional magnet.

Horizontal apertures on both sides need to be considered due to the closed bump of the trajectory created by the spectrometer and its compensation. Fig.~\ref{tau:fig:zoom_apers} shows in dashed red lines the additional apertures required to operate this scheme. The smallest aperture shown in grey at the level of the new QTL refers to the circular aperture. This considerably underestimates the available space as the beam is much larger horizontally than vertically at this location. Due to the non-circular section of the vacuum chamber in this quadrupole, the more realistic representation of the available aperture is shown in black.

\subsection{BDF dilution system}
Lastly, the dilution system discussed Sec. \ref{transf:sec:dill} will need to be strengthened. Its new location will be \SI{40}{m} closer to the BDF target which will thus require a system which is around 70\% stronger. This has not been studied in detail but the same design chosen for the nominal layout can easily be scaled up by adding more units.
\section{Target system}
\label{sec:taufv_target}

\begin{figure}[htbp]
\centering %
\includegraphics[width=0.7\linewidth]{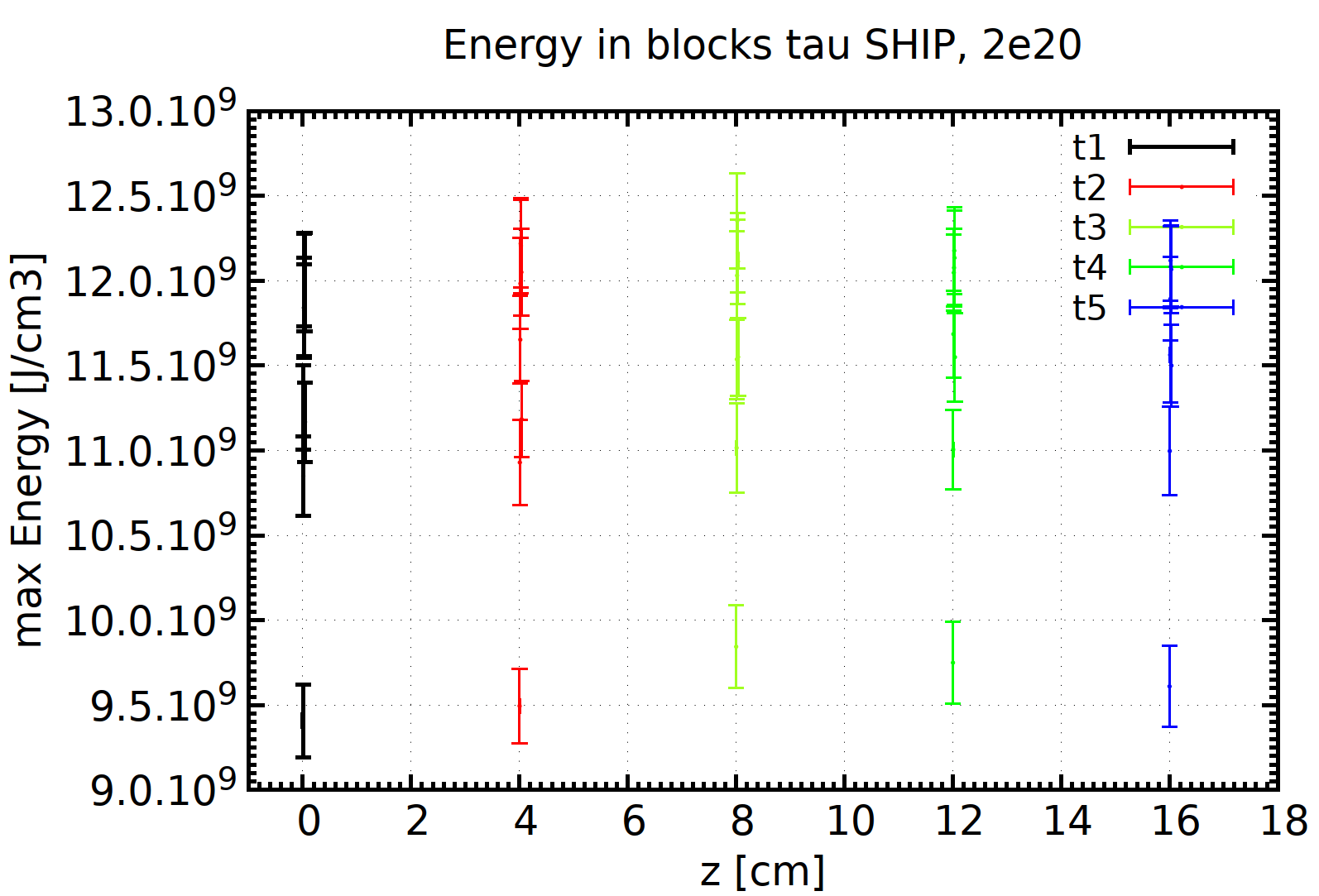}
\caption{\label{fig:TauFV:Target_fluka_Z} Energy distribution along the Z axis for the five tungsten blades assuming \num{2e20} protons on target.}
\end{figure}

The TauFV target system consists of five tungsten blades of dimensions $40\times2\times\SI{0.4}{mm^3}$. The energy deposition on the target blades has been estimated with FLUKA simulations~\cite{FLUKA_Code}. The average energy density in each blade after one pulse is expected to be of the order of $\sim$\SI{450}{J\per cm^3}. Figure~\ref{fig:TauFV:Target_fluka_Z} displays the energy density along the longitudinal axis for the different target blades, assuming \num{2e20} protons on target. 

Due to the high beam power deposited on such thin blades, and the fact that the surrounding vertex detector must operated under cooling, the target must be actively cooled\footnote{Otherwise a radiative cooled assembly under vacuum could have been proposed as refractory metals could be operated at very high temperature.}. The initial studies consider an inert gas such as helium as cooling medium for the target. Preliminary CFD simulations have been carried out in order to calculate the heat transfer coefficient that could be achieved with forced circulation of helium around the blades. 

\begin{figure}[htbp]
\centering %
\includegraphics[width=0.7\linewidth]{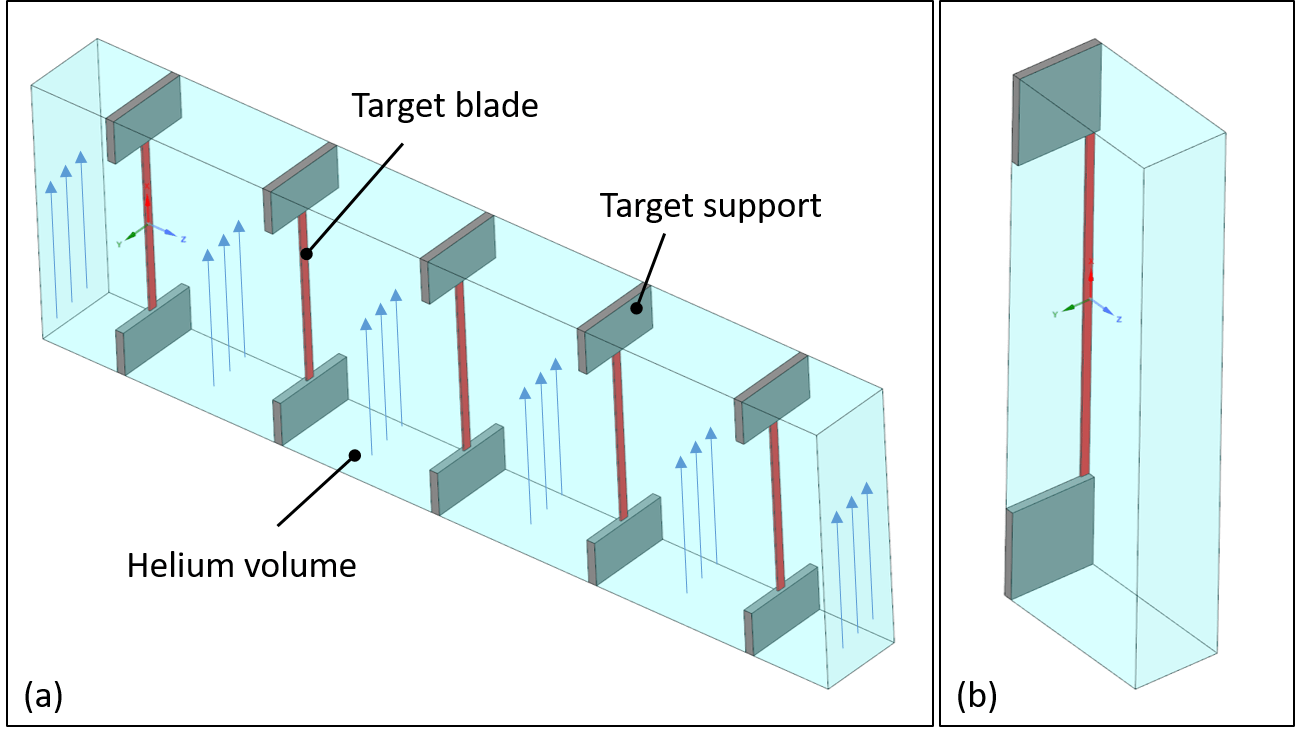}
\caption{\label{fig:TauFV:CFD_model} (a) Full 3D model of the helium cooling system design; (b) Quarter model used for the CFD calculations of the helium cooling circuit.}
\end{figure}

The preliminary study considers that the thin tungsten blades are contained in a closed-loop tank where the helium cooling circulation system will be enclosed. The horizontal cross-section of the helium volume is tentatively assumed to be $160\times\SI{40}{mm^2}$, designed in a way that the whole volume of the target blades is contained. The blades are clamped on the top and bottom edges; the supports for the blades have also been included in the 3D CFD model. Figure~\ref{fig:TauFV:CFD_model} presents the fluid-flow model of the whole helium circuit (a), and the simplified quarter model of one of the blades used for the helium cooling simulation (b).

\begin{figure}[htbp]
\centering %
\includegraphics[width=0.7\linewidth]{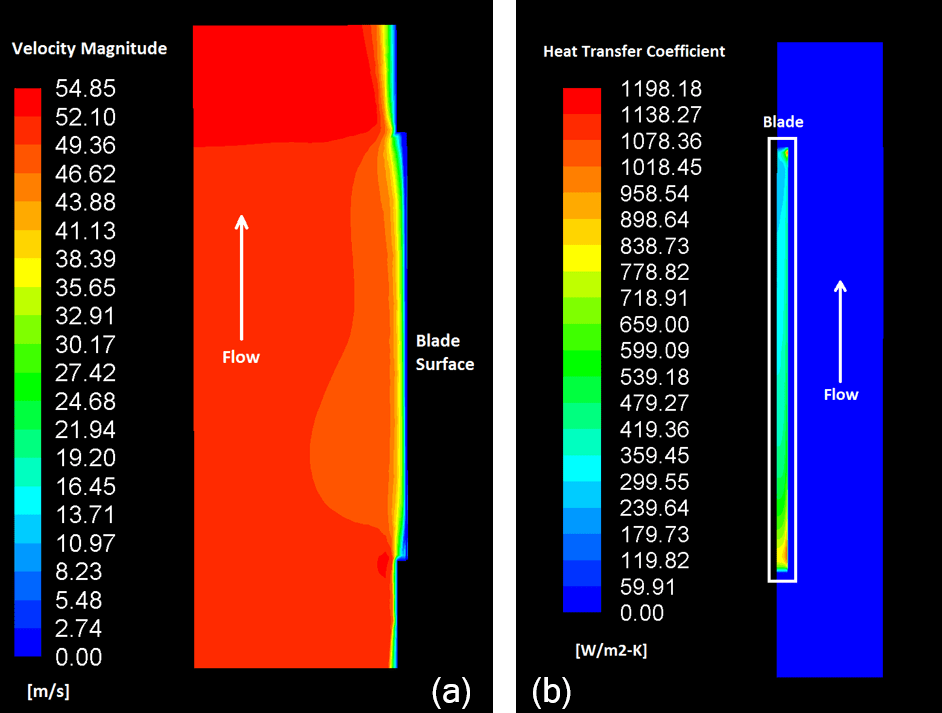}
\caption{\label{fig:TauFV:CFD_HTC} (a) Velocity magnitude contours on a plane containing the blade centre lines; (b) Heat transfer coefficient on the surface of the blade}
\end{figure}

An initial flow-rate of \SI{684}{m^3\per\hour} (equivalent to \SI{34.2}{m^3\per\hour} for the quarter case illustrated in~\ref{fig:TauFV:CFD_model}b) has been considered, which is assumed to be realistic for the cooling system of such a device. The helium rate applied leads to an average helium velocity in contact with the blades of around 50\,m/s, illustrated in Figure~\ref{fig:TauFV:CFD_HTC}(a). The average heat transfer coefficient obtained in the surface of the tungsten blades is estimated at around \SI{350}{\watt\per(m^2 \kelvin)}, as shown in Figure~\ref{fig:TauFV:CFD_HTC}(b).

\begin{figure}[htbp]
\centering %
\includegraphics[width=0.8\linewidth]{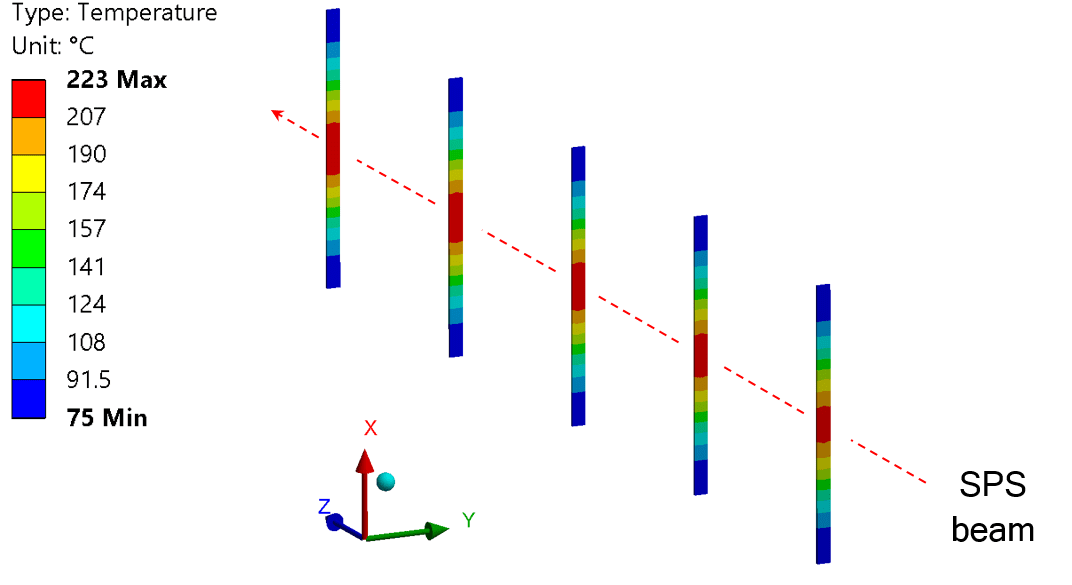}
\caption{\label{fig:TauFV:Target_temp} Temperature distribution in the target blades at the pulse peak. Maximum temperature $\approx \SI{220}{\celsius}$.}
\end{figure}

The heat transfer coefficient (HTC) value obtained can be used as boundary condition for Finite Element Model (FEM) calculations aiming at validating the current target design from a thermo-mechanical point of view. The thermal simulations performed have shown that the estimated convection through helium leads to an acceptable temperature in the target blades, around \SI{220}{\celsius} after beam impact (see Fig. \ref{fig:TauFV:Target_temp}). Given that tungsten is a refractory metal with high melting point suited for high temperature applications, the temperature reached is within the limits for safe operation of the target. The compatibility of the resulting temperature and the helium flow with the global assembly will depend on the vertex detector system and its associated cooling system that will operate close to the target assembly.

\begin{figure}[htbp]
\centering %
\includegraphics[width=0.8\linewidth]{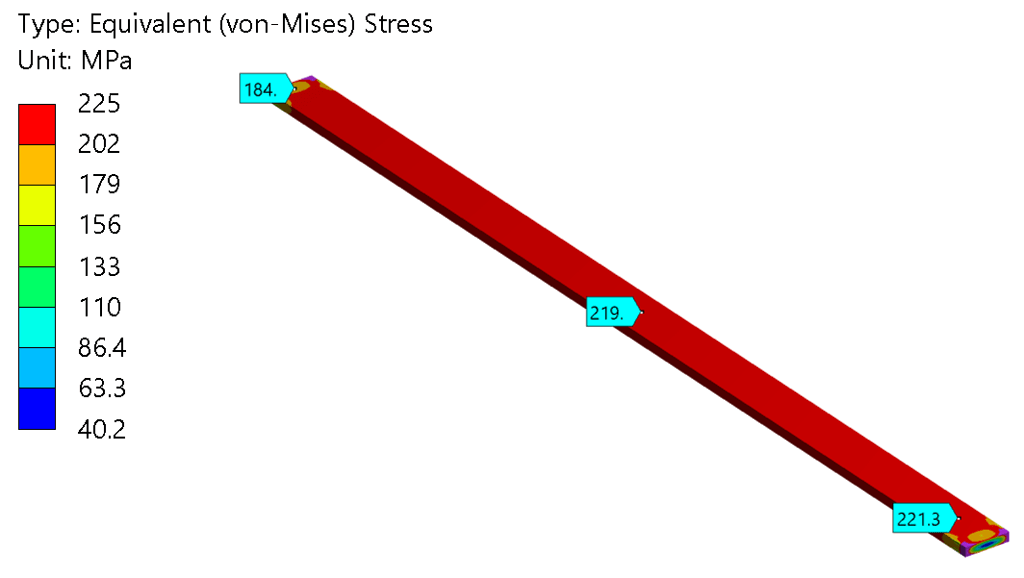}
\caption{\label{fig:TauFV:Target_stress} Von Mises equivalent stress distribution in the target blades at the pulse peak. Maximum stress $\approx \SI{220}{MPa}$.}
\end{figure}

The thermal stresses induced by the temperature increase in the target materials have also been calculated. As mentioned above, the blades are assumed to be clamped at the edges. Given the low ductility of tungsten at temperatures below \SI{300}{\celsius}, the maximum principal stress criterion is the most suited to evaluate the safety margin of the target blades against failure. However, the stresses induced by the beam are purely compressive, and the calculated maximum principal stress is negligible. The maximum von Mises equivalent stress is equal to the maximum compressive stress, and is around \SI{220}{MPa}, significantly below the tensile strength of the material at the operational temperatures (\SI{550}{MPa} at \SI{300}{\celsius} \cite{Tungsten_Schmidt}). The von Mises equivalent stress distribution in one of the target blades obtained from the structural calculations is presented in Fig.~\ref{fig:TauFV:Target_stress}. 

In conclusion, this very preliminary analysis shows that the temperatures and stresses expected in the tungsten blades are well within the material limits, thus validating the preliminary target design for the current experimental setup. The main challenges in the manufacturing of the target are associated with the production and machining of such thin tungsten blades, pure tungsten being a very brittle material at room temperature. Alternatively, a tungsten alloy with NiFe or NiCu content could be considered, allowing for a similar density but higher strength and ductility. Otherwise slightly lighter materials with much more ductility, such as tantalum or tantalum-alloys, could potentially be employed.
\section{Radiation protection}
\label{sec:taufv_rp}

\subsection{Introduction}

The main Radiation Protection (RP) challenges for TauFV arise from the high beam power, the proximity to the surface and CERN fence, as well as the goal of minimizing the impact on the rest of the BDF complex. In order to respect the applicable CERN radiation protection legislation regarding doses to personnel as well as the environmental impact~\cite{SCF}, a preliminary radiological assessment has been carried out for the proposed setup of the TauFV experiment. Specific studies of prompt and residual dose rates, as well as considerations about air and ground activation will be described. To assess the above-mentioned radiation protection aspects, extensive simulations were performed with the FLUKA Monte Carlo particle transport code~\cite{FLUKA_Code} \cite{fluka1}.

\subsection{FLUKA model}

\begin{figure}[!htb]
  \centering
  \includegraphics[width=0.72\textwidth]{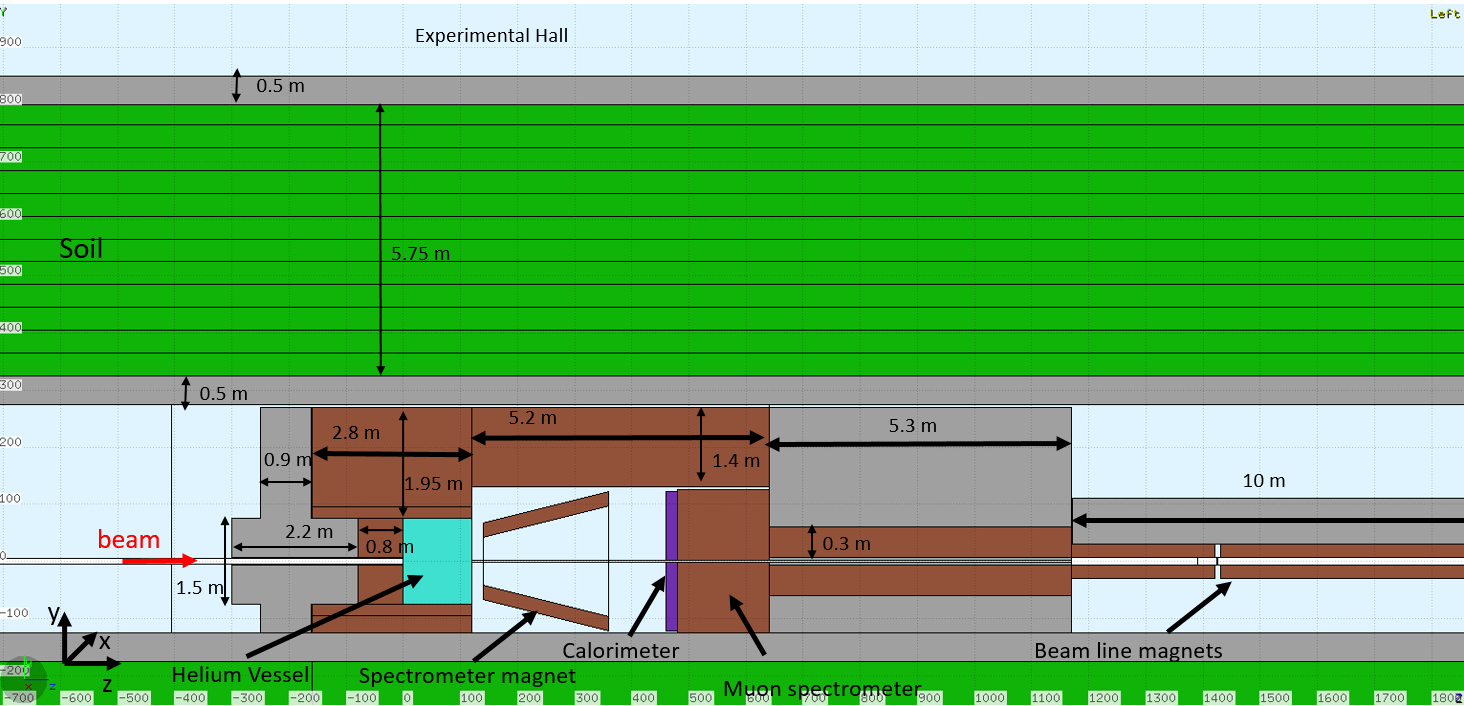}
  \includegraphics[width=0.72\textwidth]{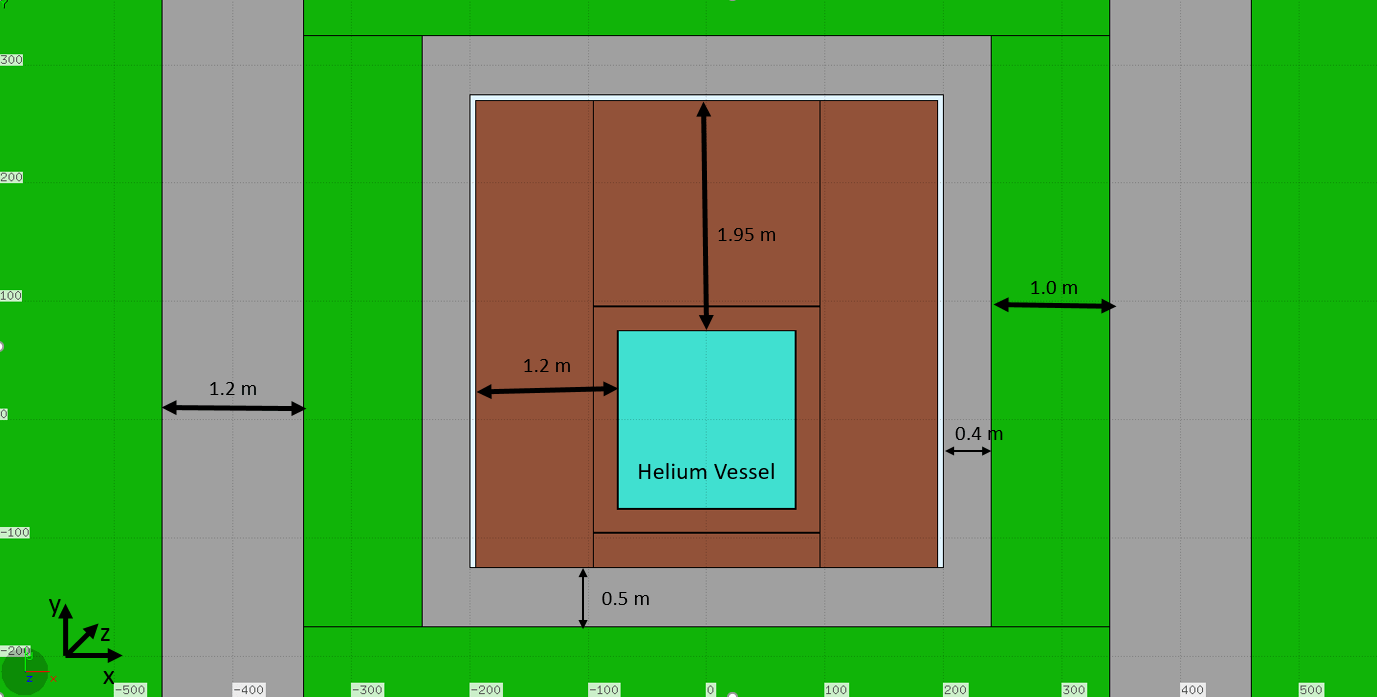}

  \captionsetup{width=0.85\textwidth} \caption{\small FLUKA model for TauFV. Vertical (top) and horizontal (bottom) views.}
\label{fig:flukageo}
\end{figure}

In the FLUKA model, for simplicity, the target was modelled as a \SI{40 x 2 x 2}{mm} single tungsten blade. The experimental setup is shown in Fig.~\ref{fig:flukageo}. A helium vessel is foreseen to host the target and the vertex locator detector (VELO) to avoid corrosion. A spectrometer magnet\footnote{Only the yoke was included in the FLUKA simulations, but not yet the coil.} with downstream tracking stations is used for momentum measurement. The spectrometer is followed by a calorimeter simulated as a GaGG\footnote{ Gadolinium Aluminium Gallium Garnet with density of \SI{}6.67}{g/cm$^{3}$} block \SI{20}{\cm} thick, and a muon detector with iron filters simulated as a \SI{1.6}{m} thick iron block. The first \SI{5.3}{m} downstream of the experiment have been reserved for shielding. The shielding is followed by a \SI{2.5}{m} long magnet correcting the trajectory of the beam with the inverse field of the spectrometer magnet. Six dipole skeletons (without magnetic field) of \SI{50 x 50 x 500}{cm} were implemented downstream of the correcting magnet. 
The beam pipe, in stainless steel, is cylindrical upstream of the detector with a radius of $\sim$\SI{5}{cm}. Throughout the experiment, and inside the downstream magnets, the beam pipe is elliptical with an aperture of \SI{20 x 1}{cm} and \SI{36 x 12}{cm}, respectively.

The shielding surrounding the helium vessel is composed of iron and concrete as follows:
\begin{itemize}
    \item iron shielding on top (\SI{195}{cm}), bottom (\SI{50}{cm}), side (\SI{120}{cm}) and upstream (\SI{80}{cm}) of the helium vessel,
    \item concrete shielding upstream of the helium vessel and around the beam pipe of \SI{150 x 150 x 220}{cm},
    \item iron (\SI{280}{cm} long) and concrete (\SI{90}{cm} long) shielding upstream the top and bottom shielding of the helium vessel.
\end{itemize}

The iron shielding is \SI{140}{cm} thick on top and \SI{65}{cm} thick on the sides of the spectrometer magnet, the calorimeter and the muon spectrometer. Downstream of the experiment, a \SI{5.3}{m} long shielding is foreseen. It consists of several layers:
\begin{itemize}
    \item an elliptical inner layer of \SI{5}{cm} tungsten with the beam pipe casted,
    \item a cylinder of iron with a radius of \SI{60}{cm} with the tungsten plug casted,
    \item a parallelepiped of concrete \SI{390 x 390 x 530}{cm} surrounding the iron cylinder.
\end{itemize}

The magnets downstream of the shielding are surrounded by \SI{80 x 240 x 160}{cm} (x,y,z) standard blocks of concrete along \SI{10}{m}.
To further improve the shielding and reduce the prompt dose to the downstream magnets, and to the tunnel and the soil:
\begin{itemize}
    \item the concrete shielding can be partially exchanged with iron,
    \item add more collimators further downstream,
    \item optimise the shape of the beam pipe inside the shielding. 
\end{itemize}  
The material compositions and densities used in the FLUKA simulation are as given in \ref{RP:Flumodel_comp}.

\subsection{Radiation protection assessment}

\begin{figure}[!htb]
  
\begin{subfigure}{0.72\textwidth}
\centering
  \includegraphics[width=0.72\textwidth]{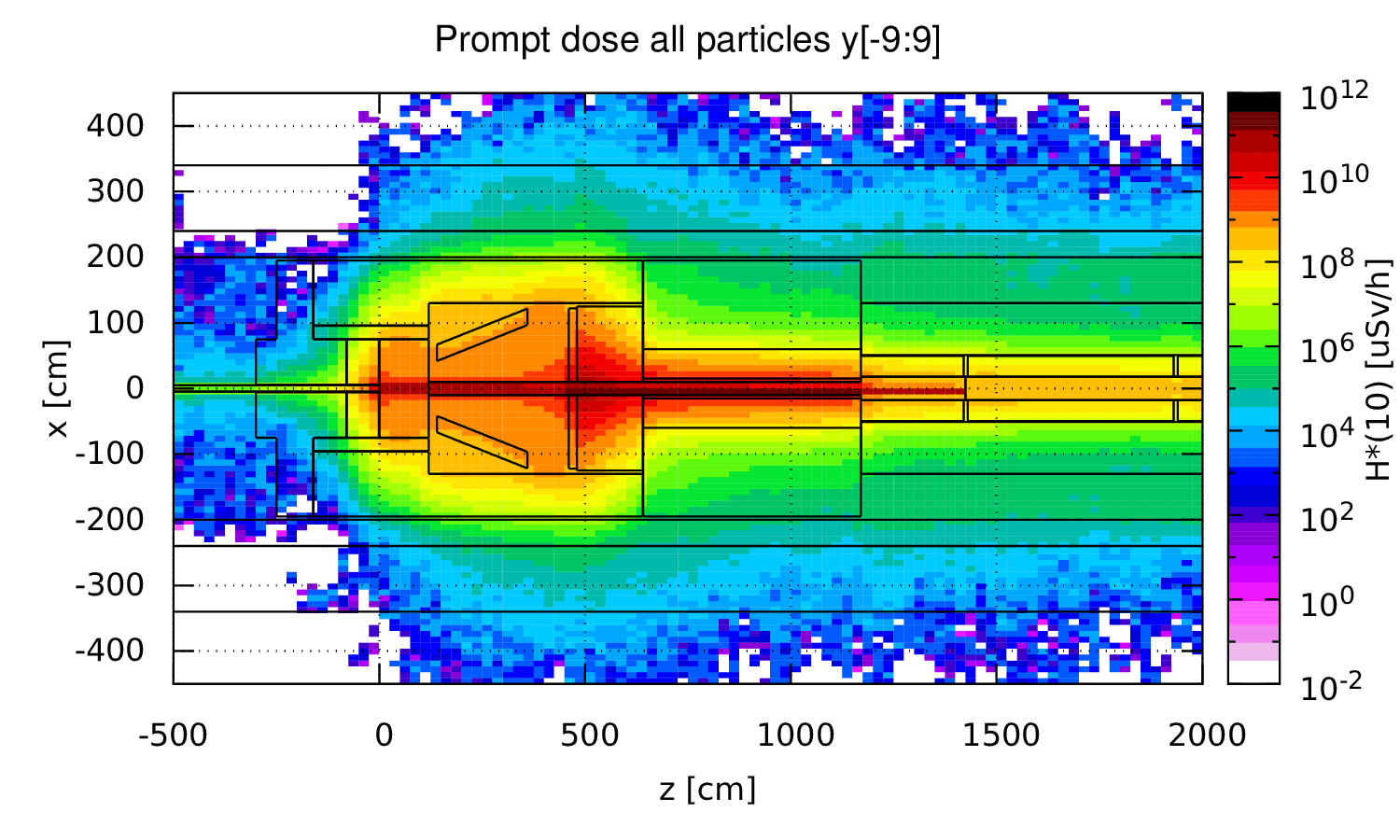}
  \label{fig:prompt1-v2}
\end{subfigure}
\begin{subfigure}{0.72\textwidth}
 \centering
  \includegraphics[width=0.72\textwidth]{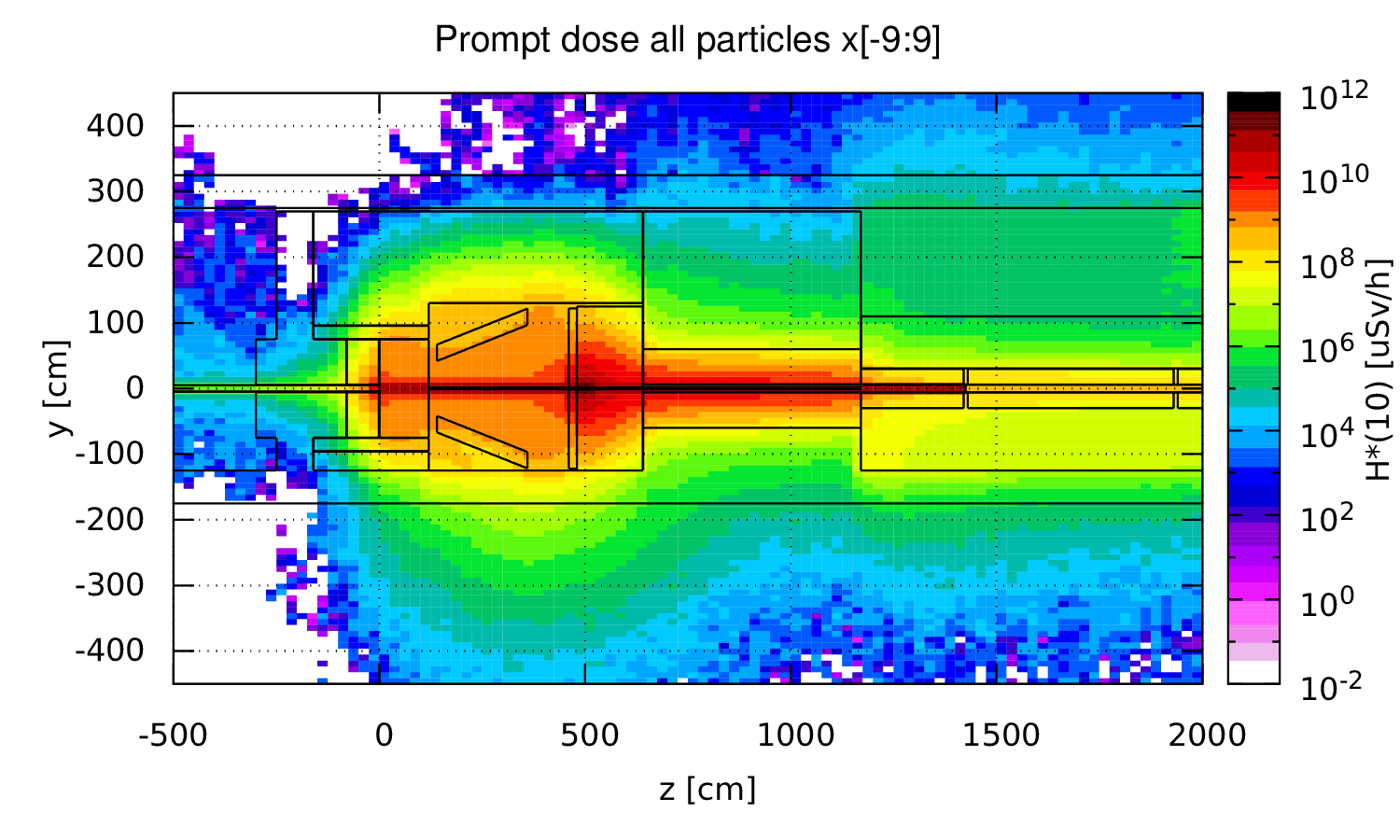}
  \label{fig:prompt3-v2}
\end{subfigure}

\captionsetup{width=0.72\textwidth} \caption{\small Prompt ambient dose equivalent rates for the horizontal (top) and vertical (bottom) cut views.}
\label{fig:tauFVpDR1}
\end{figure}

\begin{figure}[!htb]
  
\begin{subfigure}{0.72\textwidth}
\centering
  \includegraphics[width=0.72\textwidth]{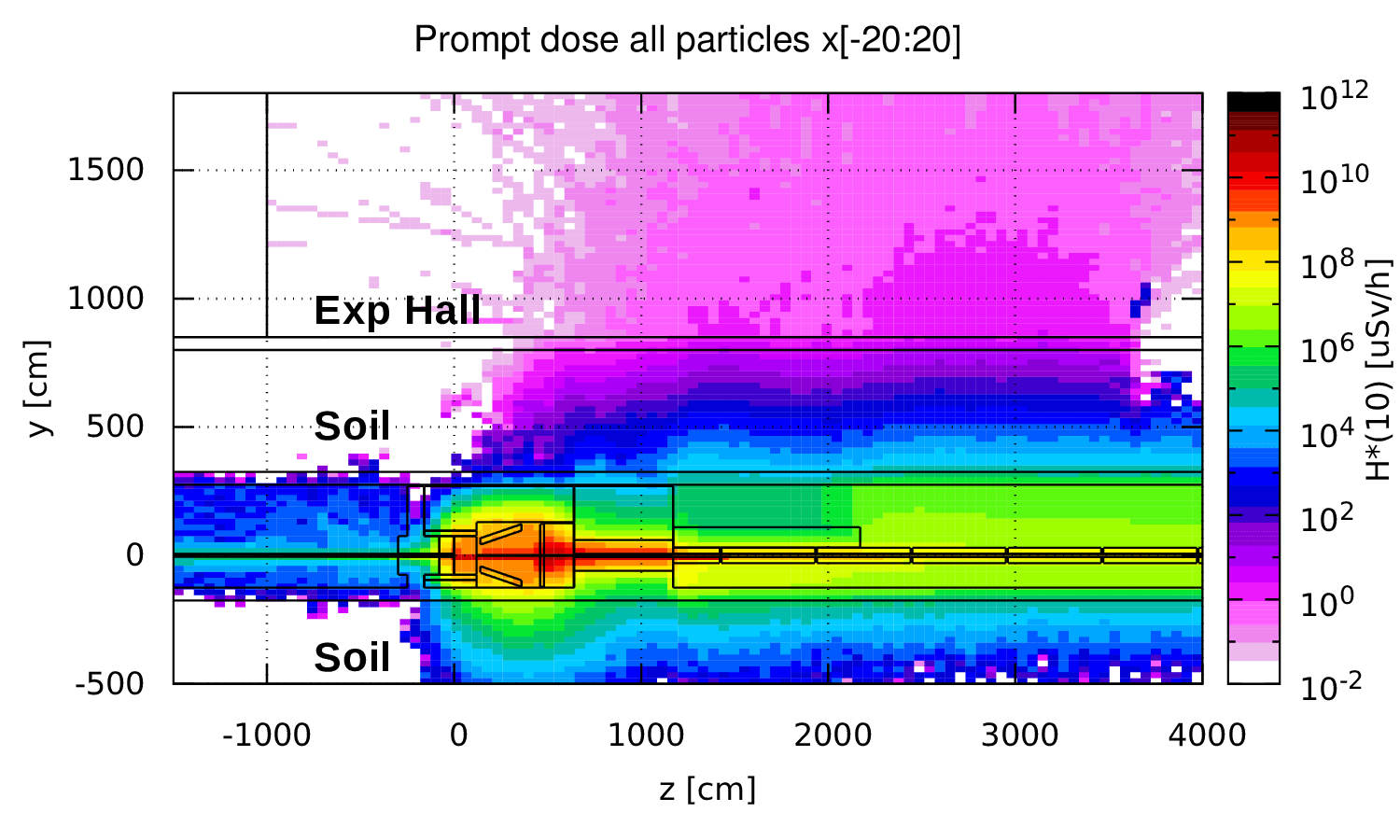}
  \label{fig:prompt2-v2}
\end{subfigure}
\begin{subfigure}{0.72\textwidth}
 \centering
  \includegraphics[width=0.72\textwidth]{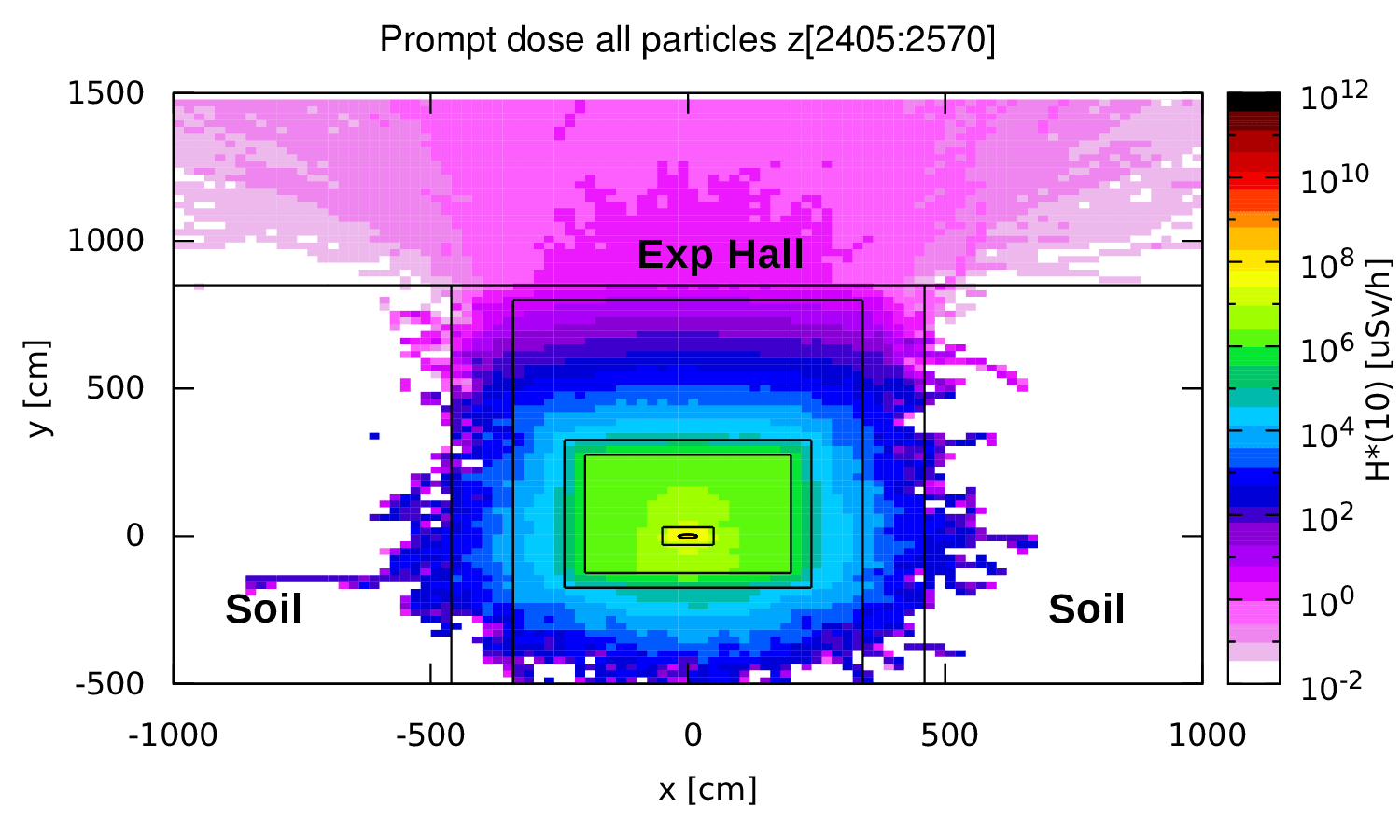}
  \label{fig:prompt4-v2}
\end{subfigure}

\captionsetup{width=0.72\textwidth} \caption{\small Prompt ambient dose equivalent rates for the vertical (top) and perpendicular (bottom) cut views.}
\label{fig:tauFVpDR2}
\end{figure}

FLUKA simulations have been performed assuming $4\times10^{13}$ protons per spill with one spill every 7.2 s. The resulting prompt ambient dose equivalent rates for the experiment and the shielding are shown in Fig.~\ref{fig:tauFVpDR1}, while Fig.~\ref{fig:tauFVpDR2} shows the prompt ambient dose equivalent rates for the tunnel, soil and experimental hall. 

As expected, the highest dose rates can be found in the region of the target, calorimeter and muon spectrometer reaching a few  
\SI{e11}{\micro\sievert/\hour}.
The levels are reduced by several orders of magnitude in the surrounding shielding. The prompt dose rates are further reduced by the soil above, such that they drop to a few \si{\micro\sievert/\hour} in the above-ground experimental hall.

\begin{figure}[!htb]
\begin{subfigure}{0.72\textwidth}
  \centering
  \includegraphics[width=0.72\textwidth]{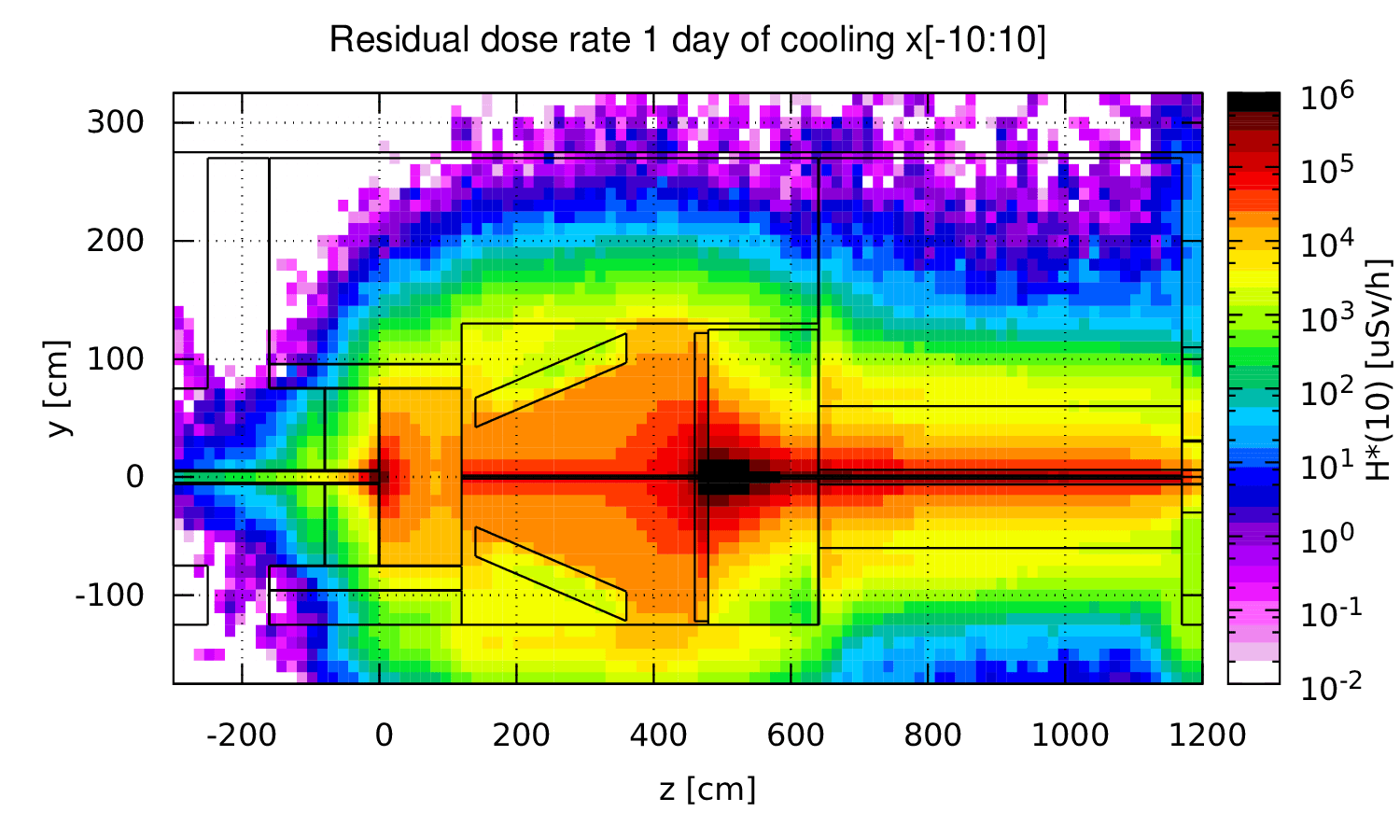}
  \label{fig:res1}
\end{subfigure}
\begin{subfigure}{0.72\textwidth}
  \centering
  \includegraphics[width=0.72\textwidth]{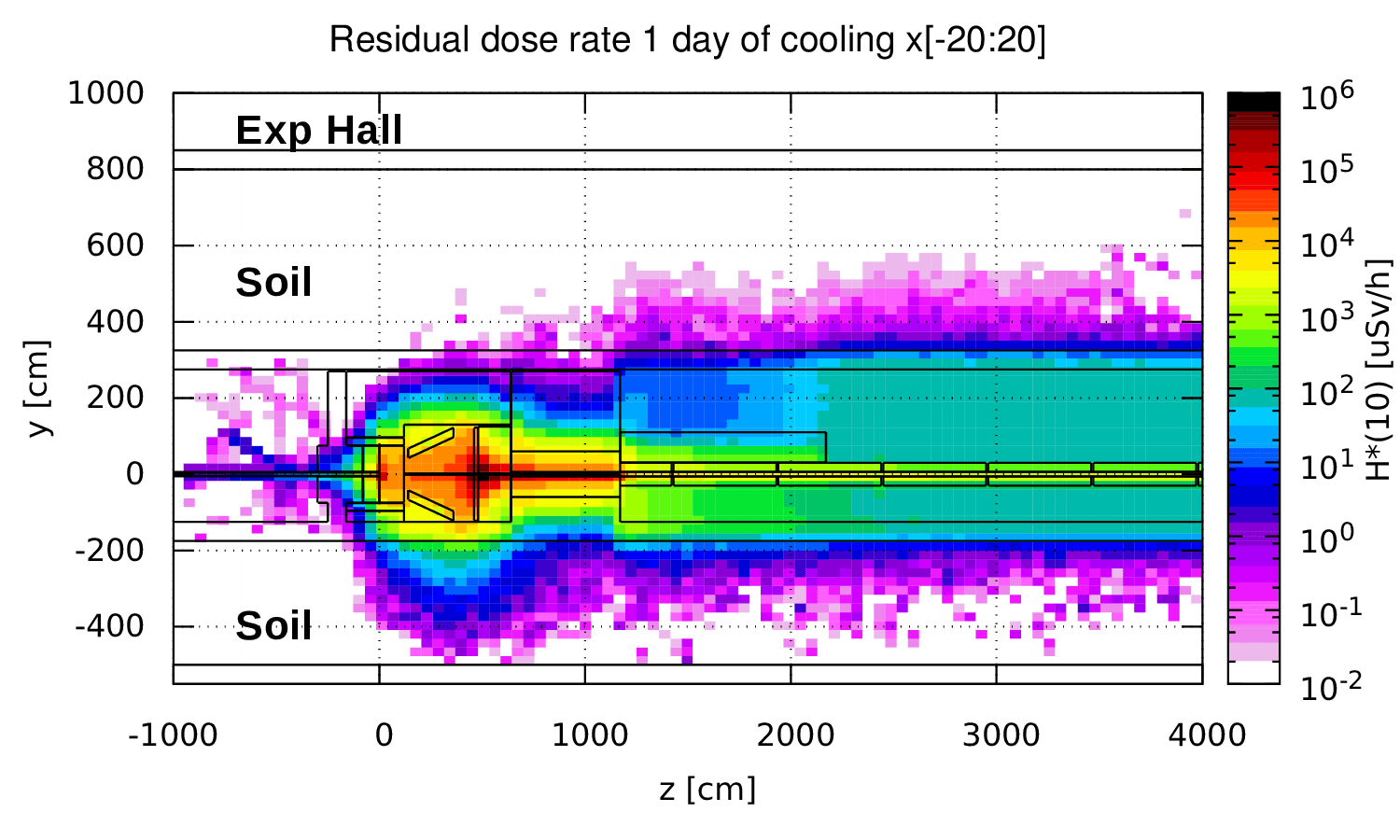}
  \label{fig:res2}
\end{subfigure}

\captionsetup{width=0.72\textwidth} \caption{\small Residual ambient dose equivalent rates after one day of cool-down for the experiment (left) and the surroundings (right).}
\label{fig:RD1}
\end{figure}

Fig.~\ref{fig:RD1} shows the expected residual dose rates after one day of cool-down time. The highest dose rates can be found in the region of the detector reaching in the order of a few \SI{e6}{\micro\sievert/\hour}. For this reason, the facility should be designed such that interventions will be performed with remote handling systems.

For the next phase of the study it is foreseen to perform optimisation of the shielding such as to further reduce the prompt and residual dose rates. The impact of the remnant beam on the activation of the BDF facility downstream should be evaluated. The environmental impact from releases of radioactive air and soil activation should also be carefully addressed in future studies. Finally, the production of radioactive waste should be evaluated.
\FloatBarrier
\section{Civil Engineering}
\label{sec:taufv_CE}

\subsection{Introduction}

The civil engineering (CE) required to implement TauFV has been studied at a conceptual level to ensure that the project is feasible. The CE involved in TauFV is no more complex than that involved in the wider BDF project. This section briefly details the CE requirements specific to TauFV and as such must be read in conjunction with the CE section for the wider BDF project for context and  detail relating to these proposals.

\subsection{CE Requirements}
TauFV is to be located on the existing beam-line within the BDF extraction tunnel as detailed in Section \ref{sec:taufv_line} of this chapter. The positioning has been driven primarily by beam dynamics constraints. The location is also favourable from a CE perspective, allowing synergy between the proposed BDF infrastructure and that for TauFV. 

The TauFV experimental arrangement would take up the full width of the planned extraction tunnel. Therefore additional space is required to allow access to the rest of the BDF transfer line:

\begin{itemize}

 \item Passage of personnel
 \item Transport of materials and equipment past the experiment 
 \item Space reservation for cable trays and  services
 \item Prevention of any dead-ends which would not allow acceptable evacuation routes from certain lengths of tunnel. 

\end{itemize}

Several options were considered to facilitate access past the TauFV detector bunker. Options reviewed include a bypass tunnel or alternative access arrangements, but the preferred option was to widen the tunnel. Tunnel enlargement would provide sufficient space to bypass the experiment with enough shielding between the experiment and access route to make the arrangement acceptable for RP. This was clearly the most cost effective option as well as providing flexibility for change in the future. 

To allow installation, operation and maintenance of the TauFV experiment, a shaft above this area will be needed over the full width and length of the experiment with direct vertical access.

A full summary of the requirements generated by TauFV is therefore:
\begin{itemize}
    
\item Tunnel widening to \SI{8.35}{\metre} at a point \SI{94}{\metre} upstream of the BDF target over a distance of approximately \SI{41}{\metre}.
\item A shaft above the proposed  location of the TauFV experiment measuring \SI{5}{\metre} wide by \SI{10}{\metre} in length. The shaft must be capable of being filled with concrete shielding blocks when the beam is in operation.

 \item An access, assembly and storage building above the experiment to house:
 
  \begin{itemize}
      \item a suitably sized crane serving both the heavy equipment shaft and TauFV shaft.
      \item sufficient shared circulation space for transport and handling detector components, magnets, beam-line equipment and shielding.
      \item flexible space to be used for temporary storage of activated equipment or assembly/ maintenance related to the detector.
      \item vehicle access to the building to enable all of the above.
    \end{itemize}
\item Concrete shielding blocks to fill the shaft and also to provide a partition between the experiment and passage way.
\end{itemize}

\subsection{CE Infrastructure}

\begin{figure}[h!]
\centering
\includegraphics[width=16cm]{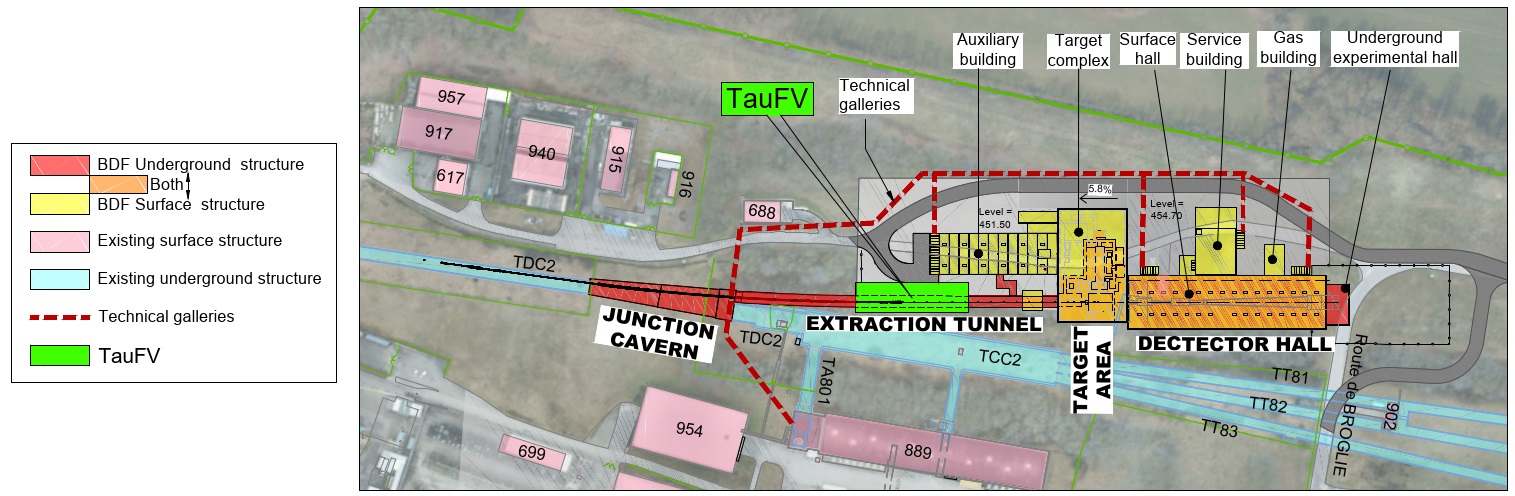}
\caption{Location plan showing the position of TauFV and the CE modifications required for implementation.}
\label{tau:fig:map}
\end{figure}

The proposed layout of CE infrastructure to meet the needs of the TauFV experiment are as shown in Fig. \ref{tau:fig:map}.

\subsubsection{Tunnel enlargement}

\begin{figure}[htbp]
\centering
\includegraphics[width=16cm]{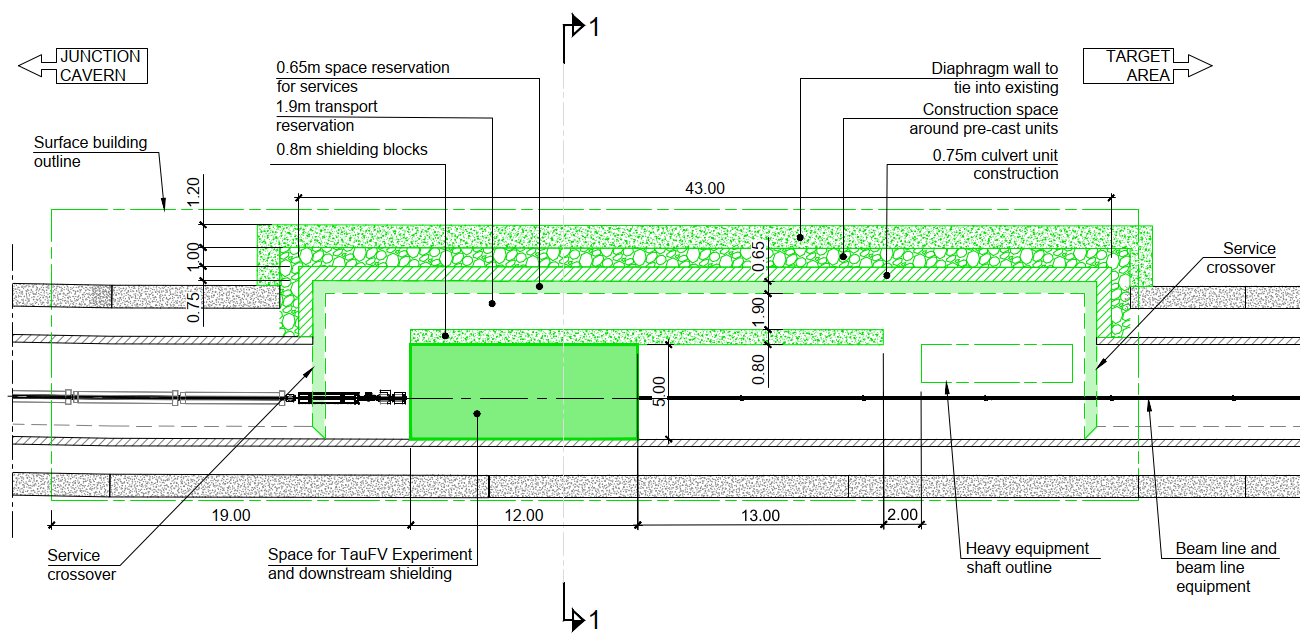}
\caption{Plan view of the proposed sub-surface CE infrastructure in green with wider BDF proposals shown in grey.}
\label{tau:fig:plan}
\end{figure}

The proposed form of construction for the tunnel enlargement will be exactly the same as that used elsewhere on BDF. The layout of the below ground infrastructure is shown in Fig. \ref{tau:fig:plan}.

Widening is only practical on the north-west side of the BDF extraction tunnel due to the proximity of the existing TCC2 tunnel and the constraints with which working close to it are associated.

Due to the CE requirements of the wider project, it is essential that a watertight perimeter is maintained around the BDF installation. Therefore, proposed diaphragm walls for the enlargement will tie into those already proposed for BDF. This will maintain continuity of the perimeter as well as benefiting from the efficiencies associated with a consistent form of construction. 

The widened tunnel itself will again be constructed from pre-cast ‘n’ units on a cast \textit{in situ} concrete base. The pre-cast units will be widened and with a deeper roof slab but in all other respects will be similar. A cast \textit{in situ} joint will need to be formed between the two sections of differing widths. The same approach to waterproofing will be adopted with a multi-layer passive system maintained through this area.

The proposed cross section for the BDF extraction tunnel includes a \SI{650}{\milli\metre} wide space reservation for cable trays and services. This extends to \SI{700}{\milli\metre} for a ventilation duct close to the tunnel roof. The enlargement for TauFV will require a \SI{650}{\milli\metre} wide strip to be maintained with the ventilation duct running outside of the required access area. This will allow continuity of the services past the TauFV experiment as well as providing acceptable levels of radiation shielding for cabling and other radiation sensitive infrastructure. A crossover from one side of the tunnel to the other will be required for the services at each end of the TauFV experiment. This should not be problematic as a crossover is already planed for BDF. The option to keep services on the Jura side of the tunnel throughout could be explored at future stages if TauFV is implemented.

\begin{figure}[htbp]
\centering
\includegraphics[width=16cm]{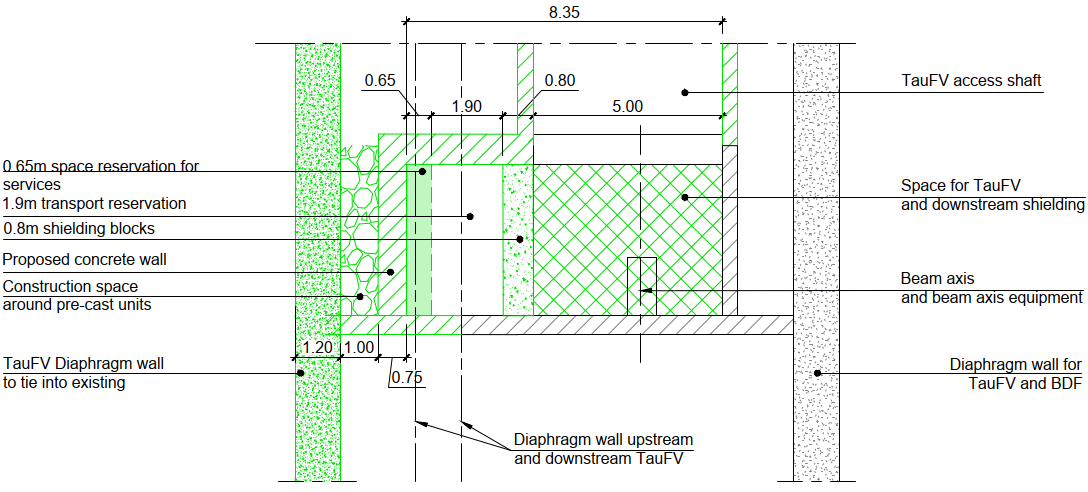}
\caption{Schematic typical section through TauFV access shaft showing proposed TauFV infrastructure in green with wider BDF proposals shown in grey. }
\label{tau:fig:section}
\end{figure}

The transport handling team have been consulted to ensure sufficient space is allowed at either end of the enlarged area for movement and turning of the longest and largest components. A minimum width of \SI{1.9}{\metre} is maintained as well as a maximum angle of \ang{30}.

\subsubsection{TauFV access shaft}

\begin{figure}[htbp]
\centering
\includegraphics[width=13cm]{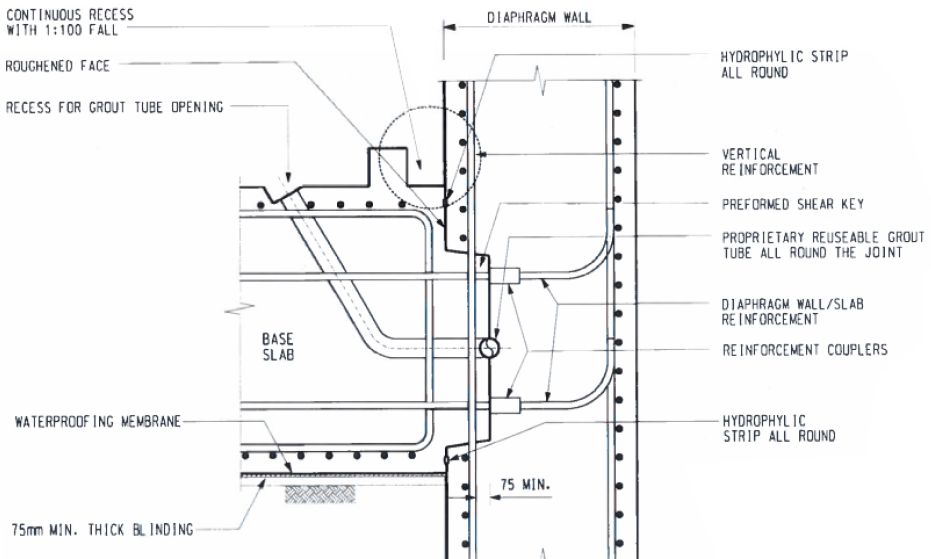}
\caption{Typical detail of diaphragm Wall to Slab coupler connection}
\label{tau:fig:joint}
\end{figure}

The experimental team has calculated the required cross section for the shaft of \SI{5}{\metre} by \SI{10}{\metre} which has been used to inform this study. The form of construction will again be similar to that used for the heavy equipment shaft. There may be scope to optimise this when the design moves past a conceptual stage and when requirements are fully understood. In the meantime, it is sufficient to say that the shaft is entirely feasible using the same techniques. Where necessary, there is also the option to use diaphragm walls to support the roof slab and the concrete shielding blocks with connections as shown in Fig.\ref{tau:fig:joint}. An interlocking arrangement will be adopted for stacking shielding.

\subsubsection{Concrete shielding blocks}
RP requirements have dictated the requirement for \SI{0.8}{\metre} thick shielding blocks to be positioned alongside the TauFV experiment and the downstream magnets for a total distance of \SI{25}{\metre}. Shielding blocks will be pre-cast concrete blocks stacked to form walls. The block walls can be dismantled to allow any future change in experimental arrangement.

\subsubsection{Access building}

An access building above the TauFV shaft is required to house a crane and space for handling, storage and work on components. The additional area required  to accommodate the needs of the experiment was agreed between transport handling and experimental area teams as 625 m$^{2}$.

\begin{figure}[htbp]
\centering
\includegraphics[width=16cm]{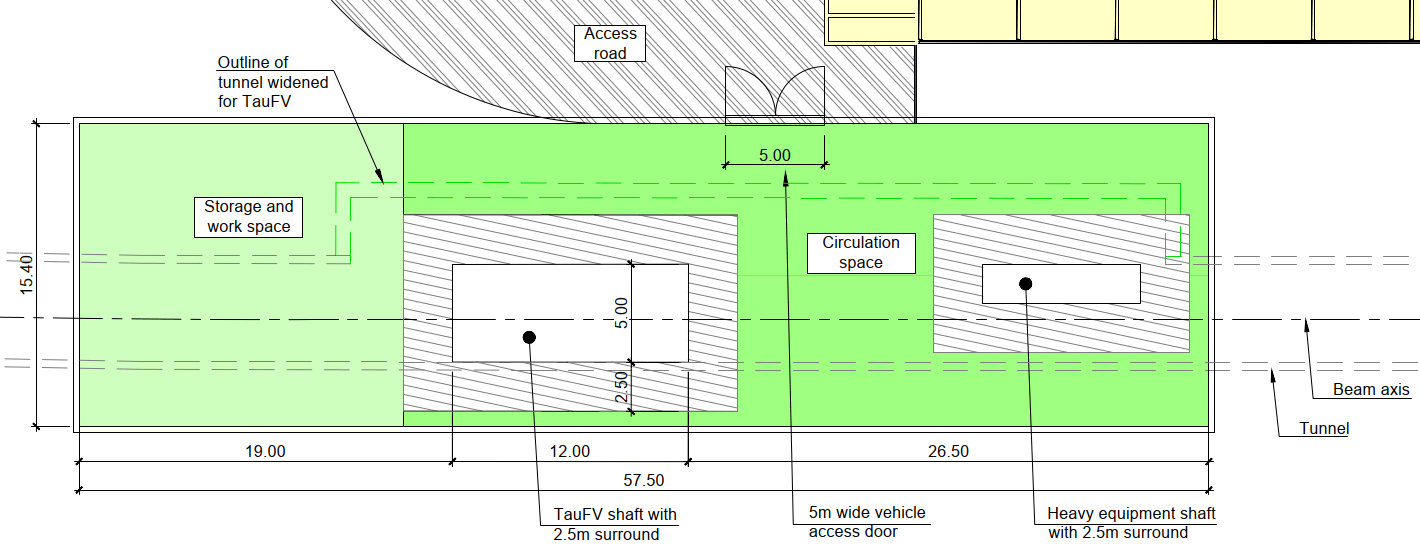}
\caption{Schematic layout of the TauFV building floor plan}
\label{tau:fig:buildingplan}
\end{figure}

Since the building is already close to the proposed BDF access building, it is more efficient to provide this space by extending the building south-west to cover both shafts. In this way, one set of crane rails and one crane will serve both shafts. Circulation space is also shared between the two, and only one access door need to be provided, again helping to reduce the overall floor space required. Initial discussions suggest a \SI{40}{t}  crane will be suitable to serve both purposes.

The expanded access building will be of the same type of construction as that for BDF. An opportunity was identified in this instance to have the building share the diaphragm walls as a foundation. The building outline has been designed to follow the same alignment as the diaphragm walls to enable this.

An indicative layout of the building floor plan is included in Fig. \ref{tau:fig:buildingplan}.

This layout has been agreed with the experimental and  transport handling teams. Further optimisation will be carried out as the design is developed. 

\subsection{Further CE Considerations}

In general, the same considerations apply to the TauFV experiment CE infrastructure as to the wider BDF proposals. One point of note, is that the scheduling of construction of these elements would need careful consideration. The TauFV infrastructure is predominantly outside of the 8m exclusion zone from TCC2, however, a substantial proportion of the access building and some of the subsurface infrastructure is not. Planning how these elements could be scheduled most effectively will be key to delivery of the BDF programme in full. TauFV construction work could be included in either work package 3 or 4 or alternatively split between both. Procurement of these works could be optimised at the next stage of project development.

\printbibliography[heading=subbibliography]

 \chapter{Applications of BDF to Nuclear Astrophysics and Material Irradiation}
\label{TargetApp}

\section{Introduction}
The unique capabilities of the particle field created by the high energy and intensity proton beam from SPS impacting on the production target could be exploited also for other purposes. In fact, owing to the large beam power on target and without perturbing the main physics goal of the installation, the facility could be parasitically exploited for other scientific and technical goals, which are further described in Section~\ref{APPX:NUC_ASTRO} and \ref{APPX:IRRADIATION}. 

The capabilities would be indeed unique in terms of total doses, particle spectra and fluences: similar values could be reached at "standard" spallation neutron sources but are not easily exploitable due to the specific physics and engineering criteria present at those facilities, including closely coupled cryogenic moderators and reflectors, which depletes significantly the high energy components of the particle spectra.

\section{Nuclear Astrophysics applications}
\label{APPX:NUC_ASTRO}

\subsection{Introduction and physics case}
Almost all elements heavier than iron are produced in the Universe by neutron-induced reactions, in particular in neutron capture processes. In the stellar environment, in particular in the Asymptotic Giant Branch (AGB) phase, characterized by relatively low temperatures and neutron densities, nuclei close to the $\beta$-stability line (i.e. stable or with long half-life) are produced in the so-called s-process (“s” standing for slow), where neutron capture times are typically much longer than competing $\beta$-decays. On the contrary, extremely high temperatures and neutron densities are reached in explosive scenarios, such as Supernovae or binary Neutron Star (NS) Merger events, leading to the production of neutron-rich nuclei in rapid capture processes (r-process).
Modelling of both s- and r-process nucleosynthesis require neutron capture cross section data for isotopes along the valley of $\beta$-stability. In fact, accurate (n,$\gamma$) cross sections are needed in order to reliably determine s-process abundances, that on the one hand provide important information on stellar evolution, and on the other hand allow to obtain the r-process counterpart by the difference with the observed abundances (N$_{r}$ = N$_{obs}$ – N$_{s}$).

Recently, studies of stellar nucleosynthesis have received a boost thanks to the enormous
progresses made by astronomical observations as well as by important developments in
modelling stellar evolution. A huge step forward in this respect has resulted from the recent multi-messenger observation of a NS-NS merger in August 2017. The simultaneous detection of gravitational waves, gamma-ray burst and electromagnetic radiation in the optical and near-optical range has unequivocally demonstrated that NS-NS merger are an important, if not the most important, site for r-process nucleosynthesis. All this is now calling for new data on neutron-induced reaction, in particular on short-lived radioactive isotopes. More details on the need of neutron data of interest for Nuclear Astrophysics can be found in Ref.~\cite{COLONNA2018177}.

An important feature of the s-process nucleosynthesis is the branching of the reaction path when isotopes with half-lives comparable with the neutron capture time are produced (for this reason called branching point isotopes). The analyses of such branchings provide crucial information on the physical conditions (temperature, neutron density, etc…) of the stellar site in which s-process nucleosynthesis takes place. A complete list of the most important branching point isotopes can be found in Ref.~\cite{RevModPhys.83.157}. Neutron cross section data on these isotopes are scarce, if any, and often discrepant between each other. Two major problems hinder reliable measurements of these reactions: on the one hand samples of sufficient mass of such radioactive isotopes are extremely difficult to procure and, on the other hand, even when samples are available, the background related to the natural radioactivity of the sample often dominates over capture count-rates. Both problems could be minimised only by the use of intense neutron beams.

Taking advantage of the high luminosity of the neutron beam, some branching point isotopes, with a relatively long half-life of tens or hundreds of years, have been measured at n\_TOF, or at new high-flux activation facilities, such as SARAF (Israel). In all these cases, sample masses of several mg (in some cases hundreds of mg) have been used. An important step forward in these measurements would require neutron fluxes orders of magnitude higher than currently available. In this case, a reasonable reaction count-rate could still be obtained with sub-mg, or even sub-$\mu$g sample masses, making feasible measurements that at present are considered unrealistic. The use of extremely low sample masses has several advantages, in particular the availability of the material and the low associated background (and safety limitations). In fact, a sufficient amount of short-lived isotopes can be obtained for example by irradiation of a stable or long-lived progenitor, or extracted from irradiated structural components in nuclear reactors or at spallation neutron sources. Most importantly, suitable samples of short-lived isotopes could be obtained by implantation of a radioactive ion beam at currently available RIB facilities. A successful attempt in this respect has recently been performed at n\_TOF, where a 90 ng sample of the short-lived $^{7}$Be isotope was produced by implantation at the CERN ISOLDE facility and irradiated with the neutron beam in the second experimental area (EAR2) at n\_TOF \cite{PhysRevLett.121.042701}. The second advantage is that an extremely small-mass sample of short-lived isotope would be characterised by a manageable total activity and related background. The only limitation in this case could be posed by the background related to neutron beam itself and its interaction with the experimental setup, as well as by the $\gamma$-rays and charged particle contamination in the beam. For these reasons, the measurements must rely on high sensitivity techniques, such as activation analysis, either off-line or on-line (i.e. in between irradiation pulses), or Accelerator Mass Spectroscopy.

\subsection{Neutron flux at the Beam Dump Facility Complex for nuclear astrophysics}
\begin{figure}[ht]
    \centering
    \includegraphics[width=0.8\linewidth]{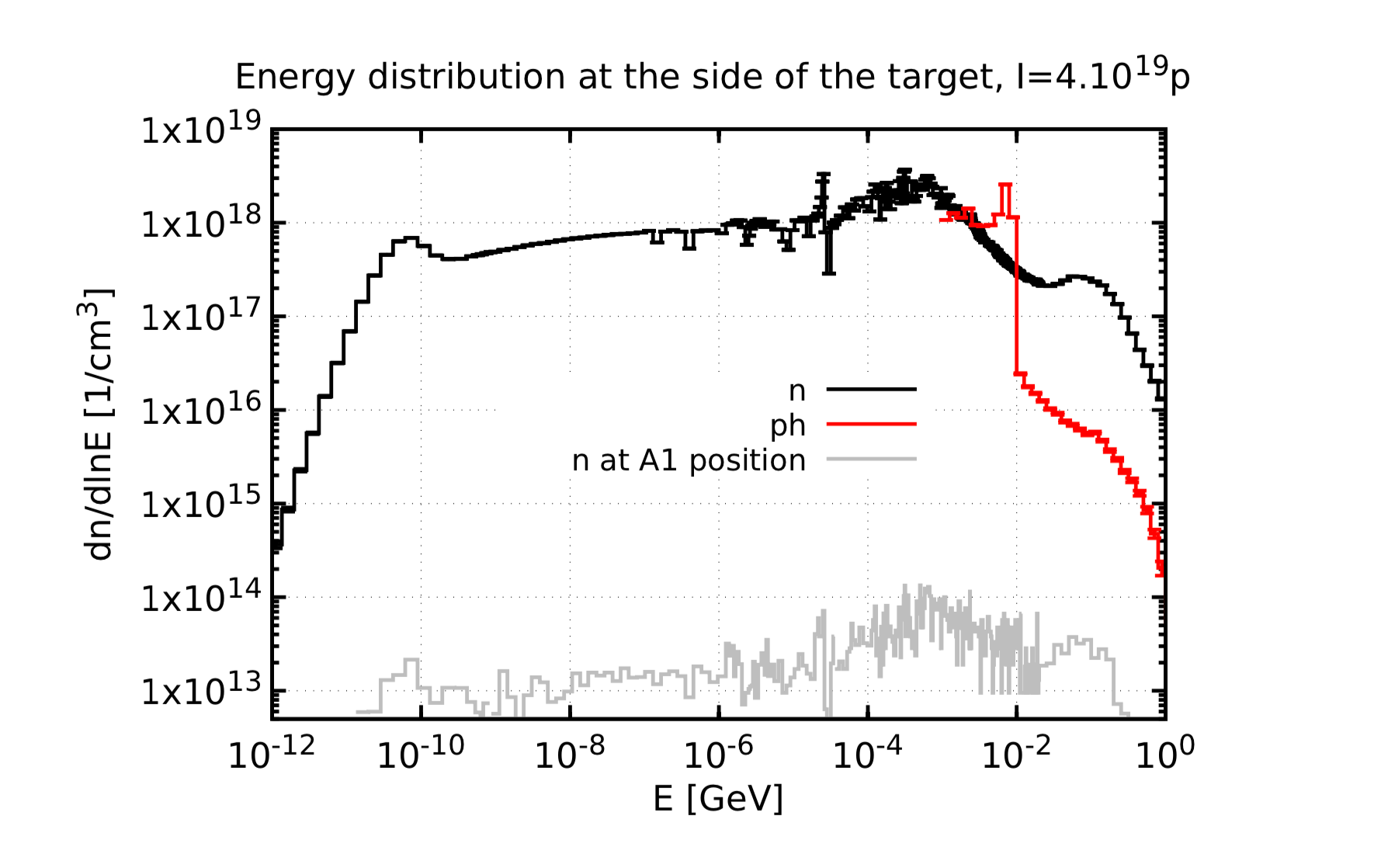}
    \caption{The figure shows the neutron spectrum on the side of the target, at roughly 70 cm from the axis of the assembly. The back line shows the neutron spectrum while the red one shows the photon background (cut at 10 MeV). The grey lines represent the neutron fluence at 3.4 meters at 90 degrees with respect to the target centre, assuming an opening through the target bunker shielding, representing a potential irradiation port.}
    \label{fig:appl:1Dfluence_close}
\end{figure}

In this context, the availability of an extremely intense neutron flux at the BDF facility would offer the unique opportunity to study neutron-induced reactions on short-lived isotopes, of interest for Nuclear Astrophysics as well as for various applications. Preliminary Monte Carlo simulations indicate that in the proximity of the BDF target a flux of around 10$^{13}$-10$^{14}$ neutrons/cm$^{2}$/pulse could be available (see Figure~\ref{fig:appl:1Dfluence_close}), with a spectrum covering from thermal energy to 100 MeV, with a peak around 1 MeV. 

Since the long duration of the proton pulse (1 s) does not allow one to use the time-of-flight technique for neutron energy determination, only integral cross section measurement can be performed. To this end, the neutron spectrum has to be suitably tailored to the different needs. In particular, for studies related to Nuclear Astrophysics, it is desirable to rely on a Maxwellian-like neutron spectrum, with kT between 10 and 100 keV. Such a spectrum can be obtained with an additional (small) moderation of the BDF neutron beam, so to lower the average neutron energy, and with the use of suitable filters (or absorbers) for suppressing the thermal and epithermal region. Studies in this sense are already undergoing within the n\_TOF Collaboration, aiming at setting up an irradiation station near the spallation target. In fact, a BDF neutron facility may greatly benefit from the experience that will be gained in the next few years at the n\_TOF near-target irradiation facility.
The huge flux envisaged for a neutron beam at BDF would surpass any other activation facility now being constructed or planned for the near future. In fact, even considering a reduction due to the moderation and filtering process of two orders of magnitude, a flux of 10$^{12}$ neutrons/cm$^{2}$/pulse would be available for the measurement, three orders of magnitude higher than expected, for example, at the mentioned near-target irradiation station at n\_TOF and two orders of magnitude higher than the most intense Maxwellian neutron source currently available, at SARAF (Israel), that is characterised by an average neutron flux of 1.2x10$^{10}$ n/cm$^{2}$/s.

The great potentiality of the BDF neutron beam can be illustrated with a few examples of measurements that could be performed at that facility. The short-lived $^{147}$Nd (t$_{1/2}$= 11 d) is an important s-process branching isotope, that constraints the neutron density of low-mass AGB stars~\cite{RevModPhys.83.157}. However, its very short half-life makes it extremely difficult at present to measure its neutron capture cross section, a key factor in the determining and modelling the nucleosynthesis path around this branching point. In fact, only a small amount of this isotope (a few tens of ng) can be produced by irradiation of the stable $^{146}$Nd isotope, for example in the core of a nuclear reactor. Such a small amount of material makes it mandatory, for measuring the $^{147}$Nd(n,$\gamma$) cross section, the use of an extremely intense neutron beam, at present not available anywhere. At BDF, the predicted flux of 10$^{12}$ neutrons/cm$^{2}$/pulse would result (assuming a realistic Maxwellian Averaged Cross Section (MACS) of 100 mb) in 10$^{4}$ neutron capture reactions per day, leading to the production of the stable $^{148}$Nd isotope. The number of reactions could later be determined off-line with sufficient accuracy by measuring the number of produced $^{148}$Nd with Accelerator Mass Spectroscopy.

Other important branching isotopes of short half-life that could be measured at BDF would be the $^{179}$Ta and $^{134}$Cs, with a t$_{1/2}$ of roughly 2 y, and $^{170}$Tm, with t$_{1/2}$ of 0.35 y. In this case as well, high-purity samples of mass of a few ng (or a few tens of ng) might reasonably be produced by implantation of a radioactive beam and later irradiated with the BDF neutron beam. This would be the case for the particularly important branching isotope $^{134}$Cs, sensitive to the s-process temperature in low-mass AGB stars. At present a measurement of the neutron capture cross section for this isotope is not considered feasible, at least for the near future. The availability of a BDF neutron beam, in combination with the ISOLDE beam, might finally make this measurement possible. The yield of this isotope at ISOLDE exceeds 10$^{9}$ ions/$\mu$C, so that a sample of 10$^{14}$ atoms (corresponding to ~10 ng) could be produced by implantation in a few days at ISOLDE. The result of the capture reaction is $^{135}$Cs, a radioactive nucleus with very long half-life (2.3x10$^{6}$ years), which again could be measured by means of Accelerator Mass Spectroscopy. Similar considerations apply also to $^{170}$Tm, that could be produced at a nuclear reactor (or at the same BDF beam) starting from the stable $^{169}$Tm isotope, with the only difference that the measurement in this case would rely on the activation technique, given the short half-life of the reaction product $^{171}$Tm. As for $^{179}$Ta, if a production mechanism could be found (for example by means of the $^{180}$Ta(n,2n) reaction), it may also become feasible at BDF. 

In summary, the availability of an extremely high-flux neutron beam at BDF would open the way to very challenging measurements of interest for Nuclear Astrophysics, both for s- and r-process nucleosynthesis, that are not feasible at neutron facilities currently operating or planned for the near future.

\section{Electronics and Material Irradiation}
\label{APPX:IRRADIATION}
\subsection{Introduction}
\begin{figure}[ht]
    \centering
    \includegraphics[width=1\linewidth]{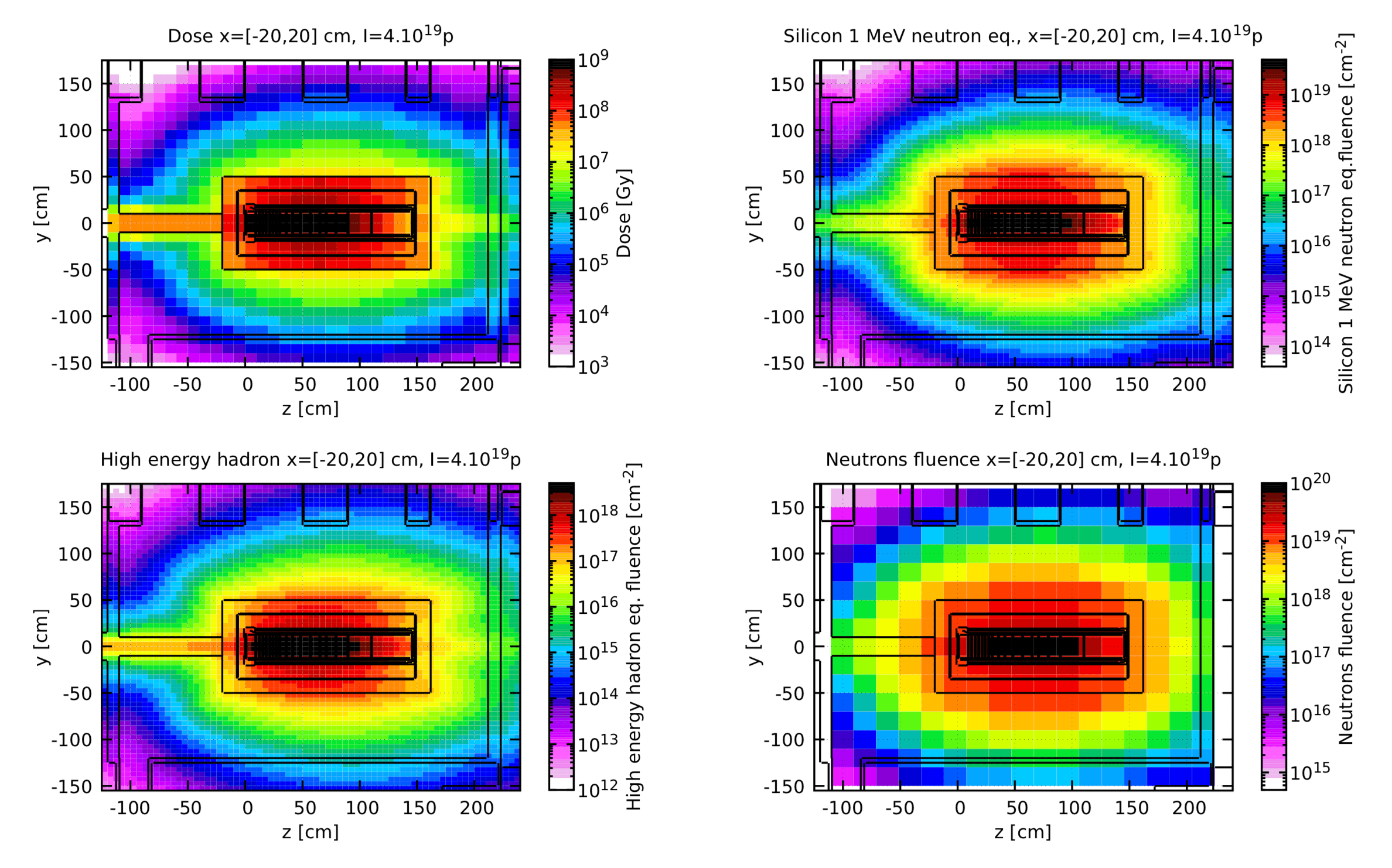}
    \caption{The figure shows a longitudinal cut of the target bunker showing relevant prompt radiation quantities averaged around the target centre line. The values are normalised per yearly protons on target.}
    \label{fig:appl:R2E_top}
\end{figure}

The Beam Dump Facility (BDF) infrastructure and operation provides a unique opportunity for hosting irradiation test facilities for material and electronics. The unparalleled mixed-field radiation levels expected near the target, reaching integral annual levels of roughly 400 MGy and 10$^{18}$ 1-MeV neutron equivalent per cm$^{2}$ (see Figure~\ref{fig:appl:R2E_top}), would allow for accelerated testing in a highly representative environment for future accelerator applications, including those for instance linked to HE-LHC, FCC-hh and potentially FCC-ee.

\subsection{Irradiation capabilities}
\subsubsection{Near-target station}
Irradiation of materials would be performed in passive mode, with a remote handling and transport system placing the samples in the irradiation areas, and being able of retrieving them in hot cell zones for post irradiation analysis. The attained levels would be sufficient to study both Total Ionising Dose (TID) (i.e. mainly affecting insulating, polymer materials) and Displacement per Atom (DPA) (mainly affecting mechanical and thermo-physical properties of metals).

Electronics would be tested in a similar way for passive, unbiased displacement damage studies, or actively (i.e. biased) for TID effects. In both cases, lifetime levels compatible with those expected in the most exposed FCC-hh detector areas would be attained, providing an excellent opportunity to evaluate their radiation tolerance, certainly beyond the limit of what present silicon-based technologies can withstand. 

An in-beam irradiation station, directly facing the diluted primary beam, could also be envisaged in order to further increase the TID and DPA levels that could be reached.

\subsubsection{Irradiation outside the proximity shielding}

In addition to the near-target irradiation stations, providing a broad range of dose rates and fluxes for ultra-high dose and fluence sample irradiation, a lower radiation level, more accessible area could be constructed of testing of electronic components, boards and systems. Such area will require neutron fluxes (>10 MeV) in the 10$^{4}$-10$^{8}$ n/cm$^{2}$/s range, and would allow for qualification of electronic equipment to be operated in the accelerator environment. Ideally, surfaces of several m$^2$ should be attained, enabling the radiation qualification of bulky systems, such as power converters. 

\begin{figure}[ht]
    \centering
    \includegraphics[width=0.8\linewidth]{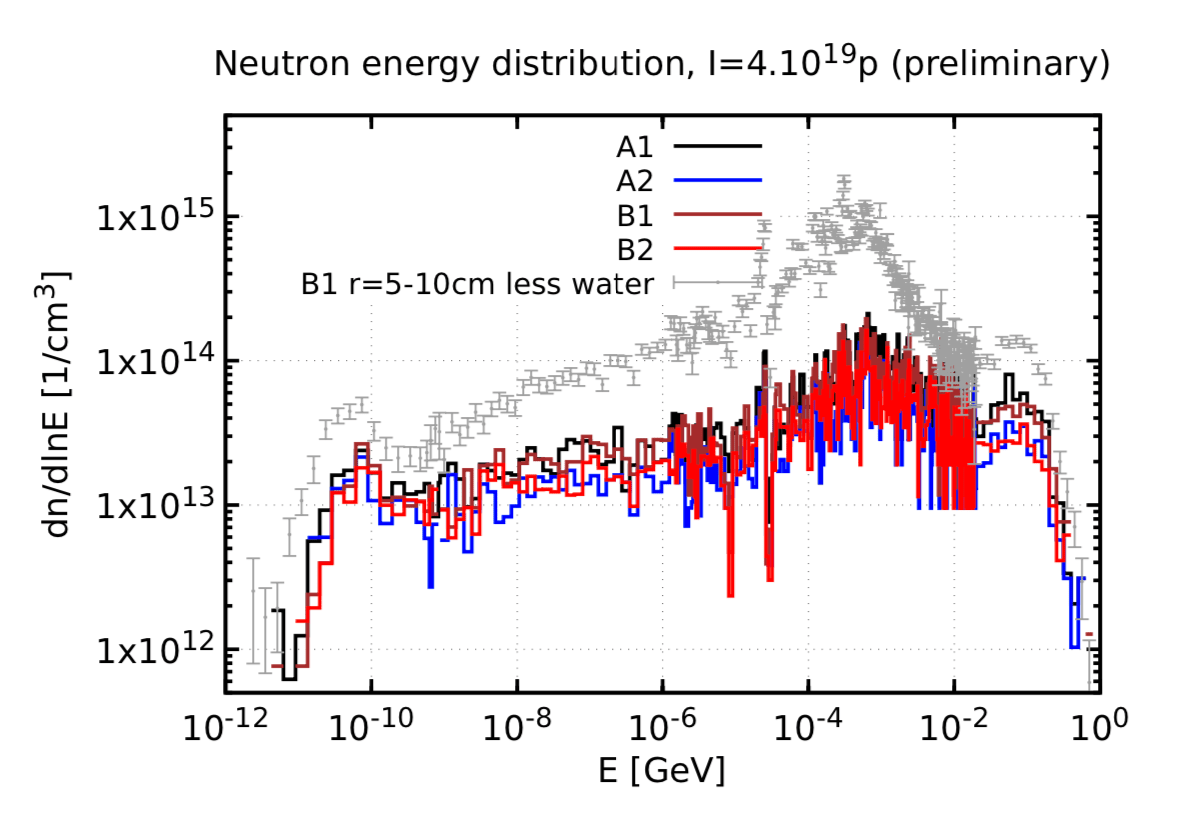}
    \caption{The figure shows the neutron spectrum on the side of the target, at different distances from the centre. Positions A1/B1 are at 3.4 meters from the centre, while A2/B2 are at 4.1 meters, respectively. Values could be further optimised with an improved configuration of the neutron collimation system: for example, the grey line shows the values at 3.4 meters with a larger collimator opening.}
    \label{fig:appl:1Dfluence_far}
\end{figure}

With the increased interest and need of using commercial-off-the-shelf electronic components and systems-on-chip, large scale irradiation is in principle the only viable solution to ensure an adequate radiation tolerance. Such conclusions does not only apply to accelerator applications, but also to ground-level (e.g. automotive, high-reliability servers), avionic and so-called new space applications. Such an area could be obtained in the BDF Target Complex, with a dedicated irradiation bunker on the side of the target, not perturbing the main physics aim of the facility. Fluxes of the same order of magnitude as those mentioned above could be reached, as evident in Figure~\ref{fig:appl:1Dfluence_far}.

Radiation protection aspects will have to be further analysed in the future to find the best suited configuration.

\subsection{Synergies with other CERN facilities}
The unique capabilities of a potential material irradiation facility could be well coupled with other facility on the CERN site, as for example HiRadMat, allowing the possibility to execute beam shock experiments on previously irradiated material. This would create world-class installation in the field of high power targets, beam windows, collimators and absorbers, which would be of interest for all high power and high intensity accelerator endeavours world-wide. Complementarity could be further explored with CHARM and Co60 irradiation facilities.

\FloatBarrier
\printbibliography[heading=subbibliography]

\end{appendices}

\end{document}